\documentclass[11pt,a4paper]{memoir}
\usepackage[utf8]{inputenc}

\usepackage[left=3cm,right=3cm,top=3cm,bottom=3cm]{geometry}

\author{Marcel Schaub}

\usepackage[english]{babel}

\usepackage{soul}
\setuldepth{0}

\usepackage{wallpaper}

\usepackage{anyfontsize}

\usepackage[absolute]{textpos}

\usepackage{upref}

\newcommand{\superscript}[1]{\ensuremath{^{\textrm{#1}}}}

\newcommand{\tho}[0]{\superscript{th}}
\renewcommand{\st}[0]{\superscript{st}}

\usepackage{enumerate}

\usepackage{amsmath}
\numberwithin{equation}{section}
\usepackage{mathtools}
\newcommand{\sm}{\smashoperator}
\renewcommand{\leq}{\leqslant}
\renewcommand{\geq}{\geqslant}

\usepackage{amsfonts}
\usepackage{amssymb}

\usepackage{amsthm}
\usepackage{thmtools, thm-restate} 

\usepackage{esvect}
\usepackage{cancel}
\usepackage{nicefrac}

\usepackage{fontawesome}
\newcommand{\speaker}{{\hspace{0.7pt}\text{\raisebox{-1.7pt}{\scalebox{1.4}{\faVolumeOff}}\hspace{0.7pt}}}}

\usepackage[T1]{fontenc}

\DeclareMathAlphabet{\mathbbs}{U}{bbold}{m}{n}
\newcommand{\Idbb}{\mathbbs 1}

\usepackage[arrow, matrix, curve]{xy}

\usepackage{ifthen}

\usepackage{oldgerm}

\usepackage{eurosym}

\usepackage{float}
\usepackage[rflt]{floatflt}
\usepackage{graphicx}
\usepackage{xcolor}

\usepackage{dlfltxbcodetips}

\newtheorem{thm}{Theorem}[section]
\newtheorem{bigthm}{Theorem}[chapter]

\newtheorem{lem}[thm]{Lemma}

\newtheorem{kor}[thm]{Corollary}

\newtheorem{prop}[thm]{Proposition}

\newtheorem{theorem}[thm]{Theorem}
\newtheorem{proposition}[thm]{Proposition}
\newtheorem{lemma}[thm]{Lemma}

\newtheoremstyle{cited}{}{}{\itshape}{}{
}{\textbf{.}}{.5em}{\textbf{#1 #2} #3}
\theoremstyle{cited}
\newtheorem{cthm}[thm]{Theorem}
\newtheorem{clem}[thm]{Lemma}
\newtheorem{cprop}[thm]{Proposition}

\theoremstyle{definition}
\newtheorem{defn}[thm]{Definition}

\newtheorem{egs}[thm]{Examples}

\newtheorem{eg}[thm]{Example}

\newtheorem{bem}[thm]{Remark}

\newtheorem*{warn}{Warning}
\newtheorem{bems}[thm]{Remarks}
\newtheorem*{varbems}{Remarks}

\newtheorem*{note}{Note}

\newtheorem{asmp}[thm]{Assumption}

\newtheorem{definition}[thm]{Definition}

\newcommand{\lra}{\longrightarrow}
\newcommand{\ra}{\rightarrow}

\newcommand{\Ra}{\Rightarrow}

\newcommand{\La}{\Leftarrow}

\newcommand{\lera}{\leftrightarrow}

\newcommand{\xra}{\xrightarrow}

\newcommand{\lk}{\left(}
\newcommand{\rk}{\right)}

\newcommand{\ov}[1]{\overline{#1}}

\newcommand{\dx}{\mathrm{d}x}

\newcommand{\dy}{\mathrm{d}y}

\newcommand{\dz}{\mathrm{d}z}

\newcommand{\ds}{\mathrm{d}s}
\newcommand{\dr}{\mathrm{d}r}

\newcommand{\du}{\mathrm{d}u}

\newcommand{\dt}{\mathrm{d}t}
\newcommand{\dd}{\mathrm{d}}

\DeclareMathOperator{\tr}{tr}
\DeclareMathOperator{\Tr}{Tr}

\DeclareMathOperator{\diam}{diam}

\DeclareMathOperator{\divv}{div}

\DeclareMathOperator{\curl}{curl}

\DeclareMathOperator{\const}{const.}
\DeclareMathOperator{\sgn}{sgn}

\DeclareMathOperator{\supp}{supp}

\DeclareMathOperator*{\esssup}{ess \, sup}
\DeclareMathOperator{\ad}{ad}

\DeclareMathOperator{\dist}{dist}
\DeclareMathOperator{\ran}{ran}

\DeclareMathOperator{\spano}{span}
\DeclareMathOperator{\spec}{spec}
\DeclareMathOperator{\rank}{rank}

\DeclareMathOperator{\Vol}{Vol}

\newcommand{\ess}{\mathrm{ess}}
\newcommand{\loc}{\mathrm{loc}}
\newcommand{\hol}{\mathrm{hol}}
\newcommand{\sa}{\mathrm{sa}}
\newcommand{\per}{\mathrm{per}}

\newcommand{\symm}{\mathrm{symm}}

\newcommand{\Imm}{\mathrm{Im}}
\newcommand{\Rem}{\mathrm{Re}}
\let\Im\undefined
\let\Re\undefined
\DeclareMathOperator{\Im}{\Imm}
\DeclareMathOperator{\Re}{\Rem}

\renewcommand{\tilde}{\widetilde}
\renewcommand{\hat}{\widehat}

\DeclareFontFamily{U}{matha}{\hyphenchar\font45}
\DeclareFontShape{U}{matha}{m}{n}{
	<5> <6> <7> <8> <9> <10> gen * matha
	<10.95> matha10 <12> <14.4> <17.28> <20.74> <24.88> matha12
}{}
\DeclareSymbolFont{matha}{U}{matha}{m}{n}
\DeclareFontSubstitution{U}{matha}{m}{n}

\DeclareFontFamily{U}{mathx}{\hyphenchar\font45}
\DeclareFontShape{U}{mathx}{m}{n}{
	<5> <6> <7> <8> <9> <10>
	<10.95> <12> <14.4> <17.28> <20.74> <24.88>
	mathx10
}{}
\DeclareSymbolFont{mathx}{U}{mathx}{m}{n}
\DeclareFontSubstitution{U}{mathx}{m}{n}

\DeclareMathDelimiter{\vvvert}{0}{matha}{"7E}{mathx}{"17}

\DeclareFontFamily{U}{mathx}{\hyphenchar\font45}
\DeclareFontShape{U}{mathx}{m}{n}{
	<5> <6> <7> <8> <9> <10>
	<10.95> <12> <14.4> <17.28> <20.74> <24.88>
	mathx10
}{}
\DeclareSymbolFont{mathx}{U}{mathx}{m}{n}
\DeclareFontSubstitution{U}{mathx}{m}{n}
\DeclareMathAccent{\widecheck}{0}{mathx}{"71}
\DeclareMathAccent{\wideparen}{0}{mathx}{"75}

\newcommand{\Cbb}{\mathbb{C}}
\newcommand{\Ebb}{\mathbb{E}}

\newcommand{\Hbb}{\mathbb{H}}

\newcommand{\Nbb}{\mathbb{N}}

\newcommand{\Rbb}{\mathbb{R}}

\newcommand{\Zbb}{\mathbb{Z}}

\newcommand{\Acal}{\mathcal{A}}
\newcommand{\Bcal}{\mathcal{B}}
\newcommand{\Ccal}{\mathcal{C}}
\newcommand{\Dcal}{\mathcal{D}}
\newcommand{\Ecal}{\mathcal{E}}
\newcommand{\Fcal}{\mathcal{F}}
\newcommand{\Gcal}{\mathcal{G}}
\newcommand{\Hcal}{\mathcal{H}}
\newcommand{\Ical}{\mathcal{I}}
\newcommand{\Jcal}{\mathcal{J}}
\newcommand{\Kcal}{\mathcal{K}}
\newcommand{\Lcal}{\mathcal{L}}
\newcommand{\Mcal}{\mathcal{M}}
\newcommand{\Ncal}{\mathcal{N}}
\newcommand{\Ocal}{\mathcal{O}}
\newcommand{\Pcal}{\mathcal{P}}
\newcommand{\Qcal}{\mathcal{Q}}
\newcommand{\Rcal}{\mathcal{R}}
\newcommand{\Scal}{\mathcal{S}}
\newcommand{\Tcal}{\mathcal{T}}
\newcommand{\Ucal}{\mathcal{U}}
\newcommand{\Vcal}{\mathcal{V}}
\newcommand{\Wcal}{\mathcal{W}}
\newcommand{\Xcal}{\mathcal{X}}

\newcommand{\Zcal}{\mathcal{Z}}
\newcommand{\Afrak}{\mathfrak{A}}

\newcommand{\Bfrak}{\mathfrak{B}}

\newcommand{\hfrak}{\mathfrak{h}}
\newcommand{\Hfrak}{\mathfrak{H}}

\newcommand{\Lloc}{L_{\mathrm{loc}}}
\newcommand{\Wloc}{W_{\mathrm{loc}}}

\newcommand{\EGL}{\mathcal E^{\mathrm{GL}}_{D, h}}
\newcommand{\EGLGSE}{E^{\mathrm{GL}}(D)}
\newcommand{\FBCS}{\mathcal F^{\mathrm{BCS}}_{
			h, T
}}
\newcommand{\Abold}{\mathbf A}
\newcommand{\Bbold}{\mathbf B}
\newcommand{\Ebold}{\mathbf E}
\newcommand{\Gbold}{\mathbf G}
\newcommand{\Sbold}{\mathbf S}
\newcommand{\Tbold}{\mathbf T}
\newcommand{\Lambdabold}{\mathbf \Lambda}

\newcommand{\Zbold}{\mathbf Z}
\newcommand{\sbold}{\mathbf s}
\newcommand{\Hmag}{H_{\mathrm{mag}}}
\newcommand{\Lmag}{L_{\mathrm{mag}}}
\newcommand{\Wper}[1]{W_{\mathrm{per}}^{#1, \infty}(Q_1)}
\newcommand{\Wperdiss}[1]{W_{\mathrm{per}}^{#1, \infty}(Q_{e_{\mathbf B}})}
\newcommand{\Lper}{L_{\mathrm{per}}^\infty(Q_1)}
\newcommand{\Lperdiss}{L_{\mathrm{per}}^\infty(Q_{e_{\mathbf B}})}
\newcommand{\Wpervec}[1]{W_{\mathrm{per}}^{#1, \infty}(Q_1; \Rbb^3)}
\newcommand{\Wpervecdiss}[1]{W_{\mathrm{per}}^{#1, \infty}(Q_{e_{\mathbf B}}; \Rbb^3)}
\newcommand{\Lpervec}{L_{\mathrm{per}}^\infty(Q_1; \Rbb^3)}
\newcommand{\Lpervecdiss}{L_{\mathrm{per}}^\infty(Q_{e_{\mathbf B}}; \Rbb^3)}
\newcommand{\Lsymm}{{L^2(Q_h \times \Rbb_{\mathrm s}^3)}}
\newcommand{\Hsymm}{{H^1(Q_h \times \Rbb_{\mathrm s}^3)}}
\newcommand{\Tc}{{T_{\mathrm{c}}}}
\newcommand{\Dc}{{D_{\mathrm{c}}}}
\newcommand{\betac}{\beta_{\mathrm{c}}}

\newcommand{\R}{\mathbb{R}}
\newcommand{\Z}{\mathbb{Z}}

\newcommand{\Curl}{\operatorname{curl}}

\newcommand{\Lat}{\mathcal{L}}

\newcommand{\Gd}{\mathcal{G}_d}
\newcommand{\Hd}{\Gamma_d}
\newcommand{\dbold}{\mathbf d}

\usepackage{esint} 

\usepackage[hidelinks]{hyperref}
\hypersetup {
	pdfpagemode = {UseNone},
	pdftitle = {BCS Theory in the Weak Magnetic Field Regime for Systems with Nonzero Flux and Exponential Estimates on the Adiabatic Theorem in Extended Quantum Lattice Systems},
	pdfauthor = {Marcel Oliver Maier},
	pdflang = {en-US}
}

\newcommand{\e}{\mathrm{e}}
\renewcommand{\i}{\mathrm{i}}

\copypagestyle{marcel}{ruled}
\makeevenfoot{marcel}{}{\thepage}{PhD Thesis}
\makeoddfoot{marcel}{\abgabetermineng}{\thepage}{Marcel Maier}
\makeheadrule{marcel}{\textwidth}{0.4pt}
\makefootrule{marcel}{\textwidth}{0.4pt}{4pt}

\usepackage{lmodern}

\usepackage[
	sorting=nyt, 
	backend=biber, 
    giveninits=true,	
	style=alphabetic,
	style=alphabetic, 
	maxbibnames=99,
	backref=true,
	eprint=true,
	isbn=false,
	date=year,
	]{biblatex}

\DeclareLabelalphaTemplate{
   \labelelement{
     \field[uppercase,final]{shorthand}     
     \field[uppercase,strwidth=1,strside=left,names=-5,noalphaothers]{labelname}
   }
   \labelelement{
     \field[strwidth=2,strside=right]{year}
   }
   \labelelement{
		\field[uppercase]{label}
   }
}
\renewbibmacro*{note+pages}{%
  \printfield{note}%
  \setunit{\bibpagespunct}%
  \printfield{pages}%
  \setunit{\bibpagespunct}
  \printfield{pagetotal}
  \newunit}

\DefineBibliographyStrings{english}{%
  backrefpage = {page}, 
  backrefpages = {pages}, 
}

\renewbibmacro{in:}{} 
\DeclareFieldFormat{pages}{#1}
\DeclareFieldFormat{pagetotal}{#1~pages}
\DeclareFieldFormat[article]{volume}{\mkbibbold{#1}}
\DeclareFieldFormat[article,inbook,incollection,inproceedings,patent,thesis,unpublished]{title}{#1} 

\AtEveryBibitem{\clearfield{number}}
\AtEveryBibitem{\clearfield{issue}}

\usepackage{pdfpages}
\usepackage{chngcntr}
\counterwithin{footnote}{section}
\counterwithout{footnote}{subsection}

\usepackage{ifthen}
\newcommand{\masterfile}{Diss}

\newcommand{\abgabetermin}{17. Juni 2022}
\newcommand{\abgabetermineng}{June 17, 2022}

\begin{document}


\ThisTileWallPaper{\paperwidth}{\paperheight}{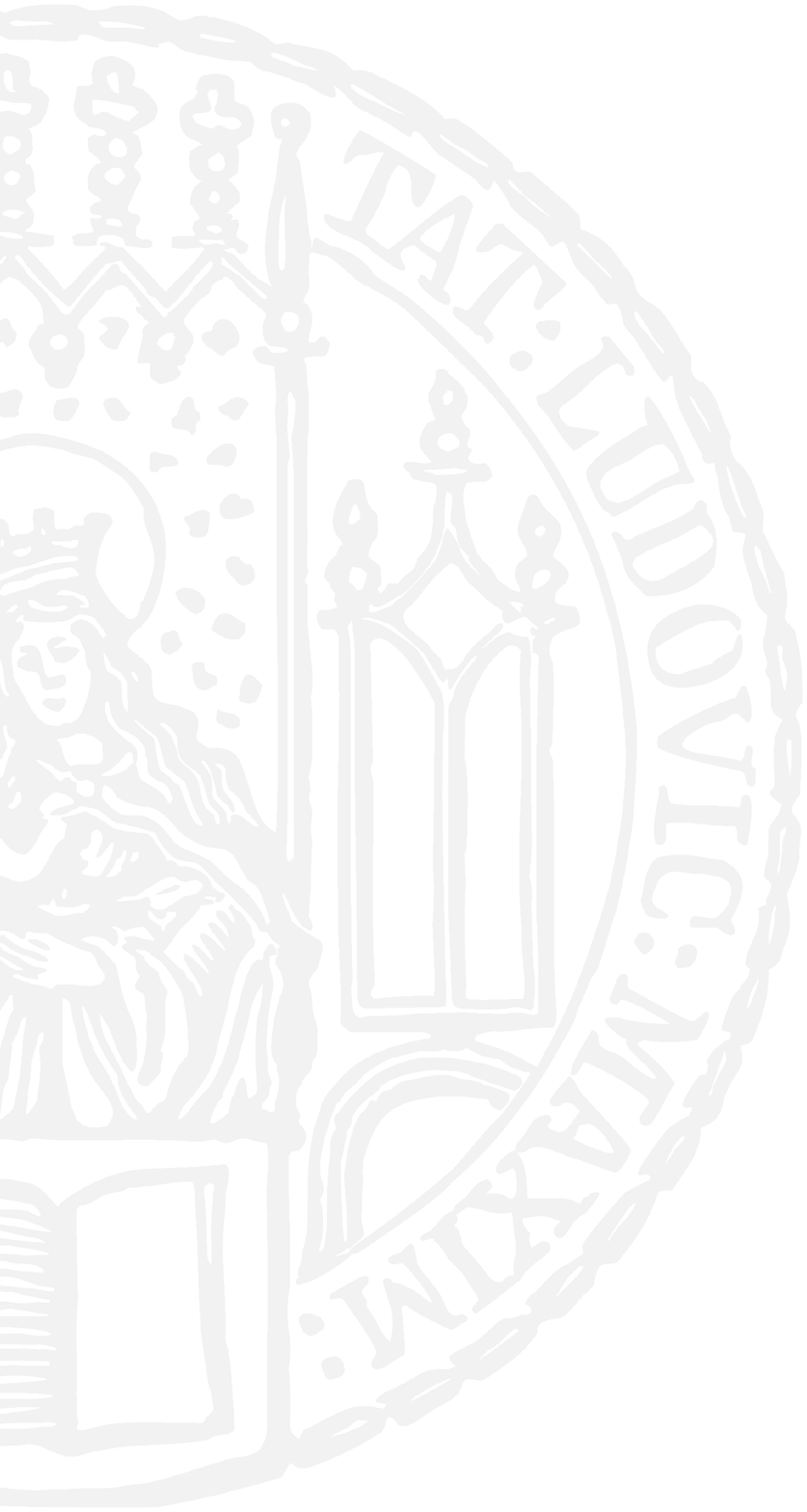}

\thispagestyle{empty}


\begin{center}
\huge 
\textsc{BCS Theory \\ in the Weak Magnetic Field Regime \\ for Systems with Nonzero Flux} \\
%
---\\
\textsc{and}\\
---\\
\textsc{Exponential Estimates \\ on the Adiabatic Theorem \\ in Extended Quantum Lattice Systems}
\end{center}

\vfill

\begin{center}
\huge \textsc{Dissertation}
\end{center}

\begin{center}
\textsc{an der Fakultät für Mathematik, Informatik und Statistik}

\textsc{der Ludwig-Maximilians-Universität München}
\end{center}

\vfill

\begin{center}
vorgelegt von 

\textsc{Marcel Oliver Maier}

\textsc{geboren Schaub}

\textsc{aus Darmstadt}

\vspace{2cm}

\textsc{München, den \abgabetermin}
\end{center}


\newpage

~
\vfill

\noindent 1. Gutachter: Prof. Dr. Christian Hainzl

\noindent 2. Gutachter: Prof. Dr. Marius Lemm

\noindent Tag der mündlichen Prüfung: 7. September 2022 

\newpage


\ThisTileWallPaper{\paperwidth}{\paperheight}{Siegelwallpaperblank.pdf}

\thispagestyle{empty}

\begin{center}
\huge 
\textsc{BCS Theory \\ in the Weak Magnetic Field Regime \\ for Systems with Nonzero Flux} \\
%
---\\
\textsc{and}\\
---\\
\textsc{Exponential Estimates \\ on the Adiabatic Theorem \\ in Extended Quantum Lattice Systems}
\end{center}

\vfill

%

\begin{center}
\huge \textsc{PhD Thesis}
\end{center}

\begin{center}
\textsc{at the Faculty of Mathematics, Informatics, and Statistics}

\textsc{Ludwig-Maximilians-Universität München}
\end{center}

\vfill

\begin{center}
by 

\textsc{Marcel Oliver Maier}

\textsc{born Schaub}

\textsc{from Darmstadt}

\vspace{2cm}

\textsc{Munich, \abgabetermineng}
\end{center}


\newpage
~\thispagestyle{empty}\newpage
%
%
%
%
%

\newpage


\def\abstractname{Abstract (Deutsch)}

\begin{abstract}
Im Hauptteil dieser Dissertation betrachten wir einen periodisch realisierten mikro\-sko\-pi\-schen Supraleiter, der durch die BCS-Theorie beschrieben wird und externen elektromagnetischen Feldern ausgesetzt ist. Wir zeigen, dass der Supraleiter im Limes makroskopischer und schwacher Magnetfelder korrekt durch die Ginzburg--Landau-Theorie beschrieben wird. Die wichtigste Neuerung unserer Ergebnisse besteht darin, dass wir einen nicht verschwindenden magnetischen Fluss durch die Einheitszelle des Periodengitters zulassen. Diese Hauptresultate werden durch verschiedene unveröffentlichte Arbeiten auf dem Gebiet der BCS-Theorie ergänzt. Außerdem stellen wir der Präsentation dieser Ergebnisse eine umfassende Einführung in die BCS-Theorie voran, die sich für Masterstudent*innen und Doktorand*innen eignet. Damit hoffen wir, einen Beitrag zur Schließung der Lücke der fehlenden Einführungsliteratur auf diesem Gebiet zu leisten.

Die Dissertation umfasst ein zweites Thema, in dem wir Ideen für den Aufbau von Quantengittersystemen liefern, um exponentielle Abschätzungen für den Adiabatensatz zu beweisen. Diese Notizen sind das Ergebnis eines Forschungsaufenthalts an der University of British Columbia (UBC) in Vancouver, Kanada.
\end{abstract}

\def\abstractname{Abstract (English)}

\begin{abstract}
In the main part of this PhD thesis, we consider a periodically realized microscopic superconductor described by BCS theory, which is subject to external electromagnetic fields. We show that the superconductor is properly described by Ginzburg--Landau theory in the macroscopic and weak magnetic field limit. The main novelty of our results is to allow for a non-vanishing magnetic flux through the unit cell of the lattice of periodicity. These main results are supplemented by various unpublished notes in the field of BCS theory. Furthermore, we preface the presentation of these results with a comprehensive introduction suitable for master's or PhD students. Thereby, we hope to contribute to filling the gap of missing introductory literature in the field.

The thesis comprises a second topic, in which we provide ideas for setting up quantum lattice systems in order to prove exponential estimates for the adiabatic theorem. These notes are the result of studies in this field, which have been conducted during a research stay at the University of British Columbia (UBC) in Vancouver, Canada.
\end{abstract}

\def\abstractname{Abstract}

\newpage
~\thispagestyle{empty}\newpage

\tableofcontents

\newpage

\listoffigures


\chapter*{Acknowledgements}
\addcontentsline{toc}{chapter}{Acknowledgements}

First and foremost, I would like to express my special thanks to my advisor Christian Hainzl, who took a great leap of faith in me when he posed me this challenging problem. I thank him for the opportunity to write this thesis and for his constant and ongoing support. I also thank him for the opportunity to work as the CRC coordinator, which allowed me to gain deep insights into the process of submitting a comprehensive proposal. I would like to thank Stefan Teufel for his commitment in creating and maintaining such a warm working environment and work group in Tübingen. I believe that this is something very special! I express my gratitude to Marius Lemm for proofreading this thesis and providing me with valuable and enlightening comments.

I am indebted to Sven Bachmann for giving me the opportunity to spend four months at the University of British Columbia in Vancouver, Canada, in 2019. I thank him for the pleasant working atmosphere we had at UBC and the continuing collaboration afterwards. For this research stay, financial support from the University of Tübingen and the German Academic Exchange Service (DAAD) is appreciated.

Further financial support from the LMU Mentoring Program for travelling to the International Congress of Mathematical Physics (ICMP) 2021 and for extensive equipment for my home office is gratefully acknowledged.

I would like to extend a thank you to the mathematical physics groups in Tübingen and Munich. It was a pleasure to be part of the numerous hilarious lunch and coffee (chocolate, rather) breaks! Special thanks go to Tim Tzaneteas, who taught me a lot about BCS theory in the early stages of my PhD and who inspired the small project on magnetic field decomposition that has found its way into this thesis (Chapter \ref{Chapter:Abrikosov_gauge}). Special thanks also go to Ruth Schulte whom I already met during our Master's studies, who continued to be a welcomed teaching partner, and who taught me the tricks of the trade when dealing with the LMU administration. Finally, I would like to thank my ``Italian office'' and the people frequently showing up there: Giovanni Antinucci, Marco Falconi, Luca Fresta, Emanuela Giacomelli, Giovanna Marcelli, and Marco Olivieri. This was great fun, which I have always missed since I moved to Munich! Cheers, everybody!

I can't express how grateful I am to my predecessor, colleague, and collaborator Andreas Deuchert. I owe almost everything I know about the physics behind the mathematics of BCS theory to him. He spent countless hours with me in discussions during my PhD, and he was the one with whom I was able to form the core team (in Oberwolfach 2019) that finally solved the problem. I would also like to thank him for allowing me to incorporate ideas from his unpublished notes into this work. Thanks for everything, Andi!

Finally, and most importantly, I would like to thank my family, my mother Jutta and my father Reiner, for their outstanding efforts to provide me with a thorough and good education, and for their invaluable and continuing support, my brother Philipp (my deep appreciation for your endurance to read Chapter \ref{Chapter:Intro}!), who continues to be a source of inspiration and a best friend, and my wife Eva for her endless support and love. 



\chapter*{Zusammenfassung}
\addcontentsline{toc}{chapter}{Zusammenfassung}

Diese Arbeit ist auf dem Gebiet der Vielteilchen-Quantensysteme angesiedelt und umfasst zwei Themen:
\begin{enumerate}[(i)]
\item BCS-Theorie der Supraleitung im Regime der schwachen magnetischen Felder,
\item Adiabatische Theorie im Kontext von Quantengittersystemen.
\end{enumerate}

Das Thema (i) bildet den Hauptteil dieser Arbeit. In diesem Teil untersuchen wir periodisch ausgedehnte Supraleiter, die schwachen und makroskopischen externen elektrischen und magnetischen Feldern ausgesetzt sind. Diese Felder sind so beschaffen, dass sie einen nicht verschwindenden magnetischen Fluss durch die Einheitszelle induzieren, was die hauptsächliche Herausforderung an der Aufgabenstellung dieser Arbeit darstellt. Der Supraleiter wird mathematisch im Rahmen der BCS-Theorie der Supraleitung beschrieben. Diese Theorie ist eine effektive Zweiteilchentheorie mit einem Paarungsmechanismus, die nach den drei Physikern John Bardeen, Leon Neil Cooper und John Robert Schrieffer benannt ist, die 1957 eine mit dem Nobelpreis ausgezeichnete attraktive Wechselwirkung im Supraleiter postulierten. Ihr berühmter Ansatz ist Inspiration für unzählige Arbeiten und seine rigorose Verifikation aus der Perspektive der Vielteilchenquantenmechanik fordert die Mathematische Physik bis heute heraus.

Unser Modell wird variationell in Form eines freien Energiefunktionals, des sogenannten BCS-Funktionals, beschrieben und wir untersuchen die Fluktuation dessen Minimums, der sogenannten BCS-Energie, sowie die Verschiebung der kritischen Temperatur des Systems, die durch die makroskopischen externen Felder im Grenzbereich schwacher Feldstärken verursacht werden. Wir zeigen, dass diese Verschiebungen in diesem Regime durch die Ginzburg--Landau-Theorie beschrieben wird und leiten geeignete asymptotische Beschreibungen her.

Unser Hauptbeitrag sind signifikante konzeptionelle Vereinfachungen in der Analysis des Funktionals auf Testzuständen im Vergleich zu den Arbeiten von Frank, Hainzl, Seiringer und Solovej aus dem Jahr 2012 bzw. 2016 sowie ein neues Zerlegungsresultat für die Cooper-Paar-Wellenfunktion, das den konstanten Anteil des Magnetfeldes einbezieht. Dieses Zerlegungsresultat ermöglicht es uns, die genannten Resultate für Systeme mit einem solchen konstanten Feldanteil zu beweisen. Letzterer ist für den nicht verschwindenden magnetischen Fluss verantwortlich und stellt die mathematische Behandlung des Problems vor erhebliche Schwierigkeiten. Der wichtigste mathematische Grund ist, dass die Komponenten des magnetischen Impulsoperators nicht kommutieren, was die Anwendung von Fourieranalysis unmöglich macht. Die Präsentation der Resultate ist in zwei Veröffentlichungen aufgeteilt, die in dieser Arbeit enthalten sind.

Darüber hinaus liefern wir mehrere Ergebnisse, die helfen können, das Projekt in Zukunft voranzubringen. Das erste Ergebnis ist ein Zerlegungsresultat für magnetische Potentiale, die im Zusammenhang mit periodischen Magnetfeldern auftreten. Diese Ergebnisse werden verwendet, um zu argumentieren, dass die magnetischen Potentiale, die in den Arbeiten in den Kapiteln \ref{Chapter:DHS1} und \ref{Chapter:DHS2} behandelt werden, erschöpfend sind. Wir liefern auch eine spektrale Zerlegung des periodischen Landau-Hamiltonians. Unser letztes Ergebnis ist eine Störungstheorie des tiefliegenden Spektrums des Operators $K_{T,\Abold} -V$, der in der BCS-Theorie eine wichtige Rolle spielt. Dies geschieht mit Hilfe einer Combes--Thomas-Abschätzung für den Resolventenkern von $K_T-V$, die von unabhängigem Interesse sein könnte.

Ein Problem in der BCS-Theorie war schon immer das Fehlen von Einsteigerliteratur. Wir versuchen, diese Lücke zu schließen, indem wir mit einer umfassenden Einführung in das Thema beginnen, die für Leser*innen geschrieben ist, die zum ersten Mal mit der BCS-Theorie in Kontakt kommen.

Im Rahmen von Punkt (ii) skizzieren wir den Beweis von exponentellen Abschätzungen für den Adiabatensatz in Quantengittersystemen. Zwar ist die Arbeit noch nicht abgeschlossen, dennoch präsentieren wir hier den aktuellen Zustand des Projekts und die Grundlagen, die wir entwickeln mussten, um dieses Problem in Zukunft lösen zu können. Diese Arbeit wurde während meines Forschungsaufenthalts an der University of British Columbia in Vancouver, Kanada, im Jahr 2019 begonnen.

\thispagestyle{plain}


\chapter*{Summary}
\addcontentsline{toc}{chapter}{Summary}

This thesis is located in the field of many-body quantum systems and covers two topics:
\begin{enumerate}[(i)]
\item BCS theory in the weak magnetic field regime,
\item Adiabatic theory for quantum lattice systems.
\end{enumerate}

The topic (i) constitutes the major part of this thesis. In this part, we investigate periodically extended superconductors that are subject to weak and macroscopic external electric and magnetic fields. These fields are such that the magnetic flux through the unit cell is non-vanishing, which opens up the main challenge for the contributions of this thesis. The superconductor is mathematically described in the framework of BCS theory of superconductivity. This theory is an effective two-particle pairing theory, which is named after the three physicists John Bardeen, Leon Neil Cooper, and John Robert Schrieffer, who postulated a Nobel prize awarded attractive interaction inside the superconductor in 1957. Their famous ansatz is inspiration for countless works and its verification from first principle quantum mechanics continues to challenge mathematical physics until the present day.

Our model is described in a variational manner in terms of a free energy functional, the so-called BCS functional, and we investigate the fluctuation of its minimum, the so-called BCS energy, and the critical temperature shift of the system, which are caused by the external fields in the weak-field limit. We show that in this regime the superconductor is described by Ginzburg--Landau theory and derive appropriate asymptotic descriptions.

We mainly contribute with significant conceptual simplifications in the trial state analysis compared to the works of Frank, Hainzl, Seiringer, and Solovej from 2012 and 2016 as well as a new decomposition result for the Cooper pair wave function which encompasses the constant magnetic field contribution. This decomposition result enables us to prove the aforementioned results for systems with such a constant magnetic field contribution. The latter is responsible for the non-vanishing magnetic flux and imposes significant difficulties to the mathematical treatment of the problem. The main mathematical reason is that the components of the magnetic momentum operator do not commute, which makes it impossible to use Fourier analysis. The work is splitted into two papers, which are included in this thesis.

In addition, we provide several results that may be helpful in continuing the project in the future. The first result is a decomposition result for magnetic potentials that arise in the context of periodic magnetic fields. These results are used to argue that the magnetic potentials covered in the works of Chapters \ref{Chapter:DHS1} and \ref{Chapter:DHS2} are exhaustive. We also provide a spectral decomposition of the periodic Landau Hamiltonian. Our last result is an asymptotic analysis of the low-lying spectrum of the operator $K_{T,\Abold} -V$, which plays a prominent role in BCS theory. This is done with the help of a Combes--Thomas  estimate for the resolvent kernel of $K_T-V$, which might be of independent interest.

A problem in BCS theory has always been the lack of introductory literature. We attempt to fill this gap by beginning with a comprehensive introduction to the subject, written for readers coming into contact with BCS theory for the first time.


In point (ii), we outline the proof of exponential estimates for the adiabatic theorem in extended quantum lattice systems. While the work is not yet complete, we present here the current state of the project and the new locality setup we needed to develop for such systems in order to solve this problem in the future. This work was started during my research stay at the University of British Columbia in Vancouver, Canada, in 2019.

\thispagestyle{plain}


\chapter*{Preface}
\addcontentsline{toc}{chapter}{Preface}

This thesis consists of three parts. In the first part, the main part, I present the results pertaining to BCS theory that have been obtained in collaboration with my advisor Christian Hainzl and my collaborator and predecessor Andreas Deuchert (University of Zurich). These results comprise two papers, which constitute the main project of my PhD studies:

\begin{enumerate}[(i)]
\item Microscopic Derivation of Ginzburg--Landau Theory and the BCS Critical Temperature Shift in a Weak Homogeneous Magnetic Field, Andreas Deuchert, Christian Hainzl, Marcel Maier (born Schaub), submitted to \emph{Probability and Mathematical Physics}, \href{https://arxiv.org/abs/2105.05623}{\texttt{ArXiv:2105.05623}}. This work will be referred to as \cite{DeHaSc2021} in this thesis and it is contained in Chapter \ref{Chapter:DHS1}. The content of Chapter \ref{Chapter:DHS1} differs from that in \cite{DeHaSc2021} insofar as we add Section \ref{DHS1:Addendum_proofs_section}, which contains slightly alternative proofs to some results, comprising more detailed descriptions than in \cite{DeHaSc2021}. 

\item Microscopic Derivation of Ginzburg--Landau Theory and the BCS Critical Temperature Shift in the Presence of Weak Macroscopic External Fields, about to be uploaded to the ArXiv. This work is the content of Chapter \ref{Chapter:DHS2}.
\end{enumerate}

In Chapters \ref{Chapter:DHS1} and \ref{Chapter:DHS2}, we have replaced references to ``M. Maier's PhD thesis'' that appear in the original works by explicit references to the corresponding passages in this document. In the affiliation list at the end of the respective chapter, the mailing address of myself has been removed since it becomes invalid in due time.

Chapters \ref{Chapter:DHS1} and \ref{Chapter:DHS2} are preceded by a comprehensive introduction to BCS theory from my own perspective, as far as it is needed to understand the material covered in this thesis. I have included this introduction in Chapter \ref{Chapter:Intro}, which is written for readers, who make their first contact with BCS theory. 

Chapter \ref{Chapter:Intro} is used to carefully introduce all the relevant components that are needed to study the BCS model as a mathematical model of superconductivity. We briefly explain the historic development and give an overview over all the techniques that are fundamental for contributing to the field. Chapter \ref{Chapter:Intro} also features several figures that have been designed for a talk at the International Congress of Mathematical Physics (ICMP) 2021 in Geneva. These aim at supporting the understanding transported by the mathematical formulas and explanations. The end of the chapter contains an overview of the status of this project and an outlook on possible follow-up projects. Chapter \ref{Chapter:Intro} is further supplemented by Chapters \ref{Chapter:Local_Traces} and \ref{Chapter:W-functions} in the appendix, which are included for the reader's convenience. The material covered there is not new: all the material is contained in the references pointed out there. However, occasionally, some additional details are found in the appendix compared to the original references, which has been a motivation to gather the content in one source and include it into the thesis. 

Part \ref{Part:Further_BCS} contains several results that did not have the place to be published within the above mentioned works (Chapter \ref{Chapter:DHS1} and \ref{Chapter:DHS2}) but might be useful for anybody who is involved in this project in the future. In Chapter \ref{Chapter:Abrikosov_gauge} we introduce a useful gauge for periodic magnetic fields, Chapter \ref{Chapter:Spectrum_Landau_Hamiltonian_Section} consists of an analysis of the spectrum of the periodic Landau Hamiltonian. Finally, Chapter \ref{Chapter:Combes-Thomas} provides the weak magnetic field asymptotics for the low-lying spectrum of the operator $K_{T, \Abold} - V$ for a wide class of magnetic fields. Our method of proof is a Combes--Thomas estimate for the resolvent kernel of $K_T - V$, which to the best of my knowledge does not exist in the literature. This analysis had been announced in \cite{DeHaSc2021} and in Chapter \ref{Chapter:DHS2}.

We remark that this thesis is, as far as BCS theory is concerned, a three-dimensional thesis. We do not cover two dimensions, although many results are valid with similar proofs also in this case.

Part \ref{Part:Adiabatic} finally consists of the unpublished and unfinished results that have been obtained in Vancouver, Canada, in collaboration with Sven Bachmann on the adiabatic theorem for extended quantum systems. The original goal was to provide exponential estimates for the results in the work \cite{SvenAdiabatic}. As this project did not come to a successful conclusion to the present date, and since several people kindly asked me to write this down somewhere, I'm happy to comply with their request and provide the corresponding results in Chapter \ref{Chapter:Adiabatic}. This chapter also contains a brief introduction to the field of quantum lattice systems.

The thesis does not contain a global introduction into its content. It is rather intended that the introduction consists of the Chapters \ref{Chapter:Intro} and the introductory Section \ref{AT:Intro}. We note that there is no global reference list at the end of this thesis. Rather, each chapter is followed by its own reference list.


\section*{Personal Contribution}
\addcontentsline{toc}{section}{Personal Contribution}

The results of Chapter \ref{Chapter:DHS1}, \ref{Chapter:DHS2}, and \ref{Chapter:Spectrum_Landau_Hamiltonian_Section} were obtained under the supervision of my advisor Christian Hainzl in collaboration with Andreas Deuchert. I was responsible for central ideas, for working out the proofs, as well as composing and formulating the manuscripts. The elaboration on the manuscripts was shared by Andreas Deuchert and me in equal parts.

The result of Chapter \ref{Chapter:Abrikosov_gauge} is a collaboration with Tim Tzaneteas. Tzaneteas was responsible for the first version of the manuscript, I contributed with the proof of Proposition \ref{GF:prop:func-eq} and a thorough revision of the manuscript.

The ideas for the results of Chapter \ref{Chapter:Combes-Thomas} originate from an unpublished note by Andreas Deuchert. I contributed by the elaboration on these, further ideas, and by composing and writing the manuscript.

The content of Chapter \ref{Chapter:Adiabatic} was obtained under the supervision of and collaboration with Sven Bachmann. I was resposible for conceptualization, central ideas, as well as for the manuscript.

\thispagestyle{plain}


\part{Main Results on BCS Theory}
\label{Part:Papers}

\begin{refsection}

\chapter{Introduction to BCS Theory}
\label{Chapter:Intro} \label{CHAPTER:INTRO}

Welcome to the introduction to BCS Theory. Both in content and writing, this introduction is intended for readers, who make their first contact with BCS theory, and who are in their advanced master's studies or beginning PhD studies --- not the experienced scientist. The style of writing in this chapter is rather informal and has the character of an overview. This means that the content is somewhat less dense and easier to read. However, it does not meet the scientific requirements of a paper. In this chapter, I occasionally switch from the common academic ``we'' to the personal form ``I'' when I want a message to be understood as a somewhat more ``personal advice'' and when I want to break the distance between the author and the reader.

The chapter contains the necessary fundamentals that are needed to understand the content of Parts \ref{Part:Papers} and \ref{Part:Further_BCS}. When further specialized knowledge is needed, I include a reference, where the reader can aquire this knowledge, if necessary. However, references are rather rare and there are only a few of them. If the reader expects further literature, they are kindly asked to consult the introductions and reference lists of Chapters \ref{Chapter:DHS1} and \ref{Chapter:DHS2}, where my collaborators and I have put a vast amount of literature. I also recommend a look at the introduction of Andreas Deuchert's PhD thesis \cite{AndiPhD} for an introduction from a slightly more physical point of view.

%




\section{Historical Development}
\label{Intro:History_Section}

\subsection{Preliminary remark}

In this historical synopsis, I recommmend to be \emph{somewhat} familiar with the lecture notes \cite{MBQM} by Jan Philip Solovej. In these notes, I have learned about many-body quantum Hamiltonians, the Friedrichs' extension, the Fock space, creation and annihilation operators, the formalism of second quantization, quadratic Hamiltonians, one- and two-particle density matrices, generalized one-particle density matrices, quasi-free states and their unitary implementation, and Bogolubov's method of approximation.

I stress the word ``somewhat'' in the preceding paragraph because for the reader to work with BCS theory \emph{only}, it is strictly speaking not necessary to know all these notions. The short explanation for this is that BCS theory is an effective two-particle theory, in which techniques like second quantization play no role anymore. A little more elaborate explanation is provided by the following example, which I would like to present because I myself found this very confusing at the beginning of my PhD studies: The mathematician working in the field of BCS theory calls a BCS state a ``one-particle density matrix'', which is a self-adjoint operator valued $2\times 2$ matrix with additional properties. We will follow this tradition below when we introduce the framework of BCS theory. However, as of today, this nomenclature can be confusing: There is a precisely defined notion of ``one-particle density matrix'' introduced for example in \cite{MBQM}, which suggests that we need to know everything about concepts like the Fock space, second quantization, and quasi-free states. However, to the best of my knowledge, there is no rigorously established connection that declares the former notion as a certain limit of the latter. In this sense, the naming ``one-particle density matrix'' in BCS theory is purely artificial. 

This has the following advantage for the inclined reader, should they be not familiar with the aforementioned notions. It opens up the possibility of reading this historical section without following the mathematical aspects I introduce and refer to (I put these mathematical terms nevertheless to make déjà vu moments possible for the reader and to make the section somewhat more vivid). If applicable, the reader may therefore confidently ignore their missing mathematical understanding of this section and step into the business when I introduce the BCS functional below in Section \ref{Intro_Section:BCS-functional}. In this way, nothing will be missed. Moreover, since it is pretty time consuming to work through the notes \cite{MBQM} (in particular, the problems, although thinking about these is enlightening!), it is certainly worth the consideration. Nevertheless, I could imagine that a little more understanding on the motivation is available if the reader has spent some time on the concepts above.

\subsection{Phenomenological description of superconductivity}

In 1911, the first experimental observation of superconductivity was achieved by Heike Kamerlingh Onnes at Leiden University, Netherlands, who was awarded the Nobel Prize in Physics for his discovery in 1913. He made experiments with a pure sample of mercury, which he obtained through repeated distillation, in liquid helium at temperatures of about 3-5 Kelvin. He had developed the apparatus to produce significant amounts of liquid helium himself the years before. The temperature is certainly a good above absolute zero, even for the state of the art at the time --- temperatures of as low as 0.00001 Kelvin had been reached and it was experimentally evident that absolute zero could not be realized. Onnes observed that the electric resistance of the mercury wire had disappeared completely. At the time, this effect had not been predictable. In fact, renowned scientists like William Thomson, 1.~Baron (Lord) Kelvin had debated beforehand whether the electric resistance should decrease linearly as the temperature decreases to zero (which had been a known behavior for higher temperatures), or if, as people including Onnes had envisioned, suddenly all motion of electrons comes to a stop, the resistance being infinitely high. In repeated experiments, Onnes' team observed and confirmed a transition from the normal conducting state of mercury to a state of zero electrical resistance at 4.2 Kelvin. Onnes himself introduced the term ``superconducting'' for this state. Later, the team found more sorts of metals which show similar transitions. The history of his experiments can be read in \cite{Onnes2, Onnes}, where the content of this paragraph is taken from.

\begin{figure}[h]
\centering
\includegraphics[width=10cm]{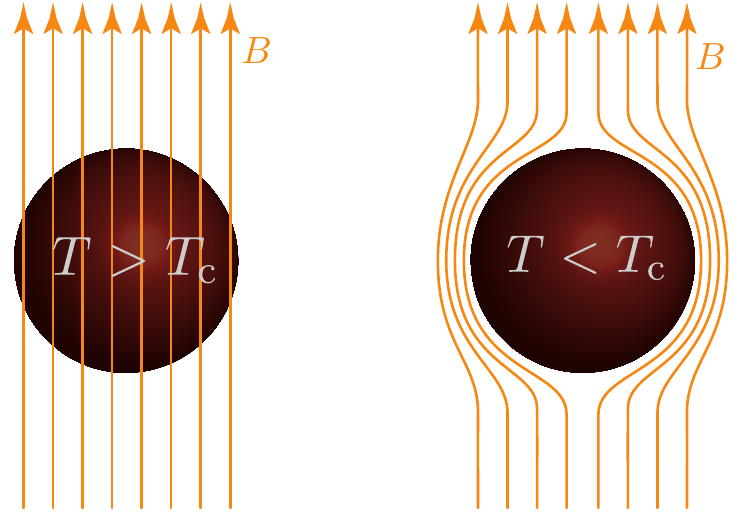}
\caption{Displacement of the magnetic field lines due to the Meißner effect.\label{Intro:Meißner}}
\end{figure}

In a short notice of one page, Walther Meißner and Robert Ochsenfeld reported on a new effect on superconductivity in 1933 \cite{Meissner}. Their experiments with the superconductors tin and lead in a weak external magnetic field induced by a coil showed that the distribution of the magnetic field lines in the exterior of the superconductor changed in such a way that is expected from a perfect diamagnet that has permeability zero. Meanwhile, in the center of the interior, the magnetic field remains almost unchanged, see Figure \ref{Intro:Meißner}. Likewise, if the exterior field was switched off in the superconducting phase, the interior field remained unchanged, while the exterior field did not vanish completely. An important parameter in this effect is the so-called \emph{penetration depth} of the exterior magnetic field into the superconductor. It describes the extent to which the interior ``reaction field'' is able to repell the penetrating exterior field.

The effect is nowadays known as the \emph{Meißner--Ochsenfeld effect} (often just \emph{Meißner effect}) and it had not been explainable with the classical physics known at the time. The authors further report on distiguished critical temperaturs (``jumping points'' as they call it) depending on whether the temperature passes the critical temperature increasingly or decreasingly. This is a phenomenon we will have to deal with also mathematically, when exterior fields are present. The rigorous microscopic justification of the Meißner effect remains a challenge for mathematicians until the present day.

The Meißner effect can be visualized by a hovering piece of superconducting material over a large electromagnet, as shown, e.g., in the video \cite{Video_Meissner}. The physical picture is that the electromagnet exposes the superconductor to a weak external magnetic field which the superconductor is able to repel from its interior by a response field that is directed opposite to the exterior field. It should be said that the Meißner effect persists in the presence of a \emph{weak} external magnetic field, which makes us be interested in the \emph{weak magnetic field regime}. After all, it is plausible that the penetration depth increases with the field strength of the exterior magnetic field so that superconducting effects will be ``destroyed'' when the field strength becomes too large. Below, we will define precisely what we mean mathematically by ``weak'' in this context. 

It took almost 40 years from the first discovery by Onnes until Vitali Ginzburg and Lev Landau presented the first theoretical, phenomenological, and macroscopic quantum description of superconductivity in 1950 \cite{GL}. The theory is based on a system of two partial differential equations for a single complex-valued function $\psi$, the so-called order parameter, and the response field, whose domains in $\Rbb^3$ cover the dimensions of the material. The function $\psi$ has the property that $|\psi(x)|$ ranges between 0 (absence of superconductivity) and 1 (presence of superconductivity) at the respective point $x$. Ginzburg--Landau theory has been highly influencial throughout the physics community and is capable of describing various macroscopic effects of superconductors. It continues to be a very active field of research until the present day, covering more and more complicated effects and domains. For us, Ginzburg--Landau theory arises as a limiting macroscopic counterpart to the microscopic BCS theory of superconductivity, which we shall discuss now.

\subsection{BCS theory in the early days of many-body quantum systems}

BCS theory is a microscopic theory of superconductivity that is named after the three physicists John Bardeen, Leon Neil Cooper, and John Robert Schrieffer. The theory was presented in their famous 1957 paper \cite{BCS}, which fits into a period marked by scientists devoting themselves to the systematic analysis of many-body quantum Hamiltonians, the latter being agreed to provide an appropriate description of microscopic physical systems.

The objective is to describe a system of $N\geq 1$ particles in a three-dimensional metallic box $Q_L = [0,L]^3$ of sidelength $L>0$.
\begin{figure}[h]
\centering
\includegraphics[width=10cm]{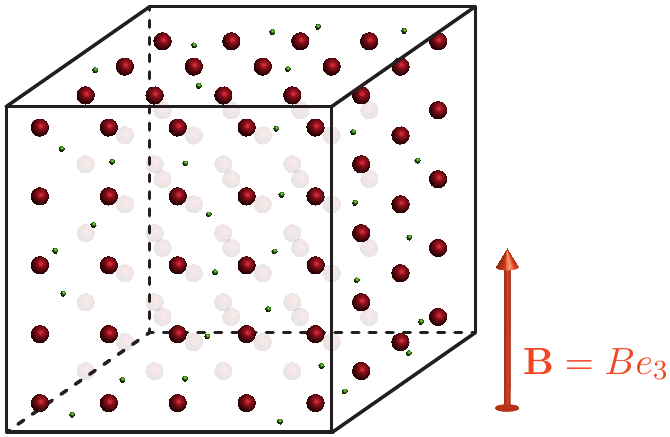}
\caption{System of fermionic particles (green balls) in a metallic box with lattice ions (red balls) subject to a constant external magnetic field $\Bbold$ (orange arrow) pointing in the $e_3$-direction.\label{Fig:fermionic_particles_in_box}}
\end{figure}
The system is subject to external fields, where the most relevant contribution is given by the constant magnetic field. It is modeled by a vector $\Bbold\in \Rbb^3$ having strength $B := |\Bbold| >0$. Such a system is displayed in Figure \ref{Fig:fermionic_particles_in_box}. In fact, we will cover more general fields in our mathematical description later.

A Hamiltonian operator which describes this situation reads, in suitable units,
\begin{align}
H_N := \sum_{i=1}^N (-\i \nabla_i + \Abold(x_i))^2 - \mu - \sum_{1 \leq i < j \leq N} V(x_i - x_j). \label{Intro:Hamiltonian_N-particles}
\end{align}
Here, $\Abold\colon \Rbb^d\ra \Rbb^d$ is a magnetic potential corresponding to the magnetic field $\curl \Abold = \Bbold$ and $V\colon \Rbb^3\ra \Rbb$ is a two-particle interaction potential. This Hamiltonian is usually realized self-adjointly in the Hilbert space $L^2(Q_L)^{\otimes N}$, the $N$-fold tensor product of $L^2(Q_L)$ with itself. Depending on the particle statistics, the domain may be restricted to the subspace $L^2(Q_L)^{\otimes_{\mathrm s} N}$ of $N$-body wave functions that are symmetric with respect to exchange of any of their coordinates (bosonic), or to the subspace $L^2(Q_L)^{\wedge N}$ of anti-symmetric wave functions (fermionic). For simplicity, we are neglecting spin throughout Parts \ref{Part:Papers} and \ref{Part:Further_BCS} of this thesis. If suitable boundary conditions are phrased, the self-adjoint realization is obtained from the corresponding differential expression on smooth functions by the Friedrichs' extension method.

We are then interested in the ground state energy, the ground state, and equilibrium properties of this Hamiltonian. Of course, these properties are given by the stationary or dynamic Schrödinger equation, so, for example, we are interested in the lowest eigenvalue of the operator $H_N$. However, it is now well known that it is extremely difficult to extract from the full Hamiltonian $H_N$ the information needed to calculate the aforementioned quantities. Undoubtedly, we are thus in need of powerful approximation methods, which allow us to consider simplified models and to carry out the computations on these, while we are able to control the errors. Unavoidably, we are then forced to restrict our predictions to limiting regimes. The most prominent examples are the thermodynamic limit $N\to \infty$, $L \to \infty$ in such a way that the particle density $\rho := N/L^3$ is kept fixed, or the mean-field and Gross--Pitaevskii limit $N\to \infty$, where the interaction strength is coupled to the interparticle distance through suitable scaling factors in the potential $V$. Even after this limit has been taken, we are then forced to further restrict our statements to limiting regimes like the ``dilute''/``low-density'' regime $\rho \ll 1$, the ``high-density'' regime $\rho \gg 1$, or the ``adiabatic'' regime of a ``slow'' dependence on time of the underlying Hamiltonian, if such a dependence is present.

In the late 1940's, physicists were able to develop the first successful and influencial ideas to systematically approximate these Hamiltonians. This endeavor had begun with the 1947 seminal paper by Nikolay Nikolayevich Bogolubov \cite{Bogolubov}, which continues to influence papers in modern mathematical quantum many-body descriptions up to the present day. In this paper, Bogolubov introduced a method to systematically approximate many-body Hamiltonians that are second-quantized with periodic boundary conditions, i.e., \eqref{Intro:Hamiltonian_N-particles} rewritten as
\begin{align}
H_N = \sum_{p \in \frac{2\pi}{L} \Zbb^3} \hfrak_p \;  a_p^* \, a_p - \frac{1}{2L^3} \sum_{k, p, q\in \frac{2\pi}{L} \Zbb^3} \hat V(k) \; a_{p+k}^* \, a_{q - k}^* \, a_q \, a_p, \label{Intro:Hamiltonian_2nd-quantized}
\end{align}
with $\hfrak_p = \langle u_p,T u_p\rangle$, $T:= (-\i \nabla + \Abold)^2-\mu$, and $u_p(x) = (2\pi L)^{-{\nicefrac 32}} \e^{-\i p\cdot x}$ the plane wave basis.

\begin{figure}[h]
\centering
\includegraphics[width=15cm]{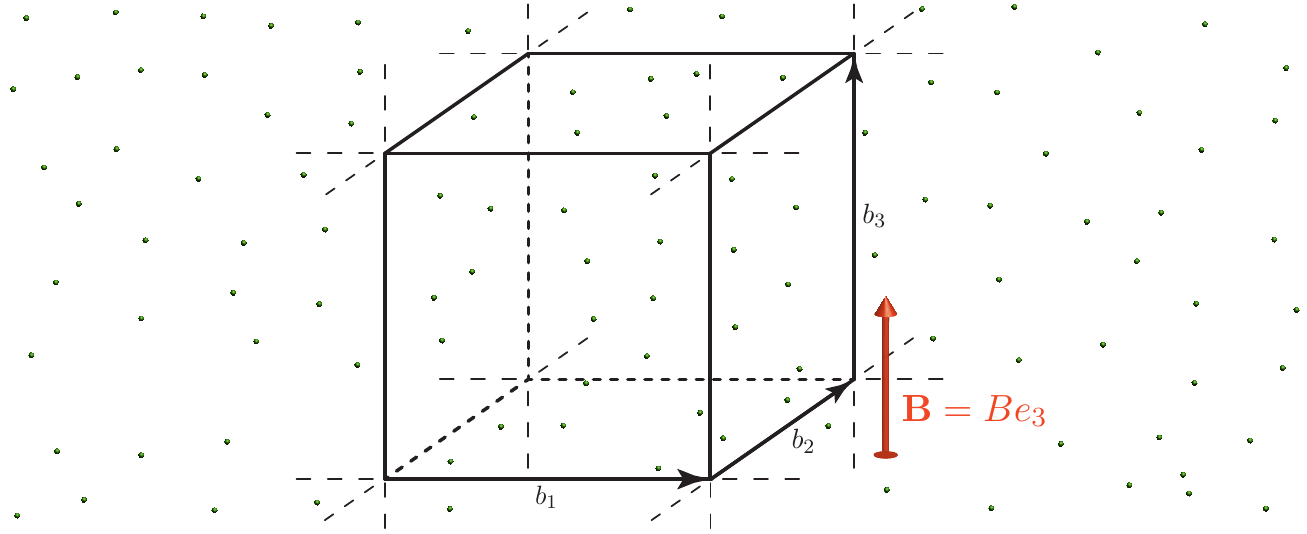}
\caption{A periodic system with a unit cell spanned by the vectors $b_1, b_2, b_3\in \Rbb^3$.}
\end{figure}

Then, it is argued that this Hamiltonian is well-approximated by what we nowadays call a ``quadratic Hamiltonian'', namely a quadratic expression in the creation and annihilation operators $a_p^*$ and $a_p$. Note that, as of \eqref{Intro:Hamiltonian_2nd-quantized}, the interaction term is quartic in these operators. The quadratic Hamiltonian has the feature of being explicitly diagonalizable so its spectrum is computable as a closed formula, see \cite[Problem 11.9]{MBQM}. The idea that paved the way for his procedure, the famous c-number substitution, came from the bosonic case. It had been known from the Albert Einstein 1925 paper \cite{Einstein} that the free (i.e., non-interacting, $V\equiv 0$) bosonic gas undergoes complete Bose--Einstein condensation in the ground state. This means that every of the $N$ particles is in its one-particle ground state (the zero momentum, i.e., constant mode) and the many-body ground state is a pure tensor product of these one-particle states. Of course, this assertion does not persist when interactions are present. However, the idea was used as an assumption for the interacting case by Bogolubov to replace the creation and annihilation operator $a_0^*$ and $a_0$ by the number of particles $\sqrt{N}$ (c-number substitution), since the occupation $a_0^*a_0$ of the ground state has an expectation that is almost equal to $N$ (i.e., almost all particles are condensed). Furthermore, he argued that every $a_p^*$ and $a_p$ with $p \neq 0$  is ``small'' compared to the large fraction of condensed particles. This enabled him to view the Hamiltonian as a ``power series'', in which he could ``neglect'' the remaining terms of ``higher order'' in the ``excitation'' operators $a_p^*$ and $a_p$ for $p \neq 0$. Nowadays, we know that this procedure is not always correct to the claimed precision (that, actually, had been pointed out by Landau soon after the publication). However, the method that modern proofs use for proving what actually holds is, in its core, inspired by \cite{Bogolubov}.

Bogolubov's method as he developed it was used by several people to compute the ground state energy of bosonic as well as fermionic systems by just computing the ground state energy densities of the respective resulting quadratic Hamiltonians. These formulas are the very well-known Lee--Huang--Yang formula in the bosonic case, see \cite{Fournais_Solovej_Bose_Gas},
\begin{align}
e_{\mathrm{B}}(\rho) &= 4\pi a \rho^2 \Bigl[  1+ \frac{128}{15 \sqrt \pi} (\rho a^3)^{\nicefrac 12}\Bigr] + o(\rho a^3)^{\nicefrac 12}, & \rho \ll 1, \label{Intro:LHY}
\end{align}
where $a$ denotes the scattering length of the interaction potential $V$, and the somewhat less-known Huang--Yang formula in the fermionic case (for spin $\nicefrac 12$ fermions)
\begin{align}
e_{\mathrm F}(\rho_\uparrow, \rho_\downarrow) &= \frac 35 (6\pi^2)^{\frac 23} \bigl( \rho_\uparrow^{\frac 53} + \rho_\downarrow^{\frac 53} \bigr) + 8\pi a \rho_\uparrow \rho_\downarrow + o(\rho^2),  & \rho := \rho_\uparrow + \rho_\downarrow \ll 1, \label{Intro:HY}
\end{align}
see \cite{FGHP20}. Here, $\rho_\uparrow$ and $\rho_\downarrow$ denote the densities of spin-up and spin-down particles, respectively. Meanwhile, these formulas have been proven in the vaccuum at temperature absolute zero. However, at the time there was no knowledge how far these quantities actually deviate from the true ground state energy in the respective regimes.

So far, we had a look on the ground state, that is, a zero temperature property. The 1957 paper \cite{BCS} analyzes what happens to fermionic systems at positive temperature $T > 0$, which amounts to studying the free energy (or, pressure)  functional
\begin{align*}
F_T(\langle \cdot \rangle) = \langle H_N\rangle - T\, S(\langle \cdot \rangle)
\end{align*}
of the system. Here $\langle \cdot \rangle$ is a state given by $\langle A\rangle = \sum_{n\in \Nbb} \lambda_n \langle \psi_n, A\psi_n\rangle$ for all observables $A$, where $0 \leq \lambda_n \leq 1$, $\sum_{n\in \Nbb} \lambda_n = 1$, $\{\psi_n\}_{n\in \Nbb}$ is an orthonormal basis of the underlying Hilbert space, and $S(\langle \cdot \rangle) := -\sum_{n\in \Nbb} \lambda_n \, \ln \lambda_n$ is the von Neumann entropy. The first observation is that the \emph{normal state} of the system at positive temperature is now the Gibbs state at temperature $T$ induced by the Hamiltonian $H_N$ and given by
\begin{align}
\langle A\rangle = \frac{\Tr [A \e^{-\beta H_N}]}{\Tr \e^{-\beta H_N}},
\end{align}
where $\beta := T^{-1}$ is the inverse temperature, see \cite{MBQM}.

Since the authors wanted to describe superconductors or superfluids they must have thought of appropriate substitutes to Bogolubov's assertion at zero temperature, which had to be combined with a suitable mechanism that allows for a notion of ``condensation'' for fermions (by the Pauli exclusion principle, a condensation of fermions is a priori impossible). Their main physical assumption, which drives the theory, is the \emph{attractive interaction} potential $-V$ which is postulated to be present due to the phonon vibrations caused by attraction of the lattice ions with the electrons. In this way, a scattering electron causes a slight displacement of the lattice ion which in turn drags the other electrons in the same direction. This induces an \emph{effective attractive interaction between electrons}.
\begin{figure}[h]
\centering
\includegraphics[width=12cm]{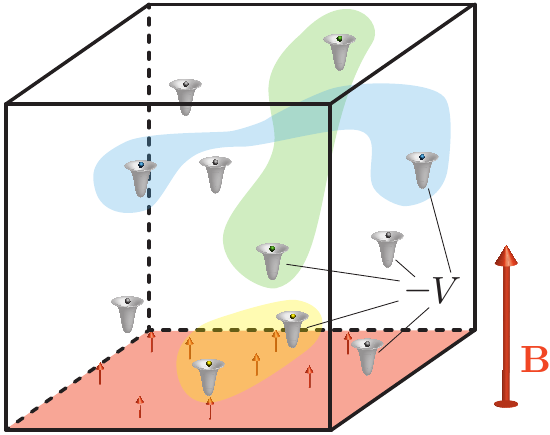}
\caption{Fermions interact via an attractive two-body potential -- Cooper pair formation indicated by the colored clouds.}
\end{figure}
With these ideas, BCS constructed a trial state for the problem that features the particles to undergo a \emph{pairing mechanism} which would then cause the pairs of fermions to behave like bosons that could condense again in the sense of Bogolubov. The pairs of electrons are called Cooper pairs nowadays due to an earlier publication \cite{Cooper} by Leon Neil Cooper. The BCS trial state could be shown to introduce a superconducting phase to the system, which means that it has strictly lower energy than the normal state. Their publication led to the significant and ongoing interest and influence that the paper is known for. BCS did not only compute energies but also made predictions concerning the penetration depth, the Meißner effect and further quantities. These predictions hold on a physical level of rigor, which means that a mathematical proof for significant parts of the theory is still missing. In particular, we lack necessity arguments for the assumed form of the trial state.

\subsection{Origin of the BCS functional}


%
%

Nowadays, we recognize the state used by BCS as a quasi-free state. Since this class of states is important for BCS theory, let us give a brief review of the basic ingredients. From now on, we fix the particle statistics to be fermionic.

A fermionic quasi-free (pure) state $\Psi$, which is a vector in the fermionic Fock space $\Fcal(\Hcal) := \bigoplus_{n\geq 0} \Hcal^{\wedge n}$ built upon a one-particle Hilbert space $\Hcal$, can be characterized by its generalized one-particle density matrix, the operator $\Gamma_\Psi \colon \Hcal \oplus \Hcal^* \ra \Hcal \oplus \Hcal^*$ defined as
\begin{align*}
\bigl\langle f_1 + Jg_1, \Gamma_\Psi (f_2 + Jg_2) \bigr\rangle_{\Hcal \oplus \Hcal^*} := \bigl\langle \Psi, \bigl( a^*(f_2) + a(g_2)\bigr) \bigl( a(f_1) + a^*(g_1)\bigr) \Psi\bigr\rangle_{\Fcal(\Hcal)}.
\end{align*}
Here, $J\colon \Hcal \ra \Hcal^*$ is the anti-linear Riesz' isomorphism given by $Jg(f) = \langle g, f\rangle$, i.e., $Jg$ is the linear functional ``taking the scalar product with $g\in \Hcal$''. Moreover, $a^*(f)$ is the creation operator corresponding to the state $f$ and $a(f)$ is the annihilation operator corresponding to the state $f$. Their definition shall not be of importance here, it is given in \cite{MBQM}. We only note that $a^*(g)$ is linear in $g$ whereas $a(f)$ is anti-linear in $f$ and the \emph{canonical anti-commutation relations}
\begin{align}
\{ a(f), a(g) \} = \{ a^*(f), a^*(g) \} &= 0 , & \{ a(f) , a^*(g) \} &= \langle f, g\rangle_\Hcal\, \Idbb, \label{Intro:CAR}
\end{align}
hold. In terms of the \emph{one-particle density matrix} $\gamma_\Psi \colon \Hcal \ra \Hcal$ and the offdiagonal operator (sometimes also called the \emph{two-particle density matrix}) $\alpha_\Psi \colon \Hcal^* \ra \Hcal$, defined by
\begin{align}
\langle f, \gamma_\Psi g\rangle_\Hcal &:= \langle \Psi, a^*(g) a(f)\Psi\rangle_{\Fcal(\Hcal)}, & \langle f, \alpha_\Psi Jg\rangle_\Hcal &:= \langle \Psi, a(g) a(f) \Psi\rangle_{\Fcal(\Hcal)}, \label{Intro:gamma_alpha_definition_Fock}
\end{align}
the generalized one-particle density matrix takes the form of a $2\times 2$ operator valued matrix of the form
\begin{align}
\Gamma_\Psi = \begin{pmatrix} \gamma_\Psi & \alpha_\Psi \\ \alpha_\Psi^* & 1 - J\gamma_\Psi J^*\end{pmatrix}, \label{Intro:One_particle_density_matrix_proper}
\end{align}
where it can be shown that $\alpha_\Psi^* = -J\alpha_\Psi J$ due to \eqref{Intro:CAR} and \eqref{Intro:gamma_alpha_definition_Fock}. We further note that $\gamma_\Psi$ is self-adjoint and $0 \leq \gamma_\Psi \leq 1$. Likewise, $\Gamma_\Psi$ is self-adjoint as well and $0 \leq \Gamma_\Psi \leq 1$. Since
\begin{align}
\Gamma_\Psi (1 - \Gamma_\Psi) &= \begin{pmatrix} \gamma_\Psi(1 - \gamma_\Psi) - \alpha_\Psi \alpha_\Psi^* & \alpha_\Psi \ov \gamma_\Psi - \gamma_\Psi \alpha_\Psi \\ \ov \gamma_\Psi \alpha_\Psi^* - \alpha_\Psi^* \gamma_\Psi & \ov \gamma_\Psi(1 - \ov \gamma_\Psi) - \alpha_\Psi^* \alpha_\Psi \end{pmatrix}, \label{Intro:Gamma(1-Gamma)}
\end{align}
the latter implies $\Gamma_\Psi (1 - \Gamma_\Psi)\geq 0$, whence $\alpha_\Psi$ and $\gamma_\Psi$ are related through the operator inequality
\begin{align}
\alpha_\Psi \alpha_\Psi^* \leq \gamma_\Psi (1 - \gamma_\Psi). \label{Intro:gamma_alpha_relation_proper}
\end{align}
Moreover, if $\Psi$ has \emph{finite particle expectation}, that is, $\Psi = \bigoplus_{N=0}^\infty \Psi^{(N)}$ with
\begin{align}
\langle \Psi, \Ncal \Psi\rangle :=\sum_{N=0}^\infty N\, \Vert \Psi^{(N)}\Vert_{\Hcal^{\wedge N}}^2 < \infty,
\end{align}
then $\gamma_\Psi$ is trace class and $\Tr \gamma_\Psi = \langle \Psi, \Ncal \Psi\rangle$. Consequently, $\alpha_\Psi \alpha_\Psi^* \colon \Hcal \ra \Hcal$ is a trace class operator by \eqref{Intro:gamma_alpha_relation_proper}. For all these facts, see \cite{MBQM}.

With these notions at hand, it is possible to follow the somewhat lengthy calculation given in \cite[Sect. 2.1]{Hainzl2015}, which we will not repeat here and which leads in a mathematically non-rigorous fashion to the BCS functional. Non-rigorous means that certain simplifications pertaining to $\mathrm{SU}(2)$-symmetry (neglection of spin) have to be made and certain infinite-volume limits have to be taken which replace sums by integrals. To the present day it is unclear how to rigorously do this. Furthermore, the so-called \emph{direct} and \emph{exchange} terms have to be neglected, which show up due to the properties of quasi-free states. See \cite{Hainzl_Braeunlich_direct_exchange} for a detailed discussion. It is sometimes argued that this neglection actually produces more accurate experimental results, which opens up the question if quasi-free states are the right type of state in the end.

A proper derivation of the functional would also have to clarify the correct type of state which allows for the desired pairing mechanism in the many-body model. In particular, we would have to understand if (and, if yes, why) quasi-free states are the appropriate class on which the superconductivity properties of the many-body quantum system are correctly displayed. This is very much related to the question why pairs are the appropriate size of clusters that are formed by electrons and why there do not exist multituples of quantum mechanically correlated particles in a significant share. In other words, a rigorous procedure that justifies BCS theory from the perspective of first principle many-body model as the correct effective microscopic theory has to understand the mechanism of suppressing multituples compared to pairs. All this is not part of this thesis and to the best of my knowledge unknown. Therefore, we will take the BCS theory of superconductivity as a given model and analyze it --- given the evidence of it arising from the many-body setting by the procedure sketched above.

To conclude this historical upshot, I once more emphasize that the ``generalized one-particle density matrix'' above is mathematically independent of what we are going to define in the next section but the roles they play are similar, hence the naming.


\section{The BCS Functional of Superconductivity}
\label{Intro_Section:BCS-functional}

For the purpose of this thesis, BCS theory is given in a variational manner by an energy functional and the physical properties of this functional are captured in the lowest energy over all admissible states and the minimizer of this problem. As usual, the minimizer will satisfy an Euler--Lagrange equation, which in this case is known as the \emph{Bogolubov--de-Gennes equation}. This equation, however, will not play a prominent role in this thesis, since we will be mostly dealing with so-called \emph{low-energy states} of the functional, which are not necessarily minimizers. In order to define the functional, we first write down a formal expression for the functional and then make sense of all quantities appearing in there step by step in the rest of this chapter. For a state $\Gamma$ of the form
\begin{align}
\Gamma &= \begin{pmatrix} \gamma & \alpha \\ \ov \alpha & 1 - \ov \gamma\end{pmatrix}, \label{Intro:Gamma_formal}
\end{align}
the formal expression reads
\begin{align}
\Tr_\Omega \bigl[ \bigl((-\i \nabla + \Abold(x))^2 + W - \mu\bigr) \gamma \bigr] - T S(\Gamma) - \fint_{\Omega} \dd X \int_{\Rbb^3} \dd r \; V(r) \, |\alpha(X, r)|^2.\label{Intro:BCS_functional_formal}
\end{align}
We refer to the first term as the \emph{kinetic energy}, which is in analogy to the many-body Hamiltonian. The second term will be the \emph{entropy term} and the last term is the \emph{interaction term} or \emph{pairing term}. The superconductivity of the system will be indicated by a non-vanishing Cooper pair wave function $\alpha$, as we will see more precisely below. In the context of energy functionals, whose states have a matrix structure as in \eqref{Intro:Gamma_formal}, the non-vaninshing off-diagonal entry $\alpha$ is sometimes also referred to as the present \emph{off-diagonal long range order (ODLRO)} of the system in the literature. This wording results from the operator $\alpha$, discussed in \eqref{Intro:gamma_alpha_definition_Fock}, which models \emph{two-particle correlations} in the system.

We should think of the functional in the following way. The kinetic term is the dominating contribution to the energy, i.e., the BCS functional is bounded from below because the entropy and the interaction can by bounded by a portion of the kinetic energy. If it was only for the kinetic energy, the state minimizing this functional would obviously be the characteristic function $\gamma = \Idbb_{(-\infty, 0]}((-\i \nabla + \Abold)^2 + W - \mu)$, which is known to be smeared out if the entropy is taken into account and $\Gamma$ is a diagonal matrix. The minimizer is then $(1 + \e^{\beta (-\i \nabla + \Abold)^2 + W - \mu)})^{-1}$. The interaction term competes against these two terms in that it ``wants to choose'' $\alpha$ as big as possible. This however, is limited by the constraint \eqref{Intro:gamma_alpha_relation_proper} (actually rather by the requirement $0 \leq \Gamma \leq 1$). Hence, $\gamma$ and $\alpha$ depend on each other in a subtle way.

Let us start with the quantities in \eqref{Intro:BCS_functional_formal} that are easily explained. First of all, $\mu\in \Rbb$ is the \emph{chemical potential} (or \emph{Fermi energy}) that plays the role of fixing the number of particles in a grand canonical quantum system. The region $\{p\in \Rbb^3: p^2 \leq \mu\}$ is often referred to as the \emph{Fermi sea}. Secondly, $T\geq 0$ is the \emph{temperature} of the system. We also have built in the attractive character of the interaction potential $V\colon \Rbb^3 \ra \Rbb$ from the beginning by writing $-V$. We may think of a nonnegative potential $V\geq 0$ so that $-V$ is indeed a negative function but for the results in this thesis to hold, this is not necessary. The \emph{mean integral} $\fint_\Omega \dd X \; f(X)$ is defined as $\frac{1}{|\Omega|} \int_\Omega \dd X \; f(X)$ for some function $f$. 

We further have to make sense of the following expressions that appear in \eqref{Intro:BCS_functional_formal}: 
\begin{itemize}
\item We want to set up the BCS model in a so-called \emph{gauge-periodic} fashion, where the domain $\Omega\subset \Rbb^3$ will be the \emph{unit cell of the lattice of periodicity}. As we will see, our notion of periodicity depends on the \emph{magnetic potential} $\Abold\colon \Rbb^3\ra \Rbb^3$, the \emph{gauge} of the magnetic field. We will therefore call the system \emph{gauge-periodic} and the size of $\Omega$ has to be linked to certain properties of $\Abold$. In particular, we need to understand different periodicity properties of the fields: The periodic function $W\colon \Rbb^3\ra \Rbb$ modeling the potential corresponding to an \emph{external electric field} will be \emph{truly} periodic (not gauge-periodic). The same holds for parts of the magnetic potential $\Abold$ and we shall clarify why.

\item The gauge-periodic state $\Gamma$ in \eqref{Intro:Gamma_formal}, which consists of the two components $\gamma$ and $\alpha$ ($\ov \gamma$ and $\ov \alpha$ denotes complex conjugation).

\item $\Tr_\Omega$ is the \emph{trace per unit volume} of the periodic operator which stands inside the argument.

\item $S(\Gamma)$ should be the usual \emph{von Neumann entropy per unit volume} of the state $\Gamma$ defined as $S(\Gamma) := -\Tr_\Omega[\Gamma \ln \Gamma]$, once $\Tr_\Omega$ is properly defined.

\item The function $\alpha(X, r)$ equals, per slight abuse of notation, $\alpha(x, y)$, which, in turn, is the kernel of $\alpha$. This kernel exists because $\alpha$ will be a Hilbert--Schmidt operator per unit volume on the natural domain of the formal expression \eqref{Intro:BCS_functional_formal}. The center-of-mass coordinate $X = \frac{x+y}{2}$ and the relative coordinate $r = x-y$ play a very important role in the regime of \emph{weak external} fields. This is the regime that we are interested in since this is the regime in which the Meißner effect takes place. 
\end{itemize}

The construction of the main items above requires substantial mathematical effort and will occupy us for the rest of this chapter. We will approach these quantities while we discuss the BCS functional in increasingly difficult settings. This discussion helps us to come across all the further relevant objects that are necessary to be known when we want to work with BCS theory. We start with the simplest situation.


\section{Translational Invariance}
\label{Intro:BCS-functional-ti_Section}

Let us first take a look at the BCS functional in the simplest situation: We look at non-interacting particles and assume that $\Abold =0$ and $W=0$. In this case, we assume that the system is translation invariant, that is, $\gamma$ and $\alpha$ are operators given by translation-invariant kernels $\gamma(x - y)$ and $\alpha(x-y)$, respectively. 

This assumption is justified since the translation-invariant BCS functional defined below in \eqref{Intro:BCS_functional_ti_definition} does not break translational invariance in the sense that the minimizer over all states indeed is translation invariant provided the temperature lies in a sufficiently small interval below $\Tc$. This result has been provided in the work \cite{DeuchertGeisinger}.

Then, the trace is simply given by evaluating the kernel of the operator in question on the diagonal $x = y$ and integrating. We can apply the Fourier transform and express everything in terms of two functions $\gamma \in L^1(\Rbb^3, (1 + p^2)\dd p)$ and $\alpha\in H^1(\Rbb^3, \dx)$ with $|\hat \alpha(p)| \leq \gamma(p) (1 - \gamma(p))$. A short calculation shows that, in this case, the BCS functional is given by the simpler expression (up to factors of $2\pi$ depending on the convention on the Fourier transform, see \cite[Eq. (3.1)-(3.6)]{Hainzl2015} or \cite{Hainzl2007})
\begin{align}
\Fcal_{\mathrm{ti}}^{\mathrm {BCS}}(\gamma, \alpha) := \int_{\Rbb^3} \dd p \; (p^2-\mu) \, \gamma(p)- T S(\gamma, \alpha) - \int_{\Rbb^3} \dd x \; V(x) \, |\alpha(x)|^2, \label{Intro:BCS_functional_ti_definition}
\end{align}
where
\begin{align}
S(\gamma, \alpha) &:=  - \int_{\Rbb^3} \dd p \; \bigl[ s_+(p) \ln s_+(p) + s_-(p) \ln(s_-(p))\bigr]
\end{align}
and $s_\pm(p)$ are the eigenvalues of the matrix
\begin{align*}
\hat \Gamma(p) = \begin{pmatrix} \gamma(p) & \hat \alpha(p) \\ \ov{\hat{\alpha} (p)} & 1 - \gamma(p) \end{pmatrix}
\end{align*}
that are determined by $s(1 - s) = \gamma(1 - \gamma) - |\hat \alpha|^2$, which means
\begin{align}
s_\pm(p) = \frac 12 \pm \frac 12 \sqrt{|\hat \alpha(p)|^2 + (1 - 2\gamma(p))^2},
\end{align}
whence, with $s(p) := s_+(p)$, we conclude
\begin{align}
S(\gamma, \alpha) &= - \int_{\Rbb^3} \dd p \; \bigl[ s(p) \ln s(p) + (1-s(p)) \ln(1-s(p))\bigr].
\end{align}

\subsection{Normal state}

The normal state is the minimizer of $\Fcal_{\mathrm{ti}}^{\mathrm{BCS}}$ in the absence of interactions, i.e., $V \equiv 0$. In this case, the functional reads
\begin{align}
\Fcal_{\mathrm{ti}}^{\mathrm{BCS}}(\gamma, \alpha) = \int_{\Rbb^3} \dd p\; (p^2 - \mu)\, \hat \gamma(p) - T S(\gamma, \alpha) \label{Intro:BCS-functional-ti}
\end{align}
and the first question to answer is whether the minimizer satisfies $\alpha \equiv 0$. To see that this is indeed the case, we define 
\begin{align}
\varphi(x) &=  x \ln(x) + (1- x) \ln(1 - x), & 0 &\leq x \leq 1. \label{Intro:varphi}
\end{align}
Note that the function is defined at the endpoints via continuous extension and it is strictly convex. Then, we need to employ the relative entropy inequality
\begin{align}
\int_{\Rbb^3} \dd p \; \varphi(s(p)) - \varphi(s'(p)) &\geq \int_{\Rbb^3} \dd p \; \varphi'(s'(p)) \, (s(p) - s'(p)), \label{Intro:Klein_invariant}
\end{align}
where $0 \leq s'(p) \leq 1$ is arbitrary. Equality holds if and only if $s = s'$. This inequality is known as Klein's inequality and it can be proven similarly to Theorem \ref{Klein's inequality}. We will discuss the general case of this trace inequality when it comes to the trace per unit volume below so we will not be detailed here.

We apply this to
\begin{align*}
s'(p) &:= \frac 12 + \frac 12 \, |1 - 2\gamma(p)| \leq s(p).
\end{align*}
Since $\varphi'(x) = \ln(\frac{x}{1-x})$, we have
\begin{align*}
\varphi'(s'(p)) = \ln \Bigl( \frac{1 + |1 - 2\gamma(p)|}{1 - |1 - 2\gamma(p)|} \Bigr) \geq 0,
\end{align*}
so that
\begin{align*}
S(\gamma, \alpha) - S(\gamma, 0) \geq 0
\end{align*}
and the inequality is strict unless $\alpha =0$.

This shows that it is energetically favorable for the minimizer to satisfy $\alpha =0$. An explicit minimization of the functional
\begin{align*}
\Fcal_0(\gamma) &:= \int_{\Rbb^3} \dd p \; (p^2-\mu) \gamma(p) - TS(\gamma, 0)
\end{align*}
proves that the well-known \emph{Fermi--Dirac distribution}
\begin{align}
\gamma_0(p) &:= \frac{1}{1 + \e^{\beta (p^2-\mu)}} \label{Intro:Fermi-Dirac}
\end{align}
minimizes \eqref{Intro:BCS-functional-ti}. To see this, we differentiate with respect to $\gamma$ and obtain the critical equation
\begin{align}
p^2-\mu + T\ln\Bigl( \frac{\gamma(p)}{1 - \gamma(p)}\Bigr) =0, \label{Intro:Normal_state_critical_equation}
\end{align}
whose solution is \eqref{Intro:Fermi-Dirac}. By the strictness of the inequality \eqref{Intro:Klein_invariant}, we have proven that \eqref{Intro:Fermi-Dirac} is also the unique minimizer.
This is the stable minimizer of the BCS functional when no external fields and no interaction are present in the model. We use the critical equation \eqref{Intro:Normal_state_critical_equation} to see that its BCS energy is given by
\begin{align}
\Fcal_{\mathrm{ti}}^{\mathrm{BCS}}(\gamma_0, 0) &= T \int_{\Rbb^3} \dd p \; \ln(1 - \gamma_0(p)) = T \int_{\Rbb^3} \dd p \; \ln(1 + \e^{-\beta (p^2 - \mu)}).
\end{align}

\subsection{Superconductivity}

The interesting features of the model become visible if we allow for an interaction potential $V\neq 0$. We used suggestive notation in the interaction term and write the interaction with a minus sign as $-V$.

We shall think of this as an attractive interaction potential as was postulated by Bardeen--Cooper--Schrieffer in their work. It is important to understand that this attractiveness is our \emph{main physical assumption} on the model. This assumption is certainly satisfied if our potential is a nonnegative function $V\geq 0$, whence $-V$ is negative. This is a good picture to keep in mind but mathematically not necessary. We will phrase later what we actually mean by an attractive interaction. 

%

Mathematically, the BCS model now becomes interesting because it gives rise to temperature regimes in which there are states with energies that are strictly lower than the energy of the normal state. This goes back to a nontrivial competition between the interaction term and the kinetic and entropic term in the BCS functional \eqref{Intro:BCS_functional_ti_definition}. A careful analysis of these effects has been provided in the work \cite{Hainzl2007} by Hainzl, Hamza, Seiringer and Solovej in 2008. Since the results obtained there are very important for this thesis, we are going to phrase and briefly discuss them here.

\begin{cthm}[{\cite[Theorem 1]{Hainzl2007}}]
\label{Intro:BCS-ti_criterion}
Let $V\in L^{\nicefrac 32}(\Rbb^3)$ and $T\geq 0$. Then the following statements are equivalent:
\begin{enumerate}[(a)]
\item The normal state $(\gamma_0, 0)$ is instable under pair formation, i.e., there is a pair $(\gamma, \alpha)$ with $\gamma\in L^1(\Rbb^3, (1 + p^2)\dd p)$, $\alpha\in H^1(\Rbb^3, \dd x)$, and $|\hat \alpha(p) |^2 \leq \gamma(p) (1 - \gamma(p))$ such that
\begin{align*}
\Fcal_{\mathrm{ti}}^{\mathrm{BCS}}(\gamma, \alpha) < \Fcal_{\mathrm{ti}}^{\mathrm{BCS}}(\gamma_0, 0).
\end{align*}

\item The linear operator $K_T - V$ has at least one negative eigenvalue. Here, $K_T$ is the pseudodifferential operator given by the symbol
\begin{align}
K_T(p) &:= \frac{p^2 - \mu}{\tanh(\frac{p^2-\mu}{2T})}. \label{Intro:KT-symbol}
\end{align}
\end{enumerate}
\end{cthm}

In \cite{Hainzl2007}, there is a third statement about nontrivial solutions to the so-called \emph{BCS gap equation}, which we leave out here because the gap equation as such does not play a prominent role in this thesis. Nevertheless, this is the object that had been first studied by physicists when they investigated the BCS model.

To briefly sketch the proof of Theorem \ref{Intro:BCS-ti_criterion}, this result is proven by the second variational test at the normal state in direction $\alpha$, where the first derivative vanishes because of the minimality of the normal state. The operator $K_T - V$ is the Hessian matrix, which makes it plausible that its sign determines the stability of the normal state as a minimum of the BCS functional. The remarkable fact is that with the interaction turned on, there is always a direction in which we can lower the energy by perturbing with $\alpha$, i.e., the normal state becomes a saddle point. The reader should keep in mind that this is not at all an expectable situation. It could very well happen that we have to first increase the energy, climb over a ``mountain range'' around the ``local minimum'' $(\gamma_0, 0)$ to be able to enter the ``valley'' in which superconductivity takes place. 

The first statement (a) of Theorem \ref{Intro:BCS-ti_criterion} is our definition for superconductivity as this is the intuitive behavior of the system lowering the energy beyond the normal state. Theorem \ref{Intro:BCS-ti_criterion} is of great significance for us since it rephrases the question of superconductivity in terms of a spectral question, which makes it relatively comfortable to handle, since a whole zoo of methods is available to treat these problems. Consequently, the spectrum of the operator $K_T$ plays a central role in the mathematical description of BCS theory and in our thesis.

\subsection{The spectrum of \texorpdfstring{$K_T - V$}{KT-V}}

On the one hand, the operator $K_T - V$ is a Schrödinger-type operator and as such shares a lot of properties that are known from $-\Delta - V$. On the other hand, the peculiar dispersion relation it has encodes both the kinetic energy of the BCS model and its entropy. Not only for this reason is it relatively clumsy and challenging to work with $K_T$ since it largely escapes explicit calculations due to the hyperbolic tangent. 
\begin{figure}[h]
\centering
\includegraphics[width=8cm]{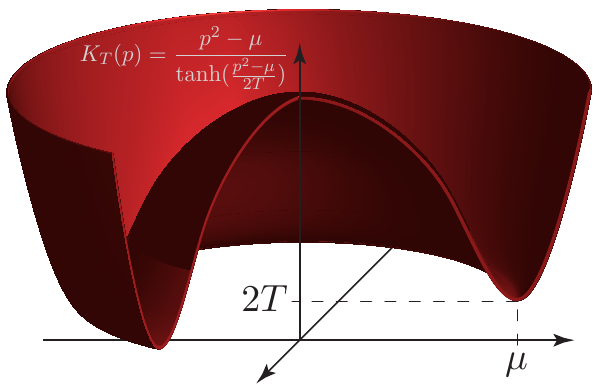}
\caption{Mexican hat shape of the operator $K_T$.\label{Intro:KT_figure}}
\end{figure}
Figure \ref{Intro:KT_figure} illustrates qualitatively the shape of the function $K_T(p)$ defined in \eqref{Intro:KT-symbol} in the case $\mu >0$. First of all, $K_T$ behaves like $p^2-\mu$ as $p \to \infty$. We also see that the minimum of the function $K_T(p)$ is attained on the so-called \emph{Fermi surface} $\{ p\in \Rbb^3 : p^2 = \mu\}$ and a short argument shows that it equals $2T$. If $\mu < 0$, however, the minimum is attained at $p =0$ and hence equals $|\mu| / \tanh(|\mu| /(2T))$. It follows that the spectrum of the operator $K_T$ equals
\begin{align}
\sigma(K_T) &= \begin{cases} [2T, \infty) , & \mu \geq 0 , \\ [|\mu| / \tanh(|\mu|/(2T)) , \infty) , & \mu < 0. \end{cases} \label{Intro:KT-spectrum}
\end{align}

When the interaction $V$ is present, in principle, anything may change in $\sigma(K_T)$. However, in practice, assumptions are made so that the essential spectrum is unchanged, e.g., if $V\in L^2(\Rbb^3)$ (also if $V\in L^{\nicefrac 32}(\Rbb^3)$ but the proof is more complicated). For, in this case the operator $VK_T^{-1}$ is Hilbert--Schmidt since $K_T(p)^{-1}$ is an $L^2(\Rbb^3)$-function of $p$. Therefore, due to Weyl's criterion, the essential spectrum $\sigma_{\ess}(K_T - V)$ is preserved under the perturbation $-V$ and thus equals \eqref{Intro:KT-spectrum}. In this situation, $V$ ``only'' adds isolated eigenvalues of finite multiplicity to the spectrum, which might, in fact, be embedded in the essential spectrum. Our interest lies in isolated eigenvalues below the bottom of the essential spectrum, i.e., below $2T$. Note that in the case $\mu < 0$ the bottom is lower than $2T$ since $\tanh(|x|) \leq |x|$. Since $K_T$ is strictly monotone in $T$, as discussed above, we have the following picture of the spectrum as a function of $T$. In this picture, we assume for simplicity that all these eigenvalues are simple, which in practice might not hold. Therefore, we are in a situation similar to the following Figure \ref{Intro:Spectrum_KT}.

\begin{figure}[h]
\centering
\includegraphics[width=8cm]{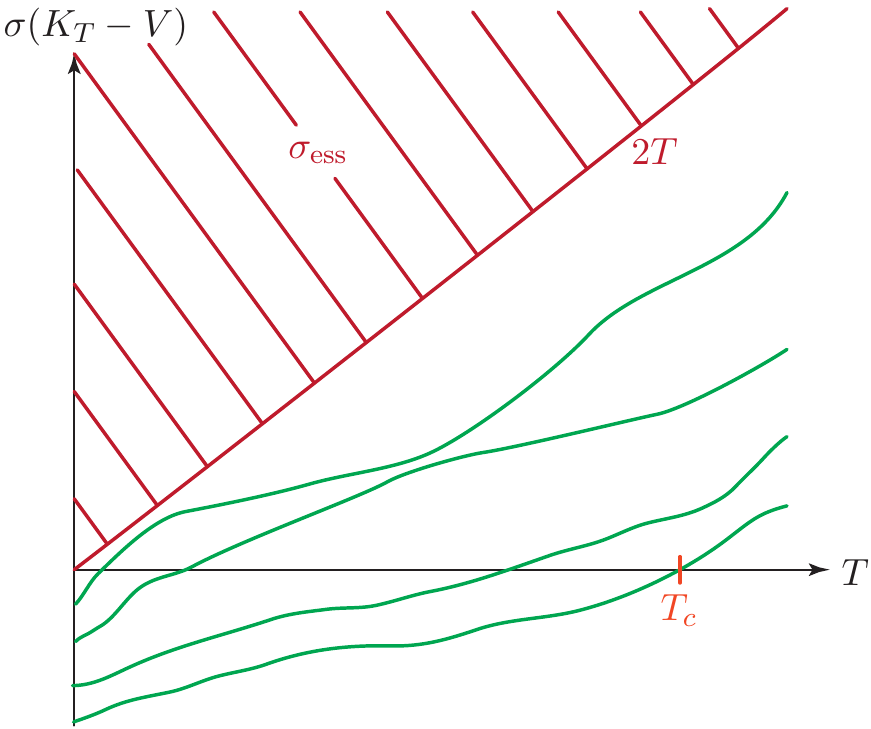}
\caption{The essential spectrum of $K_T-V$ (red) and the isolated eigenvalues (green) below it. The crossing point of the lowest eigenvalue with zero marks the critical temperature $\Tc$. \label{Intro:Spectrum_KT}}
\end{figure}

This allows for the following definition of a critical temperature.

\subsection{The critical temperature and attractive potentials}

To find a criterion for superconductivity, we use the next important notice. The function $T\mapsto K_T(p)$ is a strictly increasing function for every $p\in \Rbb^3$. This implies that the eigenvalues below $2T$ are increasing functions of $T$ as well (there might, however, be crossings in general as opposed to the impression of Figure \ref{Intro:Spectrum_KT}). This allows for the next important result.

\begin{cthm}[{\cite[Theorem 2]{Hainzl2007}}]
For any $V\in L^{\nicefrac 32}(\Rbb^3)$ there is a critical temperature $0 \leq \Tc < \infty$ such that $\inf \sigma(K_T - V) < 0$ if $T < \Tc$ and $K_T - V\geq 0$ if $T\geq \Tc$.
\end{cthm}

This result is important since it provides the first criterion for superconductivity and it clarifies the common definition
\begin{align}
\Tc := \inf\bigl\{ T \geq 0 : K_T - V \geq 0\bigr\}, \label{Intro:Tc_definition}
\end{align}
which is widely used in the literature. Furthermore, \cite{Hainzl2007} shows several sufficient conditions under which this temperature is indeed positive. As an example, we mention the following result.

\begin{cthm}[{\cite[Theorem 3 (i)]{Hainzl2007}}]
Let $V\in L^{\nicefrac 32}(\Rbb^3)$ be not identically zero and let $\mu >0$. If $V\geq 0$ then $\Tc >0$.
\end{cthm}

For us, this result determines the notion ``attractive'' for the interaction potential $V$. Namely, we understand $V$ to be \emph{attractive} if $\Tc>0$. This is a central assumption in the works we present in Chapters \ref{Chapter:DHS1} and \ref{Chapter:DHS2}. The preceding theorem provides a sufficient condition for this to be true but it is by far not necessary.

In particular, we have a critical temperature $\Tc >0$ at which the lowest eigenvalue of $K_\Tc - V$ is given by zero. Throughout this thesis, we are going to assume that this eigenvalue is simple\footnote{We will include a note on degeneracy later in Section \ref{Intro:Degeneracy_Section}, when we discuss the results of this thesis and the state of the project.} and call the corresponding eigenfunction $\alpha_*$. This function is a very central object in BCS theory and we note its eigenvalue equation
\begin{align}
K_\Tc \alpha_* - V\alpha_* = 0.\label{Intro:alphastar_ev_equation}
\end{align}
This function is well-behaved in the sense that it is a smooth function with rapid decay. For the particular case of $\alpha_*$ this has been proven in \cite{Hainzl2012} but will also be proven in this thesis in Chapter \ref{Chapter:Combes-Thomas} for all eigenfunctions of $K_T$ that belong to isolated eigenvalues below $2T$ for any temperature.

To get a feeling for the critical temperature, we should mention the result of \cite[Theorem 1]{Hainzl2007-2}, which implies that the critical temperature is exponentially small in $V$. More precisely, if $V$ is replaced by $\lambda V$, then
\begin{align*}
\Tc(\lambda V) &\sim \mu \, \e^{-\frac{1}{\lambda a_\mu(V)}}, & \lambda & \to 0,
\end{align*}
where $a_\mu(V) >0$ is a certain parameter that is not of interest here. Physically, this means that the critical temperature is pretty close to absolute zero since we explained above that the attractive interaction between the electrons is fairly weak, i.e., $0 < \lambda \ll 1$.


\section{External Fields}

\subsection{The Meißner Effect}

With these notions at hand, we can turn our attention towards the BCS functional in the presence of weak external fields. The reason why we want to understand this BCS model lies in the Meißner effect that we made contact with in Section \ref{Intro:History_Section}. Mathematically, this is modeled by a BCS functional having the formal expression
\begin{align}
&\Tr_\Omega \bigl[ \bigl((-\i \nabla + \Abold(x))^2 - \mu\bigr) \gamma \bigr] - T S(\Gamma) - \fint_{\Omega} \dd X \int_{\Rbb^3} \dd r \; V(r) \, |\alpha(X, r)|^2 \notag \\
&\hspace{150pt} + \int_\Omega \dd x \; \bigl| \curl \Abold(x) - H_{\mathrm{ext}}(x) \bigr|^2 .\label{Intro:BCS_functional_Meißner}
\end{align}
Here, $H_{\mathrm{ext}}$ models the externally applied magnetic field from the magnet and $\Abold$ is the magnetic potential of the response field of the superconductor.

This functional is hard to define rigorously --- we shall comment on this below, when more insight is available --- and there are a lot of questions to study before we would be able to analyze this functional in the superconducting phase. Unfortunately, this thesis will not be able to treat the functional in \eqref{Intro:BCS_functional_Meißner} at all. However, this is one of the ultimate goals to understand --- namely, minimize this functional both in the state $\Gamma$ and the response field $\Abold$ of the superconductor. This would prove the Meißner effect on the level of BCS theory and it is one of the main motivations to study the BCS functional with an external magnetic field.

In this thesis, we will address the minimization of the BCS functional while we are fixing the response field and dropping the magnetic field term, i.e., the functional in \eqref{Intro:BCS_functional_formal}. We will analyze the minimization problem for the BCS functional in terms of the state $\Gamma$ only, which will be a necessary preparation for the problem described in the preceding paragraph. Therefore, we need to understand the minimization problem with fixed external fields in great detail and generality.

We now turn our attention to the rigorous definition of the BCS functional with external fields as given in \eqref{Intro:BCS_functional_formal}. The first step towards this is an  understanding of periodic magnetic fields and how to model a fermionic system in the presence of these.

\subsection{Gauge-periodic systems}

We want to model a periodic system which involves periodic states and an energy functional that is compatible with this periodicity. In order to set this up, we need a notion of translations that is compatible with the magnetic field so that the BCS functional becomes invariant under these translations. It will become clear what we mean by this in due time.

We remark that it is not known whether the minimizer of the BCS model (over all admissible states) is indeed periodic as opposed to the situation described in Section \ref{Intro:BCS-functional-ti_Section}. Hence, this is a considerably simplifying assumption as of today.

When we want to describe a magnetic model, the physically relevant quantity is actually the magnetic field $\Bcal \colon \Rbb^3\ra \Rbb^3$ and \emph{not} the magnetic potential $\Abold\colon \Rbb^3\ra \Rbb^3$, which has the property that $\Bcal = \curl \Abold$. In other words, although it is the magnetic potential that arises in the magnetic Laplacian and thus in the BCS functional, the ``physical properties'' of the system remain unchanged as long as $\Abold$ is chosen such that $\curl \Abold = \Bcal$ holds. This leaves us the ``freedom of gauge'', which we should use when we choose the magnetic potential.

Since we want to describe a periodic system, we should first fix a lattice of periodicity $\Lambda \subseteq\Rbb^3$. Our ansatz for this lattice is the $\Zbb$-span of three linearly independent vectors $b_1, b_2, b_3\in \Rbb^3$ called the basis, i.e.,
\begin{align*}
\Lambda := \Bigl\{ \sum_{i=1}^3 n_i b_i \in \Rbb^3 : n_i \in \Zbb , \, i=1,2,3\Bigr\}.
\end{align*}
The \emph{unit} (or, \emph{fundamental}) \emph{cell} of the lattice $\Lambda$ is the domain
\begin{align*}
\Omega &:= \Bigl\{ \sum_{i=1}^3 a_i b_i\in \Rbb^3 : 0 \leq a_i < 1 , \;i=1,2,3 \Bigr\}.
\end{align*}

We are given a $\Lambda$-periodic magnetic field $\Bcal\colon \Rbb^3\ra \Rbb^3$ and have the task to choose a magnetic potential $\Abold$ such that $\curl \Abold = \Bcal$. Now the question is: What is the best (i.e., most convenient for us) choice of $\Abold$? There are several gauges to choose from in the literature (Coulomb or transversal gauge, Landau gauge, ...) but it has turned out that there is one particular choice of gauge that is most suitable for our purposes. For two space dimensions, this gauge is widely known and has appeared in the literature. In three dimensions, a derivation of this gauge in the ``pedestrian way'' (as presented in Chapter \ref{Chapter:Abrikosov_gauge}) is less known so we shall explain it here in some detail. This gauge is called the \emph{Abrikosov gauge} and the claim is that there is a bounded periodic magnetic potential $A\colon \Rbb^3 \ra \Rbb^3$ with $A(0) = 0$ such that the potential
\begin{align}
\Abold &:= \Abold_\Bbold + A
\end{align}
satisfies $\Bcal = \curl \Abold$, where $\Abold_\Bbold(x) := \frac 12 \, \Bbold \wedge x$ is the \emph{constant magnetic field} potential and the vector $\Bbold\in \Rbb^3$ is the \emph{average magnetic field}
\begin{align}
\Bbold := \frac{1}{|\Omega|} \int_\Omega \dd x\; \Bcal(x).
\end{align}
We remark that the condition $A(0) =0$ can be replaced by $\fint_\Omega \dx \, A(x) =0$ or any other constant shift by a simple gauge transformation.

In two dimensions, this result has been established in \cite{Tim_Abrikosov} and the three dimensional case is proven in Chapter \ref{Chapter:Abrikosov_gauge}. 

This result is important in that it limits the situations that we need to take care of to a very explicit, yet difficult, magnetic potential and a somewhat easier (bounded) but general magnetic potential. However, the potential $\Abold_\Bbold$ looks very incompatible with periodic structures (like a torus) since it grows linearly from  the origin. We will see in the following how to fix this problem.

\subsubsection{The magnetic flux}

There are several reasons why this gauge is useful to choose. The first advantage is that the constant magnetic field potential is a very explicit potential for which many algebraic identities are known. The second advantage is that the periodic magnetic potential is bounded and, more importantly, it does not contribute to the average magnetic flux $\Phi_\Bcal$ through the unit cell $\Omega$. The reason lies in Green's theorem:
\begin{align}
\Phi_\Bcal := \int_{\underline{\partial \Omega}} \dd S(x) \; \Bcal(x) \cdot \nu(x) = \oint_{\partial (\underline{\partial \Omega})}  \dd y \; \Abold(y) \cdot \ell(y).
\end{align}
Here, $\nu$ is a unit normal to the surface $\underline{\partial \Omega}$ that is determined by three perpendicularly oriented faces of $\Omega$, while $\ell$ is a unit line element along the line $\partial (\underline{\partial \Omega})$. Since $A$ is periodic, it will have the same value on two opposite edges of the cube while $\ell$ will point in different directions on opposite edges. Hence, $A$ does not contribute to $\Phi_\Bcal$, i.e., the average magnetic flux is solely determined by the constant magnetic field part.

In 2012, the work \cite{Hainzl2012} investigated the BCS functional for the periodic magnetic potential only and obtained a description of the physical behavior in the weak magnetic field regime, when the perturbation of the magnetic field is given on the macroscopic scale of the system. The result of our works (Chapters \ref{Chapter:DHS1} and \ref{Chapter:DHS2}) extends this to the general situation of periodic magnetic fields. In this case the magnetic field has a nonzero flux through the unit cell and this is responsible for many additional complications in the treatment of the model.

\begin{figure}[h]
\centering
\includegraphics[width=15cm]{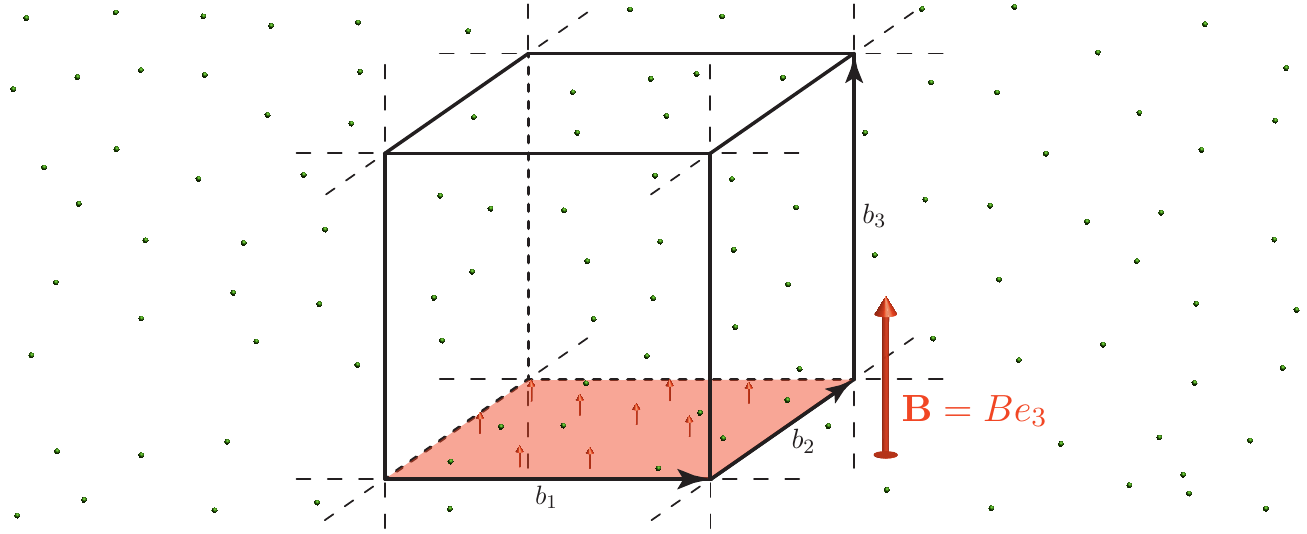}
\caption{System with nonzero flux through the unit cell. The system is exposed to a constant magnetic field, $\Bcal(x) \equiv \Bbold$.\label{Intro:System_nonzero_flux}}
\end{figure}

Figure \ref{Intro:System_nonzero_flux} illustrates a system with a constant magnetic field pointing in the $e_3$-direction. The system features a nonzero average magnetic flux through the unit cell.

\subsubsection{Our choice of lattice}

Now, we want to set up the BCS model in a periodic manner, which requires a notion of translation that is compatible with the operators that appear in \eqref{Intro:BCS_functional_formal}. In particular, the translation we choose must commute with the magnetic momentum operator 
\begin{align}
\pi_\Abold := -\i \nabla + \Abold(x). \label{Intro:Magnetic_Momentum}
\end{align}
To achieve this, we have to use the magnetic translations, which are given by
\begin{align}
T_\Bbold(v) f(x) &:= \e^{\i x \cdot \Abold_\Bbold(v)} f(x + v), & v &\in \Rbb^3 , \; x\in \Rbb^3. \label{Intro:Magnetic_Translation}
\end{align}
With this definition, we claim that for every $\lambda\in \Lambda$, we have
\begin{align}
T_\Bbold(\lambda)^* \, \pi_\Abold \, T_\Bbold(\lambda) = \pi_\Abold. \label{Intro:Magnetic_Momentum_Translation_Intertwining}
\end{align}
We call this property the \emph{gauge-periodicity} of $\pi_\Abold$ and this notion extends in an obvious way to any operator in place of $\pi_\Abold$. The term ``gauge'' refers to the fact that we use the magnetic translations instead of regular ones. Equation \eqref{Intro:Magnetic_Momentum_Translation_Intertwining} is true, since the periodicity of $A$ implies
\begin{align*}
\pi_\Abold \, T_\Bbold(\lambda) f(x) &= \bigl(-\i \nabla + \Abold_\Bbold(x) + A(x)\bigr) \e^{\i x\cdot \Abold_\Bbold(\lambda)} f(x + \lambda) \\
&= \e^{\i x\cdot \Abold_\Bbold(\lambda)} \bigl(-\i \nabla + \Abold_\Bbold(x) + A(x+\lambda) + \Abold_\Bbold(\lambda)\bigr) f(x + \lambda) \\
&= T_\Bbold(\lambda) \, \pi_\Abold f(x).
\end{align*}
In fact, this calculation even shows that
\begin{align*}
T_\Bbold(v)^* \, \pi_{\Abold_\Bbold} \, T_\Bbold(v) = \pi_{\Abold_\Bbold}
\end{align*}
for all $v\in \Rbb^3$.

At this point, we see why the magnetic potential $A$ and the electric potential $W$ are indeed periodic in the ``classical'' sense. The reason is that, as a multiplication operator, these functions commute with the lattice translations $T_\Bbold(\lambda)$.

However, the translations in \eqref{Intro:Magnetic_Translation} do not form an abelian group! Hence, it makes a difference if we translate first by a vector $v$ and then by a vector $w$ or in the opposite order, which does not allow for a sensible interpretation as \emph{lattice translations}. More precisely,
\begin{align}
T_\Bbold(v) \, T_\Bbold(w) &= \e^{-\i \Bbold \cdot (v\wedge w)} \, T_\Bbold(w) \, T_\Bbold(v), & v,w &\in \Rbb^3,
\end{align}
which yields an abelian translation group on a lattice $\Lambda$ only if $\e^{\i \Bbold \cdot (\lambda_1\wedge \lambda_2)} = 1$ for all $\lambda_1, \lambda_2\in \Lambda$. Formally, we also have the ``group law''
\begin{align}
T_\Bbold(v + w) &= \e^{\i \frac{\Bbold}{2} \cdot (v \wedge w)} \, T_\Bbold(v) \, T_\Bbold(w),
\end{align}
which is of course ill-defined since the left side is symmetric in $v$ and $w$, whereas the right side is not. In order to cure the theory from this problem as well, we must even have $\e^{\i \frac \Bbold 2\cdot (\lambda_1\wedge \lambda_2)} =1$. At this point, we opt for a lattice that is spanned by multiples of the standard basis $e_i$ in $\Rbb^3$. Then, this imposes conditions on the basis vector $b_i$, $i = 1, 2,3$ of $\Lambda$, namely for each $i,j,k=1,2,3$, we must have
\begin{align}
B_i \cdot (b_j \wedge b_k) \in 4\pi \Zbb.
\end{align}
Let us assume for simplicity that all $B_i \neq 0$, otherwise there are certain simplifications in what follows. We choose the vectors $b_i$ to be mutually orthogonal and to satisfy the conditions $B_i \, |b_j| \, |b_k| = 4\pi$, which we do by setting
\begin{align}
b_i(\Bbold) &:= \sqrt{\frac{4\pi}{B_1 \, B_2\, B_3}} \; B_i \, e_i, & i &= 1, 2, 3. \label{Intro:Basis_Vectors_definition}
\end{align}
Here, $e_i$ denotes the $i$\tho\ standard basis vector in $\Rbb^3$. These basis vectors span a $\Bbold$-dependent lattice, which we denote by
\begin{align}
\Lambda_\Bbold := \Bigl\{ \sum_{i=1}^3 n_i \, b_i(\Bbold) : n_i\in \Zbb , \, i=1,2,3\Bigr\} \label{Intro:Lattice_definition}
\end{align}
and the unit (or, fundamental) cell of $\Lambda_\Bbold$ is
\begin{align}
Q_\Bbold &:= \Bigl\{ \sum_{i=1}^3 a_i \, b_i(\Bbold) : 0 \leq a_i  < 1 , \, i =1, 2,3\Bigr\}. \label{Intro:Lattice_unit_cell_definition}
\end{align}
As a consequence of these definitions, the group $\{T_\Bbold(\lambda)\}_{\lambda \in \Lambda_\Bbold}$ is an abelian group of translations and a short computation shows that we have the expected group law
\begin{align}
T_\Bbold(\lambda_1 + \lambda_2) &= T_\Bbold(\lambda_2) \, T_\Bbold(\lambda_2), & \lambda_1, \lambda_2 &\in \Lambda_\Bbold. \label{Intro:Magnetic_group_law}
\end{align}
Finally, by our choice of basis $b_i$, $i=1, 2, 3$, we have
\begin{align}
\Bbold \cdot (\lambda_1 \wedge \lambda_2) &\in 4\pi \Zbb , & \lambda_1,\lambda_2 \in \Lambda_\Bbold. \label{Intro:Magnetic_Phase_Neutralization}
\end{align}

\begin{figure}[h]
\centering
\includegraphics[width=15cm]{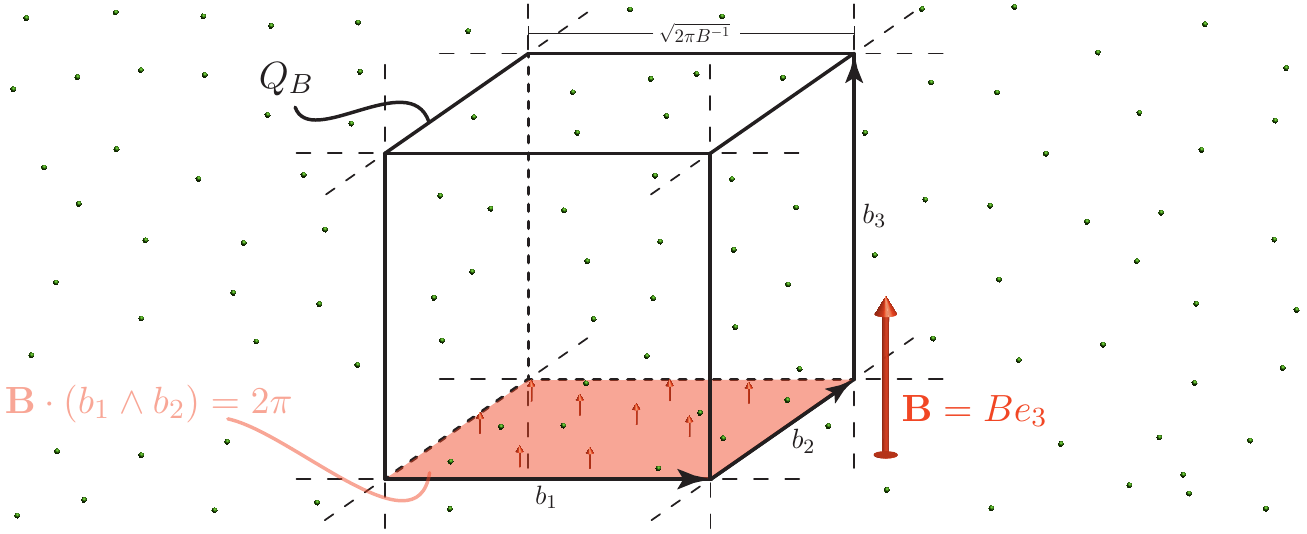}
\caption{The lattice with magnetic flux $2\pi$ in the case $\Bbold = Be_3$.\label{Intro:Lattice_with_flux}}
\end{figure}

This concludes our construction of the lattice, which we build the BCS model on. Figure \ref{Intro:Lattice_with_flux} illustrates the construction in the case $\Bbold = B e_3$ where the basis vectors are given by $b_i = \sqrt{2\pi B^{-1}} \, e_i$ and the box $Q_B$ equals per definition $Q_\Bbold$.

The next steps are to define what we mean by a periodic BCS state and to introduce the admissibility condition. In order to do this, we need to define the trace per unit volume, which is our next goal.

\subsubsection{The Meißner effect revisited}

At this point, we can look a bit closer on the difficulties that arise in connection with the Meißner effect. First of all, one problem is to even define the BCS functional in \eqref{Intro:BCS_functional_Meißner} in a rigorous fashion because the unit cell of the lattice of periodicity depends --- as we have seen --- on the \emph{average magnetic flux} of the response field through it. One could think of fixing a magnetic flux $\Phi >0$ and looking at the family of BCS functionals indexed by $\Phi$ and so that the response field has the average flux $\Phi$ through the unit cell. Then, the unit cell of the lattice has to be sized according to $\Phi$ and one is inclined to carry out the minimization problem flux-wise. However, this already spoils the reality of the model to a large extent since the superconductor should be allowed to have a ``fixed'' (independent of the flux) macroscopic size. Hence, we would have to first find a way to decouple the magnetic flux and the size of the unit cell in a sensible manner. One way that could be thought of would be to prove that the magnetic flux of the external and the response field coincide a priori. However, the possibility of this is pure speculation as of now.

Moreover, even in the situation of the non-interacting functional and without external magnetic field, we first need to prove that the normal state is the one that we already know. However, this requires proving that the response field in this case equals zero. To the best of our knowledge, this result is unknown to the present day. The reason for this is the lack of suitable magnetic field estimates for the last term in \eqref{Intro:BCS_functional_Meißner} and leading experts in the field hope to make progress in this direction within the next ten years. Therefore, there is quite some work to do before the Meißner effect can be tackled.

\subsection{Gauge-periodic operators and the Bloch--Floquet decomposition}

When we say that our system be modeled periodically with respect to the lattice $\Lambda_\Bbold$ defined in \eqref{Intro:Lattice_definition}, we mean that all the operators and observables of the system are \emph{$\Lambda_\Bbold$-gauge-periodic} (or simply \emph{gauge-periodic}) operators with respect to the magnetic lattice translations $T_\Bbold(\lambda)$ defined in \eqref{Intro:Magnetic_Translation}. Per definition, this means that an operator $S$ satisfies
\begin{align}
T_\Bbold(\lambda)^* \, S \, T_\Bbold(\lambda) &= S , & \lambda &\in \Lambda_\Bbold. \label{Intro:Periodic_operator}
\end{align}
The most prominent example of a gauge-periodic operator is the magnetic momentum operator $\pi_\Abold$, see \eqref{Intro:Magnetic_Momentum_Translation_Intertwining}. 

Gauge-periodic operators have the problem that the Hilbert space $L^2(\Rbb^3)$ is somewhat inappropriate to investigate their spectral properties and to evaluate the common energy functionals on them. These functionals are mostly given by trace functionals, including the BCS functional. Also, the spectrum of an operator depends heavily on the domain on which this operator is investigated. We only need to think of the Laplacian, which commutes with all (regular) translations, i.e., is periodic with respect to any lattice, defined on $L^2(\Rbb^3)$ versus periodic functions in $L^2([0,1]^3)$. The former has a purely continuous spectrum while the latter has the plane waves $\e^{2\pi \i k\cdot x}$ as an eigenbasis.

However, a periodic operator can never be trace class in the usual sense since, loosely speaking, its trace would equal the sum of the traces on one of the unit cells of the lattice. This, in turn, is an infinite sum of equal terms. This makes us prefer the domain of periodic functions over the space $L^2(\Rbb^3)$. In order to describe this properly, we need a notion of ``decomposing'' the operator into a family of operators. Each member of this family, the so-called \emph{fiber}, acts on a translate of the unit cell of the lattice while ``storing the information'' where the translate was located in $\Rbb^3$. A similar goal is pursued when we develop the theory of Fourier series for periodic functions and we should keep this analogy in mind. Here, the function $f$ is $\Zbb^3$-periodic, say, and the ``fiber'' is the frequency (or, momentum) $p\in \Zbb^3$ such that $f(x) = \sum_{p \in \Zbb^3} \e^{2\pi \i p\cdot x} \hat f_p$ holds. As we know, we must define
\begin{align}
\hat f_p = \int_0^1 \dd x \; \e^{-2\pi \i p \cdot x} f(x)
\end{align}
in order to achieve this. The analogue in operator theory is the so-called \emph{Bloch--Floquet decomposition}, which we discuss now.

I recommend that the reader takes a look at the Appendix \ref{Chapter:Local_Traces} or any other source that provides insight into the theory of local traces. There, I review the proof of standard inequalities like Hölder's, Peierls', and Klein's inequality. Subsequently, following \cite{Reedsimon4} to a large extent, I introduce the theory of operator valued functions and the theory of local traces. Finally, I provide local versions of the aforementioned inequalities. Nothing of this is new but might be new to the reader.

To make the long story short, we follow the publication \cite[Section 2.1]{Stefan_Peierls} and recall the construction of a magnetic Bloch--Floquet decomposition. To start with, we define the space of gauge-periodic $L^2$-functions as
\begin{align}
\Hcal_\Bbold &:= \bigl\{ f \in L_\loc^2(\Rbb^3) : T_\Bbold(\lambda) f = f , \; \lambda\in \Lambda_\Bbold\bigr\} \label{Intro:Periodic_L2}
\end{align}
and we equip it with the scalar product
\begin{align}
\langle f, g\rangle_{\Hcal_\Bbold} := \int_{Q_\Bbold} \dd x \; \ov{f(x)} \, g(x), \label{Intro:Periodic_L2_scalar_product}
\end{align}
which makes it a Hilbert space. We first note that $\Hcal_\Bbold$ is unitarily equivalent to the space $L^2(Q_\Bbold)$,
\begin{align}
\Hcal_\Bbold \cong L^2(Q_\Bbold). \label{Intro:Periodic_unitary_equivalence_box}
\end{align}
For, each $x\in \Rbb^3$ has a unique decomposition $x = \tilde x + \nu$, where $\tilde x\in Q_\Bbold$ and $\nu\in \Lambda_\Bbold$. Therefore, for $f\in L^2(Q_\Bbold)$ and $x\in \Rbb^3$, we set
\begin{align}
\Ecal_{\mathrm{mag}} f(x) &:= \e^{-\i x \cdot \Abold_\Bbold(\nu)} f(\tilde x), & x &= \tilde x + \nu, \; \tilde x\in Q_\Bbold, \; \nu \in \Lambda_\Bbold, \label{Intro:Gauge-periodic_Extension}
\end{align}
which defines a gauge-periodic extension $\Ecal_{\mathrm{mag}} f \in \Hcal_\Bbold$ of $f$. To see this, let $\lambda \in \Lambda_\Bbold$ and compute
\begin{align*}
T_\Bbold(\lambda) (\Ecal_{\mathrm{mag}}f)(x) &= \e^{\i x\cdot \Abold_\Bbold(\lambda)} \, (\Ecal_{\mathrm{mag}} f)(x + \lambda) = \e^{\i x \cdot \Abold_\Bbold(\lambda)} \, \e^{-\i (x + \lambda) \cdot \Abold_\Bbold(\nu + \lambda)} f(\tilde x) \\
&= \e^{\i x\cdot \Abold_\Bbold(\lambda)} \, \e^{-\i x \cdot \Abold_\Bbold(\nu)} \, \e^{-\i x\cdot \Abold_\Bbold(\lambda)} \, \e^{-\i \lambda\cdot \Abold_\Bbold(\nu + \lambda)} \, f(\tilde x) \\
&= \e^{-\i \lambda \cdot \Abold_\Bbold(\nu)} \; \e^{-\i x \cdot \Abold_\Bbold(\nu)} f(\tilde x).
\end{align*}
By \eqref{Intro:Magnetic_Phase_Neutralization}, we conclude that $\e^{-\i \lambda \cdot \Abold_\Bbold(\nu)}=1$, whence the right side equals $\Ecal_{\mathrm{mag}} f(x)$. This proves that $\Ecal_{\mathrm{mag}} f\in \Hcal_\Bbold$. The inverse is given by $\Ecal_{\mathrm{mag}}^{-1}f = \chi_{Q_\Bbold}f$, where $\chi_{Q_\Bbold}$ is the characteristic function of the box $Q_\Bbold$. Unitarity of the map $\Ecal_{\mathrm{mag}} \colon L^2(Q_\Bbold) \ra \Hcal_\Bbold$ is obvious.

\begin{bem}
The gauge-periodic function $\Ecal_{\mathrm{mag}}f$ is of course far from smooth even if $f$ is smooth. In fact, I do not know any ``easy to give'' nontrivial gauge-periodic function, which is smooth.
\end{bem}

We define the gauge-periodic Sobolev space
\begin{align}
\Hcal_\Bbold^m &:= \bigl\{ f\in \Hcal_\Bbold : (-\i \nabla + \Abold_\Bbold)^\nu f\in \Hcal_\Bbold, \; |\nu| \leq m \bigr\} \label{Intro:Periodic_Hm}
\end{align}
for $m \in \Nbb_0$. Here, we used multi-index notation for $\nu\in \Nbb_0^3$. It is equipped with the scalar product
\begin{align}
\langle f, g\rangle_{\Hcal_\Bbold^m} &:= \sum_{|\nu| \leq m} \bigl\langle (-\i \nabla + \Abold_\Bbold)^\nu f, (-\i \nabla + \Abold_\Bbold)^\nu g\bigr\rangle_{\Hcal_\Bbold}. \label{Intro:Periodic_Hm_scalar_product}
\end{align}
In this way, $\Hcal_\Bbold^m$ is a Hilbert space for each $m\in \Nbb_0$ with the convention that $\Hcal_\Bbold^0 := \Hcal_\Bbold$ and the magnetic momentum operator $\pi_\Abold$ in \eqref{Intro:Magnetic_Momentum} is self-adjoint on $\Hcal_\Bbold^1$.

By $\Lambda_\Bbold^*$ we denote the \emph{dual lattice} of $\Lambda_\Bbold$, which is the lattice spanned by the basis vectors $b_1^*(\Bbold) , b_2^*(\Bbold)$, and $b_3^*(\Bbold)$, uniquely determined by the condition $b_i^*(\Bbold) \cdot b_j(\Bbold) = 2\pi \delta_{ij}$. By \eqref{Intro:Basis_Vectors_definition}, this implies that
\begin{align}
b_i^*(\Bbold) &= \frac{\sqrt{\pi \, B_1 \, B_2 \, B_3}}{B_i} \; e_i, & i &= 1, 2, 3. \label{Intro:Basis_Vectors_Dual_definition}
\end{align}
In particular, we thus have
\begin{align}
\nu \cdot \lambda &\in 2\pi \Zbb,  & \lambda \in \Lambda_\Bbold , \; \nu \in \Lambda_\Bbold^*. \label{Intro:Dual_lattice_relation}
\end{align}
We also let $Q_\Bbold^*$ denote the unit cell of $\Lambda_\Bbold^*$, the so-called \emph{dual unit cell}. On $\Hcal_\Bbold$, the lattice $\Lambda_\Bbold^*$ possesses the unitary representation
\begin{align}
\begin{split} \tau \colon \Lambda_\Bbold^* &\ra \Bcal(\Hcal_\Bbold), \\ \nu &\mapsto \tau(\nu), \end{split} & \bigl( \tau(\nu) \varphi \bigr) (x) &:= \e^{\i \nu \cdot x} \, \varphi(x).
\end{align}
Here, $\Bcal(\Hcal_\Bbold)$ stands for the Banach space of bounded operators $\Hcal_\Bbold \ra \Hcal_\Bbold$. Finally, we define the space of $\tau$-equivariant $\Hcal_\Bbold$-valued functions by
\begin{align}
\Kcal_\Bbold(\tau) &:= \bigl\{ \varphi\in L^2(\Rbb^3 ; \Hcal_\Bbold) : \varphi(\vartheta - \nu) = (\tau(\nu) \varphi)(\vartheta) , \; \nu \in \Lambda_\Bbold^*\bigr\}. \label{Intro:Equivariant_K}
\end{align}
It follows from the definition that functions in $\Kcal_\Bbold(\tau)$ are fully determined by their values on $Q_\Bbold^*$. If $f\in \Kcal_\Bbold(\tau)$ and $\vartheta\in \Rbb^3$, then the $\Lambda_\Bbold$-periodic function $f(\vartheta)\in \Hcal_\Bbold$ is called the \emph{$\vartheta$\tho\ fiber} of $f$ (sometimes also $\vartheta$ itself is called the fiber). Since the target Hilbert space $\Hcal_\Bbold$ is independent of $\vartheta$, we also write $\Kcal_\Bbold(\tau)$ suggestively as a so-called \emph{constant fiber direct integral} over $\Hcal_\Bbold$:
\begin{align}
\Kcal_\Bbold(\tau) =: \int_{Q_\Bbold^*}^\oplus \dd \vartheta \; \Hcal_\Bbold. \label{Intro:Direct_Integral_definition}
\end{align}
This notation really has the interpretation of a Hilbert space $\Hcal$ being decomposed into a direct sum $\Hcal = U_1 \oplus U_2$ of two subspaces $U_1, U_2\subseteq \Hcal$. The only difference is that \eqref{Intro:Direct_Integral_definition} is a ``continuous'' version of this decomposition. In order to actually make it a Hilbert space, we equip $\Kcal_\Bbold(\tau)$ with the scalar product
\begin{align}
\langle \varphi, \psi \rangle_{\Kcal_\Bbold(\tau)} &:= \frac{1}{|Q_\Bbold^*|} \int_{Q_\Bbold^*} \dd \vartheta \; \langle \varphi(\vartheta), \psi(\vartheta) \rangle_{\Hcal_\Bbold}. \label{Intro:Equivariant_K_scalar_product}
\end{align}
where $|Q_\Bbold^*|$ denotes the Lebesgue measure of $Q_\Bbold^*$.

Now, we are in position to define the \emph{Bloch--Floquet transformation} of a smooth and compactly supported function $f\in C_c^\infty(\Rbb^3) \subseteq L^2(\Rbb^3)$ by
\begin{align}
(\Ucal_{\mathrm{BF}} f) (\vartheta)(x) := \sum_{\lambda\in \Lambda_\Bbold} \e^{-\i \vartheta \cdot (x - \lambda) } \; \bigl(T_\Bbold(\lambda) f \bigr)(x). \label{Intro:BF-Hintrafo}
\end{align}
This means that $\Ucal_{\mathrm{BF}}$ takes an $L^2(\Rbb^3)$-function $f$ and \emph{sorts} it according to its gauge-periodic fibers.  The periodicity is accomplished by the sum over $\lambda\in \Lambda_\Bbold$. If $\nu\in \Lambda_\Bbold^*$, then
\begin{align*}
(\Ucal_{\mathrm{BF}} f)(\vartheta - \nu)(x) = \e^{\i \nu \cdot x} \sum_{\lambda\in \Lambda_\Bbold} \e^{-\i \vartheta \cdot (x - \lambda)}  \e^{-\i \nu \cdot \lambda} \; \bigl( T_\Bbold(\lambda) f \bigr)(x).
\end{align*}
By \eqref{Intro:Dual_lattice_relation}, we have $\nu \cdot \lambda \in 2\pi \Zbb$ so that, indeed, $\Ucal_{\mathrm{BF}} f \in \Kcal_\Bbold(\tau)$.

A straightforward computation shows that
\begin{align*}
\Vert \Ucal_{\mathrm{BF}} f\Vert_{\Kcal_\Bbold(\tau)}^2 = \sum_{\lambda, \lambda' \in \Lambda_\Bbold} \int_{Q_\Bbold^*} \dd x \; \ov{T_\Bbold(\lambda) f(x)} \, T_\Bbold(\lambda') f(x) \; \frac{1}{|Q_\Bbold^*|} \int_{Q_\Bbold^*} \dd \vartheta \; \e^{\i \vartheta (\lambda - \lambda')},
\end{align*}
and the fact that
\begin{align*}
\frac{1}{|Q_\Bbold^*|} \int_{Q_\Bbold^*} \dd \vartheta \; \e^{\i \vartheta \cdot (\lambda - \lambda')} &= \delta_{\lambda, \lambda'}, & \lambda, \lambda' &\in \Lambda_\Bbold,
\end{align*}
shows that $\Vert \Ucal_{\mathrm{BF}} f\Vert_{\Kcal_\Bbold(\tau)} = \Vert f\Vert_{L^2(\Rbb^3)}$. Hence, $\Ucal_{\mathrm{BF}}$ extends uniquely to a unitary map $\Ucal_{\mathrm{BF}} \colon L^2 (\Rbb^3) \ra \Kcal_\Bbold(\tau)$ with inverse $\Ucal_{\mathrm{BF}}^*\colon \Kcal_\Bbold(\tau) \ra L^2 (\Rbb^3)$ given by
\begin{align}
(\Ucal_{\mathrm{BF}}^* \varphi)(x) &:= \frac{1}{|Q_\Bbold^*|} \int_{Q_\Bbold^*} \dd \vartheta \; \e^{\i \vartheta \cdot x} \, \varphi(\vartheta) (x). \label{Intro:BF-Rücktrafo}
\end{align}

Before we go on, we should invest a moment to understand how we should view the Bloch--Floquet transformation $\Ucal_{\mathrm{BF}}$. This transformation accomplishes the domain switch from $L^2(\Rbb^3)$ to the space $\Kcal_\Bbold(\tau)$ of ``families'' of $\Lambda_\Bbold$-gauge-periodic functions, which we mentioned in the beginninig of this subsection. The members (fibers) of these families are connected according to certain rules when we pass from one translate of the unit cell to another ($\tau$-equivariance) and each fiber can be viewed as a function on $Q_\Bbold$ as we saw in \eqref{Intro:Gauge-periodic_Extension}. To illustrate this further, let us investigate what $\Ucal_{\mathrm{BF}}$ implies for the magnetic momentum operator $\pi_\Abold$ in \eqref{Intro:Magnetic_Momentum}. We employ \eqref{Intro:BF-Rücktrafo} and \eqref{Intro:Magnetic_Momentum_Translation_Intertwining} to obtain
\begin{align}
\Ucal_{\mathrm{BF}} \, \pi_\Abold \, \Ucal_{\mathrm{BF}}^* \varphi(\vartheta) &= \pi_\Abold(\vartheta) \varphi(\vartheta), & \vartheta &\in \Rbb^3, \label{Intro:Magnetic_Momentum_Fibering}
\end{align}
where, actionwise,
\begin{align}
\pi_\Abold(\vartheta) := \pi_\Abold + \vartheta.
\end{align}
Since the equation \eqref{Intro:Magnetic_Momentum_Fibering} holds fiberwise without mixing, the magnetic momentum operator $\pi_\Abold$ is an example of an operator which we say to \emph{fiber in the direct integral decomposition} or be \emph{decomposable} (we can also view this as a certain sense of \emph{block diagonality} if we have invariant subspaces in mind) and we express this in the suggestive notation
\begin{align}
\Ucal_{\mathrm{BF}} \, \pi_\Abold \, \Ucal_{\mathrm{BF}}^* =: \int_{Q_\Bbold^*}^\oplus \dd \vartheta \; \pi_\Abold(\vartheta). \label{Intro:Direct_Integral_operator}
\end{align}
Furthermore, \eqref{Intro:Magnetic_Momentum_Fibering} holds in an equivariant fashion in the sense that
\begin{align}
\pi_\Abold(\vartheta - \nu) &= \tau(\nu) \, \pi_\Abold(\vartheta) \, \tau(\nu)^*, & \nu &\in \Lambda_\Bbold^*. \label{Intro:Magnetic_Momentum_Fibering_equivariant}
\end{align}
Here, $\tau(\nu)^*$ is the adjoint in $\Hcal_\Bbold$. The important point is that the operator $\pi_\Abold(\vartheta)$ acts in the Hilbert space $\Hcal_\Bbold$ for every $\vartheta\in Q_\Bbold^*$, i.e., a space of periodic functions, whereas $\pi_\Abold$ acts in $L^2(\Rbb^3)$. Moreover, \eqref{Intro:Magnetic_Momentum_Fibering_equivariant} shows that the operator $\pi_\Abold(\vartheta)$ is uniquely determined by the action on functions in $L^2(Q_\Bbold)$. Therefore, the fiber $\pi_\Abold(\vartheta)$ may be viewed as an operator on the fundamental box $Q_\Bbold$.

We will not comment further on the actual domains of the respective operators but they come out naturally, see \cite{Stefan_Peierls}. Using the $\tau$-equivariance \eqref{Intro:Magnetic_Momentum_Fibering_equivariant}, we also easily see that eigenvalues of the self-adjoint operators
\begin{align*}
&\int_{Q_\Bbold^*}^\oplus \dd \vartheta \; (\pi_\Abold(\vartheta))^2 & &\int_{Q_\Bbold^*}^\oplus \dd \vartheta \; (\pi_{\Abold_\Bbold}(\vartheta))^2
\end{align*}
are $\Lambda_\Bbold$-gauge-periodic functions. In fact, we show in Chapter \ref{Chapter:Spectrum_Landau_Hamiltonian_Section} that the eigenvalues of the latter are constant.

\subsection{The trace per unit volume}

The objective of this section is to define the \emph{local trace} of a $\Lambda_\Bbold$-gauge-periodic operator $S$ in the sense of \eqref{Intro:Periodic_operator}. With the knowledge of the preceding section, this is easy now. First of all, we assume that the operator $S$ is \emph{decomposed in the direct integral}, i.e., there are operators $S_\vartheta$, $\vartheta \in Q_\Bbold^*$ such that
\begin{align}
\Ucal_{\mathrm{BF}} \, S\, \Ucal_{\mathrm{BF}}^* = \int_{Q_\Bbold^*}^\oplus \dd \vartheta \; S_\vartheta. \label{Intro:Direct_Integral_Decomposition}
\end{align}
In this case, we define the \emph{local trace} $\Tr_{Q_\Bbold}(S)$ as the number
\begin{align}
\Tr_{Q_\Bbold}(S) &:= \frac{1}{|Q_\Bbold^*|} \int_{Q_\Bbold^*} \dd \vartheta \; \tr(S(\vartheta)), \label{Intro:Local_Trace_definition}
\end{align}
where $\tr(S)$ is the usual trace on the space $L^2(Q_\Bbold)$, see \eqref{Intro:Periodic_unitary_equivalence_box}. 

\begin{bem}
\label{Intro:Remark_on_trace}
Some sources (including the paper in Section \ref{Chapter:DHS1}) define the trace per unit volume of $S$ simply as the ``usual trace'' of $\chi_{Q_\Bbold} S$. There are two possible ways to interpret this. The first way is to utilize the unitary equivalence \eqref{Intro:Periodic_unitary_equivalence_box} and to identify the fiber $S_\vartheta$ with the restriction of $\chi_{Q_\Bbold} S$ to the space $L^2(Q_\Bbold)$ as we discussed above in \eqref{Intro:Gauge-periodic_Extension}. The second way is to consult \cite[Lemma 3]{Teufel2009}, where the authors explicitly show that, with our definition \eqref{Intro:Local_Trace_definition}, we indeed obtain $\Tr_{Q_\Bbold}(S) = \tr(\chi_{Q_\Bbold}S)$. We elaborate on this result in Appendix \ref{Chapter:Local_Traces}. It is important to understand, however, that the Bloch--Floquet transformation is not so much a transformation of the operator but rather of the space on which the operator acts, see the beginning of this section.
\end{bem}

With this at hand, we define the \emph{trace per unit volume} of $S$ as
\begin{align}
\Tr (S) &:= \frac{1}{|Q_\Bbold|} \Tr_{Q_\Bbold}(S) = \int_{Q_\Bbold^*} \frac{\dd \vartheta}{(2\pi)^3} \; \tr(S(\vartheta)). \label{Intro:Trace_per_unit_volume_definition}
\end{align}
When we denote the kernel of $S$ by $S(x,y)$, then Remark \ref{Intro:Remark_on_trace} implies
\begin{align}
\Tr(S) = \frac{1}{|Q_\Bbold|} \int_{Q_\Bbold} \dd x \; S(x,x). \label{Intro:Trace_kernel}
\end{align}

Furthermore, for $1 \leq p < \infty$, we may define the \emph{$p$\tho\ von Neumann Schatten class per unit volume} (or \emph{local von Neumann--Schatten class}) $\Scal^p$ as the space of bounded operators $S$ that obey \eqref{Intro:Direct_Integral_Decomposition} and for which $\Vert S\Vert_p^p := \Tr(|S|^p)$ is finite. The space $\Scal^\infty$ consists of all bounded periodic operators and is equipped with the usual operator norm $\Vert \cdot\Vert_\infty$. In Appendix \ref{Chapter:Local_Traces}, we prove several results on $\Vert \cdot\Vert_p$. From these, it follows that $\Scal^p$ are Banach spaces and among these are the general Hölder inequality
\begin{align}
\Vert ST\Vert_r \leq \Vert S\Vert_p \Vert T\Vert_q, \label{Intro:General_Hölder}
\end{align}
which holds for $S\in \Scal^p$ and $T\in \Scal^q$ and all $1\leq p,q,r\leq \infty$ as long as $\frac 1r = \frac 1p + \frac 1q$. We also have the inequality
\begin{align}
|\Tr S|\leq \Vert S\Vert_1. \label{Intro:Trace_Norm_Estimate}
\end{align}

\subsection{Admissible BCS states}
\label{Intro:Admissible_states_Section}

Let $\Gamma$ be a bounded self-adjoint matrix-valued operator on $L^2(\Rbb^3) \oplus L^2(\Rbb^3)$ of the form
\begin{align}
\Gamma = \begin{pmatrix} \gamma & \alpha \\ \ov \alpha & 1 - \ov \gamma \end{pmatrix}, \label{Intro:Gamma_definition}
\end{align}
where $0 \leq \Gamma \leq 1$. Here, we set $\ov \alpha := J \alpha J$ where the operator $J \colon L^2(\Rbb^3) \ra L^2(\Rbb^3)$ is the antilinear Riesz' identification, whose role is played by the complex conjugation. In the framework of BCS theory, we call an operator $\Gamma$ of the form \eqref{Intro:Gamma_definition} a \emph{generalized fermionic one-particle density matrix}. The notion is inspired by \eqref{Intro:One_particle_density_matrix_proper} but the two operators do not share any common background as explained in the beginning of Section \ref{Intro_Section:BCS-functional}.

Since $\Gamma$ is self-adjoint, $\Gamma = \Gamma^*$, we infer that $\gamma = \gamma^*$ and $\alpha^* = \ov \alpha$. The latter implies that the kernel $\alpha(x,y)$, satisfies 
\begin{align}
\alpha(x, y) = \alpha(y,x). \label{Intro:alpha_symmetry}
\end{align}
In this way, the fermionicity of $\alpha$ is deleted in comparison to the generalized one-particle density matrix introduced in \eqref{Intro:One_particle_density_matrix_proper}. Physically, this corresponds to the assumption that our system is spinless (or, in a spin singlet state), which is a simplifying assumption. The full Cooper pair wave function is then $\alpha$ times an appropriate spin operator. 

Furthermore, since $0 \leq \Gamma \leq 1$, we have $0 \leq \gamma \leq 1$ and since $\Gamma (1 - \Gamma) \geq 0$, we deduce from \eqref{Intro:Gamma(1-Gamma)} that $\alpha$ and $\gamma$ are related through the operator inequality
\begin{align}
\alpha \alpha^* \leq \gamma (1 - \gamma). \label{Intro:gamma_alpha_relation_BCS}
\end{align}

The next step is to generalize the notion of gauge-periodicity in the obvious fashion to operator valued $2\times 2$ matrices. To do this, we define the magnetic translation
\begin{align}
\Tbold_\Bbold(v) &:= \begin{pmatrix} T_\Bbold(v) & \\ & \ov{T_\Bbold(v)} \end{pmatrix} , & v &\in \Rbb^3,
\end{align}
where $T_\Bbold(v)$ is the magnetic translation in \eqref{Intro:Magnetic_Translation}, and we call an operator $\Sbold$ on the space $L^2(\Rbb^3) \oplus L^2(\Rbb^3)$ \emph{$\Lambda_\Bbold$-gauge-periodic} (or just \emph{gauge-periodic}) if 
\begin{align}
\Tbold_\Bbold (\lambda) \, \Sbold \, \Tbold_\Bbold(\lambda)^* &= \Sbold, & \lambda &\in \Lambda_\Bbold.
\end{align}
If $\Gamma$ is a gauge-periodic operator, then $\gamma$ and $\alpha$ satisfy 
\begin{align}
T_\Bbold(\lambda) \, \gamma \, T_\Bbold(\lambda)^* &= \gamma, & T_\Bbold(\lambda) \, \alpha \, \ov{T_\Bbold(\lambda)}^* &= \alpha. \label{Intro:gamma_alpha_periodicity_operator}
\end{align}
A straightforward computation shows that the kernels $\gamma(x,y)$ and $\alpha(x,y)$ then obey
\begin{align}
\gamma(x, y) &= \e^{\i \frac{\Bbold}{2} \cdot (\lambda \wedge (x-y))} \, \gamma(x + \lambda, y + \lambda), \notag\\
\alpha(x, y) &= \e^{\i \frac{\Bbold}{2} \cdot (\lambda \wedge (x+y))} \, \alpha(x + \lambda, y + \lambda). \label{Intro:gamma_alpha_periodicity_kernel}
\end{align}

\begin{defn}[BCS states]
A \emph{BCS state} is a gauge-periodic generalized one-particle density matrix $\Gamma$ of the form \eqref{Intro:Gamma_definition}. A BCS state is called \emph{admissible} if $\gamma$ and $\pi_{\Abold_\Bbold}^2 \gamma$ belong to $\Scal^1$.
\end{defn}

\begin{bem}
In many publications (including the ones in Chapter \ref{Chapter:DHS1} and \ref{Chapter:DHS2}), the admissibility condition is phrased as
\begin{align}
\Tr \bigl[ \gamma + \pi_{\Abold_\Bbold}^2 \gamma \bigr] <\infty. \label{Intro:admissibility_sloppy}
\end{align}
Now, obviously, this does not make sense because $\pi_{\Abold_\Bbold}^2\gamma$ need not be a nonnegative operator, whence the condition \eqref{Intro:admissibility_sloppy} does not help much. One way to understand it is to request $\gamma$ and $\pi_\Abold^2 \gamma$ to be of trace class, as we did above. The advantage to phrase it like in \eqref{Intro:admissibility_sloppy} is that it requires substantially less notation to be introduced (and the trace per unit volume is expectable to be familiar to the readership of a paper). The second way to understand \eqref{Intro:admissibility_sloppy} is the actually \emph{weaker} condition
\begin{align}
\Tr \bigl[ \gamma + \pi_{\Abold_\Bbold} \, \gamma \, \pi_{\Abold_\Bbold} \bigr] < \infty. \label{Intro:admissibility_weak_condition}
\end{align}
This condition does make sense as $\gamma$ is a nonnegative operator. Moreover, it is a natural condition in the sense that $\Tr [|S|] + \Tr[|\pi_{\Abold_{\Bbold}}\,  S \, \pi_{\Abold_\Bbold}|]$ defines a norm on the space $\Hcal^1(\Scal^1)$ of operators for which this expression is finite, which makes $\Hcal^1(\Scal^1)$ a Banach space and a closed subspace of $\Scal^1$. However, in order to use this definition in place of our admissibility condition, we would have to adjust the BCS functional \eqref{Intro:BCS_functional_formal} and replace $\pi_{\Abold_\Bbold}^2\gamma$ by $\pi_{\Abold_\Bbold} \gamma \pi_{\Abold_\Bbold}$ which makes the work a bit clumsy and spoils the obvious interpretation of a ``kinetic energy'' operator. It is, indeed, a matter of taste.
\end{bem}

We investigate a bit what the admissibility of $\Gamma$ implies for $\gamma$ and $\alpha$. First of all, we note that $\ov \gamma$ is trace class as well. For, we know that $\gamma = \int_{Q_\Bbold^*}^\oplus \dd \vartheta \;\gamma(\vartheta)$ with nonnegative trace class operators $\gamma(\vartheta)$ on $L^2(Q_\Bbold)$. Therefore, for some orthonormal basis $\{\varphi_n\}_{n\in \Nbb} \subseteq L^2(Q_\Bbold)$, we have  $\tr(\gamma(\vartheta)) = \sum_{n\in \Nbb_0} \langle \varphi_n, \gamma(\vartheta) \varphi_n\rangle$. Since $J = \int_{Q_\Bbold^*}^\oplus \dd \vartheta \, J(\vartheta)$ with the ($\vartheta$-independent) complex conjugation operator $J(\vartheta)$ on $L^2(Q_\Bbold)$, we conclude $\tr(\ov \gamma(\vartheta)) = \sum_{n\in \Nbb} \langle \ov{\varphi_n} , \gamma(\vartheta) \ov{\varphi_n}\rangle$. 

This has the following consequences for $\alpha$. First, we note that $\gamma\geq 0$ and \eqref{Intro:gamma_alpha_relation_BCS} imply
\begin{align}
\alpha \alpha^* \leq \gamma. \label{Intro:admissibility_alpha}
\end{align}
Therefore, admissibility of $\Gamma$ implies that $\alpha$ is a local Hilbert--Schmidt operator with a kernel $\alpha(x,y)$. Since the kernel of $\alpha \alpha^*$ equals
\begin{align}
\alpha \alpha^* (x,y) = \int_{\Rbb^3} \dd z \; \alpha(x,z) \ov{ \alpha(y,z)},
\end{align}
we conclude from \eqref{Intro:Trace_kernel} that the Hilbert--Schmidt norm per unit volume equals
\begin{align}
\Vert \alpha\Vert_2^2 &= \frac{1}{|Q_\Bbold|} \iint_{Q_\Bbold\times \Rbb^3} \dd x \dd y \; |\alpha(x,y)|^2. \label{Intro:alpha_Hilbert-Schmidt}
\end{align}
Likewise, admissibility and \eqref{Intro:admissibility_alpha} yield
\begin{align}
\Vert (-\i \nabla + \Abold_\Bbold) \alpha\Vert_2^2 = \iint_{Q_\Bbold \times \Rbb^3} \dd x \dd y \; |(-\i \nabla_x + \Abold_\Bbold(x)) \alpha(x,y)|^2 < \infty. \label{Intro:alpha_Sobolev1}
\end{align}
Finally, we often additionally require that
\begin{align}
\Vert (-\i \nabla + \Abold_\Bbold) \ov \alpha\Vert_2^2 = \iint_{Q_\Bbold \times \Rbb^3} \dd x \dd y \; |(-\i \nabla_y + \Abold_\Bbold(y)) \alpha(x,y)|^2 < \infty. \label{Intro:alpha_Sobolev2}
\end{align}

Equations \eqref{Intro:alpha_Hilbert-Schmidt}, \eqref{Intro:alpha_Sobolev1}, and \eqref{Intro:alpha_Sobolev2} imply that $\alpha$ belongs to the Sobolev space of Hilbert--Schmidt operators, for which the norm given by
\begin{align}
\Vert \alpha\Vert_2^2 + \Vert (-\i \nabla + \Abold_\Bbold) \alpha\Vert_2^2 + \Vert (-\i \nabla + \Abold_\Bbold) \ov \alpha\Vert_2^2
\end{align}
is finite.

We close this section by the introduction of center-of-mass and relative coordinates since these play a major role in the analysis of the BCS functional in the weak magnetic field regime. This can already be seen from the periodicity relation of the kernel $\alpha(x,y)$ in \eqref{Intro:gamma_alpha_periodicity_kernel}. When we introduce the center of mass of $x$ and $y$ by
\begin{align}
X := \frac{x + y}{2} \label{Intro:COM}
\end{align}
and the relative coordinate by
\begin{align}
r := x - y, \label{Intro:relative}
\end{align}
then \eqref{Intro:alpha_symmetry} and \eqref{Intro:gamma_alpha_periodicity_kernel} become
\begin{align}
\alpha(X, r) &= \e^{\i \Bbold\cdot (\lambda \wedge X)} \, \alpha(X + \lambda, r), \qquad \lambda\in \Lambda_\Bbold; & \alpha(X,r) &= \alpha(X, -r). \label{Intro:alpha_periodicity_COM}
\end{align}
Here, we abused notation slightly by writing $\alpha(X, r) \equiv \alpha(x,y)$. Therefore, let us introduce the space $L^2(Q_\Bbold \times \Rbb_{\mathrm{s}}^3)$ of functions $\alpha(x,y)$ obeying \eqref{Intro:alpha_periodicity_COM}, whose norm 
\begin{align}
\Vert \alpha\Vert_2^2 &:= \frac{1}{|Q_\Bbold|}\iint_{Q_\Bbold \times \Rbb^3} \dd x\dd y \; |\alpha(x,y)|^2
\end{align}
is finite.

A straightforward computation then shows that
\begin{align}
\pi_x &= \frac 12 \Pi + \tilde \pi, & \pi_y &= \frac 12 \Pi - \tilde \pi,
\end{align}
where $\Pi$ and $\tilde \pi$ are the center-of-mass magnetic and relative momentum operators given by
\begin{align}
\Pi &:= -\i \nabla + 2 \Abold_\Bbold, & \tilde \pi &:= -\i \nabla + \frac 12 \Abold_\Bbold. \label{Intro:Magnetic_momenta_COM}
\end{align}
Therefore, we can also let $H^1(Q_\Bbold \times \Rbb_{\mathrm{s}}^3)$ be the space of functions for which the norm
\begin{align}
\Vert \alpha\Vert_{H^1(Q_\Bbold \times \Rbb_{\mathrm s}^3)}^2 := \Vert \alpha\Vert_2^2 + \Vert \Pi\alpha\Vert_2^2 + \Vert \tilde \pi\alpha\Vert_2^2
\end{align}
is finite. 

The conclusion is that if $\Gamma$ is an admissible BCS state, then the kernel of $\alpha$ belongs to the Sobolev space $H^1(Q_\Bbold \times \Rbb_{\mathrm s}^3)$.

At this point, we can understand the expression \eqref{Intro:BCS_functional_formal} in a rigorous fashion. Before we define the BCS functional that we will work with, we need to introduce the scaling of the external fields and the weak field regime.

\subsection{The scaling of weak external fields}

\begin{figure}[h]
\centering
\includegraphics[width=15cm]{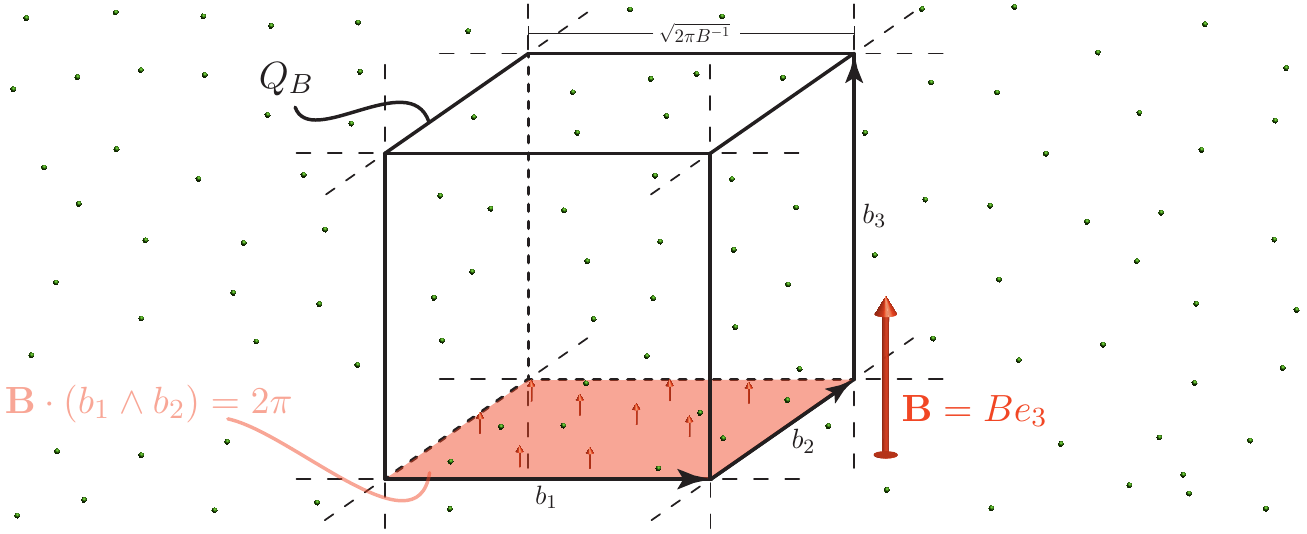}
\caption{Fermionic system with a constant nonzero magnetic flux through the unit cell. The flux is independent of the field strength $B = h^2$.\label{Intro:System_nonzero_flux_2}}
\end{figure}

We recall that we consider a system of fermions that is subject to external fields. The most relevant contribution, which also causes the most severe mathematical difficulties, stems from the constant magnetic field. It is given by a vector $\Bbold\in \Rbb^3$, which in this figure and the work presented in Chapter \ref{Chapter:DHS1}, is pointing in the $e_3$-direction and has strength $B >0$. In this case, the box $Q_\Bbold$ that we constructed in \eqref{Intro:Lattice_unit_cell_definition} is solely dependent on $B$ and is a cube of sidelength $\sqrt{2\pi B^{-1}}$, as Figure \ref{Intro:System_nonzero_flux_2} shows.
This models a system with a constant (that is, independent of $B$) but nonzero magnetic flux through the unit cell of the lattice $\Lambda_B$ of periodicity. 

Now, we define the magnetic field strength 
\begin{align*}
B := |\Bbold|
\end{align*}
and the unit vector pointing in the direction of the magnetic field
\begin{align}
e_\Bbold := (\mathfrak b_1, \mathfrak b_2, \mathfrak b_3) := B^{-1} \; \Bbold.
\end{align}
To introduce the scaling of the problem, we assume that the fixed external potentials $A$ and $W$ are periodic with respect to the lattice $\Lambda_{e_\Bbold}$. We emphasize that we say \emph{periodic}, not \emph{gauge-periodic}. This means that we have
\begin{align}
A(x + \lambda) &= A(x), & W(x + \lambda) &= W(x), & \lambda &\in \Lambda_{e_\Bbold}. \label{Intro:Periodic_fields}
\end{align}

We introduce a parameter $h >0$, which we think of as Planck's constant and which shall model the ratio between the microscopic --- order $1$ --- and macroscopic --- order $h^{-1}$ --- scale of the system. We assume $h$ to be small, $0 < h \ll 1$, and it will enable us to model the external fields to be \emph{weak} and of  \emph{macroscopic} nature. First of all, we explain the  macroscopic nature of the fields. We set the magnetic field $\Bbold$ to be of strength $B := h^2$. This extends the box $Q_\Bbold$ to be very large and the sidelength
\begin{align}
|b_i(\Bbold)| &= \sqrt{\frac{4\pi}{h^6 \, \mathfrak b_1 \, \mathfrak b_2\, \mathfrak b_3}} \, h^2 \mathfrak b_i \sim h^{-1} 
\end{align}
to be of the macroscopic order $h^{-1}$, see \eqref{Intro:Basis_Vectors_definition}. Since the constant magnetic field potential $\Abold_\Bbold(x) = \frac 12 \Bbold \wedge x$ is linear, we cannot see the difference between the macroscopic scaling and the weakness of the field, they come hand in hand. 
\begin{figure}[h]
\centering
\includegraphics[width=10cm]{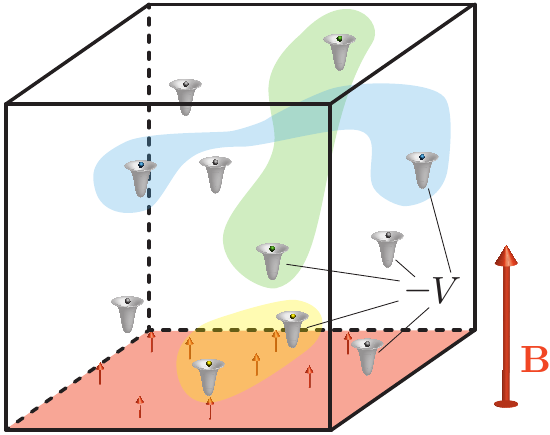}
\caption{Fermions interact via an attractive two-body potential -- Cooper pair formation indicated by the colored clouds.}
\end{figure}
This is different, however, when it comes to the periodic external fields. Here, we do see a difference. Namely, the potentials that we will insert into the functional are given by
\begin{align}
A_h(x) &:= h \, A(hx), & W_h(x) &:= h^2 \, W(hx), \label{Intro:Fields_scaling}
\end{align}
where $A$ and $W$ are the periodic fields from \eqref{Intro:Periodic_fields}. Here, the macroscopic nature of the fields results from the argument $hx$, which makes the fields ``live'' (as we say) on the large macroscopic box $Q_\Bbold$ instead of $Q_{e_\Bbold}$. In contrast, the factors of $h$ and $h^2$ in front of $A$ and $W$, respectively, indicate the weakness of the field. The macroscopic scale is complemented by the microscopic scale of order $1$ on which the interaction of the particles caused by $V$ takes place.

Since we have to deal with a separation of scales (microscopic and macroscopic), we also say that our problem is a \emph{two-scale problem}. 

The following assumptions on the external fields will guide us through the entire thesis. In order to phrase them, we introduce the Sobolev spaces $\Wperdiss m$ and $\Wpervecdiss m$ as the spaces of (vector-valued) functions $f$ such that $D^\nu f$ belongs to the space $\Lperdiss$ or $\Lpervecdiss$ of bounded periodic (vector-valued) functions, respectively, $\nu \in \Nbb_0^3$, $|\nu| \leq m$. We remark that such functions are Lipschitz continuous up to the $(m-1)$\st\ derivative and they possess a Taylor expansion in the classical sense. This can be seen from Sobolev space theory above the critical Sobolev exponent, which is maybe less common to be taught in lecture courses, but which is well explained for example in \cite{EvansPDE}. It needs knowledge on Lebesgue's Differentiation theorem, which is only stated but not proved in \cite{EvansPDE}. A proof using maximal functions can be found in \cite{Lebesgue_Differentiation_Soneji}. For the reader's convenience, we gather the relevant content in Appendix \ref{Chapter:W-functions}.

\begin{asmp}
The magnetic potential $A$ satisfies $A\in \Wpervecdiss 4$ and the electric potential satisfies $W\in \Wperdiss 1$. Additionally, we have $A(0) = 0$.
\end{asmp}

\subsection{Rigorous definition of the BCS functional}

We can now rigorously define the BCS functional as the expression which we informally introduced in \eqref{Intro:BCS_functional_formal}. For any admissible BCS state $\Gamma$ and any $h >0$, we define the BCS free energy functional at temperature $T\geq 0$ by
\begin{align}
\FBCS(\Gamma) &:= \Tr \bigl[ \bigl((-\i \nabla + \Abold_h)^2 + W_h - \mu\bigr) \gamma \bigr] - T \, S(\Gamma) - \fint_{Q_\Bbold} \dd X \int_{\Rbb^3} \dd r \; V(r) \, |\alpha(X, r)|^2.\label{Intro:BCS_functional_definition}
\end{align}
Here, $S(\Gamma) := - \Tr[\Gamma \ln(\Gamma)]$ is the von Neumann entropy per unit volume and $\Abold_h$ and $W
_h$ are as in \eqref{Intro:Fields_scaling}.

\subsection{Boundedness from below}

The first thing that we have to convince ourselves of is the fact that the BCS functional is bounded from below. Obviously, this is necessary to be able to properly set up the minimization problem, which we mentioned several times by now. The proof I present here has been conducted by Andreas Deuchert, who wrote it down in unpublished notes on the BCS functional. For the allowance to present it here, I express my gratitude to him. 

Before we start we should mention that we use the following convention in the field of analysis: $C$ denotes a generic positive constant that is allowed to change from line to line. We allow it to depend on the various fixed quantities in the theory, namely the external fields $A$ and $W$, as well as the critical temperature $\Tc$, the chemical potential $\mu$, the interaction $V$, the ground state $\alpha_*$ of $K_\Tc - V$, and so on. Of course, it does \emph{not} depend on $h$, $T$, $\gamma$, etc.

The main idea is that the kinetic energy dominates both the entropy and the interaction. First of all, we bound away the periodic potentials $A_h$ and $W_h$. Since $W$ is bounded, we obviously have $\Tr [W_h\gamma] \geq -Ch^2 \Tr \gamma$. Furthermore,
\begin{align*}
(-\i \nabla + \Abold_h)^2 = \pi_{\Abold_\Bbold}^2 + A_h \cdot \pi_{\Abold_\Bbold} + \pi_{\Abold_\Bbold} \cdot A_h + |A_h|^2.
\end{align*}
Since, for any self-adjoint operators $T$ and $S$ and $\varepsilon >0$, we have $(\sqrt \varepsilon \, T + \sqrt \varepsilon^{-1} S)^2 \geq 0$, whence
\begin{align*}
TS + ST \geq -\varepsilon \, T^2 - \varepsilon^{-1} S^2,
\end{align*}
we conclude that 
\begin{align*}
(-\i \nabla + \Abold_h)^2 \geq (1 - \varepsilon) \, \pi_{\Abold_\Bbold}^2 + \bigl( 1 - \varepsilon^{-1} \bigr) |A_h|^2 \geq (1 - \varepsilon) \, \pi_{\Abold_\Bbold}^2 - C(1 + \varepsilon^{-1}) h^2.
\end{align*}
With the choice $\varepsilon = \frac 12$, this implies that
\begin{align}
\Tr \bigl[ \bigl( (-\i \nabla + \Abold_h)^2 + W_h - \mu\bigr) \gamma\bigr] \geq \frac 12 \Tr \bigl[ (\pi_{\Abold_\Bbold}^2 - C(1+h^2)) \gamma\bigr].
\end{align}
To arrive at this inequality, we made use of the fact that $\gamma = \gamma^{\nicefrac 12} \gamma^{\nicefrac 12}$ and that we are allowed to symmetrize the operators inside the trace by cyclicity.

The next step is to bound the entropy by a portion of the kinetic energy. 
We define
\begin{align*}
\tilde \Gamma &:= \frac{1}{1 + \e^{\frac \beta 8 \tilde H}},  & \tilde H &:= \begin{pmatrix} \pi_{\Abold_\Bbold}^2 \\ & -\ov \pi_{\Abold_\Bbold}^2\end{pmatrix},
\end{align*}
and let $\tilde \gamma$ be the upper left entry of $\tilde \Gamma$. With this, a short computation shows that
\begin{align}
\frac 18 \Tr [\pi_{\Abold_\Bbold}^2\gamma] - TS(\Gamma) = \frac 18 \Tr [\pi_{\Abold_\Bbold}^2 \tilde \gamma] - TS(\tilde \Gamma) + \frac T2 \Tr \bigl[ \frac{\beta \tilde H}{8} (\Gamma - \tilde \Gamma) + \varphi(\Gamma) - \varphi(\tilde \Gamma)\bigr], \label{Intro:Boundedness_from_below_1}
\end{align}
where the function $\varphi$ is defined in \eqref{Intro:varphi}. Since $\varphi'(\tilde \Gamma) = - \frac{\beta}{8} \tilde H$, we are in position to employ the trace version of \emph{Klein's inequality}. We already came across a simplified version of this in Section \ref{Intro:BCS-functional-ti_Section}. The inequality tells us that whenever $\varphi\colon [0,1] \ra \Rbb$ is strictly convex and differentiable so that $\varphi(T)$, $\varphi(S)$, and $\varphi'(S)$ are locally trace class, then
\begin{align}
\Tr \bigl[ \varphi(T) - \varphi(S) - \varphi'(S) (T - S) \bigr] \geq 0 \label{Intro:Klein}
\end{align}
and equality holds if and only if $T = S$, see Theorem \ref{Klein_local_thm}. This inequality is the most basic trace inequality and usually the first, which one comes about. It also plays an important role in the study of the so-called relative entropy, as we shall sketch in the next section. 

Klein's inequality tells us that \eqref{Intro:Boundedness_from_below_1} is bounded from below by
\begin{align}
\frac 18 \Tr [\pi_{\Abold_\Bbold}^2\gamma] - TS(\Gamma) \geq \frac 18 \Tr [\pi_{\Abold_\Bbold}^2 \tilde \gamma] - TS(\tilde \Gamma), \label{Intro:Boundedness_from_below_2}
\end{align}
To calculate this term, we have $1 - \tilde \gamma = (1 + \e^{-\frac \beta 8 \pi_{\Abold_\Bbold}^2})^{-1}$, so that
\begin{align*}
\Tr[\tilde \Gamma \ln(\tilde \Gamma)] = \Tr [\tilde \gamma \ln (\tilde \gamma) + (1 - \tilde \gamma) \ln(1 - \tilde \gamma)].
\end{align*}
Therefore, the right side of \eqref{Intro:Boundedness_from_below_2} equals
\begin{align}
T \Tr \bigl[ \frac{\beta\pi_{\Abold_\Bbold}^2}{8} \tilde \gamma + \tilde \gamma \ln(\tilde \gamma) + (1 - \tilde \gamma) \ln(1 - \tilde \gamma)\bigr] = T \Tr \bigl[ \ln(1 + \e^{-\frac \beta 8 \pi_{\Abold_\Bbold}^2})\bigr]. \label{Intro:Boundedness_from_below_3}
\end{align}
In Chapter \ref{Chapter:Spectrum_Landau_Hamiltonian_Section}, we investigate the spectrum of the periodic Landau Hamiltonian. This shows that the right hand side of \eqref{Intro:Boundedness_from_below_3} is finite.

Next, we bound the interaction by a portion of the kinetic energy. This is true even if $V\in L^\infty(\Rbb^3)$ does not hold. For, the inequality \eqref{Intro:gamma_alpha_relation_BCS} implies $\alpha \alpha^* \leq \gamma$, whence $\Tr[\pi_{\Abold_\Bbold}^2\gamma] \geq \Tr [\pi_{\Abold_\Bbold} \alpha \alpha^* \pi_{\Abold_\Bbold}] = \Tr [\alpha \pi_{\Abold_\Bbold}^2 \alpha^*]$. The last inequality follows from the fact that $\pi_{\Abold_\Bbold} \alpha$ is Hilbert--Schmidt, which we saw in Section \ref{Intro:Admissible_states_Section}. Hence, we conclude that
\begin{align}
\frac 18 \Tr [\pi_{\Abold_\Bbold}^2 \gamma] - \fint_{Q_\Bbold} \dd X \int_{\Rbb^3} \dd r \; V(r) |\alpha(X,r)|^2 \geq \fint_{Q_\Bbold} \dd y \; \langle \alpha, (\pi_{\Abold_\Bbold}^2 - V_y) \alpha\rangle_{L^2(\Rbb^3, \dd x)}, \label{Intro:Boundedness_from_below_4}
\end{align}
where $\pi_{\Abold_\Bbold}^2 - V_y$ acts on the first coordinate of $\alpha(x,y)$ and $V_y$ is the operator acting as $(V_y\psi)(x) = V(x-y)\psi(x)$. With modest assumptions on $V$, we have that $\pi_{\Abold_\Bbold}^2 - V_y$ is bounded from below. In our case, with bounded $V$, this is trivial, since $V_y \geq -\Vert V\Vert_\infty$ and $\pi_{\Abold_\Bbold}^2 \geq 0$. The latter holds by the \emph{diamagnetic inequality}
\begin{align}
|(-\i \nabla + \Abold) \psi(x)| \geq |\nabla |\psi|(x)|, \label{Intro:Diamagnetic_inequality}
\end{align}
which holds pointwise for almost all $x\in \Rbb^3$ as long as $\Abold\in L_{\mathrm{loc}}^2(\Rbb^3)$, see \cite[Theorem 7.21]{LiebLoss} or \cite[Eq. (4.4.3)]{LiebSeiringer}. Hence, \eqref{Intro:Boundedness_from_below_4} is bounded from below by $-C \Tr[\alpha \alpha^*]$, which by the inequality $\alpha \alpha^* \leq \gamma$ shows that
\begin{align*}
\frac 18 \Tr [\pi_{\Abold_\Bbold}^2 \gamma] - \fint_{Q_\Bbold} \dd X \int_{\Rbb^3} \dd r \; V(r) \, |\alpha(X,r)|^2 \geq - C \Tr \gamma.
\end{align*}

To sum up, we have shown that
\begin{align*}
\FBCS(\Gamma) \geq \frac 14 \Tr \bigl[ (\pi_{\Abold_\Bbold}^2-C(1 + h^2))\gamma\bigr] - C.
\end{align*}
Thus, it remains to show that the first term is bounded from below. Set $D:= C(1 +h^2)$. We claim that
\begin{align}
\Tr \bigl[ (\pi_{\Abold_\Bbold}^2-D)\gamma\bigr] \geq \Tr \bigl[ (\pi_{\Abold_\Bbold}^2-D) \Idbb_{(-\infty, D]}(\pi_{\Abold_\Bbold}^2) \bigr]. \label{Intro:Boundedness_from_below_5}
\end{align}
It is clear that the right side of this is finite since the operator $\pi_{\Abold_\Bbold}^2$ has finitely many eigenvalues below the threshold $D$, see Chapter \ref{Chapter:Spectrum_Landau_Hamiltonian_Section}. To see that \eqref{Intro:Boundedness_from_below_5} holds, we write
\begin{align}
\Tr \bigl[ (\pi_{\Abold_\Bbold}^2-D)\gamma\bigr] \notag \\
&\hspace{-50pt}= \Tr \bigl[ (\pi_{\Abold_\Bbold}^2-D) \Idbb_{(-\infty, D]}(\pi_{\Abold_\Bbold}^2) \, \gamma\bigr] + \Tr \bigl[ (\pi_{\Abold_\Bbold}^2-D) \Idbb_{(D, \infty)}(\pi_{\Abold_\Bbold}^2) \, \gamma\bigr]. \label{Intro:Boundedness_from_below_6}
\end{align}
The second term is nonnegative since $\gamma\geq 0$ and
\begin{align*}
\Tr \bigl[ (\pi_{\Abold_\Bbold}^2-D) \Idbb_{(D, \infty)}(\pi_{\Abold_\Bbold}^2) \, \gamma\bigr] &\\
&\hspace{-100pt}=  \Tr \bigl[ \sqrt{(\pi_{\Abold_\Bbold}^2-D) \Idbb_{(D, \infty)}(\pi_{\Abold_\Bbold}^2)} \;\gamma \;\sqrt{(\pi_{\Abold_\Bbold}^2-D) \Idbb_{(D, \infty)}(\pi_{\Abold_\Bbold}^2)}\, \bigr].
\end{align*}
Likewise, the first term on the right side of \eqref{Intro:Boundedness_from_below_6} is bounded from below by
\begin{align*}
\Tr \bigl[ (\pi_{\Abold_\Bbold}^2-D) \Idbb_{(-\infty, D]}(\pi_{\Abold_\Bbold}^2) \, \gamma\bigr] &\\
&\hspace{-70pt} = - \Tr \bigl[ \sqrt{(D-\pi_{\Abold_\Bbold}^2) \Idbb_{(-\infty, D]}(\pi_{\Abold_\Bbold}^2)} \; \gamma \; \sqrt{(D-\pi_{\Abold_\Bbold}^2) \Idbb_{(-\infty, D]}(\pi_{\Abold_\Bbold}^2)} \, \bigr] \\
&\hspace{-70pt} \geq \Tr \bigl[ (\pi_{\Abold_\Bbold}^2-D) \Idbb_{(-\infty, D]}(\pi_{\Abold_\Bbold}^2) \bigr].
\end{align*}
This proves the claim.

\subsection{Normal state}


As in the case of the translation invariant functional, the first question that needs to be answered is the question of the normal state and its BCS energy, i.e., the minimizer of $\FBCS$ in the absense of interactions, $V =0$. The normal state is defined as the Fermi--Dirac distribution with external fields, that is,
\begin{align}
\Gamma_0 &:= \begin{pmatrix} \gamma_0  \\ & 1 - \ov \gamma_0 \end{pmatrix} , & \gamma_0 &:= \frac{1}{1 + \e^{\beta (\hfrak_{\Abold, W} - \mu)}}, & \hfrak_{\Abold, W} &:= (-\i \nabla + \Abold_h)^2 + W_h. \label{Intro:Normal_state}
\end{align}
A short calculation shows that its BCS energy is given by
\begin{align}
\FBCS(\Gamma_0) &= T \, \Tr \bigl[ \ln\bigl( 1 + \e^{-\beta ((-\i \nabla + \Abold_h^2) + W_h - \mu)}\bigr)\bigr]
\end{align}
and the trace edition of Klein's inequality \eqref{Intro:Klein}, applied to $\varphi$ as defined in \eqref{Intro:varphi}, helps us to prove that
\begin{align}
\FBCS(\Gamma) &\geq \FBCS(\Gamma_0) \label{Intro:Minimization_normal_state}
\end{align}
with equality if and only if $\Gamma = \Gamma_0$. This proves that $\Gamma_0$ is the unique minimizer of $\FBCS$ in the absence of interactions.

To see that \eqref{Intro:Minimization_normal_state} is true, we write the BCS functional as
\begin{align}
\FBCS(\Gamma) - \FBCS(\Gamma_0) &= \frac{T}{2} \Tr \bigl[ \beta H_0 (\Gamma - \Gamma_0) + \varphi(\Gamma) + \varphi(\Gamma_0) \bigr], \label{Intro:BCS_equals_relative_entropy}
\end{align}
where $\varphi$ is the function in \eqref{Intro:varphi} and
\begin{align*}
H_0 &:= \begin{pmatrix} \hfrak_{\Abold, W} \\ & -\ov{\hfrak_{\Abold, W}} \end{pmatrix}.
\end{align*}
We further make use of the identity
\begin{align}
\beta H_0 &= \ln(1- \Gamma_0) - \ln(\Gamma_0) = \varphi'(\Gamma_0).
\end{align}
From this, the claim is provided by Klein's inequality applied to $A = \Gamma$ and $B = \Gamma_0$.

Klein's inequality falls into the category of inequalities that we also refer to as \emph{relative entropy inequalities}. This is due to the fact that it implies the relative entropy of the state $\Gamma$ and a state $\Gamma_0$ of the form $\Gamma_0 = \frac{1}{1 + \e^{\beta H}}$ for a Hamiltonian $H$ given by
\begin{align}
\Hcal(\Gamma, \Gamma_0) &:= \Tr \bigl[ \Gamma (\ln(\Gamma) - \ln(\Gamma_0)) + (1 - \Gamma) (\ln(1 - \Gamma) - \ln(1 - \Gamma_0))\bigr] \label{Intro:Relative_entropy}
\end{align}
to be nonnegative and equal to zero if and only if $\Gamma = \Gamma_0$. This follows from \eqref{Intro:BCS_equals_relative_entropy} since (if $V =0$)
\begin{align*}
\FBCS(\Gamma) - \FBCS(\Gamma_0) = \Hcal(\Gamma, \Gamma_0).
\end{align*}

\subsection{Superconductivity and minimization problem}

When interactions are present, i.e., $V\not \equiv 0$, we are interested in the question whether the normal state is \emph{stable} or \emph{instable} in the sense that the minimization problem for the \emph{BCS energy}
\begin{align}
F^{\mathrm{BCS}}(h, T) := \inf \bigl\{ \FBCS(\Gamma) - \FBCS(\Gamma_0) : \Gamma \; \text{admissible BCS state} \bigr\} \label{Intro:BCS-energy}
\end{align}
has a minimizer which is different from the normal state $\Gamma_0$ in \eqref{Intro:Normal_state} or not. We call the system \emph{superconducting} if this is the case, i.e., if
\begin{align}
F^{\mathrm{BCS}} (h, T) < 0.
\end{align}
Otherwise, the system is said to be in the normal state. Note that a state $\Gamma$, which lowers the BCS energy below that of the normal state necessarily has $\alpha \not \equiv 0$ because $\Gamma_0$ is the unique minimizer among diagonal states. This is proven with the help of Klein's inequality by a similar argument to the fact that $\Gamma_0$ is the unique minimizer in the absence of interactions.

\subsection{The critical temperatures}

We next define two critical temperatures, namely the \emph{upper critical temperature}
\begin{align}
\ov{\Tc(h)} &:= \inf\bigl\{ T > 0 : \Fcal_{h, T'}^{\mathrm{BCS}}(\Gamma) > \Fcal_{h, T'}^{\mathrm{BCS}}(\Gamma_0) \text{ for all admissible } \Gamma \neq \Gamma_0 \text{ and } T' \geq T \bigr\} \label{Intro:Tc_upper_definition}
\end{align}
and the \emph{lower critical temperature}
\begin{align}
\underline{\Tc(h)} &:= \sup \bigl\{ T >0 : F^{\mathrm{BCS}} (h, T) < 0   \text{ for all } 0 \leq T' < T\bigr\}. \label{Intro:Tc_lower_definition}
\end{align}

\begin{figure}[h]
\centering
\includegraphics[width = 12cm]{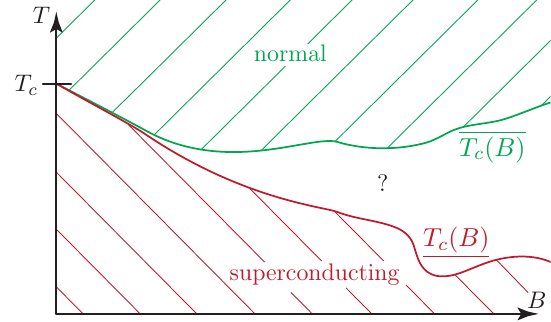}
\caption{Upper and lower critical temperature in the case of the constant magnetic field, where $B = h^2$. \label{Intro:Phase_Diagram1}}
\end{figure}

The upper critical temperature $\ov{\Tc(h)}$ is the lowest temperature above which the normal state is always stable, whereas the lower critical temperature $\underline{\Tc(h)}$ is the largest temperature below which the normal state is always instable. This situation is illustrated for the constant magnetic field strength $B = h^2$ in Figure \ref{Intro:Phase_Diagram1}, which is a hypothetic phase diagram for the BCS model.

We saw earlier in Section \ref{Intro:BCS-functional-ti_Section} that the BCS model has a unique critical temperature $\Tc$ in the absence of external fields. However, if fields are present then it might indeed be the case that the superconducting phase appears and vanishes alternatingly while we increase the temperature in between the two critical temperatures \eqref{Intro:Tc_upper_definition} and \eqref{Intro:Tc_lower_definition}. One mathematical reason for this is that the appropriate substitute for $K_T - V$ (the operator $L_{T, B}$, which will be introduced in \eqref{DHS1:LTB_definition} and analyzed thorougly in Chapter \ref{Chapter:DHS1}) is not monotone in $T$ anymore. In connection to this, recall that Meißner and Ochsenfeld already observed such a behavior, as we discussed in Section \ref{Intro:History_Section}. The question whether these temperatures coincide is not answered in this thesis. What we do investigate is how close these critical temperatures lie together and, via the definitions \eqref{Intro:Tc_upper_definition} and \eqref{Intro:Tc_lower_definition}, this is closely related to an analysis of the BCS functional at low energies.


\section{The Limiting Ginzburg--Landau Theory}

\subsection{Gauge-periodic Sobolev spaces upon the center of mass}

As a preparation, we need to define the function spaces which will accompany us throughout the thesis. The functions we are interested in are $\Lambda_\Bbold$-gauge-periodic with respect to the magnetic translations $\Tcal_\Bbold(\lambda) := T_{2\Bbold}(\lambda)$, $\lambda\in \Lambda_\Bbold$, i.e.,
\begin{align}
\Tcal_\Bbold(v) f(X) &:= \e^{\i \Bbold \cdot (v \wedge X)} f(X + v), & v &\in \Rbb^3.
\end{align}
Due to our choice of lattice, the family $\{ \Tcal_\Bbold(\lambda)\}_{\lambda\in \Lambda_\Bbold}$ is an abelian group of translations.

The reason for the factor of $2$ in comparison to the magnetic translations $T_\Bbold(\lambda)$ in \eqref{Intro:Magnetic_Translation} lies in the fact that the center of mass describes Cooper pairs which carry twice the charge of a single electron. We should note that the term representing the magnetic potential in the Hamiltonian is actually $q\Abold(X)$, where $q$ is the charge. If our units were made so that $q$ would be present, then we would have seen that $q = 2e$ for the center of mass.

For $1 \leq p \leq \infty$, we denote the $L^p$-space of $\Lambda_\Bbold$-periodic functions by
\begin{align}
\Lmag^p(Q_\Bbold) &:= \bigl\{ \Psi\in L_{\mathrm{loc}}^p(\Rbb^3) : \Tcal_\Bbold(\lambda) \Psi = \Psi , \, \lambda\in \Lambda_\Bbold \bigr\} \label{Intro:Lmag_p_definition}
\end{align}
and we equip it with the norm
\begin{align}
\Vert \Psi\Vert_p^p &:= \fint_{Q_\Bbold} \dd X \; |\Psi(X)|^p := \frac{1}{|Q_\Bbold|} \int_{Q_\Bbold} \dd X \; |\Psi(X)|^p \label{Intro:Lmag_p_norm_definition}
\end{align}
if $1\leq p < \infty$ and with the usual sup-norm if $p = \infty$. For $m\in \Nbb_0$, we further define the \emph{gauge-periodic Sobolev space} by
\begin{align}
\Hmag^m(Q_\Bbold) &:= \bigl\{ \Psi\in \Lmag^2(Q_\Bbold) : (-\i \nabla + 2\Abold_\Bbold)^\nu \in \Lmag^2(Q_\Bbold) , \; |\nu| \leq m \bigr\} \label{Intro:Hmagm_definition}
\end{align}
and we endow it with the scalar product
\begin{align}
\langle \Phi, \Psi\rangle_{\Hmag^m(Q_\Bbold)} &:= \sum_{|\nu|\leq m} h^{-2(1 + |\nu|)}\; \langle (-\i \nabla + 2\Abold_\Bbold)^\nu \Phi, (-\i \nabla + 2\Abold_\Bbold)^\nu \Psi\rangle_{\Lmag^2(Q_\Bbold)}. \label{Intro:Hmagm_norm_definition}
\end{align}
Note that if $\Psi$ is a gauge-periodic function, then $(-\i \nabla + 2\Abold) \Psi$ is gauge-periodic, too, since the magnetic momentum operator $\Pi_{\Abold_h}$ commutes with the magnetic translations $\Tcal_\Bbold(v)$, where
\begin{align}
\Pi_\Abold := -\i \nabla + 2\Abold. \label{Intro:Magnetic_momentum_COM}
\end{align}
This follows from a similar computation to \eqref{Intro:Magnetic_Momentum_Translation_Intertwining}. Moreover, the components of $\Pi_\Abold$ are self-adjoint in $\Hmag^1(Q_\Bbold)$.

\subsection{The Ginzburg--Landau scaling}

We now want to comment on the peculiar scaling that we chose for the scalar product on the space $\Hmag^m(Q_\Bbold)$, which we defined in \eqref{Intro:Hmagm_norm_definition}. To explain this, let a function $\psi\in \Lmag^2(Q_{e_\Bbold})$ be given and let us define
\begin{align}
\Psi(X) &:= h \, \psi(h X), & X &\in \Rbb^3. \label{Intro:GL-scaling}
\end{align}
We will see in a moment, when we define the Ginzburg--Landau functional, that it is invariant under this scaling, which is the reason for us to work with it. Moreover, we have
\begin{align}
\Vert \Psi\Vert_p &= h \; \Vert \psi\Vert_p, & 1 &\leq p \leq \infty, \label{Intro:Lmag_p_norm_scaling}
\end{align}
as can be verified by a short calculation. If $\psi \in \Hmag^m(Q_\Bbold)$, then $\Psi\in \Hmag^m(Q_\Bbold)$ and we have
\begin{align}
\Vert \Psi\Vert_{\Hmag^m(Q_\Bbold)} = \Vert \psi\Vert_{\Hmag^m(Q_{e_\Bbold})}. \label{Intro:Hmagm_norm_scaling}
\end{align}

This motivates the following suggestive notation that we use throughout the thesis: $\psi$ is a gauge-periodic function on the unscaled box $Q_{e_\Bbold}$ (the macroscopic ``outside-world'' perspective of the sample), whereas $\Psi$ is gauge-periodic on the large box $Q_\Bbold$ (macroscopic box from the microscopic perspective). The advantage of the incorporation of the scaling factors in the norm is that all norms can be thought of as ``order one'' with respect to $h$. Therefore, we can phrase statements in terms of $\Psi$ without saying anything about its scaling properties. However, as soon as $\Psi$ arises as a scaled version of $\psi$, we know that the norm does not contain any ``hidden'' factors of $h$. Furthermore, we never see any factors of $h$ that appear for dimensional reasons, which is an advantage since these often are a source of confusion.

\subsection{The Ginzburg--Landau functional}

When we investigate the BCS functional in the weak magnetic field limit, we need a limiting theory that describes it. Per definition, this theory is independent of $h$. 

Our limiting theory is given by an energy functional, the Ginzburg--Landau functional, too. It determines the subleading behavior of BCS theory in the weak external field limit. As explained earlier, this macroscopic theory of superconductivity is older than BCS theory. The endeavor of deriving GL theory from BCS theory has been initiated by Gor'kov \cite{Gorkov}, which then has been continued in the works \cite{Hainzl2012}, \cite{Hainzl2014}, among others, and in this thesis. Ginzburg--Landau theory is defined in terms of a single order parameter, a gauge-periodic wave function $\psi$ with respect to the lattice $\Lambda_{e_\Bbold}$. This wave function belongs to the magnetic Sobolev space $\Hmag^1(Q_{e_\Bbold})$. For positive coefficients $\Lambda_0, \Lambda_2, \Lambda_3 >0$, a real coefficient $\Lambda_1 \in \Rbb$, and a real parameter $D\in \Rbb$ the (sometimes called \emph{microscopically derived}) \emph{Ginzburg--Landau functional} is defined as
\begin{align}
\Ecal_D^{\mathrm{GL}}(\psi) &:= \int_{Q_{e_\Bbold}} \dd X\; \bigl\{ \Lambda_0 \, |(-\i \nabla + 2\Abold) \psi(X)|^2 + \Lambda_1 \, W(X) \, |\psi(X)|^2 \notag \\
&\hspace{170pt} - D \, \Lambda_2 \, |\psi(X)|^2 + \Lambda_3 \, |\psi(X)|^4 \bigr\}. \label{Intro:GL-functional_1box_definition}
\end{align}
We can define the so-called \emph{Ginzburg--Landau energy} for this functional which is simply given by
\begin{align}
\EGLGSE &:= \inf\bigl\{ \Ecal_D^{\mathrm{GL}}(\psi) : \psi\in \Hmag^1(Q_{e_\Bbold})\bigr\}. \label{Intro:GL-energy_1box_definition}
\end{align}

Let us mention that Ginzburg--Landau theory has a phase transition which can be demonstrated quite easily. Namely, depending on the parameter $D$ the functional given above can have a trivial minimizer $\psi \equiv 0$, in which case the energy is zero, or the energy is strictly negative with a nontrivial minimizer. The critical parameter for this transition is 
\begin{align}
\Dc &:= \frac{1}{\Lambda_2} \inf \spec\bigl\{ \Lambda_0\, (-\i \nabla + \Abold)^2 + \Lambda_1\, W \bigr\}. \label{Intro:Dc_definition}
\end{align}

To see that this holds, let us assume first that $D \leq \Dc$. By \eqref{Intro:Dc_definition}, this implies that the operator $\Lcal_D := \Lambda_0 (-\i \nabla + \Abold)^2 + \Lambda_1 W - D$ is nonnegative. Therefore, we may drop the positive quartic term for a lower bound and obtain
\begin{align*}
\Ecal_D^{\mathrm{GL}}(\psi) &\geq \Lambda_2 \, \langle \psi, \Lcal_D \psi\rangle \geq 0.
\end{align*}
This proves that $E^{\mathrm{GL}}(D) \geq 0$ and the test-function $\psi\equiv 0$ proves $E^{\mathrm{GL}}(D) \leq 0$. If, on the other hand, $D > \Dc$, then let $\psi$ be a ground state of the problem \eqref{Intro:Dc_definition}. Here, we have to use the fact that the quartic term vanishes faster than the quadratic part of the functional as $|\psi|$ becomes small. Therefore, for $\theta \in \Rbb$, we compute
\begin{align}
\Ecal_D^{\mathrm{GL}}(\theta \psi) &= \Lambda_2 \, (\Dc - D) \, \theta^2 \, \Vert \psi\Vert_2^2 + \Lambda_3\, \theta^4 \, \Vert \psi\Vert_4^4. \label{Intro:GL-energy_negative_1}
\end{align}
We minimize this function over $\theta$. The critical point is
\begin{align*}
\theta_{\mathrm{c}} := \frac{\Lambda_2 \, (D - \Dc) \, \Vert \psi\Vert_2^2}{2\, \Lambda_3 \, \Vert \psi\Vert_4^4}
\end{align*}
and its value is
\begin{align}
\Ecal_D^{\mathrm{GL}}(\theta_{\mathrm c}\psi) &= - \frac{\Lambda_2^2 \, (D - \Dc)^2 \, \Vert \psi\Vert_2^4}{4 \, \Lambda_3\, \Vert \psi\Vert_4^4}, \label{Intro:GL-energy_negative_2}
\end{align}
which is clearly negative. This proves that the Ginzburg--Landau functional has a phase transition at $D = \Dc$.

As Ginzburg--Landau theory is our limiting theory, it is $h$-independent. However, it is sometimes more convenient to have a scaling invariant version of the Ginzburg--Landau functional defined on the box $Q_\Bbold$ (in the microscopic perspective). Hence, for $\Psi\in \Hmag^1(Q_\Bbold)$, we set
\begin{align}
\EGL(\Psi) &:= \frac{1}{h^4} \fint_{Q_\Bbold} \dd X \; \bigl\{ \Lambda_0 \, |(-\i \nabla + 2 \Abold) \Psi(X)|^2 + \Lambda_1 \, W_h(X) \, |\Psi(X)|^2 \notag \\
&\hspace{150pt} - Dh^2 \, \Lambda_2 \, |\Psi(X)|^2 + \Lambda_3 \, |\Psi(X)|^4 \bigr\}. \label{Intro:GL-functional_definition}
\end{align}
As promised, if $\psi$ and $\Psi$ are related through \eqref{Intro:GL-scaling}, then a short calculation shows that
\begin{align}
\EGL(\Psi) = \Ecal_D^{\mathrm{GL}}(\psi),
\end{align}
whence
\begin{align}
\EGLGSE = \inf\bigl\{ \EGL(\Psi) : \Psi\in \Hmag^1(Q_\Bbold) \bigr\}. \label{Intro:GL-energy_definition}
\end{align}

\subsection{Ginzburg--Landau with magnetic field term}

We also note that the Ginzburg--Landau functional usually comes with an additional term which lets it describe the Meißner effect. In terms of an external magnetic field $H_{\mathrm{ext}}$, this functional then reads
\begin{align*}
\tilde \Ecal_D^{\mathrm{GL}} (\psi, \Abold) &:= \int_{Q_{e_\Bbold}} \dd X\; \bigl\{ \Lambda_0 \, |(-\i \nabla + 2\Abold) \psi(X)|^2 + \Lambda_1 \, W(X) \, |\psi(X)|^2 \notag \\
&\hspace{170pt} - D \, \Lambda_2 \, |\psi(X)|^2 + \Lambda_3 \, |\psi(X)|^4 \bigr\} \notag \\
&\hspace{120pt} + \int_{Q_{e_\Bbold}}  \dd X \; \bigl| \curl \Abold(X) - H_{\mathrm{ext}}(X)\bigr|^2.
\end{align*}
This functional has been investigated in the literature in great detail and it describes additional features of the superconductor like the penetration depth. In our case however, since we do not have the corresponding term in the BCS functional, the functional we obtain as a limiting functional is the one given in \eqref{Intro:GL-functional_definition}.


\section{Main Results of this Thesis}

\subsection{The BCS energy}

The results that are presented in this work address two questions. The first question is the behavior of the BCS energy $F^{\mathrm{BCS}}(h, T)$ in \eqref{Intro:BCS-energy} for small $h$ and here we want to derive a formula that describes the energy in the weak magnetic field regime. In practice, this formula will be an asymptotic expansion in powers of $h$ as $h \to 0$. Here, we want to determine the coefficients. The first coefficient we expect is already incorporated in the definition of the BCS energy, namely the energy of the normal state.

Of course this is the external-field edition of the normal state energy since we have defined it so. However, it will not surprise the reader that the energy of the normal state with external fields converges to the energy of the free normal state. The proof of this fact is not contained in this thesis but may be given with the help of Chapter \ref{Chapter:Spectrum_Landau_Hamiltonian_Section}, which allows for an evaluation of traces of a large class of functions of the Landau Hamiltonian.

The second order requires a precise temperature scaling in which we address the problem, namely
\begin{align}
T = \Tc (1 - Dh^2) \label{Intro:BCS-energy_temperature_scaling}
\end{align}
for some constant $D\in \Rbb$. In terms of the Ginzburg--Landau energy that we introduced above, we will show that the BCS energy in \eqref{Intro:BCS-energy} has the asymptotic expansion
\begin{align}
F^{\mathrm{BCS}}(h, \Tc(1 - Dh^2)) &= h^4 \, \bigl( \EGLGSE + o(1)\bigr) , & h &\to 0. \label{Intro:BCS-energy_expansion}
\end{align}
Here, the coefficients $\Lambda_i$, $i = 0, \ldots, 3$ for the Ginzburg--Landau functional are determined by the translation-invariant BCS theory. More precisely, in terms of the functions
\begin{align}
g_1(x) &:= \frac{\tanh(x/2)}{x^2} - \frac{1}{2x}\frac{1}{\cosh^2(x/2)}, & g_2(x) &:= \frac 1{2x} \frac{\tanh(x/2)}{\cosh^2(x/2)}, \label{Intro:GL-coefficients_aux-functions}
\end{align}
we have
\begin{align}
\Lambda_0 &:= \frac{\betac^2}{16} \int_{\Rbb^3} \frac{\dd p}{(2\pi)^3} \; |(-2)\hat{V\alpha_*}(p)|^2 \; \bigl( g_1 (\betac(p^2-\mu)) + \frac 23 \betac \, p^2\, g_2(\betac(p^2-\mu))\bigr), \label{Intro:GL-coefficient_1}\\
\Lambda_1 &:= \frac{\betac^2}{4} \int_{\Rbb^3} \frac{\dd p}{(2\pi)^3} \; |(-2)\hat{V\alpha_*}(p)|^2 \; g_1(\betac (p^2-\mu)), \label{Intro:GL-coefficient_W} \\
\Lambda_2 &:= \frac{\betac}{8} \int_{\Rbb^3} \frac{\dd p}{(2\pi)^3} \; \frac{|(-2)\hat{V\alpha_*}(p)|^2}{\cosh^2(\frac{\betac}{2}(p^2 -\mu))},\label{Intro:GL-coefficient_2} \\
\Lambda_3 &:= \frac{\betac^2}{16} \int_{\Rbb^3} \frac{\dd p}{(2\pi)^3} \; |(-2) \hat{V\alpha_*}(p)|^4 \;  \frac{g_1(\betac(p^2-\mu))}{p^2-\mu}.\label{Intro:GL_coefficient_3}
\end{align}
Here, $\alpha_*$ is the unique normalized ground state of $K_\Tc - V$ and we use the Fourier transform of the ``gap function'' $V\alpha_*\in L^2(\Rbb^3)$, which is defined by
\begin{align}
\hat{V\alpha_*}(p) := \int_{\Rbb^3} \dx\; \e^{-\i p\cdot x} \, V(x)\alpha_*(x). \label{Intro:Gap_function}
\end{align}
Note that we choose a non-unitary Fourier transform here.

I announced earlier that the BCS gap equation will not play a major role in this thesis. This is wrong insofar as the eigenvalue equation $K_\Tc\alpha_* = V\alpha_*$ can be viewed as one manifestation of this gap equation and especially physicists consider the so-called \emph{gap function} $-2V\alpha_*$ as the ``solution'' to the gap equation. Here, the minus sign stands for the negative charge of the particles and the factor of 2 again models the Cooper pair charge. The function $V\alpha_*$ is then the microscopic wave function which models the details inside the Cooper pair.

The coefficients $\Lambda_2$ and $\Lambda_3$ are easily seen to be positive just because of the signs of the functions $g_1(x)/x$ and $g_2$, whereas $\Lambda_1$ can have either sign and the sign depends on the sign of the derivative $\Tc$ with respect to $\mu$, see the corresponding remark below \cite[Eq. (1.21)]{Hainzl2012}. The kinetic coefficient $\Lambda_0$ is also positive but the argument is somewhat more complicated. It is given in \eqref{DHS1:GL-coefficient_1_positive} and involves calculations with the commutator of $K_\Tc$. In the work \cite{Hainzl2012}, this coefficient is actually a positive definite matrix, since the authors were able to dispense with the radiality of $V$ in the context of the fluxless model.

The result \eqref{Intro:BCS-energy_expansion} should be read in the following way. If it was not for the error $o(1)$, then we would have a criterion for superconductivity the BCS model with external fields. Indeed, in this case, the phase transition would be inherited from the Ginzburg--Landau functional, i.e., if the temperature approaches $\Tc$ in a linear fashion with a slope $D$ that is above $\Dc$, $D\geq \Dc$, then the BCS model would be superconducting, otherwise it would be in the normal state. Since we do have the error in \eqref{Intro:BCS-energy_expansion}, this interpretation is not quite possible and we need an additional result on the phase diagram of the BCS model.

It is noteworthy, however, that the first term that appears on the right side is of the order $h^4$, whereas, as we explained above, the changes in the BCS functional caused by the external fields are of the order $h^2$. The reason for this is difficult to explain without the knowledge of the proof. In some sense, however, it lies in the fact that the temperature regime we consider ranges about the critical temperature of the translation invariant model and the state which minimizes \eqref{Intro:BCS-energy} is given by the gap function $V\alpha_* = K_\Tc \alpha_*$ of that model. Therefore, the contribution that appears on the order $h^2$ cancels out.

\subsection{The critical temperature and phase diagram}

The second question we want to address is the shape of the phase diagram of BCS theory in the weak magnetic field regime, $0 < h \ll 1$. More precisely, we want to argue for small magnetic fields that, up to small errors, the upper and lower critical temperature $\ov{\Tc(h)}$ and $\underline{\Tc(h)}$ in \eqref{Intro:Tc_upper_definition} and \eqref{Intro:Tc_lower_definition} actually coincide, namely
\begin{align}
\Dc - o(1) \leq \frac{\underline{\Tc(h)} - \Tc}{\Tc h^2} \leq \frac{\ov{\Tc(h)} - \Tc}{\Tc h^2} &\leq \Dc + o(1) , & h &\to 0. \label{Intro:Tch_expansion}
\end{align}
This means that they are confined in a small cone around a linear asymptotic expansion. The linear expansion is given by
\begin{align}
\Tc(h) &:= \Tc (1 - \Dc h^2), \label{Intro:Tch_definition}
\end{align}
where $\Dc$ is the critical parameter of the Ginzburg--Landau functional, defined in \eqref{Intro:Dc_definition}. For the case of a constant magnetic field, this is illustrated in the following Figure \ref{Intro:Phase_Diagram2}.
\begin{figure}[h]
\centering
\includegraphics[width=12cm]{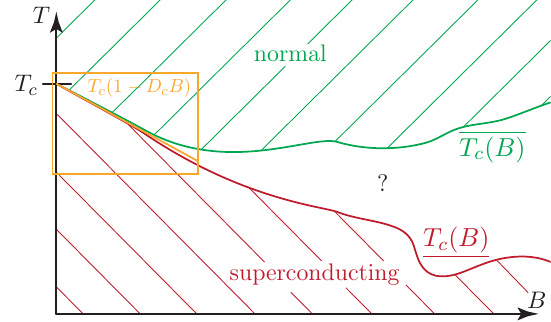}
\caption{Phase diagram of the BCS model for an external field only consisting of the constant magnetic field with strength $B = |\Bbold|$. \label{Intro:Phase_Diagram2}}
\end{figure}
As expected, the first order of the expansion \eqref{Intro:Tch_expansion} is the critical temperature $\Tc$ of the model without external fields. As we see, the next order in the expansion is the linear term given by the slope $\Dc$. This confirms the heuristics that we sketched in the previous section, namely that the phase transition of the BCS model is located at the temperature regime \eqref{Intro:Tch_definition}, i.e., it is inherited from the Ginzburg--Landau model, although, strictly speaking, we have two phase transitions that we cannot exclude to be distinct.

Figure \ref{Intro:Phase_Diagram2} suggests that the critical temperature of the model decreases with increasing strength of the external fields. This is indeed true for a purely magnetic field. When the electric field $W$ is present, however, the critical temperature might actually \emph{in}crease, depending on the lowest eigenvalue of the problem \eqref{Intro:Dc_definition}. 

The important point here is that the critical temperature of the BCS model only depends on the \emph{linear part} of Ginzburg--Landau theory. The quartic term only plays a role in the energy expansion but is not present here.

The attentive reader will have noticed that our results Theorem \ref{DHS1:Main_Result_Tc} and \ref{Main_Result_Tc} do not show \eqref{Intro:Tch_expansion}. The reason for this is that we are not able to deal with a small temperature regime close to absolute zero because our expansions deteriorate as the temperature approaches zero. I suspect that these problems are of technical nature and can be fixed in the future.

\subsection{The structure of low-energy states}

In order to understand the BCS model up to the second order, we need to understand the structure of superconducting states. Consequently, the results \eqref{Intro:BCS-energy_expansion} and \eqref{Intro:Tch_expansion} are based on the following structural result that we are going to prove. 

We assume that the temperature is not too far below the critical temperature $\Tc$, i.e.
\begin{align}
T- \Tc \geq D_0h^2 \label{Intro:Structure_Assumption1}
\end{align}
for some constant $D_0>0$. Furthermore, we assume that $\Gamma$ is an \emph{almost minimizer} (or, \emph{approximate minimizer}) of the BCS model, that is,
\begin{align}
\FBCS(\Gamma) - \FBCS(\Gamma_0) \leq 0. \label{Intro:Structure_Assumption2}
\end{align}
We can actually allow for an energy up to $D_1 h^4$ for some $D_1 \geq 0$, which is the order of the Ginzburg--Landau functional but for the sake of simplicity, we assume that $D_1 =0$ for now. 

If \eqref{Intro:Structure_Assumption1} and \eqref{Intro:Structure_Assumption2} are true, then for $h >0$ small enough there are $\Psi \in \Hmag^1(Q_\Bbold)$ and $\xi \in H^1(Q_\Bbold \times \Rbb_{\mathrm{s}}^3)$ such that the Cooper pair wave function $\alpha = \Gamma_{12}$ satisfies the decomposition
\begin{align}
\alpha(X, r) = \Psi(X)\alpha_*(r) + \xi(X, r), \label{Intro:Structure_Decomposition}
\end{align}
where
\begin{align}
\Vert \Psi\Vert_{\Hmag^1(Q_\Bbold)}^2 &\leq C, & \Vert \xi\Vert_{H^1(Q_\Bbold \times \Rbb_{\mathrm{s}}^3)}^2 \leq Ch^4 \, \Vert \Psi\Vert_{\Hmag^1(Q_\Bbold)}^2, \label{Intro:Structure_Bound}
\end{align}
and 
\begin{align}
\EGL(\Psi) \leq \EGLGSE + o(1). \label{Intro:Structure_Bound_Psi}
\end{align}
 
The important message of \eqref{Intro:Structure_Decomposition} is that the superconducting behavior of the BCS model is decoupling in center of mass and relative coordinate to leading order. The \emph{microscopic} behavior is displayed by the \emph{translation invariant} BCS theory in the relative coordinate whereas the \emph{macroscopic} behavior is displayed by the wave function $\Psi$, which only depends on the center of mass coordinate.

Mathematically, this is expressed in the fact that the Cooper pair wave function admits a product structure to leading order, where the center of mass part $\Psi$ is a \emph{macroscopic quantity}, i.e. its $\Hmag^1(Q_\Bbold)$-norm is uniformly bounded in $h$. The interpretation is that $\Psi$ is flat throughout the sample, i.e., the Cooper pairs have a uniform distribution approximately. If $\Psi$ showed oscillations, the gradient would be large. On top of that, since $\Vert \alpha_*\Vert_2 =1$, the modulus $|\Psi|^2$ of this function $\Psi$ has the interpretation of the density of Cooper pairs in the system and \eqref{Intro:Structure_Bound_Psi} shows that it is an almost minimizer of the Ginzburg--Landau model. 

Furthermore, the relative coordinate is occupied by the ground state wave function of the operator $K_\Tc - V$, which represents the translation invariant BCS theory. We should note that the $L^2(Q_\Bbold \times \Rbb_{\mathrm{s}}^3)$-norm of the leading term equals
\begin{align}
\fint_{Q_\Bbold} \dd X \int_{\Rbb^3} \dd r \; \bigl| \Psi(X) \alpha_*(r)\bigr|^2 = h^2 \, \Vert \Psi\Vert_{\Hmag^0(Q_\Bbold)}^2.
\end{align}
Therefore, it is much larger than the $H^1(Q_\Bbold \times \Rbb_{\mathrm{s}}^3)$-norm of the remainder function $\xi$.

\begin{figure}[h]
\centering
\includegraphics[width=12cm]{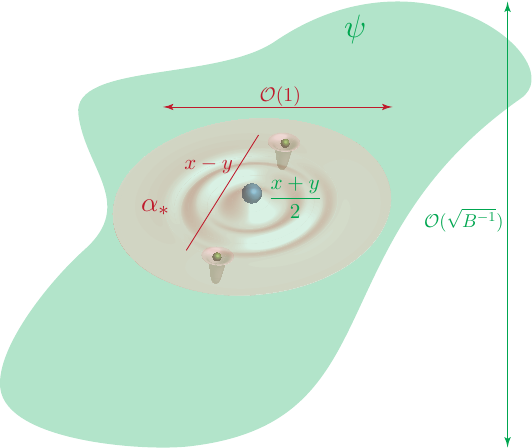}
\caption{Separation of scales \label{Intro:Separation_of_scales}}
\end{figure}

All this can be seen as a separation of scales, which is displayed in Figure \ref{Intro:Separation_of_scales}. Again, this picture is for the constant magnetic field only.

\subsection{On the proof}

With this remark, we conclude our introduction into the mathematics of BCS theory of superconductivity. The proofs of the results that appeared in this introduction are provided in the papers that are presented in Chapters \ref{Chapter:DHS1} and \ref{Chapter:DHS2}, as well as the additional Chapters~\ref{Chapter:Abrikosov_gauge}-\ref{Chapter:Combes-Thomas}. With the content of this introduction at hand, the reader should be well prepared for the reading of this content. We close this introduction by outlining the prospected future of this project.

\section{State of the Project and Outlook}

As we have discussed in detail above, the results of this thesis enable us to derive Ginzburg--Landau theory in the weak magnetic field limit for any periodic magnetic field of sufficient regularity that is applied to the BCS model. In this outlook, we want to sketch a few possibilities for the project to be continued in the future. In this section, we assume the reader to be familiar with the content and techniques used in Chapters \ref{Chapter:DHS1} and \ref{Chapter:DHS2}, as opposed to the previous sections. 

\subsection{The degenerate case}
\label{Intro:Degeneracy_Section}

The situation we describe in this thesis is s-wave superconductivity, that is, the ground state $\alpha_*$ of the operator $K_T - V$ is unique. Since the interaction potential $V$ is assumed to be radial, it follows that $\alpha_*$ is radial, too. The reason for this lies in the fact that $K_T$ is a radial symbol in Fourier space, making $K_T - V$ compatible with the angular momentum decomposition of $L^2(\Rbb^3)$, which reads
\begin{align}
L^2(\Rbb^3) &= \bigoplus_{\ell = 0}^\infty \Hfrak_\ell,
\end{align}
where
\begin{align}
\Hfrak_\ell := L^2(\Rbb_+, r^2\dd r) \otimes \spano\bigl\{ Y_m^\ell : m = -\ell, \ldots, \ell\bigr\}
\end{align}
and $Y_m^\ell$ are the spherical harmonics. If the ground state $\alpha_*$ is simple, then this implies that it lies in the angular momentum sector $\ell =0$ and is thus a radial function.

It would be interesting to generalize the results of this thesis to the case of $K_T - V$ having a degenerate ground state --- let $n := \dim(K_\Tc -V)$ denote the degeneracy of the ground state. In this case we expect that the gap function $\Delta(X, r)$ is of the form
\begin{align}
\Delta(X, r) &:= -2 \, V\alpha_*(r) \cdot \Psi(X) := -2 \sum_{i=1}^n V\alpha_*^{(i)}(r) \, \Psi_i(X),
\end{align}
where
\begin{align}
\alpha_* := (\alpha_*^{(1)}, \ldots, \alpha_*^{(n)})
\end{align}
is an orthonormal basis of $\ker(K_\Tc -V)$ and
\begin{align}
\Psi := (\Psi_1, \ldots, \Psi_n).
\end{align}
is a vector of gauge-periodic functions. Further literature on this is provided in the works \cite{FraLe2016} and \cite{DeuchertGeisinger}.

In the degenerate case, we have to change the point of view on Ginzburg--Landau theory a bit, since the gap function does not decouple into $\alpha_*(r)$ and $\Psi(X)$ so well anymore. Hence, the expected Ginzburg--Landau functional reads as follows.

\subsubsection{The Ginzburg--Landau functional in the degenerate case}

Let $\mathbf E \colon \Rbb^3 \ra \Rbb^{3\times 3}$ and $E_W, E_2, E_4\colon \Rbb^3 \ra \Rbb$ be bounded continuous functions where $E_2(p) , E_4(p) >0$ for all $p\in \Rbb^3$. For $Z\in \Cbb^n$ and $\Zcal \in \Cbb^n \otimes \Cbb^3$, the coefficients for the Ginzburg--Landau functional are defined by the quadratic terms
\begin{align}
\Lambdabold^{\alpha_*} (\Zcal) &:= \int_{\Rbb^3} \dd p \; \ov{\bigl( (-2) \hat{V\alpha_*}(p)^t \, \Zcal \bigr)} \cdot \Ebold(p) \cdot \bigl( (-2) \hat{V\alpha_*}(p)^t \, \Zcal\bigr), \\
\Lambda_W^{\alpha_*}(Z) &:= \int_{\Rbb^3} \dd p \; |(-2) \hat{V\alpha_*}(p)^t \, Z|^2 \; E_W(p), \\
\Lambda_2^{\alpha_*}(Z) &:= \int_{\Rbb^3} \dd p \; |(-2) \hat{V\alpha_*}(p)^t \, Z|^2 \; E_2(p),
\intertext{as well as the quartic term}
\Lambda_4^{\alpha_*}(Z) &:= \int_{\Rbb^3} \dd p \; |(-2) \hat{V\alpha_*}(p)^t \, Z|^4 \; E_4(p).
\end{align}
Let $\Psi \colon \Rbb^3 \ra \Cbb^n$ be a gauge-periodic function. For $h>0$ and $D\in \Rbb$, the Ginzburg--Landau functional is defined by
\begin{align}
\EGL(\alpha_* ,  \Psi) &:= \frac{1}{h^4} \fint_{Q_h} \dd X \; \Bigl\{ \Lambdabold^{\alpha_*}\bigl( (-\i \nabla + 2\Abold_h) \Psi(X) \bigr) + W_h(X)\, \Lambda_W^{\alpha_*}\bigl( \Psi(X) \bigr) \notag \\
&\hspace{140pt} - Dh^2\, \Lambda_2^{\alpha_*} \bigl( \Psi(X) \bigr) + \Lambda_4^{\alpha_*} \bigl( \Psi(X) \bigr) \Bigr\}. \label{Intro:Definition_GL-functional_degenerate}
\end{align}
Furthermore, the Ginzburg--Landau energy is defined as
\begin{align*}
E^{\mathrm{GL}}(D) := \inf \bigl\{ \EGL(\alpha_* , \Psi) : \alpha_*\in \ker(K_\Tc -V)^n\, , \, \Psi\in \Hmag^1(Q_h)^n \bigr\}
\end{align*}
and the critical parameter of the Ginzburg--Landau functional is given as follows. We define the matrix $\Ebb_2(\alpha_*) \in \Cbb^{n \times n}$ by
\begin{align}
\Ebb_2(\alpha_*) &:= \int_{\Rbb^3} \dd p \; E_2(p) \; \ov{\hat{V\alpha_*}(p)} \, V\alpha_*(p)^t
\end{align}
and the $\Cbb^{n\times n}$ matrix-valued operator
\begin{align}
\Lcal(\alpha_*) &:= \int_{\Rbb^3} \dd p \; \hat{V\alpha_*}(p) \bigl[ (-\i \nabla + 2\Abold)^* \cdot \Ebold(p) \cdot (-\i \nabla + 2\Abold) + W \, E_W(p) \bigr] \hat{V\alpha_*}(p)^t.
\end{align}
With these, we may define the critical parameter
\begin{align}
\Dc &:= \sup \bigl\{ \Dc(\alpha_*) : \alpha_* \in \ker (K_\Tc - V)\bigr\} \label{Intro:Dc_Definition_degenerate}
\end{align}
where
\begin{align}
\Dc(\alpha_*) &:= \Ebb_2(\alpha_*)^{-1} \inf \spec_{\Lmag^2(Q_{e_\Bbold};\Cbb^n)} \Lcal(\alpha_*)
\end{align}
where the infimum is taken over gauge-periodic, square integrable functions. As an example, in the case $n=2$, it is easy to see that $\Ebb_2(\alpha_*)$ is indeed a positive definite matrix whenever its upper right entry $\Ebb^{12}(\alpha_*) \neq 0$. For general $n$, we leave it to the reader to provide conditions under which $\Ebb_2(\alpha_*)$ is positive definite. We shall assume this in the following.

The definition of $\Dc$ is motivated by the fact that we again have $\EGLGSE < 0$ if $D > \Dc$ and $\EGLGSE =0$ if $D \leq \Dc$. To see this, we fix $h=1$ and let first $D > \Dc$ so that $D > \Dc(\alpha_*)$ for any $\alpha_*\in \ker(K_\Tc -V)$. Let $\psi$ be a ground state of $\Lcal(\alpha_*)$, i.e., $\Lcal(\alpha_*)\psi = \Dc(\alpha_*) \psi$. This implies that
\begin{align*}
\EGL(\alpha_*, \psi) = (\Dc(\alpha_*) - D) \langle \psi, \Ebb_2(\alpha_*) \psi\rangle + \Ebb_4(\alpha_*, \psi),
\end{align*}
where
\begin{align*}
\Ebb_4(\alpha_*, \psi) &:= \fint_{Q_{e_\Bbold}} \dd X \; \Lambda_4^{\alpha_*} \bigl( \psi(X)\bigr).
\end{align*}
By a similar argument to the one in \eqref{Intro:GL-energy_negative_1}-\eqref{Intro:GL-energy_negative_2}, we then see that
\begin{align*}
\EGL(\alpha_* , \psi) &= - ( \Dc(\alpha_*) - D)^2 \; \frac{\langle \psi, \Ebb_2(\alpha_*) \psi\rangle^2}{4 \, \Ebb_4(\alpha_*, \psi)}.
\end{align*}
In particular, this implies that $\EGLGSE$ is negative. If, on the other hand, $D \leq \Dc$, then for any $\varepsilon >0$ let $\alpha_*\in \ker(K_\Tc - V)$ be such that $\Dc(\alpha_*) \geq \Dc - \varepsilon$. Thus, the operator
\begin{align*}
\Lcal_D(\alpha_*) := \Lcal(\alpha_*) - D \, \Ebb_2(\alpha_*)
\end{align*}
satisfies the bound
\begin{align*}
\Lcal_D(\alpha_*) \geq - \varepsilon \, \Ebb_2(\alpha_*)
\end{align*}
By omitting the nonnegative quartic term of the Ginzburg--Landau functional, we thus obtain the lower bound
\begin{align*}
\EGL(\alpha_*, \psi) \geq -\varepsilon \, \langle \psi, \Ebb_2(\alpha_*)\psi\rangle.
\end{align*}
Since this is true for all $\varepsilon >0$, it follows that $\EGLGSE \geq 0$. The inequality $\EGLGSE \leq 0$ follows by testing with $\psi \equiv 0$.

\subsubsection{The main obstacle}

The degeneracy in the ground state essentially amounts to an increased difficulty in the structural result on the low-energy states, namely Theorems \ref{DHS1:Structure_of_almost_minimizers} and \ref{Structure_of_almost_minimizers}. The point is that we have to prove a bound on the $\Lsymm$-norm of the leading term and its derivative
\begin{align*}
&\fint_{Q_\Bbold} \dd X \int_{\Rbb^3} \dd r \; |\alpha_*(r)^t \, \Psi(X)|^2, & &\fint_{Q_\Bbold} \dd X \int_{\Rbb^3} \dd r \; |\alpha_*(r)^t \, \Pi \Psi(X)|^2.
\end{align*}
This cannot be decoupled as an $\Hmag^1(Q_\Bbold)$-norm bound on $\Psi$ anymore. For this to be proven, we presumably have to put more assumptions on the structure of the ground state space $\ker(K_\Tc - V)$. It is also advisable to start with a fluxless BCS model so that the external fields have been removed when it comes to the delicate analysis. The main obstacle is to find a substitute for Lemma \ref{DHS1:AAstar_Positive}.

\subsection{Towards the treatment of the Meißner effect}

When we want to set up a BCS model for the response field of a superconductor to an external field $H_{\mathrm{ext}}$, we have to solve the following problems.

\subsubsection{Regularity issues on \texorpdfstring{$\Abold$}{A}}

The derivation of Ginzburg--Landau theory has to be carried out under sufficiently low regularity conditions, which are not satisfied in the model we consider in Chapters \ref{Chapter:DHS1} and \ref{Chapter:DHS2}. In order to do this, we need to introduce an energy cut-off, similar to the one we introduce for $\Psi$ in Corollary \ref{DHS1:Structure_of_almost_minimizers_corollary} in order to regularize it. Then, we need to perform the derivation with the regularized part and control the errors obtained in this way. This is a project which should be able to be solved with modest effort.

\subsubsection{Decoupling of lattice and flux}

Since the superconductor is located in a fixed spatial region (which for us is the unit cell of the lattice of periodicity) and the magnetic response field has a magnetic flux through the unit cell which might not be rational, we need to find a way to decouple the size of the unit cell and the constant magnetic field part that we choose in our gauge. This is necessary to obtain a fluxwise definition of the BCS functional. It is, however, not even clear that this is the way to go. There might be more elegant ways to define the model.

\subsubsection{Magnetic field estimates}

We have to find a way to provide sufficiently good a priori estimates for the response fields. In particular, we have to prove that the response field is of \emph{macroscopic} nature if $H_{\mathrm{ext}}$ is. This puts up a major challenge to the business, which we do not know how to solve at the present day.

\printbibliography[heading=bibliography, title=Bibliography of Chapter \ref{Chapter:Intro}]
\end{refsection}

\def\EGL{\mathcal E^{\mathrm{GL}}_{D, B}}
\def\FBCS{\mathcal F^{\mathrm{BCS}}_{
B, T
}}

\def\Lsymm{{L^2(Q_B \times \Rbb_{\mathrm s}^3)}}
\def\Hsymm{{H^1(Q_B \times \Rbb_{\mathrm s}^3)}}

\begin{refsection}

\chapter[Microscopic Derivation of Ginzburg--Landau Theory and the BCS Critical Temperature Shift in a Weak Homogeneous Magnetic Field][BCS-Theory in a Homogeneous Magnetic Field]{
Microscopic Derivation of Ginzburg--Landau Theory and the BCS Critical Temperature Shift in a Weak Homogeneous Magnetic Field
}
 \label{Chapter:DHS1} \label{CHAPTER:DHS1}

The content of this chapter has been published in \cite{DeHaSc2021} and is co-authored by Andreas Deuchert and Christian Hainzl.

\begin{abstract}
\noindent 

Starting from the Bardeen--Cooper--Schrieffer (BCS) free energy functional, we derive the Ginzburg--Landau functional
for the case of a weak homogeneous magnetic field. We also provide an asymptotic formula for the BCS critical temperature as a function of the magnetic field. This extends the previous works \cite{Hainzl2012, Hainzl2014} of Frank, Hainzl, Seiringer and Solovej to the case of external magnetic fields with non-vanishing magnetic flux through the unit cell. 
\end{abstract}


\section{Introduction and Main Results}


\subsection{Introduction}

In 1950 Ginzburg and Landau (GL) introduced a phenomenological theory of superconductivity that is based on a system of nonlinear partial differential equations for a complex-valued wave function (the order parameter) and an effective magnetic field \cite{GL}. Their theory is \textit{macroscopic} in nature and contains no reference to a \textit{microscopic} mechanism behind the phenomenon of superconductivity. The GL equations show a rich mathematical structure, which has been investigated in great detail, see, e.g., \cite{Sigal1, Sigal2, Serfaty, SandierSerfaty, Correggi3, Correggi2, Giacomelli1, Giacomelli2} and references therein. They also inspired interesting new concepts beyond the realm of their original application.

The first generally accepted \textit{microscopic} theory of superconductivity was discovered seven years later by Bardeen, Cooper and Schrieffer (BCS) in \cite{BCS}. In a major breakthrough they realized that a pairing mechanism between the conduction electrons (formation of Cooper pairs) causes the resistance in certain materials to drop down to absolute zero if their temperature is sufficiently low. This pairing phenomenon at low temperatures is induced by an effective attraction between the electrons mediated by phonons, that is, by the quantized lattice vibrations of the crystal formed by the ion cores. In recognition of this contribution BCS were awarded the Nobel prize in physics in 1972. 

In the physics literature BCS theory is often formulated in terms of the gap equation, which, in the absence of external fields, is a nonlinear integral equation for a complex-valued function called the gap function (the order parameter of BCS theory). The name of the equation is related to the fact that its solution allows to determine the spectral gap of an effective \textit{quadratic} Hamiltonian that is open only in the superconducting phase. BCS theory also has a variational interpretation, where the gap equation arises as the Euler--Lagrange equation of the BCS free energy functional. This free energy functional can be obtained from a full quantum mechanical description of the system by restricting attention to quasi-free states, a point of view that was emphasized by Leggett in \cite{Leg1980}, see also \cite{de_Gennes}. In this formulation, the system is described in terms of a one-particle density matrix and a Cooper pair wave function. 

Although it was originally introduced to describe the phase transition from the normal to the superconducting state in metals and alloys, BCS theory can also be applied to describe the phase transition to the superfluid state in cold fermionic gases. In this case, the usual non-local phonon-induced interaction in the gap equation needs to be replaced by a local pair potential. From a mathematical point of view, the gap equation has been studied for interaction kernels suitable to describe the physics of conduction electrons in solids in \cite{Odeh1964, BilFan1968, Vanse1985, Yang1991, McLYang2000, Yang2005}. We refer to \cite{Hainzl2007, Hainzl2007-2, Hainzl2008-2, Hainzl2008-4, FreiHaiSei2012, BraHaSei2014, FraLe2016, DeuchertGeisinger} for works that investigate the translation-invariant BCS functional with a local pair interaction. BCS theory in the presence of external fields has been studied in \cite{HaSei2011, BrHaSei2016, FraLemSei2017, A2017, CheSi2020}.

A relation between the \textit{macroscopic} GL theory and the \textit{microscopic} BCS theory was established by Gor'kov in 1959 \cite{Gorkov}. He showed that, close to the critical temperature, where the order parameters of both models are expected to be small, GL theory arises from BCS theory when the free energy is expanded in powers of the gap function. The first mathematically rigorous proof of this relation was given by Frank, Hainzl, Seiringer and Solovej in 2012 \cite{Hainzl2012}. They showed that in the presence of weak and \textit{macroscopic} external fields, the \textit{macroscopic} variations of the Cooper pair wave function of the system are correctly described by GL theory if the temperature is close to the critical temperature of the sample in an appropriate sense. The precise parameter regime is as follows: The external electric field $W$ and the vector potential $A$ of the external magnetic field are given by $h^2 W(x)$ and $h A(hx)$, respectively. Here $0 < h \ll 1$ denotes the ratio between the microscopic and the macroscopic length scale of the system. Such external fields change the energy by an amount of the order $h^2$ and it is therefore natural to consider temperatures $T = \Tc (1 - D h^2)$ with $D > 0$, where $\Tc$ denotes the critical temperature of the sample in the absence of external fields. Within this setup it has been shown in \cite{Hainzl2012} that the correction to the BCS free energy on the order $h^4$ is correctly described by GL theory. Moreover, the Cooper pair wave function of the system is, to leading order in $h$, given by
\begin{equation}
	\alpha(x,y) = h\, \alpha_*(x-y) \, \psi\left( \frac{h(x+y)}{2} \right).
	\label{DHS1:eq:intro}
\end{equation}
Here, $\psi$ denotes the order parameter of GL theory and $\alpha_*(x-y)$ is related to the Cooper pair wave function in the absence of external fields. 

External electric and magnetic fields may change the critical temperature of a superconductor and this shift is expected to be described by GL theory. A justification of this claim has been provided in \cite{Hainzl2014}. More precisely, it has been shown that, within the setup of \cite{Hainzl2012} described above, the critical temperature of the sample obeys the asymptotic expansion
\begin{equation}
	\Tc(h) = \Tc(1  - \Dc h^2) + o(h^2),
	\label{DHS1:eq:intro2}
\end{equation}
where the constant $\Dc$ can be computed using linearized GL theory. 

One crucial assumption in \cite{Hainzl2012} and \cite{Hainzl2014} is that the vector potential related to the external magnetic field is periodic. In this case the magnetic flux through the unit cell equals zero. An important step towards an extension of the results in \cite{Hainzl2014} to the case of a magnetic field with non-vanishing magnetic flux through the unit cell has been provided by Frank, Hainzl and Langmann in \cite{Hainzl2017}. In this article the authors consider the problem of computing the BCS critical temperature shift in the presence of a weak homogeneous magnetic field within linearized BCS theory. Heuristically, this approximation is justified by the fact that linearized GL theory is sufficient to predict the critical temperature shift, see the discussion in the previous paragraph. In the physics literature this approximation appears in \cite{Werthamer1966, WertHelHoh1966, La1990, La1991}, for instance.

The aim of the present article is to extend the results in \cite{Hainzl2012} and \cite{Hainzl2014} to a setting with an external magnetic field having non-zero flux through the unit cell. More precisely, we consider a large periodic sample of  fermionic particles subject to a weak homogeneous magnetic field $\Bbold \in \Rbb^3$. The temperatures $T$ is chosen such that $(\Tc - T)/\Tc = D |\Bbold|$ with $D \in \Rbb$. We show that the correction of the BCS free energy of the sample at the order $|\Bbold|^2$ is given by GL theory. Moreover, to leading order in $|\Bbold|$ the Cooper pair wave function of the system is given by \eqref{DHS1:eq:intro} with $h$ replaced by $|\Bbold|^{\nicefrac{1}{2}}$. We also show that the BCS critical temperature shift caused by the external magnetic field is given by \eqref{DHS1:eq:intro2} with $\Dc$ determined by linearized GL theory. Our analysis yields the same formula that was computed within the framework of linearized BCS theory in \cite{Hainzl2017}. This can be interpreted as a justification of the approximation to use linearized BCS theory to compute the BCS critical temperature shift. The main new ingredient of our proof are a priori bounds for certain low-energy states of the BCS functional that include the magnetic field.

\subsection{Gauge-periodic samples}
\label{DHS1:Magnetically_Periodic_Samples}

We consider a $3$-dimensional sample of fermionic particles described by BCS theory that is subject to an external magnetic field $\Bbold := Be_3$ with strength $B>0$, pointing in the $e_3$-direction. We choose the magnetic vector potential $\Abold(x) := \frac 12 \Bbold \wedge x$ so that $\curl \Abold = \Bbold$, where $\Bbold \wedge x\in \Rbb^3$ denotes the cross product of two vectors. The corresponding magnetic momentum operator $\pi := -\i \nabla + \Abold$ commutes with the magnetic translations $T(v)$, defined by
\begin{align}
T(v)f(x) &:= \e^{\i \frac {\Bbold} 2\cdot (v\wedge x)} f(x+v), & v &\in \Rbb^3.\label{DHS1:Magnetic_Translation}
\end{align}
The family $\{T(v)\}_{v\in \Rbb^3}$ obeys the relation $T(v + w) = \e^{\i \frac{\Bbold}{2} \cdot (v \wedge w)} \; T(v) T(w)$, that is, it is a unitary representation of the Heisenberg group. We assume that our system is periodic with respect to the Bravais lattice $\Lambda_B = \sqrt{2\pi B^{-1}} \, \Zbb^3$ with fundamental cell
\begin{align}
Q_B &:= \bigl[0, \sqrt{2\pi B^{-1}}\bigr]^3 \subseteq \Rbb^3. \label{DHS1:Fundamental_cell}
\end{align}
The magnetic flux through the unit cell $Q_B$ equals $\Bbold \cdot (b_1\wedge b_2) =2\pi$, where $b_i = \sqrt{2\pi B^{-1}} \, e_i$ are the basis vectors spanning $\Lambda_B$. This assures that the abelian subgroup $\{T(\lambda)\}_{\lambda\in \Lambda_B}$ is a unitary representation of the lattice group. 


\subsection{The BCS functional}
\label{DHS1:BCS_functional_Section}

In BCS theory a state is described by a generalized fermionic one-particle density matrix, that is, by a self-adjoint operator $\Gamma$ on $L^2(\mathbb{R}^3) \oplus L^2(\mathbb{R}^3)$ which satisfies $0\leq \Gamma\leq 1$ and is of the form
\begin{align}
\Gamma = \begin{pmatrix} \gamma & \alpha \\ \ov \alpha & 1 - \ov \gamma \end{pmatrix}. \label{DHS1:Gamma_introduction}
\end{align}
Here, $\ov \alpha = J \alpha J$ with the Riesz identification operator $J \colon L^2(\Rbb^3) \ra L^2(\Rbb^3)$, $f\mapsto \ov f$, realized by complex conjugation. The condition $\Gamma = \Gamma^*$ implies that the one-particle density matrix $\gamma$ is a self-adjoint operator. It also implies that the Cooper pair wave function $\alpha(x,y)$, the kernel of $\alpha$, is symmetric under the exchange of its coordinates. The symmetry of $\alpha$ is due to the fact that we exclude spin variables from our description and assume that Cooper pairs are in a spin singlet state. The condition $0\leq \Gamma \leq 1$ implies $0\leq \gamma \leq 1$ as well as that $\gamma$ and $\alpha$ are related through the operator inequality 
\begin{align}
\alpha \alpha^* \leq \gamma ( 1- \gamma). \label{DHS1:gamma_alpha_fermionic_relation}
\end{align}

A BCS state $\Gamma$ is called \emph{gauge-periodic} if $\mathbf T(\lambda) \, \Gamma \, \mathbf T(\lambda)^* = \Gamma$ holds for every $\lambda\in \Lambda_B$, with the magnetic translations $\mathbf T(\lambda)$ on $L^2(\Rbb^3)\oplus L^2(\Rbb^3)$ defined by
\begin{align*}
\mathbf T(v) &:= \begin{pmatrix}
T(v) & 0 \\ 0 & \ov{T(v)}\end{pmatrix}, & v &\in \Rbb^3.
\end{align*}
For $\gamma$ and $\alpha$, this implies $T(\lambda) \gamma T(\lambda)^* = \gamma$ and   $T(\lambda)\,\alpha \,\ov{T(\lambda)}^* = \alpha$ or, in terms of their kernels,
\begin{align}
\gamma(x, y) &= \e^{\i \frac \Bbold 2 \cdot (\lambda \wedge (x-y))} \; \gamma(x+\lambda,y+ \lambda), \notag\\
\alpha(x, y) &= \e^{\i \frac \Bbold 2 \cdot (\lambda \wedge (x+y))} \; \alpha(x+\lambda,y+ \lambda), & \lambda\in \Lambda_B. \label{DHS1:alpha_periodicity}
\end{align}

\begin{bem}
Since we are interested in the situation of a constant magnetic field it seems natural to consider magnetically translation-invariant BCS states, that is, states obeying $\mathbf T(v) \, \Gamma \, \mathbf T(v)^* = \Gamma$ for every $v \in \mathbb{R}^3$. However, in this case one obtains a trivial model because the Cooper pair wave function $\alpha$ of a magnetically translation-invariant state necessarily vanishes. To see this, we note that $\alpha$ satisfies $T(v)\,\alpha \,\ov{T(v)}^* = \alpha$ for all $v \in \mathbb{R}^3$. Using this and the relation $T(v+w)\,\alpha \,\ov{T(v+w)}^* =  \e^{ \i\Bbold \cdot (v \wedge w)} \, T(v) T(w) \, \alpha \, \ov{T(w)}^* \, \ov{T(v)}^* $, we conclude that $\alpha = 0$.
\end{bem}

A gauge-periodic BCS state $\Gamma$ is said to be \emph{admissible} if 
\begin{align}
\Tr \bigl[\gamma + (-\i \nabla + \Abold)^2\gamma\bigr] < \infty \label{DHS1:Gamma_admissible}
\end{align}
holds. Here $\Tr[A]$ denotes the trace per unit volume of $A$, i.e.,
\begin{align}
\Tr [A] &:= \frac{1}{|Q_B|} \Tr_{L^2(Q_B)} [\chi A \chi],  \label{DHS1:Trace_per_unit_volume_definition}
\end{align}
with the characteristic function $\chi$ of the cube $Q_B$ in \eqref{DHS1:Fundamental_cell}. By $\Tr_{L^2(Q_B)}[A]$ we denote the usual trace of an operator $A$ on $L^2(Q_B)$. The condition in \eqref{DHS1:Gamma_admissible} is meant to say that $\gamma$ and $(-\i \nabla + \Abold)^2\gamma$ are locally trace class, that is, they are trace class with respect to the trace in \eqref{DHS1:Trace_per_unit_volume_definition}. Eq.~\eqref{DHS1:Gamma_admissible}, the same inequality with $\gamma$ replaced by $\overline{\gamma}$, and the inequality in \eqref{DHS1:gamma_alpha_fermionic_relation} imply that $\alpha$, $(-\i \nabla + \Abold)\alpha$, and $(-\i \nabla + \Abold) \ov \alpha$ are locally Hilbert--Schmidt. In Section~\ref{DHS1:Preliminaries} below we will express this property in terms of $H^1$-regularity of the kernel of $\alpha$.

For any admissible BCS state $\Gamma$, we define the Bardeen--Cooper--Schrieffer free energy functional (in the following: BCS functional) at temperature $T\geq 0$ by
\begin{align}
\FBCS(\Gamma) &:= \Tr\bigl[ \bigl( (-\i \nabla + \Abold)^2 - \mu\bigr)\gamma \bigr] - T\, S(\Gamma) - \frac{1}{|Q_B|} \int_{Q_B} \dd X \int_{\Rbb^3} \dd r\; V(r) \, |\alpha(X,r)|^2,
\label{DHS1:BCS functional}
\end{align}
with the von Neumann entropy per unit volume $S(\Gamma)= - \Tr [\Gamma \ln(\Gamma)]$ and the chemical potential $\mu\in \Rbb$. The particles interact via a two-body potential $V \in L^{\nicefrac 32}(\Rbb^3) + L_{\varepsilon}^{\infty}(\mathbb{R}^3)$. Furthermore, we introduced center-of-mass and relative coordinates $X = \frac{x+y}{2}$ and $r = x-y$. Here and in the following, we abuse notation slightly by writing $\alpha(X,r)\equiv \alpha(x,y)$. 
\begin{bem}
	We opt for the above set-up because the solution of the problem for the constant magnetic field already contains the main difficulties of the case of a general magnetic field. This is related to the fact that the vector potential of any magnetic field with non-zero flux through the unit cell can be written as a sum of a vector potential of a homogeneous magnetic field and a periodic vector potential, see e.g. \cite[Proposition~4.1]{Tim_Abrikosov}. The latter can be treated in some sense as a perturbation, see \cite{Hainzl2012,Hainzl2014}. However, this is not true for the constant magnetic field, see Remark~\ref{DHS1:Remarks_Main_Result}~(a) below. To solve the general case it is therefore crucial to understand the case of a homogeneous magnetic field. To keep the presentation to a reasonable length and to be able to convey the main ideas more clearly, we therefore decided to present this case first. We plan to extend our treatment to the case of a general magnetic field in a second paper. One motivation to treat general periodic magnetic fields with non-zero flux through the unit cell stems from the fact that it is an interesting and highly relevant problem to consider magnetic fields that are chosen self-consistently.
\end{bem}

The BCS functional is bounded from below and coercive on the set of admissible states. More precisely, it can be shown that the kinetic energy dominates the entropy and the interaction energy, i.e., there is a constant $C>0$ such that for all admissible $\Gamma$, we have
\begin{align}
\FBCS(\Gamma) &\geq \frac 12 \Tr \bigl[ \gamma + (-\i \nabla + \Abold)^2 \gamma\bigr] - C. \label{DHS1:BCS functional_bounded_from_below}
\end{align}

The unique minimizer of the BCS functional among admissible states with $\alpha =0$ is given by
\begin{align}
\Gamma_0 &:= \begin{pmatrix} \gamma_0 & 0  \\ 0 &  1-\ov\gamma_0 \end{pmatrix}, & \gamma_0 &:= \frac 1{1 + \e^{ ((-\i\nabla + \Abold)^2-\mu)/T}}.  \label{DHS1:Gamma0}
\end{align}
Since $\Gamma_0$ is also the unique minimizer of the BCS functional for sufficiently large temperatures $T$, it is called the normal state.
We define the BCS free energy by
\begin{align}
F^{\mathrm{BCS}}(B, T) := \inf\bigl\{ \FBCS(\Gamma) - \FBCS(\Gamma_0) : \Gamma \text{ admissible}\bigr\} \label{DHS1:BCS GS-energy}
\end{align}
and say that our system is superconducting if $F^{\mathrm{BCS}}(B, T) < 0$, that is, if the minimal energy is strictly smaller than that of the normal state. In this work we are interested in the regime of weak magnetic fields $0 < B \ll 1$. Our goal is to obtain an asymptotic expansion of $F^{\mathrm{BCS}}(B, T)$ in powers of $B$ that allows us to derive Ginzburg--Landau theory, and to show how the BCS critical temperature depends on the magnetic field $B$. For our main results to hold, we need the following assumptions concerning the regularity of the interaction potential $V$.
%
%
%

\begin{asmp}
\label{DHS1:Assumption_V}
We assume that the interaction potential $V$ is a nonnegative, radial function such that $(1+|\cdot|^2) V\in L^\infty(\Rbb^3)$.
\end{asmp}

\begin{bem}
Our main results Theorem~\ref{DHS1:Main_Result} and Theorem~\ref{DHS1:Main_Result_Tc} still hold if the assumption $V\geq 0$ is dropped. We only use it in Appendix \ref{DHS1:KTV_Asymptotics_of_EV_and_EF_Section} when we investigate the spectral properties of a certain linear operator involving $V$. These statements still hold in the case of potentials without a definite sign but their proof is longer. A proof of these statements in the general setting can be found in Chapter \ref{Chapter:Combes-Thomas}. We expect our results to be true also if $V$ has moderate local singularities. Furthermore, it may be possible to slightly weaken the decay assumptions of $V$. We choose to work with the assumptions above to keep the presentation at a reasonable length.
\end{bem}

\subsection{The translation-invariant BCS functional}
\label{DHS1:BCS_functional_TI_Section}

If no external fields are present, i.e. if $\Bbold =0$, we describe the system by translation-invariant states, that is, we assume that the kernels of $\gamma$ and $\alpha$ are of the form $\gamma(x-y)$ and $\alpha(x-y)$. To define the trace per unit volume we choose a cube of side length $1$. The resulting translation-invariant BCS functional and its infimum minus the free energy of the normal state are denoted by $\mathcal{F}^{\mathrm{BCS}}_{\mathrm{ti},T}$ and $F^{\mathrm{BCS}}_{\mathrm{ti}}(T)$, respectively. This functional has been studied in detail in \cite{Hainzl2007}, see also \cite{Hainzl2015} and the references therein, where it has been shown that there is a unique critical temperature $\Tc \geq 0$ such that $\mathcal{F}^{\mathrm{BCS}}_{\mathrm{ti},T}$ has a minimizer with $\alpha \neq 0$ if $T < \Tc$. For $T\geq \Tc$ the normal state in \eqref{DHS1:Gamma0} with $B=0$ is the unique minimizer. In terms of the energy, we have $F^{\mathrm{BCS}}_{\mathrm{ti}}(T) < 0$ for $T < \Tc$, while $F^{\mathrm{BCS}}_{\mathrm{ti}}(T) = 0$ if $T\geq \Tc$.

It has also been shown in \cite{Hainzl2007} that the critical temperature $\Tc$ can be characterized via a linear criterion. More precisely, the critical temperature is determined by the unique value of $T$ such that the operator 
\begin{equation*}
	K_{T} - V 
\end{equation*}
acting on $L^2_{\mathrm{sym}}(\mathbb{R}^3)$, the space of reflection-symmetric square-integrable functions, has zero as its lowest eigenvalue. Here, $K_T = K_T(-\i \nabla)$ with the symbol 
\begin{align}
	K_T(p) := \frac{p^2 - \mu}{\tanh \frac{p^2-\mu}{2T}}. \label{DHS1:KT-symbol}
\end{align}
It should be noted that the function $T \mapsto K_T(p)$ is strictly monotone increasing for fixed $p \in \mathbb{R}^3$, and that $K_T(p) \geq 2T$ if $\mu \geq 0$ and $K_T(p)\geq |\mu|/\tanh(|\mu|/(2T))$ if $\mu < 0$. Our assumptions on $V$ guarantee that the essential spectrum of the operator $K_{T} - V$ equals $[2T, \infty)$ if $\mu \geq 0$ and $[|\mu|/\tanh(|\mu|/(2T)), \infty)$ if $\mu <0$. Accordingly, an eigenvalue at zero is necessarily isolated and of finite multiplicity.

The results in \cite{Hainzl2007} have been obtained in the case where the Cooper pair wave function $\alpha(x)$ is not necessarily an even function (as opposed to our setup), which means that $K_{\Tc}(-\i \nabla) - V$ has to be understood to act on $L^2(\mathbb{R}^3)$. The results in \cite{Hainzl2007}, however, equally hold if the symmetry of $\alpha$ is enforced. 

We are interested in the situation where (a) $\Tc > 0$ and (b) the translation-invariant BCS functional has a unique minimizer with a radial Cooper pair wave function ($s$-wave Cooper pairs) for $T$ close to $\Tc$. This is implied by the following assumption. Part~(b) should be compared to \cite[Theorem~2.8]{DeuchertGeisinger}.

\begin{asmp}
\label{DHS1:Assumption_KTc}
\begin{enumerate}[(a)]
\item We assume that $\Tc >0$. If $V \geq 0$ and it does not vanish identically this is automatically implied, see \cite[Theorem 3]{Hainzl2007}. In the case of an interaction potential without a definite sign it is a separate assumption.


\item We assume that the lowest eigenvalue of $K_{\Tc} - V$ is simple.
\end{enumerate}
\end{asmp}


In the following we denote by $\alpha_*$ the unique ground state of the operator $K_{\Tc} - V$, i.e.,
\begin{align}
	K_{\Tc} \alpha_* = V\alpha_*. \label{DHS1:alpha_star_ev-equation}
\end{align}
We choose the normalization of $\alpha_*$ such that it is real-valued and $\Vert \alpha_*\Vert_{L^2(\Rbb^3)} = 1$. Since $V$ is a radial function and $\alpha_*$ is the unique solution of \eqref{DHS1:alpha_star_ev-equation} it follows that $\alpha_*$ is radial, too.


\subsection{The Ginzburg--Landau functional}
\label{DHS1:Ginzburg-Landau-functional}

We call a function $\Psi$ on $Q_B$ \textit{gauge-periodic} 
if it is left invariant by the magnetic translations of the form
\begin{align}
	T_B(\lambda)\Psi(X) &:= \e^{\i \Bbold \cdot (\lambda \wedge X)} \; \Psi(X + \lambda ), & \lambda &\in \Lambda_B. \label{DHS1:Magnetic_Translation_Charge2}
\end{align}
The operator $T(\lambda)$ in \eqref{DHS1:Magnetic_Translation} coincides with $T_B(\lambda)$  when $\Bbold$ is replaced by $2\Bbold$.

Let 
$\Lambda_0 , \Lambda_2, \Lambda_3 >0$ and $D\in\Rbb$ be given. For $B >0$ and a gauge-periodic function $\Psi$, the Ginzburg--Landau functional is defined by 
\begin{align}
\EGL(\Psi) &:= \frac{1}{B^2} \frac{1}{|Q_B|} \int_{Q_B} \dd X \; \bigl\{ \Lambda_0 \; |(-\i\nabla + 2\Abold)\Psi(X)|^2 - DB \, \Lambda_2\, |\Psi(X)|^2 + \Lambda_3\,|\Psi(X)|^4\bigr\}. \label{DHS1:Definition_GL-functional}
\end{align}
We highlight the factor of $2$ in front of the magnetic potential in \eqref{DHS1:Definition_GL-functional} and that the definition of the magnetic translation in \eqref{DHS1:Magnetic_Translation_Charge2} differs from that in  \eqref{DHS1:Magnetic_Translation} by a factor $2$. These two factors reflect the fact that $\Psi$ describes Cooper pairs, which carry twice the charge of a single particle.
The Ginzburg--Landau energy
\begin{align*}
E^{\mathrm{GL}}(D) := \inf \bigl\{ \EGL(\Psi) : \Psi\in \Hmag^1(Q_B)\bigr\}
\end{align*}
is independent of $B$ by scaling. More precisely, for given $\psi$ the function
\begin{align}
\Psi(X) &:= \sqrt{B} \; \psi\bigl( \sqrt{B} \, X\bigr), & X\in \Rbb^3, \label{DHS1:GL-rescaling}
\end{align}
satisfies
\begin{align}
\EGL(\Psi) = \mathcal E_{D,1}^{\mathrm{GL}}(\psi). \label{DHS1:EGL-scaling}
\end{align}

We also define the critical parameter
\begin{align}
\Dc &:= \frac{\Lambda_0}{\Lambda_2} \inf \spec_{\Lmag^2(Q_1)} \bigl((-\i \nabla + e_3 \wedge X)^2\bigr), \label{DHS1:Dc_Definition}
\end{align}
where the infimum is taken over gauge-periodic square-integrable functions. Its definition is motivated by the fact that $\EGLGSE < 0$ if $D > \Dc$ and $\EGLGSE =0$ if $D \leq \Dc$. This should be compared to \cite[Lemma 2.5]{Hainzl2014}. In our situation with a constant magnetic field the lowest eigenvalue of the Hamiltonian in \eqref{DHS1:Dc_Definition} equals $2$, see \cite[Eq.~(6.2)]{Tim_Abrikosov}, and $\Dc$ is explicit. In the situation of \cite{Hainzl2014}, where general external fields excluding the constant magnetic field are present, the parameter $\Dc$ is not explicit.  


\subsection{Main results}
\label{DHS1:Main_Result_Section}

Our first main result concerns the asymptotics of the BCS free energy in \eqref{DHS1:BCS GS-energy} in the regime $B \ll 1$. It also contains a statement about the asymptotics of the Cooper pair wave function of states $\Gamma$, whose energy $\FBCS(\Gamma)$ has the same asymptotic behavior as the BCS free energy (approximate minimizers). The precise statement is captured in the following theorem.

\begin{bigthm}
\label{DHS1:Main_Result}
Let Assumptions \ref{DHS1:Assumption_V} and \ref{DHS1:Assumption_KTc} hold, let $D \in \mathbb{R}$, and let the coefficients $\Lambda_0, \Lambda_2, \Lambda_3 >0$ be given by \eqref{DHS1:GL-coefficient_1}-\eqref{DHS1:GL_coefficient_3} below. Then there are constants $C>0$ and $B_0 >0$ such that for all $0 < B \leq B_0$, we have
\begin{align}
F^{\mathrm{BCS}}(B,\, \Tc(1 - DB)) = B^2 \; \bigl( \EGLGSE + R \bigr), \label{DHS1:ENERGY_ASYMPTOTICS}
\end{align}
with $R$ satisfying the estimate
\begin{align}
CB \geq R \geq - \Rcal := -C B^{\nicefrac 1{12}}. \label{DHS1:Rcal_error_Definition}
\end{align}
Moreover, for any approximate minimizer $\Gamma$ of $\FBCS$ at $T = \Tc(1 - DB)$ in the sense that
\begin{align}
\FBCS(\Gamma) - \FBCS(\Gamma_0) \leq B^2 \bigl( \EGLGSE + \rho\bigr) 
\label{DHS1:BCS_low_energy}
\end{align}
holds for some $\rho \geq 0$, we have the decomposition
\begin{align}
\alpha(X, r ) = \Psi(X) \, \alpha_*(r) + \sigma(X,r) \label{DHS1:Thm1_decomposition}
\end{align}
for the Cooper pair wave function $\alpha =\Gamma_{12}$. Here, $\sigma$ satisfies
\begin{align}
\frac{1}{|Q_B|} \int_{Q_B} \dd X \int_{\Rbb^3} \dd r \; |\sigma(X, r)|^2 &\leq C B^{\nicefrac {11}6}, \label{DHS1:Thm1_error_bound}
\end{align}
$\alpha_*$ is the normalized zero energy eigenstate of $K_{\Tc}-V$, and the function $\Psi$ obeys
\begin{align}
\EGL(\Psi) \leq \EGLGSE + \rho + \Rcal. \label{DHS1:GL-estimate_Psi}
\end{align}
\end{bigthm}

Our second main result concerns the shift of the BCS critical temperature that is caused by the external magnetic field. 

\begin{bigthm}
\label{DHS1:Main_Result_Tc} \label{DHS1:MAIN_RESULT_TC}
Let Assumptions \ref{DHS1:Assumption_V} and \ref{DHS1:Assumption_KTc} hold. Then there are constants $C>0$ and $B_0 >0$ such that for all $0 < B \leq B_0$ the following holds:
\begin{enumerate}[(a)]
	\item Let $0 < T_0 < \Tc$. If the temperature $T$ satisfies
	\begin{equation}
		T_0 \leq T \leq \Tc \, ( 1 - B \, ( \Dc + C \, B^{\nicefrac 1{2}}))
		\label{DHS1:eq:lowertemp}
	\end{equation}
	with $\Dc$ in \eqref{DHS1:Dc_Definition}, then we have
	\begin{equation*}
		F^{\mathrm{BCS}}(B,T) < 0.
	\end{equation*} 
	\item If the temperature $T$ satisfies
	\begin{equation}
		T \geq \Tc \, ( 1 - B \, ( \Dc - \Rcal ) )
		\label{DHS1:eq:uppertemp}
	\end{equation}
	with $\Dc$ in \eqref{DHS1:Dc_Definition} and $\Rcal$ in \eqref{DHS1:Rcal_error_Definition}, then we have
	\begin{equation*}
		\FBCS(\Gamma) - \FBCS(\Gamma_0) > 0
	\end{equation*}
	unless $\Gamma = \Gamma_0$.
\end{enumerate}
\end{bigthm}

\begin{bems}
\label{DHS1:Remarks_Main_Result}
\begin{enumerate}[(a)]
	
\item Theorem~\ref{DHS1:Main_Result} and Theorem~\ref{DHS1:Main_Result_Tc} extend similar results in \cite[Theorem 1]{Hainzl2012} and \cite[Theorem~2.4]{Hainzl2014} to the case of a homogeneous magnetic field. Such a magnetic field has a non-periodic vector potential and a non-zero magnetic flux through the unit cell $Q_B$. The main reason why the problem with a homogeneous magnetic field is more complicated is that it cannot be treated as a perturbation of the Laplacian. More precisely, it was possible in \cite{Hainzl2012,Hainzl2014} to work with a priori bounds for low-energy states that only involve the Laplacian and not the external fields. As noticed in \cite{Hainzl2017}, see the discussion below Remark~6, this is not possible in the case of a homogeneous magnetic field. In the proof of comparable a priori estimates involving the homogeneous magnetic field, see Theorem~\ref{DHS1:Structure_of_almost_minimizers} below, we have to deal with the fact that the components of the magnetic momentum operator do not commute, which leads to significant technical difficulties.

\item If we compare Theorem~\ref{DHS1:Main_Result} to \cite[Theorem 1]{Hainzl2012} or Theorem~\ref{DHS1:Main_Result_Tc} to \cite[Theorem~2.4]{Hainzl2014} we note the following technical differences: (1) The parameter $h$ in \cite{Hainzl2012,Hainzl2014} equals $B^{\nicefrac 12}$ in our work. (2) We use microscopic coordinates while macroscopic coordinates are used in \cite{Hainzl2012,Hainzl2014}. (3) Our free energy is normalized by a volume factor, see \eqref{DHS1:Trace_per_unit_volume_definition} and \eqref{DHS1:BCS functional}. This is not the case in \cite{Hainzl2012,Hainzl2014}. (4) The leading order of the Cooper pair wave function in \cite[Theorem 1]{Hainzl2012} is of the form 
\begin{equation}
	\frac{1}{2} \alpha_*(x-y) (\Psi(x) + \Psi(y)).
	\label{DHS1:eq:remarksA1}
\end{equation}
This should be compared to \eqref{DHS1:Thm1_decomposition}, where relative and center-of-mass coordinates are used. Using the a priori bound for the $L^2$-norm of $\nabla \Psi$ below (5.61) in \cite{Hainzl2012}, one can see that \eqref{DHS1:eq:remarksA1} equals the first term in \eqref{DHS1:Thm1_decomposition} to leading order in $h$. The analogue in our setup does not seem to be correct.

\item The Ginzburg--Landau energy appears at the order $B^2$. This should be compared to the free energy of the normal state, which is of order $1$.

\item To appreciate the bound in \eqref{DHS1:Thm1_error_bound}, we note that the first term in the decomposition of $\alpha$ in \eqref{DHS1:Thm1_decomposition} obeys
\begin{align*}
	\frac{1}{|Q_B|} \int_{Q_B} \dd X \int_{\Rbb^3} \dd r \; | \Psi(X)\alpha_*(r) |^2 &= O(B).
\end{align*}

\item We stated Theorem~\ref{DHS1:Main_Result} with fixed $D\in \Rbb$. Our explicit error bounds show that $D$ is allowed to vary with $B$ as long as there is a $B$-independent constant $D_0>0$ such that $|D| \leq D_0$ holds. 

\item Theorem~\ref{DHS1:Main_Result_Tc} gives bounds on the range of temperatures where superconductivity is present, see \eqref{DHS1:eq:lowertemp}, or absent, see \eqref{DHS1:eq:uppertemp}. The interpretation of this theorem is that for small magnetic fields $B$ the critical temperature obeys the asymptotic expansion
\begin{equation}
	\Tc(B) = \Tc(1 - \Dc B) + o(B).
	\label{DHS1:eq:BemA1}
\end{equation}
We highlight that $\Tc$ is determined by the translation-invariant problem, and that $D_{\mathrm{c}}$ is given by the macroscopic (linearized) GL theory. The same result has been obtained in \cite[Theorem~4]{Hainzl2017} in the case of linearized BCS theory. Theorem~\ref{DHS1:Main_Result_Tc} can therefore be interpreted as a justification of this approximation. Eq.~\eqref{DHS1:eq:BemA1} allows us to compute the upper critical field $B_{c2}$. That is, the magnetic field, above which, for a given temperature $T$, superconductivity is absent. In particular, it allows us to compute the derivative of $B_{c2}$ with respect to $T$ at the critical temperature from the BCS functional. For more details we refer to \cite[Appendix~A]{Hainzl2017}. 

\item We expect that the assumption $0 < T_0 \leq T$ for some arbitrary but $B$-independent constant $T_0$ in Theorem~\ref{DHS1:Main_Result_Tc}~(a) is of technical nature. We need this assumption, which similarly appears in \cite[Theorem~4]{Hainzl2017}, because our trial state analysis in Section~\ref{DHS1:Upper_Bound} breaks down when the temperature $T$ approaches zero. This is related to the fact that the Fermi distribution function $f_T(x) = (\e^{x/T}+1)^{-1}$ cannot be represented by a Cauchy-integral uniformly in the temperature. We note that there is no such restriction in Theorem~\ref{DHS1:Main_Result_Tc}~(b). It is also not needed in \cite[Theorem~2.4]{Hainzl2014}.

\end{enumerate}
\end{bems}


\subsection{Organization of the paper and strategy of proof}

In Section~\ref{DHS1:Preliminaries} we complete the introduction of our mathematical setup. We recall several properties of the trace per unit volume and introduce the relevant spaces of gauge-periodic functions. 

Section~\ref{DHS1:Upper_Bound} is dedicated to a trial state analysis. We start by introducing a class of Gibbs states, whose Cooper pair wave function is given by a product of the form $\alpha_*(r) \Psi(X)$ to leading order in $B$ with $\alpha_*$ in \eqref{DHS1:alpha_star_ev-equation} and with a gauge-periodic function $\Psi$ on $Q_B$. We state and motivate several results concerning these Gibbs states and their BCS free energy, whose proofs are deferred to Section~\ref{DHS1:Proofs}. Afterwards, these statements are used to prove the upper bound on \eqref{DHS1:ENERGY_ASYMPTOTICS} as well as Theorem~\ref{DHS1:Main_Result_Tc}~(a). As will be explained below, they are also relevant for the proofs of the lower bound in \eqref{DHS1:ENERGY_ASYMPTOTICS} and of Theorem~\ref{DHS1:Main_Result_Tc}~(b) in Section~\ref{DHS1:Lower Bound Part B}. 

Section~\ref{DHS1:Proofs} contains the proof of the results concerning the Gibbs states and their BCS free energy that have been stated without proof in Section~\ref{DHS1:Upper_Bound}. Our analysis is based on an extension of the phase approximation method, which has been pioneered in the framework of linearized BCS theory in \cite{Hainzl2017}, to our nonlinear setting. The phase approximation is a well-known tool in the physics literature, see, e.g., \cite{Werthamer1966}, and has also been used in the mathematical literature to study spectral properties of Schr\"odinger operators involving a magnetic field, for instance in \cite{NenciuCorn1998,Nenciu2002}. Our approach should be compared to the trial state analysis in \cite{Hainzl2012,Hainzl2014}, where a semi-classical expansion is used. The main novelty of our trial state analysis is Lemma~\ref{DHS1:BCS functional_identity_Lemma}, where we provide an alternative way to compute a certain trace function involving the trial state. It should be compared to the related part in the proof of \cite[Theorem~2]{Hainzl2012}. While the analysis in \cite{Hainzl2012} uses a Cauchy integral representation of the function $z \mapsto \ln(1+\e^{-z})$, our approach is based on a product expansion of the hyperbolic cosine in terms of Matsubara frequencies. In this way we obtain better decay properties in the subsequent resolvent expansion, which, in our opinion, simplifies the analysis considerably.

Section~\ref{DHS1:Lower Bound Part A} contains the proof of a priori estimates for BCS states, whose BCS free energy is smaller than or equal to that of the normal state $\Gamma_0$ in \eqref{DHS1:Gamma0} plus a correction of the order $B^2$ (low-energy states). The result is captured in Theorem~\ref{DHS1:Structure_of_almost_minimizers}, which is the main novelty of the present article. It states that the Cooper pair wave function of any low-energy state in the above sense has a Cooper pair wave function, which is, to leading order in $B$, given by a product of the form $\alpha_*(r) \Psi(X)$  with $\alpha_*(r)$ in \eqref{DHS1:alpha_star_ev-equation} and with a gauge-periodic function $\Psi(X)$ on $Q_B$. Furthermore, the function $\Psi(X)$ obeys certain bounds, which show that it is slowly varying and small in an appropriate sense. As explained in Remark~\ref{DHS1:Remarks_Main_Result}~(a), the main difficulty to overcome is that our a priori bounds involve the magnetic field. Therefore, we have to deal with the non-commutativity of the components of the magnetic momentum operator. The step where this problem appears most prominently is in the proof of Proposition~\ref{DHS1:First_Decomposition_Result}. 


The proof of the lower bound on \eqref{DHS1:ENERGY_ASYMPTOTICS} and of Theorem~\ref{DHS1:Main_Result_Tc}~(b) is provided in Section~\ref{DHS1:Lower Bound Part B}, which mostly follows the strategy in \cite[Section~6]{Hainzl2012} and \cite[Section~4.2]{Hainzl2014}. Two main ingredients for the analysis in this section are the trial state analysis in Section~\ref{DHS1:Upper_Bound} and Section~\ref{DHS1:Proofs}, and the a priori bounds for low-energy states in Section~\ref{DHS1:Lower Bound Part A}. From Theorem~\ref{DHS1:Structure_of_almost_minimizers} we know that the Cooper pair wave function of any low-energy state has a product structure to leading order in $B$. The main idea of the proof of the lower bound in \eqref{DHS1:ENERGY_ASYMPTOTICS} is to construct a Gibbs state, whose Cooper pair wave function has the same asymptotics to leading order in $B$. The precise characterization of the Cooper pair wave function of the Gibbs state in Section~\ref{DHS1:Upper_Bound} and the a priori bounds in Theorem~\ref{DHS1:Structure_of_almost_minimizers} then allow us to bound the BCS free energy of the original state from below in terms of that of the Gibbs state. The latter has been computed with sufficient precision in Section~\ref{DHS1:Upper_Bound} and Section~\ref{DHS1:Proofs}.


Throughout the paper, $c$ and $C$ denote generic positive constants that change from line to line. We allow them to depend on the various fixed quantities like $B_0$, $\mu$, $\Tc$, $V$, $\alpha_*$, etc. Further dependencies are indexed.


\section{Preliminaries}
\label{DHS1:Preliminaries}


\subsection{Schatten classes}
\label{DHS1:Schatten_Classes}

In our proofs we frequently use Schatten norms of periodic operators, which are defined with respect to the trace per unit volume in \eqref{DHS1:Trace_per_unit_volume_definition}. In this section we recall some basic facts about these norms.

A gauge-periodic operator $A$ belongs to the $p$\tho\ local von-Neumann--Schatten class $\Scal^p$ with $1\leq p < \infty$ if it has finite $p$-norm, that is, if $\Vert A\Vert_p^p := \Tr (|A|^p) <\infty$. By $\Scal^\infty$ we denote the set of bounded gauge-periodic operators and $\Vert \cdot \Vert_{\infty}$ is the usual operator norm. For the above norms the triangle inequality
\begin{align*}
\Vert A + B\Vert_p \leq \Vert A\Vert_p + \Vert B\Vert_p
\end{align*}
holds for $1\leq p \leq \infty$. Moreover, for $1 \leq p,q,r \leq \infty$ with $\frac{1}{r} = \frac{1}{p} + \frac{1}{q}$ we have the general Hölder inequality 
\begin{align}
\Vert AB\Vert_r \leq \Vert A\Vert_p \Vert B\Vert_q. \label{DHS1:Schatten-Hoelder}
\end{align}
It is important to note that the above norms are not monotone decreasing in the index $p$. This should be compared to the usual Schatten norms, where such a property holds. The familiar inequality
\begin{align*}
| \Tr A | \leq \Vert A \Vert_1
\end{align*}
is true also in the case of local Schatten norms.

The above inequalities can be reduced to the case of the usual Schatten norms, see, e.g., \cite{Simon05}, using the magnetic Bloch--Floquet decomposition. We refer to \cite[Section XIII.16]{Reedsimon4} for an introduction to the Bloch--Floquet transformation and to \cite{Stefan_Peierls} for a particular treatment of the magnetic case. More specifically, for a gauge-periodic operator $A$ we use the unitary equivalence 
\begin{align*}
A \cong \int^{\oplus}_{[0,\sqrt{ 2 \pi B}]^3} \mathrm{d} k \;  A_{k}
\end{align*}
to write the trace per unit volume as
\begin{equation}
\Tr A = \int_{[0,\sqrt{ 2 \pi B}]^3} \mathrm{d} k \; \Tr_{L^2(Q_B)} A_{k},
\label{DHS1:eq:ATPUV}
\end{equation}
where $\Tr_{L^2(Q_B)}$ denotes the usual trace over $L^2(Q_B)$. The inequalities for the trace per unit volume from above follow from the usual ones when we use that $(AB)_{k} = A_{k} B_{k}$ holds for two gauge-periodic operators $A$ and $B$.


\subsection{Gauge-periodic Sobolev spaces}
\label{DHS1:Periodic Spaces}

In this section we introduce several spaces of gauge-periodic functions, which will be used to describe the center-of-mass part of Cooper pair wave functions. 



For $1 \leq p < \infty$, the space $L_{\mathrm{mag}}^p(Q_B)$ consists of all $L_\loc^p(\Rbb^3)$-functions $\Psi$, which satisfy $T_B(\lambda)\Psi = \Psi$ for all $\lambda\in\Lambda_B$ with $T_B(\lambda)$ in \eqref{DHS1:Magnetic_Translation_Charge2}. The space is equipped with the usual $p$-norm per unit volume
\begin{align}
\Vert \Psi\Vert_{\Lmag^p(Q_B)}^p &:= \fint_{Q_B} \dd X \; |\Psi(X)|^p := \frac{1}{|Q_B|} \int_{Q_B} \dd X \; |\Psi(X)|^p, \label{DHS1:Periodic_p_Norm}
\end{align}
and we use the conventional abbreviation $\Vert \Psi\Vert_p$ when this does not lead to confusion.

For $m\in \Nbb_0$, the corresponding gauge-periodic Sobolev space is defined by
\begin{align}
\Hmag^m(Q_B) &:= \bigl\{ \Psi\in \Lmag^2(Q_B) :  (-\i\nabla + 2 \Abold)^\nu \Psi\in \Lmag^2(Q_B) \quad \forall \nu\in \Nbb_0^3, |\nu|_1\leq m\bigr\}, \label{DHS1:Periodic_Sobolev_Space}
\end{align}
where $|\nu |_1 := \sum_{i=1}^3 \nu_i$ for $\nu\in \Nbb_0^3$.
%
%
Equipped with the scalar product
\begin{align}
\langle \Phi, \Psi\rangle_{\Hmag^m(Q_B)} &:= \sum_{|\nu|_1\leq m} B^{-1 - |\nu|_1} \; \langle (-\i\nabla + 2\Abold)^\nu \Phi, (-\i \nabla + 2\Abold)^\nu \Psi\rangle_{\Lmag^2(Q_B)},  \label{DHS1:Periodic_Sobolev_Norm}
\end{align}
it is a Hilbert space. We note that $(-\i \nabla + 2 \Abold)^\nu \Psi$ is a gauge-periodic function if $\Psi$ is gauge-periodic because the magnetic momentum operator
\begin{align*}
\Pi := -\i \nabla + 2\Abold
\end{align*}
commutes with the magnetic translations $T_B(\lambda)$ in \eqref{DHS1:Magnetic_Translation_Charge2}. We also note that $\Pi$ is a self-adjoint operator on $\Hmag^1(Q_B)$. 

At this point, we shall briefly explain the scaling behavior in $B$ of the norms introduced in \eqref{DHS1:Periodic_p_Norm} and \eqref{DHS1:Periodic_Sobolev_Norm} in terms of the Ginzburg--Landau scaling in \eqref{DHS1:GL-rescaling}. First, we note that if $\psi \in \Lmag^p(Q_1)$ and $\Psi$ is as in \eqref{DHS1:GL-rescaling}, then
\begin{align}
\Vert \Psi\Vert_{\Lmag^p(Q_B)} = B^{\nicefrac 12} \, \Vert \psi\Vert_{\Lmag^p(Q_1)} \label{DHS1:Periodic_p_Norm_scaling}
\end{align}
for every $1\leq p \leq \infty$. In contrast, the scaling of the norm in \eqref{DHS1:Periodic_Sobolev_Norm} is chosen such that
\begin{align*}
\Vert \Psi\Vert_{\Hmag^m(Q_B)} = \Vert \psi\Vert_{\Hmag^m(Q_1)}.
\end{align*}
This follows from \eqref{DHS1:Periodic_p_Norm_scaling} and the fact that
$\Vert (-\i \nabla + 2\Abold)^\nu\Psi\Vert_2^2$ scales as $B^{1 + |\nu|_1}$ for $\nu\in \Nbb_0^3$.

We also mention the following magnetic Sobolev inequality because it will be used frequently in the course of the paper. For any $B>0$ and any $\Psi\in \Hmag^1(Q_B)$, we have
\begin{align}
\Vert \Psi\Vert_{\Lmag^6(Q_B)}^2 &\leq  C \, B^{-1}\, \Vert (-\i \nabla + 2\Abold)\Psi\Vert_{\Lmag^2(Q_B)}^2. \label{DHS1:Magnetic_Sobolev}
\end{align}

\begin{proof}[Proof of \eqref{DHS1:Magnetic_Sobolev}]
Since $Q_1$ satisfies the cone property, \cite[Theorem 8.8]{LiebLoss} implies 
\begin{align*}
\Vert \psi \Vert_{\Lmag^6(Q_1)}^2 &\leq C\, \bigl( \Vert \psi \Vert_{\Lmag^2(Q_1)}^2 + \Vert \nabla |\psi|\, \Vert_{\Lmag^2(Q_1)}^2\bigr).
\end{align*}
From \cite[Eq.~(6.2)]{Tim_Abrikosov} we know that the bottom of the spectrum of $(-\i \nabla + e_3 \wedge X)^2$ equals~$2$. For the first term on the right side, this implies $2\Vert \psi\Vert_2^2 \leq \Vert (-\i \nabla + e_3\wedge X)\psi\Vert_2^2$. To bound the second term, we apply the  diamagnetic inequality $|\nabla |\psi(X)|| \leq |(-\i\nabla + e_3\wedge X)\psi(X)|$, see \cite[Theorem 7.21]{LiebLoss}. This proves \eqref{DHS1:Magnetic_Sobolev} for $B = 1$ and the scaling in \eqref{DHS1:GL-rescaling} yields \eqref{DHS1:Magnetic_Sobolev} for $B >0$.
\end{proof}

As indicated below \eqref{DHS1:Trace_per_unit_volume_definition}, the Cooper pair wave function $\alpha$ related to an admissible state $\Gamma$ belongs to $\Scal^2$, the Hilbert--Schmidt class introduced in Section \ref{DHS1:Schatten_Classes}. In terms of the center-of-mass and relative coordinates, the gauge-periodicity and the symmetry of the kernel of $\alpha$ in \eqref{DHS1:alpha_periodicity} read
\begin{align}
\alpha(X,r) &= \e^{\i \Bbold \cdot (\lambda \wedge X)} \; \alpha(X+ \lambda, r), \quad \lambda\in \Lambda_B; & \alpha(X,r) &= \alpha(X, -r). \label{DHS1:alpha_periodicity_COM}
\end{align}
That is, $\alpha(X,r)$ is a gauge-periodic function of the center-of-mass coordinate $X$ and a reflection-symmetric function of the relative coordinate $r \in \mathbb{R}^3$. We make use of the isometric identification of $\Scal^2$ with the space 
\begin{align*}
L^2(Q_B \times \Rbb_{\mathrm s}^3) := \Lmag^2(Q_B) \otimes L_{\mathrm{sym}}^2(\Rbb^3),
\end{align*}
the square-integrable functions obeying \eqref{DHS1:alpha_periodicity_COM}, for which the norm
\begin{align*}
\Vert \alpha\Vert_{\Lsymm}^2 := \fint_{Q_B} \dd X\int_{\Rbb^3} \dd r \; |\alpha(X, r)|^2 = \frac{1}{|Q_B|} \int_{Q_B} \dd X\int_{\Rbb^3} \dd r \; |\alpha(X, r)|^2
\end{align*} 
is finite. By \eqref{DHS1:alpha_periodicity_COM}, the identity $\Vert \alpha\Vert_2 = \Vert \alpha\Vert_{\Lsymm}$ holds. Therefore, we do not distinguish between the scalar products $\langle \cdot, \cdot\rangle$ on $\Lsymm$ and $\Scal^2$ and identify operators in $\Scal^2$ with their kernels whenever this appears convenient. 

Finally, the Sobolev space $H^1(Q_B\times \Rbb_{\mathrm s}^3)$ consists of all functions $\alpha\in L^2(Q_B\times \Rbb_{\mathrm s}^3)$ with finite $H^1$-norm given by
\begin{align}
\Vert \alpha\Vert_{H^1(Q_B\times \Rbb_{\mathrm s}^3)}^2 &:= \Vert \alpha\Vert_2^2 + \Vert \Pi_X\alpha\Vert_2^2 + \Vert \tilde \pi_r\alpha\Vert_2^2. \label{DHS1:H1-norm}
\end{align}
Here, we used the magnetic momentum operators 
\begin{align}
\Pi_X &:= -\i\nabla_X + 2 \Abold(X), & \tilde \pi_r &:= -\i\nabla_r + \frac 12 \Abold(r), \label{DHS1:Magnetic_Momenta_COM}
\end{align}
where $\Abold(x) = \frac 12 \Bbold \wedge x$. We note that the norm in \eqref{DHS1:H1-norm} is equivalent to the norm given by $\Tr [\alpha\alpha^*] + \Tr [(-\i \nabla + \Abold)\alpha \alpha^* (-\i \nabla + \Abold)]  + \Tr [(-\i \nabla + \Abold) \alpha^* \alpha (-\i \nabla + \Abold)]$, which, in turn, is given by
$\Vert \alpha\Vert_2^2 + \Vert (-\i \nabla + \Abold)\alpha\Vert_2^2 + \Vert \alpha (-\i \nabla + \Abold)\Vert_2^2$. See also the discussion below \eqref{DHS1:Trace_per_unit_volume_definition}.

\section{Trial States and their BCS Energy}
\label{DHS1:Upper_Bound} \label{DHS1:UPPER_BOUND}

The goal of this section is to provide the upper bound on \eqref{DHS1:ENERGY_ASYMPTOTICS} and the proof of Theorem~\ref{DHS1:Main_Result_Tc}~(a). Both bounds are proved with a trial state argument using Gibbs states $\Gamma_\Delta$ that are defined via a gap function $\Delta$ in the effective Hamiltonian. In Proposition~\ref{DHS1:Structure_of_alphaDelta} we show that the Cooper pair wave function $\alpha_\Delta$ of $\Gamma_\Delta$ is a product function with respect to relative and center-of-mass coordinates to leading order provided $\Delta$ is a product function that is small in a suitable sense. A representation formula for the BCS energy in terms of the energy of these states is provided in Proposition~\ref{DHS1:BCS functional_identity}. Finally, in Theorem \ref{DHS1:Calculation_of_the_GL-energy}, we show that certain parts of the BCS energy of the trial states $\Gamma_{\Delta}$ equal the terms in the Ginzburg--Landau functional in \eqref{DHS1:Definition_GL-functional} with sufficient precision provided $T = \Tc(1 - DB)$ for some fixed $D \in \mathbb{R}$. These results, whose proofs are deferred to Section \ref{DHS1:Proofs}, are combined in Section~\ref{DHS1:Upper_Bound_Proof_Section} to give the proof of the results mentioned in the beginning of this paragraph.


\subsection{The Gibbs states \texorpdfstring{$\Gamma_\Delta$}{GammaDelta}}

For any $\Psi\in \Lmag^2(Q_B)$, let us introduce the gap function $\Delta\in L^2(Q_B\times \Rbb_{\mathrm s}^3)$, given by
\begin{align}
\Delta(X,r) := \Delta_\Psi(X, r) &:= -2 \; V\alpha_*(r) \; \Psi(X).  \label{DHS1:Delta_definition}
\end{align}
In our trial state analysis, $\Psi$ is going to be a minimizer of the Ginzburg--Landau functional in \eqref{DHS1:Definition_GL-functional}. It therefore obeys the scaling in \eqref{DHS1:GL-rescaling}, which implies that the local Hilbert-Schmidt norm $\Vert \Delta\Vert_2^2$ is of the order $B$. We highlight that the $L^2(\mathbb{R}^3)$-norm of $V\alpha_*$ is of the order $1$, that is, the size of $\Vert \Delta\Vert_2^2$ is determined by $\Psi$. In the proof of the lower bound we have less information on $\Psi$. The related difficulties are discussed in Remark~\ref{DHS1:rem:Psi} below. With
\begin{align}
\hfrak_B &:=  (-\i \nabla +\Abold )^2 - \mu, \label{DHS1:hfrakB_definition}
\end{align}
we define the Hamiltonian
\begin{align}
H_{\Delta} &:= H_0 + \delta := \begin{pmatrix}
\hfrak_B & 0 \\ 0 & -\ov{\hfrak_B}
\end{pmatrix} + \begin{pmatrix}
0 & \Delta \\ \ov \Delta & 0
\end{pmatrix} = \begin{pmatrix}
\hfrak_B & \Delta \\ \ov \Delta & -\ov {\hfrak_B}
\end{pmatrix} \label{DHS1:HDelta_definition}
\end{align}
and the corresponding Gibbs state at inverse temperature $\beta = T^{-1} >0$ as
\begin{align}
\begin{pmatrix} \gamma_\Delta & \alpha_\Delta \\ \ov{\alpha_\Delta} & 1 - \ov{\gamma_\Delta}\end{pmatrix} = \Gamma_\Delta := \frac{1}{1 + \e^{\beta H_\Delta}}. \label{DHS1:GammaDelta_definition}
\end{align}
We note that the normal state $\Gamma_0$ in \eqref{DHS1:Gamma0} corresponds to setting $\Delta =0$ in \eqref{DHS1:GammaDelta_definition}.

\begin{lem}[Admissibility of $\Gamma_\Delta$]
\label{DHS1:Gamma_Delta_admissible}
Let Assumptions \ref{DHS1:Assumption_V} and \ref{DHS1:Assumption_KTc} hold. Then, for any $B>0$, any $T>0$, and any $\Psi\in \Hmag^1(Q_B)$, the state $\Gamma_\Delta$ in \eqref{DHS1:GammaDelta_definition} is admissible, where $\Delta \equiv \Delta_\Psi$ as in \eqref{DHS1:Delta_definition}.
\end{lem}

The states $\Gamma_\Delta$ are inspired by the following observation. Via variational arguments it is straightforward to see that any minimizer of $\FBCS$ in \eqref{DHS1:BCS functional} solves the nonlinear Bogolubov--de Gennes equation
\begin{align}
\Gamma &= \frac 1{1 + \e^{\beta \, \Hbb_{V\alpha}}}, & \Hbb_{V\alpha} = \begin{pmatrix} \hfrak_B & -2\, V\alpha \\ -2\, \ov{V\alpha} & -\ov{\hfrak_B}\end{pmatrix}. \label{DHS1:BdG-equation}
\end{align}
Here, $V\alpha$ is the operator given by the kernel $V(r)\alpha(X,r)$. As we look for approximate minimizers of $\FBCS$, we choose $\Gamma_\Delta$ in order to approximately solve \eqref{DHS1:BdG-equation}. As far as the leading term of $\alpha_\Delta$ is concerned this is indeed the case, as the following result shows. It should be compared to \eqref{DHS1:Thm1_decomposition}.

\begin{prop}[Structure of $\alpha_\Delta$]
\label{DHS1:Structure_of_alphaDelta} \label{DHS1:STRUCTURE_OF_ALPHADELTA}
Let Assumption \ref{DHS1:Assumption_V} and \ref{DHS1:Assumption_KTc} (a) be satisfied and let $T_0>0$ be given. Then, there is a constant $B_0>0$ such that for any $0 < B \leq B_0$, any $T\geq T_0$, and any $\Psi\in \Hmag^2(Q_B)$ the function $\alpha_\Delta$ in \eqref{DHS1:GammaDelta_definition} with $\Delta \equiv \Delta_\Psi$ as in \eqref{DHS1:Delta_definition} has the decomposition
\begin{align}
\alpha_\Delta(X,r) &= \Psi(X) \alpha_*(r) - \eta_0(\Delta)(X,r) - \eta_{\perp}(\Delta)(X,r). \label{DHS1:alphaDelta_decomposition_eq1}
\end{align}
The remainder functions $\eta_0(\Delta)$ and $\eta_\perp(\Delta)$ have the following properties:
\begin{enumerate}[(a)]
\item The function $\eta_0$ satisfies the bound
\begin{align}
\Vert \eta_0\Vert_\Hsymm^2 &\leq  C\; \bigl( B^3 + B \, |T - \Tc|^2\bigr) \; \bigl(  \Vert \Psi\Vert_{\Hmag^1(Q_B)}^6 + \Vert \Psi\Vert_{\Hmag^1(Q_B)}^2\bigr). \label{DHS1:alphaDelta_decomposition_eq2}
\end{align}

\item The function $\eta_\perp$ satisfies the bound
\begin{align}
\Vert \eta_\perp\Vert_{\Hsymm}^2 + \Vert |r|\eta_\perp\Vert_{\Lsymm}^2 &\leq C \; B^3 \; \Vert \Psi\Vert_{\Hmag^2(Q_B)}^2. \label{DHS1:alphaDelta_decomposition_eq3}
\end{align}

\item The function $\eta_\perp$ has the explicit form
\begin{align*}
\eta_\perp(X, r) &= \int_{\Rbb^3} \dd Z \int_{\Rbb^3} \dd s \; k_T(Z, r-s) \, V\alpha_*(s) \, \bigl[ \cos(Z\cdot \Pi_X) - 1\bigr] \Psi(X)
\end{align*}
with $k_T(Z,r)$ defined in Section~\ref{DHS1:Proofs} below \eqref{DHS1:MTB_definition}. For any radial $f,g\in L^2(\Rbb^3)$ the operator
\begin{align*}
\iiint_{\Rbb^9} \dd Z \dd r \dd s \; f(r) \, k_T(Z, r-s) \, g(s) \, \bigl[ \cos(Z\cdot \Pi) - 1\bigr]
\end{align*}
commutes with $\Pi^2$, and, in particular, 
if $P$ and $Q$ are two spectral projections of $\Pi^2$ with $P Q = 0$, then $\eta_\perp$ satisfies the orthogonality property
\begin{align}
	\bigl\langle f(r) \, (P \Psi)(X), \, \eta_{\perp}(\Delta_{Q\Psi}) \bigr\rangle = 0.
	\label{DHS1:alphaDelta_decomposition_eq4}
\end{align}
\end{enumerate}
\end{prop}

\begin{bem}
	\label{DHS1:rem:Psi}
The statement of Proposition \ref{DHS1:Structure_of_alphaDelta} should be read in two different ways, depending on whether we are interested in proving the upper or the lower bound for the BCS free energy. When we prove the upper bound using trial states $\Gamma_\Delta$, part (c) is irrelevant. In this case the gap function $\Delta\equiv \Delta_\Psi$ is defined with a minimizer $\Psi$ of the GL functional, whose $\Hmag^2(Q_B)$-norm is uniformly bounded, and all remainder terms can be estimated using \eqref{DHS1:alphaDelta_decomposition_eq2} and \eqref{DHS1:alphaDelta_decomposition_eq3}.

In the proof of the lower bound for the BCS free energy in Section~\ref{DHS1:Lower Bound Part B} we are forced to work with a trial state $\Gamma_{\Delta}$, whose gap function is defined via a function $\Psi$ that is related to a low-energy state of the BCS functional, see Theorem~\ref{DHS1:Structure_of_almost_minimizers} below. For such functions we only have a bound on the $\Hmag^1(Q_B)$-norm at our disposal. To obtain a function in $\Hmag^2(Q_B)$, we introduce a regularized version of $\Psi$ as in \cite[Section~6]{Hainzl2012}, \cite[Section~6]{Hainzl2014}, and \cite[Section~7]{Hainzl2017} by $\Psi_\leq := \Idbb_{[0,\varepsilon]}(\Pi^2)\Psi$ for some $B \ll \varepsilon \ll 1$, see Corollary \ref{DHS1:Structure_of_almost_minimizers_corollary}. The $\Hmag^2(Q_B)$-norm of $\Psi_\leq$ is not uniformly bounded in $B$, see \eqref{DHS1:Psileq_bounds} below. This causes a certain error term, namely the left side of \eqref{DHS1:LBpartB_3} below, to be large, a priori. 

To overcome this problem we use part (c) of Proposition~\ref{DHS1:Structure_of_almost_minimizers}. It exploits the fact that the first term on the right side of \eqref{DHS1:LBpartB_3} has an explicit form that satisfies the orthogonality property in \eqref{DHS1:alphaDelta_decomposition_eq4}, 
which implies that the left side of \eqref{DHS1:LBpartB_3} is indeed small. This is the reason why we need to distinguish between $\eta_0$ and $\eta_\perp$. 
\end{bem}

\subsection{The BCS energy of the states \texorpdfstring{$\Gamma_\Delta$}{GammaDelta}}

This section pertains to the BCS energy of the states $\Gamma_\Delta$, which is given by the Ginzburg--Landau functional to leading order. We will see in Section \ref{DHS1:BCS functional_identity_proof_Section} that the BCS energy of $\Gamma_\Delta$ can be calculated in terms of
\begin{align}
\Tr_0\Bigl[\ln\bigl( \cosh\bigl( \frac \beta 2 H_\Delta\bigr)\bigr) - \ln\bigl( \cosh\bigl( \frac \beta 2 H_0\bigr)\bigr)\Bigr]. \label{DHS1:lncosh-operator}
\end{align}
Here, $\Tr_0$ is a weaker form of trace which will be introduced later in \eqref{DHS1:Weak_trace_definition}. The operator inside the trace is closely related to the relative entropy of $H_\Delta$ and $H_0$ but also incorporates the interaction energy of $\alpha_\Delta$. We refer to \eqref{DHS1:kinetic energy} for more details. In the following, we explain how the terms of the Ginzburg--Landau functional, which appear in the energy expansion in \eqref{DHS1:ENERGY_ASYMPTOTICS}, are obtained from the operator in \eqref{DHS1:lncosh-operator}. 

As pointed out in Remark \ref{DHS1:Remarks_Main_Result}, we should think of $\Delta$ as being small. In order to expand the term in \eqref{DHS1:lncosh-operator} in powers of $\Delta$, we use the fundamental theorem to formally write \eqref{DHS1:lncosh-operator} as
\begin{align}
\frac \beta 2 \Tr_0 \Bigl[ \int_0^1 \dd t \; \tanh\bigl( \frac \beta 2 H_{t \Delta} \bigr) \Bigl(\begin{matrix} 0 & \Delta \\ \ov \Delta & 0 \end{matrix}\Bigr) \Bigr]. \label{DHS1:lncosh-operator_fundamental_theorem}
\end{align}
This identity is not rigorous because it ignores the subtlety that $H_{t\Delta}$, $t\in [0,1]$, are unbounded operators which do not commute for distinct values of $t$. 
We present a rigorous version of \eqref{DHS1:lncosh-operator_fundamental_theorem} in Lemma \ref{DHS1:BCS functional_identity_Lemma} below. For the sake of the following discussion it is legitimate to assume that equality between \eqref{DHS1:lncosh-operator} and \eqref{DHS1:lncosh-operator_fundamental_theorem} holds. 

We use the Mittag-Leffler series expansion, see e.g. \cite[Eq. (7)]{Hainzl2017}, to write the hyperbolic tangent in \eqref{DHS1:lncosh-operator_fundamental_theorem} as
\begin{align}
\tanh\bigl( \frac \beta 2 z\bigr) &= -\frac{2}{\beta} \sum_{n\in \Zbb} \frac{1}{\i\omega_n - z} \label{DHS1:tanh_Matsubara}
\end{align}
with the Matsubara frequencies 
\begin{align}
\omega_n &:= \pi (2n+1) T, \qquad n \in \Zbb. \label{DHS1:Matsubara_frequencies}
\end{align}
The convergence of \eqref{DHS1:tanh_Matsubara} becomes manifest by combining the $+n$ and $-n$ terms. Thus,
\begin{align}
\tanh\bigl( \frac \beta 2H_{\Delta}\bigr) &= -\frac 2\beta \sum_{n\in \Zbb} \frac{1}{\i \omega_n - H_{\Delta}}. \label{DHS1:tanh-expansion}
\end{align}
We use this representation to expand the operator in \eqref{DHS1:lncosh-operator_fundamental_theorem} in powers of $\Delta$ using the resolvent equation. The first term obtained in this way is $\langle \Delta, L_{T,B}\Delta\rangle$ with the linear operator $L_{T,B} \colon \Lsymm \ra \Lsymm$, given by
\begin{align}
L_{T,B}\Delta &:= -\frac 2\beta \sum_{n\in \Zbb} (\i \omega_n - \hfrak_B)^{-1} \, \Delta \,  (\i \omega_n + \ov{\hfrak_B})^{-1}. \label{DHS1:LTB_definition}
\end{align}
In the temperature regime we are interested in, we will obtain the quadratic terms in the Ginzburg--Landau functional from $\langle \Delta, L_{T,B}\Delta\rangle$. 


The next term in the expansion of \eqref{DHS1:tanh-expansion} is the quartic term $\langle \Delta, N_{T,B}(\Delta)\rangle$ with the nonlinear map $N_{T,B}\colon \Hsymm \ra \Lsymm$ defined as
\begin{align}
N_{T,B}(\Delta) &:= \frac 2\beta \sum_{n\in \Zbb} (\i \omega_n - \hfrak_B)^{-1}\,  \Delta \,  (\i\omega_n + \ov{\hfrak_B})^{-1} \, \ov \Delta \,  (\i\omega_n - \hfrak_B)^{-1}\, \Delta \, (\i\omega_n + \ov{\hfrak_B})^{-1}. \label{DHS1:NTB_definition}
\end{align}
The expression $\langle \Delta, N_{T,B}(\Delta)\rangle$ will determine the quartic term in the Ginzburg--Landau functional. All higher order terms in the expansion of \eqref{DHS1:lncosh-operator_fundamental_theorem} in $\Delta$ will be summarized in a trace-class operator called $\Rcal_{T,B}(\Delta)$, whose local trace norm is small.



With the operators $L_{T,B}$ and $N_{T,B}$ at hand, we are in position to state a representation formula for the BCS functional. It serves as the fundamental equation, on which the proofs of Theorems \ref{DHS1:Main_Result} and \ref{DHS1:Main_Result_Tc} are based. In particular, it will be applied in the proofs of upper and lower bounds, and we therefore formulate the statement for a general state $\Gamma$ and not only for Gibbs states.

\begin{prop}[Representation formula for the BCS functional]
\label{DHS1:BCS functional_identity} \label{DHS1:BCS FUNCTIONAL_IDENTITY}
Let $\Gamma$ be an admissible state. For any $B>0$, let $\Psi\in \Hmag^1(Q_B)$ and let $\Delta \equiv \Delta_\Psi$ be as in \eqref{DHS1:Delta_definition}. For $T>0$ and if $V\alpha_*\in L^{\nicefrac 65}(\Rbb^3) \cap L^2(\Rbb^3)$, there is an operator $\Rcal_{T,B}(\Delta)\in \Scal^1$ such that
\begin{align}
\FBCS(\Gamma) - \FBCS(\Gamma_0)& \notag\\
&\hspace{-70pt}= - \frac 14 \langle \Delta, L_{T,B} \Delta\rangle + \frac 18 \langle \Delta, N_{T,B} (\Delta)\rangle + \Vert \Psi\Vert_{\Lmag^2(Q_B)}^2 \; \langle \alpha_*, V\alpha_*\rangle_{L^2(\Rbb^3)} \notag\\
&\hspace{-40pt}+ \Tr\bigl[\Rcal_{T,B}(\Delta)\bigr] \notag\\
&\hspace{-40pt}+ \frac{T}{2} \Hcal_0(\Gamma, \Gamma_\Delta) - \fint_{Q_B} \dd X \int_{\Rbb^3} \dd r \; V(r) \, \bigl| \alpha(X,r) - \alpha_*(r) \Psi(X)\bigr|^2, \label{DHS1:BCS functional_identity_eq}
\end{align}
where
\begin{align}
	\Hcal_0(\Gamma, \Gamma_\Delta) := \Tr_0\bigl[ \Gamma(\ln \Gamma - \ln \Gamma_\Delta) + (1 - \Gamma)(\ln(1-\Gamma) - \ln(1 - \Gamma_\Delta))\bigr] \label{DHS1:Relative_Entropy}
\end{align}
denotes the relative entropy of $\Gamma$ with respect to $\Gamma_\Delta$. Moreover, $\Rcal_{T,B}(\Delta)$ obeys the estimate
\begin{align*}
	\Vert\Rcal_{T,B}(\Delta) \Vert_1 \leq C \; T^{-5} \; B^3 \; \Vert \Psi\Vert_{\Hmag^1(Q_B)}^6.
\end{align*}
\end{prop}

The relative entropy defined in \eqref{DHS1:Relative_Entropy} is based on the weaker form of trace $\Tr_0$, whose introduction we postpone until \eqref{DHS1:Weak_trace_definition}.

The right side of \eqref{DHS1:BCS functional_identity_eq} should be read as follows. The first line yields the Ginzburg--Landau functional, see Theorem \ref{DHS1:Calculation_of_the_GL-energy} below. 
The second and third line consist of remainder terms. The second line is small in absolute value whereas the techniques used to bound the third line differ for upper and lower bounds. This is responsible for the different qualities of the upper and lower bounds in Theorems \ref{DHS1:Main_Result} and \ref{DHS1:Main_Result_Tc}, see \eqref{DHS1:Rcal_error_Definition}. For an upper bound, when choosing $\Gamma := \Gamma_\Delta$ as a trial state, the relative entropy term $\Hcal_0(\Gamma_\Delta, \Gamma_\Delta)=0$ drops out and the last term in \eqref{DHS1:BCS functional_identity_eq} can be estimated with the help of Proposition \ref{DHS1:Structure_of_alphaDelta}. The last term in \eqref{DHS1:BCS functional_identity_eq} is actually nonpositive by our assumptions on $V$ but we do not use this. For a lower bound, the third line needs to be bounded from below using a relative entropy estimate that we provide in Section \ref{DHS1:Lower Bound Part B}.

It remains to show that the first line of the right side of \eqref{DHS1:BCS functional_identity_eq} is indeed given by the Ginzburg--Landau functional.
In order to state the result, we need the function
\begin{align}
\hat{V\alpha_*}(p) := \int_{\Rbb^3} \dx\; \e^{-\i p\cdot x} \, V(x)\alpha_*(x), \label{DHS1:Gap_function}
\end{align}
which fixes our convention on the Fourier transform in this paper. 

\begin{thm}[Calculation of the GL energy]
\label{DHS1:Calculation_of_the_GL-energy} \label{DHS1:CALCULATION_OF_THE_GL-ENERGY}
Let Assumptions \ref{DHS1:Assumption_V} and \ref{DHS1:Assumption_KTc} (a) hold and let $D\in \Rbb$ be given. Then, there is a constant $B_0>0$ such that for any $0 < B \leq B_0$, any $\Psi\in \Hmag^2(Q_B)$, $\Delta \equiv \Delta_\Psi$ as in \eqref{DHS1:Delta_definition}, and $T = \Tc(1 - DB)$, we have
\begin{align}
- \frac 14 \langle \Delta, L_{T,B} \Delta\rangle + \frac 18 \langle \Delta, N_{T,B} (\Delta)\rangle + \Vert \Psi\Vert_{\Lmag^2(Q_B)}^2 \; \langle \alpha_*, V\alpha_*\rangle_{L^2(\Rbb^3)} & \notag\\
&\hspace{-40pt}= B^2\; \EGL(\Psi) + R(B). \label{DHS1:Calculation_of_the_GL-energy_eq}
\end{align}
Here,
\begin{align*}
|R(B)|\leq C \, B^3 \; \Vert\Psi\Vert_{\Hmag^2(Q_B)}^2 \;  \bigl[ 1 + \Vert \Psi\Vert_{\Hmag^1(Q_B)}^2 \bigr]
\end{align*}
and with the functions
%
\begin{align}
g_1(x) &:= \frac{\tanh(x/2)}{x^2} - \frac{1}{2x}\frac{1}{\cosh^2(x/2)}, & g_2(x) &:= \frac 1{2x} \frac{\tanh(x/2)}{\cosh^2(x/2)}, \label{DHS1:XiSigma}
\end{align}
the coefficients $\Lambda_0$, $\Lambda_2$, and $\Lambda_3$ in $\EGL$ are given by
\begin{align}
\Lambda_0 &:= \frac{\betac^2}{16} \int_{\Rbb^3} \frac{\dd p}{(2\pi)^3} \; |(-2)\hat{V\alpha_*}(p)|^2 \; \bigl( g_1 (\betac(p^2-\mu)) + \frac 23 \betac \, p^2\, g_2(\betac(p^2-\mu))\bigr), \label{DHS1:GL-coefficient_1}\\
\Lambda_2 &:= \frac{\betac}{8} \int_{\Rbb^3} \frac{\dd p}{(2\pi)^3} \; \frac{|(-2)\hat{V\alpha_*}(p)|^2}{\cosh^2(\frac{\betac}{2}(p^2 -\mu))},\label{DHS1:GL-coefficient_2} \\
\Lambda_3 &:= \frac{\betac^2}{16} \int_{\Rbb^3} \frac{\dd p}{(2\pi)^3} \; |(-2) \hat{V\alpha_*}(p)|^4 \;  \frac{g_1(\betac(p^2-\mu))}{p^2-\mu}.\label{DHS1:GL_coefficient_3}
\end{align}
%
%
\end{thm}


Let us comment on the positivity of the coefficients \eqref{DHS1:GL-coefficient_1}-\eqref{DHS1:GL_coefficient_3}. First, $\Lambda_2$ is trivially positive. 
Since $g_1(x)/x >0$ for all $x\in \Rbb$, the coefficient $\Lambda_3$ is positive as well. It cannot be immediately seen that $\Lambda_0$ is positive, however. In order to prove this, we introduce the positive function 
\begin{align*}
g_3(x) &:= \frac{2}{x^2} \frac{1}{\cosh^2(x/2)} - \frac{1}{x} \frac{1}{\tanh(x/2)} \frac{1}{\cosh^2(x/2)}
\end{align*}
and compute
\begin{align}
2\Re \langle \alpha_* , x_i (K_{\Tc} - V) x_i \alpha_*\rangle &= (2\pi)^{-3} \langle \hat{V\alpha_*} , K_{\Tc}(p)^{-1} [ -\i \partial_{p_i} , [K_{\Tc}(p) , -\i \partial_{p_i}]] K_{\Tc}(p)^{-1}  \hat {V\alpha_*}\rangle \notag\\
&= 8\, \Lambda_0 - 2 \betac^3 \int_{\Rbb^3} \frac{\dd p}{(2\pi)^3} \; |\hat{V\alpha_*}(p)|^2 \; p_i^2 \, g_3(\betac(p^2-\mu)). \label{DHS1:GL-coefficient_1_positive}
\end{align}
Since the left side is nonnegative, this proves that $\Lambda_0 >0$. The idea for this proof is borrowed from \cite[Eq. (1.22)]{Hainzl2012}.

Let us comment on the connection between \eqref{DHS1:Calculation_of_the_GL-energy_eq} and \cite{Hainzl2017}. The two-particle Birman--Schwinger operator $1 - V^{\nicefrac 12} L_{T,B} V^{\nicefrac 12}$ has been intensively studied in \cite{Hainzl2017} 
to identify temperature regimes, where the bottom of its spectrum is positive or negative. This operator also appears in \eqref{DHS1:Calculation_of_the_GL-energy_eq} because
\begin{align}
- \frac 14 \langle \Delta, L_{T,B} \Delta\rangle + \Vert \Psi\Vert_2^2 \, \langle \alpha_*, V\alpha_*\rangle = \bigl\langle V^{\nicefrac 12} \alpha_* \Psi, \bigl( 1 - V^{\nicefrac 12} L_{T,B} V^{\nicefrac 12} \bigr) V^{\nicefrac 12} \alpha_* \Psi\bigr\rangle. \label{DHS1:Birman-Schwinger_LTB}
\end{align}
That is, the question whether the bottom of the spectrum of $1 - V^{\nicefrac 12} L_{T,B} V^{\nicefrac 12}$ is positive or negative is intimately related to the sign of \eqref{DHS1:Calculation_of_the_GL-energy_eq}, and thus of \eqref{DHS1:BCS functional_identity_eq} and \eqref{DHS1:BCS GS-energy}. Accordingly, it is related to the question whether the systems displays superconductivity or not. We highlight that the operator on the right side of \eqref{DHS1:Birman-Schwinger_LTB} acts on functions in  $L^2(\mathbb{R}^6)$ in \cite{Hainzl2017}, while it acts on $\Lsymm$ in our case. Since the lowest eigenvalue of the operator $(-\i\nabla + 2 \Abold)^2$ equals $2B$ when understood to act on $L^2(\Rbb^3)$ or on $\Lmag^2(Q_B)$, we obtain the same asymptotic behavior of $\Tc(B)$ as in \cite[Theorem~4]{Hainzl2017}.

%

Theorem \ref{DHS1:Calculation_of_the_GL-energy} is valid for the precise temperature scaling $T = \Tc(1 - DB)$. In order to prove Theorem \ref{DHS1:Main_Result_Tc} (a), we also need to show that the system is superconducting for temperatures that are small compared to $\Tc (1- DB)$. This is guaranteed by the following proposition.

\begin{prop}[A priori bound on Theorem \ref{DHS1:Main_Result_Tc} (a)]
\label{DHS1:Lower_Tc_a_priori_bound}
Let Assumptions \ref{DHS1:Assumption_V} and \ref{DHS1:Assumption_KTc} (a) hold. Then, for every $T_0 > 0$ there are constants $B_0>0$ and $D_0>0$ such that for all $0 < B \leq B_0$ and all temperatures $T$ obeying
\begin{align*}
T_0 \leq T < \Tc (1 - D_0 B),
\end{align*}
there is an admissible BCS state $\Gamma$ with
\begin{align}
\FBCS(\Gamma) - \FBCS(\Gamma_0) < 0. \label{DHS1:Lower_critical_shift_2}
\end{align}
\end{prop}


\subsection{The upper bound on \texorpdfstring{(\ref{DHS1:ENERGY_ASYMPTOTICS})}{(\ref{DHS1:ENERGY_ASYMPTOTICS})} and proof of Theorem \ref{DHS1:Main_Result_Tc} (a)}
\label{DHS1:Upper_Bound_Proof_Section}

Using the results in the previous section, we provide the proofs of the upper bound on \eqref{DHS1:ENERGY_ASYMPTOTICS} and of Theorem \ref{DHS1:Main_Result_Tc} (a). The statements in the previous section, that is, Propositions \ref{DHS1:Structure_of_alphaDelta} and \ref{DHS1:BCS functional_identity}, as well as Theorem \ref{DHS1:Calculation_of_the_GL-energy} are proven in Section \ref{DHS1:Proofs}.

\begin{proof}[Proof of the upper bound on \eqref{DHS1:ENERGY_ASYMPTOTICS}]
Let $D\in \Rbb$ be given, let $D_0 := 1 +|D|$, and let $\Psi$ be a minimizer of the Ginzburg--Landau functional, i.e., $\mathcal E_{D,B}^{\mathrm{GL}}(\Psi) = \EGLGSE$. We note that $\Psi$ belongs to $\Hmag^2(Q_B)$ and has uniformly bounded $\Hmag^2(Q_B)$-norm. Let $\Delta \equiv\Delta_\Psi$ be as in \eqref{DHS1:Delta_definition} and let $T = \Tc(1 - DB)$. We apply Proposition \ref{DHS1:BCS functional_identity} with the choice $\Gamma = \Gamma_\Delta$ and find
\begin{align}
\FBCS(\Gamma_\Delta) - \FBCS(\Gamma_0) 
%
%
&\leq - \frac 14 \langle \Delta, L_{T,B} \Delta\rangle + \frac 18 \langle \Delta, N_{T,B} (\Delta)\rangle + \Vert \Psi\Vert_2^2 \, \langle \alpha_*, V\alpha_*\rangle \notag\\
&\hspace{-20pt} - \int_{Q_B} \dd X \int_{\Rbb^3} \dd r \; V(r) \, \bigl| \alpha_\Delta(X,r) - \alpha_*(r) \Psi(X)\bigr|^2 + C B^3.\label{DHS1:Upper_Bound_proof_1}
\end{align}
The first term in the last line is bounded by $\Vert V\Vert_\infty \Vert \eta\Vert_2^2$ and a bound for the $L^2$-norm of $\eta := \eta_0 + \eta_\perp$ is provided by \eqref{DHS1:alphaDelta_decomposition_eq2} and \eqref{DHS1:alphaDelta_decomposition_eq3}. In fact, by Assumption \ref{DHS1:Assumption_V}, this term is nonpositive but we do not need to use this here. By Theorem~\ref{DHS1:Calculation_of_the_GL-energy}, this implies
\begin{align*}
F^{\mathrm{BCS}}(\Tc(1 - DB), B) &\leq B^2 \, \EGLGSE + C B^3,
\end{align*}
which concludes the proof of the upper bound on \eqref{DHS1:ENERGY_ASYMPTOTICS}.
\end{proof}


\begin{proof}[Proof of Theorem~\ref{DHS1:Main_Result_Tc}~(a)]
Let $D_0 > 0$ be given and let us recall the definition of $\Dc$ in \eqref{DHS1:Dc_Definition}. We show that there is a constant $D_1>0$ and appropriate trial states such that \eqref{DHS1:Lower_critical_shift_2} holds for all temperatures $T$ obeying
\begin{align}
\Tc(1 - D_0 B) \leq T < \Tc (1 -  \Dc\, B - D_1 \, B^{\nicefrac 32}), \label{DHS1:Lower_critical_shift_3}
\end{align}
provided $B>0$ is small enough. Since Proposition \ref{DHS1:Lower_Tc_a_priori_bound} covers the remaining range of $T$, this proves Theorem~\ref{DHS1:Main_Result_Tc}~(a).

We define $D := \frac{\Tc - T}{B\Tc}$ and note that \eqref{DHS1:Lower_critical_shift_3} yields $D - \Dc > D_1B^{\nicefrac 12}$. Let $\psi \in \Hmag^2(Q_1)$ be a ground state of the linear operator in \eqref{DHS1:Dc_Definition} and let $\Psi$ be as in \eqref{DHS1:GL-rescaling}. Accordingly, we have $(\Lambda_0 / \Lambda_2) \Pi^2 \Psi = B\Dc \Psi$ and
\begin{align*}
\inf_{\theta \in \Rbb} \EGL(\theta \Psi) = - \frac{\Lambda_2^2(D - \Dc)^2 \Vert \psi\Vert_2^4}{4 \Lambda_3 \Vert \psi\Vert_4^4},
\end{align*}
where the optimal $\mathrm{\theta_c}$ satisfies $\Lambda_2 (D - \Dc) \Vert \psi\Vert_2^2 = 2\Lambda_3 \Vert \psi\Vert_4^4 \, \mathrm{\theta_c}^2$. We combine Proposition~\ref{DHS1:BCS functional_identity} and Theorem \ref{DHS1:Calculation_of_the_GL-energy} applied to $\Gamma = \Gamma_\Delta$ with $\Delta = \Delta_{\mathrm{\theta_c}\Psi}$, to see that \eqref{DHS1:Upper_Bound_proof_1} holds in this case as well. Let us note that \eqref{DHS1:Lower_critical_shift_3} implies $|T - \Tc| \leq CB$. Proposition \ref{DHS1:Structure_of_alphaDelta} and \eqref{DHS1:Upper_Bound_proof_1} therefore allow us to conclude that
\begin{align}
\FBCS(\Gamma_\Delta) - \FBCS(\Gamma_0) &\leq - \frac{\Lambda_2^2 \Vert \psi\Vert_2^4}{4 \Lambda_3\Vert \psi\Vert_4^4} \; (D - \Dc)^2\; B^2 + CB^3.
\end{align}
The right side is negative provided $D_1>0$ is chosen large enough since $D - \Dc > D_1 B^{\nicefrac 12}$. This shows \eqref{DHS1:Lower_critical_shift_2} for temperatures $T$ satisfying \eqref{DHS1:Lower_critical_shift_3} and completes the proof of Theorem~\ref{DHS1:Main_Result_Tc}~(a).
\end{proof}


\section{Proofs of the Results in Section \ref{DHS1:Upper_Bound}}
\label{DHS1:Proofs}




\subsection{Schatten norm estimates for operators given by product kernels}
\label{DHS1:Estimates_on_product_wave_functions_Section}

In this subsection we provide estimates for several norms of gauge-periodic operators with integral kernels given by product functions of the form $\tau(x-y) \Psi((x+y)/2)$, which will be used frequently in our proofs.

\begin{lem}
\label{DHS1:Schatten_estimate}
Let $B>0$, let $\Psi$ be a gauge-periodic function on $Q_B$ and let $\tau$ be an even and real-valued function on $\Rbb^3$. Moreover, let the operator $\alpha$ be defined via its integral kernel $\alpha(X,r) := \tau(r)\Psi(X)$, i.e., $\alpha$ acts as
\begin{align*}
\alpha f(x) &= \int_{\Rbb^3} \dd y \; \tau(x - y) \Psi\bigl(\frac{x+y}{2}\bigr) f(y), & f &\in L^2(\Rbb^3).
\end{align*}

\begin{enumerate}[(a)]
\item Let $p \in \{2,4,6\}$. If $\Psi\in \Lmag^p(Q_B)$ and $\tau \in L^{\frac {p}{p-1}}(\Rbb^3)$, then $\alpha \in \Scal^p$ and
\begin{align*}
\Vert \alpha\Vert_p \leq C \; \Vert \tau\Vert_{\frac{p}{p-1}} \; \Vert \Psi\Vert_p.
\end{align*}

\item For any $\nu > 3$, there is a $C_\nu >0$, independent of $B$, such that if $(1 +|\cdot|)^\nu \tau\in L^{\nicefrac 65}(\Rbb^3)$ and $\Psi\in \Lmag^6(Q_B)$, then $\alpha \in \Scal^\infty$ and
\begin{align*}
\Vert \alpha\Vert_\infty &\leq C_\nu \, B^{-\nicefrac 14} \; \max\{1 , B^{\nicefrac \nu 2}\} \; \Vert (1 + |\cdot|)^\nu \tau\Vert_{\nicefrac 65} \; \Vert \Psi\Vert_6.
\end{align*}
\end{enumerate}
\end{lem}

\begin{proof}
The case $p = 2$ of part (a) holds trivially with equality and $C = 1$. Since $\tau$ is even and real-valued, the kernel of $\alpha^*\alpha$ is given by
\begin{align*}
\alpha^* \alpha(x,y) &= \int_{\Rbb^3} \dd z \; \tau(x-z) \ov{\Psi\bigl( \frac{x+z}{2}\bigr)} \tau(z - y) \Psi\bigl( \frac{z+y}{2}\bigr).
\end{align*}
Using $\Vert \alpha\Vert_4^4 = \Vert \alpha^* \alpha\Vert_2^2$ and the change of variables $z \mapsto x-z$ and $y \mapsto x-y$, we see that
\begin{align*}
\Vert \alpha\Vert_4^4 &= \frac{1}{|Q_B|} \iint_{Q_B\times \Rbb^3} \dx \dy \; \Bigl| \int_{\Rbb^3} \dd z \; \tau(z) \, \tau(y-z) \,  \ov{\Psi\bigl( x - \frac z2\bigr)}\, \Psi\bigl( x-\frac{y+z}{2}\bigr)\Bigr|^2.
\end{align*}
By Hölder's inequality, Young's inequality and 
the fact that $|\Psi|$ is periodic, we see that
\begin{align*}
\Vert \alpha\Vert_4^4 &\leq \Vert \Psi\Vert_4^4 \; \int_{\Rbb^3} \dd y \, \Bigl| \int_{\Rbb^3} \dd z \; |\tau(z) \tau(y-z)|\Bigr|^2 \leq C  \; \Vert \tau\Vert_{\nicefrac 43}^4\; \Vert \Psi\Vert_4^4
\end{align*}
holds. This proves part (a) for $p =4$. If $p =6$ we use $\Vert \alpha\Vert_6^6 = \Vert \alpha \,\alpha^*\alpha\Vert_2^2$ and a similar change of variables to write
\begin{align*}
\Vert \alpha \alpha^* \alpha\Vert_2^2 &= \frac{1}{|Q_B|} \iint_{Q_B\times \Rbb^3} \dd x \dd y\; \Bigl| \iint_{\Rbb^3\times \Rbb^3} \dd z_1 \dd z_2 \; \tau(z_1) \, \tau(z_2 - z_1) \, \tau(y - z_2) \\
&\hspace{120pt} \times \Psi\bigl(x - \frac{z_1}{2}\bigr) \ov{ \Psi\bigl( x - \frac{x_1+ z_2}{2}\bigr)} \Psi\bigl( x - \frac{z_2 + y}{2}\bigr)\Bigr|^2.
\end{align*}
We thus obtain $\Vert \alpha\Vert_6^6 \leq \Vert \Psi\Vert_6^6 \; \Vert \tau * \tau * \tau \Vert_2^2$, which, in combination with Young's inequality, proves the claimed bound.

In case of part (b), we follow closely the strategy of the proof of \cite[Eq. (5.51)]{Hainzl2012}. Let $f,g\in L^2(\Rbb^3)$ and let $\chi_j$ denote the characteristic function of the cube with side length $\sqrt{2\pi B^{-1}}$ centered at $j\in \Lambda_B$. We estimate
\begin{align}
|\langle f, \alpha g\rangle| &\leq \sum_{j,k\in \Lambda_B}\iint_{\Rbb^3\times \Rbb^3} \dd x\dd y\; \bigl|\chi_j(x) f(x) \Psi\bigl( \frac{x+y}{2}\bigr) \tau(x - y)\chi_k(y)g(y)\bigr|. \label{DHS1:Schatten_estimate_1}
\end{align}
Let $|\cdot|_\infty$ and $| \cdot |$ denote the maximum norm and the euclidean norm on $\Rbb^3$, respectively. We observe that the estimates $|x - j|_\infty \leq \frac 12\sqrt{2\pi B^{-1}}$ and $|y - k|_\infty \leq \frac 12\sqrt{2\pi B^{-1}}$ imply $| \frac{x + y}{2} - \frac{j + k}{2}|_\infty \leq \frac 12\sqrt{2\pi B^{-1}}$. Accordingly, if $\chi_j(x)\chi_k(y)$ equals $1$, so does $\chi_{\frac{j+k}{2}}(\frac{x+y}{2})$ and we may replace $\Psi$ on the right side of \eqref{DHS1:Schatten_estimate_1} by $\chi_{\frac{j+k}{2}}\Psi$ without changing the term. The above bounds for $|x - j|_\infty$ and $|y - k|_\infty$ also imply $|j - k| \leq |x - y| + \sqrt{6\pi B^{-1}}$, which yields the lower bound
\begin{align}
|x - y| \geq \bigl[ |j -k| - \sqrt{6\pi B^{-1}}\bigr]_+. \label{DHS1:Schatten_estimate_2}
\end{align}
We choose $\nu >3$, insert the factor $(\sqrt{2\pi B^{-1}} + |x-y|)^\nu$ and its inverse in \eqref{DHS1:Schatten_estimate_1}, use \eqref{DHS1:Schatten_estimate_2} to estimate the inverse, apply Cauchy--Schwarz in the $x$-coordinate, and obtain
\begin{align*}
|\langle f, \alpha g\rangle| &\leq \sum_{j,k\in \Lambda_B} \bigl( \sqrt{2\pi B^{-1}} + \bigl[|j-k| - \sqrt{6\pi B^{-1}}\bigr]_+\bigr)^{-\nu} \; \Vert \chi_jf\Vert_2 \\
&\hspace{-20pt} \times \Bigl( \int_{\Rbb^3} \dd x \, \Bigl| \int_{\Rbb^3} \dd y\; \bigl| (\chi_{\frac{j+k}{2}} \Psi)\bigl( \frac{x+y}{2}\bigr) \bigl( \sqrt{2\pi B^{-1}} + |x-y| \bigr)^\nu \tau(x-y) \chi_k(y)g(y)\bigr| \; \Bigr|^2\Bigr)^{\nicefrac 12}.
\end{align*}
An application of Hölder's inequality in the $y$-coordinate shows that the second line is bounded by 
\begin{align*}
\Bigl\Vert \bigl|(\sqrt{2\pi B^{-1}} + |\cdot|)^\nu \tau \bigr|^{\nicefrac 65} * |\chi_kg|^{\nicefrac 65} \Bigr\Vert_{\nicefrac 53}^{\nicefrac 56} &\leq \bigl\Vert \bigl( \sqrt{2\pi B^{-1}} + |\cdot| \bigr)^\nu  \tau \bigr\Vert_{\nicefrac 65} \, \Vert \chi_kg \Vert_{2} 
%
\end{align*}
times $|Q_B|^{\nicefrac 16} \Vert \Psi\Vert_{\Lmag^6(Q_B)}$. We highlight that the $\Lmag^6(Q_B)$-norm is defined via a normalized integral, whence we needed to insert the factor of $|Q_B|^{-\nicefrac 16}$. 
Hence,
\begin{align*}
|\langle f, \alpha g\rangle| &\leq C B^{-\nicefrac 14} \, \Vert \Psi\Vert_6 \, \bigl\Vert \bigl( \sqrt{2\pi B^{-1}}  + |\cdot|\bigr)^\nu \tau\bigr\Vert_{\nicefrac 65} \\
&\hspace{50pt} \times \sum_{j,k\in \Lambda_B} \bigl( \sqrt{2\pi B^{-1}} + \bigl[ |j -k| - \sqrt{6\pi B^{-1}} \bigr]_+\bigr)^{-\nu} \; \Vert \chi_jf\Vert_2 \; \Vert \chi_kg\Vert_2 .
\end{align*}
For $\lambda >0$ we estimate $\Vert \chi_j f\Vert_2 \Vert \chi_k g\Vert_2 \leq \frac{\lambda}{2} \Vert \chi_j f\Vert_2^2 + \frac{1}{2 \lambda} \Vert \chi_kg\Vert_2^2$. In each term, we carry out one of the sums and optimize the resulting expression over $\lambda$. We find $\lambda = \Vert g\Vert_2 \; \Vert f\Vert_2^{-1}$ as well as
\begin{align*}
|\langle f, \alpha g\rangle| &\leq C B^{-\nicefrac 14}  \Vert f\Vert_2 \; \Vert g\Vert_2 \; \Vert \Psi\Vert_6 \; \frac{\Vert ( \sqrt{2\pi B^{-1}} +|\cdot|)^\nu \tau\Vert_{\nicefrac 65}}{(2\pi B^{-1})^{\nicefrac \nu 2}} \; \sum_{j\in \Zbb^3} \bigl( 1 + [ |j| - \sqrt{3}]_+\bigr)^{-\nu}.
\end{align*}
The fraction involving $\tau$ is bounded by $C_\nu\max\{1, B^{\nicefrac \nu 2}\} \Vert (1 + |\cdot|)^\nu \tau \Vert_{\nicefrac 65}$. This proves the claim.
\end{proof}


\subsection{Proof of Proposition \ref{DHS1:BCS functional_identity}}
\label{DHS1:BCS functional_identity_proof_Section}

We recall the definitions of $\Delta(X, r) = -2\, V\alpha_*(r)\, \Psi(X)$ in \eqref{DHS1:Delta_definition}, the Hamiltonian $H_\Delta$ in \eqref{DHS1:HDelta_definition} and $\Gamma_\Delta = (1 + \e^{\beta H_\Delta})^{-1}$ in \eqref{DHS1:GammaDelta_definition}. Throughout this section we assume that the function $\Psi$ in the definition of $\Delta$ is in $\Hmag^1(Q_B)$. From  Lemma~\ref{DHS1:Gamma_Delta_admissible}, which is proved in Section~\ref{DHS1:sec:proofofadmissibility} below, we know that $\Gamma_{\Delta}$ is an admissible BCS state in this case. We define the anti-unitary operator
\begin{align*}
\Jcal &:= \begin{pmatrix} 0 & J \\ -J & 0\end{pmatrix} 
\end{align*}
with $J$ defined below \eqref{DHS1:Gamma_introduction}. The operator $H_\Delta$ obeys the relation $\Jcal H_\Delta \Jcal^* = - H_\Delta$, which implies $\Jcal\Gamma_\Delta \Jcal^* = 1 - \Gamma_\Delta$. Using this and the cyclicity of the trace, we write the entropy of $\Gamma_{\Delta}$ as
\begin{align}
S(\Gamma_\Delta) = \frac 12 \Tr[ \varphi(\Gamma_\Delta)] \label{DHS1:entropy matrix},
\end{align}
where $\varphi(x) := -[x\ln(x) + (1 - x) \ln(1-x)]$ for $0 \leq x \leq 1$.

In order to rewrite the BCS functional, it is useful to introduce a weaker notion of trace per unit volume. More precisely, we call a gauge-periodic operator $A$ acting on $L^2(\Rbb^3)\oplus L^2(\Rbb^3)$ weakly locally trace class if $P_0AP_0$ and $Q_0AQ_0$ are locally trace class, where
\begin{align}
P_0 = \begin{pmatrix} 1 & 0 \\ 0 & 0 \end{pmatrix} \label{DHS1:P0}
\end{align}
and $Q_0 = 1-P_0$, and we define its weak trace per unit volume by
\begin{align}
\Tr_0 (A):= \Tr\bigl( P_0AP_0 + Q_0 AQ_0\bigr). \label{DHS1:Weak_trace_definition}
\end{align}
If an operator is locally trace class then it is also weakly locally trace class but the converse need not be true. It is true, however, in case of nonnegative operators. If an operator is locally trace class then its weak trace per unit volume and its usual trace per unit volume coincide. 

Before their appearance in the context of BCS theory in \cite{Hainzl2012,Hainzl2014}, weak traces of the above kind appeared in \cite{HLS05,FLLS11}. In \cite[Lemma 1]{HLS05} it has been shown that if two weak traces $\Tr_P$ and $\Tr_{P'}$ are defined via projections $P$ and $P'$ then $\Tr_{P}(A) = \Tr_{P'}(A)$ holds for appropriate $A$ if $P - P'$ is a Hilbert--Schmidt operator. 

%

Let $\Gamma$ be an admissible BCS state and recall the normal state $\Gamma_0$ in \eqref{DHS1:Gamma0}. 
In terms of the weak trace per unit volume, the BCS functional can be written as
\begin{align}
\FBCS(\Gamma) - \FBCS(\Gamma_0) \hspace{-90pt}& \notag\\
&= \frac 12 \Tr\bigl[  (H_0\Gamma - H_0\Gamma_0) - T \varphi(\Gamma) + T \varphi(\Gamma_0)\bigr] - \fint_{Q_B} \dd X \int_{\Rbb^3} \dd r \; V(r)\, |\alpha(X,r)|^2 \notag\\
&= \frac 12 \Tr_0\bigl[ (H_\Delta\Gamma_\Delta - H_0\Gamma_0) - T\varphi(\Gamma_\Delta) + T\varphi(\Gamma_0)\bigr] \label{DHS1:kinetic energy} \\
&\hspace{55pt} + \frac 12 \Tr_0\bigl[ (H_\Delta \Gamma - H_\Delta \Gamma_\Delta) - T \varphi(\Gamma) + T\varphi(\Gamma_\Delta)\bigr] \label{DHS1:relative_entropy_term}\\
&\hspace{55pt} - \frac 12 \Tr_0 \begin{pmatrix}
0 & \Delta \\ \ov \Delta & 0 \end{pmatrix} \Gamma - \fint_{Q_B} \dd X \int_{\Rbb^3} \dd r \; V(r)\, |\alpha(X,r)|^2. \label{DHS1:interaction-term}
\end{align}
Note that we added and subtracted the first term in \eqref{DHS1:kinetic energy} and that we added and subtracted the first term in \eqref{DHS1:interaction-term} to replace the Hamiltonian $H_0$ in \eqref{DHS1:relative_entropy_term} by $H_\Delta$. The operators inside the traces in \eqref{DHS1:kinetic energy} and \eqref{DHS1:relative_entropy_term} are not necessarily locally trace class, which is the reason we introduce the weak local trace. We also note that \eqref{DHS1:relative_entropy_term} equals $\frac T2$ times the relative entropy $ \Hcal_0(\Gamma, \Gamma_\Delta)$ of $\Gamma$ with respect to $\Gamma_{\Delta}$, defined in \eqref{DHS1:Relative_Entropy}. 
%
%

The first term in \eqref{DHS1:interaction-term} can be written as
\begin{align}
-\frac  12\Tr_0 \begin{pmatrix} 0 & \Delta \\ \ov \Delta & 0\end{pmatrix} \Gamma &
=2\Re \fint_{Q_B} \dd X \int_{\Rbb^3} \dd r\; (V\alpha_*)(r) \Psi(X) \; \ov \alpha(X,r). \label{DHS1:additional-term}
\end{align}
The integrands in \eqref{DHS1:interaction-term} and \eqref{DHS1:additional-term} are equal to
\begin{equation*}
-|\alpha(X,r)|^2 + 2\Re\alpha_*(r)\Psi(X) \; \ov \alpha (X,r)  = -\bigl|\alpha(X, r) - \alpha_*(r)\Psi(X)\bigr|^2 + \bigl|\alpha_*(r)\Psi(X)\bigr|^2. 
\end{equation*}
To rewrite \eqref{DHS1:kinetic energy} we need the following identities, whose proofs are straightforward computations:
\begin{align}
\Gamma_\Delta &= \frac 12 - \frac 12 \tanh\bigl( \frac \beta 2 H_\Delta\bigr), & \ln(\Gamma_\Delta) &= -\frac \beta 2 H_\Delta - \ln\bigl( 2\cosh\bigl( \frac \beta 2 H_\Delta\bigr)\bigr), \notag\\
1 - \Gamma_\Delta &= \frac 12 + \frac 12 \tanh\bigl( \frac\beta 2H_\Delta\bigr), & \ln(1 - \Gamma_\Delta) &= \frac\beta 2H_\Delta - \ln\bigl( 2\cosh\bigl( \frac \beta 2H_\Delta\bigr)\bigr).  \label{DHS1:GammaRelations}
\end{align}
Eq.~\eqref{DHS1:GammaRelations} implies
\begin{align}
\Gamma_\Delta \ln(\Gamma_\Delta) + (1-\Gamma_\Delta)\ln(1-\Gamma_\Delta) =  -\ln\bigl( 2\cosh\bigl( \frac \beta 2H_\Delta\bigr) \bigr) + \frac{\beta}{2} H_\Delta \tanh\bigl( \frac\beta 2 H_\Delta\bigr)\bigr),
\label{DHS1:GammaDeltaRelation}
\end{align}
as well as
\begin{align*}
\beta H_\Delta \Gamma_\Delta - \varphi(\Gamma_\Delta) &= \frac{\beta}{2} H_\Delta - \ln \bigl( 2 \cosh\bigl( \frac \beta 2H_\Delta\bigr)\bigr).
\end{align*}
This allows us to rewrite \eqref{DHS1:kinetic energy} as
\begin{align}
\frac 1{2\beta}\, \Tr_0\bigl[ (\beta H_\Delta\Gamma_\Delta - \beta H_0\Gamma_0) - \varphi(\Gamma_\Delta) + \varphi(\Gamma_0)\bigr] & \notag \\
&\hspace{-180pt}= \frac 14 \Tr_0 \bigl[ H_\Delta - H_0\bigr] - \frac 1{2\beta}\Tr_0 \bigl[ \ln\bigl( \cosh\bigl( \frac \beta 2H_\Delta\bigr)\bigr) - \ln\bigl( \cosh\bigl( \frac  \beta 2H_0\bigr)\bigr)\bigr]. \label{DHS1:trace-difference-ln}
\end{align}
We note that $H_\Delta - H_0$ is weakly locally trace class and that its weak trace equals $0$. This, in particular, implies that the second term on the right side of \eqref{DHS1:trace-difference-ln} is weakly locally trace class. To summarize, our intermediate result reads
\begin{align}
\FBCS(\Gamma) - \FBCS(\Gamma_0) \hspace{-60pt} & \notag\\
&= -\frac 1{2\beta}\Tr_0 \bigl[ \ln\bigl( \cosh\bigl( \frac \beta 2H_\Delta\bigr)\bigr) - \ln\bigl( \cosh\bigl( \frac  \beta 2H_0\bigr)\bigr)\bigr] \notag\\
&\hspace{30pt} + \Vert \Psi\Vert_{\Lmag^2(Q_B)}^2 \; \langle \alpha_*, V\alpha_*\rangle_{L^2(\Rbb^3)} \notag \\
&\hspace{30pt} +\frac{T}{2} \Hcal_0(\Gamma, \Gamma_\Delta) - \fint_{Q_B} \dd X\int_{\Rbb^3} \dd r\; V(r) \, \bigl|\alpha(X,r) - \alpha_*(r) \Psi(X)\bigr|^2.  \label{DHS1:BCS functional-intermediate}
\end{align}

In order to compute the first term on the right side of \eqref{DHS1:BCS functional-intermediate}, we need Lemma~\ref{DHS1:BCS functional_identity_Lemma} below. It is the main technical novelty of our trial state analysis and should be compared to the related part in the proof of \cite[Theorem~2]{Hainzl2012}. The main difference between our proof of Lemma~\ref{DHS1:BCS functional_identity_Lemma} and the relevant parts of the proof of \cite[Theorem~2]{Hainzl2012} is that we use the product representation of the hyperbolic cosine in \eqref{DHS1:cosh-Product} below instead of a Cauchy integral representation of the function $z \mapsto \ln(1+e^{-z})$. In this way we obtain better decay properties in the subsequent resolvent expansion, which simplifies the analysis considerably. 

As already noted above, the admissibility of $\Gamma_{\Delta}$ implies that the difference between the two operators in the first term on the right side of \eqref{DHS1:BCS functional-intermediate} is weakly locally trace class. We highlight that this is a nontrivial statement because each of the two operators separately does not share this property. We also highlight that our proof of Lemma~\ref{DHS1:BCS functional_identity_Lemma} does not require this as an assumption, it implies the statement independently.

In combination with \eqref{DHS1:BCS functional-intermediate}, Lemma~\ref{DHS1:BCS functional_identity_Lemma} below proves Proposition~\ref{DHS1:BCS functional_identity}.  Before we state the lemma, we recall the definitions of the operators $L_{T,B}$ and $N_{T,B}$ in \eqref{DHS1:LTB_definition} and \eqref{DHS1:NTB_definition}, respectively.


\begin{lem}
\label{DHS1:BCS functional_identity_Lemma}
Let $V\alpha_*\in L^{\nicefrac 65}(\Rbb^3)\cap L^2(\Rbb^3)$. For any $B>0$, any $\Psi\in \Hmag^1(Q_B)$, and any $T>0$, the operator
\begin{align*}
\ln\bigl( \cosh\bigl( \frac \beta 2H_\Delta\bigr)\bigr) - \ln \bigl( \cosh\bigl( \frac \beta 2H_0\bigr)\bigr) 
\end{align*}
is weakly locally trace class and its weak local trace equals
\begin{align}
-\frac 1{2\beta} \Tr_0\Bigl[ \ln\bigl( \cosh\bigl( \frac \beta 2H_\Delta\bigr)\bigr) - \ln \bigl( \cosh\bigl( \frac \beta 2H_0\bigr)\bigr) \Bigr] & \notag\\
%
%
&\hspace{-100pt}= -\frac 14\langle \Delta, L_{T,B}\Delta\rangle + \frac 18 \langle \Delta, N_{T,B}(\Delta)\rangle + \Tr\Rcal_{T,B}(\Delta). \label{DHS1:BCS functional_identity_Lemma_eq2}
\end{align}
The operator $\Rcal_{T,B}(\Delta)$ is locally trace class and its trace norm satisfies the bound
\begin{align*}
\Vert\Rcal_{T,B}(\Delta) \Vert_1 &\leq C\; T^{-5} \; B^3 \; \Vert \Psi\Vert_{\Hmag^1(Q_B)}^6.
\end{align*}
\end{lem}



\begin{proof}[Proof of Lemma \ref{DHS1:BCS functional_identity_Lemma}]
We recall the Matsubara frequencies in \eqref{DHS1:Matsubara_frequencies} and write the hyperbolic cosine in terms of the following product expansion, see \cite[Eq. (4.5.69)]{Handbook}, 
\begin{align}
\cosh\bigl(\frac \beta 2x\bigr) &= \prod_{k=0}^\infty \Bigl( 1 + \frac{x^2}{\omega_k^2}\Bigr). 
\label{DHS1:cosh-Product}
\end{align}
We have
\begin{align*}
0 &\leq \sum_{k= 0}^\infty \ln \bigl( 1 + \frac{x^2}{\omega_k^2}\bigr) = \ln \bigl( \cosh\bigl( \frac \beta 2 x\bigr)\bigr) \leq \frac\beta 2 \; |x|, & x &\in \Rbb,
\end{align*}
and accordingly
\begin{align*}
\ln\bigl( \cosh\bigl( \frac{\beta}{2} H_\Delta\bigr)\bigr) =  \sum_{k=0}^\infty \ln \bigl( 1+ \frac{H_\Delta^2}{\omega_k^2}\bigr) 
\end{align*}
holds in a strong sense on the domain of $|H_\Delta|$. Since $\Delta$ is a bounded operator by Lemma \ref{DHS1:Schatten_estimate}, the domains of $|H_\Delta|$ and $|H_0|$ coincide. The identity 
\begin{align}
\ln\bigl( \cosh\bigl( \frac{\beta}{2} H_\Delta\bigr)\bigr) - \ln\bigl( \cosh\bigl( \frac{\beta}{2} H_0\bigr)\bigr)  =  \sum_{k=0}^\infty \bigl[ \ln \bigl( \omega_k^2+ H_\Delta^2 \bigr) - \ln \bigl( \omega_k^2 + H_0^2 \bigr) \bigr] \label{DHS1:BCS functional_identity_Lemma_1}
\end{align}
therefore holds in a strong sense on the domain of $|H_0|$. Elementary arguments show that
\begin{align}
\ln\bigl( \omega^2 + H_\Delta^2\bigr) - \ln\bigl(\omega^2 + H_0^2\bigr) = -\lim_{R\to\infty} \int_\omega^R \dd u \; \Bigl[\frac{2u}{u^2 + H_\Delta^2} - \frac{2u }{u^2 + H_0^2}\Bigr] \label{DHS1:ln-integral}
\end{align}
holds for $\omega>0$ in a strong sense on the domain of $\ln(1+|H_0|)$. Therefore, by \eqref{DHS1:BCS functional_identity_Lemma_1} and \eqref{DHS1:ln-integral}, we have
\begin{align}
\ln\bigl( \cosh\bigl(\frac \beta 2 H_\Delta\bigr)\bigr) - \ln\bigl( \cosh\bigl( \frac \beta 2 H_0\bigr)\bigr) & \notag \\
&\hspace{-100pt}= -\i \sum_{k=0}^\infty \int_{\omega_k}^\infty \dd u \; \Bigl[\frac{1}{\i u - H_\Delta } - \frac{1}{\i u - H_0} + \frac{1}{\i u + H_\Delta} - \frac{1}{\i u + H_0}\Bigr] \label{DHS1:lncosh-difference}
\end{align}
in a strong sense on the domain of $|H_0|$. By a slight abuse of notation, we have incorporated the limit in \eqref{DHS1:ln-integral} into the integral.
%
%
%
%
%
%
%
%
%
%

In the next step we use the resolvent expansion
\begin{align}
(z-H_\Delta)^{-1} = (z-H_0)^{-1} + (z-H_0)^{-1} \; (H_\Delta - H_0)\; (z-H_\Delta)^{-1} \label{DHS1:Resolvent_Equation}
\end{align}
to see that the right side of \eqref{DHS1:lncosh-difference} equals
\begin{align*}
\Ocal_1 + \Dcal_2 + \Ocal_3 + \Dcal_4 + \Ocal_5 - 2\beta \,  \Rcal_{T,B}(\Delta),
\end{align*}
with two diagonal operators $\Dcal_2$ and $\Dcal_4$, three offdiagonal operators $\Ocal_1$, $\Ocal_3$ and $\Ocal_5$ and a remainder term $\Rcal_{T,B}(\Delta)$. The index of the operators reflects the number of $\delta$ matrices appearing in their definition. The diagonal operators $\Dcal_2$ and $\Dcal_4$ are given by
\begin{align*}
\Dcal_2 &:=  -\i \sum_{k=0}^\infty\int_{\omega_k}^\infty \dd u \; \Bigl[\frac{1}{\i u - H_0}\delta \frac{1}{\i u - H_0}\delta \frac{1}{\i u - H_0} + \frac{1}{\i u + H_0}\delta \frac{1}{\i u + H_0}\delta \frac{1}{\i u + H_0} \Bigr] , \\
\Dcal_4 &:=  -\i \sum_{k=0}^\infty\int_{\omega_k}^\infty \dd u \; \Bigl[\frac{1}{\i u - H_0}\delta \frac{1}{\i u - H_0}\delta \frac{1}{\i u - H_0}\delta \frac{1}{\i u - H_0}\delta \frac{1}{\i u - H_0} \\
&\hspace{120pt}+ \frac{1}{\i u + H_0}\delta \frac{1}{\i u + H_0}\delta \frac{1}{\i u + H_0}\delta \frac{1}{\i u + H_0}\delta \frac{1}{\i u + H_0}\Bigr]
\end{align*}
and the offdiagonal operators read
\begin{align*}
\Ocal_1 &:= -\i \sum_{k=0}^\infty\int_{\omega_k}^\infty \dd u \; \Bigl[\frac{1}{\i u - H_0}\delta \frac{1}{\i u - H_0} + \frac{1}{\i u + H_0}\delta \frac{1}{\i u + H_0}\Bigr], \\
\Ocal_3 &:= -\i \sum_{k=0}^\infty\int_{\omega_k}^\infty \dd u \; \Bigl[\frac{1}{\i u - H_0}\delta \frac{1}{\i u - H_0}\delta \frac{1}{\i u - H_0}\delta \frac{1}{\i u - H_0} \\
&\hspace{85pt}+ \frac{1}{\i u + H_0}\delta \frac{1}{\i u + H_0}\delta \frac{1}{\i u + H_0}\delta \frac{1}{\i u + H_0}\Bigr],\\
\Ocal_5 &:=  -\i \sum_{k=0}^\infty\int_{\omega_k}^\infty \dd u \; \Bigl[\frac{1}{\i u - H_0}\delta \frac{1}{\i u - H_0}\delta \frac{1}{\i u - H_0}\delta \frac{1}{\i u - H_0}\delta \frac{1}{\i u - H_0}\delta \frac{1}{\i u - H_0} \\
&\hspace{85pt}+ \frac{1}{\i u + H_0}\delta \frac{1}{\i u + H_0}\delta \frac{1}{\i u + H_0}\delta \frac{1}{\i u + H_0}\delta \frac{1}{\i u + H_0}\delta \frac{1}{\i u + H_0}\Bigr].
\end{align*}
Since the operators $\Ocal_1$, $\Ocal_3$, and $\Ocal_5$ are offdiagonal, they are weakly locally trace class and their weak local trace equals $0$. We also note that the operator $\Ocal_1$ is not necessarily locally trace class, which is why we need to work with the weak local trace. The operator $\Rcal_{T,B}(\Delta)$ is defined by
\begin{align*}
\Rcal_{T,B}(\Delta) & \\
&\hspace{-40pt}:= \frac{\i}{2\beta} \sum_{k=0}^\infty\int_{\omega_k}^\infty \dd u \; \Bigl[\frac{1}{\i u - H_0}\delta \frac{1}{\i u - H_0}\delta \frac{1}{\i u - H_0}\delta \frac{1}{\i u - H_\Delta}\delta \frac{1}{\i u - H_0}\delta \frac{1}{\i u - H_0}\delta \frac{1}{\i u - H_0} \\
&\hspace{50pt}+ \frac{1}{\i u + H_0}\delta \frac{1}{\i u + H_0}\delta \frac{1}{\i u + H_0}\delta \frac{1}{\i u + H_\Delta}\delta \frac{1}{\i u + H_0}\delta \frac{1}{\i u + H_0}\delta \frac{1}{\i u + H_0}\Bigr].
\end{align*}


It remains to compute the traces of $\Dcal_2$ and $\Dcal_4$, and to estimate the trace norm of $\Rcal_{T,B}(\Delta)$. We first consider $\Dcal_2$ and use Hölder's inequality in \eqref{DHS1:Schatten-Hoelder} to estimate
\begin{align}
  \Bigl\Vert \frac{1}{\i u \pm H_0} \delta \frac{1}{\i u \pm H_0} \delta \frac{1}{\i u \pm H_0}\Bigr\Vert_1 &\leq \Bigl\Vert \frac{1}{\i u \pm H_0} \Bigr\Vert_\infty^3 \; \Vert \delta\Vert_2^2 = \frac{2}{u^3} \; \Vert \Delta\Vert_2^2.
\end{align}
Therefore, Lemma \ref{DHS1:Schatten_estimate} shows that the combination of the series and the integral defining $\Dcal_2$ converges absolutely in local trace norm.
In particular, $\Dcal_2$ is locally trace class and we may arbitrarily interchange the trace, the sum, and the integral to compute its trace. We do this, use the cyclicity of the trace, and obtain
\begin{equation}
\Tr \Dcal_2 =  -\i \sum_{k=0}^\infty \int_{\omega_k}^\infty \dd u \; \Tr\Bigl[\Bigl(\frac{1}{\i u - H_0}\Bigr)^2 \delta \frac{1}{\i u - H_0}\delta + \Bigl(\frac{1}{\i u + H_0}\Bigr)^2\delta \frac{1}{\i u + H_0}\delta \Bigr]. \label{DHS1:eq:A7}
\end{equation}
Integration by parts shows
\begin{align*}
\int_{\omega_k}^\infty \dd u \; \Bigl(\frac{1}{\i u \pm H_0}\Bigr)^2 \delta \frac{1}{\i u \pm H_0}\delta &= -\i\; \frac{1}{\i \omega_k \pm H_0} \delta \frac{1}{\i \omega_k \pm H_0} \delta \\
&\hspace{80pt}- \int_{\omega_k}^\infty \dd u \; \frac{1}{\i u \pm H_0} \delta \Bigl( \frac{1}{\i u \pm H_0}\Bigr)^2 \delta,
\end{align*}
and another application of the cyclicity of the trace yields
\begin{align}
\Tr \int_{\omega_k}^\infty \dd u \; \Bigl(\frac{1}{\i u \pm H_0}\Bigr)^2 \delta \frac{1}{\i u \pm H_0}\delta &= -\frac \i 2 \; \Tr \frac{1}{\i \omega_k \pm H_0} \delta \frac{1}{\i \omega_k \pm H_0} \delta. \label{DHS1:BCS functional_identity_Lemma_2}
\end{align}
Note that
\begin{align}
\frac{1}{\i \omega_k \pm H_0}\,  \delta\,  \frac{1}{\i \omega_k \pm H_0} \, \delta 
&= \begin{pmatrix}
\frac{1}{\i \omega_k \pm \hfrak_B} \, \Delta \frac{1}{\i \omega_k \mp \ov{\hfrak_B}} \, \ov \Delta \\ & \frac{1}{\i \omega_k \mp \ov{\hfrak_B}}\,  \ov \Delta \frac{1}{\i \omega_k \pm \hfrak_B} \, \Delta
\end{pmatrix}. \label{DHS1:Calculation-entry}
\end{align}
We combine this with \eqref{DHS1:eq:A7} and \eqref{DHS1:BCS functional_identity_Lemma_2} and summarize the cases $\pm$ into a single sum over $n\in \Zbb$. This yields
\begin{align*}
-\frac{1}{2\beta}\Tr \Dcal_2 &= \frac{1}{2\beta}\sum_{n\in \Zbb} \Bigl\langle \Delta, \frac{1}{\i\omega_n - \hfrak_B} \Delta \frac{1}{\i\omega_n + \ov{\hfrak_B}}\Bigr\rangle =  - \frac 14\langle \Delta, L_{T,B}\Delta\rangle,
\end{align*}
where $L_{T,B}$ is the operator defined in \eqref{DHS1:LTB_definition}.

We argue as above to see that the integrand in the definition of $\Dcal_4$ is bounded by $C \Vert \Delta\Vert_4^4 \, u^{-5}$. Moreover, we have $\Vert \Delta\Vert_4^4 \leq CB^2 \Vert V\alpha_*\Vert_{\nicefrac 43}^4 \Vert \Psi\Vert_{\Hmag^1(Q_B)}^4$ by \eqref{DHS1:Magnetic_Sobolev} and Lemma \ref{DHS1:Schatten_estimate}.
Therefore, the integral and the sum in $\Dcal_4$ are absolutely convergent with respect to the local trace norm.
The trace of $\Dcal_4$ is computed similar to that of $\Dcal_2$. With $N_{T,B}$ defined in \eqref{DHS1:NTB_definition}, the result reads
\begin{align}
-\frac{1}{2\beta}\Tr \Dcal_4 &
= \frac 18\langle \Delta, N_{T,B}(\Delta)\rangle. \label{DHS1:NTB_size_bound_2}
\end{align} 
In case of $\Rcal_{T,B}(\Delta)$, we bound the trace norm of the operator inside the integral by $u^{-7}\Vert \Delta\Vert_6^6$. Using \eqref{DHS1:Magnetic_Sobolev} and Lemma~\ref{DHS1:Schatten_estimate}, we estimate the second factor by a constant times $\Vert V\alpha_*\Vert_{\nicefrac 65}^6 B^{-3}\Vert \Pi\Psi\Vert_2^6 \leq CB^3 \Vert \Psi\Vert_{\Hmag^1(Q_B)}^6$. Finally, integration over $u$ yields the term $6\pi^{-6} T^{-6} (2k+1)^{-6}$, which is summable in $k$. This proves the claimed bound for the trace norm of $\Rcal_{T,B}(\Delta)$.
%
\end{proof}


\subsection{Proof of Theorem \ref{DHS1:Calculation_of_the_GL-energy}}
\label{DHS1:Calculation_of_the_GL-energy_proof_Section}

\subsubsection{Magnetic resolvent estimates}
\label{DHS1:Magnetic_resolvent_estimates_Section}

In this preparatory subsection, we provide estimates for the magnetic resolvent kernel
\begin{align*}
G^z_B(x,y) &:= \frac{1}{z - \hfrak_B}(x,y), & x,y &\in \Rbb^3.
\end{align*}
We also introduce the function
\begin{align}
g_B^z(x) &:= G_B^z(x,0), &x &\in \Rbb^3. \label{DHS1:gB_definition}
\end{align}
The proof of the following statement can be found in \cite[Lemma 8]{Hainzl2017}. 

\begin{lem}
\label{DHS1:gB-identities}
For all $B\geq 0$, $z\in \Cbb \setminus [B,\infty)$ and $x,y\in \Rbb^3$ we have
\begin{enumerate}[(a)]
\item $g_B^z(-x) = g_B^z(x)$,
\item $G_B^z(x,y) = \e^{\i \frac{\Bbold}{2} \cdot (x\wedge y)} \; g_B^z(x-y)$.
\end{enumerate}
\end{lem}

We start our analysis by providing a decay estimate for the $L^1$-norm of the resolvent kernel $g^z_0$ in \eqref{DHS1:gB_definition} and its gradient in the case $B=0$. For $g_0^z$ such an estimate has been provided in \cite[Lemma~9]{Hainzl2017}. Since we additionally need an estimate for $\nabla g_0^z$, we repeat some of the arguments here.

\begin{lem}
\label{DHS1:g0_decay}
Let $a > -2$. There is a constant $C_a >0$ such that for $t,\omega\in \Rbb$, we have
\begin{align}
\left \Vert \, |\cdot|^a g_0^{\i \omega + t}\right\Vert_1 &\leq C_a \; f(t, \omega)^{1+ \frac a2}, 
\label{DHS1:g0_decay_1}
\end{align}
where
\begin{align}
f(t, \omega) := \frac{|\omega| + |t + \mu|}{(|\omega| + (t + \mu)_-)^2} \label{DHS1:g0_decay_f}
\end{align}
and $x_- := -\min\{x,0\}$. Furthermore, for any $a > -1$, there is a constant $C_a >0$ with
\begin{align}
\left \Vert \, |\cdot|^a \nabla g_0^{\i\omega + t} \right\Vert_1 \leq C_a \; f(t, \omega)^{\frac 12 + \frac a2} \; \Bigl[ 1 + \frac{|\omega| + |t+ \mu|}{|\omega| + (t + \mu)_-}\Bigr]. \label{DHS1:g0_decay_2}
\end{align}
\end{lem}


\begin{proof}
The resolvent kernel $g_0^z$ is given by
\begin{align}
g_0^{z}(x) &= -\frac{1}{4\pi |x|} \; \e^{-\sqrt{-(z+ \mu)}\; |x|}, \label{DHS1:g0_definition}
\end{align}
where $\sqrt{\cdot}$ denotes the standard branch of the square root. As long as $a > - 2$ we have
\begin{align}
\left\Vert \, |\cdot|^a g_0^z\right\Vert_1 &= \int_{\Rbb^3} \dx \; |x|^a \; \Bigl| \frac{1}{4\pi |x|}\e^{-|x|\sqrt{-(z + \mu)}} \Bigr| = \frac{\Gamma(a +2)}{(\Re \sqrt{-(z +\mu)})^{a+2}}. \label{DHS1:Resolvent_kernel_integration}
\end{align}
Moreover,
\begin{align*}
( \Re \sqrt{-z})^2 = \frac 12 (|z| - \Re z) \geq \begin{cases} \frac 14 \frac{|\Im z|^2}{\Re z + |\Im z|} & \Re z\geq 0, \\ \frac 12 |z| & \Re z < 0, \end{cases} 
\end{align*}
and hence
\begin{align*}
(\Re \sqrt{-(t +\mu + \i\omega)})^2 \geq \frac 14 \frac{(|\omega| + (t+ \mu)_-)^2}{|\omega| + |t + \mu|}.
\end{align*}
This proves \eqref{DHS1:g0_decay_1}. To prove \eqref{DHS1:g0_decay_2}, we use \eqref{DHS1:g0_definition} 
and estimate
\begin{align*}
|\nabla g_0^z(x)|\leq |z+\mu|^{\nicefrac 12} \; |g_0^z(x)| +|x|^{-1}|g_0^z(x)|.
\end{align*}
This shows the second estimate for $a > - 1$.
\end{proof}

In the next step we prove estimates for the $L^1$-norms of $g_B^z$ and $g_B^z-g_0^z$ and the gradient of these functions if $B\neq 0$. Once more, some of the arguments in \cite[Lemma~10]{Hainzl2017} reappear in our proof below, ensuring self-consistency.

\begin{lem}
\label{DHS1:gB-g_decay}
For any $a\geq 0$, there are constants $\delta_a , C_a > 0$ such that for all $t, \omega\in \Rbb$ and for all $B \geq 0$ with $f(t, \omega)^2\,  B^2 \leq \delta_a$, we have 
\begin{align}
\bigl\Vert |\cdot|^a g_B^{\i\omega+ t}\bigr\Vert_1 &\leq C_a \, f(t,\omega)^{1+\frac a2}, \notag \\
\bigl\Vert |\cdot|^a \nabla g_B^{\i\omega+ t}\bigr\Vert_1 &\leq C_a \, f(t,\omega)^{\frac 12 + \frac a2} \Bigl[ 1 + \frac{|\omega| + |t+ \mu|}{|\omega| + (t + \mu)_-}\Bigr], \label{DHS1:gB-g_decay_2}
\end{align}
and
\begin{align}
\bigl\Vert \,|\cdot|^a ( g_B^{\i\omega+ t} - g_0^{\i\omega+ t} ) \bigr\Vert_1 &\leq C_a \, B^2 \, f(t,\omega)^{3+ \frac a2} , \notag \\
\bigl\Vert \,|\cdot|^a (\nabla g_B^{\i\omega+ t} - \nabla g_0^{\i\omega+ t} ) \bigr\Vert_1 &\leq C_a \, B^2 \, f(t,\omega)^{\frac 52 + \frac a2} \Bigl[ 1 + \frac{|\omega| + |t+ \mu|}{|\omega| + (t + \mu)_-}\Bigr] \label{DHS1:gB-g_decay_1}
\end{align}
with the function $f(t,\omega)$ in \eqref{DHS1:g0_decay_f}.
\end{lem}




\begin{proof}
During the proof we use the notation $z = \i \omega + t$.
We define the function
\begin{align}
h^z(x) := \frac 14\, |e_3\wedge x|^2 \, g_0^z(x) \label{DHS1:hz_definition}
\end{align}
and choose $\delta_a$ such that $2 \delta_a D_a C_2 =1$. Here $C_2$ denotes the constant in \eqref{DHS1:g0_decay_1} and $D_a := 1$ if $0 \leq a \leq 1$ and $D_a := 2^a$ if $a>1$. Lemma~\ref{DHS1:g0_decay} and the bound $\Vert h^z\Vert_1\leq \Vert |\cdot|^2g_0^z\Vert_1$ imply
\begin{align}
B^2 D_a \Vert h^z\Vert_1 \leq \frac 12 \label{DHS1:hz_estimate}
\end{align}
for all $\omega$, $t$, and $B$ that are allowed by our assumptions. We define the operator $\tilde G_B^z$ by the kernel
\begin{align*}
\tilde G_B^z(x,y) &:= \e^{\i \frac{\Bbold}{2} (x\wedge y)} g_0^z(x-y)
\end{align*}
and note that
\begin{align}
(z - \hfrak_B) \tilde G_B^z = 1 - T_B^z, \label{DHS1:gB-g_1}
\end{align}
where $T_B^z$ is the operator given by the kernel
\begin{align*}
T_B^z(x,y) &:= \e^{\i \frac \Bbold 2(x\wedge y)} \bigl[ \Bbold \wedge (x-y) (-\i\nabla_x)g_0^z(x-y) + B^2 \; h^z(x-y)\bigr].
\end{align*}
The first term in square brackets equals $0$ because $g_0^z$ is a radial function, which implies that the vector $\nabla g_0^z(x-y)$ is perpendicular to $\Bbold \wedge (x-y)$. 
Multiplication of \eqref{DHS1:gB-g_1} with $(z - \hfrak_B)^{-1}$ from the left yields
\begin{align*}
G_B^z(x,y) - \tilde G_B^z(x,y) = \int_{\Rbb^3} \dd v \; G_B^z(x,v) T_B^z(v,y).
\end{align*}
We set $y =0$, change variables $v\mapsto x - v$, and find
\begin{align}
g_B^z(x) - g_0^z(x) = B^2  \int_{\Rbb^3} \dd v \; \e^{\i \frac \Bbold 2\cdot (v\wedge x)} \; g_B^z(v) \; h^z(x-v). \label{DHS1:gB-g_2}
\end{align}
This implies
\begin{align}
\Vert g_B^z - g_0^z\Vert_1 &\leq B^2 \Vert |g_B^z| * |h^z| \Vert_1 \leq B^2 \; \Vert g_B^z - g_0^z\Vert_1 \; \Vert h^z\Vert_1 + B^2 \; \Vert g_0^z\Vert_1 \; \Vert h^z\Vert_1. \label{DHS1:gB-g_6}
\end{align}
A straightforward calculation involving \eqref{DHS1:gB-g_1} and the Neumann series shows that $g_B^z - g_0^z$ belongs to $L^1(\Rbb^3)$. Therefore, \eqref{DHS1:hz_estimate} and \eqref{DHS1:gB-g_6} imply
\begin{align}
\Vert g_B^z - g_0^z\Vert_1 &\leq \Vert g_0^z\Vert_1 \label{DHS1:gB-g_4}
\end{align}
for all $t, \omega$, and $B$ that are allowed by our assumptions.

We use this estimate as a basis to prove the bounds claimed in the lemma and start with the first bound in \eqref{DHS1:gB-g_decay_1}. By \eqref{DHS1:gB-g_2}, we have
\begin{align*}
\Vert \, |\cdot|^a (g_B^z - g_0^z)\Vert_1 &\leq D_a B^2 \bigl[ \Vert \, |\cdot|^a g_0^z\Vert_1 \, \Vert h^z\Vert_1 + \Vert \, |\cdot|^a (g_B^z - g_0^z)\Vert_1 \, \Vert h^z\Vert_1 \\
&\hspace{80pt} + \Vert g_B^z - g_0^z\Vert_1 \, \Vert |\cdot|^a h^z\Vert_1 + \Vert g_0^z\Vert_1 \, \Vert \, |\cdot|^a h^z\Vert_1 \bigr].
\end{align*}
A similar argument to the one above \eqref{DHS1:gB-g_4} shows that $|\cdot|^a (g_B^z - g_0^z)$ belongs to $L^1(\Rbb^3)$. In combination with \eqref{DHS1:gB-g_4}, we therefore obtain
\begin{align}
\Vert \, |\cdot|^a (g_B^z - g_0^z)\Vert_1 &\leq 2D_a B^2 \; \bigl[ \Vert |\cdot|^a g_0^z\Vert_1\, \Vert h^z\Vert_1 
+ 2\, \Vert g_0^z\Vert_1 \, \Vert |\cdot|^a h^z\Vert_1\bigr]. \label{DHS1:gB-g_7}
\end{align}
With the help of Lemma~\ref{DHS1:g0_decay} and \eqref{DHS1:hz_estimate}, we read off the first bound in \eqref{DHS1:gB-g_decay_1}. Moreover, the triangle inequality, Lemma~\ref{DHS1:g0_decay}, the first bound in \eqref{DHS1:gB-g_decay_1} and the bound $f(t, \omega)^2\,  B^2 \leq \delta_a$ imply the first bound in \eqref{DHS1:gB-g_decay_2}.

Next, we consider the bounds in \eqref{DHS1:gB-g_decay_2} and \eqref{DHS1:gB-g_decay_1} involving the gradient. As a preparation, the bound $|\nabla h^z (x)| \leq |x|\, |g_0^z(x)| + |x|^2\, |\nabla g_0^z(x)|$ and Lemma~\ref{DHS1:g0_decay} show
\begin{align}
\Vert \, |\cdot|^a \nabla h^z\Vert_1 &
\leq  C_a \; f(t, \omega)^{\frac 32 + \frac a2} \Bigl[ 1 + \frac{|\omega| +  |t+\mu|}{|\omega| + (t + \mu)_-}\Bigr]. \label{DHS1:gB-g_8}
\end{align}
We use $|\nabla \e^{\i \frac \Bbold 2 \cdot (v\wedge x)}| \leq B|v|$ and \eqref{DHS1:gB-g_2} to see that
\begin{align*}
|\nabla g_B^z(x) - \nabla g_0^z(x)| &\leq B^2 \int_{\Rbb^3} \dd v \; \bigl[ B\, |v|\, |g_B^z(v)| \, |h^z(x-v)| + |g_B^z(v)| \, |\nabla h^z(x-v)|\bigr]
\end{align*}
as well as
\begin{align}
\Vert\, |\cdot|^a (\nabla g_B^z - \nabla g_0^z)\Vert_1 &\leq D_a B^2  \bigl[ B\, \Vert \, |\cdot|^{a+1} g_B^z\Vert_1 \, \Vert h^z\Vert_1 + B \, \Vert \, |\cdot| g_B^z\Vert_1 \, \Vert \, |\cdot|^a h^z\Vert_1 \notag\\
&\hspace{60pt} + \Vert \, |\cdot|^a g_B^z\Vert_1 \; \Vert \nabla h^z\Vert_1 + \Vert  g_B^z\Vert_1 \; \Vert \,|\cdot|^a\nabla h^z\Vert_1\bigr].  \label{DHS1:eq:A9}
\end{align}
When we combine \eqref{DHS1:eq:A9}, the first estimates in \eqref{DHS1:gB-g_decay_2} and \eqref{DHS1:gB-g_decay_1}, the bound in \eqref{DHS1:gB-g_8} and Lemma \ref{DHS1:g0_decay}, we see that
\begin{align*}
\Vert\, |\cdot|^a (\nabla g_B^z - \nabla g_0^z)\Vert_1 &\leq C_a \, B^2 \, f(t,\omega)^{\frac 52 + \frac a2} \bigl[ 1 + B f(t,\omega)\bigr].
\end{align*}
An application of the assumption $B^2f(t,\omega)^2 \leq \delta_a$ proves the second bound in \eqref{DHS1:gB-g_decay_1}. Finally, the triangle inequality, the second bound in \eqref{DHS1:gB-g_decay_1}, and Lemma \ref{DHS1:g0_decay} show
\begin{align*}
\Vert \, |\cdot|^a \nabla g_B^z\Vert_1 
&\leq C_a \, f(t,\omega)^{\frac 12 + \frac a2} \Bigl[ 1 + \frac{|\omega| + |t + \mu|}{|\omega| + (t + \mu)_-}\Bigr] \bigl[ 1 + B^2 f(t,\omega)^2\bigr].
\end{align*}
Another application of $B^2f(t,\omega)^2 \leq \delta_a$ on the right side proves the second bound in \eqref{DHS1:gB-g_decay_2}.
\end{proof}

\subsubsection{A representation formula for \texorpdfstring{$L_{T,B}$}{LTB} and an outlook on the quadratic terms}

In this subsection we compute the terms in \eqref{DHS1:Calculation_of_the_GL-energy_eq} involving the linear operator $L_{T,B}$ defined in \eqref{DHS1:LTB_definition}. Our starting point is the representation formula for $L_{T,B}$ in \cite[Lemma 11]{Hainzl2017},
which expresses the operator explicitly in terms of the relative and the center-of-mass coordinate.

\begin{lem}
\label{DHS1:LTB_action}
The operator $L_{T,B} \colon \Lsymm \ra \Lsymm$ in \eqref{DHS1:LTB_definition} acts as
\begin{align*}
(L_{T,B}\alpha) (X,r) &= \iint_{\Rbb^3\times \Rbb^3} \dd Z \dd s \; k_{T,B}(Z, r, s) \; (\cos(Z\cdot \Pi_X)\alpha) (X,s)
\end{align*}
with
\begin{align}
k_{T,B}(Z, r, s) &:= \frac 2\beta \sum_{n\in \Zbb} k_{T,B}^n(Z, r- s) \; \e^{\frac{\i}{4} \Bbold \cdot(r\wedge s)} \label{DHS1:kTB_definition}
\end{align}
and
\begin{align}
k_{T,B}^n(Z,r) &:= g_B^{\i\omega_n} \bigl( Z - \frac{r}{2}\bigr) \; g_B^{-\i\omega_n} \bigl( Z + \frac{r}{2}\bigr). \label{DHS1:kTBn_definition}
\end{align}
\end{lem}

We analyze the operator $L_{T,B}$ in three steps. In the first two steps we introduce two operators of increasing simplicity in their dependence on $B$:
\begin{align}
L_{T,B} = (L_{T,B} - \tilde L_{T,B}) + (\tilde L_{T,B} - M_{T,B}) + M_{T,B}, \label{DHS1:LTB_decomposition}
\end{align}
where $\tilde L_{T,B}$ and $M_{T,B}$ are defined below in \eqref{DHS1:LTBtilde_definition} and \eqref{DHS1:MTB_definition}, respectively. To obtain $\tilde L_{T,B}$ we replace the functions $g_B^z$ in the definition of $L_{T,B}$ by $g_0^z$. Moreover, $M_{T,B}$ is obtained from $\tilde L_{T,B}$ when we replace $k_{T,B}$ by $k_{T,0}$, i.e., when we additionally replace the magnetic phase $\e^{\frac{\i}{4} \Bbold \cdot(r\wedge s)}$ by $1$. In Section~\ref{DHS1:Approximation_of_LTB_Section} we prove that the terms in the brackets in \eqref{DHS1:LTB_decomposition} are small in a suitable sense. The third step consists of a careful analysis of the operator $M_{T,B}$, which takes place in Section~\ref{DHS1:Analysis_of_MTB_Section}. There, we expand the operator $\cos(Z\cdot \Pi_X)$ in powers of $Z\cdot \Pi_X$ up to second order and extract the quadratic terms of the Ginzburg--Landau functional in \eqref{DHS1:Definition_GL-functional} as well as a term that cancels the last term on the left side of \eqref{DHS1:Calculation_of_the_GL-energy_eq}. In Section~\ref{DHS1:Summary_quadratic_terms_Section} we summarize our findings.

We remark that the operator $\tilde L_{T,B}$ is called $M_{T,B}$ in \cite{Hainzl2017} and that $M_{T,B}$ is called $N_{T,B}$. The reason why we did not follow the notation in \cite{Hainzl2017} is that $N_{T,B}$ is reserved for the nonlinear term in the present paper. We note that our decomposition of $L_{T,B}$ in \eqref{DHS1:LTB_decomposition} already appeared in \cite{Hainzl2017}. Parts of our analysis follow the analysis of $L_{T,B}$ in Section~4 and Section~5 in that reference. However, we additionally need $\Hsymm$-norm bounds that are not provided in \cite{Hainzl2017}. It should also be noted that $L_{T,B}$ acts on $L^2(\mathbb{R}^6)$ in \cite{Hainzl2017}, while it acts on $\Lsymm$ in our case.

\subsubsection{Approximation of \texorpdfstring{$L_{T,B}$}{LTB}}
\label{DHS1:Approximation_of_LTB_Section}

\paragraph{The operator $\tilde L_{T,B}$.} The operator $\tilde L_{T,B}$ is defined by
\begin{align}
\tilde L_{T,B}\alpha(X,r) &:= \iint_{\Rbb^3\times \Rbb^3} \dd Z \dd s \; \tilde k_{T,B}(Z, r,s) \; (\cos(Z\cdot\Pi_X)\alpha)(X,s) \label{DHS1:LTBtilde_definition}
\end{align}
with
\begin{align}
\tilde k_{T,B} (Z, r,s) := \frac 2\beta \sum_{n\in \Zbb} k_{T,0}^n(Z, r-s) \; \e^{\frac{\i}{4} \Bbold \cdot (r\wedge s)} \label{DHS1:ktildeTB_definition}
\end{align}
and $k_{T,0}^n$ in \eqref{DHS1:kTBn_definition}. In the following proposition we provide an estimate that allows us to replace $L_{T,B}$ by $\widetilde{L}_{T,B}$ in our computations.

\begin{prop}
\label{DHS1:LTB-LtildeTB}
For any $T_0>0$ there is $B_0>0$ such that for any $0 < B \leq B_0$, any $T\geq T_0$ and whenever $|\cdot|^k V\alpha_*\in L^2(\Rbb^3)$ for $k\in \{0,1\}$, $\Psi\in \Hmag^1(Q_B)$, and $\Delta \equiv \Delta_\Psi$ as in \eqref{DHS1:Delta_definition}, we have
\begin{align*}
\Vert L_{T,B} \Delta - \tilde L_{T,B}\Delta\Vert_\Hsymm^2 \leq C \; B^5 \; \bigl(\Vert V\alpha_*\Vert_2^2 + \Vert \, |\cdot|V\alpha_*\Vert_2^2\bigr) \; \Vert \Psi\Vert_{\Hmag^1(Q_B)}^2.
\end{align*}
\end{prop}

\begin{bem}
	\label{DHS1:rem:A1}
	For the proof of Theorem~\ref{DHS1:Calculation_of_the_GL-energy} we only need a bound for $\langle \Delta, (L_{T,B} - \tilde L_{T,B})\Delta\rangle$, which is easier to obtain. This bound follows directly from Proposition~\ref{DHS1:LTB-LtildeTB}, Lemma~\ref{DHS1:Schatten_estimate} and an application of the Cauchy--Schwarz inequality. The more general bound in Proposition~\ref{DHS1:LTB-LtildeTB} is needed in the proof of Proposition~\ref{DHS1:Structure_of_alphaDelta}.
\end{bem}


Before we start with the proof of Proposition~\ref{DHS1:LTB-LtildeTB} let $a \geq 0$ and introduce the functions
\begin{align}
F_{T,B}^{a} &:= \smash{\frac 2\beta \sum_{n\in \Zbb}}  \; \bigl(|\cdot|^a \, | g_B^{\i\omega_n} - g_0^{\i\omega_n}|\bigr) * |g_B^{-\i\omega_n}| + |g_B^{\i\omega_n} - g_0^{\i\omega_n}| * \bigl(|\cdot|^a \, |g_B^{-\i\omega_n}| \bigr) \notag \\
&\hspace{50pt}+ \bigl(|\cdot|^a \, |g_0^{\i\omega_n}|\bigr) * |g_B^{-\i\omega_n} - g_0^{-\i\omega_n}| + |g_0^{\i\omega_n}| * \bigl(|\cdot|^a \, |g_B^{-\i\omega_n} -  g_0^{-\i\omega_n}|\bigr) \label{DHS1:LTB-LtildeTB_FTB_definition}
\end{align}
and
\begin{align}
G_{T,B} &:= \smash{\frac 2\beta \sum_{n\in \Zbb}}\; |\nabla g_B^{\i\omega_n} - \nabla g_0^{\i\omega_n}| * |g_B^{-\i\omega_n}| + |g_B^{\i\omega_n} - g_0^{\i\omega_n}| * |\nabla g_B^{-\i\omega_n}| \notag \\
&\hspace{80pt}+ |\nabla g_0^{\i\omega_n}| * |g_B^{-\i\omega_n} - g_0^{-\i\omega_n}| + |g_0^{\i\omega_n}| * |\nabla g_B^{-\i\omega_n} - \nabla g_0^{-\i\omega_n}|  \label{DHS1:LTB-LtildeTB_GTB_definition}
\end{align}
with the Matsubara frequencies $\omega_n$ in \eqref{DHS1:Matsubara_frequencies} and the resolvent kernel $g_B^z$ in \eqref{DHS1:gB_definition}. We claim that for any $a\geq 0$ there is a constant $B_0>0$ such that for $0\leq B\leq B_0$ we have
\begin{align}
\Vert F_{T,B}^a\Vert_1 + \Vert G_{T,B}\Vert_1 \leq C_a B^2.\label{DHS1:LTB-LtildeTB_FTBGTB}
\end{align}
To prove \eqref{DHS1:LTB-LtildeTB_FTBGTB} we note that the function $f(t, \omega)$ in \eqref{DHS1:g0_decay_f} satisfies
\begin{align}
f(0, \omega_n) &\leq C \; (T^{-1} + T^{-2}) \; |2n+1|^{-1} \label{DHS1:g0_decay_f_estimate1}
\end{align}
and that
\begin{align}
\frac{|\omega_n| + |\mu|}{|\omega_n| + \mu_-} \leq C \; (1+ T^{-1}). \label{DHS1:g0_decay_f_estimate2}
\end{align}
Since $T\geq T_0>0$, Lemmas~\ref{DHS1:g0_decay} and \ref{DHS1:gB-g_decay} prove \eqref{DHS1:LTB-LtildeTB_FTBGTB}.

\begin{proof}[Proof of Proposition \ref{DHS1:LTB-LtildeTB}]
We write
\begin{align}
\Vert L_{T,B}\Delta - \tilde L_{T,B} \Delta\Vert_\Hsymm^2 & \notag\\
&\hspace{-120pt}= \Vert L_{T,B}\Delta - \tilde L_{T,B} \Delta\Vert_2^2 + \Vert \Pi_X(L_{T,B}\Delta - \tilde L_{T,B} \Delta)\Vert_2^2 + \Vert \tilde \pi_r(L_{T,B}\Delta - \tilde L_{T,B} \Delta)\Vert_2^2 \label{DHS1:LTB-LtildeTB_8}
\end{align}
and claim that
\begin{align}
\Vert L_{T,B}\Delta - \tilde L_{T,B} \Delta\Vert_2^2 &\leq 4 \; \Vert \Psi\Vert_2^2 \; \Vert F_{T,B}^0 * |V\alpha_*| \, \Vert_2^2. \label{DHS1:LTB-LtildeTB_1}
\end{align}
If this is true, Young's inequality, \eqref{DHS1:Periodic_Sobolev_Norm}, and \eqref{DHS1:LTB-LtildeTB_FTBGTB} prove
\begin{align*}
	\Vert L_{T,B}\Delta - \tilde L_{T,B} \Delta\Vert_2^2 \leq C B^5 \; \Vert V \alpha_* \Vert_2^2 \; \Vert \Psi\Vert_{\Hmag^1(Q_B)}^2.
\end{align*}
To see that \eqref{DHS1:LTB-LtildeTB_1} holds, we expand the squared modulus in the Hilbert--Schmidt norm and obtain
\begin{align}
\Vert L_{T,B}\Delta - \tilde L_{T,B}\Delta\Vert_2^2 & \leq 4 \int_{\Rbb^3} \dd r \iint_{\Rbb^3\times \Rbb^3} \dd Z\dd Z' \iint_{\Rbb^3\times \Rbb^3} \dd s\dd s' \; |V\alpha_*(s)| \; |V\alpha_*(s')|  \notag\\
&\hspace{120pt} \times | k_{T,B}(Z, r, s) - \tilde k_{T,B} (Z, r, s)|  \notag\\
&\hspace{120pt} \times |k_{T,B}(Z', r, s') - \tilde k_{T,B} (Z', r, s')|  \notag\\
&\hspace{30pt}\times \fint_{Q_B} \dd X \; |\cos(Z\cdot \Pi_X)\Psi(X)| \; |\cos(Z'\cdot \Pi_X)\Psi(X)|. \label{DHS1:Expanding_the_square}
\end{align}
The operator $\cos(Z\cdot \Pi_X)$ is bounded by $1$ and we have
\begin{align}
\fint_{Q_B} \dd X \; |\cos(Z\cdot \Pi)\Psi(X)| \; |\cos(Z'\cdot \Pi)\Psi(X)| &\leq \Vert \Psi\Vert_2^2. \label{DHS1:LTB-LtildeTB_3}
\end{align}
By \eqref{DHS1:Expanding_the_square}, this implies
\begin{align}
\Vert L_{T,B}\Delta - \tilde L_{T,B}\Delta\Vert_2^2 & \notag\\
&\hspace{-80pt}\leq 4\; \Vert \Psi\Vert_2^2 \; \int_{\Rbb^3 }\dd r\, \Bigl| \iint_{\Rbb^3\times \Rbb^3} \dd Z\dd s \; |k_{T,B}(Z, r, s) - \tilde k_{T,B} (Z, r, s)|\; |V\alpha_*(s)|\Bigr|^2, \label{DHS1:LTB-LtildeTB_2}
\end{align}
where the integrand is bounded by
\begin{align}
|k_{T,B}(Z, r, s) - \tilde k_{T,B} (Z, r, s)| &\leq \frac 2\beta\,  \smash{\sum_{n\in \Zbb}} \, \bigl[ |g_B^{\i\omega_n} - g_0^{\i\omega_n}|\bigl( Z - \frac r2\bigr) \; |g_B^{-\i\omega_n}| \bigl( Z + \frac r2\bigr) \notag\\
&\hspace{40pt}+ |g_0^{\i\omega_n}|\bigl( Z -\frac r2\bigr)\;  |g_B^{-\i\omega_n} - g_0^{-\i\omega_n}|\bigl( Z +\frac r2\bigr)\bigr]. \label{DHS1:LTB-LtildeTB_11}
\end{align}
For $a \geq 0$, we have the estimate
\begin{align}
|Z|^a &\leq 
\bigl| Z + \frac{r}{2}\bigr|^a + \bigl| Z - \frac{r}{2}\bigr|^a. \label{DHS1:Z_estimate}
\end{align}
This, \eqref{DHS1:LTB-LtildeTB_11}, and the fact that $g_B^{\pm \i\omega_n}$ is an even function imply
\begin{align}
\int_{\Rbb^3} \dd Z \; |Z|^a\; |k_{T,B} (Z, r,s) - \tilde k_{T,B}(Z, r, s)| \leq F_{T,B}^a(r-s), \label{DHS1:LTB-LtildeTB_5}
\end{align}
where $F_{T,B}^a$ is the function in \eqref{DHS1:LTB-LtildeTB_FTB_definition}.  We apply the case $a =0$ to \eqref{DHS1:LTB-LtildeTB_2} and read off \eqref{DHS1:LTB-LtildeTB_1}.

We claim that the second term on the right side of \eqref{DHS1:LTB-LtildeTB_8} is bounded by
\begin{align}
\Vert \Pi_X(L_{T,B}\Delta - \tilde L_{T,B}\Delta)\Vert_2^2 &\leq C B^2 \; \Vert \Psi\Vert_{\Hmag^1(Q_B)}^2 \; \Vert (F_{T,B}^0 + F_{T,B}^1) * |V\alpha_*| \,\Vert_2^2. \label{DHS1:LTB-LtildeTB_6}
\end{align}
If this is true, Young's inequality and \eqref{DHS1:LTB-LtildeTB_FTBGTB} show the claimed bound for this term. To prove \eqref{DHS1:LTB-LtildeTB_6}, we use \eqref{DHS1:Expanding_the_square} with $\cos(Z\cdot\Pi_X)$ replaced by $\Pi_X\cos(Z\cdot \Pi_X)$, that is, we need to replace \eqref{DHS1:LTB-LtildeTB_3} by
\begin{align*}
\fint_{Q_B} \dd X \; |\Pi \cos(Z\cdot \Pi)\Psi(X)| \; |\Pi\cos(Z'\cdot \Pi)\Psi(X)| &\leq \Vert \Pi\cos(Z\cdot \Pi)\Psi\Vert_2\; \Vert \Pi\cos(Z'\cdot \Pi)\Psi\Vert_2.
\end{align*}
In Lemma \ref{DHS1:CommutationII} in Section~\ref{DHS1:Lower Bound Part A} we prove intertwining relations for $\cos(Z\cdot \Pi)$ with various magnetic momenta. The intertwining relation \eqref{DHS1:PiXcos} therein and \eqref{DHS1:Periodic_Sobolev_Norm} show
\begin{align}
\Vert \Pi\cos(Z\cdot \Pi)\Psi\Vert_2 &\leq \Vert \Pi\Psi\Vert_2 + 2B |Z| \; \Vert \Psi\Vert_2 \leq C\, B \; \Vert \Psi\Vert_{\Hmag^1(Q_B)} \; (1 + |Z|), \label{DHS1:PiXcos_estimate}
\end{align}
which yields
\begin{align}
\Vert \Pi_X(L_{T,B}\Delta - \tilde L_{T,B}\Delta)\Vert_2^2 &\leq C \,  B^2 \; \Vert \Psi\Vert_{\Hmag^1(Q_B)}^2  \notag\\
&\hspace{-80pt} \times  \int_{\Rbb^3} \dd r \, \Bigl| \iint_{\Rbb^3\times \Rbb^3} \dd Z\dd s \; (1+ |Z|) \; |k_{T,B} (Z, r,s) - \tilde k_{T,B}(Z, r,s)| \; |V\alpha_*(s)|\Bigr|^2. \label{DHS1:LTB-LtildeTB_9}
\end{align}
We apply the cases $a=0$ and $a=1$ of \eqref{DHS1:LTB-LtildeTB_5} to this and obtain \eqref{DHS1:LTB-LtildeTB_6}.


Concerning the third term on the right side of \eqref{DHS1:LTB-LtildeTB_8} we claim that
\begin{align}
\Vert \tilde \pi_r (L_{T,B}\Delta - \tilde L_{T,B} \Delta)\Vert_2^2 &\leq C \; \Vert \Psi\Vert_2^2 \; \bigl\Vert \bigl(G_{T,B} + F_{T,B}^1\bigr) * |V\alpha_*| + F_{T,B}^0 * |\cdot|\,|V\alpha_*| \, \bigr\Vert_2^2. \label{DHS1:LTB-LtildeTB_4}
\end{align}
If this is true, Young's inequality, \eqref{DHS1:LTB-LtildeTB_FTBGTB}, and \eqref{DHS1:Periodic_Sobolev_Norm} show the  relevant bound for this term.
To prove \eqref{DHS1:LTB-LtildeTB_4}, we estimate
\begin{align}
\Vert \tilde\pi_r (L_{T,B}\Delta - \tilde L_{T,B}\Delta)\Vert_2^2 & \notag \\
&\hspace{-90pt}\leq 4 \;\Vert \Psi\Vert_2^2  \int_{\Rbb^3} \dd r \, \Bigl| \iint_{\Rbb^3\times \Rbb^3} \dd Z\dd s \; |\tilde \pi_r k_{T,B}(Z, r, s) - \tilde \pi_r\tilde k_{T,B} (Z, r, s)| \;|V\alpha_*(s)| \Bigr|^2.  \label{DHS1:LTB-LtildeTB_10}
\end{align}
Using $\frac 14 |\Bbold \wedge r| \leq B(|r-s| + |s|)$ we see that the integrand on the right side is bounded by
\begin{align}
|\tilde \pi_r k_{T,B}(Z, r, s) - \tilde \pi_r\tilde k_{T,B} (Z, r, s)| &\leq \frac 2\beta \sum_{n\in \Zbb} \Big[ |\nabla_r k_{T,B}^n(Z, r-s) - \nabla_r k_{T,0}^n(Z, r-s)| \notag \\
&\hspace{30pt} + B |r-s|\;|k_{T,B}^n(Z, r-s) -  k_{T,0}^n(Z, r-s)| \notag \\
&\hspace{30pt} + B |s|\;|k_{T,B}^n(Z, r-s) -  k_{T,0}^n(Z, r-s)| \Big]. \label{DHS1:LTB-LtildeTB_13}
\end{align}
We also have
\begin{align*}
|\nabla_r k_{T,B}^n(Z, r) - \nabla_r k_{T,0}^n(Z, r)| &\leq |\nabla g_B^{\i\omega_n} - \nabla g_0^{\i\omega_n}|\bigl( Z + \frac r2\bigr) \;|g_B^{-\i\omega_n}| \bigl( Z - \frac r2\bigr) \\
&\hspace{30pt}+ |g_B^{\i\omega_n} - g_0^{\i\omega_n}|\bigl( Z + \frac r2\bigr)\; |\nabla g_B^{-\i\omega_n}|\bigl( Z -\frac r2\bigr) \\
&\hspace{30pt}+ |\nabla g_0^{\i\omega_n}|\bigl( Z +\frac r2\bigr)\; |g_B^{-\i\omega_n} - g_0^{-\i\omega_n}|\bigl( Z -\frac r2\bigr)\\
&\hspace{30pt}+ |g_0^{\i\omega_n}|\bigl( Z + \frac r2\bigr) |\nabla g_B^{-\i\omega_n} - \nabla g_0^{-\i\omega_n}|\bigl( Z - \frac r2\bigr).
\end{align*}
Since $g_B^{\pm \i \omega_n}$ is an even function this implies
\begin{align}
\frac 2\beta\sum_{n\in\Zbb}\int_{\Rbb^3} \dd Z \; |\nabla k_{T,B}^n(Z, r) - \nabla k_{T,0}^n(Z, r)| \leq G_{T,B}(r). \label{DHS1:LTB-LtildeTB_12}
\end{align}
Moreover, the estimate
\begin{align}
|r-s| &= \bigl| \frac{r-s}{2} + Z + \frac{r-s}{2} - Z\bigr| \leq \bigl| Z - \frac{r-s}{2}\bigr| + \bigl| Z + \frac{r-s}{2}\bigr| \label{DHS1:r-s_estimate}
\end{align}
shows that $ |r-s|\, F_{T,B}^0(r-s) \leq F_{T,B}^1(r-s)$. We conclude the estimate
\begin{align}
\smash{\int_{\Rbb^3}} \dd Z \; |\tilde \pi_r k_{T,B}(Z, r, s) - \tilde \pi_r\tilde k_{T,B} (Z, r, s)| \notag \\
&\hspace{-80pt}\leq G_{T,B}(r-s) + B\;F_{T,B}^1(r-s) + B\;F_{T,B}^0(r-s) \;|s|. \label{DHS1:LTB-LtildeTB_7}
\end{align}
From \eqref{DHS1:LTB-LtildeTB_7} we deduce \eqref{DHS1:LTB-LtildeTB_4}, which proves the claim.
\end{proof}

%

\paragraph{The operator $M_{T,B}$.} The operator $M_{T,B}$ is defined by
\begin{align}
M_{T,B} \alpha(X,r) := \iint_{\Rbb^3\times \Rbb^3} \dd Z \dd s \; k_T(Z, r-s) \;(\cos(Z\cdot \Pi_X)\alpha)(X,s), \label{DHS1:MTB_definition}
\end{align}
where $k_T(Z, r) := k_{T,0}(Z, r, 0)$ with $k_{T,0}$ in \eqref{DHS1:kTB_definition}. The following proposition allows us to replace $\tilde L_{T,B}$ by $M_{T,B}$ in our computations.

\begin{prop}
\label{DHS1:LtildeTB-MTB}
For any $T_0>0$ there is $B_0>0$ such that for any $0< B \leq B_0$, any $T\geq T_0$, and whenever  $|\cdot|^k V\alpha_*\in L^2(\Rbb^3)$ for $k\in \{0,1\}$,  $\Psi\in \Hmag^1(Q_B)$, and $\Delta \equiv \Delta_\Psi$ as in \eqref{DHS1:Delta_definition}, we have
\begin{align}
	\Vert \tilde L_{T,B}\Delta - M_{T,B}\Delta \Vert_\Hsymm^2 &\leq C\;B^3 \;\bigl( \Vert V\alpha_*\Vert_2^2 + \Vert \, |\cdot|V\alpha_*\Vert_2^2\bigr) \;\Vert \Psi\Vert_{\Hmag^1(Q_B)}^2. \label{DHS1:LtildeTB-MTB_2}
\end{align}
If instead $|\cdot|^2 V\alpha_*\in L^2(\Rbb^3)$ then
\begin{align}
	|\langle \Delta, \tilde L_{T,B}\Delta - M_{T,B}\Delta\rangle| &\leq C \;B^3 \;\Vert\,  |\cdot|^2 V\alpha_*\Vert_2^2 \;\Vert \Psi\Vert_{\Hmag^1(Q_B)}^2. \label{DHS1:LtildeTB-MTB_3}
\end{align}
\end{prop}

\begin{bem}
	The $\Hsymm$-norm bound in \eqref{DHS1:LtildeTB-MTB_2} is needed for the proof of Proposition \ref{DHS1:Structure_of_alphaDelta} and the quadratic form bound in \eqref{DHS1:LtildeTB-MTB_3} is needed for the proof of Theorem~\ref{DHS1:Calculation_of_the_GL-energy}. We highlight that the bound in \eqref{DHS1:LtildeTB-MTB_2} is insufficient as far as the proof of Theorem~\ref{DHS1:Calculation_of_the_GL-energy} is concerned. More precisely, if we apply Cauchy--Schwarz to the left side of \eqref{DHS1:LtildeTB-MTB_3}, and use \eqref{DHS1:LtildeTB-MTB_2} as well as the Lemma~\ref{DHS1:Schatten_estimate} to estimate $\Vert \Delta\Vert_2$ we obtain a bound of the order $B^2$ only. This is not good enough because $B^2$ is the order of the Ginzburg--Landau energy.

	To obtain the desired quality for the quadratic form bound \eqref{DHS1:LtildeTB-MTB_3}, we exploit the fact that $V\alpha_*$ is real-valued, which allows us to replace the magnetic phase factor $\exp(\frac{\i}{4} \Bbold (r\wedge s))$ in $\tilde k_{T,B}$ in \eqref{DHS1:ktildeTB_definition} by $\cos(\frac 14 \Bbold (r\wedge s))$. This improves the error estimate by an additional factor of $B$.
\end{bem}

Before we start with the proof of Proposition \ref{DHS1:LtildeTB-MTB}, let $a\in \Nbb_0$ and define the functions
\begin{align}
F_T^{a} := \frac 2\beta \sum_{n\in \Zbb} \sum_{b = 0}^a \binom ab \; \bigl(|\cdot|^{b}\,  |g_0^{\i\omega_n}|\bigr) * \bigl(|\cdot|^{a-b} \, |g_0^{-\i\omega_n}| \bigr) \label{DHS1:LtildeTB-MTB_FT_definition}
\end{align}
and
\begin{align}
G_T &:= \smash{\frac 2\beta \sum_{n\in \Zbb}} \bigl( |\cdot| \, |\nabla g_0^{\i \omega_n}| \bigr) * |g_0^{-\i\omega_n}| + |\nabla g_0^{\i\omega_n}| * \bigl( |\cdot| \, |g_0^{-\i\omega_n}|\bigr) \notag\\
&\hspace{110pt} + \bigl( |\cdot| \, |g_0^{\i\omega_n}| \bigr) *  |\nabla g_0^{-\i\omega_n}| + |g_0^{\i\omega_n}| * \bigl( |\cdot| \, |\nabla g_0^{-\i\omega_n}|\bigr). \label{DHS1:LtildeTB-MTB_GT_definition}
\end{align}
For $T \geq T_0 > 0$ and $a\in \Nbb_0$, by Lemma~\ref{DHS1:g0_decay},  \eqref{DHS1:g0_decay_f_estimate1}, and \eqref{DHS1:g0_decay_f_estimate2}, we have
\begin{align}
\Vert F_{T}^a\Vert_1 + \Vert G_{T}\Vert_1 &\leq C_{a}. \label{DHS1:LtildeTB-MTB_FTBGTB}
\end{align}

\begin{proof}[Proof of Proposition \ref{DHS1:LtildeTB-MTB}]
We start with the proof of \eqref{DHS1:LtildeTB-MTB_2}, which is similar to the proof of Proposition \ref{DHS1:LTB-LtildeTB}. We claim that
\begin{align}
\Vert \tilde L_{T,B}\Delta - M_{T,B}\Delta\Vert_2^2 &\leq 4 \;B^2 \;\Vert \Psi\Vert_2^2 \; \Vert F_T^1 * |\cdot|\, |V\alpha_*| \, \Vert_2^2. \label{DHS1:LtildeTB-MTB_1}
\end{align}
If this is true Young's inequality, \eqref{DHS1:Periodic_Sobolev_Norm}, and \eqref{DHS1:LtildeTB-MTB_FTBGTB} prove the claimed bound for this term.
To see that \eqref{DHS1:LtildeTB-MTB_1} holds, we argue as in \eqref{DHS1:Expanding_the_square}-\eqref{DHS1:LTB-LtildeTB_2} and find
\begin{align*}
\Vert \tilde L_{T,B}\Delta - M_{T,B}\Delta\Vert_2^2 & \\
&\hspace{-60pt}\leq 4\Vert \Psi\Vert_2^2 \int_{\Rbb^3} \dd r\,  \Bigl|\frac{2}{\beta } \sum_{n\in \Zbb} \iint_{\Rbb^3\times \Rbb^3}\dd Z \dd s \;  \bigl| k_{T,0}^n (Z, r-s) \bigl[ \e^{\frac{\i}{4} \Bbold \cdot (r\wedge s)} - 1\bigr] \bigr| \;|V\alpha_*(s)|\Bigr|^2.
\end{align*}
Since $|r\wedge s| \leq |r - s|\, |s|$, we have  $|\e^{\frac{\i}{4}\Bbold \cdot (r\wedge s)} - 1| \leq B\, |r-s|\, |s|$ as well as
\begin{align*}
|k_{T,0}^n(Z, r-s)| \bigl|\e^{\frac{\i}{4} \Bbold \cdot (r\wedge s)} - 1\bigr| &\leq B \;|g_0^{\i\omega_n}|\bigl( Z - \frac{r-s}{2}\bigr) \;|g_0^{-\i\omega_n}|\bigl( Z + \frac{r-s}{2}\bigr) \; |r-s|\; |s|.
\end{align*}
In combination with the estimate for $|r-s|$ in \eqref{DHS1:r-s_estimate} and the bound for $|Z|^a$ in \eqref{DHS1:Z_estimate}, this proves 
\begin{align}
\frac{2}{\beta} \sum_{n\in \Zbb} \int_{\Rbb^3} \dd Z \; |Z|^a \, |k_{T,0}^n(Z, r-s)| \, \bigl| \e^{\frac{\i}{4} \Bbold \cdot (r\wedge s)} - 1\bigr| &\leq B\;F_T^{a+1} (r - s)\;|s| \label{DHS1:LtildeTB-MTB_4}
\end{align}
for $a \in \Nbb_0$. The case $a = 0$ implies \eqref{DHS1:LtildeTB-MTB_1}. A computation similar to the one leading to \eqref{DHS1:LTB-LtildeTB_9} shows 
\begin{align*}
\Vert \Pi_X (\tilde L_{T,B}\Delta - M_{T,B}\Delta)\Vert_2^2 &\leq C\, B^4 \; \Vert \Psi\Vert_{\Hmag^1(Q_B)}^2 \; \Vert (F_T^1 + F_T^2) * |\cdot|\, |V\alpha_*|\, \Vert_2^2.
\end{align*}
To obtain the result we also used \eqref{DHS1:PiXcos_estimate} and \eqref{DHS1:LtildeTB-MTB_4}. We apply Young's inequality and use \eqref{DHS1:LtildeTB-MTB_FTBGTB} to prove the claimed bound for this term.
%
Finally, a computation similar to the one that leads to \eqref{DHS1:LTB-LtildeTB_10} shows
\begin{align*}
\Vert \tilde \pi_r (\tilde L_{T,B}\Delta - M_{T,B}\Delta)\Vert_2^2 & \\
&\hspace{-80pt}\leq 4\, \Vert \Psi\Vert_2^2 \int_{\Rbb^3} \dd r \, \Bigl| \iint_{\Rbb^3\times \Rbb^3}\dd Z \dd s \; \frac{2}{\beta } \sum_{n\in \Zbb} \bigl| \tilde \pi_r k_{T,0}^n (Z, r-s) \bigl[ \e^{\frac{\i}{4} \Bbold (r\wedge s)} - 1\bigr] \bigr| \; |V\alpha_*(s)|\Bigr|^2.
\end{align*}
We argue as in the proof of \eqref{DHS1:LTB-LtildeTB_7} to see that
\begin{align*}
\smash{\frac 2\beta \sum_{n\in \Zbb} \int_{\Rbb^3} \dd Z}\; \bigl| \tilde \pi_r k_{T,0}^n (Z, r-s) \bigl[ \e^{\frac{\i}{4} \Bbold (r\wedge s)} - 1\bigr] \bigr| &\\
&\hspace{-100pt}\leq C\, B \; \bigl( G_T(r-s)\; |s| + F_T^1(r-s) + F_T^0(r-s) \; |s|\bigr) .
\end{align*}
With the help of Young's inequality and \eqref{DHS1:LtildeTB-MTB_FTBGTB}, these considerations prove \eqref{DHS1:LtildeTB-MTB_2}.

It remains to prove \eqref{DHS1:LtildeTB-MTB_3}. The term we need to estimate reads
\begin{align}
\langle \Delta, \tilde L_{T,B} \Delta - M_{T,B} \Delta \rangle &= 4 \iint_{\Rbb^3\times \Rbb^3} \dd r \dd s \; \bigl( \e^{\frac \i 4 \Bbold \cdot (r\wedge s)} - 1 \bigr) V\alpha_*(r) V\alpha_*(s)  \notag\\
&\hspace{-20pt} \times \int_{\Rbb^3} \dd Z \; \frac{2}{\beta} \sum_{n\in\Zbb} k_{T,0}^n(Z, r-s)  \fint_{Q_B} \dd X \; \ov{\Psi(X)} \cos(Z\cdot \Pi_X)\Psi(X). \label{DHS1:LtildeTB-MTB_5}
\end{align}
Except for the factor $\e^{\frac \i 4 \Bbold (r\wedge s)}$, the right side is symmetric under the exchange of the coordinates $r$ and $s$. The exponential factor acquires a minus sign in the exponent when this transformation is applied. When we add the right side of \eqref{DHS1:LtildeTB-MTB_5} and the same term with the roles of $r$ and $s$ interchanged, we get
%
\begin{align}
\langle \Delta, \tilde L_{T,B} \Delta - M_{T,B} \Delta \rangle & = -8 \iint_{\Rbb^3\times \Rbb^3} \dd r \dd s \; \sin^2\bigl( \frac 1 8 \, \Bbold\cdot (r\wedge s)\bigr) \; V\alpha_*(r) V\alpha_*(s) \notag \\
&\hspace{-20pt}\times \int_{\Rbb^3} \dd Z \; \frac{2}{\beta} \sum_{n\in\Zbb} k_{T,0}^n(Z, r-s)  \fint_{Q_B} \dd X \; \ov{\Psi(X)} \cos(Z\cdot \Pi_X)\Psi(X). \label{DHS1:eq:A12}
\end{align}
To obtain \eqref{DHS1:eq:A12} we also used $\cos(x) -1 =- 2\sin^2(\frac x2)$. The operator norm of $\cos(Z\cdot\Pi_X)$ is bounded by $1$ and we have $\sin^2(\frac 18 \Bbold \cdot (r\wedge s)) \leq \frac 18 B^2 |r|^2 |s|^2$. Therefore, \eqref{DHS1:eq:A12} proves
\begin{align}
|\langle \Delta, \tilde L_{T,B}\Delta - M_{T,B}\Delta\rangle | &\leq B^2 \; \Vert \Psi\Vert_2^2 \; \bigl\Vert |\cdot|^2|V\alpha_*| \; \bigl( |\cdot |^2|V\alpha_*| * F_T^0\bigr) \bigr\Vert_1.
\label{DHS1:eq:A13}
\end{align}
Finally, we use Young's inequality, \eqref{DHS1:Periodic_Sobolev_Norm}, and \eqref{DHS1:LtildeTB-MTB_FTBGTB} and obtain \eqref{DHS1:LtildeTB-MTB_3}. This proves Proposition~\ref{DHS1:LtildeTB-MTB}.
\end{proof}

%

\subsubsection{Analysis of \texorpdfstring{$M_{T,B}$}{MTB} and calculation of the quadratic terms}
\label{DHS1:Analysis_of_MTB_Section}

We decompose $M_{T,B} = M_T^{(1)} + M_{T,B}^{(2)} + M_{T,B}^{(3)}$, where
\begin{align}
M_T^{(1)} \alpha(X,r) &:= \iint_{\Rbb^3\times \Rbb^3} \dd Z \dd s \; k_T(Z, r-s) \; \alpha(X,s), \label{DHS1:MT1_definition}\\
M_{T,B}^{(2)} \alpha(X, r) &:=  \iint_{\Rbb^3\times \Rbb^3} \dd Z \dd s\; k_T(Z, r-s) \; \bigl( -\frac 12\bigr) (Z\cdot \Pi_X)^2 \; \alpha(X, s), \label{DHS1:MTB2_definition}\\
M_{T,B}^{(3)} \alpha(X,r) &:= \iint_{\Rbb^3\times \Rbb^3} \dd Z \dd s\; k_T(Z, r-s) \; \Rcal(Z\cdot \Pi_X) \; \alpha(X, s), \label{DHS1:MTB3_definition}
\end{align}
and $\Rcal(x) = \cos(x) - 1 + \frac 12 x^2$.

\paragraph{The operator $M_T^{(1)}$.} The expression $\langle \Delta, M_T^{(1)} \Delta \rangle$ contains a term that cancels the last term on the left side of \eqref{DHS1:Calculation_of_the_GL-energy_eq} as well as the quadratic term without magnetic gradient in the Ginzburg--Landau functional in \eqref{DHS1:Definition_GL-functional}. The following result allows us to extract these terms. We recall that $\Delta \equiv \Delta_\Psi = -2 V\alpha_* \Psi$.

\begin{prop}
\label{DHS1:MT1}
Assume that $V\alpha_*\in L^2(\Rbb^3)$ and let $\Psi \in \Lmag^2(Q_B)$ and $\Delta \equiv \Delta_\Psi$ as in \eqref{DHS1:Delta_definition}.
\begin{enumerate}[(a)]
\item We have $M_{\Tc}^{(1)} \Delta (X, r) = -2\, \alpha_* (r) \Psi(X)$.

\item For any $T_0>0$ there is a constant $c>0$ such that for $T_0 \leq T \leq \Tc$ we have
\begin{align*}
\langle \Delta, M_T^{(1)} \Delta - M_{\Tc}^{(1)} \Delta \rangle \geq  c \, \frac{\Tc - T}{\Tc} \; \Vert \Psi\Vert_2^2.
\end{align*}

\item Given $D\in \Rbb$ there is $B_0>0$ such that for $0< B\leq B_0$, and $T = \Tc (1 - DB)$ we have
\begin{align*}
	\langle \Delta, M_T^{(1)} \Delta -  M_{\Tc}^{(1)} \Delta\rangle  = 4\; DB \; \Lambda_2 \; \Vert \Psi\Vert_2^2 + R(\Delta)
\end{align*}
with the coefficient $\Lambda_2$ in \eqref{DHS1:GL-coefficient_2}, and
\begin{align*}
	|R(\Delta)| &\leq C \; B^2 \; \Vert V\alpha_*\Vert_2^2\; \Vert \Psi\Vert_{2}^2.
\end{align*}

\item Assume additionally that $| \cdot | V\alpha_*\in L^2(\Rbb^3)$. There is $B_0>0$ such that for any $0< B\leq B_0$, any $\Psi\in \Hmag^1(Q_B)$, and any $T \geq T_0 > 0$ we have 
\begin{align*}
	\Vert M_T^{(1)}\Delta - M_{\Tc}^{(1)}\Delta\Vert_{\Hsymm}^2 &\leq C \, B \, | T - \Tc |^2 \,  \bigl( \Vert V\alpha_*\Vert_2^2 + \Vert \, |\cdot| V\alpha_*\Vert_2^2\bigr) \Vert\Psi\Vert_{\Hmag^1(Q_B)}^2.
\end{align*}

\end{enumerate}
\end{prop}

\begin{bem}
	The above bound for the remainder term implies
	\begin{equation*}
	| R(\Delta) | \leq C \; B^3 \; \Vert V\alpha_*\Vert_2^2\; \Vert \Psi\Vert_{\Hmag^1(Q_B)}^2.
	\end{equation*}
	Part~(b) in the Proposition is needed for the proof of Proposition~\ref{DHS1:Lower_Tc_a_priori_bound}. Part~(d) is needed in the proof of Proposition~\ref{DHS1:Structure_of_alphaDelta}.
\end{bem}

Before we give the proof of the above proposition, we introduce the function
\begin{align}
	F_{T,\Tc} &:= \frac{2}{\beta} \sum_{n\in \Zbb} |2n+1| \bigl[  |g_0^{\i\omega_n^T}| * |g_0^{\i\omega_n^{\Tc}}| * |g_0^{-\i\omega_n^T}|  + |g_0^{\i\omega_n^{\Tc}}| *  |g_0^{-\i\omega_n^T}| * |g_0^{-\i\omega_n^{\Tc}}| \bigr], \label{DHS1:MT1_FTTc_definition}
\end{align}
where we indicated the $T$-dependence of the Matusubara frequencies in \eqref{DHS1:Matsubara_frequencies} because different temperatures appear in the formula. As long as $T \geq T_0 > 0$, Lemma~\ref{DHS1:g0_decay} and \eqref{DHS1:g0_decay_f_estimate1} imply the bound  
\begin{equation}
	\Vert F_{T,\Tc} \Vert_1 \leq C.
	\label{DHS1:MT1_FTTc_bound}
\end{equation}

\begin{proof}[Proof of Proposition \ref{DHS1:MT1}]
We start with the proof of part (a). First of all, we recall that $k_T(Z, r) = k_{T,0}(Z, r, 0)$ with $k_{T,B}(Z,r,s)$ in \eqref{DHS1:kTB_definition}. In Fourier space the convolution operator $g_0^{\pm \i \omega_n}(x-y)$ equals multiplication with $(\pm\i\omega_n + \mu - k^2)^{-1}$. This allows us to write
\begin{align*}
	k_T(Z, r) &= \frac{2}{\beta} \sum_{n\in\Zbb} \iint_{\Rbb^3\times \Rbb^3} \frac{\dd k}{(2\pi)^3} \frac{\dd \ell}{(2\pi)^3} \; \frac{\e^{\i k\cdot (Z - \frac r2)}}{\i\omega_n  + \mu - k^2} \frac{\e^{\i \ell \cdot (Z + \frac r2)}}{-\i\omega_n + \mu - \ell^2} \\
	&= -\frac 2\beta \sum_{n\in \Zbb} \iint_{\Rbb^3\times \Rbb^3} \frac{\dd p}{(2\pi)^3} \frac{\dd q}{(2\pi)^3} \; \e^{\i Z\cdot q} \e^{-\i r\cdot p} \; \frac{1}{\i\omega_n + \mu - (p+\frac  q2)^2} \frac{1}{\i\omega_n - \mu + (p - \frac q 2)^2},
\end{align*}
where we applied the change of variables $q= k+\ell$ and $p = \frac{k-\ell}{2}$. We use the partial fraction expansion
\begin{align*}
	\frac{1}{E + E'} \Bigl( \frac{1}{\i\omega_n - E} - \frac{1}{\i\omega_n + E'} \Bigr) = \frac{1}{\i\omega_n - E} \frac{1}{\i\omega_n + E'}
\end{align*}
and the representation formula of the hyperbolic tangent in \eqref{DHS1:tanh_Matsubara} to see that
\begin{equation}
	k_T(Z, r) = \iint_{\Rbb^3 \times \Rbb^3} \frac{\dd p}{(2\pi)^3} \frac{\dd q}{(2\pi)^3} \; \e^{\i Z\cdot q} \e^{-\i r\cdot p} \; L_T\bigl( p+ \frac{q}{2}, p - \frac{q}{2}\bigr), \label{DHS1:kT_Fourier_representation2}
\end{equation}
where
\begin{align}
	L_T(p,q) := \frac{\tanh(\frac \beta 2(p^2-\mu)) + \tanh(\frac \beta 2 (q^2-\mu))}{p^2-\mu + q^2-\mu}. \label{DHS1:LT_definition}
\end{align}
In particular,
\begin{align}
	\int_{\Rbb^3} \dd Z \; k_T(Z, r) = \int_{\mathbb{R}^3} \frac{\dd p}{(2\pi)^3} \; \e^{-\i r\cdot p}  L_T(p, p) = \int_{\mathbb{R}^3} \frac{\dd p}{(2\pi)^3} \; \e^{-\i r\cdot p} K_T(p)^{-1} \label{DHS1:LT_integration_KT}
\end{align}
with $K_T(p)$ in \eqref{DHS1:KT-symbol}. Therefore, we have
\begin{align*}
M_T^{(1)}\Delta(X, r) = K_T^{-1} V\alpha_*(r) \, \Psi(X),
\end{align*}
which together with $K_{\Tc} \alpha_* = V \alpha_*$ proves part (a).
To prove part (b), we use \eqref{DHS1:LT_integration_KT} to write
\begin{align}
\langle \Delta, M_T^{(1)} \Delta - M_{\Tc}^{(1)} \Delta\rangle =  \int_{\Rbb^3} \frac{\dd p}{(2\pi)^3} \; \bigl[ K_T(p)^{-1} - K_{\Tc}(p)^{-1}\bigr] \, |(-2)V\alpha_*(p)|^2 \, \Vert \Psi\Vert_2^2. \label{DHS1:MT1_3}
\end{align}
With the help of the first order Taylor expansion
\begin{align}
K_T(p)^{-1} - K_{T_c}(p)^{-1} &
= \frac 12\int_{T}^{T_c} \dd T' \; \frac{1}{(T')^2} \frac{1}{\cosh^2( \frac{p^2-\mu}{2T'})} \label{DHS1:MT1_1}
\end{align}
we see that
\begin{align*}
	K_T(p)^{-1} - K_{\Tc}(p)^{-1} \geq \frac 12 \frac{\Tc - T}{\Tc^2} \, \frac{1}{\cosh^2(\frac{p^2-\mu}{2T_0})}
\end{align*}
holds for $T_0 \leq T \leq \Tc $. This and \eqref{DHS1:MT1_3} prove part (b).

To prove part (c), we expand \eqref{DHS1:MT1_1} to second order in $T-T_c$ and find
\begin{align*}
	\Bigl|\int_{\Rbb^3} \frac{\dd p}{(2\pi)^3} \; [K_T(p)^{-1} - K_{\Tc}(p)^{-1} ] \; |(-2)\hat{V\alpha_*}(p)|^2 - 4\, \Lambda_2\; \frac{\Tc - T}{\Tc}  \Bigr| \leq C\, |T- \Tc|^2 \; \Vert V\alpha_*\Vert_2^2
\end{align*}
with $\Lambda_2$ in \eqref{DHS1:GL-coefficient_2}. By \eqref{DHS1:MT1_3}, this proves part (c).

It remains to prove part (d). We use the resolvent identity to see that
\begin{align}
g_0^{\pm \i\omega_n^T} - g_0^{\pm \i \omega_n^{\Tc}}  = \mp \i (\omega_n^T - \omega_n^{\Tc}) \;  g_0^{\pm \i \omega_n^T} * g_0^{\pm \i \omega_n^{\Tc}} . \label{DHS1:gTgTc}
\end{align}
Using \eqref{DHS1:gTgTc}, it is straightforward to see that
\begin{align*}
\Vert M_T^{(1)}\Delta - M_{\Tc}^{(1)}\Delta\Vert_2^2 &\leq C \, |T - \Tc|^2 \, \Vert F_{T,\Tc} \Vert_1 \,  \Vert V\alpha_*\Vert_2^2 \, \Vert\Psi\Vert_2^2
\end{align*}
holds with $F_{T,\Tc}$ in \eqref{DHS1:MT1_FTTc_definition}. In combination with \eqref{DHS1:MT1_FTTc_bound} this proves the claimed bound for this term. The estimates for the terms $\Vert \tilde \pi_r (M_T^{(1)}\Delta - M_{\Tc}^{(1)}\Delta)\Vert_2^2$ and $\Vert \Pi_X(M_T^{(1)}\Delta - M_{\Tc}^{(1)}\Delta)\Vert_2^2$ are proved similarly. We omit the details.
\end{proof}

\paragraph{The operator $M_{T,B}^{(2)}$.} The term $\langle \Delta, M_{T,B}^{(2)} \Delta \rangle$ with $M_{T,B}^{(2)}$ in \eqref{DHS1:MTB2_definition} contains the kinetic term in the Ginzburg--Landau functional in \eqref{DHS1:Definition_GL-functional}. The following proposition allows us to compare the two.

\begin{prop}
\label{DHS1:MTB2}
Assume that the function $V\alpha_*$ is radial and belongs to $L^2(\Rbb^3)$. For any $B>0$, $\Psi\in \Hmag^1(Q_B)$, and $\Delta \equiv \Delta_\Psi$ as in \eqref{DHS1:Delta_definition}, we have
\begin{align}
\langle \Delta, M_{\Tc,B}^{(2)} \Delta\rangle = - 4\; \Lambda_0 \; \Vert \Pi\Psi\Vert_2^2 \label{DHS1:MTB2_1}
\end{align}
with $\Lambda_0$ in \eqref{DHS1:GL-coefficient_1}. Moreover, for any $T \geq T_0 > 0$ we have
\begin{align}
	|\langle \Delta,  M_{T,B}^{(2)} \Delta - M_{\Tc,B}^{(2)}\Delta\rangle| \leq C\; B^2 \; |T - \Tc| \; \Vert V\alpha_*\Vert_2^2 \; \Vert \Psi\Vert_{\Hmag^1(Q_B)}^2. \label{DHS1:MTB2_2}
\end{align}
\end{prop}

Before we give the proof of Proposition~\ref{DHS1:MTB2}, let us introduce the function
\begin{align}
F_{T,\Tc}^a &:= \smash{\frac{2}{\beta} \sum_{n\in \Zbb} \sum_{\substack{a_1,a_2,a_3\in \Nbb_0 \\ a_1+ a_2 + a_3 = a}}}  |2n+1| \bigl[ \bigl(|\cdot|^{a_1} \, |g_0^{\i\omega_n^T}|\bigr) * \bigl(|\cdot|^{a_2} \, |g_0^{\i\omega_n^{\Tc}}|\bigr) * \bigl(|\cdot|^{a_3} \, |g_0^{-\i\omega_n^T}|\bigr) \notag\\
&\hspace{110pt} + \bigl(|\cdot|^{a_1} \, |g_0^{\i\omega_n^{\Tc}}|\bigr) * \bigl(|\cdot|^{a_2} \, |g_0^{-\i\omega_n^T}|\bigr)* \bigl(|\cdot|^{a_3} \, |g_0^{-\i\omega_n^{\Tc}}|\bigr)\bigr], \label{DHS1:MTB2_FTTc_definition}
\end{align}
where $a\in \Nbb_0$ and where we indicated the $T$-dependence of the Matusubara frequencies in \eqref{DHS1:Matsubara_frequencies} because different temperatures appear in the formula. 
As long as $T \geq T_0 > 0$, Lemma~\ref{DHS1:g0_decay} and \eqref{DHS1:g0_decay_f_estimate1} imply the bound $\Vert F_{T,\Tc}^a \Vert_1 \leq C_a$.

\begin{proof}[Proof of Proposition \ref{DHS1:MTB2}]
We have
\begin{align}
\langle \Delta, M_{\Tc, B}^{(2)} \Delta\rangle &= -2 \iint_{\Rbb^3\times\Rbb^3} \dd r\dd s \; V\alpha_*(r)V\alpha_*(s) \int_{\Rbb^3} \dd Z\; k_{\Tc}(Z, r-s) \; \langle \Psi, (Z\cdot \Pi_X)^2 \Psi\rangle
\label{DHS1:eq:A10}
\end{align}
and 
\begin{align*}
\langle \Psi, (Z \cdot \Pi)^2 \Psi\rangle &= \sum_{i,j=1}^3 Z_iZ_j \; \langle \Pi^{(i)} \Psi, \Pi^{(j)} \Psi\rangle. 
\end{align*}
The integration over $Z$ in \eqref{DHS1:eq:A10} defines a $3\times 3$ matrix with matrix elements
\begin{align*}
	\int_{\Rbb^3} \dd Z \, k_{\Tc}(Z, r) \; Z_iZ_j &=   \iint_{\Rbb^3\times \Rbb^3} \frac{\dd p}{(2\pi)^3} \frac{\dd q}{(2\pi)^3} \int_{\Rbb^3} \dd Z \; \e^{-\i Z\cdot q} \e^{-\i p\cdot r} \; L_{\Tc}\bigl( p + \frac q2, p -  \frac q2\bigr)Z_iZ_j,
\end{align*}
which we have written in terms of the Fourier representation of $k_{\Tc}(Z,r)$ in \eqref{DHS1:kT_Fourier_representation2}. We use $Z_i Z_j \e^{-\i Z\cdot q} = -\partial_{q_i} \partial_{q_j} \e^{-\i Z\cdot q}$, integrate by parts twice, and find
\begin{align*}
	\int_{\Rbb^3} \frac{\dd q}{(2\pi)^3} \int_{\Rbb^3} \dd Z \; \e^{-\i Z\cdot q} \; L_{\Tc}\bigl( p + \frac q2, p - \frac q2\bigr) \; Z_iZ_j &= -\Bigl[ \frac{\partial}{\partial q_i} \frac{\partial}{\partial q_j} L_{\Tc} \big(p + \frac q2,p - \frac q2\bigr) \Bigr]_{q=0}.
\end{align*}
A tedious but straightforward computation shows that the right side of the above equation can be written in terms of the functions $g_1$ and $g_2$ in \eqref{DHS1:XiSigma} as
\begin{align*}
	-\Bigl[ \frac{\partial}{\partial q_i} \frac{\partial}{\partial q_j}  L_{\Tc} \bigl( p + \frac q2, p - \frac q2\bigr) \Bigr]_{q=0}  =  \frac{\beta_c^2}{2} \bigl[ g_1 (\beta_c(p^2-\mu)) \delta_{ij} + 2\beta_c \; g_2(\beta_c(p^2-\mu)) \; p_ip_j\bigr],
\end{align*}
and hence
\begin{equation*}
	\int_{\Rbb^3} \dd Z \; k_{\Tc}(Z, r) \, Z_iZ_j = \frac{\beta_c^2}{2}  \int_{\Rbb^3} \frac{\dd p}{(2\pi)^3} \;  \e^{-\i p \cdot r} \bigl[ g_1 (\beta_c(p^2-\mu)) \delta_{ij} + 2\beta_c \; g_2(\beta_c(p^2-\mu)) \; p_ip_j\bigr].
\end{equation*}
Let us denote the term in the bracket on the right side by $A_{ij}(p)$. When we insert the above identity into \eqref{DHS1:eq:A10} we find
\begin{equation}
	\langle \Delta, M_{\Tc, B}^{(2)} \Delta\rangle = -\frac{\beta_{\mathrm{c}}^2}{4} \sum_{i,j = 1}^3 \langle \Pi^{(i)} \Psi, \Pi^{(j)} \Psi \rangle \int_{\mathbb{R}^3} \frac{\dd p}{(2\pi)^3} \; |(-2)\hat{V\alpha_*}(p)|^2 A_{ij}(p).
\end{equation}
We use that $V\alpha_*$ is a radial function to see that the integral of the term proportional to $p_i p_j$ equals zero unless $i = j$. Since the angular average of $p_i^2$ equals $\frac 13 p^2$ this proves \eqref{DHS1:MTB2_1}.

It remains to prove \eqref{DHS1:MTB2_2}. To this end, we estimate
\begin{align}
|\langle \Delta, M_{T,B}^{(2)}\Delta - M_{\Tc,B}^{(2)}\Delta \rangle | & \notag\\
&\hspace{-100pt}\leq 2 \iiint_{\Rbb^3\times \Rbb^3\times \Rbb^3} \dd r\dd s\dd Z \; |V\alpha_*(r)|\; |V\alpha_*(s)| \; |k_T(Z, r-s)-k_{\Tc}(Z, r-s)|  \notag\\
&\hspace{160pt} \times |\langle \Psi, (Z\cdot \Pi)^2\Psi\rangle|. \label{DHS1:MTB2_3}
\end{align}
%
%
%
%
%
%
For general operators $A, B, C$, we have $|A + B + C|^2 \leq 3(|A|^2 + |B|^2 + |C|^2)$.
This implies
\begin{align}
	(Z\cdot \Pi)^2 &\leq 3 \; \bigl( Z_1^2 \; (\Pi^{(1)})^2 + Z_2^2 \; (\Pi^{(2)})^2 + Z_3^2 \; (\Pi^{(3)})^2\bigr) \leq 3 \; Z^2 \; \Pi^2, \label{DHS1:ZPiX_inequality}
\end{align}
and, in particular,
\begin{align}
|\langle \Psi, (Z\cdot \Pi)^2 \Psi\rangle | \leq 3 \, |Z|^2 \, \Vert \Pi\Psi\Vert_2^2 \leq 3\, B^2\, |Z|^2 \, \Vert \Psi\Vert_{\Hmag^1(Q_B)}^2. \label{DHS1:MTB2_4}
\end{align}
Moreover, \eqref{DHS1:gTgTc}
%
%
and the estimate for $|Z|^2$ in \eqref{DHS1:Z_estimate} show
\begin{align}
\int_{\Rbb^3} \dd Z \; |Z|^2 \, |k_T(Z, r) - k_{\Tc}(Z, r)| &\leq C\, |T-\Tc|\; F_{T,\Tc}^2(r) \label{DHS1:MTB2_5}
\end{align}
with $F_{T,\Tc}^2$ in \eqref{DHS1:MTB2_FTTc_definition}.
%
%
With the help of \eqref{DHS1:MTB2_3}, \eqref{DHS1:MTB2_4}, and \eqref{DHS1:MTB2_5}, we deduce 
\begin{align*}
|\langle \Delta, M_{T,B}^{(2)} \Delta - M_{\Tc,B}^{(2)}\Delta\rangle | &\leq C\, B^2 \; |T - \Tc| \; \Vert \Psi\Vert_{\Hmag^1(Q_B)}^2 \; \bigl\Vert |V\alpha_*| \; \bigl( |V\alpha_*| * F_{T,\Tc}^2\bigr) \bigr\Vert_1 .
%
\end{align*}
An application of Young's inequality completes the proof.
\end{proof}

\paragraph{The operator $M_{T,B}^{(3)}$.} The term $\langle \Delta, M_{T,B}^{(3)} \Delta \rangle$ with $M_{T,B}^{(3)}$ in \eqref{DHS1:MTB3_definition} is the remainder of our expansion of $\langle \Delta, M_{T,B} \Delta \rangle$ in powers of $B$. In contrast to the previous estimates, we need the $\Hmag^2(Q_B)$-norm of $\Psi$ to control its size. 

\begin{prop}
\label{DHS1:MTB3}
For any $T_0>0$ there is $B_0>0$ such that for any $0 < B \leq B_0$, any $T\geq T_0$, and whenever $V\alpha_*\in L^2(\Rbb^3)$, $\Psi \in \Hmag^2(Q_B)$, and $\Delta \equiv \Delta_\Psi$ as in \eqref{DHS1:Delta_definition}, we have
\begin{align*}
|\langle \Delta,  M_{T,B}^{(3)} \Delta\rangle| &\leq C \; B^3 \; \Vert V\alpha_*\Vert_2^2 \; \Vert \Psi\Vert_{\Hmag^2(Q_B)}^2.
\end{align*}
\end{prop}

Before we give the proof of Proposition~\ref{DHS1:MTB3}, let us introduce the function
\begin{align}
F_T(r) := \frac 1\beta \sum_{n\in \Zbb}\; \bigl(|\cdot|^4 |g_0^{\i\omega_n}| \bigr) * |g_0^{-\i\omega_n}| + |g_0^{\i\omega_n}| * \bigl(|\cdot|^4 |g_0^{-\i\omega_n}|\bigr). \label{DHS1:MTB3_FT_definition}
\end{align}
As long as $T \geq T_0 > 0$, Lemma~\ref{DHS1:g0_decay} and \eqref{DHS1:g0_decay_f_estimate1} imply $\Vert F_T\Vert_1\leq C$.

\begin{proof}[Proof of Proposition \ref{DHS1:MTB3}]
We have
\begin{align}
\langle \Delta, M_{T_B}^{(3)} \Delta \rangle &= 4 \iiint_{\Rbb^3\times \Rbb^3\times \Rbb^3} \dd r\dd s\dd Z \; V\alpha_*(r) V\alpha_*(s)\, k_T(Z, r-s) \; \langle \Psi , \Rcal(Z\cdot \Pi_X)\Psi\rangle, \label{DHS1:MTB3_1}
\end{align}
where the function $\Rcal(x) = \cos(x) - 1 + \frac{x^2}{2}$ obeys the bound $0\leq\Rcal(x) \leq \frac{1}{24} x^4$. We claim that
\begin{align}
|Z\cdot \Pi|^4 &\leq 9\; |Z|^4 \; ( \Pi^4 + 8 B^2),\label{DHS1:ZPiX_inequality_quartic}
\end{align}
which implies
\begin{align}
\langle \Psi, \Rcal(Z\cdot \Pi) \Psi\rangle &\leq \frac{9}{24} \; |Z|^4 \; \bigl( \Vert \Pi^2\Psi\Vert_2^2 + 8 B^2 \Vert \Psi\Vert_2^2\bigr) \leq C\; B^3 \; |Z|^4 \; \Vert \Psi\Vert_{\Hmag^2(Q_B)}^2. \label{DHS1:MTB3_2}
\end{align}
To see that \eqref{DHS1:ZPiX_inequality_quartic} is true, we note that $[\Pi^{(1)}, \Pi^{(2)}] = -2\i B$ implies
\begin{align}
\Pi \; \Pi^2 = \Pi^2 \; \Pi + 4\i B \;  (-\Pi^{(2)} , \Pi^{(1)} , 0)^t, \label{DHS1:PiX-operator-equality_1}
\end{align}
and hence
\begin{align}
\Pi \; \Pi^2\; \Pi = \Pi^4+ 8B^2. \label{DHS1:PiPi2Pi_equality}
\end{align}
We also have $[Z\cdot \Pi, \Pi] = -2\i \, \Bbold \wedge Z$, which implies $(Z\cdot \Pi)\Pi^2(Z\cdot \Pi)= \Pi (Z\cdot \Pi)^2\Pi$. We combine this with the operator inequality \eqref{DHS1:ZPiX_inequality} for $(Z\cdot \Pi)^2$ and \eqref{DHS1:PiPi2Pi_equality} and get $(Z\cdot \Pi)\Pi^2(Z\cdot \Pi) \leq 3|Z|^2(\Pi^4+8B^2)$. Finally, we write $|Z\cdot \Pi|^4 = (Z\cdot \Pi) (Z\cdot \Pi)^2 (Z\cdot \Pi)$, apply \eqref{DHS1:ZPiX_inequality} again, and obtain \eqref{DHS1:ZPiX_inequality_quartic}. 

Using the estimate \eqref{DHS1:Z_estimate} on $|Z|^4$ and \eqref{DHS1:MTB3_2}, we argue as in the proof of \eqref{DHS1:LTB-LtildeTB_5} to see that
\begin{align}
\int_{\Rbb^3} \dd Z\; |Z|^4 \; |k_T(Z,r)| &\leq F_T(r) \label{DHS1:MTB3_3}
\end{align}
with $F_T$ in \eqref{DHS1:MTB3_FT_definition}. The bound on the $L^1(\mathbb{R}^3)$-norm of $F_T$ below \eqref{DHS1:MTB3_FT_definition}, \eqref{DHS1:MTB3_1}, \eqref{DHS1:MTB3_2}, and \eqref{DHS1:MTB3_3} prove the claim.
\end{proof}

\subsubsection{Summary: The quadratic terms}
\label{DHS1:Summary_quadratic_terms_Section}

Let us summarize the results concerning the quadratic terms in $\Delta \equiv \Delta_\Psi$ that are relevant for the proof of Theorem~\ref{DHS1:Calculation_of_the_GL-energy} and provide an intermediate statement that is needed for the proof of Proposition \ref{DHS1:Lower_Tc_a_priori_bound}.

\begin{prop}
\label{DHS1:Rough_bound_on_BCS energy}
Given $T_0> 0$ there is a constant $B_0>0$ such that for any $T_0 \leq T\leq \Tc$, any $0 < B \leq B_0$, and whenever $|\cdot|^k V\alpha_*\in L^2(\Rbb^3)$ for $k\in \{0,1,2\}$, $\Psi \in \Hmag^1(Q_B)$, and $\Delta \equiv \Delta_\Psi$ as in \eqref{DHS1:Delta_definition}, we have
\begin{align}
- \frac 14 \langle \Delta, L_{T,B} \Delta\rangle + \Vert \Psi\Vert_2^2 \; \langle \alpha_*, V\alpha_*\rangle & \leq c  \, \frac{T - \Tc}{\Tc}\, \Vert \Psi\Vert_2^2 + C B^2  \, \Vert \Psi\Vert_{\Hmag^1(Q_B)}^2. \label{DHS1:Rough_bound_on_BCS energy_eq1}
\end{align}
\end{prop}

\begin{proof}
By Lemma \ref{DHS1:LTB_action}, the decomposition \eqref{DHS1:LTB_decomposition} of $L_{T,B}$, as well as Propositions~\ref{DHS1:LTB-LtildeTB} and \ref{DHS1:LtildeTB-MTB}, we have
\begin{align}
- \frac 14 \langle \Delta, L_{T,B} \Delta\rangle + \Vert \Psi\Vert_2^2 \; \langle \alpha_*, V\alpha_*\rangle &\notag\\
&\hspace{-100pt} = - \frac 14 \langle \Delta, M_T^{(1)}\Delta- M_{\Tc}^{(1)} \Delta\rangle - \frac 14 \langle \Delta, M_{T, B} \Delta - M_T^{(1)}\Delta\rangle + R_1(\Delta), \label{DHS1:Rough_bound_on_BCS energy_proof_1}
\end{align}
where
\begin{align*}
	| R_1(\Delta) | \leq C \; B^3 \; \Vert \Psi\Vert_{\Hmag^1(Q_B)}^2
\end{align*}
and, by Proposition \ref{DHS1:MT1}, 
\begin{align*}
-\frac 14 \langle \Delta, M_T^{(1)}\Delta- M_{\Tc}^{(1)} \Delta\rangle \leq  c \; \frac{T - \Tc}{\Tc} \; \Vert \Psi \Vert_2^2.
\end{align*}
We claim that
\begin{align}
|\langle \Delta, M_{T,B}\Delta - M_T^{(1)}\Delta\rangle| &\leq C\; B^2\; \Vert V\alpha_*\Vert_2^2 \; \Vert \Psi\Vert_{\Hmag^1(Q_B)}^2. \label{DHS1:MTB-MT1_eq1}
\end{align}
The proof of \eqref{DHS1:MTB-MT1_eq1} goes along the same lines as that of Proposition~\ref{DHS1:LtildeTB-MTB} and uses the operator inequality \eqref{DHS1:ZPiX_inequality} on $(Z\cdot \Pi)^2$ to estimate
\begin{align}
|\langle \Psi, [\cos(Z\cdot\Pi) - 1] \Psi \rangle| 
&\leq C\; B^2 \; |Z|^2 \; \Vert \Psi\Vert_{\Hmag^1(Q_B)}^2. \label{DHS1:MTB-MT1_1}
\end{align}
We omit the details. This proves \eqref{DHS1:Rough_bound_on_BCS energy_eq1}.
\end{proof}

Let the assumptions of Theorem~\ref{DHS1:Calculation_of_the_GL-energy} hold. We combine \eqref{DHS1:Rough_bound_on_BCS energy_proof_1} with the results of 
Propositions \ref{DHS1:MT1}, \ref{DHS1:MTB2}, and \ref{DHS1:MTB3} to see that for $T = \Tc(1-DB)$ with $D \in \mathbb{R}$ we have
\begin{equation}
	-\frac{1}{4} \langle \Delta, L_{T,B} \Delta \rangle + \Vert \Psi \Vert_2^2 \, \langle \alpha_*, V \alpha_* \rangle = \Lambda_0 \; \Vert \Pi\Psi\Vert_2^2 - DB \; \Lambda_2 \; \Vert \Psi\Vert_2^2 +  R_2(\Delta),
	\label{DHS1:eq:A15}
\end{equation}
where
\begin{equation*}
	| R_2(\Delta) | \leq C\,  B^3 \, \Vert \Psi\Vert_{\Hmag^2(Q_B)}^2.
\end{equation*}
This concludes our analysis of the operator $L_{T,B}$.

\subsubsection{A representation formula for the operator \texorpdfstring{$N_{T,B}$}{NTB}}


Let us introduce the notation $\Zbold$ for the vector $(Z_1,Z_2,Z_3)$ with $Z_1, Z_2, Z_3 \in \Rbb^3$. We also denote $\dd \Zbold = \dd Z_1 \dd Z_2 \dd Z_3$.
Remarkably, the strategy of the analysis we used for $L_{T,B}$ carries over to the nonlinear operator $N_{T,B}$ in \eqref{DHS1:NTB_definition}. As in the case of $L_{T,B}$, we start with a representation formula for the operator $N_{T,B}$ and note the analogy to Lemma \ref{DHS1:LTB_action}.

\begin{lem}
\label{DHS1:NTB_action}
The operator $N_{T,B} \colon \Hsymm \ra \Lsymm$ in \eqref{DHS1:NTB_definition} acts as
\begin{align*}
N_{T,B}(\alpha) (X, r) &= \iiint_{\Rbb^9} \dd \Zbold \iiint_{\Rbb^9} \dd \sbold \; \ell_{T,B}(\Zbold, r, \sbold)\; \Acal(X, \Zbold, \sbold)
\end{align*}
with
\begin{align}
\Acal(X, \Zbold, \sbold) &:= \e^{\i Z_1\cdot \Pi_X} \alpha(X, s_1) \; \ov{\e^{\i Z_2\cdot \Pi_X}\alpha(X, s_2)} \; \e^{\i Z_3\cdot \Pi_X} \alpha(X,s_3) \label{DHS1:NTB_alpha_definition}
\end{align}
and
\begin{align}
\ell_{T,B}(\Zbold, r, \sbold) &:= \frac 2\beta \sum_{n\in \Zbb} \ell_{T,B}^n(\Zbold, r, \sbold) \; \e^{\i \frac \Bbold 2 \cdot \Phi(\Zbold, r, \sbold)}, \label{DHS1:lTB_definition}
\end{align}
where
\begin{align}
\ell_{T,B}^n(\Zbold, r, \sbold) &:= g_B^{\i\omega_n} \bigl(Z_1 - \frac{r-s_1}{2}\bigr) \; g_B^{-\i\omega_n} \bigl( Z_1 - Z_2 - \frac{s_1 + s_2}{2}\bigr) \notag \\
&\hspace{50pt} \times g_B^{\i\omega_n} \bigl( Z_2 - Z_3 - \frac{s_2 + s_3}{2} \bigr) \; g_B^{-\i\omega_n} \bigl( Z_3 + \frac{r-s_3}{2}\bigr) \label{DHS1:lTBn_definition}
\end{align}
with $g_B^{\pm \i\omega_n}$ in \eqref{DHS1:gB_definition} and
\begin{align}
\Phi(\Zbold, r, \sbold) &:= \frac r2 \wedge \bigl( Z_1 - \frac{r - s_1}{2}\bigr) + \frac r2 \wedge \bigl( Z_3 + \frac{r-s_3}{2}\bigr) \notag \\
&\hspace{-35pt} + \bigl( Z_2 - Z_3 - \frac{s_2 + s_3}{2}\bigr) \wedge \bigl( Z_1 - Z_2 - \frac{s_1 + s_2}{2}\bigr) \notag \\
&\hspace{-35pt}+ \bigl( Z_3 + \frac{r - s_3}{2}\bigr) \wedge \bigl( Z_1 - Z_2 - \frac{s_1 + s_2}{2}\bigr)+ \bigl( s_2 + s_3 - \frac r2\bigr) \wedge \bigl( Z_1 - Z_2 - \frac{s_1 + s_2}{2}\bigr) \notag \\
&\hspace{-35pt}+ \bigl( Z_3 + \frac{r - s_3}{2} \bigr) \wedge \bigl( Z_3 - Z_2 + \frac{s_2 + s_3}{2}\bigr)+ \bigl( s_3 - \frac r2\bigr) \wedge \bigl( Z_3 - Z_2 + \frac{s_2 + s_3}{2}\bigr). 
%
%
%
%
\label{DHS1:PhiB_definition}
\end{align}
\end{lem}

\begin{bem}
	We highlight that the formula \eqref{DHS1:PhiB_definition} for the phase function $\Phi$ only involves the coordinates that appear in \eqref{DHS1:lTBn_definition} and the relative coordinates $r$ and $\sbold$. This structure allows us to remove the magnetic phase factor in \eqref{DHS1:lTB_definition} with techniques that are similar to the ones used in the analysis of $L_{T,B}$.
\end{bem}

\begin{proof}[Proof of Lemma \ref{DHS1:NTB_action}]
The integral kernel of $N_{T,B}$ reads
\begin{align*}
N_{T,B} (\alpha)(X,r) &= \smash{\frac 2\beta \sum_{n\in \Zbb} \iiint_{\Rbb^{9}} \dd \mathbf u \iiint_{\Rbb^9}\dd \mathbf v} \; G_B^{\i\omega_n} (\zeta_X^r, u_1)\, \alpha(u_1,v_1)\, G_{B}^{-\i\omega_n} (u_2,v_1) \, \ov{\alpha(u_2,v_2)}  \\
&\hspace{150pt} \times G_B^{\i\omega_n}(v_2,u_3)\, \alpha(u_3,v_3)\,  G_{B}^{-\i\omega_n}(\zeta_X^{-r}, v_3),
\end{align*}
where we used the short-hand notation $\zeta_X^r := X+\frac r2$. We also used that
\begin{align}
\frac{1}{\i\omega_n + \ov {\hfrak_B}} (x,y) = -G_B^{-\i\omega_n}(y,x), \label{DHS1:Kernel_of_complex_conjugate}
\end{align}
which follows from $\ov{ A^*(x,y) } = A(y,x)$ and
\begin{align*}
\frac{1}{z - \ov{\hfrak_B}} = \ov{\Bigl(\frac{1}{z - \hfrak_B}\Bigr)^*}.
\end{align*}
We hereby correct a typo in the analogue of \eqref{DHS1:Kernel_of_complex_conjugate} in the proof of \cite[Lemma~11]{Hainzl2017}.

Let us define the coordinates $\Zbold$ and $\sbold$ by
\begin{align*}
\mathbf u &= X + \Zbold + \frac \sbold 2, & \mathbf v &= X + \Zbold - \frac \sbold 2,
%
%
\end{align*}
and note that we interpret them as relative and center-of-mass coordinates. For $N_{T,B}$ this implies
\begin{align*}
N_{T,B}(\alpha) (X,r) &= \iiint_{\Rbb^9} \dd \Zbold \iiint_{\Rbb^9} \dd \sbold \; \e^{-\i \Bbold \cdot (X\wedge Z_1)} \, \e^{\i \Bbold \cdot (X \wedge Z_2)} \, \e^{-\i \Bbold \cdot (X\wedge Z_3)} \; \Acal(X,\Zbold,\sbold)  \\
&\hspace{-60pt} \times \frac{2}{\beta} \sum_{n\in \Zbb}  G_B^{\i\omega_n}(\zeta_X^r,\zeta_{Z_1 + X}^{s_1}) \, G_B^{-\i\omega_n}(\zeta_{Z_2+X}^{s_2}, \zeta_{Z_1 + X}^{-s_1}) \,  G_B^{\i\omega_n}(\zeta_{Z_2+X}^{-s_2}, \zeta_{Z_3+X}^{s_3}) \,  G_B^{-\i\omega_n}(\zeta_X^{-r}, \zeta_{Z_3+X}^{-s_3})
\end{align*}
with $\Acal(X, \Zbold, \sbold)$ in \eqref{DHS1:NTB_alpha_definition}. Here, we used $\Bbold \cdot (X\wedge Z) = Z \cdot (\Bbold \wedge X)$ and that $Z\cdot (\Bbold \wedge X)$ commutes with $Z\cdot (-\i\nabla_X)$, which implies 
\begin{align*}
\alpha(X + Z, s) = \e^{\i Z \cdot (-\i\nabla_X)} \, \alpha(X, s) = \e^{-\i \Bbold \cdot (X\wedge Z)} \; \e^{\i Z \cdot \Pi_X}\alpha(X, s).
\end{align*}
A tedious but straightforward computation that uses Lemma~\ref{DHS1:gB-identities}~(b) shows
\begin{align}
 G_B^{\i\omega_n}(\zeta_X^r,\zeta_{Z_1 + X}^{s_1}) \; G_B^{-\i\omega_n}(\zeta_{Z_2+X}^{s_2}, \zeta_{Z_1 + X}^{-s_1}) \; G_B^{\i\omega_n}(\zeta_{Z_2+X}^{-s_2}, \zeta_{Z_3+X}^{s_3}) \; G_B^{-\i\omega_n}(\zeta_X^{-r}, \zeta_{Z_3+X}^{-s_3}) &\notag \\
&\hspace{-300pt} =\e^{\i \Bbold \cdot (X\wedge Z_1)} \e^{-\i \Bbold \cdot (X \wedge Z_2)}\e^{\i \Bbold \cdot (X\wedge Z_3)} \; \e^{\i  \frac \Bbold 2 \cdot \Phi(\Zbold, r, \sbold)} \; \ell_{T,B}^n(\Zbold, r, \sbold). \label{DHS1:NTB_representation_1}
\end{align}
This proves the claim.
\end{proof}

As in the case of $L_{T,B}$, we analyze the operator $N_{T,B}$ by introducing several steps of simplification. Namely, we write
\begin{align}
N_{T,B} = (N_{T,B} - \tilde N_{T,B}) + (\tilde N_{T,B} - N_{T,B}^{(1)}) + (N_{T,B}^{(1)} - N_T^{(2)}) + N_{T}^{(2)}. \label{DHS1:NTB_decomposition}
\end{align}
with $\tilde N_{T,B}$ in \eqref{DHS1:NtildeTB_definition}, $N_{T,B}^{(1)}$ in \eqref{DHS1:NTB1_definition}, and $N_T^{(2)}$ in \eqref{DHS1:NT2_definition}. To obtain $\tilde N_{T,B}$ we replace the functions $g_B^z$ by  $g_0^z$ in $N_{T,B}$. When we replace $\ell_{T,B}$ by $\ell_{T,0}$, we obtain $N_{T,B}^{(1)}$, and $N_T^{(2)}$ is obtained from $N_{T,B}^{(1)}$ by replacing the magnetic translations $\e^{\i Z_i \cdot \Pi_X}$ by $1$. Using arguments that are comparable to the ones applied in the analysis of the operator $L_{T,B}$, we show in Section~\ref{DHS1:sec:approxNTB} below that the contributions from the terms in the parentheses in \eqref{DHS1:NTB_decomposition} can be treated as remainders. In Section~\ref{DHS1:sec:compquarticterm} we prove a proposition that allows us to extract the quartic term in the Ginzburg--Landau functional from the term $\langle \Delta, N_{T}^{(2)}(\Delta) \rangle$. Finally, we summarize our findings in Section~\ref{DHS1:sec:quarticterms}.

\subsubsection{Approximation of \texorpdfstring{$N_{T,B}$}{NTB}}
\label{DHS1:sec:approxNTB}

\paragraph{The operator $\tilde N_{T,B}$.} The operator $\tilde N_{T,B}$ is defined by
\begin{align}
\tilde N_{T,B}(\alpha) (X,r) &:= \iiint_{\Rbb^{9}} \dd \Zbold \iiint_{\Rbb^{9}} \dd \sbold \; \tilde \ell_{T,B} (\Zbold, r, \sbold) \; \Acal(X, \Zbold , \sbold) \label{DHS1:NtildeTB_definition}
\end{align}
with
\begin{align*}
\tilde \ell_{T,B}(\Zbold, r, \sbold) &:= \frac 2\beta \sum_{n\in\Zbb} \ell_{T,0}^n (\Zbold, r, \sbold) \; \e^{\i  \frac \Bbold 2 \cdot \Phi(\Zbold, r, \sbold)},
\end{align*}
$\mathcal{A}$ in \eqref{DHS1:NTB_alpha_definition}, $\ell_{T,0}^n$ in \eqref{DHS1:lTBn_definition} and $\Phi$ in \eqref{DHS1:PhiB_definition}. The following proposition quantifies the error that we make when we replace $N_{T,B}(\Delta)$ by $\tilde N_{T,B}(\Delta)$ in our computations.

\begin{prop}
\label{DHS1:NTB-NtildeTB}
Assume that $V\alpha_*\in L^{\nicefrac 43}(\Rbb^3)$. For every $T_0>0$ there is $B_0>0$ such that for any $0 < B \leq B_0$, any $T\geq T_0$, any $\Psi \in \Hmag^1(Q_B)$, and $\Delta \equiv \Delta_\Psi$ as in \eqref{DHS1:Delta_definition}, we have
\begin{align*}
|\langle \Delta,  N_{T,B}(\Delta) - \tilde N_{T,B}(\Delta)\rangle| &\leq C \; B^4 \; \Vert V\alpha_*\Vert_{\nicefrac 43}^4 \; \Vert \Psi\Vert_{\Hmag^1(Q_B)}^4.
\end{align*}
\end{prop}

Before we give the proof of Proposition~\ref{DHS1:NTB-NtildeTB}, let us introduce the function
\begin{align}
F_{T,B} &:= \smash{\frac 2\beta \sum_{n\in \Zbb}} \; |g_B^{\i\omega_n} - g_0^{\i\omega_n}| * |g_B^{-\i\omega_n} | * |g_B^{\i\omega_n}| * |g_B^{-\i\omega_n}| \notag \\
&\hspace{100pt}+|g_0^{\i\omega_n}| * |g_B^{-\i\omega_n}  - g_0^{-\i\omega_n}| * |g_B^{\i\omega_n}| * |g_B^{-\i\omega_n}| \notag \\
&\hspace{100pt} +|g_0^{\i\omega_n}| * |g_0^{-\i\omega_n} | * |g_B^{\i\omega_n} - g_0^{\i\omega_n}| * |g_B^{-\i\omega_n}| \notag \\
&\hspace{100pt} +|g_0^{\i\omega_n}| * |g_0^{-\i\omega_n} | * |g_0^{\i\omega_n}| * |g_B^{-\i\omega_n} - g_0^{-\i\omega_n}| \label{DHS1:NTB-NtildeTB_FTB_definition}.
\end{align}
By Lemmas~\ref{DHS1:g0_decay} and \ref{DHS1:gB-g_decay} as well as \eqref{DHS1:g0_decay_f_estimate1} we have
\begin{align}
	\Vert F_{T,B}\Vert_1 \leq C \, B^2
	\label{DHS1:eq:A16}
\end{align}
for $T \geq T_0 > 0$. 

\begin{proof}[Proof of Proposition \ref{DHS1:NTB-NtildeTB}]
The function $|\Psi|$ is periodic and \eqref{DHS1:Magnetic_Sobolev} therefore implies
\begin{align}
\Vert \e^{\i Z \cdot\Pi}\Psi\Vert_6^2 = \Vert \Psi\Vert_6^2 \leq C\,  B \, \Vert \Psi\Vert_{\Hmag^1(Q_B)}^2. \label{DHS1:NTB-NtildeTB_3}
\end{align}
Consequently, we have
\begin{align}
\fint_{Q_B} \dd X \; |\Psi(X)|\; \prod_{i=1}^3  |\e^{\i Z_i\cdot \Pi}\Psi(X)| &\leq \Vert \Psi\Vert_2 \; \prod_{i=1}^3  \Vert \e^{\i Z_i\cdot \Pi}\Psi\Vert_6 \leq C \, B^2 \, \Vert \Psi\Vert_{\Hmag^1(Q_B)}^4 \label{DHS1:NTB-NtildeTB_2}
\end{align}
as well as
\begin{align}
|\langle \Delta,  N_{T,B}(\Delta)- \tilde N_{T,B}(\Delta)\rangle| & \notag\\
&\hspace{-100pt}\leq C\, B^2 \, \Vert \Psi\Vert_{\Hmag^1(Q_B))}^4  \int_{\Rbb^3} \dd r \,  \iiint_{\Rbb^9} \dd \sbold\; |V\alpha_*(r)|\;  |V\alpha_*(s_1)| \; |V\alpha_*(s_2)| \; |V\alpha_*(s_3)|  \notag\\
&\hspace{60pt}\times \iiint_{\Rbb^9} \dd \Zbold \; |\ell_{T,B}(\Zbold, r,\sbold) - \tilde \ell_{T,B}(\Zbold, r, \sbold)|.  \label{DHS1:NTB-NtildeTB_1}
\end{align}
We use the change of variables 
\begin{align}
Z_1' - Z_2' &:= Z_1 - Z_2 - \frac{s_1 +s_2}{2}, & Z_2' - Z_3' &:= Z_2 - Z_3 - \frac{s_2 + s_3}{2}, & Z_3' &:= Z_3 + \frac{r-s_3}{2}, \label{DHS1:NTB_change_of_variables_1}
\end{align}
whence
\begin{align}
Z_1 - \frac{r-s_1}{2} = Z_1' - (r - s_1 - s_2 - s_3). \label{DHS1:NTB_change_of_variables_2}
\end{align}
As in the proof of \eqref{DHS1:LTB-LtildeTB_5}, we conclude
\begin{align*}
\iiint_{\Rbb^9} \dd \Zbold\; |\ell_{T,B}(\Zbold , r, \sbold) - \tilde \ell_{T,B}(\Zbold, r, \sbold)| \leq F_{T,B}(r - s_1 - s_2 - s_3)
\end{align*}
with $F_{T,B}$ in \eqref{DHS1:NTB-NtildeTB_FTB_definition}. We insert the above bound in \eqref{DHS1:NTB-NtildeTB_1} and use
\begin{align*}
\bigl\Vert V\alpha_* \; \bigl( V\alpha_* * V\alpha_* * V\alpha_* * F_{T,B}\bigr) \bigr\Vert_1 &\leq C \; \Vert V\alpha_*\Vert_{\nicefrac 43}^4 \; \Vert F_{T,B}\Vert_1
\end{align*}
as well as \eqref{DHS1:eq:A16}, which proves the claim.
\end{proof}

\paragraph{The operator $N_{T,B}^{(1)}$.} The operator $N_{T,B}^{(1)}$ is defined by
\begin{align}
N_{T,B}^{(1)}(\alpha)(X,r) &:= \iiint_{\Rbb^9} \dd \Zbold \iiint_{\Rbb^9} \dd \sbold \; \ell_{T,0} (\Zbold, r, \sbold) \; \Acal (X, \Zbold , \sbold) \label{DHS1:NTB1_definition}
\end{align}
with $\mathcal{A}$ in \eqref{DHS1:NTB_alpha_definition} and $\ell_{T,0}$ in \eqref{DHS1:lTB_definition}. The following proposition allows us to replace $\langle \Delta, \widetilde{N}_{T,B}(\Delta) \rangle$ by $\langle \Delta, N_{T,B}^{(1)}(\Delta) \rangle$ in our computations.

\begin{prop}
\label{DHS1:NtildeTB-NTB1}
Assume that $|\cdot|^kV\alpha_* \in L^{\nicefrac 43}(\Rbb^3)$ for $k\in \{0,1\}$. For every $T \geq T_0 > 0 $, every $B>0$, every $\Psi \in \Hmag^1(Q_B)$ and $\Delta \equiv \Delta_\Psi$ as in \eqref{DHS1:Delta_definition}, we have 
\begin{align*}
|\langle \Delta, \tilde N_{T,B}(\Delta) - N_{T,B}^{(1)}(\Delta)\rangle| &\leq C \; B^3 \; \bigl( \Vert V\alpha_*\Vert_{\nicefrac 43}^4 + \Vert \, |\cdot|V\alpha_*\Vert_{\nicefrac 43}^4\bigr)\; \Vert \Psi\Vert_{\Hmag^1(Q_B)}^4.
\end{align*}
\end{prop}

Before we start with the proof of Proposition~\ref{DHS1:NtildeTB-NTB1}, we introduce the functions
\begin{align}
	F_T^{(1)} &:= \smash{\frac 2\beta \sum_{n\in \Zbb}} \, |g_0^{\i\omega_n}| * \bigl(|\cdot|\, |g_0^{-\i\omega_n}|\bigr) * \bigl(|\cdot|\, |g_0^{\i\omega_n}|\bigr) * |g_0^{-\i\omega_n}| \notag \\
	&\hspace{100pt}+ |g_0^{\i\omega_n}| * \bigl(|\cdot|\, |g_0^{-\i\omega_n}|\bigr) * |g_0^{\i\omega_n}| * \bigl(|\cdot|\, |g_0^{-\i\omega_n}|\bigr) \notag \\
	&\hspace{100pt}+ |g_0^{\i\omega_n}| * |g_0^{-\i\omega_n}| * \bigl(|\cdot|\, |g_0^{\i\omega_n}|\bigr) * \bigl(|\cdot|\, |g_0^{-\i\omega_n}|\bigr) \label{DHS1:NtildeTB-NTB1_FT1_definition}
\end{align}
and
\begin{align}
	F_T^{(2)} &:= \smash{\frac 2\beta \sum_{n\in \Zbb}} \,
	\bigl( |\cdot| \, |g_0^{\i\omega_n}|\bigr) *  |g_0^{-\i\omega_n}| * |g_0^{\i\omega_n}| * |g_0^{-\i\omega_n}| 
	+ |g_0^{\i\omega_n}| * \bigl(|\cdot|\, |g_0^{-\i\omega_n}|\bigr) * |g_0^{\i\omega_n}| * |g_0^{-\i\omega_n}| \notag\\
	&\hspace{30pt}+ |g_0^{\i\omega_n}| * |g_0^{-\i\omega_n}| * \bigl(|\cdot|\, |g_0^{\i\omega_n}|\bigr) * |g_0^{-\i\omega_n}| 
	+ |g_0^{\i\omega_n}| * |g_0^{-\i\omega_n}| * |g_0^{\i\omega_n}| * \bigl(|\cdot|\, |g_0^{-\i\omega_n}|\bigr). \label{DHS1:NtildeTB-NTB1_FT2_definition}
\end{align}
As long as $T \geq T_0 > 0$, Lemma~\ref{DHS1:g0_decay} and \eqref{DHS1:g0_decay_f_estimate1} imply the bound
\begin{align}
	\Vert F_T^{(1)} \Vert_1 + \Vert F_T^{(2)} \Vert_1 \leq C. \label{DHS1:NtildeTB-NTB1_FT1-2_estimate}
\end{align}

\begin{proof}[Proof of Proposition \ref{DHS1:NtildeTB-NTB1}]
We use \eqref{DHS1:NTB-NtildeTB_2} and estimate
\begin{align}
|\langle \Delta, \tilde N_{T,B}(\Delta) - N_{T,B}^{(1)}(\Delta)\rangle| &  \notag\\
&\hspace{-100pt}\leq C\, B^2 \, \Vert \Psi\Vert_{\Hmag^1(Q_B)}^4 \int_{\Rbb^3} \dd r  \iiint_{\Rbb^9} \dd \sbold \; |V\alpha_*(r)| \; |V\alpha_*(s_1)| \; |V\alpha_*(s_2)| \; |V\alpha_*(s_3)|  \notag \\
&\hspace{30pt}\times \frac{2}{\beta} \sum_{n\in \Zbb} \iiint_{\Rbb^9} \dd \Zbold \; |\ell_{T,0}^n(\Zbold, r, \sbold)| \; \bigl| \e^{\i \Bbold \, \Phi(\Zbold, r, \sbold)} - 1\bigr|. \label{DHS1:eq:A17}
\end{align}
In terms of the coordinates in \eqref{DHS1:NTB_change_of_variables_1} and with \eqref{DHS1:NTB_change_of_variables_2}, the phase function $\Phi$ in \eqref{DHS1:PhiB_definition} can be written as
\begin{align}
\Phi(\Zbold, r,\sbold) &= (Z_2' - Z_3') \wedge (Z_1' - Z_2') + Z_3' \wedge (Z_1' - Z_2') + Z_3' \wedge (Z_3'- Z_2') \notag\\
& \hspace{50pt} + \frac r2 \wedge \bigl( Z_1' - (r - s_1 - s_2 - s_3)\bigr) + \bigl( s_2 + s_3 - \frac r2\bigr) \wedge (Z_1 ' - Z_2')\notag \\
&\hspace{50pt} + \bigl( s_3 - \frac r2\bigr) \wedge (Z_3' - Z_2') + \frac r2 \wedge Z_3'.
\end{align}
We use the estimate $|\e^{\i  \frac \Bbold 2 \cdot \Phi(\Zbold, r, \sbold)} - 1| \leq B \, |\Phi(\Zbold, r, \sbold)|$, \eqref{DHS1:eq:A17}, and argue as in the proof of \eqref{DHS1:LtildeTB-MTB_4} to see that
\begin{align*}
&\frac{2}{\beta} \sum_{n\in \Zbb} \iiint_{\Rbb^9} \dd \Zbold \; |\ell_{T,0}^n(\Zbold, r, \sbold)| \; \bigl| \e^{\i  \frac \Bbold 2 \cdot \Phi(\Zbold, r, \sbold)} - 1\bigr| \\
&\hspace{20pt} \leq CB \; \bigl[ F_T^{(1)} (r - s_1 - s_2 - s_3) + F_T^{(2)}(r - s_1 - s_2 - s_3) \; \bigl(1 + |r| + |s_1| + |s_2| + |s_3|\bigr)\bigr] 
\end{align*}
with $F_T^{(1)}$ in \eqref{DHS1:NtildeTB-NTB1_FT1_definition} and $F_T^{(2)}$ in \eqref{DHS1:NtildeTB-NTB1_FT2_definition}. Young's inequality then implies
\begin{align*}
|\langle \Delta, \tilde N_{T,B}(\Delta) - N_{T,B}^{(1)}(\Delta)\rangle| &\\
&\hspace{-100pt}\leq C\; B^3 \, \Vert \Psi\Vert_{\Hmag^1(Q_B)}^4  \; \bigl( \Vert V\alpha_*\Vert_{\nicefrac 43}^4 + \Vert \, |\cdot| V\alpha_*\Vert_{\nicefrac 43}^4 \bigr) \bigl( \Vert F_T^{(1)}\Vert_1 + \Vert F_T^{(2)}\Vert_1 \bigr).
\end{align*}
Finally, an application of \eqref{DHS1:NtildeTB-NTB1_FT1-2_estimate} 
proves the claim.
\end{proof}

\paragraph{The operator $N_T^{(2)}$.} The operator $N_{T}^{(2)}$ is defined by
\begin{align}
N_T^{(2)}(\alpha) (X, r) &:= \iiint_{\Rbb^9} \dd \Zbold \iiint_{\Rbb^9} \dd \sbold \; \ell_{T,0} (\Zbold, r, \sbold) \, \Acal(X, 0 , \sbold) \label{DHS1:NT2_definition}
\end{align}
with $\mathcal{A}$ in \eqref{DHS1:NTB_alpha_definition} and $\ell_{T,0}$ in \eqref{DHS1:lTB_definition}.

The following proposition allows us to replace $\langle \Delta, N_{T,B}^{(1)}(\Delta) \rangle$ by $\langle \Delta, N_{T}^{(2)}(\Delta) \rangle$ in our computations. We highlight that the $\Hmag^2(Q_B)$-norm of $\Psi$ is needed to bound the difference between the two terms.

\begin{prop}
\label{DHS1:NTB1-NT2}
Assume that $|\cdot|^kV\alpha_* \in L^{\nicefrac 43}(\Rbb^3)$ for $k\in \{0,2\}$. For any $T \geq T_0 >0$, any $B>0$, any $\Psi \in \Hmag^2(Q_B)$, and $\Delta\equiv \Delta_\Psi$ as in \eqref{DHS1:Delta_definition}, we have
\begin{align*}
|\langle \Delta, N_{T, B}^{(1)}(\Delta) - N_{T}^{(2)}(\Delta) \rangle|  &\leq C \; B^3 \; \bigl( \Vert V\alpha_*\Vert_{\nicefrac 43}^4 + \Vert \, |\cdot|^2 V\alpha_*\Vert_{\nicefrac 43}^4\bigr) \\
&\hspace{130pt}  \times  \Vert \Psi\Vert_{\Hmag^1(Q_B)}^3 \; \Vert \Psi\Vert_{\Hmag^2(Q_B)}.
\end{align*}
\end{prop}

Before we prove the above proposition, let us introduce the functions
\begin{align*}
	F_T^{(1)} &:= \frac 2\beta \sum_{n\in \Zbb} |g_0^{\i\omega_n}| *  |g_0^{-\i\omega_n}| * |g_0^{\i\omega_n}| * |g_0^{-\i\omega_n}|
\end{align*}
and
\begin{align*}
	F_T^{(2)} &:= \smash{\frac 2\beta \sum_{n\in \Zbb}} \; |g_0^{\i\omega_n}| *  \bigl(|\cdot|^2 \,|g_0^{-\i\omega_n}| \bigr) * |g_0^{\i\omega_n}| * |g_0^{-\i\omega_n}| \\
	&\hspace{100pt}+ |g_0^{\i\omega_n}| *  \,|g_0^{-\i\omega_n}| *   \bigl(|\cdot|^2 \, |g_0^{\i\omega_n}|\bigr) * |g_0^{-\i\omega_n}| \\
	&\hspace{100pt}+ |g_0^{\i\omega_n}| * |g_0^{-\i\omega_n}| * |g_0^{\i\omega_n}| * \bigl(|\cdot|^2 \, |g_0^{-\i\omega_n}|\bigr).
\end{align*}
For $T \geq T_0 > 0$, an application of Lemma~\ref{DHS1:g0_decay} and the estimate \eqref{DHS1:g0_decay_f_estimate1} on $f(t,\omega)$ show
\begin{align}
	\Vert F_T^{(1)} \Vert_1 + \Vert F_T^{(2)} \Vert_1 \leq C. \label{DHS1:eq:A18}
\end{align}

\begin{proof}
We have
\begin{align}
\langle \Delta, N_{T,B}^{(1)}(\Delta) - N_T^{(2)}(\Delta)\rangle &\notag\\
&\hspace{-115pt}=  16 \int_{\Rbb^3} \dd r \iiint_{\Rbb^9} \dd \sbold \; V\alpha_*(r)V\alpha_*(s_1)V\alpha_*(s_2)V\alpha_*(s_3) \iiint_{\Rbb^9} \dd \Zbold \; \ell_{T,0} (\Zbold, r,\sbold)  \notag \\
&\hspace{-115pt} \times\fint_{Q_B} \dd X \; \ov{\Psi(X)} \; \bigl( \e^{\i Z_1\cdot \Pi_X}\Psi(X) \, \ov{\e^{\i Z_2 \cdot \Pi_X} \Psi(X)} \,  \e^{\i Z_3 \cdot \Pi_X}\Psi(X) - \Psi(X)\ov{\Psi(X)} \Psi(X)\bigr). \label{DHS1:NTB1-NTB2_2}
\end{align}
Apart from the exponential factors, this expression is symmetric under the simultaneous replacement of $(Z_1,Z_2,Z_3)$ by $(-Z_1,-Z_2,-Z_3)$. 
When we expand the magnetic translations in cosine and sine functions of $Z_i \cdot \Pi_X$, $i=1,2,3$, the above symmetry implies that all terms with an odd number of sine functions vanish. Accordingly, we may replace the bracket in the last line of \eqref{DHS1:NTB1-NTB2_2} by 
\begin{align}
& \bigl(\cos(Z_1\cdot \Pi_X) \Psi(X) \; \ov{\cos(Z_2\cdot \Pi_X) \Psi(X)} \; \cos(Z_3\cdot \Pi_X) \Psi(X) - \Psi(X) \overline{ \Psi(X) } \Psi(X) \bigr) \notag \\
&\hspace{10pt}+ \cos(Z_1\cdot \Pi_X) \Psi(X) \; \ov{\i\sin(Z_2\cdot \Pi_X) \Psi(X)}  \;  \i \sin(Z_3\cdot \Pi_X) \Psi(X) \notag \\ 
&\hspace{10pt}+ \i \sin(Z_1\cdot \Pi_X) \Psi(X) \; \ov{ \cos(Z_2\cdot \Pi_X) \Psi(X)} \; \i \sin(Z_3\cdot \Pi_X) \Psi(X) \notag \\
&\hspace{10pt}+ \i \sin(Z_1\cdot \Pi_X) \Psi(X) \; \ov{\i\sin(Z_2\cdot \Pi_X) \Psi(X)}\;  \cos(Z_3\cdot \Pi_X) \Psi(X).  \label{DHS1:NTB1-NTB2_1}
\end{align}
Let us consider the first term in \eqref{DHS1:NTB1-NTB2_1}. We use $|\cos(x)  - 1|^2= 4|\sin^4(\frac x2)| \leq \frac{1}{4} |x|^4$ and the operator inequality in \eqref{DHS1:ZPiX_inequality_quartic} to see that $|\cos(Z\cdot \Pi)-1|^2 \leq C\cdot|Z|^4\; (\Pi^4 + B^2)$ holds. In particular,
\begin{align}
\Vert [ \cos(Z\cdot \Pi) - 1] \Psi \Vert_2^2 &
\leq C \; B^3 \; |Z|^4  \;  \Vert \Psi\Vert_{\Hmag^2(Q_B)}^2. \label{DHS1:NTB1-NTB2_5}
\end{align}
In combination with the estimate \eqref{DHS1:NTB-NtildeTB_3} on $\Vert \e^{\i Z\cdot \Pi}\Psi\Vert_6$, this implies
\begin{align}
&\fint_{Q_B} \dd X \; |\Psi(X)| \bigl| \cos( Z_1 \cdot \Pi)\Psi(X) \; \ov{\cos( Z_2  \cdot \Pi)\Psi(X)} \; \cos( Z_3 \cdot \Pi)\Psi(X) - \Psi(X)\ov{\Psi(X)}\Psi(X)\bigr| \notag\\
&\hspace{15pt}\leq  \Vert \Psi\Vert_6 \;  \Vert (\cos( Z_1\cdot \Pi) - 1)\Psi\Vert_2 \; \Vert \cos(Z_2\cdot \Pi)\Psi\Vert_6 \; \Vert \cos( Z_3\cdot \Pi)\Psi\Vert_6 \notag \\
&\hspace{30pt}+ \Vert \Psi\Vert_6^2 \; \Vert (\cos(Z_2\cdot \Pi) - 1)\Psi\Vert_2 \; \Vert \cos( Z_3\cdot \Pi)\Psi\Vert_6 + \Vert \Psi\Vert_6^3 \; \Vert (\cos(Z_3\cdot \Pi) - 1)\Psi\Vert_2 \notag \\
&\hspace{15pt}\leq C \, B^3 \, \Vert \Psi\Vert_{\Hmag^1(Q_B)}^3\; \Vert \Psi\Vert_{\Hmag^2(Q_B)} \; \bigl(|Z_1|^2 + |Z_2|^2 + |Z_3|^2\bigr). \label{DHS1:NTB1-NTB2_6}
%
\end{align}

To treat the other terms in \eqref{DHS1:NTB1-NTB2_1} we use the operator inequality in \eqref{DHS1:ZPiX_inequality} to see that
\begin{align*}
\Vert \sin(Z \cdot \Pi)\Psi\Vert_2^2  = \langle \Psi, \sin^2(Z\cdot \Pi) \, \Psi\rangle \leq C \, B^2\, |Z|^2 \, \Vert \Psi\Vert_{\Hmag^1(Q_B)}^2, 
%
\end{align*}
which yields 
\begin{align}
&\fint_{Q_B} \dd X \; |\Psi(X)| \; |\cos(Z_i\cdot \Pi)\Psi(X)| \; |\sin(Z_j\cdot \Pi)\Psi(X)|\; |\sin(Z_k\cdot \Pi)\Psi(X)| \notag \\
&\hspace{180pt}\leq C \, B^3 \, \bigl( |Z_j|^2 + |Z_k|^2\bigr) \; \Vert \Psi\Vert_{\Hmag^1(Q_B)}^4. \label{DHS1:NTB1-NTB2_7}
\end{align}
We gather \eqref{DHS1:NTB1-NTB2_2}, \eqref{DHS1:NTB1-NTB2_1}, \eqref{DHS1:NTB1-NTB2_6}, and \eqref{DHS1:NTB1-NTB2_7} and find
\begin{align*}
|\langle \Delta, N_{T,B}^{(1)}(\Delta)- N_T^{(2)}(\Delta)\rangle| &\leq C\, B^3 \, \Vert \Psi\Vert_{\Hmag^1(Q_B)}^3 \; \Vert \Psi\Vert_{\Hmag^2(Q_B)}  \\
&\hspace{-30pt} \times \int_{\Rbb^3} \dd r\iiint_{\Rbb^9}\dd \sbold \; |V\alpha_*(r)|\; |V\alpha_*(s_1)|\; |V\alpha_*(s_2)| \; |V\alpha_*(s_3)| \\
&\hspace{60pt} \times \iiint_{\Rbb^9} \dd \Zbold \; |\ell_{T,0}(\Zbold, r, \sbold)| \; \bigl(|Z_1|^2 + |Z_2|^2 + |Z_3|^2\bigr).
\end{align*}
When we write the coordinates $Z_i$, $i=1,2,3$, in terms of the coordinates in \eqref{DHS1:NTB_change_of_variables_1} and \eqref{DHS1:NTB_change_of_variables_2} plus linear combinations of $r$ and $s_i$, $i=1,2,3$, we see that
\begin{align*}
|Z_1| &\leq |Z_1' -Z_2'| + |Z_2' - Z_3'| + |Z_3'| + |r| + |s_1| + |s_2| + |s_3|,\\
|Z_2| &\leq |Z_2 ' - Z_3'| + |Z_3'| + |r| + |s_2| + |s_3|,\\
|Z_3| &\leq |Z_3'| + |r| + |s_3|.
\end{align*}
We use this and argue as in the proof of \eqref{DHS1:LtildeTB-MTB_4}, which yields
%
%
\begin{align*}
&\iiint_{\Rbb^9} \dd \Zbold \; |\ell_{T,0}(\Zbold, r, \sbold)| \; \bigl(|Z_1|^2 + |Z_2|^2 + |Z_3|^2\bigr)  \\
&\hspace{30pt} \leq C\bigl( F_T^{(1)}(r-s_1-s_2-s_3) \; (|r|^2 + |s_1|^2 + |s_2|^2 + |s_3|^2) + F_T^{(2)}(r - s_1 - s_2 - s_3)\bigr).
\end{align*}
In particular,
\begin{align*}
|\langle \Delta, N_{T,B}^{(1)}(\Delta)- N_T^{(2)}(\Delta)\rangle| &\leq C \;  B^3 \; \Vert \Psi\Vert_{\Hmag^1(Q_B)}^3 \; \Vert \Psi\Vert_{\Hmag^2(Q_B)} \\
&\hspace{40pt} \times \bigl( \Vert V\alpha_*\Vert_{\nicefrac 43}^4 + \Vert \, |\cdot|^2V\alpha_*\Vert_{\nicefrac 43}^4  \bigr) \bigl( \Vert F_T^{(1)}\Vert_1 + \Vert F_T^{(2)}\Vert_1\bigr).
\end{align*}
In combination with \eqref{DHS1:eq:A18}, this proves the claim.
\end{proof}

\subsubsection{Calculation of the quartic term in the Ginzburg--Landau functional}
\label{DHS1:sec:compquarticterm}

The following proposition allows us to extract the quartic term in the Ginzburg--Landau functional in \eqref{DHS1:Definition_GL-functional} from $\langle \Delta, N_T^{(2)}(\Delta) \rangle$.

\begin{prop}
\label{DHS1:NTc2}
Assume $V\alpha_* \in L^{\nicefrac 43}(\Rbb^3)$. For any $B>0$, any $\Psi\in \Hmag^1(Q_B)$, and $\Delta \equiv \Delta_\Psi$ as in \eqref{DHS1:Delta_definition}, we have
\begin{align*}
\langle \Delta, N_{\Tc}^{(2)}(\Delta)\rangle = 8\; \Lambda_3 \; \Vert \Psi\Vert_4^4
\end{align*}
with $\Lambda_3$ in \eqref{DHS1:GL_coefficient_3}. Moreover, for any $T \geq T_0 > 0$, we have
\begin{align*}
	|\langle \Delta,  N_T^{(2)}(\Delta) - N_{\Tc}^{(2)}(\Delta)\rangle| &\leq C\; B^2  \; |T - \Tc| \; \Vert V\alpha_*\Vert_{\nicefrac 43}^4 \; \Vert \Psi\Vert_{\Hmag^1(Q_B)}^4.
\end{align*}
\end{prop}

Before we prove the above proposition, let us introduce the function
\begin{align}
	F_{T,\Tc} &:= \smash{\frac 2\beta \sum_{n\in \Zbb} |2n+1| \bigl[} |g_0^{\i\omega_n^T}| * |g_0^{\i\omega_n^{\Tc}}| * |g_0^{-\i\omega_n^T}| * |g_0^{\i\omega_n^T}| * |g_0^{-\i\omega_n^T}|  \phantom {\sum^i} \notag \\
	&\hspace{120pt}+|g_0^{\i\omega_n^{\Tc}}| *  |g_0^{-\i\omega_n^T}| * |g_0^{-\i\omega_n^{\Tc}}| * |g_0^{\i\omega_n^T}| * |g_0^{-\i\omega_n^T}| \notag \\
	&\hspace{120pt}+|g_0^{\i\omega_n^{\Tc}}| *  |g_0^{-\i\omega_n^{\Tc}}| * |g_0^{\i\omega_n^T}| * |g_0^{\i\omega_n^{\Tc}}| * |g_0^{-\i\omega_n^T}| \notag \\
	&\hspace{120pt}+|g_0^{\i\omega_n^{\Tc}}| * |g_0^{-\i\omega_n^{\Tc}}| * |g_0^{\i\omega_n^{\Tc}}| * |g_0^{-\i\omega_n^T}| * |g_0^{-\i\omega_n^{\Tc}}| \bigr], \label{DHS1:eq:A19}
\end{align}
where we have included the $T$-dependence of the Matsubara frequencies in our notation once more because different temperatures appear in the formula. As long as $T \geq T_0 > 0$, Lemma~\ref{DHS1:g0_decay} and \eqref{DHS1:g0_decay_f_estimate1} imply
\begin{equation}
	\Vert F_{T,\Tc} \Vert_1 \leq C.
	\label{DHS1:eq:A21}
\end{equation}

\begin{proof}[Proof of Proposition \ref{DHS1:NTc2}]
Set
\begin{align*}
\ell_T(\Zbold, r) &:= \frac 2\beta \sum_{n\in \Zbb} g_0^{\i\omega_n}(r - Z_1) \, g_0^{-\i\omega_n}(Z_1 - Z_2) \, g_0^{\i\omega_n}(Z_2 - Z_3) \, g_0^{-\i\omega_n}(Z_3).
\end{align*}
Then, by the change of variables \eqref{DHS1:NTB_change_of_variables_1} and \eqref{DHS1:NTB_change_of_variables_2}, we have
\begin{align*}
\iiint_{\Rbb^9} \dd \Zbold \; \ell_{T,0}(\Zbold, r,\sbold) = \iiint_{\Rbb^9} \dd \Zbold \; \ell_T(\Zbold, r-s_1 - s_2 - s_3).
\end{align*}
We use that $(\pm \i \omega_n + \mu -p^2)^{-1}$ is the Fourier transform of $g_0^{\pm \i\omega_n}(x)$, which yields
\begin{align*}
	\ell_T(\Zbold, r) &= \frac 2\beta \sum_{n\in \Zbb} \iiiint_{\Rbb^{12}} \frac{\dd \mathbf p}{(2\pi)^{12}}  \; \frac{\e^{\i p_1 \cdot (r - Z_1)}}{\i\omega_n + \mu - p_1^2} \frac{\e^{\i p_2\cdot (Z_1 - Z_2)}}{-\i\omega_n + \mu - p_2^2 } \frac{\e^{\i p_3\cdot (Z_2 - Z_3)}}{\i\omega_n + \mu - p_3^2} \frac{\e^{\i p_4\cdot Z_3}}{-\i\omega_n + \mu - p_4^2}.
\end{align*}
Integration over $\Zbold$ gives
\begin{align*}
	\iiint_{\Rbb^9} \dd \Zbold \; \ell_T(\Zbold, r) = \frac 2\beta \sum_{n\in \Zbb} \int_{\Rbb^3} \frac{\dd p}{(2\pi)^3} \; \e^{\i p\cdot r} \frac{1}{(\i\omega_n + \mu - p^2)^2 (\i\omega_n - \mu + p^2)^2}. 
\end{align*}
In view of the partial fraction expansion
\begin{align*}
\frac{1}{(\i\omega_n - E)^2(\i\omega_n + E)^2} = \frac{1}{4E^2} \Bigl[ \frac{1}{(\i\omega_n - E)^2} + \frac{1}{(\i\omega_n + E)^2}\Bigr] - \frac{1}{4E^3} \Bigl[ \frac{1}{\i\omega_n - E} - \frac{1}{\i\omega_n + E}\Bigr]
\end{align*}
and the identity
\begin{align}
\frac{\beta}{2} \frac{1}{\cosh^2(\frac \beta 2z)} = \frac{\dd}{\dd z} \tanh\bigl( \frac \beta 2 z\bigr) = - \frac{2}{\beta} \sum_{n\in \Zbb} \frac{1}{(\i\omega_n - z)^2}, \label{DHS1:cosh2_Matsubara}
\end{align}
which follows from \eqref{DHS1:tanh_Matsubara}, we have
\begin{align*}
\frac{2}{\beta} \sum_{n\in \Zbb} \frac{1}{(\i\omega_n - E)^2(\i\omega_n + E)^2} = \frac{\beta^2}{2} \; \frac{g_1(\beta E)}{E}
\end{align*}
with the function $g_1$ in \eqref{DHS1:XiSigma}. We conclude that
\begin{align*}
\iiint_{\Rbb^9} \dd \Zbold \; \ell_T(\Zbold,r) &= \frac{\beta^2}{2} \int_{\Rbb^3} \frac{\dd p}{(2\pi)^3}  \; \e^{\i p\cdot r} \; \frac{g_1(\beta (p^2-\mu))}{p^2 - \mu}.
\end{align*}
For the term we are interested in, this implies
\begin{align}
\langle \Delta, N_{\Tc}^{(2)}(\Delta)\rangle &= 16\, \Vert \Psi\Vert_4^4 \; \frac{\beta_c^2}{2} \int_{\Rbb^3} \dd r \iiint_{\Rbb^9} \dd \sbold \;  V\alpha_*(r)V\alpha_*(s_1)V\alpha_*(s_2)V\alpha_*(s_3)\notag\\
&\hspace{100pt} \times \int_{\Rbb^3} \frac{\dd p}{(2\pi)^3} \; \e^{\i p\cdot (r-s_1-s_2-s_3)} \; \frac{g_1(\beta_c(p^2-\mu))}{p^2-\mu}\notag\\
&= 8 \,  \Vert \Psi\Vert_4^4 \, \frac{\beta_c^2}{16} \int_{\Rbb^3} \frac{\dd p}{(2\pi)^3} \; |(-2)\hat{V\alpha_*}(p)|^4 \; \frac{g_1(\beta_c(p^2-\mu))}{p^2-\mu} =  8\; \Lambda_3 \; \Vert \Psi\Vert_4^4 \label{DHS1:NTc2_quartic_term_result}
\end{align}
with $\Lambda_3$ in \eqref{DHS1:GL_coefficient_3}. This proves the first claim.

To prove the second claim, we note that
\begin{align}
	 \langle \Delta, N_T^{(2)}(\Delta) - N_{\Tc}^{(2)}(\Delta)& \rangle = 16 \int_{\Rbb^3} \dd r \iiint_{\Rbb^9} \dd \sbold \; V\alpha_*(r)V\alpha_*(s_1)V\alpha_*(s_2)V\alpha_*(s_3) \notag \\
	 & \times \iiint_{\Rbb^9} \dd \Zbold \; \left( \ell_{T,0} - \ell_{\Tc,0} \right) (\Zbold, r,\sbold) \fint_{Q_B} \dd X \; |\Psi(X)|^4. \label{DHS1:eq:A22}
\end{align}
Afterwards, we argue as in the proof of \eqref{DHS1:MTB2_2}, that is, we use the resolvent equation \eqref{DHS1:gTgTc} as well as the change of variables in \eqref{DHS1:NTB_change_of_variables_1} and \eqref{DHS1:NTB_change_of_variables_2} and obtain
\begin{align}
\int_{\Rbb^9} \dd \Zbold \; | \ell_{T,0}(\Zbold,r , \sbold ) - \ell_{\Tc,0}(\Zbold, r,\sbold) | &\leq C\, |T - \Tc| \; F_{T,\Tc}(r-s_1-s_2-s_3) 
\label{DHS1:eq:A20}
\end{align}
%
with the function $F_{T,\Tc}$ in \eqref{DHS1:eq:A19}. 
%
Together with \eqref{DHS1:eq:A22}, this implies
\begin{align*}
|\langle \Delta, N_T^{(2)}(\Delta) - N_{\Tc}^{(2)}(\Delta)\rangle| &\leq C \, |T-\Tc| \,  \Vert \Psi\Vert_6^3 \Vert \Psi\Vert_2 \, \bigl\Vert V\alpha_* \; \bigl( V\alpha_* * V\alpha_* * V\alpha_* * F_{T,\Tc}\bigr)\bigr\Vert_1. 
%
\end{align*}
Finally, an application of \eqref{DHS1:Magnetic_Sobolev}, Young's inequality, and \eqref{DHS1:eq:A21} concludes the proof.
\end{proof}

\subsubsection{Summary: The quartic terms and proof of Theorem~\ref{DHS1:Calculation_of_the_GL-energy}}
\label{DHS1:sec:quarticterms}

Let the assumptions of Theorem~\ref{DHS1:Calculation_of_the_GL-energy} hold. We collect the results of Lemma~\ref{DHS1:NTB_action}, as well as Propositions~\ref{DHS1:NTB-NtildeTB}, \ref{DHS1:NtildeTB-NTB1}, \ref{DHS1:NTB1-NT2}, and \ref{DHS1:NTc2}, which yield
\begin{equation}
	\frac{1}{8} \langle \Delta, N_{T,B}(\Delta) \rangle = \; \Lambda_3 \; \Vert \Psi\Vert_4^4 + R(B)
	\label{DHS1:eq:A28}
\end{equation}
with
\begin{equation*}
	| R(B) | \leq C \; B^3 \; \Vert \Psi \Vert_{\Hmag^1(Q_B)}^3 \; \Vert \Psi \Vert_{\Hmag^2(Q_B)}.
\end{equation*}
Together with \eqref{DHS1:eq:A15}, this completes the proof of  Theorem~\ref{DHS1:Calculation_of_the_GL-energy}.

\subsection{Proof of Lemma \ref{DHS1:Gamma_Delta_admissible} and Proposition \ref{DHS1:Structure_of_alphaDelta}}
\label{DHS1:sec:proofofadmissibility}

We start with the proof of Lemma \ref{DHS1:Gamma_Delta_admissible} and recall the definition of $\Gamma_\Delta$ in \eqref{DHS1:GammaDelta_definition} and that of the normal state $\Gamma_0$ in \eqref{DHS1:Gamma0}. By definition, $\Gamma_\Delta$ is a gauge-periodic generalized fermionic one-particle density matrix. Therefore, we only have to check the trace class condition \eqref{DHS1:Gamma_admissible}.

To this end, we use the expansion \eqref{DHS1:tanh-expansion} of the hyperbolic tangent in terms of the Matsubara frequencies, the first formula in \eqref{DHS1:GammaRelations}, and the resolvent equation \eqref{DHS1:Resolvent_Equation} to write
\begin{align}
\Gamma_\Delta = \frac 12 - \frac 12 \tanh\bigl( \frac \beta 2 H_\Delta\bigr) = \frac 12 + \frac{1}{\beta} \sum_{n\in \Zbb} \frac{1}{\i \omega_n - H_\Delta} = \Gamma_0 + \Ocal + \Qcal_{T,B}(\Delta), \label{DHS1:alphaDelta_decomposition_1}
\end{align}
where
\begin{align}
\Ocal &:= \frac 1\beta \sum_{n\in \Zbb} \frac{1}{ \i \omega_n - H_0} \delta \frac{1}{ \i \omega_n - H_0}, & 
\Qcal_{T,B}(\Delta) &:= \frac 1\beta \sum_{n\in \Zbb} \frac{1}{ \i \omega_n - H_0} \delta\frac{1}{ \i \omega_n - H_0} \delta \frac{1}{ \i \omega_n - H_\Delta} \label{DHS1:alphaDelta_decomposition_2}
\end{align}
with $\delta$ in \eqref{DHS1:HDelta_definition}. Since $\Ocal$ is offdiagonal, we have $[\Ocal]_{11} =0$ and the operator $(1+ \pi^2) [\Ocal]_{11}$ is locally trace class trivially. Using \eqref{DHS1:Calculation-entry}, we see that
\begin{align*}
\bigl[ \Qcal_{T,B}(\Delta)\bigr]_{11} = \frac 1\beta \sum_{n\in \Zbb} \frac{1}{\i\omega_n - \hfrak_B} \Delta \frac{1}{\i\omega_n + \ov{\hfrak_B}} \ov \Delta \Bigl[ \frac{1}{\i\omega_n - H_\Delta}\Bigr]_{11}.
\end{align*}
An application of Hölder's inequality shows that $(1+\pi^2)[\Qcal_{T,B}(\Delta)]_{11}$ is locally trace class. It remains to show that $(1+\pi^2) \gamma_0$ is locally trace class. But this follows from the bound $(1+x) (\exp(\beta (x - \mu))+1)^{-1} \leq C_{\beta,a} \e^{-\frac \beta 2 (x-\mu)}$ for $x \geq a$, the diamagnetic inequality for the magnetic heat kernel, see e.g. \cite[Theorem 4.4]{LiebSeiringer}, and the explicit formula for the heat kernel of the Laplacian. This concludes the proof of Lemma~\ref{DHS1:Gamma_Delta_admissible}.

%

Let us continue with the proof of Proposition~\ref{DHS1:Structure_of_alphaDelta}. We use $\alpha_\Delta = [\Gamma_\Delta]_{12}$, the resolvent equation \eqref{DHS1:Resolvent_Equation} and \eqref{DHS1:alphaDelta_decomposition_1} to see that
\begin{align*}
\alpha_\Delta &= [\Ocal]_{12} + [\Qcal_{T,B}(\Delta)]_{12} = [\Ocal]_{12} + \Rcal_{T,B}(\Delta), 
\end{align*}
with $\Ocal$ in \eqref{DHS1:alphaDelta_decomposition_2},
and
\begin{align*}
\Rcal_{T,B}(\Delta) &:=  \frac 1\beta \sum_{n\in \Zbb} \Bigl[ \frac{1}{ \i \omega_n - H_0} \delta\frac{1}{ \i \omega_n - H_0} \delta\frac{1}{ \i \omega_n - H_\Delta} \delta \frac{1}{ \i \omega_n - H_0}\Bigr]_{12}.
\end{align*}
The definition of $L_{T,B}$ in \eqref{DHS1:LTB_definition} implies $[\Ocal]_{12} = -\frac 12 L_{T,B}\Delta$, and we define
\begin{align}
\eta_0(\Delta) &:= \frac 12 \bigl(L_{T,B}\Delta - M_{T,B}\Delta\bigr) + \frac 12 \bigl( M_T^{(1)}\Delta - M_{\Tc}^{(1)}\Delta\bigr) + \Rcal_{T,B}(\Delta), \notag \\
\eta_\perp(\Delta) &:= \frac 12 \bigl( M_{T,B}\Delta - M_{T}^{(1)}\Delta\bigr), \label{DHS1:eta_perp_definition}
\end{align}
with $M_{T,B}$ in \eqref{DHS1:MTB_definition} and $M_T^{(1)}$ in \eqref{DHS1:MT1_definition}. Proposition~\ref{DHS1:MT1} implies that $-\frac 12 M_{\Tc}^{(1)} \Delta = \Psi\alpha_*$, so these definitions allow us to write $\alpha_{\Delta}$ as in \eqref{DHS1:alphaDelta_decomposition_eq1}. It remains to prove the properties of $\eta_0$ and $\eta_\perp$ that are listed in Proposition~\ref{DHS1:Structure_of_alphaDelta}.

We start with the proof of \eqref{DHS1:alphaDelta_decomposition_eq2}, and note that 
\begin{align*}
\Rcal_{T,B}(\Delta) &= \frac 1\beta \sum_{n\in\Zbb} \frac 1{\i\omega_n - \hfrak_B} \, \Delta \,  \frac 1{\i\omega_n + \ov{\hfrak_B}}\,  \ov \Delta\,  \Bigl[ \frac{1}{\i \omega_n - H_\Delta}\Bigr]_{11}\,  \Delta \, \frac 1{\i\omega_n + \ov{\hfrak_B}}.
\end{align*}
Using Hölder's inequality, we immediately see that $\Vert \Rcal_{T,B}(\Delta)\Vert_2 \leq C \beta^3 \Vert \Delta\Vert_6^3$. Furthermore, we estimate
\begin{equation*}
	\Vert \pi \Rcal_{T,B}(\Delta) \Vert_2 \leq \frac 1\beta \sum_{n\in\Zbb} \Bigl\Vert \pi \frac 1{\i\omega_n - \hfrak_B} \Bigr\Vert_{\infty} \Bigl\Vert \frac 1{\i\omega_n + \ov{\hfrak_B}} \Bigr\Vert_{\infty}^2 \Bigl\Vert \Bigl[ \frac{1}{\i \omega_n - H_\Delta}\Bigr]_{11} \Bigr\Vert_{\infty} \Vert \Delta \Vert_6^3.
\end{equation*}
With the help of $\Vert A\Vert_\infty^2 = \Vert A^*A\Vert_\infty$ for a general operator $A$, the first norm on the right side is bounded by
\begin{align*}
\Bigl\Vert \pi \frac 1{\i\omega_n - \hfrak_B} \Bigr\Vert_{\infty} &\leq \Bigl\Vert \frac 1{-\i\omega_n - \hfrak_B}\Bigr\Vert_\infty^{\nicefrac 12} \Bigl\Vert\pi^2 \frac 1{\i\omega_n - \hfrak_B} \Bigr\Vert_{\infty}^{\nicefrac 12} \leq C \,  |\omega_n|^{-\nicefrac 12}.
\end{align*}
Hence,
\begin{equation}
	\Vert \pi \Rcal_{T,B}(\Delta) \Vert_2 \leq C\, \beta^{\nicefrac 52}\,\Vert \Delta \Vert_6^3.
	\label{DHS1:eq:A25}
\end{equation}
With a similar argument, we see that $\Vert \Rcal_{T,B}(\Delta) \pi \Vert_2$ is bounded by the right side of \eqref{DHS1:eq:A25}, too. An application of Lemma~\ref{DHS1:Schatten_estimate} and of \eqref{DHS1:Magnetic_Sobolev} on the right side of \eqref{DHS1:eq:A25} finally shows
\begin{equation*}
	\Vert \Rcal_{T,B}(\Delta) \Vert_{H^1(Q_B \times \mathbb{R}^3_{\mathrm{s}})}^2 \leq C \; B^3 \; \Vert \Psi \Vert_{\Hmag^1(Q_B)}^6. 
\end{equation*}
The remaining terms in $\eta_0(\Delta)$ can be estimated with the help of Propositions~\ref{DHS1:LTB-LtildeTB}, \ref{DHS1:LtildeTB-MTB}, and \ref{DHS1:MT1}, which establishes \eqref{DHS1:alphaDelta_decomposition_eq2}. 

It remains to prove \eqref{DHS1:alphaDelta_decomposition_eq3} and \eqref{DHS1:alphaDelta_decomposition_eq4}. We start with the proof of \eqref{DHS1:alphaDelta_decomposition_eq3} and write
\begin{align}
	\eta_\perp(\Delta)(X,r) = \iint_{\Rbb^3\times \Rbb^3} \dd Z \dd s\; k_T(Z, r-s) \; [ \cos(Z\cdot \Pi_X) - 1 ]  \; \Delta(X, s).
	\label{DHS1:eq:A27}
\end{align}
Using \eqref{DHS1:NTB1-NTB2_5} we see that
\begin{align}
	\Vert \eta_\perp\Vert_2^2 &\leq C\; B^3  \; \Vert F_T^{(2)}\Vert_1^2 \; \Vert V\alpha_*\Vert_2^2\;  \Vert \Psi\Vert_{\Hmag^2(Q_B)}^2,
	\label{DHS1:eq:A26}
\end{align}
with the function $F_T^{(2)}$ in \eqref{DHS1:LtildeTB-MTB_FT_definition}. The $L^1(\Rbb^3$)-norm of this function was estimated in \eqref{DHS1:NtildeTB-NTB1_FT1-2_estimate}. We use this bound and conclude the claimed bound for the $L^2(Q_B \times \mathbb{R}^3_{\mathrm{s}})$-norm of $\eta_{\perp }$. Bounds for $\Vert \tilde \pi_r \eta_\perp \Vert_2$ and $\Vert |r|\eta_\perp \Vert_2$ can be proved similarly and we leave the details to the reader. 

To prove the claimed bound for $\Vert \Pi_X\eta_\perp \Vert_2$, we need to replace $[\cos(Z\cdot\Pi_X)- 1]\Psi(X)$ by $\Pi_X[\cos(Z\cdot \Pi_X) - 1]\Psi(X)$ in the proof of \eqref{DHS1:eq:A26}. Using the intertwining relation \eqref{DHS1:PiXcos} in Lemma~\ref{DHS1:CommutationII} below, the operator inquality \eqref{DHS1:ZPiX_inequality} for $(Z\cdot \Pi)^2$, and the equality \eqref{DHS1:PiPi2Pi_equality} for $\Pi \, \Pi^2\, \Pi$, we see that
\begin{align}
	\Vert \Pi [\cos(Z\cdot\Pi) - 1]\Psi \Vert^2 &\leq C \; B^3 \; |Z|^2 \; \Vert \Psi\Vert_{\Hmag^2(Q_B)}^2 \label{DHS1:alphaDelta_decomposition_4}
\end{align}
holds. The claimed bound for $\Vert \Pi_X\eta_\perp \Vert_2$ follows from \eqref{DHS1:eq:A27} and \eqref{DHS1:alphaDelta_decomposition_4}, which, in combination with the previous considerations, proves \eqref{DHS1:alphaDelta_decomposition_eq3}. 

To prove \eqref{DHS1:alphaDelta_decomposition_eq4}, we note that for any two radial functions $f,g\in L^2(\Rbb^3)$ the function
\begin{equation}
	\iint_{\Rbb^6} \dd r \dd s \; f(r) \, k_T(Z, r-s) \, g(s) 
\end{equation}
is radial in $Z$. We claim that this implies that the operator
\begin{equation}
	\iiint_{\Rbb^9} \dd Z \dd s \dd r \; f(r)  k_T(Z, r-s) g(s) [ \cos(Z\cdot \Pi) - 1 ] 
	\label{DHS1:eq:A29}
\end{equation}
equals $h( \Pi^2 )$ for some function $h \colon [0,\infty) \to \mathbb{R}$. To prove this, let us denote by $\tilde \Pi$ the same operator $\Pi$ but understood to act on $L^2(\mathbb{R}^3)$ instead of $\Lmag^2(Q_B)$. From \cite[Lemma~28]{Hainzl2017} we know that the above statement is true when $\Pi$ is replaced by $\tilde \Pi$. To reduce our claim to this case, we use the unitary Bloch--Floquet transformation 
\begin{equation}
	( \mathcal{U}_{\mathrm{BF}} \Psi )(k,X) := \sum_{\lambda \in \Lambda_B} \e^{-\i k \cdot (X-\lambda)} (T_B(\lambda) \Psi)(X) 
	\label{DHS1:eq:ABF1}
\end{equation}
with $T_B(\lambda)$ in \eqref{DHS1:Magnetic_Translation_Charge2} and inverse
\begin{equation}
	( \mathcal{U}^*_{\mathrm{BF}} \Phi )(X) = \int_{[0,\sqrt{2 \pi B}]^3} \mathrm{d} k \; \e^{\i k\cdot X} \Phi(k,X).
	\label{DHS1:eq:ABF2}
\end{equation}
The magnetic momentum operator $\tilde \Pi$ obeys the identity
\begin{equation}
	\mathcal{U}_{\mathrm{BF}}\,  \tilde \Pi\,  \mathcal{U}^*_{\mathrm{BF}} = \int^{\oplus}_{[0,\sqrt{ 2 \pi B}]^3} \mathrm{d}k \; \tilde \Pi(k)
	\label{DHS1:eq:ABF3}
\end{equation}
with $\tilde \Pi(k) = \Pi + k$ acting on $\Lmag^2(Q_B)$. The claim follows when we conjugate both sides of the equation
\begin{equation*}
	\iiint_{\Rbb^9} \dd Z \dd s \dd r \; f(r) k_T(Z, r-s) g(s) [ \cos(Z\cdot \tilde{\Pi}) - 1 ] = h(\tilde{\Pi}^2)
\end{equation*}
with the Bloch--Floquet transformation and use that $\tilde \Pi(0) = \Pi$. Eq.~\eqref{DHS1:alphaDelta_decomposition_eq4} is a direct consequence of the fact that the operator in \eqref{DHS1:eq:A29} equals $h( \Pi^2 )$. This proves Proposition~\ref{DHS1:Structure_of_alphaDelta}.

\subsection{Proof of Proposition \ref{DHS1:Lower_Tc_a_priori_bound}}
\label{DHS1:Lower_Tc_a_priori_bound_proof_Section}

Let the assumptions of Proposition~\ref{DHS1:Lower_Tc_a_priori_bound} hold. We show that there are constants $D_0>0$ and $B_0>0$ such that for $0 < B \leq B_0$ and temperatures $T$ obeying
\begin{align*}
	0 < T_0 \leq T < \Tc (1 - D_0 B)
\end{align*}
there is a function $\Psi \in \Hmag^2(Q_B)$, such that the Gibbs state $\Gamma_{\Delta}$ in \eqref{DHS1:GammaDelta_definition} built upon the gap function $\Delta(X,r) = -2 V\alpha_*(r) \Psi(X)$ obeys \eqref{DHS1:Lower_critical_shift_2}.

To prove this, we choose $\psi \in \Hmag^2(Q_1)$ with $\Vert \psi\Vert_{\Hmag^2(Q_B)}=1$ and $\Psi \in \Hmag^2(Q_B)$ as in \eqref{DHS1:GL-rescaling}. This, in particular, implies $\Vert \Psi\Vert_{\Hmag^2(Q_B)}=1$. We collect the results of Propositions~\ref{DHS1:Structure_of_alphaDelta}, \ref{DHS1:BCS functional_identity}, \ref{DHS1:Rough_bound_on_BCS energy}, as well as \eqref{DHS1:Magnetic_Sobolev} and \eqref{DHS1:eq:A28}, and conclude that
\begin{align*}
	\FBCS(\Gamma_\Delta) - \FBCS(\Gamma_0) &< B \, \bigl( - cD_0 \, \Vert \psi\Vert_2^2 + C \bigr)
\end{align*}
holds as long as $B$ is small enough. We remark that this argument can be carried out without the assumption of $\Hmag^2(Q_1)$-regularity of $\psi$ by instead using the sign of $V$. Compare this to the discussion below \eqref{DHS1:Upper_Bound_proof_1}. Choosing $D_0 = \frac{C}{c \Vert \psi\Vert_2^2}$ ends the proof of Proposition~\ref{DHS1:Lower_Tc_a_priori_bound}.

\section{The Structure of Low-Energy States}
\label{DHS1:Lower Bound Part A}

In Section~\ref{DHS1:Upper_Bound} we use a Gibbs state to show that the BCS free energy is bounded from above by the Ginzburg--Landau energy plus corrections of lower order. The Gibbs state has a Cooper pair wavefunction which is given by a product of the form $\alpha_*(r) \Psi(X)$ to leading order, where $\Psi$ is a minimizer of the Ginzburg--Landau functional in \eqref{DHS1:Definition_GL-functional} and $\alpha_*$ is the unique solution of the gap equation \eqref{DHS1:alpha_star_ev-equation}.
Moreover, close to the critical temperature the Cooper pair wave function is small in an appropriate sense, which allows us to expand the BCS functional in powers of $\Psi$ and to obtain the terms in the Ginzburg--Landau functional.

Our proof of a matching lower bound for the BCS free energy in Section~\ref{DHS1:Lower Bound Part B} is based on the fact that certain low-energy states of the BCS functional have a Cooper pair wave function with a similar structure. The precise statement is provided in Theorem~\ref{DHS1:Structure_of_almost_minimizers} below, which is the main technical novelty of this paper. This section is devoted to its proof.

%

We recall the definition of the generalized one-particle density matrix $\Gamma$ in \eqref{DHS1:Gamma_introduction}, its offdiagonal entry $\alpha$, as well as the normal state $\Gamma_0$ in \eqref{DHS1:Gamma0}.

\begin{thm}[Structure of low-energy states]
\label{DHS1:Structure_of_almost_minimizers}
Let Assumptions \ref{DHS1:Assumption_V} and \ref{DHS1:Assumption_KTc} hold. For given $D_0, D_1 \geq 0$, there is a constant $B_0>0$ such that for all $0 <B \leq B_0$ the following holds: If $T>0$ obeys $T - \Tc \geq -D_0B$ and if $\Gamma$ is a gauge-periodic state with low energy, that is,
\begin{align}
\FBCS(\Gamma) - \FBCS(\Gamma_0) \leq D_1B^2, \label{DHS1:Second_Decomposition_Gamma_Assumption}
\end{align}
then there are $\Psi\in \Hmag^1(Q_B)$ and $\xi\in \Hsymm$ such that
\begin{align}
\alpha(X,r) = \Psi(X)\alpha_*(r) + \xi(X,r), \label{DHS1:Second_Decomposition_alpha_equation}
\end{align}
where
\begin{align}
\sup_{0< B\leq B_0} \Vert \Psi\Vert_{\Hmag^1(Q_B)}^2 &\leq C, &  \Vert \xi\Vert_{\Hsymm}^2 &\leq CB^2 \bigl( \Vert \Psi\Vert_{\Hmag^1(Q_B)}^2 + D_1\bigr). \label{DHS1:Second_Decomposition_Psi_xi_estimate}
\end{align}
\end{thm}

\begin{varbems}
\begin{enumerate}[(a)]
\item Equation \eqref{DHS1:Second_Decomposition_Psi_xi_estimate}
proves that, despite $\Psi$ being dependent on $B$, it is a macroscopic quantity in the sense that its $\Hmag^1(Q_B)$-norm scales as that of the function in \eqref{DHS1:GL-rescaling}.

\item We highlight that, in contrast to the $\Hmag^1(Q_B)$-norm of $\Psi$, the $\Hsymm$-norm of $\xi$ is not scaled with additional factors of $B$, see \eqref{DHS1:H1-norm}. The unscaled $\Lmag^2(Q_B)$-norm of $\Psi$ is of the order $B^{\nicefrac 12}$, whence it is much larger than that of $\xi$.
\item Theorem~\ref{DHS1:Structure_of_almost_minimizers} should be compared to \cite[Eq.~(5.1)]{Hainzl2012} and \cite[Theorem~22]{Hainzl2017}.
\end{enumerate}
\end{varbems}

Theorem~\ref{DHS1:Structure_of_almost_minimizers} contains the natural a priori bounds for the Cooper pair wave function $\alpha$ of a low-energy state $\Gamma$ in the sense of \eqref{DHS1:Second_Decomposition_Gamma_Assumption}. However, in Section~\ref{DHS1:Lower Bound Part B} we are going to need more regularity of $\Psi$ than is provided by Theorem \ref{DHS1:Structure_of_almost_minimizers}. More precisely, we are going to use the function $\Psi$ from this decomposition to construct a Gibbs state $\Gamma_{\Delta_{\Psi}}$ and apply our trial state analysis provided by Propositions \ref{DHS1:Structure_of_alphaDelta} and \ref{DHS1:BCS functional_identity} as well as Theorem~\ref{DHS1:Calculation_of_the_GL-energy} to extract the Ginzburg--Landau energy. In order to control the errors during this analysis, we need the $\Hmag^2(Q_B)$-norm of $\Psi$. The following corollary provides us with a decomposition of $\alpha$ in terms of a center-of-mass Cooper pair wave function $\Psi_\leq$ with $\Hmag^2(Q_B)$-regularity.



\begin{kor}
\label{DHS1:Structure_of_almost_minimizers_corollary}
Let the assumptions of Theorem~\ref{DHS1:Structure_of_almost_minimizers} hold and let $\varepsilon \in [B, B_0]$. Let $\Psi$ be as in 
\eqref{DHS1:Second_Decomposition_alpha_equation} and define
\begin{align}
	\Psi_\leq &:= \Idbb_{[0,\varepsilon]}(\Pi^2) \Psi, &  \Psi_> &:= \Idbb_{(\varepsilon,\infty)}(\Pi^2) \Psi. \label{DHS1:PsileqPsi>_definition}
\end{align}
Then, we have
\begin{align}
	\Vert \Psi_\leq\Vert_{\Hmag^1(Q_B)}^2 &\leq \Vert \Psi\Vert_{\Hmag^1(Q_B)}^2, \notag \\ 
	\Vert \Psi_\leq \Vert_{\Hmag^k(Q_B)}^2 &\leq C\, (\varepsilon B^{-1})^{k-1} \, \Vert \Psi\Vert_{\Hmag^1(Q_B)}^2, \qquad k\geq 2,  \label{DHS1:Psileq_bounds}
\end{align}
as well as 
\begin{align}
\Vert \Psi_>\Vert_2^2 &\leq C \varepsilon^{-1}B^2 \, \Vert \Psi\Vert_{\Hmag^1(Q_B)}^2, & \Vert \Pi\Psi_>\Vert_2^2 &\leq CB^2 \, \Vert \Psi\Vert_{\Hmag^1(Q_B)}^2. \label{DHS1:Psi>_bound}
\end{align}
Furthermore,
\begin{align}
	\sigma_0(X,r) := \Psi_>(X)\alpha_*(r) \label{DHS1:sigma0}
\end{align}
satisfies
\begin{align}
	\Vert \sigma_0\Vert_{\Hsymm}^2 &\leq C\varepsilon^{-1}B^2 \, \Vert \Psi\Vert_{\Hmag^1(Q_B)}^2 \label{DHS1:sigma0_estimate}
\end{align}
and, with $\xi$ in \eqref{DHS1:Second_Decomposition_alpha_equation}, the function
\begin{align}
	\sigma :=  \xi + \sigma_0 \label{DHS1:sigma}
\end{align}
obeys
\begin{align}
	\Vert \sigma\Vert_{\Hsymm}^2 \leq CB^2 \bigl( \varepsilon^{-1}\Vert \Psi\Vert_{\Hmag^1(Q_B)}^2 + D_1\bigr). \label{DHS1:Second_Decomposition_sigma_estimate}
\end{align}
In terms of these functions, the Cooper pair wave function $\alpha$ of the low-energy state $\Gamma$ in \eqref{DHS1:Second_Decomposition_Gamma_Assumption} admits the decomposition
\begin{align}
	\alpha(X,r) = \Psi_\leq (X)\alpha_*(r) + \sigma(X,r). \label{DHS1:Second_Decomposition_alpha_equation_final}
\end{align}
\end{kor}


%
%

\begin{proof}
The bounds for $\Psi_\leq$ and $\Psi_>$ in \eqref{DHS1:Psileq_bounds} and \eqref{DHS1:Psi>_bound} are a direct consequence of their definition in \eqref{DHS1:PsileqPsi>_definition}.
%
The bound \eqref{DHS1:Psi>_bound} immediately implies \eqref{DHS1:sigma0_estimate}. Moreover, $\sigma$ obeys \eqref{DHS1:Second_Decomposition_sigma_estimate} by \eqref{DHS1:Second_Decomposition_Psi_xi_estimate} and \eqref{DHS1:sigma0_estimate}. Finally, \eqref{DHS1:Second_Decomposition_alpha_equation_final} follows from \eqref{DHS1:Second_Decomposition_alpha_equation}.
\end{proof}


\subsection{A lower bound for the BCS functional}

We start the proof of Theorem \ref{DHS1:Structure_of_almost_minimizers} with the following lower bound on the BCS functional.


\begin{lem}
Let $\Gamma_0$ be the normal state in \eqref{DHS1:Gamma0}. We have the lower bound
\begin{align}
\FBCS(\Gamma) - \FBCS(\Gamma_0) \geq  \Tr\bigl[ (K_{T,B} - V) \alpha  \alpha^*\bigr] + \frac{4T}{5} \Tr\bigl[ (\alpha^* \alpha)^2\bigr], \label{DHS1:Lower_Bound_A_3}
\end{align}
where $K_{T,B} = K_T(\pi)$ and $V\alpha(x,y) = V(x-y) \alpha(x,y)$.
\end{lem}

\begin{proof}
The statement follows from Eqs.~(5.3)--(5.12) in \cite{Hainzl2012} with the evident replacements. The argument uses the relative entropy inequality \cite[Lemma~1]{Hainzl2012}, which is a refinement of the bound \cite[Theorem~1]{Hainzl2008_Lewin}.
\end{proof}

In Proposition \ref{DHS1:KTV_Asymptotics_of_EV_and_EF} in Appendix~\ref{DHS1:KTV_Asymptotics_of_EV_and_EF_Section} we show that the magnetic field can lower the lowest eigenvalue zero of $K_{\Tc} - V$
at most by a constant times $B$. This information is used in the following lemma to bound $K_{T,B} - V$ from below by a nonnegative operator, up to a correction of the size $CB$. The inequality \eqref{DHS1:KTB_Lower_bound_eq} below is stated for $K_{T,B} - V$ as a one-particle operator but it holds equally for the operator $K_{T,B} - V(x-y)$ in \eqref{DHS1:Lower_Bound_A_3} because $V$ intertwines as $T(y)^* V(x) T(y) = V(x-y)$ with the magnetic translations $T(y)$ in \eqref{DHS1:Magnetic_Translation}.

\begin{lem}
\label{DHS1:KTB_Lower_bound}
Let Assumptions \ref{DHS1:Assumption_V} and \ref{DHS1:Assumption_KTc} be true.
For any $D_0 \geq 0$, there are constants $B_0>0$ and $T_0>0$ such that for $0< B\leq B_0$ and $T>0$ with $T - \Tc \geq -D_0B$, the estimate
\begin{align}
K_{T, B} - V &\geq c \; (1 - P) (1 + \pi^2) (1- P) + c \, \min \{ T_0, (T - \Tc)_+\} - CB \label{DHS1:KTB_Lower_bound_eq}
\end{align}
holds. Here, $P = |\alpha_*\rangle\langle \alpha_*|$ is the orthogonal projection onto the ground state $\alpha_*$ of $K_{\Tc} - V$.
\end{lem}

\begin{proof}
We prove two lower bounds on $K_{T,B} - V$, which we add up to etablish \eqref{DHS1:KTB_Lower_bound_eq}.

\emph{Step 1.} We claim that there are $T_0, c, C >0$ such that
\begin{align}
K_{T,B} - V \geq c \; \min \{ T_0, (T - \Tc)_+\} - CB. \label{DHS1:KTB_Lower_bound_5}
\end{align}
To prove \eqref{DHS1:KTB_Lower_bound_5}, we note that the derivative of the symbol $K_T$ in \eqref{DHS1:KT-symbol} with respect to $T$ equals
\begin{align}
\frac{\dd}{\dd T} K_T(p) = 2\;  \frac{( \frac{p^2 - \mu}{2T})^2}{\sinh^2(\frac{p^2-\mu}{2T})} \label{DHS1:KTc_bounded_derivative}
\end{align}
and is bounded from above by 2. If $T \leq \Tc$, we infer $K_{T,B} - K_{\Tc,B} \geq -2D_0B$ as an operator inequality, which, in combination with Proposition~\ref{DHS1:KTV_Asymptotics_of_EV_and_EF} in the appendix,  proves \eqref{DHS1:KTB_Lower_bound_5} in this case. To treat the case $T \geq \Tc$, we denote by $e_0^{T,B}$ and $e_1^{T,B}$ the lowest and the second lowest eigenvalue of the operator $K_{T,B} - V$, respectively. Also let $P_{T,B}$ be the spectral projection corresponding to $e_0^{T,B}$ and define $Q_{T,B} = 1 - P_{T,B}$. We have 
\begin{align*}
	K_{T,B} - V \geq e_0^{T,B} P_{T,B} + e_{1}^{T,B} Q_{T,B}.
\end{align*}
Since $K_T(p) - K_{\Tc}(p) \geq 0$ for all $p \in \mathbb{R}^3$, which follows from \eqref{DHS1:KTc_bounded_derivative}, we know the lower bound $e_{1}^{T,B} \geq e_{1}^{\Tc,B} \geq \kappa$ for some $\kappa > 0$. Here, the second inequality follows from Proposition~\ref{DHS1:KTV_Asymptotics_of_EV_and_EF}. From Proposition~\ref{DHS1:KTV_Asymptotics_of_EV_and_EF} we also know that the lowest eigenvalue of $K_{T,B} - V $ is simple. According to \eqref{DHS1:KTc_bounded_derivative}, the function $T \mapsto K_T(p)$ is increasing and has a non-vanishing derivative for each $p \in \mathbb{R}^3$. Analytic perturbation theory therefore implies the lower bound $e_0^{T,B} \geq e_0^{\Tc,B} + c (T - \Tc)$ for some $c > 0$ as long as $|T - \Tc |$ is small enough. Since Proposition~\ref{DHS1:KTV_Asymptotics_of_EV_and_EF} shows $e_0^{\Tc,B} \geq -CB$ these consideration prove \eqref{DHS1:KTB_Lower_bound_5} in the case $T \geq \Tc$.

\emph{Step 2.} We claim there are $c,C>0$ such that
\begin{align}
K_{T,B} - V \geq c \; (1 - P) (1 + \pi^2) (1- P) - CB. \label{DHS1:KTB_Lower_bound_9}
\end{align}
From the arguments in Step~1 we know that we can replace $T$ by $\Tc$ for a lower bound if we allow for a remainder of the size $-CB$. To prove \eqref{DHS1:KTB_Lower_bound_9}, we choose $0 < \eta < 1$ and write
\begin{align}
K_{\Tc,B}-V = e_0^BP_B + (1-P_B) [(1-\eta) K_{\Tc,B} -V](1-P_B) + \eta (1-P_B) K_{\Tc,B} (1-P_B), \label{DHS1:KTB_Lower_bound_1}
\end{align}
where $e_0^B$ denotes the ground state energy of $K_{\Tc, B} - V$ and $P_B = |\alpha_*^B\rangle \langle \alpha_*^B|$ is the spectral projection onto the corresponding unique ground state vector $\alpha_*^B$. From Proposition~\ref{DHS1:KTV_Asymptotics_of_EV_and_EF} we know that the first term on the right side of \eqref{DHS1:KTB_Lower_bound_1} is bounded from below by $-CB$. The lowest eigenvalue of $K_{\Tc} - V$ is simple and isolated from the rest of the spectrum. Proposition~\ref{DHS1:KTV_Asymptotics_of_EV_and_EF} therefore implies that the second term in \eqref{DHS1:KTB_Lower_bound_1} is nonnegative as long as $\eta$ is, independently of $B$, chosen small enough, and can be dropped for a lower bound.
To treat the third term, we note that the symbol $K_T(p)$ in \eqref{DHS1:KT-symbol} satisfies the inequality $K_{\Tc}(p) \geq c' (1 + p^2)$ for some constant $c'$, and hence $K_{\Tc,B} \geq c' (1 + \pi^2)$. In combination, the above considerations prove
\begin{align*}
K_{\Tc,B}-V \geq  c' \; (1-P_B)(1+\pi^2)(1-P_B) - CB.
\end{align*}
It remains to replace $P_B$ by $P = |\alpha_*\rangle\langle \alpha_*|$. To this end, we write
\begin{align}
(1-P_B)(1+\pi^2)(1-P_B) - (1-P)(1+\pi^2)(1-P) &\notag \\
&\hspace{-160pt}= (P-P_B) +  (P - P_B)\pi^2(1-P_B) + (1-P)\pi^2(P-P_B) \label{DHS1:KTB_Lower_bound_4}.
\end{align}
From Proposition \ref{DHS1:KTV_Asymptotics_of_EV_and_EF} we know that $\Vert P_B - P\Vert_\infty \leq CB$ and $\Vert \pi^2(P_B- P)\Vert_\infty \leq CB$. Hence, the norm of the operator on the right side of \eqref{DHS1:KTB_Lower_bound_4} is bounded by a constant times $B$. This shows \eqref{DHS1:KTB_Lower_bound_9} and concludes our proof.
\end{proof}

We deduce two corollaries from \eqref{DHS1:Lower_Bound_A_3} and Lemma \ref{DHS1:KTB_Lower_bound}. The first statement is an a priori bound that we use in the proof of Theorem \ref{DHS1:Main_Result_Tc} (b).

\begin{kor}
\label{DHS1:TcB_First_Upper_Bound}
Let Assumptions \ref{DHS1:Assumption_V} and \ref{DHS1:Assumption_KTc} be true. Then, there are constants $B_0>0$ and $C>0$ such that for all $0 < B \leq B_0$ and all temperatures $T\geq \Tc(1 + CB)$, we have $\FBCS(\Gamma) - \FBCS(\Gamma_0) >0$ unless $\Gamma = \Gamma_0$.
\end{kor}

\begin{proof}
Let $D_0>0$ and assume that $T \geq \Tc (1 + D_0B)$. From \eqref{DHS1:Lower_Bound_A_3} and Lemma~\ref{DHS1:KTB_Lower_bound} we know that
\begin{align}
\FBCS(\Gamma) - \FBCS(\Gamma_0) \geq (c \, \min\{T_0, \Tc D_0B\}  - CB) \, \Vert \alpha\Vert_2^2. \label{DHS1:TcB_First_Upper_Bound_1}
\end{align}
For the choice $D_0 = \frac{2 C}{ c \Tc}$ and $B_0 = \frac{T_0}{D_0 \Tc} $ the right side of \eqref{DHS1:TcB_First_Upper_Bound_1} is strictly positive unless $\alpha = 0$. We conclude that $\Gamma_0$ is the unique minimizer of $\FBCS$, which proves the claim. 
\end{proof}

The second corollary provides a bound for the Cooper pair wave functions of low-energy BCS states in the sense of \eqref{DHS1:Second_Decomposition_Gamma_Assumption}. It is based upon \eqref{DHS1:Lower_Bound_A_3} and to state it we need to introduce the operator
\begin{align}
U  &:= \e^{-\i \frac r2 \Pi_X}. \label{DHS1:U_definition}
\end{align}
We highlight that it acts on both, the relative coordinate $r = x-y$ and the center-of-mass coordinate $X = \frac{x+y}{2}$ of a function $\alpha(x,y)$. 

\begin{kor}
\label{DHS1:cor:lowerbound}
Let Assumptions \ref{DHS1:Assumption_V} and \ref{DHS1:Assumption_KTc} be true. For any $D_0, D_1 \geq 0$, there is a constant $B_0>0$ such that if $\Gamma$ satisfies \eqref{DHS1:Second_Decomposition_Gamma_Assumption}, if $0 < B\leq B_0$, and if $T$ is such that $T - \Tc \geq -D_0B$, then $\alpha = \Gamma_{12}$ obeys
\begin{align}
&\langle \alpha, [U(1 - P)(1 + \pi_r^2)(1 - P)U^* + U^*(1 - P)(1 + \pi_r^2)(1 - P)U] \alpha \rangle \notag\\
&\hspace{200pt} + \Tr\bigl[(\alpha^* \alpha)^2\bigr] \leq C B \Vert \alpha\Vert_2^2 + D_1B^2, \label{DHS1:Lower_Bound_A_2}
\end{align}
where $P = | \alpha_* \rangle \langle \alpha_* |$ and $\pi_r = -\i \nabla_r + \frac 12\Bbold \wedge r$ both act on the relative coordinate.
\end{kor}

In the statement of the corollary and in the following, we refrain from equipping the projection $P = |\alpha_*\rangle \langle \alpha_*|$ with an index $r$ although it acts on the relative coordinate. This should not lead to confusion and keeps the formulas readable.

\begin{proof}
We recall that the operator $V$ acts by multiplication with $V(x-y)$ and that $K_T(p)$ is defined in \eqref{DHS1:KT-symbol}. Using $\alpha(x,y) = \alpha(y,x)$ we write
\begin{align}
\Tr \bigl[ (K_{T,B} - V) \alpha \alpha^*\bigr] &= \frac 12 \fint_{Q_B} \dd x \int_{\Rbb^3} \dd y \; \overline{\alpha(x,y)} \bigl[ (K_T(\pi_x) - V) + (K_T(\pi_y) - V) \bigr] \alpha(x,y). \label{DHS1:Lower_Bound_A_4}
\end{align}
We note that $\pi_x = \frac 12 \Pi_X + \tilde \pi_r = U \pi_rU^*$ and $\pi_y = \frac 12 \Pi_X - \tilde \pi_r = -U^*\pi_r U$, with $\tilde \pi_r$ and $\Pi_X$ in \eqref{DHS1:Magnetic_Momenta_COM}. Using the above identities we see that
\begin{align}
K_T(\pi_x) - V(r) &= U ( K_{T}(\pi_r) - V(r) ) U^*, \notag \\ 
K_T(\pi_y) - V(r) &= U^* ( K_{T}(\pi_r) - V(r) ) U. \label{DHS1:eq:idK_T^r} 
\end{align}
The result follows from a short computation or from Lemma~\ref{DHS1:CommutationI} below.
We combine \eqref{DHS1:Second_Decomposition_Gamma_Assumption}, \eqref{DHS1:Lower_Bound_A_3}, \eqref{DHS1:Lower_Bound_A_4} and \eqref{DHS1:eq:idK_T^r} to show the inequality
\begin{align*}
\frac 12\langle \alpha, [U (K_{T}(\pi_r) - V(r))U^* + U^* (K_{T}(\pi_r) - V(r))U]\alpha\rangle + c \Tr \bigl[ (\alpha^* \alpha)^2\bigr] \leq D_1 B^2. 
\end{align*}
Finally, we apply Lemma \ref{DHS1:KTB_Lower_bound} to the first term on the left side and obtain \eqref{DHS1:Lower_Bound_A_2}.
\end{proof}


\subsection{The first decomposition result}

The proof of Theorem~\ref{DHS1:Structure_of_almost_minimizers} is based on Corollary~\ref{DHS1:cor:lowerbound} and is given in two steps. In the first step we drop the second term on the left side of \eqref{DHS1:Lower_Bound_A_2} for a lower bound, and investigate the implications of the resulting inequality for $\alpha$. The result of the corresponding analysis is summarized in Proposition~\ref{DHS1:First_Decomposition_Result} below. The second term on the left side of \eqref{DHS1:Lower_Bound_A_2} is used later in Lemma~\ref{DHS1:Bound_on_psi}. 

\begin{prop}
\label{DHS1:First_Decomposition_Result}
Given $D_0, D_1 \geq 0 $, there is $B_0>0$ with the following properties. If, for some $0< B\leq B_0$, the wave function $\alpha\in \Lsymm$ satisfies
\begin{align}
\langle \alpha , [U^* (1-P)(1 + \pi_r^2)(1-P)U + U (1-P) (1 + \pi_r^2)(1-P)U^*] \alpha \rangle & \leq D_0 B \Vert \alpha\Vert_2^2 + D_1 B^2 , \label{DHS1:First_Decomposition_Result_Assumption}
\end{align}
then there are $\Psi\in \Hmag^1(Q_B)$ and $\xi_0\in \Hsymm$ such that
\begin{align}
\alpha(X,r) =  \alpha_*(r)\cos\bigl( \frac r2 \cdot \Pi_X\bigr) \Psi(X) + \xi_0(X,r) \label{DHS1:First_Decomposition_Result_Decomp}
\end{align}
with
\begin{align}
\langle \Psi, \Pi^2 \Psi\rangle + \Vert \xi_0\Vert_{\Hsymm}^2 \leq C\bigl( B \Vert \Psi \Vert_2^2 + D_1 B^2 \bigr).  \label{DHS1:First_Decomposition_Result_Estimate}
\end{align}
\end{prop}
Before we give the proof of the a priori estimates in Proposition \ref{DHS1:First_Decomposition_Result}, we define the decomposition of $\alpha$, explain the idea behind it, and discuss relations to the existing literature. For this purpose, let the operator $A \colon \Lsymm \ra \Lmag^2(Q_B)$ be given by
\begin{align}
(A\alpha)(X) := \int_{\Rbb^3} \dd r\; \alpha_*(r) \; \cos\bigl( \frac r2\cdot \Pi_X\bigr) \alpha(X,r). \label{DHS1:Def_A}
\end{align}
A short computation shows that its adjoint $A^*\colon \Lmag^2(Q_B) \ra \Lsymm$ is given by
\begin{align}
(A^*\Psi) (X,r) = \alpha_*(r) \cos\bigl( \frac r2\cdot \Pi_X\bigr) \Psi(X). \label{DHS1:Def_Astar}
\end{align}
We highlight that this is the form of the first term in \eqref{DHS1:First_Decomposition_Result_Decomp}. For a given Cooper pair wave function $\alpha$, we use these operators to define the two functions $\Psi$ and $\xi_0$ by
\begin{align}
	\Psi &:= (AA^*)^{-1} A\alpha, & \xi_0 &:= \alpha - A^*\Psi. \label{DHS1:Def_Psixi}
\end{align}
Lemma~\ref{DHS1:AAstar_Positive} below guarantees that $AA^*$ is invertible, and we readily check that  \eqref{DHS1:First_Decomposition_Result_Decomp} holds with these definitions. Moreover, this decomposition of $\alpha$ is orthogonal in the sense that $\langle A^* \Psi, \xi_0 \rangle = 0$ holds. The claimed orthogonality follows from
\begin{align}
	A\xi_0 =0, \label{DHS1:fundamental_property}
\end{align}
which is a direct consequence of \eqref{DHS1:Def_Psixi}. In the following we motivate our choice for $\Psi$ and $\xi_0$ and comment on its appearance in the literature.

The decomposition of $\alpha$ is motivated by the minimization problem for the low-energy operator $2 -UPU^* - U^*PU$, that is, the operator in \eqref{DHS1:First_Decomposition_Result_Assumption} with $\pi_r^2$ replaced by zero. The operators $UPU^*$ and $U^*PU$ act as $A^*A$ on the space $\Lsymm$ of reflection symmetric functions in the relative coordinate. If $\Pi$ is replaced by $P_X$ in the definition of $U$ then $A^*A$ can be written as 
\begin{equation}
	A^* A \cong \int^{\oplus}_{[0,\sqrt{ 2 \pi B}]^3} \mathrm{d}P_X \; | a_{P_X} \rangle \langle a_{P_X} |, \label{DHS1:eq:directintegralA}
\end{equation}
with $| a_{P_X} \rangle \langle a_{P_X} |$ the orthogonal projection onto the function $a_{P_X}(r) = \cos(r/2 \cdot P_X) \alpha_*(r)$. Here the variable $P_X$ is the dual of the center-of-mass coordinate $X$ in the sense of Fourier transformation and $r$ denotes the relative coordinate. That is, the function $a_{P_X}(r)$ minimizes $1-A^* A$ in each fiber, whence it is the eigenfunction with respect to the lowest eigenvalue of $1 - A^* A = 1 - (UPU^* + U^*PU)/2$. This discussion should be compared to \cite[Eq. (5.47)]{Hainzl2012} and the discussion before Lemma~20 in \cite{ProceedingsSpohn}.

If we replace $P_X$ by the magnetic momentum operator $\Pi$ again the above picture changes because the components of $\Pi$ cannot be diagonalized simultaneously (they do not commute), and hence \eqref{DHS1:eq:directintegralA} has no obvious equivalent in this case. The decomposition of $\alpha$ in terms of the operators $A$ and $A^*$ above has been introduced in \cite{Hainzl2017} in order to study the operator $1 - V^{\nicefrac 12} L_{T,B} V^{\nicefrac 12}$ with $L_{T,B}$ in \eqref{DHS1:LTB_definition}, see also the discussion below Theorem~\ref{DHS1:Calculation_of_the_GL-energy}. The situation in this work is comparable to our case with $\pi_r^2$ replaced by zero in \eqref{DHS1:First_Decomposition_Result_Assumption}. Our analysis below shows that the ansatz \eqref{DHS1:Def_Psixi} is useful even if the full range of energies is considered, that is, if $\pi_r^2$ is present in \eqref{DHS1:First_Decomposition_Result_Assumption}. 

In the following lemma we collect useful properties of the operator $A A^*$. It should be compared to \cite[Lemma~27]{Hainzl2017}.
%
%
%
%
%
%

\begin{lem}
\label{DHS1:AAstar_Positive}
The operators
\begin{align*}
AA^* &= \int_{\Rbb^3} \dd r\; \alpha_*(r)^2 \cos^2\bigl( \frac{r}{2} \cdot \Pi\bigr), & 1-AA^* &= \int_{\Rbb^3} \dd r \;\alpha_*(r)^2 \sin^2\bigl( \frac r2 \cdot \Pi\bigr) 
\end{align*}
on $\Lmag^2(Q_B)$ are both bounded nonnegative functions of $\Pi^2$ and satisfy the following properties:
\begin{enumerate}[(a)]
\item $0\leq AA^*\leq 1$ and $0 \leq 1 - AA^*\leq 1$.
\item There is a constant $c>0$ such that $AA^* \geq c$ and $1 - AA^* \geq c \; \Pi^2\; (1 + \Pi^2)^{-1}$.
\end{enumerate}
In particular, $AA^*$ and $1 - AA^*$ are boundedly invertible on $\Lmag^2(Q_B)$.
\end{lem}


\begin{proof}
Part (a) is a direct consequence of the fact that $\Vert \alpha_* \Vert_2 = 1$. In the following we reduce the proof of part (b) to known results in \cite{Hainzl2017}. To this end, we introduce the operator
\begin{align}
	R := \int_{\Rbb^3} \dd r \; \alpha_*(r)^2 \; \cos( r \cdot \Pi)
	\label{DHS1:eq:R}
\end{align}
and note that
\begin{align*}
	AA^* &= \frac 12 (1 + R), & 1 - AA^* &= \frac 12 (1 - R).
\end{align*}
Let us also denote by $\widetilde{R}$ the operator in \eqref{DHS1:eq:R} but with $\Pi$ replaced by $\tilde \Pi$, which is the same operator but understood to act on $L^2(\mathbb{R}^3)$ instead of $\Lmag^2(Q_B)$. In Lemma~28 in \cite{Hainzl2017} it has been shown that $\widetilde{R}$ is a function of $\tilde \Pi^2$. Moreover, the statement of Lemma~27 in \cite{Hainzl2017} is equivalent to
\begin{align}
	1 - \widetilde{R}^2 \geq c\,  \frac{\tilde \Pi^2}{1 + \tilde \Pi^2}
	\label{DHS1:eq:R1}
\end{align}
for some $0<c<1$, and Eq.~(55) in the same reference implies
\begin{align}
	| \widetilde{R} | \leq 1 - c.
	\label{DHS1:eq:R2}
\end{align}
In combination, \eqref{DHS1:eq:R1}, \eqref{DHS1:eq:R2}, and $ \widetilde{R} \leq 1$ show
\begin{align}
	 1+\widetilde{R} &\geq c, & 1 - \widetilde{R} \geq \frac{1 - \widetilde{R}^2}{2} &\geq \frac{c}{2} \frac{\tilde \Pi^2}{1 + \tilde \Pi^2} .
	 \label{DHS1:eq:R3} 
\end{align}
It remains to argue that $R$ is a function of $\Pi^2$ and that a version of \eqref{DHS1:eq:R3} with $\tilde R$ and $\tilde \Pi$ replaced by $R$ and $\Pi$ holds. 

The fact that $R$ is a function of $\Pi^2$ follows from the argument that we used to show that the same statement is true for the operator in \eqref{DHS1:eq:A29}. To show that \eqref{DHS1:eq:R3} with $\tilde R$ and $\tilde \Pi$ replaced by $R$ and $\Pi$ holds, we conjugate both sides of the inequalities with the Bloch-Floquet transformation in \eqref{DHS1:eq:ABF1} and \eqref{DHS1:eq:ABF2}, and use \eqref{DHS1:eq:ABF3}. The inequalities in \eqref{DHS1:eq:R3} therefore hold equally in any fiber, that is, with $\tilde \Pi$ on the left and on the right sides replaced by $\Pi(k) = \Pi + k$ acting on $\Lmag^2(Q_B)$. Since $\tilde \Pi(0) = \Pi$, this proves the claim. 
\end{proof}

The remainder of this subsection is devoted to the proof of Proposition~\ref{DHS1:First_Decomposition_Result}. We start with a lower bound on the operator in \eqref{DHS1:First_Decomposition_Result_Assumption} when it acts on wave functions of the form $A^*\Psi$, see Lemma~\ref{DHS1:MainTerm} below. 

\subsubsection{Step one -- lower bound on the range of \texorpdfstring{$A^*$}{A*}}

The main result of this subsection is the following lemma.

\begin{lem}
\label{DHS1:MainTerm}
For any $\Psi\in \Lmag^2(Q_B)$, with $A$ and $A^*$ given by \eqref{DHS1:Def_A} and \eqref{DHS1:Def_Astar}, with $U$ given by \eqref{DHS1:U_definition}, and $P = |\alpha_*\rangle \langle \alpha_*|$ with $\alpha_*$ from \eqref{DHS1:alpha_star_ev-equation} acting on the relative coordinate, we have
\begin{align}
\frac 12 \langle A^* \Psi, [ U^* (1 - P) (1 + \pi_r^2) (1- P) U + U(1 - P) (1 + \pi_r^2)(1-P) U^* ] A^*\Psi\rangle \hspace{-340pt}& \notag\\
&=  \langle \Psi, AA^* (1 - AA^*)(1 + \Pi^2) \Psi\rangle \notag\\
&\hspace{10pt} +  \fint_{Q_B} \dd X \int_{\Rbb^3} \dd r \; \ov{(1 - AA^*)\Psi(X)} \; |\nabla \alpha_*(r)|^2 \; \cos^2 \bigl(\frac r2 \Pi_X\bigr) \; (1 - AA^*)\Psi(X) \notag \\
&\hspace{10pt}+ \fint_{Q_B} \dd X \int_{\Rbb^3} \dr \; \ov{AA^*\Psi(X)} \; |\nabla \alpha_*(r)|^2\; \sin^2 \bigl(\frac r2  \Pi_X\bigr) \; AA^*\Psi(X)\notag \\
&\hspace{10pt} + \frac 14 \fint_{Q_B}\dd X \int_{\Rbb^3} \dd r \; \ov{(1 - AA^*)\Psi(X)} \; |\Bbold\wedge r|^2 \alpha_*(r)^2 \; \sin^2 \bigl(\frac r2 \Pi_X\bigr) \; (1 - AA^*)\Psi(X) \notag\\
&\hspace{10pt} + \frac 14\fint_{Q_B}\dd X\int_{\Rbb^3} \dr \; \ov{AA^*\Psi(X)} \; |\Bbold\wedge r|^2 \alpha_*(r)^2 \;  \cos^2 \bigl(\frac r2 \Pi_X\bigr) \; AA^*\Psi(X). \label{DHS1:MainTerm_5}
\end{align}
In particular, we have the lower bound
\begin{align}
\frac 12 \langle A^* \Psi, [ U^* (1 - P) (1 + \pi_r^2) (1- P) U + U(1 - P) (1 + \pi_r^2)(1-P) U^* ] A^*\Psi\rangle 
%
\geq c\, \langle \Psi, \Pi^2\Psi\rangle . \label{DHS1:MainTerm_LowerBound}
\end{align}
\end{lem}


\begin{bem}
Let us replace $\pi_r^2$ on the left side of \eqref{DHS1:MainTerm_5} by zero for the moment. In this case, the substitute of \eqref{DHS1:MainTerm_5} reads
\begin{align}
\frac 12 \langle A^*\Psi, [U^*(1 - P)U + U(1 - P)U^* ] A^*\Psi\rangle = \langle \Psi , AA^*(1 - AA^*) \Psi\rangle. \label{DHS1:Low_Energy_Operator}
\end{align}
It follows from Lemma~\ref{DHS1:AAstar_Positive} that the operator $AA^*(1 - AA^*)$ is bounded from below by $\Pi^2$ only for small values of $\Pi^2$, which is not enough for the proof of Proposition \ref{DHS1:First_Decomposition_Result}. This justifies the term ``low-energy operator'' for $1 - A^*A$, which we used earlier in the discussion below \eqref{DHS1:Def_Astar}. The additional factor $1 + \Pi^2$ in the first term on the right side of \eqref{DHS1:MainTerm_5} compensates for the problematic behavior of \eqref{DHS1:Low_Energy_Operator} for high energies. The expression on the right side of \eqref{DHS1:Low_Energy_Operator} also appears in \cite{Hainzl2017}. 
\end{bem}

Before we give the proof of Lemma~\ref{DHS1:MainTerm}, we prove two technical lemmas, which provide intertwining relations for various magnetic momentum operators with $U$ and linear combinations of $U$. A part of the relations in the first lemma can be found in \cite[Lemma~24]{Hainzl2017}.

\begin{lem}
\label{DHS1:CommutationI}
Let $p_r := -\i \nabla_r$, $\pi_r = p_r + \frac 12\Bbold \wedge r$ and $\tilde \pi_r$ and $\Pi_X$ be given by \eqref{DHS1:Magnetic_Momenta_COM}. With $U$ in \eqref{DHS1:U_definition}, we have the following intertwining relations:
\begin{align*}
\begin{split}
U \Pi_X U^* &= \Pi_X - \Bbold\wedge r,\phantom{\frac 12}  \\  U^* \Pi_X U &= \Pi_X + \Bbold\wedge r, \phantom{\frac 12}
\end{split} &
\begin{split}
U \pi_r U^* &= \tilde \pi_r + \frac{1}{2}\Pi_X, \\ U^* \pi_r U &= \tilde \pi_r - \frac{1}{2}\Pi_X, \phantom{\frac 12}
\end{split} &
\begin{split}
U \tilde \pi_r U^* &= p_r + \frac{1}{2}\Pi_X, \\ U^* \tilde \pi_r U &= p_r - \frac{1}{2}\Pi_X.
\end{split}
\end{align*}
\end{lem}

\begin{proof}
Let us denote $P_X := -\i \nabla_X$. We use the fact that $r\cdot P_X$ commutes with $r\cdot (\Bbold \wedge X)$ to see that
\begin{align}
U^* = \e^{\i \frac \Bbold 2 \cdot (X\wedge r)} \; \e^{\i \frac r2P_X} \label{DHS1:representation_Ustar}
\end{align}
holds. To prove the first intertwining relation with $\Pi_X$, we compute
\begin{align*}
\Pi_XU^* &= (P_X + \Bbold\wedge X)U^* =  \e^{\i \frac \Bbold 2 \cdot (X\wedge r)} \Bigl[ P_X - \frac 12 \Bbold \wedge r + \Bbold\wedge X\Bigr] \e^{\i \frac r2P_X} \\
&= U^* \Bigl[ P_X - \frac 12 \Bbold \wedge r + \Bbold \wedge \bigl(X -\frac r2\bigr)\Bigr] = U^* [ \Pi_X - \Bbold \wedge r].
\end{align*}
Here we used that $f(X)\, \e^{\i \frac r2P_X} = \e^{\i \frac r2 P_X}\, f(X-\frac r2)$. The second intertwining relation with $\Pi_X$ is obtained by replacing $r$ by $-r$. 

Next we consider the first intertwining relation with $\pi_r$ and compute
\begin{align*}
\pi_r U^* &= \bigl( p_r + \frac 12\Bbold \wedge r\bigr) U^* = \e^{\i \frac \Bbold 2 \cdot (X\wedge r)}\Bigl[ p_r + (-\i)\i \frac 12 \Bbold \wedge X + \frac 12 \Bbold \wedge r\Bigr] \e^{\i \frac r2P_X} \\
&= U^* \Bigl[ p_r + \frac{P_X}{2} + \frac 12 \Bbold \wedge \bigl( X - \frac r2\bigr) + \frac 12 \Bbold \wedge r\Bigr] = U^* \Bigl[ \tilde \pi_r + \frac{\Pi_X}{2}\Bigr].
\end{align*}
The remaining relations can be proved similarly and we skip the details.
\end{proof}

\begin{lem}
\label{DHS1:CommutationII}
\begin{enumerate}[(a)]
\item We have the following intertwining relations for $\Pi_X$:
\begin{align}
\Pi_X \cos\bigl( \frac r2 \Pi_X\bigr) &=  \cos\bigl( \frac r2 \Pi_X\bigr)\Pi_X - \i \sin\bigl( \frac r2 \Pi_X\bigr) \Bbold\wedge r, \label{DHS1:PiXcos} \\
\Pi_X \sin\bigl( \frac r2 \Pi_X\bigr) &= \sin\bigl( \frac r2 \Pi_X\bigr) \Pi_X + \i \cos\bigl( \frac r2 \Pi_X\bigr) \Bbold\wedge r. \label{DHS1:PiXsin}
\end{align}

\item The operators $p_r$, $\tilde \pi_r$ and $\pi_r$ intertwine according to
\begin{align}
\tilde \pi_r \cos\bigl( \frac r2 \Pi_X\bigr) &= \cos\bigl( \frac r2 \Pi_X\bigr) p_r + \i \sin\bigl( \frac r2 \Pi_X\bigr) \frac{\Pi_X}{2}, \label{DHS1:pirtilde_cos_pr}\\
\tilde \pi_r \cos\bigl( \frac r2 \Pi_X\bigr) &= \cos\bigl( \frac r2 \Pi_X\bigr)\pi_r + \i \frac{\Pi_X}{2} \sin\bigl( \frac r2 \Pi_X\bigr), \label{DHS1:pirtilde_cos_pir}
\end{align}
and
\begin{align}
\tilde \pi_r \sin\bigl( \frac r2 \Pi_X\bigr)  &= \sin\bigl( \frac r2 \Pi_X\bigr) p_r - \i \cos\bigl( \frac r2 \Pi_X\bigr) \frac{\Pi_X}{2}, \label{DHS1:pirtilde_sin_pr}\\
\tilde \pi_r \sin\bigl( \frac r2 \Pi_X\bigr) &= \sin\bigl( \frac r2 \Pi_X\bigr) \pi_r - \i \frac{\Pi_X}{2} \cos\bigl( \frac r2 \Pi_X\bigr) \label{DHS1:pirtilde_sin_pir}
\end{align}
\end{enumerate}
\end{lem}

It will be useful in the proof of Lemma \ref{DHS1:MainTerm} to have displayed both, \eqref{DHS1:pirtilde_cos_pr} and \eqref{DHS1:pirtilde_cos_pir} as well as \eqref{DHS1:pirtilde_sin_pr} and \eqref{DHS1:pirtilde_sin_pir}, even though they follow trivially from each other and \eqref{DHS1:PiXcos} or \eqref{DHS1:PiXsin}.

\begin{proof}
The proof is a direct consequence of the representations
\begin{align}
	\cos\bigl( \frac r2 \Pi_X\bigr) &= \frac 12 (U^* + U), & \sin\bigl( \frac r2 \Pi_X\bigr) &= \frac 1{2\i} (U^*-U),
\end{align}
and the intertwining relations in Lemma~\ref{DHS1:CommutationII}. We omit the details.
\end{proof}

\begin{proof}[Proof of Lemma \ref{DHS1:MainTerm}]
The proof is a tedious computation that is based on the intertwining relations in Lemma \ref{DHS1:CommutationII}. 
We start by defining
\begin{align}
\Tcal_1 &:= U^* \pi_r^2 U + U\pi_r^2 U^* = 2\, \tilde \pi_r^2 + \frac 12 \, \Pi_X^2, & \Tcal_2 &:= U^*P\pi_r^2PU + UP\pi_r^2PU^*, \notag \\
\Tcal_3 &:= U^* P\pi_r^2U + UP\pi_r^2U^*, & \Tcal_4 &:= U^* \pi_r^2 PU + U\pi_r^2PU^*. \label{DHS1:MainTerm_6}
\end{align}
Then, \eqref{DHS1:MainTerm_5} can be written as
\begin{align}
\langle A^* \Psi, [ U^* (1 - P) (1 + \pi_r^2) (1- P) U + U(1 - P) (1 + \pi_r^2)(1-P) U^* ] A^*\Psi\rangle  & = \notag\\
&\hspace{-330pt}= 2\langle A^*\Psi, (1 - A^*A)A^*\Psi\rangle \notag \\
&\hspace{-300pt} + \langle A^*\Psi, \Tcal_1 A^*\Psi\rangle + \langle A^*\Psi, \Tcal_2 A^*\Psi\rangle - \langle A^*\Psi, \Tcal_3 A^*\Psi\rangle - \langle A^*\Psi, \Tcal_4 A^*\Psi\rangle. \label{DHS1:Tcal_first_line}
\end{align}
The first term on the right side equals twice the term in \eqref{DHS1:Low_Energy_Operator}, which is in its final form. 

We start by computing the $\Pi_X^2$ term in $\Tcal_1$, which reads
\begin{align*}
\langle A^* \Psi , \Pi_X^2 A^* \Psi \rangle = \fint_{Q_B} \dd X \int_{\Rbb^3} \dd r \; \ov{\Psi(X)} \alpha^*(r) \; \cos\bigl( \frac r2 \Pi_X\bigr) \Pi_X^2 \cos\bigl( \frac r2 \Pi_X\bigr) \; \alpha_*(r) \Psi(X).
\end{align*}
Our goal is to move $\Pi_X^2$ to the right. To that end, we apply \eqref{DHS1:PiXcos} twice and obtain
\begin{align*}
\Pi_X^2 \cos\bigl( \frac r2 \Pi_X\bigr) = \cos\bigl( \frac r2 \Pi_X\bigr)\Pi_X^2 - \i \sin\bigl( \frac r2 \Pi_X\bigr) \Pi_X\Bbold \wedge r - \i \Pi_X \sin\bigl( \frac r2 \Pi_X\bigr)\Bbold \wedge r.
\end{align*}
We multiply this from the left with $\cos( \frac r2 \Pi_X)$, use \eqref{DHS1:PiXcos} to commute $\Pi_X$ to the left in the last term, and find
\begin{align}
\cos\bigl( \frac r2 \Pi_X\bigr) \Pi_X^2\cos\bigl( \frac r2 \Pi_X\bigr) &= \cos^2\bigl( \frac r2 \Pi_X\bigr) \Pi_X^2 + \sin^2\bigl( \frac r2 \Pi_X\bigr) |\Bbold\wedge r|^2 \notag\\
&\hspace{-60pt}- \i \bigl[ \Pi_X\cos\bigl( \frac r2 \Pi_X\bigr)\sin\bigl( \frac r2 \Pi_X\bigr) + \cos\bigl( \frac r2 \Pi_X\bigr)\sin\bigl( \frac r2 \Pi_X\bigr) \Pi_X\bigr]\Bbold\wedge r. \label{DHS1:PiX-term_1}
\end{align}
The operator $| \Bbold \wedge r |^2$ in the second term on the right side commutes with $\sin^2( \frac r2 \Pi_X)$. The operator in square brackets is self-adjoint and commutes with $\Bbold \wedge r$. When we add \eqref{DHS1:PiX-term_1} and its own adjoint, we obtain
\begin{align}
\cos\bigl( \frac r2 \Pi_X\bigr) \Pi_X^2\cos\bigl( \frac r2 \Pi_X\bigr) & \notag\\
&\hspace{-50pt}= \frac 12 \cos^2\bigl( \frac r2 \Pi_X\bigr) \Pi_X^2 + \frac 12 \, \Pi_X^2 \cos^2\bigl( \frac r2 \Pi_X\bigr) + \sin^2\bigl( \frac r2 \Pi_X\bigr) |\Bbold\wedge r|^2. \label{DHS1:PiX-term_2}
\end{align}
We evaluate \eqref{DHS1:PiX-term_2} in the inner product with $\alpha_*\Phi$ and $\alpha_*\Psi$ on the left and right side, respectively, use the fact that $AA^*$ commutes with $\Pi^2$, see Lemma \ref{DHS1:AAstar_Positive}, and obtain
\begin{align}
\langle A^*\Phi, \Pi_X^2A^*\Psi\rangle &= \langle \Phi, AA^* \Pi^2\Psi\rangle \notag\\
&\hspace{10pt} + \fint_{Q_B}\dd X\int_{\Rbb^3} \dd r \; \ov{\Phi(X)} \; |\Bbold\wedge r|^2  \alpha_*(r)^2 \; \sin^2\bigl( \frac r2 \Pi_X\bigr)  \; \Psi(X). \label{DHS1:PiX-term}
\end{align}
When we choose $\Phi = \Psi$ we obtain the result for the term proportional to $\Pi_X^2$ in $\Tcal_1$.

Next, we investigate the term proportional to $\tilde\pi_r^2$ in $\Tcal_1$. We use \eqref{DHS1:pirtilde_cos_pir} to move the operators $\tilde \pi_r$ from the middle to the outer positions and find
\begin{align}
\langle A^*\Psi, \tilde \pi_r^2A^*\Psi\rangle &= \fint_{Q_B} \dd X\int_{\Rbb^3} \dd r\; \ov{\Psi(X)} \alpha_*(r)\left[ \pi_r\cos\bigl( \frac r2 \Pi_X\bigr) -\i \sin\bigl( \frac r2 \Pi_X\bigr) \frac{\Pi_X}{2}\right] \notag\\
&\hspace{80pt} \times \left[ \cos\bigl( \frac r2 \Pi_X\bigr)\pi_r + \i \frac{\Pi_X}{2}\sin\bigl( \frac r2 \Pi_X\bigr)\right]\alpha_*(r) \Psi(X). \label{DHS1:tildepir_eq1}
\end{align}
We multiply out the brackets and obtain four terms. The terms proportional to $\cos^2$ and $\sin^2$ read
\begin{align}
&\fint_{Q_B} \dd X \int_{\Rbb^3}\dd r\; \ov{\Psi(X)}\;  |\pi_r\alpha_*(r)|^2\;  \cos^2\bigl( \frac r2 \Pi_X\bigr) \; \Psi(X) \notag\\
&\hspace{20pt} + \fint_{Q_B}\dd X\int_{\Rbb^3} \dd r \; \ov{\Psi(X)} \; \alpha_*(r)\;  \sin\bigl( \frac r2 \Pi_X\bigr) \frac{\Pi_X^2}{4} \sin\bigl( \frac r2 \Pi_X\bigr) \; \alpha_*(r) \; \Psi(X).\label{DHS1:tildepir_eq4}
\end{align}
For the moment the second line remains untouched. It is going to be canceled by a term in \eqref{DHS1:tildepir_eq5} below. The term in the first line equals
\begin{align}
%
&\fint_{Q_B} \dd X \int_{\Rbb^3}\dd r\; \ov{\Psi(X)} \; |\nabla\alpha_*(r)|^2 \; \cos^2\bigl( \frac r2 \Pi_X\bigr) \; \Psi(X) \notag \\ 
&\hspace{50pt} + \frac 14\fint_{Q_B} \dd X \int_{\Rbb^3}\dd r\; \ov{\Psi(X)} \; |\Bbold \wedge r|^2 \alpha_*(r)^2 \; \cos^2\bigl( \frac r2 \Pi_X\bigr) \; \Psi(X). \label{DHS1:tildepir_eq2}
\end{align}
To obtain this result, we used $(\nabla \alpha_*)(r) \cdot \Bbold \wedge r =0$, which holds because $\alpha_*$ is radial. This term is in its final form.

Now we have a closer look at the terms proportional to $\sin $ times $\cos$ in \eqref{DHS1:tildepir_eq1}. The operator inside the relevant quadratic form is given by
\begin{align}
\i \pi_r\cos\bigl( \frac r2 \Pi_X\bigr) \frac{\Pi_X}{2} \sin\bigl( \frac r2 \Pi_X\bigr) - \i \sin\bigl( \frac r2 \Pi_X\bigr)\frac{\Pi_X}{2} \cos\bigl( \frac r2 \Pi_X\bigr)\pi_r. \label{DHS1:MainTerm_1}
\end{align}
We intend to interchange $\sin( \frac r2 \Pi_X)$ and $\cos( \frac r2 \Pi_X)$ in the first term. To do this, we use \eqref{DHS1:PiXsin} to move $\Pi_X$ out of the center so that the first term equals
\begin{align*}
\i \pi_r\cos\bigl( \frac r2 \Pi_X\bigr) \sin\bigl( \frac r2 \Pi_X\bigr)\frac{\Pi_X}{2} -\frac 12\pi_r\cos^2\bigl( \frac r2 \Pi_X\bigr) \Bbold \wedge r.
\end{align*}
%
In the first term we may now commute the sine and the cosine and use \eqref{DHS1:PiXcos} and \eqref{DHS1:pirtilde_sin_pir} to bring $\pi_r$ and $\Pi_X$ in the center again. We also move $\pi_r$ into the center in the second term in \eqref{DHS1:MainTerm_1}. As a result, \eqref{DHS1:MainTerm_1} equals
\begin{align}
%
&\cos\bigl( \frac r2 \Pi_X\bigr) \frac{\Pi_X^2}{4}\cos\bigl( \frac r2 \Pi_X\bigr) - \sin\bigl( \frac r2 \Pi_X\bigr) \frac{\Pi_X^2}{4}\sin\bigl( \frac r2 \Pi_X\bigr) \notag \\
&\hspace{40pt}+ \i \cos\bigl( \frac r2 \Pi_X\bigr) \frac{\Pi_X}{4} \sin\bigl( \frac r2 \Pi_X\bigr) \Bbold \wedge r \notag \\ 
&\hspace{80pt}- \frac 12 \pi_r \cos^2\bigl( \frac r2 \Pi_X\bigr)\Bbold \wedge r - \frac 12 \sin\bigl( \frac r2 \Pi_X\bigr)\tilde \pi_r \sin\bigl( \frac r2 \Pi_X\bigr)\Bbold \wedge r. \label{DHS1:MainTerm_3}
\end{align}
We use \eqref{DHS1:pirtilde_sin_pir} to move $\tilde \pi_r$ to the left in the last term in \eqref{DHS1:MainTerm_3}. One of the terms we obtain in this way cancels the third term in \eqref{DHS1:MainTerm_3}. We also use $\cos( \frac r2 \Pi_X)^2 + \sin( \frac r2 \Pi_X)^2=1$ to rewrite the fourth term in \eqref{DHS1:MainTerm_3}. In combination, these considerations imply that the terms in \eqref{DHS1:MainTerm_3} equal
\begin{align}
\begin{split}
%
\cos\bigl( \frac r2 \Pi_X\bigr) \frac{\Pi_X^2}{4}\cos\bigl( \frac r2 \Pi_X\bigr) - \sin\bigl( \frac r2 \Pi_X\bigr)\frac{\Pi_X^2}{4}\sin\bigl( \frac r2 \Pi_X\bigr) - \frac 12 \pi_r\Bbold \wedge r. \end{split} \label{DHS1:tildepir_eq5}
\end{align}
The expectation of the second term with respect to $\alpha_*(r) \Psi(X)$ cancels the second term in \eqref{DHS1:tildepir_eq4}. We multiply the last term from the left and from the right with $\alpha_*(r)$, integrate over $r$ and find 
\begin{align}
\frac 12\int_{\Rbb^3} \dd r\; \alpha_*(r) \pi_r \Bbold \wedge r \alpha_*(r) = \frac 12 \int_{\Rbb^3} \ov{p_r\alpha_*(r)} \Bbold \wedge r\alpha_*(r) + \frac 14\int_{\Rbb^3} \dd r\; |\Bbold \wedge r|^2 \alpha_*(r)^2. \label{DHS1:Tcal3_eq2}
\end{align}
The first term on the right side vanishes because $\alpha_*$ is radial, see the remark below \eqref{DHS1:tildepir_eq4}. Let us summarize where we are. We combine \eqref{DHS1:tildepir_eq1}-\eqref{DHS1:Tcal3_eq2} to see that
\begin{align}
\langle A^*\Psi, \tilde \pi_r^2A^*\Psi\rangle &= \fint_{Q_B} \dd X \int_{\Rbb^3}\dd r\; \ov{\Psi(X)} \; |\nabla\alpha_*(r)|^2 \; \cos^2\bigl( \frac r2 \Pi_X\bigr) \; \Psi(X) \notag \\
&\hspace{-30pt} + \frac 14 \fint_{Q_B} \dd X \int_{\Rbb^3}\dd r\; \ov{\Psi(X)} \; |\Bbold \wedge r|^2 \alpha_*(r)^2 \; \bigl( \cos^2 \bigl( \frac r2 \Pi_X\bigr) -1 \bigr) \; \Psi(X) \notag \\
&\hspace{-30pt} + \frac 14 \fint_{Q_B} \dd X \int_{\Rbb^3}\dd r \; \ov{\alpha_*(r) \Psi(X)} \; \cos\bigl( \frac r2 \Pi_X\bigr) \Pi_X^2 \cos\bigl( \frac r2 \Pi_X\bigr) \; \alpha_*(r) \Psi(X). \label{DHS1:eq:1}
\end{align}
The term in the last line equals $\langle A^*\Psi, \Pi_X^2A^*\Psi\rangle$ and we use \eqref{DHS1:PiX-term} to rewrite it. This yields
\begin{align*}
\langle A^*\Psi, \tilde \pi_r^2A^*\Psi\rangle &= \frac 14 \langle \Psi, AA^* \Pi^2\Psi\rangle + \fint_{Q_B} \dd X \int_{\Rbb^3} \dd r\;  \ov{\Psi(X)}\;  |\nabla \alpha_*(r)|^2\;  \cos^2\bigl( \frac r2 \Pi_X\bigr)\; \Psi(X).
\end{align*}
In combination with \eqref{DHS1:MainTerm_6} and \eqref{DHS1:PiX-term}, this yields
\begin{align}
\langle A^* \Psi, \Tcal_1 A^*\Psi\rangle &= \langle \Psi , AA^*\Pi^2\Psi\rangle \notag \\
&\hspace{10pt}+ 2\fint_{Q_B} \dd X\int_{\Rbb^3} \dd r\; \ov{\Psi(X)}\;  |\nabla \alpha_*(r)|^2 \; \cos^2\bigl( \frac r2 \Pi_X\bigr)\;  \Psi(X) \notag \\
&\hspace{10pt} + \frac 12 \fint_{Q_B}\dd X\int_{\Rbb^3} \dd r \; \ov{\Psi(X)} \; |\Bbold\wedge r|^2 \alpha_*(r)^2\; \sin^2\bigl( \frac r2 \Pi_X\bigr)  \; \Psi(X)  \label{DHS1:Tcal1-Result}
\end{align}
and completes our computation of the term involving $\Tcal_1$.

A short computation shows that
\begin{equation}
\langle A^*\Psi , \Tcal_2A^*\Psi\rangle = 2\, \langle AA^*\Psi , AA^*\Psi\rangle \Bigl[ \Vert \nabla \alpha_*\Vert_2^2 + \frac 14 \int_{\Rbb^3} \dd r \; |\Bbold \wedge r|^2 \alpha_*(r)^2\Bigr]. \label{DHS1:Tcal2_result}
\end{equation}
It remains to compute the terms in \eqref{DHS1:Tcal_first_line} involving the operators $\Tcal_3$ and $\Tcal_4$, where $\Tcal_4^* = \Tcal_3$.

In the following we compute the term with $\Tcal_3$. A short computation, which uses the fact that $\alpha_*$ is radial, shows
\begin{align}
\langle A^* \Psi, \Tcal_3A^*\Psi\rangle &= \langle \alpha_* A A^*\Psi, \pi_r^2 (U^* + U) A^*\Psi \rangle \notag\\
&\hspace{-30pt}= 2\fint_{Q_B}\dd X\int_{\Rbb^3} \dd r \; \ov{AA^*\Psi(X)} \; \ov{p_r\alpha_*(r)} \; p_r\cos^2\bigl( \frac r2 \Pi_X\bigr) \; \alpha_*(r) \; \Psi(X) \notag\\
&\hspace{-10pt} + \frac 12 \fint_{Q_B} \dd X \int_{\Rbb^3} \dd r\; \ov{AA^*\Psi(X)} \; |\Bbold \wedge r|^2 \alpha_*(r)^2\;  \cos^2\bigl( \frac r2 \Pi_X\bigr) \; \Psi(X). \label{DHS1:Tcal3_eq3}
\end{align}
%
The second term on the right side is in its final form and will be canceled by a term below. We continue with the first term, use \eqref{DHS1:pirtilde_cos_pr} twice to commute $p_r$ with the squared cosine, as well as \eqref{DHS1:PiXcos} and \eqref{DHS1:PiXsin} to commute $\Pi_X$ to the center in the emerging terms, and find
\begin{align}
p_r\cos^2\bigl( \frac r2 \Pi_X\bigr) 
&= \cos^2\bigl( \frac r2 \Pi_X\bigr)p_r + \i \sin\bigl( \frac r2 \Pi_X\bigr) \Pi_X \cos\bigl( \frac r2 \Pi_X\bigr) - \frac 12 \Bbold \wedge r. \label{DHS1:Tcal3_eq1}
\end{align}
We note that the last term, when inserted back into \eqref{DHS1:Tcal3_eq3}, vanishes because $\alpha_*$ is radial. The first term is final and its quadratic form with $p_r \alpha_* AA^*\Psi$ and $\alpha_*\Psi$ reads
\begin{align}
2 \fint_{Q_B} \dd X\int_{\Rbb^3} \dd r \; \ov{AA^*\Psi(X)} \; |\nabla \alpha_*(r)|^2 \; \cos^2\bigl( \frac r2\Pi_X\bigr) \; \Psi(X). \label{DHS1:MainTerm_4}
\end{align}
Let us continue with the second term on the right side of \eqref{DHS1:Tcal3_eq1}. We multiply it with $p_r$ from the left and use \eqref{DHS1:pirtilde_cos_pr} and \eqref{DHS1:pirtilde_sin_pr} to commute $p_r$ to the right. In the two emerging terms we bring $\Pi_X$ to the center and obtain
\begin{align*}
p_r \sin\bigl( \frac r2 \Pi_X\bigr) \Pi_X\cos\bigl( \frac r2 \Pi_X\bigr) &= \sin\bigl( \frac r2 \Pi_X\bigr)\Pi_X\cos\bigl( \frac r2 \Pi_X\bigr) p_r + \i \sin\bigl( \frac r2 \Pi_X\bigr)\frac{\Pi_X^2}{2} \sin\bigl( \frac r2 \Pi_X\bigr) \\
&\hspace{145pt}- \i \cos\bigl( \frac r2 \Pi_X\bigr) \frac{\Pi_X^2}{2}\cos\bigl( \frac r2 \Pi_X\bigr).
\end{align*}
We plug the second term of \eqref{DHS1:Tcal3_eq1}, written in this form, back into \eqref{DHS1:Tcal3_eq3} and obtain
\begin{align}
2 \i \fint_{Q_B}\dd X\int_{\Rbb^3} \dd r\; \ov{AA^*\Psi(X)} \; \ov{p_r\alpha_*(r)} \; \sin\bigl( \frac r2 \Pi_X\bigr) \Pi_X  \cos\bigl( \frac r2 \Pi_X\bigr)\; \alpha_*(r)\; \Psi(X) \hspace{-350pt}& \notag\\
&= 2 \i \fint_{Q_B}\dd X\int_{\Rbb^3} \dd r\; \ov{AA^* \Psi(X)} \; \alpha_*(r) \; \sin\bigl( \frac r2 \Pi_X\bigr) \Pi_X \cos\bigl( \frac r2 \Pi_X\bigr) \; p_r\alpha_*(r)\; \Psi(X) \notag \\
&\hspace{20pt} + \fint_{Q_B}\dd X\int_{\Rbb^3} \dd r \; \ov{AA^*\Psi(X)} \;  \alpha_*(r) \; \cos\bigl( \frac r2 \Pi_X\bigr)\Pi_X^2 \cos\bigl( \frac r2 \Pi_X\bigr) \; \alpha_*(r)\; \Psi(X) \notag \\
&\hspace{20pt} - \fint_{Q_B}\dd X\int_{\Rbb^3} \dd r \; \ov{AA^*\Psi(X)} \; \alpha_*(r) \; \sin\bigl( \frac r2 \Pi_X\bigr)\Pi_X^2 \sin\bigl( \frac r2 \Pi_X\bigr) \; \alpha_*(r) \; \Psi(X).\label{DHS1:MainTerm_7}
\end{align}
Notice that the first term on the right side equals $(-1)$ times the term on the left side. Thus, the left side equals $\frac 12$ times the third line plus the fourth line. To compute the third line of \eqref{DHS1:MainTerm_7} we use \eqref{DHS1:PiX-term} with the choice $\Phi = AA^*\Psi$. A short computation shows that \eqref{DHS1:PiX-term_2} holds equally with $\cos$ and $\sin$ interchanged. Accordingly, $(-1)$ times the fourth line of \eqref{DHS1:MainTerm_7} equals
\begin{equation*}
	\langle AA^*\Psi, (1 -AA^*) \Pi^2\Psi\rangle + \fint_{Q_B}\dd X\int_{\Rbb^3} \dd r \; \ov{AA^*\Psi(X)} \; |\Bbold\wedge r|^2  \alpha_*(r)^2 \; \cos^2\bigl( \frac r2 \Pi_X\bigr)  \; \Psi(X). 
\end{equation*}
In combination, these considerations imply that the left side of \eqref{DHS1:MainTerm_7} is given by
\begin{align}
&\frac 12 \langle AA^*\Psi, AA^* \Pi^2\Psi\rangle - \frac 12 \langle AA^*\Psi, (1 -AA^*) \Pi^2\Psi\rangle \notag \\
&\hspace{20pt}+ \frac 12 \fint_{Q_B}\dd X\int_{\Rbb^3} \dd r \; \ov{AA^*\Psi(X)} \; |\Bbold\wedge r|^2  \alpha_*(r)^2 \; \sin^2\bigl( \frac r2 \Pi_X\bigr)  \; \Psi(X) \notag\\
&\hspace{20pt} - \frac 12 \fint_{Q_B}\dd X\int_{\Rbb^3} \dd r \; \ov{AA^*\Psi(X)} \; |\Bbold\wedge r|^2  \alpha_*(r)^2 \; \cos^2\bigl( \frac r2 \Pi_X\bigr)  \; \Psi(X). \label{DHS1:Tcal3_eq4}
\end{align}
We note that the third term in \eqref{DHS1:Tcal3_eq4} cancels the second term in \eqref{DHS1:Tcal3_eq3}. Adding all this to \eqref{DHS1:MainTerm_4}, we find
\begin{align}
\langle A^*\Psi, \Tcal_3 A^*\Psi\rangle &= \frac 12 \langle \Psi, AA^* AA^* \Pi^2 \Psi\rangle - \frac 12 \langle \Psi, AA^* (1 - AA^*) \Pi^2 \Psi\rangle  \notag\\
&\hspace{-10pt} + 2 \fint_{Q_B} \dd X\int_{\Rbb^3} \dd r\; \ov{AA^*\Psi(X)} \; |\nabla \alpha_*(r)|^2 \; \cos^2\bigl( \frac r2 \Pi_X\bigr) \; \Psi(X) \notag \\ 
&\hspace{-10pt}+ \frac 12 \fint_{Q_B} \dd X \int_{\Rbb^3} \dd r \; \ov{AA^*\Psi(X)} \; |\Bbold \wedge r|^2 \alpha_*(r)^2 \;  \sin^2\bigl( \frac r2 \Pi_X\bigr)\; \Psi(X). \label{DHS1:Tcal3_result}
\end{align}
The corresponding result for $\langle A^*\Psi, \Tcal_4 A^*\Psi\rangle$ is obtained by taking the complex conjugate of the right side of \eqref{DHS1:Tcal3_result}, which amounts to interchanging the roles of $AA^*\Psi$ and $\Psi$ in the last two lines.

We are now prepared to collect our results and to provide the final formula for \eqref{DHS1:Tcal_first_line}. We need to collect the terms in \eqref{DHS1:Tcal1-Result}, \eqref{DHS1:Tcal2_result}, \eqref{DHS1:Tcal3_result} and the complex conjugate of \eqref{DHS1:Tcal3_result}. The terms involving $|\nabla \alpha_*|^2$ read
\begin{align}
&2\fint_{Q_B} \dd X\int_{\Rbb^3} \dd r\; \ov{\Psi(X)}\;  |\nabla \alpha_*(r)|^2 \; \cos^2\bigl( \frac r2 \Pi_X\bigr) \; \Psi(X) + 2\langle AA^*\Psi, AA^*\Psi\rangle \Vert \nabla \alpha_*\Vert_2^2 \notag \\
&\hspace{50pt}-2 \fint_{Q_B} \dd X\int_{\Rbb^3} \dd r\; \ov{AA^*\Psi(X)} \; |\nabla \alpha_*(r)|^2 \; \cos^2\bigl( \frac r2 \Pi_X\bigr) \; \Psi(X) \notag \\
&\hspace{50pt}- 2 \fint_{Q_B} \dd X\int_{\Rbb^3} \dd r\; \ov{\Psi(X)} \; |\nabla \alpha_*(r)|^2 \; \cos^2\bigl( \frac r2 \Pi_X\bigr) \;  AA^*\Psi(X). \label{DHS1:nablaterms}
\end{align}
When we insert the factor $1 = \cos^2(\frac r2 \Pi_X) + \sin^2(\frac r2\Pi_X)$ in the second term, we obtain the final result for the terms proportional to $|\nabla \alpha_*|^2$.


The terms proportional to $\alpha_*^2$ with magnetic fields read
\begin{align*}
&\frac 12 \fint_{Q_B}\dd X\int_{\Rbb^3} \dd r \; \ov{\Psi(X)} \; |\Bbold\wedge r|^2 \alpha_*(r)^2\; \sin^2\bigl( \frac r2 \Pi_X\bigr)  \; \Psi(X) \\
&\hspace{30pt} + \frac 12\langle AA^*\Psi, AA^*\Psi\rangle \int_{\Rbb^3} \dd r \; |\Bbold \wedge r|^2 \alpha_*(r)^2 \\
&\hspace{30pt}-  \frac 12 \fint_{Q_B} \dd X \int_{\Rbb^3} \dd r \; \ov{AA^*\Psi(X)} \; |\Bbold \wedge r|^2 \alpha_*(r)^2 \; \sin^2\bigl( \frac r2 \Pi_X\bigr) \; \Psi(X)  \\
&\hspace{30pt}-  \frac 12 \fint_{Q_B} \dd X \int_{\Rbb^3} \dd r \; \ov{\Psi(X)} \; |\Bbold \wedge r|^2\alpha_*(r)^2\; \sin^2\bigl( \frac r2 \Pi_X\bigr) \; AA^*\Psi(X).
\end{align*}
When we insert $1 = \cos^2(\frac r2 \Pi_X) + \sin^2(\frac r2\Pi_X)$ in the second term we can bring these terms in the claimed form.


Finally, we collect the terms proportional to $\alpha_*^2$ but without magnetic field. Taking into account the first term in \eqref{DHS1:Tcal_first_line}, we find
\begin{align*}
&2\langle \Psi, AA^*(1 - AA^*)\Psi\rangle + \langle \Psi, AA^*\Pi^2\Psi\rangle + \langle \Psi, AA^*(1 - AA^*)\Pi^2\Psi\rangle - \langle \Psi, AA^* AA^* \Pi^2 \Psi\rangle \\
&\hspace{50pt}= 2\langle \Psi, AA^*(1 - AA^*)(1 + \Pi^2)\Psi\rangle.
\end{align*}
To obtain the result, we used that the terms coming from $\Tcal_3$ and $\Tcal_4$ are actually the same because $AA^*$ and $1 - AA^*$ commute with $\Pi^2$, see Lemma \ref{DHS1:AAstar_Positive}. This proves \eqref{DHS1:MainTerm_5} and the lower bound \eqref{DHS1:MainTerm_LowerBound} is implied by the operator bounds in Lemma \ref{DHS1:AAstar_Positive}.
%
\end{proof}

\subsubsection{Step two -- estimating the cross terms}

In the second step of the proof of Proposition \ref{DHS1:First_Decomposition_Result} we estimate the cross terms that we obtain when the decomposition in \eqref{DHS1:First_Decomposition_Result_Decomp} with $\Psi$ and $\xi_0$ in \eqref{DHS1:Def_Psixi} is inserted into the left side of \eqref{DHS1:First_Decomposition_Result_Assumption}. 

%
\begin{lem}
\label{DHS1:Decomp_Low_Momenta_Crossterms}
Given $D_0, D_1 \geq 0$, there is $B_0>0$ with the following properties. If, for some $0< B\leq B_0$, the wave function $\alpha\in \Lsymm$ satisfies
\begin{align*}
\frac 12 \langle \alpha , [U^* (1 - P)U + U(1 - P)U^* ]\alpha \rangle \leq D_0 B\, \Vert \alpha \Vert_2^2 + D_1B^2,
\end{align*}
then $\Psi$ and $\xi_0$ in \eqref{DHS1:Def_Psixi} satisfy the estimates
\begin{align}
	\Vert \alpha \Vert_2^2 \leq C \bigl( \Vert \Psi \Vert_2^2 + D_1 B^2 \bigr)
	\label{DHS1:eq:new1}
\end{align}
and
\begin{align}
\langle \Psi, AA^*(1 - AA^*)\Psi\rangle + \Vert \xi_0\Vert_2^2 \leq C \bigl( B  \Vert \Psi \Vert_2^2 +D_1 B^2\bigr). \label{DHS1:Estimate_Low_Momenta}
\end{align}
Furthermore, for any $\eta >0$ we have
\begin{align}
|\langle \xi_0, [U(1-P)(1+\pi_r^2)(1-P)U^* + U^* (1-P)(1+\pi_r^2)(1-P)U] A^*\Psi\rangle| & \notag \\
&\hspace{-180pt}\leq \eta \, \Vert \Pi\Psi\Vert_2^2 + C \left(1+ \eta^{-1} \right)\, \bigl( B \Vert \Psi \Vert_2^2 + D_1B^2\bigr). \label{DHS1:Tcal_CrossTerms} 
\end{align}
\end{lem}

\begin{proof}
We start by noting that $A\xi_0 =0$ implies $\langle \xi_0, A^*\Psi\rangle =0$, and hence
\begin{align}
\Vert \alpha\Vert_2^2 = \Vert A^*\Psi\Vert_2^2 + \Vert \xi_0\Vert_2^2 \leq \Vert \Psi\Vert_2^2 + \Vert \xi_0\Vert_2^2. \label{DHS1:Decomp_Low_Momenta_1}
\end{align}
We use $\alpha \in \Lsymm$ and $A(1 - A^*A)A^* = AA^*(1 - AA^*)$ to see that
\begin{align}
D_0 B \Vert \alpha\Vert_2^2 + D_1 B^2 &\geq \frac 12 \langle \alpha, [U^* (1 - P)U + U(1 - P)U^* ]\alpha \rangle = \langle \alpha, (1 - A^*A) \alpha\rangle \notag\\
&= \langle A^*\Psi, (1 - A^*A)A^*\Psi\rangle + \langle \xi_0, (1- A^*A) A^*\Psi\rangle + \langle A^*\Psi, (1- A^*A)\xi_0\rangle \notag \\
&\hspace{230pt} + \langle \xi_0 , (1- A^*A)\xi_0\rangle \notag \\
&= \langle \Psi, AA^*(1 - AA^*) \Psi\rangle + \Vert \xi_0\Vert_2^2. \label{DHS1:Decomp_Low_Momenta_9}
\end{align}
From Lemma~\ref{DHS1:AAstar_Positive} we know that the first term on the right side is nonnegative and hence
\begin{equation*}
	\Vert \xi_0\Vert_2^2 \leq D_0 B \Vert \alpha\Vert_2^2 + D_1 B^2.
\end{equation*}
Together with \eqref{DHS1:Decomp_Low_Momenta_1}, this also proves $(1 - D_0B)\Vert \alpha\Vert_2^2 \leq \Vert \Psi\Vert_2^2 + D_1 B^2$, that is, \eqref{DHS1:eq:new1}. Finally, \eqref{DHS1:eq:new1} and \eqref{DHS1:Decomp_Low_Momenta_9} prove \eqref{DHS1:Estimate_Low_Momenta}.

Next we prove \eqref{DHS1:Tcal_CrossTerms}. Let us define 
\begin{align}
\Tcal := U^* (1 - P)(1 + \pi_r^2)(1 - P)U + U(1 - P)(1 + \pi_r^2)(1 - P)U^* \label{DHS1:Tcal_Op_Def}
\end{align}
and consider $\langle \xi_0, \Tcal A^*\Psi\rangle$. We note that $A\xi_0 =0$ implies $PU \xi_0 = 0 = PU^*\xi_0$, where the projection $P$ is understood to act on the relative coordinate. In combination with \eqref{DHS1:MainTerm_6} this allows us to see that
\begin{align}
\langle \xi_0, \Tcal A^*\Psi\rangle &= \Bigl\langle \xi_0, \Bigl[2\tilde \pi_r^2 + \frac{\Pi_X^2}{2} \Bigr] A^*\Psi\Bigr\rangle - \langle \xi_0, (U^* + U) \pi_r^2 \, \alpha_* AA^*\Psi\rangle \label{DHS1:Decomp_Low_Momenta_2}
\end{align}
holds. We use \eqref{DHS1:PiXcos} and \eqref{DHS1:PiXsin} to commute $\Pi_X^2$ in the first term on the right side of \eqref{DHS1:Decomp_Low_Momenta_2} to the right and find
\begin{align}
\frac 12\langle \xi_0, \Pi_X^2 A^*\Psi\rangle &= \frac 12\langle \xi_0, A^*\Pi_X^2\Psi\rangle \notag \\
&\hspace{10pt}- \i \fint_{Q_B} \dd X\int_{\Rbb^3} \dd r\; \ov{\xi_0(X,r)} \; \sin\bigl( \frac r2 \Pi_X\bigr) \; \Bbold \wedge r\; \alpha_*(r) \; \Pi_X \Psi(X) \notag \\
&\hspace{10pt}+ \frac 12\fint_{Q_B} \dd X \int_{\Rbb^3} \dd r \; \ov{\xi_0(X,r)}  \; \cos\bigl( \frac r2 \Pi_X\bigr) \;  |\Bbold\wedge r|^2 \alpha_*(r) \; \Psi(X). \label{DHS1:Decomp_Low_Momenta_4}
\end{align}
The first term on the right side vanishes because $A\xi_0 =0$. Similarly, we apply \eqref{DHS1:pirtilde_cos_pr} and \eqref{DHS1:pirtilde_sin_pr} to commute $\tilde \pi_r^2$ in the first term in \eqref{DHS1:Decomp_Low_Momenta_2} to the right and find
\begin{align}
2\langle \xi_0, \tilde \pi_r^2A^*\Psi\rangle &= 2 \fint_{Q_B} \dd X \int_{\Rbb^3} \dd r\; \ov{\xi_0(X,r)} \; \cos\bigl( \frac r2 \Pi_X\bigr) \; p^2\alpha_*(r) \; \Psi(X) \notag \\ 
&\hspace{20pt} + 2\i \fint_{Q_B} \dd X \int_{\Rbb^3} \dd r \, \ov{\xi_0(X,r)} \;  \sin\bigl( \frac r2 \Pi_X\bigr) \; p\alpha_*(r) \; \Pi_X\Psi(X). \label{DHS1:Decomp_Low_Momenta_6}
\end{align}
When we combine $\pi_r^2 \alpha_*(r) = p_r^2\alpha_*(r) + \frac 14|\Bbold \wedge r|^2\alpha_*(r)$, which holds because $\alpha_*$ is radial, \eqref{DHS1:Decomp_Low_Momenta_2}, \eqref{DHS1:Decomp_Low_Momenta_4} and \eqref{DHS1:Decomp_Low_Momenta_6}, we obtain
\begin{align*}
\langle \xi_0, \Tcal A^*\Psi\rangle &= 2 \fint_{Q_B} \dd X \int_{\Rbb^3} \dd r\; \ov{\xi_0(X,r)} \; \cos\bigl( \frac r2 \Pi_X\bigr) \; \pi_r^2\alpha_*(r) \; (1 - AA^*)\Psi(X) \\
&\hspace{30pt}+ 2\i \fint_{Q_B}\dd X\int_{\Rbb^3} \dd r\; \ov{\xi_0(X,r)} \; \sin\bigl( \frac r2 \Pi_X\bigr) \; \bigl[ p - \frac 12\Bbold \wedge r\bigr] \alpha_*(r) \; \Pi_X\Psi(X).
\end{align*}
Using Cauchy-Schwarz, we bound the absolute value of this by
\begin{align}
|\langle \xi_0, \Tcal A^*\Psi\rangle| \leq 2\Vert \xi_0\Vert_2 \; \bigl[ \Vert \pi_r^2\alpha_*\Vert_2 \; \Vert (1 - AA^*)\Psi\Vert_2 + \bigl(\Vert \nabla \alpha_*\Vert_2 + B \Vert \, | \cdot  |\alpha_*\Vert_2\bigr) \Vert \Pi\Psi\Vert_2\bigr], \label{DHS1:Decomp_Low_Momenta_7}
\end{align}
and with the decay properties of $\alpha_*$ in \eqref{DHS1:Decay_of_alphastar} we see that the norms of $\alpha_*$ on the right side are bounded uniformly in $0 \leq B \leq B_0$. Moreover, Lemma~\ref{DHS1:AAstar_Positive} and \eqref{DHS1:Estimate_Low_Momenta} imply that there is a constant $c>0$ such that
\begin{equation}
	\Vert (1 - AA^*)\Psi\Vert_2^2 \leq \langle \Psi, (1-AA^*) \Psi \rangle \leq \frac{1}{c} \langle \Psi, A A^* (1-AA^*) \Psi \rangle \leq C \bigl( B  \Vert \Psi \Vert_2^2 +D_1 B^2\bigr).
\end{equation}
For $\eta >0$ we thus obtain
\begin{align}
|\langle \xi_0, \Tcal A^*\Psi\rangle| \leq C \bigl[ \eta\,  \Vert \Pi\Psi\Vert_2^2 + \eta^{-1} \, \Vert \xi_0 \Vert_2^2 + \bigl( B \Vert \Psi \Vert_2^2 +D_1 B^2\bigr) \bigr] \label{DHS1:Decomp_Low_Momenta_8}
\end{align}
and an application of \eqref{DHS1:Estimate_Low_Momenta} proves the claim.
\end{proof}

\subsubsection{Proof of Proposition \ref{DHS1:First_Decomposition_Result}}

We recall the decomposition $\alpha = A^*\Psi + \xi_0$ with $\Psi$ and $\xi_0$ in \eqref{DHS1:Def_Psixi} as well as $\Tcal$ in \eqref{DHS1:Tcal_Op_Def}. From \eqref{DHS1:First_Decomposition_Result_Assumption} and \eqref{DHS1:eq:new1} we know that
\begin{align}
C \bigl( B \Vert\Psi\Vert_2^2 + D_1 B^2 \bigr)  \geq \langle A^*\Psi, \Tcal A^*\Psi\rangle + 2\Re\langle \xi_0, \Tcal A^*\Psi\rangle + \langle \xi_0, \Tcal\xi_0\rangle . \label{DHS1:First_Decomposition_Result_1}
\end{align}
With the help of Lemma \ref{DHS1:CommutationI}, the identities $PU\xi_0 = 0 = PU^*\xi_0$ imply
\begin{align}
\langle \xi_0, \Tcal \xi_0 \rangle = \Bigl\langle \xi_0, \bigl(2 + \frac{\Pi_X^2}{2} + 2\tilde \pi_r^2\bigr) \xi_0\Bigr\rangle \geq \frac 12 \, \Vert \xi_0\Vert_\Hsymm^2. \label{DHS1:First_Decomposition_Result_2}
\end{align}
Lemma \ref{DHS1:AAstar_Positive} guarantees the existence of a constant $\rho>0$ such that
\begin{align*}
AA^*(1 - AA^*)(1 + \Pi^2) \geq \rho\; \Pi^2.
\end{align*}
Therefore, \eqref{DHS1:MainTerm_LowerBound} implies 
\begin{align}
\langle A^* \Psi, \Tcal A^*\Psi\rangle \geq 2\, \langle \Psi, AA^*(1 - AA^*) (1 + \Pi^2)\Psi\rangle \geq 2\rho\, \langle \Psi, \Pi^2\Psi\rangle. \label{DHS1:First_Decomposition_Result_3}
\end{align}

To estimate the second term on the right side of \eqref{DHS1:First_Decomposition_Result_1}, we note that $\Tcal$ is bounded from below by $U(1 - P)U^* + U^*(1-P)U$. Therefore, we may apply Lemma \ref{DHS1:Decomp_Low_Momenta_Crossterms} with $\eta = \frac \rho2$ and find
\begin{align*}
2\Re \langle \xi_0, \Tcal A^*\Psi\rangle \geq -2\, |\langle \xi_0, \Tcal A^*\Psi\rangle| \geq - \rho \, \Vert \Pi\Psi\Vert_2^2 - C\bigl( B\Vert \Psi\Vert_2^2 + D_1 B^2\bigr).
\end{align*}
In combination with \eqref{DHS1:First_Decomposition_Result_1}, \eqref{DHS1:First_Decomposition_Result_2} and \eqref{DHS1:First_Decomposition_Result_3}, we thus obtain
\begin{align*}
C\bigl( B\Vert \Psi\Vert_2^2 + D_1B^2 \bigr)\geq \rho \, \Vert \Pi\Psi\Vert_2^2 + \frac 12 \, \Vert \xi_0\Vert_\Hsymm^2.
\end{align*}
This proves \eqref{DHS1:First_Decomposition_Result_Estimate}. 


\subsection{Uniform estimate on \texorpdfstring{$\Vert\Psi\Vert_2$}{Psi}}

Up to now we neglected the nonlinear term on the left side of \eqref{DHS1:Lower_Bound_A_2}. This term provides the inequality
\begin{align}
		\Tr\bigl[(\alpha^* \alpha)^2\bigr] \leq C \left( B \Vert \alpha\Vert_2^2 + B^2 \right). \label{DHS1:Lower_Bound_A_2b}
\end{align}
In this section we will take this term and \eqref{DHS1:Lower_Bound_A_2b} into account and show that it can be combined with Proposition~\ref{DHS1:First_Decomposition_Result} to obtain a bound for $\Vert \Psi\Vert_2$. This will afterwards allow us to prove Theorem~\ref{DHS1:Structure_of_almost_minimizers}.

\begin{lem}
\label{DHS1:Bound_on_psi}
Given $D_0\geq0$, there is $B_0>0$ such that for all $0 < B \leq B_0$ the following holds. If the wave function $\alpha\in \Lsymm$ obeys \eqref{DHS1:Lower_Bound_A_2} then $\Psi$ in \eqref{DHS1:Def_Psixi} satisfies
\begin{align}
\Vert \Psi\Vert_2^2 &\leq CB. \label{DHS1:Bound_on_psi_result}
\end{align}
\end{lem}

\begin{proof}
We recall the decomposition $\alpha = A^*\Psi + \xi_0$ with $\Psi$ and $\xi_0$ in \eqref{DHS1:Def_Psixi}. Eq.~\eqref{DHS1:Lower_Bound_A_2b} and an application of the triangle inequality imply
\begin{align}
C\bigl( B\Vert \Psi\Vert_2^2 + B^2\bigr)^{\nicefrac 14} \geq \Vert \alpha\Vert_4 \geq \Vert A^*\Psi\Vert_4 - \Vert \xi_0\Vert_4. \label{DHS1:Nonlinearity_8}
\end{align}
Thus, it suffices to prove an upper bound for $\Vert \xi_0\Vert_4$ and a lower bound for $\Vert A^*\Psi\Vert_4$. Our proof follows closely the proof of \cite[Eq. (5.48)]{Hainzl2012}. 

\emph{Step 1.} Let us start with the upper bound on $\Vert \xi_0\Vert_4$. We claim the estimate
\begin{align}
\Vert \xi_0\Vert_4 \leq C \bigl( B^{\nicefrac 14} \Vert \Psi\Vert_2^{\nicefrac 12} + B^{\nicefrac 18} \Vert \Psi\Vert_2 + B^{\nicefrac 12}\bigr).
\label{DHS1:Nonlinearity_1}
\end{align}
To see this, we first use Hölder's inequality to estimate $\Vert \xi_0\Vert_4^4 \leq \Vert \xi_0\Vert_2^2 \; \Vert \xi_0\Vert_\infty^2$. From Proposition~\ref{DHS1:First_Decomposition_Result} we know that $\Vert \xi_0\Vert_2^2 \leq C(B\Vert \Psi\Vert_2^2+ B^2)$, and it thus remains to prove a bound for $\Vert \xi_0\Vert_\infty$. We claim that for any $\nu > 3$
\begin{align}
\Vert \xi_0\Vert_\infty \leq  1 + C_\nu\, B^{-\nicefrac 14} \, \Vert (1 + |\cdot |)^\nu \alpha_*\Vert_{\nicefrac 65} \; \Vert \Psi \Vert_6, \label{DHS1:Claim_xi_infty}
\end{align}
where the right side is finite by the decay properties of $\alpha_*$ in \eqref{DHS1:Decay_of_alphastar}. To prove \eqref{DHS1:Claim_xi_infty}, we first note that \eqref{DHS1:gamma_alpha_fermionic_relation} implies $\Vert \alpha\Vert_\infty\leq 1$, and hence $\Vert \xi_0\Vert_\infty \leq 1 + \Vert A^*\Psi\Vert_\infty$. We apply Lemma \ref{DHS1:Schatten_estimate} (b) to $A^*\Psi$ and obtain \eqref{DHS1:Claim_xi_infty}. We also combine \eqref{DHS1:Magnetic_Sobolev} with Proposition \ref{DHS1:First_Decomposition_Result} and obtain $
\Vert \Psi\Vert_6 \leq C( \Vert \Psi\Vert_2 + B^{\nicefrac 12})$. In combination, these considerations imply \eqref{DHS1:Nonlinearity_1}. 

\emph{Step 2.} We claim that
\begin{align}
\Vert A^*\Psi\Vert_4^4 \geq \frac 1{16} \; \Vert \hat \alpha_* \Vert_4^4 \; \Vert \Psi\Vert_4^4 - C \bigl( B^{\nicefrac 18} \Vert \Psi\Vert_2 + B^{\nicefrac 58} \bigr)^4 \label{DHS1:Nonlinearity_5}
\end{align}
holds. To prove \eqref{DHS1:Nonlinearity_5}, we write $\Vert A^*\Psi\Vert_4^4 = \tr ((A^*\Psi)^* A^*\Psi)^2$. The fact that $\alpha_*$ is real-valued implies
\begin{align*}
\Vert A^*\Psi\Vert_4^4 &= \fint_{Q_B} \dd x \int_{\Rbb^3} \dd y \; \Bigl| \int_{\Rbb^3} \dd z \; \alpha_*(x - z)\ov{ \left[ \cos\bigl( \frac{x - z}{2}\Pi_{\frac{x+z}{2}}\bigr)  \Psi\bigl( \frac{x+z}{2}\bigr) \right] } \\
&\hspace{150pt} \times \alpha_*(z-y) \left[ \cos\bigl( \frac{z-y}{2} \Pi_{\frac{z+y}{2}}\bigr)  \Psi\bigl( \frac{z+y}{2}\bigr) \right] \Bigr|^2.
\end{align*}
We use $\cos(x) = 1 - 2\sin^2(\frac x2)$ twice and find
\begin{align}
\Vert A^*\Psi\Vert_4^4 &\geq \frac 14 \Tcal_* - C \, (\Tcal_1 + \Tcal_2), \label{DHS1:Nonlinearity_2}
\end{align}
where
\begin{align*}
\Tcal_* &:= \fint_{Q_B} \dx \int_{\Rbb^3} \dy \;\Bigl| \int_{\Rbb^3} \dd z \;  \alpha_*(x - z) \ov{\Psi\bigl(\frac{x+z}{2}\bigr)} \alpha_*(z - y) \Psi\bigl( \frac{z+y}{2}\bigr)\Bigr|^2
\end{align*}
and 
\begin{align}
\Tcal_1 &:= \fint_{Q_B} \dx \int_{\Rbb^3} \dy \;\Bigl| \int_{\Rbb^3} \dd z \;  \alpha_*(x - z) \ov{\Psi\bigl( \frac{x+z}{2}\bigr)} \notag\\
&\hspace{140pt} \times \alpha_*(z - y) \left[ \sin^2\bigl( \frac{z-y}{4} \Pi_{\frac{z+y}{2}}\bigr)  \Psi\bigl( \frac{z +y}{2}\bigr) \right] \Bigr|^2, \notag\\
\Tcal_2 &:= \fint_{Q_B} \dx \int_{\Rbb^3} \dy \; \Bigl| \int_{\Rbb^3} \dd z\; \alpha_*(x - z)\ov{ \left[ \sin^2\bigl( \frac{x -z}{4} \Pi_{\frac{x+z}{2}}\bigr) \Psi\bigl( \frac{x + z}{2}\bigr) \right]} \notag \\
&\hspace{140pt} \times\alpha_*(z - y) \left[ \cos\bigl( \frac{z - y}{2}\Pi_{\frac{z+y}{2}}\bigr)  \Psi\bigl( \frac{z+y}{2}\bigr) \right] \Bigr|^2. \label{DHS1:Nonlinearity_3}
\end{align}
In the following we derive a lower bound on $\Tcal_*$ and an upper bound on $\Tcal_1$ and $\Tcal_2$.

\emph{Lower bound on $\Tcal_*$.} We change variables $z\mapsto z + x$ and $y\mapsto y + x$ and afterwards replace $x$ by~$X$, which allows us to write
\begin{align}
\Tcal_* = \fint_{Q_B} \dd X \int_{\Rbb^3} \dy \; \Bigl| \int_{\Rbb^3} \dd z \;  \alpha_*(z) \ov{\Psi\bigl(X + \frac z2\bigr)} \alpha_*(z - y) \Psi\bigl(X + \frac{z+y}{2}\bigr)\Bigr|^2. \label{DHS1:Nonlinearity_9}
\end{align}
Next, we combine $\Psi(X + \frac z2) = \e^{\i \frac z2P_X} \Psi(X)$ and the identity $\e^{\i \frac r2P_X} = \e^{\i \frac \Bbold 2 \cdot (r\wedge X)}\e^{\i \frac r2\Pi_X}$ in \eqref{DHS1:representation_Ustar} to write $\Psi(X + \frac z2) = \e^{\i \frac \Bbold 2 \cdot (z\wedge X)} \e^{\i \frac z2\Pi_X} \Psi(X)$. 
We conclude that
\begin{align*}
\ov{\Psi\bigl( X + \frac z2\bigr)} \Psi\bigl(X + \frac{z+y}{2}\bigr) = \e^{\i \frac{\Bbold}{2} \cdot (y\wedge X)} \; \ov{\left[ \e^{\i \frac{z}{2}\Pi_X}\Psi(X) \right]} \; \left[ \e^{\i \frac{z+y}{2}\Pi_X} \Psi(X) \right],
\end{align*}
as well as
\begin{align*}
\Tcal_* = \fint_{Q_B} \dd X \int_{\Rbb^3} \dd y\; \Bigl| \int_{\Rbb^3} \dd z \; \alpha_*(-z) \ov{ \left[ \e^{\i \frac z2\Pi_X} \Psi(X) \right]} \alpha_*(z - y) \left[ \e^{\i \frac{z+y}{2}\Pi_X} \Psi(X) \right] \Bigr|^2.
\end{align*}
This also implies
\begin{align}
\Tcal_* \geq \frac 14\Tcal_*^* - C(\Tcal_*^{(1)} + \Tcal_*^{(2)})
\label{DHS1:eq:A1}
\end{align}
with
\begin{align*}
\Tcal_*^* := \fint_{Q_B}\dd X\int_{\Rbb^3}\dd y\; \Bigl| \int_{\Rbb^3} \dd z \; \alpha_*(z) \ov{\Psi(X)} \alpha_*(z - y) \Psi(X)\Bigr|^2 = \Vert \Psi\Vert_4^4 \; \Vert \alpha_* * \alpha_*\Vert_2^2
\end{align*}
and
\begin{align}
\Tcal_*^{(1)} &:= \fint_{Q_B} \dd X\int_{\Rbb^3} \dd y\; \Bigl| \int_{\Rbb^3} \dd z \; \alpha_*(z) \ov{\bigl[\e^{\i \frac{z}{2}\Pi_X}\Psi(X)\bigr]} \alpha_*(z - y) \bigl[\bigl( \e^{\i \frac{z+y}{2}\Pi_X} - 1\bigr) \Psi(X)\bigr]\Bigr|^2, \notag \\
\Tcal_*^{(2)} &:= \fint_{Q_B} \dd X\int_{\Rbb^3} \dd y\; \Bigl| \int_{\Rbb^3} \dd z \;\alpha_*(z)  \ov{\bigl[\bigl( \e^{\i \frac z2\Pi_X} - 1\bigr) \Psi(X)\bigr]} \alpha_*(z - y) \Psi(X)\Bigr|^2. \label{DHS1:Nonlinearity_10}
\end{align}

\emph{Upper bound on $\Tcal_*^{(1)}$ and $\Tcal_*^{(2)}$.} We start with $\Tcal_*^{(1)}$, expand the square and estimate
\begin{align}
&\Tcal_*^{(1)} \leq  \int_{\Rbb^3} \dd y  \int_{\Rbb^3} \dd z  \int_{\Rbb^3} \dd z' \;  | \alpha_*(z) \, \alpha_*(z') \, \alpha_*(z - y) \, \alpha_*(z' - y)| \label{DHS1:Nonlinearity_13} \\
& \times \fint_{Q_B} \dd X \; \Bigl| \bigl[ \e^{\i \frac z2\Pi_X} \Psi(X) \bigr]  \bigl[ \e^{\i \frac {z'}2\Pi_X} \Psi(X) \bigr]  \bigl[ \bigl( \e^{\i \frac{z+y}{2}\Pi_X}-1\bigr) \Psi(X) \bigr]  \bigl[ \bigl( \e^{\i \frac{z'+y}{2}\Pi_X}-1\bigr) \Psi(X) \bigr] \Bigr|. \nonumber
\end{align}
When we use Hölder's inequality, \eqref{DHS1:NTB-NtildeTB_3}, \eqref{DHS1:ZPiX_inequality}, and \eqref{DHS1:Magnetic_Sobolev}, we see that the integral in the second line can be bounded by
\begin{align}
\Vert \e^{\i \frac{z}{2}\Pi} \Psi\Vert_6^2 \; \Vert (\e^{\i \frac{z+y}{2}\Pi} - 1) \Psi\Vert_6 \; \Vert (\e^{\i \frac{z+y}{2}\Pi} - 1) \Psi\Vert_2 \leq C\; \bigl|\frac{z+y}{2}\bigr| \; B^{-\nicefrac 32} \; \Vert \Pi\Psi\Vert_2^4. \label{DHS1:Nonlinearity_11}
\end{align}
%
%
%
%
%
%
%
%
Proposition~\ref{DHS1:First_Decomposition_Result} provides us with a bound for $\Vert \Pi \Psi \Vert_2$. In combination with \eqref{DHS1:Nonlinearity_13}, \eqref{DHS1:Nonlinearity_11}, Young's inequality, and the bound $|z + y| \leq 2|z| + |z- y|$, this implies
\begin{align}
\Tcal_*^{(1)} &\leq C \, B^{-\nicefrac 32}\,  \left( B^2 \Vert \Psi \Vert_2^4 + D_1 B^4 \right) \int_{\Rbb^3} \dd y \; \Bigl| \int_{\Rbb^3} \dd z\;  |z+y|\; |  \alpha_*(z) \alpha_*(y-z)|\Bigr|^2 \nonumber \\
&\leq C \left( B^{-\nicefrac 12} \Vert \Psi \Vert_2^4 + D_1 B^{ \nicefrac 52 } \right) \, \Vert \alpha_*\Vert_{\nicefrac 43} \, \Vert \, |\cdot|\alpha_*\Vert_{\nicefrac 43}, \label{DHS1:Nonlinearity_12} 
\end{align}
where the right side is finite by \eqref{DHS1:Decay_of_alphastar}. Similarly, we see that $\Tcal_*^{(2)}$ is bounded by the right side of \eqref{DHS1:Nonlinearity_12}.

\emph{Upper bound on $\Tcal_1$ and $\Tcal_2$ in \eqref{DHS1:Nonlinearity_3}.} Bounds for $\Tcal_1$ and $\Tcal_2$ can be obtain along the same lines as the bound for $\Tcal_*^{(1)}$. We apply the same change of variables as above and use estimates similar to the ones in \eqref{DHS1:Nonlinearity_13}. In case of $\Tcal_1$, the bound in \eqref{DHS1:Nonlinearity_11} needs to be replaced by
\begin{align}
\Vert \Psi\Vert_6^2 \, \bigl\Vert \sin^2\bigl( \frac{z - y}{2}  \Pi\bigr) \Psi \bigr\Vert_6 \, \bigl\Vert \sin^2\bigl( \frac{z - y}{2}  \Pi\bigr) \Psi \bigr\Vert_2 \leq C \; \frac{|z-y|}{2} \; B^{-\nicefrac 32}\,  \Vert \Pi\Psi\Vert_2^4. \label{DHS1:Nonlinearity_14}
\end{align}
Here, we used $\sin^2(x) \leq |x|$ and the operator inequality in  \eqref{DHS1:ZPiX_inequality} to estimate the third factor. For the first and the second factor, we used
\begin{align*}
\sin^2\bigl( \frac{z-y}{4} \Pi\bigr) = - \frac{1}{4} \bigl( 2+ \e^{\i\frac{z-y}{2} \Pi} + \e^{-\i \frac{z-y}{2}\Pi}\bigr)
\end{align*}
and \eqref{DHS1:Magnetic_Sobolev} or \eqref{DHS1:NTB-NtildeTB_3}, respectively. A bound for $\Tcal_2$ can be proved analogously. The final estimate we obtain in this way reads
\begin{equation}
	\Tcal_1 + \Tcal_2 \leq C \left( B^{-\nicefrac 12} \Vert \Psi \Vert_2^4 + D_1 B^{ \nicefrac 52 } \right). 
	\label{DHS1:eq:A2}
\end{equation}
In combination with \eqref{DHS1:Nonlinearity_2}, \eqref{DHS1:eq:A1}, and \eqref{DHS1:Nonlinearity_12}, this proves \eqref{DHS1:Nonlinearity_5}.

\emph{Step 3.} We denote $c:= \frac 1{2} \Vert \hat\alpha_*\Vert_4$, insert \eqref{DHS1:Nonlinearity_1} and \eqref{DHS1:Nonlinearity_5} into \eqref{DHS1:Nonlinearity_8} and obtain
\begin{align}
C B^{\frac 14}\Vert \Psi\Vert_2^{\nicefrac 12}  \geq c \Vert \Psi\Vert_4 - CB^{\frac 18} \Vert \Psi\Vert_2 - CB^{\frac 12}, \label{DHS1:Nonlinearity_6}
\end{align}
which holds for $B$ small enough. For $\eta >0$ the left side is bounded from above by a constant times $\eta \Vert \Psi\Vert_2 + \eta^{-1} B^{\frac 12}$ and Hölder's inequality implies $\Vert \Psi\Vert_4 \geq \Vert \Psi\Vert_2$. 
Accordingly,
\begin{equation}
C \left( \eta \Vert \Psi\Vert_2 + \eta^{-1} B^{\frac 12} \right) \geq (c - CB^{\frac 18}) \Vert \Psi\Vert_2 - CB^{\frac 12}.
\label{DHS1:eq:A3}
\end{equation}
When we choose $\eta$ and $B$ in \eqref{DHS1:eq:A3} small enough, this proves the claim.
%
%
%
\end{proof}


\subsection{Proof of Theorem \ref{DHS1:Structure_of_almost_minimizers}}

The assumption \eqref{DHS1:Second_Decomposition_Gamma_Assumption} in Theorem~\ref{DHS1:Structure_of_almost_minimizers}, Corollary~\ref{DHS1:cor:lowerbound}, Proposition~\ref{DHS1:First_Decomposition_Result}, \ifthenelse{\equal\masterfile{Diss}}{as well as}{and} Lemma~\ref{DHS1:Bound_on_psi} imply the decomposition $\alpha = A^*\Psi + \xi_0$, where $\Psi$ and $\xi_0$ in \eqref{DHS1:Def_Psixi} obey
\begin{align}
\Vert \Psi\Vert_{\Hmag^1(Q_B)}^2 = B^{-1} \Vert \Psi\Vert_2^2 + B^{-2} \Vert \Pi\Psi\Vert_2^2 \leq C  \label{DHS1:Psi_estimate}
\end{align}
and
\begin{align}
\Vert \xi_0\Vert_{\Hsymm}^2 \leq C B^2 \bigl( \Vert \Psi\Vert_{\Hmag^1(Q_B)}^2 + D_1\bigr). \label{DHS1:xi0_estimate}
\end{align}
Define
\begin{align}
\xi &:= \xi_0 + \bigl( \cos\bigl( \frac r2 \Pi_X\bigr) - 1\bigr) \alpha_*(r) \Psi(X). \label{DHS1:xi_definition}
\end{align}
Then, \eqref{DHS1:Second_Decomposition_alpha_equation} holds and we claim that $\xi$ satisfies \eqref{DHS1:Second_Decomposition_Psi_xi_estimate}. To prove this, we estimate the second term in \eqref{DHS1:xi_definition}. We use $1 - \cos(x) \leq |x|$ and \eqref{DHS1:ZPiX_inequality} to bound
\begin{align*}
\bigl\Vert \bigl( \cos\bigl( \frac r2\Pi_X\bigr) - 1\bigr) \Psi \alpha_*\bigr\Vert_2^2 &\leq  C\; \Vert |\cdot|\alpha_*\Vert_2^2 \; \Vert \Pi\Psi\Vert_2^2 \leq CB^2 \, \Vert \Psi\Vert_{\Hmag^1(Q_B)}^2,
\end{align*}
where the right side is finite by the decay properties of $\alpha_*$ in \eqref{DHS1:Decay_of_alphastar}. Using additionally \eqref{DHS1:PiXcos}, we also see that
\begin{align*}
\bigl\Vert \Pi_X \bigl( \cos\bigl( \frac r2\Pi_X\bigr) -1\bigr) \alpha_* \Psi \bigr\Vert_2^2 &\leq \bigl\Vert \bigl[ \cos\bigl(\frac r2\Pi_X\bigr) - 1 \bigr] \Pi_X \alpha_* \Psi \bigr\Vert_2^2 + \bigl\Vert \sin\bigl( \frac r2\Pi_X\bigr) \Bbold \wedge r \alpha_* \Psi \bigr\Vert_2^2\\
&\leq C \bigl(\Vert \Pi\Psi\Vert_2^2 + B^2 \Vert \Psi \Vert_2^2\bigr) \leq CB^3 \, \Vert \Psi\Vert_{\Hmag^1(Q_B)}^2
\end{align*}
holds. Finally, an application of \eqref{DHS1:pirtilde_cos_pir} and \eqref{DHS1:ZPiX_inequality} allows us to estimate
\begin{align*}
\bigl\Vert \pi_r \bigl( \cos\bigl( \frac r2\Pi_X\bigr) - 1\bigr) \Psi \alpha_*\bigr\Vert_2^2 &= \bigl\Vert \bigl[ \bigl( \cos\bigl(\frac r2\Pi_X\bigr) - 1\bigr) \tilde \pi_r + \frac \i 2 \sin\bigl( \frac r2\Pi_X\bigr)\Pi_X\bigr] \Psi \alpha_*\bigr\Vert_2^2 \\
&\leq C \bigl( \Vert \Pi\Psi\Vert_2^2 + B^2 \Vert \Psi \Vert_2^2\bigr) \leq CB^2 \, \Vert \Psi\Vert_{\Hmag^1(Q_B)}^2.
\end{align*}
This proves that $\xi$ obeys \eqref{DHS1:Second_Decomposition_Psi_xi_estimate} and ends the proof of Theorem \ref{DHS1:Structure_of_almost_minimizers}.


\section{The Lower Bound on \texorpdfstring{(\ref{DHS1:ENERGY_ASYMPTOTICS})}{(\ref{DHS1:ENERGY_ASYMPTOTICS})} and Proof of Theorem \ref{DHS1:Main_Result_Tc} (b)}
\label{DHS1:Lower Bound Part B}


\subsection{The BCS energy of low-energy states}

In this section, we provide the lower bound on \eqref{DHS1:ENERGY_ASYMPTOTICS} and the proof of Theorem~\ref{DHS1:Main_Result_Tc}~(b), and thereby complete the proof of Theorems \ref{DHS1:Main_Result} and \ref{DHS1:Main_Result_Tc}. Let $D_1\geq 0$ and $D\in \Rbb$ be given and assume that $\Gamma$ is a gauge-periodic state at temperature $T = \Tc(1 - DB)$ that satisfies \eqref{DHS1:Second_Decomposition_Gamma_Assumption}.
Corollary~\ref{DHS1:Structure_of_almost_minimizers_corollary} provides us with a decomposition of the Cooper pair wave function $\alpha = [\Gamma]_{12}$ in terms of $\Psi_\leq$ in \eqref{DHS1:PsileqPsi>_definition} and $\sigma$ in \eqref{DHS1:sigma}, where $\Vert \Psi_\leq \Vert_{\Hmag^1(Q_B)} \leq C$ and where the bound
\begin{align}
	\Vert \Psi_\leq\Vert_{\Hmag^2(Q_B)}^2 &\leq C \, \varepsilon B^{-1} \, \Vert \Psi\Vert_{\Hmag^1(Q_B)}^2 \label{DHS1:Lower_Bound_B_Psileq}
\end{align}
holds in terms of the function $\Psi$ in Theorem \ref{DHS1:Structure_of_almost_minimizers}. With the function $\Psi_\leq$ we construct a Gibbs state $\Gamma_{\Delta}$ with the gap function $\Delta \equiv \Delta_{\Psi_\leq}$ as in \eqref{DHS1:Delta_definition}. Using Proposition~\ref{DHS1:BCS functional_identity}, we write the BCS free energy of $\Gamma$ as
\begin{align*}
	\FBCS(\Gamma) - \FBCS(\Gamma_0) &= - \frac 14 \langle \Delta, L_{T,B} \Delta\rangle + \frac 18 \langle \Delta, N_{T,B} (\Delta)\rangle + \Vert \Psi_\leq \Vert_2^2 \;  \langle \alpha_*, V\alpha_*\rangle \\
	&\hspace{10pt}+ \Tr\bigl[\Rcal_{T,B}(\Delta)\bigr] + \frac T2 \Hcal_0(\Gamma, \Gamma_\Delta) - \fint_{Q_B} \dd X\int_{\Rbb^3} \dd r\; V(r) \, |\sigma(X,r)|^2,
\end{align*}
where
\begin{equation*}
	\Vert \Rcal_{T,B}(\Delta) \Vert_1 \leq C \; B^3 \; \Vert \Psi \Vert_{\Hmag^1(Q_B)}^6.
\end{equation*}
We also apply Theorem~\ref{DHS1:Calculation_of_the_GL-energy} to compute the terms in the first line on the right side, and find the lower bound
\begin{align}
		\FBCS(\Gamma) - \FBCS(\Gamma_0) &\geq B^2 \; \EGL(\Psi_\leq) -C \bigl( B^3 + \varepsilon B^2 \bigr) \Vert \Psi\Vert_{\Hmag^1(Q_B)}^2 \notag \\
		&\hspace{30pt}+ \frac T2 \Hcal_0(\Gamma, \Gamma_\Delta) - \fint_{Q_B} \dd X\int_{\Rbb^3} \dd r\; V(r) \, |\sigma(X,r)|^2. \label{DHS1:Lower_Bound_B_2}
\end{align}
The relative entropy is nonnegative and the last term on the right side is nonpositive. In the next section we show that their sum is negligible.

\subsection{Estimate on the relative entropy}

In this section we prove a lower bound for the second line in \eqref{DHS1:Lower_Bound_B_2}, showing that it is negligible. We start with the following lower bound for the relative entropy.

\begin{lem}
For all admissible BCS states $\Gamma$, we have
\begin{align}
T\Hcal_0(\Gamma, \Gamma_\Delta) &\geq \Tr \Bigl[ (\Gamma - \Gamma_\Delta) \frac{H_\Delta}{\tanh(\frac \beta 2H_\Delta)} (\Gamma - \Gamma_\Delta)\Bigr]. \label{DHS1:LBpartB_1}
\end{align}
\end{lem}

\begin{proof}
The proof is given in \cite[Lemma 5]{Hainzl2012} and uses the fact that $\Gamma_\Delta$ is admissible, which follows from Lemma \ref{DHS1:Gamma_Delta_admissible}.
\end{proof}

To be able to combine the term on the right side of \eqref{DHS1:LBpartB_1} and the last term on the right side of \eqref{DHS1:Lower_Bound_B_2}, we first need to replace the operator $H_{\Delta}$ in the second factor on the right side of \eqref{DHS1:LBpartB_1} by $H_0$. To that end, we note that the estimate $H_\Delta^2\geq (1 - \delta)H_0^2 - \delta^{-1} \Vert \Delta\Vert_\infty^2$ holds for $0 < \delta < 1$ and we rewrite it as
\begin{align}
H_0 \leq (1 - \delta)^{-1} \bigl( H_\Delta^2 + \delta^{-1} \Vert \Delta\Vert_\infty^2\bigr). \label{DHS1:LBpartB_4}
\end{align}
Furthermore, we note that the series expansion 
\begin{align*}
	\frac{x}{\tanh(\frac x2)} = 2+ \sum_{k=1}^\infty \bigl( 2 - \frac{8k^2\pi^2}{x^2 + 4k^2\pi^2}\bigr),
\end{align*}
see \cite[Eq. (5.14)]{Hainzl2012}, shows that the function $x \mapsto \frac{\sqrt{x}}{\tanh(\frac \beta 2\sqrt{x})}$ is operator monotone. We use this together with \eqref{DHS1:LBpartB_4} and the monotonicity of the map $x \mapsto \tanh(x)$, which yields
 \begin{align*}
	\begin{pmatrix}
		K_{T,B} & 0 \\ 0 & \ov{K_{T,B}}
	\end{pmatrix} = \frac{H_0}{\tanh(\frac \beta 2 H_0)}
&\leq (1 - \delta)^{-\nicefrac 12} \frac{\sqrt{H_\Delta^2 + \delta^{-1}\Vert \Delta\Vert_\infty^2}}{\tanh(\frac \beta 2 \sqrt{H_\Delta^2 + \delta^{-1} \Vert\Delta\Vert_\infty^2})}.
\end{align*}
When we apply a first order Taylor expansion on the right side, the above inequality can be written as
\begin{align*}
\begin{pmatrix}
K_{T,B} & 0 \\ 0 & \ov{K_{T,B}}
\end{pmatrix} 
&\leq (1 - \delta)^{-\nicefrac 12} \Bigl[\frac{H_\Delta}{\tanh(\frac \beta 2H_\Delta)} + \frac{\beta}{4} \int_0^{\delta^{-1} \Vert \Delta\Vert_\infty^2} \dd t \; g\bigl( \frac \beta 2 \sqrt{H_\Delta^2 + t}\bigr)\Bigr]
\end{align*}
with the nonnegative function 
\begin{align*}
g(x) &:= \frac{1}{x} \frac{1}{\tanh(x)} - \frac{1}{\tanh^2(x)} \frac{1}{\cosh^2(x)}.
\end{align*}
Using $\sup_{x \geq 0} g(x) \leq 1$ and $1 \leq \frac{x}{\tanh(x)}$, we conclude that
\begin{align*}
	\begin{pmatrix}
		K_{T,B} & 0 \\ 0 & \ov{K_{T,B}}
	\end{pmatrix} \leq (1 - \delta)^{-\nicefrac 12} \bigl( 1 +  \frac{\delta^{-1} \beta^2}{8} \;  \Vert \Delta\Vert_\infty^2\bigr) \frac{H_\Delta}{\tanh(\frac \beta 2 H_\Delta)}
\end{align*}
holds. We choose $\delta := \Vert \Delta\Vert_\infty$ and note that Lemma~\ref{DHS1:Schatten_estimate} and \eqref{DHS1:Magnetic_Sobolev} imply 
\begin{equation}
	\Vert \Delta\Vert_\infty \leq C \; B^{\nicefrac 14} \; \Vert \Psi \Vert_{\Hmag^1(Q_B)}.
	\label{DHS1:eq:ABDelta}
\end{equation} 
In particular, $\delta < 1 $ as long as $B>0$ is sufficiently small, and we have
\begin{align}
	\frac{H_\Delta}{\tanh(\frac \beta 2 H_\Delta)} \geq \frac{\sqrt{1 - \Vert \Delta\Vert_\infty}}{1 + \frac{\beta^2}{8} \Vert \Delta\Vert_\infty} \; \begin{pmatrix}
		K_{T,B} & 0 \\ 0 & \ov{K_{T,B}}
	\end{pmatrix} 
	\geq (1 - C\Vert \Delta\Vert_\infty) \; \begin{pmatrix}
		K_{T,B} & 0 \\ 0 & \ov{K_{T,B}}
	\end{pmatrix}.
	\label{DHS1:LBpartB_2}
\end{align}
In combination, \eqref{DHS1:LBpartB_1} and \eqref{DHS1:LBpartB_2} prove
\begin{align*}
\frac 12 \Tr \Bigl[(\Gamma - \Gamma_\Delta)\frac{H_\Delta}{\tanh(\frac \beta 2H_\Delta)} (\Gamma - \Gamma_\Delta)\Bigr] & \\
&\hspace{-160pt} \geq 
%
(1 - C\Vert \Delta\Vert_\infty) \langle \alpha - \alpha_\Delta , K_{T,B} (\alpha - \alpha_\Delta)\rangle + (1 - C\Vert \Delta\Vert_\infty) \Tr[ (\gamma - \gamma_\Delta) K_{T,B}(\gamma - \gamma_\Delta)],
\end{align*}
where we can drop the last term for a lower bound because it is nonnegative if $B$ is sufficiently small. This is the lower bound for the relative entropy of $\Gamma$ with respect to $\Gamma_{\Delta}$ we were looking for. 

It remains to combine the first term on the right side and the interaction term on the right side of \eqref{DHS1:Lower_Bound_B_2}.
%
Let us define the function $\eta := \alpha_* \Psi_\leq - \alpha_\Delta$. By Corollary~\ref{DHS1:Structure_of_almost_minimizers_corollary} we have $\alpha - \alpha_\Delta = \sigma + \eta$ as well as
\begin{align}
	&\frac T2 \Hcal_0(\Gamma, \Gamma_\Delta) - \fint_{Q_B} \dd X \int_{\Rbb^3} \dd r \; V(r)|\sigma(X, r)|^2 \notag\\
	&\hspace{20pt} \geq (1 - C\Vert \Delta\Vert_\infty) \langle \sigma + \eta, K_{T,B} (\sigma + \eta)\rangle - \langle \sigma , V\sigma\rangle \notag\\
	&\hspace{20pt} \geq (1 - C\Vert \Delta\Vert_\infty) \langle \sigma, (K_{T,B} - V) \sigma \rangle - C\Vert \Delta\Vert_\infty \Vert V\Vert_\infty \Vert \sigma \Vert_2^2 - 2 \, | \langle\eta, K_{T,B} \sigma\rangle|. \label{DHS1:Lower_Bound_2_intermediate}
\end{align}
From \eqref{DHS1:KTB_Lower_bound_5} we know that the lowest eigenvalue of $K_{T,B} - V$ is bounded from below by $-CB$. In combination with \eqref{DHS1:Second_Decomposition_sigma_estimate} and \eqref{DHS1:eq:ABDelta}, this implies that the first two terms on the right side of \eqref{DHS1:Lower_Bound_2_intermediate} are bounded from below by 
\begin{align}
	-C \varepsilon^{-1}B^{\nicefrac 94} \Vert \Psi\Vert_{\Hmag^1(Q_B)} \; \bigl(  \Vert \Psi\Vert_{\Hmag^1(Q_B)}^2 + D_1\bigr)^{\nicefrac 12}. \label{DHS1:Lower_Bound_B_3}
\end{align}
To estimate the last term on the right side of \eqref{DHS1:Lower_Bound_2_intermediate}, we use \eqref{DHS1:KTc_bounded_derivative} to replace $K_{T, B}$ by $ K_{\Tc, B}$, which yields the estimate
\begin{align*}
	|\langle \eta ,(K_{T,B} - K_{\Tc,B} )\sigma\rangle| &\leq 2D_0 B \, \Vert \sigma\Vert_2 \, \Vert \eta\Vert_2 \leq C \, B^3 \, \Vert \Psi\Vert_{\Hmag^1(Q_B)} \; \bigl(  \Vert \Psi\Vert_{\Hmag^1(Q_B)}^2 + D_1\bigr)^{\nicefrac 12}.
\end{align*}
To obtain this result we also used \eqref{DHS1:Second_Decomposition_sigma_estimate}, 
Proposition~\ref{DHS1:Structure_of_alphaDelta} and \eqref{DHS1:Lower_Bound_B_Psileq}. Next, we decompose $\eta = \eta_0 + \eta_\perp$ with $\eta_0(\Delta)$ and $\eta_{\perp}(\Delta)$ as in Proposition~\ref{DHS1:Structure_of_alphaDelta} and write
\begin{align}
	\langle \eta, K_{\Tc,B} \sigma \rangle &= \langle \eta_0 , K_{\Tc,B} \sigma\rangle + \langle \eta_\perp, K_{\Tc,B} (\sigma - \sigma_0)\rangle + \langle \eta_\perp , K_{\Tc,B} \sigma_0\rangle. \label{DHS1:Lower_Bound_B_1}
\end{align}
Using \eqref{DHS1:alphaDelta_decomposition_eq2} and \eqref{DHS1:Second_Decomposition_sigma_estimate}, we see that the first term on the right side of \eqref{DHS1:Lower_Bound_B_1} is bounded by
\begin{align}
	|\langle \eta_0, K_{\Tc,B} \sigma \rangle| &\leq \bigl\Vert \sqrt{K_{\Tc,B}}\, \eta_0\bigr\Vert_{2} \bigl\Vert \sqrt{K_{\Tc,B}}\,\sigma\bigr\Vert_{2} \notag \\
	&\leq C \varepsilon^{-\nicefrac 12} B^{\nicefrac 52} \, \Vert \Psi\Vert_{\Hmag^1(Q_B)} \; \bigl(  \Vert \Psi\Vert_{\Hmag^1(Q_B)}^2 + D_1\bigr)^{\nicefrac 12}.
	\label{DHS1:eq:A30}
\end{align}
We note that $\sigma - \sigma_0 = \xi$ and use \eqref{DHS1:alphaDelta_decomposition_eq3}, \eqref{DHS1:Second_Decomposition_Psi_xi_estimate}, and \eqref{DHS1:Lower_Bound_B_Psileq} to estimate
\begin{align}
	|\langle \eta_\perp, K_{\Tc,B} \xi \rangle| &\leq \bigl\Vert \sqrt{K_{\Tc,B}} \, \eta_\perp\bigr\Vert_{2} \bigl\Vert \sqrt{K_{\Tc,B}} \, \xi \bigr\Vert_{2} \notag \\
	&\leq C\varepsilon^{\nicefrac 12} B^2 \, \Vert \Psi\Vert_{\Hmag^1(Q_B)} \; \bigl(  \Vert \Psi\Vert_{\Hmag^1(Q_B)}^2 + D_1\bigr)^{\nicefrac 12}. \label{DHS1:eq:A31} 
\end{align}
It remains to estimate the last term on the right side of \eqref{DHS1:Lower_Bound_B_1}, which we write as 
\begin{align}
		\langle \eta_\perp, K_{\Tc,B} \sigma_0\rangle &= \langle \eta_\perp, K_{\Tc}^r  \sigma_0\rangle + \langle \eta_\perp, [K_{\Tc,B}^r - K_{\Tc}^r ] \sigma_0\rangle + \langle \eta_\perp, (U-1)K_{\Tc,B}^r \sigma_0\rangle \notag \\
		&\hspace{190pt}  + \langle \eta_\perp, UK_{\Tc,B}^r (U^* - 1) \sigma_0\rangle. \label{DHS1:LBpartB_3}
\end{align}
with the unitary operator $U$ in \eqref{DHS1:U_definition}. We recall that the operators $K_{\Tc,B}^r$ and $K_{\Tc}^r$ act on the relative coordinate $r = x-y$.

Since $\Delta(X,r) = - 2 V(r) \alpha_*(r) \Psi_{\leq}(X)$ and $\sigma_{0}(X,r) = \alpha_*(r) \Psi_>(X)$ we know from Proposition~\ref{DHS1:Structure_of_alphaDelta}~(c) that the first term on the right side of \eqref{DHS1:LBpartB_3} vanishes. A bound for the remaining terms is provided by the following lemma. Its proof will be given in Section~\ref{DHS1:sec:A1} below.
\begin{lem}
	\label{DHS1:Lower_Bound_B_remainings}
	We have the following estimates on the remainder terms of \eqref{DHS1:LBpartB_3}:
	\begin{enumerate}[(a)]
		\item $|\langle \eta_\perp, [K_{\Tc,B}^r - K_{\Tc}^r]\sigma_0\rangle| \leq C B^3 \, \Vert \Psi\Vert_{\Hmag^1(Q_B)}^2$,
		
		\item $|\langle \eta_\perp, (U - 1) K_{\Tc,B}^r  \sigma_0\rangle| \leq C \varepsilon^{\nicefrac 12} B^2 \, \Vert \Psi\Vert_{\Hmag^1(Q_B)}^2$,
		
		\item $|\langle \eta_\perp, UK_{\Tc,B}^r (U^* - 1)\sigma_0\rangle| \leq C \varepsilon^{\nicefrac 12} B^2 \, \Vert \Psi\Vert_{\Hmag^1(Q_B)}^2$.
	\end{enumerate}
\end{lem} 

Accordingly, we have
\begin{align*}
|\langle \sigma, K_{T,B} \eta \rangle | &\leq C\bigl( \varepsilon^{-\nicefrac 12} B^{\nicefrac 52} + \varepsilon^{\nicefrac 12} B^2 \bigr) \, \Vert \Psi\Vert_{\Hmag^1(Q_B)} \; \bigl(  \Vert \Psi\Vert_{\Hmag^1(Q_B)}^2 + D_1\bigr)^{\nicefrac 12}.
\end{align*}
We combine this with \eqref{DHS1:Lower_Bound_B_2}, \eqref{DHS1:Lower_Bound_2_intermediate}, and \eqref{DHS1:Lower_Bound_B_3} to see that
\begin{align}
\FBCS(\Gamma) - \FBCS(\Gamma_0) &  \geq B^2\; \EGL(\Psi_\leq) \notag\\
&\hspace{-70pt} - C\bigl( \varepsilon^{-\nicefrac 12} B^{\nicefrac 52} + \varepsilon^{\nicefrac 12} B^2 + \varepsilon^{-1} B^{\nicefrac 94} \bigr) \Vert \Psi\Vert_{\Hmag^1(Q_B)} \bigl( \Vert \Psi\Vert_{\Hmag^1(Q_B)} + D_1 \bigr)^{\nicefrac 12}.  \label{DHS1:Lower_Bound_B_8}
\end{align}
The optimal choice $\varepsilon = B^{\nicefrac 16}$ in \eqref{DHS1:Lower_Bound_B_8} yields
\begin{align}
	&\FBCS(\Gamma) - \FBCS(\Gamma_0) \notag \\
	&\hspace{2cm}\geq B^2 \, \bigl(\EGL(\Psi_\leq) - C \,  B^{\nicefrac {1}{12}} \Vert \Psi\Vert_{\Hmag^1(Q_B)} \; \bigl(  \Vert \Psi\Vert_{\Hmag^1(Q_B)}^2 + D_1\bigr)^{\nicefrac 12} \bigr). \label{DHS1:Lower_Bound_B_5}
\end{align}
This proves the lower bound for the BCS free energy in Theorem~\ref{DHS1:Main_Result}.


\subsection{Conclusion}

Using \eqref{DHS1:Lower_Bound_B_5}, we now finish the proofs of Theorem~\ref{DHS1:Main_Result} and Theorem~\ref{DHS1:Main_Result_Tc}, and we start with the former. Let $\Gamma$ be an approximate minimizer of the BCS functional, i.e., let \eqref{DHS1:Second_Decomposition_Gamma_Assumption} hold with
\begin{align}
D_1 := \EGLGSE + \rho \label{DHS1:Lower_Bound_B_4}
\end{align}
and $\rho\geq 0$. Since $\Vert \Psi\Vert_{\Hmag^1(Q_B)} \leq C$ by \eqref{DHS1:Second_Decomposition_Psi_xi_estimate}, \eqref{DHS1:Lower_Bound_B_5} implies
\begin{align*}
B^2 \bigl(\EGLGSE + \rho\bigr) \geq \FBCS(\Gamma) - \FBCS(\Gamma_0) \geq B^2 \bigl( \EGL(\Psi_\leq) - C\, B^{\nicefrac 1{12}}\bigr).
\end{align*}
This proves the claimed bound for the Cooper pair wave function of an approximate minimizer of the BCS functional in Theorem~\ref{DHS1:Main_Result}. 


We turn to the proof of Theorem~\ref{DHS1:Main_Result_Tc}. Let the temperature $T$ obey
\begin{align}
\Tc ( 1 - B (\Dc - D_0 B^{\nicefrac 1{12}})) < T \leq \Tc(1 + CB) \label{DHS1:TcB_Upper_Fine_1}
\end{align}
with $\Dc$ in \eqref{DHS1:Dc_Definition} and $D_0 > 0$. We claim that the normal state $\Gamma_0$ minimizes the BCS functional for such temperatures $T$ if $D_0$ is chosen sufficiently large. Since Corollary \ref{DHS1:TcB_First_Upper_Bound} takes care of the remaining temperature range, this implies part (b) of Theorem \ref{DHS1:Main_Result_Tc} and completes its proof.

To see that the above claim is true, we start with the lower bound in \eqref{DHS1:Lower_Bound_B_5} and assume that \eqref{DHS1:Second_Decomposition_Gamma_Assumption} holds with $D_1 = 0$. We drop the nonnegative quartic term in the Ginzburg--Landau functional for a lower bound and obtain
\begin{align*}
	\EGL(\Psi_\leq) &\geq B^{-2}  \langle \Psi_\leq , (\Lambda_0 \, \Pi^2 - DB \Lambda_2) \Psi_\leq\rangle \geq  \Lambda_2 \, ( \Dc - D) \; B^{-1}  \Vert \Psi_\leq \Vert_2^2, 
\end{align*}
with $\Lambda_0$ in \eqref{DHS1:GL-coefficient_1}, $\Lambda_2$ in \eqref{DHS1:GL-coefficient_2}, and with $D \in \mathbb{R}$ defined by $T = \Tc(1-D B)$. We combine \eqref{DHS1:Psi>_bound} and \eqref{DHS1:First_Decomposition_Result_Estimate} and estimate
\begin{align*}
\Vert \Psi_\leq \Vert_2 \geq \Vert \Psi\Vert_2 - \Vert \Psi_>\Vert_2 \geq c \; B^{\nicefrac 12} \; \Vert \Psi \Vert_{\Hmag^1(Q_B)} ( 1 - C \, B^{\nicefrac{5}{12}} ).
\label{DHS1:eq:A32}
\end{align*}
When we insert our findings in the lower bound for the BCS energy in \eqref{DHS1:Lower_Bound_B_5}, this gives
\begin{equation}
0\geq \FBCS(\Gamma) - \FBCS(\Gamma_0) \geq c \; B^2 \; \Vert \Psi \Vert_{\Hmag^1(Q_B)}^2 \bigl( (\Dc - D)  - CB^{\nicefrac {1}{12}}\bigr), \label{DHS1:eq:A33}
\end{equation}
We note that the lower bound in \eqref{DHS1:TcB_Upper_Fine_1} is equivalent to
\begin{align}
\Dc - D > D_0 B^{\nicefrac 1{12}}. \label{DHS1:TcB_Upper_Fine_2}
\end{align}
When we choose $D_0 > C$ with $C>0$ in \eqref{DHS1:eq:A33} and use \eqref{DHS1:TcB_Upper_Fine_2} to obtain a lower bound for the right side of \eqref{DHS1:eq:A33}, we conclude that $\Psi =0$. By \eqref{DHS1:Second_Decomposition_alpha_equation} and \eqref{DHS1:Second_Decomposition_Psi_xi_estimate}, this implies that $\alpha = 0$ whence $\Gamma$ is a diagonal state. Therefore, $\Gamma_0$ is the unique minimizer of $\FBCS$ if $T$ satisfies \eqref{DHS1:TcB_Upper_Fine_1} with our choice of $D_0$. As explained below \eqref{DHS1:TcB_Upper_Fine_1}, this proves Theorem \ref{DHS1:Main_Result_Tc}.


\subsection{Proof of Lemma~\ref{DHS1:Lower_Bound_B_remainings}}
\label{DHS1:sec:A1}

In this section we prove Lemma~\ref{DHS1:Lower_Bound_B_remainings}.
%
%
Our proof of part (a) uses a Cauchy integral representation for the operator $K_{\Tc,B}- (\pi^2 -\mu)$, which is provided in Lemma~\ref{DHS1:KT_integral_rep} below. Let us start by defining the contour for the Cauchy integral.

\begin{defn}[Speaker path]
\label{DHS1:speaker path}
Let $R>0$, assume that $\mu \leq 1$ and define the following complex paths 
\begin{align*}
\begin{split}
u_1(t) &:= \frac{\pi\i}{2\betac} + (1 + \i)t\\
u_2(t) &:= \frac{\pi\i}{2\betac} - (\mu + 1)t\\
u_3(t) &:= -\frac{\pi\i}{2\betac}t - (\mu + 1)\\
u_4(t) &:= -\frac{\pi\i}{2\betac} - (\mu + 1)(1-t) \\
u_5(t) &:= -\frac{\pi\i }{2\betac} + (1 - \i)t
\end{split}
&
\begin{split}
\phantom{ \frac \i\betac }t&\in [0,R], \\
\phantom{ \frac \i\betac }t &\in [0,1], \\
\phantom{ \frac \i\betac }t&\in [-1,1],\\
\phantom{ \frac \i\betac }t &\in [0,1],\\
\phantom{ \frac \i \betac }t&\in [0,R].
\end{split}
& \begin{split} 
\text{\includegraphics[width=6cm]{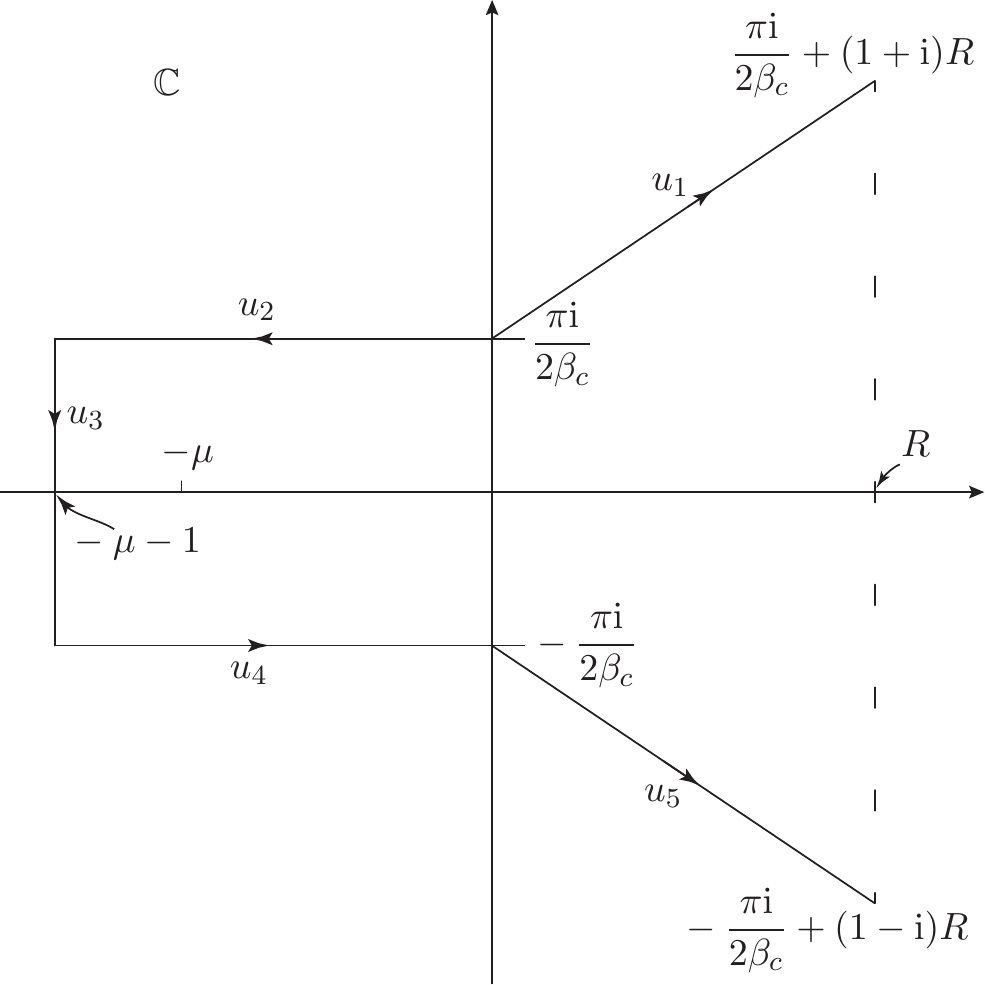}}
\end{split} 
\end{align*}
The speaker path is defined as the union of paths $u_i$, $i=1, \ldots, 5$, with $u_1$ taken in reverse direction, i.e.,
\begin{align*}
\speaker_R := \mathop{\dot -}u_1 \mathop{\dot +} u_2 \mathop{\dot +} u_3 \mathop{\dot +} u_4 \mathop{\dot +} u_5.
\end{align*}
If $\mu > 1$ we choose the same path as in the case $\mu = 1$.
\end{defn} 
This path has the property that certain norms of the resolvent kernel of $\pi^2$ are uniformly bounded for $z \in \speaker_R$ and $R > 0$. More precisely, Lemma~\ref{DHS1:gB-g_decay} implies
\begin{align}
\sup_{0\leq B\leq B_0} \sup_{R>0} \sup_{w\in \speaker_R} \bigl[ \left\Vert \, |\cdot |^a g_B^w\right\Vert_1 + \left \Vert \, |\cdot|^a \nabla g_B^w\right\Vert_1 \bigr] < \infty. \label{DHS1:g0_decay_along_speaker}
\end{align}
We could also choose a path parallel to the real axis in Lemma~\ref{DHS1:KT_integral_rep} below. In this case the above norms would depend on $R$. Although our analysis also works in this case, we decided to use the path $\speaker_R$ because of the more elegant bound in \eqref{DHS1:g0_decay_along_speaker}. 
With the above definition at hand, we are prepared to state the following lemma.

\begin{lem}
\label{DHS1:KT_integral_rep}
Let $H\colon \Dcal(H)\ra \Hcal$ be a self-adjoint operator on a separable Hilbert space~$\Hcal$ with $H\geq -\mu$ and let $\beta > 0$. Then, we have
\begin{align}
\frac{H}{\tanh(\frac{\beta}{2} H)} = H + \lim_{R\to\infty} \int_{\speaker_R} \frac{\dd z}{2\pi\i} \Bigl( \frac{z}{\tanh(\frac{\beta}{2} z)} - z \Bigr) \frac{1}{z - H}, \label{DHS1:KT_integral_rep_eq}
\end{align}
with the speaker path $\speaker_R$ in Definition~\ref{DHS1:speaker path}. The above integral including the limit is understood as an improper Riemann integral with respect to the uniform operator topology.  
\end{lem}

\begin{proof}
We have that the function $f(z) = \frac{z}{\tanh(\frac{\beta }{2}z)} - z = \frac{2z}{\e^{\beta z} - 1}$ is analytic in the open domain $\Cbb \setminus 2\pi T \i \Zbb_{\neq 0}$. The construction of the Riemann integral over the path $\speaker_R$ with respect to the uniform operator topology is standard. The fact that the limit $R \to \infty$ exists in the same topology follows from the exponential decay of the function $f(z)$ along the speaker path. To check the equality in \eqref{DHS1:KT_integral_rep_eq}, we evaluate both sides in the inner product with two vectors in $\ran \Idbb_{(-\infty, K]}(H)$ for $K > 0$, use the functional calculus, the Cauchy integral formula, and the fact that $\bigcup_{K>0} \ran \Idbb_{(-\infty, K]}(H)$ is a dense subset of $\Hcal$. This proves the claim.
\end{proof}

Henceforth, we use the symbol $\int_{\speaker}$ to denote the integral on the right side of \eqref{DHS1:KT_integral_rep_eq} including the limit and we denote $\speaker = \bigcup_{R > 0} \speaker_R$.

\begin{proof}[Proof of Lemma \ref{DHS1:Lower_Bound_B_remainings}]
We apply Cauchy-Schwarz to estimate
\begin{align}
|\langle \eta_\perp, (K_{\Tc,B}^r - K_{\Tc}^r)\sigma_0\rangle| \leq \Vert \eta_\perp\Vert_2 \, \Vert (K_{\Tc,B}^r- K_{\Tc}^r)\sigma_0\Vert_2
\label{DHS1:eq:A34}
\end{align}
and claim that 
\begin{align}
\Vert [K_{{\Tc},B}^r - K_{\Tc}^r]\sigma_0\Vert_2 \leq C \varepsilon^{-\nicefrac 12} B^2 \, \Vert \Psi\Vert_{\Hmag^1(Q_B)} \label{DHS1:Lower_Bound_B_remainings_3}
\end{align}
holds. To see this, we apply Lemma~\ref{DHS1:KT_integral_rep} and write
\begin{align}
K_{{\Tc},B}^r - K_{\Tc}^r = \pi_r^2 - p_r^2 + \int_\speaker \frac{\dd w}{2\pi\i} \; f(w) \;  \frac{1}{w + \mu - \pi_r^2} [\pi_r^2 - p_r^2] \frac{1}{w + \mu - p_r^2}, \label{DHS1:Lower-bound-final_2}
\end{align}
where $\pi_r^2 - p_r^2 = \i \, \Bbold\wedge r\;  p_r + \frac 14 |\Bbold\wedge r|^2$. Using \eqref{DHS1:Psi>_bound} and \eqref{DHS1:Decay_of_alphastar}, we estimate the first term on the right side of \eqref{DHS1:Lower-bound-final_2} by
\begin{align}
\Vert [\pi_r^2 - p_r^2]\sigma_0\Vert_2 &\leq B \, \Vert \, |\cdot|\nabla \alpha_*\Vert_2 \Vert \Psi_>\Vert_2 + B^2 \Vert \,|\cdot|^2\alpha_*\Vert_2 \Vert \Psi_>\Vert_2 \notag\\
&\leq C\varepsilon^{-\nicefrac 12} B^2 \, \Vert \Psi\Vert_{\Hmag^1(Q_B)}. \label{DHS1:Lower_Bound_B_remainings_1}
\end{align}
To estimate the second term in \eqref{DHS1:Lower-bound-final_2}, we use Hölder's inequality in \eqref{DHS1:Schatten-Hoelder} and find
\begin{align*}
\Bigl\Vert \int_\speaker \frac{\dd w}{2\pi\i} \, f(w)\;  \frac{1}{w + \mu - \pi_r^2} [\pi_r^2 - p_r^2]\frac{1}{w+\mu - p_r^2}\sigma_0\Bigr\Vert_2 &\\
&\hspace{-180pt}\leq \int_\speaker \frac{\dd |w|}{2\pi}\,  |f(w)| \, \Bigl\Vert \frac{1}{w + \mu- \pi_r^2} \Bigr\Vert_\infty \Bigl\Vert [\pi_r^2 - p_r^2]\frac{1}{w + \mu - p_r^2}\sigma_0\Bigr\Vert_2,
\end{align*}
where $\dd |w| = \dd t \; |w'(t)|$. Eq.~\eqref{DHS1:g0_decay_along_speaker} implies that the operator norm of the magnetic resolvent is uniformly bounded for $w \in \speaker$. Since the function $f$ is exponentially decaying along the speaker path it suffices to prove a bound on the last factor that is uniform for $w \in \speaker$. We have
\begin{align*}
[\pi_r^2 - p_r^2] \frac{1}{w + \mu - p_r^2} \sigma_0(X,r) &= \int_{\Rbb^3} \dd s \; [\pi_r^2 - p_r^2] g_0^w(r - s) \alpha_*(s) \Psi_>(X),
\end{align*}
which implies
\begin{align}
\Bigl\Vert [\pi_r^2 - p_r^2] \frac{1}{w + \mu - p_r^2}\sigma_0\Bigr\Vert_2^2 &
%
\leq \Vert \Psi_>\Vert_2^2 \int_{\Rbb^3} \dd r \, \Bigl| \int_{\Rbb^3} \dd s \; |[\pi_r^2 - p_r^2] g_0^w(r - s) \alpha_*(s)|\Bigr|^2. \label{DHS1:Lower_Bound_B_remainings_2}
\end{align}
Moreover,
\begin{align*}
\int_{\Rbb^3} \dd r \; \Bigl| \int_{\Rbb^3} \dd s \; |[\pi_r^2 - p_r^2] g_0^w(r - s) \alpha_*(s)|\Bigr|^2 & \\
&\hspace{-195pt} \leq CB^2 \bigl( \Vert \, |\cdot| \nabla g_0^w\Vert_1^2 \; \Vert \alpha_*\Vert_2^2 + \Vert \nabla g_0^w\Vert_1^2 \; \Vert \, |\cdot|\alpha_*\Vert_2^2 + \Vert \,  |\cdot|^2 g_0^w\Vert_1^2 \;  \Vert\alpha_*\Vert_2^2 + \Vert g_0^w\Vert_1^2 \;  \Vert \, |\cdot|^2\alpha_*\Vert_2^2\bigr).
\end{align*}
The right side is uniformly bounded for $w \in \speaker$ by \eqref{DHS1:g0_decay_along_speaker} and \eqref{DHS1:Decay_of_alphastar}. In combination with \eqref{DHS1:Psi>_bound} and \eqref{DHS1:Lower_Bound_B_remainings_2}, this implies
\begin{align*}
\Bigl\Vert [\pi_r^2 -p_r^2]\frac{1}{w + \mu - p_r^2} \sigma_0\Bigr\Vert_2^2 & \leq CB^2 \;  \Vert\Psi_>\Vert_2^2\leq C \varepsilon^{-1} B^4 \, \Vert \Psi\Vert_{\Hmag^1(Q_B)}^2.
\end{align*}
Using this and \eqref{DHS1:Lower_Bound_B_remainings_1}, we read off \eqref{DHS1:Lower_Bound_B_remainings_3}. Finally, we apply Proposition~\ref{DHS1:Structure_of_alphaDelta} to estimate $\Vert \eta_\perp\Vert_2$ in \eqref{DHS1:eq:A34}, which proves part (a).


To prove part (b), we start by noting that
\begin{align*}
|\langle \eta_\perp, (U-1) K_{\Tc,B}^r\sigma_0 \rangle| &\leq \Vert \,  |r| \eta_\perp\Vert_2 \;  \Vert\, |r|^{-1} (U - 1) K_{\Tc,B}^r \, \sigma_0 \Vert_2. 
\end{align*}
A bound for the left factor on the right side is provided by Proposition~\ref{DHS1:Structure_of_alphaDelta}. To estimate the right factor, we use \eqref{DHS1:Psi>_bound}, \eqref{DHS1:Decay_of_alphastar} and the operator inequality in \eqref{DHS1:ZPiX_inequality}, which implies $|U- 1|^2 \leq 3 r^2 \Pi_X^2$, and find
\begin{align*}
\Vert\, |r|^{-1} (U - 1) K_{\Tc,B}^r \, \sigma_0 \Vert_2 &\leq C \Vert K_{\Tc,B}^r \alpha_*\Vert_2 \; \Vert \Pi\Psi_>\Vert_2 \leq CB \, \Vert \Psi\Vert_{\Hmag^1(Q_B)}.
\end{align*}
This proves part (b).

For part (c), we estimate
\begin{align}
|\langle \eta_\perp, UK_{\Tc,B}^r (U^* - 1) \sigma_0\rangle|  &\leq \bigl\Vert \sqrt{K_{\Tc,B}^r} \, U^* \eta_\perp \bigr\Vert_2 \;  \bigl\Vert \sqrt{K_{\Tc,B}^r}\, (U^* - 1)\sigma_0\bigr\Vert_2
\label{DHS1:eq:A35}
\end{align}
and note that $K_{\Tc,B}^r \leq C(1+ \pi_r^2)$ implies
\begin{align}
\bigl\Vert \sqrt{K_{\Tc,B}^r} (U^* - 1)\sigma_0\bigr\Vert_2^2 &= \langle \sigma_0, (U - 1) K_{\Tc,B}^r (U^* - 1)\sigma_0\rangle \notag\\
&\leq C \Vert (U^* - 1)\sigma_0\Vert_2^2 + C\Vert \pi_r (U^* - 1)\sigma_0\Vert_2^2.\label{DHS1:Lower-bound-final_3}
\end{align}
Using the bound for $|U-1|^2$ in part (b), \eqref{DHS1:Psi>_bound} and \eqref{DHS1:Decay_of_alphastar}, we see that the first term is bounded by $C\Vert |r|\alpha_* \Pi_X\Psi_>\Vert^2 \leq CB^2$. Lemma~\ref{DHS1:CommutationI} allows us to write
\begin{align*}
\pi_r (U^* - 1) &= (U^* - 1) \tilde \pi_r + \frac 12 U^*\Pi_X - \frac 14 \Bbold \wedge r.
\end{align*}
Accordingly, we have
\begin{align*}
\Vert \pi_r (U^* - 1)\sigma_0\Vert_2^2 &\\
&\hspace{-50pt}\leq C\bigl( \Vert \, |r| p_r\alpha_*\Pi_X\Psi_>\Vert_2^2 + B \, \Vert \, |r|^2\alpha_* \Pi_X \Psi_> \Vert_2^2 + \Vert \alpha_*\Pi_X\Psi_>\Vert_2^2 + B \Vert \, |r|\alpha_* \Psi_>\Vert_2^2\bigr)\\
&\hspace{-50pt}\leq C \bigl( B^2 + \varepsilon^{-1} B^3 \bigr) \leq CB^2 \, \Vert \Psi\Vert_{\Hmag^1(Q_B)}^2.
\end{align*}
We conclude that the right side of \eqref{DHS1:Lower-bound-final_3} is bounded by $CB^2 \Vert \Psi\Vert_{\Hmag^1(Q_B)}^2$. 

With $K_T(p) \leq C (1+p^2)$ we see that the first factor on the right side of \eqref{DHS1:eq:A35} is bounded by
\begin{align*}
\bigl\Vert \sqrt{K_{\Tc,B}^r} \, U^* \eta_\perp\bigr\Vert_2^2 &= \langle \eta_\perp , U K_{\Tc,B}^r U^* \eta_\perp\rangle \leq C \Vert \eta_\perp\Vert_2^2 + C\Vert \pi_rU^*\eta_\perp\Vert_2^2.
\end{align*}
From Lemma~\ref{DHS1:CommutationI} we know that $\pi_r U^* = U^* [\tilde\pi_r + \frac 12 \Pi_X]$, and hence
\begin{align*}
\bigl\Vert \sqrt{K_{\Tc,B}^r} \, U^* \eta_\perp\bigr\Vert_2^2 &\leq C \bigl( \Vert \eta_\perp\Vert_2^2 + \Vert \tilde \pi_r\eta_\perp\Vert_2^2 + \Vert \Pi_X\eta_\perp\Vert_2^2\bigr) \leq C \varepsilon B^2 \, \Vert \Psi\Vert_{\Hmag^1(Q_B)}^2. 
\end{align*}
This proves part (c) and ends the proof of the Lemma~\ref{DHS1:Lower_Bound_B_remainings}.
\end{proof}

\begin{center}
\huge \textsc{--- Appendix ---}
\end{center}


\section{Estimates on Eigenvalues and Eigenfunctions of \texorpdfstring{$K_{\Tc,B}-V$}{KTcB-V}}
\label{DHS1:KTV_Asymptotics_of_EV_and_EF_Section}

In this section, we investigate the low lying eigenvalues of $K_{\Tc,B} - V$ and its ground state wave function. Our analysis is carried out at $T = \Tc$ and we omit $\Tc$ from the notation throughout the appendix. The goal is to prove the following result.

\begin{prop}
\label{DHS1:KTV_Asymptotics_of_EV_and_EF}
Let Assumptions \ref{DHS1:Assumption_V} and \ref{DHS1:Assumption_KTc} hold. There is a constant $B_0>0$ such that for any $0 \leq B\leq B_0$ the following holds. Let $e_0^B$ and $e_1^B$ denote the lowest and next to lowest eigenvalue of $K_{\Tc,B} - V$. Then:
\begin{enumerate}[(a)]
\item $|e_0^B| \leq CB$,


\item $K_{\Tc,B}- V$ has a uniform spectral gap above $e_0^B$, i.e., $e_1^B - e_0^B \geq \kappa > 0$.

\item Let $\alpha_*$ be the eigenfunction in \eqref{DHS1:alpha_star_ev-equation} and let $\alpha_*^B$ be an eigenfunction corresponding to $e_0^B$ such that $\langle \alpha_*^B , V\alpha_*\rangle$ is real and nonnegative for all $0\leq B \leq B_0$. Then,
\begin{align*}
\Vert \alpha_*^B - \alpha_*\Vert_2 + \Vert \pi^2(\alpha_*^B - \alpha_*)\Vert_2 \leq CB.
\end{align*}

\item With $P_B := |\alpha_*^B\rangle \langle \alpha_*^B|$ and $P := |\alpha_*\rangle \langle \alpha_*|$ and with $\alpha_*^B$ and $\alpha_*$ as in part (c), we have 
\begin{align*}
\Vert P_B- P\Vert_\infty  + \Vert \pi^2(P_B - P)\Vert_\infty \leq CB.
\end{align*}
\end{enumerate}
\end{prop}

\begin{bem}
	We emphasize that this appendix is the only place in the paper where the assumption $V\geq 0$ is used. It simplifies our analysis because it implies that the Birman--Schwinger operator $V^{\nicefrac 12} [K_{\Tc,B} - e]^{-1} V^{\nicefrac 12}$ is self-adjoint. However, for the statement of Proposition \ref{DHS1:KTV_Asymptotics_of_EV_and_EF} to be true, it is not necessary that $V$ has a sign. In fact, with the help of a Combes--Thomas estimate for the resolvent kernel of $K_{\Tc} - V$ it is possible to show that Proposition~\ref{DHS1:KTV_Asymptotics_of_EV_and_EF} also holds for potentials $V$ without a definite sign. This approach requires more effort, and we therefore refrain from giving a general proof here. It can be found in Chapter \ref{Chapter:Combes-Thomas}.
\end{bem}

Let us recall the decay properties of the eigenfunction $\alpha_*$ corresponding to the lowest eigenvalue of the operator $K_{\Tc}-V$. Since $\alpha_* = K_{\Tc}^{-1} V\alpha_*$ and $V\in L^\infty(\Rbb^3)$, we immediately have $\alpha_*\in H^2(\Rbb^3)$. Furthermore, for any $\nu \in \Nbb_0^3$, by \cite[Proposition 2]{Hainzl2012},
\begin{align}
\int_{\Rbb^3} \dd x \; \bigl[ |x^\nu \alpha_*(x)|^2 + |x^\nu \nabla \alpha_*(x)|^2 \bigr] < \infty. \label{DHS1:Decay_of_alphastar}
\end{align}
In fact, more regularity of $\alpha_*$ is known, see \cite[Appendix A]{Hainzl2012} but \eqref{DHS1:Decay_of_alphastar} is all we use in this paper. Before we give the proof of Proposition~\ref{DHS1:KTV_Asymptotics_of_EV_and_EF} in Section~\ref{DHS1:sec:new2} below, we prove two preparatory statements.

Let $e\in (-\infty, 2\Tc)$ and denote the kernel of the resolvent $(e - K_{\Tc})^{-1}$ by $\Gcal^e(x-y)$.

\begin{lem}
\label{DHS1:Gcal estimates}
For $e \in (-\infty, 2\Tc)$ and $k\in \Nbb_0$ the functions $|\cdot|^k\Gcal^e$ and $|\cdot|^k \nabla \Gcal^e$ belong to $L^1(\Rbb^3)$.
\end{lem}

\begin{proof}
The function $(e - K_{\Tc}(p))^{-1}$ and its derivatives belong to $L^2(\Rbb^3)$.
Therefore, we have
\begin{align}
\Vert \, |\cdot|^k \Gcal^e \Vert_1 &\leq \Bigl( \int_{\Rbb^3} \dx \; \bigl| \frac{|x|^k}{1 + |x|^{k+2}}\bigr|^2 \Bigr)^{\nicefrac 12} \; \Vert (1 + |\cdot |^{k+2})\Gcal^e\Vert_2 < \infty. \label{DHS1:Gcal_estimates_5}
\end{align}
This proves the first claim and the second follows from a similar argument.
\end{proof}

\subsection[Phase approximation]{Phase approximation for $K_{\Tc,B}$}

\begin{prop}
\label{DHS1:Resolvent_perturbation}
Let $V$ and $|\cdot|^2V$ belong to $L^\infty(\Rbb^3)$. There is $B_0>0$ such that for $0 \leq B\leq B_0$ and $e\in (-\infty, 2\Tc)$, we have
\begin{align}
\Bigl\Vert \Bigl[ \frac{1}{e - K_{\Tc,B}} - \frac{1}{e - K_{\Tc}}\Bigr] V^{\nicefrac 12} \Bigr\Vert_\infty + \Bigl\Vert \pi^2 \Bigl[ \frac{1}{e - K_{\Tc,B}} - \frac{1}{e - K_{\Tc}}\Bigr] V \Bigr\Vert_\infty \leq C_e B. \label{DHS1:Resolvent_perturbation_2}
\end{align}
\end{prop}

\begin{proof}
To prove this result, we apply a phase approximation to the operator $K_{\Tc,B}$. We pursue the strategy that we used in the proof of Lemma \ref{DHS1:gB-g_decay} and define
\begin{align}
\Scal_{B}^{e}(x,y) := \e^{\i \frac{\Bbold}{2}\cdot (x\wedge y)} \; \Gcal^{e}(x-y). \label{DHS1:Scalz_definition}
\end{align}
Let $\Scal_{B}^e$ be the operator defined by the kernel $\Scal_{B}^e(x,y)$. We claim that
\begin{align}
(e - K_{\Tc,B} )\Scal_{B}^e = \Idbb - \Tcal_{B}^{e} \label{DHS1:Interplay SAzTAz}
\end{align}
with the operator $\Tcal_{B}^e$ defined by the kernel
\begin{align}
\Tcal_{B}^e (x,y) := \e^{\i \frac \Bbold 2 \cdot (x\wedge y)} \bigl[ (K_{\Tc}(\pi_{x,y}) - K_{\Tc}) \frac{1}{e - K_{\Tc}}\bigr] (x,y) \label{DHS1:Resolvent_KT_phase_approximation}
\end{align}
and $\pi_{x,y} = -\i\nabla_x + \Abold (x-y)$. To prove \eqref{DHS1:Interplay SAzTAz}, it is sufficient to note that
\begin{align*}
K_{\Tc,B} \; \e^{\i \frac \Bbold 2 \cdot (x\wedge y)} = \e^{\i \frac \Bbold 2 \cdot (x\wedge y)} \; K_{\Tc}(\pi_{x,y}). 
\end{align*}
Using Lemma~\ref{DHS1:KT_integral_rep} and Lemma~\ref{DHS1:Gcal estimates}, a straightforward computation shows that
\begin{align}
\Vert \Tcal_B^e\Vert_\infty \leq C_e\, B 
\label{DHS1:TcalTB_bound}
\end{align}
holds for $B$ small enough.

With the operator $\Scal_B^e$ in \eqref{DHS1:Scalz_definition} we write
\begin{align}
\frac{1}{e - K_{\Tc,B}} - \frac{1}{e - K_{\Tc}} = \frac{1}{e - K_{\Tc,B}} - \Scal_B^e + \Scal_B^e - \frac{1}{e - K_{\Tc}} \label{DHS1:Resolvent_perturbation_1}
\end{align}
For the first term \eqref{DHS1:Interplay SAzTAz} implies
\begin{align*}
\frac{1}{e - K_{\Tc,B}} - \Scal_B^e = \Scal_B^e \;  \sum_{n=1}^\infty (\Tcal_B^e)^n
\end{align*}
and since $\Scal_B^e$ is a bounded operator with norm bounded by $\Vert \Gcal^e\Vert_1$, \eqref{DHS1:TcalTB_bound} yields
\begin{align*}
\Bigl\Vert \frac{1}{e - K_{\Tc,B}} - \Scal_B^e\Bigr\Vert_\infty \leq C_e   B.
\end{align*}
To estimate the second term on the right side of \eqref{DHS1:Resolvent_perturbation_1}, we use $|\e^{\i \frac \Bbold 2 \cdot (x\wedge y)} - 1|\leq B |x - y| |y|$ and bound the kernel of this term by
\begin{align*}
\bigl| \bigl[ \Scal_B^e - \frac{1}{e - K_{\Tc}}\bigr] V^{\nicefrac 12}(x,y)\bigr| &\leq B \;  |x-y| |\Gcal^e(x-y)| \; |y||V^{\nicefrac 12} (y)|.
\end{align*}
We further bound $|y||V^{\nicefrac 12}(y)| \leq \Vert \,|\cdot|^2V\Vert_\infty^{\nicefrac 12}$, which shows that
\begin{align*}
\Bigl\Vert \Bigl[\Scal_B^e - \frac{1}{e - K_{\Tc}}\Bigr] V^{\nicefrac 12}\Bigr\Vert_\infty &\leq B\; \Vert \, |\cdot|^2V\Vert_\infty^{\nicefrac 12} \; \Vert \, |\cdot|\Gcal^e\Vert_1 \leq C_e  B.
\end{align*}
This completes the proof of the first estimate in \eqref{DHS1:Resolvent_perturbation_2}. 

To prove the second estimate, we note that
\begin{align*}
\pi^2 \Bigl[ \frac{1}{e - K_{\Tc,B}} - \frac{1}{e - K_{\Tc}} \Bigr] V = \pi^2 \frac{1}{e - K_{\Tc,B}} [ K_{\Tc,B} - K_{\Tc}] \frac{1}{e - K_{\Tc}} V^{\nicefrac 12}.
\end{align*}
Since $\pi^2(e - K_{\Tc,B})^{-1}$ is a bounded function of $\pi^2$, we know that the operator norm of the operator in the above equation is uniformly bounded in $B$. Thus, it suffices to show that $[K_{\Tc,B} - K_{\Tc}] \frac{1}{e - K_{\Tc}} V$ satisfies the claimed operator norm bound. To this end, we use \eqref{DHS1:Lower-bound-final_2} and obtain two terms. Since $\pi^2 - p^2 = \Bbold \wedge x \cdot p + \frac 14|\Bbold \wedge x|^2$, the estimate for the first term reads
\begin{align*}
\bigl[ (\pi^2- p^2) \frac{1}{e - K_{\Tc}} V^{\nicefrac 12}\bigr](x,y) \leq \bigl[ B \cdot |x| |\nabla \Gcal^e(x-y)| + B^2 \; |x|^2 |\Gcal^e(x-y)| \bigr] |V(y)|.
\end{align*}
The $L^1(\Rbb^3)$-norm in $x-y$ of the right side is bounded by
\begin{align*}
C B\, \bigl[\bigl( \Vert \, |\cdot| \nabla \Gcal^e\Vert_1  + \Vert \, |\cdot|^2\Gcal^e\Vert_1 \bigr) \Vert V\Vert_\infty + \Vert \nabla  \Gcal^e\Vert_1 \, \Vert \, |\cdot| V\Vert_\infty + \Vert \Gcal^e\Vert_1 \, \Vert \, |\cdot|^2 V\Vert_\infty\bigr],
\end{align*}
which is finite by Lemma \ref{DHS1:Gcal estimates}. With the help of \eqref{DHS1:g0_decay_along_speaker} the remaining term can bounded similarly. This proves the claim.
\end{proof}


\subsection[Asymptotics]{Asymptotics for eigenvalues and eigenfunctions}
\label{DHS1:sec:new2}

We are now prepared to give the proof of Proposition~\ref{DHS1:KTV_Asymptotics_of_EV_and_EF}.

\begin{proof}[Proof of Proposition \ref{DHS1:KTV_Asymptotics_of_EV_and_EF}]
We start with the upper bound of part (a). By the variational principle for $e_0^B$ we have
\begin{align}
e_0^B \leq \langle \alpha_*, (K_{\Tc,B} - V)\alpha_*\rangle = \langle \alpha_* , (K_{\Tc} - V)\alpha_*\rangle + \langle \alpha_* , (K_{\Tc,B} - K_{\Tc})\alpha_*\rangle, \label{DHS1:KTV_Asymptotics_1}
\end{align}
where the first term on the right side equals $0$ by the definition of $\alpha_*$. We use \eqref{DHS1:g0_decay_along_speaker}, Lemma \ref{DHS1:KT_integral_rep}, and \eqref{DHS1:Decay_of_alphastar}, and argue as in the proof of \eqref{DHS1:Lower_Bound_B_remainings_1} to see that the second term is bounded by $CB$.

In the next step, we show the lower bound of part (a) and part (b) at the same time. Thus, for $n = 0,1$ we aim to show
\begin{align}
e_n^B \geq e_n^0 - C_nB \label{DHS1:KTV_Asymptotics_10}
\end{align}
for the lowest and next-to-lowest eigenvalue $e_0^B$ and $e_1^B$, respectively. For notational convenience, we give the proof for general $n\in \Nbb_0$ and we order the eigenvalues $e_n^B$ increasingly. Let $\alpha_n^B$ be the eigenfunction to $e_n^B$ for $n\geq 0$ and note that $\alpha_*^B = \alpha_0^B$. 

Now, we switch to the Birman--Schwinger picture: $e_n^B$ being the $(n+1)$-st to smallest eigenvalue of $K_{\Tc,B} - V$ is equivalent to 1 being the $(n+1)$-st to largest eigenvalue of the Birman--Schwinger operator $V^{\nicefrac 12} (K_{\Tc,B} - e_n^B)^{-1} V^{\nicefrac 12}$ corresponding to $e_n^B$. Accordingly, the min-max principle, see, e.g., \cite[Theorem 12.1 (5)]{LiebLoss}, implies
\begin{align}
1 = \max_{\substack{u_0, \ldots, u_n\in L^2(\Rbb^3) \\ u_i \perp u_j, \; i\neq j}} \min\Bigl\{ \Bigl\langle \Phi , V^{\nicefrac 12} \frac{1}{K_{\Tc,B} - e_n^B} V^{\nicefrac 12} \Phi\Bigr\rangle : \Phi \in \mathrm{span} \{u_0, \ldots, u_n\}, \; \Vert \Phi\Vert_2 =1\Bigr\}. \label{DHS1:KTV_Asymptotics_5}
\end{align}
We obtain a lower bound on \eqref{DHS1:KTV_Asymptotics_5} by choosing the functions $u_i$, $i=0, \ldots, n$ as the first $n+1$ orthonormal eigenfunctions $\varphi_i^B$ satisfying
\begin{align}
	V^{\nicefrac 12} \frac{1}{K_{\Tc,B} - e_n^B} V^{\nicefrac 12} \varphi_i^B = \eta_i^B \, \varphi_i^B, \qquad \qquad i = 0, \ldots, n, \label{DHS1:KTV_Asymptotics_8}
\end{align}
where $\eta_i^B \geq 1$ denote the first $n$ eigenvalues of the Birman--Schwinger operator in \eqref{DHS1:KTV_Asymptotics_8} ordered decreasingly. In particular, we have $\eta_n^B = 1$, as well as the relations $\varphi_n^B = V^{\nicefrac 12} \alpha_n^B$ and $\alpha_n^B = (K_{\Tc,B} - e_n^B)^{-1} V^{\nicefrac 12} \varphi_n^B$.


Denote $e_n := e_n^0$ and apply the resolvent equations to write
\begin{align}
V^{\nicefrac 12}\frac{1}{K_{\Tc,B} - e_n^B}V^{\nicefrac 12} &= V^{\nicefrac 12}\frac{1}{K_{\Tc} - e_n}V^{\nicefrac 12} + (e_n^B - e_n) \, \Qcal_n^B + \Rcal_n^B \label{DHS1:KTV_Asymptotics_4}
\end{align}
with
\begin{align*}
\Qcal_n^B &:= V^{\nicefrac 12}\frac{1}{K_{\Tc,B} - e_n} \,  \frac{1}{K_{\Tc,B} - e_n^B}V^{\nicefrac 12}, &
\Rcal_n^B &:= V^{\nicefrac 12}\Bigl[\frac{1}{K_{\Tc,B} - e_n} - \frac{1}{K_{\Tc}-e_n}\Bigr]V^{\nicefrac 12}.
\end{align*}
By Proposition \ref{DHS1:Resolvent_perturbation}, we have $\Vert \Rcal_n^B\Vert_\infty \leq C_n B$. Furthermore, we may assume without loss of generality that $e_n^B \leq e_n$, because otherwise there is nothing to prove. We combine this with \eqref{DHS1:KTV_Asymptotics_5} for $B =0$ and \eqref{DHS1:KTV_Asymptotics_4}, which yields
\begin{align}
1 
&\geq \min\Bigl\{ \Bigl\langle \Phi , V^{\nicefrac 12} \frac{1}{K_{\Tc,B} - e_n^B} V^{\nicefrac 12} \Phi\Bigr\rangle : \Phi \in \mathrm{span} \{\varphi_0^B, \ldots, \varphi_n^B\}, \; \Vert \Phi\Vert_2 =1\Bigr\} \notag\\
&\hspace{30pt}- [e_n^B - e_n] \min \bigl\{ \langle \Phi, \Qcal_n^B\Phi\rangle : \Phi \in \mathrm{span} \{\varphi_0^B, \ldots,  \varphi_n^B\}, \; \Vert \Phi\Vert_2 =1\bigr\} - C_nB.
\label{DHS1:KTV_Asymptotics_2}
\end{align}
We observe that the first term on the right side equals 1. To be able to conclude, we therefore need to show that there is a constant $c>0$, independent of $B$, such that
\begin{align}
\min \bigl\{ \langle \Phi, \Qcal_n^B\Phi\rangle : \Phi \in \mathrm{span} \{\varphi_0^B, \ldots, \varphi_n^B\}\bigr\} \geq c. \label{DHS1:KTV_Asymptotics_3}
\end{align}
Then, \eqref{DHS1:KTV_Asymptotics_2} implies $-[e_n^B-e_n] \leq C_nB$, which proves \eqref{DHS1:KTV_Asymptotics_10}.

We will prove that \eqref{DHS1:KTV_Asymptotics_3} holds with $c = \Vert V\Vert_\infty^{-1}$. To that end, we write
\begin{align*}
\langle \Phi, \Qcal_n^B \Phi\rangle &= \Bigl\langle \Phi, V^{\nicefrac 12} \frac{1}{K_{\Tc,B}- e_n^B} (K_{\Tc,B} - e_n^B) \frac{1}{K_{\Tc,B}- e_n} \frac{1}{K_{\Tc,B} - e_n^B} V^{\nicefrac 12} \Phi \Bigr\rangle,
\end{align*}
apply $-e_n^B \geq -e_n$, and infer
\begin{align*}
\langle \Phi , \Qcal_n^B\Phi\rangle &\geq \Bigl\Vert \frac{1}{K_{\Tc,B} - e_n^B} V^{\nicefrac 12} \Phi\Bigr\Vert_2^2. 
\end{align*}
Since $\Phi\in \mathrm{span}\{\varphi_0^B, \ldots, \varphi_n^B\}$, there are coefficients $c_i^B\in \Cbb$, $i = 0, \ldots, n$ such that we have $\Phi = \sum_{i=0}^n c_i^B \varphi_i^B$. We use the eigenvalue equation in \eqref{DHS1:KTV_Asymptotics_8} for $\varphi_i^B$ as well as $\langle \varphi_i^B, \varphi_j^B \rangle = \delta_{i,j}$ to see that
\begin{align*}
\Vert V\Vert_\infty \Bigl\Vert \frac{1}{K_{\Tc,B} - e_n^B} V^{\nicefrac 12} \Phi\Bigr\Vert_2^2 &\geq \Bigl\langle \frac{1}{K_{\Tc,B} - e_n^B} V^{\nicefrac 12} \Phi, V \frac{1}{K_{\Tc,B} - e_n^B} V^{\nicefrac 12} \Phi \Bigr\rangle \\
&= \sum_{i,j=0}^n \ov{c_i^B} c_j^B \, \eta_i^B\eta_j^B\, \langle \varphi_i^B, \varphi_j^B\rangle = \sum_{i=0}^n |c_i^B|^2 \; (\eta_i^B)^2 \geq 1.
\end{align*}
Here, we used that $\eta_i \geq 1$ and $\Vert \Phi\Vert_2 =1$.  This shows \eqref{DHS1:KTV_Asymptotics_3} and completes the proof of \eqref{DHS1:KTV_Asymptotics_10}.

The case $n =0$ in \eqref{DHS1:KTV_Asymptotics_10} yields the lower bound of part (a), since $e_0 = 0$. For $n = 1$, we have $e_1 = \kappa$, showing part (b) with the help of part (a).

To prove part (c), we write $\varphi_0 := \varphi_0^0$. The chosen phase of $\varphi_0^B$ in the lemma is such that $\langle \varphi_0, \varphi_0^B\rangle$ is real and nonnegative. We write $\varphi_0^B = a_B \varphi_0 + b_B \Phi$ with $\langle \Phi, \varphi_0\rangle =0$, $\Vert \Phi\Vert_2 = 1$ and $|a_B|^2 + |b_B|^2 = 1$. By construction, $a_B = \langle \varphi_0, \varphi_0^B\rangle$. Furthermore, by \eqref{DHS1:KTV_Asymptotics_4},
\begin{align}
1 = \bigl\langle \varphi_0^B , V^{\nicefrac 12} \frac{1}{K_{\Tc,B} - e_0^B} V^{\nicefrac 12} \varphi_0^B\bigr\rangle = \bigl\langle \varphi_0^B , V^{\nicefrac 12} \frac{1}{K_{\Tc}}  V^{\nicefrac 12} \varphi_0^B\bigr\rangle + \langle \varphi_0^B,\Tcal_B\varphi_0^B\rangle \label{DHS1:KTV_Asymptotics_9}
\end{align}
with $\Tcal_B := e_0^B\, Q_0^B + \Rcal_0^B$. Thus, 
\begin{align*}
1 \leq a_B^2 + |b_B|^2 \bigl\langle \Phi, V^{\nicefrac 12} \frac{1}{K_{\Tc}} V^{\nicefrac 12} \Phi\bigr\rangle + 2a_B \Re \bigl[ b_B \bigl\langle \varphi_0, V^{\nicefrac 12} \frac{1}{K_{\Tc}} V^{\nicefrac 12} \Phi\bigr\rangle \bigr] + \Vert \Tcal_B\Vert_\infty.
\end{align*}
By parts (a) and (b) and Proposition \ref{DHS1:Resolvent_perturbation}, we know that $\Vert \Tcal_B\Vert_\infty \leq CB$. Furthermore, the term in square brackets vanishes, since $V^{\nicefrac 12} K_{\Tc}^{-1} V^{\nicefrac 12} \varphi_0 = \varphi_0$ and $\langle \varphi_0, \Phi\rangle =0$. Using the orthogonality of $\Phi$ and $\varphi_0$ once more as well as the fact that 1 is the largest eigenvalue of $V^{\nicefrac 12} K_{\Tc}^{-1} V^{\nicefrac 12}$, we see that there is an $\eta < 1$ such that $\langle \Phi, V^{\nicefrac 12} K_{\Tc}^{-1} V^{\nicefrac 12} \Phi\rangle \leq \eta$. It follows that
\begin{align*}
1 \leq a_B^2 + |b_B|^2 \eta + CB.
\end{align*}
Since $a_B^2 + |b_B|^2 = 1$, this implies $|b_B|^2 \leq CB$ as well as $a_B^2 \geq 1 - CB$. Since $a_B \geq 0$, we infer $1 - a_B \leq CB$. 

The next step is to improve the estimate on $b_B$ to $|b_B|\leq CB$. To this end, we combine the two eigenvalue equations of $\varphi_0^B$ and $\varphi_0$. With $\Tcal_B$ as in \eqref{DHS1:KTV_Asymptotics_9}, we find
\begin{align*}
\varphi_0^B - \varphi_0 &= V^{\nicefrac 12} \frac{1}{K_{\Tc}} V^{\nicefrac 12} (\varphi_0^B - \varphi_0) + \Tcal_B \varphi_0^B.
\end{align*}
Testing this against $\Phi$, we obtain
\begin{align*}
b_B = \langle \Phi, \varphi_0^B - \varphi_0 \rangle = \bigl\langle V^{\nicefrac 12} \frac{1}{K_{\Tc}} V^{\nicefrac 12} \Phi, \varphi_0^B - \varphi_0\bigr\rangle + \langle \Phi, \Tcal_B \varphi_0\rangle.
\end{align*}
We apply Cauchy-Schwarz on the right side and use that $\Vert V^{\nicefrac 12} K_{\Tc}^{-1} V^{\nicefrac 12} \Phi\Vert_2 \leq\eta$, which follows from the orthogonality of $\Phi$ and $\varphi_0$. We also use $\Vert \varphi_0^B - \varphi_0\Vert_2 \leq (1 - a_B) + |b_B|$. This implies
\begin{align*}
|b_B| \leq \eta \, (1 - a_B) + \eta \, |b_B| + CB,
\end{align*}
from which we conclude that $|b_B| \leq CB$.

It remains to use these findings to prove the claimed bounds for $\alpha_*^B - \alpha_*$. According to the Birman--Schwinger correspondence, we have $\alpha_*^B = (K_{\Tc,B} - e_0^B)^{-1} V^{\nicefrac 12} \varphi_0^B$. Thus, since $\varphi_0 = V^{\nicefrac 12} \alpha_*$,
\begin{align}
\alpha_*^B - \alpha_* &= \frac{1}{K_{\Tc,B} - e_0^B} V^{\nicefrac 12} \bigl[ a_B\varphi_0 + b_B \Phi\bigr] - \frac{1}{K_{\Tc}} V^{\nicefrac 12} \varphi_0 \notag\\
&= a_B\Bigl[ \frac{1}{K_{\Tc,B} - e_0^B} - \frac{1}{K_{\Tc}}\Bigr] V \alpha_* + b_B \frac{1}{K_{\Tc,B}-e_0^B} V^{\nicefrac 12} \Phi - (1 - a_B) \frac{1}{K_{\Tc}} V \alpha_*. \label{DHS1:KTV_Asymptotics_6}
\end{align}
The proof of the norm estimates for $\alpha_*^B - \alpha$ and $\pi^2(\alpha_*^B - \alpha_*)$ is obtained from Lemma \ref{DHS1:Gcal estimates}, Proposition \ref{DHS1:Resolvent_perturbation}, and the estimates of part (a) on $e_0^B$.

Part (d) follows from part (c). This ends the proof of Proposition~\ref{DHS1:KTV_Asymptotics_of_EV_and_EF}.
\end{proof}

\begin{center}
\textsc{Acknowledgements}
\end{center}

A. D. gratefully acknowledges funding from the European Union’s Horizon 2020 research and innovation programme under the Marie Sklodowska-Curie grant agreement No 836146 and from the Swiss National Science Foundation through the Ambizione grant PZ00P2 185851. It is a pleasure for A. D. to thank Stefan Teufel and the Institute of Mathematics at the University of T\"ubingen for its warm hospitality.


\vspace{1cm}

\setlength{\parindent}{0em}

(Andreas Deuchert) \textsc{Institut für Mathematik, Universität Zürich}

\textsc{Winterthurerstrasse 190, CH-8057 Zürich}

E-mail address: \href{mailto:  andreas.deuchert@math.uzh.ch}{\texttt{andreas.deuchert@math.uzh.ch}}

\vspace{0.3cm}

(Christian Hainzl) \textsc{Mathematisches Institut der Universität München}

\textsc{Theresienstr. 39, D-80333 München}

E-mail address: \href{mailto: hainzl@math.lmu.de}{\texttt{hainzl@math.lmu.de}}

\vspace{0.3cm}

(Marcel Maier, born Schaub) \textsc{Mathematisches Institut der Universität München}

\textsc{Theresienstr. 39, D-80333 München}


\setlength{\parindent}{17pt}


\section{Addendum to the paper \texorpdfstring{\cite{DeHaSc2021}}{DHS21}}
\label{DHS1:Addendum_proofs_section}

\subsection{An addendum to the proof of Proposition \ref{DHS1:MTB2}}

The following lemma contains the computations that are necessary for the calculation of the coefficient $\Lambda_0$ in \eqref{DHS1:GL-coefficient_1}. These consist of representation formulas for the function $\tilde L_T$ and form an addendum to Proposition \ref{DHS1:MTB2}. The result is a tedious but straightforward calculation, whence it is not contained in the paper \cite{DeHaSc2021}. The content of Lemma \ref{DHS1+:LT-derivative_L1} is to provide integrability of these functions in the sense needed for justifying the integration by parts in the proof of Proposition \ref{DHS1:MTB2}. Since these functions are rapidly decaying at infinity, only their singularities have to be investigated. We leave the proof of the following result to the reader.

\begin{lem}
\label{DHS1+:LT_derivatives}
Let $p,q\in\Rbb^3$. Recall $L_T(p,q)$ from \eqref{DHS1:LT_definition} and define:
\begin{align*}
\tilde L_T(p,q) &:= L_T\bigl(p+\frac q2,p-\frac q2\bigr) = \frac{\tanh(\frac{\beta}{2}(|p+\nicefrac q2|^2-\mu)) + \tanh(\frac{\beta}{2}(|p-\nicefrac q2|^2-\mu))}{(|p+\nicefrac q2|^2 - \mu) + (|p-\nicefrac q2|^2 -\mu)}.
\end{align*}
Introduce the short-hand notations
\begin{align*}
H_T(p,q) &:= \cosh^2\Bigl( \frac{\beta}{2} \bigl( \bigl| p + \frac q2\bigr|^2 - \mu\bigr)\Bigr), & J_T(p,q) &:= \tanh\Bigl( \frac{\beta}{2} \bigl( \bigl| p + \frac q2\bigr|^2 - \mu\bigr)\Bigr), 
\end{align*}
and
\begin{align*}
\ell(p,q) &:= \bigl| p + \frac q2\bigr|^2 - \mu + \bigl| p - \frac q2\bigr|^2 - \mu,
\end{align*}
and define $H_T(p):= H_T(p,0)$, as well as $J_T(p) := J_T(p,0)$. Then, 
the following statements hold.
\begin{enumerate}[(a)]
\item For $i= 1,2,3$, we have
\begin{align*}
\partial_{q_i} \tilde L_T( p , q) &= \frac{\beta}{2\ell_\mu(p,q)} \; \Bigl[ \frac{p_i + \frac{q_i}{2}}{H_T(p,q)} - \frac{p_i - \frac{q_i}{2}}{H_T(p,-q)}\Bigr] - \frac{J_T(p,q) + J_T(p,-q)}{\ell(p,q)^2} \; q_i.
\end{align*}

\item Furthermore, for $i,j=1,2,3$, we have
\begin{align*}
\partial_{q_j}\partial_{q_i} \tilde L_T(p,q) \hspace{-20pt} &\hspace{20pt} =- \frac{\beta \; q_j}{2\ell_\mu(p,q)^2} \; \Bigl[ \frac{p_i + \frac{q_i}{2}}{H_T(p,q)} - \frac{p_i - \frac{q_i}{2}}{H_T(p,-q)}\Bigr] \\
&\hspace{130pt} - \frac{\beta \; q_i}{2\ell_\mu(p,q)^2} \; \Bigl[ \frac{p_j + \frac{q_j}{2}}{H_T(p,q)} - \frac{p_j - \frac{q_j}{2}}{H_T(p,-q)}\Bigr]\\
&+ \frac{\beta \; \delta_{ij}}{4\ell_\mu(p,q)}\Bigl[ \frac{1}{H_T(p,q)} + \frac{1}{H_T(p,-q)}\Bigr] \\
&- \frac{\beta^2}{2\ell_\mu(p,q)}\Bigl[ \frac{(p_i + \frac{q_i}{2})(p_j + \frac{q_j}{2})J_T(p,q)}{H_T(p,q)} + \frac{(p_i - \frac{q_i}{2})(p_j - \frac{q_j}{2})J_T(p,-q)}{H_T(p,-q)}\Bigr]\\
&+ [J_T(p,q) + J_T(p,-q)]\; \Bigl[ \frac{2\; q_iq_j}{\ell(p,q)^3} - \frac{\delta_{ij}}{\ell(p,q)^2}\Bigr]
\end{align*}

\item In particular,
\begin{align*}
\partial_{q_j}\partial_{q_i} \tilde L_T(p,0) &= \Bigl[\frac{\beta}{4 (p^2-\mu)} \; \frac{1}{H_T(p)} - \frac{J_T(p)}{2 (p^2-\mu)^2}\Bigr] \; \delta_{ij}  \\
&\hspace{150pt} - \frac{\beta^2}{2(p^2-\mu)} \; \frac{J_T(p)}{H_T(p)} \; p_ip_j.
\end{align*}
\end{enumerate}
\end{lem}

\begin{lem}
\label{DHS1+:LT-derivative_L1}
Let $p\in \Rbb^3$ be fixed. Then, for every $i,j=1,2,3$, both of the functions $q\mapsto \partial_{q_i} \tilde L_T(p,q)$ and $q\mapsto \partial_{q_i}\partial_{q_j} \tilde L_T(p,q)$ are elements of $L^1(\Rbb^3)$.
\end{lem}

\begin{proof}
We start by considering $\partial_{q_i} \tilde f_T(p,q)$. It suffices to investigate the singularity at the set
\begin{align}
\Bigl\{ q\in \Rbb^3 : \bigl| p + \frac q2\bigr|^2 = \mu = \bigl| p + \frac q2\bigr|^2 \Bigr\}. \label{DHS1+:fT_Singularities}
\end{align}
In all other regimes, the decay properties are sufficiently good for integrability. Separate the parts with $p_i$ and $q_i$ to obtain
\begin{align}
\partial_{q_i} \tilde L_T(p,q) &= q_i \Bigl[ \frac{\beta}{4\ell_\mu(p,q)}\bigl[\frac{1}{H_T(p,q)} + \frac{1}{H_T(p,-q)}\bigr] - \frac{J_T(p,q) + J_T(p,-q)}{\ell(p,q)^2}\Bigr]\\
&\hspace{150pt} + \frac{\beta p_i}{2\ell_\mu(p,q)}\bigl[\frac{1}{H_T(p,q)} - \frac{1}{H_T(p,-q)}\bigr] \label{DHS1+:fTprime_rewriting}
\end{align}
In the first line, we can consider the terms with $(p,q)$ and with $(p,-q)$ separately. Then, we are left with investigating
\begin{align*}
\xi_T^{(1)}(p,q) &:= \frac{\beta}{4\ell_\mu(p,q)}\frac{1}{H_T(p,q)} - \frac{J_T(p,q)}{\ell(p,q)^2} \\
\xi_T^{(2)}(p,q) &:= \frac{1}{\ell(p,q)} \bigl[ \frac{1}{H_T(p,q)} - \frac{1}{H_T(p,-q)}\bigr]
\end{align*}
Without loss, by interchanging the order if necessary, we can take one of the limits, e.g. $|p - \nicefrac q2|^2-\mu \to 0$. As $x := |p+\nicefrac q2|^2-\mu\to 0$, we then have
\begin{align*}
\frac{\beta}{4x} \frac{1}{\cosh^2(\frac\beta 2\; x)} - \frac{\tanh(\frac \beta 2\; x)}{x^2}= \frac{1}{x}\Bigl[ \frac \beta 4 \frac{1}{\cosh^2(\frac \beta 2\; x)} - \frac{\tanh(\frac \beta 2\; x)}{x}\Bigr]
\end{align*}
The term in brackets is $\Ocal(1)$ and $\frac 1x$ is integrable over the unit ball in dimensions $d\geq 2$. Proceed analogously for the term with $(p,-q)$.  Concerning $\xi_T^{(2)}(p,q)$, an application of l'Hôpital yields
\begin{align}
\lim_{x\to 0}\frac{1}{x^2} \Bigl[ \frac{1}{\cosh^2(\frac \beta 2\; x)} -1\Bigr] 
= -\frac{\beta^2}{4}. \label{DHS1+:cosh-decay}
\end{align}
Hence, this term is even linear at the singularity. The decay makes it integrable as well. Moving on to $\partial_{q_i} \partial_{q_j} \tilde f_T(p,q)$, we rewrite accordingly
\begin{align}
\partial_{q_i}\partial_{q_j}\tilde L_T(p,q) \hspace{-20pt} & \hspace{20pt} = - \frac{\beta\; (p_iq_j+p_jq_i)}{2\ell_\mu(p,q)^2} \bigl[ \frac{1}{H_T(p,q)}- \frac{1}{H_T(p,-q)}\bigr] \label{DHS1+:ft_double1} \\ 
& -\frac{\beta \; q_iq_j}{2\ell_\mu(p,q)^2} \bigl[\frac{1}{H_T(p,q)} + \frac{1}{H_T(p,-q)}\bigr] + [J_T(p,q) + J_T(p,-q)] \; \frac{2q_iq_j}{\ell(p,q)^3} \label{DHS1+:ft_double2}\\
&- \frac{\beta^2}{2\ell_\mu(p,q)}\Bigl[ \frac{(p_i + \frac{q_i}{2})(p_j + \frac{q_j}{2})J_T(p,q)}{H_T(p,q)} + \frac{(p_i - \frac{q_i}{2})(p_j - \frac{q_j}{2})J_T(p,-q)}{H_T(p,-q)}\Bigr] \label{DHS1+:ft_double3}\\
&+ \frac{\beta \; \delta_{ij}}{4\ell_\mu(p,q)}\Bigl[ \frac{1}{H_T(p,q)} + \frac{1}{H_T(p,-q)}\Bigr] - [J_T(p,q) + J_T(p,-q)]\; \frac{\delta_{ij}}{\ell(p,q)^2} \label{DHS1+:ft_double4}
\end{align}
As before, it is enough to investigate the singularity \eqref{DHS1+:fT_Singularities}. The first thing to notice is that the term in line \eqref{DHS1+:ft_double3} is bounded. Proceeding analogously to the above, the term \eqref{DHS1+:ft_double1} is bounded by \eqref{DHS1+:cosh-decay}. The term in \eqref{DHS1+:ft_double2} is $\Ocal(\frac{1}{x^2})$ by the same investigation as for $\xi_T^{(1)}$ above. This is still integrable over compacts in dimensions $d\geq 3$. The remaining term is \eqref{DHS1+:ft_double4} is $\Ocal(\frac 1x)$ by the investigation of $\xi_T^{(1)}$ above.
\end{proof}

\subsection{An alternative proof of Lemma \ref{DHS1:KTB_Lower_bound}}

This is an alternative proof for Lemma \ref{DHS1:KTB_Lower_bound}, which takes the pedestrian way without any reference to analytic perturbation theory.

\begin{lem}
\label{DHS1+:KTB_Lower_bound}
Let Assumptions \ref{DHS1:Assumption_V} and \ref{DHS1:Assumption_KTc} be true.
For any $D_0 \geq 0$, there are constants $B_0>0$ and $T_0>0$ such that for $0< B\leq B_0$ and $T>0$ with $T - \Tc \geq -D_0B$, the estimate
\begin{align}
K_{T, B} - V &\geq c \; (1 - P) (1 + \pi^2) (1- P) + c \, \min \{ T_0, (T - \Tc)_+\} - CB \label{DHS1+:KTB_Lower_bound_eq}
\end{align}
holds. Here, $P = |\alpha_*\rangle\langle \alpha_*|$ is the orthogonal projection onto the ground state $\alpha_*$ of $K_{\Tc} - V$.
\end{lem}

\begin{proof}
We prove two lower bounds on $K_{T,B} - V$, which we add up to etablish \eqref{DHS1+:KTB_Lower_bound_eq}.

\emph{Step 1.} We claim that there are $T_0>0$ such that
\begin{align}
K_{T,B} - V \geq c \; \min \{ T_0, (T - \Tc)_+\} - CB. \label{DHS1+:KTB_Lower_bound_5}
\end{align}
To prove \eqref{DHS1+:KTB_Lower_bound_5}, we note that the derivative of the symbol $K_T$ in \eqref{DHS1:KT-symbol} with respect to $T$ equals
\begin{align}
\frac{\dd}{\dd T} K_T(p) = 2\; K_T(p)^2 \frac{1}{\cosh^2(\frac{p^2-\mu}{2T})} \label{DHS1+:KTc_bounded_derivative}
\end{align}
and is bounded from above by 2. If $T \leq \Tc$, we infer $K_{T,B} - K_{\Tc,B} \geq -2D_0B$ as an operator inequality, which
proves \eqref{DHS1+:KTB_Lower_bound_5}. 
%

If on the other hand $T \geq \Tc$, then choose $T_0 := \frac \kappa 3$, where $\kappa >0$ denotes the spectral gap of the operator $K_{\Tc,B} - V$ above its lowest eigenvalue $e_0^B$. For $B$ small enough, Proposition \ref{DHS1:KTV_Asymptotics_of_EV_and_EF} ensures that $\kappa$ is independent of $B$, whence $e_0^B$ is simple. In order to prove \eqref{DHS1+:KTB_Lower_bound_5}, we may henceforth assume that $T \leq \Tc + \frac \kappa 3$ due to the monotonicity of $K_{T,B}$ in $T$, which implies $K_{T,B} \geq K_{\min\{T,\Tc+\frac \kappa 3\},B}$.

We consider the minimization problem for the operator $K_{T, B} - V$, while we assume $\Tc \leq T\leq \Tc + \frac \kappa 3$. For a normalized $\varphi \in L^2(\Rbb^3)$, we split $\varphi = a\alpha_*^B + b \zeta$ with $|a|^2 + |b|^2 = 1$ and $\langle \alpha_*^B , \zeta\rangle =0$. Here, $\alpha_*^B$ is the unique normalized ground state of $K_{\Tc,B} - V$. In fact, $a$, $b$, and $\zeta$ are dependent on $B$ but these dependencies play no role in what follows. Note that for $T \geq \Tc$, \eqref{DHS1+:KTc_bounded_derivative} implies the inequality of operators
\begin{align}
K_{T,B} - K_{\Tc, B} \leq 2(T - \Tc).
\end{align}
Therefore, an application of $uv \leq \frac \eta 2 \, u^2 + \frac{1}{2\eta} \, v^2$ for $u,v\geq 0$ and $0 < \eta < 1$ yields
\begin{align}
\langle \varphi, (K_{T,B} - V) \varphi\rangle \geq \frac 12\, |a_B|^2 \; \langle \alpha_*^B, (K_{T,B} - K_{\Tc,B}) \alpha_*^B\rangle + |b_B|^2 \; (\kappa - 2 (T-\Tc)) - |e_0^B|. \label{DHS1+:KTB_Lower_bound_8}
\end{align}
In view of \eqref{DHS1+:KTc_bounded_derivative}, it is easy to see that
\begin{align*}
\langle \alpha_*^B, (K_{T,B} - K_{\Tc, B}) \alpha_*^B\rangle &\geq \bigl\langle K_{\Tc,B} \alpha_*^B, \frac{1}{\cosh^2(\frac{\pi^2 - \mu}{2\Tc})} K_{\Tc,B} \alpha_*^B\bigr\rangle \; (T - \Tc).
\end{align*}
It remains to remove the magnetic field in every instance to show that
\begin{align}
\bigl\langle K_{\Tc,B} \alpha_*^B, \frac{1}{\cosh^2(\frac{\pi^2 - \mu}{2\Tc})} K_{\Tc,B} \alpha_*^B\bigr\rangle \geq \int_{\Rbb^3} \dd p \; |\hat{V\alpha_*}(p)|^2 \; \frac{1}{\cosh^2(\frac{p^2-\mu}{2\Tc})} - CB. \label{DHS1+:KTB_Lower_bound_6}
\end{align}
The first term on the right side equals $2\Tc \Lambda_2 >0$ with $\Lambda_2$ from \eqref{DHS1:GL-coefficient_1}.

To see that \eqref{DHS1+:KTB_Lower_bound_6} is true, we start with the eigenvalue equation $(K_{\Tc,B} - V)\alpha_*^B = e_0^B \alpha_*^B$. By Proposition \ref{DHS1:KTV_Asymptotics_of_EV_and_EF}, the estimates $|e_0^B| \leq CB$ and $\Vert \alpha_*^B - \alpha_*\Vert_2^2 \leq CB$ hold. Since $V$ and $\cosh(\frac{\pi^2-\mu}{2\Tc})^{-2}$ are bounded operators, we may therefore replace $\alpha_*^B$ by $\alpha_*$ and $e_0^B$ by zero in every occurrence for an error of $CB$, i.e.,
\begin{align}
\bigl\langle K_{\Tc,B} \alpha_*^B, \frac{1}{\cosh^2(\frac{\pi^2 - \mu}{2\Tc})} K_{\Tc,B} \alpha_*^B\bigr\rangle \geq \bigl\langle V \alpha_*, \frac{1}{\cosh^2(\frac{\pi^2 - \mu}{2\Tc})}  V\alpha_*\bigr\rangle - CB.\label{DHS1+:KTB_Lower_bound_7}
\end{align}

Furthermore, we use the identity \eqref{DHS1:cosh2_Matsubara}, which expresses $\cosh(\frac \beta 2 z)^{-2}$ in terms of the Matsubara frequencies $\omega_n$ in \eqref{DHS1:Matsubara_frequencies}. The resolvent equation
\begin{align*}
\frac{1}{(\i\omega_n - \pi^2)^2} - \frac{1}{(\i\omega_n - p^2)^2} = \frac{1}{(\i\omega_n - \pi^2)^2} \bigl[ \i\omega_n (\pi^2 - p^2) - (\pi^4 - p^4) \bigr] \frac{1}{(\i\omega_n - p^2)^2}
\end{align*}
leads us to investigate
\begin{align*}
\pi^4 - p^4 &= (\pi^2 + p^2) (\pi^2- p^2) + [\pi^2- p^2, p^2] \\ 
&\hspace{-35pt}= \bigl( 2\pi^2 - \frac 14 |\Bbold \wedge x|^2\bigr) \bigl( \Bbold \wedge x\cdot p + \frac 14 |\Bbold \wedge x|^2 \bigr) - (\Bbold \wedge x \cdot p)^2 - \frac 14 |\Bbold \wedge x|^2 (\Bbold \wedge x\cdot p) + B^2.
%
\end{align*}
Here, we used $[\pi^2 - p^2,p^2] = \frac 14 [|\Bbold \wedge x|^2,p^2] = B^2$ as well as $p^2 = \pi^2 - \Bbold \wedge x\cdot p - \frac 14 |\Bbold \wedge x|^2$ and the fact that $\Bbold \wedge x\cdot p$ commutes with $\pi^2-p^2$. Since $|\cdot|^k V\alpha_*\in L^2(\Rbb^3)$ for $k\in \Nbb$ by \eqref{DHS1:Decay_of_alphastar}, a tedious investigation of the kernels with the techniques introduced in Section~\ref{DHS1:Proofs} shows that, in \eqref{DHS1+:KTB_Lower_bound_7}, $\pi^2$ may be replaced by $p^2$ for an error of the size $CB$. This shows \eqref{DHS1+:KTB_Lower_bound_6}.

We apply \eqref{DHS1+:KTB_Lower_bound_6} to \eqref{DHS1+:KTB_Lower_bound_8}. Together with the estimates provided by Proposition \ref{DHS1:KTV_Asymptotics_of_EV_and_EF} mentioned above and $|a|^2 + |b|^2 =1$, we finally obtain
\begin{align*}
\langle \varphi, (K_{T,B} - V)\varphi\rangle &\geq \min \Bigl\{ \frac 12 \int_{\Rbb^3} \dd p \; |\hat{V\alpha_*}(p)|^2 \smash{\frac{1}{\cosh^2(\frac{p^2-\mu}{2\Tc})}} \, (T - \Tc) \; , \; \kappa - 2(T- \Tc)\Bigr\} \\
&\hspace{300pt} - CB.
\end{align*}
Since $T\leq \Tc + \frac \kappa 3$ implies $\kappa - 2(T - \Tc) \geq T - \Tc$, this proves \eqref{DHS1+:KTB_Lower_bound_5}.

\emph{Step 2.} We claim there are $c,C>0$ such that
\begin{align}
K_{T,B} - V \geq c \; (1 - P) (1 + \pi^2) (1- P) - CB. \label{DHS1+:KTB_Lower_bound_9}
\end{align}
From the arguments in Step~1 we know that we can replace $T$ by $\Tc$ for a lower bound if we allow for a remainder of the size $-CB$. To prove \eqref{DHS1+:KTB_Lower_bound_9}, we choose $0 < \eta < 1$ and write
\begin{align}
K_{\Tc,B}-V = e_0^BP_B + (1-P_B) [(1-\eta) K_{\Tc,B} -V](1-P_B) + \eta (1-P_B) K_{\Tc,B} (1-P_B), \label{DHS1+:KTB_Lower_bound_1}
\end{align}
where $e_0^B$ denotes the ground state energy of $K_{\Tc, B} - V$ and $P_B = |\alpha_*^B\rangle \langle \alpha_*^B|$ is the spectral projection onto the corresponding unique ground state vector $\alpha_*^B$. From Proposition~\ref{DHS1:KTV_Asymptotics_of_EV_and_EF} we know that the first term on the right side of \eqref{DHS1+:KTB_Lower_bound_1} is bounded from below by $-CB$. The lowest eigenvalue of $K_{\Tc} - V$ is simple and isolated from the rest of the spectrum. Proposition~\ref{DHS1:KTV_Asymptotics_of_EV_and_EF} therefore implies that the second term in \eqref{DHS1+:KTB_Lower_bound_1} is nonnegative as long as $\eta$ is, independently of $B$, chosen small enough, and can be dropped for a lower bound.
%
%
To treat the third term, we note that the symbol $K_T(p)$ in \eqref{DHS1:KT-symbol} satisfies the inequality $K_{\Tc}(p) \geq c' (1 + p^2)$ for some constant $c'$, and hence $K_{\Tc,B} \geq c' (1 + \pi^2)$. In combination, the above considerations prove
\begin{align*}
K_{\Tc,B}-V \geq  c' \; (1-P_B)(1+\pi^2)(1-P_B) - CB.
\end{align*}
It remains to replace $P_B$ by $P = |\alpha_*\rangle\langle \alpha_*|$. To this end, we write
\begin{align}
(1-P_B)(1+\pi^2)(1-P_B) - (1-P)(1+\pi^2)(1-P) &\notag \\
&\hspace{-160pt}= (P-P_B) +  (P - P_B)\pi^2(1-P_B) + (1-P)\pi^2(P-P_B) \label{DHS1+:KTB_Lower_bound_4}.
\end{align}
From Proposition \ref{DHS1:KTV_Asymptotics_of_EV_and_EF} we know that $\Vert P_B - P\Vert_\infty \leq CB$ and $\Vert \pi^2(P_B- P)\Vert_\infty \leq CB$. Hence, the norm of the operator on the right side of \eqref{DHS1+:KTB_Lower_bound_4} is bounded by a constant times $B$. This shows \eqref{DHS1+:KTB_Lower_bound_9} and concludes our proof.
\end{proof}

\subsection{An alternative proof of Lemma \ref{DHS1:KT_integral_rep}}

This is an alternative proof of Lemma \ref{DHS1:KT_integral_rep}.

\begin{defn}[Speaker path]
\label{DHS1+:speaker_path}
Let $R>0$ and $\alpha \geq 0$. Using the notation $\beta := T^{-1}$, define the following complex paths
\begin{align*}
\begin{split}
\gamma_1(t) &:= \frac{\pi\i}{2\beta} + (1 + \i)t\\
\gamma_2(t) &:= \frac{\pi\i}{2\beta} - (\alpha + 1)t\\
\gamma_3(t) &:= -\frac{\pi\i}{2\beta}t - (\alpha + 1)\\
\gamma_4(t) &:= -\frac{\pi\i}{2\beta} - (\alpha + 1)(1-t) \\
\gamma_5(t) &:= -\frac{\pi\i }{2\beta} + (1 - \i)t
\end{split}
&
\begin{split}
\phantom{ \frac \i\betac }t&\in [0,R], \\
\phantom{ \frac \i\betac }t &\in [0,1], \\
\phantom{ \frac \i\betac }t&\in [-1,1],\\
\phantom{ \frac \i\betac }t &\in [0,1],\\
\phantom{ \frac \i \betac }t&\in [0,R].
\end{split}
& \begin{split} 
\text{\includegraphics[width=6cm]{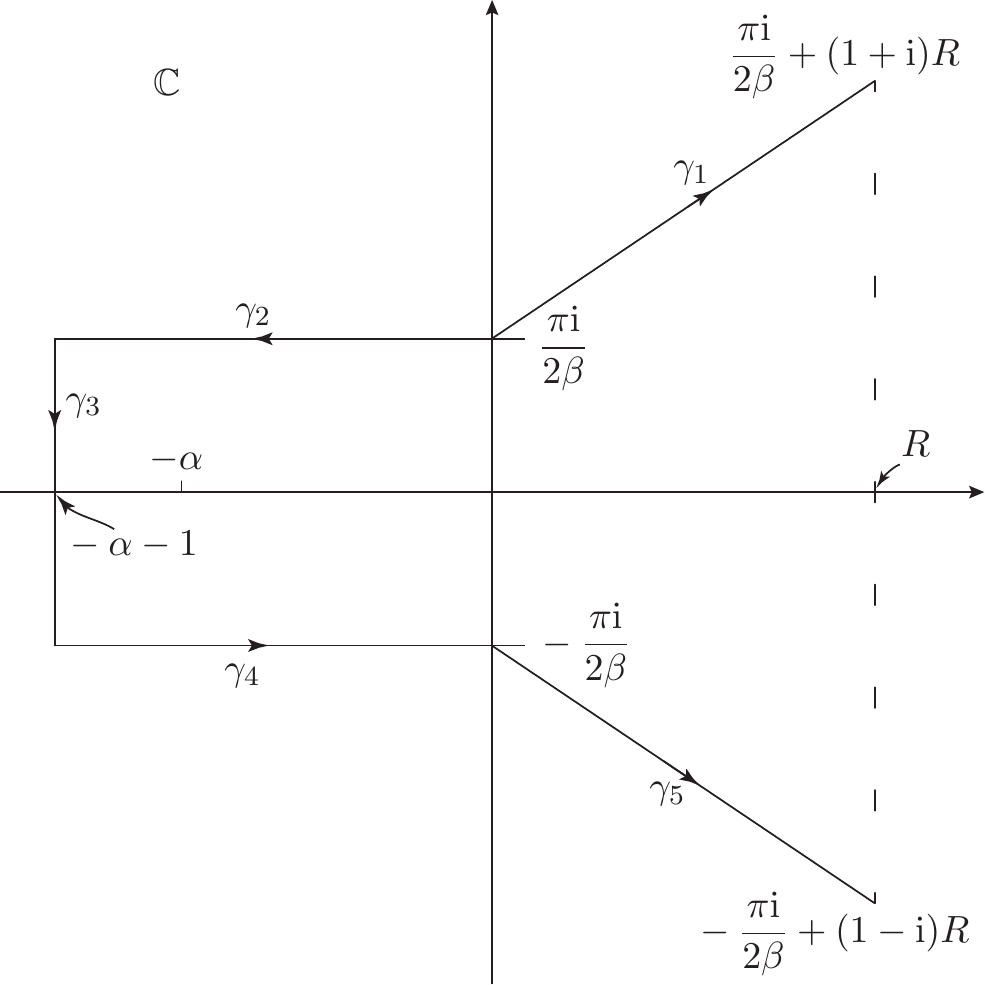}}
\end{split} 
\end{align*}
The speaker path is defined as the union of paths $\gamma_i$, $i=1, \ldots, 5$, with $\gamma_1$ taken in reverse direction, i.e.,
\begin{align*}
\speaker_{R} := \mathop{\dot -}\gamma_1 \mathop{\dot +} \gamma_2 \mathop{\dot +} \gamma_3 \mathop{\dot +} \gamma_4 \mathop{\dot +} \gamma_5.
\end{align*}
We also let $\speaker_\alpha := \bigcup_{R>0} \speaker_{\alpha,R}$.
\end{defn} 

\begin{lem}
\label{DHS1+:KT_integral_rep}
Let $\alpha \geq 0$ and let $H\colon \Dcal(H)\ra \Hcal$ be a self-adjoint operator in a separable Hilbert space $\Hcal$ with $H\geq -\alpha$. Then, we have
\begin{align*}
\frac{H}{\tanh(\frac{\beta H}{2})} = H + \lim_{R\to\infty} \int_{\speaker_{\alpha, R}} \frac{\dd z}{2\pi\i} \Bigl( \frac{z}{\tanh(\frac{\beta z}{2})} - z \Bigr) \frac{1}{z - H},
\end{align*}
where $\speaker_{\alpha, R}$ is the speaker path from Definition \ref{DHS1+:speaker_path}. The limit exists in operator norm.
\end{lem}

\begin{proof}
Call $f_T(z) = \frac{z}{\tanh(\frac{\beta z}{2})} - z = \frac{2z}{\e^{\beta z} - 1}$. Let us first prove that the limit exists and defines a bounded operator on $\Hcal$. To do this, we investigate the tails of the paths $\gamma_1$ and $\gamma_5$. For example, we have to investigate the operator norm of
\begin{align*}
\int_R^\infty \frac{\dd t}{2\pi\i} \frac{2\gamma_1(t)}{\e^{\beta \gamma_1(t)} - 1} \frac{1}{\gamma_1(t) - H} \, \gamma_1'(t).
\end{align*}
The estimates
\begin{align*}
|\gamma_1(t)| &\leq \frac{\pi}{2\beta} + \sqrt{2}t \leq Ct, & |\gamma_1'(t)| &= \sqrt{2}, & \Re \gamma_1(t) &= t,
\end{align*}
and
\begin{align*}
\Bigl\Vert \frac 1{\gamma_1(t) - H}\Bigr\Vert_\infty &\leq \frac{1}{\Im \gamma_1(t)} = \frac{1}{\frac{\pi}{2\beta} + t} \leq \frac 1t.
\end{align*}
hold for $R$ large enough, and imply
\begin{align*}
\Bigl\Vert \int_R^\infty \frac{\dd t}{2\pi\i} \frac{2\gamma_1(t)}{\e^{\beta \gamma_1(t)} - 1} \frac{1}{\gamma_1(t) - H} \, \gamma_1'(t)\Bigr\Vert_\infty &\leq C \int_R^\infty \frac{1}{\e^{\beta t} - 1} \leq C \e^{-\beta R}.
\end{align*}
The last inequality follows by taking $R$ so large that $1 \leq \frac 12 \e^{\beta t}$ for all $t\geq R$. The contribution of $\gamma_5$ is estimated in a similar fashion. This proves operator norm convergence of the limit. 

We let $K\geq 0$ and choose $\psi\in \ran(\Idbb_{(-\infty, K-1]}(H))$. Then, we take $R\geq K$ and close the speaker path by the contour $\gamma_R(t) := R + \i (R + \frac{\pi}{2\beta}) t$ where $t\in [-1,1]$. Then, for each $\varphi\in \Hcal$, by Cauchy's integral theorem, we have
\begin{align*}
\langle \varphi, f_T(H) \psi\rangle = \int_{\speaker_{\alpha,R} \dot + \gamma_R} \frac{\dd z}{2\pi\i} \; f_T(z) \langle \varphi, (z - H)\psi\rangle.
\end{align*}
When we investigate the contribution from $\gamma_R$, we have
\begin{align*}
\Bigl|\int_{\gamma_R} \frac{\dd z}{2\pi\i} \; f_T(z) \langle \varphi, (z - H)^{-1}\psi\rangle\Bigr| &\leq \int_{-1}^1 \frac{\dt}{\pi} \frac{|\gamma_R(t)|}{|\e^{\beta \gamma_R(t)}| - 1} \; |\langle \varphi, (\gamma_R(t) - H)^{-1} \psi\rangle| \; |\gamma_R'(t)|.
\end{align*}
Since $\Re \gamma_R(t) = R$, we have that
\begin{align*}
|\gamma_R(t) - (K-1)| \geq |R - K + 1| \geq 1.
\end{align*}
It follows that $\sup_{t\in [-1,1]}\Vert (\gamma_R(t) - H)^{-1}\Idbb_{(-\infty, K-1]}(H)\Vert_\infty \leq 1$. Hence, we obtain
\begin{align*}
\sup_{\Vert \varphi\Vert = 1} \Bigl|\int_{\gamma_R} \frac{\dd z}{2\pi\i} \; f_T(z) \langle \varphi, (z - H)^{-1}\psi\rangle\Bigr| &\leq C \, \frac{R^2}{\e^{\beta R} - 1} \, \Vert\psi\Vert \xra{R\to\infty} 0.
\end{align*}
This proves that
\begin{align}
f_T(H)\psi = \lim_{R\to\infty} \int_{C_{\alpha,\beta, R}} \frac{\dd z}{2\pi\i} \; f_T(z) \frac{1}{z - H} \psi. \label{DHS1+:Preliminary fTH}
\end{align}
Now, since $t\mapsto f_T(t)$, $t\geq - \alpha$ is bounded with $\Vert f_T\Vert_\infty = \frac{2\alpha}{1 - \e^{-\beta \alpha}}$, we get that $f_T(H)$ is a bounded operator and hence, \eqref{DHS1+:Preliminary fTH} extends by density to all $\psi\in \Hcal$. Hence, the claim holds for all $\psi\in \Dcal(H)$, since, then
\begin{align*}
\frac{H}{\tanh(\frac{\beta H}{2})} \psi &= H\psi + f_T(H)\psi.\qedhere
\end{align*}
\end{proof}

\printbibliography[heading=bibliography, title=Bibliography of Chapter \ref{Chapter:DHS1}]
\end{refsection}

\def\EGL{\mathcal E^{\mathrm{GL}}_{D, h}}
\def\FBCS{\mathcal F^{\mathrm{BCS}}_{
h, T
}}

\def\Lsymm{{L^2(Q_h \times \Rbb_{\mathrm s}^3)}}
\def\Hsymm{{H^1(Q_h \times \Rbb_{\mathrm s}^3)}}

\begin{refsection}

\chapter[Microscopic Derivation of Ginzburg--Landau Theory and the BCS Critical Temperature Shift in the Presence of Weak Macroscopic External Fields][BCS-Theory in the Presence of Weak External Fields]{
Microscopic Derivation of Ginzburg--Landau Theory and the BCS Critical Temperature Shift in the Presence of Weak Macroscopic External Fields
}
\label{Chapter:DHS2} \label{CHAPTER:DHS2}

\begin{abstract}
\noindent 
We consider the Bardeen--Cooper--Schrieffer (BCS) free energy functional with weak and macroscopic external electric and magnetic fields and derive the Ginzburg--Landau functional. We also provide an asymptotic formula for the BCS critical temperature as a function of the external fields. This extends our previous results in \cite{DeHaSc2021} for the constant magnetic field to general magnetic fields with a nonzero magnetic flux through the unit cell.
\end{abstract}


\section{Introduction and Main Results}


\subsection{Introduction}

Ginzburg--Landau (GL) theory has been introduced as the first macroscopic and phenomenlogical description of superconductivity in 1950 \cite{GL}. The theory comprises a system of partial differential equations for a complex-valued function, the order parameter, and an effective magnetic field. Ginzburg--Landau theory has been highly influencial and investigated in numerous works, among which are \cite{Sigal1, Sigal2, Serfaty, SandierSerfaty, Correggi3, Correggi2, Giacomelli1, Giacomelli2, Giacomelli3} and references therein.

Bardeen--Cooper--Schrieffer (BCS) theory of superconductivity is the first commonly accepted and Nobel prize awarded microscopic theory of superconductivity \cite{BCS}. As a major breakthrough, the theory features a pairing mechanism between the electrons below a certain critical temperature, which causes the electrical resistance in the system to drop to zero in the superconducting phase. This effect is due to an effective attraction between the electrons, which arises as a consequence of the phonon vibrations of the lattice ions in the superconductor.

As Leggett pointed out \cite{Leg1980}, BCS theory can be formulated variationally in terms of an energy functional, see also \cite{de_Gennes}. This free energy functonal can be obtained from the first principle quantum mechanical description via a restriction to quasi-free states. Such states are determined by their one-particle density matrix and the Cooper pair wave function. The BCS functional has be studied intensively from a mathematical point of view in the absence of external fields \cite{Hainzl2007, Hainzl2007-2, Hainzl2008-2, Hainzl2008-4, FreiHaiSei2012, BraHaSei2014, FraLe2016, DeuchertGeisinger} and in the presence of external fields \cite{HaSei2011, BrHaSei2016, FraLemSei2017, A2017, CheSi2020}. The BCS gap equation arises as the Euler--Lagrange equation of the BCS functional and its solution is used to compute the spectral gap of an effective Hamiltonian, which is open in the superconducting phase. BCS theory from the point of view of its gap equation is studied in \cite{Odeh1964, BilFan1968, Vanse1985, Yang1991, McLYang2000, Yang2005}.

The present article continues a series of works, in which the \emph{macroscopic} GL theory is derived from the \emph{microscopic} BCS theory in the weak external field regime. This endeavor has been initiated by Gor'kov in 1959 \cite{Gorkov}. The first mathematically rigorous derivation of the GL functional from the BCS functional has been provided by Frank, Hainzl, Seiringer, and Solovej for periodic external electric and fluxless magnetic fields in 2012 \cite{Hainzl2012}. Recently, their work has been extended to systems exposed to a homogeneous magnetic field by Deuchert, Hainzl, and Maier \cite{DeHaSc2021}. This extends the derivation of GL theory to the case of a system with a nonzero magnetic flux through the unit cell of periodicity. The present work unites these results and provides the derivation of GL theory for general external fields. GL theory arises from BCS theory when the temperature is sufficiently close to the critical temperature. More precisely, if $0 < h \ll 1$ denotes the ratio between the microscopic and the macroscopic length scale, then the external electric field $W$ and the magnetic vector potential $\Abold$ are given by $h^2 W(hx)$ and $h\Abold(hx)$, respectively. Furthermore, the temperature regime is such that $T - \Tc = -\Tc Dh^2$ for some constant $D >0$, where $\Tc$ is the critical temperature in absence of external fields. When this scaling is in effect, it is shown in \cite{Hainzl2012} and \cite{DeHaSc2021} that the Cooper pair wave function $\alpha(x,y)$ is given by
\begin{align}
\alpha(x,y) = h\, \alpha_*(x - y) \, \psi \bigl( \frac{h(x+y)}{2}\bigr) \label{eq:intro1}
\end{align}
to leading order in $h$. Here, $\alpha_*$ is the microscopic Cooper pair wave function in the absence of external fields and $\psi$ is the GL order parameter.

Moreover, the influence of the external fields causes a shift in the critical temperature of the BCS model, which is described by linearized GL theory in the same scaling regime. More precisely, it has been shown in \cite{Hainzl2014} and \cite{DeHaSc2021} that the critical temperature shift in BCS theory is given by
\begin{align}
\Tc(h) = \Tc (1 - \Dc h^2) \label{eq:intro2}
\end{align}
to leading order, where $\Dc$ denotes a critical parameter that can be computed using linearized GL theory.

The present work is an extension of the paper \cite{DeHaSc2021}, in which we proved the expansions \eqref{eq:intro1} and \eqref{eq:intro2} for systems exposed to a constant magnetic field. In this article, we incorporate general periodic electric fields $W$ and magnetic vector potentials $\Abold$ that give rise to periodic magnetic fields. We show that within the scaling introduced above, the Ginzburg--Landau energy arises as leading order correction on the order $h^4$. Furthermore, we show that the Cooper pair wave function admits the leading order term \eqref{eq:intro1} and that the critical temperature shift is given by \eqref{eq:intro2} to leading order. The proof of these results relies to a large extent on a priori bounds for certain low-energy BCS states that include the magnetic field and have been proved in \cite{DeHaSc2021}. The main technical novelty of this article is a further development of the phase approximation method, which has been pioneered in the framework of BCS theory for the case of the constant magnetic field in \cite{Hainzl2017} and \cite{DeHaSc2021}, which allows us to extend the trial state analysis in \cite{DeHaSc2021} to the situation of general external fields.

\subsection{Gauge-periodic samples}
\label{Magnetically_Periodic_Samples}

Our objective is to study a system of three-dimensional fermionic particles that is subject to weak and slowly varying external electromagnetic fields within the framework of BCS theory. Let us define the magnetic field $\Bbold \coloneqq h^2 e_3$. It can be written in terms of the vector potential $\Abold_{\Bbold}(x) \coloneqq \frac{1}{2} \Bbold \wedge x$, where $x \wedge y$ denotes the cross product of two vectors $x,y \in \mathbb{R}^3$, as $\Bbold = \curl \Abold_\Bbold$. To the vector potential $\Abold_{\Bbold}$ we associate the magnetic translations
\begin{align}
	T(v)f(x) &:= \e^{\i \frac {\Bbold} 2\cdot (v\wedge x)} f(x+v), & v &\in \Rbb^3, \label{Magnetic_Translation}
\end{align}
which commute with the magnetic momentum operator
\begin{align}
	\pi := -\i \nabla + \Abold_\Bbold. \label{pi_definition}
\end{align}
The family $\{ T(v) \}_{v\in \mathbb{R}^3}$ satisfies $T(v+w) = \e^{\i \frac{\Bbold}{2} \cdot (v \wedge w)} T(v) T(w)$ and is therefore a unitary representation of the Heisenberg group. We assume that our system is periodic with respect to the Bravais lattice $\Lambda_h \coloneqq \sqrt{2\pi} \, h^{-1} \, \Zbb^3$ with fundamental cell
\begin{align}
Q_h &\coloneqq \bigl[0, \sqrt{2\pi} \, h^{-1}\bigr]^3 \subseteq \Rbb^3. \label{Fundamental_cell}
\end{align}
Let $b_i = \sqrt{2\pi} \, h^{-1} \, e_i$ denote the basis vectors that span $\Lambda_h$. The magnetic flux through the face of the unit cell spanned spanned by $b_1$ and $b_2$ equals $2 \pi$, and hence the abelian subgroup $\{ T(\lambda) \}_{\lambda \in \Lambda_h}$ is a unitary representation of the lattice group. 

Our system is subject to an external electric field $W_h(x) = h^2W(hx)$ with a fixed function $W \colon \Rbb^3 \ra \Rbb$, as well as a magnetic field defined in terms of the vector potential $\Abold_h(x) = h \Abold(hx)$, which admits the form $\Abold := \Abold_{e_3} + A$ with $A \colon \Rbb^3\ra \Rbb^3$ and $\Abold_{e_3}$ as defined above. We assume that $A$ and $W$ are periodic with respect to $\Lambda_1$. The total magnetic momentum operator of the system is $\pi_{\Abold_h}$, where
\begin{align}
\pi_\Abold \coloneqq -\i \nabla + \Abold. \label{piA_definition}
\end{align}
Since $A$ is a periodic function we know that $\pi_{\Abold_h}$ commutes with $T(\lambda)$ as long as $\lambda\in \Lambda_h$.

The flux of the magnetic field $\curl A_h$ through all faces of the unit cell $Q_h$ vanishes because $A_h$ is a periodic function. Accordingly, the magnetic field $\curl \Abold_h$ has the same fluxes through the faces of the unit cell as $\Bbold$. 

The above representation of $\Abold_h$ is general in the sense that any periodic magnetic field field $B(x)$ that satisfies the Maxwell equation $\divv B = 0$ can be written as the curl of a vector potential $A_B$ of the form $A_B(x)= \frac{1}{2} b \wedge x + A_{\mathrm{per}}(x)$, where $b$ denotes the vector with components given by the average magnetic flux of $B$ through the faces of $Q_h$ and $A_{\mathrm{per}}$ is a periodic vector potential. For more information concerning this decomposition we refer to Chapter \ref{Chapter:Abrikosov_gauge}. For a treatment of the two-dimensional case, see \cite{Tim_Abrikosov}.


\subsection{The BCS functional}
\label{BCS_functional_Section}

In BCS theory a state is conveniently described by its generalized one-particle density matrix, that is, by a self-adjoint operator $\Gamma$ on $L^2(\Rbb^3) \oplus L^2(\Rbb^3)$, which obeys $0 \leq \Gamma \leq 1$ and is of the form
\begin{align}
\Gamma = \begin{pmatrix} \gamma & \alpha \\ \ov \alpha & 1 - \ov \gamma \end{pmatrix}. \label{Gamma_introduction}
\end{align}
Here, $\ov \alpha$ denotes the operator $\alpha$ with the complex conjugate integral kernel in the position space representation. Since $\Gamma$ is self-adjoint we know that $\gamma$ is self-adjoint and that $\alpha$ is symmetric in the sense that its integral kernel satisfies $\alpha(x,y) = \alpha(y,x)$. This symmetry is related to the fact that we exclude spin degrees of freedom from our description and assume that all Cooper pair wave functions are in a spin singlet state. The condition $0 \leq \Gamma \leq 1$ implies that the one-particle density matrix $\gamma$ satisfies $0 \leq \gamma \leq 1$ and that $\alpha$ and $\gamma$ are related through the inequality
\begin{align}
\alpha \alpha^* \leq \gamma ( 1- \gamma). \label{gamma_alpha_fermionic_relation}
\end{align}

Let us define the magnetic translations $\mathbf T(\lambda)$ on $L^2(\Rbb^3)\oplus L^2(\Rbb^3)$ by
\begin{align*}
	\mathbf T(v) &:= \begin{pmatrix}
		T(v) & 0 \\ 0 & \ov{T(v)}\end{pmatrix}, & v &\in \Rbb^3.
\end{align*}
We say that a BCS state $\Gamma$ is \emph{gauge-periodic} provided $\mathbf T(\lambda) \, \Gamma \, \mathbf T(\lambda)^* = \Gamma$ holds for any $\lambda\in \Lambda_h$. This implies the relations $T(\lambda) \, \gamma \, T(\lambda)^* = \gamma$ and   $T(\lambda)\,\alpha \,\ov{T(\lambda)}^* = \alpha$, or, in terms of integral kernels,
\begin{align}
\gamma(x, y) &= \e^{\i \frac \Bbold 2 \cdot (\lambda \wedge (x-y))} \; \gamma(x+\lambda,y+ \lambda), \notag\\
\alpha(x, y) &= \e^{\i \frac \Bbold 2 \cdot (\lambda \wedge (x+y))} \; \alpha(x+\lambda,y+ \lambda), & \lambda\in \Lambda_h. \label{alpha_periodicity}
\end{align}

We further say that a gauge-periodic BCS state $\Gamma$ is \emph{admissible} if 
\begin{align}
\Tr \bigl[\gamma + (-\i \nabla + \Abold_\Bbold)^2\gamma\bigr] < \infty \label{Gamma_admissible}
\end{align}
holds. Here $\Tr[\Rcal]$ denotes the trace per unit volume of an operator $\Rcal$ defined by
\begin{align}
\Tr [\Rcal] &:= \frac{1}{|Q_h|} \Tr_{L^2(Q_h)} [\chi \Rcal \chi],  \label{Trace_per_unit_volume_definition}
\end{align}
where $\chi$ denotes the characteristic function of the cube $Q_h$ in \eqref{Fundamental_cell} and $\Tr_{L^2(Q_h)}[\cdot]$ is the usual trace over an operator on $L^2(Q_h)$. By the condition in \eqref{Gamma_admissible}, we mean that $\chi \gamma \chi$ and $\chi (-\i \nabla + \Abold_\Bbold)^2 \gamma \chi$ are trace-class operators. 
Eq.~ \eqref{gamma_alpha_fermionic_relation}, \eqref{Gamma_admissible}, and the same inequality with $\gamma$ replaced by $\overline{\gamma}$ imply that $\alpha$, $(-\i \nabla + \Abold_\Bbold)\alpha$, and $(-\i \nabla + \Abold_\Bbold) \ov \alpha$ are locally Hilbert--Schmidt. We will rephrase this property as a notion of $H^1$-regularity for the kernel of $\alpha$ in Section~\ref{Preliminaries} below.

Let $\Gamma$ be an admissible BCS state. We define the Bardeen--Cooper--Schrieffer free energy functional, or BCS functional for short, at temperature $T\geq 0$ by the formula
\begin{align}
\FBCS(\Gamma) &:= \Tr\bigl[ \bigl( (-\i \nabla + \Abold_h)^2 - \mu + W_h(x) \bigr)\gamma \bigr] - T\, S(\Gamma)  \notag \\
&\hspace{120pt} - \frac{1}{|Q_h|} \int_{Q_h} \dd X \int_{\Rbb^3} \dd r\; V(r) \, |\alpha(X,r)|^2,
\label{BCS functional}
\end{align}
where  $S(\Gamma)= - \Tr [\Gamma \ln(\Gamma)]$ denotes the von Neumann entropy per unit volume and $\mu\in \Rbb$ is a chemical potential. The interaction energy is written in terms of the center-of-mass and relative coordinates $X = \frac{x+y}{2}$ and $r = x-y$. Throughout this paper, we write, by a slight abuse of notation, $\alpha(x,y) \equiv \alpha(X,r)$. That is, we use the same symbol for the function depending on the original coordinates and for the one depending on $X$ and $r$.

The natural space for the interaction potential guaranteeing that the BCS functional is bounded from below is $V \in L^{\nicefrac 32}(\Rbb^3) + L_{\varepsilon}^{\infty}(\mathbb{R}^3)$. Under these assumptions it can be shown that the BCS functional satisfies the lower bound
\begin{align}
\FBCS(\Gamma) &\geq \frac 12 \Tr \bigl[ \gamma + (-\i \nabla + \Abold_\Bbold)^2 \gamma\bigr] - C \label{BCS functional_bounded_from_below}
\end{align}
for some constant $C>0$. In other words, the BCS functional is bounded from below and coercive on the set of admissible BCS states. 

The normal state $\Gamma_0$ is the unique minimizer of the BCS functional when restricted to admissible states with $\alpha = 0$ and reads
\begin{align}
\Gamma_0 &:= \begin{pmatrix} \gamma_0 & 0  \\ 0 &  1-\ov\gamma_0 \end{pmatrix}, & \gamma_0 &:= \frac 1{1 + \e^{ ((-\i\nabla + \Abold_h)^2 + W_h-\mu)/T}}.  \label{Gamma0}
\end{align}
Its name is motivated by the fact that it is also the unique minimizer of the BCS functional if the temperature $T$ is chosen sufficiently large.
We define the BCS free energy by
\begin{align}
F^{\mathrm{BCS}}(h, T) := \inf\bigl\{ \FBCS(\Gamma) - \FBCS(\Gamma_0) : \Gamma \text{ admissible}\bigr\} \label{BCS GS-energy}
\end{align}
and say that the system is superconducting at temperature $T$ if $F^{\mathrm{BCS}}(h, T) < 0$. Although it is not difficult to prove that the BCS functional has a minimizer, we refrain from giving a proof here. If we assume that the BCS functional has a minimizer $\Gamma$ then the condition $F^{\mathrm{BCS}}(h, T) < 0$ implies $\alpha = \Gamma_{12} \neq 0$. 

The goal of this paper is to derive an asymptotic formula for $F^{\mathrm{BCS}}(h, T)$ for small $h > 0$. This will allow us to derive Ginzburg--Landau theory and to show how the critical temperature depends on the external electric and magnetic field and on $h$. For our main results to hold, we need the following assumptions.

\begin{asmp}
\label{Assumption_V}
We assume that the interaction potential $V$ is a radial function that satisfies $(1+|\cdot|^2) V\in L^2(\mathbb{R}^3) \cap L^\infty(\Rbb^3)$. Moreover, the electric and the magnetic potentials $W\in W^{1, \infty}(\Rbb^3)$ and $A\in W^{3, \infty}(\Rbb^3; \Rbb^3)$ are $\Lambda_1$-periodic functions, i.e.  $W(x + \lambda) = W(x)$ and $A(x + \lambda) = W(x)$ for $\lambda\in \Lambda_1$ and a.e. $x\in \Rbb^3$. We also assume that $A(0) = 0$.
\end{asmp}

\subsection{The translation-invariant BCS functional}
\label{BCS_functional_TI_Section}

In the absence of external fields we describe the system by translation-invariant states, that is, we assume that the integral kernels of $\gamma$ and $\alpha$ are of the form $\gamma(x-y)$ and $\alpha(x-y)$. The trace per unit volume is in this case defined with respect to a cube with sidelength $1$. We denote the resulting translation-invariant BCS functional by $\mathcal{F}^{\mathrm{BCS}}_{\mathrm{ti},T}$ and the difference between its infimum and the energy of the normal state by $F^{\mathrm{BCS}}_{\mathrm{ti}}(T)$. The translation-invariant BCS functional is analyzed in detail in \cite{Hainzl2007}, see also the review article \cite{Hainzl2015}. In \cite{Hainzl2007} it has been shown that there exists a unique critical temperature $\Tc \geq 0$ such that $\mathcal{F}^{\mathrm{BCS}}_{\mathrm{ti},T}$ has a minimizer with $\alpha \neq 0$ for $T < \Tc$. The normal state in \eqref{Gamma0} with $h=0$ is the unique minimizer if $T\geq \Tc$. Moreover, the critical temperature $\Tc$ can be characterized by a linear criterion: It equals the unique temperature $T$ such that the linear operator
\begin{align*}
K_{T} - V 
\end{align*}
has zero as its lowest eigenvalue. Here $K_T = K_T(-\i \nabla)$ with the symbol 
\begin{align}
	K_T(p) := \frac{p^2 - \mu}{\tanh \frac{p^2-\mu}{2T}}. \label{KT-symbol}
\end{align}
The operator $K_{T} - V $ is understood to act on the space $L_{\mathrm{sym}}^2(\Rbb^3)$ of reflection-symmetric square-integrable functions on $\mathbb{R}^3$. To be precise, the results in \cite{Hainzl2007} have been proven without the assumption $\alpha(-x) = \alpha(x)$ for a.e. $x\in \mathbb{R}^3$. In this case, the operator $K_\Tc - V$ acts in the Hilbert space $L^2(\Rbb^3)$ instead of $L_{\mathrm{sym}}^2(\Rbb^3)$. The results in \cite{Hainzl2007}, however, equally hold in the case of symmetric Cooper pair wave functions.

We note that the function $K_T(p)$ satisfies the inequalities $K_T(p) \geq 2T$ for $\mu \geq 0$, as well as $K_T(p)\geq |\mu|/\tanh(|\mu|/(2T))$ for $\mu < 0$. Our assumptions on $V$ guarantee that the essential spectrum of $K_T-V$ equals that of $K_T$, and hence an eigenvalue below $2T$ for $\mu \geq 0$ or below $|\mu|/\tanh(|\mu|/(2T))$ for $\mu < 0$ is necessarily isolated and of finite multiplicity. This, in particular, applies to an eigenvalue of $K_T - V$ at $0$.

We are interested in the situation, where $\Tc > 0$ and where the translation-invariant BCS functional has a unique minimizer with a radial Cooper pair wave function (s-wave Cooper pairs) for $T$ close to $\Tc$. The following assumptions guarantee that we are in such a situation. Part (b) should be compared to \cite[Theorem~2.8]{DeuchertGeisinger}. 

\begin{asmp}
\label{Assumption_KTc}
We assume that the interaction potential $V$ is such that the following holds:
\begin{enumerate}[(a)]
\item We have $\Tc >0$. 

\item The lowest eigenvalue of $K_{\Tc} - V$ is simple.
\end{enumerate}
\end{asmp}

We refer to \cite[Theorem 3]{Hainzl2007} for a sufficient condition for $V$ that implies our first assumption. Throughout this paper we denote by $\alpha_*$ the unique solution to the equation
\begin{align}
K_\Tc \alpha_* = V\alpha_*. \label{alpha_star_ev-equation}
\end{align}
We will assume that $\alpha_*$ is real-valued and satisfies $\Vert \alpha_*\Vert_{L^2(\Rbb^3)} = 1$. If we write the above equation as $\alpha_* = K_{\Tc}^{-1} V\alpha_*$, we see that $V\in L^\infty(\Rbb^3)$ implies $\alpha_*\in H^2(\Rbb^3)$. Moreover, we know from \cite[Proposition 2]{Hainzl2012} that
\begin{align}
\int_{\Rbb^3} \dd x \; \bigl[ |x^\nu \alpha_*(x)|^2 + |x^\nu \nabla \alpha_*(x)|^2 \bigr] < \infty \label{Decay_of_alphastar}
\end{align}
holds for $\nu \in \Nbb_0^3$.


\subsection{The Ginzburg--Landau functional}
\label{Ginzburg-Landau-functional_Section}

We say that a function $\Psi$ on $Q_h$ is \textit{gauge-periodic} if the magnetic translations of the form
\begin{align}
	T_h(\lambda)\Psi(X) &:= \e^{\i \Bbold \cdot (\lambda \wedge X)} \; \Psi(X + \lambda ), & \lambda &\in \Lambda_h, \label{Magnetic_Translation_Charge2}
\end{align}
leave $\Psi$ invariant. We highlight that $T(\lambda)$ in \eqref{Magnetic_Translation} equals $T_h(\lambda)$ provided we replace $\Bbold$ by $2\Bbold$. Let $\Lambda_0, \Lambda_2, \Lambda_3 >0$, $\Lambda_1, D \in \Rbb$, and let $\Psi$ be a gauge-periodic function. The Ginzburg--Landau functional is defined by
\begin{align}
\EGL(\Psi) &:= \frac{1}{h^4} \frac{1}{|Q_h|} \int_{Q_h} \dd X \; \bigl\{ \Lambda_0 \; |(-\i\nabla + 2\Abold_h)\Psi(X)|^2 + \Lambda_1 \, W_h(X)\, |\Psi(X)|^2 \notag \\
&\hspace{150pt} - Dh^2 \, \Lambda_2\, |\Psi(X)|^2 + \Lambda_3\,|\Psi(X)|^4\bigr\}. \label{Definition_GL-functional}
\end{align}
We emphasize the factor $2$ in front of the magnetic vector potential in \eqref{Definition_GL-functional}. Its appearance is due to the fact that $\Psi$ describes the center-of-mass motion of Cooper pairs carrying twice the charge of a single fermion.

The Ginzburg--Landau energy is defined by
\begin{align*}
E^{\mathrm{GL}}(D) := \inf \bigl\{ \EGL(\Psi) : \Psi\in \Hmag^1(Q_h)\bigr\}.
\end{align*}
By scaling, it is independent of $h$. More precisely, for given $\psi$ the function
\begin{align}
\Psi(X) &:= h \; \psi\bigl(h \, X\bigr), & X\in \Rbb^3, \label{GL-rescaling}
\end{align}
obeys
\begin{align}
\EGL(\Psi) = \mathcal E_{D,1}^{\mathrm{GL}}(\psi). \label{EGL-scaling}
\end{align}
We also define the critical parameter
\begin{align}
\Dc &:= \frac 1{\Lambda_2} \inf \spec_{\Lmag^2(Q_1)} \bigl(\Lambda_0 \, (-\i \nabla + \Abold)^2 + \Lambda_1 \; W\bigr). \label{Dc_Definition}
\end{align}
As has been shown in \cite[Lemma 2.5]{Hainzl2014}, we have $\EGLGSE < 0$ if $D > \Dc$ and $\EGLGSE =0$ if $D \leq \Dc$.


\subsection{Main results}
\label{Main_Result_Section}

Our first main result concerns an asymptotic expansion of the BCS free energy in the small parameter $h>0$. The precise statement is captured in the following theorem.

\begin{bigthm}
\label{Main_Result}
Let Assumptions \ref{Assumption_V} and \ref{Assumption_KTc} hold, let $D \in \Rbb$, and let the coefficients $\Lambda_0, \Lambda_1, \Lambda_2$, and $\Lambda_3$ be given by \eqref{GL-coefficient_1}-\eqref{GL_coefficient_3} below. Then there are constants $C>0$ and $h_0 >0$ such that for all $0 < h \leq h_0$, we have
\begin{align}
F^{\mathrm{BCS}}(h,\, \Tc(1 - Dh^2)) = h^4 \; \bigl( \EGLGSE + R \bigr), \label{ENERGY_ASYMPTOTICS}
\end{align}
with $R$ satisfying the estimate
\begin{align}
Ch \geq R \geq - \Rcal := -C h^{\nicefrac 1{6}}. \label{Rcal_error_Definition}
\end{align}
Moreover, for any approximate minimizer $\Gamma$ of $\FBCS$ at $T = \Tc(1 - Dh^2)$ in the sense that
\begin{align}
\FBCS(\Gamma) - \FBCS(\Gamma_0) \leq h^4 \bigl( \EGLGSE + \rho\bigr) 
\label{BCS_low_energy}
\end{align}
holds for some $\rho \geq 0$, we have the decomposition
\begin{align}
\alpha(X, r ) = \alpha_*(r) \Psi(X)  + \sigma(X,r) \label{Thm1_decomposition}
\end{align}
for the Cooper pair wave function $\alpha = \Gamma_{12}$. Here, $\sigma$ satisfies
\begin{align}
\frac{1}{|Q_h|} \int_{Q_h} \dd X \int_{\Rbb^3} \dd r \; |\sigma(X, r)|^2 &\leq C h^{\nicefrac {11}3}, \label{Thm1_error_bound}
\end{align}
$\alpha_*$ is the normalized zero energy eigenstate of $K_{\Tc}-V$, and the function $\Psi$ obeys
\begin{align}
\EGL(\Psi) \leq \EGLGSE + \rho + \Rcal. \label{GL-estimate_Psi}
\end{align}
\end{bigthm}

Our second main result is a statement about the dependence of the critical temperature of the BCS functional on $h>0$ and on the external fields.

\begin{bigthm}
\label{Main_Result_Tc} \label{MAIN_RESULT_TC}
Let Assumptions \ref{Assumption_V} and \ref{Assumption_KTc} hold. Then there are constants $C>0$ and $h_0 >0$ such that for all $0 < h \leq h_0$ the following holds:
\begin{enumerate}[(a)]
	\item Let $0 < T_0 < \Tc$. If the temperature $T$ satisfies
	\begin{equation}
		T_0 \leq T \leq \Tc \, ( 1 - h^2 \, ( \Dc + C \, h^{\nicefrac 12}))
		\label{eq:lowertemp}
	\end{equation}
	with $\Dc$ in \eqref{Dc_Definition}, then we have
	\begin{equation*}
		F^{\mathrm{BCS}}(h,T) < 0.
	\end{equation*} 
	\item If the temperature $T$ satisfies
	\begin{equation}
		T \geq \Tc \, ( 1 - h^2 \, ( \Dc - \Rcal ) )
		\label{eq:uppertemp}
	\end{equation}
	with $\Dc$ in \eqref{Dc_Definition} and $\Rcal$ in \eqref{Rcal_error_Definition}, then we have
	\begin{equation*}
		\FBCS(\Gamma) - \FBCS(\Gamma_0) > 0
	\end{equation*}
	unless $\Gamma = \Gamma_0$.
\end{enumerate}
\end{bigthm}

\begin{bems}
\label{Remarks_Main_Result}
\begin{enumerate}[(a)]
\item Theorem~\ref{Main_Result} and Theorem~\ref{Main_Result_Tc} extend similar results in \cite{Hainzl2012} and \cite{Hainzl2014} to the case of general external electric and magnetic fields. In these references the main restriction is that the vector potential is assumed to be periodic, that is, the corresponding magnetic field has vanishing flux through the faces of the unit cell $Q_h$, compare with the discussion in Section~\ref{Magnetically_Periodic_Samples}. Removing this restriction causes major mathematical difficulties because the vector potential of a constant magnetic field cannot be treated as a perturbation of the Laplacian. More precisely, it was possible in \cite{Hainzl2012,Hainzl2014} to work with a priori bounds for low-energy states that do not involve the externel magnetic field. As noticed in the discussion below Remark~6 in \cite{Hainzl2017}, this is not possible if the magnetic field has nonzero flux through the faces of the unit cell. To prove a priori bounds that involve a constant magnetic field one has to deal with the fact that the components of the magnetic momentum operator do not commute, which leads to significant technical difficulties. These difficulties have been overcome in \cite{DeHaSc2021}, which allowed us to extend the results \cite{Hainzl2012,Hainzl2014} to the case of a system in a constant magnetic field. Our proof of Theorem~\ref{Main_Result} and Theorem~\ref{Main_Result_Tc} uses these results, and should therefore be interpreted as an extension of the methods in \cite{DeHaSc2021} to the case of general external electric and magnetic fields.

\item When we compare our result in Theorem~\ref{Main_Result} to the main Theorem in \cite{Hainzl2012}, we notice two differences. The first is that our BCS energy is normalized by a volume factor while this is not the case in \cite{Hainzl2012}. The second difference is that the Cooper pair wave function of an approximate minimizer of the BCS functional is decomposed in \cite{Hainzl2012} as
\begin{equation*}
	\alpha(x,y) = \frac{1}{2} \alpha_*(x-y) \left( \Psi(x) + \Psi(y) \right) + \xi(x,y)
\end{equation*}
for some function $\xi$, which should be compared to \eqref{Thm1_decomposition}. When we use the a priori bound for $\Vert \nabla \Psi \Vert_2$ below Eq.~(5.61) in \cite{Hainzl2012}, we see that this decomposition equals that in \eqref{Thm1_decomposition} to leading order in $h$. The analogue in our setting does not seem to be correct.

\item The Ginzburg--Landau energy appears at the order $h^4$. This needs to be compared to the energy of the normal state, which is of order $1$ in $h$.

\item The size of the remainder in \eqref{Thm1_error_bound} should be compared to the $L^2$-norm per unit volume of the leading order part of the Cooper pair wave function in \eqref{Thm1_decomposition}, which satisfies
\begin{equation*}
	\frac{1}{| Q_h |} \int_{Q_h} \dd X \int_{\mathbb{R}^3} \dd r \; | \alpha_*(x) \Psi(X) |^2 = O(h^2).
\end{equation*}

\item Our bounds show that $D$ in Theorem~\ref{Main_Result} can be chosen as a function of $h$ as long as $|D| \leq D_0$ holds for some constant $D_0>0$.

\item The upper bound for the error in \eqref{Rcal_error_Definition} is worse than the corresponding bound in \cite{DeHaSc2021} by the factor $h^{-1}$. It is of the same size as the comparable error term in \cite[Theorem~1]{Hainzl2012}.

\item Theorem~\ref{MAIN_RESULT_TC} gives bounds on the temperature regions where superconductivity is present or absent. The interpretation of the theorem is that the critical temperature of the full model satisfies
\begin{equation*}
	\Tc(h) = \Tc \left( 1 - D_{\mathrm{c}} h^2 \right) + o(h^2),
\end{equation*}
with the critical temperature $\Tc$ of the translation-invariant problem. The coefficient $D_{\mathrm{c}}$ is determined by linearized Ginzburg--Landau theory, see \eqref{Magnetic_Translation_Charge2}. The above equation allows us to compute the upper critical field $B_{\mathrm{c}2}$, above which superconductivity is absent. It also allows to to compute the derivative of $B_{\mathrm{c}2}$ with respect to $T$ at $\Tc$, see \cite[Appendix~A]{Hainzl2017}.

\item We expect that the assumption $0 < T_0 < \Tc$ in part (a) of Theorem~\ref{MAIN_RESULT_TC}, which also appeared in \cite{Hainzl2017}, is only of technical nature. We need it because our trial state analysis breaks down as $T$ approaches zero. We note that there is no such restriction in part (b) of Theorem~\ref{MAIN_RESULT_TC} or in Theorem~\ref{Main_Result}.
\end{enumerate}
\end{bems}


\subsection{Organization of the paper and strategy of proof}

To a large extent, our proof relies on the same strategy as has been pursued in \cite{DeHaSc2021} and the numbering of the sections is identical in order to make comparison easy.

In Section~\ref{Preliminaries}, the introduction of our mathematical setup is completed. We collect useful properties of the trace per unit volume and introduce the relevant spaces of gauge-periodic functions. 

Section~\ref{Upper_Bound} presents the extension of the trial state analysis in \cite{DeHaSc2021}. We introduce a class of Gibbs states having Cooper pair wave functions, which admit a product structure of the form $\alpha_*(r) \Psi(X)$ to leading order in $h$. Here, $\alpha_*$ is the ground state in \eqref{alpha_star_ev-equation} and $\Psi$ is a gauge-periodic function. We state the results that pertain to the structure and the BCS energy of these states. With the help of these results, we provide the proofs of the upper bound on \eqref{ENERGY_ASYMPTOTICS} as well as Theorem \ref{Main_Result_Tc} (a). We will also need these results in Section \ref{Lower Bound Part B}, where we give the proofs of the lower bound on \eqref{ENERGY_ASYMPTOTICS} and of Theorem \ref{Main_Result_Tc} (b). 

Section~\ref{Proofs} marks the main part of this paper, in which we present the proof of the trial state analysis in Section~\ref{Upper_Bound}. Our method of proof is based on an extensive use of the phase approximation method for general magnetic fields. In the case of a constant magnetic field, this method has been pioneered within the framework of BCS theory in \cite{Hainzl2017, DeHaSc2021}, whereas the case of magnetic fields with zero flux through the unit cell is contained in the unpublished notes \cite{BdGtoGL}. We extend these results to our setting with general external fields and provide the relevant weak magnetic field estimates for the proof of the results in Section \ref{Upper_Bound}. Within this general framework, it should be noted that these estimates require considerable additional effort in comparison to \cite{DeHaSc2021}, which is also reflected in the length of the proofs. In contrast, the effect of the electric field can be studied by straightforward perturbation theory. Some of the auxiliary results, which have been proven in \cite{DeHaSc2021}, hold in this more general case as well and require only minimal adjustments in the proof. Therefore, we refrain from giving these proofs a second time and instead refer to \cite{DeHaSc2021}.

In Section~\ref{Lower Bound Part A}, we prove a priori estimates for BCS states, whose BCS free energy is low in the sense that it is less than or equal to that of the normal state $\Gamma_0$ in \eqref{Gamma0} plus a correction of the order $h^4$. The proof of this result in the case of a constant magnetic field has been the main novelty of the work \cite{DeHaSc2021}, and we adapt it to the more general situation in this paper. Our method relies on the perturbative removal of the additional external fields $A$ and $W$ and a subsequent use of the results in \cite{DeHaSc2021}. In this way, we obtain a decomposition result for the Cooper pair wave function of the same form as that of the Gibbs states in Section~\ref{Upper_Bound}.

The proof of the lower bound on \eqref{ENERGY_ASYMPTOTICS} and of Theorem~\ref{Main_Result_Tc}~(b) is provided in Section~\ref{Lower Bound Part B} and follows the same lines as that presented in \cite{DeHaSc2021}. This, in particular, completes the proofs of Theorem \ref{Main_Result} and \ref{Main_Result_Tc}. The a priori estimates in Section \ref{Lower Bound Part A} allow us to replace a low-energy state by a suitable Gibbs state and to compute the BCS energy of the latter, since the leading behavior of the Cooper pair wave functions is the same. The errors are controlled by our trial state analysis of Section \ref{Upper_Bound}. Because of the large overlap in content with \cite{DeHaSc2021}, we will shorten the proofs to a minimal length here.

Throughout the paper, $c$ and $C$ denote generic positive constants that change from line to line. We allow them to depend on the various fixed quantities like $h_0$, $D_0$, $\mu$, $\Tc$, $V$, $A$, $W$, $\alpha_*$, etc. Further dependencies are indexed.


\section{Preliminaries}
\label{Preliminaries}


\subsection{Schatten classes}
\label{Schatten_Classes}

The trace per unit volume in \eqref{Trace_per_unit_volume_definition} gives rise to periodic Schatten classes and the Schatten norms of periodic operators play an important role during our proofs. In the following lines, we recall several facts about these classes.

For $1 \leq p < \infty$, the $p$\tho\ local von-Neumann--Schatten class $\Scal^p$ consists of all gauge-periodic operators $A$ having finite $p$-norm, that is, $\Vert A\Vert_p^p := \Tr (|A|^p) <\infty$. The space of bounded gauge-periodic operators $\Scal^\infty$ is equipped with the usual operator norm. We note that the $p$-norm is not monotone decreasing in the index $p$. This should be compared to the usual Schatten norms, where such a property holds, see the discussion below \cite[Eq. (3.9)]{Hainzl2012}.

We recall that the triangle inequality 
\begin{align*}
\Vert A + B\Vert_p \leq \Vert A\Vert_p + \Vert B\Vert_p
\end{align*}
holds on $\Scal^p$ for $1 \leq p \leq \infty$. We also have the generalized version of Hölder's inequality
\begin{align}
\Vert AB\Vert_r \leq \Vert A\Vert_p \Vert B\Vert_q, \label{Schatten-Hoelder}
\end{align}
which holds for $1 \leq p,q,r \leq \infty$ with $\frac{1}{r} = \frac{1}{p} + \frac{1}{q}$. The familiar inequality
\begin{align*}
| \Tr A | \leq \Vert A \Vert_1
\end{align*}
holds in the case of local Schatten norms as well.

The above inequalities can be deduced from their versions in the case of usual Schatten norms, see, e.g., \cite{Simon05}, with the help of the magnetic Bloch--Floquet decomposition. We refer to \cite[Section XIII.16]{Reedsimon4} for an introduction to the Bloch--Floquet transformation and to \cite{Stefan_Peierls} for a particular treatment of the magnetic case. To be more precise, a gauge-periodic operator $A$ satisfies the unitary equivalence 
\begin{align*}
A \cong \int^{\oplus}_{[0,\sqrt{ 2 \pi }\, h]^3} \mathrm{d} k \;  A_{k},
\end{align*}
which we use to write the trace per unit volume as
\begin{align}
\Tr A = \int_{[0,\sqrt{ 2 \pi }\, h]^3} \frac{\dd k}{(2\pi)^3} \; \Tr_{L^2(Q_h)} A_{k}. \label{eq:ATPUV}
\end{align}
Here, $\Tr_{L^2(Q_h)}$ denotes the usual trace over $L^2(Q_h)$. When we use that $(AB)_k = A_k B_k$ holds for operators $A$ and $B$, the above mentioned inequalities for the trace per unit volume are implied by their usual versions.


\subsection{Gauge-periodic Sobolev spaces}
\label{Periodic Spaces}

The center-of-mass part of Cooper pair wave functions is described by gauge-periodic functions. In the following we introduce the relevant spaces for these. 

An $L_\loc^p(\Rbb^3)$-functions $\Psi$ belongs to the space $L_{\mathrm{mag}}^p(Q_h)$, where $1 \leq p < \infty$, provided $T_h(\lambda)\Psi = \Psi$ holds for all $\lambda\in\Lambda_h$ (with $T_h(\lambda)$ in \eqref{Magnetic_Translation_Charge2}). We endow $L_{\mathrm{mag}}^p(Q_h)$ with the usual $p$-norm per unit volume
\begin{align}
\Vert \Psi\Vert_{\Lmag^p(Q_h)}^p &:= \fint_{Q_h} \dd X \; |\Psi(X)|^p := \frac{1}{|Q_h|} \int_{Q_h} \dd X \; |\Psi(X)|^p. \label{Periodic_p_Norm}
\end{align}
As usual, we use the abbreviation $\Vert \Psi\Vert_p$.

Analogously, for $m\in \Nbb_0$, we define the Sobolev spaces of gauge-periodic functions corresponding to the constant magnetic field part as
\begin{align}
\Hmag^m(Q_h) &:= \bigl\{ \Psi\in \Lmag^2(Q_h) :  (-\i\nabla + 2\Abold_\Bbold)^\nu \Psi\in \Lmag^2(Q_h) \quad \forall \nu\in \Nbb_0^3, |\nu|_1\leq m\bigr\}, \label{Periodic_Sobolev_Space}
\end{align}
where $|\nu |_1 := \sum_{i=1}^3 \nu_i$ for $\nu\in \Nbb_0^3$. It is a Hilbert space endowed with the scalar product
\begin{align}
\langle \Phi, \Psi\rangle_{\Hmag^m(Q_h)} &:= \sum_{|\nu|_1\leq m} h^{-2 - 2|\nu|_1} \; \langle (-\i\nabla + 2\Abold_\Bbold)^\nu \Phi, (-\i \nabla + 2\Abold_\Bbold)^\nu \Psi\rangle_{\Lmag^2(Q_h)}.  \label{Periodic_Sobolev_Norm}
\end{align}
It is noteworthy that if $\Psi$ is a gauge-periodic function then so is $(-\i \nabla + 2\Abold_\Bbold)^\nu \Psi$, since the magnetic momentum operator
\begin{align}
\Pi := -\i \nabla + 2\Abold_\Bbold \label{Pi_definition}
\end{align}
commutes with the magnetic translations $T_h(\lambda)$ in \eqref{Magnetic_Translation_Charge2}. Furthermore, $\Pi$ is a self-adjoint operator on $\Hmag^1(Q_h)$. The full magnetic momentum operator reads $\Pi_{\Abold_h}$ with
\begin{align*}
\Pi_\Abold := -\i \nabla + 2 \Abold. 
\end{align*}

The norms introduced in \eqref{Periodic_p_Norm} and \eqref{Periodic_Sobolev_Norm} feature a scaling behavior in $h$, which is motivated by the Ginzburg--Landau scaling in \eqref{GL-rescaling}. More precisely, if $\psi \in \Lmag^p(Q_1)$ and $\Psi$ is as in \eqref{GL-rescaling}, then
\begin{align}
\Vert \Psi\Vert_{\Lmag^p(Q_h)} = h \, \Vert \psi\Vert_{\Lmag^p(Q_1)} \label{Periodic_p_Norm_scaling}
\end{align}
for every $1\leq p \leq \infty$. Meanwhile, the scaling of the norm in \eqref{Periodic_Sobolev_Norm} is
\begin{align*}
\Vert \Psi\Vert_{\Hmag^m(Q_h)} = \Vert \psi\Vert_{\Hmag^m(Q_1)}.
\end{align*}
This follows from \eqref{Periodic_p_Norm_scaling} and the fact that $\Vert (-\i \nabla + 2\Abold_\Bbold)^\nu\Psi\Vert_2^2$ scales as $h^{2 + 2|\nu|_1}$ for $\nu\in \Nbb_0^3$.

For the sake of completeness, let us record the following magnetic Sobolev inequality, which we will make use of several times throughout the paper. For any $h>0$ and any $\Psi\in \Hmag^1(Q_h)$, we have
\begin{align}
\Vert \Psi\Vert_{\Lmag^6(Q_h)}^2 &\leq  C \, h^{-2}\, \Vert (-\i \nabla + 2\Abold_\Bbold)\Psi\Vert_{\Lmag^2(Q_h)}^2. \label{Magnetic_Sobolev}
\end{align}
The proof can be found in \cite[Eq. (2.7)]{DeHaSc2021}.


The Cooper pair wave function $\alpha$, which is the offdiagonal entry of an admissible state $\Gamma$, belongs to the Hilbert--Schmidt class $\Scal^2$, defined in Section \ref{Schatten_Classes}, see the discussion below \eqref{Trace_per_unit_volume_definition}. The symmetry and the gauge-periodicity of the kernel of $\alpha$ in \eqref{alpha_periodicity} can be reformlated as
\begin{align}
\alpha(X,r) &= \e^{\i \Bbold \cdot (\lambda \wedge X)} \; \alpha(X+ \lambda, r), \quad \lambda\in \Lambda_h; & \alpha(X,r) &= \alpha(X, -r) \label{alpha_periodicity_COM}
\end{align}
in terms of center-of-mass and relative coordinates. In other words, $\alpha(X,r)$ is a gauge-periodic function of the center-of-mass coordinate $X$ and a reflection-symmetric function of the relative coordinate $r \in \mathbb{R}^3$. We make use of the unitary equivalence of $\Scal^2$ to the space
\begin{align*}
\Lsymm := \Lmag^2(Q_h) \otimes L_{\mathrm{sym}}^2(\Rbb^3),
\end{align*}
which consists of the square-integrable functions satisfying \eqref{alpha_periodicity_COM}, with finite norm given by
\begin{align*}
\Vert \alpha\Vert_{\Lsymm}^2 := \fint_{Q_h} \dd X\int_{\Rbb^3} \dd r \; |\alpha(X, r)|^2 = \frac{1}{|Q_h|} \int_{Q_h} \dd X\int_{\Rbb^3} \dd r \; |\alpha(X, r)|^2.
\end{align*} 
The identity $\Vert \alpha\Vert_2 = \Vert \alpha\Vert_{\Lsymm}$ follows from \eqref{alpha_periodicity_COM}. Therefore, we identify the scalar products $\langle \cdot, \cdot\rangle$ on $\Lsymm$ and $\Scal^2$ with each other and we do not distinguish operators in $\Scal^2$ and their kernels as this does not lead to confusion. 

We define the Sobolev space $H^1(Q_h\times \Rbb_{\mathrm s}^3)$ of all functions $\alpha\in L^2(Q_h\times \Rbb_{\mathrm s}^3)$, which have finite $H^1$-norm defined as
\begin{align}
\Vert \alpha\Vert_{H^1(Q_h\times \Rbb_{\mathrm s}^3)}^2 &:= \Vert \alpha\Vert_2^2 + \Vert \Pi\alpha\Vert_2^2 + \Vert \tilde \pi\alpha\Vert_2^2. \label{H1-norm}
\end{align}
The magnetic momentum operators used in this definition are given by
\begin{align}
\Pi &:= -\i\nabla_X + 2 \Abold_\Bbold(X), & \tilde \pi &:= -\i\nabla_r + \frac 12  \Abold_\Bbold(r). \label{Magnetic_Momenta_COM}
\end{align}
The full magnetic momentum operators are given by $\Pi_{\Abold_h, X}$ and $\tilde \pi_{\Abold_h, r}$ with
\begin{align}
\Pi_{\Abold} &:= -\i\nabla_X + 2 \Abold, & \tilde \pi_{\Abold} &:= -\i\nabla_r + \frac 12 \Abold(r). \label{Magnetic_Momenta_full_COM}
\end{align}
We highlight that the norm in \eqref{H1-norm} is equivalent to the norm defined as
\begin{align}
\Tr [\alpha\alpha^*] + \Tr [(-\i \nabla + \Abold_\Bbold)\alpha \alpha^* (-\i \nabla + \Abold_\Bbold)]  + \Tr [(-\i \nabla + \Abold_\Bbold) \alpha^* \alpha (-\i \nabla + \Abold_\Bbold)]. \label{Norm_equivalence_1}
\end{align}
The latter is further equivalent to the norm defined as
\begin{align}
\Vert \alpha\Vert_2^2 + \Vert (-\i \nabla + \Abold_\Bbold)\alpha\Vert_2^2 + \Vert \alpha (-\i \nabla + \Abold_\Bbold)\Vert_2^2, \label{Norm_equivalence_2}
\end{align}
compare also with the discussion below \eqref{Trace_per_unit_volume_definition}.


\subsection{Periodic Sobolev spaces}

For our external fields $A$ and $W$, we define the spaces of periodic functions
\begin{align}
\Lper := \{ f\in L^\infty(\Rbb^3) : f(x+\lambda) = f(x) \text{ a.e. in } \Rbb^3, \lambda\in \Lambda_1 \} \label{Lper_definition}
\end{align}
and
\begin{align}
\Lpervec := \{ f\in L^\infty(\Rbb^3;\Rbb^3) : f(x+\lambda) = f(x) \text{ a.e. in } \Rbb^3, \lambda\in \Lambda_1 \}. \label{Lpervec_definition}
\end{align}
Likewise, for $m\in \Nbb$, we define the periodic Sobolev spaces
\begin{align}
\Wper m := \{ f\in \Lper : (-\i \nabla)^\nu f\in \Lper, \quad \forall \nu\in \Nbb_0^3, |\nu|_1\leq mm\} \label{Wper_definition}
\end{align}
and
\begin{align}
\Wpervec m := \{f\in \Lpervec : (-\i \nabla)^\nu f\in \Lper, \quad \forall \nu\in \Nbb_0^3, |\nu|_1\leq m\}. \label{Wpervec_definition}
\end{align}
These spaces are endowed with their usual sup-norms over $Q_1$.

For the external fields $A$ and $W$, Assumption \ref{Assumption_V} is equivalent to $A \in \Wpervec 3$, while $W\in \Wper 1$.


\section{Trial States and their BCS Energy}
\label{Upper_Bound} \label{UPPER_BOUND}

In this section, we extend the trial state analysis presented in \cite{DeHaSc2021} to the case of our general external field setting. In order to do this, we state the extensions of the results presented in \cite[Section 3]{DeHaSc2021}, which are needed to prove the upper bound on \eqref{ENERGY_ASYMPTOTICS} and Theorem~\ref{Main_Result_Tc}~(a). For these proofs, we refer to \cite[Section 3.3]{DeHaSc2021}.

Our trial state analysis involves the Gibbs states $\Gamma_\Delta$. These are constructed upon a gap function $\Delta$ in terms of the effective Hamiltonian. Our first structural result, Proposition~\ref{Structure_of_alphaDelta}, shows that, if $\Delta$ is a product function in terms of the center-of-mass and relative coordinates which is small in a suitable sense, then the Cooper pair wave function $\alpha_\Delta$ of $\Gamma_\Delta$ admits a product structure to leading order as well. This, in particular, implies that $\Gamma_{\Delta}$ approximately solves the Euler--Lagrange equation of the BCS functional in the vicinity of the critical temperature. Therefore, $\Gamma_\Delta$ is a good candidate for an approximate minimizer. In order to compute the BCS energy of the trial states $\Gamma_\Delta$, we prove a representation formula for the BCS functional in Proposition~\ref{BCS functional_identity}. Parts of the expression we provide are shown to equal the Ginzburg--Landau functional in \eqref{Definition_GL-functional} up to a small error along the temperatures $T = \Tc (1 - Dh^2)$ for some $D\in\Rbb$. We postpone the proof of these results until Section \ref{Proofs}.


\subsection{The Gibbs states \texorpdfstring{$\Gamma_\Delta$}{GammaDelta}}

The gap function $\Delta\in \Lsymm$ is defined upon a $\Psi\in \Lmag^2(Q_h)$ as
\begin{align}
\Delta(X,r) := \Delta_\Psi(X, r) &:= -2 \; V\alpha_*(r) \Psi(X).  \label{Delta_definition}
\end{align}
The gauge-periodic function $\Psi$ is a minimizer of the Ginzburg--Landau functional in \eqref{Definition_GL-functional} as far as the trial state analysis of this section is concerned. Since $\Psi$ satisfies the scaling in \eqref{GL-rescaling}, the local Hilbert--Schmidt norm of $\Delta$ is of the order $h$, while the $L^2(\Rbb^3)$-norm of $V\alpha_*$ is of the order $1$. Therefore, $\Psi$ fully determines the size of $\Vert \Delta\Vert_2$. The notation
\begin{align}
\hfrak_{\Abold, W} &:= \hfrak_\Abold + W := (-\i \nabla +\Abold_h )^2 + W_h - \mu \label{hfrakAW_definition}
\end{align}
allows us to introduce the Hamiltonian
\begin{align}
H_{\Delta} &:= H_0 + \delta := \begin{pmatrix}
\hfrak_{\Abold, W} & 0 \\ 0 & -\ov{\hfrak_{\Abold, W}}
\end{pmatrix} + \begin{pmatrix}
0 & \Delta \\ \ov \Delta & 0
\end{pmatrix} = \begin{pmatrix}
\hfrak_{\Abold, W} & \Delta \\ \ov \Delta & -\ov {\hfrak_{\Abold, W}}
\end{pmatrix}. \label{HDelta_definition}
\end{align}
This enables us to define the corresponding Gibbs state at inverse temperature $\beta = T^{-1} >0$ by
\begin{align}
\begin{pmatrix} \gamma_\Delta & \alpha_\Delta \\ \ov{\alpha_\Delta} & 1 - \ov{\gamma_\Delta}\end{pmatrix} = \Gamma_\Delta := \frac{1}{1 + \e^{\beta H_\Delta}}. \label{GammaDelta_definition}
\end{align}
In this way, setting $\Delta =0$ yields the normal state $\Gamma_0$ in \eqref{Gamma0}.

\begin{lem}[Admissibility of $\Gamma_\Delta$]
\label{Gamma_Delta_admissible}
Let Assumptions \ref{Assumption_V} and \ref{Assumption_KTc} hold. Then, for any $h>0$, any $T>0$, and any $\Psi\in \Hmag^1(Q_h)$, the state $\Gamma_\Delta$ in \eqref{GammaDelta_definition} is admissible.
\end{lem}

The states $\Gamma_\Delta$ are inspired by the solution $\Gamma$ of the nonlinear Bogolubov--de Gennes equation, the Euler--Lagrange equation of the BCS functional,
\begin{align}
\Gamma &= \frac 1{1 + \e^{\beta \, \Hbb_{V\alpha}}}, & \Hbb_{V\alpha} = \begin{pmatrix} \hfrak_{\Abold, W} & -2\, V\alpha \\ -2\, \ov{V\alpha} & -\ov{\hfrak_{\Abold, W}}\end{pmatrix}. \label{BdG-equation}
\end{align}
Here, $V\alpha$ is the operator given by the kernel $V(r)\alpha(X,r)$. The following result shows that our candidate $\Gamma_\Delta$ for an approximate solution to \eqref{BdG-equation} is indeed appropriate as far as the leading term is concerned, compare this to \eqref{Thm1_decomposition}.


\begin{prop}[Structure of $\alpha_\Delta$]
\label{Structure_of_alphaDelta} \label{STRUCTURE_OF_ALPHADELTA}
Let Assumption \ref{Assumption_V} and \ref{Assumption_KTc} (a) be satisfied and let $T_0>0$ be given. Then, there is a constant $h_0>0$ such that for any $0 < h \leq h_0$, any $T\geq T_0$, and any $\Psi\in \Hmag^2(Q_h)$ the function $\alpha_\Delta$ in \eqref{GammaDelta_definition} with $\Delta \equiv \Delta_\Psi$ as in \eqref{Delta_definition} has the decomposition
\begin{align}
\alpha_\Delta(X,r) &= \Psi(X) \alpha_*(r) - \eta_0(\Delta)(X,r) - \eta_{\perp}(\Delta)(X,r). \label{alphaDelta_decomposition_eq1}
\end{align}
The remainder functions $\eta_0(\Delta)$ and $\eta_\perp(\Delta)$ have the following properties:
\begin{enumerate}[(a)]
\item The function $\eta_0$ satisfies the bound
\begin{align}
\Vert \eta_0\Vert_\Hsymm^2 &\leq  C\; \bigl( h^5 + h^2 \, |T - \Tc|^2\bigr) \; \bigl(  \Vert \Psi\Vert_{\Hmag^1(Q_h)}^6 + \Vert \Psi\Vert_{\Hmag^1(Q_h)}^2\bigr). \label{alphaDelta_decomposition_eq2}
\end{align}

\item The function $\eta_\perp$ satisfies the bound
\begin{align}
\Vert \eta_\perp\Vert_{\Hsymm}^2 + \Vert |r|\eta_\perp\Vert_{\Lsymm}^2 &\leq C \; h^6 \; \Vert \Psi\Vert_{\Hmag^2(Q_h)}^2. \label{alphaDelta_decomposition_eq3}
\end{align}

\item The function $\eta_\perp$ has the explicit form
\begin{align*}
\eta_\perp(X, r) &= \int_{\Rbb^3} \dd Z \int_{\Rbb^3} \dd s \; k_T(Z, r-s) \, V\alpha_*(s) \, \bigl[ \cos(Z\cdot \Pi) - 1\bigr] \Psi(X)
\end{align*}
with $k_T(Z,r)$ defined in Section~\ref{Proofs} below \eqref{MTA_definition}. For any radial $f,g\in L^2(\Rbb^3)$ the operator
\begin{align*}
\iiint_{\Rbb^9} \dd Z \dd r \dd s \; f(r) \, k_T(Z, r-s) \, g(s) \, \bigl[ \cos(Z\cdot \Pi) - 1\bigr]
\end{align*}
commutes with $\Pi^2$, and, in particular, 
if $P$ and $Q$ are two spectral projections of $\Pi^2$ with $P Q = 0$, then $\eta_\perp$ satisfies the orthogonality property
\begin{align}
	\bigl\langle f(r) \, (P \Psi)(X), \, \eta_{\perp}(\Delta_{Q\Psi}) \bigr\rangle = 0.
	\label{alphaDelta_decomposition_eq4}
\end{align}
\end{enumerate}
\end{prop}

\begin{bem}
The reason why the statement of Proposition \ref{Structure_of_alphaDelta} may seem somewhat complicated lies in the different techniques of proof for the upper and lower bounds on the BCS functional. The detailed reasoning can be found in \cite[Remark 3.3]{DeHaSc2021}.

\end{bem}


\subsection{The BCS energy of the states \texorpdfstring{$\Gamma_\Delta$}{GammaDelta}}

In this section, we analyze the BCS energy of the states $\Gamma_\Delta$, and we are going to show that it is determined by the Ginzburg--Landau functional to leading order in the appropriate temperature scaling. In \cite[Section 3.2]{DeHaSc2021}, it has been argued that the BCS energy of $\Gamma_\Delta$ can be calculated in terms of the operators $L_{T, \Abold, W}$ and $N_{T, \Abold, W}$, which we define in the following lines. 
With the Matsubara frequencies 
\begin{align}
\omega_n &:= \pi (2n+1) T, \qquad n \in \Zbb, \label{Matsubara_frequencies}
\end{align}
we define the linear operator $L_{T, \Abold, W} \colon \Lsymm \ra \Lsymm$ given by
\begin{align}
L_{T, \Abold, W}\Delta &:= -\frac 2\beta \sum_{n\in \Zbb} \frac 1{\i \omega_n - \hfrak_{\Abold, W}} \, \Delta \,  \frac 1 {\i \omega_n + \ov{\hfrak_{\Abold, W}}}. \label{LTAW_definition}
\end{align}
In the temperature regime we are interested in, we will obtain the quadratic terms in the Ginzburg--Landau functional from $\langle \Delta, L_{T, \Abold, W}\Delta\rangle$. 


We also define the nonlinear map $N_{T, \Abold, W}\colon \Hsymm \ra \Lsymm$ as
\begin{align}
N_{T, \Abold, W}(\Delta) &:= \frac 2\beta \sum_{n\in \Zbb} \frac 1{\i \omega_n - \hfrak_{\Abold, W}}\,  \Delta \,  \frac 1{\i\omega_n + \ov{\hfrak_{\Abold, W}}} \, \ov \Delta \,  \frac 1{\i\omega_n - \hfrak_{\Abold, W}}\, \Delta \, \frac 1{\i\omega_n + \ov{\hfrak_{\Abold, W}}}. \label{NTAW_definition}
\end{align}
The expression $\langle \Delta, N_{T, \Abold, W}(\Delta)\rangle$ will determine the quartic term in the Ginzburg--Landau functional. 



The operators $L_{T, \Abold, W}$ and $N_{T, \Abold, W}$ enable us to formulate a representation formula for the BCS functional. This result is the basis for the proofs of Theorems \ref{Main_Result} and \ref{Main_Result_Tc}. As this result is relevant for upper and lower bounds on the BCS functional, the statement is phrased not only for Gibbs states but for a general state $\Gamma$. The proof is given in \cite[Proposition 3.4]{DeHaSc2021}.

\begin{prop}[Representation formula for the BCS functional]
\label{BCS functional_identity} \label{BCS FUNCTIONAL_IDENTITY}
Let $\Gamma$ be an admissible state. For any $h>0$, let $\Psi\in \Hmag^1(Q_h)$ and let $\Delta \equiv \Delta_\Psi$ be as in \eqref{Delta_definition}. For $T>0$ and if $V\alpha_*\in L^{\nicefrac 65}(\Rbb^3) \cap L^2(\Rbb^3)$, there is an operator $\Rcal_{T, \Abold, W}^{(1)}(\Delta)\in \Scal^1$ such that
\begin{align}
\FBCS(\Gamma) - \FBCS(\Gamma_0)& \notag\\
&\hspace{-70pt}= - \frac 14 \langle \Delta, L_{T, \Abold, W} \Delta\rangle + \frac 18 \langle \Delta, N_{T, \Abold, W} (\Delta)\rangle + \Vert\Psi \Vert_{\Lmag^2(Q_h)} \, \langle \alpha_*, V\alpha_*\rangle_{L^2(\Rbb^3)} \notag \\
&\hspace{-40pt}+ \Tr\bigl[\Rcal_{T, \Abold, W}^{(1)}(\Delta)\bigr] \notag\\
&\hspace{-40pt}+ \frac{T}{2} \Hcal_0(\Gamma, \Gamma_\Delta) - \fint_{Q_h} \dd X \int_{\Rbb^3} \dd r \; V(r) \, \bigl| \alpha(X,r) - \alpha_*(r) \Psi(X)\bigr|^2, \label{BCS functional_identity_eq}
\end{align}
where
\begin{align}
	\Hcal_0(\Gamma, \Gamma_\Delta) := \Tr_0\bigl[ \Gamma(\ln \Gamma - \ln \Gamma_\Delta) + (1 - \Gamma)(\ln(1-\Gamma) - \ln(1 - \Gamma_\Delta))\bigr] \label{Relative_Entropy}
\end{align}
denotes the relative entropy of $\Gamma$ with respect to $\Gamma_\Delta$. Moreover, $\Rcal_{T, \Abold, W}^{(1)}(\Delta)$ obeys the estimate
\begin{align*}
	\Vert\Rcal_{T, \Abold, W}^{(1)}(\Delta) \Vert_1 \leq C \; T^{-5} \; h^6 \; \Vert \Psi\Vert_{\Hmag^1(Q_h)}^6.
\end{align*}
\end{prop}

The definition \eqref{Relative_Entropy} of relative entropy uses a weaker form of trace $\Tr_0$, which is defined as follows. A gauge-periodic operator $A$, which acts on $L^2(\Rbb^3)\oplus L^2(\Rbb^3)$, is called weakly locally trace class if $P_0AP_0$ and $Q_0AQ_0$ are locally trace class, where
\begin{align}
P_0 = \begin{pmatrix} 1 & 0 \\ 0 & 0 \end{pmatrix} \label{P0}
\end{align}
and $Q_0 = 1-P_0$. Its weak trace per unit volume is defined by
\begin{align}
\Tr_0 (A):= \Tr\bigl( P_0AP_0 + Q_0 AQ_0\bigr). \label{Weak_trace_definition}
\end{align}
We remark that an operator $A$ is weakly locally trace class provided it is locally trace class but the converse may not hold. However, if $A$ is nonnegative, then the converse does hold. Moreover, the trace per unit volume and the weak trace per unit volume coincide for operators which are locally trace class.


An explanation of the roles played by the different terms on the right side of \eqref{BCS functional_identity_eq} can be found below \cite[Proposition 3.4]{DeHaSc2021}. Here, we focus on the first line and show that, within the appropriate temperature scaling $T = \Tc (1 - Dh^2)$, it is determined by the Ginzburg--Landau functional. The statement of the result involves the function
\begin{align}
\hat{V\alpha_*}(p) := \int_{\Rbb^3} \dx\; \e^{-\i p\cdot x} \, V(x)\alpha_*(x) \label{Gap_function}
\end{align}
and, hereby, we fix our convention on the Fourier transform in the present article.

\begin{thm}[Calculation of the GL energy]
\label{Calculation_of_the_GL-energy} \label{CALCULATION_OF_THE_GL-ENERGY}
Let Assumptions \ref{Assumption_V} and \ref{Assumption_KTc} (a) hold and let $D\in \Rbb$ be given. Then, there is a constant $h_0>0$ such that for any $0 < h \leq h_0$, any $\Psi\in \Hmag^2(Q_h)$, $\Delta \equiv \Delta_\Psi$ as in \eqref{Delta_definition}, and $T = \Tc(1 - Dh^2)$, we have
\begin{align}
- \frac 14 \langle \Delta, L_{T, \Abold, W} \Delta\rangle + \frac 18 \langle \Delta, N_{T, \Abold, W} (\Delta)\rangle + \Vert \Psi\Vert_{\Lmag^2(Q_h)}^2 \; \langle \alpha_*, V\alpha_*\rangle_{L^2(\Rbb^3)} & \notag\\
&\hspace{-60pt}= h^4\; \EGL(\Psi) + R(h). \label{Calculation_of_the_GL-energy_eq}
\end{align}
Here,
\begin{align*}
|R(h)|\leq C \, \bigl[ h^5 \, \Vert \Psi\Vert_{\Hmag^1(Q_h)}^2 + h^6 \, \Vert\Psi\Vert_{\Hmag^2(Q_h)}^2 \bigr] \, \bigl[ 1 + \Vert \Psi\Vert_{\Hmag^1(Q_h)}^2 \bigr]
\end{align*}
and with the functions
%
\begin{align}
g_1(x) &:= \frac{\tanh(x/2)}{x^2} - \frac{1}{2x}\frac{1}{\cosh^2(x/2)}, & g_2(x) &:= \frac 1{2x} \frac{\tanh(x/2)}{\cosh^2(x/2)}, \label{XiSigma}
\end{align}
the coefficients $\Lambda_0$, $\Lambda_2$, and $\Lambda_3$ in $\EGL$ are given by
\begin{align}
\Lambda_0 &:= \frac{\beta_c^2}{16} \int_{\Rbb^3} \frac{\dd p}{(2\pi)^3} \; |(-2)\hat{V\alpha_*}(p)|^2 \; \bigl( g_1 (\beta_c(p^2-\mu)) + \frac 23 \beta_c \, p^2\, g_2(\beta_c(p^2-\mu))\bigr), \label{GL-coefficient_1}\\
\Lambda_1 &:= \frac{\betac^2}{4} \int_{\Rbb^3} \frac{\dd p}{(2\pi)^3} \; |(-2)\hat{V\alpha_*}(p)|^2 \; g_1(\betac (p^2-\mu)), \label{GL-coefficient_W} \\
\Lambda_2 &:= \frac{\beta_c}{8} \int_{\Rbb^3} \frac{\dd p}{(2\pi)^3} \; \frac{|(-2)\hat{V\alpha_*}(p)|^2}{\cosh^2(\frac{\beta_c}{2}(p^2 -\mu))},\label{GL-coefficient_2} \\
\Lambda_3 &:= \frac{\beta_c^2}{16} \int_{\Rbb^3} \frac{\dd p}{(2\pi)^3} \; |(-2) \hat{V\alpha_*}(p)|^4 \;  \frac{g_1(\beta_c(p^2-\mu))}{p^2-\mu}.\label{GL_coefficient_3}
\end{align}
%
%
\end{thm}



It has been argued in \cite{DeHaSc2021} that the coefficients $\Lambda_0$, $\Lambda_2$, and $\Lambda_3$ are positive. We note that the coefficient $\Lambda_1$ can, in principle, have either sign, see also the remark below \cite[Eq. (1.21)]{Hainzl2012}. 


Let us note that the error estimate $h^5$ is worse than the estimate given in \cite[Theorem 3.5]{DeHaSc2021} due to the presence of the bounded periodic potential $A$. However, this does not affect the error in our main results Theorems \ref{Main_Result} and \ref{Main_Result_Tc} because the largest contribution stems from the term involving the $H^2$-norm of $\Psi$. This is in accordance with the works \cite{Hainzl2012} and \cite{Hainzl2014}. 

Theorem \ref{Calculation_of_the_GL-energy} analyzes the BCS energy in the temperature scaling $T = \Tc(1 - Dh^2)$. However, we also need a preliminary result, which proves that our system is superconducting for temperatures below this regime, in order to prove Theorem \ref{Main_Result_Tc} (a). The statement reads as follows.

\begin{prop}[A priori bound on Theorem \ref{Main_Result_Tc} (a)]
\label{Lower_Tc_a_priori_bound}
Let Assumptions \ref{Assumption_V} and \ref{Assumption_KTc} (a) hold and let $T_0>0$. Then, there are constants $h_0>0$ and $D_0>0$ such that for all $0 < h \leq h_0$ and all temperatures $T$ obeying
\begin{align*}
 T_0 \leq T < \Tc (1 - D_0 h^2),
\end{align*}
there is a BCS state $\Gamma$ with
\begin{align}
\FBCS(\Gamma) - \FBCS(\Gamma_0) < 0. \label{Lower_critical_shift_2}
\end{align}
\end{prop}


\subsection{The upper bound on \texorpdfstring{(\ref{ENERGY_ASYMPTOTICS})}{(\ref{ENERGY_ASYMPTOTICS})} and proof of Theorem \ref{Main_Result_Tc} (a)}
\label{Upper_Bound_Proof_Section}

The results of the previous section enable us to give the proof of the upper bound on \eqref{ENERGY_ASYMPTOTICS} and of Theorem \ref{Main_Result_Tc} (a). We refer to \cite[Section 3.3]{DeHaSc2021} for the detailed presentation.

\section{Proofs of the Results in Section \ref{Upper_Bound}}
\label{Proofs}




\subsection{Schatten norm estimates for operators given by product kernels}
\label{Estimates_on_product_wave_functions_Section}

In the following, we frequently need Schatten norm estimates for several gauge-periodic operators, whose kernels have the product structure $\tau(x-y) \Psi((x+y)/2)$. The following result is proven in \cite[Lemma 4.1 (a)]{DeHaSc2021}.

\begin{lem}
\label{Schatten_estimate}
Let $h>0$, let $\Psi$ be a gauge-periodic function on $Q_h$ and let $\tau$ be an even and real-valued function on $\Rbb^3$. Moreover, let the operator $\alpha$ be defined via its integral kernel $\alpha(X,r) := \tau(r)\Psi(X)$, i.e., $\alpha$ acts as
\begin{align*}
\alpha f(x) &= \int_{\Rbb^3} \dd y \; \tau(x - y) \Psi\bigl(\frac{x+y}{2}\bigr) f(y), & f &\in L^2(\Rbb^3).
\end{align*}

\begin{enumerate}[(a)]
\item Let $p \in \{2,4,6\}$. If $\Psi\in \Lmag^p(Q_h)$ and $\tau \in L^{\frac {p}{p-1}}(\Rbb^3)$, then $\alpha \in \Scal^p$ and
\begin{align*}
\Vert \alpha\Vert_p \leq C \; \Vert \tau\Vert_{\frac{p}{p-1}} \; \Vert \Psi\Vert_p.
\end{align*}

\item For any $\nu > 3$, there is a $C_\nu >0$, independent of $h$, such that if $(1 +|\cdot|)^\nu \tau\in L^{\nicefrac 65}(\Rbb^3)$ and $\Psi\in \Lmag^6(Q_h)$, then $\alpha \in \Scal^\infty$ and
\begin{align*}
\Vert \alpha\Vert_\infty &\leq C_\nu \, h^{-\nicefrac 12} \; \max\{1 , h^\nu\} \; \Vert (1 + |\cdot|)^\nu \tau\Vert_{\nicefrac 65} \; \Vert \Psi\Vert_6.
\end{align*}
\end{enumerate}
\end{lem}

\subsection{Proof of Theorem \ref{Calculation_of_the_GL-energy}}
\label{Calculation_of_the_GL-energy_proof_Section}


\subsubsection{Technical preparation --- the phase approximation method}
\label{DHS2:Phase_approximation_method_Section}

In this subsection, we analyze the resolvent kernel 
\begin{align}
G^z_\Abold(x,y) &:= \frac{1}{z - (-\i \nabla + \Abold)^2 + \mu}(x,y), & x,y &\in \Rbb^3, \label{GAz_definition}
\end{align}
of the magnetic Laplacian and use the phase approximation method to prove perturbative estimates in the weak magnetic field regime. The phase approximation method presented here follows the gauge-invariant perturbation theory in \cite[pp. 1290]{Nenciu2002} and we note that versions of several results of this section are proven in \cite[Chapter 6]{AlissaPhD}.

The first result, which we will frequently use in this context pertains to the free resolvent kernel
\begin{align}
g^z := \frac{1}{z - (-\i \nabla)^2 + \mu} \label{g_definition}
\end{align}
and its gradient. The proof of the following decay estimate on their $L^1$-norms can be found in \cite[Lemma 4.4]{DeHaSc2021}.

\begin{lem}
\label{g_decay}
Let $a > -2$. There is a constant $C_a >0$ such that for $t,\omega\in \Rbb$, we have
\begin{align}
\left \Vert \, |\cdot|^a g^{\i \omega + t}\right\Vert_1 &\leq C_a \; f(t, \omega)^{1+ \frac a2}, 
\label{g0_decay_1}
\end{align}
where
\begin{align}
f(t, \omega) := \frac{|\omega| + |t + \mu|}{(|\omega| + (t + \mu)_-)^2} \label{g0_decay_f}
\end{align}
and $x_- := -\min\{x,0\}$. Furthermore, for any $a > -1$, there is a constant $C_a >0$ with
\begin{align}
\left \Vert \, |\cdot|^a \nabla g^{\i\omega + t} \right\Vert_1 \leq C_a \; f(t, \omega)^{\frac 12 + \frac a2} \; \Bigl[ 1 + \frac{|\omega| + |t+ \mu|}{|\omega| + (t + \mu)_-}\Bigr]. \label{g0_decay_2}
\end{align}
\end{lem}

The core of the phase approximation method is the nonintegrable phase factor, sometimes also called the Wilson line, defined by
\begin{align}
\Phi_\Abold(x,y) := -\int_y^x \Abold(u) \cdot \dd u := -\int_0^1 \dd t\; \Abold(y + t(x-y))\cdot (x-y). \label{PhiA}
\end{align}

Here and in the following, in order to avoid having to write $\Abold_h$ in every instance, we assume more generally that $\Abold = \Abold_\Bbold + A$ for a fixed but arbitrary vector $\Bbold \in \Rbb^3$ and a periodic potential $A$ of appropriate regularity. We note that the $(n-1)\st$ derivative of a function in $\Wpervec{n}$ is Lipschitz continuous, whence line integrals are well defined.

\begin{lem}
\label{PhiA_derivative}
Let $\Bbold\in\Rbb^3$ and $A\in \Wpervec{2}$. For $\Abold = \Abold_\Bbold + A$, we have
\begin{align}
\nabla_x \Phi_\Abold(x,y) = -\Abold(x) + \tilde \Abold(x,y), \label{PhiA_derivative_eq1}
\end{align}
where
\begin{align}
\tilde \Abold (x,y) := \int_0^1 \dt \; t \curl \Abold(y + t(x - y)) \wedge (x-y) \label{Atilde_definition}
\end{align}
is the transversal Poincar\'e gauge relative to $y\in \Rbb^3$.
\end{lem}


\begin{proof}
By assumption, $\curl A$ is a Lipschitz continuous function. Therefore, the line integral over $\curl A$ is well defined. We make use of the formula
\begin{align}
\nabla (v \cdot w) = (v\cdot \nabla)w + (w\cdot \nabla) v + v\wedge \curl w + w\wedge \curl v \label{PhiA_derivative_4}
\end{align}
for arbitrary vector fields $v$ and $w$. For fixed $y\in \Rbb^3$, we apply this to
\begin{align*}
v(x) &:= \int_0^1\, \dt \; \Abold(y + t(x-y)), & w(x) &:= x-y,
\end{align*}
whence $\curl w =0$ so that
\begin{align}
-\nabla_x \Phi_\Abold(x,y) &= \Bigl( \int_0^1 \dt \; \Abold(y + t(x-y)) \cdot \nabla_x\Bigr) (x-y) \notag \\
&\hspace{-60pt} + \lk (x-y) \cdot \nabla_x\rk \int_0^1 \dt \; \Abold(y + t(x-y)) + (x-y) \wedge \int_0^1 \dt \; t \, \curl \Abold(y + t(x-y)). \label{PhiA_derivative_1}
\end{align}
The first term on the right side equals
\begin{align}
\sum_{i=1}^3 \int_0^1\dt \; \Abold_i(y + t(x-y)) \, \partial_i (x-y) &= \int_0^1 \dt \; \Abold(y + t(x-y)). \label{PhiA_derivative_2}
\end{align}
For the second term on the right side of \eqref{PhiA_derivative_1} we use integration by parts to get
\begin{align}
\left( (x-y)\cdot \nabla_x\right) \int_0^1 \dt \; \Abold(y + t(x-y)) &= \int_0^1 \dt \; t \, \frac{\dd}{\dt} \Abold(y + t(x-y)) \notag \\
&\hspace{-50pt} = t \, \Abold(y + t(x-y))\Big|_0^1 - \int_0^1 \dt \; \Abold(y + t(x-y)). \label{PhiA_derivative_3}
\end{align}
Therefore, the sum of \eqref{PhiA_derivative_2} and \eqref{PhiA_derivative_3} equals $\Abold(x)$. Since the last term on the right side of \eqref{PhiA_derivative_1} equals equals $-\tilde \Abold(x,y)$, this proves \eqref{PhiA_derivative_eq1}.
\end{proof}

We define the gauge-invariant version of the free resolvent kernel $g^z$ in \eqref{g_definition} by
\begin{align}
\tilde G_\Abold^z(x,y) := \e^{\i \Phi_\Abold(x,y)} \, g^z(x-y). \label{GtildeAz}
\end{align}
In the following lines, we investigate the intertwining relation between $(-\i \nabla  + \Abold)^2$ and the operator $\tilde G_\Abold^z$ associated to the kernel in \eqref{GtildeAz}. First of all, \eqref{PhiA_derivative_eq1} implies
\begin{align}
(-\i \nabla_x + \Abold(x)) \; \e^{\i \Phi_\Abold(x,y)} = \e^{\i \Phi_\Abold(x,y)}\; (-\i\nabla_x + \tilde \Abold(x,y)) \label{PhiA_Magnetic_Momentum_Action}
\end{align}
where $\tilde \Abold(x,y)$ is the Poincaré gauge in \eqref{Atilde_definition}. Furthermore, a short computation shows that \eqref{PhiA_Magnetic_Momentum_Action} implies the operator equation
\begin{align}
(z - (-\i \nabla + \Abold)^2 + \mu) \tilde G_\Abold^z = \Idbb - T_\Abold^z,  \label{GAz_TAz_relation}
\end{align}
where $T_\Abold^z$ is the operator associated to the kernel
\begin{align}
T_\Abold^z (x,y) := \e^{\i \Phi_\Abold(x,y)} \bigl( 2 \, \tilde \Abold(x,y) (-\i \nabla_x) -\i \divv_x \tilde \Abold(x,y) + |\tilde \Abold(x,y)|^2 \bigr) g^z(x-y). \label{TAz_definition}
\end{align}
Since $\tilde \Abold(x,y)$ is perpendicular to $x-y$, the first term in brackets vanishes by the radiality of $g^z$. The following result shows that $T_\Abold$ is a bounded operator, whose norm is small if $\Abold$ is replaced by $\Abold_h$.

\begin{lem}
\label{TAz_boundedness}
Let $\Bbold \in \Rbb^3$, $A \in \Wpervec 3$, and let $\Abold := \Abold_\Bbold + A$. Introduce the constant
\begin{align}
M_\Abold := \max\bigl\{ \Vert \curl(\curl \Abold) \Vert_{L^\infty(\Rbb^3)} \, , \, \Vert \curl \Abold \Vert_{L^\infty(\Rbb^3)}^{\nicefrac 32} \bigr\}. \label{MA_definition}
\end{align}
Then, we have
\begin{align}
|T_\Abold^z(x,y) | &\leq M_\Abold \; \rho_\Abold^z(x-y), \label{TAz_boundedness_eq1}
\end{align}
where
\begin{align}
\rho_\Abold^z(x) &:= \bigl( |x| + \Vert \curl \Abold \Vert_\infty^{\nicefrac 12} \, |x|^2\bigr) \, |g^z(x)|. \label{rhoAz_definition}
\end{align}
In particular, the operator $T_\Abold^z$ corresponding to the kernel $T_\Abold^z(x,y)$ in \eqref{TAz_definition} is bounded and
\begin{align}
\Vert T_\Abold^z\Vert_\infty \leq M_\Abold \; \Vert \rho_\Abold^z \Vert_{L^1(\Rbb^3)}. \label{TAz_boundedness_eq2}
\end{align}
\end{lem}

\begin{proof}
To see that \eqref{TAz_boundedness_eq1} is true, we need to compute the terms $\divv_x \tilde \Abold(x,y)$ and $|\tilde \Abold (x,y)|$ and show that they satisfy appropriate bounds. For general vector fields $v$ and $w$, we make use of the formula
\begin{align*}
\divv (v \wedge w) = \curl (v) \cdot w - \curl (w) \cdot v
\end{align*}
and apply it to $v(x) := \curl \Abold(y + t(x-y))$ and $w(x) = x-y$. Since $\curl w =0$, this shows
\begin{align*}
\divv_x \bigl(\curl \Abold(y+t(x-y)) \wedge (x-y)\bigr) &= t \; \curl(\curl \Abold) (y + t (x-y)) \cdot (x-y).
\end{align*}
Therefore, since $\curl(\curl \Abold)$ is a Lipschitz continuous function, we conclude
\begin{align}
|\divv_x \tilde \Abold(x,y)| &\leq \int_0^1\dt \; t^2\, | \curl (\curl \Abold)(y + t(x-y))\cdot (x-y)| \notag\\
&\leq \Vert \curl (\curl \Abold) \Vert_\infty \; |x-y|. \label{TAz_boundedness_1}
\end{align}
Furthermore, we have
\begin{align}
|\tilde \Abold(x,y)|^2 &= \int_0^1 \dt \int_0^1 \ds\; ts \; \bigl(\curl \Abold(y + t(x-y)) \wedge (x-y)\bigr) \notag \\
& \hspace{100pt} \cdot \bigl(\curl \Abold (y+s(x-y)) \wedge (x-y)\bigr)\notag\\
&\leq \Vert \curl \Abold\Vert_\infty^2 \, |x-y|^2. \label{TAz_boundedness_2}
\end{align}
This proves \eqref{TAz_boundedness_eq1}. The estimate \eqref{TAz_boundedness_eq2} follows from \eqref{TAz_boundedness_eq1} and Young's inequality.
\end{proof}

In the next step, we investigate the relation between the full magnetic resolvent kernel $G_\Abold^z$ in \eqref{GAz_definition} and the gauge-invariant version of the free resolvent kernel $\tilde G_\Abold^z$ in \eqref{g_definition}. We do this by analyzing the tilted resolvent kernels
\begin{align}
\Gbold_{\Bbold, \Abold}^z (x,y) &:= \e^{-\i \Phi_{\Abold_\Bbold}(x,y)} \, G_\Abold^z(x, y) \label{GboldBAz_definition}
\end{align}
and
\begin{align}
\tilde \Gbold_{\Bbold, \Abold}^z (x,y) &:= \e^{-\i \Phi_{\Abold_\Bbold}(x,y)} \, \tilde G_\Abold^z(x, y).  \label{GboldtildeBAz_definition}
\end{align}
As long as $\Abold = \Abold_\Bbold + A$, we have
\begin{align}
\tilde \Gbold_{\Bbold, \Abold}^z (x,y) = \e^{\i \Phi_A(x,y)} \, g^z(x-y). \label{GboldtildeBAz_first_identity}
\end{align}
The reason that we analyze the kernels $\Gbold_{\Bbold, \Abold}^z$ and $\tilde \Gbold_{\Bbold, \Abold}^z$ --- and not $G_\Abold^z$ directly --- lies in the fact that our phase approximation needs to cover derivatives, which spoil the gauge invariance if we do not correct by the leading phase of the (unbounded) constant magnetic field potential $\Abold_\Bbold$. A simpler version without the constant magnetic field present has been carried out in \cite[Chapter 6]{AlissaPhD}. The following result is the analogue of \cite[Lemma 4.5]{DeHaSc2021}.

\begin{lem}
\label{GAz-GtildeAz_decay}
There is $\delta >0$ such that for any $a\geq 0$, any $t, \omega\in \Rbb$, and for all vector potentials $\Abold = \Abold_\Bbold + A$ with $\Bbold\in \Rbb^3$ and $A\in \Wpervec{3}$, which satisfy
\begin{align}
f(t, \omega)^{\frac 32} \, M_{\Abold} +  f(t,\omega) \, \Vert \curl \Abold\Vert_\infty+  f(t,w) \, \Vert A\Vert_\infty^2 \leq \delta \label{GAz-GtildeAz_decay_assumption}
\end{align}
with $f(t, \omega)$ in \eqref{g0_decay_f}, there are even $L^1(\Rbb^3)$-functions
\begin{align}
&\Gcal_\Abold^{\i\omega + t}, && \Gcal_{\nabla, \Abold, A}^{\i\omega + t}, && \Hcal_\Abold^{\i\omega + t}, && \Hcal_{\nabla, \Abold, A}^{\i\omega + t}, && \Ical_A^{\i\omega + t}, \label{Gcals_Hcals_definition}
\end{align}
such that 
\begin{align}
|\Gbold_{\Bbold, \Abold}^{\i\omega + t}(x,y)| &\leq \Gcal_\Abold^{\i\omega + t} (x-y),   \notag\\
|\nabla_x \Gbold_{\Bbold, \Abold}^{\i\omega + t}(x,y)| &\leq \Gcal_{\nabla, \Abold, A}^{\i\omega + t} (x-y), \notag \\
|\nabla_y \Gbold_{\Bbold, \Abold}^{\i\omega + t}(x,y)| &\leq \Gcal_{\nabla, \Abold, A}^{-\i\omega + t} (x-y) , \label{GAz-GtildeAz_decay_eq1}
\end{align}
as well as
\begin{align}
|\Gbold_{\Bbold, \Abold}^{\i\omega + t}(x,y) - \tilde \Gbold_{\Bbold, \Abold}^{\i\omega + t}(x,y)| &\leq \Hcal_\Abold^{\i\omega + t} (x-y), \notag \\
| \nabla_x \Gbold_{\Bbold, \Abold}^{\i\omega + t}(x,y) - \nabla_x \tilde \Gbold_{\Bbold, \Abold}^{\i \omega + t}(x,y)| &\leq \Hcal_{\nabla, \Abold, A}^{\i\omega + t} (x-y) , \notag \\
|\nabla_y \Gbold_{\Bbold, \Abold}^{\i\omega + t}(x,y) - \nabla_y \tilde \Gbold_{\Bbold, \Abold}^{\i \omega + t}(x,y)| &\leq \Hcal_{\nabla, \Abold, A}^{-\i\omega + t} (x-y) , \label{GAz-GtildeAz_decay_eq2}
\end{align}
and
\begin{align}
|\nabla_x \tilde \Gbold_{\Abold}^{\i\omega + t}(x,y)| &\leq \Ical_A^{\i\omega + t}(x-y), \notag \\
|\nabla_y \tilde \Gbold_{\Abold}^{\i\omega + t}(x,y)| &\leq \Ical_A^{-\i\omega + t}(x-y). \label{GAz-GtildeAz_decay_eq3}
\end{align}
Furthermore, we have the estimates
\begin{align}
\Vert \, |\cdot|^a \Gcal_\Abold^{\i\omega + t} \Vert_1 &\leq C_a \, f(t, \omega)^{1 + \frac a2}, \notag \\
\Vert \, |\cdot|^a \Gcal_{\nabla, \Abold, A}^{\i\omega + t} \Vert_1 &\leq C_a \, f(t,\omega)^{\frac 12 + \frac a2} \, \Bigl[ 1 + \frac{|\omega| + |t - \mu|}{|\omega| + (t - \mu)_-} \Bigr], \label{GAz-GtildeAz_decay_eq4}
\end{align}
as well as
\begin{align}
\Vert \, |\cdot|^a \Hcal_\Abold^{\i\omega + t} \Vert_1 &\leq C_a \, M_\Abold \, f(t, \omega)^{\frac 52 + \frac a2}, \notag \\
\Vert \, |\cdot|^a \Hcal_{\nabla, \Abold, A}^{\i\omega + t} \Vert_1 &\leq C_a \, M_\Abold \, f(t, \omega)^{2 + \frac a2} \Bigl[ 1 + \frac{|\omega| + |t - \mu|}{|\omega| + (t - \mu)_-} \Bigr], \label{GAz-GtildeAz_decay_eq5}
\end{align}
and
\begin{align}
\Vert \, |\cdot|^a \Ical_A^{\i\omega + t}\Vert_1 &\leq C_a \; f(t,\omega)^{\frac 12 + \frac a2} \Bigl[ 1 + \frac{|\omega| + |t - \mu|}{|\omega| + (t -\mu)_-}\Bigr]. \label{GAz-GtildeAz_decay_eq6}
\end{align}
\end{lem}

\begin{bem}
In comparison to \cite[Lemma 4.5]{DeHaSc2021}, we lose a power of $h$ in the estimate since the constant $M_{\Abold_h}$ in \eqref{MA_definition} is of the order $h^3$, while the estimate \cite[Eq. (4.32)]{DeHaSc2021} is of the order $h^4$. This is due to the second term in the bracket of \eqref{TAz_definition} and in accordance with the works \cite{Hainzl2012} and \cite{Hainzl2014}.
\end{bem}

\begin{proof}[Proof of Lemma \ref{GAz-GtildeAz_decay}]
We use the abbreviation $z = \i \omega + t$ throughout the proof. First of all, for the function $\rho_\Abold^z$ in \eqref{rhoAz_definition}, by Lemma \ref{g_decay} and the assumption \eqref{GAz-GtildeAz_decay_assumption}, we have
\begin{align*}
\Vert \, |\cdot|^a \rho_\Abold^z\Vert_1 \leq C_a \; f(t, \omega)^{\frac 32 + \frac a2} \bigl[ 1 + \Vert \curl \Abold\Vert_\infty^{\nicefrac 12} \, f(t,\omega)^{\nicefrac 12}\bigr] \leq C_a \, f(t, \omega)^{\frac 32 + \frac a2},
\end{align*}
which by \eqref{GAz-GtildeAz_decay_assumption} yields
\begin{align}
M_\Abold \, \Vert \rho_\Abold^z\Vert_1 \leq  C\, M_\Abold \, f(t,\omega)^{\frac 32} \leq \frac 12 \label{GAz-GtildeAz_decay_2}
\end{align}
for all $t, \omega$ and $\Abold$ in the lemma provided $\delta$ is chosen suitably small. Let $(\rho_\Abold^z)^{*j}$ denote the $j$-fold convolution of $\rho_\Abold^z$ with itself. Then, the symmetric function
\begin{align}
\tilde \Hcal_\Abold^z := \sum_{j=1}^\infty M_\Abold^j \, (\rho_\Abold^z)^{*j}, \label{HAz}
\end{align}
satisfies
\begin{align*} 
|\cdot|^a \Hcal_\Abold^z \leq \sum_{j=1}^\infty M_\Abold^j \sum_{m=1}^j j^{(a-1)_+} \, \rho_\Abold^z * \cdots * \bigl( |\cdot|^a\rho_\Abold^z\bigr) * \cdots * \rho_\Abold^z.
\end{align*}
Here, $|\cdot|^a \rho_\Abold^z$ appears in the $m$\tho\ slot and $j^{a_+}$ is the constant in the estimate 
\begin{align}
|x_1 + \cdots + x_j|^a \leq j^{(a-1)_+} \bigl(|x_1|^a + \cdots + |x_j|^a \bigr). \label{DHS2:convexity}
\end{align}
Therefore, by \eqref{GAz-GtildeAz_decay_2} and Young's inequality, we have
\begin{align}
\Vert \, |\cdot|^a \tilde \Hcal_\Abold^z \Vert_1 &
\leq M_\Abold \, \Vert \, |\cdot|^a \rho_\Abold^z\Vert_1 \sum_{j=1}^\infty \frac{j^{1 + (a-1)_+}}{2^{j-1}} \leq C_a \; M_\Abold \; f(t, \omega)^{\frac 32 + \frac a2}. \label{GAz-GtildeAz_decay_1}
\end{align}
Furthermore, by \eqref{GAz_TAz_relation}, \eqref{GAz-GtildeAz_decay_2} and Lemma \ref{TAz_boundedness}, the Neumann-series
\begin{align}
\frac 1{z - (-\i \nabla + \Abold)^2 + \mu} = \tilde G_{\Abold}^z\; \sum_{j=0}^\infty \bigl(T_\Abold^z\bigr)^j \label{Neumann-series}
\end{align}
converges in terms of the operators defined by the kernels $\tilde G_\Abold^z$ in \eqref{GtildeAz} and $T_\Abold^z$ in \eqref{TAz_definition}. When we denote the kernel of $\sum_{j=1}^\infty (T_\Abold^z)^j$ by $\Scal_\Abold^z$, this implies $|\Scal_\Abold^z(x, y)| \leq \tilde \Hcal_\Abold^z(x-y)$ and
\begin{align}
G_\Abold^z(x,y) - \tilde G_{\Abold}^z(x,y) = \int_{\Rbb^3} \dd u \; \e^{\i \Phi_\Abold(x, u)} g^z(x-u) \; \Scal_\Abold^z(u,y). \label{GAz-GtildeAz_decay_3}
\end{align}
Therefore, the functions $\Hcal_\Abold^z := |g^z| * \tilde \Hcal_\Abold^z$ and $\Gcal_\Abold^z := \Hcal_\Abold^z + |g^z|$ obey the first estimates in \eqref{GAz-GtildeAz_decay_eq1} and \eqref{GAz-GtildeAz_decay_eq2} and, by Lemma \ref{g_decay} and \eqref{GAz-GtildeAz_decay_1}, we obtain the first $L^1$-norm estimates claimed in \eqref{GAz-GtildeAz_decay_eq4} and \eqref{GAz-GtildeAz_decay_eq5}, respectively.

To proceed with the derivatives, the function
\begin{align*}
\Ical_A^z := C \, \bigl( \Vert A\Vert_\infty \, |g^z| + \Vert \curl A\Vert_\infty \, |\cdot|\,  |g^z| + |\nabla g^z| \bigr) 
\end{align*}
satisfies \eqref{GAz-GtildeAz_decay_eq6} by the assumption \eqref{GAz-GtildeAz_decay_assumption} and, by Lemma \ref{PhiA_derivative}, we have
\begin{align*}
|\nabla_x\tilde \Gbold_{\Bbold, \Abold}^z(x,y)| &=  |\nabla_x \e^{\i \Phi_A(x,y)} g^z(x-y)| \leq \Ical_A^z(x-y),
\end{align*}
which proves the first estimate of \eqref{GAz-GtildeAz_decay_eq3}. Furthermore, \eqref{GAz-GtildeAz_decay_3} implies
\begin{align}
\Gbold_{\Bbold, \Abold}^z(x,y) - \tilde \Gbold_{\Bbold, \Abold}^z(x,y) = \int_{\Rbb^3} \dd u \; \tilde \Gbold_{\Bbold, \Abold}^z(x,u) \; \e^{\i \Phi_{\Abold_\Bbold}(x, u-y)} \Scal_\Abold^z(u,y), \label{GAz-GtildeAz_decay_4}
\end{align}
which yields
\begin{align*}
\bigl|\nabla_x \Gbold_{\Bbold, \Abold}^z(x,y) - \nabla_x \tilde \Gbold_{\Bbold, \Abold}^z(x,y)\bigr| &\leq \Ical_A * \tilde \Hcal_\Abold^z (x-y) + |\Bbold|\, |g^z| * \bigl( |\cdot| \, \tilde \Hcal_\Abold^z\bigr)(x-y).
\end{align*}
Hence, the function
\begin{align*}
\Hcal_{\nabla, \Abold, A}^z := \Ical_A * \tilde \Hcal_\Abold^z + \Vert \curl \Abold\Vert_\infty \, |g^z| * \bigl( |\cdot|\, \tilde \Hcal_\Abold^z \bigr) 
\end{align*}
satisfies the second estimate of \eqref{GAz-GtildeAz_decay_eq2} and obeys the second estimate of \eqref{GAz-GtildeAz_decay_eq5}. If we define $\Gcal_{\nabla, \Abold, A}^z := \Hcal_{\nabla, \Abold, A}^z + \Ical_\Abold^z$, then the second estimate of \eqref{GAz-GtildeAz_decay_eq1} holds and the second estimate of \eqref{GAz-GtildeAz_decay_eq4} follows in a straightforward manner.

Finally, since $A(x,y) = \ov{A^*(y,x)}$, we have $G_\Abold^z(x,y) = \ov{G_\Abold^{\ov z}(y,x)}$ whence \eqref{GAz-GtildeAz_decay_4} is equivalent to
\begin{align*}
\Gbold_{\Bbold, \Abold}^z(x,y) - \tilde \Gbold_{\Bbold, \Abold}^z(x,y) = \int_{\Rbb^3} \dd u \; \ov{ \tilde \Gbold_{\Bbold, \Abold}^{\ov z}(y,u)} \; \ov{ \e^{\i \Phi_{\Abold_\Bbold}(y, u-x)} \Scal_\Abold^{\ov z}(u,x)}.
\end{align*}
Differentiating this with respect to $y$ proves the last estimates of \eqref{GAz-GtildeAz_decay_eq1}-\eqref{GAz-GtildeAz_decay_eq3}.
\end{proof}


\subsubsection{Decomposition of \texorpdfstring{$L_{T, \Abold, W}$}{LTAW} --- separation of \texorpdfstring{$W$}{W}}

We use the resolvent equation
\begin{align}
(z- T)^{-1} = (z-S)^{-1} + (z-T)^{-1} \; (S - T)\; (z- S)^{-1} \label{Resolvent_Equation}
\end{align}
to decompose the operator $L_{T, \Abold, W}$ in \eqref{LTAW_definition} as
\begin{align}
L_{T, \Abold, W} = L_{T, \Abold} + L_{T, \Abold}^W + \Rcal_{T, \Abold,W}^{(2)} \label{LTAW_decomposition}
\end{align}
with
\begin{align}
L_{T, \Abold} := L_{T, \Abold, 0} \label{LTA_definition}
\end{align}
as well as
\begin{align}
L_{T, \Abold}^W \Delta &:= -\frac 2 \beta \sum_{n\in \Zbb} \Bigl[ \frac 1{\i \omega_n - \hfrak_\Abold} \, W_h \, \frac 1{\i \omega_n - \hfrak_\Abold} \, \Delta \, \frac{1}{\i \omega_n + \ov{\hfrak_\Abold}} \notag \\
&\hspace{100pt} - \frac{1}{\i\omega_n - \hfrak_\Abold} \, \Delta \, \frac{1}{\i \omega_n + \ov{\hfrak_\Abold}} \, W_h \, \frac{1}{\i\omega_n + \ov{\hfrak_\Abold}}\Bigr] \label{LTA^W_definition} 
\end{align}
and
\begin{align}
\Rcal_{T, \Abold, W}^{(2)}\Delta &:= -\frac 2\beta \sum_{n\in \Zbb} \Bigl[ \frac{1}{\i\omega_n - \hfrak_\Abold} \, W_h \, \frac{1}{\i \omega_n - \hfrak_\Abold} \, W_h \, \frac{1}{\i\omega_n - \hfrak_{\Abold, W}} \, \Delta \, \frac{1}{\i \omega_n + \ov{\hfrak_\Abold}} \notag \\
&\hspace{60pt} + \frac{1}{\i\omega_n - \hfrak_\Abold} \, \Delta \, \frac{1}{\i\omega_n + \ov{\hfrak_\Abold}} \, W_h \, \frac{1}{\i\omega_n + \ov{\hfrak_{\Abold, W}}} \, W_h \, \frac{1}{\i \omega_n + \ov{\hfrak_\Abold}} \notag \\
&\hspace{60pt} -\frac{1}{\i \omega_n - \hfrak_\Abold} \, W_h \, \frac{1}{\i\omega_n - \hfrak_{\Abold, W}} \, \Delta \, \frac{1}{\i\omega_n + \ov{\hfrak_{\Abold, W}}} \, W_h \, \frac 1{\i\omega_n + \ov{\hfrak_\Abold}}. \label{RTAW2_definition}
\end{align}

\begin{lem}
\label{RTAW2_estimate}
For any $A \in \Lpervec$ and $W\in \Lper$ there is $h_0>0$ such that for any $T>0$, any $0 < h \leq h_0$, and whenever $V\alpha_*\in L^2(\Rbb^3)$, $\Psi\in \Hmag^1(Q_h)$, and $\Delta \equiv \Delta_\Psi$ as in \eqref{Delta_definition}, we have
\begin{align*}
\Vert L_{T, \Abold}^W \Delta \Vert_{\Hsymm}^2 &\leq C \, ( \beta^4 + \beta^3 ) \, h^6 \, \Vert \Psi\Vert_{\Hmag^1(Q_h)}^2
\end{align*}
and
\begin{align*}
\Vert \Rcal_{T, \Abold, W}^{(2)} \Delta\Vert_{\Hsymm}^2 &\leq C\, (\beta^6 + \beta^5) \, h^{10} \, \Vert \Psi\Vert_{\Hmag^1(Q_h)}^2.
\end{align*}
\end{lem}

\begin{proof}
According to \eqref{Norm_equivalence_2} we need to bound three terms and we start with the first term. Hölder's inequality shows that the Hilbert--Schmidt norm per unit volume of the terms in the sum in \eqref{LTA^W_definition} is bounded by $C\, |\omega_n|^{-3} \, \Vert W_h\Vert_\infty\,  \Vert \Delta\Vert_2$, which by Lemma \ref{Schatten_estimate} is bounded by a constant times $\beta^3\, h^3 \Vert \Psi\Vert_{\Hmag^1(Q_h)}$.

For the second and third term recall the notations $\pi$ and $\pi_\Abold$ from \eqref{pi_definition} and \eqref{piA_definition}. We use $\Vert S\Vert_\infty^2 = \Vert S^*S\Vert_\infty$ for a general operator $S$ and obtain
\begin{align*}
\Bigl\Vert \pi\, \frac 1{\i\omega_n - \hfrak_\Abold} \Bigr\Vert_{\infty} &\leq \Bigl\Vert \frac 1{\i\omega_n + \hfrak_\Abold}\Bigr\Vert_\infty^{\nicefrac 12} \; \Bigl\Vert \pi_\Abold^2 \, \frac 1{\i\omega_n - \hfrak_\Abold} \Bigr\Vert_{\infty}^{\nicefrac 12} + \Bigl\Vert A_h \, \frac{1}{\i\omega_n - \hfrak_\Abold}\Bigr\Vert_\infty \\
&\leq C \, \bigl( |\omega_n|^{-\nicefrac 12} + \Vert A_h\Vert_\infty \, |\omega_n|^{-1} \bigr).
\end{align*}
A similar estimate holds for $(\i \omega_n + \ov{\hfrak_\Abold})^{-1} \ov \pi$. This proves the claim for $L_{T, \Abold}^W$. The proof for $\Rcal_{T, \Abold, W}^{(2)}$ is analogous.
\end{proof}



\subsubsection{A representation formula for \texorpdfstring{$L_{T,\Abold}$}{LTA} and an outlook on the quadratic terms}

This following subsections are devoted to the computation of the terms in \eqref{Calculation_of_the_GL-energy_eq}, which are given by the linear operator $L_{T, \Abold}$ defined in \eqref{LTA_definition}. The following representation formula expresses the operator in terms of the center-of-mass and relative coordinate and is our starting point for the analysis.


\begin{lem}
\label{LTA_action}
The operator $L_{T,\Abold} \colon \Lsymm \ra \Lsymm$ in \eqref{LTA_definition} acts as
\begin{align*}
(L_{T,\Abold}\alpha) (X,r) &= \iint_{\Rbb^3\times \Rbb^3} \dd Z \dd s \; k_{T, \Bbold, \Abold}(X, Z, r, s) \; (\e^{\i Z \cdot \Pi}\alpha) (X,s)
\end{align*}
with
\begin{align}
k_{T,\Bbold, \Abold}(X, Z, r, s) := \frac 2\beta \sum_{n\in \Zbb} k_{T, \Bbold,  \Abold}^n(X, Z, r, s) \; \e^{\i \frac{\Bbold}{4} \cdot (r \wedge s)} \label{kTBA_definition}
\end{align}
and
\begin{align}
k_{T, \Bbold, \Abold}^n(X, Z, r, s) &:= \Gbold_{\Bbold, \Abold_h}^{\i\omega_n} \bigl(X + \frac r2, X + Z + \frac s2\bigr) \; \Gbold_{\Bbold, \Abold_h}^{-\i\omega_n}\bigl(X - \frac r2, X + Z - \frac s2\bigr), \label{kTBAn_definition}
\end{align}
where $\Gbold_{\Bbold, \Abold}^z$ is defined in \eqref{GboldBAz_definition}.
\end{lem}

This result should be compared to \cite[Lemma 4.6]{DeHaSc2021} for the operator $L_{T, B}$. The tilted resolvent kernels $\Gbold_{\Bbold, \Abold}^z$ play the role of the translation-invariant part (called $g_B$) of the magnetic resolvent kernel in \cite{DeHaSc2021}. The more complicated structure of $k_{T, \Bbold, \Abold}^n$, which, in particular, becomes evident in its $X$-dependence, stems from the periodic magnetic field $A$. Moreover, this expression is not symmetric in $Z$, whence the operator $\cos(Z\cdot \Pi)$ in \cite{DeHaSc2021} is replaced by $\e^{\i Z\cdot \Pi}$. Since $\Pi$ is missing the periodic magnetic potential, we have to recover this from a hidden term in $k_{T, \Bbold, \Abold}$ at a later stage.

\begin{proof}[Proof of Lemma \ref{LTA_action}]
The integral kernel of $L_{T, \Abold}$ reads
\begin{align*}
L_{T,\Abold} \alpha(X, r) &= \frac 2\beta \sum_{n\in \Zbb} \iint_{\Rbb^3\times \Rbb^3} \dd u\dd v\; G_{\Abold_h}^{\i\omega_n} (\zeta_X^r, u) \, G_{\Abold_h}^{-\i\omega_n}(\zeta_X^{-r}, v) \, \alpha(u,v),
\end{align*}
where we write $\zeta_X^r := X + \frac r2$ with $X =\frac{x+y}{2}$ and $r=x-y$ for short. We also used the fact that the magnetic resolvent kernel $G_\Abold^z$ in \eqref{GAz_definition} satisfies
\begin{align}
\frac{1}{\i\omega_n + \ov {(-\i \nabla + \Abold)^2} - \mu} (x,y) = -G_\Abold^{-\i\omega_n}(y,x), \label{GAz_Kernel_of_complex_conjugate}
\end{align}
which follows from $\ov{ A^*(x,y) } = A(y,x)$ and
\begin{align*}
\frac{1}{z - \ov{(-\i \nabla + \Abold)^2} + \mu} = \ov{\Bigl(\frac{1}{z - (-\i \nabla + \Abold)^2 + \mu} \Bigr)^*}.
\end{align*}
We define the coordinates $Z$ and $s$ by
\begin{align}
u &= X + Z + \frac s2, & v = X + Z - \frac s2,
\end{align}
which implies $\alpha(u,v) = \e^{\i Z\cdot (-\i \nabla_X)} \alpha(X, s)$, and we multiply and divide by the factor
\begin{align*}
\e^{\i \Phi_{\Abold_\Bbold}(X + \frac r2, X + Z + \frac s2)} \; \e^{\i \Phi_{\Abold_\Bbold}(X - \frac r2, X + Z - \frac s2)} = \e^{\i \Bbold \cdot (X\wedge Z)} \, \e^{\i \frac{\Bbold}{4} \cdot (r\wedge s)}.
\end{align*}
Since $Z\cdot (-\i \nabla_X)$ and $\Bbold \cdot (X\wedge Z)= Z\cdot (\Bbold \wedge X)$ commute, this implies the claimed formula.
\end{proof}

In the following, the operator $L_{T,\Abold}$ in \eqref{LTA_definition} is analyzed in four steps, where the global strategy is similar to the one pursued in \cite{DeHaSc2021}. The first three steps consist of introducing a chain of operators of increasing simplicity in their dependence on $\Abold$ and decomposing $L_{T, \Abold}$ according to
\begin{align}
L_{T,\Abold} = (L_{T,\Abold} - \tilde L_{T, \Bbold, A}) + (\tilde L_{T, \Bbold, A} - \tilde M_{T,\Abold}) + (\tilde M_{T,\Abold} - M_{T,\Abold}) + M_{T, \Abold}, \label{LTA_decomposition}
\end{align}
where $\tilde L_{T, \Bbold, A}$, $\tilde M_{T, \Abold}$, and $M_{T,\Abold}$ are defined in \eqref{LtildeTA_definition}, \eqref{MtildeTA_definition}, and \eqref{MTA_definition} below, respectively. Due to the periodic contribution of the magnetic field, we need one additional step in comparison to \cite{DeHaSc2021}, in which we extract the leading magnetic phase contribution from the periodic magnetic field. This is an algebraic identity in \cite{DeHaSc2021}. In this work, it is the result of the  approximation of the tilted resolvent kernels $\Gbold_{\Bbold, \Abold}^z$ in \eqref{GboldBAz_definition} by the functions $\tilde \Gbold_{\Bbold, \Abold}^z$ in \eqref{GboldtildeBAz_definition}, which leads to the operator $\tilde L_{T, \Bbold, A}$. 
In the next step, we separate the magnetic phase from $\tilde \Gbold_{\Bbold, \Abold}^z$ and, by approximation, obtain the magnetic field term that is missing in $\e^{\i Z\cdot\Pi}$. Since the resulting integral kernel is symmetric in $Z$, we recover the operator $\cos(Z\cdot \Pi_\Abold)$, which leads us to the operator $\tilde M_{T, \Abold}$.
The operator $\tilde M_{T, \Abold}$ is the analogue of $\tilde L_{T, B}$ in \cite{DeHaSc2021}, since it contains a similar residual magnetic phase factor. We obtain the operator $M_{T,\Abold}$ from $\tilde M_{T, \Abold}$ by replacing this factor by $1$. 
The careful analysis of the operator $M_{T, \Abold}$, which takes place in Section~\ref{Analysis_of_MTA_Section} follows the strategy presented in \cite{DeHaSc2021} (see the remark below \cite[Lemma 4.6]{DeHaSc2021}) and leads to the quadratic terms of the Ginzburg--Landau functional in \eqref{Definition_GL-functional} (except for the $W$-term) as well as a term that cancels the last term on the left side of \eqref{Calculation_of_the_GL-energy_eq}.
We carry out a similar analysis for the operator $L_{T, \Abold}^W$ in \eqref{LTA^W_definition} in Sections \ref{LTAW_action_Section}-\ref{Analysis_of_MTW_Section} and extract the quadratic term involving $W$ in the Ginzburg--Landau functional. We conclude by a summary of our results in Section~\ref{Summary_quadratic_terms_Section}.

%
%
%
%
%
%
%
%
%
%


\subsubsection{Approximation of \texorpdfstring{$L_{T,\Abold}$}{LTA}}
\label{Approximation_of_LTAW_Section}

\paragraph{The operator $\tilde L_{T, \Bbold, A}$.}

We define the operator $\tilde L_{T, \Bbold, A}$ by
\begin{align}
\tilde L_{T, \Bbold, A}\alpha(X,r) &:= \iint_{\Rbb^3\times \Rbb^3} \dd Z \dd s \; \tilde k_{T, \Bbold, A}(X, Z, r,s) \; ( \e^{\i Z \cdot \Pi} \alpha)(X,s) \label{LtildeTA_definition}
\end{align}
with
\begin{align}
\tilde k_{T, \Bbold, A} (X, Z, r,s) := \frac 2\beta \sum_{n\in \Zbb} k_T^n(Z, r-s) \; \e^{\i \tilde \Phi_{A_h}(X, Z, r, s)} \, \e^{\i \frac{\Bbold}{4} \cdot (r\wedge s)}, \label{ktildeTA_definition}
\end{align}
where
\begin{align}
k_T^n(Z, r) := g^{\i\omega_n}\bigl(Z - \frac{r}{2} \bigr) \, g^{-\i \omega_n}\bigl( Z + \frac{r}{2} \bigr)
\end{align}
and
\begin{align}
\tilde \Phi_A(X, Z, r, s) &:= \Phi_A\bigl(X + \frac r2, X + Z + \frac s2\bigr) + \Phi_A\bigl(X - \frac r2, X + Z - \frac s2\bigr). \label{LtildeTA_PhitildeA_definition}
\end{align}

\begin{prop}
\label{LTA-LtildeTA}
For any $T_0>0$ and $A\in \Wpervec 3$ there is $h_0>0$ such that for any $0 < h \leq h_0$, any $T\geq T_0$ and whenever $V\alpha_*\in L^2(\Rbb^3)$, $\Psi\in \Hmag^1(Q_h)$, and $\Delta \equiv \Delta_\Psi$ as in \eqref{Delta_definition}, we have
\begin{align*}
\Vert L_{T,\Abold} \Delta - \tilde L_{T, \Bbold, A} \Delta\Vert_\Hsymm^2 \leq C \; h^8 \; \Vert V\alpha_*\Vert_2^2 \; \Vert \Psi\Vert_{\Hmag^1(Q_h)}^2.
\end{align*}
\end{prop}

\begin{bem}
\label{rem:A1}
We only need to prove a bound on $\langle \Delta, (L_{T,\Abold} - \tilde L_{T, \Bbold, A})\Delta\rangle$, when we are interested proving Theorem \ref{Calculation_of_the_GL-energy}. We prove an $\Hsymm$-norm bound here in preparation for the proof of Proposition \ref{Structure_of_alphaDelta} and the bound needed for Theorem \ref{Calculation_of_the_GL-energy} follows from Proposition \ref{LTA-LtildeTA}, Lemma~\ref{Schatten_estimate}, and the Cauchy--Schwarz inequality.
%
\end{bem}

In the proof of Proposition~\ref{LTA-LtildeTA}, the functions
\begin{align}
F_{T,\Abold}^{a} &:= \smash{\frac 2\beta \sum_{n\in \Zbb}}  \; \bigl(|\cdot|^a \, \Hcal_\Abold^{\i \omega_n} \bigr) * \Gcal_\Abold^{-\i\omega_n} + \Hcal_\Abold^{\i\omega_n} * \bigl(|\cdot|^a \, \Gcal_\Abold^{-\i\omega_n} \bigr) \notag \\
&\hspace{80pt}+ \bigl(|\cdot|^a \, |g^{\i\omega_n}|\bigr) * \Hcal_\Abold^{-\i\omega_n} + |g^{\i\omega_n}| * \bigl(|\cdot|^a \, \Hcal_\Abold^{-\i\omega_n} \bigr), \label{LTA-LtildeTA_FTA_definition}
\end{align}
where $a \geq 0$, and
\begin{align}
G_{T, \Abold, A}^\pm &:= \frac 2\beta \sum_{n\in \Zbb} \Hcal_{\nabla, \Abold, A}^{\pm \i\omega_n} * \Gcal_\Abold^{-\i\omega_n} + \Hcal_\Abold^{\i\omega_n} * \Gcal_{\nabla , \Abold, A}^{\mp \i \omega_n}  + \Ical_A^{\pm \i\omega_n} * \Hcal_\Abold^{-\i\omega_n} + |g^{\i\omega_n}| * \Hcal_{\nabla, \Abold, A}^{\mp \i\omega_n}  \label{LTA-LtildeTA_GTA_definition}
\end{align}
play a prominent role. Here, $\omega_n$ are the Matsubara frequencies in \eqref{Matsubara_frequencies}, $g^z$ is the resolvent kernel in \eqref{g_definition}, and we used the functions in \eqref{Gcals_Hcals_definition}. For any $a\geq 0$, we prove momentarily that there is a constant $h_0>0$ such that for $0\leq h\leq h_0$ we have
\begin{align}
\Vert F_{T,\Abold_h}^a \Vert_1 + \Vert G_{T,\Abold_h, A_h}^\pm \Vert_1 \leq C_a \, M_{\Abold_h} \leq C_a \; h^3, \label{LTA-LtildeTA_FTAGTA}
\end{align}
where $M_\Abold$ is the constant in \eqref{MA_definition}. To see that \eqref{LTA-LtildeTA_FTAGTA} is true we note that the function $f(t, \omega)$ in \eqref{g0_decay_f} obeys the estimate
\begin{align}
f(0, \omega_n) &\leq C \; (T^{-1} + T^{-2}) \; |2n+1|^{-1} \label{g0_decay_f_estimate1}
\end{align}
and that further
\begin{align}
\frac{|\omega_n| + |\mu|}{|\omega_n| + \mu_-} \leq C \; (1+ T^{-1}). \label{g0_decay_f_estimate2}
\end{align}
Since $T\geq T_0>0$, the assumption \eqref{GAz-GtildeAz_decay_assumption} in Lemma \ref{GAz-GtildeAz_decay} is satisfied for all $0 < h \leq h_0$ provided $h_0>0$ is small enough. Thus, Lemmas~\ref{g_decay} and \ref{GAz-GtildeAz_decay} imply \eqref{LTA-LtildeTA_FTAGTA}.

\begin{proof}[Proof of Proposition \ref{LTA-LtildeTA}]
By definition, we have
\begin{align}
\Vert L_{T, \Abold}\Delta - \tilde L_{T, \Bbold, A} \Delta\Vert_\Hsymm^2 &= \Vert L_{T, \Abold}\Delta - \tilde L_{T, \Bbold, A} \Delta\Vert_2^2 \notag \\
&\hspace{-50pt} + \Vert \Pi (L_{T, \Abold}\Delta - \tilde L_{T, \Bbold, A} \Delta)\Vert_2^2 + \Vert \tilde \pi(L_{T, \Abold}\Delta - \tilde L_{T, \Bbold, A} \Delta)\Vert_2^2 \label{LTA-LtildeTBA_8}
\end{align}
and we claim that the first term on the right side satisfies
\begin{align}
\Vert L_{T, \Abold}\Delta - \tilde L_{T, \Bbold, A} \Delta\Vert_2^2 &\leq 4 \; \Vert \Psi\Vert_2^2 \; \Vert F_{T, \Abold_h}^0 * |V\alpha_*| \, \Vert_2^2. \label{LTA-LtildeTBA_1}
\end{align}
If this holds, the desired estimate for this term follows from Young's inequality, \eqref{Periodic_Sobolev_Norm}, and \eqref{LTA-LtildeTA_FTAGTA}. To prove \eqref{LTA-LtildeTBA_1}, an expansion of the squared modulus in the Hilbert--Schmidt norm yields
\begin{align}
\Vert L_{T, \Abold}\Delta - \tilde L_{T, \Bbold, A}\Delta\Vert_2^2 & \leq 4 \int_{\Rbb^3} \dd r \iint_{\Rbb^3\times \Rbb^3} \dd Z\dd Z' \iint_{\Rbb^3\times \Rbb^3} \dd s\dd s' \; |V\alpha_*(s)| \; |V\alpha_*(s')|  \notag\\
&\hspace{60pt} \times \esssup_{X\in \Rbb^3} | (k_{T, \Bbold, \Abold} - \tilde k_{T, \Bbold, A}) (X, Z, r, s)| \notag\\
&\hspace{60pt} \times \esssup_{X\in \Rbb^3} |(k_{T, \Bbold, \Abold} - \tilde k_{T, \Bbold, A}) (X, Z', r, s')|  \notag\\
&\hspace{30pt}\times \fint_{Q_h} \dd X \; |\e^{\i Z \cdot \Pi} \Psi(X)| \; | \e^{\i Z'\cdot \Pi}\Psi(X)|. \label{Expanding_the_square}
\end{align}
Since the operator $\e^{\i Z\cdot \Pi}$ is bounded by $1$, we have
\begin{align}
\fint_{Q_h} \dd X \; |\e^{\i Z \cdot \Pi} \Psi(X)| \; | \e^{\i Z'\cdot \Pi}\Psi(X)| &\leq \Vert \Psi\Vert_2^2. \label{LTA-LtildeTBA_3}
\end{align}
Consequently, \eqref{Expanding_the_square} yields
\begin{align}
\Vert L_{T, \Abold}\Delta - \tilde L_{T, \Bbold, A}\Delta\Vert_2^2 & \leq 4\; \Vert \Psi\Vert_2^2 \notag\\
&\hspace{-80pt} \times \int_{\Rbb^3 }\dd r\, \Bigl| \iint_{\Rbb^3\times \Rbb^3} \dd Z\dd s \; \esssup_{X\in \Rbb^3} |(k_{T, \Bbold, \Abold} - \tilde k_{T, \Bbold, A}) (X, Z, r, s)|\; |V\alpha_*(s)|\Bigr|^2. \label{LTA-LtildeTBA_2}
\end{align}
We note that
\begin{align}
k_T^n(Z, r - s) \, \e^{\i \tilde \Phi_{A_h}(X, Z,r, s)} &= \tilde G_{A_h}^{\i \omega_n}\bigl( X + \frac r2 , X + Z + \frac s2\bigr) \; \tilde G_{A_h}^{-\i \omega_n} \bigl( X - \frac r2, X + Z - \frac s2 \bigr) \notag \\
&\hspace{-50pt} = \tilde \Gbold_{\Bbold, \Abold_h}^{\i\omega_n} \bigl( X + \frac r2 , X + Z + \frac s2\bigr) \;  \tilde \Gbold_{\Bbold, \Abold_h}^{-\i\omega_n} \bigl( X - \frac r2, X + Z - \frac s2 \bigr) \notag \\
&\hspace{-50pt}=: \tilde k_{T, \Bbold, \Abold}^n(X, Z, r, s) \label{LTA-LtildeTBA_14}
\end{align}
with $\tilde \Gbold_{\Bbold, \Abold_h}^z$ in \eqref{GboldtildeBAz_definition}, whence
\begin{align}
(k_{T, \Bbold, \Abold}^n - \tilde k_{T, \Bbold, A}^n)(X, Z, r, s) & \notag \\
&\hspace{-120pt} = \bigl( \Gbold_{\Bbold, \Abold_h}^{\i \omega_n} - \tilde \Gbold_{\Bbold, \Abold_h}^{\i \omega_n}\bigr) \bigl( X+ \frac r2, X + Z+ \frac s2\bigr) \, \Gbold_{\Bbold, \Abold_h}^{-\i \omega_n} \bigl( X - \frac r2, X + Z - \frac s2\bigr) \notag \\
&\hspace{-100pt} + \tilde \Gbold_{\Bbold, \Abold_h}^{\i \omega_n} \bigl( X + \frac r2, X + Z + \frac s2\bigr) \, \bigl( \Gbold_{\Bbold, \Abold_h}^{-\i \omega_n} - \tilde \Gbold_{\Bbold, \Abold_h}^{-\i \omega_n} \bigr) \bigl( X - \frac r2 , X + Z - \frac s2\bigr) \label{LTA-LtildeTBA_15}
\end{align}
so that, by Lemma \ref{GAz-GtildeAz_decay}, the integrand in \eqref{LTA-LtildeTBA_2} is bounded by
\begin{align}
\bigl|(k_{T, \Bbold, \Abold} - \tilde k_{T, \Bbold, A}) (X, Z, r, s)\bigr| &\leq  \frac 2\beta\,  \smash{\sum_{n\in \Zbb}} \, \bigl[ \Hcal_{\Abold_h}^{\i\omega_n} \bigl( Z - \frac {r-s}2\bigr) \; \Gcal_{\Abold_h}^{-\i\omega_n} \bigl( Z + \frac {r-s}2\bigr) \notag\\
&\hspace{30pt}+ |g^{\i\omega_n}|\bigl( Z -\frac {r-s}2\bigr)\;  \Hcal_{\Abold_h}^{-\i\omega_n} \bigl( Z +\frac {r-s}2\bigr)\bigr]. \label{LTA-LtildeTBA_11}
\end{align}
We combine the bound
\begin{align}
|Z|^a &\leq 
\bigl| Z + \frac{r}{2}\bigr|^a + \bigl| Z - \frac{r}{2}\bigr|^a, \label{Z_estimate}
\end{align}
which holds for $a \geq 0$, \eqref{LTA-LtildeTBA_11}, and the fact that the functions in \eqref{LTA-LtildeTBA_11} are even (see Lemma \ref{GAz-GtildeAz_decay}), which implies
\begin{align}
\int_{\Rbb^3} \dd Z \; |Z|^a\; \esssup_{X\in \Rbb^3} |(k_{T, \Bbold, \Abold} - \tilde k_{T, \Bbold, A})(X, Z, r, s)| \leq F_{T, \Abold_h}^a(r-s), \label{LTA-LtildeTBA_5}
\end{align}
where $F_{T, \Abold}^a$ is the function in \eqref{LTA-LtildeTA_FTA_definition}. When we apply the case $a =0$ to \eqref{LTA-LtildeTBA_2}, we obtain \eqref{LTA-LtildeTBA_1}.

For the second term on the right side of \eqref{LTA-LtildeTBA_8}, we claim that
\begin{align}
\Vert \Pi(L_{T, \Abold}\Delta - \tilde L_{T, \Bbold, A}\Delta)\Vert_2^2 &\leq C \, h^2 \; \Vert \Psi\Vert_{\Hmag^1(Q_h)}^2 \notag\\
& \hspace{-20pt} \times \Vert (F_{T, \Abold_h}^0 + F_{T, \Abold_h}^1 + G_{T, \Abold_h, A_h}^{+} + G_{T, \Abold_h, A_h}^{-}) * |V\alpha_*| \,\Vert_2^2. \label{LTA-LtildeTBA_6}
\end{align}
If this holds, the desired estimate for this term follows from Young's inequality and \eqref{LTA-LtildeTA_FTAGTA}. To see that \eqref{LTA-LtildeTBA_6} is true, we apply \eqref{Expanding_the_square} with $\e^{\i Z \cdot \Pi}$ replaced by $\Pi \, \e^{\i Z \cdot \Pi}$. This amounts to replacing \eqref{LTA-LtildeTBA_3} by
\begin{align*}
\fint_{Q_h} \dd X \; |\Pi \, \e^{\i Z \cdot \Pi}\Psi(X)| \; |\Pi \, \e^{\i Z' \cdot \Pi}\Psi(X)| &\leq \Vert \Pi \, \e^{\i Z \cdot \Pi}\Psi\Vert_2\; \Vert \Pi \, \e^{\i Z' \cdot \Pi}\Psi\Vert_2.
\end{align*}
From \cite[Lemma 5.11]{DeHaSc2021} or from a direct computation, we know that
\begin{align}
\Pi\, \e^{\i Z\cdot \Pi} = \e^{\i Z \cdot \Pi} \bigl[\Pi - 2 \, \Bbold \wedge Z\bigr]. \label{PiU_intertwining}
\end{align}
Using \eqref{Periodic_Sobolev_Norm}, this yields
\begin{align}
\Vert \Pi \, \e^{\i Z\cdot \Pi} \Psi\Vert_2 &\leq \Vert \Pi\Psi\Vert_2 + |\Bbold| \, |Z| \; \Vert \Psi\Vert_2 \leq C\, h^2 \; \Vert \Psi\Vert_{\Hmag^1(Q_h)} \; (1 + |Z|), \label{PiXcos_estimate}
\end{align}
which subsequently proves
\begin{align}
\Vert \Pi(L_{T, \Abold}\Delta - \tilde L_{T, \Bbold, A}\Delta)\Vert_2^2 &\leq C \, h^2 \; \Vert \Psi\Vert_{\Hmag^1(Q_h)}^2  \notag\\
&\hspace{-110pt} \times  \int_{\Rbb^3} \dd r \, \Bigl| \iint_{\Rbb^3\times \Rbb^3} \dd Z\dd s \; \bigl[h\, (1+ |Z|) \; \esssup_{X\in \Rbb^3} |(k_{T, \Bbold, \Abold} - \tilde k_{T, \Bbold, A})(X, Z, r,s)|  \notag \\
&\hspace{-20pt} + \esssup_{X\in \Rbb^3} |(\nabla_X k_{T, \Bbold, \Abold} - \nabla_X \tilde k_{T, \Bbold, A})(X, Z, r, s)|\bigr] \, |V\alpha_*(s)|\Bigr|^2. \label{LTA-LtildeTBA_9}
\end{align}
With the help of Lemma \ref{GAz-GtildeAz_decay} and \eqref{LTA-LtildeTBA_15} a straightforward computation shows that
\begin{align*}
\int_{\Rbb^3} \dd Z \; \esssup_{X\in \Rbb^3} |(\nabla_X k_{T, \Bbold, \Abold} - \nabla_X \tilde k_{T, \Bbold, A})(X, Z, r, s)| \leq \bigl( G_{T, \Abold_h, A_h}^+ + G_{T, \Abold_h, A_h}^-\bigr) (r-s).
\end{align*}
Consequently, an application of $a=0$ and $a=1$ in \eqref{LTA-LtildeTBA_5} to \eqref{LTA-LtildeTBA_9} shows \eqref{LTA-LtildeTBA_6}.

We claim that the third term on the right side of \eqref{LTA-LtildeTBA_8} satisfies
\begin{align}
\Vert \tilde \pi (L_{T, \Abold}\Delta - \tilde L_{T, \Bbold, A} \Delta)\Vert_2^2 \leq C \; \Vert \Psi\Vert_2^2 \; \Vert (F_{T, \Abold_h}^1 + G_{T, \Abold_h, A_h}^+) * |V\alpha_*| \, \Vert_2^2. \label{LTA-LtildeTBA_4}
\end{align}
If this holds, the desired bound for this term follows from Young's inequality, \eqref{LTA-LtildeTA_FTAGTA}, and \eqref{Periodic_Sobolev_Norm}. To see that \eqref{LTA-LtildeTBA_4} is true, we estimate
\begin{align}
\Vert \tilde \pi (L_{T, \Abold}\Delta - \tilde L_{T, \Bbold, A}\Delta)\Vert_2^2 & \leq 4 \;\Vert \Psi\Vert_2^2 \notag \\
&\hspace{-100pt} \times  \int_{\Rbb^3} \dd r \, \Bigl| \iint_{\Rbb^3\times \Rbb^3} \dd Z\dd s \; \esssup_{X\in \Rbb^3} |(\tilde \pi k_{T, \Bbold, \Abold} - \tilde \pi\tilde k_{T, \Bbold, A} )(X, Z, r, s)| \;|V\alpha_*(s)| \Bigr|^2.  \label{LTA-LtildeTBA_10}
\end{align}
We use $\nabla_r \Bbold\cdot (r\wedge s) = -\Bbold \wedge s$ and
\begin{align}
\tilde \pi \, \e^{\i \frac{\Bbold}{4} \cdot (r\wedge s)} = \e^{\i \frac{\Bbold}{4} \cdot (r\wedge s)} \bigl[ -\i \nabla_r + \frac 14 \, \Bbold \wedge (r-s)\bigr], \label{tildepi_magnetic_phase}
\end{align}
which implies that the integrand on the right side of \eqref{LTA-LtildeTBA_10} obeys
\begin{align}
|\tilde \pi k_{T, \Bbold, \Abold} - \tilde \pi\tilde k_{T, \Bbold, A}| &\leq |\nabla_r k_{T, \Bbold, \Abold} - \nabla_r \tilde k_{T, \Bbold, \Abold}| + |\Bbold| \,  |r-s| \, |k_{T, \Bbold, \Abold} - \tilde k_{T, \Bbold, \Abold}|. \label{LTA-LtildeTBA_13}
\end{align}
By \eqref{LTA-LtildeTBA_15} and Lemma \ref{GAz-GtildeAz_decay}, a straightforward computation shows
\begin{align}
\int_{\Rbb^3} \dd Z \; \esssup_{X\in \Rbb^3} |(\nabla_r k_{T, \Bbold, \Abold} - \nabla_r \tilde k_{T, \Bbold, A})(X, Z, r, s)| \leq G_{T, \Abold_h, A_h}^+(r-s). \label{LTA-LtildeTBA_12}
\end{align}
Furthermore, for $a\geq 0$, we have the bound
\begin{align}
|r-s|^a &= \bigl| \frac{r-s}{2} + Z + \frac{r-s}{2} - Z\bigr|^a \leq 2^{(a-1)_+} \Bigl( \bigl| Z - \frac{r-s}{2}\bigr|^a + \bigl| Z + \frac{r-s}{2}\bigr|^a\Bigr), \label{r-s_estimate}
\end{align}
which proves that $|r-s|\, F_{T, \Abold}^0(r-s) \leq F_{T, \Abold}^1(r-s)$. Therefore, \eqref{LTA-LtildeTBA_13} implies the estimate
\begin{align}
\int_{\Rbb^3} \dd Z \; \esssup_{X\in \Rbb^3} |(\tilde \pi k_{T, \Bbold, \Abold} - \tilde \pi\tilde k_{T, \Bbold, A}) (X, Z, r, s)| \leq \bigl(F_{T, \Abold_h}^1 + G_{T, \Abold_h, A_h}^+\bigr)(r-s). \label{LTA-LtildeTBA_7}
\end{align}
Finally, \eqref{LTA-LtildeTBA_7} implies \eqref{LTA-LtildeTBA_4}, which finishes the proof.
\end{proof}

\paragraph{The operator $\tilde M_{T,\Abold}$.}

We define $\tilde M_{T,\Abold}$ by
\begin{align}
\tilde M_{T,\Abold}\alpha(X,r) := \iint_{\Rbb^3 \times \Rbb^3} \dd Z\dd s \; k_T(Z , r-s) \, \e^{-\i \frac{r-s}{4} \cdot D\Abold_h (X)(r+s)} \, \bigl(\cos(Z \cdot \Pi_{\Abold_h}) \alpha\bigr) (X,s), \label{MtildeTA_definition}
\end{align}
where $k_T(Z, r) := k_{T,0,0}(0, Z, r, 0)$ with $k_{T,0,0}$ in \eqref{kTBA_definition} and where $(D\Abold)_{ij} := \partial_j \Abold_i$ is the Jacobian matrix of $\Abold$. In our calculation, we may replace $\tilde L_{T, \Bbold, A}$ by $\tilde M_{T, \Abold}$ due to the following error bound.

\begin{prop}
\label{LtildeTA-MtildeTA}
For any $T_0>0$ and $A\in \Wpervec 2$ there is $h_0>0$ such that for any $0< h \leq h_0$, any $T\geq T_0$, and whenever  $|\cdot|^k V\alpha_*\in L^2(\Rbb^3)$ for $k\in \{0,1,2\}$,  $\Psi\in \Hmag^1(Q_h)$, and $\Delta \equiv \Delta_\Psi$ as in \eqref{Delta_definition}, we have
\begin{align}
\Vert \tilde L_{T, \Bbold, A}\Delta - \tilde M_{T,\Abold}\Delta \Vert_\Hsymm^2 &\leq C\; h^8 \; \max_{k =0,1,2} \Vert \, |\cdot|^k V\alpha_*\Vert_2^2 \;\Vert \Psi\Vert_{\Hmag^1(Q_h)}^2. \label{LtildeTA-MtildeTA_eq1}
\end{align}
\end{prop}

For the proof, we need several preparatory lemmas. The first result extracts the magnetic field contribution $\Phi_{2A_h}(X, X + Z)$, which completes the operator $\e^{\i Z\cdot \Pi_{\Abold_h}}$ in $\tilde L_{T, \Bbold, \Abold}$, see \eqref{Phi_Approximation_eq1}. In \cite{DeHaSc2021}, where only the constant magnetic field is present, many of these approximations all hold as algebraic identities. Parts of the following result are proven in \cite[Chapter 6]{AlissaPhD}.

\begin{lem}
\label{Phi_Approximation}
For all magnetic potentials $A\in \Wpervec{3}$ and all vectors $Z,r,s\in \Rbb^3$, we have
\begin{align}
\esssup_{X\in \Rbb^3} \bigl| \tilde \Phi_A(X, Z, r, s) - \Phi_{2A}(X,X+Z) + \frac 14 (r-s)\cdot D A(X)(r+s)\bigr| & \notag\\
&\hspace{-200pt} \leq C \; \Vert D^2A\Vert_\infty \; \bigl( |Z| +  |r-s|\bigr)  \bigl( |s|^2 + |r-s|^2\bigr), \label{Phi_Approximation_eq1}
\end{align}
where $\Phi_A$ is defined in \eqref{PhiA} and $\tilde \Phi_A$ is defined in \eqref{LtildeTA_PhitildeA_definition}. Furthermore, $(DA)_{ij} = \partial_j A_i$ is the Jacobian matrix of $A$ and $(D^2A)_{ijk} = \partial_k \partial_j A_i$ is the Hessian of $A$. We also have
\begin{align}
\esssup_{X\in \Rbb^3} |\nabla_X \tilde\Phi_{A}(X, Z, r, s) | \leq C \, \Vert DA\Vert_\infty \Bigl( \bigl| Z + \frac{r-s}{2}\bigr| + \bigl| Z - \frac{r-s}{2}\bigr|\Bigr) \label{Phi_Approximation_eq2}
\end{align}
and
\begin{align}
\esssup_{X\in \Rbb^3} \bigl| \nabla_X \tilde \Phi_A(X, Z, r, s) - \nabla_X \Phi_{2A} (X, X + Z) \bigr| & \notag\\
&\hspace{-100pt} \leq C \; \Vert D^2A \Vert_\infty\; \bigl( |Z| + |r-s|\bigr)  \bigl(|s| + |r-s|\bigr) \label{Phi_Approximation_eq3}
\end{align}
as well as
\begin{align}
\esssup_{X\in \Rbb^3} \bigl| \nabla_X \frac{r-s}{4} DA(X) (r+s) \bigr| \leq C \, \Vert D^2A\Vert_\infty \, |r-s| \, \bigl( |s| + |r-s|\bigr) \label{Phi_Approximation_eq4}
\end{align}
and
\begin{align}
\esssup_{X\in \Rbb^3} |\nabla_r \tilde \Phi_{A}(X, Z, r, s)| &\leq \Vert DA\Vert_\infty \, \bigl( |Z| + |r-s| \bigr) + \Vert A\Vert_\infty \label{Phi_Approximation_eq5}
\end{align}
and
\begin{align}
\esssup_{X\in \Rbb^3} \bigl| \nabla_r \tilde \Phi_A(X, Z, r, s) + \frac 14 \nabla_r (r-s)\cdot D A(X)(r+s) \bigr| &\notag\\
&\hspace{-100pt}\leq C\, \Vert D^2A\Vert_\infty \, \bigl( |s|^2 + |r-s|^2 + |Z|^2\bigr). \label{Phi_Approximation_eq6}
\end{align}
\end{lem}


\begin{proof}[Proof of Lemma \ref{Phi_Approximation}]
In this proof, we use the short-hand notation $\zeta_X^r := X + \frac r2$. We insert the definition, sort together the terms with $Z$ and $\frac{r-s}{2}$, and obtain
\begin{align}
\tilde \Phi_A(X, Z, r, s) &= \int_0^1 \dt\; \bigl[A\bigl( \zeta_{X + Z - tZ}^{s + t(r-s)} \bigr) + A\bigl(\zeta_{X + Z - tZ}^{-s - t(r-s)}\bigr) \bigr] \cdot Z \notag \\
& \hspace{50pt} -  \int_0^1 \dt\; \bigl[A\bigl(\zeta_{X + Z - tZ}^{s + t(r-s)}\bigr) - A\bigl( \zeta_{X + Z - tZ}^{-s - t(r-s)} \bigr) \bigr] \cdot \frac{r-s}{2}. \label{Phi_Approximation_1}
\end{align}
We consider the terms in square brackets and perform a second order Taylor expansion in the variable $\pm \frac 12 (s + t(r-s))$, that is, we have the estimate
\begin{align}
\Bigl| A\bigl(X + Z- tZ \pm \frac{s + t(r-s)}{2}\bigr) - A(X + Z-tZ) \mp \frac 12 DA(X + Z-tZ) (s + t(r-s)) \Bigr|   & \notag\\
&\hspace{-220pt} \leq C\; \Vert D^2A\Vert_\infty \; \bigl(|s|^2 +|r-s|^2\bigr). \label{Phi_Approximation_2}
\end{align}
For the first term on the right side of \eqref{Phi_Approximation_1}, this implies
\begin{align}
\Bigl| \int_0^1 \dt\; \bigl[A\bigl( \zeta_{X + Z - tZ}^{s + t(r-s)} \bigr) + A\bigl( \zeta_{X + Z - tZ}^{-s - t(r-s)} \bigr) \bigr] \cdot Z  -\Phi_{2A}(X,X+Z) \Bigr| & \notag\\
&\hspace{-100pt}  \leq C \, \Vert D^2A\Vert_\infty \, |Z| \, \bigl(|s|^2 + |r-s|^2\bigr), \label{Phi_Approximation_4}
\end{align}
while for the second term of \eqref{Phi_Approximation_1}, \eqref{Phi_Approximation_2} implies
\begin{align}
\Bigl| - \int_0^1 \dt\; \bigl[A\bigl( \zeta_{X + Z - tZ}^{s + t(r-s)} \bigr) - A\bigl(\zeta_{X + Z - tZ}^{-s - t(r-s)}\bigr) \bigr] \cdot \frac{r-s}{2}+ \frac 14 (r-s) \cdot DA(X) (r+s) \Bigr| & \notag\\
&\hspace{-350pt} \leq \Bigl| -\frac{r-s}2 \cdot \int_0^1 \dd t \; \bigl[ DA(X + Z - tZ) - DA(X) \bigr](s + t(r-s)) \Bigr| \notag \\
&\hspace{-150pt} + C\, \Vert D^2A\Vert_\infty \, \bigl( |s|^2 + |r-s|^2\bigr) \, |r-s| \notag \\
&\hspace{-350pt} \leq C\, \Vert D^2A \Vert_\infty \, |r-s| \, \bigl[|s|^2 + |r-s|^2 + \bigl(|s| + |r-s|\bigr) |Z| \bigr] . \label{Phi_Approximation_3}
\end{align}
Adding up \eqref{Phi_Approximation_4} and \eqref{Phi_Approximation_3} proves \eqref{Phi_Approximation_eq1}.

Differentiating \eqref{Phi_Approximation_1} with respect to $X$, we infer
\begin{align*}
\nabla_X \tilde \Phi_A(X, Z,r , s) &= \int_0^1 \bigl[ DA\bigl( \zeta_{X + Z - tZ}^{s + t(r-s)} \bigr)^t + DA\bigl(\zeta_{X + Z - tZ}^{-s - t(r-s)}\bigr)^t \bigr] Z \notag \\
& \hspace{50pt} -  \int_0^1 \dt\; \bigl[ DA\bigl(\zeta_{X + Z - tZ}^{s + t(r-s)}\bigr)^t - DA\bigl( \zeta_{X + Z - tZ}^{-s - t(r-s)} \bigr)^t \bigr] \frac{r-s}{2}
\end{align*}
and deduce \eqref{Phi_Approximation_eq2}. When we approximate
\begin{align}
\Bigl| DA \bigl( X + Z- tZ \pm \frac{s + t(r-s)}{2} \bigr)^t - DA(X + Z - tZ)^t\Bigr| \leq \Vert D^2A\Vert_\infty\, \bigl( |s| + |r-s| \bigr), \label{Phi_Approximation_6} 
\end{align}
this further proves \eqref{Phi_Approximation_eq3}. The proof of \eqref{Phi_Approximation_eq4} is a straightforward computation.

The estimate \eqref{Phi_Approximation_eq5} follows from differentiating \eqref{Phi_Approximation_1} with respect to $r$, which reads
\begin{align}
\nabla_r \tilde \Phi_A(X, Z, r, s) &= \int_0^1 \dt\; \frac t2\bigl[DA\bigl( \zeta_{X + Z - tZ}^{s + t(r-s)} \bigr)^t - DA\bigl(\zeta_{X + Z - tZ}^{-s - t(r-s)}\bigr)^t \bigr] Z \notag \\
& \hspace{50pt} -  \int_0^1 \dt\; \frac t2 \bigl[DA\bigl(\zeta_{X + Z - tZ}^{s + t(r-s)} \bigr)^t + DA\bigl(\zeta_{X + Z - tZ}^{-s - t(r-s)} \bigr)^t \bigr] \frac{r-s}{2} \notag \\
& \hspace{50pt} - \frac 12 \int_0^1 \dt\; \bigl[A\bigl(\zeta_{X + Z - tZ}^{s + t(r-s)} \bigr) - A\bigl(\zeta_{X + Z - tZ}^{-s - t(r-s)} \bigr) \bigr]. \label{Phi_Approximation_7}
\end{align}
It remains to prove \eqref{Phi_Approximation_eq6}. We use \eqref{Phi_Approximation_6} to see that the first term of \eqref{Phi_Approximation_7} is bounded by $C\Vert D^2A\Vert_\infty (|s| + |r-s|) |Z|$. For the second and third term of \eqref{Phi_Approximation_7} we note
\begin{align}
\frac 14 \nabla_r (r - s) \cdot DA(X) (r + s) = \frac 14 DA(X)^t (r-s) + \frac 14 DA(X) (r+s). \label{Phi_Approximation_9}
\end{align}
We use the first term of this on the second term of \eqref{Phi_Approximation_7}, employ \eqref{Phi_Approximation_6}, and obtain
\begin{align}
\Bigl| -  \int_0^1 \dt\; \frac t2 \bigl[DA\bigl(\zeta_{X + Z - tZ}^{s + t(r-s)} \bigr)^t + DA\bigl(\zeta_{X + Z - tZ}^{-s - t(r-s)} \bigr)^t \bigr] \frac{r-s}{2} + \frac 14 DA(X)^t (r-s) \bigr| & \notag \\
&\hspace{-360pt} \leq \Bigl| \int_0^1 \dd t \; \frac{t}{2} \bigl[ DA(X + Z - tZ)^t - DA(X)^t\bigr]  (r-s) \Bigr| + C \, \Vert D^2A\Vert_\infty \bigl( |s| + |r-s|\bigr) |r-s| \notag\\
&\hspace{-360pt} \leq C \, \Vert D^2A\Vert_\infty \, \bigl[ |r-s| \, |Z| + \bigl( |s| + |r-s|\bigr) |r-s|\bigr] . \label{Phi_Approximation_8}
\end{align}
For the third term of \eqref{Phi_Approximation_7}, we use the second term of \eqref{Phi_Approximation_9} and get
\begin{align*}
\Bigl| - \frac 12 \int_0^1 \dt\; \bigl[A\bigl(\zeta_{X + Z - tZ}^{s + t(r-s)} \bigr) - A\bigl(\zeta_{X + Z - tZ}^{-s - t(r-s)} \bigr) \bigr] + \frac 14  DA(X) (r+ s)\Bigr| \leq \Tcal_+ + \Tcal_- + \Tcal,
\end{align*}
where
\begin{align*}
\Tcal_\pm &:= \Bigl| -\frac 12 \int_0^1 \dd t \; \bigl[ A\bigl( \zeta_{X + Z - tZ}^{\pm s \pm t(r-s)}\bigr) \mp A(X + Z - tZ) - \frac 12 DA(X + Z- tZ) (s+ t(r-s)) \bigr]\Bigr|
\end{align*}
and
\begin{align*}
\Tcal &:= \Bigl| -\frac 12 \int_0^1 \dd t \; DA(X + Z- tZ) (s + t(r-s)) + \frac 14 DA(X) (r+s)\Bigr|.
\end{align*}
By \eqref{Phi_Approximation_2}, we have
\begin{align*}
\Tcal_\pm &\leq C \, \Vert D^2 A\Vert_\infty \, \bigl( |s|^2 + |r-s|^2\bigr), & \Tcal &\leq C\, \Vert D^2A\Vert_\infty \, \bigl( |s| + |r-s|\bigr ) \, |Z|.
\end{align*}
In combination with \eqref{Phi_Approximation_8}, these considerations imply \eqref{Phi_Approximation_eq6}.
\end{proof}

The next result is the substitute for the identity \cite[Eq. (5.39)]{DeHaSc2021} for the more general magnetic field we consider here.

\begin{lem}
\label{Magnetic_Translation_Representation}
Let $Z\in \Rbb^3$, $A\in \Lper$, and $h>0$. Then, on $\Lmag^2(Q_h)$, we have the operator equation
\begin{align}
\e^{\i \Phi_{2A_h}(X , X +Z)} \, \e^{\i Z\cdot \Pi} = \e^{\i Z\cdot \Pi_{\Abold_h}}. 
\end{align}
\end{lem}

This is a consequence of the following abstract proposition, whose proof can be found in \cite[p. 290]{Werthamer1966} and which is included here for the sake of completeness.

\begin{prop}
\label{Operator_Equality_Abstract}
Let $\Hcal$ be a Hilbert space, $P\colon \Dcal(P)\ra \Hcal$ self-adjoint and let $Q$ be bounded and self-adjoint. Assume that $[\e^{\i tP} \, Q \, \e^{-\i tP} , \e^{\i sP} \, Q \, \e^{-\i sP} ] =0$ for every $t,s\in [0,1]$. Then, we have
\begin{align*}
\exp\Bigl(\i \int_0^1\dt \; \e^{\i tP} \, Q\, \e^{-\i tP}\Bigr) \, \e^{\i P} = \e^{\i(P + Q)}.
\end{align*}
\end{prop}

\begin{proof}
For $s\in \Rbb$, set $Q(s) := \e^{\i sP} \, Q \, \e^{-\i sP}$ and $W(s) := \e^{\i s(P+Q)} \, \e^{-\i sP}$. On $\Dcal(P)$, we may differentiate $W$ to get
\begin{align*}
-\i W'(s) = \e^{\i s(P+Q)} (P+Q) \, \e^{-\i sP} - \e^{\i s(P+Q)} \, P \, \e^{-\i sP} 
=  \e^{\i s(P+Q)} \, Q \, \e^{-\i sP} = W(s) \, Q(s).
\end{align*}
This equation can be extended to all of $\Hcal$, since both sides are continuous. Hence, $W$ satisfies the linear differential equation $W'(s) = \i W(s) \, Q(s)$ of which the unique solution is given by
\begin{align*}
\tilde W(s) := \exp\Bigl(\i \int_0^s \dt\; Q(t)\Bigr).
\end{align*}
Since $[Q(t),Q(s)] =0$ by assumption, we conclude that $\tilde W$ indeed is continuously differentiable and satisfies $\tilde W'(s) = \i \tilde W(s) \, Q(s)$ for each $s\in [0,1]$. By uniqueness, we obtain
\begin{align*}
\exp\Bigl(\i \int_0^s \dt\; Q(t)\Bigr) = \e^{\i s(P+Q)}\e^{-\i sP}.
\end{align*}
Setting $s =1$ gives the claim.
\end{proof}

\begin{proof}[Proof of Lemma \ref{Magnetic_Translation_Representation}]
We first show that $\e^{\i \Phi_{2A_h}(X, X+Z)} \, \e^{\i Z \cdot (-\i \nabla_X)} = \e^{\i Z \cdot (-\i \nabla_X + 2A_h)}$. The full statement is then clear because $Z \cdot (\Bbold \wedge X)$ commutes with $Z\cdot (-\i \nabla_X)$ as well as with $Z\cdot A_h$. We apply Proposition \ref{Operator_Equality_Abstract} to $P = Z\cdot (-\i \nabla)$ and $Q = Z\cdot 2A_h$. When we define $Q(t) := \e^{\i tP}\, Q\, \e^{-\i tP}$, we have $Q(t) = Z \cdot 2A_h(X + tZ)$ and, in particular, $[Q(s),Q(t)]=0$ for all $s,t\in \Rbb$. Furthermore, a change of variables $t\mapsto 1 -t$ shows
\begin{align*}
\int_0^1 \dt \; Q(t) &= \Phi_{2A_h}(X,X+Z).
\end{align*}
Hence, Proposition \ref{Operator_Equality_Abstract} gives the claim.
\end{proof}

We will momentarily start with the proof of Proposition \ref{LtildeTA-MtildeTA}, in which the functions
\begin{align}
F_T^{a} := \frac 2\beta \sum_{n\in \Zbb} \sum_{b = 0}^a \binom ab \; \bigl(|\cdot|^{b}\,  |g^{\i\omega_n}|\bigr) * \bigl(|\cdot|^{a-b} \, |g^{-\i\omega_n}| \bigr) \label{LtildeTA-MtildeTA_FT_definition}
\end{align}
and
\begin{align}
G_T^a &:= \smash{\frac 2\beta \sum_{n\in \Zbb} \sum_{b=0}^a \binom ab} \; \bigl( |\cdot|^b \, |\nabla g^{\i \omega_n}| \bigr) * \bigl( |\cdot|^{a-b} \, |g^{-\i\omega_n}|\bigr) \notag \\
&\hspace{150pt} + \bigl( |\cdot|^b \, |g^{\i\omega_n}| \bigr) * \bigl( |\cdot|^{a-b} \, |\nabla g^{-\i\omega_n}| \bigr) \label{LtildeTA-MtildeTA_GT_definition}
\end{align}
play a prominent role, where $a\in \Nbb_0$. An application of Lemma~\ref{g_decay},  \eqref{g0_decay_f_estimate1}, and \eqref{g0_decay_f_estimate2} shows that for $T \geq T_0 > 0$ and $a\in \Nbb_0$, we have
\begin{align}
\Vert F_{T}^a\Vert_1 + \Vert G_T^a \Vert_1 &\leq C_{a}. \label{LtildeTA-MtildeTA_FTGT}
\end{align}

\begin{proof}[Proof of Proposition \ref{LtildeTA-MtildeTA}]
We start by claiming
\begin{align}
\Vert \tilde L_{T, \Bbold, A} \Delta - \tilde M_{T, \Abold} \Delta \Vert_2^2 &\leq C \, \Vert \Psi\Vert_2^2 \; \Vert D^2A_h\Vert_\infty^2 \; \Vert F_T^3 * |V\alpha_*| + F_T^1 * |\cdot|^2\, |V\alpha_*|  \, \Vert_2^2. \label{LtildeTA-MtildeTA_6}
\end{align}
If this holds, the desired bound for this term follows from \eqref{LtildeTA-MtildeTA_FTGT}, \eqref{Periodic_Sobolev_Norm}, and Young's inequality. To prove \eqref{LtildeTA-MtildeTA_6}, we note that for any $X, r\in \Rbb^3$ we have $D\Abold_\Bbold(X) \, r = \frac 12\Bbold \wedge r$. Hence, a short computation shows
\begin{align*}
- \frac{r-s}{4} \, D\Abold_\Bbold(X) \, (r + s) = \frac{\Bbold}{4} \cdot (r\wedge s).
\end{align*}
Since the integrand of $\tilde M_{T, \Abold}$ is symmetric in $Z$, by Lemma \ref{Magnetic_Translation_Representation}, we observe
\begin{align*}
\tilde M_{T, \Abold} \alpha(X, r) &= \smash{\iint_{\Rbb^3\times \Rbb^3}} \dd Z \dd s \; k_T(Z, r-s) \, \e^{\i \Phi_{2A_h} (X, X+Z)} \, \e^{-\i \frac{r-s}{4} DA_h(X) (r + s)} \, \e^{\i \frac{\Bbold}{4} \cdot (r\wedge s)} \\
&\hspace{230pt} \times \e^{\i Z \cdot \Pi} \alpha (X, s)
\end{align*}
so that
\begin{align}
\bigl(\tilde L_{T, \Bbold, A} \Delta - \tilde M_{T, \Abold} \Delta\bigr) (X, r) &= -2 \iint_{\Rbb^3 \times \Rbb^3} \dd Z \dd s \; k_T(Z , r-s) \, V\alpha_*(s) \, \e^{\i \frac \Bbold 4 \cdot (r\wedge s)} \; \e^{\i Z \cdot \Pi} \Psi(X)\notag \\
&\hspace{-20pt} \times \bigl[ \e^{\i \tilde \Phi_{A_h}(X, Z, r, s)} - \e^{\i \Phi_{2A_h}(X, X + Z)}\e^{-\i \frac{r-s}{4} DA_h(X)(r + s)} \bigr]. \label{LtildeTA-MtildeTA_1}
\end{align}
Using this as well as the techniques of the estimate on $\Vert \e^{\i Z\cdot \Pi} \Psi\Vert_2$ in \eqref{LTA-LtildeTBA_3} and of the expansion of the squared modulus in \eqref{LTA-LtildeTBA_2}, we obtain
\begin{align}
\Vert \tilde L_{T, \Bbold, A} \Delta - \tilde M_{T, \Abold} \Delta \Vert_2^2 &\leq 4 \, \Vert \Psi\Vert_2^2 \int_{\Rbb^3} \dd r \, \Bigl| \iint_{\Rbb^3\times \Rbb^3} \dd Z \dd s \; |k_T(Z, r-s)| \, |V\alpha_*(s)| \notag \\
&\hspace{-30pt} \times \esssup_{X\in \Rbb^3} \bigl| \e^{\i \tilde \Phi_{A_h}(X, Z, r, s) - \i \Phi_{2A_h}(X, X+ Z) + \i \frac{r-s}{4} DA_h(X) (r + s)} -1\bigr| \Bigr|^2. \label{LtildeTA-MtildeTA_7}
\end{align}
Furthermore, Lemma \ref{Phi_Approximation}, as well as the estimate \eqref{Z_estimate} on $|Z|^a$ and \eqref{r-s_estimate} on $|r-s|^a$ imply
\begin{align}
\int_{\Rbb^3} \dd Z \; |Z|^a \, |k_T(Z, r-s)| \,  \esssup_{X\in \Rbb^3} \bigl| \e^{\i \tilde \Phi_A(X, Z, r, s) - \i \Phi_{2A}(X, X+ Z) + \i \frac{r-s}{4} DA(X) (r + s)} -1\bigr| & \notag \\
&\hspace{-250pt} \leq C \; \Vert D^2A\Vert_\infty \bigl[  F_T^{3+a}(r-s) + F_T^{1+a}(r-s)\; |s|^2\bigr] \label{LtildeTA-MtildeTA_2}
\end{align}
with the function $F_T^a$ in \eqref{LtildeTA-MtildeTA_FT_definition}. In combination with \eqref{LtildeTA-MtildeTA_7}, we deduce \eqref{LtildeTA-MtildeTA_6}.

We claim that the term involving $\Pi$ is bounded by
\begin{align}
\Vert \Pi (\tilde L_{T, \Bbold, A}\Delta - \tilde M_{T, \Abold}\Delta)\Vert_2^2 &\leq C\,  h^2 \, \Vert \Psi\Vert_{\Hmag^1(Q_h)}^2 \, \Vert D^2 A_h\Vert_\infty^2 \, \bigl( 1 +  \Vert DA_h\Vert_\infty^2\bigr) \notag \\
&\hspace{-120pt} \times \bigl[ \Vert (F_T^2 + F_T^3 + F_T^4) * |V\alpha_*|\, \Vert_2^2 + \Vert F_T^1 * |\cdot| \, |V\alpha_*| \, \Vert_2^2 + \Vert (F_T^1 + F_T^2)* |\cdot|^2|V\alpha_*| \, \Vert_2^2\bigr]. \label{LtildeTA-MtildeTA_3}
\end{align}
If this holds, the desired bound for this term follows from Young's inequality and \eqref{LtildeTA-MtildeTA_FTGT}. To prove \eqref{LtildeTA-MtildeTA_3}, we use the techniques to prove \eqref{LTA-LtildeTBA_9},  combine these with \eqref{LtildeTA-MtildeTA_1}, and obtain
\begin{align}
\Vert \Pi (\tilde L_{T, \Bbold, A}\Delta - \tilde M_{T, \Abold}\Delta)\Vert_2^2 &\leq C \, h^2 \, \Vert \Psi\Vert_{\Hmag^1(Q_h)}^2 \int_{\Rbb^3} \dd r\, \Bigl| \iint_{\Rbb^3\times \Rbb^3} \dd Z \dd s \;  \, |V\alpha_*(s)| \notag \\
&\hspace{-120pt} \times |k_T(Z, r-s)| \Bigl[ h\,  (1 + |Z|)\,   \esssup_{X\in\Rbb^3} \bigl| \e^{\i \tilde \Phi_{A_h}(X, Z, r, s) - \i \Phi_{2A_h}(X, X+ Z) + \i \frac{r-s}{4} DA_h(X) (r + s)} -1\bigr| \notag \\
&\hspace{-70pt}+ \esssup_{X\in \Rbb^3} \bigl| \nabla_X \e^{\i \tilde \Phi_{A_h}(X, Z, r, s)} - \nabla_X \e^{\i \Phi_{2A_h}(X, X + Z)} \e^{-\i \frac{r-s}{4} DA_h(X)(r + s)} \bigr| \Bigr] \Bigr|^2. \label{LtildeTA-MtildeTA_4}
\end{align}
A straightforward computation shows that
\begin{align*}
\bigl| \nabla_X \e^{\i \tilde \Phi_{A}(X, Z, r, s)} - \nabla_X \e^{\i \Phi_{2A}(X, X + Z)} \e^{-\i \frac{r-s}{4} DA(X)(r + s)} \bigr| & \notag \\
&\hspace{-220pt} \leq \bigl| \e^{\i \tilde \Phi_{A}(X, Z, r, s) - \i \Phi_{2A}(X, X+Z) + \i \frac{r-s}{4} DA(X) (r + s)} - 1\bigr| |\nabla_X \tilde \Phi_{A}(X, Z, r, s)| \notag\\
&\hspace{-200pt} + \bigl| \nabla_X \tilde \Phi_{A}(X, Z, r, s) - \nabla_X \Phi_{2A}(X, X + Z) \bigr| + \bigl|\nabla_X \frac{r-s}{4} DA(X) (r+s)\bigr|, 
\end{align*}
which, by Lemma \ref{Phi_Approximation}, is bounded by
\begin{align*}
\bigl| \nabla_X \e^{\i \tilde \Phi_{A}(X, Z, r, s)} - \nabla_X \e^{\i \Phi_{2A}(X, X + Z)} \e^{-\i \frac{r-s}{4} DA(X)(r + s)} \bigr| & \leq C \, \Vert D^2A\Vert_\infty \, \bigl(1 + \Vert DA\Vert_\infty\bigr) \\
&\hspace{-180pt} \times \bigl[ \bigl( |s|^2 + |r-s|^2 \bigr) \bigl( |Z|^2 + |r-s|^2\bigr) + \bigl( |s| + |r-s|\bigr) \bigl( |Z| + |r-s|\bigr)\bigr].
\end{align*}
With the help of \eqref{LtildeTA-MtildeTA_2}, we obtain
\begin{align*}
&\int_{\Rbb^3} \dd Z \; |k_T(Z, r-s)| \, \Bigl[ h\,  (1 + |Z|)  \esssup_{X\in\Rbb^3} \bigl| \e^{\i \tilde \Phi_{A}(X, Z, r, s) - \i \Phi_{2A}(X, X+ Z) + \i \frac{r-s}{4} DA(X) (r + s)} -1\bigr| \\
&\hspace{80pt}+ \esssup_{X\in \Rbb^3} \bigl| \nabla_X \e^{\i \tilde \Phi_{A}(X, Z, r, s)} - \nabla_X \e^{\i \Phi_{2A}(X, X + Z)} \e^{-\i \frac{r-s}{4} DA(X)(r + s)} \bigr| \Bigr]\\
&\hspace{40pt} \leq C \, \Vert D^2A\Vert_\infty \bigl( 1 + \Vert DA\Vert_\infty\bigr) \\
&\hspace{60pt} \times \bigl[ (F_T^2 + F_T^3 + F_T^4)(r-s) + F_T^1(r-s) \, |s| + (F_T^1 +  F_T^2)(r-s) \, |s|^2 \bigr],
\end{align*}
which in combination with \eqref{LtildeTA-MtildeTA_4} proves \eqref{LtildeTA-MtildeTA_3}.

We claim that the term involving $\tilde \pi$ is bounded by
\begin{align}
\Vert \tilde \pi (\tilde L_{T, \Bbold, A} \Delta - \tilde M_{T, \Abold}\Delta)\Vert_2^2 &\leq C \, h^2 \, \Vert \Psi\Vert_{\Hmag^1(Q_h)}^2 \, \Vert D^2A_h\Vert_\infty^2 \, \bigl( 1 + \Vert A_h\Vert_\infty^2 + \Vert DA_h\Vert_\infty^2\bigr) \notag\\
&\hspace{-110pt} \times \bigl \Vert (F_T^2 + F_T^3 + F_T^4+ G_T^2) * |V\alpha_*|  + (F_T^0 + F_T^1 + F_T^2 + G_T^0) * |\cdot|^2 |V\alpha_*| \bigr\Vert_2^2. \label{LtildeTA-MtildeTA_5}
\end{align}
If this holds, the desired bound for this term follows from Young's inequality and \eqref{LtildeTA-MtildeTA_FTGT}. To start out, by \eqref{LtildeTA-MtildeTA_1}, we have
\begin{align*}
\Vert \tilde \pi (\tilde L_{T, \Bbold, A}\Delta - \tilde M_{T, \Abold}\Delta)\Vert_2^2 &\leq C \, h^2 \, \Vert \Psi\Vert_{\Hmag^1(Q_h)}^2 \, \int_{\Rbb^3} \dd r \, \Bigl| \iint_{\Rbb^3 \times \Rbb^3} \dd Z \dd s \; |V\alpha_*(s)| \\
&\hspace{-100pt} \times \bigl|\tilde \pi k_T(Z, r-s) \e^{\i \frac{\Bbold}{4} \cdot (r\wedge s)}\bigr| \esssup_{X\in \Rbb^3} \bigl| \e^{\i \tilde \Phi_{A_h}(X, Z, r, s) - \i \Phi_{2A_h}(X, X+ Z) + \i \frac{r-s}{4} DA_h(X) (r + s)} -1\bigr| \\
&\hspace{-80pt} + |k_T(Z, r-s)| \, \esssup_{X\in \Rbb^3} \bigl| \nabla_r \e^{\i \tilde \Phi_{A_h}(X, Z, r, s)} - \nabla_r \e^{\i \Phi_{2A_h}(X, X+ Z)} \e^{-\i \frac{r-s}{4} DA_h(X) (r + s)}\bigr|\Bigr|^2.
\end{align*}
By the intertwining relation \eqref{tildepi_magnetic_phase} of $\tilde \pi$ with $\e^{\i \frac \Bbold 4 \cdot (r\wedge s)}$, we have
\begin{align*}
|\tilde \pi k_T(Z,r-s) \e^{\i \frac{\Bbold}{4} (r\wedge s)}| &\leq |\nabla_r k_T(Z, r-s)| + |\Bbold| \, |r-s| \, |k_T(Z,r-s)|.
\end{align*}
Hence, a computation similar to \eqref{LtildeTA-MtildeTA_2} shows that
\begin{align}
\int_{\Rbb^3} \dd Z \; |\tilde \pi k_T(Z, r-s) \e^{\i \frac{\Bbold}{4} \cdot (r\wedge s)}| \,  \esssup_{X\in \Rbb^3} \bigl| \e^{\i \tilde \Phi_A(X, Z, r, s) - \i \Phi_{2A}(X, X+ Z) + \i \frac{r-s}{4} DA(X) (r + s)} -1\bigr| & \notag \\
&\hspace{-350pt} \leq C \; \Vert D^2A\Vert_\infty \bigl[ (F_T^3 + G_T^2)(r-s) + (F_T^1 + G_T^0)(r-s)\; |s|^2\bigr].  \label{LtildeTA-MtildeTA_9}
\end{align}
Furthermore, we have
\begin{align*}
\bigl| \nabla_r \e^{\i \tilde \Phi_{A}(X, Z, r, s)} - \nabla_r \e^{\i \Phi_{2A}(X, X+ Z) - \i \frac{r-s}{4} DA(X) (r + s)}\bigr| & \\
&\hspace{-200pt} \leq \bigl| \e^{\i \tilde \Phi_{A} (X, Z, r, s) - \i \Phi_{2A}(X, X + Z)+ \i \frac{r-s}{4} DA(X) (r + s)} - 1\bigr| |\nabla_r \tilde \Phi_{A}(X, Z, r, s)| \\
&\hspace{-150pt} + \bigl| \nabla_r \tilde \Phi_{A}(X, Z,r, s) + \nabla_r \frac{r-s}{4} DA(X) (r+s)\bigr|,
\end{align*}
which by Lemma \ref{Phi_Approximation} is bounded by
\begin{align*}
\bigl| \nabla_r \e^{\i \tilde \Phi_{A}(X, Z, r, s)} - \nabla_r \e^{\i \Phi_{2A}(X, X+ Z)} \e^{-\i \frac{r-s}{4} DA(X) (r + s)}\bigr| & \leq  C \, \Vert D^2A\Vert_\infty \bigl( 1+ \Vert A\Vert_\infty + \Vert D A\Vert_\infty \bigr)\\
&\hspace{-220pt} \times \bigl[ \bigl( |s|^2 + |r-s|^2\bigr) \bigl( |Z| + |r-s|\bigr) \bigl( |Z| + |r-s| + 1\bigr) + \bigl( |s|^2 + |r-s|^2 + |Z|^2\bigr)\bigr].
\end{align*}
Therefore,
\begin{align*}
\int_{\Rbb^3} \dd Z \; |k_T(Z,r-s)| \, \esssup_{X\in \Rbb^3} \bigl| \nabla_r \e^{\i \tilde \Phi_{A}(X, Z, r, s)} - \nabla_r \e^{\i \Phi_{2A}(X, X+ Z)} \e^{-\i \frac{r-s}{4} DA(X) (r + s)}\bigr| & \\
&\hspace{-350pt} \leq C\, \Vert D^2A\Vert_\infty \bigl( 1 + \Vert A\Vert_\infty + \Vert DA\Vert_\infty \bigr) \\
&\hspace{-300pt} \times \bigl[ (F_T^2 + F_T^3 + F_T^4)(r-s) + (F_T^0 + F_T^1 + F_T^2)(r-s) \, |s|^2\bigr].
\end{align*}
We combine this with \eqref{LtildeTA-MtildeTA_9} and obtain \eqref{LtildeTA-MtildeTA_5}.
\end{proof}

\paragraph{The operator $M_{T,\Abold}$.} We define the operator $M_{T,\Abold}$ by
\begin{align}
M_{T,\Abold} \alpha(X,r) := \iint_{\Rbb^3\times \Rbb^3} \dd Z \dd s \; k_T(Z, r-s) \;(\cos(Z\cdot \Pi_{\Abold_h})\alpha)(X,s), \label{MTA_definition}
\end{align}
where $k_T$ is defined below \eqref{MtildeTA_definition}. In our calculation, we may replace $\tilde M_{T, \Abold}$ by $M_{T,\Abold}$ due to the following error bound.

\begin{prop}
\label{MtildeTA-MTA}
For any $T_0>0$ and $A\in \Wpervec 2$ there is $h_0>0$ such that for any $0< h \leq h_0$, any $T\geq T_0$, and whenever  $|\cdot|^k V\alpha_*\in L^2(\Rbb^3)$ for $k\in \{0,1\}$,  $\Psi\in \Hmag^1(Q_h)$, and $\Delta \equiv \Delta_\Psi$ as in \eqref{Delta_definition}, we have
\begin{align}
	\Vert \tilde M_{T,\Abold}\Delta - M_{T,\Abold}\Delta \Vert_\Hsymm^2 &\leq C\;h^6 \;\bigl( \Vert V\alpha_*\Vert_2^2 + \Vert \, |\cdot|V\alpha_*\Vert_2^2\bigr) \;\Vert \Psi\Vert_{\Hmag^1(Q_h)}^2. \label{MtildeTA-MTA_eq1}
\end{align}
If instead $|\cdot|^k V\alpha_*\in L^2(\Rbb^3)$ for $k\in \{0,2\}$ then
\begin{align}
	|\langle \Delta, \tilde M_{T,\Abold}\Delta - M_{T,\Abold}\Delta\rangle| &\leq C \;h^6 \; \bigl( \Vert V\alpha_*\Vert_2^2 + \Vert\,  |\cdot|^2 V\alpha_*\Vert_2^2 \bigr) \;\Vert \Psi\Vert_{\Hmag^1(Q_h)}^2. \label{MtildeTA-MTA_eq2}
\end{align}
\end{prop}

\begin{bem}
We highlight that we need the two bounds \eqref{MtildeTA-MTA_eq1} and \eqref{MtildeTA-MTA_eq2}. The bound \eqref{MtildeTA-MTA_eq1} is insufficient for the proof of Theorem \ref{Calculation_of_the_GL-energy} but is needed for the proof of Proposition~\ref{Structure_of_alphaDelta}. The bound \eqref{MtildeTA-MTA_eq2} exploits the fact that $V\alpha_*$ is real-valued, which allows for the replacement of $\exp(-\i \frac{r-s}{4} D\Abold_h(X)(r+s))$ by $\cos(\frac{r-s}{4} D\Abold_h(X)(r+s))$ in $\tilde M_{T, \Abold}$. For a detailed explanation, we refer to \cite[Remark 4.10]{DeHaSc2021}.
%
%
\end{bem}

\begin{proof}[Proof of Proposition \ref{MtildeTA-MTA}]
The proof is similar to that of Proposition \ref{LTA-LtildeTA} and we begin by proving \eqref{MtildeTA-MTA_eq1}. We claim that
\begin{align}
\Vert \tilde M_{T, \Abold}\Delta - M_{T, \Abold}\Delta\Vert_2^2 &\leq 4 \, \Vert \Psi\Vert_2^2 \, \Vert D\Abold_h\Vert_\infty^2 \, \Vert F_T^2 * |V\alpha_*| + F_T^1 * |\cdot|\, |V\alpha_*|\, \Vert_2^2 \label{MtildeTA-MTA_6}
\end{align}
with the function $F_T^a$ in \eqref{LtildeTA-MtildeTA_FT_definition}. If this holds, the desired bound for this term follows from Young's inequality, \eqref{Periodic_Sobolev_Norm}, and the $L^1$-norm estimate \eqref{LtildeTA-MtildeTA_FTGT} on $F_T^a$. To prove \eqref{MtildeTA-MTA_6}, we repeat the arguments of \eqref{Expanding_the_square}-\eqref{LTA-LtildeTBA_2} and obtain
\begin{align}
\Vert \tilde M_{T, \Abold}\Delta - M_{T, \Abold}\Delta\Vert_2^2 & \leq 4 \, \Vert \Psi\Vert_2^2 \notag \\
&\hspace{-90pt} \times \int_{\Rbb^3} \dd r\,  \Bigl| \iint_{\Rbb^3\times \Rbb^3}\dd Z \dd s \; |k_T (Z, r-s)| \, \esssup_{X\in \Rbb^3} \bigl| \e^{-\i \frac{r-s}{4} D\Abold_h(X) (r+s)}  - 1\bigr| \;|V\alpha_*(s)|\Bigr|^2. \label{MtildeTA-MTA_4}
\end{align}
We combine
\begin{align}
\bigl| (r-s) D\Abold_h(X) (r+s)\bigr| &\leq \Vert D\Abold_h \Vert_\infty \, |r-s| \, \bigl( |s| + |r-s|\bigr) \label{MtildeTA-MTA_1}
\end{align}
with the estimate for $|r-s|$ in \eqref{r-s_estimate} and obtain
\begin{align}
\int_{\Rbb^3} \dd Z \; |k_T (Z, r-s)| \, \esssup_{X\in \Rbb^3} \bigl| \e^{-\i \frac{r-s}{4} D\Abold_h(X) (r+s)}  - 1\bigr| & \notag \\
&\hspace{-100pt} \leq C\, \Vert D\Abold_h\Vert_\infty \bigl[ F_T^2(r-s) + F_T^1 (r - s)\;|s|\bigr]. \label{MtildeTA-MTA_5}
\end{align}
In combination with \eqref{MtildeTA-MTA_4}, we get \eqref{MtildeTA-MTA_6}.

We claim that the term involving $\Pi$ is bounded by
\begin{align}
\Vert \Pi (\tilde M_{T, \Abold}\Delta - M_{T, \Abold}\Delta)\Vert_2^2 &\leq C\, h^2 \; \Vert \Psi\Vert_{\Hmag^1(Q_h)}^2 \notag \\
&\hspace{-70pt} \times \bigl[ \Vert D^2A_h\Vert_\infty^2 \, \Vert F_T^2 *|V\alpha_*| + F_T^1 * |\cdot| \, |V\alpha_*| \, \Vert_2^2 \notag \\
&\hspace{-50pt} + h^2\, \Vert D\Abold_h\Vert_\infty^2\, \Vert (F_T^2 + F_T^3)* |V\alpha_*| + (F_T^1 + F_T^2) * |\cdot|\, |V\alpha_*|\, \Vert_2^2 \bigr], \label{MtildeTA-MTA_2}
\end{align}
which shows the desired bound for this term by Young's inequality and \eqref{LtildeTA-MtildeTA_FTGT}. To prove \eqref{MtildeTA-MTA_2}, we first claim that 
\begin{align}
\Vert \Pi \cos(Z\cdot \Pi_{\Abold_h}) \Psi\Vert_2 \leq C \, h^2 \, (1 + |Z|) \, \Vert \Psi\Vert_{\Hmag^1(Q_h)}. \label{MtildeTA-MTA_Lemma}
\end{align}
If this holds, \eqref{MtildeTA-MTA_2} follows from a computation similar to the one leading to \eqref{LTA-LtildeTBA_9}, using Lemma \ref{Phi_Approximation},  \eqref{MtildeTA-MTA_5}, and \eqref{MtildeTA-MTA_Lemma}. To see that \eqref{MtildeTA-MTA_Lemma} holds, we utilize Lemma \ref{Magnetic_Translation_Representation} to note that
\begin{align*}
\e^{\pm \i Z \cdot \Pi_{\Abold_h}} = \e^{\pm \i Z \cdot \Pi} \e^{-\i \Phi_{2A_h}(X, X \mp Z)}.
\end{align*}
Therefore, by the intertwining relation \eqref{PiU_intertwining} of $\Pi$ with $\e^{\i Z\cdot \Pi}$,
\begin{align*}
[\Pi, \e^{\pm \i Z\cdot \Pi_{\Abold_h}}] = \e^{\pm \i Z\cdot \Pi_{\Abold_h}} \bigl[ \mp 2\, \Bbold\wedge Z - \nabla_X \Phi_{2A_h}(X, X \mp Z)\bigr].
\end{align*}
By the definition \eqref{PhiA}, we have
\begin{align}
\Phi_{2A}(X, X \pm Z) = \pm 2\int_0^1 \dt \; A(X \pm tZ) \cdot Z, \label{MtildeTA-MTA_3}
\end{align}
whence the vector analysis identity \eqref{PhiA_derivative_4} implies
\begin{align*}
\nabla_X \Phi_{2A}(X, X \mp Z) = 2A(X\mp Z) - 2A(X) \pm 2\int_0^1 \dt \; \curl A(X \mp tZ) \wedge Z.
\end{align*}
We conclude $|\nabla_X\Phi_{2A}(X, X \mp Z)|\leq C\, \Vert DA\Vert_\infty |Z|$, which implies
\begin{align}
\Vert [\Pi, \cos(Z\cdot \Pi_{\Abold_h})]\Psi\Vert_2 \leq C \, (|\Bbold| + \Vert DA_h\Vert_\infty) \, |Z| \, \Vert \Psi\Vert_2 \label{PiXcosPiA_estimate}
\end{align}
and \eqref{MtildeTA-MTA_Lemma}.

%
Finally, we use the strategy that leads to \eqref{LTA-LtildeTBA_10} and obtain
\begin{align*}
\Vert \tilde \pi (\tilde M_{T, \Abold}\Delta - M_{T, \Abold}\Delta)\Vert_2^2 & \leq 4\, \Vert \Psi\Vert_2^2 \\
&\hspace{-100pt} \times \int_{\Rbb^3} \dd r \, \Bigl| \iint_{\Rbb^3\times \Rbb^3}\dd Z \dd s \; \bigl| \tilde \pi k_T (Z, r-s) \bigl[ \e^{-\i \frac{r-s}{4} D\Abold_h(X)(r + s)} - 1\bigr] \bigr| \; |V\alpha_*(s)|\Bigr|^2.
\end{align*}
By \eqref{Phi_Approximation_9}, we have
\begin{align*}
\bigl|\nabla_r (r - s) D\Abold(X) (r+s) \bigr| \leq C \, \Vert D\Abold\Vert_\infty \bigl( |s| + |r-s|\bigr).
\end{align*}
Therefore, we use \eqref{MtildeTA-MTA_1} and the estimate $|\Bbold \wedge r| \leq |s| + |r-s|$ to see that
\begin{align*}
\int_{\Rbb^3} \dd Z\; \bigl| \tilde \pi k_T (Z, r-s) \bigl[ \e^{-\i \frac{r-s}{4} D\Abold(X) (r+s)} - 1\bigr] \bigr| &\\
&\hspace{-150pt}\leq C\, \Vert D\Abold\Vert_\infty \; \bigl( (F_T^1 + G_T^2)(r-s) + (F_T^0 + G_T^1)(r-s) \; |s|  \bigr) .
\end{align*}
We apply Young's inequality and \eqref{LtildeTA-MtildeTA_FTGT} and obtain \eqref{MtildeTA-MTA_eq1} from these considerations.

We are left with proving \eqref{MtildeTA-MTA_eq2}. The term that has to be bounded reads
\begin{align}
\langle \Delta, \tilde M_{T, \Abold} \Delta - M_{T, \Abold} \Delta \rangle &= 4 \iint_{\Rbb^3\times \Rbb^3} \dd r \dd s \; \bigl( \e^{-\i \frac{r-s}{4} D\Abold_h(X) (r+s)} - 1 \bigr) V\alpha_*(r) V\alpha_*(s)  \notag\\
&\hspace{-20pt} \times \int_{\Rbb^3} \dd Z \; k_T(Z, r-s)  \fint_{Q_h} \dd X \; \ov{\Psi(X)} \cos(Z\cdot \Pi_{\Abold_h})\Psi(X). \label{MtildeTA-MTA_7}
\end{align}
When we exchange the coordinates $r$ and $s$, the right side remains unchanged, except for the factor $\e^{-\i \frac{r-s}{4} D\Abold_h(X) (r+s)}$, which is complex conjugated when we apply this transformation. Thus, we may add up the right side of \eqref{MtildeTA-MTA_7} with the transformed version, and get
%
\begin{align}
\langle \Delta, \tilde M_{T, \Abold} \Delta - M_{T, \Abold} \Delta \rangle & = -8 \iint_{\Rbb^3\times \Rbb^3} \dd r \dd s \; \sin^2\bigl( \frac 1 8 \, (r-s)D\Abold_h(X)(r+s)\bigr) \; V\alpha_*(r) V\alpha_*(s) \notag \\
&\hspace{-20pt}\times \int_{\Rbb^3} \dd Z \; k_T(Z, r-s)  \fint_{Q_h} \dd X \; \ov{\Psi(X)} \cos(Z\cdot \Pi_{\Abold_h})\Psi(X). \label{MtildeTA-MTA_8}
\end{align}
Here, we used $\cos(x) -1 =- 2\sin^2(\frac x2)$. Since $\cos(Z\cdot\Pi_{\Abold_h})$ is bounded by $1$ in operator norm, \eqref{MtildeTA-MTA_1} implies
\begin{align*}
\sin^2\bigl(\frac 18 (r-s)D\Abold_h(X)(r+s) \bigr) \leq \frac 18 \, \Vert D\Abold_h\Vert_\infty^2\,  |r-s|^2 \, \bigl( |s|^2+ |r-s|^2\bigr).
\end{align*}
By, \eqref{MtildeTA-MTA_8} we thus have
\begin{align*}
|\langle \Delta, \tilde M_{T, \Abold}\Delta - M_{T, \Abold}\Delta\rangle | &\leq \Vert D\Abold_h\Vert_\infty^2 \; \Vert \Psi\Vert_2^2 \; \bigl\Vert |V\alpha_*| \; \bigl( F_T^4 * |V\alpha_*| + F_T^2 * |\cdot |^2 |V\alpha_*| \bigr) \bigr\Vert_1.
\end{align*}
The desired bound \eqref{MtildeTA-MTA_eq2} now follows from Young's inequality, \eqref{Periodic_Sobolev_Norm}, as well as the $L^1$-norm estimate \eqref{LtildeTA-MtildeTA_FTGT} on $F_T^a$. The proof of Proposition~\ref{MtildeTA-MTA} is thus completed.
\end{proof}

%


\subsubsection{Analysis of \texorpdfstring{$M_{T, \Abold}$}{MTA} and calculation of two quadratic terms}
\label{Analysis_of_MTA_Section}

We decompose $M_{T, \Abold} = M_T^{(1)} + M_{T, \Abold}^{(2)} + M_{T, \Abold}^{(3)}$, where
\begin{align}
M_T^{(1)} \alpha(X,r) &:= \iint_{\Rbb^3\times \Rbb^3} \dd Z \dd s \; k_T(Z, r-s) \; \alpha(X,s), \label{MT1_definition}\\
M_{T, \Abold}^{(2)} \alpha(X, r) &:=  \iint_{\Rbb^3\times \Rbb^3} \dd Z \dd s\; k_T(Z, r-s) \; \bigl( -\frac 12\bigr) (Z\cdot \Pi_{\Abold_h})^2 \; \alpha(X, s), \label{MTB2_definition}\\
M_{T, \Abold}^{(3)} \alpha(X,r) &:= \iint_{\Rbb^3\times \Rbb^3} \dd Z \dd s\; k_T(Z, r-s) \; \Rcal(Z\cdot \Pi_{\Abold_h}) \; \alpha(X, s), \label{MTA3_definition}
\end{align}
and $\Rcal(x) = \cos(x) - 1 + \frac 12 x^2$.

\paragraph{The operator $M_T^{(1)}$.} From the quadratic form $\langle \Delta, M_T^{(1)} \Delta \rangle$, we extract the quadratic term without magnetic gradient and $W$-field in the Ginzburg--Landau functional in \eqref{Definition_GL-functional}. We also extract a term which cancels the last term on the left side of \eqref{Calculation_of_the_GL-energy}. 
%
%
The result, which performs the extraction, is proven in \cite[Proposition 4.11]{DeHaSc2021} and we repeat the statement in the next proposition. We recall that $\Delta \equiv \Delta_\Psi = -2 V\alpha_* \Psi$ as in \eqref{Delta_definition}.

\begin{prop}
\label{MT1}
Assume that $V\alpha_*\in L^2(\Rbb^3)$ and let $\Psi \in \Lmag^2(Q_h)$ and $\Delta \equiv \Delta_\Psi$ as in \eqref{Delta_definition}.
\begin{enumerate}[(a)]
\item We have $M_{\Tc}^{(1)} \Delta (X, r) = -2\, \alpha_* (r) \Psi(X)$.

\item For any $T_0>0$ there is a constant $c>0$ such that for $T_0 \leq T \leq \Tc$ we have
\begin{align*}
\langle \Delta, M_T^{(1)} \Delta - M_{\Tc}^{(1)} \Delta \rangle \geq  c \, \frac{\Tc - T}{\Tc} \; \Vert \Psi\Vert_2^2.
\end{align*}

\item Given $D\in \Rbb$ there is $h_0>0$ such that for $0< h\leq h_0$, and $T = \Tc (1 - Dh^2)$ we have
\begin{align*}
	\langle \Delta, M_T^{(1)} \Delta -  M_{\Tc}^{(1)} \Delta\rangle  = 4\; Dh^2 \; \Lambda_2 \; \Vert \Psi\Vert_2^2 + R(\Delta)
\end{align*}
with the coefficient $\Lambda_2$ in \eqref{GL-coefficient_2}, and
\begin{align*}
	|R(\Delta)| &\leq C \; h^6 \; \Vert V\alpha_*\Vert_2^2\; \Vert \Psi\Vert_{\Hmag^1(Q_h)}^2.
\end{align*}

\item Assume additionally that $| \cdot | V\alpha_*\in L^2(\Rbb^3)$. Then, there is $h_0>0$ such that for any $0< h \leq h_0$, any $\Psi\in \Hmag^1(Q_h)$, and any $T \geq T_0 > 0$ we have 
\begin{align*}
	\Vert M_T^{(1)}\Delta - M_{\Tc}^{(1)}\Delta\Vert_{\Hsymm}^2 &\leq C \, h^2 \, | T - \Tc |^2 \,  \bigl( \Vert V\alpha_*\Vert_2^2 + \Vert \, |\cdot| V\alpha_*\Vert_2^2\bigr) \Vert\Psi\Vert_{\Hmag^1(Q_h)}^2.
\end{align*}
\end{enumerate}
\end{prop}

\paragraph{The operator $M_{T,\Abold}^{(2)}$.} The kinetic contribution in the Ginzburg--Landau functional in \eqref{Definition_GL-functional} is contained in the term $\langle \Delta, M_{T,\Abold}^{(2)} \Delta \rangle$, where $M_{T,\Abold}^{(2)}$ is defined in \eqref{MTB2_definition}. In the following proposition, we extract this term.

\begin{prop}
\label{MTB2}
Assume that the function $V\alpha_*$ is radial and belongs to $L^2(\Rbb^3)$. For any $A\in \Wpervec 1$, $h>0$, $\Psi\in \Hmag^1(Q_h)$, and $\Delta \equiv \Delta_\Psi$ as in \eqref{Delta_definition}, we have
\begin{align}
\langle \Delta, M_{\Tc,\Abold}^{(2)} \Delta\rangle = - 4\; \Lambda_0 \; \Vert \Pi_{\Abold_h} \Psi\Vert_2^2 \label{MTA2_1}
\end{align}
with $\Lambda_0$ in \eqref{GL-coefficient_1}. Moreover, for any $T \geq T_0 > 0$ we have
\begin{align}
	|\langle \Delta,  M_{T,\Abold}^{(2)} \Delta - M_{\Tc,\Abold}^{(2)}\Delta\rangle| \leq C\; h^4 \; |T - \Tc| \; \Vert V\alpha_*\Vert_2^2 \; \Vert \Psi\Vert_{\Hmag^1(Q_h)}^2. \label{MTA2_2}
\end{align}
\end{prop}


\begin{proof}
The proof is analogous to the proof of \cite[Proposition 4.13]{DeHaSc2021} with the obvious replacements.
\end{proof}

\paragraph{The operator $M_{T,\Abold}^{(3)}$.} The remainder of the expansion of $\langle \Delta, M_{T,\Abold} \Delta \rangle$ is given by the term $\langle \Delta, M_{T,\Abold}^{(3)} \Delta \rangle$, where $M_{T,\Abold}^{(3)}$ is defined in \eqref{MTA3_definition}. As in the work \cite{DeHaSc2021}, the $\Hmag^2(Q_h)$-norm of $\Psi$ is required to bound its size. 


\begin{prop}
\label{MTB3}
For any $T_0>0$ and $A\in \Wpervec 2$ there is $h_0>0$ such that for any $0 < h \leq h_0$, any $T\geq T_0$, and whenever $V\alpha_*\in L^2(\Rbb^3)$, $\Psi \in \Hmag^2(Q_h)$, and $\Delta \equiv \Delta_\Psi$ as in \eqref{Delta_definition}, we have
\begin{align*}
|\langle \Delta,  M_{T,\Abold}^{(3)} \Delta\rangle| &\leq C \; h^6 \; \Vert V\alpha_*\Vert_2^2 \; \Vert \Psi\Vert_{\Hmag^2(Q_h)}^2.
\end{align*}
\end{prop}

We need the following auxiliary result on the operator $|Z\cdot \Pi_\Abold|^4$.

\begin{lem}
\label{ZPiX_inequality_quartic}
\begin{enumerate}[(a)]
\item Let $A\in \Wpervec 2$. For any $Z\in \Rbb^3$ and any $\varepsilon >0$, we have
\begin{align}
|Z\cdot \Pi_\Abold|^4 &\leq 9\; |Z|^4 \; \bigl( \Pi_\Abold^4 + \varepsilon \, \Pi_\Abold^2 + 2 \, |\curl \Abold|^2 + \varepsilon^{-1} \, |\curl(\curl \Abold)|^2 \bigr). \label{ZPiX_inequality_quartic_eq}
\end{align}

\item There is a constant $h_0>0$ such that for any $0 < h \leq h_0$, any $\Psi\in \Hmag^2(Q_h)$, and any $Z\in \Rbb^3$, we have
\begin{align*}
\langle \Psi, \, |Z\cdot \Pi_{\Abold_h}|^4 \, \Psi\rangle &\leq C\; h^6 \; |Z|^4 \; \Vert \Psi\Vert_{\Hmag^2(Q_h)}^2.
\end{align*}
\end{enumerate}
\end{lem}

\begin{proof}
We note that 
\begin{align}
\bigl[\Pi_\Abold^{(i)}, \Pi_\Abold^{(j)}\bigr] = -\i \sum_{k=1}^3 \varepsilon_{ijk} \; (\curl \Abold)_k \label{ZPiX_inequality_quartic_3}
\end{align}
with the Levi--Civita symbol $\varepsilon_{ijk}$, which is defined as $1$ if $(i, j, k)$ is a cyclic permutation of $\{1, 2, 3\}$, as $-1$ if it is an anticyclic permutation, and zero if at least two indices coincide. We claim that
\begin{align}
\Pi_\Abold \; \Pi_\Abold^2\; \Pi_\Abold = \Pi_\Abold^4 + 2 \, |\curl \Abold|^2 - \curl (\curl \Abold) \cdot \Pi_\Abold. \label{PiPi2Pi_equality}
\end{align}
In particular, since all terms except the last are self-adjoint, this implies 
\begin{align}
\bigl[ \curl (\curl \Abold) , \Pi_\Abold \bigr] = 0. \label{ZPiX_inequality_quartic_4}
\end{align}
To prove \eqref{PiPi2Pi_equality}, we note that
\begin{align}
\Pi_\Abold \; \Pi_\Abold^2 &= \Pi_\Abold^2 \; \Pi_\Abold + 2 \, [\Pi_\Abold, \Pi_\Abold^{(1)}] \, \Pi_\Abold^{(1)}  + 2 \, [\Pi_\Abold, \Pi_\Abold^{(2)}] \, \Pi_\Abold^{(2)}  + 2 \, [\Pi_\Abold, \Pi_\Abold^{(3)}] \, \Pi_\Abold^{(3)} \notag \\
&\hspace{50pt} + \bigl[\Pi_\Abold^{(1)} , [\Pi_\Abold, \Pi_\Abold^{(1)}] \bigr] + \bigl[\Pi_\Abold^{(2)} , [\Pi_\Abold, \Pi_\Abold^{(2)}] \bigr] + \bigl[\Pi_\Abold^{(3)} , [\Pi_\Abold, \Pi_\Abold^{(3)}] \bigr]. \label{ZPiX_inequality_quartic_6}
\end{align}
Furthermore, with the help of \eqref{ZPiX_inequality_quartic_3}, a straightforward computation shows that
\begin{align}
[\Pi_\Abold, \Pi_\Abold^{(1)}] \, \Pi_\Abold^{(1)}  + [\Pi_\Abold, \Pi_\Abold^{(2)}] \, \Pi_\Abold^{(2)}  + [\Pi_\Abold, \Pi_\Abold^{(3)}] \, \Pi_\Abold^{(3)} = \i \, (\curl \Abold) \wedge \Pi_\Abold \label{ZPiX_inequality_quartic_7}
\end{align}
and
\begin{align}
\bigl[\Pi_\Abold^{(1)} , [\Pi_\Abold, \Pi_\Abold^{(1)}] \bigr] + \bigl[\Pi_\Abold^{(2)} , [\Pi_\Abold, \Pi_\Abold^{(2)}] \bigr] + \bigl[\Pi_\Abold^{(3)} , [\Pi_\Abold, \Pi_\Abold^{(3)}] \bigr] = - \curl(\curl \Abold). \label{ZPiX_inequality_quartic_8}
\end{align}
Therefore, the combination of \eqref{ZPiX_inequality_quartic_6}-\eqref{ZPiX_inequality_quartic_8} yields
\begin{align*}
\Pi_\Abold \; \Pi_\Abold^2 \; \Pi_\Abold = \Pi_\Abold^4 + 2\i \, \bigl( (\curl \Abold) \wedge \Pi_\Abold\bigr) \cdot \Pi_\Abold - \curl (\curl \Abold) \cdot \Pi_\Abold
\end{align*}
Since $\i\,  ( (\curl \Abold) \wedge \Pi_\Abold) \cdot \Pi_\Abold = |\curl \Abold|^2$, we conclude \eqref{PiPi2Pi_equality}.

We further claim that
\begin{align}
(Z\cdot \Pi_\Abold) \, \Pi_\Abold^2\, (Z\cdot \Pi_\Abold)= \Pi_\Abold \, (Z\cdot \Pi_\Abold)^2 \, \Pi_\Abold + (Z \cdot \curl(\curl \Abold)) \; (Z \cdot \Pi_\Abold), \label{ZPiX_inequality_quartic_1}
\end{align}
which likewise implies
\begin{align}
\bigl[ Z \cdot \curl(\curl \Abold), Z\cdot \Pi_\Abold \bigr] =0. \label{ZPiX_inequality_quartic_5}
\end{align}
To see that \eqref{ZPiX_inequality_quartic_1} holds, we note that
\begin{align}
(Z\cdot \Pi_\Abold) \, \Pi_\Abold^2\, (Z\cdot \Pi_\Abold) = \sum_{i,j =1}^3 Z_iZ_j \; \Pi_\Abold^{(i)} \, \Pi_\Abold^2 \, \Pi_\Abold^{(j)}  \label{ZPiX_inequality_quartic_2}
\end{align}
as well as
\begin{align*}
\Pi_\Abold^{(i)} \, \Pi_\Abold^2 \, \Pi_\Abold^{(j)} = \Pi_\Abold \, \Pi_\Abold^{(i)} \, \Pi_\Abold^{(j)} \, \Pi_\Abold + \Pi_\Abold \, \Pi_\Abold^{(i)} \, [ \Pi_\Abold, \Pi_\Abold^{(j)}] + [\Pi_\Abold^{(i)}, \Pi_\Abold] \, \Pi_\Abold \, \Pi_\Abold^{(j)}.
\end{align*}
Since the sum in \eqref{ZPiX_inequality_quartic_2} is manifest symmetric with respect to the exchange of $i$ and $j$, we combine the terms as $\frac 12$ times
\begin{align*}
\Pi_\Abold^{(i)} \, \Pi_\Abold^2 \, \Pi_\Abold^{(j)} + \Pi_\Abold^{(j)} \, \Pi_\Abold^2 \, \Pi_\Abold^{(i)} &= \Pi_\Abold \, \Pi_\Abold^{(i)} \, \Pi_\Abold^{(j)} \, \Pi_\Abold + \Pi_\Abold \, \Pi_\Abold^{(j)} \, \Pi_\Abold^{(i)} \, \Pi_\Abold \\
&\hspace{20pt} + \bigl[ [\Pi_\Abold^{(i)}, \Pi_\Abold] , \Pi_\Abold \, \Pi_\Abold^{(j)}\bigr] + \bigl[ [\Pi_\Abold^{(j)}, \Pi_\Abold] , \Pi_\Abold \, \Pi_\Abold^{(i)}\bigr].
\end{align*}
We further use $[A, BC] = [A, B] \, C + B\, [A, C]$ on the last two terms. For the third term, this implies
\begin{align*}
\bigl[ [\Pi_\Abold^{(i)}, \Pi_\Abold] , \Pi_\Abold \, \Pi_\Abold^{(j)}\bigr] &= \bigl[ [\Pi_\Abold^{(i)}, \Pi_\Abold] , \Pi_\Abold \bigr] \, \Pi_\Abold^{(j)} + \Pi_\Abold \bigl[ [\Pi_\Abold^{(i)}, \Pi_\Abold] , \Pi_\Abold^{(j)}\bigr]
\end{align*}
and likewise for $i$ and $j$ interchanged. The last term drops out because of symmetry. It remains to analyze the first term. A straightforward computation using \eqref{ZPiX_inequality_quartic_3} shows
\begin{align*}
\bigl[ [\Pi_\Abold^{(i)} , \Pi_\Abold] , \Pi_\Abold\bigr] = (\curl (\curl \Abold))_i.
\end{align*}
Hence, 
\begin{align*}
\frac 12 \sum_{i,j =1}^3 Z_iZ_j \Bigl( \bigl[ [\Pi_\Abold^{(i)} , \Pi_\Abold] , \Pi_\Abold\bigr] \Pi_\Abold^{(j)} + \bigl[ [\Pi_\Abold^{(j)} , \Pi_\Abold] , \Pi_\Abold\bigr] \Pi_\Abold^{(i)}\Bigr) = (Z \cdot \curl(\curl \Abold) )\; (Z\cdot \Pi_\Abold),
\end{align*}
which proves \eqref{ZPiX_inequality_quartic_1}.

We turn to the proof of \eqref{ZPiX_inequality_quartic_eq} and start by noting that for general operators $A, B, C$, we have $|A + B + C|^2 \leq 3(|A|^2 + |B|^2 + |C|^2)$, which implies
\begin{align}
(Z\cdot \Pi_\Abold)^2 &\leq 3 \; \bigl( Z_1^2 \; (\Pi_\Abold^{(1)})^2 + Z_2^2 \; (\Pi_\Abold^{(2)})^2 + Z_3^2 \; (\Pi_\Abold^{(3)})^2\bigr) \leq 3 \; |Z|^2 \; \Pi_\Abold^2. \label{ZPiX_inequality}
\end{align}
We use \eqref{ZPiX_inequality_quartic_1}, apply \eqref{ZPiX_inequality} to $(Z\cdot \Pi_\Abold)^2$, and use \eqref{PiPi2Pi_equality} to obtain
\begin{align*}
(Z\cdot \Pi_\Abold) \, \Pi_\Abold^2\, (Z\cdot \Pi_\Abold) &\leq 3\, |Z|^2 \, \bigl(\Pi_\Abold^4+ 2\, |\curl \Abold|^2 - \curl(\curl \Abold) \cdot \Pi_\Abold\bigr) \\
&\hspace{100pt} + (Z \cdot \curl(\curl \Abold)) \, (Z\cdot \Pi_\Abold). 
\end{align*}
Then, we write $|Z\cdot \Pi_\Abold|^4 = (Z\cdot \Pi_\Abold)  (Z\cdot \Pi_\Abold)^2 (Z\cdot \Pi_\Abold)$, apply \eqref{ZPiX_inequality} again, and obtain 
\begin{align*}
|Z\cdot \Pi_\Abold|^4 &\leq 9 \, |Z|^4 \bigl( \Pi_\Abold^4 + 2\, |\curl \Abold|^2 - \curl (\curl \Abold) \cdot \Pi_\Abold \bigr) \\
&\hspace{150pt} + 3 \, |Z|^2 \, (Z \cdot \curl(\curl \Abold)) \, (Z\cdot \Pi_\Abold).
\end{align*}
Finally, we use $AB \leq \frac \varepsilon 2 \, A^2 + \frac{1}{2\varepsilon} \, B^2$ for $\varepsilon>0$ and commuting self-adjoint operators $A$ and $B$. By \eqref{ZPiX_inequality_quartic_4} and \eqref{ZPiX_inequality_quartic_5} as well as \eqref{ZPiX_inequality} and the regular Cauchy-Schwarz inequality, this implies
\begin{align*}
- 9 \, |Z|^4 \, \curl(\curl \Abold) \cdot \Pi_\Abold + 3 \, |Z|^2 \, (Z\cdot \curl(\curl\Abold))\, (Z\cdot \Pi_\Abold) \\
&\hspace{-70pt} \leq 9 \, |Z|^4 \bigl( \varepsilon \, \Pi_\Abold^2 + \varepsilon^{-1} \, |\curl(\curl \Abold)|^2\bigr),
\end{align*}
which proves \eqref{ZPiX_inequality_quartic_eq} and part (a). Part (b) follows from the expansion
\begin{align*}
\Pi_\Abold^4 &= \Pi^4 + (\Pi^2 + |A|^2) \, (\Pi \cdot A + A \cdot \Pi ) + (\Pi \cdot A + A\cdot \Pi) \, (\Pi^2 + |A|^2) \\
&\hspace{200pt} + \Pi^2 \, |A|^2 + |A|^2 \, \Pi^2 + |A|^4,
\end{align*}
a similar expansion for $\Pi_\Abold^2$, the Cauchy-Schwarz inequality, the fact that the estimate $\Vert D^k A_h\Vert_\infty \leq Ch^{k+1}$ holds for $k\in \{0,1,2\}$, the choice $\varepsilon = h^2$, and part (a). This completes the proof.
\end{proof}


\begin{proof}[Proof of Proposition \ref{MTB3}]
We start with the identity
\begin{align}
\langle \Delta, M_{T, \Abold}^{(3)} \Delta \rangle &= 4 \iiint_{\Rbb^3\times \Rbb^3\times \Rbb^3} \dd r\dd s\dd Z \; V\alpha_*(r) V\alpha_*(s)\, k_T(Z, r-s) \; \langle \Psi , \Rcal(Z\cdot \Pi_{\Abold_h})\Psi\rangle, \label{MTB3_1}
\end{align}
where the function $\Rcal(x) = \cos(x) - 1 + \frac{x^2}{2}$ satisfies $0\leq\Rcal(x) \leq \frac{1}{24} x^4$. By Lemma \ref{ZPiX_inequality_quartic} we infer
\begin{align}
\langle \Psi, \Rcal(Z\cdot \Pi_{\Abold_h}) \Psi\rangle &\leq C\; h^6 \; |Z|^4 \; \Vert \Psi\Vert_{\Hmag^2(Q_h)}^2. \label{MTB3_2}
\end{align}
We use the estimate \eqref{Z_estimate} on $|Z|^4$, repeat the arguments that lead to \eqref{LtildeTA-MtildeTA_2}, and obtain
\begin{align}
\int_{\Rbb^3} \dd Z\; |Z|^4 \; |k_T(Z,r)| &\leq F_T^4(r), \label{MTB3_3}
\end{align}
where $F_T^4$ is the function defined in \eqref{LtildeTA-MtildeTA_FT_definition}, whose $L^1(\Rbb^3)$-norm has been bounded in \eqref{LtildeTA-MtildeTA_FTGT}. The proof is finished by an application of \eqref{MTB3_1}, \eqref{MTB3_2}, and \eqref{MTB3_3}.
\end{proof}


\subsubsection{A representation formula for the operator \texorpdfstring{$L_{T,\Abold}^W$}{LTAW}}
\label{LTAW_action_Section}

As in the case of $L_{T, \Abold}$ we start our analysis with a representation formula for the operator $L_{T, \Abold}^W$.

\begin{lem}
\label{LTAW_action}
The operator $L_{T, \Abold}^W \colon \Lsymm \ra \Lsymm$ in \eqref{LTA^W_definition} acts as
\begin{align}
L_{T, \Abold}^W \alpha(X, r) &= \iint_{\Rbb^3 \times \Rbb^3} \dd Z \dd s \; k_{T, \Bbold, \Abold}^W(X, Z, r, s) \; (\e^{\i Z \cdot \Pi} \alpha)(X, s),
\end{align}
where
\begin{align}
k_{T, \Bbold, \Abold}^W (X, Z, r, s) &:= \smash{\frac 2\beta \sum_{n\in \Zbb} \int_{\Rbb^3} \dd Y} \; W_h(X + Y) \bigl[  k_{T, \Bbold, \Abold, +}^{W,n}(X, Y, Z, r, s) \; \e^{\i \frac \Bbold 2 \cdot \Phi_+^W(Y, Z, r, s)} \notag \\
&\hspace{80pt} + k_{T, \Bbold, \Abold, -}^{W,n}(X, Y, Z, r, s) \; \e^{\i \frac \Bbold 2 \cdot \Phi_-^W(Y, Z, r, s)} \bigr] \label{kTAW_definition}
\end{align}
as well as
\begin{align}
k_{T, \Bbold, \Abold, \pm}^{W,n}(X, Y, Z, r, s) &:= \Gbold_{\Bbold, \Abold_h}^{\pm \i \omega_n} \bigl( X \pm \frac r2, X + Y\bigr) \; \Gbold_{\Bbold, \Abold_h}^{\pm \i \omega_n} \bigl( X + Y, X + Z \pm \frac s2\bigr)\notag \\
&\hspace{120pt}  \times \Gbold_{\Bbold, \Abold_h}^{\mp \i \omega_n} \bigl( X \mp \frac r2, X + Z \mp \frac s2\bigr), \label{kTAWn_definition}
\end{align}
where $\Gbold_{\Bbold, \Abold}^z$ is defined in \eqref{GboldBAz_definition}, and
\begin{align}
\Phi_\pm^W (Y, Z,r, s) &:= \pm \frac r2\wedge \bigl(Y \mp \frac r2\bigr) + \bigl( Y \mp \frac r2\bigr) \wedge \bigl( Z - Y \pm \frac s2\bigr) \notag\\
&\hspace{60pt} \pm \frac r2\wedge \bigl( Z - Y\pm \frac s2\bigr) \mp \frac r2 \wedge \bigl( Z\pm \frac{r-s}{2}\bigr). \label{LTA^W_Phi_definition}
\end{align}
\end{lem}

\begin{proof}
The proof is analogous to the proof of Lemma \ref{LTA_action}. We employ \eqref{GAz_Kernel_of_complex_conjugate} to get
\begin{align*}
L_{T, \Abold}^W \alpha(x, y) &=  \frac 2\beta \sum_{n\in \Zbb} \iiint_{\Rbb^9} \dd u \dd v \dd w \; \bigl[ G_{\Abold_h}^{\i \omega_n} (x, u) \, W_h(u) \, G_{\Abold_h}^{\i\omega_n} (u,v) \, \alpha(v, w) \, G_{\Abold_h}^{-\i\omega_n}(y, w) \\
&\hspace{100pt} + G_{\Abold_h}^{\i\omega_n} (x, v) \, \alpha(v, w) \, G_{\Abold_h}^{-\i\omega_n} (u,w) \, W_h(u) \, G_{\Abold_h}^{-\i\omega_n}(y, u)\bigr].
\end{align*}
In terms of the coordinates $X = \frac{x + y}{2}$ and $r = x-y$, the change of variables
\begin{align*}
u &= X + Y, & v &= X + Z + \frac s2, & w &= X + Z - \frac s2,
\end{align*}
and the multiplication and division by the factor
\begin{align*}
\e^{\i \Phi_{\Abold_\Bbold}(X \pm \frac r2, X + Y)} \, \e^{\i \Phi_{\Abold_\Bbold}(X + Y, X + Z \pm \frac s2)} \, \e^{\i \Phi_{\Abold_\Bbold}(X \mp \frac r2 , X + Z \mp \frac s2)} = \e^{\i \Bbold \cdot (X\wedge Z)} \, \e^{\i \frac{\Bbold}{2} \cdot \Phi_\pm^W(Y, Z, r, s)}
\end{align*}
lead to the claimed formula.
\end{proof}

\subsubsection{Approximation of the operator \texorpdfstring{$L_{T, \Abold}^W$}{LTAW}}
\label{Approximation_of_LTA^W_Section}

We apply a similar four step analysis to the operator $L_{T, \Abold}^W$ and ecompose
\begin{align}
L_{T, \Abold}^W = \bigl( L_{T, \Abold}^W - \tilde L_{T, \Bbold, A}^W \bigr) + \bigl( \tilde L_{T, \Bbold, A}^W - \tilde M_{T, \Bbold}^W\bigr) + \bigl( \tilde M_{T, \Bbold}^W - M_T^W\bigr) + M_T^W, 
\end{align}
where $\tilde L_{T, \Bbold, A}^W$, $\tilde M_{T, \Bbold}^W$, and $M_T^W$ are operators of increasing simplicity in their dependence on $\Abold$. These operators are defined below in \eqref{LtildeTAW_definition}, \eqref{MtildeTAW_definition}, and \eqref{MTW_definition}. As in the case of $L_{T, \Abold}$, we show that the terms in brackets are small in a suitable sense and we extract the $W$-field term in the Ginzburg--Landau functional in \eqref{Definition_GL-functional} from $M_T^W$.

\paragraph{The operator $\tilde L_{T, \Bbold, A}^W$.}

We define the operator $\tilde L_{T, \Bbold, A}^W$ by
\begin{align}
\tilde L_{T, \Bbold, A}^W \alpha(X, r) &:= \iint_{\Rbb^3 \times \Rbb^3} \dd Z \dd s \; \tilde k_{T, \Bbold, A}^W(X, Z, r, s) \; (\e^{\i Z \cdot \Pi} \alpha)(X, s), \label{LtildeTAW_definition}
\end{align}
where
\begin{align}
\tilde k_{T, \Bbold, A}^W(X, Z, r, s) &:= W_h(X) \; \smash{\frac 2\beta \sum_{n\in \Zbb} \int_{\Rbb^3} \dd Y} \; \bigl[ \tilde k_{T,+}^{W, n}(Y, Z, r, s) \; \e^{\i \tilde \Phi_{A_h, +}^W(X, Y, Z, r, s)} \; \e^{\i \frac \Bbold 2 \cdot \Phi_+^W(Y, Z, r, s)} \notag \\
&\hspace{100pt}  + \tilde k_{T,-}^{W,n}(Y, Z, r, s) \; \e^{\i \tilde \Phi_{A_h, -}^W(X, Y, Z, r, s) } \; \e^{\i \frac \Bbold 2 \cdot \Phi_-^W(Y, Z, r, s) } \bigr]
\end{align}
with $\Phi_\pm^W$ in \eqref{LTA^W_Phi_definition},
\begin{align}
\tilde k_{T, \pm}^{W, n} (Y, Z, r, s) &:= g^{\pm \i\omega_n} \bigl( Y \mp \frac r2\bigr) \, g^{\pm \i \omega_n} \bigl( Z - Y \pm \frac s2\bigr) \; g^{\mp \i \omega_n} \bigl( Z \pm \frac{r-s}{2}\bigr), \label{ktildeTnW_definition}
\end{align}
and
\begin{align}
\tilde \Phi_{A, \pm}^W(X, Y, Z, r, s) &:= \Phi_A \bigl( X \pm \frac r2, X + Y\bigr) + \Phi_A \bigl( X + Y, X + Z \pm \frac s2\bigr)\notag \\
&\hspace{140pt}  + \Phi_A \bigl( X \mp \frac r2, X + Z \mp \frac s2\bigr). \label{LTA^W_PhitildeA_definition}
\end{align}

\begin{prop}
\label{LTAW-LtildeTAW}
Given $T_0>0$, $A\in \Wpervec 3$, and $W\in \Wper 1$ there is $h_0>0$ such that for any $0  < h \leq h_0$, any $T\geq T_0$, and whenever $|\cdot|^k V\alpha_*\in L^2(\Rbb^3)$ for $k\in \{0,1\}$, $\Psi\in \Hmag^1(Q_h)$, and $\Delta \equiv \Delta_\Psi$ as in \eqref{Delta_definition}, we have
\begin{align*}
|\langle \Delta, L_{T, \Abold}^W \Delta - \tilde L_{T, \Bbold, A}^W\Delta \rangle| \leq C\,  h^5\,  \bigl( \Vert V\alpha_*\Vert_2^2 + \Vert \, |\cdot| V\alpha_*\Vert_2^2 \bigr) \, \Vert \Psi\Vert_{\Hmag^1(Q_h)}^2.
\end{align*}
\end{prop}

\begin{proof}
The proof is analogous to the proof of Proposition \ref{LTA-LtildeTA}. Therefore, we are less detailed in our description. We have
\begin{align}
|\langle \Delta, L_{T, \Abold}^W\Delta- \tilde L_{T, \Bbold, A}^W \Delta \rangle | & \notag \\
&\hspace{-110pt}\leq 4 \, \Vert \Psi\Vert_2^2 \iiint_{\Rbb^9} \dd Z \dd r \dd s \; \esssup_{X\in \Rbb^3} |(k_{T, \Bbold, \Abold}^W - \tilde k_{T, \Bbold, A}^W)(X, Z, r, s)| \; |V\alpha_*(r)| \, |V\alpha_*(s)|, \label{LTAW-LtildeTAW_2}
\end{align}
and since
\begin{align*}
k_{T, \pm}^{W, n}(Y, Z, r, s) \, \e^{\i \tilde \Phi_{A_h, \pm}^W(X, Y, Z, r, s)} &\\
&\hspace{-130pt} = \tilde \Gbold_{\Bbold, \Abold_h}^{\pm \i \omega_n} \bigl( X \pm \frac r2, X + Y\bigr) \, \tilde \Gbold_{\Bbold, \Abold_h}^{\pm \i \omega_n} \bigl( X + Y, X + Z \pm \frac s2\bigr) \, \tilde \Gbold_{\Bbold, \Abold_h}^{\mp \i \omega_n} \bigl( X \mp \frac r2, X + Z \mp \frac s2\bigr)\\
&\hspace{-130pt} =: \tilde k_{T, \Bbold, \Abold, \pm}^{W, n}(X, Y, Z, r, s),
\end{align*}
where $\tilde \Gbold_{\Bbold, \Abold}^z$ is defined in \eqref{GboldtildeBAz_definition}, the integrand in \eqref{LTAW-LtildeTAW_2} is bounded by
\begin{align}
|(k_{T, \Bbold, \Abold}^W - \tilde k_{T, \Bbold, A}^W)(X, Z, r, s)| & \notag\\
&\hspace{-100pt} \leq \smash{\frac 2\beta \sum_{n\in \Zbb} \int_{\Rbb^3} \dd Y} \; |W_h(X + Y) - W_h(X)| \; \bigl|\bigl(k_{T, \Bbold, \Abold, +}^{W, n} + k_{T, \Bbold, \Abold, -}^{W, n}\bigr)(X, Y, Z, r, s)\bigr| \notag \\
&\hspace{-10pt} + |W_h(X)| \; \bigl|\bigl(k_{T, \Bbold, \Abold, +}^{W, n} - \tilde k_{T, \Bbold, \Abold, +}^{W, n}\bigr)(X, Y, Z, r, s)\bigr| \notag \\
&\hspace{-10pt}+ |W_h(X)| \; \bigl|\bigl(k_{T, \Bbold, \Abold, -}^{W, n} - \tilde k_{T, \Bbold, \Abold, -}^{W, n}\bigr)(X, Y, Z, r, s)\bigr|. \label{LTAW-LtildeTAW_1}
\end{align}
For the first term, we note that $|W(X + Y) - W(X)|\leq \Vert DW\Vert_\infty \, |Y|$ and with the bound $|Y| \leq |Y \pm \frac r2| + r$, we obtain
\begin{align*}
\smash{\frac 2\beta \sum_{n\in \Zbb} \iint_{\Rbb^3\times \Rbb^3} \dd Y \dd Z} \; |Y| \; \esssup_{X\in \Rbb^3} \bigl| k_{T, \Bbold, \Abold, \pm}^{W,n}(X, Y, Z, r, s)\bigr| & \\
&\hspace{-70pt} \leq F_{T, \Abold_h, \pm}^{(\mathrm a), 0}(r-s)\, |r| + F_{T, \Abold_h, \pm}^{(\mathrm a), 1}(r-s), 
\end{align*}
where
\begin{align*}
F_{T, \Abold,\pm}^{(\mathrm a), a} := \frac 2\beta \sum_{n\in \Zbb} \bigl( |\cdot|^a \, \Gcal_\Abold^{\pm \i\omega_n}\bigr) * \Gcal_\Abold^{\pm \i\omega_n} * \Gcal_\Abold^{\mp \i \omega_n}.
\end{align*}
Since $\Vert F_{T, \Abold, \pm}^{(a), a}\Vert_1 \leq C$ by Lemma \ref{GAz-GtildeAz_decay} and \eqref{g0_decay_f_estimate1} and since $\Vert DW_h\Vert_\infty \leq Ch^3$, we conclude the claimed estimate for this term.

For the second and third term in \eqref{LTAW-LtildeTAW_1}, we bound $|W_h(X)| \leq \Vert W_h\Vert_\infty \leq Ch^2$ and
\begin{align*}
\frac 2\beta \sum_{n\in \Rbb^3} \iint_{\Rbb^3\times \Rbb^3} \dd Y\dd Z\; \esssup_{X\in \Rbb^3} \bigl|\bigl(k_{T, \Bbold, \Abold, \pm}^{W,n} - \tilde k_{T, \Bbold, \Abold, \pm}^{W,n}\bigr)(X, Y, Z, r, s)\bigr| & \leq F_{T, \Abold_h,\pm}^{(\mathrm b)}(r - s)
\end{align*}
with
\begin{align*}
F_{T, \Abold,\pm}^{(\mathrm b)} &:= \smash{\frac 2\beta \sum_{n\in \Zbb}} \; \Hcal_\Abold^{\pm \i \omega_n} * \Gcal_\Abold^{\pm \i\omega_n} * \Gcal_\Abold^{\mp \i\omega_n} + |g^{\pm \i \omega_n}| * \Hcal_\Abold^{\pm \i\omega_n} * \Gcal_\Abold^{\mp \i\omega_n} \\
&\hspace{180pt} + |g^{\pm \i \omega_n}| * |g^{\pm \i \omega_n}| * \Hcal_\Abold^{\mp \i\omega_n}.
\end{align*}
Likewise, we have $\Vert F_{T, \Abold, \pm}^{(\mathrm b)} \Vert_1 \leq C M_\Abold$ by Lemmas \ref{g_decay}, \ref{GAz-GtildeAz_decay}, and \eqref{g0_decay_f_estimate1}. Since $M_{\Abold_h} \leq Ch^3$, we conclude the claimed estimate for this term.
\end{proof}

\paragraph{The operator $\tilde M_{T, \Bbold}^W$.}

We define the operator $\tilde M_{T, \Bbold}^W$ by
\begin{align}
\tilde M_{T, \Bbold}^W \alpha(X, r) &:= W_h(X) \iint_{\Rbb^3\times \Rbb^3} \dd Z \dd s \; k_T^W(Z, r-s) \; \bigl( \cos(Z \cdot \Pi) \alpha\bigr) (X, s), \label{MtildeTAW_definition}
\end{align}
where
\begin{align}
k_T^W(Z, r) &:= \frac 2\beta \sum_{n\in \Zbb} k_{T, +}^n (Z, r) + k_{T,-}^n(Z, r) \label{kTW_definition}
\end{align}
and
\begin{align}
k_{T, \pm}^{W,n}(Z, r) &:= (g^{\pm \i \omega_n} * g^{\pm \i \omega_n})\bigl( Z \mp \frac r2\bigr)\;  g^{\mp \i \omega_n} \bigl( Z \pm \frac r2\bigr). \label{kTWn_definition}
\end{align}

\begin{prop}
\label{LtildeTAW-MtildeTAW}
Given $T_0>0$, $A\in \Wpervec 1$, and $W\in \Lper$ there is $h_0>0$ such that for any $0  < h \leq h_0$, any $T\geq T_0$, and whenever $|\cdot|^k V\alpha_*\in L^2(\Rbb^3)$ for $k\in \{0,1\}$, $\Psi\in \Hmag^1(Q_h)$, and $\Delta \equiv \Delta_\Psi$ as in \eqref{Delta_definition}, we have
\begin{align*}
|\langle \Delta, \tilde L_{T, \Bbold, A}^W \Delta - \tilde M_{T, \Bbold}^W\Delta \rangle| \leq C\,  h^5\,  \bigl( \Vert V\alpha_*\Vert_2^2 + \Vert \, |\cdot| V\alpha_*\Vert_2^2 \bigr) \, \Vert \Psi\Vert_{\Hmag^1(Q_h)}^2.
\end{align*}
\end{prop}

\begin{proof}
Since $g^{\i\omega_n}$ is a symmetric function, we have
\begin{align*}
\tilde M_{T, \Bbold}^W \alpha(X, r) &= W_h(X) \iint_{\Rbb^3 \times \Rbb^3} \dd Z \dd s \; k_T^W(Z, r-s) \; \bigl( \e^{\i Z\cdot \Pi}\alpha\bigr)(X, s).
\end{align*}
It follows that
\begin{align*}
|\langle \Delta, \tilde L_{T, \Bbold, A}^W \Delta - \tilde M_{T, \Bbold}^W \Delta\rangle | &\leq 4 \, \Vert \Psi\Vert_2^2 \, \Vert W_h\Vert_\infty \, \frac 2\beta \sum_{n\in \Zbb} \iiiint_{\Rbb^{12}} \dd Y\dd Z \dd r \dd s \; |V\alpha_*(r)| \, |V\alpha_*(s)| \\
& \hspace{-30pt} \times \esssup_{X\in \Rbb^3} \Bigl[ \bigl| \tilde k_{T,+}^{W, n} (Y, Z, r, s)\bigr| \; \bigl| \e^{\i \tilde \Phi_{A_h, +}^W(X, Y, Z, r, s)} \, \e^{\i \frac{\Bbold}{2} \cdot \Phi_+^W(Y, Z, r, s)} - 1 \bigr| \\
&\hspace{10pt} + \bigl|\tilde k_{T,-}^{W, n} (Y, Z,r, s)\bigr| \; \bigl| \e^{\i \Phi_{\Abold_h, -}^W(X, Y, Z,r, s)} \, \e^{\i \frac{\Bbold}{2} \cdot \Phi_-^W(Y, Z, r, s)} - 1\bigr| \Bigr], 
\end{align*}
where $\tilde k_{T, \pm}^{W, n}$ is defined in \eqref{ktildeTnW_definition}. In view of the estimate
\begin{align}
\bigl| \e^{\i \tilde \Phi_{A_h, \pm}^W(X, Y, Z, r, s)} \, \e^{\i \frac{\Bbold}{2} \cdot \Phi_\pm^W(Y, Z, r, s)} - 1 \bigr| &\leq |\tilde \Phi_{A_h, \pm}^W(X, Y, Z, r, s)| + |\Bbold \cdot \Phi_\pm^W(Y, Z, r, s)|, \label{LtildeTAW-MtildeTAW_2}
\end{align}
it remains to bound the two terms on the right side separately. We start by the first term and note that $|\Phi_A(x,y)| \leq \Vert A\Vert_\infty \, |x-y|$. Therefore,
\begin{align*}
|\tilde \Phi_A(X, Y, Z, r, s)| &\leq \Vert A\Vert_\infty \Bigl[ \bigl| Y \mp \frac r2\bigr| + \bigl| Z - Y \pm \frac s2\bigr| + \bigl| Z \pm \frac{r-s}{2}\bigr| \Bigr],
\end{align*}
which implies that
\begin{align*}
\smash{\frac 2\beta \sum_{n\in \Zbb} \iint_{\Rbb^3\times \Rbb^3} \dd Y \dd Z} \; \bigl|\tilde k_{T, \pm}^{W, n} (Y, Z, r, s)\bigr| \; \esssup_{X\in \Rbb^3} \bigl| \tilde \Phi_{A, \pm}^W(X, Y, Z, r, s)\bigr| & \\
&\hspace{-50pt} \leq C \, \Vert A\Vert_\infty \, F_{T, \pm}^{(\mathrm a)}(r-s) , 
\end{align*}
where 
\begin{align}
F_{T, \pm}^{(\mathrm a)} &:= \smash{\frac 2\beta \sum_{n\in \Zbb}} \; \bigl( |\cdot| \, |g^{\pm \i \omega_n}|\bigr) * |g^{\pm \i \omega_n}| * |g^{\mp \i\omega_n}| +  |g^{\pm \i \omega_n}| * \bigl( |\cdot| \, |g^{\pm \i \omega_n}| \bigr) * |g^{\mp \i\omega_n}| \notag \\
&\hspace{160pt} +  |g^{\pm \i \omega_n}| * |g^{\pm \i \omega_n}|* \bigl( |\cdot| \, |g^{\mp \i\omega_n}| \bigr). \label{LtildeTAW-MtildeTAW_3}
\end{align}
Since the $L^1$-norm of this function is uniformly bounded, since $\Vert W_h\Vert_\infty \leq Ch^2$, and since $\Vert A_h\Vert_\infty \leq Ch$ this proves the claim for the first term on the right side of \eqref{LtildeTAW-MtildeTAW_2}. We turn to the second term. Using the definition \eqref{LTA^W_Phi_definition} of $\Phi_\pm^W$, a straightforward computation shows
\begin{align*}
\smash{\frac 2\beta \sum_{n\in \Zbb} \iint_{\Rbb^3\times \Rbb^3} \dd Y \dd Z} \;  \esssup_{X\in \Rbb^3}  \bigl| \tilde k_{T, \pm}^{W, n} (Y, Z, r, s)\bigr| \; \bigl| \Bbold \cdot \Phi_\pm^W(Y, Z, r, s) \bigr| &\\
&\hspace{-150pt}\leq C \, |\Bbold| \, \bigl( F_{T, \pm}^{(\mathrm a)}(r-s) \, |r| + F_{T, \pm}^{(\mathrm b)}(r-s)\bigr), 
\end{align*}
where $F_{T, \pm}^{(\mathrm a)}$ is as in \eqref{LtildeTAW-MtildeTAW_3} and
\begin{align*}
F_{T, \pm}^{(\mathrm b)} &:= \smash{\frac 2\beta \sum_{n\in \Zbb}} \bigl( |\cdot| \, |g^{\pm \i \omega_n}|\bigr) * \bigl( |\cdot| \, |g^{\pm \i \omega_n}| \bigr) * |g^{\mp \i\omega_n}|.
\end{align*}
Since $\Vert F_T^{(\mathrm b)}\Vert_1 \leq C$ as before, we conclude the claimed estimate for the second term on the right side of \eqref{LtildeTAW-MtildeTAW_2}, which finishes the proof.
\end{proof}

\paragraph{The operator $M_T^W$.}

We define the operator $M_T^W$ by
\begin{align}
M_T^W \alpha(X, r) &:= W_h(X) \iint_{\Rbb^3\times \Rbb^3} \dd Z \dd s \; k_T^W(Z, r-s) \; \alpha(X, s) \label{MTW_definition}
\end{align}
with $k_T^W$ in \eqref{kTW_definition}.

\begin{prop}
\label{MtildeTAW-MTW}
For any $h >0$, any $W\in \Lper$, any $T\geq T_0>0$, and whenever $V\alpha_*\in L^2(\Rbb^3)$, $\Psi\in \Hmag^1(Q_h)$, and $\Delta \equiv \Delta_\Psi$ as in \eqref{Delta_definition}, we have
\begin{align*}
|\langle \Delta, \tilde M_{T, \Bbold}^W \Delta - M_T^W\Delta \rangle| \leq C \,  h^6\, \Vert V\alpha_*\Vert_2^2 \; \Vert \Psi\Vert_{\Hmag^1(Q_h)}^2.
\end{align*}
\end{prop}

\begin{proof}
We combine the estimate $\cos(x) - 1 = -2\sin^2(\frac x2) \leq \frac 12 |x|^2$ with the operator inequality in \eqref{ZPiX_inequality} (for $A=0$) and obtain
\begin{align*}
|\langle \Delta, \tilde M_{T, \Bbold}^W\Delta - M_T^W\Delta\rangle| &\\
&\hspace{-70pt} \leq 6 \, \Vert W_h\Vert_\infty \, \Vert \Pi\Psi\Vert_2^2 \iiint_{\Rbb^9} \dd Z \dd r \dd s \; |Z|^2 \, |k_T^W(Z, r-s)| \; |V\alpha_*(r)|\, |V\alpha_*(s)|.
\end{align*}
An application of the estimate on $|Z|^2$ in \eqref{Z_estimate} yields
\begin{align*}
\int_{\Rbb^3} \dd Z \; |Z|^2 \, |k_T^W(Z, r)| \leq C\, F_T(r),
\end{align*}
where $F_T := F_{T, +} + F_{T, -}$ and
\begin{align*}
F_{T, \pm} &:= \smash{\frac 2\beta \sum_{n\in \Zbb}} \bigl( |\cdot|^2 \, |g^{\pm \i \omega_n}| \bigr) * |g^{\pm \i\omega_n}| * |g^{\mp \i \omega_n}| + |g^{\pm \i\omega_n}| * \bigl( |\cdot|^2 \, |g^{\pm \i \omega_n}| \bigr) * |g^{\mp \i \omega_n}| \\
&\hspace{200pt} + |g^{\pm \i\omega_n}| * |g^{\pm \i\omega_n}| * \bigl( |\cdot|^2 \, |g^{\mp \i\omega_n}| \bigr).
\end{align*}
Since $\Vert F_{T, \pm}\Vert_1 \leq C$ by Lemma \ref{g_decay}, $\Vert \Pi\Psi\Vert_2 \leq Ch^2 \Vert \Psi\Vert_{\Hmag^1(Q_h)}$, and $\Vert W_h\Vert_\infty \leq Ch^2$, we conclude the claimed estimate. This finishes the proof.
\end{proof}

\subsubsection{Analysis of \texorpdfstring{$M_T^W$}{MTW} and calculation of the quadratic \texorpdfstring{$W$}{W}-term}
\label{Analysis_of_MTW_Section}

\begin{prop}
\label{MTW}
Assume that $V\alpha_* \in L^2(\Rbb^3)$ and that $W\in \Lper$. For any $h>0$, $\Psi\in \Hmag^1(Q_h)$, and $\Delta \equiv \Delta_\Psi$ as in \eqref{Delta_definition}, we have
\begin{align}
\langle \Delta, M_\Tc^W \Delta\rangle = -4\; \Lambda_1 \; \langle \Psi, W_h \Psi\rangle  \label{MTW_1}
\end{align}
with $\Lambda_1$ in \eqref{GL-coefficient_W}. Moreover, for any $T \geq T_0 > 0$ we have
\begin{align}
|\langle \Delta,  M_T^W \Delta - M_\Tc^W \Delta\rangle| \leq C\; h^4 \; |T - \Tc| \; \Vert V\alpha_*\Vert_2^2 \; \Vert \Psi\Vert_{\Hmag^1(Q_h)}^2. \label{MTW_2}
\end{align}
\end{prop}

\begin{proof}
By the definition in \eqref{kTW_definition}, we have
\begin{align*}
k_T(Z, r) &= - \frac 2\beta \sum_{n\in \Zbb} \int_{\Rbb^3} \frac{\dd p}{(2\pi)^3} \int_{\Rbb^3} \frac{\dd q}{(2\pi)^3} \; \Bigl[ \frac{\e^{\i Z\cdot (p+q)} \e^{\i \frac r2\cdot (q-p)}}{(\i \omega_n + \mu - p^2)^2 (\i\omega_n - \mu + p^2)} \\
&\hspace{150pt} - \frac{\e^{\i Z\cdot (p+q)} \e^{\i \frac r2\cdot (p-q)}}{(\i\omega_n - \mu + p^2)^2 (\i\omega_n + \mu - p^2)}\Bigr]
\end{align*}
so that
\begin{align*}
\int_{\Rbb^3} \dd Z \; k_T(Z, r) &= -\frac 4\beta \sum_{n\in \Zbb} \int_{\Rbb^3} \frac{\dd p}{(2\pi)^3} \; \e^{\i r \cdot p} \, \frac{p^2 -\mu }{(\i \omega_n + \mu - p^2)^2 (\i\omega_n - \mu + p^2)^2}.
\end{align*}
In view of the partial fraction expansion
\begin{align*}
\frac{1}{(\i\omega_n - E)^2(\i\omega_n + E)^2} = \frac{1}{4E^2} \Bigl[ \frac{1}{(\i\omega_n - E)^2} + \frac{1}{(\i\omega_n + E)^2}\Bigr] - \frac{1}{4E^3} \Bigl[ \frac{1}{\i\omega_n - E} - \frac{1}{\i\omega_n + E}\Bigr]
\end{align*}
and the identity
\begin{align}
\frac{\beta}{2} \frac{1}{\cosh^2(\frac \beta 2z)} = \frac{\dd}{\dd z} \tanh\bigl( \frac \beta 2 z\bigr) = - \frac{2}{\beta} \sum_{n\in \Zbb} \frac{1}{(\i\omega_n - z)^2}, \label{cosh2_Matsubara}
\end{align}
which follows from the Mittag-Leffler series expansion, see e.g. \cite[Eq. (3.12)]{DeHaSc2021} (its convergence becomes manifest by combining the $+n$ and $-n$ terms)
\begin{align}
\tanh\bigl( \frac \beta 2 z\bigr) &= -\frac{2}{\beta} \sum_{n\in \Zbb} \frac{1}{\i\omega_n - z}, \label{tanh_Matsubara}
\end{align}
we have
\begin{align*}
\frac{4}{\beta} \sum_{n\in \Zbb} \frac{E}{(\i\omega_n - E)^2(\i\omega_n + E)^2} = \beta^2 \; g_1(\beta E)
\end{align*}
with the function $g_1$ in \eqref{XiSigma}. Therefore,
\begin{align*}
\langle \Delta, M_\Tc^W \Delta\rangle &= -\betac^2 \int_{\Rbb^3} \frac{\dd p}{(2\pi)^3} \; |(-2)\hat{V\alpha_*}(p)|^2 \, g_1(\betac (p^2-\mu)) \; \langle \Psi, W_h \Psi\rangle \\
&= - 4\, \Lambda_1 \, \langle \Psi, W_h\Psi\rangle.
\end{align*}
The estimate in \eqref{MTW_2} follows in a straightforward manner analogously to the proof of \eqref{MTA2_2}, using \cite[Eq. (4.91)]{DeHaSc2021}.
\end{proof}


\subsubsection{Summary: The quadratic terms}
\label{Summary_quadratic_terms_Section}

In this section, we give a summary of the results pertaining to the quadratic terms in $\Delta \equiv \Delta_\Psi$ that are relevant for the proof of Theorem~\ref{Calculation_of_the_GL-energy}. We also prove a preparatory result, which we will use in the proof of Proposition \ref{Lower_Tc_a_priori_bound}.

\begin{prop}
\label{Rough_bound_on_BCS energy}
Given $T_0> 0$, $A\in \Wpervec 3$, and $W\in \Wper 1$ there is a constant $h_0>0$ such that for any $T_0 \leq T\leq \Tc$, any $0 < h \leq h_0$, and whenever $|\cdot|^k V\alpha_*\in L^2(\Rbb^3)$ for $k\in \{0,1,2\}$, $\Psi \in \Hmag^1(Q_h)$, and $\Delta \equiv \Delta_\Psi$ as in \eqref{Delta_definition}, we have
\begin{align}
- \frac 14 \langle \Delta, L_{T,\Abold, W} \Delta\rangle + \Vert \Psi\Vert_2^2 \; \langle \alpha_*, V\alpha_*\rangle & \leq c  \, \frac{T - \Tc}{\Tc}\, \Vert \Psi\Vert_2^2 + C h^4  \, \Vert \Psi\Vert_{\Hmag^1(Q_h)}^2. \label{Rough_bound_on_BCS energy_eq1}
\end{align}
\end{prop}

\begin{proof}
By the decomposition \eqref{LTAW_decomposition} of $L_{T, \Abold, W}$, we have
\begin{align}
-\frac 14 \langle \Delta, L_{T, \Abold, W} \Delta \rangle  = - \frac 14 \langle \Delta, L_{T,\Abold} \Delta\rangle - \frac 14 \langle \Delta, L_{T, \Abold}^W \Delta \rangle - \frac 14 \langle \Delta, \Rcal_{T, \Abold, W}^{(2)} \Delta\rangle \label{Rough_bound_on_BCS energy_proof_2}
\end{align}
with $L_{T, \Abold}^W$ and $R_{T, \Abold, W}^{(2)}$ in \eqref{LTA^W_definition} and \eqref{RTAW2_definition}, respectively. By Lemma \ref{RTAW2_estimate} as well as Propositions \ref{LTAW-LtildeTAW}, \ref{LtildeTAW-MtildeTAW}, \ref{MtildeTAW-MTW}, and \ref{MTW}, we have
\begin{align*}
|\langle \Delta, L_{T, \Abold}^W \Delta \rangle| + |\langle \Delta, \Rcal_{T, \Abold, W}^{(2)} \Delta\rangle| \leq C \, h^4 \, \Vert \Psi\Vert_{\Hmag^1(Q_h)}^2.
\end{align*}
Furthermore, by Lemma \ref{LTA_action}, the decomposition \eqref{LTA_decomposition} of $L_{T,\Abold}$, as well as Propositions~\ref{LTA-LtildeTA}, \ref{LtildeTA-MtildeTA}, and \ref{MtildeTA-MTA}, we have
\begin{align}
- \frac 14 \langle \Delta, L_{T,\Abold} \Delta\rangle + \Vert \Psi\Vert_2^2 \; \langle \alpha_*, V\alpha_*\rangle &\notag\\
&\hspace{-100pt} = - \frac 14 \langle \Delta, M_T^{(1)}\Delta- M_{\Tc}^{(1)} \Delta\rangle - \frac 14 \langle \Delta, M_{T,\Abold} \Delta - M_T^{(1)}\Delta\rangle + R_1(\Delta), \label{Rough_bound_on_BCS energy_proof_1}
\end{align}
with a remainder $R_1(\Delta)$ obeying the bound
\begin{align*}
|R_1(\Delta) | \leq C \; h^5 \; \Vert \Psi\Vert_{\Hmag^1(Q_h)}^2,
\end{align*}
and where Proposition \ref{MT1} implies 
\begin{align*}
-\frac 14 \langle \Delta, M_T^{(1)}\Delta- M_{\Tc}^{(1)} \Delta\rangle \leq  c \; \frac{T - \Tc}{\Tc} \; \Vert \Psi \Vert_2^2.
\end{align*}
The proof of the bound
\begin{align}
|\langle \Delta, M_{T,\Abold}\Delta - M_T^{(1)}\Delta\rangle| &\leq C\; h^4 \; \Vert V\alpha_*\Vert_2^2 \; \Vert \Psi\Vert_{\Hmag^1(Q_h)}^2 \label{MTB-MT1_eq1}
\end{align}
follows the same strategy as that of Proposition~\ref{MtildeTA-MTA} and it employs the operator inequality \eqref{ZPiX_inequality} on $(Z\cdot \Pi_\Abold)^2$ to bound
\begin{align}
|\langle \Psi, [\cos(Z\cdot\Pi_\Abold) - 1] \Psi \rangle| &\leq C\; h^4 \; |Z|^2 \; \Vert \Psi\Vert_{\Hmag^1(Q_h)}^2. \label{MTB-MT1_1}
\end{align}
The details are left to the reader. This completes the proof of \eqref{Rough_bound_on_BCS energy_eq1}.
\end{proof}

Under the assumptions of Theorem~\ref{Calculation_of_the_GL-energy}, we combine \eqref{Rough_bound_on_BCS energy_proof_2} and \eqref{Rough_bound_on_BCS energy_proof_1} with the results of Propositions~\ref{MT1}, \ref{MTB2}, \ref{MTB3}, \ref{LTAW-LtildeTAW}, \ref{LtildeTAW-MtildeTAW}, \ref{MtildeTAW-MTW}, and \ref{MTW}, which shows that, within the temperature regime $T = \Tc(1-Dh^2)$ with $D \in \Rbb$, the identity
\begin{align}
-\frac{1}{4} \langle \Delta, L_{T,\Abold, W} \Delta \rangle + \Vert \Psi \Vert_2^2 \, \langle \alpha_*, V \alpha_* \rangle & \notag \\
&\hspace{-80pt} = \Lambda_0 \; \Vert \Pi_{\Abold_h}\Psi\Vert_2^2 - Dh^2 \; \Lambda_2 \; \Vert \Psi\Vert_2^2 + \Lambda_1 \; \langle \Psi, W_h\Psi\rangle +  R_2(\Delta) \label{eq:A15}
\end{align}
holds. Here, the error term $R_2(\Delta)$ obeys the estimate
\begin{align*}
	| R_2(\Delta) | \leq C\;  \bigl( h^5 \; \Vert \Psi\Vert_{\Hmag^1(Q_h)}^2 + h^6 \; \Vert \Psi\Vert_{\Hmag^2(Q_h)}^2\bigr).
\end{align*}
This concludes the extraction of the quadratic terms in the Ginzburg--Landau functional and finishes the analysis of the operator $L_{T,\Abold, W}$.


\subsubsection{Decomposition of \texorpdfstring{$N_{T, \Abold, W}$}{NTAW} --- perturbation of \texorpdfstring{$W$}{W}}

We use the resolvent equation \eqref{Resolvent_Equation} to decompose the operator $N_{T, \Abold, W}$ in \eqref{NTAW_definition} as
\begin{align}
N_{T, \Abold, W} = N_{T, \Abold} + \Rcal_{T, \Abold,W}^{(3)} \label{NTAW_decomposition}
\end{align}
with
\begin{align}
N_{T, \Abold} := N_{T, \Abold, 0} \label{NTA_definition}
\end{align}
as well as
\begin{align}
\Rcal_{T, \Abold, W}^{(3)}(\Delta) & \notag\\
&\hspace{-40pt} := \frac 2\beta \sum_{n\in \Zbb} \Bigl[ \frac{1}{\i\omega_n - \hfrak_\Abold} \, W_h \,  \frac{1}{\i \omega_n -\hfrak_{\Abold, W}} \, \Delta \, \frac{1}{\i\omega_n + \ov{\hfrak_{\Abold, W}}} \, \ov \Delta \, \frac{1}{\i \omega_n - \hfrak_{\Abold, W}} \, \Delta \, \frac{1}{\i \omega_n + \ov{\hfrak_{\Abold, W}}} \notag \\
&\hspace{10pt} - \frac{1}{\i\omega_n - \hfrak_\Abold} \, \Delta \, \frac{1}{\i \omega_n + \ov{\hfrak_\Abold}} \, W_h \, \frac{1}{\i\omega_n + \ov{\hfrak_{\Abold, W}}} \, \ov \Delta \,  \frac{1}{\i \omega_n - \hfrak_{\Abold, W}} \, \Delta \, \frac{1}{\i \omega_n + \ov{\hfrak_{\Abold, W}}} \notag \\
&\hspace{10pt} + \frac{1}{\i\omega_n - \hfrak_\Abold} \, \Delta \, \frac{1}{\i \omega_n + \ov{\hfrak_\Abold}} \, \ov \Delta \, \frac{1}{\i\omega_n - \hfrak_\Abold} \, W_h \,  \frac{1}{\i \omega_n - \hfrak_{\Abold, W}} \, \Delta \, \frac{1}{\i \omega_n + \ov{\hfrak_{\Abold, W}}} \notag \\
&\hspace{10pt} - \frac{1}{\i\omega_n - \hfrak_\Abold} \, \Delta \, \frac{1}{\i \omega_n + \ov{\hfrak_\Abold}} \, \ov \Delta \, \frac{1}{\i\omega_n - \hfrak_\Abold} \, \Delta \,  \frac{1}{\i \omega_n + \ov{\hfrak_\Abold}} \, W_h \, \frac{1}{\i \omega_n + \ov{\hfrak_{\Abold, W}}}. \label{RTAW3_definition}
\end{align}

\begin{lem}
\label{RTAW3_estimate}
Assume that $V\alpha_*\in L^{\nicefrac 43}(\Rbb^3)$. For any $T>0$, any $A\in \Wpervec 1$, any $W\in \Lper$, any $h>0$, and whenever $\Psi\in \Hmag^1(Q_h)$ and $\Delta \equiv \Delta_\Psi$ as in \eqref{Delta_definition}, we have
\begin{align*}
\Vert \Rcal_{T, \Abold, W}^{(3)} (\Delta)\Vert_{\Lsymm}^2 &\leq C\, \beta^8 \, h^{10} \, \Vert \Psi\Vert_{\Hmag^1(Q_h)}^6.
\end{align*}
\end{lem}

\begin{proof}
Hölder's inequality shows that the Hilbert-Schmidt norm per unit volume of the terms in the sum in \eqref{RTAW3_definition} are bounded by $C\, |\omega_n|^{-5} \, \Vert W_h\Vert_\infty \,  \Vert \Delta\Vert_6^3$, which by Lemma \ref{Schatten_estimate} and \eqref{Magnetic_Sobolev} is bounded by a constant times $\beta^5\, h^5 \Vert \Psi\Vert_{\Hmag^1(Q_h)}^3$. This proves the claim.
\end{proof}

\subsubsection{A representation formula for the operator \texorpdfstring{$N_{T, \Abold}$}{NTA}}

From now on in this Subsection \ref{Calculation_of_the_GL-energy_proof_Section}, the notation $\Zbold$ is short for the vector $(Z_1,Z_2,Z_3)$ with $Z_1, Z_2, Z_3 \in \Rbb^3$ and we abbreviate $\dd \Zbold = \dd Z_1 \dd Z_2 \dd Z_3$ to restrict the length of formulas.
%
%
The analysis we used for the case of $L_{T, \Abold}$ applies to the nonlinear operator $N_{T, \Abold}$ in \eqref{NTA_definition} as well and we start with a representation formula for $N_{T, \Abold}$.

\begin{lem}
\label{NTA_action}
The operator $N_{T, \Abold} \colon \Hsymm \ra \Lsymm$ in \eqref{NTAW_definition} acts as
\begin{align*}
N_{T, \Abold}(\alpha) (X, r) &= \iiint_{\Rbb^9} \dd \Zbold \iiint_{\Rbb^9} \dd \sbold \; \ell_{T, \Bbold, \Abold}(X, \Zbold, r, \sbold)\; \Acal(X, \Zbold, \sbold)
\end{align*}
with
\begin{align}
\Acal(X, \Zbold, \sbold) &:= \e^{\i Z_1\cdot \Pi} \alpha(X, s_1) \; \ov{\e^{\i Z_2\cdot \Pi} \alpha(X, s_2)} \; \e^{\i Z_3\cdot \Pi} \alpha(X,s_3) \label{NTA_alpha_definition}
\end{align}
as well as
\begin{align}
\ell_{T, \Bbold, \Abold}(X, \Zbold, r, \sbold) &:= \frac{2}{\beta} \sum_{n\in \Zbb} \ell_{T, \Bbold, \Abold}^n(X, \Zbold, r, \sbold) \; \e^{\i \frac \Bbold 2 \cdot \Phi(\Zbold, r, \sbold)} \label{lTA_definition}
\end{align}
and
\begin{align}
\ell_{T, \Bbold, \Abold}^n(X, \Zbold, r, \sbold) &:=  \Gbold_{\Bbold, \Abold_h}^{\i\omega_n}\bigl(X + \frac r2 \, , \, X + Z_1 + \frac{s_1} 2\bigr) \, \Gbold_{\Bbold, \Abold_h}^{-\i\omega_n} \bigl( X + Z_2 - \frac{s_2}{2} \, , \, X + Z_1 - \frac{s_1}{2}\bigr)  \notag\\
&\hspace{-50pt} \times \Gbold_{\Bbold, \Abold_h}^{\i\omega_n}\bigl( X + Z_2 + \frac{s_2}{2} \, , \, X + Z_3 + \frac{s_3}{2}\bigr) \, \Gbold_{\Bbold, \Abold_h}^{-\i\omega_n}\bigl(X - \frac{r}{2} \, , \, X + Z_3 - \frac{s_3}{2}\bigr), \label{lTAn_definition}
\end{align}
where $\Gbold_{\Bbold, \Abold}^z$ is defined in \eqref{GboldBAz_definition}, and
\begin{align}
\Phi(\Zbold, r, \sbold) &:= \frac r2 \wedge \bigl( Z_1 - \frac{r - s_1}{2}\bigr) + \frac r2 \wedge \bigl( Z_3 + \frac{r-s_3}{2}\bigr) \notag \\
&\hspace{-35pt} + \bigl( Z_2 - Z_3 - \frac{s_2 + s_3}{2}\bigr) \wedge \bigl( Z_1 - Z_2 - \frac{s_1 + s_2}{2}\bigr) \notag \\
&\hspace{-35pt}+ \bigl( Z_3 + \frac{r - s_3}{2}\bigr) \wedge \bigl( Z_1 - Z_2 - \frac{s_1 + s_2}{2}\bigr)+ \bigl( s_2 + s_3 - \frac r2\bigr) \wedge \bigl( Z_1 - Z_2 - \frac{s_1 + s_2}{2}\bigr) \notag \\
&\hspace{-35pt}+ \bigl( Z_3 + \frac{r - s_3}{2} \bigr) \wedge \bigl( Z_3 - Z_2 + \frac{s_2 + s_3}{2}\bigr)+ \bigl( s_3 - \frac r2\bigr) \wedge \bigl( Z_3 - Z_2 + \frac{s_2 + s_3}{2}\bigr). \label{PhiB_definition}
\end{align}
\end{lem}

\begin{proof}
When we compute theintegral kernel of $N_{T, \Abold}$, we get
\begin{align*}
N_{T, \Abold} (\alpha)(X,r) &= \smash{\frac 2\beta \sum_{n\in \Zbb} \iiint_{\Rbb^{9}} \dd \mathbf u \iiint_{\Rbb^9}\dd \mathbf v} \; G_{\Abold_h}^{\i\omega_n} (\zeta_X^r, u_1)\, \alpha(u_1,v_1)\, G_{\Abold_h}^{-\i\omega_n} (u_2,v_1) \, \ov{\alpha(u_2,v_2)}  \\
&\hspace{150pt} \times G_{\Abold_h}^{\i\omega_n}(v_2,u_3)\, \alpha(u_3,v_3)\,  G_{\Abold_h}^{-\i\omega_n}(\zeta_X^{-r}, v_3).
\end{align*}
Here, we denoted $\zeta_X^r := X+\frac r2$ for short and used \eqref{GAz_Kernel_of_complex_conjugate}. We introduce the center-of-mass coordinate $\Zbold$ and the relative coordinate $\sbold$ determined by
\begin{align*}
\mathbf u &= X + \Zbold + \frac \sbold 2, & \mathbf v &= X + \Zbold - \frac \sbold 2.
\end{align*}
Furthermore, we multiply and divide by the factor
\begin{align*}
\e^{\i \Phi_{\Abold_\Bbold}(\zeta_X^r,\zeta_{Z_1 + X}^{s_1})} \, \e^{\i \Phi_{\Abold_\Bbold}(\zeta_{Z_2+X}^{s_2}, \zeta_{Z_1 + X}^{-s_1})} \, \e^{\i \Phi_{\Abold_\Bbold}(\zeta_{Z_2+X}^{-s_2}, \zeta_{Z_3+X}^{s_3})} \, \e^{\i \Phi_{\Abold_\Bbold}(\zeta_X^{-r}, \zeta_{Z_3+X}^{-s_3})} & \\
&\hspace{-200pt} = \e^{\i \Bbold \cdot (X\wedge Z_1)} \, \e^{-\i \Bbold \cdot (X \wedge Z_2)} \, \e^{\i \Bbold \cdot (X\wedge Z_3)} \; \e^{\i \frac \Bbold 2 \cdot \Phi(\Zbold, r, \sbold)},
\end{align*}
where the equality is a tedious but straightforward computation. Using the fact that $\Bbold \cdot (X\wedge Z) = Z \cdot (\Bbold \wedge X)$ and that $Z\cdot (\Bbold \wedge X)$ commutes with $Z\cdot (-\i\nabla_X)$, this implies the claimed formula.
\end{proof}

We analyze the operator $N_{T, \Abold}$ with the by now familiar four-step analysis, as in the case of $L_{T,\Abold}$. In order to do this, we decompose $N_{T, \Abold}$ according to
\begin{align}
N_{T, \Abold} = (N_{T, \Abold} - \tilde N_{T, \Bbold, A}) + (\tilde N_{T, \Bbold, A} - N_{T, \Bbold}^{(1)}) + (N_{T, \Bbold}^{(1)} - N_T^{(2)}) + N_{T}^{(2)}. \label{NTB_decomposition}
\end{align}
where $\tilde N_{T, \Bbold, A}$ is defined in \eqref{NtildeTB_definition}, $N_{T, \Bbold}^{(1)}$ in \eqref{NTB1_definition}, and $N_T^{(2)}$ in \eqref{NT2_definition} below. 
After we have proven that the terms in brackets are small in Section~\ref{sec:approxNTA}, the extraction of the quartic term in the Ginzburg--Landau functional from the term $\langle \Delta, N_{T}^{(2)}(\Delta) \rangle$ is achieved by a result in Section~\ref{sec:compquarticterm}. Section~\ref{sec:quarticterms} summarizes and concludes the analysis.

\subsubsection{Approximation of \texorpdfstring{$N_{T, \Abold}$}{NTA}}
\label{sec:approxNTA}

\paragraph{The operator $\tilde N_{T, \Bbold, A}$.} We define the operator $\tilde N_{T, \Bbold, A}$ by
\begin{align}
\tilde N_{T, \Bbold, A}(\alpha) (X,r) &:= \iiint_{\Rbb^{9}} \dd \Zbold \iiint_{\Rbb^{9}} \dd \sbold \; \tilde \ell_{T, \Bbold, A} (X, \Zbold, r, \sbold) \; \Acal(X, \Zbold , \sbold) \label{NtildeTB_definition}
\end{align}
with
\begin{align*}
\tilde \ell_{T, \Bbold, A}(X, \Zbold, r, \sbold) &:= \frac 2\beta \sum_{n\in\Zbb} \ell_T^n (\Zbold, r, \sbold) \; \e^{\i \tilde \Phi_{A_h}(X, \Zbold, r, \sbold)} \; \e^{\i \frac \Bbold 2 \cdot \Phi(\Zbold, r, \sbold)},
\end{align*}
where
\begin{align}
\ell_T^n(\Zbold, r, \sbold) &:= g^{\i\omega_n} \bigl(Z_1 - \frac{r-s_1}{2}\bigr) \; g^{-\i\omega_n} \bigl( Z_1 - Z_2 - \frac{s_1 + s_2}{2}\bigr) \notag \\
&\hspace{50pt} \times g^{\i\omega_n} \bigl( Z_2 - Z_3 - \frac{s_2 + s_3}{2} \bigr) \; g^{-\i\omega_n} \bigl( Z_3 + \frac{r-s_3}{2}\bigr) \label{lTn_definition}
\end{align}
and
\begin{align}
\tilde \Phi_A(X, \Zbold, r, \sbold) &:= \Phi_A\bigl(X + \frac r2 \, , \, X + Z_1 + \frac{s_1} 2\bigr) + \Phi_A\bigl( X + Z_2 - \frac{s_2}{2} \, , \, X + Z_1 - \frac{s_1}{2}\bigr) \notag\\
&\hspace{-30pt} + \Phi_A\bigl( X + Z_2 + \frac{s_2}{2} \, , \, X + Z_3 + \frac{s_3}{2}\bigr) + \Phi_A\bigl(X - \frac{r}{2} \, , \, X + Z_3 - \frac{s_3}{2}\bigr). \label{NtildeTA_PhitildeA_definition}
\end{align}
In our calculation, we may replace $N_{T, \Abold}(\Delta)$ by $\tilde N_{T, \Bbold, A}(\Delta)$ due to the following error bound.

\begin{prop}
\label{NTA-NtildeTA}
Assume that $V\alpha_*\in L^{\nicefrac 43}(\Rbb^3)$ and that $A\in \Wpervec{3}$. For every $T_0>0$ there is $h_0>0$ such that for any $0 < h \leq h_0$, any $T\geq T_0$, any $\Psi \in \Hmag^1(Q_h)$, and $\Delta \equiv \Delta_\Psi$ as in \eqref{Delta_definition}, we have
\begin{align*}
|\langle \Delta,  N_{T, \Abold}(\Delta) - \tilde N_{T, \Bbold, A}(\Delta)\rangle| &\leq C \; h^6 \; \Vert V\alpha_*\Vert_{\nicefrac 43}^4 \; \Vert \Psi\Vert_{\Hmag^1(Q_h)}^4.
\end{align*}
\end{prop}

The function
\begin{align}
F_{T,\Abold} &:= \smash{\frac 2\beta \sum_{n\in \Zbb}} \; \Hcal_\Abold^{\i\omega_n} * \Gcal_\Abold^{-\i \omega_n} * \Gcal_\Abold^{\i \omega_n} * \Gcal_\Abold^{-\i \omega_n} + |g^{\i\omega_n}| * \Hcal_\Abold^{-\i \omega_n} * \Gcal_\Abold^{\i \omega_n} * \Gcal_\Abold^{-\i \omega_n} \notag \\
&\hspace{50pt} + |g^{\i\omega_n}| * |g^{-\i\omega_n} | * \Hcal_\Abold^{\i \omega_n} * \Gcal_\Abold^{-\i \omega_n} + |g^{\i\omega_n}| * |g^{-\i\omega_n} | * |g^{\i\omega_n}| * \Hcal_\Abold^{-\i \omega_n}. \label{NTA-NtildeTA_FTA_definition}
\end{align}
plays a prominent role in the proof of Proposition \ref{NTA-NtildeTA}. For any $T \geq T_0 > 0$, Lemmas~\ref{g_decay} and \ref{GAz-GtildeAz_decay} as well as \eqref{g0_decay_f_estimate1} show that its $L^1(\Rbb^3)$-norm is bounded by
\begin{align}
	\Vert F_{T, \Abold_h}\Vert_1 \leq C \; h^3. \label{NTA-NtildeTA_FTA_estimate}
\end{align}

\begin{proof}[Proof of Proposition \ref{NTA-NtildeTA}]
Since the function $|\Psi|$ is periodic, \eqref{Magnetic_Sobolev} yields
\begin{align}
\Vert \e^{\i Z \cdot\Pi}\Psi\Vert_6^2 = \Vert \Psi\Vert_6^2 \leq C\,  h^2 \, \Vert \Psi\Vert_{\Hmag^1(Q_h)}^2, \label{NTB-NtildeTB_3}
\end{align}
which further implies
\begin{align}
\fint_{Q_h} \dd X \; |\Psi(X)|\; \prod_{i=1}^3  |\e^{\i Z_i\cdot \Pi}\Psi(X)| &\leq \Vert \Psi\Vert_2 \; \prod_{i=1}^3  \Vert \e^{\i Z_i\cdot \Pi}\Psi\Vert_6 \leq C \, h^4 \, \Vert \Psi\Vert_{\Hmag^1(Q_h)}^4 \label{NTA-NtildeTBA_2}
\end{align}
and
\begin{align}
|\langle \Delta,  N_{T, \Abold}(\Delta)- \tilde N_{T, \Bbold, A}(\Delta)\rangle| & \notag\\
&\hspace{-100pt}\leq C\, h^4 \, \Vert \Psi\Vert_{\Hmag^1(Q_h))}^4  \int_{\Rbb^3} \dd r \,  \iiint_{\Rbb^9} \dd \sbold\; |V\alpha_*(r)|\;  |V\alpha_*(s_1)| \; |V\alpha_*(s_2)| \; |V\alpha_*(s_3)|  \notag\\
&\hspace{-20pt}\times \iiint_{\Rbb^9} \dd \Zbold \; \esssup_{X\in \Rbb^3} \bigl|(\ell_{T, \Bbold, \Abold} - \tilde \ell_{T, \Bbold, A})(X, \Zbold, r, \sbold)\bigr|.  \label{NTB-NtildeTB_1}
\end{align}
Furthermore, we have
\begin{align*}
\tilde \ell_T^n (\Zbold, r, \sbold) \, \e^{\i \tilde \Phi_A(X, \Zbold, r, \sbold)} \\
&\hspace{-80pt} = \tilde \Gbold_{\Bbold, \Abold}^{\i\omega_n}\bigl(X + \frac r2 \, , \, X + Z_1 + \frac{s_1} 2\bigr) \, \tilde \Gbold_{\Bbold, \Abold}^{-\i\omega_n} \bigl( X + Z_2 - \frac{s_2}{2} \, , \, X + Z_1 - \frac{s_1}{2}\bigr)  \notag\\
&\hspace{-50pt} \times \tilde \Gbold_{\Bbold, \Abold}^{\i\omega_n}\bigl( X + Z_2 + \frac{s_2}{2} \, , \, X + Z_3 + \frac{s_3}{2}\bigr) \, \tilde \Gbold_{\Bbold, \Abold}^{-\i\omega_n}\bigl(X - \frac{r}{2} \, , \, X + Z_3 - \frac{s_3}{2}\bigr),
\end{align*}
where $\tilde \Gbold_{\Bbold, \Abold}^z$ is defined in \eqref{GboldtildeBAz_definition}. Let us employ the change of variables 
\begin{align}
Z_1' - Z_2' &:= Z_1 - Z_2 - \frac{s_1 +s_2}{2}, & Z_2' - Z_3' &:= Z_2 - Z_3 - \frac{s_2 + s_3}{2}, & Z_3' &:= Z_3 + \frac{r-s_3}{2}, \label{NTA_change_of_variables_1}
\end{align}
which implies
\begin{align}
Z_1 - \frac{r-s_1}{2} = Z_1' - (r - s_1 - s_2 - s_3). \label{NTA_change_of_variables_2}
\end{align}
We repeat the argument that leads to \eqref{LTA-LtildeTBA_5}, and deduce
\begin{align*}
\iiint_{\Rbb^9} \dd \Zbold\; \esssup_{X\in \Rbb^3} \bigl|(\ell_{T, \Bbold, \Abold} - \tilde \ell_{T, \Bbold, A})(X, \Zbold, r, \sbold)\bigr| \leq F_{T, \Abold_h}(r - s_1 - s_2 - s_3),
\end{align*}
where $F_{T, \Abold}$ is the function in \eqref{NTA-NtildeTA_FTA_definition}.  Plugging this into \eqref{NTB-NtildeTB_1}, applying the estimate
\begin{align*}
\bigl\Vert V\alpha_* \; \bigl( V\alpha_* * V\alpha_* * V\alpha_* * F_{T, \Abold}\bigr) \bigr\Vert_1 &\leq C \; \Vert V\alpha_*\Vert_{\nicefrac 43}^4 \; \Vert F_{T, \Abold}\Vert_1,
\end{align*}
and using \eqref{NTA-NtildeTA_FTA_estimate} finishes the proof.
\end{proof}

\paragraph{The operator $N_{T, \Bbold}^{(1)}$.} We define the operator $N_{T, \Bbold}^{(1)}$ by
\begin{align}
N_{T, \Bbold}^{(1)}(\alpha)(X,r) &:= \iiint_{\Rbb^9} \dd \Zbold \iiint_{\Rbb^9} \dd \sbold \; \ell_T (\Zbold, r, \sbold) \; \Acal (X, \Zbold , \sbold) \label{NTB1_definition}
\end{align}
with
\begin{align}
\ell_T(\Zbold, r, \sbold) &:= \ell_{T,0,0}(0, \Zbold, r, \sbold) \label{lT_definition}
\end{align}
where $\Acal$ is defined in \eqref{NTA_alpha_definition} and $\ell_{T,0,0}$ in \eqref{lTA_definition}. In our calculations, we may replace $\langle \Delta, \tilde N_{T, \Bbold, A}(\Delta) \rangle$ by $\langle \Delta, N_{T, \Bbold}^{(1)}(\Delta) \rangle$ due to the following error bound.

\begin{prop}
\label{NtildeTBA-NTB1}
Assume that $|\cdot|^kV\alpha_* \in L^{\nicefrac 43}(\Rbb^3)$ for $k\in \{0,1\}$ and $A\in\Lpervec$. For every $T \geq T_0 > 0 $, every $h>0$, every $\Psi \in \Hmag^1(Q_h)$ and $\Delta \equiv \Delta_\Psi$ as in \eqref{Delta_definition}, we have 
\begin{align*}
|\langle \Delta, \tilde N_{T, \Bbold, A}(\Delta) - N_{T, \Bbold}^{(1)}(\Delta)\rangle| &\leq C \; h^5 \; \bigl( \Vert V\alpha_*\Vert_{\nicefrac 43}^4 + \Vert \, |\cdot|V\alpha_*\Vert_{\nicefrac 43}^4\bigr)\; \Vert \Psi\Vert_{\Hmag^1(Q_h)}^4.
\end{align*}
\end{prop}

The functions
\begin{align}
	F_T^{(1)} &:= \smash{\frac 2\beta \sum_{n\in \Zbb}} \, |g^{\i\omega_n}| * \bigl(|\cdot|\, |g^{-\i\omega_n}|\bigr) * \bigl(|\cdot|\, |g^{\i\omega_n}|\bigr) * |g^{-\i\omega_n}| \notag \\
	&\hspace{100pt}+ |g^{\i\omega_n}| * \bigl(|\cdot|\, |g^{-\i\omega_n}|\bigr) * |g^{\i\omega_n}| * \bigl(|\cdot|\, |g^{-\i\omega_n}|\bigr) \notag \\
	&\hspace{100pt}+ |g^{\i\omega_n}| * |g^{-\i\omega_n}| * \bigl(|\cdot|\, |g^{\i\omega_n}|\bigr) * \bigl(|\cdot|\, |g^{-\i\omega_n}|\bigr) \label{NtildeTB-NTB1_FT1_definition}
\end{align}
and
\begin{align}
F_T^{(2)} &:= \smash{\frac 2\beta \sum_{n\in \Zbb}} \, \bigl( |\cdot| \, |g^{\i\omega_n}|\bigr) *  |g^{-\i\omega_n}| * |g^{\i\omega_n}| * |g^{-\i\omega_n}| + |g^{\i\omega_n}| * \bigl(|\cdot|\, |g^{-\i\omega_n}|\bigr) * |g^{\i\omega_n}| * |g^{-\i\omega_n}| \notag\\
&\hspace{30pt}+ |g^{\i\omega_n}| * |g^{-\i\omega_n}| * \bigl(|\cdot|\, |g^{\i\omega_n}|\bigr) * |g^{-\i\omega_n}| + |g^{\i\omega_n}| * |g^{-\i\omega_n}| * |g^{\i\omega_n}| * \bigl(|\cdot|\, |g^{-\i\omega_n}|\bigr). \label{NtildeTB-NTB1_FT2_definition}
\end{align}
play a prominent role in the proof of Proposition~\ref{NtildeTBA-NTB1}. For any $T \geq T_0 > 0$, Lemma~\ref{g_decay} and \eqref{g0_decay_f_estimate1} show that their $L^1(\Rbb^3)$-norms are bounded by
\begin{align}
	\Vert F_T^{(1)} \Vert_1 + \Vert F_T^{(2)} \Vert_1 \leq C. \label{NtildeTB-NTB1_FT1-2_estimate}
\end{align}

\begin{proof}[Proof of Proposition \ref{NtildeTBA-NTB1}]
The estimate \eqref{NTA-NtildeTBA_2} implies the bound
\begin{align}
|\langle \Delta, \tilde N_{T, \Bbold, A}(\Delta) - N_{T, \Bbold}^{(1)}(\Delta)\rangle| &  \notag\\
&\hspace{-120pt}\leq C\; h^4 \, \Vert \Psi\Vert_{\Hmag^1(Q_h)}^4 \int_{\Rbb^3} \dd r  \iiint_{\Rbb^9} \dd \sbold \; |V\alpha_*(r)| \; |V\alpha_*(s_1)| \; |V\alpha_*(s_2)| \; |V\alpha_*(s_3)|  \notag \\
&\hspace{-100pt}\times \frac{2}{\beta} \sum_{n\in \Zbb} \iiint_{\Rbb^9} \dd \Zbold \; |\ell_T^n(\Zbold, r, \sbold)| \; \esssup_{X\in \Rbb^3} \bigl| \e^{\i\, \tilde \Phi_{A_h}(X, \Zbold, r, \sbold)} \, \e^{\i \frac \Bbold 2 \cdot \Phi(\Zbold, r,\sbold)} - 1\bigr| \label{NtildeTBA-NTB1_1}
\end{align}
with $\tilde \Phi_A$ in \eqref{NtildeTA_PhitildeA_definition} and $\Phi$ in \eqref{PhiB_definition}. We rewrite the phase function $\Phi$ in terms of the coordinates in \eqref{NTA_change_of_variables_1} and \eqref{NTA_change_of_variables_2}, which yields
\begin{align}
\Phi(\Zbold, r,\sbold) &=  (Z_1' - Z_2') \wedge (Z_2' - Z_3') + (Z_1' - Z_2') \wedge Z_3' + (Z_3'- Z_2') \wedge Z_3' \notag\\
& \hspace{50pt} + \frac r2 \wedge \bigl( Z_1' - (r - s_1 - s_2 - s_3)\bigr) + \bigl( s_2 + s_3 - \frac r2\bigr) \wedge (Z_1 ' - Z_2')\notag \\
&\hspace{50pt} + \bigl( s_3 - \frac r2\bigr) \wedge (Z_3' - Z_2') + \frac r2 \wedge Z_3'.\label{NtildeTBA-NTB1_2}
\end{align}
Furthermore, we bound
\begin{align*}
|\tilde \Phi_A(X, \Zbold, r, \sbold)| &\leq \Vert A\Vert_\infty \bigl( (Z_1' - (r - s_1 - s_2 - s_3)) + (Z_1' - Z_2') + (Z_2' - Z_3') + Z_3'\bigr),
\end{align*}
which follows from the definition \eqref{PhiA} of $\Phi_A$. Then, we combine this with \eqref{NtildeTBA-NTB1_1} and \eqref{NtildeTBA-NTB1_2}, repeat the argument leading to \eqref{MtildeTA-MTA_5}, and obtain
\begin{align*}
&\frac{2}{\beta} \sum_{n\in \Zbb} \iiint_{\Rbb^9} \dd \Zbold \; |\ell_{T,0}^n(\Zbold, r, \sbold)| \; \bigl| \e^{\i \tilde \Phi_{A_h}(X, \Zbold, r, \sbold)}\, \e^{\i \frac \Bbold 2 \cdot \Phi(\Zbold, r, \sbold)} - 1\bigr| \\
&\hspace{20pt} \leq Ch \; \bigl[ F_T^{(1)} (r - s_1 - s_2 - s_3) + F_T^{(2)}(r - s_1 - s_2 - s_3) \; \bigl(1 + |r| + |s_1| + |s_2| + |s_3|\bigr)\bigr],
\end{align*}
where the functions $F_T^{(1)}$ and $F_T^{(2)}$ are defined in \eqref{NtildeTB-NTB1_FT1_definition} and \eqref{NtildeTB-NTB1_FT2_definition}, respectively. We apply Young's inequality to this and conclude that
\begin{align*}
|\langle \Delta, \tilde N_{T, \Bbold, A}(\Delta) - N_{T, \Bbold}^{(1)}(\Delta)\rangle| &\\
&\hspace{-100pt}\leq C\; h^5 \, \Vert \Psi\Vert_{\Hmag^1(Q_h)}^4  \; \bigl( \Vert V\alpha_*\Vert_{\nicefrac 43}^4 + \Vert \, |\cdot| V\alpha_*\Vert_{\nicefrac 43}^4 \bigr) \bigl( \Vert F_T^{(1)}\Vert_1 + \Vert F_T^{(2)}\Vert_1 \bigr).
\end{align*}
Finally, the claim follows from \eqref{NtildeTB-NTB1_FT1-2_estimate}.
\end{proof}

\paragraph{The operator $N_T^{(2)}$.} We define the operator $N_{T}^{(2)}$ by
\begin{align}
N_T^{(2)}(\alpha) (X, r) &:= \iiint_{\Rbb^9} \dd \Zbold \iiint_{\Rbb^9} \dd \sbold \; \ell_{T} (\Zbold, r, \sbold) \, \Acal(X, 0 , \sbold), \label{NT2_definition}
\end{align}
where $\Acal$ is defined in \eqref{NTA_alpha_definition} and $\ell_{T}$ in \eqref{lT_definition}.

In our calculations, we may replace $\langle \Delta, N_{T, \Bbold}^{(1)}(\Delta) \rangle$ by $\langle \Delta, N_{T}^{(2)}(\Delta) \rangle$ due to the following error bound. Its proof can be found in \cite[Proposition 4.20]{DeHaSc2021}. In order to estimate the size of the error, the $\Hmag^2(Q_h)$-norm of $\Psi$ is needed once more.

\begin{prop}
\label{NTB1-NT2}
Assume that $|\cdot|^kV\alpha_* \in L^{\nicefrac 43}(\Rbb^3)$ for $k\in \{0,2\}$. For any $T \geq T_0 >0$, any $h>0$, any $\Psi \in \Hmag^2(Q_h)$, and $\Delta\equiv \Delta_\Psi$ as in \eqref{Delta_definition}, we have
\begin{align*}
|\langle \Delta, N_{T, \Bbold}^{(1)}(\Delta) - N_{T}^{(2)}(\Delta) \rangle|  &\leq C \; h^6 \; \bigl( \Vert V\alpha_*\Vert_{\nicefrac 43}^4 + \Vert \, |\cdot|^2 V\alpha_*\Vert_{\nicefrac 43}^4\bigr)  \\
&\hspace{130pt}  \times  \Vert \Psi\Vert_{\Hmag^1(Q_h)}^3 \; \Vert \Psi\Vert_{\Hmag^2(Q_h)}.
\end{align*}
\end{prop}

\subsubsection{Calculation of the quartic term in the Ginzburg--Landau functional}
\label{sec:compquarticterm}

The quartic term in the Ginzburg--Landau functional in \eqref{Definition_GL-functional} is contained in $\langle \Delta, N_T^{(2)}(\Delta) \rangle$. It is extracted by the following proposition whose proof can be found in \cite[Proposition 4.21]{DeHaSc2021}.

\begin{prop}
\label{NTc2}
Assume $V\alpha_* \in L^{\nicefrac 43}(\Rbb^3)$. For any $h>0$, any $\Psi\in \Hmag^1(Q_h)$, and $\Delta \equiv \Delta_\Psi$ as in \eqref{Delta_definition}, we have
\begin{align*}
\langle \Delta, N_{\Tc}^{(2)}(\Delta)\rangle = 8\; \Lambda_3 \; \Vert \Psi\Vert_4^4
\end{align*}
with $\Lambda_3$ in \eqref{GL_coefficient_3}. Moreover, for any $T \geq T_0 > 0$, we have
\begin{align*}
|\langle \Delta,  N_T^{(2)}(\Delta) - N_{\Tc}^{(2)}(\Delta)\rangle| &\leq C\; h^4  \; |T - \Tc| \; \Vert V\alpha_*\Vert_{\nicefrac 43}^4 \; \Vert \Psi\Vert_{\Hmag^1(Q_h)}^4.
\end{align*}
\end{prop}

\subsubsection{Summary: The quartic terms and proof of Theorem~\ref{Calculation_of_the_GL-energy}}
\label{sec:quarticterms}

We work under the assumptions of Theorem~\ref{Calculation_of_the_GL-energy}. In view of the decomposition \eqref{NTAW_decomposition} of $N_{T, \Abold, W}$, the results of Lemmas~\ref{RTAW3_estimate} and \ref{NTA_action}, and those of Propositions~\ref{NTA-NtildeTA}, \ref{NtildeTBA-NTB1}, \ref{NTB1-NT2}, and \ref{NTc2}, we have
\begin{align}
\frac{1}{8} \langle \Delta, N_{T, \Abold, W}(\Delta) \rangle = \; \Lambda_3 \; \Vert \Psi\Vert_4^4 + R(h) \label{eq:A28}
\end{align}
where the remainder $R(h)$ is bounded by
\begin{align*}
| R(h) | \leq C \; \Vert \Psi \Vert_{\Hmag^1(Q_h)}^3 \, \bigl( h^5 \; \Vert \Psi \Vert_{\Hmag^1(Q_h)} + h^6 \; \Vert \Psi \Vert_{\Hmag^2(Q_h)}\bigr).
\end{align*}
Theorem~\ref{Calculation_of_the_GL-energy} follows from this and \eqref{eq:A15}. This finishes the proof. 


\subsection{Proof of Lemma \ref{Gamma_Delta_admissible} and Proposition \ref{Structure_of_alphaDelta}}
\label{sec:proofofadmissibility}

We begin this section by proving Lemma \ref{Gamma_Delta_admissible}. First of all, we recall the definition of $\Gamma_\Delta$ in \eqref{GammaDelta_definition}, which is a gauge-periodic generalized fermionic one-particle density matrix per definition. Consequently, only the trace class condition \eqref{Gamma_admissible} needs to be verified.

In terms of the expansion
\begin{align}
\tanh\bigl( \frac \beta 2 H_\Delta \bigr) &= -\frac 2\beta \sum_{n\in \Zbb} \frac{1}{\i \omega_n - H_{\Delta}}, \label{tanh-expansion}
\end{align}
which follows from \eqref{tanh_Matsubara}, the expression
\begin{align}
\Gamma_\Delta &= \frac 12 - \frac 12 \tanh\bigl( \frac \beta 2 H_\Delta\bigr), \label{GammaDelta_tanh_relation}
\end{align}
and the resolvent equation \eqref{Resolvent_Equation}, $\Gamma_\Delta$ can be rewritten as
\begin{align}
\Gamma_\Delta = \frac 12 - \frac 12 \tanh\bigl( \frac \beta 2 H_\Delta\bigr) = \frac 12 + \frac{1}{\beta} \sum_{n\in \Zbb} \frac{1}{\i \omega_n - H_\Delta} = \Gamma_0 + \Ocal + \Qcal_{T,\Abold, W}(\Delta), \label{alphaDelta_decomposition_1}
\end{align}
where $\Gamma_0$ is the normal state in \eqref{Gamma0} and, in terms of $\delta$ in \eqref{HDelta_definition},
\begin{align}
\Ocal &:= \frac 1\beta \sum_{n\in \Zbb} \frac{1}{ \i \omega_n - H_0} \delta \frac{1}{ \i \omega_n - H_0}, & 
\Qcal_{T,\Abold, W}(\Delta) &:= \frac 1\beta \sum_{n\in \Zbb} \frac{1}{ \i \omega_n - H_0} \delta\frac{1}{ \i \omega_n - H_0} \delta \frac{1}{ \i \omega_n - H_\Delta}. \label{alphaDelta_decomposition_2}
\end{align}
Since $\Ocal$ is offdiagonal, the entry $[\Ocal]_{11}$ vanishes, whence the operator $(1+ \pi^2) [\Ocal]_{11}$ is locally trace class trivially. To see that $(1 + \pi^2) [\Qcal_{T, \Abold, W}(\Delta)]_{11}$ is locally trace class, we use
\begin{align*}
\frac{1}{\i \omega_n \pm H_0}\,  \delta\,  \frac{1}{\i \omega_n \pm H_0} \, \delta 
&= \begin{pmatrix}
\frac{1}{\i \omega_n \pm \hfrak_{\Abold, W}} \, \Delta \frac{1}{\i \omega_n \mp \ov{\hfrak_{\Abold, W}}} \, \ov \Delta \\ & \frac{1}{\i \omega_n \mp \ov{\hfrak_{\Abold, W}}}\,  \ov \Delta \frac{1}{\i \omega_n \pm \hfrak_{\Abold, W}} \, \Delta
\end{pmatrix} 
\end{align*}
and obtain
\begin{align*}
\bigl[ \Qcal_{T, \Abold, W}(\Delta)\bigr]_{11} = \frac 1\beta \sum_{n\in \Zbb} \frac{1}{\i\omega_n - \hfrak_{\Abold, W}} \, \Delta \, \frac{1}{\i\omega_n + \ov{\hfrak_{\Abold, W}}} \, \ov \Delta \, \Bigl[ \frac{1}{\i\omega_n - H_\Delta}\Bigr]_{11}.
\end{align*}
By Hölder's inequality in \eqref{Schatten-Hoelder}, the terms in the sum of $(1+\pi^2)[\Qcal_{T,\Abold, W}(\Delta)]_{11}$ are bounded in local trace norm by
\begin{align*}
\Bigl\Vert (1 + \pi^2) \frac{1}{\i \omega_n - \hfrak_{\Abold, W}}\Bigr\Vert_\infty  \; \frac{1}{|\omega_n|^2} \; \Vert \Delta\Vert_2^2.
\end{align*}
To see that the first factor is bounded, we combine
\begin{align}
\pi^2 = (\pi_\Abold - A)^2 = (\pi_\Abold^2 + W) + \pi_\Abold \cdot A + A\cdot \pi_\Abold + |A|^2 - W \label{alphaDelta_decomposition_3}
\end{align}
with the inequality
\begin{align}
\pi_\Abold \cdot A + A\cdot \pi_\Abold \leq \frac 12 \bigl( (\pi_\Abold^2 + W) + |A|^2 - W\bigr), \label{alphaDelta_decomposition_4}
\end{align}
which proves that $(1 + \pi^2) (\i \omega_n - \hfrak_{\Abold, W})^{-1}$ is uniformly bounded in $n$. We conclude that $(1 + \pi^2) [\Qcal_{T, \Abold, W}(\Delta)]_{11}$ is locally trace class.

The following result completes the proof of Lemma \ref{Gamma_Delta_admissible}.

\begin{lem}
$(1 + \pi^2) \gamma_0$ is locally trace class for $h>0$ small enough.
\end{lem}

\begin{proof}
By the diamagnetic inequality \cite[Eq. (4.4.3)]{LiebSeiringer}, the operator $1+ \pi_{\Abold_h}^2 +W_h$ is nonnegative for $h$ small enough. We first show that the operator $(1 + \pi_{\Abold_h}^2+W_h) \gamma_0$ is locally trace class. Since this is a nonnegative function of $\pi_{\Abold_h}^2+W_h$, we apply the bound $(1+x)(\exp(\beta (x - \mu))+1)^{-1} \leq C_{\beta} \e^{-\frac \beta 2 (x-\mu)}$ for $x \geq 0$. Therefore, it suffices to show that the local trace of $\e^{-\frac \beta 2(\pi_{\Abold_h}^2 + W_h-\mu)}$ is finite. 
Since $\pi_{\Abold_h}^2 + W_h \geq \pi_{\Abold_h}^2 - C$, the corresponding inequality holds for the eigenvalues of the respective operators so that
\begin{align*}
\Tr \bigl[ \e^{-\frac \beta 2 (\pi_{\Abold_h}^2 + W_h - \mu)}\bigr] \leq C_\beta \Tr \bigl[ \e^{-\frac \beta 2 (\pi_{\Abold_h}^2 - \mu)}\bigr].
\end{align*}
Therefore, the diamagnetic inequality for the magnetic heat kernel, see e.g. \cite[Theorem 4.4]{LiebSeiringer}, and the explicit formula for the heat kernel of the Laplacian prove that $\e^{-\frac \beta 2(\pi_{\Abold_h}^2 + W_h-\mu)}$ is locally trace class. 
In particular, $\gamma_0$ and, since $W$ is a bounded operator, $\pi_{\Abold_h}^2 \gamma_0$ are locally trace class.

To prove the claim of the lemma, we employ \eqref{alphaDelta_decomposition_3}. Since $\pi_\Abold \cdot A = -\i \divv A + A\cdot \pi_\Abold$ and since $\divv A$, $|A|^2$, and $W$ are bounded operators, it remains to show that $(A_h\cdot \pi_{\Abold_h}) \gamma_0$ is locally trace class. To see this, we write
\begin{align*}
(A_h\cdot \pi_{\Abold_h}) \gamma_0 = (A_h\cdot \pi_{\Abold_h}) \, \frac{1}{1 + \pi_{\Abold_h}^2} \; (1 + \pi_{\Abold_h}^2) \gamma_0
\end{align*}
and use \eqref{ZPiX_inequality} to conclude that
\begin{align*}
\Bigl\Vert (A\cdot \pi_\Abold) \; \frac{1}{1 + \pi_\Abold^2} \, f \Bigr\Vert_2^2 = \Bigl\langle f, \frac{1}{1 + \pi_\Abold^2} \, (A\cdot \pi_\Abold)^2 \, \frac{1}{1 + \pi_\Abold^2} f \Bigr\rangle \leq C \, \Vert A\Vert_\infty^2 \, \Bigl\Vert \frac{|\pi_\Abold|}{1 + \pi_\Abold^2} \Bigr\Vert_\infty^2 \, \Vert f\Vert_2^2.
\end{align*}
This finishes the proof.
\end{proof}


%

We now proceed with proving Proposition~\ref{Structure_of_alphaDelta}. To his end, we combine the fact that $\alpha_\Delta = [\Gamma_\Delta]_{12}$, the resolvent equation \eqref{Resolvent_Equation}, and \eqref{alphaDelta_decomposition_1}, which yields
\begin{align*}
\alpha_\Delta &= [\Ocal]_{12} + [\Qcal_{T, \Abold, W}(\Delta)]_{12} = [\Ocal]_{12} + \Rcal_{T, \Abold, W}^{(4)}(\Delta).
\end{align*}
Here, $\Ocal$ is the operator in \eqref{alphaDelta_decomposition_2} and
\begin{align}
\Rcal_{T,\Abold, W}^{(4)}(\Delta) &:=  \frac 1\beta \sum_{n\in \Zbb} \Bigl[ \frac{1}{ \i \omega_n - H_0} \delta\frac{1}{ \i \omega_n - H_0} \delta\frac{1}{ \i \omega_n - H_\Delta} \delta \frac{1}{ \i \omega_n - H_0}\Bigr]_{12}. \label{RTAW4_definition}
\end{align}
We note that $[\Ocal]_{12} = -\frac 12 L_{T, \Abold, W}\Delta$ with $L_{T, \Abold, W}$ in \eqref{LTAW_definition}. In view of the decomposition \eqref{LTAW_decomposition} of $L_{T, \Abold, W}$, we define
\begin{align}
\eta_0(\Delta) &:= \frac 12 \bigl(L_{T, \Abold}\Delta - M_{T, \Abold}\Delta\bigr) + \frac 12 L_{T, \Abold}^W \Delta + \frac 12 \bigl( M_T^{(1)}\Delta - M_{\Tc}^{(1)}\Delta\bigr) \notag \\
&\hspace{60pt} + \frac 12 \bigl( M_{T, \Abold}\Delta - M_{T, \Abold_{e_3}}\Delta\bigr) + \Rcal_{T, \Abold, W}^{(2)}\Delta + \Rcal_{T, \Abold, W}^{(4)}(\Delta), \notag \\
\eta_\perp(\Delta) &:= \frac 12 \bigl( M_{T, \Abold_{e_3}}\Delta - M_{T}^{(1)}\Delta\bigr), \label{eta_perp_definition}
\end{align}
with $M_{T, \Abold}$ in \eqref{MTA_definition}, $L_{T, \Abold}^W$ in \eqref{LTA^W_definition}, and $M_T^{(1)}$ in \eqref{MT1_definition}. The remainders $\Rcal_{T, \Abold, W}^{(2)}$ and $\Rcal_{T, \Abold, W}^{(4)}$ are defined in \eqref{RTAW2_definition} and \eqref{RTAW4_definition}, respectively. Proposition~\ref{MT1} implies that $-\frac 12 M_{\Tc}^{(1)} \Delta = \Psi\alpha_*$, so these definitions allow us to write $\alpha_{\Delta}$ as in \eqref{alphaDelta_decomposition_eq1}. The new operator $M_{T, \Abold_{e_3}}$ is $M_{T, \Abold}$ in \eqref{MTA_definition} with the constant magnetic field potential only, which needs to be isolated to maintain the orthogonality property \eqref{alphaDelta_decomposition_eq4}, compare this to \cite[Eq. (3.9)]{DeHaSc2021}. 

The rest of the proof is devoted to the properties, which we claim $\eta_0$ and $\eta_\perp$ to have in Proposition~\ref{Structure_of_alphaDelta}. First, we prove \eqref{alphaDelta_decomposition_eq2}. To this end, we use the definition \eqref{RTAW4_definition} to write
\begin{align*}
\Rcal_{T, \Abold, W}^{(4)}(\Delta) &= \frac 1\beta \sum_{n\in\Zbb} \frac 1{\i\omega_n - \hfrak_{\Abold, W}} \, \Delta \,  \frac 1{\i\omega_n + \ov{\hfrak_{\Abold, W}}}\,  \ov \Delta\,  \Bigl[ \frac{1}{\i \omega_n - H_\Delta}\Bigr]_{11}\,  \Delta \, \frac 1{\i\omega_n + \ov{\hfrak_{\Abold, W}}}.
\end{align*}
From this and Hölder's inequality, the estimate $\Vert \Rcal_{T, \Abold, W}^{(4)}(\Delta)\Vert_2 \leq C \beta^3 \Vert \Delta\Vert_6^3$ is immediate. Likewise, we have the bound
\begin{equation*}
	\Vert \pi \Rcal_{T, \Abold, W}^{(4)}(\Delta) \Vert_2 \leq \frac 1\beta \sum_{n\in\Zbb} \Bigl\Vert \pi \frac 1{\i\omega_n - \hfrak_{\Abold, W}} \Bigr\Vert_{\infty} \Bigl\Vert \frac 1{\i\omega_n + \ov{\hfrak_{\Abold, W}}} \Bigr\Vert_{\infty}^2 \Bigl\Vert \Bigl[ \frac{1}{\i \omega_n - H_\Delta}\Bigr]_{11} \Bigr\Vert_{\infty} \Vert \Delta \Vert_6^3.
\end{equation*}
For a general operator $A$, we have $\Vert A\Vert_\infty^2 = \Vert A^*A\Vert_\infty$, which enables us to bound the first factor in the sum by
\begin{align*}
\Bigl\Vert \pi \frac 1{\i\omega_n - \hfrak_{\Abold, W}} \Bigr\Vert_{\infty} &\leq \Vert A_h\Vert_\infty \Bigl\Vert \frac{1}{\i \omega_n - \hfrak_{\Abold, W}}\Bigr\Vert_\infty \\
&\hspace{50pt} +  \Bigl\Vert \frac 1{\i\omega_n + \hfrak_{\Abold, W}}\Bigr\Vert_\infty^{\nicefrac 12} \Bigl[ \Bigl\Vert  \frac {\pi_{\Abold_h}^2 + W_h}{\i\omega_n - \hfrak_{\Abold, W}} \Bigr\Vert_{\infty} + \Vert W_h\Vert_\infty\Bigr]^{\nicefrac 12} \\
&\leq C \, \bigl( |\omega_n|^{-\nicefrac 12} + |\omega_n|^{-1}\bigr).
\end{align*}
It follows that
\begin{align}
\Vert \pi \Rcal_{T, \Abold, W}^{(4)}(\Delta) \Vert_2 \leq C\, ( \beta^{\nicefrac 52} + \beta^3) \,\Vert \Delta \Vert_6^3. \label{eq:A25}
\end{align}
In a similar manner, the right side of \eqref{eq:A25} bounds $\Vert \Rcal_{T, \Abold, W}^{(4)}(\Delta)\pi \Vert_2$ as well. By the reformulation \eqref{Norm_equivalence_2} of the $\Hsymm$-norm, we may therefore apply Lemma~\ref{Schatten_estimate} and \eqref{Magnetic_Sobolev} on the right side of \eqref{eq:A25} to conclude that
\begin{align*}
\Vert \Rcal_{T, \Abold, W}^{(4)}(\Delta) \Vert_{\Hsymm}^2 \leq C \; h^6 \; \Vert \Psi \Vert_{\Hmag^1(Q_h)}^6. 
\end{align*}

The next result shows that the newly obtained error term $M_{T, \Abold}\Delta - M_{T, \Abold_{e_3}}\Delta$ satisfies an appropriate bound.

\begin{prop}
\label{MTA-MTAe3}
For any $T_0>0$ and $A\in \Wpervec 1$ there is $h_0>0$ such that for any $0 < h \leq h_0$, any $T\geq T_0$, and whenever $|\cdot|^k V\alpha_*\in L^2(\Rbb^3)$ for $k \in \{0,1\}$, $\Psi\in \Hmag^1(Q_h)$, and $\Delta \equiv \Delta_\Psi$ as in \eqref{Delta_definition}, we have
\begin{align}
\Vert M_{T, \Abold} \Delta - M_{T, \Abold_{e_3}} \Delta\Vert_\Hsymm^2 &\leq C \, h^5 \, \bigl( \Vert V\alpha_*\Vert_2^2 + \Vert \, |\cdot|V\alpha_*\Vert_2^2 \bigr) \, \Vert \Psi\Vert_{\Hmag^1(Q_h)}^2. \label{MTA-MTAe3_eq1}
\end{align}
\end{prop}

\begin{proof}
Define the operator
\begin{align*}
\Qcal_{T, \Bbold, A}\alpha(X, r) := \iint_{\Rbb^3\times \Rbb^3} \dd Z \dd s \; k_T(Z, r-s) \; \e^{\i A_h(X) \cdot Z} \; (\e^{\i Z\cdot \Pi} \alpha)(X,s),
\end{align*}
where $k_T(Z,r) := k_{T, 0, 0}(0,Z, r, 0)$  with $k_{T, 0, 0}$ in \eqref{kTBA_definition}, and decompose 
\begin{align}
M_{T, \Abold} \Delta - M_{T, \Abold_{e_3}} \Delta = \bigl( M_{T, \Abold} \Delta - \Qcal_{T, \Bbold, A} \Delta \bigr) + \bigl( Q_{T, \Bbold, A} \Delta - M_{T, \Abold_{e_3}} \Delta\bigr). \label{MTA-MTAe3_1}
\end{align}
We are going to show that both terms in brackets of the right side are bounded in $\Hsymm$-norm by the right side of \eqref{MTA-MTAe3_eq1}. Since the integrand of $M_{T, \Abold}$ is symmetric with respect to $Z$, by Lemma \ref{Magnetic_Translation_Representation}, we have
\begin{align*}
(M_{T, \Abold} \alpha - \Qcal_{T, \Bbold, A} \alpha)(X, r) &\\
&\hspace{-100pt} = \iint_{\Rbb^3\times \Rbb^3} \dd Z \dd s\; k_T(Z, r-s) \, \bigl[ \e^{\i \Phi_{A_h}(X, X+Z)} - \e^{\i A_h(X) \cdot Z}\bigr] \, (\e^{\i Z\cdot \Pi} \alpha)(X, s).
\end{align*}
By \eqref{MtildeTA-MTA_3}, we have
\begin{align*}
\Phi_A(X, X+ Z) - A(X)\cdot Z = \int_0^1 \dd t \; \bigl[ A(X + tZ) - A(X)\bigr] \cdot Z
\end{align*}
so that
\begin{align}
\bigl| \e^{\i \Phi_A(X, X+Z)} - \e^{\i A(X)\cdot Z} \bigr| \leq \Vert DA\Vert_\infty \; |Z|^2. \label{MTA-MTAe3_2}
\end{align}
It follows that
\begin{align*}
\Vert M_{T, \Abold}\Delta - \Qcal_{T, \Bbold, A}\Delta\Vert_2^2 &\leq C\, \Vert \Psi\Vert_2^2 \, \Vert DA_h\Vert_\infty^2 \, \Vert F_T^{(2)} \Vert_1 \, \Vert V\alpha_*\Vert_2^2,
\end{align*}
where $F_T^{(2)}$ is defined in \eqref{LtildeTA-MtildeTA_FT_definition}. Since its $L^1$-norm is uniformly bounded, see \eqref{LtildeTA-MtildeTA_FTGT}, we conclude the claimed estimate for this term. In a similar manner, we bound
\begin{align*}
\Vert \Pi (M_{T, \Abold}\Delta - \Qcal_{T, \Bbold, A}\Delta )\Vert_2^2 &\leq C \, \int_{\Rbb^3} \dd r \; \Bigl| \iint_{\Rbb^3\times \Rbb^3} \dd Z \dd s\; |k_T(Z, r-s)| \, |V\alpha_*(s)| \\
&\hspace{20pt} \times \Bigl[ \esssup_{X\in \Rbb^3} \bigl| \nabla_X \e^{\i \Phi_{A_h}(X, X + Z)} - \nabla_X\e^{\i A_h(X)\cdot Z} \bigr| \; \Vert \Psi\Vert_2 \\
&\hspace{40pt}+ \esssup_{X\in \Rbb^3} \bigl| \e^{\i \Phi_{A_h}(X, X + Z)} - \e^{\i A_h(X)\cdot Z}\bigr| \; \Vert \Pi \e^{\i Z\cdot \Pi} \Psi\Vert_2 \Bigr] \Bigr|^2.
\end{align*}
Since
\begin{align*}
\bigl| \nabla_X \e^{\i \Phi_A(X, X + Z)} - \nabla_X\e^{\i A(X)\cdot Z} \bigr| &\leq \bigl| \nabla_X \Phi_A(X, X + Z) - \nabla_X A(X)\cdot Z \bigr| \\
&\hspace{20pt} + \bigl|\Phi_A(X, X+Z) - A(X)\cdot Z\bigr| \, |\nabla_XA(X)\cdot Z| \\
&\leq \bigl[ \Vert D^2A\Vert_\infty + \Vert DA\Vert_\infty^2 \bigr] \bigl[ |Z|^2 + |Z|^3\bigr],
\end{align*}
and since $\Vert D^2A_h\Vert_\infty \leq Ch^3$ and $\Vert DA_h\Vert_\infty^2 \leq Ch^4$, we use \eqref{PiXcosPiA_estimate} and infer
\begin{align*}
\Vert \Pi (M_{T, \Abold} \Delta - \Qcal_{T, \Bbold, A} \Delta)\Vert_2^2 &\leq C \, h^5 \, \Vert \Psi\Vert_{\Hmag^1(Q_h)}^2 \, \Vert V\alpha_*\Vert_2^2 \, \Vert F_T^{(2)} + F_T^{(3)} \Vert_1^2,
\end{align*}
which together with \eqref{LtildeTA-MtildeTA_FTGT} proves the claim for this term. Finally, we have
\begin{align*}
\Vert \tilde \pi (M_{T, \Abold} \Delta - \Qcal_{T, \Bbold, A}\Delta) \Vert_2^2 &\leq C \, \Vert \Psi\Vert_2^2 \\
&\hspace{-100pt} \times \int_{\Rbb^3} \dd r \, \Bigl| \iint_{\Rbb^3\times \Rbb^3} \dd Z\dd s\; |\tilde \pi k_T(Z, r-s)| \, |V\alpha_*(s)| \, \esssup_{X\in \Rbb^3} \bigl| \e^{\i \Phi_{A_h}(X, X + Z)} - \e^{\i A_h(X)\cdot Z}\bigr|\Bigr|^2.
\end{align*}
We employ $\frac 14 |\Bbold \wedge r| \leq |r-s| + |s|$ and the estimate \eqref{Z_estimate} to see that
\begin{align}
\int_{\Rbb^3} \dd Z \; |\tilde \pi k_T(Z, r-s)| \, |Z|^2 \leq G_T^2(r-s) + F_T^3(r-s) + F_T^2(r-s) \, |s| \label{MTA-MTAe3_3}
\end{align}
with $F_T^a$ and $G_T^a$ in \eqref{LtildeTA-MtildeTA_FT_definition} and \eqref{LtildeTA-MtildeTA_GT_definition}, respectively. The combination with \eqref{MTA-MTAe3_2} proves the claim for this term and concludes the proof for the first term on the right side of \eqref{MTA-MTAe3_1}.

We move on to the second term on the right side of \eqref{MTA-MTAe3_1} and start by the identity
\begin{align*}
(\Qcal_{T, \Bbold, A}\alpha - M_{T, \Abold_{e_3}}\alpha)(X, r) &\\
&\hspace{-60pt} = -2 \iint_{\Rbb^3\times \Rbb^3} \dd Z \dd s \; k_T(Z, r-s) \, \sin^2\bigl(\frac 12\,  A_h(X)\cdot Z \bigr) (\cos(Z\cdot \Pi) \alpha)(X, s) \\
&\hspace{-30pt}- \int_{\Rbb^3\times \Rbb^3} \dd Z \dd s \; k_T(Z, r-s) \, \sin(A_h(X) \cdot Z) \, (\sin(Z\cdot \Pi) \alpha)(X, s),
\end{align*}
which follows from a straightforward expansion of the complex exponential and the identity $\cos(x) - 1 = -2\sin^2(\frac x2)$. From this, it is easy to see that
\begin{align*}
\Vert \Qcal_{T, \Bbold, A}\Delta - M_{T, \Abold_{e_3}}\Delta\Vert_2^2 &\leq C\, \bigl[ \Vert\Psi\Vert_2^2 \, \Vert A_h\Vert_\infty^4 + \Vert \Pi\Psi\Vert_2^2 \Vert A_h\Vert_\infty^2 \bigr] \, \Vert F_T^{(2)} \Vert_1^2 \, \Vert V\alpha_*\Vert_2^2,
\end{align*}
which proves the claim for the first term. Secondly,
\begin{align}
\Vert \Pi ( \Qcal_{T, \Bbold, A}\Delta - M_{T, \Abold_{e_3}}\Delta)\Vert_2^2 &\leq C \, \int_{\Rbb^3} \dd r \, \Bigl| \iint_{\Rbb^3\times \Rbb^3} \dd Z\dd s \; |k_T(Z, r-s)| \, |V\alpha_*(s)| \notag \\%
& \hspace{-120pt} \times \esssup_{X\in \Rbb^3} \Bigl[ \bigl|\nabla_X \sin^2 \bigl( \frac 12\, A_h(X)\cdot Z\bigr)\bigr| \, \Vert \cos(Z\cdot \Pi)\Psi\Vert_2 \notag \\
&\hspace{50pt} + \bigl|\sin^2\bigl( \frac 12 \, A_h(X)\cdot Z\bigr)\bigr| \, \Vert \Pi \cos(Z\cdot\Pi)\Psi\Vert_2 \notag \\
&\hspace{-100pt}   + |\nabla_X \sin(A_h(X)\cdot Z)| \, \Vert \sin(Z\cdot \Pi) \Psi\Vert_2 + |\sin(A_h(X)\cdot Z)| \, \Vert \Pi\sin(Z\cdot \Pi) \Psi\Vert_2\Bigr]\Bigr|^2 \label{alphaDelta_decomposition_5}
\end{align}
We have
\begin{align*}
\Vert \sin(Z\cdot \Pi)\Psi\Vert_2 \leq C \, h^2 \, \Vert \Psi\Vert_{\Hmag^1(Q_h)} \, |Z|
\end{align*}
Furthermore, from a direct computation or from \cite[Lemma 5.12]{DeHaSc2021}, we know that
\begin{align*}
\Pi \, \sin(Z\cdot \Pi) = \sin(Z\cdot \Pi) \;\Pi + 2\i \, \cos (Z\cdot \Pi) \; \Bbold \wedge Z,
\end{align*}
which implies
\begin{align*}
\Vert \Pi \sin(Z\cdot \Pi)\Psi\Vert_2 &\leq C \, h^2 \Vert \Psi\Vert_{\Hmag^1(Q_h)} \, (1 + |Z|)
\end{align*}
In a similar way, we have the estimate \eqref{MtildeTA-MTA_Lemma} for $\Pi\cos(Z\cdot \Pi)\Psi$. From these estimates and \eqref{alphaDelta_decomposition_5}, the claimed bound for $\Vert \Pi ( \Qcal_{T, \Bbold, A}\Delta - M_{T, \Abold_{e_3}}\Delta)\Vert_2^2$ is easily shown by using the functions $F_T^a$ in \eqref{LtildeTA-MtildeTA_FT_definition}. Finally, a straightforward computation shows that
\begin{align*}
\Vert \tilde \pi(\Qcal_{T, \Bbold, A}\Delta - M_{T, \Abold_{e_3}}\Delta)\Vert_2^2 &\leq C\bigl[ \Vert A_h\Vert_\infty^4 \Vert \Psi\Vert_2^2 + \Vert A_h\Vert_\infty^2 \Vert \Pi\Psi\Vert_2^2 \bigr] \\
&\hspace{20pt} \times \int_{\Rbb^3} \dd r \, \Bigl| \iint_{\Rbb^3\times \Rbb^3} \dd Z \dd s \; |\tilde \pi k_T(Z, r-s)| \, |V\alpha_*(s)| \, |Z|^2\Bigr|^2,
\end{align*}
whence the claimed bound follows from \eqref{MTA-MTAe3_3}. This finishes the proof.
\end{proof}

In order to conclude the proof of \eqref{alphaDelta_decomposition_eq2}, we note that Lemma \ref{RTAW2_estimate} as well as Propositions~\ref{LTA-LtildeTA}, \ref{MtildeTA-MTA}, and \ref{MT1} provide the appropriate bounds on $\eta_0(\Delta)$. Parts (b) and (c) of Proposition \ref{Structure_of_alphaDelta} are proven in \cite[Proposition 3.2 (b)-(c)]{DeHaSc2021}. This ends the proof of Proposition \ref{Structure_of_alphaDelta}.

\subsection{Proof of Proposition \ref{Lower_Tc_a_priori_bound}}
\label{Lower_Tc_a_priori_bound_proof_Section}

Under the assumptions of Proposition~\ref{Lower_Tc_a_priori_bound}, we prove the existence of constants $D_0>0$ and $h_0>0$ such that for $0 < h \leq h_0$ the following holds. Within the temperature regime
\begin{align*}
	0 < T_0 \leq T < \Tc (1 - D_0 h^2)
\end{align*}
there is a function $\Psi \in \Hmag^2(Q_h)$, such that the gap function $\Delta(X,r) = -2 V\alpha_*(r) \Psi(X)$ constitutes a Gibbs state $\Gamma_\Delta$ as in \eqref{GammaDelta_definition}, which satisfies \eqref{Lower_critical_shift_2}.

This is proven by choosing an arbitrary $\psi \in \Hmag^2(Q_1)$ with $\Vert \psi\Vert_{\Hmag^2(Q_h)}=1$, whence $\Psi$ defined as in \eqref{GL-rescaling} belongs to $\Hmag^2(Q_h)$ and satisfies $\Vert \Psi\Vert_{\Hmag^2(Q_h)}=1$. Applying Propositions~\ref{Structure_of_alphaDelta}, \ref{BCS functional_identity}, \ref{Rough_bound_on_BCS energy}, as well as \eqref{Magnetic_Sobolev} and \eqref{eq:A28} yields
\begin{align*}
	\FBCS(\Gamma_\Delta) - \FBCS(\Gamma_0) &< h^2 \, \bigl( - cD_0 \, \Vert \psi\Vert_2^2 + C \bigr)
\end{align*}
as long as $h$ is small enough.
The proof of Proposition~\ref{Lower_Tc_a_priori_bound} is completed by the choice $D_0 = \frac{C}{c \Vert \psi\Vert_2^2}$.

\section{The Structure of Low-Energy States}
\label{Lower Bound Part A}

In this section we prove a priori bounds for low-energy states of the BCS functional in the sense of \eqref{Second_Decomposition_Gamma_Assumption} below. The goal is to show that their Cooper pair wave function has a structure similar to that of the trial state we use in the proof of the upper bound in Section~\ref{Upper_Bound}. These bounds and the trial state analysis in Section~\ref{Upper_Bound} are the main technical ingredients for the proof of the lower bound in Section~\ref{Lower Bound Part B}. To prove the a priori bounds, we show that $W_h$ and $A_h$ can be treated as a perturbation, which reduces the problem to proving the same bounds as for the case of a constant magnetic field. A solution to this problem has been obtained in \cite[Theorem~5.1]{DeHaSc2021} and we use it here.

%
%
We recall the definition of the generalized one-particle density matrix $\Gamma$ in \eqref{Gamma_introduction}, its Cooper pair wave function $\alpha = \Gamma_{12}$, as well as the normal state $\Gamma_0$ in \eqref{Gamma0}.

\begin{thm}[Structure of low-energy states]
\label{Structure_of_almost_minimizers}
Let Assumptions \ref{Assumption_V} and \ref{Assumption_KTc} hold. For given $D_0, D_1 \geq 0$, there is a constant $h_0>0$ such that for all $0 <h \leq h_0$ the following holds: If $T>0$ obeys $T - \Tc \geq -D_0h^2$ and if $\Gamma$ is a gauge-periodic state with low energy, that is,
\begin{align}
\FBCS(\Gamma) - \FBCS(\Gamma_0) \leq D_1h^4, \label{Second_Decomposition_Gamma_Assumption}
\end{align}
then there are $\Psi\in \Hmag^1(Q_h)$ and $\xi\in \Hsymm$ such that
\begin{align}
\alpha(X,r) = \alpha_*(r) \Psi(X) + \xi(X,r), \label{Second_Decomposition_alpha_equation}
\end{align}
where
\begin{align}
\sup_{0< h\leq h_0} \Vert \Psi\Vert_{\Hmag^1(Q_h)}^2 &\leq C, &  \Vert \xi\Vert_\Hsymm^2 &\leq Ch^4 \bigl( \Vert \Psi\Vert_{\Hmag^1(Q_h)}^2 + D_1\bigr). \label{Second_Decomposition_Psi_xi_estimate}
\end{align}
\end{thm}

\begin{varbems}
\begin{enumerate}[(a)]
\item Equation~\eqref{Second_Decomposition_Psi_xi_estimate} shows that $\Psi$ is a macroscopic quantity in the sense that its $\Hmag^1(Q_h)$-norm scales as that of the function in \eqref{GL-rescaling}. It is important to note that this norm is scaled with $h$, see \eqref{Periodic_Sobolev_Norm}. The unscaled $\Lmag^2(Q_h)$-norm of $\Psi$ is of the order $h$, and therefore much larger than that of $\xi$, see \eqref{Second_Decomposition_Psi_xi_estimate}.



\item Theorem~\ref{Structure_of_almost_minimizers} has been proven in \cite[Theorem 5.1]{DeHaSc2021} for the case of a constant external magnetic field, where $A_h =0$ and $W_h = 0$. Our proof of Theorem~\ref{Structure_of_almost_minimizers} for general external fields uses the main part of this proof.
\end{enumerate}
\end{varbems}

Although Theorem~\ref{Structure_of_almost_minimizers} contains the natural a priori bounds for low-energy states, we need a slightly different version of it in our proof of the lower bound for the BCS free energy in Section~\ref{Lower Bound Part B}. The main reason is that we intend to use the function $\Psi$ from the decomposition of the Cooper pair wave function of a low-energy state in \eqref{Second_Decomposition_alpha_equation} to construct a Gibbs state $\Gamma_{\Delta}$ as in \eqref{GammaDelta_definition}. In order to be able to justify the relevant computations with this state, we need $\Psi \in \Hmag^2(Q_h)$, which is not guaranteed by Theorem~\ref{Structure_of_almost_minimizers} above, see also \cite[Remark 3.3]{DeHaSc2021}. The following corollary provides us with a decomposition of $\alpha$, where the center-of-mass wave function $\Psi_\leq$ has the required $\Hmag^2(Q_h)$-regularity.



\begin{kor}
\label{Structure_of_almost_minimizers_corollary}
Let the assumptions of Theorem~\ref{Structure_of_almost_minimizers} hold and let $\varepsilon \in [h^2, h_0^2]$. Let $\Psi$ be as in 
\eqref{Second_Decomposition_alpha_equation} and define
\begin{align}
	\Psi_\leq &:= \Idbb_{[0,\varepsilon]}(\Pi^2) \Psi, &  \Psi_> &:= \Idbb_{(\varepsilon,\infty)}(\Pi^2) \Psi. \label{PsileqPsi>_definition}
\end{align}
Then, we have
\begin{align}
	\Vert \Psi_\leq\Vert_{\Hmag^1(Q_h)}^2 &\leq \Vert \Psi\Vert_{\Hmag^1(Q_h)}^2, \notag \\ 
	\Vert \Psi_\leq \Vert_{\Hmag^k(Q_h)}^2 &\leq C\, (\varepsilon h^{-2})^{k-1} \, \Vert \Psi\Vert_{\Hmag^1(Q_h)}^2, \qquad k\geq 2,  \label{Psileq_bounds}
\end{align}
as well as 
\begin{align}
\Vert \Psi_>\Vert_2^2 &\leq C \varepsilon^{-1}h^4 \, \Vert \Psi\Vert_{\Hmag^1(Q_h)}^2, & \Vert \Pi\Psi_>\Vert_2^2 &\leq Ch^4 \, \Vert \Psi\Vert_{\Hmag^1(Q_h)}^2. \label{Psi>_bound}
\end{align}
Furthermore,
\begin{align}
	\sigma_0(X,r) := \alpha_*(r) \Psi_>(X) \label{sigma0}
\end{align}
satisfies
\begin{align}
	\Vert \sigma_0\Vert_{H^1_\symm(Q_h\times \Rbb^3)}^2 &\leq C\varepsilon^{-1}h^4 \, \Vert \Psi\Vert_{\Hmag^1(Q_h)}^2 \label{sigma0_estimate}
\end{align}
and, with $\xi$ in \eqref{Second_Decomposition_alpha_equation}, the function
\begin{align}
	\sigma :=  \xi + \sigma_0 \label{sigma}
\end{align}
obeys
\begin{align}
	\Vert \sigma\Vert_{H^1_\symm(Q_h\times \Rbb^3)}^2 \leq Ch^4 \bigl( \varepsilon^{-1}\Vert \Psi\Vert_{\Hmag^1(Q_h)}^2 + D_1\bigr). \label{Second_Decomposition_sigma_estimate}
\end{align}
In terms of these functions, the Cooper pair wave function $\alpha$ of the low-energy state $\Gamma$ in \eqref{Second_Decomposition_Gamma_Assumption} admits the decomposition
\begin{align}
	\alpha(X,r) = \alpha_*(r) \Psi_\leq (X) + \sigma(X,r). \label{Second_Decomposition_alpha_equation_final}
\end{align}
\end{kor}

For a proof of the Corollary we refer to the proof of Corollary~5.2 in \cite{DeHaSc2021}.


\subsection{A lower bound for the BCS functional}

We start the proof of Theorem \ref{Structure_of_almost_minimizers} with the following lower bound on the BCS functional, whose proof is literally the same as that of Lemma~5.3 in \cite{DeHaSc2021}.


\begin{lem}
Let $\Gamma_0$ be the normal state in \eqref{Gamma0}. We have the lower bound
\begin{align}
\FBCS(\Gamma) - \FBCS(\Gamma_0) \geq  \Tr\bigl[ (K_{T,\Abold, W} - V) \alpha  \alpha^*\bigr] + \frac{4T}{5} \Tr\bigl[ (\alpha^* \alpha)^2\bigr], \label{Lower_Bound_A_3}
\end{align}
where
\begin{align}
K_{T, \Abold, W} = \frac{(-\i \nabla + \Abold_h)^2 + W_h- \mu}{\tanh (\frac{(-\i \nabla + \Abold_h)^2 + W_h - \mu}{2T})} \label{KTAW_definition}
\end{align}
and $V\alpha(x,y) = V(x-y) \alpha(x,y)$.
\end{lem}

In Proposition~\ref{prop:gauge_invariant_perturbation_theory} in Appendix~\ref{KTV_Asymptotics_of_EV_and_EF_Section} we show that the external electric and magnetic fields can lower the lowest eigenvalue zero of $K_{\Tc} - V$ at most by a constant times $h^2$. We use this in the next lemma to show that $K_{T, \Abold, W} - V$ is bounded from below by a nonnegative operator, up to a correction of the size $Ch^2$. We state the inequality \eqref{KTB_Lower_bound_eq} below for the one-particle operator $K_{T, \Abold, W} - V$ but it holds for the operator $K_{T, \Abold, W} - V(x-y)$ in \eqref{Lower_Bound_A_3} as well. This is due to the fact that 
\begin{equation*}
	T(y)^* \left[ K_{T, \Abold_y, W_y} - V(x) \right] T(y) = K_{T, \Abold, W} - V(x-y),
\end{equation*}
where $W_y(x) = h^2 W(h(x+y))$ and $\Abold_y(x) = \Abold_{\mathbf{B}}(x) + h A(h(x+y))$. By $T(y)$ we denoted the magnetic translations in \eqref{Magnetic_Translation}. We note that $W(x+y)$ and $A(x+y)$ are periodic functions of $x$ because $A(x)$ and $W(x)$ are. 
\begin{lem}
\label{KTB_Lower_bound}
Let Assumptions \ref{Assumption_V} and \ref{Assumption_KTc} be true. For any $D_0 \geq 0$, there are constants $h_0>0$ and $T_0>0$ such that for $0< h\leq h_0$ and $T>0$ with $T - \Tc \geq -D_0h^2$, the estimate
\begin{align}
K_{T, \Abold, W} - V &\geq c \; (1 - P) (1 + \pi^2) (1- P) + c \, \min \{ T_0, (T - \Tc)_+\} - Ch^2 \label{KTB_Lower_bound_eq}
\end{align}
holds. Here, $P = |\alpha_*\rangle\langle \alpha_*|$ is the orthogonal projection onto the ground state $\alpha_*$ of $K_{\Tc} - V$.
\end{lem}

\begin{proof}
Since $W \in L^{\infty}(\mathbb{R}^3)$ we can use \cite[Lemma 6.4]{DeHaSc2021} to show that $K_{T, \Abold, W} \geq K_{T, \Abold, 0} - C h^2$ holds. The rest of the proof goes along the same lines as that of \cite[Lemma 5.4]{DeHaSc2021} with the obvious replacements. In particular, \cite[Proposition~A.1]{DeHaSc2021} needs to be replaced by Proposition~\ref{prop:gauge_invariant_perturbation_theory}. We omit the details.
\end{proof}

We deduce two corollaries from \eqref{Lower_Bound_A_3} and Lemma \ref{KTB_Lower_bound}. The first statement is an a priori bound that will be used in the proof of Theorem \ref{Main_Result_Tc}~(b). Its proof goes along the same lines as that of \cite[Corollary 5.5]{DeHaSc2021}.
\begin{kor}
\label{TcB_First_Upper_Bound}
Let Assumptions \ref{Assumption_V} and \ref{Assumption_KTc} be true. Then, there are constants $h_0>0$ and $C>0$ such that for all $0 < h \leq h_0$ and all temperatures $T\geq \Tc(1 + Ch^2)$, we have $\FBCS(\Gamma) - \FBCS(\Gamma_0) >0$ unless $\Gamma = \Gamma_0$.
\end{kor}


The second corollary provides us with an inequality for Cooper pair wave functions of low-energy BCS states in the sense of \eqref{Second_Decomposition_Gamma_Assumption}. Before we state it, let us define the operator
\begin{align}
U  &:= \e^{-\i \frac r2 \Pi}, \label{U_definition}
\end{align}
with $\Pi$ in \eqref{Magnetic_Momenta_COM}, which acts on the relative coordinate $r = x-y$ as well as on the center-of-mass coordinate $X = \frac{x+y}{2}$ of a function $\alpha(x,y)$.

\begin{kor}
\label{cor:lowerbound}
Let Assumptions \ref{Assumption_V} and \ref{Assumption_KTc} be true. For any $D_0, D_1 \geq 0$, there is a constant $h_0>0$ such that if $\Gamma$ satisfies \eqref{Second_Decomposition_Gamma_Assumption}, if $0 < h\leq h_0$, and if $T$ is such that $T - \Tc \geq -D_0h^2$, then $\alpha = \Gamma_{12}$ obeys
\begin{align}
&\langle \alpha, [U(1 - P)(1 + \pi^2)(1 - P)U^* + U^*(1 - P)(1 + \pi^2)(1 - P)U] \alpha \rangle \notag\\
&\hspace{200pt} + \Tr\bigl[(\alpha^* \alpha)^2\bigr] \leq C h^2 \Vert \alpha\Vert_2^2 + D_1h^4, \label{Lower_Bound_A_2}
\end{align}
where $P = | \alpha_* \rangle \langle \alpha_* |$ and $\pi = -\i \nabla_r + \Abold_\Bbold(r)$ both act on the relative coordinate.
\end{kor}

In the statement of the corollary and in the following, we refrain from equipping the operator $\pi$ and the projection $P = |\alpha_*\rangle \langle \alpha_*|$ with an index $r$ although it acts on the relative coordinate. This does not lead to confusion and keeps the formulas readable.

\begin{proof}
The proof follows the same strategy as that of \cite[Corollary 5.6]{DeHaSc2021}. Let us denote by $\pi_{\Abold}^x$ and $\pi_{\Abold}^y$ the magnetic momentum operators acting on $x$ and $y$, respectively. We claim that
\begin{align}
\pi_{\Abold_h}^x &= U \pi_{\Abold^+_h}^r U^*, & -\pi_{\Abold_h}^y &= U^* \pi_{\Abold^-_h}^r U, \label{cor:lowerbound_1}
\end{align}
where $\pi_{\Abold^\pm}^r = -\mathrm{i} \nabla_r + \Abold^{\pm}(r)$ with
\begin{align*}
\Abold^\pm(r) &\coloneqq \Abold_{e_3}(r) \pm A(X \pm r).
\end{align*}
To obtain \eqref{cor:lowerbound_1}, we denote $P_X = -\mathrm{i} \nabla_X$ and note that $[r \cdot P_X, r \cdot (\Bbold \wedge X)] = 0$ implies $U = \e^{\i \frac{\Bbold}{2} (X \wedge r)} \e^{- \i \frac{r}{2} P_X}$. Using this identity we conclude
\begin{align*}
	U \; (-\mathrm{i} \nabla_r + \Abold_{\Bbold}(r)) \,  U^* &= -\mathrm{i} \nabla_r + \frac{1}{2} \Abold_{\Bbold}(r) - \frac{1}{2} \Pi_X, \\
	U^* (-\mathrm{i} \nabla_r + \Abold_{\Bbold}(r)) \, U \; &= -\mathrm{i} \nabla_r + \frac{1}{2} \Abold_{\Bbold}(r) + \frac{1}{2} \Pi_X.
\end{align*}  
Eq.~\eqref{cor:lowerbound_1} is a direct consequence of these two identities. 

We also have
\begin{align*}
W(x) &= U \, W^+(r) \, U^*, & W(y) &= U^* \, W^-(r) \, U, & W^\pm(r) &:= W(X \pm r).
\end{align*}
Consequently, if $K_{T, \Abold, W}^x$ and $K_{T, \Abold, W}^y$ denote the operators $K_{T, \Abold, W}$ acting on the $x$ and $y$ coordinate, respectively, we infer
\begin{align}
K_{T, \Abold, W}^x - V(r) &= U^* ( K_{T, \Abold^+, W^+}^r - V(r) ) \, U, \notag \\ 
K_{T, \Abold, W}^y - V(r) &= U \; ( K_{T, \Abold^-, W^-}^r - V(r) ) \, U^*. \label{eq:idK_T^r} 
\end{align}
Here $K_{T, \Abold^\pm, W^\pm}^r$ denotes the operator $K_{T, \Abold^\pm, W^\pm}$ with magnetic potential $\Abold^\pm$ and electric potential $W^\pm$, which depend on $X$, acting on the relative coordinate $r$.

The operator $V$ in \eqref{Lower_Bound_A_3} acts as by multiplication with the function $V(x-y)$. We apply the symmetry $\alpha(x,y) = \alpha(y,x)$ and deduce
\begin{align}
\Tr \bigl[ (K_{T, \Abold, W} - V) \alpha \alpha^*\bigr] & \notag\\
&\hspace{-80pt} = \frac 12 \fint_{Q_h} \dd x \int_{\Rbb^3} \dd y \; \overline{\alpha(x,y)} \bigl[ (K_{T, \Abold, W}^x - V) + (K_{T, \Abold, W}^y - V) \bigr] \alpha(x,y). \label{Lower_Bound_A_4}
\end{align}
In combination, \eqref{Second_Decomposition_Gamma_Assumption}, \eqref{Lower_Bound_A_3}, \eqref{eq:idK_T^r}, and \eqref{Lower_Bound_A_4} therefore prove the bound
\begin{align*}
\frac 12\langle \alpha, [U^* ( K_{T, \Abold^+, W^+}^r - V(r) ) \, U + U \; ( K_{T, \Abold^-, W^-}^r - V(r) ) \, U^*]\alpha\rangle + c \Tr \bigl[ (\alpha^* \alpha)^2\bigr] \leq D_1 h^4.
\end{align*}
An application of Lemma \ref{KTB_Lower_bound} on the left side establishes \eqref{Lower_Bound_A_2}.
\end{proof}

%
%

\subsection{Proof of Theorem \ref{Structure_of_almost_minimizers}}

The statement in Theorem \ref{Structure_of_almost_minimizers} is a direct consequence of Corollary~\ref{cor:lowerbound} above and the results in \cite{DeHaSc2021}. More precisely, we need to combine Corollary~\ref{cor:lowerbound} and \cite[Proposition~5.7]{DeHaSc2021}, \cite[Lemma~5.14]{DeHaSc2021}, and the arguments in \cite[Section~5.4]{DeHaSc2021}.

\section{The Lower Bound on \texorpdfstring{(\ref{ENERGY_ASYMPTOTICS})}{(\ref{ENERGY_ASYMPTOTICS})} and Proof of Theorem \ref{Main_Result_Tc} (b)}
\label{Lower Bound Part B}

\subsection{The BCS energy of low-energy states}

In this section, we complete the proofs of Theorems \ref{Main_Result} and \ref{Main_Result_Tc}, which amounts to providing the lower bound on \eqref{ENERGY_ASYMPTOTICS}, the bound in \eqref{GL-estimate_Psi}, and the proof of Theorem~\ref{Main_Result_Tc}~(b). Let $D_1\geq 0$ and $D\in \Rbb$ be given, choose $T = \Tc(1 - Dh^2)$, and assume that $\Gamma$ is a gauge-periodic state that satisfies \eqref{Second_Decomposition_Gamma_Assumption}. 

Corollary~\ref{Structure_of_almost_minimizers_corollary} guarantees a decomposition of the Cooper pair wave function $\alpha = [\Gamma]_{12}$ in terms of $\Psi_{\leq}$ in \eqref{PsileqPsi>_definition} and $\sigma$ in \eqref{sigma}. The function $\Psi_{\leq}$ satisfies the bounds
\begin{equation}
	\Vert \Psi_{\leq} \Vert_{\Hmag^1(Q_h)}^2 \leq \Vert \Psi \Vert_{\Hmag^1(Q_h)}^2 \leq C \quad \text{ and } \quad \Vert \Psi_{\leq} \Vert_{\Hmag^2(Q_h)}^2 \leq C \varepsilon h^{-2} \Vert \Psi \Vert_{\Hmag^1(Q_h)}^2, \label{eq:lowerboundB1}
\end{equation}
with $\Psi$ in \eqref{Second_Decomposition_alpha_equation}. Let us define the state $\Gamma_{\Delta}$ as in \eqref{GammaDelta_definition} with $\Delta(X,r) = -2 V \alpha_* (r) \Psi_{\leq}(X)$. We apply Proposition~\ref{BCS FUNCTIONAL_IDENTITY} and Theorem~\ref{CALCULATION_OF_THE_GL-ENERGY} to obtain the following lower bound for the BCS energy of $\Gamma$:
\begin{align}
	\FBCS(\Gamma) - \FBCS(\Gamma_0) &\geq h^4\; \EGL(\Psi_{\leq}) - C \left( h^5 + \varepsilon h^4 \right) \Vert \Psi \Vert_{\Hmag^1(Q_h)}^2 \label{eq:lowerboundB2} \\
	&\hspace{20pt}+ \frac{T}{2} \Hcal_0(\Gamma, \Gamma_\Delta) - \fint_{Q_h} \dd X \int_{\Rbb^3} \dd r \; V(r) \, | \sigma(X,r) |^2. \nonumber
\end{align}
In the next section we prove a lower bound for the terms in the second line of \eqref{eq:lowerboundB2}.

\subsection{Estimate on the relative entropy}

The arguments in \cite[Eqs.~(6.1)-(6.14)]{DeHaSc2021} apply in literally the same way here, too. We obtain the correct bounds when we replace $B$ by $h^2$ in all formulas. This, in particular, applies to the statement of \cite[Lemma~6.2]{DeHaSc2021}. The only difference is that \cite[Eq.~(6.10)]{DeHaSc2021} is now given by
\begin{equation*}
	| \langle \eta_{0},  K_{T_{\mathrm{c}},\Abold,W} \sigma \rangle | \leq \ C \varepsilon^{-\nicefrac{1}{2}} h^{\nicefrac{9}{2}} \Vert \Psi \Vert_{\Hmag^1(Q_h)} \bigl( \Vert \Psi\Vert_{\Hmag^1(Q_h)}^2 + D_1\bigr)^{\nicefrac{1}{2}},
\end{equation*}
which is due to the reason that the bound for the $L^2$-norm of $\eta_0$ in Proposition~\ref{Structure_of_alphaDelta} is worse than the comparable bound we obtained in \cite[Proposition~3.2]{DeHaSc2021}. This, however, does not change the size of the remainder in the final bound because other error terms come with a worse rate. 

With the choice $\varepsilon = h^{1/3}$ we therefore obtain the bound
\begin{align}
	&\FBCS(\Gamma) - \FBCS(\Gamma_0) \notag \\
	&\hspace{2cm}\geq h^4 \, \bigl(\EGL(\Psi_\leq) - C \,  h^{\nicefrac {1}{6}} \Vert \Psi\Vert_{\Hmag^1(Q_h)} \; \bigl(  \Vert \Psi\Vert_{\Hmag^1(Q_h)}^2 + D_1\bigr)^{\nicefrac 12} \bigr), \label{Lower_Bound_B_5}
\end{align}
which is the equivalent of \cite[Eq~(6.14)]{DeHaSc2021}.

\subsection{Conclusion}

The arguments in \cite[Section~6.3]{DeHaSc2021} apply in literally the same way also here and we obtain the correct formulas when we replace $B^{\nicefrac{1}{2}}$ by $h$. This concludes the proof of Theorem~\ref{Main_Result} and Theorem~\ref{MAIN_RESULT_TC}.

\subsection{Proof of the equivalent of \texorpdfstring{\cite[Lemma~6.2]{DeHaSc2021}}{Lemma~6.2 in Deuchert, Hainzl, Schaub (2021)} in our setting}

To obtain a proof of the equivalent of \cite[Lemma~6.2]{DeHaSc2021} in our setting, we follow the proof strategy in \cite{DeHaSc2021}. The additional terms coming from the external electric potential are not difficult to bound because $W$ is a bounded function. To obtain bounds of the correct size in $h$ for the terms involving the periodic vector potential $A_h$, we need to use that $A(0) = 0$, which is guaranteed by Assumption~\ref{Assumption_V}. This is relevant for example when we estimate our equivalent of the term on the left side of \cite[Eq.~(6.24)]{DeHaSc2021}, that is, of
\begin{equation*}
	\Vert [ \pi_{\Abold_h}^2 + W_h(r) - p_r^2 ] \sigma_0 \Vert_2   
\end{equation*}
with $p_r = - \mathrm{i} \nabla_r$ and $\sigma_0$ in \eqref{sigma0}. We make use of the decomposition \eqref{alphaDelta_decomposition_3}. This leaves us with bounding the term $\Vert [\pi^2 - p_r^2]\sigma_0\Vert_2$, which has been done in \cite[Eq.~(6.24)]{DeHaSc2021}, and we find that it is bounded by a constant times $\varepsilon^{-\nicefrac{1}{2}} h^4 \Vert \Psi\Vert_{\Hmag^1(Q_h)}$.
%
%
%
%
%
%
%
%
We use \eqref{Psi>_bound} to see that the terms involving $|A_h|^2$ and $W_h$ are bounded by 
\begin{align}
	\bigl( \Vert A_h\Vert_\infty^2 + \Vert W_h \Vert_\infty \bigr)  \ \Vert \sigma_0 \Vert_2 \leq C \varepsilon^{-\nicefrac{1}{2}} h^4 \Vert \Psi\Vert_{\Hmag^1(Q_h)}.
	\label{eq:proofoflemma32b}
\end{align}
With the same argument we see that the contribution from the first three terms on the right side of
\begin{align}
\pi_\Abold \cdot A + A \cdot \pi_\Abold = \divv A \cdot A + \divv A\cdot \pi + 2 \, |A|^2 + 2A \cdot \pi  \label{eq:proofoflemma32a}
\end{align}
are bounded by the term on the right side of \eqref{eq:proofoflemma32b}. To obtain a bound for the contribution from the fourth term on the right side of \eqref{eq:proofoflemma32a} we write
\begin{equation*}
	A_h(r) = h^2 \int_0^1 \dd t \ (D A)(h r t) \cdot r, 
\end{equation*}
where $DA$ denotes the Jacobian matrix of $A$. We conclude
\begin{equation*}
	\Vert A_h(r) \cdot \pi \, \sigma_0 \Vert_2 \leq \ h^2 \Vert DA \Vert_\infty \, \Vert \ | \cdot | \pi \alpha_* \Vert_2 \, \Vert \Psi_> \Vert_2 \leq C h^4 \varepsilon^{-\nicefrac{1}{2}} \Vert \Psi\Vert_{\Hmag^1(Q_h)}
\end{equation*}
as well as 
\begin{equation*}
	\Vert [ \pi_{\Abold_h}^2 + W_h(r) - p_r^2 ] \sigma_0 \Vert_2 \leq C \varepsilon^{-\nicefrac{1}{2}} h^4 \Vert \Psi\Vert_{\Hmag^1(Q_h)}.
\end{equation*}
All other bounds in the proof of the equivalent of \cite[Lemma~6.2]{DeHaSc2021} in our setting that involve $W_h$ or $A_h$ can be estimated with similar ideas. We therefore omit the details.

\begin{center}
\huge \textsc{--- Appendix ---}
\end{center}


\section{Gauge-Invariant Perturbation Theory for \texorpdfstring{$K_{\Tc, \Abold} -V$}{KTcA-V}}
\label{KTV_Asymptotics_of_EV_and_EF_Section}

In this appendix, we discuss the behavior of the eigenvalues below the essential spectrum of the operator $K_{\Tc, \Abold} - V$ for small $h >0$, where $K_{\Tc, \Abold} := K_{\Tc, \Abold, 0}$ with $K_{T, \Abold, 0}$ defined in \eqref{KTAW_definition}. Recall that the magnetic potential is composed of the constant magnetic field potential $\Abold_{e_3}(x) = \frac 12 e_3\wedge x$ and a bounded potential $A\in W^{3, \infty}(\Rbb^3; \Rbb^3)$, which satisfies $A(0) =0$. The full magnetic potential is then given by $\Abold := \Abold_{e_3} + A$ and $\Abold_h (x) := h\Abold(hx)$. The aim of this appendix is to prove the following proposition. 

\begin{prop}
	\label{prop:gauge_invariant_perturbation_theory}
	Assume $(1+|\cdot|^2) V \in L^{2}(\mathbb{R}^3) \cap L^{\infty}(\mathbb{R}^3)$. Then there is an $h_0>0$ such that for $0 < h \leq h_0$ the following statements hold:
	\begin{enumerate}[(a)]
		\item Let $\lambda$ be an isolated eigenvalue of multiplicity $m \in \mathbb{N}$ of the operator $K_\Tc - V$ with spectral projection $P$. Then there are $m$ eigenvalues $\lambda_1(h), \ldots ,\lambda_m(h)$ of the operator $K_{\Tc, \Abold_h}-V$ with spectral projection $P(h)$ such that  
		\begin{align}
			\max_{i=1,\ldots ,m} | \lambda_i(h) - \lambda | \leq C h^2 \quad \text{ and } \quad \Vert P(h) - P \Vert_{\infty} \leq  C h^2.
			\label{eq:app1}
		\end{align}
		\item Assume that $K_\Tc-V$ has a simple lowest eigenvalue with eigenfunction $\alpha_*$ and denote by $\alpha_*^{\Abold_h}$ the eigenfunction to the lowest eigenvalue of $K_{\Tc, \Abold} -V$, which is normalized such that $\langle \alpha_*, \alpha_*^{\Abold_h} \rangle \geq 0$ holds. Then we have the bound
		\begin{align}
			\Vert (1+\pi^2) (\alpha_*^{\Abold_h} - \alpha_*) \Vert_2 \leq C h^2,
			\label{eq:app2}
		\end{align}
		where $\pi^2$ is the magnetic Laplacian defined in \eqref{pi_definition}.
	\end{enumerate}
\end{prop}

We denote the resolvent of $K_{T, \Abold} - V$ at $z\in \rho(K_{T, \Abold} - V)$ by $\Rcal_\Abold^{z, V} := (z - (K_{\Tc, \Abold} - V))^{-1}$. The integral kernel of $\Rcal_\Abold^{z, V}$ is denoted by $\Gcal_\Abold^{z, V}(x,y)$. If $\Abold =0$, we write $\Rcal^{z, V}$ for $\Rcal_0^{z, V}$ and $\Gcal^{z, V}$ for $\Gcal_0^{z, V}$. Similarly, $\Gcal^z$ stands for $\Gcal^{z, 0}$. Before we give the proof of the above proposition, we state and prove three preparatory lemmas.

\subsection{Preparatory lemmas}

The first lemma concerns the regularity of the kernel $\Gcal^z$.

\begin{lem}
	\label{lem:App1}
	There is a continuous function $a \colon \mathbb{C} \backslash [2\Tc,\infty) \to \mathbb{R}_+$ such that
	\begin{align}
		\int_{\mathbb{R}^3} \dd x \ (1+x^2) | \nabla \Gcal^z(x) |  \leq a(z).
		\label{eq:Prep1}
	\end{align}
\end{lem}
\begin{proof}
	We use the resolvent identity $\Rcal^z = \Rcal^0 + z \Rcal^0 \Rcal^z$ to write $\Gcal^z$ as
	\begin{align}
		\Gcal^z(x) = \Gcal^0(x) + z \int_{\mathbb{R}^3} \dd v \; \Gcal^0(x-v) \, \Gcal^z(v),
		\label{eq:Prep2}
	\end{align}
	which implies
	\begin{align}
		\Vert (1+|\cdot|^2) \nabla \Gcal^z \Vert_1 \leq \Vert (1+2 |\cdot|^2) \nabla \Gcal^0 \Vert_1 \left( 1 + |z|\,  \Vert (1+ |\cdot|^2) \Gcal^z \Vert_1 \right). 
		\label{eq:Prep3}
	\end{align}
	The second $L^1(\mathbb{R}^3)$-norm on the right side of \eqref{eq:Prep3} is bounded by
	\begin{align}
		\Vert (1+|\cdot|^2) \Gcal^z \Vert_1 \leq C \ \Vert (1+|\cdot|^2)^2 \Gcal^z \Vert_2 = C \, \Bigl( \int_{\mathbb{R}^3} \dd p \ \Bigl| (1-\Delta_p)^2 \frac{1}{z - K_\Tc(p)} \Bigr|^2 \Bigr)^{\nicefrac 12}, 
		\label{eq:Prep4}
	\end{align}
	which, when multiplied with $z$, meets the requirements of the lemma. It therefore remains to consider $\Vert (1+|\cdot|^2) \nabla \Gcal^0 \Vert_1$.
	
	From \cite[Eq.~(A.6)]{Hainzl2012} and \cite[Theorem~6.23]{LiebLoss} we know that $\Gcal^0(x)$ can be written as
	\begin{align}
		\Gcal^0(x) &= \frac{2}{\pi} \sum_{n=1}^{\infty} \frac{1}{n - \frac{1}{2}} \text{Im} \ g^{\i (n-\frac{1}{2}) 2 \pi \Tc}(x) \label{eq:Prep5} \\
		&= \frac{1}{2 \pi^2 |x|} \sum_{n=1}^{\infty} \frac{1}{n - \frac{1}{2}} \text{Im} \exp \Bigl( \i \sqrt{\mu+\i \left(n-\frac{1}{2} \right) 2 \pi \Tc} \, |x| \Bigr), \nonumber
	\end{align}
	where $g^z$ is the free resolvent kernel in \eqref{g_definition} and $\sqrt{\cdot}$ denotes the principal square root. We use $\Im \e^{\i z} = \sin(\Re z) \exp(- \Im z)$ for $z \in \mathbb{C}$ and 
	\begin{align}
		\sqrt{a+ \i b} = \frac{1}{\sqrt{2}} \sqrt{\sqrt{a^2+b^2}+a} +  \frac{\i \sgn(b)}{\sqrt{2}} \sqrt{\sqrt{a^2+b^2}-a} 
		\label{eq:Prep5b} 
	\end{align}
	for $a, b \in \mathbb{R}$ to see that
	\begin{align}
		\text{Im} \exp\Bigl( \i \sqrt{\mu+\i \left(n-\frac{1}{2} \right) 2 \pi \Tc} \, |x| \Bigr) = \sin\left( |x| c_n^{+} \right)  \exp\left( -|x| c_n^{-} \right), \label{eq:Prep6}
	\end{align}
	where 
	\begin{align}
		c_n^{\pm} = \frac{ 1 }{ \sqrt{2} } \sqrt{ \sqrt{\mu^2 + ((n-1/2) 2 \pi \Tc )^2} \pm \mu}.
		\label{eq:Prep6b}
	\end{align}
	In particular,
	\begin{align}
		\nabla \Gcal^0(x) &= -\frac{x}{2 \pi^2 |x|^3} \sum_{n=1}^{\infty} \frac{1}{n - \frac{1}{2}} \sin\left( |x| c_n^{+} \right)  \exp\left( -|x| c_n^{-} \right) \nonumber\\
		&\hspace{50pt}+ \frac{x}{2 \pi^2 |x|^2} \sum_{n=1}^{\infty} \frac{c_n^{+}}{n - \frac{1}{2}} \cos\left( |x| c_n^{+} \right)  \exp\left( -|x| c_n^{-} \right)  \nonumber \\
		&\hspace{50pt}- \frac{x}{2 \pi^2 |x|^2} \sum_{n=1}^{\infty} \frac{c_n^{-}}{n - \frac{1}{2}} \sin\left( |x| c_n^{+} \right)  \exp\left( -|x| c_n^{-} \right).  \label{eq:Prep6c}
	\end{align}
	The above formula implies the bound
	\begin{align}
		\Vert (1+|\cdot|^2) \nabla \Gcal^0 \Vert_1 &\leq \sum_{n=1}^{\infty} \frac{C}{n} \int_{0}^{\infty} \dd r \left( 1+r^2 + (r+r^3) \sqrt{1+n} \right) \exp\left( -r c_n^{-} \right) \leq C \sum_{n=1}^{\infty} n^{-\frac 32}. \label{eq:Prep6d} 
	\end{align}
	This proves the claim of the lemma.
\end{proof}

The second lemma concerns bounds for the operator norm of $(z- (K_\Tc - V))^{-1}$ and commutators of this operator with $x$, when viewed as maps from $L^2(\mathbb{R}^3)$ to $L^{\infty}(\mathbb{R}^3)$. Here and in the following we denote by $\Vert A \Vert_{2;\infty}$ the norm of a bounded operator $A$ from $L^2(\mathbb{R}^3) \ra L^{\infty}(\mathbb{R}^3))$. From \cite[Corollary~A.1.2]{Simon82} we know that such an operator, which is bounded also from $L^2(\Rbb^3)$ to itself, has an integral kernel given by a measurable function $A(x,y)$, which obeys
\begin{align}
	\esssup_{x \in \mathbb{R}^3} \left( \int_{\mathbb{R}^3} \dd y \; | A(x,y) |^2 \right)^{\nicefrac 12} < \infty.	\label{eq:Prep6e}
\end{align}
The norm $\Vert A \Vert_{2;\infty}$ equals the norm of the integral kernel of $A$ in \eqref{eq:Prep6e}.

The following Lemma \ref{lem:App2} shows, in particular, that the resolvent kernel $\Gcal^{z, V}$ satisfies
\begin{align}
\esssup_{x\in \Rbb^3} \Bigl( \int_{\Rbb^3} \dd y \; \bigl( 1 + |x-y|^4\bigr) |\Gcal^{z, V}(x,y)|^2 \Bigr)^{\nicefrac 12} < \infty. \label{eq:App_decay_resolvent_kernel}
\end{align}
We remark that our assumptions on $V$ would allow for more: it can be shown that $\Gcal^{z, V}$ is exponentially decaying in the sense that \eqref{eq:App_decay_resolvent_kernel} holds with $1 + |x-y|^4$ replaced by $\e^{\delta |x-y|}$ for some $\delta >0$ depending on the distance of $z$ to the spectrum of $K_\Tc- V$. Since this result is not necessary for the proof of Proposition \ref{prop:gauge_invariant_perturbation_theory} and requires substantially more effort, we refrain from giving the proof here. It follows from a Combes--Thomas estimate for the operator $K_\Tc -V$ and can be found in Chapter \ref{Chapter:Combes-Thomas}.

\begin{lem}
	\label{lem:App2}
	If $V$ belongs to the space $L^{\infty}_{\varepsilon}(\mathbb{R}^3)$ of bounded functions that vanish at infinity, then there is a continuous function $a \colon \rho(K_\Tc-V) \to \mathbb{R}_+$ such that
	\begin{align}
		\Vert \Rcal^{z, V}  \Vert_{2;\infty} + \bigl\Vert [x, \Rcal^{z, V} ] \bigr\Vert_{2;\infty} + \bigl\Vert [x,[x, \Rcal^{z, V} ]] \bigr\Vert_{2;\infty} \leq a(z). 
		\label{eq:Prep7}
	\end{align}
\end{lem}

\begin{proof}
	We start by proving the bound for the first term on the right side of \eqref{eq:Prep7}. We use the fact that $(1-\Delta)^{-1}$ is a bounded linear map from $L^2(\mathbb{R}^3)$ to $L^{\infty}(\mathbb{R}^3)$ and the resolvent identity to estimate
	\begin{align}
		\Vert \Rcal^{z, V} \Vert_{2;\infty} \leq C \Vert (1-\Delta) \Rcal^{z, V} \Vert_\infty \leq C \Vert (1-\Delta) \Rcal^z \Vert_\infty \left( 1 + \Vert V \Vert_\infty \Vert \Rcal^{z, V} \Vert_\infty \right). 
		\label{eq:Prep8}
	\end{align}
	Since $V\in L_\varepsilon^\infty(\Rbb^3)$, we have $\rho(K_\Tc) \subseteq \rho(K_\Tc - V)$, whence both $z$-dependent terms on the right side meet the requirements of the lemma.
	
	To obtain a bound for the second term on the right side of \eqref{eq:Prep7}, we note that
	\begin{align}
		[x, \Rcal^{z, V} ]  = \Rcal^{z, V} [K_\Tc,x] \Rcal^{z, V} \quad \text{as well as} \quad [K_\Tc,x] = -\i (\nabla f)(-\i\nabla),
		\label{eq:Prep9}
	\end{align}
	where $f(p) := K_\Tc(p)$ is the symbol in \eqref{KT-symbol}. Using this, we estimate
	\begin{align}
		\Vert [x,\Rcal^{z, V}] \Vert_{2;\infty} \leq \Vert \Rcal^{z, V} \Vert_{2;\infty} \, \Vert (\nabla f)(-\i\nabla) \Rcal^{z, V} \Vert_\infty.
	\end{align}
	A bound for the first factor on the right side was obtained in \eqref{eq:Prep8}. Using the resolvent identity again, we bound the second factor by
	\begin{align}
		\Vert (\nabla f)(-\i\nabla) \Rcal^{z, V} \Vert_\infty \leq \Vert (\nabla f)(-\i\nabla) \Rcal^z \Vert_\infty \left( 1 + \Vert V \Vert_\infty \Vert \Rcal^{z, V} \Vert_\infty \right),
	\end{align}
	which proves the claim for the second term on the right side of \eqref{eq:Prep7}. A bound for the third term can be derived similarly, and we therefore leave the remaining details to the reader. This proves the claim.
\end{proof}

\begin{lem}
	\label{lem:regEF}
	Assume $(1+|\cdot|^2)V \in L^{2}(\mathbb{R}^{3}) \cap L^{\infty}(\mathbb{R}^{3})$ and let $\alpha$ be an eigenfunction of the operator $K_\Tc - V$ with eigenvalue $\lambda < 2 \Tc$. Then we have
	\begin{align}
		\Vert \, |\cdot| \nabla \alpha \Vert_2 + \Vert \, |\cdot|^2 \alpha \Vert_2 < \infty.
		\label{eq:reg1}
	\end{align}
\end{lem}

\begin{proof}
	We use the eigenvalue equation to write the Fourier transform of $\alpha$ as 
	\begin{align}
		\hat{\alpha}(p) = -\frac{1}{\lambda - K_\Tc(p)} \,  (\hat{V} \ast \hat{\alpha})(p).
		\label{eq:reg2}
	\end{align}
	Using Young's inequality, we see that this implies
	\begin{align}
		\Vert \, |\cdot|^2 \alpha \Vert &= \Bigl( \int_{\mathbb{R}^3} \dd p \; \Bigl| \Delta_p \frac{1}{\lambda - K_\Tc(p)} \, (\hat{V} \ast \hat{\alpha}) (p) \Bigr|^2  \Bigr)^{\nicefrac 12} \nonumber \\
		&\leq C \left( \Vert \hat{V} \ast \hat{\alpha} \Vert_\infty + \Vert \Delta (\hat{V} \ast \hat{\alpha}) \Vert_\infty \right) \leq C \, \Vert (1+|\cdot|^2) V \Vert_\infty. \label{eq:reg3} 
	\end{align}
	To prove the other bound, we use the resolvent identity to write \eqref{eq:reg2} as
	\begin{align}
		\hat{\alpha}(p) = - \frac{1}{K_\Tc(p)} \, (\hat{V} \ast \hat{\alpha})(p) + \frac{\lambda}{K_\Tc(p)} \frac{1}{\lambda - K_\Tc(p)} \, (\hat{V} \ast \hat{\alpha})(p). \label{eq:reg4}
	\end{align}
	We argue as in \eqref{eq:reg3} to see that the $L^2(\mathbb{R}^3)$-norm of $\nabla_p p \frac{\lambda}{K_\Tc(p)} \frac{1}{\lambda - K_\Tc(p)} (\hat{V} \ast \hat{\alpha})(p)$ is bounded by a constant times $\Vert \, |\cdot| V \Vert_\infty$. To treat the other term, we go back to position space and note that
	\begin{align}
		\Bigl( \int_{\mathbb{R}^3} \dd x \left| \int_{\mathbb{R}^3} \dd y \; |x| \, |\nabla \Gcal^0(x-y)| \, |V\alpha(y)| \right|^2 \Bigr)^{\nicefrac 12} \leq \Vert (1+|\cdot|) \nabla \Gcal^0 \Vert_1 \, \Vert (1+|\cdot|) V \Vert_{\infty}. 
		\label{eq:reg5}
	\end{align} 
In combination with Lemma~\ref{lem:App1}, these considerations prove the claim.
\end{proof}

\subsection{Proof of Proposition~\ref{prop:gauge_invariant_perturbation_theory}}

The proof of Proposition~\ref{prop:gauge_invariant_perturbation_theory} is based on an adaption of gauge-invariant perturbation theory for Schr\"odinger operators as introduced in \cite{Nenciu2002} to our setting.  The core of the argument is contained in the following lemma.

\begin{lem}
	\label{lem:Appmain}
	Assume that $(1+|\cdot|^2) V \in L^{2}(\mathbb{R}^3) \cap L^{\infty}(\mathbb{R}^3)$. There is a continuous function $a \colon  \rho(K_\Tc-V) \to \mathbb{R}_+$ such that the following holds: For every compact set $K \subset \rho(K_\Tc-V)$ there is a constant $h(K) > 0$ such that for $0 < h < h(K)$ we have
	\begin{align}
		\Rcal_\Abold^{z, V} = \Scal_{\Abold_h}^{z, V} + h^2 \, \eta_h(z) \quad \text{ with } \quad \Vert (1+\pi^2) \eta_h(z) \Vert_\infty \leq a(z).
	\end{align} 
	Here $\Scal_\Abold^{z, V}$ is the operator defined by the kernel
	\begin{align}
		\Scal_\Abold^{z, V}(x,y) &:= \e^{\i \Phi_{\Abold}(x,y)} \, \Gcal^{z, V}(x,y) \label{eq:app4}
	\end{align}
	with the phase factor $\Phi_\Abold(x, y)$ in \eqref{PhiA}.
\end{lem}

\begin{proof}
We employ \eqref{PhiA_Magnetic_Momentum_Action} and the integral representation \cite[Lemma 6.4]{DeHaSc2021} for $K_{\Tc, \Abold}$ to see that
\begin{align}
	K_{\Tc, \Abold}^x \, \e^{\i \Phi_{\Abold_h}(x,y)} = \e^{\i \Phi_{\Abold_h}(x,y)} \, K_{\Tc, \Abold_y}^x,
	\label{eq:app6}
\end{align}
where $K_{\Tc, \Abold}^x$ is understood to act on the $x$-coordinate and $\Abold_y(x) := \tilde \Abold(x,y)$ denotes the vector potential in transversal Poincar\'e gauge relative to the point $y$, defined in \eqref{Atilde_definition}. The result \cite[Lemma~6.4]{DeHaSc2021} also implies the identity
\begin{align}
	K_{\Tc, \Abold_y} - K_\Tc &= \left( -2\i (\Abold_h)_y(x) \nabla_x -\i \divv (\Abold_h)_y(x) + |(\Abold_h)_y(x)|^2 \right) \label{eq:app8} \\
	&\hspace{20pt} + \int_{\speaker} \frac{\dd z}{2 \pi \i} \; \varphi(z) \; \frac{1}{z + (-\i \nabla + (\Abold_h)_y)^2 + \mu} \notag \\
&\hspace{40pt} \times \left( -2\i (\Abold_h)_y(x) \nabla_x -\i \divv (\Abold_h)_y(x) + |(\Abold_h)_y(x)|^2 \right) \frac{1}{z + \Delta + \mu}, \nonumber
\end{align}
where $\varphi(z) = (z/\tanh( z/(2\Tc) ) - z )$. Let us define the operator $\Tcal_\Abold^{z, V}$ via the equation
\begin{align}
	\left( z - (K_{\Tc, \Abold} - V) \right) \Scal_{\Abold_h}^{z, V} = 1 - \Tcal_{\Abold_h}^{z, V}.
	\label{eq:app5}
\end{align}
Using \eqref{eq:app6}, \eqref{eq:app8}, and \eqref{eq:app5}, we write the integral kernel of $\Tcal_\Abold^{z, V}$ as
\begin{align}
	\Tcal_\Abold^{z, V}(x,y) &= \e^{\i \Phi_{\Abold}(x,y)} \Big[ \left( -2 \i \Abold_y(x) \nabla_x - \i \divv \Abold_y(x) + |\Abold_y(x)|^2 \right) \Gcal^{z, V}(x,y) \nonumber \\
	&\hspace{-20pt}+ \int_{\speaker} \frac{\dd \zeta}{2 \pi \i} \; \varphi(\zeta) \int_{\mathbb{R}^6} \dd v \dd w \; G_{\Abold_y}^\zeta(x,v) \left( -2\i \Abold_y(v) \nabla_v -\i \divv \Abold_y(v) + |\Abold_y(v)|^2 \right)  \nonumber \\
	&\hspace{200pt} \times g^\zeta(v-w) \, \Gcal^{z, V}(v, y) \Big], \label{eq:app9}
\end{align}
where $G_\Abold^z$ is the magnetic resolvent kernel defined in \eqref{GAz_definition}. In the next step we use this formula to prove a bound for the operator norm of $\Tcal_\Abold^{z, V}$.

Let us denote the first and the second term on the right side of \eqref{eq:app9} by $\Tcal_\Abold^{(1)}(x,y)$ and $\Tcal_\Abold^{(2)}(x,y)$, respectively. Using \eqref{TAz_boundedness_1} and \eqref{TAz_boundedness_2}, we see that
\begin{align}
	| \Tcal_{\Abold_h}^{(1)}(x,y) | \leq Ch^2 \bigl( |x-y| \, | \nabla_x \Gcal^{z, V}(x,y) | + \bigl( |x - y|  + | x-y |^2   \bigr) | \Gcal^{z, V}(x,y) | \bigr).
	\label{eq:app10}
\end{align} 
Using the resolvent identity $\Rcal^{z, V} = \Rcal^z + \Rcal^z V \Rcal^{z, V}$, we estimate the first term on the right side of \eqref{eq:app10} by
\begin{align}
	\left| x-y \right| \left| \nabla_x \Gcal^{z, V}(x,y) \right| \leq& \left| x-y \right| \left| \nabla \Gcal^z( x-y) \right| \nonumber \\
	&+ \int_{\mathbb{R}^3} \dd w \ \left| |x-w| \, \nabla \Gcal^z (x-w) \, V( w) \, \Gcal^{z, V}(w,y) \right| \nonumber \\
	&+ \int_{\mathbb{R}^3} \dd w \ \left| \nabla \Gcal^z (x-w) \, V(w) \, |w-y| \, \Gcal^{z, V}(w,y) \right|. \label{eq:app11}
\end{align}
Eq.~\eqref{eq:app11} allows us to obtain the following bound for the operator norm of $\Tcal_\Abold^{(1)}$: 
\begin{align}
	\Vert \Tcal_{\Abold_h}^{(1)} \Vert_{\infty} &\leq Ch^2 \big[ \left\Vert |\cdot| \nabla \Gcal^z \right\Vert_1 \left( 1 + \left\Vert V \right\Vert_2 \Vert \Rcal^{z, V} \Vert_{2;\infty} \right) \nonumber \\
	&\hspace{20pt} + \left( 1+ \Vert \nabla \Gcal^z \Vert_1 \left\Vert V \right\Vert_2  \right) \Vert [x, \Rcal^{z, V} ] \Vert_{2;\infty} +  \Vert [x,[x, \Rcal^{z, V} ]] \Vert_{2;\infty} \big]. \label{eq:app12}
\end{align}
From Lemma~\ref{lem:App1}~and~\ref{lem:App2}, we know that there is a continuous $a \colon \rho(K_\Tc-V) \to \mathbb{R}_+$ such that the right side of \eqref{eq:app12} is bounded by $a(z)$. In the following we will denote by $a(z)$ a generic function with these properties whose precise form may change from line to line.

To obtain a bound for the operator norm of $\Tcal_\Abold^{(2)}(z)$, we first estimate its kernel by
\begin{align}
	&| \Tcal_{\Abold_h}^{(2)}(x,y) | \leq Ch^2 \Bigl( \int_{\speaker} \text{d}|\zeta| \; \left| \varphi(\zeta) \right|  \Bigr) \sup_{\zeta \in \speaker} \int_{\mathbb{R}^6} \dd v \dd w \; | G_{(\Abold_h)_y}^\zeta(x,v) | \nonumber\\
	&\hspace{0.5cm} \times \big[ | v-y | \ | \nabla g^{\zeta}(v-w)| + \left( |v-y| + |v-y|^2 \right) | g^\zeta(v-w) | \big]  |\Gcal^{z, V}(w, y)|. \label{eq:app13}
\end{align}
From Lemma~\ref{GAz-GtildeAz_decay} we know that the absolute value of the resolvent kernel of the magnetic Laplacian is bounded from above by a function only depending on $x-v$, whose $L^1(\mathbb{R}^3)$-norm is bounded by a constant times  $f( \Re \zeta, \Im \zeta)$ with $f$ in \eqref{g0_decay_f}. This, in particular, implies that this $L^1(\mathbb{R}^3)$-norm is uniformly bounded in $\zeta \in \speaker$ and $h$ as long as the latter is small enough, compare this to \cite[Eq. (6.19)]{DeHaSc2021}. We use this bound, $| v-y | \leq | v-w | + | w - y |$, and the resolvent identity for $\Rcal^{z, V}$ to bound the operator norm of $\Tcal_\Abold^{(2)}$ by
\begin{align}
	\Vert \Tcal_{\Abold_h}^{(2)} \Vert_{\infty} &\leq h^2 C \big(  \sup_{\zeta \in \speaker} \Vert (1+|\cdot|) \nabla g^\zeta \Vert_1 \; , \; \sup_{\zeta \in \speaker} \Vert (1+|\cdot|^2) g^{\zeta} \Vert_1 \; , \; \Vert (1+|\cdot|^2) \Gcal^z \Vert_1 \; , \; \nonumber\\
	&\hspace{50pt} \Vert V \Vert_2 \; , \; \Vert \Rcal^{z, V} \Vert_{2;\infty} \; , \; \Vert [x, \Rcal^{z, V} ] \Vert_{2,\infty} \; , \; \Vert [x, [x, \Rcal^{z, V}] ] \Vert_{2,\infty} \big). \label{eq:app14}
\end{align}
The constant on the right side is an affine function of each of its arguments. From Lemma~\ref{g_decay} we know that the norms involving $g^{\zeta}$ are finite. \ifthenelse{\equal\masterfile{Diss}}{With}{In combination with} Lemmas~\ref{lem:App1} and \ref{lem:App2}, this implies that the right side of \eqref{eq:app14} is bounded by $a(z)$. We conclude that
\begin{align}
	\Vert \Tcal_{\Abold_h}^{z, V} \Vert_\infty \leq a(z)\, h^2
	\label{eq:app15}
\end{align}
holds.

Let $K \subset \rho(K_\Tc-V)$ be compact. The above bounds allow us to find $h_0(K) > 0$ such that for $z \in K$ and as long as $0 < h < h_0(K)$ we can write the resolvent of $K_{\Tc, \Abold}-V$ as
\begin{align}
	\frac{1}{z - (K_{\Tc, \Abold} - V)} = \Scal_{\Abold_h}^{z, V} + h^2 \,  \eta_h(z) \quad \text{ with } \quad \eta_h(z) := h^{-2} \Scal_{\Abold_h}^{z, V} \sum_{n=1}^{\infty} (\Tcal_{\Abold_h}^{z, V})^n.
	\label{eq:app16}
\end{align}
To show that the operator norm of $(1 + \pi^2) \eta_h(z)$ is bounded by $a(z)$, we use 
\begin{align}
	\Vert (1 + \pi^2) \eta_h(z) \Vert_\infty \leq \Vert (1 + \pi^2) \Scal_{\Abold_h}^{z, V} \Vert_\infty \; \sum_{n=1}^{\infty} h^{2(n-1)} a(z)^n. \label{eq:app17}
\end{align} 
With the resolvent identity for $\Rcal^{z, V}$ and Lemma~\ref{lem:App1}, we easily see that the operator norm of $(1 + \pi^2) \Scal_{\Abold_h}^{z, V}$ is bounded by $a(z)$. This proves the claim.
\end{proof}

With the resolvent estimates in Lemma~\ref{lem:Appmain} at hand, we turn to the proof of Proposition~\ref{prop:gauge_invariant_perturbation_theory}. Let $\lambda < 2\Tc$ be an eigenvalue of the operator $K_\Tc - V$. Our assumption on $V$ guarantees that it has finite multiplicity $m \in \mathbb{N}$. We choose $\varepsilon > 0$ such that the ball $B_{\varepsilon}(\lambda) \subseteq \mathbb{C}$ contains no other point of the spectrum of $K_\Tc - V$ than $\lambda$ and define 
\begin{align}
	P(h) :=  \int_{\partial B_{\varepsilon}(\lambda)} \frac{\dd z}{2 \pi \i}  \; \Rcal_\Abold^{z, V} =  \int_{\partial B_{\varepsilon}(\lambda)} \frac{\dd z}{2 \pi \i} \; \Scal_{\Abold_h}^{z, V} + h^2 \int_{\partial B_{\varepsilon}(\lambda)} \frac{\dd z}{2 \pi \i}  \; \eta_h(z). \label{eq:app19}
\end{align}
From Lemma~\ref{lem:Appmain} we know that the operator norm of the second term on the right side is bounded by a constant times $h^2$ provided $h$ is small enough. The integral kernel of the first term is given by
\begin{align}
	\Bigl( \int_{\partial B_{\varepsilon}(\lambda)} \frac{\dd z}{2 \pi \i} \; \Scal_\Abold^{z, V} \Bigr)(x,y) = \e^{\i \Phi_{\Abold}(x,y)} \sum_{i=1}^m u_i(x) \overline{ u_i(y) },
	\label{eq:app20}
\end{align}
where the vectors $\{ u_i \}_{i=1}^m$ span the eigenspace of $\lambda$. Let us denote by $P$ the projection onto that linear space. Using \eqref{eq:app19}, \eqref{eq:app20}, and $|\Phi_{\Abold_h}(x,y)| \leq C h^2 (|x|^2 + |y|^2)$, we obtain the bound
\begin{align}
	\Vert P(h) - P \Vert_{\infty} \leq C h^2 \, \max_{i=1, \ldots ,m} \Vert \, |\cdot|^2 u_i \Vert_2.
	\label{eq:app21}
\end{align}
In combination with Lemma~\ref{lem:regEF}, this proves $\rank P(h) = m$ for $h$ small enough as well as the second bound in \eqref{eq:app1}.

To prove the bounds for the eigenvalues we use the identity 
\begin{align}
	(K_{\Tc, \Abold} - V) P(h) =  \int_{\partial B_{\varepsilon}(\lambda)} \frac{\dd z}{2 \pi \i} \; z \, \Rcal_\Abold^{z, V} . \label{eq:app22}
\end{align}
As long as $h$ is small enough, the rank of this operator equals $m$ and its eigenvalues are given by $\lambda_1(h), \ldots , \lambda_m(h)$. Similar arguments to the above for the spectral projections allow us to conclude that
\begin{align}
	\Vert (K_{\Tc, \Abold} - V) P(h) - (K_\Tc - V) P \Vert_{\infty} \leq  C h^2 \, \max_{i=1, \ldots ,m} \Vert \, |\cdot|^2 u_i \Vert_2 \label{eq:app23}
\end{align}
holds. This proves the claimed bound for the eigenvalues. It remains to prove \eqref{eq:app2}.

Let us write $\alpha_*^{\Abold_h} = a(h) \alpha_* + b(h) \phi_h$ with $\langle \alpha_*, \phi_h \rangle = 0$ and $|a(h)|^2 + |b(h)|^2 = 1$. Our assumptions imply $a(h) = \langle \alpha_* , \alpha_*^{\Abold_h} \rangle \geq 0$. We rewrite the equation $P(h) \alpha_*^{\Abold_h} = \alpha_*^{\Abold_h}$ to see that $b(h) \phi_h = (P(h)-P) \alpha_*^{\Abold_h}$. An application of \eqref{eq:app1} thus implies $|b(h)| \leq C h^2$. Using this, $|a(h)|^2 + |b(h)|^2 = 1$, and the fact that $a(h) \geq 0$, we see that $| a(h) - 1 | \leq C h^2$. This allows us to conclude that
\begin{align}
	\Vert (1+\pi^2) ( \alpha_*^{\Abold_h} - \alpha_* ) \Vert_2 \leq |a(h)-1| \ \Vert (1 + \pi^2) \alpha_* \Vert_2 + | b(h) | \ \Vert (1 + \pi^2) \phi_h \Vert_2 \leq C h^2.
	\label{eq:app24}
\end{align}
To obtain the result we used Lemma~\ref{lem:regEF} to see that $\Vert (1 + \pi^2) \alpha_* \Vert_2 < \infty$, as well as $\Vert (1 + \pi^2) \phi_h \Vert_2 \leq \Vert (1 + \pi^2) \alpha_* \Vert_2 + \Vert (1 + \pi^2) \alpha_*^{\Abold_h} \Vert_2$, and 
\begin{align}
	\Vert (1+\pi^2) \alpha_*^{\Abold_h} \Vert_2 \leq C \Vert K_{\Tc, \Abold} \alpha_*^{\Abold_h} \Vert_2 \leq C \left( | \lambda(h) | + \Vert V \alpha_*^{\Abold_h} \Vert_2 \right) \leq C \left( | \lambda(h) | + \Vert V \Vert_\infty \right).
\end{align}
This proves \eqref{eq:app2} and also finishes the proof of Proposition~\ref{prop:gauge_invariant_perturbation_theory}.


\begin{center}
\textsc{Acknowledgements}
\end{center}

A.~D. gratefully acknowledges funding from the Swiss National Science Foundation through the Ambizione grant PZ00P2 185851.


\vspace{1cm}

\setlength{\parindent}{0em}

(Andreas Deuchert) \textsc{Institut für Mathematik, Universität Zürich}

\textsc{Winterthurerstrasse 190, CH-8057 Zürich}

E-mail address: \href{mailto:  andreas.deuchert@math.uzh.ch}{\texttt{andreas.deuchert@math.uzh.ch}}

\vspace{0.3cm}

(Christian Hainzl) \textsc{Mathematisches Institut der Universität München}

\textsc{Theresienstr. 39, D-80333 München}

E-mail address: \href{mailto: hainzl@math.lmu.de}{\texttt{hainzl@math.lmu.de}}

\vspace{0.3cm}

(Marcel Maier, born Schaub) \textsc{Mathematisches Institut der Universität München}

\textsc{Theresienstr. 39, D-80333 München}


\printbibliography[heading=bibliography, title=Bibliography of Chapter \ref{Chapter:DHS2}]
\end{refsection}


\part{Further Results on BCS Theory}
\label{Part:Further_BCS}


\begin{refsection}

\chapter{The Abrikosov Gauge for Periodic Magnetic Fields}
\label{Chapter:Abrikosov_gauge} \label{CHAPTER:ABRIKOSOV_GAUGE}

This chapter is devoted to the question whether the magnetic potentials that are covered in the works presented in Chapters \ref{Chapter:DHS1} and \ref{Chapter:DHS2} are exhaustive to understand the periodic BCS model in full generality --- given that the external magnetic field is fixed. Therefore, we have to analyze the possible choices of magnetic potentials corresponding to external periodic magnetic fields. The answer is that, as long as the magnetic field is smooth, the magnetic potential can always be chosen such that it admits the form discussed in Chapter~\ref{Chapter:DHS2}.


\section{Introduction}

There are a number of gauges for magnetic fields, and this note concerns a useful gauge for periodic magnetic fields, which does not seem to appear in the literature as such, at least in three dimensions. The key characteristic of this gauge is that the magnetic potential is the sum of two terms, the first corresponding to a constant magnetic field whose strength represents the average magnetic field per unit cell, and the second term being a periodic perturbation.

In this note, we will assume, for the sake of simplicity, that all functions are smooth. We now state our main result.

\begin{theorem}
\label{Thm:Main_Result}
Consider a magnetic potential $A \colon \R^d \to \R^d$, $d = 2, 3$, such that the
corresponding magnetic field $B = \Curl A$ is $\Lat$-periodic,
where $\Lat = r \Z^d$ for some $r > 0$.
Let $b$ be the average magnetic flux per unit cell, i.e.,
\begin{align}
    b = r^{-d} \int_{\Omega} \dd x \; B(x), \label{GF:eq:Main_Result_avg_flux_vector}
\end{align}
where $\Omega = [0, r]^d$ is the unit cell of $\Lat$.

Then $A$ is gauge equivalent to $A_b + a$, where
\begin{align}
A_b(x) = 
\begin{cases}
    \frac{b}{2} x^\perp, & d = 2, \\
    \frac{1}{2} b \wedge x, & d = 3,
\end{cases} \label{GF:eq:Main_Result_normal-A}
\end{align}
and $a \colon \R^d \to \R^d$ is $\Lat$-periodic.
More precisely, there is an $\eta \colon \R^d \to \R$ such that
\begin{align}
    A(x) + \nabla\eta(x) = A_b(x) + a(x). \label{GF:eq:Main_Result_eta}
\end{align}
\end{theorem}

In two dimensions this gauge is useful for the study of Abrikosov lattice solutions of the Ginzburg-Landau equations and there are number of proofs (see \cite{bgt, dutour, ts} and references therein). The proofs and details presented here are very similar to those in \cite{bgt, dutour}, although the overall point of view is somewhat different and the results are generalized to three dimensions. There are also many parallels with the theory of complex line bundles over tori and Chern classes (this is a very large field but see \cite{gunning, mumford} for an introduction to this area).

It should be noted that this gauge is compatible with the Coulomb gauge, and indeed it can be proven that a further gauge transformation allows to assume that the periodic perturbation $a$ is divergence-free. It is also possible to use a change of coordinates (via translation) to assume that $a$ has mean zero. These statements are proven in the Abrikosov lattice papers mentioned above.

As mentioned above we assume all functions are smooth but the definitions and proofs below can be extended to the case where less regularity is required.


\section{Admissible families of gauge transformations}
\label{sec:adm-fam}

Our main focus is to study certain families of gauge transformations, which arise from magnetic potentials $A \colon \R^d \to \R^d$, whose magnetic field $\Curl A$ is $\Lat$-periodic for some lattice $\Lat$. 

To begin with, the periodicity of $\Curl A$ implies
\begin{align}
    \Curl \left( A(x + t) - A(x) \right) = 0, \label{GF:eq:Motivation_1}
\end{align}
for all $t \in \Lat$, which in turn implies that there is a function
$g_t \colon \R^d \to \R$ such that
\begin{align}
    A(x + t) = A(x) + \nabla g_t(x). \label{GF:eq:Motivation_2}
\end{align}
Let $t, s \in \Lat$. On the one hand, \eqref{GF:eq:Motivation_2} implies
\begin{align*}
    A(x + t + s) = A(x) + \nabla g_{t+s}(x),
\end{align*}
and, on the other hand, \eqref{GF:eq:Motivation_2} also gives
\begin{align*}
    A(x + t + s) = A(x + t) + \nabla g_s(x + t)
    = A(x) + \nabla g_t(x) + \nabla g_s(x + t),
\end{align*}
Therefore,
\begin{align}
	\nabla g_{t+s}(x) - \nabla g_t(x) - \nabla g_s(x + t) = 0. \label{GF:eq:Motivation_3}
\end{align}
This property of the family of maps $g_t$ is, as we shall see,
very important to understanding the properties of the magnetic field
and motivates the following definition.

\begin{definition}
\label{Def:Admissible}
We call a family $g = \{g_t\}_{t\in \Lat}$ of gauge transformations $g_t \colon \R^d \to \R$ \emph{admissible} if for all $t,s \in \Lat$ there are constants $\Ical_g(t, s) \in \R$ such that for all $x \in \R^d$,
\begin{align}
    g_{t+s}(x) - g_t(x) - g_s(x + t) = \Ical_g(t, s). \label{GF:eq:cocyle} 
\end{align}
The set of admissible families is denoted by $\Gd$.
\end{definition}

As an immediate consequence of the Definition, we note that
\begin{align}
\Ical_{\alpha g + h}(t, s) = \alpha \, \Ical_g(t, s) + \Ical_h(t, s) \label{GF:eq:I_linearity}
\end{align}
for $\alpha\in \R$, $g,h\in \Gd$, and $t,s\in \Lat$.


The most important basic fact about $\Gd$ is that it admits a real vector space structure via the usual pointwise operations. A further important property of $\Gd$ is that it is closed under the following operation.

\begin{lemma}
\label{GF:Lemma:Perturbation_by_eta}
Let $g \in \Gd$ and let $\eta\colon \R^d \to \R$ be any function. Then the family $h = \{h_t\}_{t\in \Lat}$ defined by $h_t(x) := g_t(x) + \eta(x+t) - \eta(x)$ belongs to $\Gd$.
\end{lemma}

\begin{proof}
Fix $t, s \in \Lat$ and calculate
\begin{align*}
h_{t+s}(x) - h_t(x) - h_s(x+t) &= g_{t+s}(x) + \eta(x+t+s) - \eta(x) \\
&\qquad - g_t(x) - \eta(x+t) + \eta(x) \\
&\qquad - g_s(x+t) - \eta(x+t+s) + \eta(x+t) \\
&= g_{t+s}(x) - g_t(x) - g_s(x+t) = \Ical_g(t,s).
\end{align*}
This proves the statement.
\end{proof}

In view of Lemma \ref{GF:Lemma:Perturbation_by_eta}, we may now introduce a notion of gauge-equivalence that is related to the gauge transformation of magnetic potentials as we shall see below.

\begin{definition}
\label{GF:Def:Equivalence}
We say that two admissible families $g, h\in \Gd$ are gauge-equivalent and write $g \sim h$ if there is a function $\eta \colon \R^d \to \R$ and a family $\alpha = \{\alpha_t\}_{t\in \Lat}$ of constants $\alpha_t\in \R$ such that
\begin{align}
h_t(x) = g_t(x) + \eta(x+t) - \eta(x) + \alpha_t \label{GF:Def:Equivalence_1}
\end{align}
holds for all $x\in \R^d$ and $t\in \Lat$.
\end{definition}

It is easy to verify that $\sim$ is an equivalence relation and that it is compatible with the vector space structure on $\Gd$. We can therefore study the quotient space
\begin{align}
\Hd := \Gd|_\sim \label{GF:eq:Quotient_space}
\end{align}
In order to do this we study the associated bilinear form described in the following proposition.

\begin{proposition}
\label{GF:Prop:Bilinear_form}
\begin{enumerate}[(a)]
\item For any $g \in \Gd$, the expression
\begin{align}
    \Dcal_g(t,s) := \Ical_g(t, s) - \Ical_g(s, t) \label{GF:eq:Bilinear_form}
\end{align}
defines an antisymmetric bilinear form on $\Lat \times \Lat$.

\item If $g,h\in \Gd$ with $g\sim h$, then $\Dcal_g = \Dcal_h$.
\end{enumerate}
\end{proposition}

\begin{proof}
It is immediate from the definition that $\Dcal_g$ is antisymmetric. Therefore, it suffices to show linearity in the second slot to conclude bilinearity.

To prove linearity in the second slot, we first prove
\begin{align}
\Dcal_g(t, -s) = -\Dcal_g(t, s). \label{GF:eq:Bilinear_form_proof3}
\end{align}
To see this, we apply \eqref{GF:eq:cocyle} with $x = s$ to $\Ical_g(t, -s)$ and with $x = t+s$ to $\Ical_g(-s, t)$. This implies
\begin{align}
\Dcal_g(t, -s) &= g_{t-s}(s) - g_t(s) - g_{t-s}(t+s) + g_t(t). \label{GF:eq:Bilinear_form_proof1}
\end{align}
Furthermore, \eqref{GF:eq:cocyle} with $t \equiv s$ and $s \equiv t-s$ implies
\begin{align*}
g_t(x) - g_s(x) - g_{t-s}(x+s) = \Ical_g(s, t-s).
\end{align*}
Applying this with $x = 0$ and $x = t$ yields
\begin{align*}
g_{t-s}(s) - g_{t-s}(t+s) = g_t(0) - g_s(0) - g_t(t) + g_s(t).
\end{align*}
We insert this into \eqref{GF:eq:Bilinear_form_proof1}, add and subtract $g_{t+s}(0)$, and obtain
\begin{align*}
\Dcal_g(t, -s) = - \bigl( g_{t+s}(0) - g_t(0) - g_s(t) \bigr) + g_{t+s}(0) - g_s(0) - g_t(s),
\end{align*}
which proves \eqref{GF:eq:Bilinear_form_proof3}, see \eqref{GF:eq:cocyle}. 

In the next step, we show that 
\begin{align}
\Dcal_g(t, s+r) = \Dcal_g(t,s) + \Dcal_g(t, r). \label{GF:eq:Bilinear_form_proof4}
\end{align}
To see this, we first note that for any $x\in \R^d$, \eqref{GF:eq:cocyle} implies
\begin{align}
\Dcal_g(t, s) = g_{t}(x + s) - g_t(x) - g_s(x + t) + g_s(x). \label{GF:eq:Bilinear_form_proof5}
\end{align}
Therefore,
\begin{align}
\Dcal_g(t, s+r) &= g_t(s + r) - g_t(0) - g_{s+r}(t) + g_{s+r}(0) \label{GF:eq:Bilinear_form_proof2}
\end{align}
When we apply \eqref{GF:eq:cocyle} once more, we obtain
\begin{align*}
g_{s+r}(0) - g_{s+r}(t) = g_r(s) + g_s(0) - g_s(t) - g_r(t + s) 
\end{align*}
We insert this into \eqref{GF:eq:Bilinear_form_proof2} and add and subtract $g_t(s)$. This yields
\begin{align*}
\Dcal_g(t, s+r) &= g_t(s) - g_t(0) - g_s(t) + g_s(0) \\
&\hspace{40pt} + g_t(s + r) - g_t(s) - g_r(s + t) + g_r(s),
\end{align*}
which proves \eqref{GF:eq:Bilinear_form_proof4}. An induction argument on \eqref{GF:eq:Bilinear_form_proof3} and \eqref{GF:eq:Bilinear_form_proof4} then shows that $\Dcal_g$ is linear in the second slot. This proves part (a).

To prove part (b), we have $h_t(x) = g_t(x) + \eta(x + t) - \eta(x) + \alpha_t$ by hypothesis for some $\eta\colon \R^d \ra \R$ and $\alpha_t\in \R$. When we use \eqref{GF:eq:Bilinear_form_proof5} with $x =0$, we see that all occurrences of $\alpha$ drop out and all terms with $\eta$ cancel. Therefore, $\Dcal_h = \Dcal_g$. This completes the proof.
\end{proof}

We let $\Lambda_d$ denote the vector space of real antisymmetric bilinear forms on $\Lat \times \Lat$. It is well known that $\Lambda_d$ is isomorphic to $\R^n$, where $n =\binom{d}{2} = \frac 12 d (d-1)$. Moreover, a straightforward computation shows that any antisymmetric bilinear form $\Dcal$ on $\Lat \times \Lat$ satisfies
\begin{align}
\Dcal(t, s) = \begin{cases} \mathbf d \; (t\wedge s) , & d = 2 , \\ \dbold \cdot (t\wedge s) , & d =3, \end{cases} \label{GF:eq:Bilinear_form_representation}
\end{align}
where
\begin{align}
\dbold := \begin{cases} \Dcal(\tau_1, \tau_2), & d = 2, \\ (\Dcal(\tau_2, \tau_3) , \Dcal(\tau_3, \tau_1), \Dcal(\tau_1, \tau_2))^t, & d = 3. \end{cases} \label{GF:eq:Bilinear_form_generator}
\end{align}
Here, $\tau_i$, $i = 1, \ldots, d$ are the basis vectors spanning $\Lat$.

We can now prove the main tool needed for the proof of Theorem \ref{Thm:Main_Result}.

\begin{theorem}
\label{GF:Thm:Isomorphic}
We have $\Hd \cong \Lambda_d$, where an isomorphism is given by the map $[g] \mapsto \Dcal_g$, and therefore
\begin{align*}
\Hd \cong \R^{\binom{d}{2}}.
\end{align*}
\end{theorem}

\begin{proof}
By Proposition \ref{GF:Prop:Bilinear_form} the map $[g] \mapsto \Dcal_g$ is well-defined and, by \eqref{GF:eq:I_linearity}, it is linear. Therefore we only need to show that it is injective and surjective. We begin with the latter.

Let $\Dcal$ be an antisymmetric bilinear form on $\Lat \times \Lat$ and define
\begin{align*}
g_t(x) := \frac 12 \, \Dcal(t, x),
\end{align*}
where $\Dcal(t, x)$ is understood to be defined via linear extension, see \eqref{GF:eq:Bilinear_form_representation}. Then,
\begin{align*}
g_{t+s}(x) - g_t(x) - g_s(x + t) &= \frac 12 \bigl( \Dcal(t + s, x) - \Dcal(t, x) -  \Dcal(s, x + t)\bigr) \\
&= \frac 12 \, \Dcal(t, s),
\end{align*}
which shows that $g\in \Gd$ with $\Ical_g(t,s) = \frac 12 \Dcal(t,s)$. It follows that $\Dcal_g = \Dcal$, since $\Dcal_g(t, s) = \frac 12 \Dcal(t, s) - \frac 12 \Dcal(s, t) = \Dcal(t, s)$.

We turn to the injectivity of the map $[g] \mapsto \Dcal_g$.
Let $g\in \Gd$ be such that $\Dcal_g = 0$. We need to show that $g$ is gauge-equivalent to $0$, i.e., that there are $\eta \colon \R^d \to \R$
and constants $\alpha_t \in \R$ such that
\begin{align}
\eta(x+t) - \eta(x) + \alpha_t &= g_t(x), & t &\in \Lat , \; x\in \Rbb^d. \label{GF:eq:Isomorphic_proof1}
\end{align}
To start out with, we let $\Lat$ be equipped with the graph norm, i.e., if $t = \sum_{i=1}^d n_i \tau_i$, then $\Vert t\Vert := \sum_{i=1}^d |n_i|$. We prove \eqref{GF:eq:Isomorphic_proof1} by induction in $n = \Vert t\Vert$. Let first $n =0$, i.e., $t=0$. In this case, \eqref{GF:eq:cocyle} shows that $g_0$ is constantly equal to $\Ical_g(0,0)$, so \eqref{GF:eq:Isomorphic_proof1} holds with $\alpha_0 := \Ical_g(0,0)$. For the case $\Vert t \Vert = 1$ we note that the functions $g_{\tau_1}, \ldots, g_{\tau_n}$ satisfy the hypothesis of Proposition~\ref{GF:prop:func-eq} in \ifthenelse{\equal\masterfile{Diss}}{Section}{Appendix}~\ref{GF:AppendixA} since $\Dcal_g =0$. Therefore, there is a function $\eta\colon \R^d \ra \R$ such that
\begin{align}
  \eta(x + \tau_i) - \eta(x) &= g_{\tau_i}(x), & i &= 1, \dots, d, \label{GF:eq:Isomorphic_proof2}
\end{align}
where $\tau_i = re_i$ are the basis vectors spanning $\Lat$. Hence, \eqref{GF:eq:Isomorphic_proof1} holds with $\alpha_{\tau_i} = 0$. Likewise, we have
\begin{align*}
\eta(x - \tau_i) - \eta(x) = -\bigl( \eta(x - \tau_i + \tau_i) - \eta(x - \tau_i) \bigr) = - g_{\tau_i}(x - \tau_i)
\end{align*}
and, by \eqref{GF:eq:cocyle} applied to $t=-\tau_i$ and $s = \tau_i$, we obtain
\begin{align*}
- g_{\tau_i}(x - \tau_i) = g_{-\tau_i}(x) + \Ical_g(-\tau_i, \tau_i) - g_0(x).
\end{align*}
Since $g_0(x) = -\Ical_g(0,0)$ for all $x$ as argued above, we see that \eqref{GF:eq:Isomorphic_proof1} holds with
\begin{align*}
\alpha_{-\tau_i} := -\Ical_g(0,0) - \Ical_g(-\tau_i, \tau_i).
\end{align*}
This proves \eqref{GF:eq:Isomorphic_proof1} for all vectors $t\in \Lat$ with $\Vert t\Vert = 1$. By induction we assume that \eqref{GF:eq:Isomorphic_proof1} holds for all vectors $t\in \Lat$ with $\Vert t \Vert \leq n$. Let $s \in \Lat$ with $\Vert s\Vert = n+1$. Then, there is $t\in \Lat$ with $\Vert t\Vert = n$ and $i\in \{1, \ldots, d\}$ such that $s = t \pm \tau_i$ and, by \eqref{GF:eq:cocyle} and \eqref{GF:eq:Bilinear_form_proof5}, a short computation shows
\begin{align}
\eta(x + s) - \eta(x) + C_{t, \pm \tau_i} = g_s(x), \label{GF:eq:Isomorphic_proof3}
\end{align}
with
\begin{align*}
C_{t, \tau_i} &:= \alpha_t + \Ical_g(\tau_i, t), & C_{t, -\tau_i} &:= \alpha_t + \alpha_{-\tau_i} + \Ical_g(-\tau_i, t).
\end{align*}
In order to prove \eqref{GF:eq:Isomorphic_proof1}, it remains to show that $C_{t, \pm \tau_i}$ does not depend on the representation $s = t \pm \tau_i$ but only on $s$. However, this follows from the argument above because if $s = t' \pm \tau_j$ is another such decomposition, then \eqref{GF:eq:Isomorphic_proof3} implies that $C_{t, \pm \tau_i} = C_{t', \pm \tau_j}$. Hence, when we set
\begin{align*}
\alpha_s := C_{t, \pm \tau_i}
\end{align*}
then \eqref{GF:eq:Isomorphic_proof1} holds and the induction is complete.
\end{proof}


\section{Magnetic fields and admissible families}

We can now apply the previous section to magnetic fields and prove the main theorem. We start with the following result.

\begin{proposition}
\label{GF:Prop:Existence_of_g}
Suppose that $A$ is a magnetic potential such that $\Curl A$ is $\Lat$-periodic. Then there is an admissible family $g\in \Gd$ such that for all $t \in \Lat$,
\begin{align}
    A(x + t) = A(x) + \nabla g_t(x). \label{GF:eq:Existence_of_g_eq1}
\end{align}
Moreover, the corresponding antisymmetric bilinear form $\Dcal_g$ satisfies \eqref{GF:eq:Bilinear_form_representation} with $\dbold = b$, where $b$ is the average magnetic flux defined in \eqref{GF:eq:Main_Result_avg_flux_vector}.
\end{proposition}

\begin{proof}
We have already seen the proof of \eqref{GF:eq:Existence_of_g_eq1} in \eqref{GF:eq:Motivation_1}-\eqref{GF:eq:Motivation_3}. 
%
It remains to show that $\dbold = b$. When $d = 2$, we use \eqref{GF:eq:Main_Result_avg_flux_vector} to see that $b r^2$ equals
\begin{align*}
&\int_0^r \dd x_1\, \int_0^r \dd x_2 \; \left( \partial_1 A_2 (x_1, x_2) - \partial_2 A_1(x_1, x_2) \right) \\
&\hspace{10pt}= \int_0^r \dd x_2\; \left( A_2(r, x_2) - A_2(0, x_2) \right) - \int_0^r \dd x_1\; \left( A_1(x_1, r) - A_1(x_1, 0) \right) \\
&\hspace{30pt}= \int_0^r \dd x_2\; \partial_2 g_{\tau_1} (0, x_2) - \int_0^r \dd x_1\; \partial_1 g_{\tau_2}(x_1, 0).
\end{align*}
The last equality follows from \eqref{GF:eq:Existence_of_g_eq1}. We integrate this and use \eqref{GF:eq:cocyle} to conclude that
\begin{align*}
b\, r^2 = g_{\tau_1}(\tau_2) - g_{\tau_1}(0) - g_{\tau_2}(\tau_1) + g_{\tau_2}(0) = \Dcal_g(\tau_1, \tau_2) = \dbold \, (\tau_1\wedge \tau_2).
\end{align*}
Since $\tau_1\wedge \tau_2 = r^2$, the claim follows.

When $d = 3$, a similar calculation shows that $b_1r^3$ equals
\begin{align*}
&\int_0^r \dd x_1  \int_0^r \dd x_2 \int_0^r \dd x_3\; \left( \partial_2 A_3 (x_1, x_2, x_3) - \partial_3 A_2(x_1, x_2, x_3) \right) \\
%
%
%
%
%
&\hspace{10pt} = r \int_0^1 \dd x_1\, \left( g_{\tau_2}(x_1 \tau_1 + \tau_3) - g_{\tau_2}(x_1 \tau_1) - g_{\tau_3}(x_1 \tau_1 + \tau_2) + g_{\tau_3}(x_1 \tau_1) \right) \\
&\hspace{10pt} = r\int_0^1 \dd x_1\, \Dcal_g(\tau_2, \tau_3).
\end{align*}
This proves that $b_1 = \dbold_1$. The proof for $b_2=\dbold_2$ and $b_3=\dbold_3$ is analogous. This completes the proof of the proposition.
\end{proof}

%

We are in position to prove Theorem \ref{Thm:Main_Result}. Let $A$ be as in the theorem and let $g\in \Gd$ be a corresponding admissible family. Then, we know that the bilinar form $\Dcal_g$ satisfies \eqref{GF:eq:Bilinear_form_representation} with $\dbold$ replaced by
\begin{align}
\dbold_g := \begin{cases} \Dcal_g(\tau_1, \tau_2), & d = 2, \\ (\Dcal_g(\tau_2, \tau_3) , \Dcal_g(\tau_3, \tau_1), \Dcal_g(\tau_1, \tau_2))^t, & d = 3. \end{cases} \label{GF:eq:Bilinear_form_generator_g}
\end{align}
By Proposition \ref{GF:Prop:Existence_of_g}, we have $\dbold_g = b$, where $b$ is defined in \eqref{GF:eq:Main_Result_avg_flux_vector}. On the other hand, $\Curl A_b = b$, which means that the corresponding magnetic field is constant and therefore periodic. If $h\in \Gd$ denotes an admissible family corresponding to $A_b$, then the antisymmetric bilinear form $\Dcal_h$ satisfies \eqref{GF:eq:Bilinear_form_representation} with $\dbold_h = b$, too, because of \eqref{GF:eq:Main_Result_avg_flux_vector}.
By Theorem \ref{GF:Thm:Isomorphic}, it follows that $g$ and $h$ are gauge-equivalent, i.e., there are a function $\eta \colon \R^d \to \R$ and constants $\alpha_t\in \Rbb$ and such that \eqref{GF:Def:Equivalence_1} holds. We consider the magnetic potential $\tilde{A} = A + \nabla\eta$ and see that
\begin{align*}
    \tilde{A}(x + t) = \tilde{A}(x) + \nabla g_t(x) + \nabla \eta(x+t)
    - \nabla \eta(x) = \tilde A(x) + \nabla h_t(x),
\end{align*}
which means that $h$ is an admissible family associated to $\tilde{A}$. Therefore, the magnetic potential $a := A_b - \tilde{A}$ is $\Lat$-periodic by Proposition \ref{GF:Prop:Existence_of_g} applied to $A_b$. This proves Theorem \ref{Thm:Main_Result}.


\section{Functional align}
\label{GF:AppendixA}

In this appendix we solve the following functional align. The idea of the proof has been sketched in \cite{dutour}.

\begin{proposition}
\label{GF:prop:func-eq}
Let $\Lat = r\Z^d$ and let $\Omega = [0, r]^d$. Suppose that $g_1, \ldots, g_d$ are smooth functions satisfying
\begin{align}
g_i(x + \tau_j) - g_i(x) - g_j(x + \tau_i) - g_j(x) &= 0 , & i,j &= 1, \ldots, d, \; x\in \Rbb^d, \label{GF:eq:functional-equality_assumption}
\end{align}
where $\tau_i = re_i \in \Lat$. Then, the problem
\begin{align*}
  \eta(x + \tau_i) - \eta(x) &= g_{\tau_i}(x), & i &= 1, \dots, d,
\end{align*}
has a smooth solution $\eta$.
\end{proposition}

In order to prove this proposition we need the following result. For a set $M\subseteq \Rbb^d$ and $\varepsilon >0$, we define the open $\varepsilon$-fattening of $M$ as
\begin{align}
M_\varepsilon := \bigl\{ x\in \Rbb^d : \dist(x, M) < \varepsilon\bigr\}.
\end{align}

\begin{theorem}
\label{GF:lem:reduced-functional-equality}
Let $\Lat = r\Z^d$ and let $\Omega = [0, r)^d$. Suppose that $g_1, \ldots, g_d$ are smooth functions satisfying
\begin{align}
g_i(x + \tau_j) - g_i(x) - g_j(x + \tau_i) - g_j(x) &= 0 , & i,j &= 1, \ldots, d, \; x\in \Rbb^d. \label{GF:eq:reduced-functional-equality_assumption1}
\end{align}
Assume further that there is an $x_0 \in \R^d$ and $\varepsilon >0$ such that for all $i =1, \ldots, d$,
\begin{align}
g_i \big|_{x_0 + (\partial\Omega)_\varepsilon + \Lat} \equiv 0. \label{GF:eq:reduced-functional-equality_assumption2}
\end{align}
Then the problem
\begin{align}
\eta(x + \tau_i) - \eta(x) &= g_i(x), & i &= 1, \dots, d, \; x\in \Rbb^d, \label{GF:eq:reduced-functional-equality}
\end{align}
has a smooth solution $\eta$.
\end{theorem}

\begin{proof}
For $n = 1, \ldots, d$, we set
\begin{align*}
U_n := \mathrm{span}\{ \tau_i : i = 1, \ldots, d , \; i \neq n\}
\end{align*}
for the subspace of $\Rbb^d$ generated by all the $\tau_i$'s but $\tau_n$ and
\begin{align}
S_n := U_n + [0,1) \tau_n
\label{GF:lem:reduced-functional-equality_2}
\end{align}
for the corresponding strip. 

Our strategy is to define functions $\eta_n$, $n = 1, \ldots, d$, on $\Rbb^d$ recursively, show that we have
\begin{align}
\eta_n(x + \tau_i) - \eta_n(x) &= g_i(x), & i = 1, \ldots, n , \; x\in \Rbb^d, \label{GF:lem:reduced-functional-equality_3}
\end{align}
and prove that $\eta_n$ is smooth. Then, \eqref{GF:eq:reduced-functional-equality} is satisfied by the function $\eta := \eta_d$. We prove \eqref{GF:lem:reduced-functional-equality_3} by induction in $n$. 

To start out with, we put
\begin{align}
\eta_1\big|_{x_0 + S_1} := 0. \label{GF:lem:reduced-functional-equality_1}
\end{align}
For $x\in \Rbb^d$, we uniquely decompose $x = y + \ell \tau_1$, where $y\in x_0 + S_1$ and $\ell \in \Zbb$. Then, we recursively define
\begin{align}
\eta_1(x) &:= \begin{cases} \eta_1(x - \tau_1) + g_1(x - \tau_1), & \ell > 0, \\ \eta_1(x + \tau_1) - g_1(x), & \ell < 0.\end{cases} \label{GF:lem:reduced-functional-equality_4}
%
\end{align}
With these definitions, we claim that \eqref{GF:lem:reduced-functional-equality_3} holds for $n = 1$. To see this, decompose $x\in S_1$ uniquely as $x = y + \ell\tau_1$ with $y\in x_0 + S_1$ and $\ell\in \Zbb$. If $\ell \geq 0$, then, by \eqref{GF:lem:reduced-functional-equality_4}, we have
\begin{align}
\eta_1(x + \tau_1) - \eta_1(x) &= \eta_1(y + (\ell + 1) \tau_1) - \eta_1(x) \notag \\
&= \eta_1(y + \ell \tau_1) + g_1(y + \ell \tau_1) - \eta_1(x) = g_1(x). \label{GF:lem:reduced-functional-equality_9}
\end{align}
If $\ell < 0$, then \eqref{GF:lem:reduced-functional-equality_4} implies
\begin{align}
\eta_1(x +  \tau_1) - \eta_1(x) &= \eta_1(x + \tau_1) - \eta_1(y + \ell \tau_1) \notag \\
&= \eta_1(x + \tau_1) - \bigl( \eta_1(y + (\ell + 1)\tau_1) - g_1(y + \ell \tau_1) \bigr) = g_1(x). \label{GF:lem:reduced-functional-equality_10}
\end{align}
This proves \eqref{GF:lem:reduced-functional-equality_3} for $n = 1$.

The next step is to show that $\eta_1$ is smooth on $\Rbb^d$. It is clear that $\eta_1$ smooth everywhere except on $x_0 + U_1 +  \tau_1 \Zbb$. To prove this as well, we show that for all $x\in x_0 + U_1$, all $m\in \Nbb_0$ such that $g_1^{(m)}$ exists and is continuous, and all $\ell \in \Zbb$, we have
\begin{align}
\lim_{a \nearrow \ell} \eta_1^{(m)}(x + a \tau_1) = 0 = \lim_{a \searrow \ell} \eta_1^{(m)}(x + a \tau_1 ). \label{GF:lem:reduced-functional-equality_13}
\end{align}
We prove this by induction in both directions and start with the direction $\ell \in \Nbb$. Let first $\ell = 1$. By \eqref{GF:lem:reduced-functional-equality_1}, we trivially have $\lim_{a   \nearrow 1} \eta_1^{(m)} (x + a \tau_1) =0$, since $x + a\tau_1 \in x_0 + S_1$ for all $a\in [0,1)$. By using \eqref{GF:lem:reduced-functional-equality_3}, we also have
\begin{align*}
\lim_{a\searrow 1} \eta_1^{(m)}(x + a\tau_1) &= \lim_{a\searrow 0} \eta_1^{(m)}(x + (a + 1)\tau_1) = \lim_{a\searrow 0} \eta_1^{(m)}(x + a\tau_1) + \lim_{a\searrow 0} g_1^{(m)}(x + a\tau_1).
\end{align*}
The first limit is zero because of \eqref{GF:lem:reduced-functional-equality_1}. We claim that the second limit vanishes as well. To see this, we note that $U_1 \subseteq \partial \Omega + \Lat$ as well as $(\partial \Omega + \Lat)_\varepsilon = (\partial \Omega)_\varepsilon + \Lat$. It follows that
\begin{align}
x + (a + k)\tau_1 &\in x_0 + (\partial \Omega)_\varepsilon + \Lat, & a &\in [-\varepsilon, \varepsilon], \; k\in \Zbb. \label{GF:lem:reduced-functional-equality_14}
\end{align}
whence, by \eqref{GF:eq:reduced-functional-equality_assumption2}, $g_1(x + a\tau_1) = 0$ for $0 \leq a \leq \varepsilon$.

By induction, we assume that \eqref{GF:lem:reduced-functional-equality_13} holds for $\ell \in \Nbb$. Then, using \eqref{GF:lem:reduced-functional-equality_3}, we have
\begin{align*}
\lim_{a\nearrow \ell + 1} \eta_1^{(m)}(x + a\tau_1) = \lim_{a\nearrow \ell} \eta_1^{(m)}(x + (a+1)\tau_1) = \lim_{a\nearrow \ell} \eta_1^{(m)}(x + a\tau_1) +  \lim_{a\nearrow \ell} g_1^{(m)}(x + a\tau_1)
\end{align*}
The first limit vanishes by induction and the second vanishes because of \eqref{GF:lem:reduced-functional-equality_14} and \eqref{GF:eq:reduced-functional-equality_assumption2}. In the same manner, we see that $\lim_{a\searrow \ell + 1} \eta_1^{(m)} (x + a\tau_1) =0$. Analogously, the reader may prove \eqref{GF:lem:reduced-functional-equality_13} for $\ell =0$ and $\ell\in -\Nbb$. This completes the construction of $\eta_1$.

Suppose by induction that smooth functions $\eta_1, \ldots, \eta_{n-1}$ have been constructed on $\Rbb^n$ such that \eqref{GF:lem:reduced-functional-equality_3} holds for $n-1$. We are going to define $\eta_n$ on $\Rbb^d$ and show \eqref{GF:lem:reduced-functional-equality_3} for $n$. To do this, we first define
\begin{align}
\eta_n\big|_{x_0 + S_n} := \eta_{n-1}\big|_{x_0 + S_n}. \label{GF:lem:reduced-functional-equality_5}
\end{align}
Furthermore, let $x\in \Rbb^d$ and decompose uniquely $x = y + \ell \tau_n$, where $y\in x_0 + S_n$ and $\ell\in \Zbb$. Then, we define recursively,
\begin{align}
\eta_n(x) &:= \begin{cases} \eta_n(x - \tau_1) + g_n(x - \tau_n), & \ell > 0, \\ \eta_n(x + \tau_n) - g_n(x), & \ell < 0.\end{cases} \label{GF:lem:reduced-functional-equality_6}
\end{align}
With these definitions, a simple computation similar to \eqref{GF:lem:reduced-functional-equality_9} and \eqref{GF:lem:reduced-functional-equality_10} shows that
\begin{align}
\eta_n(x + \tau_n) - \eta_n(x) &= g_n(x), & x &\in \Rbb^d. \label{GF:lem:reduced-functional-equality_8}
\end{align}
We claim that this implies
\begin{align}
\eta_n(x + \ell \tau_n) - \eta_n(x) &= \begin{dcases}
\sum_{j=1}^\ell g_n(x + (j-1)\tau_n), & \ell \geq 0, \\ - \sum_{j=1}^{-\ell} g_n(x - j \tau_n), & \ell < 0.
\end{dcases} \label{GF:lem:reduced-functional-equality_11}
\end{align}
The proof of \eqref{GF:lem:reduced-functional-equality_11} is a simple induction argument using \eqref{GF:lem:reduced-functional-equality_8}. 

With this, we are in position to prove that
\begin{align}
\eta_n(x + \tau_i) - \eta_n(x) &= g_i(x), & i &= 1, \ldots, n-1 , \; x\in \Rbb^d. \label{GF:lem:reduced-functional-equality_12}
\end{align}
If this holds, then the proof of \eqref{GF:lem:reduced-functional-equality_3} is completed. To prove \eqref{GF:lem:reduced-functional-equality_12}, let $x\in \Rbb^d$ be given and, once more, choose unique $y\in x_0 + S_n$ and $\ell\in \Zbb$ such that $x = y + \ell \tau_n$. Then, we have $y, y + \tau_i \in x_0 + S_n$, whence $\eta_n(y + \tau_i) - \eta_n(y) = \eta_{n-1}(y + \tau_i) - \eta_{n-1}(y)$ by \eqref{GF:lem:reduced-functional-equality_5}. Therefore, the induction hypothesis \eqref{GF:lem:reduced-functional-equality_3} implies $\eta_{n-1}(y + \tau_i) - \eta(y) = g_i(y)$, whence by \eqref{GF:lem:reduced-functional-equality_11}, we obtain
\begin{align*}
\eta_n(x + \tau_i) - \eta_n(x) &= \eta_{n-1}(y + \tau_i) - \eta_{n-1}(y) + \eta_n(y + \tau_i + \ell \tau_n) - \eta_n(y + \tau_i) \\
&\hspace{180pt}- \bigl( \eta_n(y + \ell \tau_n) - \eta_n(y)\bigr)\\
&\hspace{-50pt}= g_i(y) + \begin{dcases} \sum_{j=1}^\ell g_n(y + \tau_i + (j - 1) \tau_n) - g_n(y + (j-1)\tau_n), & \ell \geq 0, \\ \sum_{j=1}^{-\ell} g_n(y + \tau_i - j \tau_n) - g_n(y -j \tau_n+ \tau_i), & \ell < 0. \end{dcases}
\end{align*}
A simple induction argument using the hypothesis \eqref{GF:eq:reduced-functional-equality_assumption1} shows that
\begin{align*}
g_i(x) + \sum_{j=1}^\ell g_n(x + \tau_i + (j - 1) \tau_n) - g_n(x + (j-1)\tau_n) = g_i(x + \ell \tau_n)
\end{align*}
for $\ell \geq 0$, as well as
\begin{align*}
g_i(x) + \sum_{j=1}^{-\ell} g_n(x - j\tau_n) - g_n(x - j\tau_n + \tau_i) = g_i(x + \ell \tau_n)
\end{align*}
for $\ell < 0$. This completes the proof of \eqref{GF:lem:reduced-functional-equality_12}.

It remains to show that $\eta_n$ is smooth. Like in the case $n = 1$ it is clear that $\eta_n$ is smooth everywhere but on $x_0 + U_n + \tau_n \Zbb$.  To prove this as well, we show that for all $x\in x_0 + U_n$, all $m\in \Nbb_0$ such that $g_n^{(m)}$ exists and is continuous, and all $\ell \in \Zbb$, we have
\begin{align}
\lim_{a \nearrow \ell} \eta_n^{(m)}(x + a \tau_1)  = \lim_{a \searrow \ell} \eta_n^{(m)}(x + a \tau_1). \label{GF:lem:reduced-functional-equality_15}
\end{align}
We prove this by induction in $\ell$ and start with $\ell =1$. Then since $x + a\tau_n \in x_0 + S_n$ for $a\in [0,1)$, we conclude that
\begin{align*}
\lim_{a \nearrow 1} \eta_n^{(m)}(x + a\tau_n) = \lim_{a \nearrow 1} \eta_{n-1}^{(m)}(x + a\tau_n) = \lim_{a \searrow 1} \eta_{n-1}^{(m)}(x + a\tau_n) = \lim_{a\searrow 1} \eta_n^{(m)}(x + a\tau_n).
\end{align*}
Here, we used \eqref{GF:lem:reduced-functional-equality_5} twice and that $\eta_{n-1}$ is smooth by induction. This proves \eqref{GF:lem:reduced-functional-equality_15} for $\ell = 1$. Similarly,
\begin{align*}
\lim_{a\nearrow \ell + 1} \eta_n^{(m)} (x + a\tau_n) &= \lim_{a\nearrow \ell} \eta_n^{(m)}(x + (a + 1)\tau_n) = \lim_{a\nearrow \ell} \eta_n^{(m)} (x + a\tau_n) + \lim_{a\nearrow \ell} g_n^{(m)}(x + a\tau_n).
\end{align*}
The last limit vanishes by a similar argument to the one leading to \eqref{GF:lem:reduced-functional-equality_14}. Applying this argument again, we see that the right hand side equals
\begin{align*}
\lim_{a\searrow \ell} \eta_n^{(m)} (x + a\tau_n) = \lim_{a\searrow \ell} g_n^{(m)} (x + a\tau_n) = \lim_{a\searrow \ell + 1} \eta_n^{(m)} (x + a \tau_n).
\end{align*}
The induction argument for $\ell = 0$ and $\ell \in -\Nbb$ is similar and left to the reader. This proves \eqref{GF:lem:reduced-functional-equality_15} for $\ell \in \Zbb$ and concludes the proof of the theorem.
\end{proof}

\begin{proof}[Proof of Proposition \ref{GF:prop:func-eq}]
We construct a partition of unity of $\Rbb^d$ as follows. Choose two $r$-periodic $\chi_0, \chi_1\in C^\infty(\Rbb)$ such that $0\leq \chi_0, \chi_1\leq 1$, $\chi_0 + \chi_1 =1$, and
\begin{align*}
\chi_0|_{[-\nicefrac r8, \nicefrac r8] + r\Zbb} &\equiv 0, &  \chi_1|_{[\nicefrac {3r}8, \nicefrac {5r}8] + r\Zbb} & \equiv 0.
\end{align*}

\begin{center}
\ifthenelse{\equal\masterfile{Diss}}
{\includegraphics[width=15cm]{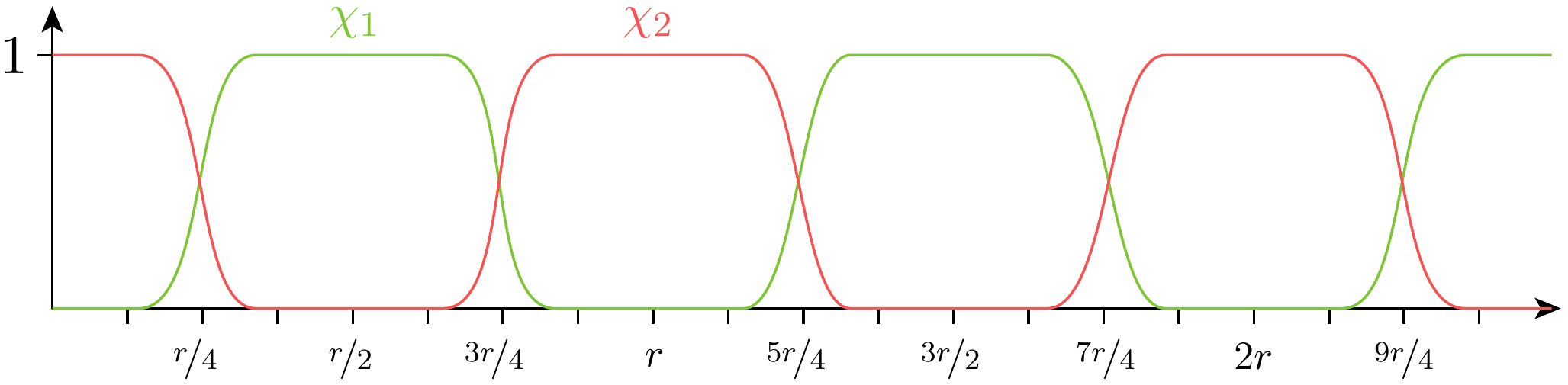}}
{\includegraphics[width=13cm]{source/gauge-fixing/Cut-Offs.pdf}}
\end{center}
Furthermore, for each $n = (n_1, \ldots, n_d)\in \{0,1\}^d$, we set
\begin{align*}
\varphi_n(x) := \chi_{n_1}(x_1) \cdots \chi_{n_d}(x_d),
\end{align*}
where $n_i = 1, 2$ for $i = 1, \ldots, d$. It is clear that 
\begin{align}
\sum_{n\in \{0,1\}^d} \varphi_n(x) =1
\end{align}
holds and for every $n\in \{0,1\}^d$, we have that
\begin{align}
\varphi_n\big|_{\frac r2n + \partial \Omega + \Lat} \equiv 0. \label{GF:prop:functional-equality_1}
\end{align}

Fix $n\in \{0,1\}^d$. For every $i=1, \ldots, d$ we define the function $g_i^n := \varphi_n g_i$. Since $\varphi_n$ is $\Lat$-periodic, the hypothesis \eqref{GF:eq:functional-equality_assumption} implies that \eqref{GF:eq:reduced-functional-equality_assumption1} holds. Furthermore, \eqref{GF:prop:functional-equality_1} implies that $g_n^i$ satisfies \eqref{GF:eq:reduced-functional-equality_assumption2}. Therefore, there is a smooth function $\eta_i$ such that
\begin{align*}
\eta_n(x + \tau_i) - \eta_n(x) &= g_i^n(x) , & i &= 1, \ldots, d.
\end{align*}
Then, a straightforward calculation shows that $\eta := \sum_{n\in \{0,1\}^d} \eta_n$ satisfies the statement of the proposition.
\end{proof}

\printbibliography[heading=bibliography, title=Bibliography of Chapter \ref{Chapter:Abrikosov_gauge}]
\end{refsection}

\begin{refsection}

\chapter{The Eigenvalues of the Periodic Landau Hamiltonian}
\label{Chapter:Spectrum_Landau_Hamiltonian_Section} \label{CHAPTER:SPECTRUM_LANDAU_HAMILTONIAN_SECTION}

In this chapter, we investigate the eigenvalues of the gauge periodic Landau Hamiltonian and their multiplicity. The chapter is written in the setting of Chapter \ref{Chapter:DHS1}. We shall fix an arbitrary charge $q\in \Nbb$ and we consider the space $\Lmag^{q,2}(Q_B)$ of $L^2_{\mathrm{loc}}(\Rbb^3)$-functions $\Psi$, which are gauge-perodic with respect to the magnetic translations
\begin{align}
T_{B,q}(v)\Psi(x) &:= \e^{\i \frac{q\Bbold}{2}\cdot (v \wedge x)} \Psi(x+v), & v &\in \Rbb^3, \label{DHS1:Magnetic_Translation_charge-q}
\end{align}
of the lattice $\Lambda_B$ defined above \eqref{DHS1:Fundamental_cell}, that is, these functions satisfy $T_{B, q}(\lambda)\Psi = \Psi$ for every $\lambda\in \Lambda_B$. The magnetic translations obey $T_{B,q}(v+w) = \e^{\i \frac{q\Bbold}{2} \cdot (v\wedge w)} T_{B, q}(v) T_{B,q}(w)$, whence the group $\{T_{B,q}(\lambda)\}_{\lambda\in \Lambda_B}$ is abelian.

On the Sobolev space $\Hmag^{q,2}(Q_B)$ of gauge-periodic functions, where
\begin{align}
\Hmag^{q,m}(Q_B) &:= \bigl\{ \Psi\in \Lmag^{q,2}(Q_B) :  \Pi^\nu \Psi\in \Lmag^{q, 2} (Q_B) \quad \forall \nu\in \Nbb_0^3, |\nu|_1\leq m\bigr\} \label{DHS1:Periodic_Sobolev_Space_charge-q}
\end{align}
for $m\in \Nbb_0$, we consider the Landau Hamiltonian $\Pi_q^2$ with magnetic momentum given by $\Pi_q :=-\i \nabla + q \Abold$. This operator commutes with the translations in \eqref{DHS1:Magnetic_Translation_charge-q} and the magnetic flux through the unit cell $Q_B$ is equal to $2\pi q$, see \eqref{DHS1:Fundamental_cell} and the discussion below \eqref{DHS1:Fundamental_cell}. In this respect, Sections \ref{DHS1:Magnetically_Periodic_Samples} and \ref{DHS1:Periodic Spaces} correspond to the special cases $q=1$ and $q=2$, respectively.

We choose a Bloch--Floquet decomposition $\Ucal_{\mathrm{BF}}$ (see also Section \ref{DHS1:Schatten_Classes}) such that $\Pi_q$ fibers according to
\begin{align}
\Ucal_{\mathrm{BF}} \, \Pi_q \, \Ucal_{\mathrm{BF}}^* =  \int_{[0,1)^3}^\oplus \dd \vartheta \; \Pi_q(\vartheta) \label{DHS1:Bloch-Floquet-decomposition}
\end{align}
with fiber momentum operators
\begin{align*}
\Pi_q(\vartheta) := -\i \nabla + q\Abold + \sqrt{2\pi B} \, \vartheta
\end{align*}
acting on the magnetic Sobolev space $\Hmag^{q,2}(Q_B)$ in \eqref{DHS1:Periodic_Sobolev_Space_charge-q}.

\begin{prop}
\label{DHS1:Spectrum_Landau_Hamiltonian}
For every $B >0$, $q\in \Nbb$, and $\vartheta\in [0, 1)^3$, the spectrum of $\Pi_q(\vartheta)^2$ consists of the isolated eigenvalues
\begin{align}
E_{q,B, \vartheta}(k, p) &:= q \, B \, (2k+1) + 2\pi \, B \, (p + \vartheta_3)^2, & k\in \Nbb_0, \; p\in \Zbb. \label{DHS1:Spectrum_Landau_Hamiltonian_eq1}
\end{align}
Furthermore, their multiplicity is finite and equals
\begin{align}
\dim \ker ( \Pi_q(\vartheta)^2 - E_{q,B, \vartheta}(k, p)) = q.
\label{DHS1:Spectrum_Landau_Hamiltonian_eq2}
\end{align}
\end{prop}

In preparation for the proof, we first note that,  by rescaling, $\Pi_q(\vartheta)^2$ is isospectral to $B \, (-\i \nabla + \frac q2 e_3\wedge x+ \sqrt{2\pi}\, \vartheta )^2$. We henceforth assume that $B =1$.

Furthermore, we introduce the notation $x = (x_\perp , x_3)^t$ and define the two-dimensional operator $\Pi_{\perp, q}(\vartheta) := (\Pi_q^{(1)}(\vartheta), \Pi_q^{(2)}(\vartheta))^t$. This operator acts on functions $\psi_\perp$ satisfying the gauge-periodic condition $T_{\perp,q} (\lambda) \psi_\perp = \psi_\perp$ for all $\lambda \in \sqrt{2\pi}\, \Zbb^2$ with
\begin{align}
T_{\perp,q} (v) \psi_\perp(x) &:= \e^{\i \frac{q}{2} (v_1x_2 - v_2x_1)}  \psi_\perp(x + v),  & v &\in \Rbb^2.
\end{align}

The following result is well known, even for more general lattices, see for example \cite[Proposition 6.1]{Tim_Abrikosov}. We include the proof for the sake of completeness, adding the treatment of the perturbation by $\vartheta$.

\begin{lem}
\label{DHS1:Spectrum_Landau_Hamiltonian_Lemma}
For every $q\in \Nbb$, the spectrum of the operator $\Pi_{\perp, q}(\vartheta)^2$ consists of the isolated eigenvalues $E_q(k) := (2k+1)q$, $k\in \Nbb_0$. Each of $E_q(k)$ is $q$-fold degenerate.
\end{lem}

\begin{proof}
Since $[\Pi_q^{(1)}(\vartheta), \Pi_q^{(2)}(\vartheta)] = -\i q$, the creation and annihilation operators
\begin{align}
a(\vartheta) &:= \frac{1}{\sqrt{2q}} \bigl( \Pi_q^{(1)}(\vartheta) - \i \Pi_q^{(2)}(\vartheta)\bigr), & a^*(\vartheta) &:= \frac{1}{\sqrt{2q}} \bigl( \Pi_q^{(1)}(\vartheta) + \i \Pi_q^{(2)}(\vartheta)\bigr) \label{DHS1:Creation-Annihilation}
\end{align}
satisfy $[a(\vartheta), a^*(\vartheta)] = 1$ and it is easy to show that
\begin{align}
\Pi_q(\vartheta)^2 = q\, (2\,  a^*(\vartheta) a(\vartheta) + 1). \label{DHS1:Spectrum_Landau_Hamiltonian_eq3}
\end{align}
From this, we read off the formula for $E_q(k)$.

The rest of the proof is devoted to the statement about the degeneracy. First, with the help of the creation and annihilation operators, it is easy to show that the degeneracy of $E_q(k)$ is equal to that of $E_q(0)$ for all $k\in \Nbb_0$. Therefore, it is sufficient to determine the degeneracy of $E_q(0)$. By \eqref{DHS1:Spectrum_Landau_Hamiltonian_eq3},  $\ker(\Pi_{\perp, q}^2(\vartheta) - q)$ equals $\ker(a(\vartheta))$ so it suffices to determine the latter. A straightforward calculation shows that
\begin{align*}
\e^{\frac q4 |x_\perp - \frac 2q \sqrt{2\pi} J\vartheta|^2} \; a(\vartheta) \; \e^{-\frac q4|x_\perp - \frac 2q \sqrt{2\pi} J\vartheta|^2} &= - \frac{\i}{\sqrt{2q}} \, [  \partial_{x_1} - \i \partial_{x_2}], & J := \bigl(\begin{matrix} & -1 \\ 1\end{matrix}\bigr).
\end{align*}
Therefore, the property $\psi_\perp \in\ker a(\vartheta)$ is equivalent to the function $\xi := \e^{\frac q4 |x_\perp - \frac 2q \sqrt{2\pi} J\vartheta|^2} \psi_\perp$ satisfying $\partial_{x_1}\xi - \i \partial_{x_2}\xi =0$. If we identify $z = x_1 + \i x_2\in \Cbb$, then $J\vartheta = \i (\vartheta_1 + \i \vartheta_2)$ and this immediately implies that the complex conjugate function $\ov \xi$ solves the Cauchy-Riemann differential equations, whence it is entire. We define the entire function
\begin{align*}
\Theta(z) := \e^{-2\i z \Re \vartheta} \, \e^{-\frac{q}{2\pi} (z - \frac{2\pi \i}{q}   \vartheta)^2} \; \ov {\xi \Bigl( \sqrt{\frac 2\pi} \, z\Bigr)}. 
\end{align*}
A tedious calculation shows that the gauge-periodicity of $\psi_\perp$ is equivalent to the relations
\begin{align}
\Theta(z + \pi) &= \Theta(z), \label{DHS1:Spectrum_Landau_1}\\ 
\Theta(z + \i \pi) &= \e^{-2\pi \vartheta} \, \e^{-2\i q z}\,  \e^{q\pi} \,  \Theta(z). \label{DHS1:Spectrum_Landau_2}
\end{align}
Therefore, it suffices to show that the space of entire functions $\Theta$ which obey \eqref{DHS1:Spectrum_Landau_1} and \eqref{DHS1:Spectrum_Landau_2} is a vector space of dimension $q$. We claim that \eqref{DHS1:Spectrum_Landau_1} implies that $\Theta$ has an absolutely convergent Fourier series expansion of the form
\begin{align}
\Theta(z) = \sum_{k\in \Zbb} c_k \; \e^{2\i kz}. \label{DHS1:Spectrum_Landau_3}
\end{align}
To prove this, we first note that, for fixed imaginary part $x_2$, we may expand $\Theta$ in an absolutely convergent series $\Theta(z) = \sum_{k\in \Zbb} a_k(x_2) \e^{2\i kx_1}$ with
\begin{align*}
a_k(x_2) = \frac 1\pi \int_0^\pi \dd x_1 \; \e^{-2\i k x_1} \, \Theta(x_1 + \i x_2).
\end{align*}
By the Cauchy-Riemann equations, it is easy to verify that $a_k' = -2k \, a_k$. Therefore, the number $c_k := \e^{2k x_2} \, a_k(x_2)$ is independent of $x_2$ and provides the expansion \eqref{DHS1:Spectrum_Landau_3}. Furthermore, \eqref{DHS1:Spectrum_Landau_2} implies that $c_{k+q} = \e^{-\pi (2k+q)} \e^{2\pi \vartheta} c_k$. Therefore, the series \eqref{DHS1:Spectrum_Landau_3} is fully determined by the values of $c_0, \ldots, c_{q-1}$ and we conclude that $\ker a(\vartheta)$ is a $q$-dimensional vector space.
\end{proof}

\begin{proof}[Proof of Proposition \ref{DHS1:Spectrum_Landau_Hamiltonian}]
As mentioned before, it suffices to prove the proposition for $B = 1$. It is easy to verify that for any $\vartheta \in [0,1)^3$ the spectrum of $(\Pi_q(\vartheta)^{(3)})^2$ consists of the simple eigenvalues $2\pi\, (p+ \vartheta_3)^2$ with $p\in \Zbb$. Since $\Pi_q(\vartheta)^{(3)}$ and $\Pi_{\perp,q}(\vartheta)$ commute, Lemma \ref{DHS1:Spectrum_Landau_Hamiltonian_Lemma} implies the existence of an orthonormal basis of eigenvectors for $\Pi_q(\vartheta)^2$ of the form $\psi_\perp^{k, m}(x_\perp) \psi_3^{\vartheta, p}(x_3)$ with $k\in \Nbb_0$, $m = 1, \ldots, q$ and $p\in \Zbb$, corresponding to the eigenvalue $E_{q,1, \vartheta}(k,p)$. This proves the formulas \eqref{DHS1:Spectrum_Landau_Hamiltonian_eq1} and  \eqref{DHS1:Spectrum_Landau_Hamiltonian_eq2}.
\end{proof}

\printbibliography[heading=bibliography, title=Bibliography of Chapter \ref{Chapter:Spectrum_Landau_Hamiltonian_Section}]

\end{refsection}

\begin{refsection}

\chapter[The Low Lying Spectrum of \texorpdfstring{$K_{T,\Abold} - V$}{KTA-V} via a Combes--Thomas Estimate][The Low Lying Spectrum of $K_{T,\Abold} - V$]{The Low Lying Spectrum of \texorpdfstring{$K_{T,\Abold} - V$}{KTA-V} via a Combes--Thomas Estimate}
\label{Chapter:Combes-Thomas}
\label{CHAPTER:COMBES-THOMAS}


\section{Introduction}

In this chapter, we are going to perform a Combes--Thomas estimate for the resolvent kernel of the operator $K_T - V$ and thereby prove that its resolvent kernel is exponentially decaying in an integral sense. Since this analysis requires quite some effort, we should spend a few lines discussing the purpose of this chapter.

First and foremost, this extends the analysis presented in the work \cite{DeHaSc2021}, which is included in this thesis in Chapter \ref{Chapter:DHS1}. In Section \ref{DHS1:KTV_Asymptotics_of_EV_and_EF_Section} of that chapter (which is \cite[Appendix A]{DeHaSc2021}), we prove asymptotic formulas for the lowest eigenvalue of the operator $K_{T, \Abold} - V$, the corresponding eigenfunction, and the spectral gap above the ground state in the case of the constant magnetic field potential $\Abold(x) = \frac 12 \Bbold \wedge x$. We emphasize that the proof of Proposition \ref{DHS1:KTV_Asymptotics_of_EV_and_EF} presented in Chapter \ref{Chapter:DHS1} is valid only if $V$ has a sign. More precisely, we assume there that $V\geq 0$. The reason is that the analysis is based on the Birman--Schwinger correspondence for the Birman--Schwinger operator $V^{\nicefrac 12} (K_T - \lambda)^{-1}V^{\nicefrac 12}$ at the eigenvalue $\lambda$ of $K_T - V$. This operator is self-adjoint only if $V$ has a sign and the self-adjointness is essential for the proof presented there. For example, it allows for a variational characterization of eigenvalues, which is heavily used. In Section \ref{DHS1:KTV_Asymptotics_of_EV_and_EF_Section}, we state an explicit reference to this thesis and, in fact, to this chapter, and announce that the result of Proposition \ref{DHS1:KTV_Asymptotics_of_EV_and_EF} does also hold if the assumption $V\geq 0$ is dropped. The proof of this claim is the present chapter.

Secondly, the Combes--Thomas estimate enables us to prove exponential localization of general eigenfunctions corresponding to isolated eigenvalues of $K_T - V$, which is a generalization of what has been proven in \cite[Appendix A]{Hainzl2012}. There, this result is shown for the eigenfunction of $K_{\Tc} - V$ corresponding to the eigenvalue zero. In the situation of dealing with a zero eigenvalue, the analysis is considerably simpler since one can use explicit expansion formulas for the hyperbolic tangent to explicitly obtain bounds on the resolvent kernel of $K_T$ in terms of the resolvent kernel of the Laplacian, which in $d =3$ is explicit. It should be noted that it would suffice to have an exponential bound for the resolvent kernel of $K_T$ (without $V$). However, unless this is proven with a completely different method (other than a Combes--Thomas estimate), I don't expect the analysis to simplify much.

The third reason why I think this analysis is worth to be written up is that there is still only a poor amount of mathematical literature about the BCS theory of superconductivity and to the best of my knowledge there is no Combes--Thomas estimate executed for the operator $K_T - V$ anywhere in the literature. In particular, most of the Combes--Thomas estimates one finds in the literature are proven for the (magnetic) Laplacian and in trace ideal sense --- not with kernels --- so we present a technique rarely used. Of course, in the end, $K_T - V$ is a Schrödinger-type operator like $-\Delta - V$ and many results carry over to the case of $K_T$ ``just'' because of this fact. However, in practice, it turns out that the slight difference of dividing $p^2-\mu$ by the hyperbolic tangent does cause technical difficulties and headaches to the one having to deal with them. On top of that, we have to face the difficulty that arises when dealing with magnetic fields, as is the case if anybody in the future wants to come closer to proving the Meißner effect. In this chapter, we perform a phase approximation for the resolvent of $K_{T, \Abold} - V$ and thereby develop a detailed understanding of the method that we already discussed for the Laplacian in Sections \ref{DHS1:Magnetic_resolvent_estimates_Section} and \ref{DHS2:Phase_approximation_method_Section} for different magnetic fields. In my opinion, this chapter is therefore a good occasion to get familiar with this very central operator $K_{T, \Abold} - V$ and gain some confidence in
\begin{enumerate}
\item dealing with the (pseudo-)differential operator $K_T$ with symbol
\begin{align*}
K_T(p) := \frac{p^2 - \mu}{\tanh(\frac{p^2-\mu}{2T})}.
\end{align*}
Recall that the spectrum of $K_T$ equals $[2T, \infty)$ if $\mu\geq 0$ and $[|\mu|/\tanh(|\mu|/(2T)), \infty)$ if $\mu < 0$. Figure \ref{CT:fig_KT} shows the shape of the symbol $K_T(p)$ for $\mu >0$.

\begin{figure}[h]
\label{CT:fig_KT}
\centering
\includegraphics[width = 8cm]{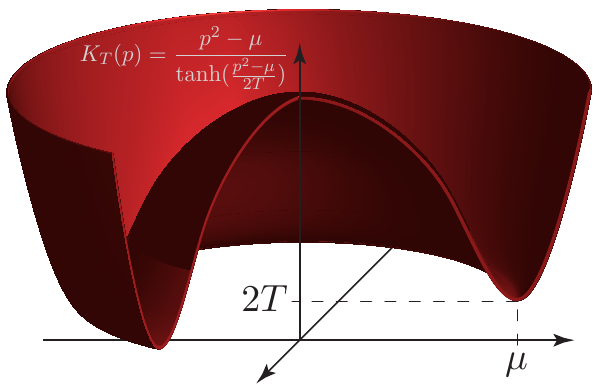}
\caption{The shape of the differential symbol $K_T(p)$.}
\end{figure}

\item performing magnetic field approximations with the phase approximation method in the spirit of \cite[Section 5]{Nenciu2002}.
\end{enumerate}

Some ideas and concepts of what will be presented here are contained in the  unpublished notes \cite{Deuchert} by Andreas Deuchert, to whom I once more express my gratitude.

We work under the following assumptions:
\begin{itemize}
\item The chemical potential $\mu$ is an arbitrary real number, $\mu\in \Rbb$.

\item $T$ is a fixed positive temperature, $T>0$.

\item The interaction potential $V\in L^2(\Rbb^3)$ satisfies $(1 + |\cdot|^2) V\in L^\infty(\Rbb^3)$.

We remark that such a potential in particular preserves the essential spectrum. The spectrum of $K_T - V$ typically looks like Figure \ref{CT:fig_spectrum} shows.

\begin{figure}[h]
\label{CT:fig_spectrum}
\centering
\includegraphics[width = 8cm]{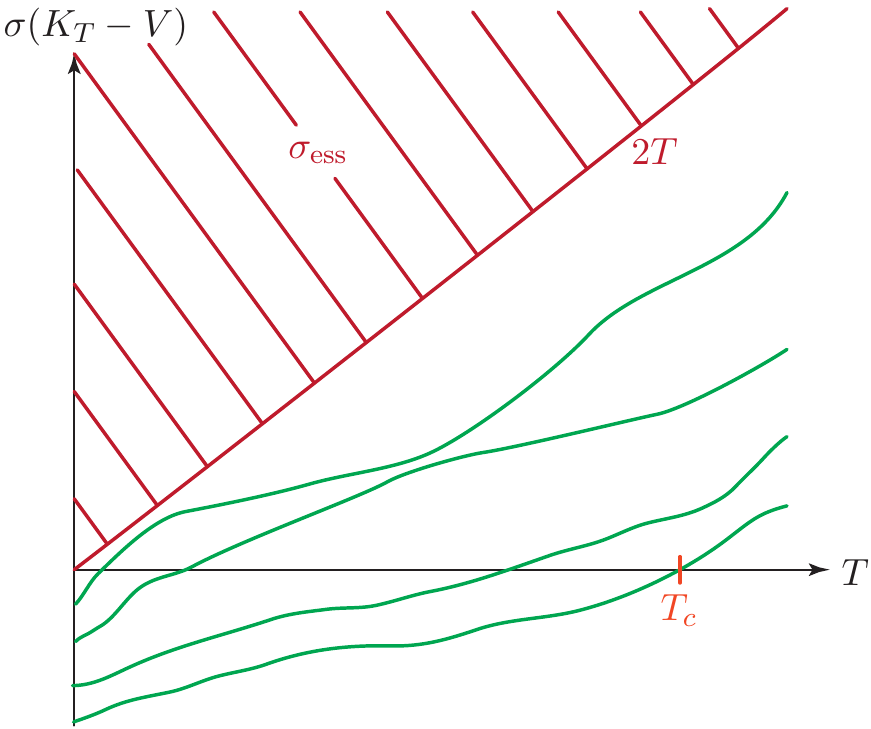}
\caption{The spectrum of $K_T - V$ as a function of $T$.}
\end{figure}

\item $\Abold$ satisfies:
\end{itemize}

\begin{asmp}
\label{CT:Assumption_A}
$\Abold\colon \Rbb^3\ra \Rbb^3$ is a measurable magnetic vector potential, which is three times weakly differentiable and whose derivatives $D^k\Abold$ --- not $\Abold$ itself --- belong to $L^\infty(\Rbb^3;\Rbb^3)$, $k = 1, 2, 3$. We note that such an $\Abold$ is Lipschitz continuous. We also assume that $\Abold(0) = 0$.
\end{asmp}

\begin{bem}
The situation we have in mind here is the one in Chapter \ref{Chapter:DHS2}, namely that the magnetic potential consists of a sum of a bounded periodic potential and a potential corresponding to the constant magnetic field. There is no assumption on periodicity since the analysis presented here would be applied in the relative coordinate of BCS theory, where periodicity is irrelevant. Let us comment a bit on the last assumption $\Abold(0) =0$. Of course, this assumption is satisfied for the constant magnetic field potential $\Abold(x) = \frac 12 \Bbold \wedge x$, where $\Bbold\in \Rbb^3$ is fixed. Moreover, in the periodic setting of BCS theory in Chapter \ref{Chapter:DHS2}, we can always find a gauge such that the periodic part $A$ has mean zero, i.e.
\begin{align*}
\frac{1}{|Q|} \int_Q \dd y \; A(y) =0.
\end{align*}
This means that we can estimate
\begin{align*}
|A(x)| \leq \frac{1}{|Q|} \int_Q \dd y \; |A(x) - A(y)| \leq \Vert DA\Vert_\infty \, \frac{1}{|Q|} \int_Q \dd y \; |x-y| \leq C \, \Vert DA\Vert_\infty (1 + |x|).
\end{align*}
A similar behavior of $\Abold$ is achieved by our assumption $\Abold(0) =0$.
\end{bem}

In this section, we also denote the free momentum operator by $p := -\i \nabla$. The magnetic operator $K_{T, \Abold}$ is then defined as $K_{T, \Abold} := K_T(-\i \nabla + \Abold)$. 

The operator $K_{T, \Abold} - V$ acts in the usual magnetic Soboblev space $H_\Abold^2(\Rbb^3 ; \Rbb^3)$, where
\begin{align*}
H_\Abold^m(\Rbb^3) := \bigl\{ f\in L^2(\Rbb^3) : (-\i \nabla + \Abold)^\nu f\in L^2(\Rbb^3) \; \forall \nu\in \Nbb_0^3, \, |\nu|_1 \leq m\bigr\}.
\end{align*}
The resolvent of $K_{T, \Abold} - V$ at $z\in \rho(K_{T, \Abold} - V)$ is denoted by
\begin{align}
\Rcal_{T, \Abold}^{z, V} := \frac{1}{z - (K_{T, \Abold} - V)}
\end{align}
and is a bounded operator $L^2(\Rbb^3) \ra H_\Abold^2(\Rbb^3)$. If either $\Abold =0$ or $V =0$, we omit the corresponding index, i.e.,  we write $\Rcal_T^{z, V} := \Rcal_{T, 0}^{z, V}$, as well as $\Rcal_{T, \Abold}^z := \Rcal_{T, \Abold}^{z,0}$ and $\Rcal_T^z := \Rcal_{T, 0}^{z, 0}$. We show in the next section that its kernel exists and we denote it by 
\begin{align}
\Gcal_{T, \Abold}^{z, V}(x,y) &:= \frac{1}{z - (K_{T, \Abold} - V)}(x,y), & x,y&\in \Rbb^3. \label{CT:GcalTAz_definition}
\end{align}
If either $\Abold =0$ or $V =0$, we likewise abbreviate $\Gcal_T^{z, V}$, as well as $\Gcal_{T, \Abold}^z$ and $\Gcal_T^z$.

The main results we will prove in this chapter are 
\begin{enumerate}
\item Theorem \ref{CT:Gcal_exponential_decay}, which proves the exponential decay estimate on the resolvent kernel of $K_T - V$,

\item Proposition \ref{CT:Exponential_localization}, which provides the exponential localization of eigenfunctions corresponding to isolated eigenvalues of $K_T - V$ below $2T$,

\item Theorem \ref{CT:KTV_Perturbation_Theorem}, which finally proves the stability and asymptotic expansions for the eigenvalues and spectral projections of $K_{T, \Abold} - V$.
\end{enumerate}

Our strategy of proof relies on a Combes--Thomas estimate for the resolvent kernel of $K_T - V$, which means that we are going to prove that there is a $\delta>0$ such that
\begin{align}
\sup_{x\in \Rbb^3} \Bigl(\int_{\Rbb^3} \dd y \; \e^{\delta \, |x-y|} \, |\Gcal_T^{z, V}(x,y)|^2 \Bigr)^{\nicefrac 12} < \infty. \label{CT:Intro1}
\end{align}
We refer to this result as exponential decay of the resolvent kernel in $(2,\infty)$-norm sense. Roughly speaking, the idea of such a result comes from the analogy of the Fourier transform $\hat f$ of a function $f$ being exponentially decaying if $f$ has an analytic extension to a complex strip around the real axis. In a similar spirit, we will extend the symbol $K_T(p)$ to certain momenta $p + \lambda a$ where $\lambda\in \Cbb$ and $a\in \Rbb^3$ is a unit vector. Working through all technical difficulties that arise, this enables us to prove \eqref{CT:Intro1}.

On the basis of the exponential decay of the resolvent kernel, we can use the technique that has already been used in \cite[]{Hainzl2012} to prove exponential localizaton of an arbitrary eigenfunction of $K_T - V$. Furthermore, we use the exponential estimate for the resolvent kernel to prove an asymptotic estimate for the eigenvalues of $K_{T, \Abold} - V$ and their corresponding spectral projections.

As usual $C$ denotes a generic positive constant that is allowed to change from line to line. We allow it to depend on the fixed quantities like $\mu$, $V$, and $T$. Further dependencies are indexed. It is needless to say that the constants in our theory deteriorate if $T$ approaches zero. In particular, in this chapter, we regard $\Abold$ as a variable, whose influence we shall keep track of precisely.

\begin{bem}
I should say that the mathematics of this chapter could (and would, in a paper) be carried out in a smarter and more efficient way. The style in which it is written is the ``students way'' so to speak. I decided against optimizing for the most efficient presentation for the sake of clarity.
\end{bem}


\section{Combes--Thomas Estimate}

The Combes--Thomas estimate consists of several steps. First, we are going to prove an estimate on the resolvent kernel of $K_T - V$. In the second step, we want to extend this to the analytic family $K_T^{\lambda a} - V$ for certain $\lambda\in\Cbb$ and $a\in \Rbb^3$ a unit vector. This requires several steps of preparation. First, we need to prove a resolvent estimate for $(p +  \lambda a)^2 -\mu$. Then, we need to make sense of the operator $K_T^{\lambda a} - V$ as an analytic family of type (A) on $H^2(\Rbb^3)$. To prove a resolvent estimate for this analytic family, we need to extend the integral representation in Lemma \ref{DHS1:KT_integral_rep} to non-self-adjoint operators. Using this, we finally are able to provide the desired resolvent estimate for $K_T^{\lambda a} -V$.

The third step is then to interpret the perturbation by $\lambda a$ of $p$ as a ``non-unitary translation operator'' in Fourier space, which is an exponential factor with a real exponent. In this way, we can relate the resolvent kernel of the analytic family back to the original resolvent kernel, which essentially amounts to an exponential tilt of the resolvent kernel. Then, for suitably chosen $\lambda$ and $a$ the aforementioned estimate on $K_T^{\lambda a} - V$ provides us with an exponential bound for the resolvent kernel of $K_T - V$ in the sense of \eqref{CT:Intro1}.

\subsection{Explicit resolvent estimates for \texorpdfstring{$K_T - V$}{KT-V}}

Our starting point is the result \cite[Corollary A.1.2]{Simon82}, which for our case reads as follows.

\begin{lem}
\label{CT:Simon}
Let $1\leq p < \infty$. If $A$ is a bounded operator on $L^p(\Rbb^3)$ and $A$ is bounded also from $L^p(\Rbb^3)$ to $L^\infty(\Rbb^3)$, then there is a measurable function $K_A$ on $\Rbb^3\times \Rbb^3$ obeying
\begin{align}
\Vert A\Vert_{p, \infty} := \sup_{x\in \Rbb^3} \Bigl( \int_{\Rbb^3} \dd y \; |K_A(x, y)|^q \Bigr)^{\nicefrac 1q} < \infty, \label{CT:Simon_1}
\end{align}
where $q = \frac p{p-1}$ is the Hölder conjugate of $p$, so that, for any $f\in L^p(\Rbb^3)$,
\begin{align}
(Af)(x) = \int_{\Rbb^3} \dd y \; K_A(x,y)\, f(y). \label{CT:Simon_2}
\end{align}
Conversely, if $A\colon L^p(\Rbb^3) \ra L^p(\Rbb^3)$ has an integral kernel $K_A$ in the sense of \eqref{CT:Simon_2} obeying \eqref{CT:Simon_1}, then $A$ is a bounded map from $L^p(\Rbb^3)$ to $L^\infty(\Rbb^3)$
\end{lem}

We also point out the references given in \cite{Simon82} to Korotkov as well as Dunford and Pettis for this result. Operators satisfying Lemma \ref{CT:Simon} are also called \emph{Carleman} operators.

For us, Lemma \ref{CT:Simon} and the fact that $H^2(\Rbb^3)$ embeds continuously into $L^\infty(\Rbb^3)$ imply the existence of the resolvent kernel $\Gcal_{T, \Abold}^{z, V}$ in \eqref{CT:GcalTAz_definition}. This is enough at this point since all further estimates and dependencies are computed explicitly.

\begin{lem}
\label{Resolvent_KTV_bounded_uniformly}
Let $V\in L^\infty(\Rbb^3)$. For any $z\in \rho(K_T-V)$, the resolvent $\Rcal_T^{z,V}$ of $K_T - V$ at $z$ is a bounded operator from $L^2(\Rbb^3)$ to $L^\infty(\Rbb^3)$ with
\begin{align*}
\Vert \Rcal_T^{z,V}\Vert_{2,\infty} \leq C \; \Bigl[ 1 + \frac{1}{\dist(z, \sigma(K_T-V))} \Bigr].
\end{align*}
\end{lem}

\begin{proof}
Let us write $\ell := \dist(z,\sigma(K_T -V))$. By the resolvent equation, we have
\begin{align*}
\Rcal_T^{z,V} = \Rcal_T^z - \Rcal_T^z \, V \, \Rcal_T^{z,V}.
\end{align*}
Since $\Vert \Rcal_T^{z,V}\Vert_\infty = \ell^{-1}$ and $V\in L^\infty(\Rbb^3)$, it is enough to show that $\Rcal_T^z$ is a bounded operator from $L^2(\Rbb^3)$ to $L^\infty(\Rbb^3)$ with a suitable norm bound. To show this, let us utilize the first resolvent equation for $\nu\in \rho(K_T)$ arbitrary but fixed, which implies
\begin{align*}
\Rcal_T^z = \Rcal_T^\nu + (z - \nu) \, \Rcal_T^\nu \, \Rcal_T^z.
\end{align*}
We note that $\sigma(K_T)\subseteq \sigma(K_T-V)$, since the essential spectrum is preserved by our assumptions on $V$, and this implies
\begin{align*}
\dist(z,\sigma(K_T)) = \inf\bigl\{ |z - \eta| : \eta\in \sigma(K_T)\bigr\} \geq \dist(z,\sigma(K_T-V)) = \ell.
\end{align*}
Hence, $\Vert \Rcal_T^z\Vert_\infty \leq \ell^{-1}$. It remains to provide a bound on the $(2,\infty)$-norm of $\Rcal_T^\nu$. For this, use the resolvent equation again to get that
\begin{align*}
\Rcal_T^\nu = (\nu - (p^2-\mu))^{-1} + (\nu - (p^2-\mu))^{-1} \bigl[ K_T - (p^2-\mu)\bigr] \Rcal_T^{\nu}
\end{align*}
As we already know, $\Vert \Rcal_T^{\nu}\Vert_\infty \leq \dist(\nu, \sigma(K_T))^{-1}$. We claim $\Vert K_T - (p^2-\mu)\Vert_\infty = \frac{2\mu}{1 - \e^{-\beta\mu}}$. To see this, we show that the function $f(t) = \frac{t}{\tanh(\frac{t}{2T})} - t = \frac{2t}{\e^{\nicefrac tT} - 1}$ is monotonically decreasing on $t\geq -\mu$. To see this, we calculate its derivative
\begin{align*}
f'(t) = \frac{2}{(\e^{\nicefrac tT} - 1)^2} \bigl[\e^{\nicefrac tT} - 1 - \frac tT \cdot \e^{\nicefrac tT}\bigr].
\end{align*}
By l'Hôpital applied two times, we see that $f'(0) = -1 < 0$. Outside $t=0$, it suffices to consider $g(t) = \e^{\nicefrac tT} - 1 - \frac tT\e^{\nicefrac tT}$ and show that it is nonpositive. We have $g(0) =0$ and $g'(t) = -\frac{t}{T^2}\e^{\nicefrac tT}$. Hence, $g'(t) >0$ if $t<0$ and $g'(t) <0$ if $t>0$. We conclude that $g(t) \leq 0$ and thus $f'(t)\geq 0$ for all $t\geq - \mu$. It follows that $f$ takes its maximum at the left boundary, proving the claim.

It remains to provide a bound on the $(2,\infty)$-norm of $(\nu - (p^2-\mu))^{-1}$. Since $\nu < \mu$, we see that $(\nu - (p^2-\mu))^{-1}$ is given by the convolution with its $L^2$-symbol. Hence, its $(2,\infty)$-norm is given by
\begin{align*}
\Bigl\Vert \frac 1{\nu - (p^2 - \mu)} \Bigr\Vert_{2,\infty} ^2 &= \int_{\Rbb^3} \frac{1}{|\nu + \mu - |p|^2|^2} \, \dd p < \infty.\qedhere
\end{align*}
\end{proof}

\subsection{Analytic extension}

\subsubsection{Explicit resolvent estimate for \texorpdfstring{$(p+ \lambda a)^2-\mu$}{(p+lambda)2-mu}}

\begin{lem}
\label{Laplace_Analytic_estimate}
Let $z\in \rho(p^2)$, $a\in \Rbb^3$ with $|a|=1$ and let $\lambda\in \ov B_{r_z}(0)\subseteq \Cbb$ with 
\begin{align*}
r_z = \min \Bigl\{ 1, \; \frac 12 \Bigl[ \sqrt{\frac{2}{|z| - \Re z}} + \frac{1}{\dist(z, \sigma(p^2))}\Bigr]^{-1}\Bigr\}.
\end{align*}
Then, $z\in \rho((p+\lambda a)^2)$ and
\begin{align*}
\Bigl\Vert \frac 1{z - (p+\lambda a)^2}\Bigr\Vert_\infty \leq \frac{2}{\dist(z, \sigma(p^2))}.
\end{align*}
\end{lem}

\begin{proof}
We have
\begin{align*}
z - (p + \lambda a)^2 = \bigl(1 - (2\lambda a p + \lambda^2 )(z - p^2)^{-1} \bigr) (z - p^2)
\end{align*}
and claim that
\begin{align}
\Bigl\Vert (2 \, a   p + \lambda) \, \frac 1{z - p^2} \Bigr\Vert_\infty &\leq \sqrt{\frac{2}{|z| - \Re z}} + |\lambda| \, \frac{1}{\dist(z, \sigma(p^2))} \leq \frac{1}{2r_z}. \label{Laplace_analytic_1}
\end{align}
If this is true, then $|\lambda| \, \Vert (2\, a p + \lambda)(z - p^2)^{-1}\Vert_\infty \leq \frac 12$ so that, by the Neumann series,
\begin{align*}
\Bigl \Vert \frac 1 {1 - (2\,a  p\lambda + \lambda^2)(z - p^2)^{-1}} \Bigr\Vert_\infty \leq 2,
\end{align*}
whence the lemma is proven. It remains to show \eqref{Laplace_analytic_1}. Here, the only difficulty is to estimate, for some $\psi\in L^2(\Rbb^3)$,
\begin{align*}
\Vert a p (z - p^2)^{-1}\psi\Vert_2^2 &= \langle (z - p^2), (a  p)^2 \; (z - p^2)^{-1}\psi\rangle \\
&\leq \langle \psi, (\ov z - p^2)^{-1} p^2 (z - p^2) \psi\rangle =  \int_{\sigma(p^2)} \dd \mu_\psi(t) \; \frac{t}{|z - t|^2} .
\end{align*}
Here, we have estimated
$|a p|^2 \leq p^2$ by Cauchy--Schwarz for $p\in \Rbb^3$ in Fourier space. We claim that the function $f(t) = \frac{t}{(\Re z - t)^2 + (\Im z)^2}$, $t\geq 0$ has a unique maximum at $t = |z|$ with value
\begin{align*}
f(|z|) = \frac{|z|}{(\Re z - |z|)^2 + (\Im z)^2} = \frac{1}{2(|z| - \Re z)}.
\end{align*}
To see this note that $f(t) \to 0$ as $t\to \infty$, $f(|z|) >0$ and $f(0) =0$. Hence, there must exist a maximum. Since $f'(t) =0$ if and only if $t=|z|$, it must be located at $t=|z|$. We readily conclude
\begin{align}
\Vert a  p \, (z - p^2)^{-1}\Vert_\infty \leq \sqrt{f(|z|)} = \sqrt{\frac{1}{2 \, (|z| - \Re z)}}. \label{Laplace_analytic_2}
\end{align}
This proves \eqref{Laplace_analytic_1}.
\end{proof}

\subsubsection{Defining \texorpdfstring{$K_T^{\lambda a}$}{KTlambdaa} for nonreal \texorpdfstring{$\lambda$}{lambda}}

\begin{lem}
\label{CT:Analytic_family_well-def}
Let $a\in\Rbb^3$ with $|a| = 1$ and let 
\begin{align}
S_{\mu, T} := \Bigl\{z\in \Cbb : |\Im z|^2 <\frac 12 \bigl[\sqrt{\mu^2 + (2\pi T)^2} - \mu\bigr]\Bigr\}\subseteq \Cbb.\label{CT:Analytic_family_strip}
\end{align}
Then, for each $\lambda\in S_{\mu, T}$, the operator
\begin{align}
K_T^{\lambda a} := \frac{(-\i \nabla + \lambda a)^2 - \mu}{\tanh\bigl( \frac{(-\i \nabla +\lambda a)^2 - \mu}{2T}\bigr)} \label{Analytic family}
\end{align}
is well-defined on $H^2(\Rbb^3)$ as a Fourier multiplier.
\end{lem}

\begin{proof}
We have to verify that the hyperbolic tangent has no zero in the claimed domain when the numerator has not. As we know, the zeros of $\tanh$ are those of $\sinh$, namely $0 = \sinh(\frac{z}{2T}) = \frac 12 \e^{-\nicefrac zT} ( \e^{\nicefrac zT} - 1)$ if and only if $z_n = 2\pi\i nT$ for some $n\in\Zbb$. Note that we need to exclude $z = 0$ since here the numerator vanishes as well and the symbol is bounded. Hence, we have to verify that the equation
\begin{align}
(p + \lambda a)^2 - \mu = 2\pi\i nT \label{Zero-equation of tanh}
\end{align}
has no solution $\lambda$ in $S_{\mu, T}$ for any $n\in\Zbb\setminus \{0\}$ and any $p\in\Rbb^3$. By choosing a suitable basis on $\Rbb^3$, we may assume that $a = e_1$. If we assume for contradiction that \eqref{Zero-equation of tanh} had a solution, then separating real and imaginary parts yields the equations
\begin{align}
p^2 + 2x p_1 + x^2 - y^2 - \mu &=0, & 2yp_1 + 2xy - 2\pi n T &=0, \label{CT:Analytic_family_well-def_1}
\end{align}
where $\lambda = x + \i y$. Without loss, we may assume that $y \neq 0$ since otherwise the second equation in \eqref{CT:Analytic_family_well-def_1} simplifies to $2\pi n T =0$, which has no solution at all and we have finished. We solve the second equation in \eqref{CT:Analytic_family_well-def_1} for $p_1$, insert this into the first equation, and find
\begin{align*}
\bigl[ \frac{\pi n T}{y} - x\bigr]^2 + p_2^2 + p_3^2 + 2x\bigl[ \frac{\pi n T}{y} - x\bigr] + x ^2 - y^2 = \mu
\end{align*}
or, put differently,
\begin{align*}
\bigl(\frac{\pi n T}{y}\bigr)^2 + p_2^2  + p_3^2 - y^2 - \mu =0.
\end{align*}
This equation cannot have any solution provided we can guarantee for
\begin{align*}
\bigl(\frac{\pi nT}{y}\bigr)^2 - y^2 -\mu \geq \bigl(\frac{\pi T}{y}\bigr)^2 - y^2 -\mu >0.
\end{align*}
But this is true on $S_{\mu, T}$, since here $y^4 + \mu y^2 - (\pi T)^2 < 0$.
\end{proof}

\subsubsection{Integral representation for \texorpdfstring{$K_T^{\lambda a}$}{KTlambdaa}}

In Lemma \ref{DHS1:KT_integral_rep}, we have proven an integral representation for the operator $K_T$ that exploits the fact that the Laplacian is bounded from below.  We need to extend this result now to also hold for $K_T^{\lambda a}$ with nonreal $\lambda$. The path for the integral representation has already been introduced in Definition \ref{DHS1:speaker path}. We restate it here for convenience.

\begin{defn}[Speaker path]
\label{CT:speaker_path}
Let $R>0$ and $\alpha \geq 0$. Using the notation $\beta := T^{-1}$, define the following complex paths
\begin{align*}
\begin{split}
\gamma_1(t) &:= \frac{\pi\i}{2\beta} + (1 + \i)t,\\
\gamma_2(t) &:= \frac{\pi\i}{2\beta} - (\alpha + 1)t,\\
\gamma_3(t) &:= -\frac{\pi\i}{2\beta}t - (\alpha + 1),\\
\gamma_4(t) &:= -\frac{\pi\i}{2\beta} - (\alpha + 1)(1-t), \\
\gamma_5(t) &:= -\frac{\pi\i }{2\beta} + (1 - \i)t,
\end{split}
&
\begin{split}
\phantom{ \frac \i\betac }t&\in [0,R], \\
\phantom{ \frac \i\betac }t &\in [0,1], \\
\phantom{ \frac \i\betac }t&\in [-1,1],\\
\phantom{ \frac \i\betac }t &\in [0,1],\\
\phantom{ \frac \i \betac }t&\in [0,R].
\end{split}
& \begin{split} 
\text{\includegraphics[width=6cm]{3_Perturbation_Theory/Speaker_path.pdf}}
\end{split} 
\end{align*}
The speaker path is defined as the union of paths $\gamma_i$, $i=1, \ldots, 5$, with $\gamma_1$ taken in reverse direction, i.e.,
\begin{align*}
\speaker_{\alpha,R} := \mathop{\dot -}\gamma_1 \mathop{\dot +} \gamma_2 \mathop{\dot +} \gamma_3 \mathop{\dot +} \gamma_4 \mathop{\dot +} \gamma_5.
\end{align*}
We also let $\speaker_\alpha := \bigcup_{R>0} \speaker_{\alpha,R}$.
\end{defn} 

\begin{kor}
\label{CT:Speaker_Path_Estimate}
Define
\begin{align}
r_\speaker := \inf_{z\in \speaker_{\mu_+}} \min \Bigl\{ 1, \frac 12 \Bigl[ \sqrt{\frac{2}{|z+\mu| - \Re (z+\mu)}} + \frac{1}{\dist(z + \mu, \sigma(p^2))}\Bigr]^{-1} \Bigr\}. \label{CT:Speaker_Path_Estimate_Assumption}
\end{align}
Then, for any $a\in \Rbb^3$ with $|a|=1$ and any $\lambda\in \ov B_{r_\speaker}(0)$, we have $\speaker_{\mu_+} \subseteq \rho( (p+\lambda a)^2 -\mu)$ and
\begin{align*}
\sup_{z\in \speaker_{\mu_+}} \Bigl\Vert \frac 1{z - (p+\lambda a)^2} \Bigr\Vert_\infty \leq C.
\end{align*}
\end{kor}

\begin{proof}
By construction of the speaker path, $r_\speaker>0$. Hence, we have the conclusion of Lemma \ref{Laplace_Analytic_estimate} for every $z\in \speaker_{\mu_+}$. Furthermore, on the speaker path, we always have $\Im z \geq \frac{\pi T}{2}$ or $\Re z+\mu_+ = -1$, i.e., $\dist(\speaker_{\mu_+}, \sigma(p^2))^{-1} \leq C$. Hence, the bound follows.
\end{proof}

For the sake of convenience, we restate Lemma \ref{DHS1:KT_integral_rep} here, whose extended proof we have given in Lemma \ref{DHS1+:KT_integral_rep}.

\begin{lem}
\label{CT:KT_integral_rep}
Let $\alpha \geq 0$ and let $H\colon \Dcal(H)\ra \Hcal$ be a self-adjoint operator in a separable Hilbert space $\Hcal$ with $H\geq -\alpha$. Then, we have
\begin{align*}
\frac{H}{\tanh(\frac{\beta H}{2})} = H + \lim_{R\to\infty} \int_{\speaker_{\alpha, R}} \frac{\dd z}{2\pi\i} \Bigl( \frac{z}{\tanh(\frac{\beta z}{2})} - z \Bigr) \frac{1}{z - H},
\end{align*}
where $\speaker_{\alpha, R}$ is the speaker path from Definition \ref{DHS1+:speaker_path}. The limit exists in operator norm.
\end{lem}

We now provide an analytic version of Lemma \ref{CT:KT_integral_rep} The proof of the following Lemma \ref{CT:KTlambdaa_integral_representation} is a bit more complicated since $(p +\lambda a)^2-\mu$ is not self-adjoint but only normal. The drawback is that we restrict to the operator $(p + \lambda a)^2-\mu$ instead of a general operator, as had been the case in the results mentioned.

\begin{lem}
\label{CT:KTlambdaa_integral_representation}
Let $\lambda\in S_{\mu, T} \cap \ov B_{r_\speaker}(0)$ with $r_\speaker$ from \eqref{CT:Speaker_Path_Estimate_Assumption} and $S_{\mu, T}$ from \eqref{CT:Analytic_family_strip}. Then, for any $a\in \Rbb^3$ with $|a|=1$, the identity
\begin{align*}
\frac{(p+\lambda a)^2 - \mu}{\tanh(\frac{(p+\lambda a)^2-\mu}{2T})} - (p + \lambda a)^2-\mu = \lim_{R\to\infty} \int_{\speaker_{\mu_+, R}} \frac{\dd z}{2\pi\i} \Bigl( \frac{z}{\tanh(\frac{z}{2T})} - z \Bigr) \frac{1}{z+\mu - (p + \lambda a)^2 },
\end{align*}
holds in the operator norm topology. Here $\speaker_{\mu_+, R}$ is the speaker path from Definition \ref{CT:speaker_path}. The limit exists in operator norm and defines a uniformly bounded operator in $\lambda$.
\end{lem}

\begin{proof}
The function $f_T(z) := \frac{z}{\tanh(\frac{z}{2T})} - z = \frac{2z}{\e^{\nicefrac zT} - 1}$ is an analytic function in the open domain $\Cbb \setminus 2\pi T\i \Zbb_{\neq 0}$. Let us write $H_\lambda := (p+\lambda a)^2-\mu$ for short. We first prove that the limit 
\begin{align*}
g_T(H_\lambda) := \lim_{R\to\infty} \int_{\speaker_{\mu_+,R}} \frac{\dd z}{2\pi\i} \; f_T(z) \, \frac{1}{z - H_\lambda}
\end{align*}
exists in operator norm and defines a bounded operator.  

To do this, we investigate the tails of the paths $\gamma_1$ and $\gamma_5$. For example, we have to investigate the operator norm of
\begin{align*}
\int_R^\infty \frac{\dd t}{2\pi\i} \frac{2\gamma_1(t)}{\e^{\beta \gamma_1(t)} - 1} \frac{1}{\gamma_1(t) - H_\lambda} \, \gamma_1'(t).
\end{align*}
The following simple estimates
\begin{align}
|\gamma_1(t)| &\leq \frac{\pi}{2\beta} + \sqrt{2}\, t \leq C\, t, & |\gamma_1'(t)| &= \sqrt{2}, & \Re \gamma_1(t) &= t, \label{CT:KTlambdaa_integral_representation_1}
\end{align}
hold for $R$ large enough. Furthermore, since $|\lambda|  \leq r_\speaker$, \eqref{CT:KTlambdaa_integral_representation_1} and Corollary \ref{CT:Speaker_Path_Estimate} imply
\begin{align*}
\Bigl\Vert \int_R^\infty \frac{\dd t}{2\pi\i} \frac{2\gamma_1(t)}{\e^{\beta \gamma_1(t)} - 1} \frac{1}{\gamma_1(t) - H_\lambda} \; \gamma_1'(t)\Bigr\Vert_\infty &\leq C \int_R^\infty \dt \; \frac{t}{\e^{\beta t} - 1} \leq C \, \e^{-\frac \beta  2 R}.
\end{align*}
The last inequality follows by taking $R$ so large that $1 \leq \frac 12 \e^{\beta t}$ and $t\e^{-\frac\beta 2t} \leq \frac 12$ for all $t\geq R$. The contribution of $\gamma_5$ is estimated in a similar fashion. This proves operator norm convergence of the limit and the fact that $g_T(H_\lambda)$ is a bounded operator with uniform norm bound in $\lambda$.

Let $K\geq 1$ and choose $\psi\in \ran(\Idbb_{S_K}(H_\lambda))$, where $S_K := \{z\in \Cbb : \Re z \leq K\}$. Take $R\geq K+1$ and close the speaker path by the contour $\gamma_R(t) := R + (R + \frac{\pi}{2\beta})\i t$, where $t\in [-1,1]$. We recall that $f_T$ is an analytic function in the open domain $\Cbb \setminus 2\pi T\i \Zbb_{\neq 0}$, in particular in the interior of the closed path $\speaker_{\mu_+,R} \dot + \gamma_R$. Hence, for each $\varphi\in L^2(\Rbb^3)$, by Cauchy's integral theorem and the spectral theorem for normal operators, we obtain
\begin{align}
\langle \varphi, [K_T^{\lambda a} - H_\lambda] \psi\rangle = \langle \varphi, f_T(H_\lambda)\psi\rangle = \int_{\speaker_{\mu_+, R}\dot + \gamma_R} \frac{\dd z}{2\pi\i} \; f_T(z) \; \langle \varphi, (z - H_\lambda)\psi\rangle. \label{KTlambdaa_First_identity}
\end{align}
When we investigate the contribution from $\gamma_R$, we have
\begin{align}
\Bigl|\int_{\gamma_R} \frac{\dd z}{2\pi\i} \; f_T(z) \langle \varphi, (z - H_\lambda)^{-1}\psi\rangle\Bigr| &\leq \int_{-1}^1 \frac{\dt}{\pi} \frac{|\gamma_R(t)|}{|\e^{\beta \gamma_R(t)}| - 1} \; |\langle \varphi, (\gamma_R(t) - H_\lambda)^{-1} \psi\rangle|\, |\gamma_R'(t)|. \label{CT:KTlambdaa_integral_representation_2}
\end{align}
First of all, since $\psi\in \ran(\Idbb_{S_K}(H_\lambda))$, we have
\begin{align*}
|\langle \varphi, (\gamma_R(t) - H_\lambda)^{-1} \psi\rangle| \leq \Bigl\Vert \frac{1}{\gamma_R(t) - H_\lambda} \, \Idbb_{S_K}(H_\lambda) \Bigr\Vert_\infty \, \Vert \varphi\Vert_2 \, \Vert \psi\Vert_2.
\end{align*}
Let us give a bound on $\Vert(\gamma_R(t) - H_\lambda)^{-1}\Idbb_{S_K}(H_\lambda)\Vert_\infty$. For $\eta\in L^2(\Rbb^3)$ consider
\begin{align*}
\Bigl\Vert \frac 1{\gamma_R(t) - H_\lambda} \, \Idbb_{S_K}(H_\lambda) \, \eta\Bigr\Vert_2^2 = \int_{\sigma((p+\lambda a)^2-\mu)} \frac{1}{|\gamma_R(t) - s|^2} \; \Idbb_{S_K}(s) \, \dd \mu_\eta(s),
\end{align*}
where the integral has to be understood as a Lebesgue integral on $\Cbb$. Now, the function of the integrand is bounded as follows
\begin{align*}
\frac{1}{|\gamma_R(t) - s|^2} = \frac{1}{(R- \Re s)^2 + (\Im \gamma_R(t) - \Im s)^2} \leq \frac{1}{1 + (\Im \gamma_R(t) - \Im s)^2} \leq 1.
\end{align*}
Hence, $\sup_{t\in [-1,1]} \Vert (\gamma_R(t) - H_\lambda)^{-1} \Idbb_{S_K}(H_\lambda)\Vert_\infty \leq 1$. In combination with \eqref{CT:KTlambdaa_integral_representation_2}, we obtain
\begin{align*}
\sup_{\Vert \varphi\Vert_2 = 1} \Bigl|\int_{\gamma_R} \frac{\dd z}{2\pi\i} \; f_T(z) \langle \varphi, (z - H_\lambda)^{-1}\psi\rangle\Bigr| &\leq C \, \frac{R^2}{\e^{\beta R} - 1} \; \Vert\psi\Vert_2 \xra{R\to\infty} 0.
\end{align*}
This proves that \eqref{KTlambdaa_First_identity} converges as $R\to\infty$ uniformly in $\varphi\in L^2(\Rbb^3)$ and we obtain
\begin{align}
f_T(H_\lambda)\psi = g_T(H_\lambda)\psi. \label{Preliminary fTH}
\end{align}
In particular, $f_T(H_\lambda)$ is a bounded operator on every subspace $\ran(\Idbb_{S_K}(H_\lambda))$ with uniform norm bound in $K$.

Since $\Idbb_{S_K}(H_\lambda) \to \Idbb$ strongly and $\bigcup_{K>0} \ran(\Idbb_{S_K}(H_\lambda))$ is dense in $L^2(\Rbb^3)$, by standard methods for unique continuation of bounded operators, we get \eqref{Preliminary fTH} for all $\psi\in L^2(\Rbb^3)$. 
%
\end{proof}

\subsubsection{\texorpdfstring{$K_{T}-V$}{KT-V} extends to an analytic family}

\begin{thm}
Let $a\in \Rbb^3$ with $|a| = 1$ and let $V\in L^2(\Rbb^3)$. Then the operator family $\{K_T^{\lambda a}-V\}_{\lambda}$, defined on $H^2(\Rbb^3)$, is strongly continuous on $S_{\mu, T}\cap \ov B_{r_\speaker}(0)$ and analytic of type (A) on $S_{\mu, T} \cap B_{r_\speaker}(0)$. Here, $K_T^{\lambda a}$ is given by \eqref{Analytic family}, $S_{\mu, T}$ by \eqref{CT:Analytic_family_strip}, and $r_\speaker$ by \eqref{CT:Speaker_Path_Estimate_Assumption}.
\end{thm}

\begin{note}
As above in the case of $(p + \lambda a)^2$, the operator $K_T^{\lambda a}$ is not self-adjoint for nonreal $\lambda$, but only normal. However, the operator $K_T^{\lambda a} - V$ has none of these properties.
\end{note}

\begin{proof}
We first show that $K_T^{\lambda a}- V$ is a well-defined closed operator on $H^2(\Rbb^3)$ for every $\lambda\in S_{\mu, T} \cap B_{r_\speaker}(0)$. By Lemma \ref{CT:Analytic_family_well-def}, we know that $K_T^{\lambda a}$ is a well-defined operator on $H^2(\Rbb^3)$ for every $\lambda\in S_{\mu, T}$. Call $f_T(z) := \frac{z}{\tanh(\frac{z}{2T})} - z = \frac{2z}{\e^{\nicefrac zT}-1}$. Then, we rewrite
\begin{align*}
K_T^{\lambda a} = -\Delta - \mu + 2\lambda \, a \, (-\i \nabla) + \lambda^2 + M(\lambda)
\end{align*}
with $M(\lambda) := f_T((-\i\nabla + \lambda a)^2-\mu)$. From Lemma \ref{CT:KTlambdaa_integral_representation}, we know that $M(\lambda)$ is a uniformly bounded operator for $\lambda \in S_{\mu, T} \cap \ov B_{r_\speaker}(0)$. Hence, when we define
\begin{align*}
W(\lambda) := -\mu + 2\lambda \, a p + \lambda^2 + M(\lambda) - V,
\end{align*}
we get
\begin{align}
K_T^{\lambda a} - V = -\Delta + W(\lambda). \label{KTV_Analytic_Family_1}
\end{align}
We claim that $p$ and $V$ are $-\Delta$-bounded with $-\Delta$-bound zero. This is true for $V$ because $V(1 + p^2)^{-1}$ is a Hilbert-Schmidt operator, thus compact and hence $p^2$-bounded with $p^2$-bound $0$. For any $\varepsilon>0$ and $\psi\in H^2(\Rbb^3)$, we also have $\Vert p\psi\Vert_2^2 \leq \varepsilon \Vert p^2\psi\Vert_2^2 + (4\varepsilon)^{-1}\Vert \psi\Vert_2^2$, since $p$ is self-adjoint. Hence, $W(\lambda)$ is $-\Delta$-bounded with $-\Delta$-bound $0$ as well. In particular, $\Dcal(K_T^{\lambda a} - V) = H^2(\Rbb^3)$. It follows that $K_T^{\lambda a}- V$ is a well-defined densely defined operator on $H^2 (\Rbb^3)$.

We claim that $K_T^{\lambda a} - V$ is closed for each $\lambda\in S_{\mu, T} \cap \ov B_{r_\speaker}(0)$. To prove this let $\{\psi_n\}_n \subseteq \Dcal(K_T^{\lambda a} - V) = H^2(\Rbb^3)$ be convergent in $L^2(\Rbb^3)$ to some $\psi\in L^2(\Rbb^3)$ and assume that there is $\eta\in L^2(\Rbb^3)$ such that $(K_T^{\lambda a } - V)\psi_n\to \eta$ in $L^2(\Rbb^3)$. Then, for every $\varepsilon>0$,
\begin{align*}
\Vert (-\Delta)(\psi_n - \psi_m)\Vert_2 &\leq \Vert (-\Delta + W(\lambda))(\psi_m - \psi_n)\Vert_2 + \varepsilon\,  \Vert (-\Delta)(\psi_n - \psi_m)\Vert_2 \\
&\hspace{180pt} + C_\varepsilon \,  \Vert \psi_m - \psi_n\Vert_2
\end{align*}
Taking $\varepsilon < 1$, this shows that $(1 - \varepsilon)\Vert (-\Delta)(\psi_n - \psi_m)\Vert$ tends to $0$ as $m, n\to\infty$. We conclude that $\{\psi_n\}_n$ converges in $H^2(\Rbb^3)$, which implies that $\psi\in H^2(\Rbb^3) = \Dcal(K_T^{\lambda a} - V)$.
Furthermore, using $-\Delta$-boundedness of $W(\lambda)$ again, we obtain
\begin{align*}
\Vert \eta - (-\Delta + W(\lambda))\psi\Vert_2 &= \lim_{n\to\infty} \Vert (-\Delta + W(\lambda))(\psi_n - \psi)\Vert_2 \\
&\leq (1 + \varepsilon) \lim_{n\to\infty}\Vert (-\Delta)(\psi_n - \psi)\Vert_2 + C_\varepsilon \lim_{n\to\infty} \Vert \psi_n - \psi\Vert_2 =0
\end{align*}
so that $\eta = (-\Delta + W(\lambda))\psi$. This proves that $K_T^{\lambda a} - V$ is closed.

To prove that $-\Delta +W(\lambda)$ is an analytic family of type (A), we have to further prove that (see \cite[p.16]{Reedsimon4} or \cite[p. 375]{Kato}) 
\begin{enumerate}[(i)]
\item $-\Delta + W(\lambda)$ has non-empty resolvent set\footnote{We remark that closedness is necessary but not sufficient for this condition.} for each $\lambda\in S_{\mu, T} \cap B_{r_\speaker}(0)$,
\item $\{-\Delta + W(\lambda)\}_\lambda$ is strongly analytic on $S_{\mu, T} \cap B_{r_\speaker}(0)$.
\end{enumerate}
To prove (i), we write, for some $z\in \Cbb \setminus [0,\infty)$:
\begin{align*}
z +\Delta - W(\lambda) = \bigl( 1- W(\lambda) \frac{1}{z + \Delta} \bigr) (z+\Delta).
\end{align*}
We know that $z + \Delta$ is invertible from $H^2(\Rbb^3)$ to $L^2(\Rbb^3)$. Hence, it suffices to prove that $\Vert W(\lambda)(z +\Delta)^{-1}\Vert_\infty \leq \frac 12$ for suitably chosen $z$. To see this, use that $W(\lambda)$ is $-\Delta$-bounded with $-\Delta$-bound $0$. This means that for all $\psi\in H^2(\Rbb^3)$, we have
\begin{align*}
\Vert W(\lambda)\psi\Vert_2 \leq \frac 14 \, \Vert (-\Delta)\psi\Vert_2 + C \, \Vert \psi\Vert_2.
\end{align*}
Hence,
\begin{align}
\bigl\Vert W(\lambda) \frac{1}{z + \Delta} \bigr\Vert_\infty \leq \frac 14 \, \bigl\Vert (-\Delta)\frac{1}{z + \Delta}\bigr\Vert_\infty + C \, \bigl\Vert \frac{1}{z + \Delta}\bigr\Vert_\infty. \label{KTV_Analytic_Family_2}
\end{align}
The function $t\mapsto \frac{t}{z - t}$ on $[0,\infty)$ is bounded by $1$ in absolute value provided $\Re z \leq  0$. For, the modulus function $t\, [(\Re z - t)^2+(\Im z)^2]^{-\nicefrac 12}$ tends to $1$ as $t\to\infty$, is $0$ at $t =0$ and has a maximum at $t = \frac{|z|^2}{\Re z}$ if and only if $\Re z > 0$. For $\Re z =0$, the bound is trivial. Hence, choosing $\Im z \geq 4C$ ensures that $\Vert W(\lambda)(z + \Delta)^{-1}\Vert_\infty \leq \frac 12$. This proves (i). Part (ii) is trivial for all terms in $W(\lambda)$ except for $M(\lambda)$. By \cite[Theorem VI.4]{Reedsimon1}, it suffices to prove that $M(\lambda)$ is weakly analytic. The expectation value reads
\begin{align*}
\langle \psi, M(\lambda)\psi\rangle = \int_{\Rbb^3} \dd p \; f_T((p + \lambda a)^2-\mu) \, |\hat \psi(p)|^2.
\end{align*}
Now, the claim follows from Morera's theorem and Fubini's theorem provided we can show that $\langle \psi, M(\lambda)\psi\rangle$ is continuous. Let $\psi\in H^2(\Rbb^3)$ and let $\lambda, \lambda_0\in S_{\mu, T} \cap B_{r_\speaker}(0)$ be given. Then, Lemma \ref{CT:KTlambdaa_integral_representation} implies
\begin{align*}
\langle \psi, M(\lambda)\psi\rangle - \langle \psi, M(\lambda_0)\psi\rangle &= \\
&\hspace{-110pt}= \int_{\Rbb^3} \dd p \int_{\speaker_{\mu_+}} \frac{\dd z}{2\pi\i } \; f_T(z) \, \frac{1}{z - H_\lambda(p)} \, \bigl[ (p + \lambda a)^2 - (p + \lambda_0 a)^2\bigr] \, \frac{1}{z - H_{\lambda_0}(p)} \, |\hat \psi(p)|^2.
\end{align*}
Here, we used again the notation $H_\lambda(p) = (p + \lambda a)^2 -\mu$. Since
\begin{align*}
(p + \lambda a)^2 - (p + \lambda_0a)^2 = 2(\lambda - \lambda_0)a p + (\lambda - \lambda_0) (\lambda + \lambda_0),
\end{align*}
and $|ap|\leq |p|$ by the Cauchy-Schwarz inequality on $\Rbb^3$, as well as $|\lambda + \lambda_0 |\leq 2r_\speaker$, we infer
\begin{align*}
|\langle \psi, M(\lambda)\psi\rangle - \langle \psi, M(\lambda_0)\psi\rangle| &\leq 2 \, |\lambda - \lambda_0| \sup_{z\in \speaker_{\mu_+}} \Bigl\Vert \frac{1}{z - H_\lambda}\Bigr\Vert_\infty \, \sup_{z\in \speaker_{\mu_+}} \Bigl\Vert \frac{1}{z - H_{\lambda_0}}\Bigr\Vert_\infty \\
&\hspace{50pt}\times \int_{\speaker_{\mu_+}} \frac{\dd |z|}{2\pi} \; |f_T(z)| \int_{\Rbb^3} \dd p \; (|p| + r_\speaker) |\hat\psi(p)|^2.
\end{align*}
Here, $\dd |z| = \dt\, |z'(t)|$. Since $\psi\in H^2(\Rbb^3)$, the latter integral is finite. The uniform operator norm bounds are finite by Corollary \ref{CT:Speaker_Path_Estimate}, and since the contour integral converges, we conclude continuity of $\lambda \mapsto \langle \psi, M(\lambda)\psi\rangle$.
\end{proof}

\subsubsection{Explicit resolvent estimates for \texorpdfstring{$K_T^{\lambda a} - V$}{KTlambdaa}}

\begin{thm}
\label{KTlambdaV_Analytic_Estimate}
Let $V\in L^\infty(\Rbb^3)$, $z\in \rho(K_T - V)$. For all $\nu \in \Rbb$, $\nu < -\mu$, define the function
\begin{align*}
f_{\nu,V}(z) &:= \frac{1}{\dist(z, \sigma(K_T - V))} + \frac{2}{\sqrt{|\nu + \mu|}} \Bigl[ 1+  \frac{2\mu_+}{1 - \e^{-\beta \mu_+}}\Bigr]\Bigl[ 1 + \frac{|\nu - z|}{\dist(z, \sigma(K_T - V))}\Bigr] \\
&\hspace{200pt}\times \Bigl[ 1 + \frac{\Vert V\Vert_\infty}{\dist(z, \sigma(K_T - V))}\Bigr] \\
&\hspace{30pt} + \int_{\speaker_{\mu_+}} \frac{\dd |w|}{2\pi}\; |f_T(w)|  \sup_{w\in \speaker_{\mu_+}} \frac{2}{\dist(w+\mu, \sigma(p^2))} \\
&\hspace{100pt} \times \sup_{w\in \speaker_{\mu_+}} \biggl[ \sqrt{\frac{2}{|w + \mu| - \Re (w + \mu)}} + \frac{1}{\dist(w + \mu, \sigma(p^2))}\biggr].
\end{align*}
Then, for any $\lambda \in S_{\mu, T} \cap \ov B_{r_\speaker}(0) \cap \ov B_{r_z}(0)$ with $r_\speaker>0$ from \eqref{CT:Speaker_Path_Estimate_Assumption} and
\begin{align*}
r_z := \frac 12 \, \frac 1{f_{\nu, V}(z)}
\end{align*}
and $a\in \Rbb^3$ with $|a|=1$, the following statements are true:
\begin{enumerate}[(a)]
\item $z\in \rho(K_T^{\lambda a} - V)$ holds and the resolvent $\Rcal_T^{z,V}(\lambda a)$ of $K_T^{\lambda a} - V$ at $z$ satisfies
\begin{align*}
\Vert \Rcal_T^{z,V}(\lambda a)\Vert_\infty \leq \frac{2}{\dist(z, \sigma(K_T - V))}.
\end{align*}

\item $\Rcal_T^{z,V}(\lambda a)$ is a bounded operator from $L^2(\Rbb^3)$ to $L^\infty(\Rbb^3)$ with norm bound
\begin{align*}
\Vert \Rcal_T^{z,V}(\lambda a)\Vert_{2,\infty} &\leq 2\, \Vert \Rcal_T^{z,V}\Vert_{2,\infty}
\end{align*}
\end{enumerate}
\end{thm}

\begin{proof}
We start with the identity
\begin{align}
\begin{split}
z - (K_T^{\lambda a} - V) &= z - (K_T - V) + K_T - K_T^{\lambda a}\\
&= \bigl[ 1 - (K_T^{\lambda a} - K_T) \Rcal_T^{z,V} \bigr] (z - (K_T - V)).
\end{split} \label{KTlambdaV_Analytic_Estimate_2}
\end{align}
Hence, to prove part (a), we need to show that
\begin{align}
\bigl\Vert (K_T^{\lambda a} - K_T) \Rcal_T^{z,V}\bigr\Vert_\infty \leq \frac 12 \label{KTlambdaV_Analytic_Estimate_1}
\end{align}
for all $\lambda \in S_{\mu, T} \cap \ov B_{r_z}(0) \cap \ov B_{r_\speaker}(0)$. If this is true, then the Neumann series implies
\begin{align}
\Bigl\Vert \frac 1{1 - (K_T^{\lambda a} - K_T) \Rcal_T^{z,V}} \Bigr\Vert_\infty \leq 2 \label{KTlambdaV_Analytic_Estimate_3}
\end{align}
and the claim follows. To see that \eqref{KTlambdaV_Analytic_Estimate_1} is true, let us employ the integral representation Lemma \ref{CT:KTlambdaa_integral_representation} to infer
\begin{align*}
K_T^{\lambda a} - K_T = 2\lambda \, ap + \lambda^2 + \int_{\speaker_{\mu_+}} \frac{\dd w}{2\pi\i } \; f_T(w) \Bigl[ \frac{1}{w+\mu - (p+\lambda a)^2} \, (2\lambda ap+\lambda^2) \,  \frac{1}{w+\mu - p^2}\Bigr].
\end{align*}
We begin by bounding the last term. Recall equation \eqref{Laplace_analytic_1} in the proof of Corollary \ref{Laplace_Analytic_estimate}. Since the speaker path stays away from $\sigma(p^2)$ uniformly for $\lambda\in \ov B_{r_\speaker}(0)$ (in particular $|\lambda|\leq 1$), we get that
\begin{align*}
\Bigl\Vert\int_{\speaker_{\mu_+}} \frac{\dd w}{2\pi\i } \; f_T(w) \Bigl[ \frac{1}{w+\mu - (p+\lambda a)^2} \, (2 ap+\lambda) \, \frac{1}{w+\mu - p^2}\Bigr]\Bigr\Vert_\infty &\\
&\hspace{-250pt} \leq \int_{\speaker_{\mu_+}} \frac{\dd |w|}{2\pi} \; |f_T(w)|   \sup_{w\in \speaker_{\mu_+}} \frac{2}{\dist(w+\mu, \sigma(p^2))} \\
&\hspace{-160pt} \times \sup_{w\in \speaker_{\mu_+}} \biggl[ \sqrt{\frac{2}{|w+\mu|-\Re (w+\mu)}}+\frac{1}{\dist(w+\mu, \sigma(p^2))}\biggr].
\end{align*}
Here, $\dd |w|= \dt\, |w'(t)|$. It remains to bound $ap\, \Rcal_T^{z,V}$. To do this, we use the resolvent equations multiple times to arrive at
\begin{align*}
ap \, \Rcal_T^{z,V} = ap \, (\nu - p^2 + \mu)^{-1} \bigl[ 1 + [K_T - p^2 + \mu] \, \Rcal_T^\nu\bigr] \bigl[ 1 + (\nu - z)\,  \Rcal_T^z\bigr] \bigl[ 1 + V \, \Rcal_T^{z,V}\bigr].
\end{align*}
Recall the estimate in \eqref{Laplace_analytic_2}, whence, using $\nu + \mu <0$, we read off the bound
\begin{align*}
\Vert ap \, \Rcal_T^{z,V}\Vert_\infty &\leq \frac{1}{2 \sqrt{|\nu+\mu|}} \Bigl[ 1 + \frac{2\mu_+}{1 -\e^{-\beta \mu_+}}\Bigr] \Bigl[ 1 + \frac{|\nu - z|}{\dist(z, \sigma(K_T - V))}\Bigr] \\
&\hspace{200pt} \times \Bigl[1 + \frac{\Vert V\Vert_\infty}{\dist(z, \sigma(K_T - V))}\Bigr]. 
\end{align*}
Multiplying this by $2$ and adding $\Vert \lambda \, \Rcal_T^{z,V}\Vert_\infty \leq \dist(z, \sigma(K_T - V))^{-1}$, proves that 
\begin{align*}
\Vert (K_T^{\lambda a} - K_T) \Rcal_T^{z,V}\Vert_\infty \leq |\lambda| \, f_{\nu,V}(z).
\end{align*}
Hence, for $\lambda\in \ov B_{r_z}(0)$, we conclude \eqref{KTlambdaV_Analytic_Estimate_1}. This proves part (a). For part (b), we utilize \eqref{KTlambdaV_Analytic_Estimate_2} to note that
\begin{align*}
\Rcal_T^{z,V}(\lambda a) = \Rcal_T^{z,V} \, \bigl[ 1 - (K_T^{\lambda a} - K_T)\Rcal_T^{z,V}\bigr]^{-1}.
\end{align*}
It follows that the $(2,\infty)$-norm of $\Rcal_T^{z,V}(\lambda a)$ is bounded by $\Vert \Rcal_T^{z,V}\Vert_{2,\infty}$ times $2$, see \eqref{KTlambdaV_Analytic_Estimate_3}.
\end{proof}

\subsection{Exponential estimate for \texorpdfstring{$K_T - V$}{KT-V}}

\begin{thm}
\label{CT:Gcal_exponential_decay}
Let $V\in L^2(\Rbb^3) \cap L^\infty(\Rbb^3)$ and $z\in \rho(K_T - V)$. Let 
\begin{align}
\delta_0(z) := \sup \bigl\{ r>0 : B_r(0) \subseteq S_{\mu, T} \cap B_{r_\speaker}(0) \cap B_{r_z}(0) \bigr\} \label{CT:Gcal_exponential_decay_eq1}
\end{align}
with $r_\speaker$ and $r_z$ from Theorem \ref{KTlambdaV_Analytic_Estimate}. Then, the resolvent kernel $\Gcal_T^{z,V}(x,y)$ in \eqref{CT:GcalTAz_definition} satisfies
\begin{align*}
\Vert \Gcal_T^{z,V}\Vert_\delta := \sup_{x\in\Rbb^3} \Bigl( \int_{\Rbb^3} \dd y\;  \e^{\delta \, |x - y|} \, |\Gcal_T^{z,V}(x,y)|^2\Bigr)^{\nicefrac 12} \leq 12 \cdot \Vert \Rcal_T^{z,V}\Vert_{2,\infty}.
\end{align*}
for every $0 \leq \delta < \delta_0(z)$. 
\end{thm}

\begin{proof}
By Lemma \ref{CT:Simon}, $\Rcal_T^{z,V}(\lambda a)$ possesses an integral kernel $\Gcal_T^{z,V}(\lambda a;x,y)$, which satisfies
\begin{align*}
\Vert \Rcal_T^{z,V}(\lambda a)\Vert_{2,\infty} = \sup_{x\in \Rbb^3} \Bigl( \int_{\Rbb^3} \dy\; |\Gcal_T^{z,V}(\lambda a; x,y)|^2 \Bigr)^{\nicefrac 12} < \infty .
\end{align*}
As a first step, let us prove that for a.e. $x,y\in \Rbb^3$ and $\lambda \in S_{\mu, T} \cap \ov B_{r_\speaker}(0) \cap \ov B_{r_z}(0)$, we have
\begin{align}
\Gcal_T^{z,V}(\lambda a;x,y) = \e^{\i \lambda a \cdot (x - y)} \, \Gcal_T^{z,V}(x,y). \label{CT:Rz-decay_1}
\end{align}
To see this, let $\varphi, \psi\in C_c^\infty(\Rbb^3)$ and we claim that 
\begin{align}
\langle \varphi, \Rcal_T^{z,V}(\lambda a) \psi\rangle = \langle \varphi, \e^{\i \lambda a \mathop \cdot} \, \Rcal_T^{z,V} \, \e^{-\i \lambda a \mathop \cdot} \psi \rangle \label{CT:Rz-decay_3}
\end{align}
holds for all $\lambda \in S_{\mu, T} \cap B_{r_\speaker}(0) \cap B_{r_z}(0)$. Since $K_T^{\lambda a} - V$ is analytic of type (A), we infer that $\langle \varphi, \Rcal_T^{z,V}(\lambda a)\psi\rangle$ is analytic\footnote{This follows from Theorem XII.7 together with Problem 10 of that section in \cite{Reedsimon4}.} in the open domain $S_{\mu, T} \cap B_{r_\speaker}(0) \cap B_{r_z}(0)$. The right-hand side of \eqref{CT:Rz-decay_3} is an entire function of $\lambda$, since $\psi$ and $\varphi$ have compact support, by the mean value theorem, and by dominated convergence. Since left and right side coincide for $\lambda\in \Rbb$, we conclude by the identity theorem that equality holds in the whole domain $S_{\mu, T} \cap B_{r_\speaker}(0) \cap B_{r_z}(0)$. On the level of kernels, \eqref{CT:Rz-decay_3} translates to
\begin{align*}
\int_{\Rbb^3} \dx \int_{\Rbb^3} \dy \;  \Gcal_T^{z,V}(\lambda a;x,y)  \, \ov{\varphi(x)} \psi(y) &= \int_{\Rbb^3} \dd x \int_{\Rbb^3} \dd y\; \e^{\i \lambda a \cdot (x-y)}  \,  \Gcal_T^{z,V}(x,y) \, \ov{\varphi(x)}\psi(y).
\end{align*}
We apply the fundamental lemma of the calculus of variations twice and deduce \eqref{CT:Rz-decay_1}.

Let $\delta_0$ be as in \eqref{CT:Gcal_exponential_decay_eq1} and $0< \delta  < \delta_0$ (for $\delta =0$, the claim is Lemma \ref{Resolvent_KTV_bounded_uniformly}). Let $b\in \Rbb^3$ be given with $|b| = \delta$. Apply \eqref{CT:Rz-decay_1} to $a := \delta^{-1} b$ and $\lambda = -\i \delta$. This implies $|a| = 1$ so that (by a slight abuse of notation)
\begin{align}
\Gcal_T^{z,V}(b; x,y) &= \e^{b\cdot (x-y)} \, \Gcal_T^{z,V}(x,y). \label{CT:Rz-decay_2}
\end{align}

Let us decompose $\Rbb^3$ into six disjoint subsets. We define the top and bottom spherical cap (cone, rather) by, respectively,
\begin{align*}
\Ccal_+ &:= \bigl\{ x\in \Rbb^3 : x = r(\sin \theta \cos \varphi, \sin\theta \sin\varphi , \cos\theta) :  r>0, \theta \in [0 , \nicefrac \pi 4],\varphi\in [0,2\pi) \bigr\},\\
\Ccal_- &:= \bigl\{ x\in \Rbb^3 : x = r(\sin \theta \cos \varphi, \sin\theta \sin\varphi , \cos\theta) : r>0, \theta \in [\nicefrac {3\pi}4 , \pi], \varphi\in [0,2\pi)\bigr\}.
\end{align*}
Furthermore, we define the sector
\begin{align*}
\Scal_1 := \bigl\{ x\in \Rbb^3 : x = r(\sin \theta \cos \varphi, \sin\theta \sin\varphi , \cos\theta) : r>0,\theta \in (\nicefrac \pi 4, \nicefrac {3\pi} 4), \varphi \in [-\nicefrac \pi 4, \nicefrac \pi 4)\bigr\}
\end{align*}
as well as the sectors $\Scal_2$, $\Scal_3$, and $\Scal_4$ by successive counterclockwise rotation of $\Scal_1$ in $\varphi$ by $\nicefrac \pi 2$. Now, let $b_1 := \delta e_1$. If $x-y\in \Scal_1$, then
\begin{align*}
b_1 \cdot (x-y) =  \delta\, |x-y| \, \sin \theta \, \cos \varphi \geq \frac 12\,  \delta \, |x-y|.
\end{align*}
We have similar estimates for $b_2 := \delta e_2$, $b_3 = -\delta e_1$ and $b_4 := -\delta e_2$ and if $x-y\in \Scal_2$ (respectively, $\Scal_3$ and $\Scal_4$) as well as the estimates $b_\pm \cdot (x-y) \geq \frac{1}{\sqrt{2}}\, \delta\,  |x-y|\geq \frac 12 \delta |x-y|$ for $b_\pm = \pm \delta e_3$ if $x-y\in \Ccal_\pm$. 

From \eqref{CT:Rz-decay_2}, we therefore obtain
\begin{align*}
\e^{\frac 12 \delta |x-y|} \, \bigl| \Gcal_T^{z,V}(x,y)\bigr| \leq \bigl| \Gcal_T^{z,V}(b_i;x,y) \bigr|,
\end{align*}
where $i\in \{1,2,3,4,\pm\}$ is chosen so that $x-y\in \Scal_i$ or $\Ccal_\pm$, respectively. We finally conclude
\begin{align*}
\sup_{x\in \Rbb^3} \Bigl(\int_{\Rbb^3} \dd y\; \e^{\delta |x-y|} \, |\Gcal_T^{z,V}(x,y)|^2 \Bigr)^{\nicefrac 12} &\leq \sum_{i\in \{1,\cdots, 4,\pm\}} \Vert \Rcal_T^{z,V}(b_i)\Vert_{2,\infty}.
\end{align*}
The bound on the last line follows from Theorem \ref{KTlambdaV_Analytic_Estimate}.
\end{proof}


\section{Exponential Localization of Non-Embedded Eigenfunctions}

\begin{prop}
\label{CT:Exponential_localization}
Let $\lambda < 2T$ be an eigenvalue of the operator $K_T - V$ and let $\alpha$ be a normalized eigenfunction corresponding to $\lambda$, i.e.,
\begin{align*}
(K_T - V)\alpha = \lambda \, \alpha.
\end{align*}
Then, the following statements are true:
\begin{enumerate}[(a)]
\item There is a $\delta_0(\lambda) >0$ such that for every $0 \leq \delta \leq \delta_0(\lambda)$, we have $\e^{\delta \, |\cdot|} \sqrt{|V|} \, \alpha \in L^2(\Rbb^3)$.

\item For any $\nu\in \Nbb_0^3$, we have
\begin{align*}
\int_{\Rbb^3} \dd x\; \bigl( |x^\nu \, \nabla \alpha|^2 + |x^\nu \, \alpha(x)|^2\bigr)  < \infty.
\end{align*}
\end{enumerate}
\end{prop}

\begin{proof}
The proof is analogous to the proof of Proposition 1 in Appendix A of \cite{Hainzl2012}. Since $V\in L^\infty(\Rbb^3)$, the function $ \phi := \sqrt{|V|} \,  \alpha$ belongs to $L^2(\Rbb^3)$ and satisfies
\begin{align*}
\phi = - \sqrt{|V|} \, \frac{1}{\lambda -K_T} \, \sqrt{V} \, \phi,
\end{align*}
where $\sqrt{V} := \frac{V}{\sqrt{|V|}}$. For fixed $R>0$, we decompose $\phi = \phi_1 + \phi_2$ with $\phi_2 := \chi_{[R,\infty)}(|\cdot |) \phi$. Then, it is clear that $\e^{\delta |\cdot|} \phi_1 \in L^2(\Rbb^3)$ for any $\delta >0$. We also set $U_1 := \chi_{[R, \infty)}(|\cdot|) \sqrt{|V|}$ and $U_2 := \chi_{[R,\infty)}(|\cdot|) \sqrt{V}$. Then, we have
\begin{align*}
\phi_2 &= -U_1 \, (\lambda - K_T)^{-1} \, U_2 \phi_2 + f, & f &:= -U_1 \, (\lambda - K_T)^{-1} \, \sqrt{V}  \, \phi_1.
\end{align*}
Let us first show that $\e^{\delta |\cdot|} f\in L^2(\Rbb^3)$. This amounts to showing that the operator 
\begin{align*}
\Tcal := -\e^{\delta |\cdot|} U_1 \, (\lambda - K_T)^{-1} \, \sqrt{V} \e^{-\delta |\cdot|}
\end{align*}
is bounded. To see this, let $\psi\in L^2(\Rbb^3)$ and estimate
\begin{align*}
\Vert \Tcal \psi\Vert_2^2 &= \int_{\Rbb^3} \dd x \, \Bigl| \int_{\Rbb^3} \dd y \; \chi_{[R, \infty)}(x) \, \sqrt{|V(x)|} \, \sqrt{V(y)} \, \e^{\delta(|x| - |y|)} \Gcal_T^\lambda(x-y) \psi(y)\Bigr|^2 \\
&\leq \Vert V\Vert_\infty \, \Vert \psi\Vert_2^2 \int_{\Rbb^3} \dd x \; \chi_{[R, \infty)}(|x|) \, |V(x)| \; \int_{\Rbb^3} \dd y \; \e^{2\delta (|x| - |y|)} \, |\Gcal_T^\lambda(x-y)|^2.
\end{align*}
By Theorem \ref{CT:Gcal_exponential_decay}, the last factor is bounded by $\Vert \Gcal_T^\lambda\Vert_{2\delta}$ as long as $\delta$ is small enough, while the third factor is bounded by $\Vert (1 + |\cdot|^2)V\Vert_2$ times
\begin{align*}
\Bigl(\int_{\Rbb^3} \dd x \; \Bigl|\frac{1}{1 + |x|^2} \Bigr|^2 \; \chi_{[R, \infty)}(|x|) \Bigr)^{\nicefrac 12},
\end{align*}
which is bounded independently of $R$ (it actually tends to zero as $R\to\infty)$. We conclude that
\begin{align*}
\Vert \Tcal\Vert_\infty \leq C \, \Vert V\Vert_\infty^{\nicefrac 12} \, \Vert (1 + |\cdot|^2)V\Vert_\infty^{\nicefrac 12} \, \Vert \Gcal_T^\lambda\Vert_{2\delta}.
\end{align*}
Since $\e^{\delta |\cdot|} \phi_1\in L^2(\Rbb^3)$, this proves that $\e^{\delta|\cdot|}f\in L^2(\Rbb^3)$. In a similar manner, we see that
\begin{align*}
\Vert \e^{\delta |\cdot|} U_1 \, (K_T -\lambda)^{-1} U_2 \e^{-\delta |\cdot|} \Vert_\infty &\\
& \hspace{-70pt} \leq \Vert V\Vert_\infty^{\nicefrac 12} \, \Vert (1 + |\cdot|^2) V\Vert_\infty^{\nicefrac 12} \, \Vert \Gcal_T^\lambda\Vert_{2\delta} \, \Bigl(\int_{\Rbb^3} \dd x \; \Bigl|\frac{1}{1 + |x|^2} \Bigr|^2 \; \chi_{[R, \infty)}(|x|) \Bigr)^{\nicefrac 14} \leq \frac 12
\end{align*}
for $R$ large enough. This implies
\begin{align*}
\e^{\delta |\cdot|} \phi_2 = \Bigl( 1 + \e^{\delta |\cdot|} U_1 \, (\lambda - K_T)^{-1} U_2 \phi_2\Bigr)^{-1} \e^{\delta |\cdot|} f.
\end{align*}
Part (a) now follows from the Neumann-series. To prove part (b), we note that since $V\in L^\infty(\Rbb^3)$, we have $|\cdot|^\nu \alpha \in L^2(\Rbb^3)$. The claim for the gradient term follows from integration by parts and the fact that $\alpha\in H^2(\Rbb^3)$.
\end{proof}


\section{The Phase Approximation Method for \texorpdfstring{$K_{T, \Abold} - V$}{KTA-V}}

The exponential decay estimate of Theorem \ref{CT:Gcal_exponential_decay} enables us to set up a phase approximation for the resolvent of $K_{T, \Abold} - V$. This, in turn, helps us to prove the asymptotic expansions for eigenvalues and spectral projetions.

\subsection{Preliminary estimates}

As a preparation, we need to recall following result, which is proven in Lemma \ref{g_decay} (or \ref{DHS1:g0_decay}) for the free resolvent kernel
\begin{align}
g^z = \frac{1}{z - (-\i \nabla)^2 + \mu}, \label{CT:g_definition}
\end{align}
see \eqref{g_definition}, and we restate it here for the readers convenience. 

\begin{lem}
\label{CT:g_decay}
Let $a > -2$. There is a constant $C_a >0$ such that for $t,\omega\in \Rbb$, we have
\begin{align}
\left \Vert \, |\cdot|^a g^{\i \omega + t}\right\Vert_1 &\leq C_a \; f(t, \omega)^{1+ \frac a2}, 
\label{CT:g0_decay_1}
\end{align}
where
\begin{align}
f(t, \omega) := \frac{|\omega| + |t + \mu|}{(|\omega| + (t + \mu)_-)^2} \label{CT:g0_decay_f}
\end{align}
and $x_- := -\min\{x,0\}$. Furthermore, for any $a > -1$, there is a constant $C_a >0$ with
\begin{align}
\left \Vert \, |\cdot|^a \nabla g^{\i\omega + t} \right\Vert_1 \leq C_a \; f(t, \omega)^{\frac 12 + \frac a2} \; \Bigl[ 1 + \frac{|\omega| + |t+ \mu|}{|\omega| + (t + \mu)_-}\Bigr]. \label{CT:g0_decay_2}
\end{align}
\end{lem}

In particular, Lemma~\ref{CT:g_decay} implies
\begin{align}
\sup_{w\in \speaker_{\mu_+}} \bigl[ \left\Vert \, |\cdot |^a g^w\right\Vert_1 + \left \Vert \, |\cdot|^a \nabla g^w\right\Vert_1 \bigr] < \infty. \label{CT:g_decay_along_speaker}
\end{align}

Furthermore, we need to provide an estimate on the $L^1$-norm of $\Gcal_T^z$ and $\nabla \Gcal_T^z$, which we do now.

\begin{lem}
\label{CT:GcalTz_estimates}
Let $z\in \rho(K_T)$. For any $k\in \Nbb_0$, we have
\begin{align}
\Vert \, |\cdot|^k \Gcal_T^z\Vert_{L^1(\Rbb^3)} &\leq C_k \, \Vert \Gcal_T^z\Vert_2. \label{CT:GcalTz_estimates_eq1}
\end{align}
Furthermore,
\begin{align}
\Vert\, |\cdot|^k \nabla \Gcal_{T}^z\Vert_{L^1(\Rbb^3)} &\leq C_{k, \delta} \, \bigl( 1 + \Vert \Gcal_T^z\Vert_2 \bigr). \label{CT:GcalTz_estimates_eq2}
\end{align}
\end{lem}

\begin{proof}
As long as $0 \leq \delta < \frac 12 \delta' < \frac 12 \delta_0(z)$ (with $\delta_0(z)$ from Theorem \ref{CT:Gcal_exponential_decay}), we have
\begin{align*}
\int_{\Rbb^3} \dx \; \e^{\delta |x|} |\Gcal_T^z(x)| &= \int_{\Rbb^3} \dx \; \e^{-(\frac 12\delta' - \delta)|x|} \cdot \e^{\frac 12 \delta' |x|} |\Gcal_T^z(x)| \\
&\leq \Bigl( \int_{\Rbb^3} \dx \; \e^{-(\delta' - 2\delta)|x|} \Bigr)^{\nicefrac 12} \Bigl( \int_{\Rbb^3} \dx \; \e^{\delta' |x|} |\Gcal_T^z(x)|^2 \Bigr)^{\nicefrac 12}
\end{align*}
Since $\Vert \Gcal_T^z\Vert_\delta \leq C \Vert \Gcal_T^z\Vert_2$ by Theorem \ref{CT:Gcal_exponential_decay}, this proves \eqref{CT:GcalTz_estimates_eq1}.

Furthermore, Lemma \ref{CT:KT_integral_rep} shows that
\begin{align*}
K_T - p^2 + \mu = \int_{\speaker_{\mu_+}} \frac{\dd w}{2\pi\i}\; f_T(w) \frac{1}{w + \mu - p^2},
\end{align*}
whence $K_T - p^2 +\mu$ has an exponentially decaying integral kernel $\Kcal$ and, by \eqref{CT:g_decay_along_speaker}, we have $\Vert |\cdot|^k \Kcal\Vert_1 <\infty$ for all $k\in \Nbb_0$. For suitably chosen $\nu\in \rho(p^2-\mu)$ (for example $\nu = -\mu - 1$), the resolvent equations therefore imply
\begin{align*}
\Gcal_T^z &= \Gcal_T^\nu + (z - \nu) \, \Gcal_T^\nu * \Gcal_T^z\\
&= g^\nu + g^\nu * \Kcal * \Gcal_T^\nu + (z-\nu) \, g^\nu * \Gcal_T^z + (z-\nu) \, g^\nu * \Kcal * \Gcal_T^\nu * \Gcal_T^z,
\end{align*}
where $g^\nu$ is the resolvent kernel of the Laplacian in \eqref{CT:g_definition}.
When we differentiate, the derivative falls on $g^\nu$. Furthermore, the factor $|\cdot|^k$ can be distributed via the inequality \eqref{DHS2:convexity} among the terms in the convolution. An application of Lemma \ref{CT:g_decay} and \eqref{CT:GcalTz_estimates_eq1} shows \eqref{CT:GcalTz_estimates_eq2} and completes the proof.
\end{proof}

Finally, we incorporate $V$ into the estimates.

\begin{thm}
\label{CT:GcalTzV_estimates}
Let $V\in L^2(\Rbb^3) \cap L^\infty(\Rbb^3)$ and let $z\in\rho(K_T - V)$. Then, for every $k\in \Nbb_0$, the operators $\Zcal_T^{z, V}(k)$ and $\Zcal_{T, \nabla}^{z, V}(k)$ associated to the kernels
\begin{align}
\Zcal_T^{z, V}(k; x,y) &:= |x-y|^k \, |\Gcal_T^{z, V}(x,y)|, & \Zcal_{T, \nabla}^{z, V}(k; x,y) &:= |x-y|^k \, |\nabla_x \Gcal_T^{z, V}(x,y)|,
\end{align}
respectively, are bounded operators which satisfy the estimate
\begin{align}
\Vert \Zcal_T^{z, V}(k)\Vert_\infty &\leq C_k \, \Vert \Gcal_T^z\Vert_2 \, \bigl(1 + \Vert \Gcal_T^{z, V}\Vert_{2, \infty}\bigr), \label{CT:GcalTzV_estimates_eq1}\\
\Vert \Zcal_{T, \nabla}^{z, V}(k)\Vert_\infty &\leq C_k \, \bigl( 1 + \Vert \Gcal_T^z\Vert_2 \bigr)  \, \bigl(1 + \Vert \Gcal_T^{z, V}\Vert_{2, \infty}\bigr). \label{CT:GcalTzV_estimates_eq2} 
\end{align}
\end{thm}

\begin{proof}
By the resolvent equation
\begin{align*}
\Rcal_T^{z, V} = \Rcal_T^z  + \Rcal_T^z \, V\, \Rcal_T^{z, V},
\end{align*}
we have $\Zcal_T^{z, V}(k) = \Zcal_{T, 1}^z(k) + \Zcal_{T, 2}^{z, V}(k)$ with
\begin{align*}
\Zcal_{T, 1}^z(k; x-y) &:= |x-y|^k \, \Gcal_T^z(x-y) \\
\Zcal_{T, 2}^{z, V}(k; x,y) &:= |x-y|^k \, \int_{\Rbb^3} \dd u \; \Gcal_T^z(x-u) \, V(u) \, \Gcal_T^{z, V}(x,y).
\end{align*}
By Lemma \ref{CT:GcalTz_estimates}, $\Zcal_{T, 1}^z(k, \cdot)$ is an $L^1$-function whence, by Young's inequality, we conclude $\Vert \Zcal_{T, 1}^z(k)\Vert_\infty \leq C_k \Vert \Gcal_T^z\Vert_2$.

We claim that $\Zcal_{T, 2}^{z, V}(k)$ is a Hilbert-Schmidt operator with a suitable norm bound. To see this, we estimate
\begin{align}
\Zcal_{T, 2}^{z, V}(k; x,y) &\leq 2^{(k-1)_+}\int_{\Rbb^3} \dd u \; |x-u|^k \, \Gcal_T^z(x-u) \, V(u) \, \Gcal_T^{z, V}(u,y) \notag \\
&\hspace{50pt} + 2^{(k-1)_+} \int_{\Rbb^3} \dd u \; \Gcal_T^z(x-u) \, V(u) \, |u-y|^k \, \Gcal_T^{z, V}(u,y). \label{CT:GcalTzV_estimates_1}
\end{align}
The Hilbert-Schmidt norm squared of the first term is bounded as
\begin{align*}
\iint_{\Rbb^3\times \Rbb^3} \dd x\dd y \, \Bigl| \int_{\Rbb^3} \dd u \; |x-u|^k \Gcal_T^z(x-u) \, V(u) \, \Gcal_T^{z, V}(u,y) \Bigr|^2 \leq \Vert \, |\cdot|^k \Gcal_T^z\Vert_1^2 \, \Vert V\Vert_2^2 \, \Vert \Gcal_T^{z, V}\Vert_{2, \infty}^2.
\end{align*}
Since $\Vert \Gcal_T^z\Vert_1 \leq C_k \Vert \Gcal_T^z\Vert_2$ by Lemma \ref{CT:GcalTz_estimates}, we conclude the claimed estimate for this term. Similarly, the Hilbert-Schmidt norm squared of the second term in \eqref{CT:GcalTzV_estimates_1} is bounded as
\begin{align*}
\iint_{\Rbb^3\times \Rbb^3} \dd x\dd y \, \Bigl| \int_{\Rbb^3} \dd u \; \Gcal_T^z(x-u) \, V(u) \, |u-y|^k\Gcal_T^{z, V}(u,y) \Bigr|^2 \leq \Vert \Gcal_T^z\Vert_1^2 \, \Vert V\Vert_2^2 \, \Vert \, |\cdot|^k \Gcal_T^{z, V}\Vert_{2,\infty}^2.
\end{align*}
Since $\Vert \, |\cdot|^k \Gcal_T^{z, V}\Vert_{2, \infty} \leq C_K \Vert \Gcal_T^{z, V} \Vert_\delta \leq C_k \Vert \Gcal_T^{z, V} \Vert_{2, \infty}$, this term satisfies the same estimate.

The estimate for $\Zcal_{T, \nabla}^{z, V}(k)$ goes along the same lines, except that we have to replace $\Vert \, |\cdot|^k \Gcal_T^z\Vert_1$ by $\Vert \, |\cdot|^k \nabla \Gcal_T^z\Vert_1 \leq C_k (1 + \Vert \Gcal_T^z\Vert_2)$, see Lemma \ref{CT:GcalTz_estimates}.
\end{proof}

\subsection{The comprehensive phase approximation method for \texorpdfstring{$K_{T, \Abold}-V$}{KTA-V}}

We are now in position to set up a phase approximation for the operator $K_{T, \Abold} - V$. Recall that $\Gcal_{T, \Abold}^{z,V}(x,y)$ in \eqref{CT:GcalTAz_definition} denotes the kernel of the resolvent $\Rcal_{T, \Abold}^{z,V} := (z - (K_{T, \Abold}-V))^{-1}$. 

The core of the phase approximation method due to \cite[pp. 1290]{Nenciu2002} is the nonintegrable phase factor, sometimes also called the Wilson line, defined by
\begin{align}
\Phi_\Abold(x,y) := -\int_y^x \Abold(u) \cdot \dd u := -\int_0^1 \dd t\; \Abold(y + t(x-y))\cdot (x-y). \label{CT:PhiA}
\end{align}

\begin{lem}
\label{CT:PhiA_derivative}
Let $\Abold$ be a vector-valued, and weakly differentiable function such that $D\Abold\in W^{1,\infty}(\Rbb^3; \Rbb^3)$. Then, we have
\begin{align}
\nabla_x \Phi_\Abold(x,y) = -\Abold(x) + \tilde \Abold(x,y), \label{CT:PhiA_derivative_eq1}
\end{align}
where
\begin{align}
\Abold_y (x) := \int_0^1 \dt \; t \curl \Abold(y + t(x - y)) \wedge (x-y) \label{CT:Atilde_definition}
\end{align}
is the transversal Poincar\'e gauge relative to $y\in \Rbb^3$.
\end{lem}

\begin{proof}
It is easy to see that the proof of Lemma \ref{PhiA_derivative} is valid here.
\end{proof}

We define the gauge-invariant version of the free resolvent kernel $\Gcal_T^{z, V}$ in \eqref{CT:GcalTAz_definition} by
\begin{align}
\Scal_{T,\Abold}^{z,V}(x,y) := \e^{\i \Phi_\Abold(x,y)} \, \Gcal_T^{z,V}(x,y). \label{CT:ScalTAz_definition}
\end{align}
In the following lines, we investigate the intertwining relation between $K_{T, \Abold}$ and the operator $\Scal_{T, \Abold}^{z, V}$ associated to the kernel \eqref{CT:ScalTAz_definition}. First of all, we recall \eqref{PhiA_Magnetic_Momentum_Action}, which reads
\begin{align}
(-\i \nabla_x + \Abold(x)) \; \e^{\i \Phi_\Abold(x,y)} = \e^{\i \Phi_\Abold(x,y)}\; (-\i\nabla_x + \Abold_y(x)) \label{CT:PhiA_Magnetic_Momentum_Action}
\end{align}
and which follows from \eqref{CT:PhiA_derivative_eq1}, where $\Abold_y(x)$ is the Poincaré gauge in \eqref{CT:Atilde_definition}. We use the notation
\begin{align}
\pi_\Abold := -\i \nabla + \Abold. \label{CT:piA_definition}
\end{align}
Since $\pi_\Abold^2 \geq 0$ (diamagnetic inequality \cite[Theorem 7.21]{LiebLoss} or \cite[Eq. (4.4.3)]{LiebSeiringer}), we find that the speaker path $\speaker_{\mu_+}$ lies in the resolvent set of $\pi_{\Abold_y}^2 -\mu$ as well as $\pi_\Abold^2-\mu$, see Definition \ref{CT:speaker_path}. Hence, for each $w\in\speaker_{\mu_+}$, we infer from \eqref{CT:PhiA_Magnetic_Momentum_Action} (by multiplying with the respective resolvents from the left and the right) that
\begin{align*}
(w - \pi_\Abold^2-\mu)^{-1} \, \e^{\i \Phi_\Abold(x,y)} =\e^{\i \Phi_\Abold(x,y)} \, (w - \pi_{\Abold_y}^2 - \mu)^{-1}.
\end{align*}
Hence, by Lemma \ref{CT:KT_integral_rep}, we conclude that
\begin{align*}
K_{T,\Abold} \, \e^{\i \Phi_\Abold(x,y)} = \e^{\i \Phi_\Abold(x,y)} \, K_{T,\Abold_y}.
\end{align*}
Thus, a straightforward computation shows that
\begin{align}
(z - (K_{T,\Abold} - V)) \, \Scal_{T,\Abold}^{z,V} = \Idbb - \Tcal_{T,\Abold}^{z,V}, \label{CT:Interplay_SAzTAz}
\end{align}
where $\Tcal_{T, \Abold}^{z, V}$ is the operator associated to the kernel
\begin{align}
\Tcal_{T, \Abold}^{z,V} (x,y) := \e^{\i \Phi_\Abold(x,y)} \bigl[ (K_{T,\Abold_y} - K_T)\Rcal_T^{z,V}\bigr] (x,y). \label{TcalAboldzV}
\end{align}
The next result shows that this operator is bounded with a suitable norm bound.

\begin{lem}
\label{CT:TAVz_decay}
Let $V\in L^2(\Rbb^3) \cap L^\infty(\Rbb^3)$ and let $\Abold$ satisfy Assumption \ref{CT:Assumption_A}. Define
\begin{align}
\Mcal(\Abold) := \max \Bigl\{ \Vert \curl \Abold\Vert_{L^\infty(\Rbb^3)} \; , \; \Vert \curl(\curl \Abold)\Vert_{L^\infty(\Rbb^3)} \; , \; \Vert \curl \Abold\Vert_{L^\infty(\Rbb^3)}^2 \Bigr\}. \label{CT:McalA_definition}
\end{align}
Then, there is a continuous function
\begin{align}
\Dcal \colon \rho(K_T - V) \ra \Rbb_+ \label{CT:TAVz_decay_functionDcal}
\end{align}
such that the operator $\Tcal_{T,\Abold}^{z,V}$ corresponding to the kernel in \eqref{TcalAboldzV} is bounded by
\begin{align}
\bigl\Vert \Tcal_{T,\Abold}^{z,V} \bigr\Vert_\infty \leq C\; \Dcal(z) \; \Mcal(\Abold). \label{CT:TAVz_decay_eq1}
\end{align}
\end{lem}

\begin{proof}
By Lemma \ref{CT:KT_integral_rep}, we have
\begin{align*}
[K_{T, \Abold_y} - K_T] \, \Rcal_T^{z,V} = [\pi_{\Abold_y}^2 - p^2] \, \Rcal_T^{z,V} + \int_{\speaker_{\mu_+}} \frac{\dd w}{2\pi\i} \, f_T(w) \, \Bigl[ \frac 1{w - \pi_{\Abold_y}^2} - \frac 1 {w - p^2}\Bigr] \, \Rcal_T^{z,V}.
\end{align*}
Here, $\speaker_{\mu_+}$ is the speaker path in Definition \ref{DHS1+:speaker_path}. We call $\Tcal_1$ the first term and $\Tcal_2$ the second. Let us start by estimating the term $\Tcal_1$. Writing it out, we obtain
\begin{align}
\Tcal_1(x,y) = \Bigl[ -\i \divv \Abold_y(x) - 2\i \, \Abold_y(x) \cdot \nabla + |\Abold_y(x)|^2 \Bigr] \, \Gcal_T^{z,V}(x,y). \label{TAVz decay_1}
\end{align}
We use \eqref{TAz_boundedness_1} and \eqref{TAz_boundedness_2} so see that
\begin{align*}
|\Tcal_1(x,y)|\leq \Mcal(\Abold) \, \Bigl[ \bigl( |x-y| + |x-y|^2\bigr) |\Gcal_T^{z, V}(x,y)| + |x-y| \, |\nabla_x\Gcal_T^{z, V}(x,y)| \Bigr].
\end{align*}
Estimates for these terms have been provided in Theorem \ref{CT:GcalTzV_estimates}.

Let us move on to the term $\Tcal_2$. It reads
\begin{align*}
\Tcal_2(x,y) = \int_{\speaker_{\mu_+}} \frac{\dd w}{2\pi\i}\; f_T(w) \,  \Bigl[ \bigl( (w-k_{\Abold_y})^{-1} - (w - k_0)^{-1} \bigr) \, \Rcal_T^{z,V} \Bigr] \, (x,y).
\end{align*}
Using the resolvent equation, we write this out as
\begin{align*}
(w - k_{\Abold_y})^{-1} - (w- k_0)^{-1} & \\
&\hspace{-70pt} = (w - k_{\Abold_y})^{-1} \bigl[ -\i \divv \Abold_y - 2\i \, \Abold_y \cdot \nabla + |\Abold_y|^2 \bigr] (w - k_0)^{-1}.
\end{align*}
Hence, we obtain
\begin{align*}
|\Tcal_2(x,y)| \leq C \;  \Mcal(\Abold)  \int_{\speaker_{\mu_+}}\frac{\dd |w|}{2\pi} \; |f_T(w)| \cdot \bigl[ \Tcal_2^1(x,y;w) + \Tcal_2^2(x,y;w)\bigr].
\end{align*}
Here, $\dd |w| = \dd t \cdot |w'(t)|$ and
\begin{align*}
\Tcal_2^1(x,y;w) &:= \int_{\Rbb^3} \dd u \int_{\Rbb^3} \dd v \; |G_{\Abold_y}^w(x,u)| \, |u-y| \,  |\nabla g^w(u - v)| \, |\Gcal_T^{z,V}(v,y)| \\
\Tcal_2^2(x,y;w) &:= \int_{\Rbb^3} \dd u \int_{\Rbb^3} \dd v \; |G_{\Abold_y}^w(x,u)| \,  \bigl( |u - y| + |u-y|^2\bigr) \, |g^w(u - v)| \, |\Gcal_T^{z,V}(v,y)|.
\end{align*}
Estimates for these functions are provided in Lemmas\footnote{We concretely use the first estimates of \eqref{GAz-GtildeAz_decay_1} and \eqref{GAz-GtildeAz_decay_2}, which do hold for $\Gbold_{\Bbold, \Abold}^z$ replaced by $G_\Abold^z$ and $\Abold_y$ being the Poincaré gauge of $\Abold$ satisfying Assumption \ref{CT:Assumption_A}. The reader may consult the proof of Lemma \ref{GAz-GtildeAz_decay} to verify this claim.} \ref{GAz-GtildeAz_decay}, \ref{CT:g_decay}, and Theorem \ref{CT:GcalTzV_estimates}. This finishes the proof.
\end{proof}

\begin{kor}
\label{CT:Magnetic_Resolvent_expansion}
Let $V\in L^2(\Rbb^3)\cap L^\infty(\Rbb^3)$, assume that $\Abold$ satisfies Assumption \ref{CT:Assumption_A}. Then, there is a continuous function $\Mcal_0 \colon \rho(K_T - V) \ra \Rbb_+$ such that the following holds. If $z\in \rho(K_T - V)$ and $0 \leq \Mcal(\Abold) \leq \Mcal_0(z)$, then $z\in \rho(K_{T, \Abold} - V)$ and there is a bounded linear operator $\tilde\Tcal_{T,\Abold}^{z,V}$ such that
\begin{align*}
\Rcal_{T,\Abold}^{z,V} = \Scal_{T,\Abold}^{z,V} + \tilde \Tcal_{T,\Abold}^{z,V}.
\end{align*}
Furthermore, there is a continuous function $\Dcal \colon \rho(K_T - V) \ra \Rbb_+$ such that $\tilde\Tcal_{T,\Abold}^{z,V}$ satisfies the estimate
\begin{align*}
\bigl\Vert \tilde \Tcal_{T,\Abold}^{z,V}\bigr\Vert_\infty &\leq C \; \Dcal(z) \; \Mcal(\Abold).
\end{align*}
\end{kor}

\begin{proof}
Define the function $\Mcal_0$ by
\begin{align}
C \, \Dcal(z) \; \Mcal_0(z) = \frac 12, \label{CT:Magnetic_Resolvent_expansion_Assumption}
\end{align}
where $C\, \Dcal(z)$ is from \eqref{CT:TAVz_decay_eq1}. Then $\Mcal_0$ is continuous. Furthermore, by hypothesis and Lemma \ref{CT:TAVz_decay}, the operator $\Idbb + \Tcal_{T,\Abold}^{z,V}$ is invertible and we may solve \eqref{CT:Interplay_SAzTAz} for $\Rcal_{T,\Abold}^{z,V}$ to get
\begin{align*}
\Rcal_{T,\Abold}^{z,V} = \Scal_{T,\Abold}^{z,V} + \sum_{n=1}^\infty \Scal_{T,\Abold}^{z,V}  \bigl( \Tcal_{T,\Abold}^{z,V} \bigr)^n.
\end{align*}
The bound on the operator norm of the second term is given by the Neumann series and reads
\begin{align*}
\Bigl\Vert \sum_{n=1}^\infty \Scal_{T,\Abold}^{z,V} \bigl( \Tcal_{T,\Abold}^{z,V} \bigr)^n \Bigr\Vert_\infty &\leq C \; \Dcal(z) \; \Mcal(\Abold) \; \Vert \Rcal_T^{z, V}\Vert_\infty \sum_{n=1}^\infty \frac 1{2^n}.
\end{align*}
Here, we used, by a pointwise estimate on the kernel, that $\Scal_{T,\Abold}^{z,V}$ is a bounded operator with norm bounded by $\Vert \Rcal_T^{z,V}\Vert_\infty$. We also used Lemma \ref{CT:TAVz_decay} and \eqref{CT:Magnetic_Resolvent_expansion_Assumption}. This completes the proof.
\end{proof}


\section{Asymptotics of Spectral Projections and Eigenvalues of \texorpdfstring{$K_{T,\Abold}-V$}{KTA-V}}

The goal of this section is to prove stability of the spectral projections corresponding to eigenvalues of finite multiplicity of $K_T -V$ under the perturbation $\Abold$, as long as
\begin{align}
\ov \Mcal(\Abold) := \max \bigl\{ \Mcal(\Abold) \, , \, \Vert D\Abold\Vert_{L^\infty(\Rbb^3)}\bigr\} \label{CT:McalAbar_definition}
\end{align}
is small, where $\Mcal(\Abold)$ is from \eqref{CT:McalA_definition}.

\begin{thm}
\label{CT:KTV_Perturbation_Theorem}
Let $V\in L^2(\Rbb^3)$ such that $|\cdot|^k V\in L^\infty(\Rbb^3)$ for $k\in \{0,1,2\}$. Let $\lambda$ be an isolated eigenvalue of finite multiplicity $m\in \Nbb$ of $K_T-V$ and let $\Pcal_T^V$ be the corresponding spectral projection. Then, the following statements are true:
\begin{enumerate}[(a)]
\item $\lambda$ is stable in the sense of Kato \cite[Section VIII.1.4]{Kato}, i.e.,
\begin{enumerate}[(i)]
\item There is $\varepsilon>0$ such that for every $z\in B_{2\varepsilon}(\lambda)\setminus \{\lambda\}$ the following holds. There is $\Mcal_0(z) >0$ such that whenever $0\leq \Mcal(\Abold)\leq \Mcal_0(z)$, we have $z\in \rho(K_{T,\Abold}-V)$.

\item $\Rcal_{T,\Abold}^{z,V} \to \Rcal_T^{z,V}$ strongly as $\ov \Mcal(\Abold) \to 0$ and as $\Phi_\Abold(x,y) \to 0$ pointwise in $x,y\in \Rbb^3$ for all $z\in B_{2\varepsilon}(\lambda) \setminus \{\lambda\}$.

\item There is $\Mcal_0> 0$ such that if $0\leq \ov \Mcal(\Abold) \leq \Mcal_0$, we have
\begin{align*}
\rank \Pcal_{T,\Abold}^V = m,
\end{align*}
where
\begin{align*}
\Pcal_{T, \Abold}^V := \int_{\partial B_\varepsilon(\lambda)} \frac{\dd z}{2\pi \i}\; \Rcal_{T,\Abold}^{z,V}.
\end{align*}
The integral has to be understood as a complex contour integral with a positively oriented contour along $\partial B_\varepsilon(\lambda)$.

\end{enumerate}

\item There is $\Mcal_0>0$ such that if $0\leq \ov \Mcal(\Abold)\leq \Mcal_0$, we have
\begin{align*}
\Vert \Pcal_{T,\Abold}^V - \Pcal_T^V \Vert _\infty + \Vert \pi_\Abold^2 (\Pcal_{T, \Abold}^V - \Pcal_T^V)\Vert_\infty &\leq C\, \ov \Mcal(\Abold).
\end{align*}

\item Let $\lambda_1(\Abold), \ldots, \lambda_m(\Abold)$ be the (not necessarily distinct) eigenvalues of $(K_{T,\Abold} - V)\Pcal_{T, \Abold}^V$. Then, there is $\Mcal_0>0$ such that for all $0\leq \ov \Mcal(\Abold)\leq \Mcal_0$, we have
\begin{align*}
|\lambda_i(\Abold) - \lambda| &\leq C\, \ov \Mcal(\Abold), & i &=1, \ldots, m.
\end{align*}

\item Assume that the lowest eigenvalue $\eta_T$ of $K_T - V$ is simple and denote the spectral gap above $\eta_T$ by $\kappa_T >0$. Then, there is $\Mcal_0>0$ such that whenever $0 \leq \ov \Mcal(\Abold) \leq \Mcal_0$ the lowest eigenvalue of $K_{T,\Abold}-V$ is simple and there is a uniform spectral gap above it. In particular, if $\underline \Pcal_T^V$ denotes the ground state projection corresponding to $\eta_T$, then
\begin{align}
K_{T,\Abold} - V \geq \eta_T \, \underline \Pcal_T^V + \frac 12\, \kappa_T \, \bigl( 1 - \underline \Pcal_T^V \bigr) - C \; \ov \Mcal(\Abold). \label{CT:KTV_Perturbation_Theorem_eq1}
\end{align}
\end{enumerate}
\end{thm}

For the proof, we need the following auxiliary statement.

\begin{lem}
\label{Projections have same rank}
Let $(P_n)_{n\in \Nbb}$ be a sequence of projections in a separable Hilbert space $\Hcal$ and let $P$ be a projection in $\Hcal$ with finite rank $m\in \Nbb$. If $P_n \to P$ in operator norm, then $\rank P_n = m$ for all sufficiently large $n$.
\end{lem}

\begin{proof}
Assume for contradiction that, for each $k\in \Nbb$, there is $n_k\geq k$ such that
\begin{enumerate}[(a)]
\item $\rank P_{n_k} \geq m+1$ or
\item $\rank P_{n_k} \leq m-1$.
\end{enumerate}
Since one of the cases (a) or (b) is admitted infinitely often, we may rule out both of them separately. Let $\psi_i\in \ran P$, $i=1, \ldots, m$ form an orthonormal basis for $\ran P$. In case (a), let $\varphi_{n_k} \in \ran P_{n_k}$ with $\Vert \varphi_{n_k}\Vert = 1$ and $\langle \varphi_{n_k} ,\psi_i\rangle =0$ for all $i=1, \ldots, m$ and all $k\in \Nbb$. It follows that $P\psi_{n_k} =0$ and thus
\begin{align*}
1 = \Vert \psi_{n_k}\Vert = \Vert P_{n_k}\psi_{n_k}\Vert = \Vert (P_{n_k} - P)\psi_{n_k}\Vert \leq \Vert P_{n_k} - P\Vert \xra{n\to\infty} 0,
\end{align*}
a contradiction. In case (b), for each $k\in \Nbb$, there is an $i\in \{1, \ldots, m\}$ such that $P_{n_k} \psi_i =0$. Hence, one of $i\in \{1,\ldots, m\}$ is hit infinitely often. Without loss, assume that $i=1$. This means that $P_{n_k} \psi_1 =0$ for all $k\in \Nbb$. As above, we get
\begin{align*}
1 = \Vert \psi_1\Vert = \Vert P\psi_1\Vert = \Vert (P - P_{n_k})\psi_1\Vert \leq \Vert P - P_{n_k}\Vert \xra{n\to\infty} 0,
\end{align*}
again a contradiction. It follows that there is $N\in \Nbb$ such that for all $n\geq N$, we have $\rank P_n = m$.
\end{proof}

\begin{proof}[Proof of Theorem \ref{CT:KTV_Perturbation_Theorem}]
Let us start by proving the three statements in (a). First of all, since $\lambda$ is an isolated eigenvalue of $K_T -V$, we may pick $\varepsilon>0$ such that $z\in \rho(K_T-V)$ for all $z\in B_{2\varepsilon}(\lambda) \setminus \{\lambda\}$. Corollary \ref{CT:Magnetic_Resolvent_expansion} shows the existence if $\Mcal_0(z) >0$ such that such $z$ also belong to $\rho(K_{T,\Abold} -V)$ provided $\Abold$ obeys $0 \leq \Mcal(\Abold) \leq \Mcal_0(z)$, whence (i) is proved. Let $\psi\in L^2(\Rbb^3)$ be arbitrary. Then
\begin{align*}
\bigl\Vert \Rcal_{T,\Abold}^{z,V}\psi - \Rcal_T^{z,V}\psi\bigr\Vert_2 &\leq \bigl\Vert \Rcal_{T,\Abold}^{z,V}\psi - \Scal_{T,\Abold}^{z,V}\psi\bigr\Vert_2 + \bigl\Vert \Scal_{T,\Abold}^{z,V} \psi - \Rcal_T^{z,V}\psi\bigr\Vert_2.
\end{align*}
The first term can be bounded by $C\, \Mcal(\Abold)\, \Vert \psi\Vert$ using Corollary \ref{CT:Magnetic_Resolvent_expansion}. For the second term, we have
\begin{align*}
\bigl\Vert \Scal_{T,\Abold}^{z,V} \psi - \Rcal_T^{z,V}\psi\bigr\Vert_2^2 = \int_{\Rbb^3} \dx \; \Bigl|\int_{\Rbb^3} \dy\; \bigl( \e^{\i \Phi_\Abold(x,y)} - 1\bigr) \Gcal_T^{z,V}(x,y)\psi(y)\Bigr|^2.
\end{align*}
Hence, we may bound the factor with the exponential by 2 to obtain an integrable dominant. Pointwise convergence of the integrand follows from the hypothesis. The dominated convergence theorem then yields (ii). 

To show part (iii), we first prove that $V\Rcal_{T,\Abold}^{z,V} \to V\Rcal_T^{z,V}$ and $\Rcal_{T,\Abold}^{z,V} V \to \Rcal_T^{z,V} V$ in norm as $ \Mcal(\Abold) \to 0$ uniformly for $z\in \partial B_\varepsilon(\lambda)$. We write the proof only for the second convergence, the first is analogous. To start out with, let $\Mcal_0^1$ be defined by
\begin{align}
\sup_{z\in \partial B_\varepsilon(\lambda)} C \, \Dcal(z) \; \Mcal_0^1 = \frac 12, \label{Projection-sup-constraint}
\end{align}
where $C\, \Dcal(z)$ is from \eqref{CT:Magnetic_Resolvent_expansion_Assumption}. Then, $\Mcal_0^1 >0$ because $\partial B_\varepsilon(\lambda)$ is compact and $\Dcal(z)$ is continuous. 

We write in a similar fashion as above
\begin{align*}
\bigl\Vert \Rcal_{T,\Abold}^{z,V}V - \Rcal_T^{z,V}V\bigr \Vert_\infty \leq\bigl \Vert \Rcal_{T,\Abold}^{z,V}V - \Scal_{T,\Abold}^{z,V}V\bigr\Vert_\infty + \bigl\Vert \Scal_{T,\Abold}^{z,V}V - \Rcal_T^{z,V}V\bigr\Vert_\infty.
\end{align*}
The first term can be bounded by $C\Vert V\Vert_\infty \Mcal(\Abold)$, using Corollary \ref{CT:Magnetic_Resolvent_expansion}. For the second term, we investigate the kernel to get
\begin{align*}
\bigl| V(y) \Scal_{T,\Abold}^{z,V}(x,y) - V(y) \Gcal_T^{z,V}(x,y)\bigr| &\leq |V(y)| \, |\e^{\i \Phi_\Abold(x,y)}-1| \, |\Gcal_T^{z,V}(x,y)|.
\end{align*}
Now, we use the bound
\begin{align}
|\Phi_\Abold(x,y)|\leq C \, \Vert D\Abold\Vert_\infty \, \bigl( \min\{|x| \, , \, |x|\} \, |x-y| + |x-y|^2\bigr),\label{CT:PhiA_estimate}
\end{align}
which follows from the assumption $\Abold(0) =0$, and obtain
\begin{align*}
\bigl| V(y) \Scal_{T,\Abold}^{z,V}(x,y) - V(y) \Gcal_T^{z,V}\bigr| & \\
&\hspace{-120pt} \leq C \, \Vert D\Abold\Vert_\infty \bigl( \Vert \,|\cdot|V\Vert_\infty \, |x - y| \, |\Gcal_T^{z,V}(x,y)| + \Vert V\Vert_\infty \, |x-y|^2 \, |\Gcal_T^{z, V}(x,y)|\bigr). 
\end{align*}
The estimates for these functions provided by Lemma \ref{CT:GcalTzV_estimates} are continuous in $z$. Therefore, there is $\Mcal_0^2 >0$ such that if $0 \leq \Vert D\Abold\Vert_\infty \leq \Mcal_0^2$, we have
\begin{align*}
\sup_{z\in \partial B_\varepsilon(\lambda)} \bigr\Vert \Rcal_{T,\Abold}^{z,V}V - \Scal_{T,\Abold}^{z,V}V\bigr\Vert_\infty \leq C \, \Vert D\Abold\Vert_\infty.
\end{align*}
We conclude that
\begin{align}
\sup_{z \in \partial B_\varepsilon(\lambda)} \bigl\Vert \Rcal_{T,\Abold}^{z,V}V - \Rcal_T^{z,V}V\bigr \Vert_\infty + \sup_{z \in \partial B_\varepsilon(\lambda)} \bigl\Vert V\Rcal_{T,\Abold}^{z,V} - V\Rcal_T^{z,V}\bigr \Vert_\infty \leq C \; \ov \Mcal(\Abold) \label{CT:KTV_Perturbation_Theorem_2}
\end{align}
provided $0 \leq \ov \Mcal(\Abold) \leq \Mcal_0 := \min \{\Mcal_0^1\, ,\,  \Mcal_0^2\}$. In particular, uniform operator norm convergence as $\ov \Mcal(\Abold)\to 0$ follows on $\partial B_\varepsilon(\lambda)$.

We use \eqref{CT:KTV_Perturbation_Theorem_2} to prove estimates on the spectral projections. Since $K_{T,\Abold}$ has no eigenvalues, we may write
\begin{align*}
\Pcal_{T,\Abold}^V &= \int_{\partial B_\varepsilon(\lambda)} \frac{\dd z}{2\pi\i}\; \Rcal_{T,\Abold}^{z,V} = \int_{\partial B_\varepsilon(\lambda)} \frac{\dz}{2\pi\i} \; \Rcal_{T,\Abold}^{z,V} - \Rcal_{T,\Abold}^z = \int_{\partial B_\varepsilon(\lambda)} \frac{\dz}{2\pi\i}\; \Rcal_{T,\Abold}^{z,V} \,  V \,  \Rcal_{T,\Abold}^z.
\end{align*}
We claim that the operator in the integrand converges to $\Rcal_T^{z,V} V \Rcal_T^z$ as $\ov \Mcal(\Abold)\to 0$. For,
\begin{align*}
\Rcal_{T,\Abold}^{z,V} \, V \, \Rcal_{T,\Abold}^z - \Rcal_T^{z,V} \, V \, \Rcal_T^z &= \bigl[ \Rcal_{T,\Abold}^{z,V} \, V - \Rcal_T^{z,V} \, V \bigr] \Rcal_{T,\Abold}^z + \Rcal_T^{z,V} \bigl[ V \, \Rcal_{T,\Abold}^z - V \, \Rcal_T^z\bigr].
\end{align*}
Since $\Rcal_{T,\Abold}^z$ is bounded uniformly in $\Abold$, see Corollary \ref{CT:Magnetic_Resolvent_expansion}, we conclude convergence of the integrand in operator norm by \eqref{CT:KTV_Perturbation_Theorem_2}. This proves the first part of (b) and (iii) of (a) by Lemma \ref{Projections have same rank}.

The next part of the proof is devoted to the second part of (b). Similarly to the above, our starting point is
\begin{align}
\pi_\Abold^2 (\Pcal_{T,\Abold}^V - \Pcal_T^V) &= \int_{\partial B_\varepsilon(\lambda)} \frac{\dd z}{2\pi\i} \;\bigl(\pi_\Abold^2 \, \Rcal_{T,\Abold}^{z,V} \, V \, \Rcal_{T,\Abold}^z - \pi_\Abold^2 \, \Rcal_T^{z,V} \, V \, \Rcal_T^z\bigr) \notag \\
&= \int_{\partial B_\varepsilon(\lambda)} \frac{\dd z}{2\pi\i} \; \pi_\Abold^2 \,  \Rcal_{T,\Abold}^{z,V} \, \bigl[ V \, \Rcal_{T,\Abold}^z - V\, \Rcal_T^z\bigr] \notag \\
&\hspace{80pt} + \int_{\partial B_\varepsilon(\lambda)} \frac{\dd z}{2\pi\i} \; \bigl[ \pi_\Abold^2 \, \Rcal_{T,\Abold}^{z,V} \, V - \pi_\Abold^2 \,  \Rcal_T^{z,V} \, V \bigr] \Rcal_T^z. \label{CT:KTV_Perturbation_Theorem_1}
\end{align}
By expanding with the resolvent equation, we have
\begin{align*}
\pi_\Abold^2\Rcal_{T,\Abold}^{z,V} = \pi_\Abold^2\,  \Rcal_{T, \Abold}^z\,  \bigl[ 1 + V\Rcal_{T,\Abold}^{z,V}\bigr].
\end{align*}
The first operator is bounded by $1$ since it is a function of $\pi_\Abold^2$. Hence, the first term of \eqref{CT:KTV_Perturbation_Theorem_1} is bounded of order $\ov \Mcal(\Abold)$ according to \eqref{CT:KTV_Perturbation_Theorem_2}. The delicate term is the second. Here, we apply the resolvent equation to get
\begin{align*}
\pi_\Abold^2 \, (\Rcal_{T,\Abold}^{z,V} - \Rcal_T^{z,V}) \, V = \pi_\Abold^2 \Rcal_{T,\Abold}^{z,V} \, (K_{T,\Abold} - K_T) \, \Rcal_T^{z,V} \, V.
\end{align*}
We are going to show that
\begin{align}
\sup_{z \in \partial B_\varepsilon(\lambda)} \Vert(K_{T, \Abold} - K_T) \Rcal_T^{z,V} \,  V \Vert_\infty \leq C \; \ov \Mcal(\Abold). \label{CT:KTV_Perturbation_Theorem_3}
\end{align}
To see this, apply the integral representation Lemma \ref{CT:KT_integral_rep} to $K_{T, \Abold} - K_T$ once more. This gives two terms, one is 
\begin{align*}
\pi_\Abold^2-p^2 = -\i \divv \Abold + \Abold \cdot p + |\Abold|^2.
\end{align*}
Since $\Vert \divv \Abold\Vert_\infty \leq\Mcal(\Abold)$, there is nothing left to prove for this term. For the other terms, we use $\Abold(x) \leq \Vert D\Abold\Vert_\infty |x|$ as well as the triangle inequality $|x| \leq |x-y| + |y|$. This implies that we need to provide a bound on
\begin{align*}
&(|y| + |x - y|) \, |\nabla \Gcal_T^{z,V}(x,y)| \, |V(y)| + (|y|^2 + |x - y|^2) \, |\Gcal_T^{z,V}(x,y)| \, |V(y)| \\
&\hspace{50pt} \leq \bigl[\Vert \,|\cdot | V\Vert_\infty + \Vert \, |\cdot|^2V\Vert_\infty\bigr] \, \bigl(|\nabla \Gcal_T^{z,V}(x,y)| + |\Gcal_T^{z,V}(x,y)|\bigr) \\
&\hspace{100pt} + \Vert V\Vert_\infty \bigl(|x -y|\,  |\nabla \Gcal_T^{z,V}(x,y)| + |x - y|^2 \, |\Gcal_T^{z,V}(x,y)|\bigr),
\end{align*}
which gives rise to a bounded operator, whose bound is uniform on $\partial B_\varepsilon(\lambda)$, as can be deduced from Theorem \ref{CT:GcalTzV_estimates}. As for the second term in the integral representation. The resolvents in the integrand of the contour integral read
\begin{align*}
\frac{1}{w +\mu - \pi_\Abold^2} - \frac{1}{w + \mu - p^2} = \frac{1}{w +\mu - \pi_\Abold^2} [\pi_\Abold^2 - p^2] \frac{1}{w + \mu - p^2}.
\end{align*}
Here, we proceed similarly. However, we need to use the triangle inequality once more to reach $V$. Once we have estimated
\begin{align*}
|x| &\leq |x - u| + |u - y| + |y| & |x|^2 \leq 2|x - u|^2 + 2|u - y|^2 + 2|y|^2,
\end{align*}
we are subject to showing that
\begin{align*}
&\Bigl| \frac{1}{w + \mu - \pi_\Abold^2} [\pi_\Abold^2-p^2] \frac{1}{w + \mu - p^2} \; \Rcal_T^{z,V}\; V\; (x,y)\bigr| \\
&\hspace{30pt}\leq \ov \Mcal(\Abold) \, \Bigl[ \Vert V\Vert_\infty \iint_{\Rbb^3\times \Rbb^3} \dd u_1\dd u_2 \; |G_\Abold^w(x,u_1)| \, |g^w(u_1 - u_2)| \, |\Gcal_T^{z, V}(u_2, y)| \,  \\
&\hspace{50pt}+ \big[ \Vert \, |\cdot|V\Vert_\infty + \Vert \, |\cdot|^2V\Vert_\infty\bigr] \iint_{\Rbb^3\times \Rbb^3} \dd u_1\dd u_2\; |G_\Abold^w(x,u_1)| \\
&\hspace{150pt} \times\bigl[ |\nabla G^w(u_1 - u_2)| + |G^w(u_1 - u_2)|\bigr] |\Gcal_T^{z,V}(u_2,y)| \\
&\hspace{50pt}+ \Vert V\Vert_\infty \iint_{\Rbb^3\times \Rbb^3} \dd u_1 \dd u_2 \; |G_\Abold^w(x,u_1)| \bigl[ (|u_1 - u_2| + |u_2 - y|)|\nabla G^w(u_1 - u_2)| \\
&\hspace{150pt} + (|u_1 - u_2|^2 + |u_2 - y|^2) |G^w(u_1 - u_2)|\bigr] |\Gcal_T^{z,V}(u_2,y)|\Bigr] 
\end{align*}
gives rise to a bounded operator with uniform bound on $\partial B_\varepsilon(\lambda)$. The estimates for this to be proven are provided by Lemmas \ref{GAz-GtildeAz_decay} and Theorem \ref{CT:GcalTzV_estimates}. This proves (b).

Now, we prove part (c). By the residue theorem and Corollary \ref{CT:Magnetic_Resolvent_expansion},
\begin{align*}
(K_{T, \Abold} - V)\Pcal_{T,\Abold}^V &= \int_{\partial B_\varepsilon(\lambda)} \frac{\dd z}{2\pi\i }\; z \, \Rcal_{T,\Abold}^{z,V} = \int_{\partial B_\varepsilon(\lambda)} \frac{\dd z}{2\pi\i }\; z \,  \Scal_{T,\Abold}^{z,V}  + \int_{\partial B_\varepsilon(\lambda)} \frac{\dd z}{2\pi\i }\; z \, \tilde \Tcal_{T, \Abold}^{z,V}.
\end{align*}
Since the last operator is bounded of order $\Mcal(\Abold)$, we are left with investigating the first term. If $\alpha_i$, $i=1, \ldots, m$ are the orthonormal eigenfunctions corresponding to $\lambda$, then it has a kernel given by
\begin{align*}
\int_{\partial B_\varepsilon(\lambda)} \frac{\dd z}{2\pi\i }\; z \, \Scal_{T,\Abold}^{z,V}(x,y) &= \lambda \,  \e^{\i \Phi_{\Abold}(x,y)} \sum_{i=1}^m \alpha_i(x) \ov{\alpha_i(y)}\\
&= \lambda \, \Pcal_T^V(x,y) + \lambda  \, \bigl( \e^{\i \Phi_\Abold(x,y)} - 1\bigr) \sum_{i=1}^m \alpha_i(x)\ov{\alpha_i(y)}.
\end{align*}
It remains to use the estimate \eqref{CT:PhiA_estimate}, so that $|\Phi_\Abold(x,y)| \leq C \Vert D\Abold\Vert_\infty (|x|^2 + |y|^2)$ and
\begin{align*}
|\e^{\i \Phi_\Abold (x,y)} - 1| \sum_{i=1}^m |\alpha_i(x)| \, |\alpha_i(y)| \leq C\, \Vert D\Abold\Vert_\infty \sum_{i=1}^m\bigl[ |x|^2 |\alpha_i(x)| \, |\alpha_i(y)| +  |\alpha_i(x)| \, |y|^2|\alpha_i(y)|\bigr].
\end{align*}
We infer that the Hilbert-Schmidt norm of this kernel is bounded by
\begin{align*}
\iint_{\Rbb^3\times \Rbb^3} \dd x \dd y\; \Bigl| \sum_{i=1}^m |x|^2\, |\alpha_i(x)|\,  |\alpha_i(y)|\Bigr|^2 & \\
&\hspace{-170pt} = \sum_{i,j=1}^m \iint_{\Rbb^3\times \Rbb^3} \dd x\dd y\; |x|^2\, |\alpha_i(x)|\, |x|^2\, |\alpha_j(x)| \, |\alpha_i(y)| \, |\alpha_j(y)| \leq \Bigl( \sum_{i=1}^m \Vert \, |\cdot|^2 \alpha_i\Vert_2 \, \Vert \alpha_i\Vert_2\Bigr)^2.
\end{align*}
These norms are finite by Proposition \ref{CT:Exponential_localization}. Hence, we have shown the asymptotics
\begin{align}
\bigl\Vert (K_{T,\Abold}-V)\Pcal_{T, \Abold}^V - \lambda \, \Pcal_T^V\bigr\Vert_\infty \leq C \; \ov \Mcal(\Abold).
\end{align}
By the min-max principle \cite[Theorem 4.12]{MBQM}, if $\lambda_1(\Abold) \leq \cdots \leq \lambda_m(\Abold)$ are ordered increasingly, we have the characterization
\begin{align*}
\lambda_i(\Abold) = \inf \bigl\{ \max_{\varphi\in M, \Vert \varphi\Vert = 1} \langle \varphi, (K_{T,\Abold} - V)\Pcal_{T,\Abold}^V \varphi\rangle : M\subseteq L^2(\Rbb^3), \dim M = i\bigr\}.
\end{align*}
From this and \eqref{CT:KTV_Perturbation_Theorem_3}, we immediately deduce part (c).

To obtain part (d), let $\kappa >0$ denote the spectral gap of $K_T  -V$ above its lowest eigenvalue $\eta_T$, i.e., $\kappa = \eta_T^1 - \eta_T$, where $\eta_T^1$ is the next-to-lowest eigenvalue. Also let $e_0^\Abold$ and $e_1^\Abold$ denote the lowest and next-to-lowest eigenvalue of $K_{T, \Abold} - V$. Then, part (c) shows that, for $\ov \Mcal(\Abold)>0$ small enough,
\begin{align*}
e_1^\Abold - e_0^\Abold &= \kappa + (e_1^\Abold - \eta_T^1) - (e_0^\Abold - \eta_T) \\
&\geq \kappa - |e_1^\Abold - \eta_T^1| - |e_0^\Abold - \eta_T| \geq \kappa - C \; \ov \Mcal(\Abold) \geq \frac 12 \kappa.
\end{align*}
If $\underline \Pcal_{T, \Abold}^V$ denotes the ground state projection of $K_{T, \Abold} - V$, we conclude
\begin{align*}
K_{T, \Abold} - V \geq \eta_T \, \underline \Pcal_{T, \Abold}^V + \frac 12\, \kappa \bigl( 1 - \underline \Pcal_{T, \Abold}^V\bigr) + (e_0^\Abold - \eta_T) \, \underline \Pcal_{T, \Abold}^V.
\end{align*}
An application of part (b) now shows \eqref{CT:KTV_Perturbation_Theorem_eq1}.
\end{proof}

\printbibliography[heading=bibliography, title=Bibliography of Chapter \ref{Chapter:Combes-Thomas}]
\end{refsection}


\part{On the Adiabatic Theorem in Extended Quantum Systems}
\label{Part:Adiabatic}

\begin{refsection}

\chapter[Preparation of Exponential Estimates on the Adiabatic Theorem in Extended Quantum Lattice Systems][Exponential Estimates for the Adiabatic Theorem]{Preparation of Exponential Estimates on the Adiabatic Theorem in Extended Quantum Lattice Systems}
\label{Chapter:Adiabatic} \label{CHAPTER:ADIABATIC}


\section{Introduction}
\label{AT:Intro}

\subsection{How it came to this chapter}

In spring 2019, I had the opportunity to travel to the University of British Columbia (UBC) in Vancouver, Canada, for a research visit, which lasted four months. During the stay, my guest advisor Sven Bachmann and I started the project on proving exponential estimates for the adiabatic theorem in \cite{SvenAdiabatic}. Unfortunately, we could not finish the project in due time and planned to complete it in a second visit in 2020, which was then canceled due to the COVID-19 pandemic. Consequently, this project is still lying on my desk waiting to be completed and the progress we made so far forms the content of this chapter. As the time of my PhD studies is coming to an end, I feel that this thesis is a good opportunity to present the state of the project. I am also happy to hereby comply with a corresponding request by Stefan Teufel in Tübingen.

Since discussions on a regular basis are hard to organize between Sven and me due to the physical distance and the time shift, I am grateful for several fruitful discussions with the postdoc researcher Amanda Young, who is currently on the leave from Technische Universität München.

It is needless to say that I plan to undertake another attempt to bring the project to a successful conclusion in the remaining time after the submission of this thesis but I have to see if time permits enough commitment.

\subsection{Exponential estimates via optimal truncation}

Mathematically, the intention has been to prove exponential error estimates for the adiabatic theorem that has been proved in the work \cite{SvenAdiabatic}. We have in mind using a similar strategy to the one presented in the paper \cite{HagedornJoye}, namely to recursively provide an explicit estimate for the adiabatic error in the expansion after the $n$\tho\ step and then perform an optimal truncation argument. We will be somewhat more detailed below on what we mean by an optimal truncation argument. This goal has not been reached yet but we believe that the content of the chapter is helpful to understand the problem and needed to be applied in one way or the other by anybody in the future, who wants to complete the work.

Let us explain a bit further why we did not yet manage to prove the aimed result. When we began the project at the UBC, it soon became clear that the desired estimate would not be possible to prove within the by now standard locality setup that has already been used in \cite{AutomorphicEquivalence} and has worked its way through countless publications in the business of quantum lattice systems in the meantime. We only mention the recent publications \cite{BdRFL2021, BrunoQuasilocal, NWY2020} and point out to the reader the references therein as examples for communities using this framework. The reason why this setup fails in the context of our problem are the badly behaved ``continuity'' estimates, as we shall call them. Namely, when an operation like the commutator is applied to a pair of \emph{local Hamiltonians} (sometimes also called \emph{quasilocal operators}), the question is: What are the locality properties of the commutator provided the corresponding properties of the initial Hamiltonians are known and can we relate them quantitatively? In other words, if $G_1, G_2$ are the initial Hamiltonians, such a quantitative relation would be given by an estimate of the type
\begin{align}
\Vert \, [G_1, G_2]\, \Vert_{\mathrm{fin}} \leq C_{\mathrm{in}, \mathrm {fin}}([\cdot, \cdot]) \; \Vert G_1\Vert_{\mathrm {in}} \; \Vert G_2\Vert_{\mathrm {in}}. \label{AT:Continuity_estimate}
\end{align}
Here, $\Vert \cdot\Vert_{\mathrm{fin}}$ and $\Vert \cdot\Vert_{\mathrm{in}}$ are two possibly different initial and final norms, which should best be related in a convenient way and $C_{\mathrm{in}, \mathrm{fin}}([\cdot, \cdot]) >0$ is a constant independent of $G_1$ and $G_2$. We shall call \eqref{AT:Continuity_estimate} a ``continuity estimate'' for the operation $[\cdot, \cdot]$ for obvious reasons. The problem is often not to find such a constant but rather to gain precise control of it in that we have a formula at hand which is easy to work with. Additionally, the constant needs to behave well enough for the operation to be applied over and over again in an iterative procedure like the adiabatic expansion.

We remark that one is tempted to choose the operator norm $\Vert \cdot\Vert$ for $\Vert \cdot\Vert_{\mathrm{in}}$ and $\Vert \cdot\Vert_{\mathrm{fin}}$ but the reader familiar to extended quantum systems will immediately realize the problem of the volume dependence of this choice and the necessity of a norm that measures the ``locality'' of the local Hamiltonian. We will comment on this problem further below.

It turns out that the standard machinery mentioned above behaves too badly under repeated application of such operations and the constants that are stacking on the way grow much too fast. It is difficult to explain this phenomenon more precisely at this point without any notation at hand but we will provide some details below in Subsection \ref{AT:Norms_Section}.

When we sat in front of the problem in Vancouver, we realized that before we could even think about actually performing the adiabatic expansion in the manner of \cite[Lemma 4.3]{SvenAdiabatic}, we were forced to go back to the very beginning of the theory and tidy up the setup of locality within our quantum lattice system. After we defined the new norm, the first item on the list was then to provide a Lieb--Robinson bound for Hamiltonians that have a finite local norm. We realized that, within this new setup, it is actually possible to gain a pretty simple proof of a Lieb--Robinson bound even for tree graphs, which we shall present in Section \ref{AT:LR_bounds_Section}.

In Section \ref{AT:I_Section} we will introduce the map $\Ical$, which is the ``generator of the spectral flow'', ``local inverse of the Liouvillian'', or ``weighted integral operator'', depending on the convention within the respective community. This operator is a central tool in solving the Schrödinger equation in a locality preserving fashion and it requires the fact that the underlying Hamiltonian has a spectral gap. Under the assumption of a gap, we show that it provides an inverse of the map $G\mapsto [H, G]$, where $H$ is the Hamiltonian. In order to do this, we have to construct a new weight function which depends on the gap and satisfies an estimate in terms of our explicit decay functions. We are aware of and will not participate in the debate whether the restriction of an open gap is too severe for realistic models to be true. Instead we will just stick to this assumption and see how far we get with it. It seems that already the (unsolved, after all) problem of this chapter in the presence of a gap is difficult enough and I encourage the reader to provide improvements in the gapless direction.

The continuity estimates are contained in Section \ref{AT:Estimating_Operations_Section}, starting with the (multi-) commutator. We further provide a well-behaved explicit estimate for the operation $\Ical$, showing the locality preserving property. For us, the main technical announcement is that we are able to provide the first explicit locality estimate for arbitrary high derivatives of $\Ical(G)$, provided the Hamiltonian for the underlying time evolution and $G$ are smooth. To the best of my knowledge, such an estimate is not known in the existing literature.

As an addendum, we also point out in Section \ref{AT:Analytic_Interactions_Section} how an analyticity property of the Hamiltonian $H$ allows to transport estimates on $H$ to higher derivatives of $H$. 

Unfortunately, to the present day, we were not able to perform the adiabatic expansion in its full generality in the sense of providing suitable estimates for the operators that are constructed iteratively. However, we present a sketch of a concept that in our opinion has the potential to provide the desired result when all ingredients are taken into account. We demonstrate this in Section \ref{AT:Concept_Section} with a toy estimate that discards the involvement of derivatives.

The reader should keep in mind that this chapter has the character of a working note, which means that the proofs may look a bit overdetailed sometimes. In particular, it should be noted that, in contrast to the previous chapters on BCS theory, one of the central interests in the endeavor here is the size and dependency of constants. For this reason, statements are always as explicit as possible and there is no ``constant $C$ that is allowed to change from line to line'' in this chapter.

\subsection{Basics of quantum spin systems}

\subsubsection{Lattice systems and their shape}

The prototype of a lattice system is, as the name suggests, the lattice $\Zbb^d$ with its graph norm\footnote{A graph is the tuple $(\Gamma, \Ecal)$ of $\Gamma$ with a set of edges $\Ecal$ where an element $e\in \Ecal$ connects two elements $x,y\in \Gamma$. The graph metric $d(x,y)$ is then defined as the cardinality of the smallest subset of $\Ecal$, which connects $x$ and $y$. If $\Gamma = \Zbb^d$, the standard graph metric is given by the 1-norm $|\cdot|_1$.} $|x|_1 := \sum_{i=1}^d |x_i|$. The reason why I don't like this picture so much is that it suggests the notion of a ``\emph{dimension}'' describing the shape of the lattice, namely $d$. However, it turns out that the dimension is not so relevant in this chapter in the sense that we never choose a basis for any space. Furthermore, this notion is not available for tree graphs and these are the reasons why I would rather like to think of the more general notion of ``\emph{growth relations}'' of the cardinality of a set in terms of its diameter. In this spirit, $\Zbb^d$ would be of polynomial growth with the power $d$ and a tree graph would be of exponential growth. We will comment on this a bit further below. 

That is why, in general, a lattice system for us will be a countable metric space $(\Gamma, d)$ with the graph metric. The ball of radius $n\in \Nbb_0$ about the point $x\in \Gamma$ is denoted by
\begin{align}
B_n(x) &:= \bigl\{ z\in \Gamma : d(x,z) \leq n\bigr\} \label{AT:Ball_definition}
\end{align}
and a set $Z\subset \Gamma$ has a (possibly infinite) \emph{diameter}, which we denote by
\begin{align}
\Dcal(Z) &:= \diam(Z) := \sup \bigl\{ d(x,y) : x,y\in Z\bigr\}. \label{AT:Diameter}
\end{align}
The set of finite and nonempty subsets
\begin{align}
\Fcal(\Gamma) &:= \bigl\{ \Lambda \subset \Gamma : 0 < |\Lambda| < \infty\bigr\} \label{AT:Finite_subsets_definition}
\end{align}
plays an important role throughout this chapter, where $|\Lambda|$ is the cardinality of the set $\Lambda$. Typically, to fix the shape of the lattice to be polynomial, we make assumptions of the following kind: There are constants $\kappa >0$ and $d \in \Nbb_0$ such that for every $Z\in \Fcal(\Gamma)$, we have
\begin{align}
|Z| \leq \kappa \, (1 + \Dcal(Z))^d. \label{AT:Growth_condition}
\end{align}
This means that $\Gamma$ has a polynomial growth of degree $d$, or is $d$-dimensional. Obviously, a tree graph violates \eqref{AT:Growth_condition}.

\subsubsection{The quantum setup}

We raise a quantum system upon $\Gamma$ by associating to each \emph{vertex} or \emph{lattice site} $x\in \Gamma$ a so-called \emph{on-site} Hilbert space $\Hcal_x$. Furthermore, to any subset $\Lambda \subset \Gamma$ (finite or infinite), the Hilbert space $\Hcal_\Lambda := \bigotimes_{x\in \Lambda} \Hcal_x$ is associated. Then, we also define the algebra of observables $\Acal_\Lambda := \Bcal(\Hcal_\Lambda)$, which consists of the bounded operators $\Hcal_\Lambda \ra \Hcal_\Lambda$, eqipped with the usual operator norm $\Vert \cdot \Vert$. There is a canonical embedding $\Acal_\Lambda \hookrightarrow \Acal_{\Lambda'}$ if $\Lambda \subset \Lambda'$, which is defined by $A \mapsto A \otimes \bigotimes_{x\in \Lambda'\setminus \Lambda} \Idbb \in \Acal_{\Lambda'}$ and which we use without comment. We also define the set of local observables by $\Acal_{\mathrm{loc}} := \bigcup_{\Lambda \in \Fcal(\Gamma)} \Acal_{\Lambda}$. For a local observable $A\in \Acal_{\mathrm{loc}}$, we may define its \emph{support} by the smallest set $X \in \Fcal(\Gamma)$ such that $A\in \Acal_X$, i.e.,
\begin{align}
\supp(A) &:= \bigcap \bigl\{ X\in \Fcal(\Gamma) : A\in \Acal_X \bigr\}.
\end{align}

Interesting quantum effects start becoming visible in the infinite volume limit, that is, when operators are considered that have an infinite support. This fact can be dealt with in two different ways:
\begin{enumerate}[(1)]
\item We can try to define all objects in infinite volume as limiting objects of their finite volume correspondents.
\item We try to restrict to a finite volume and execute the theory in such a way that the quantities of interest survive the infinite volume limit.
\end{enumerate}
It has turned out that option (1) suffers from the fact that several central objects such as the Hamiltonian do not possess a nice representation in the infinite volume and can only (if at all) be abstractly extended. However, we should not overlook that several other quantities can indeed be extended. Nevertheless, option (2) is the method we will pursue. Therefore, all objects are introduced in finite volume and the theory is never allowed to depend on this finite volume.

To get started, an interaction is a map $\Phi \colon \Fcal(\Gamma) \ra \Acal_{\mathrm{loc}}$, which assigns $Z\mapsto \Phi(Z)$ with $\Phi(Z) \in \Acal_Z$. It is easy to see that the space of interactions inherits a complex vector space structure from the spaces of bounded operators via the canonical embedding. A sum of local terms
\begin{align}
\sum_{Z \in \Fcal(\Gamma)} \Phi(Z) \label{AT:Sum_of_local_terms}
\end{align}
is then an operator which has infinite operator norm unless $\Phi$ is \emph{compactly supported}, meaning that there is an $R>0$ such that $\Phi(Z) =0$ if $|Z| \geq R$ (example: nearest neighbor interactions). This raises the need for a locality setup, since the operator norm of a term $\Phi(Z)$ in \eqref{AT:Sum_of_local_terms} typically grows linearly in $|Z|$. Therefore, we need a different measure for the ``size'' of such an object, which encodes the fact that realistic interactions in quantum lattice systems merely feature nearest neighbor interactions in the most probable case and interactions of multituples of sites are suppressed with growing number of involved sites.

\subsubsection{Norms}
\label{AT:Norms_Section}

Mathematically speaking, the ``size'' of an object like \eqref{AT:Sum_of_local_terms} is expressed in terms of a norm. This is the point where is becomes technically delicate because the whole theory is very sensitive to the number of symbols and notions that are used in the definition of the norm. Therefore, there is inevitable need to keep this formula as concise as possible. To express the suppression of terms coming from large sets in the sum \eqref{AT:Sum_of_local_terms}, we need a bounded, non-increasing, positive function $\zeta\colon [0,\infty) \ra (0,\infty)$ that is logarithmically superadditive, i.e., for $x,y\geq 0$, we have
\begin{align}
\zeta(x + y) &\geq \zeta(x) \, \zeta(y).
\end{align}
In practice, $\zeta$ is a rapidly decaying function. The norm that we propose then reads
\begin{align}
\Vert \Phi\Vert_\zeta &:= \sup_{x\in \Gamma} \sum_{\substack{Z\subset \Gamma \\ x\in Z}} \frac{\Vert \Phi(Z)\Vert}{\zeta(\Dcal(Z))}. \label{AT:new_norm}
\end{align}
It should be noted that a norm which needs to be uniform in the lattice $\Gamma$ must have a point $x\in \Gamma$ which it is ``attached to'' and which is then suped over. Otherwise, we would sum over \emph{all} subsets in the lattice and even if we restrict to sets $Z\in \Fcal(\Gamma)$, we immediately realize that the number of sets of a given fixed diameter is heavily dependent on the shape of the lattice then.

Let us compare our definition to the one in \cite{SvenAdiabatic}, which reads
\begin{align}
\Vert \Phi\Vert_{\zeta, N} &:= \sup_{x, y\in \Gamma}  \frac{1}{F(d(x,y)) \, \zeta(d(x,y))} \sum_{\substack{Z \subset \Gamma \\ x,y\in Z}} |Z|^N \; \Vert \Phi(Z)\Vert, & F(r) &:= (1 + r)^{-(d+1)} \label{AT:old_norm}
\end{align}
In our opinion, \eqref{AT:old_norm} suffers from three facts that trigger a lot of problems:
\begin{enumerate}
\item The measure for locality is determined by two different parameters that are not comparable: a natural number $N\in \Nbb_0$, which indexes a volume moment $|Z|^N$ for a finite subset $Z$ of the lattice, as well as a rapidly decaying function $\zeta$, which by the factor $\zeta(d(x,y))^{-1}$ in the norm likewise indicates decay. It turns out that there is need to unify these decay parameters so that the local norm would only be determined by just one parameter. Since $|Z|$ and $d(x,y)$ are incomparable, however, we decided to replace $d(x,y)$ by the diameter $\Dcal(Z)$. In this way, the quantities $|Z|$ and $\Dcal(Z)$ become comparable --- provided we make a suitable assumption on shape of the lattice, like \eqref{AT:Growth_condition} --- and we do not need the volume factor $|Z|^N$ anymore, which had kept arising in each operation in the setup of \cite{SvenAdiabatic}. This turns out to cure the theory to a large extent and further makes it possible to leave away one point of $x,y\in Z$, which even more simplifies the business.

\item The existence of the decay function $F$, whose shape limits the theory to polynomially growing lattices from the very beginning, and whose so-called ``convolution property'' is used to prove a Lieb--Robinson bound, is a big problem in practice. We get rid of this function completely. In the case of interest, the reader may have a look at Section \ref{AT:Old_norm_estimate_Section}, where we present a continuity estimate for $\Ical$ in the framework of the old norm in \cite{SvenAdiabatic}, which we obtained before we realized that iterating $\Ical$ with the help of this estimate is highly problematic on its own. This illustrates very well the problematic behavior of the function $F$, when it comes to precise continuity estimates and the proof shows that it is somewhat clumsy to work with the norm in \eqref{AT:old_norm}.

\item The abstract class of rapidly decaying functions $\zeta$ makes it difficult to prove concise estimates. We choose an ``explicit'' class of rapidly decaying functions $\zeta$, namely the stretched exponential decay functions $\e^{-b x^s}$ for $b>0$ and $0 < s \leq 1$. Since $s$ is basically fixed throughout the theory, the norm effectively becomes determined by a single positive number $b >0$, which in our opinion purges the theory further and enables us to prove short, handy, and well-behaved continuity estimates for the relevant operations.
\end{enumerate}

\subsubsection{Preliminary note on the adiabatic theorem and optimal truncation}

So far, we have discussed locality setups for general quantum spin systems. Before we introduce the setup we will work with in the context of the adiabatic theorem, we take a somewhat closer look at the adiabatic theorem itself. The adiabatic theorem is a statement about the approximation of the time evolution of a state $\rho(u)$ given by a Hamiltonian $H(\varepsilon t)$, where $\varepsilon >0$ is the small \emph{adiabatic parameter} and $\rho(u)$ is initially given by a state $\rho_0$. The adiabatic parameter $\varepsilon$ measures the ``slowness'' of the variation of $H$ in time compared to the hopping it induces on the physical space, i.e., the system has a ``slow'' and a ``fast'' degree of freedom that are separated by the scale $\varepsilon$. The time evolution takes place according to the (adiabatic) Schrödinger equation
\begin{align}
\i \varepsilon \frac{\dd}{\dd u} \rho(u) &= [H(u), \rho(u)], & \rho(0) &= \rho_0. \label{AT:Schroedinger}
\end{align}
Here, we have already performed a rescaling of times $u = \varepsilon t$. The rescaled time is thus $u\in [0,1]$, which corresponds to the physical time $t \in [0 , \varepsilon^{-1}]$. The theorem has a long standing history which we will not repeat here (see \cite{SvenAdiabatic} for details), and has been proven in various different contexts and classes of problems.

Let $\rho_0 = P_0$ be the ground state projection of $H(0)$ and let $P(u)$ be the ground state projection of $H(u)$. A type of question one would then typically like to address is how much $\rho(u)$ differs from $P(u)$. In other words, are the operations ``take the ground state projection'' and ``evolve according to \eqref{AT:Schroedinger}'' commutative? Or, if we start with an instantaneous ground state and evolve it according to \eqref{AT:Schroedinger} for a time $u$, is the resulting state equal to $P(u)$ and, if not, how large is the error? The ``classical'' adiabatic theorem usually should answer this question in the following manner: We have
\begin{align}
\sup_{s\in [0,1]} \Vert \rho(u) - P(u) \Vert \leq C \varepsilon. \label{AT:Adiabatic_Thm_classic}
\end{align}
As pointed out earlier, the problem is that the constant $C$ in \eqref{AT:Adiabatic_Thm_classic} typically grows linearly with the number of lattice sites involved in the support of $H$ so \eqref{AT:Adiabatic_Thm_classic} becomes meaningless in the infinite volume limit, as long as $\varepsilon$ is nonzero, even if we restrict to finite volume initially. Therefore, to be able to prove a reasonable analogon for quantum spin systems, we have to replace the left side of \eqref{AT:Adiabatic_Thm_classic} by a volume independent quantity.
The revolutionary idea that has been introduced successfully in \cite{SvenAdiabatic} is to 
\begin{enumerate}
\item consider a finite volume $\Lambda$ and carry out the theory on the finite volume, proving uniform estimates in $|\Lambda|$.

\item do this by relaxing the topology from the operator norm to expectation values of observables with finite support within $\Lambda$. These do not probe the whole quantum system but only a small region.
\end{enumerate}
The reader is referred to \cite{SvenAdiabatic} for a thorough introduction into the business of the adiabatic theorem in the context of extended quantum systems and to \cite{SvenAdiabatic_Note} for further explanations around the topic.

Let us briefly sketch what we call an optimal truncation argument. Typically, under the assumption of a Hamiltonian that has an analytic extension to a complex strip around the real axis, \eqref{AT:Adiabatic_Thm_classic} can be improved to
\begin{align}
\Vert \rho(1) - P(1) \Vert &\leq C_m \varepsilon^m , & m&\in \Nbb, \label{AT:Intro_Optimal_truncation_1}
\end{align}
in terms of a constant $C_m$ that depends on $m$ as $C_m \lesssim m^m$. The function $f_\varepsilon(x) := x^x \varepsilon^x$, which can be written as $f_\varepsilon(x) = \exp( x \log(x\varepsilon))$, then has a unique global minimum at $x(\varepsilon) := (\e\varepsilon)^{-1}$ with value
\begin{align}
f_\varepsilon( x(\varepsilon)) = \e^{- \frac{\e^{-1}}{\varepsilon}},
\end{align}
which is exponentially decaying as $\varepsilon \to 0$. When we successfully deal with the subtlety that $x(\varepsilon)$ is not an integer in general, we can therefore choose $m\in \Nbb$ in \eqref{AT:Intro_Optimal_truncation_1} optimally in terms of $\varepsilon$ to improve the error to exponential decay in $\varepsilon^{-1}$.

\subsection{The setup of locality for this chapter}

We assume that the dimension of the on-site Hilbert spaces $\Hcal_x$ is uniformly bounded, that is,
\begin{align*}
\sup_{x\in \Gamma} \dim \Hcal_x < \infty.
\end{align*}

\subsubsection{Interactions and the new norm}

We are in position to introduce the locality setup suitable for the adiabatic theorem.

\paragraph{Time-independent setting.} An \emph{interaction} on $\Gamma$ is a family $\Phi = \{\Phi^\Lambda\}_{\Lambda \in \Fcal(\Gamma)}$ of maps $\Phi^\Lambda \colon \Fcal(\Lambda) \ra \Acal_\Lambda$, which assign $Z \mapsto \Phi^\Lambda(Z)$, where $\Phi^\Lambda(Z) \in \Acal_Z$. We denote the set of interactions by $\Bcal$. An interaction $\Phi\in \Bcal$ is \emph{self-adjoint} if $\Phi^\Lambda(Z)$ is self-adjoint in $\Acal_Z$ for all $\Lambda\in \Fcal(\Gamma)$ and $Z\subset \Lambda$.

A family of operators $G = \{G^\Lambda\}_{\Lambda\in \Fcal(\Gamma)}$ is called a \emph{local Hamiltonian} if there is an interaction $\Phi_G \in \Bcal$ such that for all $\Lambda\in \Fcal(\Gamma)$, we have
\begin{align}
G^\Lambda = \sum_{Z\subset \Lambda} \Phi_G^\Lambda (Z). \label{AT:Local_Hamiltonian_decomposition}
\end{align}
The set of local Hamiltonians is denoted by $\Lcal$. Note that $\Phi_G\in \Bcal$ such that \eqref{AT:Local_Hamiltonian_decomposition} holds is not unique. A local Hamiltonian $H\in \Lcal$ is \emph{self-adjoint} if every interaction $\Phi_H \in \Bcal$ such that \eqref{AT:Local_Hamiltonian_decomposition} holds is self-adjoint. For the sake of readability, we will mostly omit the dependence on $\Lambda$ in the rest of this chapter, since this does not cause any confusion. 

It is easy to verify that $\Bcal$ and $\Lcal$ are complex vector spaces with the usual operations that are inherited from the space of bounded operators.

With the help of the stretched exponential decay functions
\begin{align}
\chi_{s, b}(x) &:= \e^{-b \, x^s} , & 0 < s \leq 1, \quad  b\geq 0, \quad x \geq 0, \label{AT:chi_definition}
\end{align}
we define the \emph{local norm} of an interaction $\Phi\in \Bcal$ as
\begin{align}
\Vert \Phi\Vert_{s, b} &:= \sup_{\Lambda \in \Fcal(\Gamma)} \sup_{x\in \Lambda} \sum_{\substack{Z\subset \Lambda \\ Z \ni x}} \frac{\Vert \Phi^\Lambda(Z)\Vert}{\chi_{s, b}(\Dcal(Z))}.
\end{align}
Then, let us define the set of \emph{$(s, b)$-localized interactions} by
\begin{align*}
\Bcal_{s,b} := \bigl\{ \Phi \in \Bcal, \; \Vert \Phi \Vert_{s,b} < \infty\bigr\}.
\end{align*}
If $G\in \Lcal$ such that \eqref{AT:Definition_gapped_Hamiltonian} holds for some $\Phi\in \Bcal_{s, b}$, then $G$ is called \emph{$(s, b)$-localized} and the set of $(s, b)$-localized Hamiltonians is denoted by $\Lcal_{s, b}$. Since $\Phi_G$ is not unique, the term $\Vert G\Vert_{s, b}$ remains undefined but equals, per convention, $\Vert \Phi_G\Vert_{s, b}$ provided an interaction has been specified previously or is clear from the context.

\begin{lem}
For each $0 < s \leq 1$ and $b\geq 0$, the functional $\Vert \cdot \Vert_{s, b}$ is a norm on $\Bcal_{s, b}$, and $(\Bcal_{s, b}, \Vert \cdot\Vert_{s, b})$ is a Banach space. 
\end{lem}

\begin{proof}
Homogeneity and triangle inequality are inherited from the operator norm. Assume that $\Vert \Phi\Vert_{s, b} =0$ for some interaction $\Phi$. Then, for all $x\in \Lambda$, we have $\Vert \Phi^\Lambda(Z)\Vert =0$ for every set $Z\subset \Lambda$ that contains $x$. If a nonempty $Z'\subset \Lambda$ is given, choose an arbitrary point $x'\in Z'$ and obtain $\Vert \Phi^\Lambda(Z') \Vert =0$. This proves that $\Phi^\Lambda(Z') =0$ for every $Z'\subset \Lambda$, whence $\Phi =0$. The completeness property of $\Bcal_{s, b}$ is inherited from the completeness of $\Bcal(\Hcal_\Lambda)$.
\end{proof}

\begin{bem}
Since $\chi_{s, b_1} > \chi_{s, b_2}$ whenever $b_1 < b_2$, we have that $b \mapsto \Vert \Phi\Vert_{s, b}$ is a strictly increasing function on $[0, \infty)$ with values in $\Rbb \cup \{\infty\}$. Therefore, if there is an $0 < s \leq 1$ and a $b>0$ such that $\Vert \Phi\Vert_{s, b} < \infty$ then, either there is $b_0>0$ such that $\lim_{b \nearrow b_0}\Vert \Phi\Vert_{s, b} = \infty$ or we have $\lim_{b\nearrow \infty} \Vert \Phi\Vert_{s, b} = \infty$. In any case, we may assume without loss of generality that
\begin{align}
\Vert \Phi\Vert_{s, b} \geq 1. \label{AT:Hamiltonian_norm_greater_than_1}
\end{align}
\end{bem}

Let us denote the class of $s$-localized interactions by
\begin{align}
\Bcal_s &:= \bigcup_{b >0 } \Bcal_{s, b} , & 0 &< s \leq 1, \label{AT:Interactions_s-localized_definition}
\end{align}
and denote the corresponding classes of local Hamiltonians by $\Lcal_s$. Furthermore, define the classes of (stretched) exponentially localized interactions by
\begin{align}
\Ecal &:= \Bcal_1, & \Scal &:= \bigcup_{0 < s < 1} \Bcal_s, \label{AT:Interactions_stretched_exponentially_localized_definition}
\end{align}
respectively, and write $\Lcal_\Ecal$ and $\Lcal_\Scal$ for the corresponding sets of local Hamiltonians. 

\paragraph{Time-dependent setting.} Since we will be concerned with analytic interactions of time, we consider an open subset $I \subset \Cbb$. A \emph{time-dependent interaction} is a map $\Phi \colon I \ra \Bcal$ and we denote the set of time-dependent interactions by $\Bcal(I)$. A time-dependent local Hamiltonian is a map $G\colon I \ra \Lcal$ such that \eqref{AT:Local_Hamiltonian_decomposition} holds for $G(u)$ for every $u\in I$. The set of time-dependent local Hamiltonians is denoted by $\Lcal(I)$.

For $0 < s \leq 1$ and $b \geq 0$, we define the \emph{local norm} of an interaction $\Phi\in \Bcal(I)$ by
\begin{align}
\vvvert \Phi\vvvert_{I, s, b} := \sup_{u\in I} \Vert \Phi(u)\Vert_{s, b}
\end{align}
and the set of $(s, b)$-localized time-dependent interactions is defined by
\begin{align*}
\Bcal_{s, b}(I) := \bigl\{ \Phi\in \Bcal(I), \; \vvvert \Phi\vvvert_{I, s, b} < \infty\bigr\}.
\end{align*}
Likewise, we conclude that $\vvvert\cdot\vvvert_{I, s, b}$ is a norm on $\Bcal_{s, b}(I)$ and $(\Bcal_{s, b}(I), \vvvert \cdot \vvvert_{I, s, b})$ is a Banach space. We also define the symbols $\Bcal_s(I)$, $\Lcal_s(I)$, $\Ecal(I)$, $\Scal(I)$, $\Lcal_\Ecal(I)$, and $\Lcal_\Scal(I)$ as the obvious analoga to the symbols defined in \eqref{AT:Interactions_s-localized_definition} and \eqref{AT:Interactions_stretched_exponentially_localized_definition}.

An interaction $\Phi \in \Bcal(I)$ is called \emph{holomorphic} if for every $\Lambda \in \Fcal(\Gamma)$ and $Z\subset \Lambda$ the map
\begin{align*}
u \longmapsto \Phi(u)^\Lambda(Z),
\end{align*}
is holomorphic on $I$. For any set $\Xcal(I)$ of time-dependent interactions (or local Hamiltonians), we write $\Xcal^\hol(I)$ for the corresponding set of holomorphic interactions (or local Hamiltonians).

\section{Technical Preparations --- growth conditions}
\label{AT:Technical_Preparations_Section}

In order to prove the adiabatic theorem, we will need a lattice system with polynomial growth. However, there are many results which hold in higher generality. Therefore, we present here various growth conditions on the lattice $\Gamma$ that will play an important role throughout the chapter.

\subsection{Abstract growth conditions}

\begin{asmp}[Abstract volume growth]
\label{AT:Assumption_abstract_V}
For $0 < s \leq 1$ and $b>0$, we assume that the constant
\begin{align}
V_s(b) := \sup_{Z\in \Fcal(\Gamma)} |Z| \; \chi_{s,b}(\Dcal(Z)) \label{AT:Vb_definition}
\end{align}
is finite. Here, $|Z|$ is the cardinality of $Z\in \Fcal(\Gamma)$ and $\chi_{s, b}$ is from \eqref{AT:chi_definition}.
\end{asmp}

Let us remark several facts about the constants in \eqref{AT:Vb_definition}. First of all, we have
\begin{align}
V_s(b) \geq 1, \label{AT:V_bigger_than_1}
\end{align}
since $V_s(b) \geq |\{x\}| \, \chi_{s, b}(\Dcal(\{x\})) = 1$ for every $x\in \Gamma$. We also have $V_{s}(b) \to \infty$ as $b\to 0$.

For $k >0$, we put
\begin{align}
V_{s, k}(b) &:= V_s\bigl( \frac bk\bigr) = \sup_{Z\in \Fcal(\Gamma)} |Z| \; \chi_{s, \frac bk}(\Dcal(Z)). \label{AT:Vkb_definition}
\end{align}
For $b>0$ and $k\leq k'$, these constants satisfy $V_{s,k}(b) \leq V_{s,k'}(b)$ and, by \eqref{AT:V_bigger_than_1},
\begin{align}
V_k(b)^k \leq V_{k'}(b)^{k'}. \label{AT:Vkb-kleqkprime_estimate}
\end{align}

The next growth condition is somewhat similar to the summability of the function $F$ in \cite{SvenAdiabatic}.

\begin{asmp}[Abstract $F$-norm growth]
\label{AT:Assumption_abstract_F}
For $0 < s \leq 1$ and $b>0$, we assume that the constant
\begin{align}
F_s(b) &:= \sup_{z\in \Gamma} \sum_{n=0}^\infty |B_n(z)| \; \chi_{s,b}(2n) \label{AT:Fb_definition}
\end{align}
is finite. Here, $B_n(z)$ is the ball of radius $n$ about $z$ defined in \eqref{AT:Ball_definition} and $\chi_{s, b}$ is from \eqref{AT:chi_definition}.
\end{asmp}

We note that
\begin{align}
F_s(b) \geq 1 , \label{AT:F_bigger_than_1}
\end{align}
since $F_s(b)$ dominates the term $n =0$ in the sum.

\subsection{The decay functions \texorpdfstring{$\chi_{s,b}$}{chisb}}

\begin{lem}
\label{AT:chi_estimate}
For any $0 < s \leq 1$ and $b \geq 0$ the function $\chi_{s, b}$ in \eqref{AT:chi_definition} satisfies the following properties:
\begin{enumerate}[(a)]

\item 
\begin{enumerate}[(i)]
\item $\chi_{s,b}$ is bounded, positive and monotonically decreasing. If $b>0$ then $\chi_{s,b}$ is strictly decreasing.

\item $\chi_{s, b}$ is logarithmically superadditive, that is, we have $\chi_{s,b}(x+y) \geq \chi_{s,b}(x)\chi_{s,b}(y)$ for every $x,y\geq 0$.
\end{enumerate}

\item For all $k\geq 0$, $b>0$ and $x\geq 0$, define $f_{s,k,b}(x) := x^k\chi_{s,b}(x)$. Then, with the convention $0^0 = 1$, we have
\begin{align*}
\Vert f_{s,k,b}\Vert_{L^\infty(\Rbb_+)} = f_{s,k,b}\bigl( \bigl( \frac{k}{bs}\bigr)^{\frac 1s}\bigr) = \bigl( \frac{k}{bs\, \e}\bigr)^{\frac ks}.
\end{align*}

\item For all $k\geq 0$, any $b'>b$, and $t\geq 0$, we have
\begin{align*}
\int_t^\infty \dx \; x^k \; \chi_{s,b'}(x) \leq \Gamma\bigl(1 + \frac 1s\bigr)\; \bigl( \frac{k}{s\e}\bigr)^{\frac ks} \; \bigl( \frac{2}{b'-b}\bigr)^{\frac{k+1}{s}} \; \chi_{s,b}(t).
%
\end{align*}
Here, $\Gamma$ is the standard $\Gamma$-function.

\item For every $d \in \Nbb_0$, we have
\begin{align*}
\sup_{n\in \Nbb_0} \max\{1, n\}^d \, \chi_{s, b}(n) &\leq \begin{dcases} \bigl( \frac{d}{s\, \e \, b}\bigr)^{\frac{d}{s}} & bs \leq d, \\ 1 & \text{otherwise}. \end{dcases}
\end{align*}

\item For every $t\geq 0$, we have
\begin{align*}
\e^{-bt} \leq E_{s,b} \; \chi_{s,b}(t)
\end{align*}
with $E_{s,b} := \exp(b(1-s)\; s^{\frac{s}{1-s}})$.
\end{enumerate}
\end{lem}

\begin{proof}
Part (a) (i) is clear. To see logarithmic superadditivity, use that $f(x) = x^s$ is concave. We may deduce $f(x) \geq \frac{y}{x+y}f(0)+\frac{x}{x+y}f(x+y) \geq \frac{x}{x+y}f(x+y)$ for $x+y\geq 0$ with $x+y>0$ (if $x=y=0$, then clear). Adding up these inequalities for $x$ and $y$, we obtain $f(x) + f(y) \geq f(x+y)$ so that superadditivity of $\chi_{s,b}$ follows. We continue with part (b). It is easy to see that $f_{s,k,b}$ has a unique maximum, which, by the first derivative test, is located at $(\nicefrac k{bs})^{\nicefrac 1s}$. To prove part (c), write
\begin{align*}
\int_t^\infty \dx \; x^k \; \chi_{s,b'}(x) = \int_t^\infty \dx \; x^k \; \chi_{\frac{b'-b}{2}}(x) \; \chi_{\frac{b'-b}{2}}(x) \; \chi_{s,b}(x).
\end{align*}
We estimate as the last factor as $\chi_{s,b}(x)\leq \chi_{s,b}(t)$. For the first two factors, we apply part (b). We are thus left with the integral, which we carry out on $[0,\infty)$. Perform a substitution $\tau(x) = ax^s$ so that $\dx = \frac 1s \frac{1}{a^{\nicefrac 1s}}\; \tau^{\frac 1s-1}\dd \tau$. Then,
\begin{align*}
\int_0^\infty \dd x \; \chi_{s,a}(x) = \frac 1s \frac{1}{a^{\nicefrac 1s}} \int_0^\infty \dd \tau \; \tau^{\frac{1}{s}-1} \e^{-\tau} = \frac{\Gamma(\nicefrac 1s)}{s} \; \frac{1}{a^{\frac 1s}} = \Gamma\bigl(1 + \frac 1s\bigr) \; \frac{1}{a^{\frac 1s}}.
\end{align*}
Apply this for $a = \frac{b'-b}{2}$ to get the claim. Part (d) follows from maximizing the function $\max\{1, x\}^d \chi_{s, b}(x)$ for $x\in [0,\infty)$. For the last part (e), we write $\e^{-bx} = \e^{-b(x - x^s)} \; \e^{-bx^s}$ and estimate the maximum of $g_s(x) = \e^{-b(x -x^s)}$. Consequently, $g_s'(x) = -b g_s(x) (1 - s \; x^{s-1})$ so that the maximal value is $g_s(s^{\frac{1}{1-s}}) = \exp(b(1-s)s^{\frac{s}{1-s}})$.
\end{proof}

\subsection{Concrete growth condition}

\begin{asmp}[Concrete polynomial growth condition]
\label{AT:Assumption_concrete_polynomial_growth}
Suppose that there are $\kappa>0$ and $d\in \Nbb_0$ such that for every $Z\in \Fcal(\Gamma)$, we have
\begin{align*}
|Z| \leq \kappa \, \max\{1, \Dcal(Z)\}^d.
\end{align*}
\end{asmp}

\begin{lem}[Concrete estimate on $V_s(b)$]
\label{AT:Vb_concrete_estimate}
Suppose that Assumption \ref{AT:Assumption_concrete_polynomial_growth} holds. Then, Assumption \ref{AT:Assumption_abstract_V} holds and if $bs \leq d$, we have
\begin{align*}
V_s(b) \leq \kappa \, \bigl( \frac{d}{s\, \e \, b}\bigr)^{\frac ds}.
\end{align*}
\end{lem}

\begin{proof}
Let $Z\in \Fcal(\Gamma)$. By assumption, we have
\begin{align*}
|Z| \; \chi_{s,b}(\Dcal(Z)) \leq \kappa \, \max\{1, \Dcal(Z)\}^d \, \chi_{s, b}(\Dcal(Z))
\end{align*}
We take the supremum on both sides and arrive at
\begin{align*}
V_s(b) \leq \kappa \sup_{n\in \Nbb_0} \max\{1, n\}^d \, \chi_{s, b}(n).
\end{align*}
By Lemma \ref{AT:chi_estimate}, the right side is finite and if $bs \leq d$, we have
\begin{align*}
\sup_{n\in \Nbb_0} \max\{1, n\}^d \, \chi_{s, b}(n) \leq \bigl(\frac{d}{s\, \e \, b}\bigr)^{\frac ds}.
\end{align*}
This finishes the proof.
\end{proof}

\begin{lem}[Concrete estimate on $F_s(b)$]
\label{AT:F_concrete_estimate}
Suppose that Assumption \ref{AT:Assumption_concrete_polynomial_growth} holds. Then, Assumption \ref{AT:Assumption_abstract_F} is satisfied and if $b < 4$ and $bs < 2d$, then
\begin{align*}
F_s(b) \leq \kappa \; \Gamma\bigl(1+ \frac 1s\bigr) \; \bigl( \frac 4b \bigr)^{\frac 1s} \; \bigl( \frac{2d}{s\, \e \, b} \bigr)^{\frac ds}.
\end{align*}
\end{lem}

\begin{proof}
By assumption, we have
\begin{align*}
F_s(b) &\leq \kappa \sum_{n =0}^\infty \chi_{s, \frac b2}(2n) \; \max\{ 1, 2n\}^d \, \chi_{s, \frac b2}(2n).
\end{align*}
Therefore, Lemma \ref{AT:chi_estimate} implies
\begin{align*}
F_s(b) &\leq \kappa \, \bigl( \frac{2d}{s \, \e \, b}\bigr)^{\frac{d}{s}} \sum_{n=0}^\infty \chi_{s, \frac b2}(2n).
\end{align*}
Since $\chi_{s, b}$ is a monotonically decreasing function, we have
\begin{align*}
\sum_{n=0}^\infty \chi_{s, \frac b2}(2n) \leq 1 + \frac 12 \int_0^\infty \dd x \; \chi_{s, \frac b2}(x),
\end{align*}
which in combination with Lemma \ref{AT:chi_estimate} implies
\begin{align*}
F_s(b) &\leq \kappa \, \bigl[ 1 + \frac 12 \, \Gamma\bigl( 1 + \frac 1s\bigr) \, \bigl( \frac 4b\bigr)^{\frac 1s}\bigr] \, \bigl( \frac{2d}{s \, \e \, b}\bigr)^{\frac ds}
\end{align*}
since $bs \leq 2d$. Since $1 + \nicefrac 1s \geq 2$, we have $\Gamma(1 + \nicefrac 1s) \geq 1$. By the assumption $b \leq 4$, we infer
\begin{align*}
1 \leq \frac 12\, \Gamma\big(1 + \frac 1s\bigr) \bigl( \frac 4b\bigr)^{\frac 1s},
\end{align*}
from which the claim follows.
\end{proof}


\section{Lieb--Robinson bound}
\label{AT:LR_bounds_Section}

\subsection{Time-independent setting}

The first thing we want to prove with the new norm is a Lieb--Robinson type bound for the evolution
\begin{align}
\tau_t(A) = \e^{\i t H} \, A \, \e^{-\i tH}, \label{AT:taut_definition}
\end{align}
where $A\in \Acal_X$ and $H\in \Lcal$ is a local Hamiltonian.

\begin{thm}
\label{AT:LR-bound}
Let Assumption \ref{AT:Assumption_abstract_V} be true, let $X,Y\subset \Lambda$ and let $A\in \Acal_X$, $B\in \Acal_Y$. Let also $0 < s \leq 1$ and $b'>0$. Assume that $H\in \Lcal_{s, b'}$ is self-adjoint. Then, for any $\Phi\in \Bcal_{s, b'}$ such that \eqref{AT:Local_Hamiltonian_decomposition} holds, any $t\in \Rbb$, and any $0 \leq b < b'$, we have
\begin{align*}
\Vert \, [\tau_t(A), B] \, \Vert &\leq 2 \; \Vert A\Vert \; \Vert B\Vert \; \Bigl[\delta_{X,Y} +\frac{1}{V_{s,1}(b'-b)}\bigl(   \e^{b \; \Vcal_{s,b,1}(b'-b) \;  |t|} - 1\bigr) \; g_{s,b}(X,Y)\Bigr].
\end{align*}
Here, $\delta_{X,Y} = 1$ if $X\cap Y\neq\emptyset$ and $\delta_{X,Y} = 0$ otherwise. Furthermore, for $k\in \Nbb$ and $a>0$, we set the Lieb--Robinson velocity to be
\begin{align}
\Vcal_{s,a,k}(b) &:= \frac{2 \, \Vert \Phi\Vert_{s, b'}}{a} \; V_{s,k}(b), \label{AT:LR-velocity_definition}
\end{align}
where $V_{s, k}(b)$ is defined in \eqref{AT:Vkb_definition}. Finally,
\begin{align}
g_{s,b}(X,Y) := \begin{dcases} \sum_{x\in X} \chi_{s,b}(d(x,Y)) & |X| \leq |Y|, \\ \sum_{y\in Y} \chi_{s,b}(d(X,y)) & |X| > |Y|.\end{dcases} \label{AT:LR-g_definition}
\end{align}
\end{thm}

\begin{proof}
The method of proof for this result is inspired by the proof of \cite[Theorem 3.1]{BrunoQuasilocal}. Let us agree on the decomposition $H = \sum_{Z\subset \Lambda} \Phi(Z)$ and call $f(t) := [\tau_t(A),B]$. Then, we have
\begin{align*}
f'(t) &= \i [\tau_t([H,A]), B] = -\i \sum_{\substack{ Z\subset \Lambda \\ Z\cap X\neq \emptyset}} \bigl[B, [\tau_t(\Phi(Z)), \tau_t(A)]\bigr].
\end{align*}
Now, make use of the Jacobi identity $[A,[B,C]] + [B,[C,A]] + [C,[A,B]] = 0$ to get
\begin{align*}
f'(t) = -\i  \Bigl[ f(t), \sum_{Z\cap X\neq \emptyset}\tau_t(\Phi(Z))\Bigr] + \i \sum_{Z\cap X\neq\emptyset} \bigl[ \tau_t(A), [B, \tau_t(\Phi(Z))]\bigr].
\end{align*}
Solving the homogeneous equation amounts to evaluating the Dyson series for the new time-dependent Hamiltonian $\Kcal(t) := \sum_{Z\cap X\neq \emptyset} \tau_t(\Phi(Z))$. This gives a unitary solution operator $\Ucal(t,s)$ so that with $g(t) := \i \sum_{Z\cap X\neq \emptyset} [\tau_t(A),[B,\tau_t(\Phi(Z))]]$ we have
\begin{align*}
f(t) - f(0) = \int_0^t \ds \; \Ucal(t,s)^* \; g(s) \;\Ucal(t,s).
\end{align*}
We conclude that
\begin{align*}
\Vert f(t) \Vert &\leq \Vert f(0)\Vert + \int_0^{|t|} \ds \, \Vert g(s)\Vert \\
&\leq\Vert \, [A,B] \, \Vert + 2\sum_{Z\cap X\neq \emptyset} \Vert \Phi(Z)\Vert \int_0^{|t|} \dd s \; \Vert A\Vert \; \frac{\Vert \, [B,\tau_s(\Phi(Z))] \, \Vert}{\Vert\Phi(Z)\Vert}.
\end{align*}
With the definition
\begin{align*}
C_B(X, t) := \sup_{A\in \Acal_X} \frac{\Vert f(t)\Vert}{\Vert A\Vert} = \sup_{A\in \Acal_X} \frac{\Vert \, [\tau_t(A), B] \, \Vert}{\Vert A\Vert},
\end{align*}
we obtain
\begin{align*}
C_B(X,t) \leq C_B(X,0) + 2 \sum_{Z\cap X\neq \emptyset} \Vert \Phi(Z)\Vert \int_0^{|t|} \ds\;  C_B(Z,s).
\end{align*}
Iterating this gives the expansion
\begin{align*}
C_B(X,t) &\leq C_B(X,0) + \sum_{k=1}^\infty \frac{(2|t|)^k}{k!} \; \sum_{Z_1\cap X\neq \emptyset} \sum_{Z_2\cap Z_1\neq \emptyset} \cdots \sum_{Z_k\cap Z_{k-1}\neq \emptyset} C_B(Z_k,0) \; \prod_{i=1}^k \Vert \Phi(Z_i)\Vert. 
\end{align*}
We observe
\begin{align*}
C_B(X,0) &\leq \begin{cases} 2\Vert B\Vert & X\cap Y\neq \emptyset, \\ 0 & \text{otherwise}.\end{cases}
\end{align*}
Hence, when we define
\begin{align*}
a_k(X,Y) &:= \sum_{Z_1\cap X\neq \emptyset} \sum_{Z_2\cap Z_1\neq \emptyset} \cdots \sum_{\substack{Z_k\cap Z_{k-1}\neq \emptyset \\ Z_k\cap Y \neq \emptyset}}\prod_{i=1}^k \Vert \Phi(Z_i)\Vert,
\end{align*}
we obtain
\begin{align}
C_B(X,t) \leq 2\Vert B\Vert \; \delta_{X,Y} + 2\Vert B\Vert \sum_{k=1}^\infty \frac{(2|t|)^k}{k!} \; a_k(X,Y).\label{LR-bound new norm eq1}
\end{align}
We omit the index $s$ from the notation in the following and claim that
\begin{align}
a_k(X,Y) \leq V_1(b'-b)^{k-1} \; \Vert \Phi\Vert_{b'}^k \; g_b(X,Y). \label{LR-bound new norm eq2}
\end{align}
To prove this, let us assume without loss that $|X| \leq |Y|$ (otherwise interchange the roles of $X$ and $Y$ in what follows). Then, going for induction, we bound the case $k=1$ as
\begin{align*}
a_1(X,Y) &= \sum_{\substack{Z\cap X\neq \emptyset \\ Z\cap Y\neq \emptyset}} \Vert \Phi(Z)\Vert \leq \sum_{x\in X} \sum_{\substack{Z\subset \Lambda \\ Z\ni x\\ Z \cap Y\neq \emptyset}} \frac{\Vert \Phi(Z)\Vert}{\chi_b(\Dcal(Z))} \; \chi_b(\Dcal(Z)).
\end{align*}
Since $x\in Z$ and $Z\cap Y \neq \emptyset$, we conclude that $\Dcal(Z) \geq d(x,Y)$. Hence,
\begin{align*}
a_1(X,Y) \leq \sum_{x\in X} \chi_b(d(x,Y)) \sum_{\substack{Z\subset \Lambda \\ Z\ni x}} \frac{\Vert \Phi(Z)\Vert}{\chi_b(\Dcal(Z))} \leq \Vert \Phi\Vert_b \; \sum_{x\in X} \chi_b(d(x,Y)),
\end{align*}
which is bounded by \eqref{LR-bound new norm eq2} since $\Vert \Phi\Vert_b \leq \Vert \Phi\Vert_{b'}$. Furthermore, we have
\begin{align*}
a_{k+1}(X,Y) \leq \sum_{x\in X} \sum_{\substack{Z_1\subset \Lambda \\ Z_1\ni x}} \frac{\Vert \Phi(Z_1)\Vert}{\chi_{b'}(\Dcal(Z_1))} \; \chi_{b'}(\Dcal(Z_1)) \sum_{Z_2\cap Z_1\neq \emptyset} \cdots \sum_{\substack{Z_{k+1} \cap Z_k\neq \emptyset \\ Z_{k+1} \cap Y\neq \emptyset}} \prod_{i=2}^{k+1} \Vert \Phi(Z_i)\Vert.
\end{align*}
Since the last factor is exactly $a_k(Z_1,Y)$, we may apply the induction hypothesis and relabel $Z_1$ by $Z$ to obtain
\begin{align*}
a_{k+1}(X,Y) &\leq \Vert \Phi\Vert_{b'}^{k} \; V_1(b'-b)^{k-1}  \sum_{x\in X} \sum_{\substack{Z\subset \Lambda \\ Z \ni x}} \frac{\Vert \Phi(Z)\Vert}{\chi_{b'}(\Dcal(Z))} \; \chi_{b'-b}(\Dcal(Z)) \\
&\hspace{200pt} \times  \sum_{z\in Z} \chi_{b}(\Dcal(Z)) \; \chi_b(d(z,Y)).
\end{align*}
Since $x,z\in Z$, we obtain $\Dcal(Z) \geq d(x,z)$ and subsequently $d(x,z) + d(z,Y)\geq d(x,Y)$ which, by the logarithmic superadditivity, implies that
\begin{align*}
\chi_b(\Dcal(Z)) \; \chi_b(d(z,Y))\leq \chi_b(d(x,z) + d(z,Y)) \leq \chi_b(d(x,Y)).
\end{align*}
Then, the sum over $z\in Z$ yields a factor of $|Z|$, which, together with $\chi_{b'-b}(\Dcal(Z))$ gives $V_1(b'-b)$ for an upper bound. All in all, we conclude,
\begin{align*}
a_{k+1}(X,Y) \leq \Vert \Phi\Vert_{b'}^{k} \;  V_1(b'-b)^k \; \sum_{x\in X} \chi_b(d(x,Y))  \sum_{\substack{ Z\subset \Lambda \\ Z\ni x}} \frac{\Vert \Phi(Z)\Vert}{\chi_{b'}(\Dcal(Z))}.
\end{align*}
Bounding the last term as the $(k+1)$\st\ copy of $\Vert \Phi\Vert_{b'}$ the induction is complete and \eqref{LR-bound new norm eq2} is proven. Looking back at \eqref{LR-bound new norm eq1}, we have shown that
\begin{align*}
C_B(X,t) &\leq 2\Vert B\Vert \Bigl[ \delta_{X,Y} + \frac{1}{V_{s,1}(b'-b)}\sum_{k=1}^\infty \frac{1}{k!} \bigl[ 2|t| \Vert \Phi\Vert_{s,b'} V_{s,1}(b'-b)\bigr]^k g_{s,b}(X,Y)\Bigr],
\end{align*}
which by definition of $C_B(X,t)$ yields the claim.
\end{proof}

\begin{kor}
\label{AT:LR-bound_kor}
Let Assumption \ref{AT:Assumption_abstract_V} be true, let $X,Y\subset \Lambda$ with $d(X,Y)>0$ and let $A\in \Acal_X$ and $B\in \Acal_Y$. Let also $0 < s\leq 1$ and $b'>0$ and assume that $H\in \Lcal_{s, b'}$ is self-adjoint. Then, for any $\Phi\in \Bcal_{s, b'}$ such that \eqref{AT:Local_Hamiltonian_decomposition} holds, any $t\in \Rbb$, and any $0\leq b < b'$, we have 
\begin{align*}
\Vert \, [\tau_t(A), B] \, \Vert &\leq \frac{2}{V_{s,1}(b'-b)} \; \min\{|X|,|Y|\} \; \Vert A\Vert \; \Vert B\Vert \; (\e^{b \; \Vcal_{s,b,1}(b'-b) \; |t|} - 1) \; \chi_{s,b}(d(X,Y)),
\end{align*}
where $\Vcal_{s,b,1}(b'-b)$ is given in \eqref{AT:LR-velocity_definition}.
\end{kor}

\begin{proof}
Obviously, $g_{s,b}(X,Y) \leq \min\{|X|,|Y|\} \; \chi_b(d(X,Y))$. Apply Theorem \ref{AT:LR-bound}.
\end{proof}

\subsection{Time-dependent setting}

To start out with, let $I\subset \Rbb$ be an interval. For a local Hamiltonian $H \in \Lcal(I)$, let $U(t,s)$ be the unique strong unitary solution \cite[Proposition 2.2]{BrunoQuasilocal} to
\begin{align}
\frac{\dd}{\dd u} U(u,v) &= -\i \, H(u) \, U(u,v), & U(v,v) &= \Idbb. \label{Heisenberg-evolution-eq}
\end{align}
Consequently, we have $U(u,v)^{-1} = U(u,v)^* = U(v,u)$ for all $u,v\in I$ and
\begin{align*}
\frac{\dd}{\dd u} U(u,v)^* = \i \, U(u,v)^* \, H(u)
\end{align*}
Let $A\in \Acal_X$. With this, we define the Heisenberg dynamics $\tau_{u,v}(A)$ as
\begin{align*}
\tau_{u,v}(A) := U(u,v)^* \, A \, U(u,v).
\end{align*}

\begin{thm}
\label{LR-bound for Heisenberg}
Let Assumption \ref{AT:Assumption_abstract_V} be true, let $X,Y\subset \Lambda$ and let $A\in \Acal_X$, $B\in \Acal_Y$. Let also $0 < s \leq 1$ and $b'>0$. Given an interval $I\subset \Rbb$, assume that $H\in \Lcal_{s,b'}(I)$ is self-adjoint. Then, for any $\Phi\in \Bcal_{s,b'}(I)$ such that \eqref{AT:Local_Hamiltonian_decomposition} holds, any $0 \leq b < b'$, and any $u,v\in I$, we have
\begin{align*}
\Vert \, [\tau_{u,v}(A),B] \, \Vert \leq 2 \; \Vert A\Vert \; \Vert B\Vert \; \Bigl[ \delta_{X,Y} + \frac{1}{V_{s,1}(b'-b)} \bigl( \e^{b \; \Vcal_{s,b,1}(b'-b) \; |u-v|} - 1\bigr) \; g_{s,b}(X,Y)\Bigr].
\end{align*}
Here, $\delta_{X,Y} = 1$ if $X\cap Y\neq \emptyset$ and $\delta_{X,Y} =0$ otherwise, and $g_{s, b}$ is defined in \eqref{AT:LR-g_definition}. Furthermore, for $k\in \Nbb$ and $a>0$, the Lieb--Robinson velocity is defined as
\begin{align}
\Vcal_{s,a,k}(b) &:= \frac{2 \, \vvvert \Phi\vvvert_{I, s, b'}}{a} \; V_{s,k}(b), \label{AT:LR-velocity_time-dependent_definition}
\end{align}
where $V_{s, k}(b)$ is defined in \eqref{AT:Vkb_definition}.
\end{thm}

\begin{proof}
The method of proof for this result is inspired by the proof of \cite[Theorem 3.1]{BrunoQuasilocal}. Let $\Phi$ be a time-dependent interaction such that the  decomposition $H(u) = \sum_{Z\subset \Lambda} \Phi(Z,u)$ holds, where $\Phi(Z, u) := \Phi(u)(Z)$, and for fixed $v\in I$ define $f_v(u) := [\tau_{u,v}(A),B]$. Then, by \eqref{Heisenberg-evolution-eq} and a short calculation, we have
\begin{align*}
f_v'(u) &= \i [\tau_{u,v}([H,A]), B] = -\i \sum_{\substack{ Z\subset \Lambda \\ Z\cap X\neq \emptyset}} \bigl[B, [\tau_{u,v}(\Phi(Z,u)), \tau_{u,v}(A)]\bigr].
\end{align*}
Now, make use of the Jacobi identity $[A,[B,C]] + [B,[C,A]] + [C,[A,B]] = 0$ to get
\begin{align*}
f_v'(u) = -\i  \Bigl[ f_v(u), \sum_{Z\cap X\neq \emptyset}\tau_{u,v}(\Phi(Z,u))\Bigr] + \i \sum_{Z\cap X\neq\emptyset} \bigl[ \tau_{u,v}(A), [\tau_{u,v}(\Phi(Z,u)),B]\bigr].
\end{align*}
Solving the homogeneous equation amounts to evaluating the Dyson series for the new time-dependent Hamiltonian $\Kcal_v(u) := \sum_{Z\cap X\neq \emptyset} \tau_{u,v}(\Phi(Z,u))$. This gives a unitary solution operator $\Ucal_v(u,w)$ so that
\begin{align*}
f_v(u) - f_v(w) = \int_w^u \dr \; \Ucal_v(r,w)^* \; g_v(r) \; \Ucal_v(r,w) \qquad \qquad \forall u,w\in I,
\end{align*}
with
\begin{align*}
g_v(u) := \i \sum_{\substack{Z\subset \Lambda \\ Z\cap X\neq \emptyset}} \bigl[ \tau_{u,v}(A), [\tau_{u,v}(\Phi(Z,u)), B]\bigr].
\end{align*}
Setting $w = v$, we get
\begin{align*}
\Vert f_v(u) \Vert &\leq \Vert f_v(v)\Vert + \int_{\min\{u,v\}}^{\max\{u,v\}} \dd r \, \Vert g_v(r)\Vert,
\end{align*}
or, in other words,
\begin{align*}
\Vert \, [\tau_{u,v}(A),B] \, \Vert \leq\Vert \, [A,B] \, \Vert + 2\, \Vert A\Vert \int_{\min\{u,v\}}^{\max\{u,v\}} \dd r \sum_{\substack{Z\subset \Lambda \\ Z\cap X\neq \emptyset}} \Vert \, [B,\tau_{r,v}(\Phi(Z,r))] \, \Vert.
\end{align*}
Iterating this up to finite order $n\in \Nbb$ gives the expansion
\begin{align}
\Vert \, [\tau_{u,v}(A),B] \, \Vert &\leq \Vert \, [A,B] \, \Vert + 2 \, \Vert A\Vert \; \Vert B\Vert \sum_{k=1}^n 2^k \; a_k(u,v,X,Y) + R_{n}(u,v,X) \label{LR for Heisenberg eq1}
\end{align}
with
\begin{align*}
a_k(u,v,X,Y) := \int_{u\vee v}^{u\wedge v} \dr_1 \cdots \int_{r_{k-1}\vee v}^{r_{k-1}\wedge v} \dr_k \sum_{\substack{Z_1\subset \Lambda \\ Z_1\cap X\neq \emptyset}} \sum_{\substack{Z_2\subset \Lambda \\ Z_2 \cap Z_1 \neq \emptyset}} \cdots \sum_{\substack{ Z_k\subset \Lambda \\ Z_k\cap Z_{k-1} \neq \emptyset \\ Z_k \cap Y\neq \emptyset}} \prod_{i=1}^k \Vert \Phi(Z_i,r_i)\Vert,
\end{align*}
where $u \wedge v := \max\{u, v\}$ and $u\vee v := \min\{u, v\}$, and
\begin{align*}
R_{n}(u,v,X) &:= 2^{n+2} \Vert A\Vert \int_{u\vee v}^{u\wedge v} \dr_1 \int_{r_1\vee v}^{r_1 \wedge v} \dr_2 \cdots \int_{r_{n}\vee v}^{r_{n}\wedge v} \dr_{n+1}   \\
&\hspace{-30pt} \times \sum_{\substack{Z_1\subset \Lambda \\ Z_1\cap X\neq \emptyset}}\sum_{\substack{Z_2\subset \Lambda \\ Z_2 \cap Z_1 \neq \emptyset}} \cdots \sum_{\substack{ Z_n\subset \Lambda \\ Z_{n+1}\cap Z_{n} \neq \emptyset}} \Vert \, [\tau_{r_{n+1},v}(\Phi(Z_{n+1},r_{n+1}), B] \, \Vert \, \prod_{i=1}^n \Vert \Phi(Z_i,r_i)\Vert.
\end{align*}

Now, first we have the bound $\Vert [A,B]\Vert \leq 2\, \Vert A\Vert \, \Vert B\Vert \, \delta_{X,Y}$. Next, concerning $R_{n}(u,v,X)$, we have
\begin{align*}
\Vert \, [\tau_{r_{k+1},v}(\Phi(Z_{n+1}r_{n+1})),B] \, \Vert \leq 2\, \Vert \Phi(Z_{n+1},r_{n+1})\Vert \; \Vert B\Vert.
\end{align*}
This allows us to bound the integrand of the error term $R_{n}$ in the following way. We omit the index $s$ from the notation in the following and claim that
\begin{align}
\sum_{\substack{Z_1\subset \Lambda \\ Z_1 \cap X \neq \emptyset}} \cdots \sum_{\substack{ Z_n\subset \Lambda \\ Z_{n+1}\cap Z_{n} \neq \emptyset}}  \prod_{i=1}^{n+1} \Vert \Phi(Z_i,r_i)\Vert &\leq \vvvert \Phi\vvvert_{I, b}^{n+1} \; V_1(b)^n \; |X|. \label{LR for Heisenberg eq2}
\end{align}
Going for induction, the case $n=1$ is bounded as
\begin{align*}
\sum_{Z_1\cap X\neq \emptyset}  \sum_{Z_2\cap Z_1\neq \emptyset} \Vert \Phi(Z_1,r_1)\Vert \; \Vert \Phi(Z_2,r_2)\Vert &\\
&\hspace{-70pt} \leq \sum_{x\in X} \sum_{\substack{Z_1\subset \Lambda \\ Z_1\ni x}} \frac{\Vert \Phi(Z_1,r_1)\Vert}{\chi_{b'}(\Dcal(Z_1))} \; \chi_{b'}(\Dcal(Z_1)) \sum_{z_1\in Z_1} \sum_{\substack{Z_2\subset \Lambda \\ z_1\in Z_2}} \frac{\Vert\Phi(Z_2,r_2)\Vert}{\chi_{b'}(\Dcal(Z_2))} .
\end{align*}
On the right, we may take away the norm $\vvvert \Phi\vvvert_{I, b}$. Then, the factor $\chi_b(\Dcal(Z_1)) \; |Z_1|$ is bounded by $V_1(b)$. Evaluating the second norm, we arrive at the claim \eqref{LR for Heisenberg eq2} in the case $n = 1$. Inductively, we have
\begin{align*}
\sum_{\substack{Z_1\subset \Lambda \\ Z_1 \cap X \neq \emptyset}} \cdots \sum_{\substack{ Z_n\subset \Lambda \\ Z_{n+1}\cap Z_{n} \neq \emptyset}}  \prod_{i=1}^{n+1} \Vert \Phi(Z_i,r_i)\Vert &\\
&\hspace{-50pt} \leq \sum_{x\in X} \sum_{\substack{Z_1\subset \Lambda \\ Z_1 \ni x}} \frac{\Vert \Phi(Z_1,r_1)\Vert}{\chi_b(\Dcal(Z_1))} \; \chi_b(\Dcal(Z_1)) \; \vvvert \Phi\vvvert_{I, b}^n \; V_1(b)^{n-1} |Z_1|.
\end{align*}
The similar strategy to the case $n=1$ readily yields \eqref{LR for Heisenberg eq2}. All in all, we conclude the bound of the remainder term to be
\begin{align*}
R_{n}(u,v,X) &\leq \frac{2}{V_1(b)} \; |X| \; \Vert A\Vert \; \Vert B\Vert \; \frac{(2 \; \vvvert \Phi\vvvert_{I, b} \; V_1(b) \; |u-v|)^{n+1}}{(n+1)!},
\end{align*}
which vanishes as $n\to\infty$. Similarly, we bound the $k$\tho\ coefficient $a_k(u,v, X,Y)$. Here, the only difference is that we have the additional constraint of $Z_k \cap Y\neq \emptyset$. Assume without loss that $|X|\leq |Y|$, otherwise interchange the roles of $X$ and $Y$ in what follows. We claim that
\begin{align}
a_k(u,v,X,Y) \leq \vvvert \Phi\vvvert_{I, b'}^k \; V_1(b'-b)^{k-1}  \; \frac{|u-v|^k}{k!} \; \sum_{x\in X} \chi_b(d(x,Y)). \label{LR for Heisenberg eq3}
\end{align}
We start the induction again by the case $k=1$ and
\begin{align*}
a_1(u,v,X,Y) =\int_{u\vee v}^{u\wedge v} \dr \sum_{\substack{Z\subset \Lambda \\ Z\cap X\neq \emptyset \\ Z\cap Y\neq\emptyset}} \Vert \Phi(Z,r)\Vert.
\end{align*}
 Then,
\begin{align*}
a_1(u,v,X,Y) \leq \int_{u\vee v}^{u\wedge v} \dr \sum_{x\in X} \sum_{\substack{Z\subset \Lambda \\ Z\ni x \\ Z\cap Y\neq \emptyset}} \frac{\Vert \Phi(Z,r)\Vert}{\chi_{b}(\Dcal(Z))} \; \chi_{b}(\Dcal(Z)).
\end{align*}
Since $\Dcal(Z) \geq d(x,y)$ for any point $y\in Y$, we have $\Dcal(Z) \geq d(x,Y)$, so
\begin{align*}
a_1(u,v,X,Y) \leq \int_{u\vee v}^{u\wedge v} \dr \; \vvvert \Phi\vvvert_{I, b} \; \sum_{x\in X} \chi_b(d(x,Y)) \leq \vvvert \Phi\vvvert_{I, b'} \;  |u-v| \sum_{x\in X} \chi_b(d(x,Y)).
\end{align*}
The case $k=1$ is proven. Supposing the claim is true for $k$, we get that
\begin{align*}
a_{k+1}(u,v,X,Y) &\leq \int_{u\vee v}^{u\wedge v} \dr_1 \sum_{x\in X} \sum_{\substack{Z_1\subset \Lambda \\ Z_1 \ni x}} \Vert \Phi(Z_1,r_1)\Vert \\
&\hspace{20pt} \times\int_{r_1\vee v}^{r_1\wedge v}\dr_2 \cdots \int_{r_{k} \vee v}^{r_{k}\wedge v}\dr_{k +1} \sum_{\substack{Z_2\subset \Lambda \\ Z_2 \cap Z_1\neq \emptyset}} \cdots \sum_{\substack{Z_{k+1}\subset \Lambda \\ Z_{k+1}\cap Z_{k} \neq \emptyset}} \prod_{i=2}^{k+1} \Vert \Phi(Z_i,r_i)\Vert .
\end{align*}
Since the second row is equal to $a_k(r_1,v,Z_1,Y)$, we apply the induction hypothesis and obtain
\begin{align*}
a_{k+1}(u,v,X,Y) &\leq \vvvert \Phi\vvvert_{I, b'}^k \; V_{s,1}(b'-b)^{k-1}  \int_{u\vee v}^{u\wedge v} \dr_1 \; \frac{|r_1-v|^{k}}{k!}  \\
&\hspace{25pt} \times \sum_{x\in X} \sum_{\substack{Z_1\subset \Lambda \\ Z_1 \ni x}} \frac{\Vert \Phi(Z_1,r_1)\Vert}{\chi_{b'}(\Dcal(Z_1))} \; \chi_{b'-b}(\Dcal(Z_1)) \sum_{z_1\in Z_1} \, \chi_b(\Dcal(Z_1)) \, \chi_b(d(z_1,Y)).
\end{align*}
Since $x,z_1\in Z_1$, we get $\Dcal(Z_1)\geq d(x,z_1)$. Then, the logarithmic superadditivity, together with the triangle inequality $d(x,z_1) + d(z_1,Y) \geq d(x, Y)$ yields
\begin{align*}
\chi_b(\Dcal(Z_1)) \, \chi_b(d(z_1,Y)) \leq \chi_b(d(x,z_1) + d(z_1,Y)) \leq \chi_b(d(x,Y)).
\end{align*}
The right-most sum gives a factor of $|Z_1|$, which, together with $\chi_{b'-b}(\Dcal(Z_1))$ yields an upper bound by a copy of $V_1(b'-b)$. Bounding away the norm $\vvvert \Phi\vvvert_{I, b'}$, we arrive at the bound
\begin{align*}
a_{k+1}(u,v,X,Y) &\leq \vvvert \Phi\vvvert_{I, b'}^{k+1} \; V_1(b'-b)^k \; \int_{u\vee v}^{u\wedge v} \dr \;\frac{|r - v|^k}{k!} \; \sum_{x\in X} \chi_b(d(x,Y)).
\end{align*}
Integrating out the last integral finally gives \eqref{LR for Heisenberg eq3} for $k+1$. Looking back at \eqref{LR for Heisenberg eq1}, we have shown that
\begin{align*}
\Vert \, [\tau_{u,v}(A),B] \, \Vert &\leq 2\Vert A\Vert \Vert B\Vert \Bigl[ \delta_{X,Y} + \frac{1}{V_1(b'-b)}\sum_{k=1}^\infty \frac{[b \; \Vcal_{b,1}(b'-b) |u-v|]^k}{k!} \; g_b(X,Y)\Bigr],
\end{align*}
which yields the claim.
\end{proof}

\begin{kor}
\label{LR-bound for Heisenberg 2}
Let Assumption \ref{AT:Assumption_abstract_V} be true, let $X,Y\subset \Lambda$ with $d(X,Y)>0$ and let $A\in \Acal_X$ and $B\in \Acal_Y$. Let also $0 < s\leq 1$ and $b'>0$. For an interval $I\subset \Rbb$, assume that $H\in \Lcal_{s, b'}(I)$ is self-adjoint. Then, for any $\Phi\in \Bcal_{s, b'}(I)$ such that \eqref{AT:Local_Hamiltonian_decomposition} holds, any $0 < b < b'$, and any $u,v\in I$, we have 
\begin{align*}
\Vert \, [\tau_{u,v}(A), B]\, \Vert &\leq \frac{2 \; \min\{|X|,|Y|\}}{V_{s,1}(b'-b)} \; \Vert A\Vert \; \Vert B\Vert \; (\e^{b \; \Vcal_{s,b,1}(b'-b) \; |u-v|} - 1) \; \chi_{s,b}(d(X,Y)),
\end{align*}
where $\Vcal_{s,b,1}(b'-b)$ is given in \eqref{AT:LR-velocity_time-dependent_definition}.
\end{kor}

\begin{proof}
Obviously, $g_{s,b}(X,Y) \leq \min\{|X|,|Y|\} \; \chi_{s, b}(d(X,Y))$. Apply Theorem \ref{LR-bound for Heisenberg}.
\end{proof}


\section{The map \texorpdfstring{$\Ical$}{I}}
\label{AT:I_Section}

In order to solve the adiabatic Schrödinger equation \eqref{AT:Schroedinger} given by the Hamiltonian~$H$, it is unavoidable to have a tool at hand which inverts the operation $G\mapsto [H, G]$ for certain operators $G$. Note that such a property cannot hold for \emph{all} operators $G$ since every function of $H$ lies in the kernel of this map. The inverse property should be compared to the work \cite{HagedornJoye}, where the resolvent of $H$ at the lowest eigenvalue was considered as a bounded operator on a proper subspace of the underlying Hilbert space --- the orthogonal complement of the eigenfunction corresponding to the eigenvalue. The important point is that we need to invert $G\mapsto [H, G]$ in a locality perserving manner. This is done by the map $\Ical_{s, \gamma}$ as we shall see now.

\subsection{Definition of \texorpdfstring{$\Ical_{s, \gamma}$}{I}}

Let $G, H\in \Lcal$, assume that $H$ is self-adjoint, and for $0 < s < 1$ and $\gamma>0$ let a function $W_{s, \gamma}\in L^1(\Rbb)$ be given. Then, we define
\begin{align}
\Ical_{s, \gamma}(G) &:= \int_\Rbb \dd t \; W_{s, \gamma}(t) \; \e^{\i t H} \, G \, \e^{-\i t H}. \label{AT:I_definition}
\end{align}
We claim that if $G \in \Lcal_s$ and $H \in \Lcal_\Ecal$, then $\Ical_{s, \gamma}(G)\in \Lcal_s$. In other words, \eqref{AT:I_definition} defines a map $\Ical_{s, \gamma} \colon \Lcal_s \ra \Lcal_s$. We shall show this in Section \ref{AT:I_Estimate_Section}.

For an interval $I\subset \Rbb$ and time-dependent Hamiltonians $G, H\in \Lcal(I)$, the map $\Ical_{s, \gamma}$ extends in a natural way to a map $\Ical_{s, \gamma} \colon \Lcal_s(I) \ra \Lcal_s(I)$ via the pointwise definition by the same formula \eqref{AT:I_definition}.

Therefore, $\Ical_{s, \gamma}$ is a ``locality preserving'' map, which is a very important property in the business of quantum lattice systems. 
%

%

\subsection{Inverse property of \texorpdfstring{$\Ical_{s, \gamma}$}{I}}

The second important property of $\Ical_{s, \gamma}$ is that it provides an inverse of the commutator. This depends on the parameter $\gamma$ that we did not use so far and that plays the role of the spectral gap of $H$.

\begin{defn}[Gapped Hamiltonian]
\label{AT:Definition_gapped_Hamiltonian}
Let $I\subset \Rbb$ be an interval. Let $H\in \Lcal(I)$ be a self-adjoint time-dependent local Hamiltonian. We call $H$ \emph{gapped} if and only if for every $\Lambda\in \Fcal(\Gamma)$ and $u\in I$ the spectrum $\sigma(H^\Lambda(u))$ admits the decomposition 
\begin{align}
\sigma(H^\Lambda(u)) &= \Sigma_0^\Lambda(u) \cup \Sigma_1^\Lambda(u), \label{AT:H_spectrum_decomposition}
\end{align}
such that $\Sigma_0^\Lambda(u)$ and $\Sigma_1^\Lambda(u)$ are separated by a \emph{uniform spectral gap} $\gamma$, that is,
\begin{align}
\gamma := \inf\bigl\{ \dist\bigl( \Sigma_0^\Lambda(u) \, , \, \Sigma_1^\Lambda(u) \bigr): \Lambda \in \Fcal(\Gamma) , \; u\in I\bigr\} >0. \label{AT:H_spectral_gap}
\end{align}
\end{defn}

\begin{prop}
Let $I\subset \Rbb$ be an interval, let $G, H\in \Lcal(I)$ be two time-dependent local Hamiltonians and assume that $H$ is gapped with uniform spectral gap $\gamma >0$. Let $P$ denote the spectral projection of $H$ onto the spectral patch $\Sigma_0$. Then, for any $0 < s < 1$, there is a function $W_{s, \gamma} \in L^1(\Rbb)$ such that the following statements hold pointwise for all $u\in I$:
\begin{enumerate}[(a)]
\item If $G$ satisfies the offdiagonal condition
\begin{align}
G = P\, G\, (1 - P) + (1 - P) \, G\, P, \label{AT:Offdiagonal_condition}
\end{align}
then
\begin{align}
G = -\i [H , \, \Ical_{s, \gamma}(G)\, ]. \label{AT:I_inverse_property}
\end{align}

\item We have
\begin{align*}
[G, P] - \i \bigl[ [\, \Ical_{s, \gamma}(G) \, , H], P\bigr] =0.
\end{align*}
\end{enumerate}
\end{prop}

\begin{proof}
The proof is given in \cite[Proposition 4.1 (a), (b)]{SvenAdiabatic} and uses that $W_{s, \gamma}$ is a function that satisfies $\hat W_{s, \gamma}(\xi) = \frac{-\i }{\sqrt{2\pi} \, \xi}$ if $|\xi| \geq \gamma$. We explicitly construct such a function in the following. For the required property, see Lemma \ref{AT:W_gamma_estimate} (b).
\end{proof}

\subsection{Construction of the function \texorpdfstring{$W_{s,\gamma}$}{Wsgamma}}

In a first step, we are going to construct a function $w_{s, \gamma}$ for $0 < s < 1$ with rapid decay and whose Fourier transform\ifthenelse{\equal\masterfile{Diss}}{\footnote{Note that, in contrast to the papers on BCS theory in Chapters \ref{Chapter:DHS1} and \ref{Chapter:DHS2}, we choose here the unitary Fourier transform.}}{}
\begin{align}
\hat w_{s, \gamma}(\xi) &:= \frac{1}{\sqrt{2\pi}} \int_\Rbb \dd t \; \e^{-\i \xi t} \, w_{s, \gamma} (t) \label{AT:Fourier_transform}
\end{align}
is compactly supported in the interval $[-\gamma, \gamma]$. A comprehensive method to construct such functions with a desired bound by a given rapidly decaying function has been used in \cite[Lemma 2.3]{AutomorphicEquivalence} and we utilize this method to construct such a function for our $\chi_{b, s}$ for every $0 < s < 1$. We point out to the reader that such a function cannot exist for $s = 1$ since an exponentially decaying function has a Fourier transform with an analytic continuation to a complex strip around the real axis. Since we require the Fourier transform to have compact support as well, it thus vanishes identically by the identity theorem.

\begin{lem}
\label{AT:w_gamm_estimate}
Let $0 < s < 1$ and define $a(s) := \frac{\gamma}{2\zeta(2-s)}$. Set $a_n(s) := \frac{a(s)}{n^{2-s}}$ for each $n\geq 1$. Then, we have $\sum_{n\geq 1} a_n(s) = \frac \gamma 2$ and the infinite product
\begin{align*}
w_{s,\gamma}(t) &:= c_{s,\gamma} \prod_{n=1}^\infty \Bigl( \frac{\sin(a_n(s) \; t)}{a_n(s) \; t}\Bigr)^2
\end{align*}
defines a nonnegative, even function $w_{s,\gamma}\in L^1(\Rbb)$. We choose $c_{s,\gamma}$ so that $\Vert w_{s,\gamma}\Vert_{L^1(\Rbb)} = 1$. Furthermore, the following statements are true:
\begin{enumerate}[(a)]
\item Consider the function
\begin{align*}
f_s(t) := \exp\bigl[\log(t) \; \bigl(2+2s-s^2-\frac{(1-s)^2}{\zeta(2-s)} \; t^s\bigr)\bigr].
\end{align*}
and let $\xi(s)\in (0,\infty)$ be the unique solution of
\begin{align*}
\log(\xi(s)) = \frac{(2+2s-s^2)\zeta(2-s)}{(1-s)^2} \; \frac{1}{\xi(s)} -1.
\end{align*}
Then, for all $t\in \Rbb$, we have the estimate
\begin{align*}
w_{s,\gamma}(t) &\leq c_{s,\gamma} \; D_s \;  \chi_{s, \mu_0(s)}(\gamma t).
\end{align*}
where
\begin{align*}
D_s &:= \frac{(4\pi)^{2-s}\e^{2(2-s)}}{(2\zeta(2-s))^{4-s}} \; f_s(\xi(s)^s), & \mu_0(s) &:= \frac{2-s+(1-s)\log(2\zeta(2-s))}{\zeta(2-s)}.
\end{align*}

\item The support of $\hat w_{s,\gamma}$ is contained in the interval $[-\gamma,\gamma]$.
\end{enumerate}
\end{lem}

\begin{proof}
First, note that $\frac{a(s)}{\gamma} = \frac{1}{2\zeta(2-s)}$. Hence, it will be convenient to express all quantities in terms of this ratio. The product converges pointwise (actually uniformly on compact subsets of $\Rbb$) as each factor is a member of $[0,1]$. Hence, the sequence of partial products is monotone decreasing and bounded from below by $0$. For every $N\in \Nbb$, we have the estimate
\begin{align*}
w_{s,\gamma}(t) \leq c_{s,\gamma} \prod_{n=1}^N \Bigl(\frac{ \sin(a_n(s) \; t)}{a_n(s) \; t}\Bigr)^2 \leq c_{s,\gamma} \prod_{n=1}^N \frac{n^{4-2s}}{(a(s) \; t)^2} = c_{s,\gamma} \frac{(N!)^{4-2s}}{(a(s) \; t)^{2N}}.
\end{align*}
Employing Stirling's formula $N! \leq \sqrt{4\pi N} (\frac N\e)^N$ (the $4$ in the square root instead of $2$ is just to get the upper bound), we get
\begin{align*}
w_{s,\gamma}(t) &\leq (4\pi)^{2-s} \; c_{s,\gamma} \; N^{2-s} \; N^{(4-2s)N} \; \e^{-(4-2s)N} \; (a(s) \; t)^{-2N}.
\end{align*}
Without loss, assume that $t\geq 0$, otherwise consider $|t|$. Now, choose $N := \lfloor \frac{a(s) t}{(\gamma t)^{1-s}}\rfloor $ and use that $\frac{(\gamma t)^s}{2\zeta(2-s)}-1 \leq N\leq \frac{(\gamma t)^s}{2\zeta(2-s)}$ to get
\begin{align*}
\frac{w_{s,\gamma}(t)}{(4\pi)^{2-s} c_{s,\gamma}} &\leq \e^{(2-s) \log(\frac{(\gamma t)^s}{2\zeta(2-s)})} 
 \; \e^{2(2-s) \frac{(\gamma t)^s}{2\zeta(2-s)} \log(\frac{(\gamma t)^s}{2\zeta(2-s)})} 
 \; \e^{-2(2-s)(\frac{(\gamma t)^s}{2\zeta(2-s)} - 1)} \\
&\hspace{220pt}\times \e^{-2(\frac{(\gamma t)^s}{2\zeta(2-s)} -1)\log(a(s) \; t)}. 
\end{align*}
Multiplying and dividing the last factor by $\e^{2(\frac{(\gamma t)^s}{2\zeta(2-s)} -1)\log((\gamma t)^{1-s})}$, and rearranging the right-hand side, we obtain
\begin{align*}
&\e^{(4-s)s\log(\gamma t)} \; \e^{-(4-s)\log(2\zeta(2-s))} \; \e^{-\frac{(1-s)^2}{\zeta(2-s)} (\gamma t)^s\log(\gamma t)} \; \e^{-\frac{1-s}{\zeta(2-s)} (\gamma t)^s \log(2\zeta(2-s))} \\
&\hspace{240pt} \; \e^{-\frac{2-s}{\zeta(2-s)}(\gamma t)^s} \; \e^{2(2-s)} \; \e^{2(1-s)\log(\gamma t)}\\
&= \smash{\frac{\e^{2(2-s)}}{(2\zeta(2-s))^{4-s}}} \; \e^{[(4-s)s+2(1-s)]\log(\gamma t)} \; \e^{-\frac{(1-s)^2}{\zeta(2-s)} (\gamma t)^s \log(\gamma t)} \\
&\hspace{250pt} \times \e^{-[\frac{2-s}{\zeta(2-s)} + \frac{1-s}{\zeta(2-s)} \log(2\zeta(2-s))] \; (\gamma t)^s}.
\end{align*}
From this, we conclude that
\begin{align*}
\frac{w_{s,\gamma}(t)}{(4\pi)^{2-s} c_{s,\gamma}} &\leq  \frac{\e^{2(2-s)}}{(2\zeta(2-s))^{4-s}} \; f_s(\gamma t) \;  \chi_{s,\mu_0(s)}(\gamma t).
\end{align*}
It remains to estimate the maximal value of $f_s$. Since the exponential is monotone, $f_s$ attains its maximum at $\e^{g_\mathrm{max}}$ where $g_{\mathrm{max}}$ is the maximal value of
\begin{align*}
g_s(t) &:= \log(t) \bigl(2+2s-s^2 - \frac{(1-s)^2}{\zeta(2-s)} \; t^s\bigr)
\end{align*}
Since $g(t) \to -\infty$ for $t\to 0$ as well as for $t\to\infty$, it has a maximum $\xi(s)\in (0,\infty)$. However, the critical equation
\begin{align}
\log(\xi(s)^s) = \frac{(2+2s-s^2)\zeta(2-s)}{(1-s)^2} \; \frac{1}{\xi(s)^s} - 1. \label{fgamma critical eq}
\end{align}
has at most one solution since the left-hand side is strictly increasing in $\xi(s)^s$, whereas the right-hand side is strictly decreasing. This proves part (a). To prove part (b), note that
\begin{align*}
\frac{1}{\sqrt{2\pi}}\int_{-\infty}^\infty \dx \; \e^{\i tx} \sqrt{\frac{\pi}{2a^2}} \; \Idbb_{[-a,a]}(x) = \frac{\sin(at)}{at}.
\end{align*}
Hence, the Fourier transform of $\frac{\sin(ax)}{ax}$ is an amplified indicator function on the interval $[-a,a]$. By the lemma below, the support of $\hat w_{s,\gamma}$ is thus contained in $[-2S,2S]$ with $S = \sum_{n=1}^\infty a_n(s) = \frac \gamma 2$.
\end{proof}

\begin{lem}
Let $0 < a \leq b$ and let $f_b\in L^\infty(\Rbb)$ with $\supp f_b \subseteq [-b,b]$. Then $\Idbb_{[-a,a]} * f_b$ has support contained in $[-(a+b),a+b]$.
\end{lem}

\begin{proof}
Compute
\begin{align*}
(\Idbb_{[-a,a]}* f_b )(x) &= \int_{-b}^b \dy\; \Idbb_{[-a,a]}(x-y) f_b(y).
\end{align*}
If $x < -(a+b)$ and $y\in [-b,b]$, then $x-y < -a-b + b = -a$. Hence, $\Idbb_{[-a,a]}(x-y) =0$ for all $y$ in the integration range. So, the convolution is $0$. Likewise for $x>a+b$.
\end{proof}

The following is our version of \cite[Lemma 2.6]{AutomorphicEquivalence}.

\begin{lem}
\label{AT:W_gamma_estimate}
Define
\begin{align*}
W_{s,\gamma}(t) &:= \begin{dcases} \int_t^\infty \dd r \; w_{s,\gamma}(r) &t\geq 0, \\ -\int_{-\infty}^t \dd r\; w_{s,\gamma}(r) & t < 0.\end{dcases}
\end{align*}
Then, the following statements hold:
\begin{enumerate}[(a)]
\item $W_{s,\gamma}$ is a bounded, odd function with
\begin{align*}
\Vert W_{s,\gamma}\Vert_\infty  = W_{s,\gamma}(0) = \frac 12.
\end{align*}

\item If $|\xi| \geq \gamma$, then
\begin{align*}
\hat W_{s, \gamma} (\xi) = - \frac{\i }{\sqrt{2\pi} \, \xi}.
\end{align*}

\item For every $0< \mu <\mu_0(s)$ and any $t\in \Rbb$, we have
\begin{align*}
|W_{s,\gamma}(t)| &\leq \frac{c_{s,\gamma}}{\gamma} \; \frac{D_s \; \Gamma(\nicefrac 1s)}{s} \; \bigl(\frac{2}{\mu_0(s) - \mu}\bigr)^{\nicefrac 1s} \; \chi_{s,\mu}(\gamma \, |t|),
\end{align*}
where $c_{s,\gamma}$ and $D_s$ are from Lemma \ref{AT:W_gamma_estimate}.

\item For $r\geq 0$ and $T\geq 0$, define
\begin{align*}
I_{s,\gamma, r}(T) &:= \int_T^\infty \dt \; t^r \; W_{s,\gamma}(t).
\end{align*}
Then, for each $0 < \mu < \mu_0(s)$ and $T\geq 0$, the estimate
\begin{align*}
I_{s,\gamma,r}(T) &\leq D_{I_{s,\gamma,r}}(\mu) \; \chi_{s,\mu}(\gamma T)
\end{align*}
holds, where
\begin{align}
D_{I_{s,\gamma,r}}(\mu) &:= \frac{c_{s,\gamma}}{\gamma^{2+r}} \; \frac{D_s \; \Gamma(\nicefrac 1s)^2}{s^2} \; \bigl( \frac{r}{s\e}\bigr)^{\frac rs} \; \Bigl(\frac{4}{\mu_0(s)-\mu}\Bigr)^{\frac{2+r}{s}}. \label{AT:W_estimate_constant_definition}
\end{align}
\end{enumerate}
\end{lem}

\begin{proof}
Part (a) is proven by $2W_{s,\gamma}(0) = \Vert w_{s,\gamma}\Vert_1 = 1$. Part (b) follows from an integration-by-parts argument. We have
\begin{align*}
\hat W_{s, \gamma}(\xi) &= - \frac{1}{\sqrt{2\pi}} \frac{1}{\i \xi} \Bigl[  \e^{-\i \xi t}\int_t^\infty \dd r \; w_{s, \gamma}(r)\Bigr]_0^\infty + \frac{1}{\sqrt{2\pi}} \frac{1}{\i \xi} \Bigl[ \e^{-\i \xi t} \int_{-\infty}^t \dd r \; w_{s, \gamma}(r) \Bigr]_{-\infty}^0 \\
&\hspace{-20pt} + \frac{1}{\sqrt{2\pi}} \frac{1}{\i \xi} \int_0^\infty \dd t \; \e^{-\i \xi t} \frac{\dd}{\dd t} \int_t^\infty \dd r \; w_{s, \gamma}(r) - \frac{1}{\sqrt{2\pi}} \frac{1}{\i \xi} \int_{-\infty}^0 \dd t \; \e^{-\i \xi t} \frac{\dd}{\dd t} \int_{-\infty}^t \dd r \; w_{s, \gamma}(r).
\end{align*}
Since $w_{s, \gamma}\in L^1(\Rbb)$ with $\Vert w_{s, \gamma} \Vert_1 =1$, we conclude that
\begin{align*}
\hat W_{s, \gamma}(\xi) = - \frac{\i}{\sqrt{2\pi} \, \xi} + \frac{\i}{\sqrt{2\pi} \, \xi} \, \hat w_{s,\gamma}(\xi).
\end{align*}
Since $\hat w_{s, \gamma} \equiv 0$ outside $[-\gamma,\gamma]$, see Lemma \ref{AT:w_gamm_estimate} (b), the claim follows.

We prove part (c). Inserting Lemma \ref{AT:w_gamm_estimate} (a), we obtain
\begin{align*}
|W_{s,\gamma}(t)| &\leq \int_{|t|}^\infty \dd \xi \; w_{s,\gamma}(\xi) \leq c_{s,\gamma} D_s \int_{|t|}^\infty \dd \xi \; \chi_{s, \mu_0(s) \; \gamma^s}(\xi).
\end{align*}
Applying Lemma \ref{AT:chi_estimate} (c) with $r =0$, $b' = \mu_0(s)\gamma^s$, and $b =\mu \gamma^s$ gives the claim. Likewise, we obtain part (d) by applying part (c) to $\mu' = \mu + \frac{\mu_0(s)-\mu}{2}$. We get
\begin{align*}
I_{s,\gamma, r}(T) &\leq \frac{c_{s,\gamma}}{\gamma} \; \frac{D_s \; \Gamma(\nicefrac 1s)}{s} \; \bigl(\frac{2}{\mu_0(s)-\mu'}\bigr)^{\nicefrac 1s} \; \int_T^\infty \dt \; t^r \; \chi_{s,\mu'\gamma^s}(t).
\end{align*}
Applying Lemma \ref{AT:chi_estimate} (c) with $b' = \gamma^s \mu'$ and $b = \gamma^s \mu$ and evaluating the rates, we get $\mu'-\mu = \frac{\mu_0(s)-\mu}{2} = \mu_0(s)-\mu'$ and the claim.
\end{proof}


\section{Continuity estimates for several operations}
\label{AT:Estimating_Operations_Section}


\subsection{Commutators}

Let $G_0,G_1\in \Lcal$ be two local Hamiltonians and let $\Phi_{G_i}\in \Bcal$, $i =0,1$ be given such that \eqref{AT:Local_Hamiltonian_decomposition} holds. Then, we define an interaction for $[G_1,G_0]$ on the set $W\subset \Lambda$ by
\begin{align}
\Phi_{[G_0,G_1]}(W) := \sum_{\substack{Z_0, Z_1 \subset\Lambda \\ Z_0 \cap Z_1 \neq \emptyset \\ Z_0\cup Z_1 = W}} [\Phi_{G_1}(Z_1), \Phi_{G_0}(Z_0)]. \label{AT:Interaction_Commutator}
\end{align}
This implies $\Phi_{[G_0, G_1]}\in \Bcal$, whence $[G_1,G_0] \in \Lcal$. By induction, this implies an interaction for multi-commutators as well. For local Hamiltonians $G_0, \ldots, G_k\in \Lcal$, we put
\begin{align}
\Phi_{\ad_{G_k} \cdots \ad_{G_1}(G_0)}(W) = \sum_{\substack{Z_0, \ldots, Z_k\subset \Lambda \\ Z_i \cap \bigcup_{\ell = 0}^{i-1} Z_\ell \neq \emptyset \; \forall i\geq 1 \\ \bigcup_{i=0}^k Z_i = W}} \ad_{\Phi_{G_k}(Z_k)} \cdots \ad_{\Phi_{G_1}(Z_1)}(\Phi_{G_0}(Z_0)). \label{AT:Interaction_Multicommutator}
\end{align}
Analogously, we then have $\Phi_{\ad_{G_k} \cdots \ad_{G_1}(G_0)}\in \Bcal$ and $\ad_{G_k} \cdots \ad_{G_1}(G_0)\in \Lcal$.

\begin{bem}
We can think of much more complicated configurations of commutators of $G_0, \ldots, G_k$ than the operator
\begin{align}
\ad_{G_k} \cdots \ad_{G_1}(G_0) \label{AT:Multicommutator_operator}
\end{align}
and we will, indeed, encounter and deal with them when we estimate the higher derivatives of $\Ical_{s, \gamma}$ in local norm in Subsection \ref{AT:I_Higher_derivatives_Section} below. However, the commutators there appear with additional operations that have to be dealt with simultaneously so that the commutator estimate we present here would not be applicable there. Therefore, an estimate on the configurations like in \eqref{AT:Multicommutator_operator} is sufficient for our purposes and we keep the digression readable by restricting to such configurations. It becomes clear from the proof that Theorem \ref{AT:Commutator_Estimate} also applies to every other configuration of $k$ commutators.
\end{bem}

\begin{thm}
\label{AT:Commutator_Estimate}
Let Assumption \ref{AT:Assumption_abstract_V} be true, let $k\in \Nbb$, $0 < s \leq 1$, and $b'> 0$. Assume that $G_0, \ldots, G_{k} \in \Lcal_{s, b'}$. Then, for any $0 < b < b'$, we have $\ad_{G_k}\cdots \ad_{G_1}(G_0)\in \Lcal_{s, b}$ and the interaction in \eqref{AT:Interaction_Multicommutator} satisfies the estimate
\begin{align*}
\Vert\Phi_{\ad_{G_k} \cdots \ad_{G_1}(G_0)}\Vert_{s, b} &\leq 4^k \; V_{s,k}(b'-b)^k \; \prod_{\ell=0}^{k} \Vert \Phi_{G_\ell}\Vert_{s, b'}, 
\end{align*}
where $V_{s,k}(b)$ is from \eqref{AT:Vkb_definition}.
\end{thm}

\begin{proof}
We omit the index $s$ throughout, since it does not play any role. Let $b_0, \ldots, b_k>0$ be such that $0 < b_k < b_{k-1} < \cdots < b_1 < b_0 \leq b'$. We claim that
\begin{align}
\Vert \Phi_{\ad_{G_k} \cdots \ad_{G_1}(G_0)}\Vert_{b_k} \leq 4^k \; \Vert \Phi_{G_0}\Vert_{b_0} \; \prod_{i=1}^k V_1(b_i - b_{i-1}) \; \Vert \Phi_{G_i}\Vert_{b_{i-1}}. \label{Commutator proof 1}
\end{align}
If this is true, we may choose $b = b_k$, as well as $b_0 = b'$, and $b_i - b_{i-1} = \frac{b'-b}{k}$. Consequently, we have $V_1(b_i-b_{i-1}) = V_{k}(b'-b)$. Using that $b_{i-1} \leq b'$ for all $i=1,\ldots, k$, we conclude the theorem.

It remains to prove \eqref{Commutator proof 1} per induction. We start with $k = 1$. Let $x\in \Lambda$ be given and estimate
\begin{align}
\sum_{\substack{Z\subset \Lambda \\Z\ni x}} \frac{\Vert \Phi_{\ad_{G_1}(G_0)}(Z)\Vert}{\chi_b(\Dcal(Z))} \leq 2\sum_{\substack{Z_0,Z_1\subset \Lambda \\ Z_0\cap Z_1\neq \emptyset \\ Z_0\cup Z_1 \ni x}} \frac{\Vert \Phi_{G_1}(Z_1)\Vert \; \Vert \Phi_{G_0}(Z_0)\Vert}{\chi_b(\Dcal(Z_1)) \;  \chi_b(\Dcal(Z_0))}. \label{Commutator proof 2}
\end{align}
Here, we used the logarithmic superadditivity of $\chi_b$ as well as $\Dcal(Z) \leq \Dcal(Z_0) + \Dcal(Z_1)$. Now, we get two terms, the terms with $x\in Z_0$ and the ones with $x\in Z_1$. For the case $x\in Z_0$, we get the upper bound
\begin{align*}
\sum_{\substack{Z_0,Z_1\subset \Lambda \\ Z_0\cap Z_1\neq \emptyset \\ Z_0 \ni x}} \frac{\Vert \Phi_{G_1}(Z_1)\Vert \; \Vert \Phi_{G_0}(Z_0)\Vert}{\chi_b(\Dcal(Z_1)) \;  \chi_b(\Dcal(Z_0))} &\leq \sum_{\substack{Z_0\subset \Lambda \\ Z_0\ni x}} \frac{\Vert \Phi_{G_0}(Z_0)\Vert}{\chi_{b'}(\Dcal(Z_0))} \; \chi_{b'-b}(\Dcal(Z_0))\sum_{z\in Z_0} \sum_{\substack{Z_1\subset \Lambda \\ Z_1 \ni z}} \frac{\Vert \Phi_{G_1}(Z_1)\Vert}{\chi_b(\Dcal(Z_1))} .
\end{align*}
Bounding away the norm $\Vert \Phi_{G_1}\Vert_b \leq \Vert \Phi_{G_1}\Vert_{b'}$, we get a factor of $|Z_0|$, which, together with $\chi_{b'-b}(\Dcal(Z_0))$ is bounded by $V_1(b'-b)$. We are left with the norm $\Vert \Phi_{G_0}\Vert_{b'}$. Hence, the total bound for this case is $2 V_1(b'-b) \Vert \Phi_{G_1}\Vert_{b'} \Vert \Phi_{G_0}\Vert_{b'}$. The case $x\in Z_1$ produces the same bound again. We arrive at \eqref{Commutator proof 1} for the case $k=1$ with $b_0 = b'$ and $b_1 = b$.

The induction argument is now straightforward. Assume that \eqref{Commutator proof 1} is true for $k -1$ and let $b_k < b_{k-1} < \cdots < b_1 < b_0\leq b'$ be given. Then, by case $k = 1$, we get
\begin{align*}
\Vert \Phi_{\ad_{G_k}\cdots \ad_{G_1}(G_0)}\Vert_{s,b_k}\leq 4 \; V_1(b_k-b_{k-1}) \; \Vert \Phi_{G_k}\Vert_{b_{k-1}} \; \Vert \Phi_{\ad_{G_{k-1}}\cdots \ad_{G_1}(G_0)}\Vert_{b_{k-1}}.
\end{align*}
Applying the induction hypothesis proves \eqref{Commutator proof 1}.
\end{proof}

\subsection{An estimate for \texorpdfstring{$\Ical_{s, \gamma}$}{I}}
\label{AT:I_Estimate_Section}

In this section, we prove the locality estimate for the map $\Ical_{s, \gamma}$ defined in \eqref{AT:I_definition}. We do this in two steps, the first of which is an estimate on local observables. In the second step, we extend this to local Hamiltonians.

\subsubsection{The map \texorpdfstring{$\Ical_{s, \gamma}$}{I} on local observables}

For each $Z\subset \Lambda$ and $n\in \Nbb_0$, define the $n$\tho\ fattening of $Z$ to be
\begin{align}
Z_n := \bigl\{ z\in \Lambda : \dist(z,Z) \leq n \bigr\}. \label{AT:Fattening_definition}
\end{align}
For a local observable $A\in \Acal_X$, where $X\subset \Lambda$, define
\begin{align}
\Delta_{s, \gamma}^0(A) &:= \int_\Rbb \dt \; W_{s,\gamma}( t) \; \Ebb_{X}(\tau_t(A)) \label{AT:I_Delta0_definition}
\end{align}
where $\Ebb_Z(B) = \frac{\tr_{Z^{\mathrm c}}(B)}{\dim \Hcal_Z}$ is the normalized partial trace. Furthermore, for $n\geq 1$, set
\begin{align}
\Delta_{s, \gamma}^n(A) &:= \int_\Rbb \dt\; W_{s,\gamma}(t) \bigl( \Ebb_{X_n}(\tau_t(A)) - \Ebb_{X_{n-1}}(\tau_t(A))\bigr) \label{AT:I_Deltan_definition}
\end{align}
Then, $\Delta_{s, \gamma}^n(A) =0$ for all $n$ sufficiently large, since $\Lambda$ is finite, and $\supp(\Delta_{s, \gamma}^n(A))\subseteq X_n\cap \Lambda$. Also,
\begin{align}
\Ical_{s, \gamma}(A) = \sum_{n=0}^\infty \Delta_{s, \gamma}^n(A), \label{AT:I_decomposition_local_observable}
\end{align}
where the sum is indeed finite.

\begin{lem}
\label{AT:I_Delta-estimate_Lemma}
Let Assumption \ref{AT:Assumption_abstract_V} be true, let $a' > a > 0$ and assume that $H\in \Lcal_{1, a'}$. For any $X\subset \Lambda$ and $A\in \Acal_X$, we have
\begin{align*}
\Vert \Delta_{s, \gamma}^0(A)\Vert &\leq \Vert W_{s,\gamma}\Vert_1 \; \Vert A\Vert.
\end{align*}
Let $0 < s < 1$, $\gamma>0$, and for any $0 < \mu < \mu_0(s)$ (with $\mu_0(s)$ from Lemma \ref{AT:w_gamm_estimate}) define
\begin{align}
\eta_{s, \gamma}(a,\mu) &:= \min \Bigl\{ \frac a{4^s} \, , \, \frac{\mu \; \gamma^s}{(4 \; \Vcal_{1,a,2}(a'-a))^s}\Bigr\}. \label{AT:eta_definition}
\end{align}
Then, for any integer $n\geq 1$, we have the estimate
\begin{align*}
\Vert \Delta_{s, \gamma}^n (A)\Vert \leq C_{s, \gamma}^\Delta(a,\mu) \; |X| \; \Vert A\Vert \; \chi_{s, \eta_{s, \gamma}(a,\mu)}(2n),
\end{align*}
where $C_\Delta$ is given by
\begin{align*}
C_{s, \gamma}^\Delta(a,\mu) &:= \frac{4 \, \e^a \; E_{s,a}}{a \; \Vcal_{1,a,2}(a'-a)} + 8 \, D_{I_{s,\gamma,0}}(\mu)
\end{align*}
with $D_{I_{s,\gamma,0}}$ from Lemma \ref{AT:W_gamma_estimate}, $E_{s,a}$ from Lemma \ref{AT:chi_estimate}, and $\Vcal_{1,a,2}(a'-a)$ from \eqref{AT:LR-velocity_definition}.
\end{lem}

For the proof, we need the following auxiliary result, whose proof can be found in \cite[Lemma 3.1]{AutomorphicEquivalence}.

\begin{lem}
\label{AT:Partial_Trace_Lemma}
Let $\Hcal_1$ and $\Hcal_2$ be finite-dimensional Hilbert spaces and suppose $\varepsilon \geq 0$ and a bounded operator $A$ on $\Hcal_1\otimes \Hcal_2$ are such that
\begin{align*}
\Vert [ A, \Idbb \otimes B]\Vert \leq \varepsilon \; \Vert B\Vert
\end{align*}
for all bounded operators $B$ on $\Hcal_2$. Then
\begin{align*}
\Vert \Ebb(A) \otimes \Idbb - A \Vert \leq \varepsilon,
\end{align*}
where $\Ebb(A) := \frac{1}{\dim \Hcal_2} \Tr_{\Hcal_2}(A)$ is the partial trace.
\end{lem}

\begin{proof}[Proof of Lemma \ref{AT:I_Delta-estimate_Lemma}]
The estimate for $\Delta_{s, \gamma}^0(A)$ is trivial. Let $n\geq 1$ and decompose
\begin{align}
\Delta_{s, \gamma}^n(A) = \tilde \Delta_{s, \gamma}^n(A) - \tilde \Delta_{s, \gamma}^{n-1}(A) \label{Deltan-decomp}
\end{align}
with
\begin{align*}
\tilde \Delta_{s, \gamma}^n(A) &:= \int_\Rbb \dt\; W_{s,\gamma}(t) \; \bigl( \Ebb_{X_n}(\tau_t(A)) - \tau_t(A)\bigr).
\end{align*}
Now, for $T>0$ to be chosen, we have
\begin{align*}
\Vert \tilde \Delta_{s, \gamma}^n(A)\Vert &\leq \Vert W_{s,\gamma}\Vert_\infty \int_{-T}^T\dt\; \bigl\Vert \Ebb_{X_n}(\tau_t(A)) - \tau_t(A)\bigr\Vert + 4\Vert A\Vert \; I_{s,\gamma,0}(T). 
\end{align*}
For the first term -- let us call it $\Tcal$ --, we use the Lieb-Robinson bound Corollary \ref{AT:LR-bound_kor} with $v := \Vcal_{1,a,1}(a'-a)$. Note that $\dist(X,X_n) \geq n$. Hence, by Lemma \ref{AT:Partial_Trace_Lemma} (and using $\Vert W_{s,\gamma}\Vert_\infty = \frac 12$, Lemma \ref{AT:W_gamma_estimate} (a)), the first term  $\Tcal$ is bounded by
\begin{align*}
\Tcal &\leq 2 \; |X| \; \Vert A\Vert \;  \e^{-an} \int_0^T \dt\; \e^{av t} \leq \frac{2}{av} \; |X| \; \Vert A\Vert  \; \e^{-a(n - vT)}.
\end{align*}
Now choose $vT = \frac{n+1}{2}$ to get
\begin{align*}
\Tcal &\leq \frac{2\e^{\frac a2}}{av} \; |X| \; \Vert A\Vert \;  \e^{-\frac a2 (n+1)}.
\end{align*}
Make use of Lemma \ref{AT:chi_estimate} (e) to obtain $\e^{-a \frac{2(n+1)}{4}} \leq E_{s,a} \, \chi_{\eta_{s, \gamma}}(2(n+1))$ with $\eta_{s, \gamma}$ in \eqref{AT:eta_definition}. Finally, again by \eqref{AT:eta_definition}, we have that 
\begin{align*}
I_{s,\gamma,0}(T) &= I_{s,\gamma,0}\bigl( \frac{n+1}{2v}\bigr) \leq D_{I_{s,\gamma,0}} \; \chi_{s,\mu} \bigl( \frac{\gamma}{4v} \; 2(n+1)\bigr) \leq D_{I_{s,\gamma, 0}}  \; \chi_{\eta_{s, \gamma}}(2(n+1)).
\end{align*}
Putting everything together, we conclude that
\begin{align*}
\Vert \tilde \Delta_{s, \gamma}^n(A)\Vert &\leq \frac 12 \; C_{s, \gamma}^\Delta \; |X| \; \Vert A\Vert \; \chi_{\eta_{s, \gamma}}(2(n+1)).
\end{align*}
From this and a triangle inequality, the bound on $\Delta_{s, \gamma}^n(A)$ follows, see \eqref{Deltan-decomp}.
\end{proof}

\begin{kor}
\label{AT:I_Delta-estimate_united}
Let Assumption \ref{AT:Assumption_abstract_V} be true, let $a' > a>0$ and assume that $H\in \Lcal_{1,a'}$. Let $0 < s < 1$, $0 < \mu < \mu_0(s)$, and let $\gamma>0$. For any $X\subset \Lambda$, any $A\in \Acal_X$, and any integer $n\geq 0$, we have
\begin{align*}
\Vert \Delta_{s, \gamma}^n(A)\Vert \leq D_{s, \gamma}^\Delta(a,\mu) \; |X| \; \Vert A\Vert \; \chi_{s, \eta_{s, \gamma}(a,\mu)}(n).
\end{align*}
Here,
\begin{align}
D_{s, \gamma}^\Delta(a, \mu) := \max \bigl \{C_{s, \gamma}^{\Delta}(a,\mu) \, , \, \Vert W_{s,\gamma}\Vert_1 \bigr\}, \label{AT:I_DDelta_definition}
\end{align}
where $C_{s, \gamma}^{\Delta}$ and $\eta_{s, \gamma}$ are taken from Lemma \ref{AT:I_Delta-estimate_Lemma}.
\end{kor}

\begin{proof}
This is unifying the estimates in Lemma \ref{AT:I_Delta-estimate_Lemma}.
\end{proof}

\subsubsection{The map \texorpdfstring{$\Ical_{s, \gamma}$}{I} on local Hamiltonians}

Let $G$ be a local Hamiltonian so that $G = \sum_{Y\subset \Lambda} \Phi_G(Y)$. To prove that $\Ical_{s, \gamma}(G)$ is still local, define
\begin{align}
\Phi_{\Ical_{s, \gamma}(G)}(Z) &:= \sum_{n=0}^\infty \sum_{\substack{Y\subset \Lambda \\ Y_n = Z}} \Delta_{s, \gamma}^n(\Phi_G(Y)). \label{AT:I_interaction_definition}
\end{align}
Then, we claim that
\begin{align}
\sum_{Z\subset \Lambda} \Phi_{\Ical_{s, \gamma}(G)}(Z) = \Ical_{s, \gamma}(G) \label{AT:I_decomposition}
\end{align}
holds. To see this, let us use \eqref{AT:I_decomposition_local_observable} so that
\begin{align*}
\Ical_{s,\gamma}(G) = \sum_{Y\subset \Lambda} \sum_{n=0}^\infty \Delta_{s, \gamma}^n(\Phi_G(Y)).
\end{align*}
We insert $1 = \sum_{Z\subset \Lambda} \Idbb_{Y_n}(Z)$ and interchange the order of sums. This is allowed since only finitely mans terms in the sum over $n$ are nonzero, see \eqref{AT:I_decomposition_local_observable}. This implies
\begin{align*}
\Ical_{s, \gamma}(G) &= \sum_{Z\subset \Lambda} \sum_{n=0}^\infty \sum_{Y\subset \Lambda} \Idbb_Z(Y_n)  \; \Delta_{s, \gamma}^n(\Phi_G(Y)).
\end{align*}
This proves \eqref{AT:I_decomposition}.

\begin{thm}
\label{AT:I_estimate_local_Hamiltonian}
Let Assumptions \ref{AT:Assumption_abstract_V} and \ref{AT:Assumption_abstract_F} be true, let $a'>0$ and assume that $H\in \Lcal_{1,a'}$. Let $0 < s < 1$ and $b'>0$ such that $G\in \Lcal_{s,b'}$. Let $0 < \mu < \mu_0(s)$, $\gamma>0$, and $0 < a < a'$. Then, for any $0 \leq b < \min\{ b', \eta_{s, \gamma}(a,\mu)\}$ with $\eta_{s, \gamma}$ in \eqref{AT:eta_definition}, we have $\Ical_{s, \gamma}(G) \in \Lcal_{s, b}$ and the interaction in \eqref{AT:I_interaction_definition} satisfies the estimate
\begin{align*}
\Vert \Phi_{\Ical_{s, \gamma}(G)} \Vert_{s,b} &\leq D_{s, \gamma}^\Delta(a,\mu) \; V_{s,1}(b'-b) \; F_{s}(\eta_{s, \gamma}(a,\mu) - b) \; \Vert \Phi_G\Vert_{s, b'},
\end{align*}
where $D_{s, \gamma}^\Delta(a, \mu)$ is from \eqref{AT:I_DDelta_definition}.
\end{thm}

\begin{proof}
Let $x\in \Lambda$ be given. Again, we suppress the dependence on $s$. The object to be estimated is bounded by the logarithmic superadditivity and monotonicity of $\chi_b$, as well as the inequality\footnote{Let $x,y\in Y$ with $d(x,y) = \Dcal(Y)$. For any $x'\in B_n(x)$ and $y'\in B_n(y)$, it follows that $x',y'\in Y_n$. Furthermore, $d(x',y') \leq d(x',x) + d(x,y) + d(y,y') \leq \Dcal(Y) + 2n$. Maximizing over $x',y'\in Y_n$ shows $\Dcal(Y_n) \leq \Dcal(Y) + 2n$.} $\Dcal(Y_n) \leq \Dcal(Y) + 2n$. We get
\begin{align*}
\sum_{\substack{Z\subset \Lambda \\ Z \ni x}}\frac{ \Vert \Phi_{\Ical_{s, \gamma}(G)}(Z)\Vert}{\chi_b(\Dcal(Z))} &\leq \sum_{Z\ni x} \sum_{n=0}^\infty \sum_{\substack{Y\subset \Lambda \\ Y_n = Z}} \frac{\Vert \Delta^n(\Phi_G(Y))\Vert}{\chi_b(\Dcal(Y)+2n)} \leq \sum_{n=0}^\infty \frac{1}{\chi_b(2n)} \sum_{\substack{Y\subset \Lambda \\ Y_n \ni x}} \frac{\Vert \Delta^n(\Phi_G(Y))\Vert}{\chi_b(\Dcal(Y))}.
\end{align*}
For the resummation, we fix $Y$ and $n\in\Nbb_0$. Then there is a point $\tilde x\in B_n(x)\cap Y$. Hence, $Y$ is hit if we sum over all $\tilde x\in B_n(x)$ and $Y'\subset \Lambda$ containing $\tilde x$. Using Corollary \ref{AT:I_Delta-estimate_united}, we obtain the upper bound
\begin{align*}
\sum_{\substack{Z\subset \Lambda \\ Z \ni x}} \frac{\Vert \Phi_{\Ical_{s, \gamma}(G)}(Z)\Vert}{\chi_b(\Dcal(Z))} &\leq D^\Delta \sum_{n=0}^\infty \chi_{\eta-b}(2n)\sum_{\tilde x\in B_n(x)} \sum_{\substack{Y \subset \Lambda \\ Y\ni\tilde x}} |Y| \; \frac{\Vert \Phi_G(Y)\Vert}{\chi_{b'}(\Dcal(Y))} \; \chi_{b'-b}(\Dcal(Y)).
\end{align*}
At this point, we estimate $|Y| \, \chi_{b'-b}(\Dcal(Y)) \leq V_1(b'-b)$. Subsequently, we can take away the norm $\Vert \Phi_G\Vert_{b'}$ to conclude that
\begin{align*}
\sum_{\substack{Z\subset \Lambda \\ Z \ni x}} \frac{\Vert \Phi_{\Ical_{s, \gamma}(G)}(Z)\Vert}{\chi_b(\Dcal(Z))}  &\leq  D^\Delta \; V_1(b'-b) \; \Vert \Phi_G\Vert_{b'} \; \sum_{n=0}^\infty |B_n(x)| \; \chi_{\eta-b}(2n).
\end{align*}
The last term is bounded by $F(\eta-b)$ in \eqref{AT:Fb_definition}, which finishes the proof.
\end{proof}

\subsection{Outlook on the first derivative of \texorpdfstring{$\Ical_{s, \gamma}$}{I}}

In this section, we take a look at the derivative of the map $\Ical_{s, \gamma}(G)$ with respect to $u$. For this purpose, we need to introduce the following map. For two local Hamiltonians $G_1, G_2$, define
\begin{align}
\Jcal_{s, \gamma}(G_1,G_2) := \i \int_\Rbb \dt\; W_{s,\gamma}(t) \int_0^t \dr \; \bigl[\tau_r(G_1),\tau_t(G_2)\bigr]. \label{AT:J_definition}
\end{align}
To compute the derivative, we make use of Duhamel's formula (see \cite{Duhamel})
\begin{align}
\frac{\dd}{\du} \e^{\i tH(u)} = \i t\int_0^1 \dd \lambda \; \e^{\i \lambda tH(u)} \, \dot H(u)  \, \e^{\i(1-\lambda)tH(u)}.\label{AT:Duhamels_formula}
\end{align}
Note that this expression carries a $\lambda \lera 1-\lambda$ symmetry. With this, we compute
\begin{align}
\smash{\frac{\dd}{\du} \Ical_{s, \gamma}(G)}  &= \Ical_{s, \gamma}(\dot G) +\smash{\i\int_\Rbb \dd t \; t \; W_{s,\gamma}(t) \int_0^1 \dd \lambda} \; \bigl[ \e^{\i \lambda t H(u)} \, \dot H(u) \, \e^{\i (1-\lambda)t H(u)} \, G(u) \, \e^{-\i tH(u)}\notag\\
&\hspace{160pt} - \e^{\i t H(u)} \, G(u) \, \e^{-\i (1-\lambda)tH^u} \, \dot H(u) \, \e^{-\i \lambda tH(u)}\bigr] \notag\\
&
= \Ical_{s, \gamma}(\dot G) + \Jcal_{s, \gamma}(\dot H, G). \label{Ical first derivative}
\end{align}

Now, a local norm bound for $\Jcal_{s, \gamma}$ would imply that $\frac \dd \du \Ical_{s, \gamma}(G)$ is a local Hamiltonian if $G$ is and we would have a local norm bound in terms of $G$ and its derivative. 

To prove this, we could actually decompose $\Jcal_{s, \gamma}$ in a rather straightforward fashion and prove a local norm bound. However, to understand higher derivatives, more is necessary. Namely, we need to understand the $n$\tho\ derivative of $\tau_t(G)$, which is complicated in general. With each derivative, there is a chain of integrals of derivatives of $H$ coming into the game, which need to be estimated. Therefore, we need a general procedure to write down the terms that arise in the $n$\tho\ derivative of $\Ical_{s, \gamma}(G)$ and we set this up the next section. This allows us to prove that all derivatives of $\Ical_{s, \gamma}(G)$ are local Hamiltonians and we prove a local norm estimate.

\subsection{The derivatives of \texorpdfstring{$\Ical_{s, \gamma}$}{I}}
\label{AT:I_Higher_derivatives_Section}

When it comes to higher derivatives of $\Ical_{s, \gamma}$, we need a system of notation to phrase what this derivative looks like. It should be pointed out that it is screamingly clear how to --- order by order --- compute this derivative from the previous section. But for our purpose it is necessary to develop a framework that is sufficiently close to a closed formula. Eventually, we investigate the form of terms arising and use a rather rough upper bound to the number of these terms to arrive at a norm estimate.

\subsubsection{Occurring terms in higher derivatives}

In this section, we develop a notation that describes the form of the terms that arise in the higher derivatives. This means that the order $\beta\in \Nbb_0$ of derivative is fixed and that we investigate one of the terms occurring in the formula for the derivative.

As a motivation, let us briefly describe how the terms that we look at below come about. If we look closer to the computation we did in \eqref{Ical first derivative}, it is not hard to believe the general building strategy for the terms. Namely, either we derive the local Hamiltonian inside or we insert a commutator with an integral over $\dot H$ in front of the local Hamiltonian in question. The integral defining $\Ical_{s, \gamma}$ plays a minor role here, it is rather the $\tau_t(G)$ that is responsible for the complex structure of the terms. Hence, we have to be able to deal with different types of integral chains, mixed with commutators.

For an interval $J\subset \Rbb$ we denote the space of continuous sections $J\ra \Lcal$ by $\Scal_J(\Lcal)$. We will decompose the construction of a term in the derivative into small building bricks and every operator that we define in the following can be realized as a map $\Lcal \ra \Lcal$ or $\Scal_J(\Lcal) \ra \Scal_J(\Lcal)$. However, the way our proof is written, we cannot use the locality bounds of the intermediate steps elegantly because we need to decompose the integral operator with the weight function $W_{s, \gamma}$ and the time evolution $\tau$ simultaneously. This may be improved in the future. Therefore, we give a locality bound only for the final operator, which encodes one term in the derivative. The decomposition into local observables and the locality bound for the intermediate steps are left as an exercise to the reader.


\begin{defn}[Primitive operator]
Define the linear operator $\Jcal \colon \Scal_J(\Lcal) \lra \Scal_J(\Lcal)$ by
\begin{align*}
\Jcal(\Gcal) (t) &:= \int_0^t \dd t' \; \Gcal(t'). 
\end{align*}
\end{defn}


\begin{defn}[Time evolution]
\begin{enumerate}[(a)]
\item For $G, H \in \Lcal$, we define the time evolution operator $\tau \colon \Lcal \ra \Scal_J(\Lcal)$ by
\begin{align*}
\tau(G)(t) := \tau_t(G) := \e^{\i t H} \, G \, \e^{-\i t H}.
\end{align*}

\item For $m\in \Nbb$, we define the pullback of the time evolution as
\begin{align*}
\tau^* \colon \bigl( \Scal_J(\Lcal)^m \ra \Scal_J(\Lcal) \bigr) \lra \bigl( \Lcal^m \ra \Scal_J(\Lcal) \bigr)
\end{align*}
by
\begin{align*}
\tau^* (\Rcal) (G_1, \ldots, G_m) &:= \Rcal\bigl( \tau(G_1), \ldots, \tau(G_m)\bigr).
\end{align*}
\end{enumerate}
\end{defn}

\begin{eg}
$\tau^* (\Jcal) \colon \Lcal \ra \Scal_J(\Lcal)$ with
\begin{align*}
\tau^*(\Jcal) (G) = \int_0^\bullet \dd t' \; \tau_{t'} (G).
\end{align*}
Note that this operator appears in the definition of $\Jcal_{s, \gamma}$ in \eqref{AT:J_definition}.
\end{eg}

\begin{defn}[The valley operator]
\begin{enumerate}[(a)]
\item For $\ell \in \Nbb_0$ define the ``valley'' operator
\begin{align*}
\Vcal_\ell \colon \Scal_J(\Lcal)^{\ell + 1} \lra \Scal_J(\Lcal) 
\end{align*}
by
\begin{align*}
\Vcal_0 (\Gcal_0) := \Gcal_0
\end{align*}
and
\begin{align*}
\Vcal_\ell (\Gcal_0 , \ldots, \Gcal_\ell) := \Vcal_{\ell - 1} \bigl( [\Jcal(\Gcal_0), \Gcal_1] , \Gcal_2, \ldots, \Gcal_\ell \bigr).
\end{align*}

\item For $m\in \Nbb$, $1\leq p\leq m$ and $\ell\in \Nbb_0$, we define the pullback valley operator
\begin{align*}
\Vcal_{p, \ell}^* \colon \bigl( \Scal_J(\Lcal)^m \ra \Scal_J(\Lcal) \bigr) \lra \bigl( \Scal_J(\Lcal)^{m + \ell} \ra \Scal_J(\Lcal) \bigr)
\end{align*}
by
\begin{align*}
\Vcal_{p, \ell}^*(\Rcal) (\Gcal_1, \ldots, \Gcal_{p-1} , \Gcal_1' , \ldots, \Gcal_\ell' , \Gcal_p, \ldots, \Gcal_m) & \\
&\hspace{-100pt} := \Rcal\bigl(\Gcal_1, \ldots, \Gcal_{p-1}, \Vcal_\ell(\Gcal_1', \ldots, \Gcal_\ell', \Gcal_p) , \Gcal_{p+1}, \ldots, \Gcal_m\bigr).
\end{align*}
\end{enumerate}
\end{defn}

\begin{egs}
\begin{enumerate}[(a)]
\item $\Vcal_4 \colon \Scal_J(\Lcal)^5 \ra \Scal_J(\Lcal)$ with
\begin{align*}
\Vcal_4(\Gcal_0, \Gcal_1, \Gcal_2, \Gcal_3, \Gcal_4) = \ad_{\int_0^\bullet \dd t_1 \; \ad_{ \int_0^{t_1} \dd t_2 \; \ad_{ \int_0^{t_2} \dd t_3 \; \ad_{ \int_0^{t_3} \dd t_4 \; \Gcal_0(t_4)} \Gcal_1(t_3)} \Gcal_2(t_2)} \Gcal_3(t_1)} \Gcal_4(\bullet)
\end{align*}
The interpretation of the valley operator is that we have a stacking chain of commutators and integrals of length $4$.

\item $\Vcal_{2, 2}^*(\Vcal_2) \colon \Scal_J(\Lcal)^5 \ra \Scal_J(\Lcal)$ with
\begin{align*}
(\Vcal_{2, 2}^* \circ \Vcal_2) (\Gcal_0, \ldots, \Gcal_4) &= \ad_{\int_0^\bullet \dd t_1 \; \ad_{\int_0^{t_1} \dd t_2 \; \Gcal_0(t_2) } \ad_{\int_0^{t_1} \dd t_2 \; \ad_{\int_0^{t_2} \dd t_3 \; \Gcal_1(t_3) } \Gcal_2(t_2) } \Gcal_3(t_1) } \Gcal_4(\bullet)
\end{align*}
The pullback valley operator inserts a valley of depth 2 at position 2 of the valley $\Vcal_2$.

\item $\tau^* (\Vcal_3) \colon \Lcal^4 \ra \Scal_J(\Lcal)$ with
\begin{align*}
\tau^*(\Vcal_3)(G_0, G_1, G_2, G_3) = \ad_{\int_0^\bullet \dd t_1 \; \ad_{ \int_0^{t_1} \dd t_2 \; \ad_{ \int_0^{t_2} \dd t_3 \; \tau_{t_3}(G_0)} \tau_{t_2}(G_1)} \tau_{t_1}(G_2)} \tau_\bullet(G_3).
\end{align*}

\item $\tau^* (\Vcal_{2, 2}^*\circ \Vcal_1) \colon \Lcal^4 \ra \Scal_J(\Lcal)$ with
\begin{align*}
\tau^*(\Vcal_{2, 2}^*\circ \Vcal_1)(G_0, G_1, G_2, G_3) = \ad_{\int_0^\bullet \dd t_1 \; \ad_{ \int_0^{t_1} \dd t_2 \; \tau_{t_2}(G_0) } \ad_{ \int_0^{t_1} \dd t_2 \; \tau_{t_2}(G_1) } \tau_{t_1}(G_2)} \tau_\bullet(G_3).
\end{align*}
\end{enumerate}
\end{egs}

\begin{defn}[The mountain range operator]
Let $\Fcal(\Nbb_0)$ denote the set of sequences $\ell = (\ell_m)_{m\in \Nbb_0}$ with entries $\ell_m\in \Nbb_0$, where only finitely many entries are nonzero, and let $|\ell| := \sum_{m=1}^\infty \ell_m$ denote the $\ell^1$-norm of $\ell\in \Fcal(\Nbb_0)$. Let also
\begin{align}
m_\ell := \begin{cases} -1 & \ell = 0, \\ \max \{ m \in \Nbb : \ell_m \neq 0\}, & \ell \neq 0, \end{cases} \label{AT:mell_definition}
\end{align}
i.e., $\ell = ( \ell_0, \ldots, \ell_{m_\ell}, 0, \ldots) \equiv (\ell_0, \ldots, \ell_{m_\ell})$.

For $\ell \in \Fcal(\Nbb_0)$, we define the mountain range operator
\begin{align*}
\Wcal^\ell \colon \Scal_J(\Lcal)^{|\ell| + 1} \lra \Scal_J(\Lcal)
\end{align*}
by
\begin{align*}
\Wcal^0(\Gcal) := \Gcal
\end{align*}
and
\begin{align*}
\Wcal^\ell := \Vcal_{m_\ell , \ell_{m_\ell}}^* \bigl( \Wcal^{(\ell_1, \ldots, \ell_{m_\ell-1})}\bigr).
\end{align*}
\end{defn}

\begin{egs}
\begin{enumerate}[(a)]
\item We claim that $\Wcal^{(1,2,1)} \colon \Scal_J(\Lcal)^5 \ra \Scal_J(\Lcal)$ is given by
\begin{align*}
\Wcal^{(1, 2, 1)}(\Gcal_0, \ldots , \Gcal_4) &= \ad_{\int_0^\bullet \dd t_1\; \Gcal_0(t_1) } \ad_{\int_0^\bullet \dd t_1 \; \ad_{ \int_0^{t_2} \; \Gcal_1(t_2)} \ad_{\int_0^{t_1} \dd t_2 \; \Gcal_2(t_2)} \Gcal_3(t_1) } \Gcal_4(\bullet).
\end{align*}
We read this as follows from the left to the right. At position $0$, we have a starting chain of commutators and integrals of length $\ell_0 = 1$, at position $1$ we have a chain of length $\ell_1 = 2$, and at position $2$ we have a chain of length $\ell_2 = 1$. The remaining operators $\Gcal_3$ and $\Gcal_4$ have to be inserted into the remaining slots. Alternatively, we can think of this as $\ell = (1, 2, 1, 0, 0)$ because we may always attach zeros.

The above formula is true because
\begin{align*}
\Wcal^{(1,2,1)} (\Gcal_0, \ldots, \Gcal_4) = \Vcal_1 \bigl( \Gcal_0, \Vcal_2 \bigl( \Gcal_1, \Vcal_1(\Gcal_2, \Gcal_3), \Gcal_4 \bigr)\bigr)
\end{align*}
and
\begin{align*}
\Vcal_1 \bigl( \Gcal_0, \Vcal_2 \bigl( \Gcal_1, \Vcal_1(\Gcal_2, \Gcal_3), \Gcal_4\bigr)\bigr) &= \Bigl[ \Jcal(\Gcal_0) \, , \, \Jcal\Bigl( \bigl[ \Jcal(\Gcal_1) \, , \,  [\Jcal(\Gcal_2), \Gcal_3] \bigr] \Bigr) \, , \, \Gcal_4 \Bigr].
\end{align*}

\item $\ell := (2,2,0,1)$. We have
\begin{align*}
\Wcal^{(2,2,0,1)}(\Gcal_0, \ldots, \Gcal_5) &= \Vcal_2\bigl( \Gcal_0, \Vcal_2( \Gcal_1, \Gcal_2, \Vcal_1(\Gcal_3, \Gcal_4)), \Gcal_5 \bigr)
\end{align*}
which equals
\begin{align*}
\Bigl[ \Jcal \Bigl( \bigl[\Jcal(\Gcal_0) \, , \, \Jcal \bigl( [ \Jcal(\Gcal_1) \, , \, \Gcal_2 ] \bigr) , [ \Jcal(\Gcal_3), \Gcal_4] \bigr] \Bigr) \, , \, \Gcal_5 \Bigr].
\end{align*}
Therefore,
\begin{align*}
\Wcal^{(2,2,0,1)}(\Gcal_0, \ldots, \Gcal_5) & \\
&\hspace{-50pt} = \ad_{ \int_0^\bullet \dd t_1 \; \ad_{\int_0^{t_1} \dd t_2 \; \Gcal_0(t_2) } \ad_{ \int_0^{t_1} \dd t_2 \; \ad_{ \int_0^{t_2} \dd t_3 \; \Gcal_1(t_3) } \Gcal_2(t_2) } \ad_{ \int_0^{t_1} \dd t_2 \; \Gcal_3(t_1) } \Gcal_4(t_1) } \Gcal_5(\bullet).
\end{align*}

\end{enumerate}
\end{egs}

\begin{defn}[Generalized weighted integral operator]
For $0 < s < 1$, $\gamma>0$, and $\ell \in \Fcal(\Nbb_0)$, we define $\Ical_{s, \gamma}^\ell \colon \Lcal^{|\ell| + 1} \ra \Lcal$ by
\begin{align*}
\Ical_{s, \gamma}^\ell(G_0, \ldots, G_{|\ell|}) := \i^{|\ell|} \int_\Rbb \dd t \; W_{s, \gamma} (t) \; \bigl( \tau^* \circ \Wcal^\ell\bigr)(G_0 \, \ldots, G_{|\ell|}) (t). 
\end{align*}
\end{defn}

\begin{egs}[Consistency check]
\label{Jcalell example}
\begin{enumerate}[(a)]
\item For $\ell =0$, we have
\begin{align*}
\Ical_{s, \gamma}^0 (G) &= \int_\Rbb \dd t \; W_{s, \gamma}(t) \; \tau_t(G) = \Ical_{s, \gamma}(G).
\end{align*}


\item For $\ell = 1 = (1,0)$, we have
\begin{align*}
\Wcal^1(\Gcal_0, \Gcal_1) = \Vcal_1(\Gcal_0, \Gcal_1) = \ad_{ \int_0^\bullet \dd t \; \Gcal_0(t) } \Gcal_1(\bullet).
\end{align*}
Therefore, comparing with the definition \eqref{AT:J_definition} of $\Jcal_{s, \gamma}$, we see that $\Ical_{s,\gamma}^1 = \Jcal_{s, \gamma}$. This term appears in the first derivative of $\Ical_{s, \gamma}$.
\end{enumerate}
\end{egs}

\subsubsection{Estimating \texorpdfstring{$\Ical_{s, \gamma}^\ell$}{Il} on local observables}

We fix $\ell \in \Fcal(\Nbb_0)$ and let $A_i \in \Acal_{X_i}$ for $i=0, \ldots, |\ell|$. For $n\in \Nbb_0^{|\ell| + 1}$ define
\begin{align*}
\Omega^{\ell, n}(A_0, \ldots, A_{|\ell|}) := \i^{|\ell|} \int_\Rbb \dt \; W_{s,\gamma}(t) \; \Wcal^\ell \bigl( \Theta_{n_0}(A_0)(t) , \ldots, \Theta_{n_{|\ell|}}(A_{|\ell|})(t) \bigr), 
\end{align*}
where $\Theta_0(A) := \Ebb_X(\tau(A))$ and
\begin{align*}
\Theta_n(A) &:= \Ebb_{X_n}(\tau(A)) - \Ebb_{X_{n-1}}(\tau(A)), &n & \geq 1,
\end{align*}
and where $X_n$ is the $n$\tho\ fattening of $X$, as defined in \eqref{AT:Fattening_definition}. This implies that 
\begin{align*}
\tau(A) = \sum_{n=0}^\infty \Theta_n(A)
\end{align*}
and this sum is finite, in fact, since each $\Theta_n(A)$ has support in $X_n \cap \Lambda$. It follows that $\Omega^{\ell, n} (A_0,\ldots, A_{|\ell|})$ has support in $\bigcup_{i=0}^{|\ell|} X_{i,n_i}\cap \Lambda$ and
\begin{align*}
\Ical_{s, \gamma}^\ell(A_0, \ldots, A_{|\ell|}) = \sum_{n\in \Nbb_0^{|\ell| + 1}} \Omega^{\ell, n}(A_0, \ldots, A_{|\ell|}),
\end{align*}
where in fact only finitely many terms in the sum are nonzero.

\begin{lem}
\label{AT:Omegaelln_estimate}
Let Assumption \ref{AT:Assumption_abstract_V} be true, let $a'>0$ and assume that $H\in \Lcal_{1,a'}$. Let $0 < s < 1$, $0 < \mu < \mu_0(s)$, and $\gamma>0$. For $0 < a < a'$, we put
\begin{align}
\eta_{s, \gamma}^k(a,\mu) &:= \min \bigl\{ \frac{a}{4^s}, \; \frac{\mu}{k} \bigl(\frac{\gamma}{4 \; \Vcal_{1,a,2}(a'-a)}\bigr)^s\bigr\}, & k&\in \Nbb, \label{AT:etak_definition}
\end{align}
where $\Vcal_{1,a,2}(a'-a)$ is from \eqref{AT:LR-velocity_definition}. Let $\ell \in \Fcal(\Nbb_0)$ and let $A_i\in \Acal_{X_i}$ for $i=0, \ldots, |\ell|$. Then, for every $n\in \Nbb_0^{|\ell| + 1}$, the estimate
\begin{align*}
\Vert \Omega^{\ell, n}(A_0,\ldots, A_{|\ell|})\Vert &\leq \frac{1}{\ell!} \; D_{s, \gamma}^{\Omega, |\ell| + 1}(a,\mu) \; \prod_{i=0}^{|\ell|} |X_i| \; \Vert A_i\Vert \; \chi_{s,\eta_{s, \gamma}^{|\ell|+1}(a,\mu)}(2n_i)
\end{align*}
holds, where $\ell! := \ell_0! \cdots \ell_{m_\ell}!$ with $m_\ell$ from \eqref{AT:mell_definition}, and 
\begin{align}
D_{s, \gamma}^{\Omega, k}(a,\mu) &:= \frac{16^k \; \e^{a'k}}{a \; k \; \Vcal_{1,a,2}(a'-a)^k} \; \frac{2 \; E_{s,a}^k}{\e^k (a'-a)^k} + 4^{k+1} \; D_{I_{s,\gamma,k-1}}, & k &\in\Nbb. \label{AT:Omegaelln_estimate_DOmega_definition}
\end{align}
Here, $E_{s,a}$ is from Lemma \ref{AT:chi_estimate}, and $D_{I_{s,\gamma,k-1}}$ is from Lemma \ref{AT:W_gamma_estimate}.
\end{lem}

\begin{proof}
Let us start with an estimate on the norms of $\Theta_n$ for $A\in \Acal_X$ and $t\in\Rbb$. We decompose according to $\Theta_n(A) = \tilde \Theta_n(A) - \tilde \Theta_{n-1}(A)$ for $n\geq 1$ where
\begin{align*}
\tilde\Theta_n(A) := \Ebb_{X_n}(\tau(A)) - \tau(A).
\end{align*}
For $n=0$, we have $\Vert \Theta_0(A)\Vert \leq \Vert A\Vert$. For $n\geq 1$, we use the fact that $d(X, X_n) \geq n$ and set $a'' := a + \frac{a'-a}{2}$. By the Lieb--Robinson bound Corollary \ref{AT:LR-bound_kor}, the hypothesis of Lemma \ref{AT:Partial_Trace_Lemma} holds with
\begin{align*}
\varepsilon := \frac{2}{V_{1, 1}(a'- a'')} \; |X| \; \Vert A\Vert \; (\e^{a'' \Vcal_{1, a'', 1}(a' - a'') |t|} - 1) \; \chi_{1,a''}(n).
\end{align*}
Since $V_{1,1}(a' - a'') \geq 1$, see \eqref{AT:V_bigger_than_1}, and with $\Vcal_{1, a'', 1}(a' - a'') \leq \Vcal_{1,a,2}(a'-a)=: v$, Lemma \ref{AT:Partial_Trace_Lemma} therefore implies
\begin{align*}
\Vert \tilde \Theta_n(A)(t)\Vert 
&\leq 2 \, \e^{a''} \; |X| \; \Vert A\Vert \; \e^{-a''((n+1)-v|t|)}.
\end{align*}
Since $1 \leq 4\e^{a'}$, combining these estimates yields
\begin{align}
\Vert \Theta_n(A)(t)\Vert \leq 4 \; \e^{a'} \; |X| \; \Vert A\Vert \; \e^{-a''(n-v|t|)}, \label{Thetant final for Jcalell}
\end{align}
for any $n\geq 0$. To begin the estimate for the integral $\Omega^{\ell, n}(A_0, \ldots,A_{|\ell|})$, let us use Hölder to get
\begin{align*}
\Vert \Omega^{\ell, n}(A_0, \ldots, A_{|\ell|})\Vert &\leq \frac{2^{|\ell| + 1}}{\ell!} \int_\Rbb \dd t\; |W_{s,\gamma}(t)| \; |t|^{|\ell|} \prod_{i=0}^{|\ell|} \sup_{r\in [-|t|,|t|]} \Vert \Theta_{n_i}(A_i)(r)\Vert \\
&\leq \frac{2^{|\ell|+1}}{\ell!} \prod_{i=0}^{|\ell|} \Bigl( \int_\Rbb \dt \; |W_{s,\gamma}(t)| \; |t|^{|\ell|} \; \bigl(\sup_{r\in [-|t|,|t|]} \Vert \Theta_{n_i}(A_i)(r)\Vert \bigr)^{|\ell|+1}\Bigr)^{\nicefrac 1{|\ell|+1}}.
\end{align*}
The $\ell!$ arises from integrating over the simplices inside $\Wcal^\ell$ and the $2^{|\ell|+1}$ comes from the number of terms the commutator generates. Let $i\in \{0, \ldots, |\ell|\}$. Breaking the integral for $T\geq 0$ to be chosen, we obtain the upper bound
\begin{align*}
\int_\Rbb \dt \; |W_{s,\gamma}(t)| \; |t|^{|\ell|} \; \bigl(\sup_{r\in [-|t|,|t|]} \Vert \Theta_{n_i}(A_i)(r)\Vert \bigr)^{|\ell| + 1} &\leq \Tcal_1(T) + \Tcal_2(T),
\end{align*}
where, with $I_{s, \gamma, |\ell|}(T)$ from Lemma \ref{AT:W_gamma_estimate},
\begin{align*}
\Tcal_1(T) &:= \Vert W_{s,\gamma}\Vert_\infty \int_{-T}^T \dt \; |t|^{|\ell|} \; \sup_{r\in [-|t|,|t|]} \Vert \Theta_{n_i}(A_i)(r)\Vert^{|\ell| + 1}, \\\
\Tcal_2(T) &:= 2^{|\ell|+2} \;  \Vert A_i\Vert^{|\ell|+1} \;  I_{s,\gamma,{|\ell|}}(T).
\end{align*}
Let us start by estimating $\Tcal_1$. Applying \eqref{Thetant final for Jcalell} and using $\Vert W_{s,\gamma}\Vert_\infty = \nicefrac 12$ by Lemma \ref{AT:W_gamma_estimate} (a), we obtain
\begin{align*}
\Tcal_1(T) &\leq 4^{|\ell|+1} \; \e^{a'(|\ell|+1)} \; |X_i|^{|\ell|+1} \; \Vert A_i\Vert^{|\ell|+1} \; \int_0^T \dt\; t^{|\ell|} \; \e^{-a'' (|\ell|+1) (n_i - vt)}.
\end{align*}
Now, for any $k\in \Nbb$ we estimate the integral to get
\begin{align*}
\int_0^T \dt \; t^k \; \e^{-a''(k + 1)(n_i- vt)} &\leq \frac{T^k \e^{-a''(k + 1) n_i}}{a''(k + 1)v} \; \e^{a''(k + 1)vt}\Big|_0^T \leq \frac{T^k}{a''(k + 1)v} \; \e^{-a''(k + 1)(n_i - vT)}.
\end{align*}
Choose $vT := \frac {n_i}2$ to get that
\begin{align*}
\int_0^T \dt \; t^k \; \e^{-a''(k+1)(n_i- vt)} &\leq \frac{1}{a''(k+1)v^{k + 1}} \; \bigl( \frac{n_i}{2}\bigr)^k \; \e^{-a''(k+1) \frac{n_i}{2}} \\
&\leq \frac{2}{a''(k+1)v^{k + 1}} \; \bigl( \frac{n_i}{2}\bigr)^{(k+1)} \; \e^{-a''(k+1) \frac{n_i}{2}}.
\end{align*}
We consider the function $f(t) := t \, \e^{-\varepsilon t}$. A short computation shows that $f(\varepsilon^{-1}) = (\varepsilon\e)^{-1}$ is the maximal value of $f$. Applying this to $\varepsilon = a''-a = \frac{a'-a}{2}$, this implies that
\begin{align*}
\bigl(\frac{n_i}{2} \; \e^{-a'' \, \frac{n_i}{2}}\bigr)^k &\leq  \frac{2^k}{(a'-a)^k\e^k} \, \e^{-a \, k \, \frac{n_i}2}.
\end{align*}
Finally, we estimate $\e^{-a \, k \, \frac{n_i}{2}} \leq E_{s,ka} \, \chi_{s,\frac{ka}{4^s}}(2 n_i)$ by Lemma \ref{AT:chi_estimate} (d) and note that we have $E_{s,ka} = E_{s,a}^k$. By the choice of $\eta_{s, \gamma}^k(a,\mu)$, we conclude that
\begin{align*}
\Tcal_1\bigl( \frac{n_i}{2v}\bigr) &\leq \frac{8^{|\ell|+1} \; \e^{a'(|\ell|+1)}}{a \, (|\ell|+1) \, v^{|\ell|+1}} \; \frac{2 \,  E_{s,a}^{|\ell|+1}}{(a'-a)^{|\ell|+1} \, \e^{|\ell|+1}} \; |X_i|^{|\ell|+1} \; \Vert A_i\Vert^{|\ell|+1} \; \chi_{s,(|\ell|+1) \eta_{s, \gamma}^{|\ell|+1}}(2n_i).
\end{align*}
Finally, apply Lemma \ref{AT:W_gamma_estimate} for $r = |\ell|$ and obtain
\begin{align*}
I_{s,\gamma,|\ell|}\bigl( \frac{n_i}{2v}\bigr) &\leq D_{I_{s,\gamma,|\ell|}}(\mu) \; \chi_{s,(|\ell| + 1) \, \frac\mu {|\ell|+1}} \bigl(\frac{\gamma}{4v} \; 2n_i\bigr) \leq D_{I_{s,\gamma,|\ell|}}(\mu) \; \chi_{s,(|\ell|+1) \, \eta_{s, \gamma}^{|\ell|+1}}(2n_i).
\end{align*}
It follows that 
\begin{align*}
\Tcal_2\bigl(\frac{n_i}{2v}\bigr) &\leq 2^{|\ell|+2} \; D_{I_{s,\gamma,|\ell|}}(\mu) \; \Vert A_i\Vert^{|\ell| + 1} \; \chi_{s,(|\ell|+1) \, \eta_{s, \gamma}^{|\ell|+1}}(2n_i).
\end{align*}
Collecting the two terms $\Tcal_1$ and $\Tcal_2$ gives the final estimate
\begin{align*}
\Bigl(\int_\Rbb \dt \; |W_{s,\gamma}(t)| \; |t|^{|\ell|} \; \bigl(\sup_{r\in [-|t|,|t|]} \Vert \Theta_{n_i}(A_i,r)\Vert \bigr)^{|\ell|+1}\Bigr)^{\frac 1{|\ell|+1}} & \\
&\hspace{-80pt} \leq \bigl( D_{s, \gamma}^{\Omega,|\ell|+1}\bigr)^{\frac 1{|\ell|+1}} \; |X_i| \; \Vert A_i\Vert \; \chi_{s,\eta_{s, \gamma}^{|\ell|+1}}(2n_i).
\end{align*}
Multiplying all the bounds completes the proof.
\end{proof}

\subsubsection{Estimating \texorpdfstring{$\Ical_{s, \gamma}^\ell$}{Il} on local Hamiltonians}

Let us agree on the following interaction for $\Ical_{s, \gamma}^\ell(G_0, \ldots, G_{|\ell|})$, where $G_0, \ldots, G_{|\ell|} \in \Lcal$:
\begin{align}
\Phi_{\Ical_{s, \gamma}^\ell(G_0, \ldots, G_{|\ell|})}(W) := \sum_{\substack{Z_0, \ldots, Z_{|\ell|}\subset \Lambda \\ Z_i \cap \bigcup_{m=0}^{i-1} Z_m\neq\emptyset \, \forall i\geq 1\\ \bigcup_{i=0}^{|\ell|} Z_i = W}} \sum_{n\in \Nbb_0^{|\ell|+1}} \sum_{\substack{Y_0, \ldots, Y_{|\ell|} \subset \Lambda \\ Y_{i,n_i} = Z_i}} \Omega^{\ell, n} \bigl(\Phi_{G_0}(Y_0), \ldots, \Phi_{G_{|\ell|}}(Y_{|\ell|})\bigr). \label{AT:Iell_interaction_definition}
\end{align}
As we convinced ourselves a lot of times by now, we then get
\begin{align*}
\sum_{W\subset \Lambda} \Phi_{\Ical_{s, \gamma}^\ell(G_0, \ldots, G_{|\ell|})}(W) = \Ical_{s, \gamma}^\ell(G_0, \ldots, G_{|\ell|}).
\end{align*}

\begin{thm}
\label{AT:I_generalized_estimate}
Let Assumptions \ref{AT:Assumption_abstract_V} and \ref{AT:Assumption_abstract_F} be true, let $a'>0$ and $H\in \Lcal_{1,a'}$. Let $\ell \in \Fcal(\Nbb_0)$, $0 < s < 1$, and $0 < \mu < \mu_0(s)$. Let $\gamma >0$ and $0 < b' < \eta_{s, \gamma}^{|\ell|+1}(a,\mu)$, where $\eta_{s, \gamma}^k(a,\mu)$ is from \eqref{AT:etak_definition}. Assume that $G_0,\ldots, G_{|\ell|}\in \Lcal_{s,b'}$. Then $\Ical_{s, \gamma}^\ell (G_0,\ldots, G_{|\ell|})\in \Lcal_{s,b}$ for any $0 < b < b'$ and the interaction in \eqref{AT:Iell_interaction_definition} obeys the estimate
\begin{align*}
\Vert \Phi_{\Ical_{s, \gamma}^\ell(G_0,\ldots, G_{|\ell|})}\Vert_{s,b} & \leq  \frac{(|\ell|+1)!}{\ell!} \; D_{s, \gamma}^{\Omega, |\ell| + 1}(a,\mu)\\
&\hspace{20pt} \times V_{s,2(|\ell|+1)}(b'-b)^{2(|\ell|+1)} \; F_{s}\bigl(\eta_{s, \gamma}^{|\ell|+1}(a, \mu) - b'\bigr)^{|\ell|+1} \; \prod_{i=0}^{|\ell|}\Vert \Phi_{G_i}\Vert_{s,b'},
\end{align*}
where $D_{s, \gamma}^{\Omega, k}(a,\mu)$ is from Lemma \ref{AT:Omegaelln_estimate}, $V_{s,2k}(b)$ is from \eqref{AT:Vkb_definition}, and $F_{s}(b)$ is from \eqref{AT:Fb_definition}.
\end{thm}

\begin{proof}
We omit the index $s$ throughout. Let $x\in \Lambda$. Then, we need to estimate
\begin{align}
\sum_{W\ni x} \frac{\Vert\Phi_{\Ical^\ell(G_0, \ldots, G_{|\ell|})}(W)\Vert}{\chi_b(\Dcal(W))} &\leq \frac{D^{\Omega,|\ell|+1}}{\ell!} \; \smash{\sm[l]{\sum_{\substack{Z_0, \ldots, Z_{|\ell|}\subset \Lambda \\ Z_i\cap \bigcup_{m=0}^{i-1} Z_m \neq \emptyset\;\forall i\geq 1 \\ \bigcup_{i=0}^{|\ell|} Z_i=: W \ni x}}}} \frac{1}{\chi_b(\Dcal(W))} \prod_{i=0}^{|\ell|}  \sum_{n_i\in \Nbb_0} \chi_{\eta^{|\ell|+1}}(n_i) \notag \\
&\hspace{130pt} \times \sum_{\substack{Y_i\subset \Lambda \\ Y_{i,n_i} = Z_i}} |Y_i| \; \Vert \Phi_{G_i}(Y_i)\Vert. \label{AT:I_generalized_estimate_1}
\end{align}
Let us explain the strategy. We are going to resum this in the following way. Since $x$ lies in the union of all $Z$'s, there is a permutation of $|\ell|+1$ elements, $\sigma\in S_{|\ell|+1}$, such that the following holds. There is an index $\sigma(0)\in \{0, \ldots, |\ell|\}$ such that $x\in Z_{\sigma(0)}$. Then, since $Z_{\sigma(0)}$ has nonempty intersection with a new set $Z_{\sigma(1)}$, this set is hit by summing over all points in $Z_{\sigma(0)}$ and all sets $Z_{\sigma(1)}$ containing that point. The next set $Z_{\sigma(2)}$ is attached to the union $Z_{\sigma(0)}\cup Z_{\sigma(1)}$, so we sum over all points therein and sets $Z_{\sigma(2)}$ that contain this point. Continuing this procedure and relabeling the $Z_{\sigma(i)}$ as $Z_i$ for all $i$, we arrive at the following upper bound for \eqref{AT:I_generalized_estimate_1}, leaving out the constants in front:
\begin{align}
&\sum_{\sigma\in S_{|\ell|+1}} \sum_{n_0\in \Nbb_0} \chi_{\eta^{|\ell|+1}}(2n_0) \sum_{\substack{Z_0\subset \Lambda \\ Z_0\ni x}} \sum_{\substack{Y_0\subset \Lambda \\ Y_{0,n_0} = Z_0}} |Y_0| \; \Vert \Phi_{G_{\sigma(0)}}(Y_0)\Vert \notag \\
&\hspace{20pt}\sum_{z_1\in Z_0} \sum_{n_1\in \Nbb_0} \chi_{\eta^{|\ell|+1}}(2n_1) \sum_{\substack{Z_1\subset \Lambda \\ Z_1\ni z_1}} \sum_{\substack{Y_1\subset \Lambda \\ Y_{1,n_1} = Z_1}} |Y_1| \; \Vert \Phi_{G_{\sigma(1)}}(Y_1)\Vert\notag\\
&\hspace{30pt}\sum_{z_2\in Z_0\cup Z_1} \sum_{n_2\in \Nbb_0} \chi_{\eta^{|\ell|+1}}(2n_2) \sum_{\substack{Z_2\subset \Lambda \\ Z_2\ni z_2}} \sum_{\substack{Y_2\subset \Lambda \\ Y_{2,n_2} = Z_2}} |Y_2| \; \Vert \Phi_{G_{\sigma(2)}}(Y_2)\Vert \label{AT:I_generalized_estimate_2} \\
&\hspace{80pt}\ddots\notag\\
&\hspace{10pt}\sum_{z_{|\ell|}\in \bigcup_{j=0}^{|\ell|} Z_j} \sum_{n_{|\ell|}\in \Nbb_0} \chi_{\eta^{|\ell|+1}}(2n_{|\ell|}) \sum_{\substack{Z_{|\ell|}\subset \Lambda \\ Z_{|\ell|}\ni z_{|\ell|}}} \sum_{\substack{Y_{|\ell|}\subset \Lambda \\ Y_{|\ell|,n_{|\ell|}} = Z_{|\ell|}}} |Y_{|\ell|}| \; \Vert \Phi_{G_{\sigma(|\ell|)}}(Y_{|\ell|})\Vert  \; \frac{1}{\chi_b(\Dcal(\bigcup_{j=0}^{|\ell|} Z_j))}.\notag 
\end{align}
Now, we are in need of decay factors in terms of the increasing unions. These are constructed as follows:
\begin{align*}
\frac{1}{\chi_b(\Dcal(\bigcup_{j=0}^{|\ell|} Z_j))} &= \frac{1}{\chi_{b'}(\Dcal(\bigcup_{j=0}^{|\ell|} Z_j))} \; \chi_{\frac{b'-b}{|\ell|+1}}\Bigl(\Dcal\Bigl(\bigcup_{j=0}^{|\ell|} Z_j\Bigr)\Bigr)^{|\ell|+1} \\
&\leq \prod_{i=0}^{|\ell|} \frac{1}{\chi_{b'}(\Dcal(Y_i)) \chi_{b'}(2n_i)} \; \chi_{\frac{b'-b}{|\ell|+1}} \Bigl( \Dcal\Bigl( \bigcup_{j=0}^i Z_i\Bigr)\Bigr).
\end{align*}
Here, we used the logarithmic superadditivity and $\Dcal(\bigcup_{j=0}^{|\ell|} Z_j) \leq \sum_{i=0}^{|\ell|} \Dcal(Z_i)$ as well as $\Dcal(Z_i)\leq \Dcal(Y_i) + 2n_i$. For the bound on the last factor, we just omitted the sets that are of no interest anymore.

The next step consists of resumming the fattenings. We know how to do this\ifthenelse{\equal\masterfile{Exponential}}{\footnote{Careful, the estimate of $\Ical_{s, \gamma}$ will of course not appear in a potential paper, since everything will be in this proof. Hence, have to motivate idea again...}}{} by just summing over all points in the $n$-ball of points that lie in the fattened set and sets that contain this new point. This yields the following upper bound to \eqref{AT:I_generalized_estimate_2}
\begin{align}
&\sum_{\sigma\in S_{|\ell| +1}} \sum_{n_0\in \Nbb_0} \chi_{\eta^{|\ell|+1} - b'}(2n_0) \sum_{\tilde x\in B_{n_0}(x)} \sum_{\substack{Y_0\subset \Lambda \\ Y_0 \ni \tilde x}} \frac{|Y_0| \; \Vert \Phi_{G_{\sigma(0)}}(Y_0)\Vert}{\chi_{b'}(\Dcal(Y_0))} \; \chi_{\frac{b'-b}{|\ell|+1}}(\Dcal(Y_{0,n_0}))\notag\\
&\hspace{10pt} \sum_{z_1\in Y_{0,n_0}} \sum_{n_1\in \Nbb_0} \chi_{\eta^{|\ell|+1} - b'}(2n_1) \sum_{\tilde z_1\in B_{n_1}(z_1)} \sum_{\substack{Y_1\subset \Lambda \\ Y_1 \ni \tilde z_1}} \frac{|Y_1| \; \Vert \Phi_{G_{\sigma(1)}}(Y_1)\Vert}{\chi_{b'}(\Dcal(Y_1))} \; \chi_{\frac{b'-b}{|\ell|+1}}(\Dcal(Y_{0,n_0}\cup Y_{1,n_1}))  \notag\\
&\hspace{80pt}\ddots \label{AT:I_generalized_estimate_3}\\
&\hspace{10pt} \sum_{z_{|\ell|-1}\in \bigcup_{j=0}^{|\ell|-1} Y_{j,n_j}} \sum_{n_{|\ell|-1}\in \Nbb_0} \chi_{\eta^{|\ell|+1} - b'}(2n_{|\ell|-1}) \notag\\
&\hspace{30pt} \times \sum_{\tilde z_{|\ell|-1}\in B_{n_{|\ell|-1}}(z_{|\ell|-1})} \sum_{\substack{Y_{|\ell|-1}\subset \Lambda \\ Y_{|\ell|-1} \ni \tilde z_{|\ell|-1}}} \frac{|Y_{|\ell|-1}| \; \Vert \Phi_{G_{\sigma(|\ell|-1)}}(Y_{|\ell|-1})\Vert}{\chi_{b'}(\Dcal(Y_{|\ell|-1}))} \chi_{\frac{b'-b}{|\ell|+1}}\bigl(\Dcal\bigl(\bigcup_{j=0}^{|\ell|-1} Y_{j,n_j}\bigr)\bigr) \notag \\
&\hspace{40pt} \sum_{z_{|\ell|}\in \bigcup_{j=0}^{|\ell|-1} Y_{j,n_j}} \sum_{n_{|\ell|}\in \Nbb_0} \chi_{\eta^{|\ell|+1} - b'}(2n_{|\ell|}) \sum_{\tilde z_{|\ell|}\in B_{n_{|\ell|}}(z_{|\ell|})} \sum_{\substack{Y_{|\ell|}\subset \Lambda \\ Y_{|\ell|} \ni \tilde z_{|\ell|}}} \frac{|Y_{|\ell|}| \; \Vert \Phi_{G_{\sigma(|\ell|)}}(Y_{|\ell|})\Vert}{\chi_{b'}(\Dcal(Y_{|\ell|}))} \notag\\ 
&\hspace{300pt} \times \chi_{\frac{b'-b}{|\ell|+1}}\bigl( \Dcal\bigl( \bigcup_{j=0}^{|\ell|} Y_{j,n_{j}}\bigr)\bigr).\notag 
\end{align}
At this point, we start estimating the last row of \eqref{AT:I_generalized_estimate_3}. First,
\begin{align*}
|Y_{|\ell|}| \; \chi_{\frac{b'-b}{|\ell|+1}} \Bigl(\Dcal\bigl( \bigcup_{j=0}^{|\ell|} Y_{j,n_j}\bigr)\Bigr) \leq V_{|\ell|+1}(b'-b).
\end{align*}
Also, note that $V_{|\ell|+1}(b'-b)\leq V_{2(|\ell|+1)}(b'-b)^2$, see \eqref{AT:Vkb-kleqkprime_estimate}. Taking away the norm $\Vert \Phi_{G_{\sigma(|\ell|)}}\Vert_{b'}$, the volume $|B_{n_{|\ell|}}(z_{|\ell|})|$, together with the sum over $n_{|\ell|}$ gives $F(\eta^{|\ell|+1} - b')$. The total bound for the last row of \eqref{AT:I_generalized_estimate_3} is thus
\begin{align*}
\Bigl| \bigcup_{j=0}^{|\ell|-1} Y_{j,n_j}\Bigr| \; F(\eta^{|\ell|+1} - b') \; V_{2(|\ell|+1)}(b'-b)^2 \; \Vert \Phi_{\sigma(|\ell|)}\Vert_{b'}.
\end{align*}
The first factor is passed on to the second to last row and we obtain
\begin{align*}
|Y_{|\ell|-1}| \; \Bigl| \bigcup_{j=0}^{|\ell| - 1} Y_{j,n_j}\Bigr| \; \chi_{\frac{b'-b}{|\ell|+1}} \Bigl( \Dcal\Bigl( \bigcup_{j=0}^{|\ell|-1} Y_{j,n_j}\Bigr)\Bigr)\leq V_{2(|\ell|+1)}(b'-b)^2.
\end{align*}
Now, the procedure continues, so that the final bound for the second to last row is
\begin{align*}
\Bigl| \bigcup_{j=0}^{|\ell|-2} Y_{j,n_j}\Bigr| \;  F(\eta^{|\ell|+1} - b') \; \Vert \Phi_{G_{\sigma(|\ell|-1)}}\Vert_{b'} \; V_{2(|\ell|+1)}(b'-b)^2.
\end{align*}
Collapsing \eqref{AT:I_generalized_estimate_3} in this way to the first line, this provides $|\ell|+1$ copies of $V_{2(|\ell|+1)}(b'-b)^2$, as well as $|\ell|+1$ copies of $F(\eta^{|\ell|+1} - b')$, and all the norms. The last step is to note that $|S_{|\ell|+1}| = (|\ell|+1)!$. This proves the claim.
\end{proof}

\subsubsection{Estimate on the derivatives of \texorpdfstring{$\Ical_{s, \gamma}$}{I}}

As a Corollary to Theorem \ref{AT:I_generalized_estimate}, we finally obtain an estimate on the norm of the derivatives of $\Ical_{s, \gamma}$. We start with a representation formula for the derivative.

\begin{lem}
\label{AT:I_Derivative_formula}
Let $\beta \in \Nbb$, $0 < s\leq 1$, and $\gamma >0$. Let $G, H \in \Lcal(I)$ be $(\beta - 1)$-fold differentiable time-dependent local Hamiltonians. Then
\begin{align}
\frac{\dd^{\beta - 1}}{\dd u^{\beta - 1}} \Ical_{s, \gamma}(G)(u) = \sum_{\ell\in \Mcal_{\beta- 1}} \sum_{\substack{ q\in \Nbb^{|\ell| + 1} \\ |q| = \beta}} N_{\beta, \ell, q} \; \Ical_{s, \gamma}^\ell \bigl(H^{(q_1)}(u) \, ,  \ldots, \, H^{(q_{|\ell|})}(u) \, , \, G^{(q_0 - 1)}(u)\bigr), \label{AT:I_Derivative_formula_eq1}
\end{align}
where $N_{\beta, \ell, q} \in \Nbb$ and
\begin{align}
\Mcal_\beta := \Bigl\{ \ell\in \Fcal(\Nbb_0) : |\ell| \leq \beta \; , \; \sum_{m=0}^i \ell_m \geq 1 + i , \; i = -1, \ldots, m_\ell \Bigr\}.
\end{align}
\end{lem}

We point out that if we wanted to give a formula for $N_{\beta, \ell, q}$, we would need to understand more precisely how the terms in the derivative are built. Certainly, the multinomial coefficient
\begin{align*}
\binom{\beta - 1}{q_0 - 1 \, , \, q_1 \, , \, \ldots\, , \, q_{|\ell|}} := \frac{(\beta - 1)!}{(q_0 - 1)! \,  q_1! \, \cdots \, q_{|\ell|}!}
\end{align*}
plays a role but there are other mechanics that would need to be captured. We are not capable of keeping track of this at the moment. However, we do not expect the bound on the derivative given below to improve much by taking this into account.

\begin{proof}[Proof of Lemma \ref{AT:I_Derivative_formula}]
We prove this by induction and we note that the case $\beta =1$ trivially holds with $N_{0,0,1} = 1$. As a preparation for the induction step, we note that \eqref{AT:Duhamels_formula} implies
\begin{align}
\frac{\dd}{\dd u}\tau(G) = \tau(\dot G) + \Vcal_1(\tau(\dot H), \tau(G)). \label{AT:I_Derivative_formula_1}
\end{align}
Furthermore, for $\ell\in \Fcal(\Nbb_0)$, $m = 0, \ldots, |\ell|$, and $G, G_0, \ldots, G_{|\ell|}\in \Lcal$, we have
\begin{align*}
(\Vcal_{m, 1}^* \circ \Wcal^\ell)(G_0, \ldots, G_{m-1}, G, G_m , \ldots, G_{|\ell|}) &\\
&\hspace{-80pt} = \Wcal^\ell(G_0, \ldots, G_{m-1} \, ,  \, \Vcal_1(G, G_m) \,  , \, G_{m+1} , \ldots,G_{|\ell|}) \\
&\hspace{-80pt} = \Wcal^{\varphi_m(\ell)} (G_0, \ldots, G_{|\ell|}),
\end{align*}
where
\begin{align*}
\varphi_m(\ell) := \bigl( \ell_0 \, , \, \ldots \, , \, \ell_{m-1} \, , \, \ell_m + 1 \, , \, 0 \, , \, \ell_{m+1} \, , \, \ldots \, , \, \ell_{|\ell|}\bigr).
\end{align*}
For differentiable Hamiltonians $G_0, \ldots, G_{|\ell|}$, \eqref{AT:I_Derivative_formula_1} therefore implies
\begin{align*}
\frac{\dd}{\dd u} (\tau^* \circ \Wcal^\ell)(G_0, \ldots, G_{|\ell|}) &= \sum_{m=0}^{|\ell|} (\tau^* \circ \Wcal^\ell)(G_0, \ldots, \dot G_m , \ldots G_{|\ell|}) \\
&\hspace{50pt} + (\tau^* \circ \Wcal^{\varphi_m(\ell)})(G_1, \ldots, G_{m-1}, \dot H, G_m, \ldots, G_{|\ell|})
\end{align*}
and this implies 
\begin{align}
\frac{\dd}{\dd u} \Ical_{s, \gamma}^\ell(G_0, \ldots, G_{|\ell|}) &= \smash{\sum_{m=0}^{|\ell|}} \; \Ical_{s, \gamma}^\ell (G_0, \ldots, \dot G_m , \ldots, G_{|\ell|}) \notag \\
&\hspace{30pt} + \Ical_{s, \gamma}^{\varphi_m(\ell)} (G_0, \ldots, G_{m-1}, \dot H, G_m, \ldots, G_{|\ell|}). \label{AT:I_Derivative_formula_2}
\end{align}

Suppose \eqref{AT:I_Derivative_formula} holds for some $\beta\in \Nbb$. In view of \eqref{AT:I_Derivative_formula_2}, it is sufficient to show that if $\ell\in \Mcal_\beta$, then $\varphi_m(\ell)\in \Mcal_{\beta + 1}$ for all $m = 0, \ldots, |\ell|$. It is obvious that $|\varphi_m(\ell)|\leq \beta + 1$. We have to verify that
\begin{align*}
\sum_{j=0}^i \varphi_m(\ell)_j &\geq 1 + i , & i &= -1, \ldots, m_{\varphi_m(\ell)}.
\end{align*}
For $i = 0, \ldots, m-1$, this is clear per definition of $\varphi_m(\ell)$. Since $m_{\varphi_m(d)} = m_\ell + 1$, for $i = m , \ldots, m_\ell + 1$, we have
\begin{align}
\sum_{j=0}^i \varphi_m(\ell)_j = 1 + \sum_{j=0}^m \ell_j + \sum_{j=m+2}^i \ell_{j-1} = 1 + \sum_{j=0}^{i-1} \ell_j \geq 1 + (i - 1) = 1 + i.
\end{align}
Thus, $\varphi_m(\ell) \in \Mcal_{\beta + 1}$ and this completes the proof.
\end{proof}

%

We agree on the interaction for $\frac{\dd^{\beta- 1}}{\dd u^{\beta - 1}} \Ical_{s, \gamma}(G)$ which is given by the decomposition \eqref{AT:I_Derivative_formula_eq1} and \eqref{AT:Iell_interaction_definition}.

\begin{thm}
\label{AT:I_betath_derivative_estimate}
Let Assumptions \ref{AT:Assumption_abstract_V} and \ref{AT:Assumption_abstract_F} be true, $a'> a > 0$, and $\beta\in \Nbb$. For an open interval $I\subset \Rbb$, assume that $H\in \Lcal_{1,a'}(I)$ is a $(\beta - 1)$-fold differentiable local Hamiltonian. Let also $0 < s < 1$, $0 < \mu < \mu_0(s)$, $\gamma>0$, and $0 < b' < \eta_{s, \gamma}^\beta(a,\mu)$, where $\eta_{s, \gamma}^\beta(a, \mu)$ is from Lemma \ref{AT:Omegaelln_estimate}. Let $G$ be a $(\beta-1)$-fold differentiable local Hamiltonian with $G^{(j)}\in \Lcal_{s,b'}(I)$ for all $0\leq j \leq \beta-1$ and assume that $H^{(j)}\in \Lcal_{s,b'}(I)$ for $1\leq j \leq \beta-1$. Then, $\frac{\dd^{\beta-1}}{\dd u^{\beta-1}} \Ical_{s, \gamma}(G) \in \Lcal_{s,b}(I)$ for any $0 < b < b'$ and the estimate
\begin{align*}
\bigl\Vert \Phi_{\frac{\dd^{\beta-1}}{\dd u^{\beta-1}} \; \Ical_{s, \gamma}(G)}(u)\bigr\Vert_{s,b} \leq D_{s, \gamma}^{\dd \Ical , (1)}(a,\mu)^\beta \; D_s^{\dd \Ical, (2),\beta}(b'-b) \; F_s(\eta^\beta - b')^\beta \; \Ncal_{s,b',\beta}(H,G)(u)
\end{align*}
holds for all $u\in I$, where 
\begin{align*}
D_{s, \gamma}^{\dd\Ical, (1)}(a,\mu) &:= 2 + 64 \; \sup_{k\in \Nbb} \Bigl[ \frac{\e^{a'-1}}{a^{\frac 1k} \, \Vcal_{1, a, 2}(a' - a)} \; \frac{E_{s,a}}{a'-a}\\
&\hspace{70pt} + \frac{1}{\gamma^{1+\frac 1k}} \Bigl( \frac{c_{s,\gamma} D_s\Gamma(\nicefrac 1s)^2}{s^2}\Bigr)^{\frac{1}{k}} \Bigl( \frac{4}{\mu_0(s)-\mu}\Bigr)^{\frac 1s+\frac{1}{sk}} \Bigl(\frac{1}{s\e}\Bigr)^{\frac 1s-\frac 1{sk}}\Bigr]
\end{align*}
with $\Vcal_{1,a,2}(a'-a)$ from \eqref{AT:LR-velocity_definition}, $E_{s,a}$ from Lemma \ref{AT:chi_estimate}, $c_{s,\gamma}$, $D_s$, and $\mu_0(s)$ from Lemma \ref{AT:w_gamm_estimate},
\begin{align*}
D_s^{\dd\Ical, (2), \beta} (b'-b) &:= \beta! \; (\beta -1)! \; (\beta-1)^{\frac{\beta-1}{s}} \; V_{s, 2\beta}(b'-b)^{2\beta},
\end{align*}
and
\begin{align}
\Ncal_{s,b,\beta} (H,G)(u) &:= \sup_{0\leq j\leq \beta-1} \sup_{\substack{q\in \Nbb^{j+1} \\ |q| = \beta}} \Vert G^{(q_0 -1)}(u) \Vert_{s, b} \; \Vert H^{(q_1)}(u) \Vert_{s, b} \cdots \Vert H^{(q_j)}(u) \Vert_{s, b}. \label{AT:Ncal_definition}
\end{align}
\end{thm}

\begin{proof}
The goal is to obtain an upper bound on the number of terms in the derivative in \eqref{AT:I_Derivative_formula_eq1} and to apply Theorem \ref{AT:I_generalized_estimate} in the worst possible case. Let us start by counting the number of terms \eqref{AT:I_Derivative_formula_eq1}. If $T_{\beta-1}$ symbolizes the total number of terms in \eqref{AT:I_Derivative_formula_eq1}, the first observation is the fact that each derivative falling on a $\tau_t(G)$ causes two new terms, see \eqref{AT:I_Derivative_formula_1}. Pretending that each term in the $(\beta-1)$\st\ derivative has the form $\Ical_{s, \gamma}^\ell(G_1, \ldots, G_{|\ell| + 1})$ with $|\ell| + 1 = \beta$ (which is not the case, many $\ell$'s have $|\ell| + 1 < \beta$, see \eqref{AT:I_Derivative_formula_eq1}), this leads to the upper bound $T_{\beta} \leq T_{\beta-1} \; 2\beta$. Obviously, we have $T_0 = 1$. Define $S_\beta := 2^\beta \beta!$. We claim that $T_{\beta-1} \leq S_{\beta-1}$ for each $\beta\in \Nbb$. To see this, note that $S_0 = 2^0 \; 0! = 1$, i.e. $T_0 \leq S_0$. Assume that $T_{\beta-1} \leq S_{\beta-1}$ for some $\beta\in \Nbb$. Then, 
\begin{align*}
T_{\beta} \leq T_{\beta-1} \, 2\beta \leq S_{\beta-1} \; 2\beta = 2^{\beta-1} \; (\beta-1)! \; 2\beta = 2^{\beta} \, \beta! = S_{\beta}.
\end{align*}

It remains to apply Theorem \ref{AT:I_generalized_estimate} to all of the terms in \eqref{AT:I_Derivative_formula_eq1}.
This gives an upper bound of the form
\begin{align*}
\bigl\Vert \Phi_{\Ical_{s, \gamma}^\ell(H^{(q_1)}, \ldots,  H^{(q_{|\ell|})},G^{(q_0-1)})}\bigr\Vert_b & \\
&\hspace{-100pt} \leq (|\ell|+1)! \;  D_{s, \gamma}^{\Omega,|\ell|+1}(a,\mu) \; V_{2(|\ell|+1)}(b'-b)^{2(|\ell|+1)} \; F(\eta_{s, \gamma}^{|\ell|+1} - b')^j \; \Ncal_{b',\beta}(H,G).
\end{align*}
Now, a straightforward computation using the definition of $D_{s, \gamma}^{\Omega,\beta}$, of $D_{I_{s,\gamma,\beta-1}}$ and the subadditivity of the $\beta$\tho\ root shows that
\begin{align*}
D_{s, \gamma}^{\Omega,\beta}(a,\mu)^{\frac 1\beta} \leq \frac 12 \; (\beta-1)^{\frac{\beta - 1}{s\beta}} \; D_{s, \gamma}^{\dd \Ical, (1)}(a,\mu).
\end{align*}
Since, by definition, $D_{s, \gamma}^{\dd \Ical, (1)} \geq 2$, we may estimate
\begin{align*}
D_{s, \gamma}^{\Omega, \beta}(a,\mu) \leq \frac{1}{2^\beta} \; (\beta-1)^{\frac{\beta - 1}{s}} \; D_{s, \gamma}^{\dd \Ical, (1)} (a,\mu)^\beta.
\end{align*}
Together with the factor $2^{\beta - 1} (\beta - 1)!$, we get the total claimed bound.
\end{proof}


\section[Analytic interactions]{Analytic time-dependent interactions}
\label{AT:Analytic_Interactions_Section}

\subsection{Derivatives of analytic Hamiltonians}

In this subsection, we show how analyticity of an interaction enables us to relate local norms of derivatives of a Hamiltonian to the norm of the Hamiltonian itself. To do this, fix an interval $I\subset \Rbb$, let $\delta >0$ and we consider the complex fattening
\begin{align}
I_\delta := \{u\in \Cbb : \dist(u, I) < \delta \}. \label{AT:Sdelta_definition}
\end{align}
We recall that $\Bcal_{s, a}^\hol(I_\delta)$ is the space of holomorphic interactions in $\Bcal_{s, a}(I_\delta)$ and $\Lcal_{s, a}^\hol(I_\delta)$ is the space of local Hamiltonians with interactions in $\Bcal_{s, a}^\hol(I_\delta)$, where $0 < s \leq 1$, $a>0$. In the following, we abbreviate the norm of $\Phi\in \Bcal_{s, a}^\hol(I_\delta)$ as
\begin{align*}
\vvvert \Phi\vvvert_{\delta, s, a} := \vvvert \Phi\vvvert_{I_\delta, s, a}.
\end{align*}

A version of the following Lemma in the one-particle case has been proven in \cite[Lemma 3.1]{HagedornJoye}. 

\begin{lem}
\label{hochhangeln}
Define $B(0) = 1$ and $B(k) = k^k$ for any integer $k\geq 1$. Let $0 < s \leq 1$ and $a>0$. For $\delta >0$ suppose $\Phi \in \Bcal_{s,a}^\hol(I_\delta)$. If there are $k\in \Nbb_0$ and a constant $C_{s, a}>0$ such that $\Phi$ satisfies
\begin{align*}
\Vert \Phi(u)\Vert_{s,a} &\leq C_{s,a} \; B(k) \; (\delta - \dist(u, I))^{-k}, & u &\in I_\delta,
\end{align*}
then
\begin{align*}
\Vert \dot \Phi(u)\Vert_{s,a} &\leq C_{s,a} \; B(k+1) \; (\delta - \dist(u, I))^{-k-1}, & u &\in I_\delta.
\end{align*}
In particular, $\dot \Phi\in \Bcal_{s,a}^\hol(I_\delta)$.
\end{lem}

\begin{proof}
We omit the index $s$ and write $\Phi(Z, u) := \Phi(u)(Z)$. First of all, for all $Z\in\Fcal(\Gamma)$ we have that $\dot \Phi(Z, u)$ has support in $Z$, since by continuity of $\Ebb_Z$, we have
\begin{align*}
\Ebb_Z[\dot \Phi(Z, u)] = \lim_{h\to 0 } \frac{\Ebb_Z[\Phi(Z, u+h) - \Phi(Z, u)]}{h} = \lim_{h\to 0} \frac{\Phi(Z, u+h) - \Phi(Z, u)}{h} = \dot \Phi(Z, u).
\end{align*}

Let first $k\geq 1$. By Cauchy's integral formula, we have
\begin{align*}
\dot \Phi(Z, u) = \frac{1}{2\pi\i} \int_\eta \dd v \; \frac{\Phi(Z, v)}{(u-v)^2},
\end{align*}
where $\eta(t) = u + r\e^{\i t}$ is the circle with center $u$ and radius $r = \frac{1}{k+1} (\delta - \dist(u, I))$. For $v \in \eta$, we have
\begin{align*}
\delta - \dist(v, I) \geq \delta - \dist(u, I) - r = k r.
\end{align*}
Thus, for $v\in \eta$,
\begin{align*}
\Vert \Phi(v)\Vert_a &\leq C_a \; k^k \; (\delta - \dist(v, I))^{-k} \leq C_a\; k^k \; (kr)^{-k} = C_a \; r^{-k}
\end{align*}
It follows that  for $x\in \Lambda$
\begin{align*}
\sum_{\substack{Z \subset \Lambda \\ Z \ni x}} \frac{\Vert \dot \Phi(Z, u)\Vert}{\chi_a(\Dcal(Z))} &\leq \frac{1}{2\pi} \int_0^{2\pi} \dt \; \sum_{\substack{Z\subset \Lambda \\ Z \ni x}} \frac{\Vert \Phi(Z, \eta(t))\Vert}{\chi_a(\Dcal(Z))} \; \frac{1}{|u - u - r\e^{\i t}|^2} \; |r\i \; \e^{\i t}| \\
&\leq C_a \; r^{-k} \; \frac{1}{r^2} \; r = C_a \; r^{-k-1} = C_a\; (k+1)^{k+1} \; (\delta - \dist(u, I))^{-(k+1)}.
\end{align*}
This proves the claim for $k\geq 1$. For $k =0$, we use the same argument with radius $r = \alpha (\delta - \dist(u, I))$ for any $\alpha < 1$. Then,
\begin{align*}
\sum_{\substack{Z\subset \Lambda \\ Z \ni x}} \frac{\Vert \dot \Phi(Z, u)\Vert}{\chi_a(\Dcal(Z))} \leq \frac{1}{2\pi} \dt \; \int_0^{2\pi} \frac{\Vert \Phi(\eta(t))\Vert_a}{r^2} \; r = C_a \; \frac{1}{\alpha }\; (\delta - \dist(u, I))^{-1} .
\end{align*}  
Since this is true for all $\alpha < 1$, infing over $\alpha$ yields the claim.
\end{proof}

\begin{kor}
\label{AT:Hamiltonian_Analytic_derivatives}
Let $a >0$, $0 < s \leq 1$, and $\delta_0>0$. Let $H\in \Lcal_{s,a}^\hol(I_{\delta_0})$. Then, for all $k\in \Nbb_0$, we have that $H^{(k)}\in \Lcal_{s,a}^\hol(I_{\delta_0})$ with $\Phi_{H^{(k)}} = \Phi_H^{(k)}$ and, for all $0 < \delta \leq \delta_0$, and $u\in I_\delta$, we have
\begin{align*}
\Vert \Phi_H^{(k)}(u)\Vert_{s,a} \leq \vvvert\Phi_H \vvvert_{\delta_0, s,a} \; B(k)\; (\delta - \dist(u, I))^{-k}.
\end{align*}
\end{kor}

\begin{proof}
By definition, $\Vert \Phi_H(u)\Vert_{s,a} \leq \vvvert \Phi_H\vvvert_{\delta_0, s,a}$. Hence, the claim holds for $k =0$. The induction is now driven by Lemma \ref{hochhangeln}.
\end{proof}

\begin{kor}
Let $\beta \in \Nbb$, $0 < s \leq 1$, $a>0$, and let $H\in \Lcal_{s,a}^\hol(I_{\delta_0})$. Then, for each $0 < \delta \leq \delta_0$ and $u\in I_\delta$, we have
\begin{align*}
\Ncal_{s, a,\beta}(H, \dot H)(u) \leq (\delta - \dist(u, I))^{-\beta} \; B(\beta) \; \vvvert \Phi_H\vvvert_{\delta_0, s,a}^\beta,
\end{align*}
where $\Ncal_{s, a,\beta}(H, \dot H)$ is from \eqref{AT:Ncal_definition}.
\end{kor}

\begin{proof}
We have
\begin{align*}
\Ncal_{s, a,\beta}(H , \dot H)(u) = \sup_{0\leq j \leq \beta - 1} \sup_{\substack{q\in \Nbb^{j+1} \\ |q| = \beta}} \Vert H^{(q_0)}(u) \Vert_{s, a} \; \Vert H^{(q_1)}(u) \Vert_{s, a} \cdots \Vert H^{(q_j)}(u) \Vert_{s, a}.
\end{align*}
Without loss, we may assume that $\vvvert \Phi_H\vvvert_{\delta_0, s, a} \geq 1$, see \eqref{AT:Hamiltonian_norm_greater_than_1}. By Corollary \ref{AT:Hamiltonian_Analytic_derivatives}, we therefore obtain
\begin{align*}
\Ncal_{s, a,\beta}(H , \dot H)(u) \leq \sup_{0\leq j \leq \beta - 1} \sup_{\substack{q\in \Nbb^{j+1} \\ |q| = \beta}} \vvvert \Phi_H\vvvert_{\delta_0, s, a}^{j+1} \; (\delta - \dist(u, I))^{-|q|} \; \prod_{i=1}^j B(q_i).
\end{align*}
Since $B(q_i) \leq \beta^{q_i}$, we have $\prod_{i=0}^j B(q_i) \leq \beta^{|q|} = B(\beta)$. The bound readily follows.
\end{proof}

%
%


\section[Expansion Concept]{Concept for the adiabatic expansion}
\label{AT:Concept_Section}

In the following, we present the concept of the adiabatic expansion that should be good enough for an optimal truncation.
\begin{center}

\vspace{0.5cm}

\begin{huge}
--- \textsc{Warning} ---
\end{huge}

\vspace{0.5cm} 

\textbf{In this section, the digression is mostly informal and by lacks mathematical rigor.}
\end{center}

\subsection{From \texorpdfstring{$\varepsilon$}{epsilon} to \texorpdfstring{$\delta$}{delta}}

For this explanation, let us fix a local Hamiltonian $H\in \Lcal_{1,a}$ that determines our time evolution and let us fix $0 < b < a$. This is the decay parameter for the class $\Bcal_b$, which the counter-diabatic driving unitary $U_n$ is aimed to belong to. Indeed, the interaction for $U_n$ will depend on $b$ and $U_{n,b} \in \Lcal_{s,b}$ is the condition that we want to be satisfied, together with a traceable bound on the $A_\alpha$'s. 

The main observation is that the adiabatic expansion is dependent on the shape of the lattice whereas the adiabatic parameter $\varepsilon >0$ is not. However, in disguise, $\varepsilon$ models the step size that we go from order to order so that this is a problem when it comes to error estimates. The solution is that we go from the given adiabatic parameter $\varepsilon>0$ to another parameter $\delta(\varepsilon) \geq \varepsilon$, which depends on the dimensionality of the lattice, such that $\delta(\varepsilon) \to 0$ as $\varepsilon \to 0$ but not as fast. The adiabatic expansion will feature the new parameter $\delta$ instead of $\varepsilon$.

We want to prove an optimal truncation result, i.e., we want to perform the adiabatic expansion for arbitrary $n\in \Nbb$ and then choose the optimal $n$ depending on $\varepsilon$. In the situation of a polynomially growing lattice, we can expect the optimal truncation to be roughly located at
\begin{align}
n_{\mathrm{opt}} := \frac{1}{\delta^{\nicefrac 1{\ell(d)}}}, \label{AT:nopt_definition}
\end{align}
where $\ell(d)$ depends on the dimensionality of the lattice.

Suppose $\Delta_{a - b}\colon \Rbb_+ \ra\Rbb_+$ is a sufficiently smooth and monotonically increasing function that models the shape of the lattice. Then, the quotient
\begin{align*}
\frac{\delta}{\Delta_{a-b}( \delta^{-\nicefrac 1{\ell(d)}})}
\end{align*}
vanishes in the limit $\delta \to 0$ and by the inverse function theorem, there is a $\delta_\varepsilon >0$ such that
\begin{align}
\frac{\delta_\varepsilon}{\Delta_{a-b}(\delta_\varepsilon^{-\nicefrac 1{\ell(d)}})} = \varepsilon. \label{AT:delta_epsilon_equation}
\end{align}

For this reason, our goal is therefore to provide the adiabatic expansion for $n\in \Nbb$ and an arbitrary adiabatic parameter $\delta>0$ with the quotient
\begin{align*}
\frac{\delta}{\Delta_{a-b}(n)}
\end{align*}
and eventually choose the optimal $n$ so that by \eqref{AT:delta_epsilon_equation}, we actually solve the adiabatic equation.

Before we state the result, let us perform a toy calculation, which makes the concept a little more concrete. Assuming that
\begin{align*}
\Delta_{a-b}(x) \sim V_{x^{\nicefrac 1{\ell(d)}}}(a -b) \sim x^{\frac{d}{s\ell(d)}},
\end{align*}
which holds by Lemma \ref{AT:Vb_concrete_estimate}, we have
\begin{align*}
\frac{\delta}{\Delta_{a-b}(\delta^{-\nicefrac 1{\ell(d)}})} \sim \delta^{1 + \frac{d}{s\ell(d)}},
\end{align*}
which by \eqref{AT:delta_epsilon_equation} implies that
\begin{align}
\delta \sim \varepsilon^{ \frac{s\ell(d)}{d + s\ell(d)}} .
\end{align}
This means that we lose a little bit of decay in our end result, since $\delta$ tends to $0$ slower than $\varepsilon$.

\subsection{A new Ansatz for the expansion}

With the decomposition \eqref{AT:delta_epsilon_equation}, we formulate a new lemma, which features the constant $\Delta_{a-b}$ in the counter-diabatic driving. We remark again that we do not solve the adiabatic equation unless $n$ is equal to the optimally chosen $n_{\mathrm{opt}}$ in \eqref{AT:nopt_definition}.

\begin{lem}
\label{AT:Adiabatic_expansion}
Let $\delta >0$ and for $n\in \Nbb$ let $\Delta(n) > 0$ be given. Let $I$ be an interval, let $H\in \Lcal_1(I)$ be a gapped self-adjoint local Hamiltonian in the sense of Definition \ref{AT:Definition_gapped_Hamiltonian} and let $P$ denote the spectral projection onto the spectral patch $\Sigma_0$. Then, there are self-adjoint operators $A_\alpha$, $1\leq  \alpha\leq n$, such that
\begin{align*}
\Pi_{n,\delta} &:= U_{n,\delta} \, P \, U_{n,\delta}^*, &  U_{n,\delta} := \exp\Bigl( \frac{\i}{\Delta(n)} \sum_{\alpha = 1}^n \delta ^\alpha A_\alpha\Bigr),
\end{align*}
solves 
\begin{align*}
\i \, \frac{\delta}{\Delta(n)} \, \dot \Pi_{n,\delta} = [ H + R_{n,\delta}, \Pi_{n,\delta}],
\end{align*}
where $R_{n, \delta}$ is of the order $\delta^{n+1}$.
\end{lem}

\begin{proof}
We essentially repeat the computation in \cite[Lemma 4.3]{SvenAdiabatic}. Drop $\delta$ from the notation. Then
\begin{align*}
\i \frac{\delta}{\Delta(n)} \dot \Pi_n &= \i \frac{\delta}{\Delta(n)} \, \dot U_n \, P \, U_n^* + \i \frac{\delta}{\Delta(n)} \,  U_n P \,\dot U_n - \frac{\delta}{\Delta(n)} \, U_n \, [K,P] \, U_n \\
&= [H,\Pi_n] + \Bigl[ \i \frac{\delta}{\Delta(n)} \, \dot U_n \, U_n^* - \frac{\delta}{\Delta(n)} \, U_n \, K \, U_n^* + (U_nHU_n^* - H),\Pi_n\Bigr].
\end{align*}
Here, we used that $U_n \dot U_n^* = -\dot U_n U_n^*$ and $[U_nHU_n,\Pi_n] = U_n[H,P] U_n = 0$ as well as $\dot P = \i[K,P]$, where $K := \Ical_{s, \gamma}(\dot H)$. Next, we write the second commutator as
\begin{align*}
\Bigl[ \i \frac{\delta}{\Delta(n)} \dot U_n U_n^* - \frac{\delta}{\Delta(n)} U_nKU_n^* + (U_nHU_n^* - H),\Pi_n\Bigr] & \\
&\hspace{-100pt} = U_n \Bigl[ \i \frac{\delta}{\Delta(n)} U_n^* \dot U_n - \frac{\delta}{\Delta(n)} K + H - U_n^* HU_n, P\Bigr] U_n^*
\end{align*}
Let us make use of the following expansion: For self-adjoint operators $S$ and $T$, $\lambda\in \Rbb$, and $n \in \Nbb_0$, we have
\begin{align}
\e^{-\i \lambda S} \, T \, \e^{\i \lambda S} &= \sum_{k=0}^n \frac{(-\i)^k \lambda^k}{k!} \ad_S^k(T) \notag \\
&\hspace{30pt} + (-\i)^{n+1} \int_0^\lambda \dd \lambda_1 \; \cdots \int_0^{\lambda_n} \dd \lambda_{n+1} \; \e^{-\i \lambda_{n+1}S} \, \ad_S^{n+1}(T) \, \e^{-\i \lambda_{n+1}S}. \label{AT:Adiabatic_expansion_1}
\end{align}
The proof an induction argument and left to the reader. Consequently, we have
\begin{align*}
U^* H U = \sum_{k=0}^n \frac{(-\i)^k}{k!} \ad_S^k(H) + h_{n+1}(\delta)
\end{align*}
with
\begin{align}
h_{n+1}(\delta) = (-\i)^{n+1} \int_0^1\dd\lambda_1 \int_0^{\lambda_1} \dd \lambda_2 \; \cdots \int_0^{\lambda_n} \dd \lambda_{n+1} \; \e^{-\i \lambda_{n+1} S} \ad_S^{n+1}(H) \, \e^{\i \lambda_{n+1} S}. \label{AT:Adiabatic_expansion_h_error_definition}
\end{align}
Furthermore, Duhamel's formula (see \cite{Duhamel}) reads
\begin{align*}
\frac \dd {\dd u} \e^{-\i S(u)} = \i \int_0^1 \dd \lambda \; \e^{\i ( 1 - \lambda) S(u)} \, \dot S(u) \, \e^{\i \lambda S(u)},
\end{align*}
whence
\begin{align*}
\i \, U_n^* \dot U_n = - \int_0^1 \dd \lambda \e^{-\i \lambda S} \, \dot S \, \e^{\i \lambda S}
\end{align*}
and, by \eqref{AT:Adiabatic_expansion_1} applied to $n-1$, we obtain
\begin{align*}
\i \, U_n^*\dot U_n = \sum_{k=0}^{n-1} \frac{\i^k(-1)^{k+1}}{(k+1)!} \ad_S^k(\dot S) + \delta^{-1} \, q_{n+1}(\delta)
\end{align*}
with 
\begin{align}
q_{n+1}(\delta) = - (-\i)^{n+1} \delta \int_0^1\dd\lambda_1 \; \cdots \int_0^{\lambda_n} \dd \lambda_{n+1} \; \e^{-\i \lambda_{n+1} S} \ad_S^n(\dot S) \, \e^{\i \lambda_{n+1} S} \label{AT:Adiabatic_expansion_q_error_definition}
\end{align}
Then, inserting $S = \frac{1}{\Delta(n)} \sum_{\alpha =1}^n \delta^\alpha A_\alpha$, we get
\begin{align*}
U_n^* H U_n = \sum_{k=0}^n \frac{(-\i)^k}{k! \, \Delta(n)^k} \Bigl( \sum_{\alpha =1}^n \delta^\alpha \ad_{A_\alpha}\Bigr)^k (H) + h_{n+1}(\delta) =: \sum_{\alpha=0}^n \delta^\alpha H_\alpha + h_{n+1}(\delta) + \tilde h_{n+1}(\delta),
\end{align*}
where $h_{n+1}(\delta)$ is defined in \eqref{AT:Adiabatic_expansion_h_error_definition}, 
\begin{align*}
H_\alpha &:= \sum_{k=0}^\alpha \frac{(-\i)^k}{k! \, \Delta(n)^k} \sum_{\substack{j\in \Nbb^k \\ |j| =\alpha}} \ad_{A_{j_k}} \cdots \ad_{A_{j_1}}(H), & 0 &\leq \alpha \leq n,
\end{align*}
and
\begin{align*}
\tilde h_{n+1}(\delta) &:= \sum_{k=0}^n \frac{(-\i)^k}{k! \, \Delta(n)^k} \sum_{\substack{j\in \Nbb^k \\ |j| \geq n+1}} \delta^{|j|}\, \ad_{A_{j_k}} \cdots \ad_{A_{j_1}}(H).
\end{align*}
In the same manner, we obtain
\begin{align*}
\i \, U_n^* \dot U_n &= \sum_{k=0}^{n-1} \frac{\i^k(-1)^{k+1}}{(k+1)! \, \Delta(n)^{k+1}} \Bigl( \sum_{\alpha=1}^n \delta^\alpha \ad_{A_\alpha} \Bigr)^k \Bigl( \sum_{j=1}^n \delta^j \dot A_j\Bigr) + \delta^{-1} \, q_{n+1} (\delta)  \\
&=: \sum_{\alpha=1}^{n-1} \delta^\alpha Q_\alpha + \delta^{-1} \, q_{n+1}(\delta) + \delta^{-1} \, \tilde q_{n+1}(\delta),
\end{align*}
where $q_{n+1}(\delta)$ is defined in \eqref{AT:Adiabatic_expansion_q_error_definition},
\begin{align}
Q_\alpha &:= -\i \sum_{k=1}^n \frac{(-\i)^k}{k! \, \Delta(n)^{k}} \sum_{\substack{j\in \Nbb^k \\ |j| = \alpha}} \ad_{A_{j_k}} \cdots \ad_{A_{j_2}} (\dot A_{j_1}), & 1 &\leq  \alpha \leq n, \label{AT:Qalpha_definition}
\end{align}
and 
\begin{align*}
\tilde q_{n+1}(\delta) := -\i \delta \sum_{k=1}^n \frac{(-\i)^k}{k! \, \Delta(n)^{k}} \sum_{\substack{j\in \Nbb^k \\ |j| \geq n}} \delta^{|j|}\, \ad_{A_{j_k}} \cdots \ad_{A_{j_2}} (\dot A_{j_1}).
\end{align*}
With the definition $Q_0 := -K$, we obtain
\begin{align*}
\i  \frac{\delta}{\Delta(n)} U_n^* \dot U_n - \frac{\delta}{\Delta(n)} K + H - U_n^* H U_n & \\
&\hspace{-120pt} =  \sum_{\alpha =1}^{n} \delta^\alpha \bigl(\frac{ Q_{\alpha-1}}{\Delta(n)} - H_\alpha\bigr) - h_{n+1}(\delta) - \tilde h_{n+1}(\delta) + q_{n+1}(\delta) + \tilde q_{n+1}(\delta).
\end{align*}
Hence, the lemma follows if we can choose $A_\alpha$ in such a way that it solves the equation
\begin{align*}
\Bigl[ \frac{Q_{\alpha-1}}{\Delta(n)} - H_\alpha, P \Bigr] =0,
\end{align*}
for all $\alpha = 1, \ldots, n$. Let us prove this by induction. The case $\alpha = 1$ reads
\begin{align*}
0=\Bigl[ - \frac{K}{\Delta(n)} - \bigl( -\frac{\i}{\Delta(n)} [A_1,H]\bigr) , P\Bigr] = -\frac{1}{\Delta(n)} \bigl[ K - \i [A_1,H],P\bigr]
\end{align*}
This is solved by choosing $A_1 := \Ical_{s, \gamma}(K)$ for any $0 < s < 1$. Suppose now that $A_1,\ldots, A_{\alpha-1}$ have been constructed. Then, isolating the dependence on $A_\alpha$, we have
\begin{align*}
H_\alpha = \frac{\i}{\Delta(n)} [A_\alpha,H] + \frac 1 {\Delta(n)} L_\alpha
\end{align*}
with
\begin{align}
L_\alpha := \sum_{k=2}^n \frac{(-\i)^k}{k! \, \Delta(n)^{k-1}} \sum_{\substack{j\in \Nbb^k \\ |j| = \alpha }} \ad_{A_{j_k}} \cdots \ad_{A_{j_1}} (H). \label{AT:Lalpha_definition}
\end{align}
Then, we need to solve the equation
\begin{align*}
0 = \Bigl[ \frac{Q_{\alpha-1}}{\Delta(n)} - \frac{\i}{\Delta(n)} [A_\alpha,H] - \frac{1}{\Delta(n)} L_\alpha, P\Bigr]  = \frac{1}{\Delta(n)} \bigl[ Q_{\alpha - 1} - L_\alpha - \i[A_\alpha,H], P\bigr],
\end{align*}
which is solved by choosing
\begin{align*}
A_\alpha &:= \Ical_{s, \gamma}(Q_{\alpha-1} - L_\alpha). \qedhere
\end{align*}
\end{proof}

\subsection{Inductive estimate on \texorpdfstring{$A_\alpha$}{Aalpha} pretending that \texorpdfstring{$Q_\alpha =0$}{Qalpha=0}}

In the preceding section, we have constructed self-adjoint operators $A_1, \ldots, A_n$ in an inductive procedure. We assume that the Hamiltonian $H$ has an interaction $\Phi_H$ that belongs to the decay class $\Bcal_{1, a'}$ for some $a'>0$ and a spectral gap $\gamma>0$ as in Definition \ref{AT:Definition_gapped_Hamiltonian}. In the following, we want to show that this further implies that $A_1, \ldots, A_n$ are local Hamiltonians, which satisfy a certain decay estimate.

To do this, let us make some preparations. Fix the following parameters arbitrarily:
\begin{align}
0 &< s < 1, & 0 &< \mu < \mu_0(s), & 0 &< a < a'. \label{AT:Induction_Parameter_choice}
\end{align}
Since, by Theorem \ref{AT:I_estimate_local_Hamiltonian}, $\Ical_{s, \gamma}$ maps into the space $\Bcal_{s, b}$, where $b < \eta_{s, \gamma}(a, \mu)$ with $\eta_{s, \gamma}(a, \mu)$ in \eqref{AT:eta_definition}, the decay classes of the $A_\alpha$'s are capped at $\eta_{s, \gamma}(a,\mu)$.

The goal we should have in mind is to prove an estimate for $A_n$ at a given decay rate $b_n < \eta_{s, \gamma}(a, \mu)$. In order to achieve this, we throw in points $b_\alpha$, $\alpha = 1, \ldots, n$, according to
\begin{align*}
b_n < b_{n-1} < \cdots < b_2 < b_1 < b_0 := \eta_{s, \gamma}(a, \mu),
\end{align*}
see also Figure~\ref{AT:Figure_bs} below, where the $b_i$ are equidistantly distributed, i.e., if
\begin{align}
\Theta_{s, \gamma, n}(a, \mu) &:= \frac{\eta_{s, \gamma}(a, \mu) - b_n}{n}, \label{AT:Induction_Theta_definition_cheat}
\end{align}
then 
\begin{align}
b_{\alpha-1} - b_\alpha &= \Theta_{s, \gamma, n}(a, \mu), & \frac{\eta_{s, \gamma}(a, \mu) - b_\alpha}{\alpha} &= \Theta_{s, \gamma, n}(a,\mu), & \alpha &= 1, \ldots, n. \label{AT:Induction_balpha_distance_cheat}
\end{align}
Furthermore, we introduce a number $q \in \Nbb$, which is representing the number of operations that it takes to construct $A_\alpha$ from $A_{\alpha - 1}$. Therefore, we insert
\begin{align}
b_\alpha = b_\alpha^{(0)} < b_\alpha^{(1)} < b_\alpha^{(2)} < \cdots < b_\alpha^{(q)} = b_{\alpha - 1} \label{AT:Induction_balpha_definition_cheat}
\end{align}
in an equidistant fashion, i.e., we have
\begin{align}
b_\alpha^{(i)} - b_\alpha^{(i-1)} = \frac{b_{\alpha - 1} - b_\alpha}{q} &= \frac{\Theta_{s, \gamma, n}(a, \mu)}{q}, & i &= 1, \ldots, q. \label{AT:Induction_balphamu_distance_cheat}
\end{align}
In this digression, we have $q = 2$ but this may change when we take $Q_{\alpha - 1}$ into account in the future. The construction is summarized in Figure \ref{AT:Figure_bs}.
\begin{figure}[h]
\centering
\ifthenelse{\equal\masterfile{Diss}}
{
\includegraphics[width=15cm]{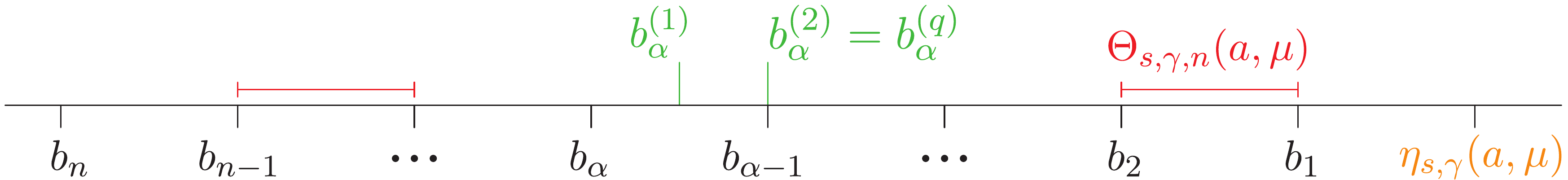}
}{
\includegraphics[width=14cm]{img/Induktion_gecheated.pdf}
}
\caption{The differently chosen $b_\alpha$'s.\label{AT:Figure_bs}}
\end{figure}
From now on, we write $\eta$ and $\Theta_n$ in place of \eqref{AT:eta_definition} and \eqref{AT:Induction_Theta_definition_cheat} for the sake of readability.

For constants $C_1, C_2, C_3, C_4, C_5 >0$ to be chosen, we assume that $b_n$ is such that
\begin{align}
\Theta_n &\leq \min \bigl\{ \frac d{s\, \e} \, , C_2 \, , \, C_4\bigr\} \label{AT:Induction_hypothesis_Theta}
\end{align}
and we want to show inductively that then, for all $\alpha = 1, \ldots, n$ and $0 < b \leq b_\alpha$, we have
\begin{align}
\Vert A_\alpha\Vert_b &\leq C_1^{\alpha + 1} \; \bigl( \frac{C_2 \, \alpha}{\Theta_n}\bigr)^{C_3 (\alpha + 2)} \, \bigl( \frac{C_4}{\Theta_n}\bigr)^{C_5(\alpha + 1)}.  \label{AT:Induction_hypothesis_cheat}
\end{align}
We are going to show \eqref{AT:Induction_hypothesis_cheat} for $\alpha = 1$ at the end. Let us assume that \eqref{AT:Induction_hypothesis_cheat} holds for every $1 \leq j \leq \alpha - 1$ and we are going to show \eqref{AT:Induction_hypothesis_cheat} for $b = b_\alpha$ since it then trivially holds for $0 < b \leq b_\alpha$ as well by the monotonicity of the norm $\Vert \cdot\Vert_b$. We emphasize that we do this under the assumption that $Q_\alpha = 0$, where $Q_\alpha$ is defined in \eqref{AT:Qalpha_definition}, so that derivatives are not going to play any role. This amounts to the assumption $A_\alpha = -\Ical_{s, \gamma}(L_\alpha)$, where $L_\alpha$ is defined in \eqref{AT:Lalpha_definition}. Furthermore, we work under the polynomial growth condition Assumption \ref{AT:Assumption_concrete_polynomial_growth} as well as the following proposition.

\begin{prop}
\label{AT:Induction_Function_estimate_cheated}
Let $p \geq 1$. Then, there is a constant $D_p>0$ such that for any $\alpha\in \Nbb$, we have
\begin{align*}
\sum_{k=2}^\alpha \sum_{\substack{j\in \Nbb^k \\ |j| = \alpha}} j_1^{p j_1} \cdots j_k^{pj_k} \leq D_p \; \alpha^{p\alpha}. 
\end{align*}
\end{prop}

We postpone the proof of Proposition \ref{AT:Induction_Function_estimate_cheated} to the end of this section and start with the induction argument. We note that Theorem \ref{AT:I_estimate_local_Hamiltonian} as well as \eqref{AT:Induction_balpha_distance_cheat} and \eqref{AT:Induction_balphamu_distance_cheat} imply
\begin{align*}
\Vert A_\alpha\Vert_{b_\alpha} = \Vert \Ical_{s, \gamma}(L_\alpha)\Vert_{b_\alpha} &\leq D^\Delta \; F(\eta - b_\alpha) \; V_1(b_\alpha^{(1)} - b_\alpha) \; \Vert L_\alpha\Vert_{b_\alpha^{(1)}} \\
&= D^\Delta \; F(\alpha \, \Theta_n) \; V_q(\Theta_n) \; \Vert L_\alpha\Vert_{b_\alpha^{(1)}}.
\end{align*}
Here, $D^\Delta$ is from \eqref{AT:I_DDelta_definition}. Furthermore, by the triangle inequality and Theorem \ref{AT:Commutator_Estimate}, we have
\begin{align*}
\Vert L_\alpha\Vert_{b_\alpha^{(1)}} &\leq \sum_{k=2}^n \frac{1}{k! \, \Delta(n)^{k-1}} \sum_{\substack{j\in \Nbb^k \\ |j| = \alpha}} \Vert \ad_{A_{j_k}} \cdots \ad_{A_{j_1}} (H)\Vert_{b_\alpha^{(1)}} \\
&\leq \sum_{k=2}^\alpha \frac{4^k}{k!} \frac{V_k(b_\alpha^{(2)} - b_\alpha^{(1)})^k}{ \Delta(n)^{k-1}} \sum_{\substack{j\in \Nbb^k \\ |j| = \alpha}} \Vert A_{j_k}\Vert_{b_\alpha^{(2)}} \; \cdots \; \Vert A_{j_1}\Vert_{b_\alpha^{(2)}} \; \Vert H\Vert_{b_\alpha^{(2)}}.
\end{align*} 
We note that $V_k(b_\alpha^{(2)} - b_\alpha^{(1)}) = V_{qk}(\Theta_n)$ and use $\Vert H\Vert_{b_\alpha^{(2)}} \leq \Vert H\Vert_a$. Then, we apply the induction hypothesis \eqref{AT:Induction_hypothesis_cheat} and obtain
\begin{align}
\Vert A_\alpha\Vert_{b_\alpha} &\leq D^\Delta \; F(\alpha \, \Theta_n) \; V_q(\Theta_n) \; C_1^\alpha \; \Vert H\Vert_a \; \sum_{k=2}^\alpha \frac{4^k}{k!} \frac{(C_1 \, V_{q k}(\Theta_n))^k}{\Delta(n)^{k-1}} \notag \\
&\hspace{-30pt} \times \sum_{\substack{j\in \Nbb^k \\ |j| = \alpha}} \bigl( \frac{C_2 \, j_1}{\Theta_n}\bigr)^{C_3(j_1 + 2)} \cdots \bigl( \frac{C_2 \, j_k}{\Theta_n}\bigr)^{C_3(j_k + 2)} \; \bigl( \frac{C_4}{\Theta_n}\bigr)^{C_5(j_1 + 1)} \cdots \bigl( \frac{C_4}{\Theta_n}\bigr)^{C_5(j_k + 1)}. \label{AT:Induction_2}
\end{align}
The term in the sum of the second line in \eqref{AT:Induction_2} equals
\begin{align}
&\bigl( \frac{C_2}{\Theta_n}\bigr)^{C_3(\alpha + 2k)} \bigl( \frac{C_4}{\Theta_n}\bigr)^{C_5(\alpha + k)} \; j_1^{C_3j_1} \cdots j_k^{C_3j_k} \; (j_1 \cdots j_k)^{2C_3}. \label{AT:Induction_4}
\end{align}
The last factor is bounded by $\alpha^{2C_3k}$, whence \eqref{AT:Induction_2} and \eqref{AT:Induction_4} yield
\begin{align}
\Vert A_\alpha\Vert_{b_\alpha} &\leq D^\Delta \; F(\alpha \, \Theta_n) \; V_{q}(\Theta_n) \; C_1^\alpha \; \Vert H\Vert_a \; \bigl( \frac{C_2}{\Theta_n}\bigr)^{C_3\alpha} \; \bigl( \frac{C_4}{\Theta_n} \bigr)^{C_5\alpha} \notag \\
&\hspace{10pt} \times \sum_{k=2}^\alpha \frac{4^k}{k!} \frac{(C_1 \,   \, V_{q k}(\Theta_n))^k}{\Delta(n)^{k-1}} \bigl( \frac{C_2 \, \alpha}{\Theta_n} \bigr)^{2C_3k} \bigl( \frac{C_4}{\Theta_n}\bigr)^{C_5k} \sum_{\substack{j\in \Nbb^k \\ |j| = \alpha}} j_1^{C_3j_1} \cdots j_k^{C_3 j_k} . \label{AT:Induction_1}
\end{align}
Since $\Theta_n \leq \frac d{s\, \e}$, see \eqref{AT:Induction_hypothesis_Theta}, 
Lemma \ref{AT:Vb_concrete_estimate} implies
\begin{align}
V_{q k}(\Theta_n) \leq \kappa \bigl( \frac{d q \, k }{s \e \, \Theta_n}\bigr)^{\frac{d}{s}} \leq \kappa \bigl( \frac{C_2 \, k}{\Theta_n}\bigr)^{C_3} 
\label{AT:Induction_Vmunk_estimate_cheat}
\end{align}
provided that $C_2$ and $C_3$ are chosen such that
\begin{align}
C_2 &:= \frac{d q}{s \e} , & C_3 &\geq \frac{d}{s} \geq 1. \label{AT:Induction_choice_C2C3_cheat}
\end{align}
For $\rho \geq 4$ to be chosen, we define
\begin{align}
\Delta(n) &:= \Bigl( C_1 \; \kappa \; \bigl( \frac{C_2\, n}{\Theta_n}\bigr)^{2C_3} \; \bigl( \frac{C_4}{\Theta_n} \bigr)^{C_5} \Bigr)^\rho. \label{AT:Induction_choice_Deltan_cheat}
\end{align}
Then, a short calculation using that $\rho(k-1) \geq k$ for $k\geq 2$ shows that \eqref{AT:Induction_Vmunk_estimate_cheat}, \eqref{AT:Induction_choice_C2C3_cheat}, and \eqref{AT:Induction_choice_Deltan_cheat} and the hypothesis \eqref{AT:Induction_hypothesis_Theta} imply
\begin{align*}
\frac{(C_1 \, V_{qk}(\Theta_n))^k}{\Delta(n)^{k-1}} \; \bigl( \frac{C_2\, \alpha}{\Theta_n} \bigr)^{2C_3k} \; \bigl( \frac{C_4}{\Theta_n}\bigr)^{C_5k} \leq 1,
\end{align*}
since $k\leq \alpha \leq n$. We combine \eqref{AT:Induction_1} with Proposition \ref{AT:Induction_Function_estimate_cheated} and obtain
\begin{align}
\Vert A_\alpha\Vert_{b_\alpha} &\leq D_{C_3} \; D^\Delta \; F(\alpha\, \Theta_n) \; V_q(\Theta_n) \; C_1^\alpha \, \bigl( \sup_{k\in \Nbb} \frac{4^k}{k!}\bigr) \, \Vert H\Vert_a \, \bigl( \frac{C_2 \, \alpha}{\Theta_n}\bigr)^{C_3\alpha} \; \bigl( \frac{C_4}{\Theta_n}\bigr)^{C_5\alpha} . \label{AT:Induction_7}
\end{align}
We have $F(\alpha \Theta_n) \leq F(\Theta_n)$ and, by Lemma \ref{AT:F_concrete_estimate} and \eqref{AT:Induction_hypothesis_Theta}, $F(\Theta_n)$ is bounded by
\begin{align}
F\bigl( \frac{\Theta_n} q\bigr) &\leq 2^{\frac ds} \kappa \; \Gamma\bigl(1 + \frac 1s\bigr) \; \bigl( \frac{d q}{s\e \, \Theta_n}\bigr)^{\frac ds} \; \bigl( \frac{4q}{\Theta_n} \bigr)^{\frac 1s} \leq 2^{\frac ds} \kappa \; \Gamma\bigl(1 + \frac 1s\bigr) \; \bigl( \frac{C_2}{\Theta_n}\bigr)^{C_3} \; \bigl( \frac{C_4}{\Theta_n} \bigr)^{C_5} \label{AT:Induction_F_estimate}
\end{align}
with the choice
\begin{align}
C_4 &:= 4q , & C_5 &:= \frac 1s. \label{AT:Induction_choice_C4C5_cheat}
\end{align}
When we apply \eqref{AT:Induction_Vmunk_estimate_cheat} with $k =1$, \eqref{AT:Induction_F_estimate} implies
\begin{align}
\Vert A_\alpha\Vert_{b_\alpha} &\leq 2^{\frac ds} \; \kappa^2 \; \Gamma\bigl( 1 + \frac 1s\bigr) \; D_{C_3} \; D^\Delta \; \bigl( \sup_{k\in \Nbb} \frac{4^k}{k!}\bigr) \; \Vert H\Vert_a \; C_1^\alpha \notag\\
&\hspace{140pt} \times \bigl( \frac{C_2 \, \alpha}{\Theta_n} \bigr)^{C_3(\alpha + 2)} \; \bigl( \frac{C_4}{\Theta_n} \bigr)^{C_5 (\alpha + 1)} \label{AT:Induction_5}
\end{align}
Therefore, if $C_1$ is chosen such that
\begin{align}
C_1 &\geq 2^{\frac ds} \; \kappa^2 \; \Gamma\bigl( 1 + \frac 1s\bigr) \;  D_{C_3} \; D^\Delta \; \bigl( \sup_{k\in \Nbb} \frac{4^k}{k!} \bigr) \; \Vert H\Vert_a, \label{AT:Induction_choice_C1_cheat}
\end{align}
then \eqref{AT:Induction_5} implies \eqref{AT:Induction_hypothesis_cheat} for $\alpha$.

We close the induction by showing that \eqref{AT:Induction_hypothesis_cheat} holds for $\alpha = 1$. By Theorem \ref{AT:I_estimate_local_Hamiltonian} applied twice, we obtain
\begin{align*}
\Vert A_1\Vert_{b_1} = \Vert \Ical_{s, \gamma}(K)\Vert_{b_1} \leq D^\Delta \, V_1(b_1^{(1)} - b_1) \, F(\eta - b_1) \, \Vert K\Vert_{b_1^{(1)}}
\end{align*}
as well as
\begin{align*}
\Vert K\Vert_{b_1^{(1)}} = \Vert \Ical_{s, \gamma}(\dot H)\Vert_{b_1^{(1)}} \leq D^\Delta \, V_1(b_1^{(1)} - b_1^{(2)}) \, F(\eta - b_1^{(1)}) \, \Vert \dot H\Vert_{b_1^{(2)}}.
\end{align*}
Combining these estimates, we have
\begin{align*}
\Vert A_1\Vert_{b_1} \leq (D^\Delta)^2 \, \Vert \dot H\Vert_\eta \, V_q(\Theta_n)^2 \, F\bigl( \frac{\Theta_n}{q}\bigr)^2
\end{align*}
When we apply \eqref{AT:Induction_Vmunk_estimate_cheat} with $k =1$ and \eqref{AT:Induction_F_estimate} with $\alpha = 1$, we arrive at
\begin{align}
\Vert A_1\Vert_{b_1} \leq \bigl(2^{\frac ds} \, \kappa^2 \, D^\Delta \, \Gamma\bigl( 1 + \frac 1s\bigr)  \bigr)^2 \, \Vert \dot H\Vert_\eta \, \bigl( \frac{C_2}{\Theta_n}\bigr)^{\frac{4d}s} \bigl( \frac{C_4}{\Theta_n} \bigr)^{\frac 2s}. \label{AT:Induction_6}
\end{align}
Thus, in light of \eqref{AT:Induction_choice_C2C3_cheat} and \eqref{AT:Induction_choice_C1_cheat}, we make the choices
\begin{align*}
C_1 &:= 2^{\frac ds} \, \kappa^2 \, \Gamma\bigl( 1 + \frac 1s\bigr) D^\Delta \; \max\Bigl\{ \sqrt{\Vert \dot H\Vert_\eta} \; , \; D_{C_3} \, \bigl( \sup_{k\in \Nbb} \frac{4^k}{k!}\bigr) \, \Vert H\Vert_a\Bigr\}
\end{align*}
and
\begin{align*}
C_3 &:= \frac{4d}{3s},
\end{align*}
whence \eqref{AT:Induction_6} implies \eqref{AT:Induction_hypothesis_cheat} for $\alpha = 1$.

\begin{proof}[Proof of Proposition \ref{AT:Induction_Function_estimate_cheated}]
Consider the function
\begin{align}
f_\alpha^p(x) &:= x^{px} \, (\alpha - x)^{p(\alpha- x)}, & 0 &\leq x \leq \alpha.
\end{align}
This function is symmetric about $x = \nicefrac \alpha 2$ and since $(f_\alpha^p)'(x) = f_\alpha^p(x) \, p \, [\log(x) - \log(\alpha - x)]$ for $0 < x < \alpha$, we conclude that $f_\alpha^p$ is strictly decreasing on $[0, \nicefrac \alpha 2)$. We claim that
\begin{align}
\sum_{j=1}^{\alpha - 1} f_\alpha^p(j) \leq \min \bigl\{ 2 + 4^p \, , \, \alpha\bigr\} \, f_\alpha^p(1). \label{AT:Induction_Function_estimate_cheated_1}
\end{align}
To see this, the left side is bounded by $2 \, f_\alpha^p(1) + (\alpha - 3) f_\alpha^p(2)$ due to the monotonicity of $f_\alpha^p$. On the one hand, we further have $(\alpha - 2) f_\alpha^p(2) \leq 4^p f_\alpha^p(1)$, which proves the first bound, while on the other hand, the monotonicity implies
\begin{align*}
2 \, f_\alpha^p(1) + (\alpha - 2) f_\alpha^p(2) = \alpha \, f_\alpha^p(1) - (\alpha - 2) \bigl( f_\alpha^p(1) - f_\alpha^p(2)\bigr) \leq \alpha \, f_\alpha^p(1).
\end{align*}
This proves \eqref{AT:Induction_Function_estimate_cheated_1}. For $2 \leq k \leq \alpha$ we claim that
\begin{align}
S_{\alpha, k}^p := \sum_{\substack{j\in \Nbb^k \\ |j| = \alpha}} j_1^{pj_1} \cdots j_k^{pj_k} \leq \min \bigl\{ 2+4^p, \alpha\bigr\} \, f_\alpha^p(1). \label{AT:Induction_Function_estimate_cheated_2}
\end{align}
We prove this by induction. Since $S_{\alpha, 2}^p = \sum_{j=1}^{\alpha - 1} f_\alpha^p(j)$, the case $k =2$ is \eqref{AT:Induction_Function_estimate_cheated_1}. By the second case of the induction hypothesis \eqref{AT:Induction_Function_estimate_cheated_2}, we further have
\begin{align*}
S_{\alpha, k+1}^p = \sum_{j = 1}^{\alpha - k} j^{pj} S_{k, \alpha -j}^p \leq \sum_{j=1}^{\alpha - k} j^{pj}\, (\alpha- j) \, f_{\alpha - j}^p(1) \leq \sum_{j=1}^{\alpha - 1} f_{\alpha}^p(j).
\end{align*}
Here, we used that $(\alpha - j) f_{\alpha - j}^p(1) \leq (\alpha - j)^{p(\alpha -j)}$. A further application of \eqref{AT:Induction_Function_estimate_cheated_1} proves \eqref{AT:Induction_Function_estimate_cheated_2}. Summing \eqref{AT:Induction_Function_estimate_cheated_2} over $k$ completes the proof.
\end{proof}

\ifthenelse{\equal\masterfile{Diss}}{}{

\subsection{Induction with derivatives}

In the preceding section, we have constructed $n\in \Nbb$ self-adjoint operators $A_1, \ldots, A_n$ in an inductive procedure. We assume that the Hamiltonian $H$ has an interaction $\Phi_H$ that belongs to the decay class $\Bcal_{1, a'}$ for some $a'>0$ and a spectral gap $\gamma>0$ as in Definition \ref{AT:Definition_gapped_Hamiltonian}. We also assume that $\Phi_H$ is setwise analytic, which, by Corollary \ref{AT:Hamiltonian_Analytic_derivatives}, implies that $\Phi_H^{(k)} \in \Bcal_{s, a'}$ for all $k\in \Nbb$. In the following, we want to show that this further implies that $A_1, \ldots, A_n$ are local Hamiltonians, which satisfy a certain decay estimate.

To do this, let us make some preparations. Fix the following parameters arbitrarily:
\begin{align}
0 &< s < 1, & 0 &< \mu < \mu_0(s), & 0 &< a < a'. \label{AT:Induction_Parameter_choice}
\end{align}

Let $1 \leq \beta \leq n$ denote the order of derivative. Since, by Theorem \ref{AT:I_betath_derivative_estimate}, $\frac{\dd^{\beta - 1}}{\dd u^{\beta - 1}}\Ical_{s, \gamma}$ maps into the space $\Bcal_{s, b}$, where $b < \eta_{s, \gamma}^n(a, \mu)$ for all $1 \leq \beta \leq n$ with $\eta_{s, \gamma}^n(a, \mu)$ in \eqref{AT:etak_definition}, the decay classes of the $A_\alpha^{(\beta)}$'s are capped at $\eta_{s, \gamma}^n(a,\mu)$ (since $n$ derivatives fall on $A_1$ in the scope of the induction process).

The goal we should have in mind is to prove an estimate for $A_n$ at a given decay rate $b_n < \eta_{s, \gamma}^n(a, \mu)$. In order to achieve this, we throw in points $b_\alpha$, $\tilde b_\alpha$, $\alpha = 1, \ldots, n$, according to
\begin{align*}
b_n < b_{n-1} < \cdots < b_2 < b_1 < b_0 := \eta_{s, \gamma}^n(a, \mu),
\end{align*}
see also Figure~\ref{AT:Figure_bs} below, where the $b_i$ are equidistantly distributed, i.e., if
\begin{align}
\Theta_{s, \gamma, n}(a, \mu) &:= \frac{\eta_{s, \gamma}^n(a, \mu) - b_n}{2n}, \label{AT:Induction_Theta_definition}
\end{align}
then 
\begin{align}
b_{\alpha-1} - b_\alpha &= 2\Theta_{s, \gamma, n}(a, \mu), & \frac{\eta_{s, \gamma}^n(a, \mu) - b_\alpha}{\alpha} &= 2\Theta_{s, \gamma, n}(a,\mu), & \alpha &= 1, \ldots, n. \label{AT:Induction_balpha_distance}
\end{align}
Furthermore, we introduce an intermediate point $\tilde b_\alpha$ according to
\begin{align}
\tilde b_\alpha := \frac{b_{\alpha - 1} - b_\alpha}{2}, \label{AT:Induction_btildealpha_definition}
\end{align}
since it takes two steps to construct $A_\alpha$ from $A_{\alpha - 1}$. Therefore, we have
\begin{align}
b_{\alpha - 1} - \tilde b_\alpha = \tilde b_\alpha - b_\alpha = \frac{b_{\alpha - 1} - b_\alpha}{2} &= \Theta_{s, \gamma, n}(a, \mu). \label{AT:Induction_btildealpha_distance}
\end{align}
In this digression, we have $q = 2$ but this may change when we take $Q_{\alpha - 1}$ into account in the future. The construction is summarized in Figure \ref{AT:Figure_bs}.
\begin{figure}[h]
\centering
\ifthenelse{\equal\masterfile{Diss}}
{
\includegraphics[width=15cm]{../Exponential_estimates_for_the_adiabatic_theorem/Notes/img/Induktion.pdf}
}{
\includegraphics[width=14cm]{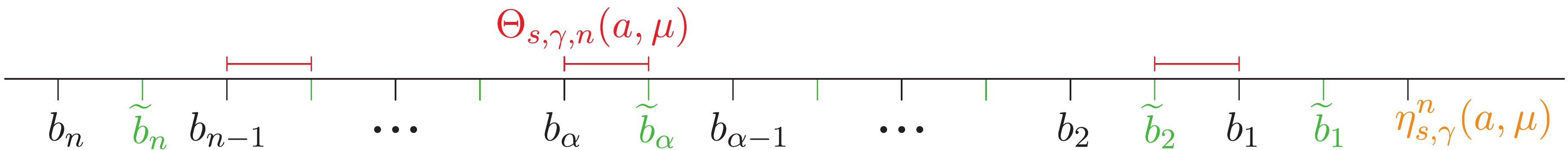}
}
\caption{The differently chosen $b_\alpha$'s.\label{AT:Figure_bs}}
\end{figure}
From now on, we write $\eta^n$ and $\Theta_n$ in place of \eqref{AT:eta_definition} and \eqref{AT:Induction_Theta_definition} for the sake of readability.

For constants $C_1, C_2, C_3, C_4, C_5 >0$ to be chosen, we assume that $b_n$ is such that
\begin{align}
\Theta_n &\leq \min \bigl\{ \frac d{s\, \e} \, , C_2 \, , \, C_4\bigr\}. \label{AT:Induction_hypothesis_Theta}
\end{align}
In the following, we want to show for fixed $1 \leq \beta \leq n$ by induction in $\alpha$ that then, for all $\alpha = 1, \ldots, n$, $\beta = 1, \ldots, n$ such that $\alpha + \beta \leq n$ and $0 < b \leq b_{\alpha + \beta}$, we have ($\nu$ to be chosen)
\begin{align}
\Vert A_\alpha^{(\beta-1)}\Vert_b &\leq C_1^{\alpha + 1} \; \bigl( \frac{C_2 \, (\alpha + \beta)}{\Theta_n}\bigr)^{C_3 (\alpha + 2\beta)} \, \bigl( \frac{C_4}{\Theta_n}\bigr)^{C_5(\alpha + \beta)} \, \beta^{(\beta + 1)\nu} \, \delta^{-\beta} \, \vvvert H\vvvert_{\eta^n}.  \label{AT:Induction_hypothesis}
\end{align}
We are going to show \eqref{AT:Induction_hypothesis} for $\alpha = 1$ at the end. Let us assume that \eqref{AT:Induction_hypothesis} holds for every $1 \leq j \leq \alpha - 1$ and we are going to show \eqref{AT:Induction_hypothesis} for $b = b_{\alpha + \beta}$ since it then trivially holds for $0 < b \leq b_{\alpha + \beta}$ as well by the monotonicity of the norm $\Vert \cdot\Vert_b$. 
Furthermore, we work under the polynomial growth condition Assumption \ref{AT:Assumption_concrete_polynomial_growth} as well as the following assumption.

\begin{asmp}
\label{AT:Induction_Function_estimateed}
Let $p \geq 1$. Then, there is a constant $D_p>0$ such that for any $\alpha\in \Nbb$ and $\beta\in \Nbb$, we have
\begin{align*}
\sum_{k=2}^{\alpha} \smash{\sum_{\substack{j\in \Nbb^k \\ |j| = \alpha}} \sum_{\substack{\ell \in \Nbb_0^k \\ |\ell| = \beta - 1}}} \binom{\beta - 1}\ell (j_1+\ell_1)^{p (j_1+2\ell_1)} \, (j_2+\ell_2)^{p(j_2+2\ell_2)} \cdots (j_k + \ell_k)^{p(j_k + 2\ell_k)} &\\
&\hspace{-100pt}\leq D_p \; (\alpha + \beta)^{p(\alpha + \beta)} 
\end{align*}
as well as
\begin{align*}
\sum_{k=1}^{\alpha - 1} \smash{\sum_{\substack{j\in \Nbb^k \\ |j| = \alpha - 1}} \sum_{\substack{\ell\in \Nbb_0^k \\ |\ell| = \beta - 1}}} \binom{\beta - 1}\ell (j_1 + \ell_1 + 1)^{C_3(j_1 + 2(\ell_1 + 1))} & \\
&\hspace{-150pt} \times (j_2 + \ell_2)^{C_3(j_2 + 2\ell_2)} \cdots (j_k + 2\ell_k)^{C_3(j_k + 2\ell_k)} \leq D_p \, (\alpha + \beta)^{p(\alpha + \beta)}.
\end{align*}
\end{asmp}

Now, we start with the induction argument. We note that Theorem \ref{AT:I_betath_derivative_estimate} as well as \eqref{AT:Induction_balpha_distance} and \eqref{AT:Induction_btildealpha_distance} imply
\begin{align*}
\Vert A_\alpha^{(\beta - 1)}\Vert_{b_{\alpha + \beta}} &= \Bigl\Vert \frac{\dd^{\beta - 1}}{\dd u^{\beta - 1}}\Ical_{s, \gamma}(L_\alpha - Q_{\alpha - 1}) \Bigr\Vert_{b_\alpha} \\
&\leq (D^{\dd \Ical, (1)})^\beta \, D^{\dd I, (2), \beta}(\tilde b_\alpha - b_\alpha) \, F(\eta^\beta - \tilde b_\alpha)^\beta \, \Ncal_{\tilde b_\alpha, \beta}(H, L_\alpha - Q_{\alpha - 1}) \\
&\leq (D^{\dd \Ical, (1)})^\beta \, D^{\dd I, (2), \beta}(\Theta_n) \, F(\alpha \, \Theta_n)^\beta \, \Ncal_{\tilde b_\alpha, \beta}(H, L_\alpha - Q_{\alpha - 1}),
\end{align*}
where the last inequality follows from $\eta^\beta - \tilde b_\alpha \geq \Theta_n$, see \eqref{AT:Induction_balpha_distance} and \eqref{AT:Induction_btildealpha_definition}. Furthermore,
\begin{align*}
\Ncal_{\tilde b_\alpha, \beta}(H, L_\alpha - Q_{\alpha - 1}) &= \sup_{0\leq j \leq \beta - 1} \sup_{\substack{q\in \Nbb^{j+1} \\ |q| = \beta}} \Vert (L_\alpha - Q_{\alpha - 1})^{(q_0 - 1)}\Vert_{\tilde b_\alpha} \, \Vert H^{(q_1)}\Vert_{\tilde b_\alpha} \cdots \Vert H^{(q_j)}\Vert_{\tilde b_\alpha}.
\end{align*}
In the following, we are concerned with an estimate on the two terms on the right side of
\begin{align}
\Vert (L_\alpha - Q_{\alpha - 1})^{(q_0 - 1)}\Vert_{\tilde b_\alpha} \leq \Vert L_\alpha^{(q_0 - 1)}\Vert_{\tilde b_\alpha} + \Vert Q_{\alpha - 1}^{(q_0 - 1)}\Vert_{\tilde b_\alpha} \label{AT:Induction_two_terms_to_estimate}
\end{align}
We start by an estimate on $L_\alpha^{(q_0 - 1)}$, use the triangle inequality and Theorem \ref{AT:Commutator_Estimate}, and obtain
\begin{align}
\Vert L_\alpha^{(q_0 - 1)} \Vert_{\tilde b_\alpha} &\leq \sum_{k=2}^n \frac{1}{k! \, \Delta(n)^{k-1}} \sum_{\substack{j\in \Nbb^k \\ |j| = \alpha}} \; \Bigl\Vert \frac{\dd^{q_0 - 1}}{\dd u^{q_0 - 1}}\ad_{A_{j_k}} \cdots \ad_{A_{j_1}} (H) \Bigr\Vert_{\tilde b_\alpha} \label{AT:Induction_Lalpha_1}
\end{align}
and
\begin{align}
\Bigl\Vert \frac{\dd^{q_0 - 1}}{\dd u^{q_0 - 1}}\ad_{A_{j_k}} \cdots \ad_{A_{j_1}} (H) \Bigr\Vert_{\tilde b_\alpha} &\leq \sum_{\substack{\ell \in \Nbb_0^{k+1} \\ |\ell| = q_0 - 1}} \binom{q_0 - 1}\ell \, \bigl\Vert \ad_{A_{j_k}^{(\ell_k)}} \cdots \ad_{A_{j_1}^{(\ell_1)}}(H^{(\ell_0)})\bigr\Vert_{\tilde b_\alpha} \notag \\
&\hspace{-120pt} \leq 4^k \, V_k(\Theta_n)^k \sum_{\substack{\ell \in \Nbb_0^{k+1} \\ |\ell| = q_0 - 1}} \binom{q_0 - 1}\ell \Vert A_{j_k}^{(\ell_k)}\Vert_{b_{\alpha - 1}} \cdots \Vert A_{j_1}^{(\ell_1)}\Vert_{b_{\alpha - 1}} \Vert H^{(\ell_0)}\Vert_{b_{\alpha - 1}}. \label{AT:Induction_Lalpha_2}
\end{align} 
Here, we used \eqref{AT:Induction_btildealpha_distance}. We apply Corollary \ref{AT:Hamiltonian_Analytic_derivatives} and the induction hypothesis \eqref{AT:Induction_hypothesis}, which yields
\begin{align}
\Vert A_{j_1}^{(\ell_1)}\Vert_{b_{\alpha - 1}} \cdots \Vert A_{j_k}^{(\ell_k)} \Vert_{b_{\alpha - 1}} \, \Vert H^{(\ell_0)}\Vert_{b_{\alpha - 1}} & \notag\\
&\hspace{-150pt} \leq C_1^{j_1 + 1} \, \bigl( \frac{C_2 (j_1 + \ell_1)}{\Theta_n}\bigr)^{C_3(j_1 + 2\ell_1)} \, \bigl( \frac{C_4}{\Theta_n}\bigr)^{C_5(j_1 + \ell_1)} \, \ell_1^{(\ell_1 + 1)\nu} \, \delta^{-\ell_1} \, \vvvert H\vvvert_{\eta^n} \notag \\
&\hspace{-100pt} \ddots \notag \\
&\hspace{-120pt} \times C_1^{j_k + 1} \, \bigl( \frac{C_2 (j_k + \ell_k)}{\Theta_n}\bigr)^{C_3(j_k + 2\ell_k)} \, \bigl( \frac{C_4}{\Theta_n}\bigr)^{C_5(j_k + \ell_k)} \, \ell_k^{(\ell_k + 1)\nu} \, \delta^{-\ell_k} \, \vvvert H\vvvert_{\eta^n} \notag \\
&\hspace{110pt} \times \ell_0^{\ell_0} \, \delta^{-\ell_0} \, \vvvert H\vvvert_{\eta^n}. \label{AT:Induction_Lalpha_3}
\end{align}
We have $\ell_0^{\ell_0} \ell_1^{(\ell_1 + 1)\nu} \cdots \ell_k^{(\ell_k + 1)\nu} \leq (q_0 - 1)^{(q_0 - 1 + k)\nu}$. Therefore, the right side of \eqref{AT:Induction_Lalpha_3} is further bounded by
\begin{align*}
&C_1^{\alpha + k} \, \vvvert H\vvvert_{\eta^n}^{k+1} \, \delta^{-(q_0  -1)} \, (q_0 - 1)^{(q_0 - 1 + k)\nu} \, \bigl( \frac{C_2}{\Theta_n}\bigr)^{C_3(\alpha + 2(q_0 - 1))} \, \bigl( \frac{C_4}{\Theta_n}\bigr)^{C_5(\alpha + q_0-1)} \\
&\hspace{150pt} \times (j_1 + \ell_1)^{C_3(j_1 + 2\ell_1)} \cdots (j_k + \ell_k)^{C_3(j_k + 2\ell_k)}.
\end{align*}
When we combine \eqref{AT:Induction_Lalpha_1}, \eqref{AT:Induction_Lalpha_2}, and \eqref{AT:Induction_Lalpha_2}, we obtain
\begin{align}
\Vert L_\alpha^{(q_0 - 1)}\Vert_{\tilde b_\alpha} &\leq C_1^\alpha \, \vvvert H\vvvert_{\eta^n} \, \delta^{-(q_0 - 1)} \, (q_0 - 1)^{(q_0 - 1)\nu} \, \bigl( \frac{C_2}{\Theta_n}\bigr)^{C_3(\alpha + 2(q_0 - 1))} \, \bigl( \frac{C_4}{\Theta_n}\bigr)^{C_5(\alpha + q_0-1)} \notag \\
&\hspace{10pt} \times \sum_{k=2}^\alpha \frac{4^k}{k!} \frac{(C_1 \, \vvvert H\vvvert_{\eta_n} \, (q_0 - 1)^\nu \, V_k(\Theta_n))^k}{\Delta(n)^{k-1}} \notag \\
&\hspace{10pt} \times \sum_{\substack{j\in \Nbb^k \\ |j| = \alpha}} \sum_{\substack{\ell \in \Nbb_0^{k+1} \\ |\ell| = q_0 - 1}} \binom{q_0 - 1}\ell (j_1 + \ell_1)^{C_3(j_1 + 2\ell_1)} \cdots (j_k + \ell_k)^{C_3(j_k + 2\ell_k)}. \label{AT:Induction_Lalpha_4}
\end{align}
In a similar manner, we bound $Q_{\alpha - 1}$ as
\begin{align}
\Vert Q_{\alpha - 1}^{(q_0 - 1)}\Vert_{\tilde b_\alpha} &\leq \sum_{k=1}^{\alpha - 1} \frac{1}{k! \, \Delta(n)^k} \sum_{\substack{j\in \Nbb^k \\ |j| = \alpha - 1}} \; \Bigl\Vert \frac{\dd^{q_0 -1}}{\dd u^{q_0 -1}} \ad_{A_{j_k}} \cdots \ad_{A_{j_2}}(\dot A_{j_1})\Bigr\Vert_{\tilde b_\alpha} \label{AT:Induction_Qalpha_1}
\end{align}
and
\begin{align}
\Bigl\Vert \frac{\dd^{q_0 -1}}{\dd u^{q_0 -1}} \ad_{A_{j_k}} \cdots \ad_{A_{j_2}}(\dot A_{j_1})\Bigr\Vert_{\tilde b_\alpha} &\leq \sum_{\substack{\ell\in \Nbb_0^k \\ |\ell| = q_0 - 1}} \binom{q_0 - 1}\ell \, \bigl\Vert \ad_{A_{j_k}^{(\ell_k)}} \cdots \ad_{A_{j_2}^{(\ell_2)}} (A_{j_1}^{(\ell_1 + 1)}) \bigr\Vert_{\tilde b_\alpha} \notag \\
&\hspace{-130pt} \leq 4^{k-1} \, V_{k-1}(\Theta_n)^{k-1} \sum_{\substack{\ell \in \Nbb_0^k \\ |\ell| = q_0 - 1}} \binom{q_0 - 1}\ell \, \Vert A_{j_k}^{(\ell_k)} \Vert_{b_{\alpha - 1}} \cdots \Vert A_{j_2}^{(\ell_2)}\Vert_{b_{\alpha - 1}} \Vert A_{j_1}^{\ell_1 + 1}\Vert_{b_{\alpha - 1}}. \label{AT:Induction_Qalpha_2}
\end{align}
Likewise, we apply the induction hypothesis \eqref{AT:Induction_hypothesis} and obtain
\begin{align}
\Vert A_{j_k}^{(\ell_k)} \Vert_{b_{\alpha - 1}} \cdots \Vert A_{j_2}^{(\ell_2)}\Vert_{b_{\alpha - 1}} \Vert A_{j_1}^{(\ell_1 + 1)}\Vert_{b_{\alpha - 1}} & \notag \\
&\hspace{-150pt} \leq C_1^{j_k + 1} \, \bigl( \frac{C_2 (j_k + \ell_k)}{\Theta_n}\bigr)^{C_3(j_k + 2\ell_k)} \, \bigl( \frac{C_4}{\Theta_n}\bigr)^{C_5(j_k + \ell_k)} \ell_k^{(\ell_k + 1)\nu} \, \delta^{-\ell_k} \, \vvvert H\vvvert_{\eta^n} \notag \\
&\hspace{-110pt} \ddots \notag \\
&\hspace{-130pt} \times C_1^{j_2 + 1} \, \bigl( \frac{C_2 (j_2 + \ell_2)}{\Theta_n}\bigr)^{C_3(j_2 + 2\ell_2)} \, \bigl( \frac{C_4}{\Theta_n}\bigr)^{C_5(j_2 + \ell_2)} \ell_2^{(\ell_2 + 1)\nu} \, \delta^{-\ell_2} \, \vvvert H\vvvert_{\eta^n} \notag \\
&\hspace{-110pt} \times C_1^{j_1 + 1} \, \bigl( \frac{C_2 (j_1 + \ell_1 + 1)}{\Theta_n}\bigr)^{C_3(j_1 + 2(\ell_1 + 1))} \, \bigl( \frac{C_4}{\Theta_n}\bigr)^{C_5(j_1 + \ell_1 + 1)} \notag \\
&\hspace{40pt} \times (\ell_1 + 1)^{(\ell_1 + 2)\nu} \, \delta^{-(\ell_1 +1)} \, \vvvert H\vvvert_{\eta^n}. \label{AT:Induction_Qalpha_3}
\end{align}
We have $(\ell_1 + 1)^{(\ell_1 + 2)\nu} \ell_2^{(\ell_2 + 1)\nu} \cdots \ell_k^{(\ell_k + 1)\nu} \leq q_0^{(q_0 + k)\nu}$. Therefore, the right side of \eqref{AT:Induction_Qalpha_3} is further bounded by
\begin{align*}
&C_1^{\alpha + k} \, \bigl( \frac{C_2}{\Theta_n}\bigr)^{C_3(\alpha + 2q_0)} \, \bigl( \frac{C_4}{\Theta_n} \bigr)^{C_5(\alpha + q_0)} \, \delta^{-q_0} \, q_0^{(q_0 + k)\nu} \, \vvvert H\vvvert_{\eta^n}^k \\
&\hspace{70pt} (j_1 + \ell_1 + 1)^{C_3(j_1 + 2(\ell_1 + 1))} \, (j_2 + \ell_2)^{C_3(j_2 + 2\ell_2)} \cdots (j_k + 2\ell_k)^{C_3(j_k + 2\ell_k)}
\end{align*}
When we combine \eqref{AT:Induction_Qalpha_1}, \eqref{AT:Induction_Qalpha_2}, and \eqref{AT:Induction_Qalpha_3}, we obtain
\begin{align}
\Vert Q_{\alpha - 1}^{(q_0 - 1)}\Vert_{\tilde b_\alpha} &\leq C_1^\alpha \bigl( \frac{C_2}{\Theta_n} \bigr)^{C_3(\alpha + 2q_0)} \bigl( \frac{C_4}{\Theta_n}\bigr)^{C_5(\alpha + q_0)} \, \delta^{-q_0} \, q_0^{q_0\nu} \notag \\
&\hspace{10pt} \times \sum_{k=1}^{\alpha - 1} \frac{4^{k-1}}{k!} \frac{V_{k-1}(\Theta_n)^{k-1}}{\Delta(n)^k} \, C_1^k \, \vvvert H\vvvert_{\eta^n}^k \, q_0^{k\nu} \notag \\
&\hspace{10pt} \times \smash{\sum_{\substack{j\in \Nbb^k \\ |j| = \alpha - 1}} \sum_{\substack{\ell\in \Nbb_0^k \\ |\ell| = q_0 - 1}} \binom{q_0 - 1}\ell} (j_1 + \ell_1 + 1)^{C_3(j_1 + 2(\ell_1 + 1))} \notag \\
&\hspace{110pt} \times (j_2 + \ell_2)^{C_3(j_2 + 2\ell_2)} \cdots (j_k + 2\ell_k)^{C_3(j_k + 2\ell_k)}. \label{AT:Induction_Qalpha_4}
\end{align}

Since $\Theta_n \leq \frac d{s\, \e}$, see \eqref{AT:Induction_hypothesis_Theta}, 
Lemma \ref{AT:Vb_concrete_estimate} implies
\begin{align}
V_k(\Theta_n) \leq \kappa \bigl( \frac{d \, k }{s \e \, \Theta_n}\bigr)^{\frac{d}{s}} \leq \kappa \bigl( \frac{C_2 \, k}{\Theta_n}\bigr)^{C_3} 
\label{AT:Induction_Vmunk_estimate}
\end{align}
provided that $C_2$ and $C_3$ are chosen such that
\begin{align}
C_2 &:= \frac{d}{s \e} , & C_3 &\geq \frac{d}{s} \geq 1. \label{AT:Induction_choice_C2C3}
\end{align}
For $\rho \geq 4$ to be chosen, we define
\begin{align}
\Delta(n) &:= \Bigl( C_1 \; \kappa \; \bigl( \frac{C_2\, n}{\Theta_n}\bigr)^{2C_3} \; \bigl( \frac{C_4}{\Theta_n} \bigr)^{C_5} \Bigr)^\rho. \label{AT:Induction_choice_Deltan}
\end{align}
Then, a short calculation using that $\rho(k-1) \geq k$ for $k\geq 2$ shows that \eqref{AT:Induction_Vmunk_estimate}, \eqref{AT:Induction_choice_C2C3}, and \eqref{AT:Induction_choice_Deltan} and the hypothesis \eqref{AT:Induction_hypothesis_Theta} imply
\begin{align*}
\frac{(C_1 \, \vvvert H\vvvert_{\eta_n} \, (q_0 - 1)^\nu \, V_k(\Theta_n))^k}{\Delta(n)^{k-1}} &\leq 1, & \frac{(C_1 \, V_{qk}(\Theta_n))^k}{\Delta(n)^{k-1}} \; \bigl( \frac{C_2\, \alpha}{\Theta_n} \bigr)^{2C_3k} \; \bigl( \frac{C_4}{\Theta_n}\bigr)^{C_5k} &\leq 1.
\end{align*}

\vspace{10cm}
\begin{align}
\Vert A_\alpha\Vert_{b_\alpha} &\leq D^\Delta \; F(\alpha \, \Theta_n) \; V_q(\Theta_n) \; C_1^\alpha \; \Vert H\Vert_a \; \sum_{k=2}^\alpha \frac{4^k}{k!} \frac{(C_1 \, V_{q k}(\Theta_n))^k}{\Delta(n)^{k-1}} \notag \\
&\hspace{-30pt} \times \sum_{\substack{j\in \Nbb^k \\ |j| = \alpha}} \bigl( \frac{C_2 \, j_1}{\Theta_n}\bigr)^{C_3(j_1 + 2)} \cdots \bigl( \frac{C_2 \, j_k}{\Theta_n}\bigr)^{C_3(j_k + 2)} \; \bigl( \frac{C_4}{\Theta_n}\bigr)^{C_5(j_1 + 1)} \cdots \bigl( \frac{C_4}{\Theta_n}\bigr)^{C_5(j_k + 1)}. \label{AT:Induction_2}
\end{align}
The term in the sum of the second line in \eqref{AT:Induction_2} equals
\begin{align}
&\bigl( \frac{C_2}{\Theta_n}\bigr)^{C_3(\alpha + 2k)} \bigl( \frac{C_4}{\Theta_n}\bigr)^{C_5(\alpha + k)} \; j_1^{C_3j_1} \cdots j_k^{C_3j_k} \; (j_1 \cdots j_k)^{2C_3}. \label{AT:Induction_4}
\end{align}
The last factor is bounded by $\alpha^{2C_3k}$, whence \eqref{AT:Induction_2} and \eqref{AT:Induction_4} yield
\begin{align}
\Vert A_\alpha\Vert_{b_\alpha} &\leq D^\Delta \; F(\alpha \, \Theta_n) \; V_{q}(\Theta_n) \; C_1^\alpha \; \Vert H\Vert_a \; \bigl( \frac{C_2}{\Theta_n}\bigr)^{C_3\alpha} \; \bigl( \frac{C_4}{\Theta_n} \bigr)^{C_5\alpha} \notag \\
&\hspace{10pt} \times \sum_{k=2}^\alpha \frac{4^k}{k!} \frac{(C_1 \,   \, V_{q k}(\Theta_n))^k}{\Delta(n)^{k-1}} \bigl( \frac{C_2 \, \alpha}{\Theta_n} \bigr)^{2C_3k} \bigl( \frac{C_4}{\Theta_n}\bigr)^{C_5k} \sum_{\substack{j\in \Nbb^k \\ |j| = \alpha}} j_1^{C_3j_1} \cdots j_k^{C_3 j_k} . \label{AT:Induction_1}
\end{align}

since $k\leq \alpha \leq n$. We combine \eqref{AT:Induction_1} with Proposition \ref{AT:Induction_Function_estimateed} and obtain
\begin{align}
\Vert A_\alpha\Vert_{b_\alpha} &\leq D_{C_3} \; D^\Delta \; F(\alpha\, \Theta_n) \; V_q(\Theta_n) \; C_1^\alpha \, \bigl( \sup_{k\in \Nbb} \frac{4^k}{k!}\bigr) \, \Vert H\Vert_a \, \bigl( \frac{C_2 \, \alpha}{\Theta_n}\bigr)^{C_3\alpha} \; \bigl( \frac{C_4}{\Theta_n}\bigr)^{C_5\alpha} . \label{AT:Induction_7}
\end{align}
We have $F(\alpha \Theta_n) \leq F(\Theta_n)$ and, by Lemma \ref{AT:F_concrete_estimate} and \eqref{AT:Induction_hypothesis_Theta}, $F(\Theta_n)$ is bounded by
\begin{align}
F\bigl( \frac{\Theta_n} q\bigr) &\leq 2^{\frac ds} \kappa \; \Gamma\bigl(1 + \frac 1s\bigr) \; \bigl( \frac{d q}{s\e \, \Theta_n}\bigr)^{\frac ds} \; \bigl( \frac{4q}{\Theta_n} \bigr)^{\frac 1s} \leq 2^{\frac ds} \kappa \; \Gamma\bigl(1 + \frac 1s\bigr) \; \bigl( \frac{C_2}{\Theta_n}\bigr)^{C_3} \; \bigl( \frac{C_4}{\Theta_n} \bigr)^{C_5} \label{AT:Induction_F_estimate}
\end{align}
with the choice
\begin{align}
C_4 &:= 4q , & C_5 &:= \frac 1s. \label{AT:Induction_choice_C4C5}
\end{align}
When we apply \eqref{AT:Induction_Vmunk_estimate} with $k =1$, \eqref{AT:Induction_F_estimate} implies
\begin{align}
\Vert A_\alpha\Vert_{b_\alpha} &\leq 2^{\frac ds} \; \kappa^2 \; \Gamma\bigl( 1 + \frac 1s\bigr) \; D_{C_3} \; D^\Delta \; \bigl( \sup_{k\in \Nbb} \frac{4^k}{k!}\bigr) \; \Vert H\Vert_a \; C_1^\alpha \notag\\
&\hspace{140pt} \times \bigl( \frac{C_2 \, \alpha}{\Theta_n} \bigr)^{C_3(\alpha + 2)} \; \bigl( \frac{C_4}{\Theta_n} \bigr)^{C_5 (\alpha + 1)} \label{AT:Induction_5}
\end{align}
Therefore, if $C_1$ is chosen such that
\begin{align}
C_1 &\geq 2^{\frac ds} \; \kappa^2 \; \Gamma\bigl( 1 + \frac 1s\bigr) \;  D_{C_3} \; D^\Delta \; \bigl( \sup_{k\in \Nbb} \frac{4^k}{k!} \bigr) \; \Vert H\Vert_a, \label{AT:Induction_choice_C1}
\end{align}
then \eqref{AT:Induction_5} implies \eqref{AT:Induction_hypothesis} for $\alpha$.

We close the induction by showing that \eqref{AT:Induction_hypothesis} holds for $\alpha = 1$. By Theorem \ref{AT:I_estimate_local_Hamiltonian} applied twice, we obtain
\begin{align*}
\Vert A_1\Vert_{b_1} = \Vert \Ical_{s, \gamma}(K)\Vert_{b_1} \leq D^\Delta \, V_1(b_1^{(1)} - b_1) \, F(\eta - b_1) \, \Vert K\Vert_{b_1^{(1)}}
\end{align*}
as well as
\begin{align*}
\Vert K\Vert_{b_1^{(1)}} = \Vert \Ical_{s, \gamma}(\dot H)\Vert_{b_1^{(1)}} \leq D^\Delta \, V_1(b_1^{(1)} - b_1^{(2)}) \, F(\eta - b_1^{(1)}) \, \Vert \dot H\Vert_{b_1^{(2)}}.
\end{align*}
Combining these estimates, we have
\begin{align*}
\Vert A_1\Vert_{b_1} \leq (D^\Delta)^2 \, \Vert \dot H\Vert_\eta \, V_q(\Theta_n)^2 \, F\bigl( \frac{\Theta_n}{q}\bigr)^2
\end{align*}
When we apply \eqref{AT:Induction_Vmunk_estimate} with $k =1$ and \eqref{AT:Induction_F_estimate} with $\alpha = 1$, we arrive at
\begin{align}
\Vert A_1\Vert_{b_1} \leq \bigl(2^{\frac ds} \, \kappa^2 \, D^\Delta \, \Gamma\bigl( 1 + \frac 1s\bigr)  \bigr)^2 \, \Vert \dot H\Vert_\eta \, \bigl( \frac{C_2}{\Theta_n}\bigr)^{\frac{4d}s} \bigl( \frac{C_4}{\Theta_n} \bigr)^{\frac 2s}. \label{AT:Induction_6}
\end{align}
Thus, in light of \eqref{AT:Induction_choice_C2C3} and \eqref{AT:Induction_choice_C1}, we make the choices
\begin{align*}
C_1 &:= 2^{\frac ds} \, \kappa^2 \, \Gamma\bigl( 1 + \frac 1s\bigr) D^\Delta \; \max\Bigl\{ \sqrt{\Vert \dot H\Vert_\eta} \; , \; D_{C_3} \, \bigl( \sup_{k\in \Nbb} \frac{4^k}{k!}\bigr) \, \Vert H\Vert_a\Bigr\}
\end{align*}
and
\begin{align*}
C_3 &:= \frac{4d}{3s},
\end{align*}
whence \eqref{AT:Induction_6} implies \eqref{AT:Induction_hypothesis} for $\alpha = 1$.

\begin{proof}[Proof of Proposition \ref{AT:Induction_Function_estimate}]
Consider the function
\begin{align}
f_\alpha^p(x) &:= x^{px} \, (\alpha - x)^{p(\alpha- x)}, & 0 &\leq x \leq \alpha.
\end{align}
This function is symmetric about $x = \nicefrac \alpha 2$ and since $(f_\alpha^p)'(x) = f_\alpha^p(x) \, p \, [\log(x) - \log(\alpha - x)]$ for $0 < x < \alpha$, we conclude that $f_\alpha^p$ is strictly decreasing on $[0, \nicefrac \alpha 2)$. We claim that
\begin{align}
\sum_{j=1}^{\alpha - 1} f_\alpha^p(j) \leq \min \bigl\{ 2 + 4^p \, , \, \alpha\bigr\} \, f_\alpha^p(1). \label{AT:Induction_Function_estimate_1}
\end{align}
To see this, the left side is bounded by $2 \, f_\alpha^p(1) + (\alpha - 3) f_\alpha^p(2)$ due to the monotonicity of $f_\alpha^p$. On the one hand, we further have $(\alpha - 2) f_\alpha^p(2) \leq 4^p f_\alpha^p(1)$, which proves the first bound, while on the other hand, the monotonicity implies
\begin{align*}
2 \, f_\alpha^p(1) + (\alpha - 2) f_\alpha^p(2) = \alpha \, f_\alpha^p(1) - (\alpha - 2) \bigl( f_\alpha^p(1) - f_\alpha^p(2)\bigr) \leq \alpha \, f_\alpha^p(1).
\end{align*}
This proves \eqref{AT:Induction_Function_estimate_1}. For $2 \leq k \leq \alpha$ we claim that
\begin{align}
S_{\alpha, k}^p := \sum_{\substack{j\in \Nbb^k \\ |j| = \alpha}} j_1^{pj_1} \cdots j_k^{pj_k} \leq \min \bigl\{ 2+4^p, \alpha\bigr\} \, f_\alpha^p(1). \label{AT:Induction_Function_estimate_2}
\end{align}
We prove this by induction. Since $S_{\alpha, 2}^p = \sum_{j=1}^{\alpha - 1} f_\alpha^p(j)$, the case $k =2$ is \eqref{AT:Induction_Function_estimate_1}. By the second case of the induction hypothesis \eqref{AT:Induction_Function_estimate_2}, we further have
\begin{align*}
S_{\alpha, k+1}^p = \sum_{j = 1}^{\alpha - k} j^{pj} S_{k, \alpha -j}^p \leq \sum_{j=1}^{\alpha - k} j^{pj}\, (\alpha- j) \, f_{\alpha - j}^p(1) \leq \sum_{j=1}^{\alpha - 1} f_{\alpha}^p(j).
\end{align*}
Here, we used that $(\alpha - j) f_{\alpha - j}^p(1) \leq (\alpha - j)^{p(\alpha -j)}$. A further application of \eqref{AT:Induction_Function_estimate_1} proves \eqref{AT:Induction_Function_estimate_2}. Summing \eqref{AT:Induction_Function_estimate_2} over $k$ completes the proof.
\end{proof}
}

\ifthenelse{\equal\masterfile{Diss}}{}{

\newpage
\subsection{Expected result by optimal truncation}

\subsubsection{What is the error estimate?}

We assume that we can, in this way, prove a result of the form
\begin{align*}
\Vert A_n\Vert \leq \const\; C^n (\ell n+m)^{\ell n + m}
\end{align*}
for some $\ell\geq 1$, $m\geq 0$ and $C>0$ (all depending on $d$). Note that we can always assume $m=0$ at the expense of an increase of $\ell$. Now, consider the function
\begin{align*}
f_\delta(x) := (C\delta)^x (\ell x+m)^{\ell x+m}.
\end{align*}
We have
\begin{align*}
f_\delta'(x) &= f_\delta(x) \bigl( \log C\delta + \ell\cdot \log(\ell x+m) + \ell\bigr)\\ f_\delta''(x) &= f_\delta(x) \bigl( \log C\delta + \ell \log(\ell x+m) + \ell\bigr)^2 + f_\delta(x) \cdot \frac{\ell^2}{\ell x+m}
\end{align*}
The term in square brackets is always nonnegative and the last term is strictly positive. Hence, any solution of $f_\delta'(x) =0$ is a nondegenerate minimum. However, we have that $\log C\delta + \ell \log (\ell x+m) + \ell =0$ iff  $(\ell x+m) \cdot \e = \frac{1}{\sqrt[\ell]{\delta}}$. Or
\begin{align}
x_{\min}(\delta) = \frac{1}{\ell}\bigl(\frac{1}{\e\cdot \sqrt[\ell]{C\delta}}-m\bigr). \label{Optimal choice}
\end{align}

\begin{bem}
Note that, in this way, we really solve the equation $\i \varepsilon \dot \Pi =[H + R_n,\Pi]$ in Lemma \ref{AT:Adiabatic_expansion} because the ratio in front becomes $\varepsilon$ by construction. Not really because we were sloppy, but ultimately...
\end{bem}

Plugging this into $f_\delta$ yields the minimal value in dependence of $\delta$
\begin{align*}
f_\delta(x_{\min}(\delta)) = (C\delta)^{-\frac m\ell} \bigl( \frac{\sqrt[\ell]{C\delta}}{\e \sqrt[\ell]{C\delta}}\bigr)^{\frac{1}{\e \sqrt[\ell]{C\delta}}}  = (C\delta)^{-\frac m\ell} \cdot \exp\bigl(-\frac{1}{\sqrt[\ell]{C} \e} \cdot \frac{1}{\sqrt[\ell]{\delta}}\bigr).
\end{align*}
Now, if $m\neq 0$, we can sacrifice a little bit of the decay rate to get
\begin{align*}
f_\delta(x_{\min}(\delta) &= (C\delta)^{-\frac m\ell} \exp\bigl( -\frac{\rho}{\e} \cdot \frac{1}{\sqrt[\ell]{C\delta}}\bigr) \cdot \exp\bigl( -\frac{1-\rho}{\sqrt[\ell]{C}\e} \cdot \frac{1}{\sqrt[\ell]{\delta}}\bigr) \\
&\leq C_{\rho, m}\cdot \exp\bigl( -\frac{1-\rho}{\sqrt[\ell]{C}\e} \cdot \frac{1}{\sqrt[\ell]{\delta}}\bigr).
\end{align*}

Recall that if $m =0$, we have $\delta \sim C' \varepsilon^{\frac{s\ell}{s\ell+2d}}$. Thus, we obtain the ``final result''
\begin{align*}
\exp\bigl(-\frac{C'}{\sqrt[\ell]{C}\e} \cdot \varepsilon^{- \frac{s}{s\ell+2d} }\bigr).
\end{align*}

\subsubsection{Reproduce the ``old'' result?}

We can obtain a ``finite order'' result if we first choose $n$ and then set $\delta := \varepsilon \Delta(n)$. If we write up the splitting according to this instead, we get a finite order result in terms of full powers of $\varepsilon$.

\subsection{The ``ultimate'' problem: Catching the local norm of the unitaries}

The ultimate problem that I always feared is the local norm estimate on the unitaries in the error term. Because of the local norm, we cannot make use of its super nice operator norm bound 1. Instead, it grows pretty quickly. Omitting the dimensional dependence, we have something like
\begin{align*}
\Vert U_n\Vert_{\tilde b} \leq \exp \Bigl( \frac{C}{\Delta(n)} \sum_{\alpha =1}^n f_\delta(\alpha)\Bigr)
\end{align*}
Now, as we know, setting $m=0$, the optimal truncation is $n = \frac{1}{\ell \e \sqrt[\ell]{C\delta}}$. Hence, the question is, if the sum
\begin{align*}
\sum_{\alpha =1}^{\frac{1}{\ell\e \sqrt[\ell]{C\delta}}} f_\delta(\alpha)
\end{align*}
is uniformly bounded in $\delta \to 0$. Since the function $f_\delta(\alpha) = (C\delta)^\alpha (\ell\alpha)^{\ell\alpha}$ is monotonically decreasing on its way to the minimum $\frac{1}{\ell\e \sqrt[\ell]{C\delta}}$, we may investigate the integral instead. Since $f_\delta(0) =1$ and $f_\delta(\nicefrac 1\ell) = \sqrt[\ell]{C\delta}$, we obtain
\begin{align*}
\int_0^{\frac{1}{\delta \e}} f_\delta(\alpha) \,\dd \alpha = \int_0^{\nicefrac 1\ell} f_\delta(\alpha) \, \dd \alpha + \int_{\nicefrac 1\ell}^{\frac{1}{\ell \e \sqrt[\ell]{C\delta}}} f_\delta(\alpha) \dd \alpha \leq f_\delta(0) + \frac{1}{\ell \e \sqrt[\ell]{C\delta}} \cdot f_\delta(1) = 1 + \frac{1}{\ell\e}.
\end{align*}
Hence, we indeed obtain
\begin{align*}
\sup_{\delta >0} \int_0^{\frac{1}{\delta \e}} f_\delta(\alpha) \, \dd \alpha \leq 1 + (\ell\e)^{-1} < \infty.
\end{align*}
This means that the optimally truncated norm of $U_n$ is indeed uniformly bounded on the way $\delta \to 0$.
}



\section{An Estimate on \texorpdfstring{$\Ical$}{I} with the Old Norm}
\label{AT:Old_norm_estimate_Section}

In this section, we want to demonstrate a fairly sharp estimate for $\Ical$ on a local Hamiltonian, while using the ``old'' norm defined in \cite{SvenAdiabatic}, see also \eqref{AT:old_norm}. The estimate is based on working out the precise constants in \cite[Lemma 4.7 \& Theorem 4.8]{AutomorphicEquivalence}, while using the stretched exponential decay functions $\chi_{\nicefrac 12,b}$ defined in \eqref{AT:chi_definition} (as opposed to the decay class that had been used in \cite{AutomorphicEquivalence}). This estimate is not usable for us for the business of proving exponential estimates. However, since, to the best of my knowledge, there is no such explicit estimate in the literature, I found it instructive to put it in. It should be said, however, that one does not have to go as far as to estimate $\Ical$ to realize that, with the norm in \eqref{AT:old_norm}, exponential estimates are out of reach. Already the commutator estimate presented in \cite{SvenAdiabatic} are far from good enough. I am aware of unpublished notes by Felix Rexze\footnote{Felix Rexze used to be a Master's student under the supervision of Stefan Teufel at the University of Tübingen in 2018.}  on commutator bounds which look somewhat better but are still not good enough to satisfy our needs.

\begin{warn}
The notation is not streamlined to the previous section, since it refers to a version of the notes where $s = \nicefrac 12$ was fixed! No proof-reading was done in this section.
\end{warn}

We denote the Banach spaces that are defined upon the norm \eqref{AT:old_norm} by $\Bcal_{b,N}$ or $\Lcal_{b,N}$ for the respective classes decaying according to $F_b(r) = F(r) \chi_b(r)$ with the decay function $F(r) = (1 + r)^{-(d+1)}$.

For a local observable $A\in \Acal_X$, we define the operators $\Delta_s^n$, $n \in \Nbb_0$ as in \eqref{AT:I_Delta0_definition} and \eqref{AT:I_Deltan_definition}, whence $\Ical(A)$ satisfies the decomposition \eqref{AT:I_decomposition_local_observable}. The interaction $\Phi_{\Ical(G)}$ for a local Hamiltonian $G$ is chosen as in \eqref{AT:I_interaction_definition}, so that $\sum_{Z\subset \Lambda} \Phi_{\Ical(G)}(Z) = \Ical(G)$.


We start with estimates on the local observables $\Delta^n$ and we derive them under the assumption of a Lieb--Robinson bound.

\begin{asmp}[Lieb--Robinson bound]
\label{LR-bound old norm}
We assume that there is an $a>0$ and $K_a, v_a>0$ such that the following holds. For each $A\in \Acal_X$, each $B\in \Acal_Y$ and each $t\in\Rbb$: If $\dist(X,Y) >0$, then
\begin{align*}
\Vert [\tau_t(A),B]\Vert &\leq K_a\;  \Vert F\Vert_1\;  \Vert A\Vert \; |X|\; \e^{a(v_a|t|-d(X,Y))}\; \Vert B\Vert,
\end{align*}
where
\begin{align*}
\Vert F\Vert_1 &:= \sup_{x\in \Gamma} \sum_{z\in \Gamma} F(d(x,z)).
\end{align*}
\end{asmp}

\begin{lem}
\label{Delta-estimate old norm}
For any $X\subset \Lambda$, $A\in \Acal_X$ and integer $n\geq 0$, we have the estimate
\begin{align*}
\Vert \Delta^n (A)\Vert \leq C_\Delta \; \Vert A\Vert \;|X| \; \chi_\eta(n).
\end{align*}
Here, $C_\Delta := \max\{ \tilde C_\Delta, \Vert W_\gamma\Vert_1\}$ with
\begin{align*}
\eta &:= \frac{1}{\sqrt{2}} \min \Bigl\{ a, \frac{3}{16}\sqrt{\frac{\gamma}{v_a}}\Bigr\}, & \tilde C_\Delta &:= \max\bigl\{ \frac{2K_a\Vert F\Vert_1 \; \e^{\frac{5a}{4}}}{a\; v_a}\; , \; 8C_{I_\gamma}\bigr\}.
\end{align*}
\end{lem}

\begin{proof}
The estimate for $\Delta^0(A)$ is trivial. Let $n> 0$ and decompose
\begin{align}
\Delta^n(A) = \tilde \Delta^n(A) - \tilde \Delta^{n-1}(A) \label{Deltan-decomp old norm}
\end{align}
with
\begin{align*}
\tilde \Delta^n(A) &:= \int_\Rbb \dt\; W_\gamma(t) \bigl( \Ebb_{X_n}(\tau_t(A)) - \tau_t(A)\bigr).
\end{align*}
Now, for $T>0$ to be chosen, we have
\begin{align*}
\Vert \tilde \Delta^n(A)\Vert &\leq \Vert W_\gamma\Vert_\infty \int_{-T}^T\dt\; \bigl\Vert \Tr_{X_n}(\tau_t(A)) - \tau_t(A)\bigr\Vert + 4\Vert A\Vert\; I_\gamma(T). 
\end{align*}
For the first term -- let us call it $\Tcal$ --, we use the Lieb-Robinson bound Assumption \ref{LR-bound old norm}. Note that $\dist(X,X_n) \geq n$. Hence, by Lemma 3.2 in the equivalence paper (with $\varepsilon$ -- the finite dimensional case) (and using $\Vert W_\gamma\Vert_\infty = \frac 12$, Lemma \ref{AT:W_gamma_estimate} (a)), the first term  $\Tcal$ is bounded by
\begin{align*}
\Tcal &\leq K_a \; D_F\; \Vert A\Vert \; |X|\; \e^{-an} \int_0^T \dt\; \e^{av_at} \leq \frac{K_a\Vert F\Vert_1}{av_a} \; \Vert A\Vert \; |X| \; \e^{-a(n - v_aT)}.
\end{align*}
Now choose $v_aT = \frac{n+1}{2}$ to get that
\begin{align*}
\Tcal &\leq \frac{K_a \Vert F\Vert_1 \e^a}{av_a} \; \Vert A\Vert\; |X| \; \e^{-\frac a2 \; (n+1)} .
\end{align*}
Utilizing Lemma \ref{AT:chi_estimate} (d), we may bound $\e^{-a\frac{(n+1)}2} \leq \e^{\frac{a}{4}} \chi_{\frac a{\sqrt{2}}}(n+1)$ and conclude that by the choice of $\eta$
\begin{align*}
\Tcal &\leq \frac{K_a \Vert F\Vert_1 \; \e^{\frac{5a}{4}}}{av_a} \; |X|\; \Vert A\Vert \; \chi_\eta(n+1).
\end{align*}
Finally, again by the choice of $\eta$, we have that 
\begin{align*}
I_\gamma(T) &= I_\gamma\bigl( \frac{n+1}{2v_a}\bigr) \leq C_{I_\gamma} \; \chi_{\frac{3}{16}} \bigl( \frac{\gamma}{2v_a} \; (n+1)\bigr) \leq C_{I_\gamma} \;\chi_\eta(n+1).
\end{align*}
Putting everything together, we conclude that
\begin{align*}
\Vert \tilde \Delta^n(A)\Vert &\leq \frac 12 \; C_\Delta \; |X|\; \Vert A\Vert \; \chi_\eta(n+1).
\end{align*}
From this and a triangle inequality, the bound on $\Delta^n(A)$ follows, see \eqref{Deltan-decomp old norm}.
\end{proof}

\begin{thm}
\label{AT:old_norm_Iestimate}
Let $\beta := \sqrt{3}$ and let $b \geq 0$ such that $\beta b < \eta$. Let $N\in \Nbb$ be given. If there is $b'\in (b,b+1)$ such that $\beta b' \leq \eta$ and $G\in \Lcal_{\beta b', N+1}$, then $\Ical(G) \in \Lcal_{b,N}$ and the estimate
\begin{align*}
\Vert \Phi_{\Ical(G)}\Vert_{b,N} &\leq C_\Ical(d,C_\Delta) \; D_\Ical(d,N,b'-b) \; \Vert \Phi_G\Vert_{\beta b',N+1}
\end{align*}
holds. Here, $F$ is the fixed function $F(r) = (1+r)^{-(d+1)}$ and the two constants $C_\Ical$ and $\Dcal_\Ical$ are given by
\begin{align*}
C_\Ical(d,C_\Delta) &:= C_\Delta \; 66 \; 3^{d+1} \; \max\{\kappa^2, \kappa \Vert F\Vert_1\}
\end{align*}
and
\begin{align*}
D_\Ical(d,N,b'-b) &:= \kappa^N \; \Bigl( \frac{2\sqrt{6} \; d(N+2)}{\e}\Bigr)^{2d(N+2)} \; \frac{1}{(b' - b)^{2d(N+2)+4}}.
\end{align*}
\end{thm}

\begin{lem}
\label{F-upper-estimate}
For $r\geq 0$ let $F(r) = (1+r)^{-(d+1)}$. Then, for any $0 < \varepsilon \leq 1$ and $r\geq 0$, we have the inequality
\begin{align*}
F(\varepsilon r) \leq \frac{1}{\varepsilon^{d+1}} \; F(r).
\end{align*}
\end{lem}

\begin{proof}
We have
\begin{align*}
\frac{F(\varepsilon r)}{F(r)} &= \bigl( \frac{1+r}{1 + \varepsilon r}\bigr)^{d+1} = \bigl( \frac{1}{\varepsilon} \; \frac{1+r}{\varepsilon^{-1} + r}\bigr)^{d+1} \leq \frac{1}{\varepsilon^{d+1}}. \qedhere
\end{align*}
\end{proof}

\begin{lem}
\label{f-l-eps-2}
Let $\varepsilon >0$ and $\ell\geq 0$. For $t\geq 0$, define
\begin{align*}
f_{\ell,\varepsilon}(t) := (1+t)^\ell \;  \e^{-\varepsilon \sqrt{t}}.
\end{align*}
Then
\begin{align*}
\Vert f_{\ell,\varepsilon}\Vert_{L^\infty([0,\infty))} &\leq \begin{dcases} \bigl( \frac{2\ell}{\e \;\varepsilon}\bigr)^{2\ell} & \ell > \varepsilon, \\ 1 & \ell \leq \varepsilon.\end{dcases}
\end{align*}
\end{lem}

\begin{proof}
The first derivative reads
\begin{align}
f_{\ell,\varepsilon}'(t) &= (1 + t)^{\ell-1} \e^{-\varepsilon\sqrt{t}} \bigl[ \ell - \varepsilon(1+t) \;\frac{1}{2\sqrt{t}}\bigr]. \label{f-l-eps-2.1}
\end{align}
This has a zero at
\begin{align*}
\sqrt{t_0} &= \frac{\ell}{\varepsilon} + \sqrt{\bigl(\frac{\ell}{\varepsilon}\bigr)^2-1}
\end{align*}
provided that $\nicefrac \ell \varepsilon \geq 1$. If this is violated, then one can easily check that $f_{\ell,\varepsilon}'(1) < 0$ so that $f_{\ell,\varepsilon}$ is monotone decreasing and has its maximum at $0$. On the other hand, if $\nicefrac \ell\varepsilon > 1$ holds with strict inequality, we immediately see that $\sqrt{t_0} > 1$. Taking the derivative of the term of $f_{\ell,\varepsilon}'$ that stands in brackets, we get
\begin{align*}
f_{\ell,\varepsilon}''(t_0) = \frac{1}{\sqrt{t_0}}\; f_{\ell-1,\varepsilon}(t_0)\; \bigl[-\frac{\varepsilon}{2} + \frac{\varepsilon}{4} \bigl( 1 + \frac 1{t_0}\bigr)\bigr] <0.
\end{align*}
In the borderline case, the solution $t_0 = 1$ to \eqref{f-l-eps-2.1} becomes unique and inserting $4$ and $4^{-1}$ into $f_{\ell,\varepsilon}'$ shows that $t_0$ is a saddle and that $f_{\ell,\varepsilon}$ is monotone decreasing. The maximum is thus located at $0$.
\end{proof}

\begin{asmp}[Abel partial sum formula]
\label{Abel-partial-sum}
Let $(a_n)_{n\geq 0}$ be a sequence of real numbers and for any $t\in \Rbb$ define
\begin{align*}
A(t) := \sum_{n=0}^{\lfloor t\rfloor} a_n.
\end{align*}
Let $x < y$ be real and $\phi\in C^1[x,y]$. Then
\begin{align*}
\sum_{n = \lceil x\rceil}^{\lfloor y\rfloor} a_n \phi(n) = A(y)\phi(y) - A(x)\phi(x) - \int_x^y \du \; A(u) \phi'(u).
\end{align*}
\end{asmp}

\begin{lem}
\label{chi-n sum estimate}
Under Assumption \ref{Abel-partial-sum}, the following holds. Let $k\in \Nbb_0$, $b>0$ and $t\geq 0$. Then
\begin{align*}
\sum_{n = \lfloor t\rfloor +1}^\infty n^k \; \chi_b(n) & \leq \frac{1}{b^{2k+2}} \; (2k+3)! \; \max\{ 1, \, b^2t\}^{k+1} \; \chi_{b}(t).
\end{align*}
\end{lem}

\begin{proof}
Let $a_n = n^k$ for $n\geq 0$. It follows that
\begin{align*}
0 \leq A(t) = \sum_{n=1}^{\lfloor t\rfloor} n^k \leq t^{k+1}.
\end{align*}
Let $\phi(t) = \e^{-b \sqrt{t}}$. We have $\phi'(t) = -b\e^{-b\sqrt{t}}\; \frac{1}{2\sqrt{t}}$. Then, since $A(t) \phi(t)\geq 0$, by Assumption \ref{Abel-partial-sum} for $t<y$, we get
\begin{align*}
\sum_{n=\lceil t\rceil}^{\lfloor y\rfloor} n^k \; \e^{-b\sqrt{n}} \leq y^{k+1} \; \e^{-b\sqrt{y}} + \frac{b}{2} \int_t^y \dd u \; u^{k+\frac 12} \; \chi_b(u).
\end{align*}
Taking the limit $y\to \infty$, we have $y^{k+1}\e^{-b\sqrt{y}} \to 0$ and applying Lemma \ref{AT:chi_estimate} (a) yields
\begin{align*}
\int_t^\infty \du \; u^{k+\frac 12}\; \chi_b(u) &\leq \frac{2}{b^{2k+3}}\; (2k+3)! \; \max\{1, b^2t\}^{k+1} \; \chi_b(t).
\end{align*}
Multiplying by $\frac b2$ gives the claim.
\end{proof}

We are now in position to give the proof of Theorem \ref{AT:old_norm_Iestimate} and remark that this is essentially a simplicitation of the proof of \cite[Theorem 4.8]{AutomorphicEquivalence}, where we work out all the constants explicitly.

\begin{proof}[Proof of Theorem \ref{AT:old_norm_Iestimate}]
Let $x,y\in \Lambda$ be given. The object to be estimated is
\begin{align*}
\sum_{Z\supset \{x,y\}} |Z|^N\; \Vert \Phi_{\Ical(G)}(Z)\Vert &\leq \sum_{Z\supset \{x,y\}} \sum_{n=0}^\infty \sum_{\substack{Y\subset \Lambda \\ Y_n = Z}} |Y_n|^N\; \Vert \Delta^n(\Phi_G(Y))\Vert \\
&\leq \kappa^N \sum_{n=0}^\infty n^{dN}\sum_{\substack{Y\subset \Lambda \\ Y_n \supset \{x,y\}}} |Y|^N \; \Vert \Delta^n(\Phi_G(Y))\Vert.
\end{align*}
Now perform the resummation. If $Y$ and $n\in\Nbb_0$ are fixed, then there are points $\tilde x\in B_n(x)\cap Y$ and $\tilde y\in B_n(y)\cap Y$. Hence, $Y$ is hit if we sum over all $\tilde x\in B_n(x)$ and $\tilde y\in B_n(y)$ and $Y'\subset \Lambda$ containing $\tilde x$ and $\tilde y$. Using Lemma \ref{Delta-estimate old norm}, we obtain the upper bound
\begin{align}
\sum_{Z\supset \{x,y\}} \Vert \Phi_{\Ical(G)}(Z)\Vert \leq  & \notag\\
&\hspace{-30pt} \leq \kappa^N C_\Delta \sum_{n=0}^\infty n^{dN} \chi_\eta(n)\sum_{\substack{\tilde x\in B_n(x) \\ \tilde y\in B_n(y)}} F_{\beta b'}(d(\tilde x, \tilde y)) \sum_{\substack{Y\subset \Lambda \\ Y\supset \{\tilde x,\tilde y\}}} |Y|^{N+1} \; \frac{\Vert \Phi_G(Y)\Vert}{F_{\beta b'}(d(\tilde x,\tilde y))}\notag\\
&\hspace{-30pt} \leq \kappa^N C_\Delta \Vert \Phi_G\Vert_{\beta b',N+1} \; \sum_{n=0}^\infty n^{dN} \chi_\eta(n) \sum_{\substack{\tilde x\in B_n(x) \\ \tilde y\in B_n(y)}} F_{\beta b'}(d(\tilde x,\tilde y)). \label{Ical estimate 1 old norm}
\end{align}
Now split the sum over $n$ at $n_0 := \lfloor \varepsilon\; \frac{d(x,y)}{2}\rfloor$ for some $0 < \varepsilon < 1$ to be chosen. Then, for $0 \leq n \leq n_0$ and $\tilde x\in B_n(x)$ and $\tilde y\in B_n(y)$, we have the estimate 
\begin{align*}
d(x,y) \leq d(x,\tilde x) + d(\tilde x,\tilde y) + d(\tilde y,y) \leq \varepsilon d(x,y) + d(\tilde x,\tilde y).
\end{align*}
Hence, $d(\tilde x,\tilde y) \geq (1-\varepsilon) d(x,y)$. The part $0 \leq n \leq n_0$ in \eqref{Ical estimate 1 old norm} is thus bounded by
\begin{align*}
\kappa^2\sum_{n=0}^{n_0} n^{d(N+2)} \; F_{\beta b'}((1-\varepsilon)d(x,y)) \leq \kappa^2\;  \Bigl(1+ \frac{d(x,y)}{2}\Bigr)^{d(N+2)+1}\;  F_{\beta b'}((1-\varepsilon) \; d(x,y)).
\end{align*}
Now, use Lemma \ref{F-upper-estimate} to bound $F((1-\varepsilon)d(x,y)) \leq (1-\varepsilon)^{-(d+1)} F(d(x,y))$ and decompose $\chi$ according to $\chi_{\beta b'}((1-\varepsilon)r) =\chi_{\sqrt{1-\varepsilon}\; \beta b}(r) \; \chi_{\sqrt{\frac{1-\varepsilon}{2}}\;  \beta(b'-b)}(\frac r2) \; \chi_{\sqrt{\frac{1-\varepsilon}{2}}\;  \beta(b'-b)}(\frac r2)$. Then utilize Lemma \ref{f-l-eps-2} with $b' - b < 1$ to estimate
\begin{align*}
\bigl( 1+ \frac r2\bigr)^{d(N+2)} \; \chi_{\sqrt{\frac{1 - \varepsilon}2}\; ( b-b')}\bigl( \frac r2\bigr) &\leq \Bigl( \frac{2\sqrt 2\; d(N+2)}{\sqrt{1-\varepsilon}\; \e\;  (b'-b)}\Bigr)^{2d(N+2)},
\end{align*}
as well as
\begin{align*}
\bigl( 1 + \frac r2\bigr) \chi_{\sqrt{\frac{1 - \varepsilon}2}\; ( b-b')}\bigl( \frac r2\bigr) &\leq \Bigl( \frac{2\sqrt{2}}{\sqrt{1 - \varepsilon}\; \e \; (b'-b)}\Bigr)^2.
\end{align*}
It follows that the part $0\leq n\leq n_0$ from \eqref{Ical estimate 1 old norm} is bounded by
\begin{align*}
&\frac{\kappa^2}{(1-\varepsilon)^{d+1}} \; \Bigl( \frac{2\sqrt 2\; d(N+2)}{\sqrt{1-\varepsilon}\; \e \;(b'-b)}\Bigr)^{2d(N+2)} \Bigl( \frac{2\sqrt{2}}{\sqrt{1 - \varepsilon}\; \e\; (b'-b)}\Bigr)^2 \;  F_{\sqrt{1-\varepsilon}\; \beta b}(d(x,y)). 
\end{align*}

For $n\geq n_0+1$ in \eqref{Ical estimate 1 old norm}, using $\eta \geq\beta b'$, we read off the bound
\begin{align}
\sum_{n=n_0+1}^\infty  \chi_{\beta b'}(n)\sum_{\substack{\tilde x\in B_n(x)\\ \tilde y\in B_n(y)}} F_{\beta b'}(d(\tilde x,\tilde y)) &\leq \Vert F\Vert_1 \kappa  \sum_{n=n_0 + 1}^\infty n^{d(N+2)+1} F_{\beta b}(n) \chi_{\beta(b'-b)}(n). \label{Ical tail1 old norm}
\end{align}
Now, we have $F_{\beta b}(n) \leq (\nicefrac 2\varepsilon)^{d+1} F_{\sqrt{\frac \varepsilon 2} \; \beta b}(d(x,y))$ by Lemma \ref{F-upper-estimate}. It follows that \eqref{Ical tail1 old norm} is bounded by
\begin{align*}
\bigl( \frac{2}{\varepsilon}\bigr)^{d+1} F_{\sqrt{\frac \varepsilon 2} \; \beta b}(d(x,y))\; \sum_{n=n_0+1}^\infty (1+n)^{d(N+2)} \chi_{\frac{\beta(b'- b)}{3}}(n) \; (1+n)\chi_{\frac{\beta(b'- b)}{3}}(n)\; \chi_{\frac{\beta(b'- b)}{3}}(n)
\end{align*}
Again, we employ Lemma \ref{f-l-eps-2} with $b-b' < 1$ to estimate
\begin{align*}
(1+n)^{d(N+2)} \chi_{\frac{\beta(b'- b)}{3}}(n) &\leq \Bigl( \frac{6d(N+2)}{\e \beta(b'-b)}\Bigr)^{2d(N+2)}
\end{align*}
and
\begin{align*}
(1+n)\chi_{\frac{\beta(b'- b)}{3}}(n) &\leq \Bigl( \frac{2\; 3}{\e \beta(b'-b)}\Bigr)^2.
\end{align*}
Then, we make use of Lemma \ref{chi-n sum estimate} with $k =0$ to bound
\begin{align*}
\sum_{n=n_0+1}^\infty \chi_{\frac{\beta(b'-b)}{3}}(n) \leq 6\bigl(\frac{3}{\beta(b'-b)}\bigr)^2 \bigl( 1 + \bigl( \frac{\beta(b'-b) \sqrt{n_0}}{3}\bigr)^2\bigr) \e^{-\frac{\beta(b-b')\sqrt{n_0}}{3}} \leq \frac{2^2\; 3^3}{\e  \beta^2 (b'-b)^2}.
\end{align*}
In the last step, we used that the function $f(t) = (1+t^2)\e^{-t}$ is bounded by $\nicefrac 2\e$. Finally, equating the two decay rates $\sqrt{\nicefrac \varepsilon 2} = \sqrt{1 - \varepsilon}$, we obtain the optimal choice $\varepsilon = \nicefrac 23$. Collecting everything, since $\frac{1}{1-\varepsilon} = \frac 2\varepsilon = 3$, we obtain
\begin{align*}
\Vert \Phi_{\Ical(G)}\Vert_{b,N} &\leq \kappa^N C_\Delta 3^{d+1}\Vert \Phi_G\Vert_{\beta b', N+1} \Bigl[ \kappa^2 \Bigl( \frac{2\sqrt{6}\; d(N+2)}{\e (b'-b)}\Bigr)^{2d(N+2)} \Bigl( \frac{2\sqrt{6}}{\e\; (b'-b)}\Bigr)^2 \\
&\hspace{50pt} + \Vert F\Vert_1 \kappa \Bigl( \frac{6d(N+2)}{\e\beta (b'-b)}\Bigr)^{2d(N+2)} \Bigl( \frac{2\; 3}{\e\beta(b'-b)}\Bigr)^2 \frac{2^2\; 3^3}{\e\beta^2(b'-b)^2}\Bigr]
\end{align*}
Since $2\sqrt{6} > \beta^{-1}\; 6$ and $b'- b < 1$, we may extract $\Dcal_\Ical(d,N,b'-b)$ and obtain
\begin{align*}
\Vert \Phi_{\Ical(G)}\Vert_{b,N} &\leq \Dcal_\Ical \; C_\Delta 3^{d+1} \Vert \Phi_G\Vert_{\beta b',N+1} \Bigl[ \bigl( \frac{2\sqrt{6}}{\e}\bigr)^2 + \bigl(\frac{2 \; 3}{\beta\e}\bigr)^2\; \frac{2^23^3}{\e\beta^2}\Bigr] \max\{\kappa^2, \Vert F\Vert_1\kappa\}.
\end{align*}
Elementary estimates using $\e < 3$ and $\beta^2 = 3$ provide $C_\Ical$ and the theorem.
\end{proof}

\printbibliography[heading=bibliography, title=Bibliography of Chapter \ref{Chapter:Adiabatic}]
\end{refsection}


\appendix

\part{Appendix}



\chapter{Local Trace Theory}
\label{Chapter:Local_Traces} \label{CHAPTER:LOCAL_TRACES}

\begin{refsection}

In this appendix, I want to present the results and methods, which helped me to understand traces per unit volume. Nothing in this appendix is new and sources are explicitly referenced. The chapter is intended to be a source to look up elementary facts for the reader, who is new to the topic of local traces. I expect the reader to be familiar with basis notions of von Neumann--Schatten classes. In case of need, the reader may consult a comprehensive course on mathematical quantum mechanics I have taken, including an introduction to trace ideals \cite{MQM2}.


\section{Standard Traces}

We start by a proof of Hölder's inequality, which is robust enough to be applied for the local trace case as well. Furthermore, we review Klein's inequality, which is a basic but widely used and important relative entropy inequality.

The set of smooth and compactly supported functions is denoted by $C_c^\infty(\Rbb^d)$. The Schwartz space $\Scal(\Rbb^d)$ consists of all smooth functions, which, together with their derivatives, decay faster than any inverse power at infinity.

Let $\Hcal$ be a Hilbert space. The space of bounded linear operators $\Hcal \ra \Hcal$ is denote by $\Bcal(\Hcal)$. We denote the space of compact operators by $\Scal^\infty(\Hcal)$, equipped with the usual operator norm. For $1 \leq p < \infty$ we define the $p$\tho\ von Neumann--Schatten class $\Scal^p(\Hcal)$ as the space of compact operators $A$ for which
\begin{align*}
\Vert A\Vert_p^p := \tr(|A|^p) < \infty.
\end{align*}
$\Scal^p(\Hcal)$ is equipped with the norm $\Vert \cdot\Vert_p$. It is shown in \cite[Section A.3]{MQM2} that any operator which belongs to $\Scal^p(\Hcal$ is compact) and in \cite[Section A.4]{MQM2} that $\Scal^p(\Hcal)$ is a Banach space. The latter is surprisingly hard to show.

We assume the reader to be familiar with basic facts on compact operators like existence of polar decompositions and the Schmidt decomposition.

\subsection{Hölder's inequality}

First, we prove Hadamard's three line theorem from \cite[Appendix to IX.4]{Reedsimon2}.

\begin{lem}[Hadamard's three line theorem]
Let $S :=\{z\in \Cbb : 0<\Re z< 1\}$ be the open strip and $\varphi\colon \ov S\lra \Cbb$ be bounded and continuous, analytic in $S$ with
\begin{align*}
|\varphi(z)| &\leq M_0 & \Re z &= 0
\end{align*}
and
\begin{align*}
|\varphi(z)| &\leq M_1 & \Re z &= 1.
\end{align*}
Then $|\varphi(z)| \leq M_0^{1 - \Re z} M_1^{\Re z}$ for all $z\in \ov S$.
\end{lem}

\begin{proof}
Since $\tilde \varphi(z) := \varphi(z) M_0^{z-1} M_1^{-z}$ satisfies the hypothesis with the bounds
\begin{align}
M_0 = 1 = M_1. \label{Threeline-wlog}
\end{align}
we may as well assume \eqref{Threeline-wlog}. If $\varphi(z) \to 0$ as $|z| \to\infty$, $z\in \ov S$, we choose a compact set $K\subseteq \ov S$ so that $|\varphi|\leq \frac 12$ outside $K$. Then $|\varphi|\leq 1$ in $K$ follows from the maximum modulus principle. Otherwise, consider $\varphi_n(z) =\varphi(z) \e^{\frac{z^2-1}{n}}$. Then, for $\Re z \in \{0,1\}$,
\begin{align*}
|\e^{\frac{z^2 - 1}{n}}| &= \bigl| \exp \bigl[\frac 1n \left[ (\Re z)^2 - 1 - (\Im z)^2 + 2\i \Im z \Re z\right]\bigr]\bigr| = \e^{\frac 1n ((\Re z)^2 - 1 - (\Im z)^2)}\\
&= \begin{cases} \e^{-\frac 1n (1 + (\Im z)^2)} & \Re z =0 \\ \e^{-\frac 1n (\Im z)}  & \Re z = 1 \end{cases} \\ &\leq 1.
\end{align*}
Hence, $|\varphi_n(z)|\leq 1$ on $\partial S$ and $\varphi_n(z) \to 0$ as $|z|\to\infty$, $z\in \ov S$. We conclude that $|\varphi_n(z)|\leq 1$ for all $z\in \ov S$ and since $\e^{\frac{z^2-1}{n}}\to 1$ as $n\to\infty$, we conclude that $|\varphi(z)| \leq 1$ for all $z\in \ov S$.
\end{proof}

\begin{cprop}[{\cite[Appendix to IX.4, Prop. 5]{Reedsimon2}}]
Let $1\leq p,q\leq \infty$ and $p,q$ be Hölder conjugate. If $A\in \Scal^p(\Hcal)$ and $B\in\Scal^q(\Hcal)$, then $AB\in \Scal^1(\Hcal)$ and $\Vert AB\Vert_1 \leq \Vert A\Vert_p\cdot\Vert B\Vert_q$
\end{cprop}

\begin{proof}
Let $A = U|A|$ and $B = V|B|$ be the polar decompositions. Let $S = \{z\in \Cbb : 0 < \Re z < 1\}$ and, for $z\in \ov S$, define
\begin{align*}
F(z) := \tr ( U |A|^{pz} V |B|^{q(1-z)}).
\end{align*}
Then, $F$ is well-defined and bounded since
\begin{align*}
|F(z)| \leq \Vert U\Vert \cdot \Vert \,|A|^{pz}\, \Vert \cdot \Vert V\Vert \cdot \tr (|B|^q) \cdot\Vert \, |B|^{-qz}\, \Vert.
\end{align*}
Now,
\begin{align*}
\Vert \, |A|^{qz}\, \Vert \leq \Vert \,|A|^{p\Re z} \, \Vert\cdot \Vert \, |A|^{\i p\Im z}\,\Vert \leq \sup_{0\leq s\leq 1} \sup_{t\in [-\Vert A\Vert ,\Vert A\Vert]} |t^{ps}| <\infty.
\end{align*}
Similarly for $B$. Continuity and analyticity in the strip are clear. Furthermore, for $y\in \Rbb$,
\begin{align*}
|F(\i y)| = |\tr( U|A|^{\i p y} V|B|^{-\i qy} |B|^q)| \leq \tr (|B|^q) = \Vert B\Vert_q^q
\end{align*}
and
\begin{align*}
|F(1 + \i y)| = \tr (U |A|^{\i py}|A|^p V|B|^{-\i qy})| \leq \tr(|A|^p) = \Vert A\Vert_p^p,
\end{align*}
since $\Vert AB\Vert_p \leq \Vert A\Vert_p \Vert B\Vert_\infty$ for any $1\leq p\leq \infty$ (here used $p=1$) by the min-max-principle. By Hadamard's three line theorem, we infer that
\begin{align}
|\tr(AB)| = | F(\nicefrac 1p)| \leq \Vert A\Vert_p\Vert B\Vert_q. \label{little-hoelder}
\end{align}
Finally, by cyclicity of the trace and \eqref{little-hoelder},
\begin{align*}
\tr(|AB|) &= \tr(\sqrt{V|B|^2V^* |A|^2}) = \tr(\sqrt{V|B|^2V^*}\cdot |A|) \\
&\leq \Vert \sqrt{V|B|^2V^*}\Vert_q \Vert A\Vert_p = \Vert A\Vert_p \Vert B\Vert_q. \qedhere
\end{align*}
\end{proof}

\begin{kor}[Generalized Hölder's inequality]
Let $1 \leq p,q,r \leq \infty$ such that we have $p^{-1} + q^{-1} = r^{-1}$. Let $A\in \Scal^p(\Hcal)$ and $B\in \Scal^q(\Hcal)$. Then $AB\in \Scal^r(\Hcal)$ and
\begin{align}
\Vert AB\Vert_r \leq \Vert A\Vert_p \Vert B\Vert_q \label{Generalized Hölder}
\end{align}
holds.
\end{kor}

\subsection{Peierl's and Klein's inequality}

For an interval $I\subseteq \Rbb$ and bounded and measurable a function $f\colon I \lra \Rbb$ define
\begin{align*}
\Scal_f^1(\Hcal) := \{ A\in \Scal^\infty(\Hcal) : A=A^*, \; \sigma(A)\subseteq I, \; f(A)\in \Scal^1(\Hcal)\}.
\end{align*}

\begin{thm}[{\cite[Theorem 2.9]{TraceInequalities}}, Peierl's inequality]
\label{Peierl's inequality}
Let $I\subseteq \Rbb$ be an interval, $f\colon I\ra \Rbb$ convex and $A\in \Scal_f^1(\Hcal)$. Let $\{u_n\}_{n\in \Nbb}$ be any ONB of $\Hcal$. Then
\begin{align}
\sum_{n=1}^\infty f(\langle u_n, Au_n\rangle) \leq \tr(f(A)) \label{Peierl}
\end{align}
and equality holds in \eqref{Peierl} if and only if $u_n$ is an eigenvector of $A$ for all $n\in \Nbb$. If $f$ is strictly convex, then equality in \eqref{Peierl} holds only in this case.
\end{thm}

\begin{proof}
Let $A = \sum_{n=1}^\infty \lambda_n |\varphi_n\rangle\langle \varphi_n|$ be the Schmidt-decomposition of $A$. Then
\begin{align*}
\tr(f(A)) &= \sum_{n=1}^\infty \langle u_n, f(A)u_n\rangle = \sum_{n=1}^\infty \sum_{i=1}^\infty f(\lambda_i) |\langle \varphi_i, u_n\rangle|^2\\
&\geq \sum_{n=1}^\infty f\Bigl( \sum_{i=1}^\infty \lambda_i |\langle \varphi_i, u_n\rangle|^2\Bigr) = \sum_{n=1}^\infty f\Bigl( \sum_{i=1}^\infty \langle \varphi_i, u_n\rangle \langle u_n, A\varphi_i\rangle\Bigr) \\
&= \sum_{n=1}^\infty f\Bigl( \Bigl\langle u_n, \sum_{i=1}^\infty \langle \varphi_i, u_n\rangle A\varphi_i\Bigr\rangle\Bigr) = \sum_{n=1}^\infty f(\langle u_n, Au_n\rangle).
\end{align*}
Moreover,
\begin{align*}
\sum_{i=1}^\infty f( \langle \varphi_i,A\varphi_i\rangle) = \sum_{i=1}^\infty f(\lambda_i) = \sum_{i=1}^\infty f(\lambda_i) \langle \varphi_i,\varphi_i\rangle = \sum_{i=1}^\infty \langle \varphi_i, f(A)\varphi_i\rangle = \tr(f(A)).
\end{align*}
Suppose that $f$ is strictly convex. Then, since $\lambda_i\neq 0$ for all $i$, $u_i = \varphi_i$ is the only possibility for equality.
\end{proof}

\begin{kor}
\label{Convex-Tracefunction}
Let $I\subseteq \Rbb$ be an interval and $f\colon I\lra \Rbb$ convex. Then
\begin{align*}
\Phi_f \colon \Scal_f^1(\Hcal) &\lra \Rbb \\
A &\longmapsto \tr(f(A))
\end{align*}
is convex and $\Phi_f$ is strictly convex if and only if $f$ is strictly convex.
\end{kor}

\begin{proof}
Let $t,s\in [0,1]$ with $t + s = 1$ and $A,B\in \Scal_f^1(\Hcal)$. Let $(\varphi_n)_n$ be the eigenbasis for $tA+sB$. Replacing $f$ by $|f|$ (convex!) in the following computation shows that $tA + sB\in \Scal_f^1(\Hcal)$. By Peierl's inequality \eqref{Peierl}, we get
\begin{align}
\tr(f(t A + sB)) &= \sum_{n=1}^\infty f(\langle \varphi_n, (tA+sB) \varphi_n\rangle) = \sum_{n=1}^\infty f(t\langle \varphi_n, A\varphi_n\rangle + s\langle \varphi_n, B\varphi_n\rangle) \notag \\
&\leq \sum_{n=1}^\infty t f(\langle \varphi_n, A\varphi_n\rangle) + s f(\langle \varphi_n,B\varphi_n\rangle) \label{convex-star}\\
&\leq t \tr(f(A)) + s \tr (f(B)). \notag
\end{align}
Furthermore, \eqref{convex-star} is strict if and only if $f$ is strictly convex.
\end{proof}

\begin{lem}
Let $f\colon I\lra \Rbb$ be a convex function. Then for all $s,t,u\in I$ with $s<t<u$, we have
\begin{align}
\frac{f(t) - f(s)}{t-s} \leq \frac{f(u)- f(s)}{u-s} \leq \frac{f(u) - f(t)}{u-t} \label{convex-monotonicity}
\end{align}
\end{lem}

\begin{proof}
Set $x = u -s$ and let $\eta >0$ such that $s + \eta x = t$. Equivalently, $t = u\eta + s(1 - \eta)$ or $\eta = \frac{t-s}{u-s}$. Then, the definition of convexity gives
\begin{align*}
f(t) \leq \eta f(u) + (1 - \eta) f(s)
\end{align*}
so that
\begin{align*}
f(t) - f(s) \leq \eta f(u) - \eta f(s)
\end{align*}
which, in turn, is
\begin{align*}
\frac{f(t) - f(s)}{t - s} \leq \frac{f(u) - f(s)}{u- s}.
\end{align*}
In the same manner, we obtain
\begin{align*}
f(u) - f(t) &\geq (1 - \eta) f(u) - (1 - \eta) f(s) = \frac{u-t}{u-s} [f(u) - f(s)]. \qedhere
\end{align*}
\end{proof}

\begin{thm}[{\cite[Theorem 2.11]{TraceInequalities}}, Klein's inequality]
\label{Klein's inequality}
Let $I\subseteq \Rbb$ be an interval and $f\colon I\lra \Rbb$ convex. Let $A,B\in \Scal_f^1(\Hcal)$. Assume that the right-sided derivative $f_+'$ is bounded on $\sigma(B)$ and that $A - B\in \Scal^1(\Hcal)$. Then
\begin{align}
\tr (f(A) - f(B) - f'_+(B) (A - B))\geq 0. \label{Klein}
\end{align}
If $f$ is strictly convex, then equality holds in \eqref{Klein} if and only if $A =B$.
\end{thm}

\begin{proof}
Let $C := A - B$ so that for $0\leq t \leq 1$, we have that
\begin{align*}
B + tC = (1-t) B + tA.
\end{align*}
Since $|f|$ is convex, $B + tC\in \Scal_f^1(\Hcal)$ for $0\leq t \leq 1$. Define $\varphi(t) := \tr(f(B + tC))$. By Corollary \ref{Convex-Tracefunction}, $\varphi$ is convex. Since $t = 0\cdot (1 - t) + 1\cdot t$, we infer that
\begin{align}
\varphi(t) \leq \varphi(0) (1-t) + \varphi(1) t \label{pre-Klein-2}
\end{align}
or, equivalently,
\begin{align}
\varphi(1) - \varphi(0) \geq \frac{\varphi(t) - \varphi(0)}{t}. \label{pre-Klein}
\end{align}
Applying \eqref{convex-monotonicity} with $s =0$ and $t < u$, we see that
\begin{align*}
\frac{\varphi(t) - \varphi(0)}{t} \leq \frac{\varphi(u) - \varphi(0)}{u}
\end{align*}
so that the right-hand side of \eqref{pre-Klein} monotonously decreases to $\varphi_+'(0)$ as $t\searrow 0$. But $\varphi_+'(0) = \tr(f_+'(B)(B-A))$. Plugging in, we get \eqref{Klein}. If $f$ is strictly convex, then $\varphi$ is strictly convex if and only if $C\neq 0$ (see Corollary \ref{Convex-Tracefunction}). This is equivalent to \eqref{pre-Klein-2} being strict for all $0 < t < 1$. By the monotonicity of the right-hand side of the \eqref{pre-Klein}, this, in turn, is equivalent to \eqref{pre-Klein} being strict in the limit $t \searrow 0$. Hence, \eqref{Klein} is strict if and only if $A \neq B$. 
\end{proof}

%


\section{Periodic Operators and Bloch--Floquet Direct Integrals}

\subsection{Hilbert space valued functions}

In this section, we follow \cite{Reedsimon1}, Section II.1, Example 6 (p. 40, 41), as well as Problem 12 (p. 64).

\begin{defn}
Let $\Hcal'$ be a separable Hilbert space and $(M,\mu)$ a measure space. A function $f\colon M\lra \Hcal'$ is called measurable iff $m\mapsto \langle y,f(m)\rangle_{\Hcal'}$ is measurable for all $y\in \Hcal'$.
\end{defn}

\begin{lem}
\label{lem2.2.2}
Suppose that $f,g\colon M\lra \Hcal'$ are measurable. Then $m\mapsto \Vert f(m)\Vert_{\Hcal'}^2$ and $m\mapsto \langle f(m), g(m)\rangle_{\Hcal'}$ are measurable.
\end{lem}

\begin{proof}
Let $(\varphi_n)_n$ be an ONB of $\Hcal'$ and for a.e. $m\in M$ write 
\begin{align*}
\langle f(m), g(m)\rangle_{\Hcal'} &= \sum_{n=1}^\infty \langle f(m), \varphi_n\rangle  \langle \varphi_n, g(m)\rangle.
\end{align*}
For a.e. $m\in M$ set
\begin{align*}
\Psi_N(m) := \sum_{n=1}^N \langle f(m), \varphi_n\rangle \langle \varphi_n, g(m)\rangle.
\end{align*}
Then $\Psi_N$ is measurable $M\ra \Rbb$ as usual. Furthermore, for a.e. $m\in M$, by Hölder,
\begin{align*}
|\langle f(m), g(m)\rangle - \Psi_N(m)| &\leq \sum_{n=N+1}^\infty |\langle f(m),\varphi_n\rangle \langle \varphi_n,g(m)\rangle| \\
&\leq \Bigl( \sum_{n=N+1}^\infty |\langle f(m) , \varphi_n\rangle|^2\Bigr)^{\nicefrac 12} \Bigl( \sum_{n=N+1} ^\infty |\langle \varphi_n, g(m)\rangle|^2\Bigr)^{\nicefrac 12} .
\end{align*}
The expressions on the right-hand side are bounded by $\Vert f(m)\Vert_{\Hcal'}$ and $\Vert g(m)\Vert_{\Hcal'}$, respectively. These, in turn, are finite for a.e. $m\in M$ since $f(m), g(m)\in \Hcal'$ for a.e. $m\in M$. Hence, for a.e. $m\in M$, we conclude
\begin{align*}
|\langle f(m),g(m)\rangle - \Psi_N(m)| \xra{N\to\infty} 0.
\end{align*}
So that $m\mapsto \langle f(m),g(m)\rangle$ is a pointwise limit of measurable functions and thus measurable. Set $f = g$ to conclude for $m\mapsto \Vert f(m)\Vert_{\Hcal'}^2$.
\end{proof}

\begin{defn}
Let $\Hcal'$ be a separable Hilbert space and $(M,\mu)$ a measure space. Then, we define $\Hcal := L^2(M,\mu; \Hcal')$ as the set of all (equivalence classes of $\mu$-a.e. equal) measurable functions $f\colon M\lra \Hcal'$ such that
\begin{align*}
\Vert f\Vert^2_{\Hcal} := \int_M \Vert f(m)\Vert_{\Hcal'}^2\, \dd \mu(m) < \infty.
\end{align*}
We also write
\begin{align*}
\Hcal = \int_M^\oplus \Hcal' \, \dd \mu.
\end{align*}
\end{defn}

The set $\Hcal$ becomes a Hilbert space via
\begin{align*}
\langle f,g\rangle_\Hcal := \int_M \langle f(m),g(m)\rangle_{\Hcal'} \, \dd \mu(m).
\end{align*}

\begin{clem}[{\cite[Problem 12, p. 64]{Reedsimon1}}]
\label{lem2.2.4}
\begin{enumerate}[(a)]
\item Let $(\varphi_k)_{k\in \Nbb}$ be an ONB. Let $g\in \Hcal$. Then,
\begin{align*}
\sum_{k=1}^N \langle \varphi_k,g(\cdot) )\varphi_k \xra{N\to\infty} g
\end{align*}
in $\Hcal$ and if $f\in \Hcal$ is another function, then
\begin{align*}
\langle f,g\rangle_\Hcal = \sum_{k=1}^\infty \int_M \langle f(m),\varphi_k\rangle_{\Hcal'} \langle \varphi_k,g(m)\rangle_{\Hcal'} \, \dd \mu(m).
\end{align*}
In particular,
\begin{align}
\Vert f\Vert_{\Hcal}^2 = \sum_{k=1}^\infty \Vert \langle \varphi_k, f(\cdot)\rangle_{\Hcal'}\Vert_{L^2(M,\mu)}^2. \label{Pythagoras}
\end{align}

\item If $L^2(M,\mu)$ is separable, then so is $\Hcal$.
\end{enumerate}
\end{clem}

\begin{proof}
\begin{enumerate}[(a)]
\item We have that
\begin{align*}
\Bigl\Vert g - \sum_{k=1}^N \langle \varphi_k,g(\cdot)\rangle \varphi_k\Bigr\Vert_\Hcal^2 = \int_M \Bigl\Vert g(m) - \sum_{k=1}^N \langle \varphi_k, g(m)\rangle_{\Hcal'} \varphi_k\Bigr\Vert_{\Hcal'}^2 \, \dd \mu(m).
\end{align*}
The integrand is bounded by $4 \Vert g(m)\Vert_{\Hcal'}^2$ which is integrable and since pointwise convergence holds by the usual Hilbert space techniques, we conclude by dominated convergence. Furthermore, we have
\begin{align*}
\langle f,g\rangle_\Hcal = \int_M \sum_{k=1}^\infty \langle f(m),\varphi_k\rangle\langle \varphi_k,g(m)\rangle \, \dd\mu(m).
\end{align*}
Call $\Psi_N(m)$ the partial sum in the integrand for a.e. $m\in M$. We intend to apply dominated convergence to prove that
\begin{align*}
\langle f,g\rangle_\Hcal = \int_M \lim_{N\to\infty} \Psi_N(m)\, \dd \mu(m) = \lim_{N\to\infty} \int_M \Psi_N(m) \, \dd \mu(m).
\end{align*}
We must provide an $N$-independent integrable dominant for $\Psi_N$. We have
\begin{align*}
|\Psi_N(m)| \leq \Bigl( \sum_{k=1}^N |\langle f(m), \varphi_k\rangle|^2 \Bigr)^{\nicefrac 12} \Bigl( \sum_{k=1}^N |\langle \varphi_k,g(m)\rangle|^2\Bigr)^{\nicefrac 12} \leq \Vert f(m)\Vert_{\Hcal'}^2\cdot \Vert g(m)\Vert_{\Hcal'}^2.
\end{align*}
The right-hand side is integrable since $f,g\in \Hcal$. Thus dominated convergence applies.

\item Let $(f_n)_n\subseteq L^2(M,\mu)$ be an ONB and let $(\varphi_k)_k\subseteq \Hcal'$ be an ONB. We claim that $\{f_n\varphi_k\}_{(m,k)\in \Nbb^2}$ is an ONB for $\Hcal$. First of all, note that for $(m,k),(n,\ell)\in \Nbb^2$, we have
\begin{align*}
\langle f_m\varphi_k, f_n\varphi_\ell\rangle_\Hcal &= \int_M \langle f_m(m)\varphi_k,f_n(m)\varphi_\ell\rangle \dd \mu(m) = \int_M \ov{f_m(m)}f_n(m) \dd \mu(m) \langle  \varphi_k,\varphi_\ell\rangle \\
&= \langle f_m,f_n \rangle_{L^2(M,\mu)} \langle \varphi_k,\varphi_\ell\rangle_{\Hcal'} = \delta_{(m,k),(n,\ell)}.
\end{align*}
Hence, $\{f_m\varphi_k\}_{(m,k)\in \Nbb^2}$ is an orthonormal set. Furthermore, for any $f\in \Hcal$, we have
\begin{align*}
f(m) = \sum_{k\in \Nbb} \langle \varphi_k, f(m)\rangle_{\Hcal'} \varphi_k = \sum_{k\in \Nbb} \sum_{m\in \Nbb} \bigl \langle f_n , \langle \varphi_k, f(\cdot)\rangle_{\Hcal'}\bigr\rangle_{L^2(M,\mu)} f_n(m)\varphi_k,
\end{align*}
i.e.,
\begin{align*}
f = \sum_{k\in \Nbb} \sum_{m\in \Nbb} \bigl \langle f_n , \langle \varphi_k, f(\cdot)\rangle_{\Hcal'}\bigr\rangle_{L^2(M,\mu)} f_n\varphi_k.
\end{align*}
Let's show that the right-hand side converges in $\Hcal$. By \eqref{Pythagoras}, we have
\begin{align*}
\Bigl\Vert \sum_{k=1}^\infty \sum_{m=1}^\infty \bigl\langle f_m, \langle \varphi_k,f(\cdot)\rangle_{\Hcal'} \bigr\rangle_{L^2(M,\mu)} f_m\varphi_k\Bigr\Vert^2_\Hcal = \hspace{-202pt} & \\
&= \sum_{k,k'\in \Nbb} \sum_{m,m'\in \Nbb} \ov{ \bigl\langle f_m,\langle \varphi_k, f(\cdot)\rangle_{\Hcal'}\bigr\rangle_{L^2(M,\mu)}} \bigl\langle f_{m'}, \langle \varphi_{k'} , f(\cdot)\rangle_{\Hcal'} \bigr\rangle_{L^2(M,\mu)} \langle f_m\varphi_k, f_{m'} \varphi_{k'} \rangle_{\Hcal}\\
&= \sum_{k=1}^\infty \sum_{m=1}^\infty |\langle f_m, \langle \varphi_k,f(\cdot)\rangle_{\Hcal'}\rangle_{L^2(M,\mu)}|^2 = \sum_{k=1}^\infty \Vert \langle \varphi_k,f(\cdot)\rangle_{\Hcal'}\Vert_{L^2(M,\mu)}^2 = \Vert f\Vert_\Hcal^2 <\infty.
\end{align*}
This completes the proof.
\end{enumerate}
\end{proof}

\subsection{Different notions of measurability}

We follow \cite{Reedsimon1}, Appendix to IV.5 (pp. 115).

\begin{defn}
Let $(M, \Acal)$ be a measurable space and let $E$ be a Banach space. Let $f\colon M\lra E$ be a function.
\begin{enumerate}[(a)]
\item $f$ is called strongly measurable if and only if there is a sequence of functions $f_n\colon M\lra E$ such that $f_n(m) \to f(m)$ in norm for a.e. $m\in M$ and each $f_n$ takes only finitely many values, each value being taken on a set in $\Acal$.

\item $f$ is called Borel measurable if  and only if $f^{-1}(C)\in \Acal$ for each open set $C$ in $E$ (in the metric space topology on $E$).

\item $f$ is called weakly measurable if and only if $\ell(f(m))$ is a complex-valued measurable function for each $\ell\in E'$.
\end{enumerate}
\end{defn}

\begin{prop}
\label{MeasurabilitiesBanach}
Let $f\colon M\lra E$ be a function.
\begin{enumerate}[(a)]
\item Let $(f_n)_n$ be a sequence of Borel measurable functions such that $f_n(m) \to f(m)$ in norm as $n\to\infty$. Then $f$ is Borel measurable.

\item If $f$ is strongly measurable, then $f$ is Borel measurable.

\item If $f$ is Borel measurable, then $f$ is weakly measurable.
\end{enumerate}
\end{prop}

\begin{proof}
\begin{enumerate}[(a)]
\item Let $C\subseteq E$ be open. Define $C_k := \{e\in E : B_{\frac 1k}(e) \subseteq C\}$ for $k\in \Nbb$. Then we have $C =\bigcup_{k\in \Nbb} C_k$. Furthermore, for a.e. $m\in M$ and all $k\in \Nbb$, there is $N_k\in \Nbb$ such that for all $n\geq N_k$, we have $f_n(m) \in B_{\frac 1k}(f(m))$. Hence
\begin{align*}
f^{-1}(C) = \bigcup_{k\in \Nbb} \bigcup_{N\in \Nbb} \bigcap_{n\geq N} f_n^{-1}(C_k),
\end{align*}
which is measurable.

\item The approximating sequence is Borel measurable. Then apply (a).

\item The natural Borel $\sigma$-algebra on $E$ (generated by open sets) makes $\ell\in E'$ Borel measurable in the above sense, since $\ell$ is continuous. Then, the composition of Borel functions is a Borel function.
\end{enumerate}
\end{proof}

\begin{lem}
\label{ApproxSimpleFunctions}
Let $(M,\Acal)$ be a measurable space and $f\colon M\lra\Cbb$ be measurable. Then, there is a sequence $(f_n)_n$ of simple functions with $|f_n(m)|\leq |f(m)|$ and $f_n(m)\to f(m)$ for a.e. $m\in M$ as $n\to\infty$.
\end{lem}

\begin{proof}
By treating real and imaginary part and positive and negative part separately, we may assume that $f$ is real-valued and $f\geq 0$. For given $m\in \Nbb$ and $\varepsilon>0$ choose $n = n(m,\varepsilon)\in \Nbb$ so large that $\eta := \frac{m}{n} < \varepsilon$. For $i = 1, \ldots, n$ set $A_i^n := f^{-1}([(i-1)\eta, i\eta))$ and $\alpha_i^n := (i-1)\eta$. Define $f_n := \sum_{i=1}^n \alpha_i^n \Idbb_{A_i^n}$. For almost every $m\in M$, there is $m\in \Nbb$ such that $|f(m)|< m$ which means $m\in A_i^n$ for some $i = 1, \ldots, n$ so that $(i - 1)\eta \leq f(m)\leq  i\eta$. Hence, $f_n(m) = \alpha_i^n = \eta(i-1) \leq f(m)$. Furthermore, 
\begin{align*}
|f(m) - f_n(m)| = f(m) - \eta(i-1) \leq i \eta - i\eta + \eta = \eta < \varepsilon,
\end{align*}
which concludes pointwise convergence.
\end{proof}

\begin{cthm}[{\cite[Theorem IV.22]{Reedsimon1}}]
\label{MeasurabilitiesHilbert}
Let $\Hcal$ be a separable Hilbert space and $(M,\Acal)$ a measurable space. Let $f\colon M\lra \Hcal$. Then the following statements are equivalent:
\begin{enumerate}[(a)]
\item $f$ is strongly measurable.

\item $f$ is Borel measurable.

\item $f$ is weakly measurable.
\end{enumerate}
\end{cthm}

\begin{proof}
By Proposition \ref{MeasurabilitiesBanach}, it suffices to prove that (c) implies (a). Assuming (c), fix an ONB $(\psi_n)_n\subseteq \Hcal$ and define $a_n := (\psi_n, f(\cdot))_{\Hcal}$. By hypothesis, the $a_n$'s are measurable complex-valued functions. By Lemma \ref{ApproxSimpleFunctions}, there is a sequence $(a_{n,m})_m$ of simple functions with $|a_{n,m}(m)| \leq |a_n(m)|$ and $a_{n,m}(m)\to a_n(m)$ for all $n\in \Nbb$ and a.e. $m\in M$. Define $f_N := \sum_{n=1}^N a_{n,N} \psi_n$. For given $\varepsilon>0$ and $n\in\Nbb$ a.e. $m\in M$ choose $N \in \Nbb$ so large that $|a_{n}(m) - a_{n,N}(m)|^2\leq \frac{\varepsilon}{2^n}$. Then, for a.e. $m\in M$, we have
\begin{align*}
\Vert f(m) - f_N(m)\Vert^2 &= \sum_{n=1}^N |a_n(m) - a_{n,N}(m)|^2 + \sum_{n=N+1}^\infty |a_n(m)|^2 \\
&\leq \varepsilon \sum_{n=1}^\infty 2^{-n} + \sum_{n=N+1}^\infty |a_n(m)|^2 \xra{N\to\infty} \varepsilon. \qedhere
\end{align*}
\end{proof}

\subsection{Decomposable operators}

We follow Section XIII.16 in \cite{Reedsimon4} (pp. 279). Let $(M,\mu)$ be a measure space. A function $A\colon M \lra \Bcal(\Hcal')$ is called measurable if and only if $m\mapsto \langle \varphi, A(m)\psi\rangle$ is measurable for all $\varphi,\psi\in \Hcal'$ (i.e. weakly measurable). By $L^\infty(M,\mu,\Bcal(\Hcal'))$ denote the space of measurable functions $A\colon M\lra \Bcal(\Hcal')$ with
\begin{align*}
\Vert A\Vert_\infty := \esssup_{m\in M} \Vert A(m)\Vert_{\Bcal(\Hcal')} < \infty.
\end{align*}

\begin{defn}
A bounded operator $A$ on $\Hcal = \int_M^\oplus \Hcal' \dd\mu$ is decomposed by the direct integral decomposition if and only if there is $\Afrak \in L^\infty(M, \mu, \Bcal(\Hcal))$ such that
\begin{align*}
(A\psi)(m) = \Afrak(m) \psi(m)
\end{align*}
for almost every $m\in M$ and all $\psi\in\Hcal$. In this case, we call $A$ decomposable and write
\begin{align*}
A =\int_M^\oplus \Afrak(m) \, \dd \mu(m).
\end{align*}
\end{defn}

\begin{cthm}[{\cite[Theorem XIII.83]{Reedsimon4}}]
\label{Definition direct integral}
If $\Afrak\in L^\infty(M,\mu, \Bcal(\Hcal'))$, then there is a unique decomposable operator $A\in \Bcal(\Hcal)$ such that $(A\psi)(m) = \Afrak(m)\psi(m)$ holds for all $\psi\in \Hcal$ and a.e. $m\in M$.
\end{cthm}

\begin{proof}
For uniqueness note that if $A,B\in\Bcal(\Hcal)$ are two such operators, then we have $(A\psi)(m) - (B\psi)(m) =0$ for all $\psi\in \Hcal$ and a.e. $m\in M$. Thus $A = B$. For existence let $\psi\in L^2(M,\mu, \Hcal')$ and let $\{\eta_k\}_{k \in \Nbb}$ be an ONB for $\Hcal'$. Then
\begin{align}
\Afrak(m) \psi(m) = \sum_{k=1}^\infty \langle \eta_k, \psi(m)\rangle_{\Hcal'} \Afrak(m)\eta_k \label{XIII.83.1}
\end{align}
for a.e. $m\in M$ since $\Afrak(m)$ is a bounded operator for a.e. $m\in M$. Now, $\Afrak(m)\eta_k$ is weakly measurable. Hence, for all $N\in \Nbb$,
\begin{align*}
\varphi_N(m) := \sum_{k=1}^N \langle \eta_k  , \psi(m)\rangle \Afrak(m) \eta_k
\end{align*}
is weakly (and hence, strongly) measurable by Theorem \ref{MeasurabilitiesHilbert}. Moreover, for $N,K\in \Nbb$, we have
\begin{align*}
\int_M \Vert \varphi_N(m) - \varphi_K(m)\Vert^2 \, \dd \mu &= \int_M \Bigl\Vert \Afrak(m) \sum_{k=K+1}^N \langle \eta_k, \psi(m)\rangle \eta_k \Bigr\Vert^2\, \dd \mu  \\
&\leq \Vert \Afrak\Vert_\infty^2 \int_M \Bigl\Vert \sum_{k=K+1}^N \langle \eta_k, \psi(m)\rangle\eta_k\Bigr\Vert^2 \, \dd \mu \leq \Vert \Afrak\Vert_\infty^2 \cdot \Vert \psi\Vert^2
\end{align*}
since
\begin{align}
\Bigl\Vert \sum_{k=K+1}^N \langle \eta_k, \psi(m)\rangle \eta_k\Bigr\Vert^2 = \sum_{k=K+1}^N |\langle \eta_k, \psi(m)\rangle|^2 \leq \Vert \psi(m)\Vert^2 \label{XIII.83 Dominant}
\end{align}
for almost all $m\in M$ and all $N\in \Nbb$ by Bessel. Now, $\sum_{k=K+1}^N \langle \eta_k, \psi(m)\rangle \eta_k \to 0$ as $K,N\to\infty$ by \eqref{XIII.83 Dominant} and it is in $L^1(M,\mu)$. Hence, by dominated convergence, we conclude that $(\varphi_N)_{N\in \Nbb}$ is Cauchy and hence, converges to a limit $\varphi\in \Hcal$. But that limit is equal to $\Afrak(m)\psi(m)$ for almost all $m\in M$ by \eqref{XIII.83.1}. Hence, $A\psi(m):= \Afrak(m)\psi(m)$ defines an $L^2$-function $M\to \Hcal'$, that is, an element of $\Hcal$. Also, using dominated convergence again,
\begin{align*}
\Vert A\psi\Vert^2 = \int_M \lim_{N\to\infty} \Vert \varphi_N(m)\Vert^2 \, \dd \mu \leq \Vert \Afrak\Vert_\infty^2 \Vert \psi\Vert^2
\end{align*}
So, $A$ is bounded and $\Vert A \Vert \leq \Vert \Afrak\Vert_\infty$. To prove the converse inequality, let $\alpha,\beta\in \Hcal'$ and let $f\in L^1(M,\mu)$. Decompose\footnote{For example, set $g := |f|^{\nicefrac 12}$ and $h := \nicefrac fg \cdot \Idbb_{\{f\neq 0\}}$.} $f$ as $f = gh$ with $g,h\in L^2(M,\mu)$ and $\Vert f\Vert_2^2 =\Vert h\Vert_2^2 = \Vert f\Vert_1$. Set $\psi := \ov g \alpha$, $\varphi = h\beta$. Then
\begin{align*}
\Bigl| \int_M f(m) \langle \alpha, \Afrak(m) \beta\rangle \, \dd \mu\Bigr| &= |\langle \psi, A\varphi\rangle| \leq \Vert A \Vert\cdot \Vert \psi\Vert_{\Hcal'}  \Vert \varphi\Vert_{\Hcal'} \\
&=\Vert A\Vert \Bigl( \int_M \Vert \ov g(m) \alpha \Vert^2 \, \dd \mu\Bigr)^{\nicefrac 12} \Bigl( \int_M \Vert h(m)\beta\Vert_{\Hcal'}^2 \, \dd \mu\Bigr)^{\nicefrac 12}\\
&= \Vert A\Vert \cdot \Vert \alpha\Vert \cdot \Vert \beta\Vert \cdot \Vert f\Vert_1
\end{align*}
Since $L^\infty(M)$ is the dual of $L^1(M)$, it follows that
\begin{align*}
|\langle \alpha, \Afrak(m)\beta\rangle| &\leq \Vert A\Vert \cdot \Vert \alpha\Vert \cdot \Vert \beta\Vert. \qedhere
\end{align*}
\end{proof}

\begin{cthm}[{\cite[Theorem XIII.84]{Reedsimon4}}]
Let $\Hcal = \int_M^\oplus \Hcal' \, \dd \mu$ where $(M,\mu)$ is a $\sigma$-finite separable measure space and $\Hcal'$ is a separable Hilbert space. Let $\Acal$ be the algebra of decomposable operators whose fibers are all multiples of the identity. Then $A\in \Bcal(\Hcal)$ is decomposable if and only if $A$ commutes with each operator in $\Acal$.
\end{cthm}

\begin{proof}
\begin{itemize}
\item[($\Ra$)] trivial.

\item[($\La$)] Since $\mu$ is $\sigma$-finite, we can finde a strictly positive $F\in L^1(M,\mu)$ so that $\nu := F\mu$ has unit mass. To see this, let $(A_n)_n\subseteq M$ with $\bigcup_{n\in \Nbb} A_n = M$ and $0 < \mu(A_n)< \infty$. Define $F|_{A_n} := 2^{-n} \mu(A_n)^{-1}$. Then,
\begin{align*}
\int_M \dd \nu := \int_M F\, \dd \mu = \sum_{n\in \Nbb} \int_{A_n} 2^{-n} \mu(A_n)^{-1} \dd \mu = \sum_{n\in \Nbb} 2^{-n} = 1.
\end{align*}
Let $\tilde \Hcal := \int_M \Hcal'\, \dd \nu$. Then, the map $U\colon \Hcal \lra \tilde \Hcal$, $Ug = F^{-\nicefrac 12} g$ is unitary since
\begin{align*}
\int_M \Vert Ug(m)\Vert^2 \, \dd \nu = \int_M |F^{-\nicefrac 12}|^2 \cdot \Vert g(m)\Vert^2 \cdot F\, \dd \mu = \Vert g\Vert^2_{\Hcal}
\end{align*}
and $U\Acal U^{-1} = \tilde \Acal$. Hence, we may suppose without loss that $\int_M \dd \mu = 1$. Choose an ONB $\{\eta_k\}_{k\in \Nbb}$ for $\Hcal'$ and let $F_k$ be the element of $\Hcal$ with $F_k(m) := \eta_k$ for almost all $m\in M$. The $F_k$ are orthonormal since $\int_M \dd \mu = 1$. Moreover, any $\psi\in \Hcal$ has an expansion $\psi = \sum_{k=1}^\infty f_k F_k$ with $f_k\in L^2(M, \mu; \Cbb)$ and $\Vert \psi\Vert^2 = \sum_{k=1}^\infty \Vert f_k\Vert^2$ (see Lemma \ref{lem2.2.4} (a)). Define functions $a_{km} \colon M\lra \Cbb$ by $AF_k(m) = \sum_{\ell = 1}^\infty a_{k\ell}(m) F_\ell$ for almost all $m\in M$. Choose a countable dense subset $\Dcal$ in $\Hcal'$ of vectors $\varphi$ of the form $\varphi = \sum_{k=1}^N \alpha_k \eta_k$ and define $\Phi := \sum_{k=1}^N \alpha_k F_k$ (whence $\Vert \Phi\Vert_{\Hcal} = \Vert \varphi\Vert_{\Hcal'}$). Then, for any $f\in L^\infty(M, \mu, \Cbb)$, setting $\Afrak(m)\varphi := \sum_{\ell=1}^\infty \sum_{k=1}^N \alpha_k a_{k\ell}(m)\eta_\ell$, we obtain
\begin{align*}
A(f\Phi)(m) = f(A\Phi)(m) &= \sum_{k=1}^N f(m) \alpha_k AF_k(m) = \sum_{k=1}^N \sum_{\ell = 1}^\infty f(m) \alpha_k a_{k\ell}(m) F_\ell(m) \\
&= \sum_{\ell=1}^\infty \sum_{k=1}^N f(m) \alpha_k a_{k\ell}(m)\eta_\ell = f(m) \Afrak(m)\varphi\\
&= f(m)\Afrak(m)\Phi(m),
\end{align*}
since $f\Idbb_{\Hcal}\in \Acal$. Moreover, we get
\begin{align*}
\Vert A(f\varphi)(m)\Vert^2 = \Bigl\Vert \sum_{k=1}^N \sum_{\ell = 1}^\infty \alpha_k a_{k\ell}(m)\eta_\ell\Bigr\Vert^2 &= \int_M |f(m)|^2 \sum_{\ell\in \Nbb} \Bigl| \sum_{k=1}^N \alpha_k a_{k\ell}(m)\Bigr|^2 \, \dd \mu(m) \\
&\leq \Vert A\Vert^2 \int_M |f(m)|^2 \cdot \Vert \varphi\Vert_{\Hcal'}^2 \, \dd \mu(m) \\
&=\Vert A\Vert^2 \int_M |f(m)|^2 \, \dd \mu(m) \cdot \sum_{k=1}^N |\alpha_k|^2.
\end{align*}
Hence, by duality, using $\Vert \Phi\Vert^2_{\Hcal} = \Vert \varphi\Vert_{\Hcal'}^2 = \sum_{k=1}^N |\alpha_k|^2$, we have
\begin{align*}
\Vert \Afrak(m)\varphi\Vert \leq \Vert A\Vert \cdot \Vert \Phi\Vert = \Vert A\Vert \cdot \Vert \varphi\Vert
\end{align*}
for almost all $m\in M$. Hence, $\Afrak(m)$ may be extended to a bounded operator on $\Hcal'$ for almost every $m\in M$ and $\Afrak\in L^\infty(M, \Bcal(\Hcal')$. Then, let $B$ be the corresponding decomposable operator and let $\psi\in \Hcal$ have the form $\psi = \sum_{k=1}^N f_k F_k$ with $f_k\in L^2(M,\mu)$. Then
\begin{align*}
(A\psi)(m) &= \sum_{k=1}^N f_k(m) (AF_k)(m) = \sum_{k=1}^N f_k(m) \Afrak(m)\eta_k = \Afrak(m) \sum_{k=1}^N f_k(m) \eta_k \\
&= \Afrak(m) \psi(m) = (B\psi)(m).
\end{align*}
Since such $\psi$'s are dense, $B = A$. To see density, let $\psi\in \Hcal$. Then, $\psi(m) = \sum_{k=1}^\infty f_k(m) \eta_k$ for certain $f_k(m)\in \Cbb$ and a.e. $m\in M$. For all $\ell\in \Nbb$, we have $\langle \eta_\ell , \psi(m)\rangle = f_\ell(m)$. Hence,
\begin{align*}
\Vert f_\ell\Vert^2 = \int_M |f_\ell(m)|^2\, \dd \mu = \int_M |\langle \eta_\ell, \psi(m)\rangle|^2\, \dd\mu \leq \int_M \Vert \psi(m)\Vert^2 \, \dd \mu = \Vert \psi\Vert^2.
\end{align*}
Thus, $f_\ell\in L^2(M, \mu)$ for all $\ell\in \Nbb$ and $\sum_{\ell\in \Nbb} \Vert f_\ell\Vert^2 = \Vert\psi\Vert^2$ (monotone convergence and Parseval). Finally, for $N\in \Nbb$, we have
\begin{align*}
\Bigl\Vert \psi - \sum_{k=1}^N f_k F_k\Bigr\Vert^2 = \int_M \Bigl\Vert \sum_{k=N+1}^\infty f_k(m)F_k(m)\Bigr\Vert^2 \, \dd \mu.
\end{align*}
Here, the integrand goes pointwise to $0$ as $N\to\infty$ and is bounded by $\Vert \psi(m)\Vert^2$ which is integrable. We conclude by dominated convergence.\qedhere
\end{itemize}
\end{proof}

\begin{defn}
Let $\Lcal_\mathrm{sa}(\Hcal'):= \{A\colon \Dcal(A) \ra \Hcal': A = A^*\}$ denote the set of self-adjoint operators in $\Hcal'$. A function $\Afrak\colon M\lra \Lcal_{\mathrm{sa}}(\Hcal')$ is called measurable if and only if the function $m\mapsto (\Afrak(m) + \i)^{-1}$ is measurable. Given such a function, we define an operator $A$ on $\Hcal = \int_M^\oplus \Hcal'\, \dd\mu$ with the domain
\begin{align*}
\Dcal(A) = \Bigl\{ \psi\in \Hcal: \psi(m) \in \Dcal(\Afrak(m)) \text{ a.e. \& } \int_M \Vert \Afrak(m)\psi(m) \Vert^2 \, \dd\mu < \infty\Bigr\}.
\end{align*}
\end{defn}

\begin{bem}
The foregoing definition makes the following sense. If $\Afrak\colon M\lra \Lcal_\sa(\Hcal')$ is measurable and $\psi(m)\in \ran((\Afrak(m) + \i)^{-1})$ for almost every $m\in M$, so that there is an $\eta(m)\in \Hcal'$ with $(\Afrak(m) +\i)^{-1}\eta(m) = \psi(m)$ a.e., then
\begin{align*}
\Afrak(m) \psi(m) &= \Afrak(m) (\Afrak(m) + \i)^{-1} \eta(m) \\
&= (\Afrak(m) + \i)(\Afrak(m) + \i)^{-1} \eta(m) - \i (\Afrak(m) + \i)^{-1} \eta(m) \\
&= \eta(m) + \i(\Afrak(m) + \i)^{-1} \eta(m)
\end{align*}
is measurable. Thus, the definition of $\Dcal(A)$ makes sense.
\end{bem}

\begin{cthm}[{\cite[Theorem XIII.85 (b)]{Reedsimon4}}]
\label{Unbounded operator decomp}
A self-adjoint operator $A$ on $\Hcal$ has the form $\int_M^\oplus \Afrak(m)\, \dd \mu$ if and only if $(A + \i)^{-1}$ is a bounded decomposable operator.
\end{cthm}

\begin{proof}
\begin{itemize}
\item[($\Ra$)] By assumption, there is a function $\Afrak \colon M \lra \Lcal_\sa(\Hcal')$ such that $g(m) := (\Afrak(m) + \i)^{-1}$ is measurable. We may define
\begin{align*}
G := \int_M^\oplus g(m)\, \dd\mu.
\end{align*}
To show is that $G = (A + \i)^{-1}$. Let $\psi\in \Hcal$, then $(G\psi)(m) = (\Afrak(m) + \i)^{-1} \psi(m)$ i.e. $(G\psi)(m) \in \Dcal(\Afrak(m))$ a.e. Furthermore, since
\begin{align*}
\Afrak(m) (G\psi)(m) &= \Afrak(m) (\Afrak(m) + \i)^{-1} \psi(m) = \psi(m) - \i (\Afrak(m) + \i)^{-1} \psi(m) \\
&= \psi(m) - \i(G\psi)(m),
\end{align*}
we have that $\Afrak(m) (G\psi)(m)$ is square integrable (since $\psi\in \Hcal$ and $G$ is bounded (by 1)) so that $G\psi\in \Dcal(A)$. Since $A= \int_M^\oplus \Afrak(m) \, \dd\mu$, we obtain
\begin{align*}
((A + \i)G\psi)(m) = (\Afrak(m) + \i)(\Afrak(m) + \i)^{-1} \psi(m) = \psi(m).
\end{align*}
Thus, $G$ is a right-sided inverse of $A + \i$. Furthermore, let $\psi\in \Dcal(A)$ and compute
\begin{align*}
(G(A + \i)\psi)(m) = (\Afrak(m) + \i)^{-1} (\Afrak(m) + \i)\psi(m) = \psi(m).
\end{align*}
Thus, $G$ is also a left-sided inverse and hence the inverse of $A+ \i$.

\item[($\La$)] By assumption, there is a measurable function $g\in L^\infty(M,\mu; \Bcal(\Hcal'))$ such that
\begin{align*}
(A+  \i)^{-1} = \int_M^\oplus g(m)\,\dd \mu.
\end{align*}
First, we note that $(A - \i)^{-1} = \int_M^\oplus g(m)^*\, \dd \mu$. This follows from a standard computation. We claim that $g(m)$ has dense range and is injective almost everywhere. To see this, let $\varphi\in \Hcal$ with $\varphi(m)\in \ker(g(m))$ a.e. Then $0 = g(m) \varphi(m) = ((A + \i)^{-1}\varphi)(m)$ so that $(A + \i)^{-1}\varphi =0$ which means $\varphi =0$ since $(A + \i)^{-1}$ is injective. Hence, $g(m)$ is injective almost everywhere. If $\varphi\in \Hcal'$ with $\varphi(m)\in \ran(g(m))^\perp$ almost everywhere, then, for all $\psi\in \Hcal$:
\begin{align*}
0 = \langle \varphi(m), g(m) \psi(m)\rangle = \langle g(m)^* \varphi(m) , \psi(m)\rangle = \langle (A -\i)^{-1} \varphi, \psi\rangle(m)
\end{align*}
so that $\varphi\in \ker((A - \i)^{-1})$. This implies $\varphi =0$ so that $\varphi(m) =0$ almost everywhere. Hence $g(m)$ has dense range in $\Hcal'$. Hence, define $\Dcal(\Afrak(m)) := \ran(g(m))$ and 
\begin{align*}
\Afrak(m) := g(m)^{-1} - \i
\end{align*}
on $\Dcal(\Afrak(m))$. Then $\Afrak(m)$ is densely defined. We claim that $\Afrak(m)$ is self-adjoint for almost every $m\in M$. To see this, first note that $\Afrak(m)$ is symmetric: For a.e. $m\in M$ pick $\varphi(m),\psi(m)\in \Dcal(\Afrak(m)) = \ran(g(m))$ so there is some $\eta(m)\in \Hcal'$ with $\psi(m) = g(m) \eta(m) = ((A + \i)^{-1} \eta)(m)$. This implies that $\psi\in \ran(A + \i)^{-1} = \Dcal(A)$. Analogously, $\varphi(m) = g(m)\xi(m)$ for some $\xi(m)\in \Hcal'$. Hence,
\begin{align*}
\langle \varphi(m), \Afrak(m)\psi(m)\rangle \hspace{-70pt} &\hspace{70pt} = \langle \xi(m) , g^*(m)[g(m)^{-1} - \i] g(m)\eta(m)\rangle\\
&= \langle \xi(m), g^*(m) [\Idbb - \i g(m)]\eta(m)\rangle = \langle \xi, (A - \i)^{-1} [\Idbb - \i(A + \i)^{-1}]\eta\rangle(m)\\
&= \langle \xi, (A - \i)^{-1} [A + \i - \i](A + \i)^{-1}\eta\rangle(m) \\
&= \langle \xi, (A - \i)^{-1} [A - \i + \i](A + \i)^{-1}\eta\rangle(m) \\
&= \langle \xi, [\Idbb + \i(A - \i)^{-1}](A + \i)^{-1} \eta\rangle(m) = \langle \xi(m), [\Idbb + \i g^*(m)]\psi(m)\rangle \\
&= \langle [\Idbb - \i g(m)]\xi(m), \psi(m)\rangle = \langle [g(m)^{-1} - \i] \varphi(m), \psi(m) \rangle \\
&= \langle \Afrak(m)\varphi(m), \psi(m)\rangle.
\end{align*}

To show that $\Afrak(m)$ is even self-adjoint for a.e. $m\in M$, let $\eta(m)\in \Hcal'$ and set $\psi(m) := g(m)\eta(m)$. Per definition, $\psi(m)\in \Dcal(\Afrak(m))$ and
\begin{align*}
(\Afrak(m) + \i )\psi(m) = (g(m)^{-1} - \i + \i) g(m) \eta(m) = \eta(m).
\end{align*}
Hence, $\ran(\Afrak(m) + \i) = \Hcal'$. For $\ran(\Afrak(m) -\i)$, we proceed analogously, using that
\begin{align*}
(A + \i) (A -\i)^{-1} = (A - \i + 2\i)(A - \i)^{-1} = \Idbb + 2\i (A - \i)^{-1}
\end{align*}
implies that
\begin{align}
g(m)^* = g(m) [\Idbb + 2\i g(m)^*]. \label{2.2.14.1}
\end{align}
If $\psi(m)\in \ran(g(m))$ so that $\psi(m) = g(m) \eta(m)$, we have that
\begin{align}
g(m)^* [\Idbb - 2\i g(m)] \eta(m) = g(m) [\Idbb + 2\i g(m)^*][1 - 2\i g(m)]\eta(m) = \psi(m). \label{2.2.14.2}
\end{align}
The last equality follows from an easy computation involving the first resolvent equation showing
\begin{align*}
[\Idbb - 2\i g(m)] [\Idbb + 2\i g(m)^*] = [\Idbb - 2\i (A + \i)^{-1}][\Idbb + 2\i (A - \i)^{-1}](m) = \Idbb(m).
\end{align*}
Hence, $\psi(m)\in \ran(g(m)^*)$. Thus, interchanging the roles of $\i$ and $-\i$ and the stars, we obtain $\ran(g(m)) = \ran(g(m)^*)$. Hence, if $\xi(m)\in \Hcal'$ and $\varphi(m) := g(m)^*\xi(m)$, we have $\varphi(m)\in \Dcal(\Afrak(m))$ and
\begin{align*}
(\Afrak(m) - \i)\varphi(m) &= [g(m)^{-1} - 2\i] g(m)^* \xi(m) \\
&= [g(m)^{-1} - 2\i] g(m) [\Idbb + 2\i g(m)^*] \xi(m) \\
&= [\Idbb - 2\i g(m)] [\Idbb + 2\i g(m)] \xi (m) = \xi(m).
\end{align*}
Thus, $\Afrak(m)$ is self-adjoint for almost all $m\in M$. This now enables us to define $\tilde A := \int_M^\oplus \Afrak(m)\, \dd \mu$. Let us show that $\Dcal(\tilde A) = \Dcal (A)$. To see this let $\psi\in \Dcal(\tilde A)$. Then $\psi(m)\in \Dcal(\Afrak(m)) = \ran(g(m))$ a.e. so $\psi(m) = g(m) \eta(m)$ for a certain $\eta(m)\in \Hcal'$. We obtain that
\begin{align}
\Afrak(m) \psi(m) = [g(m)^{-1} - \i]g(m) \eta(m) = \eta(m) - \i \psi(m). \label{2.2.14bstar}
\end{align} 
Since $\psi\in \Hcal$ and $\int_M \Vert\Afrak(m)\psi(m)\Vert^2 \, \dd\mu <\infty$ by definition of $\Dcal(\tilde A)$, we infer that $\eta\in \Hcal$. Hence, $\psi = (A + \i)^{-1}\eta\in \Dcal(A)$. Reversely, let $\psi\in\Dcal(A)$ so that $\psi = (A + \i)^{-1}\eta$ for some $\eta\in \Hcal$. Then $\psi(m) = g(m)\eta(m)$ so that $\psi(m)\in \Dcal(\Afrak(m))$ for almost all $m\in M$. Furthermore, \eqref{2.2.14bstar} says that $\Afrak(m)\psi(m)$ is square integrable over $M$. Hence, $\psi\in \Dcal(\tilde A)$. Finally, we have that
\begin{align*}
(A\psi)(m) &= (A(A + \i)^{-1}\eta)(m)= (\eta - \i (A - \i)^{-1}\eta)(m) \\
&= \eta(m) - \i g(m) \eta(m) = g(m)^{-1} \psi(m) - \i \psi(m) = \Afrak(m)\psi(m) \\
&= (\tilde A\psi)(m). \qedhere
\end{align*}
\end{itemize}
\end{proof}

\begin{cthm}[{\cite[Theorem XIII.85]{Reedsimon4}}]
Let $\Afrak\colon M\lra \Lcal_\sa(\Hcal')$ be measurable and assume that $A = \int_M^\oplus \Afrak(m) \, \dd \mu$. Then
\begin{enumerate}[(a)]
\item The operator $A$ is self-adjoint.

\item For any bounded Borel function $F\colon \Rbb \lra \Cbb$,
\begin{align*}
F(A) =\int_M^\oplus F(\Afrak(m))\, \dd \mu.
\end{align*}

\item $\lambda\in \sigma(A)$ if and only if for all $\varepsilon>0$,
\begin{align*}
\mu(\{m\in M : \sigma(\Afrak(m))\cap (\lambda - \varepsilon, \lambda + \varepsilon)\neq\emptyset \})>0.
\end{align*}

\item $\lambda$ is an eigenvalue of $A$ if and only if
\begin{align*}
\mu(\{m\in M : \lambda \text{ is an eigenvalue of $\Afrak(m)$}\})>0.
\end{align*}


\item Suppose that $\Bfrak\colon M \lra \Lcal_\sa(\Hcal')$ is measurable and $B = \int_M^\oplus \Bfrak(m) \, \dd \mu$. If $B$ is $A$-bounded with $A$-bound $a$, then, almost everywhere, $\Bfrak(m)$ is $\Afrak(m)$-bounded with $\Afrak(m)$-bound $a(m) \leq a$. If $a < 1$, then
\begin{align*}
A + B = \int_M^\oplus (\Afrak(m) + \Bfrak(m))\, \dd \mu
\end{align*}
is self-adjoint on $\Dcal(A)$.
\end{enumerate}
\end{cthm}

\begin{proof}
\begin{enumerate}[(a)]
\item First, $A$ is symmetric so it suffices to prove that $\ran(A \pm \i) = \Hcal$. Let $g(m) := (\Afrak(m) + \i)^{-1}$. Then $g(m)$ is measurable and $\Vert g(m)\Vert \leq 1$ almost everywhere. Hence, we may define $G := \int_M^\oplus g(m)\, \dd \mu$ by Theorem \ref{Definition direct integral}. Let $\eta\in\Hcal$ and set $\psi := G\eta$. Then $\psi(m)\in \ran(g(m)) = \Dcal(\Afrak(m))$ a.e. and
\begin{align*}
\Vert \Afrak(m)\psi(m)\Vert = \Vert A(m)g(m) \eta(m)\Vert \leq \Vert \eta(m)\Vert
\end{align*}
is square integrable since $\Vert A(m)g(m)\Vert \leq 1$ a.e. Hence, $\psi\in \Dcal(A)$ and $(A + \i) \psi =\eta$ whence $\ran(A + \i)= \Hcal$. Similarly, $g(m)^* = (A(m) - \i)^{-1}$ is measurable (compare \eqref{2.2.14.1} and \eqref{2.2.14.2}) and thus $\ran(A - \i)= \Hcal$.

\item By Theorem \ref{Unbounded operator decomp}, we have
\begin{align*}
(A + \i)^{-1} = \int_M^\oplus (\Afrak(m) + \i)^{-1} \,\dd \mu.
\end{align*}
The first goal is to extend this to resolvents at points $\lambda\in \Cbb$ with $\Im \lambda \neq 0$. Set $g_\lambda(m) := (\Afrak(m) - \lambda)^{-1}$. As a preparation, we note that
\begin{align*}
(A + \i)(A -\lambda)^{-1} = (A - \lambda + (\i + \lambda))(A - \lambda)^{-1} = \Idbb + (\i + \lambda)(A - \lambda)^{-1}
\end{align*}
and we get an analogous equation for $\lambda$ and $-\i$ interchanged. We obtain
\begin{align*}
[\Idbb + (\i + \lambda)(A - \lambda)^{-1}][\Idbb - (\i + \lambda)(A + \i)^{-1}] = (A + \i)(A - \lambda)^{-1}(A - \lambda)(A + \i)^{-1} = \Idbb.
\end{align*}
By interchanging again, we see that $\Idbb - (\i + \lambda)(A + \i)^{-1}$ is invertible and the inverse is $\Idbb + (\i + \lambda)(A + \lambda)^{-1}$. Exactly the same computation holds for $g_\lambda(m)$ in place of $(A -\lambda)^{-1}$ for all $\lambda\in \Cbb\setminus \Rbb$. Thus, $\Idbb - (\i + \lambda) g_\i(m)$ is invertible. Then, a straightforward computation using that
\begin{align*}
\Idbb - (\i + \lambda)(A + \i)^{-1} =\int_M^\oplus \Idbb - (\i + \lambda)g_\i(m)\, \dd\mu
\end{align*}
shows that
\begin{align*}
[\Idbb - (\i + \lambda) (A +\i)^{-1}]^{-1} = \int_M^\oplus [\Idbb + (\i + \lambda) g_\i(m)]^{-1}\, \dd \mu.
\end{align*}
This implies that
\begin{align*}
(A -\lambda)^{-1} = \int_M^\oplus g_\lambda(m)\, \dd\mu.
\end{align*}
for every $\lambda\in \Cbb\setminus \Rbb$. Now, for $\varepsilon>0$, $a,b\in \Rbb$ with $a < b$ and $t\in \Rbb$ we set
\begin{align*}
h_{\varepsilon, a,b}(t) := \frac{1}{\pi \i} \int_a^b \frac{1}{t - \lambda - \i \varepsilon} - \frac{1}{t - \lambda + \i \varepsilon} \, \dd \lambda.
\end{align*}
A straightforward computation shows that $\sup_{\varepsilon >0} \Vert h_{\varepsilon, a,b}\Vert_\infty = 2$ for all $a,b\in \Rbb$, $a < b$ and $\varepsilon>0$ and that
\begin{align*}
h_{\varepsilon, a,b}(t)\xra{\varepsilon\to 0} \Idbb_{[a,b]}(t) + \Idbb_{(a,b)}(t)
\end{align*}
for all $t\in \Rbb$. Together with 
\begin{align}
\Idbb_{[a - \delta, b-\delta]}(t) &\xra{\delta \to 0} \Idbb_{[a,b)}(t) & \Idbb_{(a-  \delta, b - \delta)}(t) &\xra{\delta \to 0} \Idbb_{[a,b)}(t) \label{2.2.15.1}
\end{align}
we see that, using dominated convergence, for all $a,b\in \Rbb$, $a<b$ and all $\psi\in \Hcal$:
\begin{align*}
\lim_{\delta \to 0}\lim_{\varepsilon\to 0}\int_M \Bigl\Vert \Idbb_{[a,b)}(\Afrak(m)) \psi(m)  - \frac 12 h_{\varepsilon, a-\delta, b - \delta}(\Afrak(m))\psi(m) \Bigr\Vert^2 \, \dd \mu = 0.
\end{align*}
Hence, define $G_{a,b}$ by
\begin{align*}
G_{a,b}\psi := \int_M^\oplus \Idbb_{[a,b)}(\Afrak(m)) \psi(m)\, \dd \mu.
\end{align*}
Then, \eqref{2.2.15.1} implies that
\begin{align*}
\Idbb_{[a,b)}(A)\psi &= \frac 12\lim_{\delta \to 0} \lim_{\varepsilon\to 0} h_{\varepsilon, a - \delta, b - \delta}(A)\psi = \frac 12 \lim_{\delta \to 0} \lim_{\varepsilon\to 0} \int_M^\oplus h_{\varepsilon, a-\delta, b - \delta}(\Afrak(m))\psi(m)\, \dd \mu \\
&= \int_M^\oplus \Idbb_{[a,b)}(\Afrak(m)\psi(m)\, \dd \mu = G_{a,b}\psi
\end{align*}
for each $\psi\in \Hcal$. Hence, $\Idbb_{[a,b)}(A) = G_{a,b}\psi$. Since half-open intervals generate the Borel $\sigma$-algebra $\Bcal(\Rbb)$, this proves the claim for $\Idbb_U$ for each $U\in \Bcal(\Rbb)$. Finally, for $F$ bounded and measurable, choose a sequence of simple functions $F_n$ with $\Vert F - F_n\Vert \to 0$ and use dominated convergence again using that $\sup_n \Vert F_n\Vert_\infty < \infty$.

\item We apply part (b) to $\Idbb_{(\lambda - \varepsilon, \lambda + \varepsilon)}$. Then, the claim follows from the fact that we have $\int_M^\oplus T(m)\, \dd \mu =0$ if and only if $T(m) =0$ a.e.

\item This is similar to part (c) by using $\Idbb_{\{\lambda\}}$.

\item If $\Vert B\psi\Vert \leq a \Vert A\psi\Vert + b\Vert\psi\Vert$, then
\begin{align*}
\Vert B(A +\i k)^{-1} \Vert \leq a\Vert A (A + \i k)^{-1}\Vert + b \Vert (A + \i k)^{-1}\Vert = a+ bk^{-1}.
\end{align*}
Hence, $\Vert \Bfrak(m) (\Afrak(m) + \i k)^{-1}\Vert \leq a + bk^{-1}$ for a.e. $m\in M$. For given $\varepsilon>0$ choose $k(m)>0$ so large that $bk(m)^{-1} < \varepsilon$. Then, for $\psi\in \Dcal(\Afrak(m))$:
\begin{align*}
\Vert \Bfrak(m)\psi\Vert &= \Vert \Bfrak(m) (\Afrak(m) + \i k)^{-1} (\Afrak(m) + \i k ) \psi\Vert \\
&\leq (a + \varepsilon) \Vert \Afrak(m)\psi\Vert + (a + \varepsilon)k(m)\Vert \psi\Vert.
\end{align*}
Hence, since $\varepsilon >0$ was arbitrary, $\Bcal(m)$ is $\Afrak(m)$-bounded with $\Afrak(m)$-bound $a(m) \leq a$. If $a < 1$, then Kato-Rellich implies that $\Afrak(m) + \Bfrak(m)$ is self-adjoint on $\Dcal(\Afrak(m))$ almost everywhere. Since
\begin{align}
\frac{1}{\Afrak(m) + \Bfrak(m) + 2\i} = \frac{1}{\Afrak(m) + \i} \frac{1}{(\Bfrak(m) + \i)^{-1} + (\Afrak(m) + \i)^{-1}} \frac{1}{\Bfrak(m) + \i} \label{2.2.15 (e)}
\end{align}
we infer that $m\mapsto \Afrak(m) + \Bfrak(m)$ is measurable by using part (b) for $\Afrak(m)$ and $\Bcal(m)$ with $f(t) = (t + \frac \i 2)^{-1}$. Thus, we may define
\begin{align*}
G := \int_M^\oplus \Afrak(m) + \Bfrak(m) \, \dd\mu.
\end{align*}
By Kato-Rellich, we know that $A + B$ is self-adjoint on $\Dcal(A)$. This implies that $\Dcal(A) = \Dcal(G)$. Linearity of the direct integral then implies $G = A + B$.\qedhere
\end{enumerate}
\end{proof}

\subsection{Direct integral decomposition of \texorpdfstring{$L^2(\Rbb^d)$}{L2Rd}}

In this section, we want to decompose the space $L^2(\Rbb^d)$ into a (constant fiber) direct integral with respect to a certain lattice (the lattice of periodicity of the Hamiltonian), where the fibers consist of the $L^2$-space over the unit cell of the lattice. Consider a basis $\{a_1, \ldots, a_d\}\subseteq \Rbb^d$ and let
\begin{align*}
\Gamma := \Gamma(a_1, \ldots, a_d) := \Bigl\{ \sum_{i=1}^d n_i a_i : n_i \in \Zbb , ~i = 1, \ldots, d\Bigr\}\subseteq \Rbb^d
\end{align*}
denote the lattice spanned by the vectors $a_1, \ldots, a_d$ as well as its closed unit cell
\begin{align*}
\mathcal C := \Ccal(a_1, \ldots, a_d) := \Bigl\{ \sum_{i=1}^d \lambda_i a_i : 0\leq \lambda_i < 1 , ~i =1, \ldots, d\Bigr\} \subseteq \Rbb^d.
\end{align*}
Introduce the corresponding dual lattice
\begin{align*}
\Gamma^* = \{ x\in \Rbb^d : \langle x,\eta\rangle \in 2\pi \Zbb ~\forall \eta\in \Gamma\}.
\end{align*}
It is known that $\Gamma^* = \Gamma(a_1^*, \ldots, a_d^*)$ with $a_1^*, \ldots, a_d^*\in \Rbb^d$ chosen in such a way that $\langle a_i^*, a_j\rangle = 2\pi \delta_{ij}$ for all $i,j=1, \ldots, d$. The dual unit cell is given by $\Ccal^* := \Ccal(a_1^*, \ldots, a_d^*)$.

\begin{clem}[{\cite[Lemma to Thm XIII.88]{Reedsimon4}}]
Let $\{a_1, \ldots, a_d\}$ and $\{b_1, \ldots, b_d\}$ be two bases in $\Rbb^d$ and let $\Gamma = \Gamma(a_1, \ldots, a_d)$ and $\Lambda = \Gamma(b_1, \ldots, b_d)$ be the two lattices in $\Rbb^d$ generated by these bases. Denote the unit cell of $\Gamma$ by $\Ccal$ and the dual unit cell of $\Lambda$ by $\Lcal^*$. Let $\Hcal' = L^2(\Ccal)$ and
\begin{align*}
\Hcal := \int_{\Lcal^*}^\oplus \Hcal' \, \frac{\dd \theta}{\Vol(\Lcal^*)}.
\end{align*}
Then, $U\colon L^2(\Rbb^d) \lra \Hcal$ given, for $f\in \Scal(\Rbb^d)$, by
\begin{align}
(Uf)_\theta (x) := \sum_{n\in \Zbb^d} \e^{-\i \langle \theta , \sum_{i=1}^d n_i b_i\rangle} f\Bigl(x + \sum_{i=1}^d n_i a_i\Bigr), \label{DID}
\end{align}
for $\theta\in \Lcal^*$ and $x\in \Ccal$, is unitary. The inverse $U^*\colon \Hcal\lra L^2(\Rbb^d)$ is given by
\begin{align}
(U^*g) \Bigl( x + \sum_{i=1}^d n_i a_i\Bigr) = \int_{\Lcal^*} \e^{\i \langle \theta, \sum_{i=1}^d n_i b_i\rangle} g_\theta(x) \, \frac{\dd \theta}{\Vol(\Lcal^*)} \label{DID_back}
\end{align}
for $x\in \Ccal$ and $n= (n_1, \ldots , n_d)\in \Zbb^d$. Moreover,
\begin{align*}
U ( - \Delta) U^* = \int_{\Lcal^*}^\oplus (-\Delta)_\theta \, \frac{\dd \theta}{\Vcal}.
\end{align*}
where $(-\Delta)_\theta$ is $-\Delta$ on $L^2(\Ccal)$ with boundary conditions
\begin{align*}
\varphi(x + a_i) &= \e^{\i \theta_i} \varphi(x), & \nabla \varphi (x + a_i) &= \e^{\i \theta_i} \nabla \varphi (x).
\end{align*}
\end{clem}

\begin{bem}
\begin{enumerate}[(a)]
\item In \cite{Reedsimon4}, the authors have set $\Lambda := \Zbb^d$ so that $\Lcal^* = [0,2\pi)^d$. One then gets the statement with $b_n = n$ for all $n\in \Zbb^d$. We wanted to stress that one may choose both lattices independently. The fact that $[0,2\pi)^d$ is the parameter set for the decomposition has nothing to do with the periodicity of the original operator one wants to decompose. This is only encoded in $\Gamma$.

\item In view of the Fourier transform for periodic functions, one may also set $\Lambda := \Gamma$.
\end{enumerate}
\end{bem}

\begin{proof}
For $n\in \Zbb^d$ set $a_n := \sum_{i=1}^d n_i a_i$ and $b_n := \sum_{i=1}^d n_i b_i$. Since $f\in \Scal(\Rbb^d)$, we have that $\sup_{x\in \Rbb^d} |x^2 f(x) |\leq C$ for some constant $C>0$. If $x\in \Gamma$, then there is $n_x\in \Zbb^d$ with $a_{n_x} + x = 0$. Thus,
\begin{align}
\Bigl|\sum_{n\in \Zbb^d} \e^{-\i \langle \theta , b_n \rangle} f(x + a_n)\Bigr| &\leq \sum_{n\in \Zbb^d\setminus \{n_x\}} |x + a_n|^{-2} \cdot |x  + a_n|^2 \cdot |f(x + b_n)| + |f(0)| \notag\\
&\leq C \sum_{n\in \Zbb^d\setminus \{n_x\}} |x + a_n|^{-2}. \label{Lemma 2.2.12.1}
\end{align}
Otherwise, we have
\begin{align}
\Bigl|\sum_{n\in \Zbb^d} \e^{-\i \langle \theta , b_n \rangle} f(x + a_n)\Bigr| &\leq \sum_{n\in \Zbb^d} |x + a_n|^{-2} \cdot |x  + a_n|^2 \cdot |f(x + b_n)| \notag\\
&\leq C \sum_{n\in \Zbb^d} |x + a_n|^{-2}. \label{Lemma 2.2.12.2}
\end{align}
Now, let $x\in \Rbb^d$ and choose $n\in \Zbb^d$ so that $|n_i| \geq 2|x_i|$. Since all norms in $\Rbb^d$ are equivalent, we have $|x|_2\geq c_d |x|_\infty$ for some $c_d>0$. Writing $x = \sum_{i=1}^d x_i a_i$ yields:
\begin{align*}
|x + a_n| = \Bigl| \sum_{i=1}^d (x_i - n_i)a_i\Bigr| \geq c_d \max_{i=1, \ldots, d} |x_i - n_i| \geq \frac{c_d}{2} \max_{i=1, \ldots, d} |n_i|.
\end{align*}
Using the equivalence of norms again, we have $|n|_\infty \geq \tilde c_d |n|_2$ for all $n\in \Zbb^d$ and some $\tilde c_d >0$. Hence, \eqref{Lemma 2.2.12.1} and \eqref{Lemma 2.2.12.2} converge. Next, we prove that $U$ is isometric. Set $\Vcal := \Vol(\Lcal^*)$ and compute
\begin{align*}
\Vert Uf\Vert_\Hcal^2 &= \int_{\Lcal^*} \frac{\dd \theta}{\Vcal} \int_\Ccal \Bigl| \sum_{n\in \Zbb^d} \e^{-\i \langle \theta , b_n\rangle} f(x + a_n)\Bigr|^2 \, \dx\\
&= \sum_{m,n\in \Zbb^d} \int_\Ccal \dx \,  \ov{ f(x + a_n)} f(x+a_m) \int_{\Lcal^*} \frac{\dd \theta}{\Vcal} \, \e^{-\i \langle \theta , b_m - b_n\rangle}\\
&= \sum_{n\in \Zbb^d} \int_\Ccal |f(x + a_n)|^2 = \int_{\Rbb^d} |f(x)|^2 \, \dx.
\end{align*}
Here, we used Fubini and the fact that
\begin{align*}
\int_{\Lcal^*} \frac{\dd \theta}{\Vcal} \, \e^{-\i \langle \theta, b_m - b_n\rangle} = \delta_{n,m},
\end{align*}
which can be proven using the transformation rule. Let us verify that $U^*$ is indeed the adjoint of $U$. We compute
\begin{align*}
\langle g, Uf\rangle_\Hcal &= \int_{\Lcal^*} \frac{\dd \theta}{\Vcal} \langle g_\theta, (Uf)_\theta\rangle_{L^2(\Ccal)} = \int_{\Lcal^*}  \frac{\dd \theta}{\Vcal} \int_\Ccal \dx \, \ov{g_\theta (x)} \sum_{n\in \Zbb^d} \e^{-\i \langle \theta, b_n\rangle} f(x + a_n)\\
&= \sum_{n\in \Zbb^d} \int_\Ccal \dx \ov{ \int_{\Lcal^*} \frac{\dd \theta}{\Vcal} \, g_\theta(x) \e^{\i\langle \theta, b_n\rangle}} \cdot f(x + a_n) = \langle U^* g, f\rangle_{L^2(\Rbb^d)}.
\end{align*}
Thus, we finish the proof by showing that also $U^*$ is isometric. We have
\begin{align*}
\Vert U^* g\Vert^2 &= \int_{\Rbb^d} |U^*g(y)|^2\, \dy = \sum_{n\in \Zbb^d} \int_\Ccal |U^*g(y + a_n)|^2\, \dy = \sum_{n\in\Zbb^d} \int_\Ccal \Bigl| \int_{\Lcal^*} \frac{\dd \theta}{\Vcal} \, \e^{\i \langle \theta, b_n\rangle} g_\theta (y) \Bigr|^2 \, \dy \\
&= \sum_{n\in \Zbb^d} \int_\Ccal \dy \int_{\Lcal^*} \frac{\dd \theta}{\Vcal} \int_{\Lcal^*} \frac{\dd \theta'}{\Vcal} \, \e^{\i \langle  \theta - \theta' , b_n\rangle} g_\theta(y)\ov{g_{\theta'}(y)} = \int_{\Lcal^*}\int_\Ccal |g_\theta(y)|^2 \, \dy \, \frac{\dd \theta}{\Vcal} = \Vert g\Vert_\Hcal^2.
\end{align*}
Here, we used that
\begin{align*}
\sum_{n\in \Zbb^d} \e^{\i \langle \theta - \theta', b_n\rangle} = \Vcal \delta(\theta -\theta'),
\end{align*}
which may be verified by testing against a function. It remains to show the decomposition of the Laplacian. Let $A := \int_{\Lcal^*}^\oplus (-\Delta)_\theta \frac{\dd \theta}{\Vcal}$ and let $f\in \Scal(\Rbb^d)$. Then $(Uf)_\theta\in C^\infty(\Ccal)$ for all $\theta\in \Lcal^*$ and for $j,k= 1, \ldots, d$ we have
\begin{align*}
(U\partial_k f)_\theta(x + a_j) &= \sum_{n\in \Zbb^d} \e^{-\i \langle \theta, b_n\rangle} \partial_k f(x + a_n + a_j) = \sum_{n\in \Zbb^d} \e^{-\i \langle \theta, b_n\rangle} \partial_k \e^{-\i \theta_j} f(x + a_n) \\
&= \e^{\i \theta_j} (U \partial_k f)_\theta(x).
\end{align*}
Hence, $(Uf)_\theta\in \Dcal((-\Delta)_\theta)$ for all $\theta\in \Lcal^*$ and
\begin{align*}
U(-\Delta f))_\theta (x) = \sum_{n\in \Zbb^d} \e^{-\i \langle \theta, b_n\rangle} (-\Delta f)(x + a_n) = (-\Delta)_\theta (Uf)_\theta (x).
\end{align*}
In particular,
\begin{align*}
\int_{\Lcal*} \Vert (-\Delta)_\theta (Uf)_\theta\Vert_{L^2(\Ccal)}^2 \, \frac{\dd \theta}{\Vcal} = \Vert (-\Delta)f\Vert_{L^2(\Rbb^d)}
\end{align*}
so that $Uf\in \Dcal(A)$. Reversely, let $\psi\in \Dcal(A)$, then $\psi_\theta\in \Dcal((-\Delta)_\theta)$ and
\begin{align*}
\infty > \int_{\Lcal^*} \Vert U^* (-\Delta)_\theta \psi_\theta \Vert^2\, \frac{\dd \theta}{\Vcal} = \int_{\Rbb^d} |(-\Delta) U^* \psi|^2\, \dx.
\end{align*}
Hence, $U^*\psi\in \Dcal(-\Delta)$.
\end{proof}

\subsubsection{Direct integral decomposition of bounded periodic operators}

We follow  Now let $A$ be a bounded operator on $L^2(\Rbb^d)$ which is periodic with respect to the lattice $\Gamma$. Per definiton, this means that $A T_\eta = T_\eta A$ for all $\eta\in \Gamma$. Here, $T_\eta$ denotes the translation operator by $\eta$. Writing \eqref{DID} a bit differently as
\begin{align*}
(Uf)_\theta = \sum_{n\in \Zbb^d} \e^{-\i \langle \theta , \sum_{i=1}^d n_i b_i\rangle} T_{a_n}^* f 
\end{align*}
we can compute, using the periodicity of $A$:
\begin{align*}
(UAf)_\theta &= \sum_{n\in \Zbb^d} \e^{-\i \langle \theta, b_n\rangle} T_{a_n}^* Af = \sum_{n\in \Zbb^d} \e^{-\i \langle \theta, b_n\rangle} AT_{a_n}^*f = A  \cdot \sum_{n\in \Zbb^d} \e^{-\i \langle \theta, b_n\rangle} T_{a_n}^* f.
\end{align*}
Define the ($\theta$-independent) $\theta$\tho\ fiber of $A$ by
\begin{align}
\begin{split}
A_\theta \colon L^2(\Ccal) &\lra L^2(\Ccal) \\
f &\longmapsto A|_{L^2(\Ccal)} f.\end{split} \label{Fiber-Def}
\end{align}
Being a little sloppy, we obtain $(UAf)_\theta = A_\theta (Uf)_\theta$ and so the direct integral decomposition
\begin{align}
UAU^* = \int_{\Lcal^*}^\oplus A_\theta \, \frac{\dd\theta}{\Vol(\Lcal^*)}. \label{Decomp of periodic ops}
\end{align}
This definition is not correct, since it suggests that one could restrict the operator $A$ successfully to the ``invariant subspace'' $L^2(\Ccal)$, which we cannot. However, the picture is correct and in the following, we only want to give the proper proof of the Bloch-Floquet decomposition of periodic operators. To do this, we use the Bloch-Floquet-Zac transform instead. We use the same lattice $\Gamma =\Lambda$ from now on. Introduce
\begin{align*}
L_\per^2(\Rbb^d) := \{ f\in \Lloc^2(\Rbb^d) : f(x-\gamma) = f(x)\;  \forall \gamma\in \Gamma, \text{ a.e. } x\in\Rbb^d \}.
\end{align*}
This becomes a Hilbert space via the inner product
\begin{align*}
\langle f, g\rangle_{L_\per^2(\Rbb^d)} := \int_\Ccal \ov{f(x)}g(x)\, \dx.
\end{align*}
We want to define ``$A$'' on $L^2_\per(\Rbb^d)$. To do this, let $f\in L_\per^2(\Rbb^d)$. Then, a simple argument by dominated convergence shows that $f = \sum_{\eta\in \Gamma} T_\eta \Idbb_\Ccal f$ where $T_\eta$ is the notation for translation by $\eta$. We have that $\Idbb_\Ccal f\in L^2(\Rbb^d)$, since $f\in \Lloc^2(\Rbb^d)$. Define
\begin{align*}
A_\per f := \sum_{\eta\in\Gamma } T_\eta A\Idbb_\Ccal f.
\end{align*}
This definition is independent of the chosen cube $\Ccal$ due to the periodicity of $f$ and $A$. For each $\eta\in \Gamma$, we immediately get that
\begin{align*}
T_\eta A_\per f = \sum_{\nu\in \Gamma} T_\eta T_\nu A\Idbb_\Ccal f = \sum_{\nu\in \Gamma} T_\nu A\Idbb_\Ccal f = A_\per f 
\end{align*}
so that $A_\per f$ is periodic. Furthermore, $A_\per$ is bounded, since
\begin{align*}
\Vert A_\per f\Vert_{L_\per^2}^2 &= \Bigl\Vert \sum_{\eta\in \Gamma} T_\eta A\Idbb_\Ccal f\Bigr\Vert_{L^2_\per}^2 = \sum_{\eta,\nu\in \Gamma} \langle T_\eta A\Idbb_\Ccal f, T_\nu A\Idbb_\Ccal f\rangle_{L^2(\Ccal)} \\
&= \sum_{\eta,\nu \in \Gamma} \langle AT_\eta \Idbb_\Ccal f, AT_\nu \Idbb_\Ccal f\rangle \\
&\leq \sum_{\eta,\nu\in \Gamma} \Vert A\Vert^2 \cdot \Vert T_\eta \Idbb_\Ccal f\Vert_{L^2(\Ccal)} \cdot \Vert T_\nu \Idbb_\Ccal f\Vert_{L^2(\Ccal)} =\Vert A\Vert \Vert f\Vert_{L^2_\per}^2.
\end{align*}
The last equality follows from
\begin{align*}
\sum_{\eta\in \Gamma} \Vert T_\eta \Idbb_\Ccal f\Vert_{L^2(\Ccal)} &= \sum_{\eta\in \Gamma} \Bigl( \int_\Ccal |T_\eta \Idbb_\Ccal f|^2 \, \dx\Bigr)^{\nicefrac 12} = \sum_{\eta\in \Gamma} \Bigl( \int_\Ccal |\underbrace{\Idbb_\Ccal(x-\eta)}_{= \delta_{\eta,0}} f(x-\eta)|^2\, \dx \Bigr)^{\nicefrac 12} \\
&= \Vert f\Vert_{L^2_\per}.
\end{align*}
This means that $A_\per$ is bounded as an operator $L_\per^2 (\Rbb^d)\lra L^2_\per(\Rbb^d)$ with $\Vert A_\per\Vert \leq \Vert A\Vert$. To prove that the Bloch-Floquet theory really decomposes $A$, we use a slight variant of the decomposition above. The Zac transform has a little better periodicity behavior and is defined on $\Scal(\Rbb^d)\subseteq L^2(\Rbb^d)$ by
\begin{align}
(Uf)_\theta(x)  := \sum_{\eta\in \Gamma} \e^{\i \theta(x +\eta)} f(x + \eta). \label{Zac}
\end{align}
Writing $x = c - \nu$ uniquely by $c\in \Ccal$ and $\nu\in \Gamma$, we obtain
\begin{align*}
(UAf)_\theta(x) &= \sum_{\eta\in \Gamma} \e^{\i \theta(x + \eta)} Af(x + \eta) = \sum_{\eta\in \Gamma} A\e^{\i \theta (x +\eta)} f(x + \eta) = \sum_{\eta\in \Gamma} A \e^{\i \eta(c - \nu + \eta)} f(c - \nu + \eta) \\
&= \sum_{\eta\in \Gamma} AT_{-\eta} T_\nu \Idbb_\Ccal (\e^{\i \theta \cdot} f)(x) = \sum_{\eta\in \Gamma} T_{-\eta} A \Idbb_\Ccal (\e^{\i \theta\cdot} f)(x) \\
&= A_\per \Bigl( \sum_{\eta\in \Gamma} T_{-\eta}  \Idbb_\Ccal \e^{\i \theta \cdot} f\Bigr) (x) = A_\per (Uf)_\theta(x).
\end{align*}

Hence, setting $A_\theta = A_\per$ for each $\theta$ we obtain that $A$ is indeed decomposable by the direct integral decomposition. In this sense, we will canonically identify $L_\per^2(\Rbb^d)$ with $L^2(\Ccal)$ and consider $A$ as decomposable as in \eqref{Decomp of periodic ops}.


\section{Local Traces}

Let $(M, \mu)$ be a measure space, $\Hcal'$ be a separable Hilbert space and $\Hcal = \int_M^\oplus \Hcal' \, \dd\mu$. Let $A\in\Bcal(\Hcal)$ be decomposable. Note that the function $m\longmapsto \tr(\Afrak(m))$ is measurable as a limit of measurable functions, see Lemma \ref{lem2.2.2}. We say that $A$ is locally compact, $A\in \Scal_\loc^\infty(\Hcal)$ if and only if $\Afrak(m)\in \Scal^\infty(\Hcal')$ for almost all $m\in M$. We equip $\Scal_\loc^\infty(\Hcal)$ with the norm $\Vert A\Vert_{\Bcal(\Hcal)} := \Vert \Afrak\Vert_\infty$. Let $1 \leq p < \infty$. We say that $A$ is locally Schatten-$p$, symbolically $A\in\Scal_\loc^p(\Hcal)$, if and only if its local Schatten-$p$ norm
\begin{align*}
\Vert A\Vert_{p,\loc}^p = \int_M \Vert \Afrak(m)\Vert_p^p \,  \dd \mu < \infty.
\end{align*}
If this is the case for $p = 1$, we say that $A$ is locally trace-class and define the local trace of $A$ by
\begin{align*}
\tr_\loc (A) := \int_M \tr (\Afrak(m))\, \dd \mu.
\end{align*}

\begin{thm}
Let $(M, \mu)$ be a measure space and $\Hcal = \int_M^\oplus \Hcal' \, \dd \mu$. Let $A,B\in \Bcal(\Hcal)$ be decomposable.
\begin{enumerate}[(a)]
\item Assume that $AB$ and $BA$ are locally trace class. Then, $\tr_\loc(AB) = \tr_\loc(BA)$.

\item The generalized Hölder's inequality holds: Let $1 \leq p,q,r\leq \infty$ such that $p^{-1} + q^{-1} = r^{-1}$ and assume that $A\in \Scal_\loc^p(\Hcal)$ and $B\in \Scal_\loc^q(\Hcal)$. Then $AB\in \Scal_\loc^r(\Hcal)$ and we have
\begin{align*}
\Vert AB\Vert_{r,\loc} \leq \Vert A\Vert_{p,\loc} \cdot \Vert B\Vert_{q,\loc}.
\end{align*}
\end{enumerate}
\end{thm}

\begin{proof}
\begin{enumerate}[(a)]
\item If $AB$ is locally trace class, then $\tr(\Afrak(m)\Bfrak(m)) <\infty$ for almost all $m\in M$. This makes the result follow from the one about standard traces.

\item Apply \eqref{Generalized Hölder} pointwise almost everywhere and the usual generalized Hölder's inequality.
\end{enumerate}
\end{proof}
Analogously to the case of standard traces, for an interval $I\subseteq \Rbb$ and a function $f\colon I \lra \Rbb$, we define
\begin{align*}
\Scal_{f, \loc}^1(\Hcal) := \{ A\in\Scal_\loc^1(\Hcal) : A=A^*, \; \sigma(A)\subseteq I, \; f(A)\in \Scal_\loc^1(\Hcal)\}.
\end{align*}

\begin{thm}[Peierl's inequality -- local version]
\label{Peierl's inequality local}
Let $(M, \mu)$ be a measure space and $\Hcal = \int_M^\oplus \Hcal'\, \dd \mu$. Let $A = \int_M^\oplus \Afrak(m) \, \dd \mu\in \Scal^1_\loc(\Hcal)$ and $f\colon \Rbb \lra\Rbb$ convex. For a.e. $m\in M$ let $\{u_n(m)\}_{n\in \Nbb}$ be any ONB of $\Hcal'$. Then
\begin{align}
\int_M \sum_{n=1}^\infty f\lk\langle u_n(m), \Afrak(m)u_n(m)\rangle\rk \, \dd \mu(m) \leq \tr_\loc(f(A)) \label{Peierl local}
\end{align}
and equality holds in \eqref{Peierl local} if and only if $u_n(m)$ is an eigenvector of $\Afrak(m)$ for all $n\in \Nbb$ and almost all $m\in M$. If $f$ is strictly convex, then equality in \eqref{Peierl local} holds only in this case.
\end{thm}

\begin{proof}
Apply Theorem \ref{Peierl's inequality} for a.e. $m\in M$ and use monotonicity of the integral. If $f$ is strictly convex then equality holds only if equality holds pointwise for almost all $m\in M$. By \ref{Peierl's inequality}, this is true only if $u_n(m)$ is an eigenvector of $\Afrak(m)$ for a.e. $m\in M$.
\end{proof}

\begin{kor}
\label{Convex-Tracefunction local}
Let $f\colon \Rbb \lra \Rbb$ be convex.
Then
\begin{align*}
\Phi_f \colon \Scal_{f,\loc}^1(\Hcal) &\lra \Rbb \\
A &\longmapsto \tr_\loc(f(A))
\end{align*}
is convex and $\Phi_f$ is strictly convex if and only if $f$ is strictly convex.
\end{kor}

\begin{proof}
Follows from the corresponding elementwise statement \ref{Convex-Tracefunction}.
\end{proof}

\begin{thm}[Klein's inequality -- local version]
\label{Klein_local_thm}
Let $I\subseteq \Rbb$ be an interval and $f\colon I\lra \Rbb$ convex. Let $A,B\in \Scal_{f,\loc}^1(\Hcal)$. Assume that the right-sided derivative $f_+'$ is bounded on $\sigma(B)$ and that $A - B\in \Scal_\loc^1(\Hcal)$. Then
\begin{align}
\tr_\loc (f(A) - f(B) - f'_+(B) (A - B))\geq 0. \label{Klein local}
\end{align}
If $f$ is strictly convex, then equality holds in \eqref{Klein local} if and only if $A =B$.
\end{thm}

\begin{proof}
Follows from the corresponding elementwise inequality \ref{Klein's inequality}.
\end{proof}

\subsection{Application to periodic operators}

Let us come back to the periodic operator $A\in \Bcal(L^2(\Rbb^d))$ from the previous section. Assume that it is locally trace-class. Following \cite{Teufel2009}, we intend to compute its local trace and claim that it is equal to $\tr(\Idbb_\Ccal A)$. To do this, we need to fix an ONB of $L^2(\Rbb^d)$ and see how it transforms under the Zac transform $U$ from \eqref{Zac}. Recall that the plane wave basis $g_{\gamma^*}(x) := |\Ccal|^{-\nicefrac 12}\cdot \Idbb_\Ccal(x) \e^{\i \gamma^* x}$, where $\gamma^*\in \Gamma^*$, forms an ONB of $\ran(\Idbb_\Ccal)\subseteq L^2(\Rbb^d)$. For $\alpha\in \Gamma$, consider
\begin{align*}
g_{\gamma^*}^\alpha(x) := g_{\gamma^*} (x - \alpha) = |\Ccal|^{-\nicefrac 12}\Idbb_{\Ccal + \alpha}(x) \e^{\i \gamma^*x}
\end{align*}
This defines an ONB $\{g_{\gamma^*}^\alpha\}_{\gamma^*\in \Gamma^*, \alpha\in \Gamma}$ of $L^2(\Rbb^d)$ because of the following. Suppose that $\varepsilon >0$, $f\in L^2(\Rbb^d)$ and for each $\alpha\in \Gamma$, we find a function $g_\alpha\in \ran(\Idbb_{\Ccal + \alpha})$ in the span of the $\{g_{\gamma^*}^\alpha\}_{\gamma^*\in \Gamma^*}$ such that $\Vert \Idbb_{\Ccal + \alpha}f  - f_\alpha\Vert \leq \frac{\varepsilon}{2^{|\alpha|}}$. If this is the case, we may define $g := \sum_{\alpha\in \Gamma} f_\alpha$. Then
\begin{align*}
\Vert f - g\Vert_{L^2(\Rbb^d)} \leq \sum_{\alpha\in \Gamma} \Vert \Idbb_{\Ccal + \alpha} (f - f_\alpha)\Vert_{L^2(\Rbb^d)} \leq \sum_{\alpha \in \Gamma} \frac{\varepsilon}{2^{|\alpha|}} \leq C_d \cdot \varepsilon
\end{align*}
The Zac transform of $g_{\gamma^*}^\alpha$ is
\begin{align*}
(Ug_{\gamma^*}^\alpha)_\theta (x) &= \sum_{\eta\in \Gamma} \e^{\i \theta(x + \eta)} g_{\gamma^*}^\alpha (x + \eta) = |\Ccal|^{-\nicefrac 12} \sum_{\eta\in \Gamma} \e^{\i \theta(x + \eta)} \Idbb_{\Ccal +\alpha}(x + \eta) \e^{\i \gamma^* (x + \eta)} \\
&= |\Ccal|^{-\nicefrac 12} \e^{\i (\theta + \gamma^*)(x + \eta)} \Idbb_{\Ccal + \alpha}(x + \eta)
\end{align*}
If $x\in \Ccal$, we obtain that $Ug_{\gamma^*}^\alpha = e_{\gamma^*}^\alpha$ with $(e_{\gamma^*}^\alpha)_\theta (x) = |\Ccal|^{-\nicefrac 12}\cdot  \e^{\i (\theta + \gamma^*)(x + \alpha)}$. We obtain that
\begin{align*}
\tr (A\Idbb_\Ccal) = \sum_{\gamma^*\in \Gamma^*,\alpha\in \Gamma} \langle g_{\gamma^*}^\alpha, A\Idbb_\Ccal g_{\gamma^*}^\alpha\rangle_{L^2(\Rbb^d)}
\end{align*}
Recall that $\Idbb_\Ccal g_{\gamma^*}^\alpha = \delta_{\alpha,0} g_{\gamma^*}^0$. This implies that
\begin{align*}
\tr(A \Idbb_\Ccal) = \sum_{\gamma^*\in \Gamma^*} \langle g_{\gamma^*}^0 , A g_{\gamma^*}^0 \rangle_{L^2(\Rbb^d)} = \frac{1}{|\Ccal^*|} \int_\Ccal \dd \theta \sum_{\gamma^*\in \Gamma^*} \langle (Ug_{\gamma^*}^0)_\theta, A_\theta (Ug_{\gamma^*}^0)_\theta\rangle_{L^2(\Ccal)}
\end{align*}

Note that, since $Ug_{\gamma^*}^0 = e_{\gamma^*}^0$, and since $(e^{\gamma^*})_\theta)_{\gamma^*\in \Gamma^*}\subseteq L^2(\Ccal)$ forms an ONB, we obtain that
\begin{align*}
\tr(A\Idbb_\Ccal) = \frac{1}{|\Ccal^*|} \int_\Ccal \dd \theta \tr(A_\theta) = \tr_\loc(A)
\end{align*}

All this is copied from \cite{Teufel2009}. It follows that for two periodic operators $A,B$, the local trace is given by $\tr (\Idbb_\Ccal AB)$. Should $A$ and $B$ have kernels $K_A(x,y)$ and $K_B(x,y)$, then we obtain
\begin{align*}
\tr_\loc (AB) = \tr(\Idbb_\Ccal AB) = \int_\Ccal \dx \int_{\Rbb^d} \, \dy \;  K_A(x,y)K_B(y,x).
\end{align*}

\printbibliography[heading=bibliography, title=Bibliography of Appendix \ref{Chapter:Local_Traces}]

\end{refsection}


\chapter{Differential Calculus for \texorpdfstring{$W^{2,\infty}$}{W2infty}-Functions}
\label{Chapter:W-functions} \label{CHAPTER:W-FUNCTIONS}

\begin{refsection}

All we treat here is taken from \cite{EvansPDE} and \cite{Lebesgue_Differentiation_Soneji}.

\section{Lebesgue's Differentiation Theorem}

\begin{lem}[Vitali's Covering lemma, \cite{Lebesgue_Differentiation_Soneji}]
\label{Vitali}
Let $x_1, \ldots, x_n\in \Rbb^d$ and $r_1, \ldots, r_n>0$ and let $E\subseteq \Rbb^d$ be such that $E \subseteq \bigcup_{i=1}^n B_{r_i}(x_i)$. Then, there is a disjoint subfamily $I\subseteq \{1, \ldots, n\}$, i.e. $B_{r_i}(x_i)\cap B_{r_j}(x_j) =\emptyset$ for all $i,j\in I$, $i\neq j$ and we have
\begin{align*}
E \subseteq \bigcup_{i\in I} B_{3r_i}(x_i)
\end{align*}
\end{lem}

\begin{bem}
One can prove this in more generality, but the constant ``$3$'' does not remain true. Often it is proved with $5$ instead but possibly anything bigger than $3$ would do.
\end{bem}

\begin{proof}
Without loss assume that $r_1\geq r_2 \geq \cdots \geq r_n\geq 0$. Choose $D_1 := B_{r_1}(x_1)$. Assume recursively that $D_1, \ldots, D_{k-1}\in \{B_{r_i}(x_i)\}_{i=1}^n$ are already chosen. If
\begin{align*}
I_k := \Bigl\{1\leq j\leq n: B_{r_j}(x_j) \cap \bigcup_{i=1}^{k-1} D_i =\emptyset\Bigr\} = \emptyset
\end{align*}
then set $m = k-1$ and terminate. Otherwise, let $\ell_k := \min I_k$ and choose $D_k:= B_{r_{\ell_k}}(x_{\ell_k})$. If $B_{r_i}(x_i)\notin \{D_j\}_{j=1}^m$ for some $i\in \{1, \ldots, n\}$, then, by definition, there is $j< i$ such that $B_{r_j}(x_j) \cap B_{r_i}(x_i)\neq \emptyset$ (if not, then $i$ would be the minimum of $I_k$ for some $k$ and $B_{r_i}(x_i)$ would have been chosen). Since $r_j\geq r_i$, we get $B_{r_i}(x_i)\subseteq B_{3r_j}(x_j)$ because for $y\in B_{r_i}(x_i)$ and $z\in B_{r_i}(x_i)\cap B_{r_j}(x_j)$ arbitrary, we have
\begin{align*}
|y-x_j|\leq |y-x_i| + |x_i - z| + |z - x_j|  < r_i + r_i + r_j \leq 3r_j
\end{align*}
Hence, we get $D_k \cap D_\ell = \emptyset$ for $1\leq k,\ell\leq m$ with $k\neq \ell$ by construction and
\begin{align*}
\bigcup_{i=1}^n B_{r_i}(x_i) &\subseteq \bigcup_{j=1}^m B_{3r_{\ell_j}}(x_{\ell_j}) \qedhere
\end{align*}
\end{proof}

Let $f\in L^1(\Rbb^d)$ and define the Hardy-Littlewood maximal function by
\begin{align*}
(Mf)(x) := \sup_{r>0} \frac{1}{|B_r(x)|} \int_{B_r(x)} |f(y)|\, \dy
\end{align*}
Here, $|B|$ denotes the Lebesgue measure of $B\subseteq \Rbb^n$.

\begin{lem}
If $f\in L^1(\Rbb^d)$, then $Mf$ is lower semi-continuous and thus measurable.
\end{lem}

\begin{proof}
We want to prove that for $0< t <\infty$; the set 
\begin{align*}
E_t := \{x\in \Rbb^d : (Mf)(x) > t\} = (Mf)^{-1}((t,\infty))
\end{align*}
is open. For $t\leq 0$, $E_t = \Rbb^n$ so the claim there is clear. Let $x\in E_t$. Then, by definition of $Mf$, there is $r>0$ such that
\begin{align*}
\frac{1}{|B_r(x)|} \int_{B_r(x)} |f(y)| \, \dy > t
\end{align*}
Choose $r' > r$ such that still
\begin{align*}
\frac 1{|B_{r'}(x)|} \int_{B_r(x)} |f(y)| \, \dy > t
\end{align*}
Let $x'\in \Rbb^d$ with $|x - x'| < r'-r$. Then $B_r(x)\subseteq B_{r'}(x')$ since for each $y\in B_r(x)$, we have
\begin{align*}
|y-x'| \leq |y-x| + |x'-x|  < r + r' -r = r'
\end{align*}
Hence, by the translational invariance of the Lebesgue measure, we get
\begin{align*}
t &< \frac{1}{|B_{r'}(x)|} \int_{B_r(x)} |f(y)|\, \dy = \frac{1}{|B_{r'}(x')|} \int_{B_r(x)} |f(y)|\, \dy \\
&\leq \frac{1}{|B_{r'}(x')|} \int_{B_{r'}(x')} |f(y)|\, \dy \leq (Mf)(x')
\end{align*}
Hence, $x'\in E_t$ and $B_{r'-r}(x)\subseteq E_t$.
\end{proof}

\begin{thm}[Weak type maximal inequality]
\label{Maximal inequality}
If $f\in L^1(\Rbb^d)$, and $t>0$, we have that
\begin{align*}
|\{x\in \Rbb^n : (Mf)(x) > t\}| \leq \frac{3^n}{t} \cdot \Vert f\Vert_{L^1(\Rbb^n)}
\end{align*}
\end{thm}

\begin{proof}
Let $K\subseteq E_t$ be compact and consider the trivial cover $\{B_{r_x}(x)\}_{x\in E_t}$ of $K$. By compactness, there is a finite subcover. By Vitali's covering lemma \ref{Vitali}, there is a disjoint subfamily $\{B_{r_i}(x_i)\}_{i=1}^k$ so that $K \subseteq \bigcup_{i=1}^k B_{3r_i}(x_i)$. For each $x\in E_t$ there is an $r>0$ such that
\begin{align*}
\frac{1}{|B_r(x)|} \int_{B_r(x)} |f(y)| \, \dy > t
\end{align*}
In other words,
\begin{align*}
|B_r(x)| \leq \frac 1t \int_{B_r(x)} |f(y)|\, \dy
\end{align*}
Hence,
\begin{align*}
|K| \leq 3^n \sum_{i=1}^k |B_{r_i}(x_i)| \leq \frac{3^n}{t} \sum_{i=1}^k \int_{B_{r_i}(x_i)} |f(y)|\, \dy \leq \frac{3^n}{t} \cdot \Vert f\Vert_{L^1(\Rbb^n)}
\end{align*}
By the inner regularity of the Lebesgue measure, we conclude that
\begin{align*}
|E_t| &= \sup_{\substack{K\subseteq E_t\\ K\text{ compact}}} |K| \leq \frac{3^n}{t} \cdot \Vert f\Vert_{L^1(\Rbb^d)} \qedhere
\end{align*}
\end{proof}

\begin{thm}[Lebesgue's Differentiation Theorem]
\label{Lebesgues differentiation theorem}
Let $1\leq p < \infty$ and $f\in \Lloc^p(U)$, where $U\subseteq \Rbb^d$ is open. Then, for almost all $x\in U$, we have
\begin{align*}
\limsup_{r\searrow 0} \fint_{B_r(x)} |f(y) - f(x)|^p \, \dy =0
\end{align*}
\end{thm}

\begin{proof}
Let $g\in C_c(\Rbb^d)$. Then $g$ is uniformly continuous. Hence, given $\varepsilon>0$, there is $\delta >0$ such that $|f(x) - f(y)| < \varepsilon^{\nicefrac 1p}$ whenever $|x - y|< \delta$. Fix $x\in \Rbb^d$ and let $r < \delta$. Then
\begin{align*}
\frac{1}{|B_r(x)|} \int_{B_r(x)} |f(y) - f(x)|^p \, \dy < \frac{1}{|B_r(x)|} \int_{B_r(x)} \, \dy \cdot\varepsilon = \varepsilon
\end{align*}
Hence,
\begin{align}
\limsup_{r\searrow 0} \fint_{B_r(x)} |f(y) - f(x)|^p \, \dy = 0 \label{Differentiation continuous}
\end{align}
Let $f\in \Lloc^p(U)$. By extending to all of $\Rbb^d$ with $0$, we may assume that $f\in \Lloc^p(\Rbb^d)$. Without loss, assume that $f\in L^p(\Rbb^d)$. If not, then for $k\in \Nbb$, the function $f_k = f\Idbb_{B_k(0)}$ is in $L^p(\Rbb^d)$. Assume the result holds for $f_k$ except for a null set $E_k\subseteq \Rbb^n$. Then, it holds for $f$ on the complement of the null set $\bigcup_{k\in \Nbb} E_k$. For, if $x\in \Rbb^d\setminus \bigcup_{k\in \Nbb} E_k$, and $r>0$ is small, then there is $k_0\in \Nbb$ with $B_r(x) \subseteq  B_{k_0}(0)$. Then, since $x\notin E_{k_0}$, apply the result to $f_{k_0}$, since $f_{k_0}|_{B_r(x)} = f|_{B_r(x)}$. Now, for an arbitrary $g\in C_c(\Rbb^d)$ and $x\in \Rbb^d$, use \eqref{Differentiation continuous} and convexity of $t\mapsto t^p$ to get
\begin{align*}
\limsup_{r\searrow 0} \fint_{B_r(x)} |f(y) - f(x)|^p \, \dy &\leq 2^{p-1} \limsup_{r\searrow 0} \fint_{B_r(x)} |g(y) - g(x)|^p \, \dy \\
& \hspace{30pt} + 4^{p-1} \limsup_{r\searrow 0} \fint_{B_r(x)} |f(y) - g(y)|^p \, \dy \\
&\hspace{30pt} + 4^{p-1}|f(x) - g(x)|^p\\
&\leq 4^{p-1} \sup_{r>0} \fint_{B_r(x)} |f(y) - g(y)|^p  + 4^{p-1} |f(x) - g(x)|^p \\
&= 4^{p-1}\bigl[ M(|f-g|^p) (x) + |f(x) - g(x)|^p\bigr]
\end{align*}
Hence, if, for given $\varepsilon>0$, we have
\begin{align*}
\limsup_{r\searrow 0} \fint_{B_r(x)} |f(y) - f(x)|^p\, \dy > \varepsilon
\end{align*}
Then, $M(|f - g|^p) (x) > 2^{2p-3} \varepsilon$ or $|f(x)- g(x)| > 2^{2p-3} \varepsilon$. Now, using the Tshebyshev inequality, we get
\begin{align*}
|X_1^\varepsilon|:=\bigl|\bigl\{ x\in \Rbb^d : |f(x) - g(x)|^p > 2^{2p-3} \varepsilon\bigr\}\bigr| \leq \frac 1{2^{2p-3} \varepsilon} \cdot \Vert f-g\Vert_{L^p(\Rbb^d)}^p
\end{align*}
and using the maximal inequality \ref{Maximal inequality} we obtain
\begin{align*}
|X_2^\varepsilon|:=  \bigl|\bigl\{ x\in \Rbb^d: M(|f-g|^p)(x) > 2^{2p-3}\varepsilon\bigr\}\bigr| \leq \frac{3^d}{2^{2p-3} \varepsilon} \cdot \Vert f - g\Vert_{L^p(\Rbb^d)}^p
\end{align*}
Putting everything together, we see that
\begin{align*}
|A_\varepsilon| := \Bigl|\Bigl\{ x\in \Rbb^d: \limsup_{r\searrow 0} \fint_{B_r(x)} |f(y) - f(x)|^p \, \dy > \varepsilon\Bigr\} \Bigr| &\leq |X_1^\varepsilon| + |X_2^\varepsilon| \\
&\leq \frac{3^d + 1}{2^{2p-3}\varepsilon} \cdot \Vert f - g\Vert_{L^p(\Rbb^d)}^p
\end{align*}
By density, we may choose a sequence $(g_n)_n\subseteq C_c(\Rbb^d)$ with $g_n \to f$ in $L^p(\Rbb^d)$ as $n\to\infty$. It follows that $A_\varepsilon$ has zero measure for every $\varepsilon>0$. Hence,
\begin{align*}
0 = \Bigl| \bigcup_{k\in \Nbb} A_{\frac 1k}\Bigr| = \Bigl|\Bigl\{ x\in \Rbb^d: \limsup_{r\searrow 0} \fint_{B_r(x)} |f(y) - f(x)|^p \, \dy > 0\Bigr\}\Bigr|
\end{align*}
so that the claim holds for almost each $x\in \Rbb^d$.
\end{proof}

\section{Taylor Expansion}

First, we prove the remark on \cite[p. 268]{EvansPDE} by using elements of the foregoing proof.\\

First, let $u\in C^1(\Rbb^d)$ and $x\in \Rbb^d$, $r>0$. We claim that
\begin{align}
\fint_{B_r(x)} |u(x) - u(y)| \, \dy \leq \frac 1d \int_{B_r(x)} \frac{|Du(y)|}{|y-x|^{d-1}} \,\dy \label{Taylor-auxilliary}
\end{align}

\begin{proof}
Fix $w\in \partial B_1(0)$ and let $0 < s < r$. Then
\begin{align*}
|u(x + sw) - u(x)| &= \Bigl| \int_0^s \frac{\dd}{\dt} u(x + tw)\, \dt \Bigr| = \Bigl| \int_0^s Du(x + tw) \cdot w\, \dt\Bigr| \\
&\leq \int_0^s |Du(x + tw)| \, \dt
\end{align*}
Hence,
\begin{align*}
\int_{\partial B_1(0)} |u(x + sw) - u(x)|\, \dd S(w) &\leq  \int_0^s \int_{\partial B_1(0)} |Du(x + tw)|\, \dd S(w) \dt \\
&= \int_0^s t^{d-1}\int_{\partial B_1(0)} \cdot \frac{|Du(x + tw)|}{t^{d-1}} \, \dd S(w) \dt
\end{align*}
Letting $y = x + tw$ so that $t = |x-y$, we obtain that
\begin{align*}
\int_{\partial B_1(0)} |u(x + sw) - u(x) | \, \dd S(w) \leq \int_{B_s(x)} \frac{|Du(y)|}{|x-y|^{d-1}}\, \dy \leq \int_{B_r(x)} \frac{|Du(y)|}{|x-y|^{d-1}} \, \dy
\end{align*}
On the other hand, multiplying the left hand side with $s^{d-1}$ and integrating over $[0,r]$, we obtain
\begin{align*}
\int_0^r s^{d-1} \int_{\partial B_1(0)} |u(x + sw) - u(x)|\, \dd S(w)\ds = \int_{B_r(x)} |u(y) - u(x)|\, \dy
\end{align*}
so that all in all, we get
\begin{align*}
\int_{B_r(x)} |u(y) - u(x)| \, \dy &\leq \frac{r^d}{d} \int_{B_r(x)} \frac{|Du(y)|}{|x-y|^{d-1}} \, \dy\qedhere
\end{align*}
\end{proof}

We are going to use this to prove a Morrey-type inequality: There is a constant $C$, depending only on $p$ and $d$ such that for $y\in B_r(x)$, we have
\begin{align}
|u(x) - u(y)| \leq C r^{1-\nicefrac dp} \Bigl( \int_{B_{2r}(x)} |Du(y)|^p\Bigr)^{\nicefrac 1p} \label{Morrey-type inequality}
\end{align}

\begin{proof}
Let $x,y\in \Rbb^d$, $x\neq y$ (otherwise there is nothing to prove) and call $r := |x-y|>0$. Then set $W := B_r(x) \cap B_r(y)$. We start by noting that
\begin{align*}
|u(x) - u(y)| &= \fint_W |u(x) - u(y) + u(z) - u(z)|\, \dz \\
&\leq \fint_W |u(x) - u(z)|\, \dz + \fint_W |u(y) - u(z)| \, \dz
\end{align*}
Now, $W\supseteq B_{\frac r2} (\frac 12 (x+y))$ so that $|W| \geq \frac{1}{2^d} |B_r(x)|$ by translational invariance. Hence, applying \eqref{Taylor-auxilliary} and H\"older, we obtain
\begin{align*}
\fint_W |u(x) - u(z)| \, \dz &\leq 2^d \fint_{B_r(x)} |u(x) - u(z)|\,\dz \leq \frac{2^d}{d} \int_{B_r(z)} \frac{|Du(z)|}{|x-z|^{d-1}}\, \dz \\
&\leq \frac{2^d}{d} \Bigl( \int_{B_r(x)} |Du(z)|^p\, \dz\Bigr)^{\nicefrac 1p} \Bigl( \int_{B_r(x)} \frac{1}{|x-z|^{(d-1)\frac{p}{p-1}}}\, \dz\Bigr)^{\frac{p-1}{p}}
\end{align*}
An explicit calculation shows that
\begin{align*}
\Bigl( \int_{B_r(x)} \frac{1}{|x-z|^{(d-1)\frac{p}{p-1}}}\, \dz\Bigr)^{\frac{p-1}{p}} = C_{d,p}\cdot  r^{1 - \nicefrac dp}
\end{align*}
To summarize, we have
\begin{align*}
\fint_W |u(x) - u(z)|\, \dz \leq C r^{1 - \nicefrac dp} \Bigl( \int_{B_r(x)} |Du(z)|\, \dz\Bigr)^{\nicefrac 1p}
\end{align*}
The same bound applies for $y$ with $B_r(y)$ instead of $B_r(x)$. Now, note that $B_r(x) \cup B_r(y) \subseteq B_{2r}(x)$. This implies that one may replace $B_r(x)$ (respectively $B_r(y)$) in both estimates by $B_{2r}(x)$ and we are done.
\end{proof}

Next, we prove a variant of Theorem 5 on \cite[p. 280]{EvansPDE}.

\begin{thm}
\label{Taylor for Winfty}
Assume that $u\in \Wloc^{2,p}(U)$ where $U\subseteq \Rbb^d$ is open and $d< p \leq \infty$. Then $u$ is twice differentiable almost everywhere in $U$ and its derivatives coincide with its weak derivatives a.e.
\end{thm}

\begin{proof}
First, assume that $n < p < \infty$. We note that the inequality \eqref{Morrey-type inequality} extends to all $u\in W^{1,p}(U)$ by interior approximation. Let $x\in U$ and $r>0$ small enough such that $B_r(x)\subseteq U$. For $y\in B_r(x)$, set
\begin{align}
v(y) := u(y) - u(x) - Du(x) (y-x) - \frac 12 (y-x) \cdot D^2u(x)  (y-x) \label{Second order Taylor expansion}
\end{align}
Note that $v(x) =0$. Then, by \eqref{Morrey-type inequality}, we get
\begin{align*}
|v(y)| = |v(y) - v(x)| &\leq Cr^{1-\nicefrac dp}\Bigl( \int_{B_{2r}(x)} |Dv(z) |^p\, \dz\Bigr)^{\nicefrac 1p} \\
&= Cr^{1-\nicefrac dp} \Bigl( \int_{B_{2r}(x)} |Du(z) - Du(x) - D^2u(x) (z-x)|^p\,\dz\Bigr)^{\nicefrac 1p}
\end{align*}
For $z\in B_{2r}(x)$, let now $\tilde v(z) := Du(z) - Du(x) - D^2u(x) (z-x)$. Again, $\tilde v(x) =0$. We obtain by \eqref{Morrey-type inequality}:
\begin{align*}
|\tilde v(z)| = |\tilde v(z) - \tilde v(x)| &\leq C (2r)^{1 - \nicefrac dp} \Bigl( \int_{B_{4r}(x)} |D\tilde v(w)|^p\, \dd w\Bigr)^{\nicefrac 1p} \\
&= Cr \Bigl( \fint_{B_{4r}(x)} |D^2u(w) - D^2u(x)|^p\Bigr)^{\nicefrac 1p}
\end{align*}
Note that we incorporated $r^{-\nicefrac dp}$ into the integral in cost of a constant. Inserting this above, we get
\begin{align*}
|v(z)| &\leq Cr \Bigl( \fint_{B_{2r}(x)} \dz \cdot Cr \fint_{B_{4r}(x)} |D^2u(w) - D^2u(x)|^p\Bigr)^{\nicefrac 1p} \\
&= Cr^2 \Bigl( \fint_{B_{4r}(x)} |D^2u(w) - D^2u(x)|^p\Bigr)^{\nicefrac 1p}
\end{align*}
The last expression is $o(r^2)$ by Lebesgue's differentiation theorem \ref{Lebesgues differentiation theorem}, since $|D^2u|\in \Lloc^p(U)$. This shows that $u$ is twice differentiable at $x$ and $D^2u(x)$ is the second derivative. Now, it is easy to see that $Du(x)$ is the first derivative. In the case $p = \infty$, note that $\Wloc^{2,\infty}(U)\subseteq \Wloc^{2,p}(U)$ for all $n < p\leq\infty$. Thus, apply the reasoning above.
\end{proof}

\section{Weak Differential Calculus in \texorpdfstring{$\Rbb^d$}{Rd}}

\begin{lem}
\label{Curl identity}
Let $v,w\colon \Rbb^3\lra \Rbb^3$ be differentiable a.e. Then
\begin{align*}
\nabla (v \cdot w) = (v\cdot \nabla)w + (w\cdot \nabla) v + v\wedge \curl w + w\wedge \curl v \qquad \text{a.e.}
\end{align*}
\end{lem}

\begin{proof}
This is a tedious but straightforward computation. Let us call the components $v = \sum_{i=1}^3 v_i e_i$ and $w = \sum_{i=1}^3 w_i e_i$. The right-hand side is given by
\begin{align*}
w \wedge \curl v + v\wedge \curl w + (v \cdot \nabla ) w + (w\cdot \nabla)v = \hspace{-150pt}&\\ 
&= w \wedge 
\begin{pmatrix} 
\partial_2 v_3 - \partial_3 v_2 \\ 
\partial_3 v_1 - \partial_1 v_3 \\ 
\partial_1 v_2 - \partial_2 v_1 
\end{pmatrix} + v\wedge 
\begin{pmatrix} 
\partial_2 w_3 - \partial_3 w_2 \\ 
\partial_3 w_1 - \partial_1 w_3 \\ 
\partial_1 w_2 - \partial_2 w_1 
\end{pmatrix} 
+ \sum_{i=1}^3 v_i \partial_i w + \sum_{i=1}^3 w_i \partial_i v \\
&= \begin{pmatrix}
w_2 \partial_1 v_2 + w_3 \partial_1 v_3 + v_2 \partial_1 w_2 + v_3 \partial_1 w_3 + v_1 \partial_1 w_1 + w_1 \partial_1 v_1 \\
w_3 \partial_2 v_3 + w_1 \partial_2 v_1 + v_3 \partial_2 w_3 + v_1 \partial_2 w_1 + v_2 \partial_2 w_2 + w_2 \partial_2 v_2 \\
w_1 \partial_3 v_1 + w_2 \partial_3 v_2 + v_1 \partial_3 w_1 + v_2 \partial_3 w_2 + v_3 \partial_3 w_3 + w_3 \partial_3 v_3
\end{pmatrix}\\
&= \begin{pmatrix}
w_1 \partial_1 v_1 + w_2 \partial_1 v_2 + w_3 \partial_1 v_3 \\
w_1 \partial_2 v_1 + w_2 \partial_2 v_2 + w_3 \partial_2 v_3 \\
w_1 \partial_3 v_1 + w_2 \partial_3 v_2 + w_3 \partial_3 v_3
\end{pmatrix} + \begin{pmatrix}
v_1 \partial_1 w_1 + v_2 \partial_1 w_2 + v_3 \partial_1 w_3 \\
v_1 \partial_2 w_1 + v_2 \partial_2 w_2 + v_3 \partial_2 w_3 \\
v_1 \partial_3 w_1 + v_2 \partial_3 w_2 + v_3 \partial_3 w_3
\end{pmatrix} \\
&= \begin{pmatrix}
\partial_1 (v_1 w_1 + v_2 w_2 + v_3 w_3) \\
\partial_2 (v_1 w_1 + v_2 w_2 + v_3 w_3) \\
\partial_3 (v_1 w_1 + v_2 w_2 + v_3 w_3) 
\end{pmatrix} = \nabla (v\cdot w)\qedhere
\end{align*}
\end{proof}

\begin{prop}[Product rule for $W^{1,\infty}(\Rbb^d)$]
\label{Product rule for Winfty}
Let $f\in W^{1,\infty}(\Rbb^d)$ and $g\in H^1(\Rbb^d)$. Then $f\cdot g\in H^1(\Rbb^d)$ with weak derivative
\begin{align*}
\nabla (f g) = (\nabla f)\cdot g + f\cdot (\nabla g).
\end{align*}
\end{prop}

\begin{proof}
Since $f$ is weakly differentiable, for every $v\in C_c^\infty(\Rbb^3)$, we have
\begin{align}
\int_{\Rbb^3} g\cdot \nabla v \, \dx = -\int_{\Rbb^3} (\nabla g) \cdot  v\, \dx. \label{Product rule weak diffbility}
\end{align}
By density, this extends to all $v\in H_0^1(\Rbb^d) = H^1(\Rbb^d)$. Now, since $f\in W^{1,\infty}(\Rbb^d)$, $f$ is almost everywhere classically differentiable and we have $\nabla (fv) = \nabla f \cdot v + f\nabla v$ almost everywhere. Since $v$ has compact support and $f,\nabla f$ are bounded, we conclude that $fv\in H^1(\Rbb^3)$. Thus, $fv$ is a valid test function in \eqref{Product rule weak diffbility}. Therefore, we obtain
\begin{align*}
-\int_{\Rbb^d} (\nabla g)\cdot fv\, \dx &=\int_{\Rbb^d} g\cdot \nabla (fv)\, \dx= \int_{\Rbb^d} g\cdot (\nabla f)\cdot v\, \dx +\int_{\Rbb^d} g\cdot f\cdot (\nabla v)\, \dx.
\end{align*} 
Thus,
\begin{align*}
\int_{\Rbb^d} fg\cdot \nabla v \, \dx = - \int_{\Rbb^d} (f\cdot \nabla g + \nabla f \cdot g)\, \dx.
\end{align*}
Hence, $fg$ is weakly differentiable with $\nabla (fg) = f\nabla g + \nabla f\cdot g$. Furthermore, since $f, \nabla f\in L^\infty(\Rbb^d)$ and $g,\nabla g\in L^2(\Rbb^d)$, we have that $fg, \nabla (fg)\in L^2(\Rbb^d)$. Thus, $fg\in H^1(\Rbb^d)$.
\end{proof}

\begin{prop}[Chain rule for $W^{1,\infty}(\Rbb^d)$]
\label{Chain rule for Winfty}
Let $f\in W^{1,\infty}(\Rbb^d)$ and $g\in C^1(\Rbb)$ such that $g'\in L^\infty(\Rbb)$. Then $g\circ f \in W^{1,\infty}(\Rbb^d)$ with weak derivative
\begin{align*}
\nabla (g\circ f) = (g'\circ f) \cdot \nabla f.
\end{align*}
\end{prop}

\begin{proof}
Since $g\in C^1(\Rbb)$, for every $t\in \Rbb$, we have
\begin{align*}
g(t) = g(0) + \int_0^t g'(s)\, \ds.
\end{align*}
Denoting $F := g\circ f$, it follows that for almost all $x\in \Rbb^d$:
\begin{align*}
|F(x)| &\leq |g(0)| + \Bigl| \int_0^{f(x)} g'(s)\, \ds\Bigr| \leq |g(0)| + \Vert g'\Vert_\infty \cdot |f(x)| \leq |g(0)| + \Vert g'\Vert_\infty \cdot \Vert f\Vert_\infty
%
\end{align*}
Hence, $F\in L^\infty(\Rbb^d)$. Since $f$ is classically differentiable almost everywhere, we obtain $\nabla F(x) = g'(f(x))\cdot \nabla f(x)$ for a.e. $x\in \Rbb^d$ by the classical chain rule. Then $\nabla F$ coincides with the weak derivative of $F$ a.e. Furthermore, $|\nabla F(x)|\leq \Vert g'\Vert_\infty \cdot \Vert \nabla f\Vert_\infty < \infty$. Hence, $F\in W^{1,\infty}(\Rbb^d)$.
\end{proof}

\printbibliography[heading=bibliography, title=Bibliography of Appendix \ref{Chapter:W-functions}]

@Article{FGHP20,
  author  = {{Falconi}, Marco and {Giacomelli}, Emanuela L. and {Hainzl}, Christian and {Porta}, Marcello},
  title   = {The dilute Fermi gas via Bogoliubov theory},
  journal = {Annales Henri Poincar\'e},
  year    = {2021},
  volume  = {22},
  pages   = {2283--2353},
  month   = may,
  doi     = {10.1007/s00023-021-01031-6},
}

@Article{Bogolubov,
  author  = {Nikolai Nikolajewitsch Bogoljubow},
  title   = {On the theory of superfluidity},
  journal = {Jornal of Physics (USSR)},
  year    = {1947},
  number  = {11},
  pages   = {p. 23},
  url     = {http://ufn.ru/pdf/jphysussr/1947/11_1/3jphysussr19471101.pdf},
}

@Article{Einstein,
  author  = {Albert Einstein},
  title   = {Quantentheorie des einatomigen idealen Gases. Zweite Abhandlung},
  journal = {Sitzungsberichte der Preussischen Akademie der Wissenschaften zu Berlin},
  year    = {1925},
  pages   = {p. 3-14},
  url     = {http://echo.mpiwg-berlin.mpg.de/ECHOdocuView?url=/permanent/echo/einstein/sitzungsberichte/R1Y4X6GP/index.meta&start=1&pn=1},
}

@Book{EvansPDE,
  title         = {Partial Differential Equations},
  publisher     = {American Mathematical Society},
  year          = {2010},
  author        = {Evans, L.C.},
  series        = {Graduate studies in mathematics},
  __markedentry = {[Marcel:]},
  isbn          = {9780821849743},
  lccn          = {2009044716},
}

@Online{MBQM,
  author    = {Jan Philip Solovej},
  title     = {Many body quantum mechanics},
  year      = {2014},
  note      = {Lecture Notes},
  url       = {http://web.math.ku.dk/~solovej/MANYBODY/mbnotes-ptn-5-3-14.pdf},
  urldate   = {2021-11-04},
}

@Unpublished{TraceInequalities,
  author  = {Eric A. Carlen},
  title   = {Trace inequalities and quantum entropy},
  note    = {An introductory course},
  year    = {2009},
  url     = {http://www.ueltschi.org/AZschool/notes/EricCarlen.pdf},
  urldate = {2021-12-02},
}

@Misc{Onnes,
  author       = {Dirk van Delft and Peter Kes},
  title        = {The discovery of superconductivity},
  howpublished = {University of Leiden, Netherlands},
  month        = sep,
  year         = {2010},
  doi          = {10.1063/1.3490499},
  url          = {https://www.lorentz.leidenuniv.nl/history/cold/DelftKes_HKO_PT.pdf},
  urldate      = {2021-12-01},
}

@Article{Onnes2,
  author    = {Rudolf de Bruyn Ouboter},
  title     = {Heike Kamerlingh Onnes’s Discovery of Superconductivity},
  journal   = {Scientific American},
  year      = {1997},
  volume    = {276},
  number    = {3},
  pages     = {98--103},
  issn      = {00368733, 19467087},
  publisher = {Scientific American, a division of Nature America, Inc.},
  url       = {https://web.njit.edu/~tyson/supercon_papers/Onnes.pdf},
  urldate   = {2021-12-01},
}

@Unpublished{MQM2,
  author = {Marcel Schaub},
  title  = {Mathematical Quantum Mechanics II},
  note   = {Unpublished lecture notes at the Ludwig-Maximilians-Universit\"at M\"unchen, lecturer: Prof. Dr. Peter M\"uller},
  year   = {2016},
}

@Unpublished{Lebesgue_Differentiation_Soneji,
  author  = {Parth Soneji},
  title   = {Lebesgue's Differentation Theorem via Maximal Functions},
  note    = {Slides of a talk to a Hüttenseminar},
  month   = dec,
  year    = {2013},
  url     = {https://www.mathematik.uni-muenchen.de/~diening/ws13/huette/vortraege/soneji.pdf},
  urldate = {2021-12-11},
}

@Misc{Video_Meissner,
  title        = {Meissner Effect},
  howpublished = {Harvard Natural Sciences Lecture Demonstrations (Youtube channel)},
  month        = aug,
  year         = {2015},
  url          = {https://www.youtube.com/watch?v=HRLvVkkq5GE},
  urldate      = {2021-12-14},
}

@PhdThesis{AndiPhD,
  author  = {Andreas Deuchert},
  title   = {Contributions to the mathematical study of BCS theory},
  school  = {University of Tübingen},
  year    = {2016},
  url     = {https://publikationen.uni-tuebingen.de/xmlui/bitstream/handle/10900/72407/dissertation.pdf?sequence=1&isAllowed=y},
  urldate = {2022-01-12},
}

@Article{Cooper,
  author    = {Cooper, Leon N.},
  title     = {Bound Electron Pairs in a Degenerate Fermi Gas},
  journal   = {Phys. Rev.},
  year      = {1956},
  volume    = {104},
  pages     = {1189--1190},
  month     = nov,
  doi       = {10.1103/PhysRev.104.1189},
  issue     = {4},
  numpages  = {0},
  publisher = {American Physical Society},
  url       = {https://link.aps.org/doi/10.1103/PhysRev.104.1189},
}

@Article{Meissner,
  author  = {Walther Meißner and Robert Ochsenfeld},
  title   = {Ein neuer Effekt bei Eintritt der Supraleitfähigkeit},
  journal = {Naturwissenschaften},
  year    = {1933},
  volume  = {21},
  pages   = {787--788},
}

@Article{Hainzl_Braeunlich_direct_exchange,
  author  = {Br\"{a}unlich, Gerhard and Hainzl, Christian and Seiringer, Robert},
  title   = {Translation-invariant quasi-free states for fermionic systems and the BCS approximation},
  journal = {Reviews in Mathematical Physics},
  year    = {2014},
  volume  = {26},
  number  = {07},
  pages   = {1450012},
  doi     = {10.1142/S0129055X14500123},
}

@Article{Fournais_Solovej_Bose_Gas,
  author    = {S{\o}ren Fournais and Jan Philip Solovej},
  title     = {{The energy of dilute Bose gases}},
  journal   = {Annals of Mathematics},
  year      = {2020},
  volume    = {192},
  number    = {3},
  pages     = {893 -- 976},
  doi       = {10.4007/annals.2020.192.3.5},
  publisher = {Department of Mathematics of Princeton University},
  url       = {https://doi.org/10.4007/annals.2020.192.3.5},
}

\end{refsection}




\chapter*{Eidesstattliche Versicherung}
\addcontentsline{toc}{chapter}{Eidesstattliche Versicherung}

(Siehe Promotionsordnung vom 12.07.11, §8 Abs. 2 Pkt. 5)

\vspace{3cm}

\noindent Hiermit erkläre ich an Eidesstatt, dass die Dissertation von mir selbstständig, ohne unerlaubte Hilfe angefertigt wurde.

\vspace{1cm}

\noindent München, den \abgabetermin

\vspace{2cm}

\noindent\makebox[15cm]{\dotfill}

\noindent (Marcel Oliver Maier)


%

\newpage

~\thispagestyle{empty}

\end{document}